\newif\ifabstract
\abstracttrue
\newif\iffull
\ifabstract \fullfalse \else \fulltrue \fi

\documentclass[11pt]{article}
\usepackage{amsfonts}
\usepackage{amssymb}
\usepackage{amstext}
\usepackage{amsmath}
\usepackage{xspace}
\usepackage{ntheorem}
\usepackage{graphicx}
\usepackage{url}
\usepackage{graphics}
\usepackage{colordvi}
\usepackage{hyperref}
\usepackage{subfigure}
\usepackage{xcolor}
\usepackage{bm}
\usepackage{cleveref}
\advance \oddsidemargin by -0.8in
\newcommand{\myparskip}{3pt}
\parskip \myparskip
\newif\ifnocomments

\ifnocomments

\else

\fi
\newcommand{\halpha}{\hat \alpha}
\newcommand{\talpha}{\tilde \alpha}
\newcommand{\vol}{\operatorname{vol}}

\newcommand{\tT}{\tilde T}
\newcommand{\tk}{\tilde k}
\newcommand{\hH}{\hat H}

\renewcommand{\hm}{\hat m}

\newcommand{\band}{\wedge}

\newcommand{\thecore}{J}

\newcommand{\algsc}{\ensuremath{{\mathcal{A}}_{\mbox{\textup{\scriptsize{ARV}}}}}\xspace}

\newcommand{\BCD}{\textsf{Basic Cluster Disengagement}\xspace}
\newcommand{\ACD}{\textsf{Advanced Cluster Disengagement}\xspace}
\newcommand{\algdecompose}{\ensuremath{\mathsf{AlgDecompose}}\xspace}
\newcommand{\algfindguiding}{\ensuremath{\mathsf{AlgFindGuiding}}\xspace}

\newcommand{\alginitpartition}{\ensuremath{\mathsf{AlgInitPartition}}\xspace}
\newcommand{\procsplit}{\ensuremath{\mathsf{ProcSplit}}\xspace}
\newcommand{\algbasicdisengagement}{\ensuremath{\mathsf{AlgBasicDisengagement}}\xspace}
\newcommand{\algadvanceddisengagement}{\ensuremath{\mathsf{AlgAdvancedDisengagement}}\xspace}
\newcommand{\algclassifycluster}{\ensuremath{\mathsf{AlgClassifyCluster}}\xspace}

\newcommand{\alg}{\ensuremath{\mathsf{Alg}}\xspace}

\newcommand{\parent}{\mathsf{parent}}
\newcommand{\lchild}{\mathsf{lchild}}
\newcommand{\rchild}{\mathsf{rchild}}
\newcommand{\cjset}{\check{\mathcal{J}}}
\newcommand{\cJ}{\check J}

\newcommand{\cK}{\check K}
\newcommand{\cpsi}{\check \psi}
\newcommand{\notS}{\overline{S}}
\newcommand{\notU}{\overline{U}}
\newcommand{\alphasc}{\ensuremath{\beta_{\mbox{\tiny{\sc ARV}}}}}
\newcommand{\cCMG}{\ensuremath{c_{\mbox{\tiny{\sc CMG}}}}}

\newcommand{\cro}{\mathsf{cr}}
\newcommand{\out}{\mathsf{out}}
\newcommand{\dep}{\mathsf{dep}}
\newcommand{\inn}{\mathsf{in}}

\newcommand{\algcombine}{\ensuremath{\mathsf{AlgCombineDrawings}}\xspace}
\newcommand{\algrec}{\ensuremath{\mathsf{AlgRecursiveCNwRS}}\xspace}
\newcommand{\cnwrs}{\ensuremath{\mathsf{MCNwRS}}\xspace}

\newcommand{\vstrip}{\mathsf{VStrip}}
\newcommand{\hstrip}{\mathsf{HStrip}}
\newcommand{\portals}{\mathsf{Portals}}
\newcommand{\entryportals}{\mathsf{EntryPortals}}
\newcommand{\exitportals}{\mathsf{ExitPortals}}
\newcommand{\intpairs}{\mathsf{IntPairs}}
\newcommand{\cellr}{\mathsf{CellRegion}}

\newcommand{\guided}{\operatorname{guided}}
\newcommand{\gd}{\operatorname{light}}
\newcommand{\light}{\operatorname{light}}
\newcommand{\down}{\operatorname{down}}
\newcommand{\up}{\operatorname{up}}
\newcommand{\bad}{\operatorname{bad}}
\newcommand{\forbidden}{\times}
\newcommand{\con}{\operatorname{concise}}
\newcommand{\cellset}{\mathsf{CellSet}}
\newcommand{\cell}{\mathsf{Cell}}

\newcommand{\good}{\operatorname{good}}

\newcommand{\mcn}{\textsf{Minimum Crossing Number}\xspace}

\newcommand{\CNwRS}{\textnormal{\textsf{MCNwRS}}\xspace}
\newcommand{\MCN}{\textsf{MCN}\xspace}

\newcommand{\del}{\mathsf{del}}
\newcommand{\dirty}{\mathsf{dirty}}

\newcommand{\lef}{\operatorname{left}}

\newcommand{\rig}{\operatorname{right}}
\newcommand{\spann}{\operatorname{span}}
\newcommand{\straight}{\operatorname{straight}}
\newcommand{\through}{\operatorname{over}}
\newcommand{\thr}{\textsf{thr}}
\newcommand{\temp}{\beta}
\newcommand{\ntcost}{\mathsf{cost}_{\mathsf {NT}}}
\newenvironment{prog}[1]{
	\begin{minipage}{6.5 in}
		\begin{center}
			{\sc #1}
		\end{center}
	}
	{
	\end{minipage}
}

\newcommand{\program}[3]{\begin{figure} \fbox{\vspace{2mm}\begin{prog}{#1} #3 \end{prog}\vspace{2mm}} 
		\caption{#1 \label{#2}} \end{figure}}

\newcommand{\cost}{\operatorname{cost}}
\newcommand{\inE}{E^{\textsf{out}}}
\newcommand{\Eout}{E^{\textnormal{\textsf{out}}}}
\newcommand{\heout}{\hat E^{\textnormal{\textsf{out}}}}

\newcommand{\innerE}{E^{\textsf{in}}}

\newcommand{\tfset}{\tilde{\mathcal{F}}}
\newcommand{\tG}{\textbf{G}}

\newcommand{\cH}{\check H}
\newcommand{\tE}{\textbf{E}'}

\newcommand{\tpsi}{\bm{\psi}}

\newcommand{\NP}{\mbox{\sf NP}}

\newcommand{\opt}{\mathsf{OPT}}
\newcommand{\optcro}{\mathsf{OPT}_{\mathsf{cr}}}
\newcommand{\optcrors}{\mathsf{OPT}_{\mathsf{cnwrs}}}
\newcommand{\set}[1]{\left\{ #1 \right\}}
\newcommand{\sse}{\subseteq}

\newcommand{\tset}{{\mathcal T}}
\newcommand{\uset}{{\mathcal U}}
\newcommand{\iset}{{\mathcal{I}}}
\newcommand{\pset}{{\mathcal{P}}}
\newcommand{\hpset}{{\hat{\mathcal{P}}}}

\newcommand{\dset}{{\mathcal{D}}}

\newcommand{\qset}{{\mathcal{Q}}}

\newcommand{\lset}{{\mathcal{L}}}
\newcommand{\bset}{{\mathcal{B}}}

\newcommand{\aset}{{\mathcal{A}}}
\newcommand{\cset}{{\mathcal{C}}}
\newcommand{\fset}{{\mathcal{F}}}

\newcommand{\jset}{{\mathcal{J}}}
\newcommand{\kset}{{\mathcal K}}
\newcommand{\xset}{{\mathcal{X}}}
\newcommand{\wset}{{\mathcal{W}}}
\newcommand{\gset}{{\mathcal{G}}}
\newcommand{\oset}{{\mathcal{O}}}
\newcommand{\yset}{{\mathcal{Y}}}
\newcommand{\rset}{{\mathcal{R}}}

\newcommand{\hset}{{\mathcal{H}}}
\newcommand{\sset}{{\mathcal{S}}}

\newcommand{\event}{{\cal{E}}}
\newcommand{\floor}[1]{\ensuremath{\left\lfloor#1\right\rfloor}}
\newcommand{\ceil}[1]{\ensuremath{\left\lceil#1\right\rceil}}
\newtheorem{theorem}{Theorem}[section]
\newtheorem{lemma}[theorem]{Lemma}

\newtheorem{observation}[theorem]{Observation}
\newtheorem{corollary}[theorem]{Corollary}
\newtheorem{claim}[theorem]{Claim}

\newtheorem{definition}[theorem]{Definition}
\newenvironment{proof}{\par \smallskip{\bf Proof:}}{\hfill\stopproof}
\def\stopproof{\square}
\def\square{\vbox{\hrule height.2pt\hbox{\vrule width.2pt height5pt \kern5pt
\vrule width.2pt} \hrule height.2pt}}

\newenvironment{proofof}[1]{\noindent{\bf Proof of #1.}}
{\hspace*{\fill}\stopproof}

\renewcommand{\phi}{\varphi}
\newcommand{\eps}{\epsilon}

\newcommand{\half}{\ensuremath{\frac{1}{2}}}
\newcommand{\poly}{\operatorname{poly}}
\newcommand{\dist}{\mbox{\sf dist}}
\newcommand{\reals}{{\mathbb R}}

\newcommand{\expect}[2][]{\text{\bf E}_{#1}\left [#2\right]}
\newcommand{\prob}[2][]{\text{\bf Pr}_{#1}\left [#2\right]}

\newenvironment{properties}[2][0]
{
\begin{enumerate} \setcounter{enumi}{#1}}{\end{enumerate}}

\renewcommand{\cong}{\operatorname{cong}}
\newcommand{\dem}{\operatorname{Dem}}

\newcommand{\bc}{\mathsf{BC}}

\newcommand{\interestconst}{{50}}

\begin{document}
\bibliographystyle{alpha}
	

\begin{titlepage}
	
	\title{A Subpolynomial Approximation Algorithm for Graph Crossing Number in Low-Degree Graphs\footnote{Extended Abstract to appear in STOC 2022.}}
	\author{Julia Chuzhoy\thanks{Toyota Technological Institute at Chicago. Email: {\tt cjulia@ttic.edu}. Supported in part by NSF grant CCF-2006464.}  \and Zihan Tan\thanks{Computer Science Department, University of Chicago. Email: {\tt zihantan@uchicago.edu}. Supported in part by NSF grant CCF-2006464.}}
	\maketitle

	\thispagestyle{empty}
	\begin{abstract}
We consider the classical Minimum Crossing Number problem: given an $n$-vertex graph $G$, compute a drawing of $G$ in the plane, while minimizing the number of crossings between the images of its edges. This is a fundamental and extensively studied problem, whose approximability status is widely open. In all currently known approximation algorithms, the approximation factor depends polynomially on $\Delta$ -- the maximum vertex degree in $G$. The best current approximation algorithm achieves an $O(n^{1/2-\eps}\cdot \poly(\Delta\cdot\log n))$-approximation, for a small fixed constant $\epsilon$, while the best negative result is APX-hardness, leaving a large gap in our understanding of this basic problem. In this paper we design a randomized $O\left(2^{O((\log n)^{7/8}\log\log n)}\cdot\poly(\Delta)\right )$-approximation algorithm for Minimum Crossing Number. This is 
the first approximation algorithm for the problem that achieves a subpolynomial in $n$ approximation factor (albeit only in graphs whose maximum vertex degree is  subpolynomial in $n$). 

In order to achieve this approximation factor, we design a new algorithm for a closely related problem called Crossing Number with Rotation System, in which, for every vertex $v\in V(G)$, the circular ordering, in which the images of the edges incident to $v$ must enter the image of $v$ in the drawing is fixed as part of input. Combining this result with the recent reduction of [Chuzhoy, Mahabadi, Tan '20] immediately yields the improved approximation algorithm for Minimum Crossing Number. 
We introduce several new technical tools, that we hope will be helpful in obtaining better algorithms for the problem in the future.

\end{abstract}
\end{titlepage}

\pagenumbering{gobble}
\tableofcontents
\newpage 
\pagenumbering{arabic}

\newpage

\section{Introduction}

We study the classical Minimum Crossing Number (\MCN) problem: given an $n$-vertex graph $G$, compute a drawing of $G$ in the plane while minimizing the number of its crossings. Here, a drawing $\phi$ of a graph $G$ is a mapping, that maps every vertex $v\in V(G)$ to some point $\phi(v)$ in the plane, and every edge $e=(u,v)\in E(G)$ to a continuous simple curve $\phi(e)$, whose endpoints are $\phi(u)$ and $\phi(v)$. For a vertex $v\in V(G)$ and an edge $e\in E(G)$, we refer to $\phi(v)$ and to $\phi(e)$ as the \emph{images} of $v$ and of $e$, respectively. We require that, for every vertex $v$ and edge $e$, $\phi(v)\in \phi(e)$ only if $v$ is an endpoint of $e$. We also require that, if some point $p$ belongs to the images of three or more edges, then it must be the image of a shared endpoint of these edges. 
A \emph{crossing} in a drawing $\phi$ of $G$ is a point that belongs to the images of two edges of $G$, and is not their common endpoint. The \emph{crossing number} of a graph $G$, denoted by $\optcro(G)$, is the minimum number of crossings in any drawing of $G$ in the plane.

The \MCN problem was initially introduced by Tur\'an \cite{turan_first} in 1944, and has been extensively studied since then  (see, e.g., \cite{chuzhoy2011algorithm, chuzhoy2011graph, chimani2011tighter, chekuri2013approximation, KawarabayashiSidi17, kawarabayashi2019polylogarithmic,chuzhoy2020towards}, and also \cite{richter_survey, pach_survey, matousek_book, vrto_biblio, schaefer2012graph} for excellent surveys). The problem is of interest to several communities, including, for example, graph theory and algorithms, and graph drawing.  
As such, much effort was invested into studying it  from different angles. 
But despite all this work, most aspects of the problem are still poorly understood. 

In this paper we focus on the algorithmic aspect of \MCN. Since the problem is  \NP-hard \cite{crossing_np_complete}, and it remains \NP-hard even in cubic graphs \cite{Hlineny06a, cabello2013hardness}, it is natural to consider approximation algorithms for it. 
Unfortunately, the approximation ratios of all currently known algorithms depend polynomially on $\Delta$, the maximum vertex degree of the input graph. To the best of our knowledge, no non-trivial approximation algorithms are known for the general setting, where $\Delta$ may be arbitrarily large.
One of the most famous results in this area, the Crossing Number Inequality, by Ajtai, Chv\'atal, Newborn and Szemer\'edi \cite{ajtai82} and by Leighton \cite{leighton_book}, shows that, for every graph $G$ with $|E(G)|\geq 4|V(G)|$, the crossing number of $G$ is  $\Omega(|E(G)|^3/|V(G)|^2)$.
Since the problem is most interesting when the crossing number of the input graph is low, it is reasonable to focus on low-degree graphs, where the maximum vertex degree $\Delta$ is bounded by either a constant, or a slowly-growing (e.g. subpolynomial) function of $n$. While we do not make such an assumption explicitly, like in all previous work, the approximation factor that we achieve also depends polynomially on $\Delta$. 

Even in this setting, there is still a large gap in our understanding of the problem's approximability, and the progress in closing this gap has been slow. On the negative side, only APX-hardness is known \cite{cabello2013hardness,ambuhl2007inapproximability},
that holds even in cubic graphs. 
On the positive side, the first non-trivial approximation algorithm for \MCN was obtained by Leighton and Rao in their seminal paper \cite{leighton1999multicommodity}.
Given as input an $n$-vertex graph $G$, the algorithm computes a drawing of $G$ with at most $O((n+\optcro(G)) \cdot \Delta^{O(1)} \log^4n)$ crossings.
This bound was later improved to $O((n+\optcro(G))\cdot \Delta^{O(1)} \log^3n)$ by \cite{even2002improved}, and then to $O((n+\optcro(G)) \cdot \Delta^{O(1)} \log^2n)$   following the improved approximation algorithm of  \cite{ARV} for Sparsest Cut. Note that all these algorithms only achieve an $O(n \poly(\Delta\log n)))$-approximation factor. However, their performance improves significantly when the crossing number of the input graph is large.
A sequence of papers~\cite{chuzhoy2011graph,chuzhoy2011algorithm} provided an improved $\tilde O(n^{0.9}\cdot\Delta^{O(1)})$-approximation algorithm for \MCN, followed by a more recent sequence of papers  by Kawarabayashi and Sidiropoulos \cite{KawarabayashiSidi17, kawarabayashi2019polylogarithmic}, who obtained an $\tilde O\left(\sqrt{n}\cdot \Delta^{O(1)}\right )$-approximation algorithm. All of the above results follow the same high-level algorithmic framework, and it was shown by Chuzhoy, Madan and Mahabadi \cite{chuzhoy-lowerbound}  (see \cite{chuzhoy2016improved} for an exposition) that this framework is unlikely to yield a better than  $O(\sqrt{n})$-approximation.
The most recent result, by 
%
%
%
Chuzhoy, Mahabadi and Tan \cite{chuzhoy2020towards}, obtained an $\tilde O(n^{1/2-\eps}\cdot \poly(\Delta))$-approximation  algorithm for some small fixed constant $\eps>0$. This result was achieved by proposing a new algorithmic framework for the problem, that departs from the previous approach. Specifically, \cite{chuzhoy2020towards} reduced the \MCN problem to another problem, called Minimum Crossing Number with Rotation System  (\CNwRS) that we discuss below, which appears somewhat easier than the \MCN problem, and then provided an algorithm for approximately solving the \cnwrs problem.

Our main result is  a randomized  $O\left(2^{O((\log n)^{7/8}\log\log n)}\cdot\Delta^{O(1)}\right )$-approximation algorithm for \MCN. In order to achieve this result, we design a new algorithm for the \CNwRS problem that achieves significantly stronger guarantees than those of  \cite{chuzhoy2020towards}. This algorithm, combined with the reduction of  \cite{chuzhoy2020towards}, immediately implies the improved approximation for the \MCN problem. We also design several new technical tools that we hope will eventually lead to further improvements. We now turn to discuss the \CNwRS problem.

In the Minimun Crossing Number with Rotation System (\CNwRS) problem, the input consists of a multigraph $G$, and, for every vertex $v\in V(G)$, a circular ordering $\oset_v$ of edges that are incident to $v$, that we call a \emph{rotation} for vertex $v$.
The set $\Sigma=\set{\oset_v}_{v\in V(G)}$ of all such rotations is called a \emph{rotation system} for graph $G$. We say that a drawing $\phi$ of $G$ \emph{obeys} the rotation system $\Sigma$, if, for every vertex $v\in V(G)$, the images of the edges in $\delta_G(v)$ enter the image of $v$ in the order $\oset_v$ (but the \emph{orientation} of the ordering can be either clock-wise or counter-clock-wise). In the \CNwRS problem, given a graph $G$ and a rotation system $\Sigma$ for $G$, the goal is to compute a drawing $\phi$ of $G$ that obeys the rotation system $\Sigma$ and minimizes the number of edge crossings. For an instance $I=(G,\Sigma)$ of the \CNwRS problem, we denote by $\optcrors(I)$ the value of the optimal solution for $I$, that is, the smallest number of crossings in any drawing of $G$ that obeys $\Sigma$. The results of \cite{chuzhoy2020towards} show the following reduction from \MCN to \CNwRS: suppose there is an efficient (possibly randomized) algorithm for the \CNwRS problem, that, for every instance $I=(G,\Sigma)$, produces a solution whose expected cost is at most $\alpha(m)\cdot (\optcrors(I)+m)$, where $m=|E(G)|$. Then there is a randomized $O(\alpha(n)\cdot \poly(\Delta\cdot\log n))$-approximation algorithm for the \MCN problem.  Our main technical result is a randomized algorithm, that, given an instance $I=(G,\Sigma)$ of \cnwrs, with high probability produces a solution to instance $I$ with at most $2^{O((\log m)^{7/8}\log\log m)}\cdot \left(\optcrors(G,\Sigma)+m\right)$ crossings, where $m=|E(G)|$. Combining this with the result of \cite{chuzhoy2020towards}, we immediately obtain a randomized $O\left(2^{O((\log n)^{7/8}\log\log n)}\cdot\poly(\Delta)\right )$-approximation algorithm for the \MCN problem.


The best previous algorithm for the \cnwrs problem, due to \cite{chuzhoy2020towards}, is a randomized algorithm, that, given an instance $I=(G,\Sigma)$ of the problem, with high probability produces a solution with at most $\tilde O\left((\optcrors(G,\Sigma)+m)^{2-\eps}\right)$ crossings, where $\eps=1/20$.
A variant of \CNwRS was previously studied by Pelsmajer et al.~\cite{pelsmajer2011crossing}, where for each vertex $v$ of the input graph $G$, both the rotation $\oset_v$ of its incident edges, and the orientation of this rotation (say clock-wise) are fixed. 
They showed that this variant of the problem is also \NP-hard, and provided an $O(n^4)$-approximation algorithm with running time $O(m^n\log m)$, where $n=|V(G)|$ and $m=|E(G)|$.
They also obtained approximation algorithms with improved guarantees for some special families of graphs.

We introduce a number of new technical tools, that we discuss in more detail in \Cref{subsec: techniques}. Some of these tools require long and technically involved proofs, which resulted in the large length of the paper. We view these tools as laying a pathway towards obtaining better algorithms for the Minimum Crossing Number problem, and it is our hope that these tools will eventually be streamlined and that their proofs will be simplified, leading to a better understanding of the problem and cleaner and simpler algorithms. We believe that some of these tools are interesting in their own right.


\subsection{Our Results}

Throughout this paper, we allow graphs to have parallel edges (but not self-loops); graphs with no parallel edges are explicitly called simple graphs. For convenience, we will assume that the input to the \MCN problem is a simple graph, while graphs serving as inputs to the \cnwrs problem may have parallel edges. The latter is necessary in order to use the reduction of \cite{chuzhoy2020towards} between the two problems. Note that the number of edges in a graph with parallel edges may be much higher than the number of vertices.
Our main technical contribution is an algorithm for the \cnwrs problem, that is summarized in the following theorem.

\begin{theorem}
\label{thm: main_rotation_system}
There is an efficient randomized algorithm, that, given an instance $I=(G,\Sigma)$ of \CNwRS with $|E(G)|=m$, computes a drawing of $G$ that obeys the rotation system $\Sigma$. The number of crossings in the drawing is w.h.p. bounded by $2^{O((\log m)^{7/8}\log\log m)}\cdot \left(\optcrors(I)+m\right)$. 
\end{theorem}

We rely on the following theorem from \cite{chuzhoy2020towards} in order to obtain an approximation algorithm for the \MCN problem.

\begin{theorem}[Theorem 1.3 in~\cite{chuzhoy2020towards}]
\label{thm: MCN_to_rotation_system}
There is an efficient algorithm, that, given an $n$-vertex graph $G$ with maximum vertex degree $\Delta$, computes an instance $I=(G',\Sigma)$ of the \CNwRS problem, with $|E(G')|\leq O\left(\optcro(G)\cdot \poly(\Delta\cdot\log n)\right)$, and   $\optcrors(I)\leq O\left(\optcro(G)\cdot \poly(\Delta\cdot\log n)\right )$. Moreover, there is an efficient algorithm that, given any solution  of value $X$ to instance $I$ of \CNwRS, computes a drawing of $G$ with the number of crossings bounded by  $O\left ((X+\optcro(G))\cdot \poly(\Delta\cdot\log n)\right )$.
\end{theorem}

Combining \Cref{thm: main_rotation_system} and \Cref{thm: MCN_to_rotation_system}, we immediately obtain the following corollary, whose proof appears in Section \ref{sec: proof of main theorem} of Appendix.

\begin{corollary}
\label{thm: main_result}
There is an efficient randomized algorithm, that, given a simple $n$-vertex  graph $G$ with maximum vertex degree $\Delta$, computes a drawing of $G$, such that, w.h.p., the number of crossings in the drawing is at most $O\left( 2^{O((\log n)^{7/8}\log\log n)}\cdot\poly(\Delta)\cdot \optcro(G) \right )$.
\end{corollary}

\subsection{Our Techniques}
\label{subsec: techniques}

In this subsection we provide an overview of the techniques used in the proof of our main technical result, \Cref{thm: main_rotation_system}. For the sake of clarity of exposition, some of the discussion here is somewhat imprecise. Our algorithm relies on the divide-and-conquer technique. Given an instance $I=(G,\Sigma)$ of the \cnwrs problem, we compute a collection $\iset$ of new instances, whose corresponding graphs are significantly smaller than $G$, and then solve each of the resulting new instances separately. Collection $\iset$ of instances is called a \emph{decomposition of $I$}. We require that the decomposition has several useful properties that will allow us to use it in order to obtain the guarantees from \Cref{thm: main_rotation_system}, by solving the instances in $\iset$ recursively. Before we define the notion of decomposition of an instance, we need the notion of a \emph{contracted graph}, that we use throughout the paper. Suppose $G$ is a graph, and let $\rset=\set{R_1,\ldots,R_q}$ be a collection of disjoint subsets of vertices of $G$. The contracted graph of $G$ with respect to $\rset$, that we denote by $G_{|\rset}$, is a graph that is obtained from $G$, by contracting, for all $1\leq i\leq q$, the vertices of $R_i$ into a supernode $u_i$. Note that every edge of the resulting graph $G_{|\rset}$ corresponds to some edge of $G$, and we do not distinguish between them. 
The vertices in set $V(G_{|\rset})\setminus \set{u_1,\ldots,u_q}$ are called \emph{regular vertices}. Each such vertex $v$ also lies in $G$, and moreover, $\delta_{G_{|\rset}}(v)=\delta_G(v)$. Abusing the notation, given a collection $\cset=\set{C_1,\ldots,C_r}$ of disjoint subgraphs of $G$, we denote by $G_{|\cset}$ the contracted graph of $G$ with respect to the collection $\set{V(C_1),\ldots,V(C_r)}$ of subsets of vertices of $G$.
Given a graph $G$ and its drawing $\phi$, we denote by $\cro(\phi)$ the number of crossings in $\phi$. 

\noindent{\bf Decomposition of an Instance.}
Given an instance $I=(G,\Sigma)$ of the \cnwrs problem, we will informally refer to $|E(G)|$ as the \emph{size of the instance}.
Assume that we are given an instance $I=(G,\Sigma)$ of \cnwrs with $|E(G)|=m$, and some parameter $\eta$ (we will generally use $\eta=2^{O((\log m)^{3/4}\log\log m)}$). Assume further that we are given another collection $\iset$ of instances of \cnwrs. We say that $\iset$ is an \emph{$\eta$-decomposition of $I$}, if $\sum_{I'=(G',\Sigma')\in \iset}|E(G')|\leq m\poly\log m$, and $\sum_{I'\in \iset}\optcrors(I')\le \left (\optcrors(I)+|E(G)|\right )\cdot \eta$. Additionally, we require that there is an efficient algorithm $\alg(I)$, that, given a feasible solution $\phi(I')$ to every instance $I'\in \iset$, computes a feasible solution $\phi$ for instance $I$, with at most $O\left (\sum_{I'\in \iset}\cro(\phi(I'))\right )$ crossings.

At a high level, our algorithm starts with the input instance $I^*=(G^*,\Sigma^*)$ of the \cnwrs problem. Throughout the algorithm, we denote $m^*=|E(G^*)|$, and we use a parameter $\mu=2^{O((\log m^*)^{7/8}\log\log m^*)}$. Over the course of the algorithm, we  consider various other instances $I$ of \cnwrs, but parameters $m^*$ and $\mu$ remain unchanged, and they are defined with respect to the original input instance $I^*$. The main subroutine of the algorithm, that we call \algdecompose, receives as input an instance $I=(G,\Sigma)$ of \cnwrs, and computes an $\eta$-decomposition $\iset$ of $I$, for  $\eta=2^{O((\log m)^{3/4}\log\log m)}$, where $m=|E(G)|$. The subroutine additionally ensures that every instance in the decomposition is sufficiently small compared to $I$, that is, for each instance $I'=(G',\Sigma')\in \iset$, $|E(G')|\leq |E(G)|/\mu$. 
We note that this subroutine is in fact randomized, and, instead of ensuring that $\sum_{I'\in \iset}\optcrors(I')\le \left (\optcrors(I)+|E(G)|\right )\cdot \eta$, it only ensures this in expectation. We will ignore this minor technicality in this high-level exposition.

It is now easy to complete the proof of \Cref{thm: main_rotation_system} using Algorithm $\algdecompose$: we simply apply Algorithm $\algdecompose$ to the input instance $I^*$, obtaining a collection $\iset$ of new  instances. We recursively solve each instance in $\iset$, and then combine the resulting solutions using Algorithm $\alg(I^*)$, in order to obtain the final solution to instance $I^*$. Since the sizes of the instances decrease by the factor of at least $\mu$ with each application of the algorithm, the depth of the recursion is bounded by $O\left ((\log m^*)^{1/8}\right )$. At each recursive level, the sum of the optimal solution costs and of the number of edges in all instances at that recursive level increases by at most factor $2^{O((\log m^*)^{3/4}\log\log m^*)}$, leading to the final bound of $2^{O((\log m^*)^{7/8}\log\log m^*)}\cdot \left(\optcrors(I^*)+m^*\right)$ on the solution cost. 

From now on we focus on the description of Algorithm \algdecompose. 
We start by describing several technical tools that this algorithm builds on. Throughout, given a graph $G$, we refer to connected vertex-induced subgraphs of $G$ as \emph{clusters}. Given a collection $\cset$ of disjoint clusters of $G$, we denote by $\Eout_G(\cset)$ the set of all edges $e\in E(G)$, such that the endpoints of $e$ do not lie in the same cluster of $\cset$. 
We will also use the notion of \emph{subinstances} that we define next.

\paragraph{Subinstances.} Suppose we are given two instances $I=(G,\Sigma)$ and $I'=(G',\Sigma')$ of \cnwrs. We say that $I'$ is a \emph{subinstance} of instance $I$, if the following hold. First, graph $G'$ must be a graph that is obtained from a subgraph of $G$ by contracting some subsets of its vertices into supernodes. Formally\footnote{We note that this definition closely resembles the notion of graph minors, but, in contrast to the definition of minors, we do not require that the induced subgraphs $\set{G[R_i]}_{1\leq i\leq q}$ are connected.}, there must be a graph $G''\subseteq G$, and a collection $\rset=\set{R_1,\ldots,R_q}$ of disjoint subsets of vertices of $G''$, such that $G'=G''_{|\rset}$. For every regular vertex $v$ of $G'$, the rotation $\oset_v\in \Sigma'$ must be consistent with the rotation $\oset_v\in \Sigma$ (recall that $\delta_{G'}(v)\subseteq \delta_G(v)$). For every supernode $u_i$ of $G'$, its rotation $\oset_{u_i}\in \Sigma'$ can be chosen arbitrarily.
Note that the notion of subinstances is transitive: if $I'$ is a subinstance of $I$ and $I''$ is a subinstance of $I'$, then $I''$ is a subinstance of $I$. 


The main tool that we use is \emph{disengagement of clusters}. 
Intuitively, given an instance $I=(G,\Sigma)$ of \cnwrs, and a collection $\cset$ of disjoint clusters of $G$, the goal is to compute an $\eta$-decomposition $\iset$ of $I$, such that every instance $I'=(G',\Sigma')\in \iset$ is a subinstance of $I$, and moreover, there is at most one cluster $C\in \cset$ that is contained in $G'$, and all edges of $G'$ that do not lie in $C$ must belong to $\Eout_G(\cset)$. Assume for now that we can design an efficient algorithm for computing such a decomposition. In this case, the high-level plan for implementing Algorithm $\algdecompose$ would be as follows. First, we compute a collection $\cset$ of disjoint clusters of graph $G$, such that, for each cluster $C\in \cset$, $|E(C)|\leq |E(G)|/(2\mu)$, and $|\Eout_G(\cset)|\leq |E(G)|/(2\mu)$. Then we perform disengagement of clusters in $\cset$, obtaining an $\eta$-decomposition of the input instance $I$. We are then guaranteed that each resulting instance in $\iset$ is sufficiently small. We note that it is not immediately clear how to compute the desired collection $\cset$ of disjoint clusters of $G$; we discuss this later. For now we focus on algorithms for computing disengagement of clusters. We do not currently have an algorithm to compute the disengagement of clusters in the most general setting described above. In this paper, we design a number of algorithms for computing disengagement of clusters, under some conditions. We start with the simplest algorithm that only works in some restricted settings, and then generalize it to more advanced algorithms that work in more and more general settings. In order to describe the disengagement algorithm for the most basic setting, we need the notion of congestion, and of internal and external routers, that we use throughout the paper, and describe next.

\noindent{\bf Congestion, Internal Routers, and External Routers.}
Given a graph $G$ and a set $\pset$ of paths in $G$, the \emph{congestion} that the set $\pset$ of paths causes on an edge $e\in E(G)$, that we denote by $\cong_G(\pset,e)$, is the number of paths in $\pset$ containing $e$. The total congestion caused by the set $\pset$ of paths in $G$ is $\cong_G(\pset)=\max_{e\in E(G)}\set{\cong_G(\pset,e)}$.

Consider now a graph $G$ and a cluster $C\subseteq G$. We denote by $\delta_G(C)$ the set of all edges $e\in E(G)$, such that exactly one endpoint of $e$ lies in $C$. An \emph{internal $C$-router} is a collection $\qset(C)=\set{Q(e)\mid e\in \delta_G(C)}$ of paths, such that, for each edge $e\in \delta_G(C)$, path $Q(e)$ has $e$ as its first edge, and all its inner vertices lie in $C$. We additionally require that all paths in $\qset(C)$ terminate at a single vertex of $C$, that we call the \emph{center vertex of the router}. Similarly, an \emph{external $C$-router} 
 is a collection $\qset'(C)=\set{Q'(e)\mid e\in \delta_G(C)}$ of paths, such that, for each edge $e\in \delta_G(C)$, path $Q'(e)$ has $e$ as its first edge, and all its inner vertices lie in $V(G)\setminus V(C)$. We additionally require that all paths in $\qset'(C)$ terminate at a single vertex of $V(G)\setminus V(C)$, that we call the \emph{center vertex of the router}. For a cluster $C\subseteq G$, we denote by $\Lambda_G(C)$ and $\Lambda'_G(C)$ the sets of all internal and all external $C$-routers, respectively.
 
 \noindent{\bf Basic Cluster Disengagement.}
 In the most basic setting for cluster disengagement, we are given an instance $I=(G,\Sigma)$ of the \cnwrs problem, and a collection $\cset$ of disjoint clusters of $G$. Additionally, for each cluster $C\in \cset$, we are given an internal $C$-router $Q(C)$, whose center vertex we denote by $u(C)$, and an external $C$-router $Q'(C)$, whose center vertex we denote by $u'(C)$. The output of the disengagement procedure is a collection $\iset$ of subinstances of $I$, that consists of a single global instance $\hat I=(\hat G,\hat \Sigma)$, and, for every cluster $C\in \cset$, an instance $I_C=(G_C,\Sigma_C)$ associated with it. Graph $\hat G$ is the contracted graph of $G$ with respect to $\cset$; that is, it is obtained from $G$ by contracting every cluster $C\in \cset$ into a supernode $v_C$. For each cluster $C\in \cset$, graph $G_C$ is obtained from $G$ by contracting the vertices of $V(G)\setminus V(C)$ into a supernode $v^*_C$. For every cluster $C\in \cset$, the rotation $\oset_{v_C}\in \hat \Sigma$ of the supernode $v_C$ in instance $\hat I$ and the rotation $\oset_{v^*_C}\in \Sigma_C$ of the supernode $v^*_C$ in instance $I_C$ need to be defined carefully, in order to ensure that the sum of the optimal solution costs of all resulting instances is low, and that we can combine the solutions to these instances to obtain a solution to $I$. Observe that the set of edges incident to vertex $v_C$ in $\hat G$ and the set of edges incident to vertex $v^*_C$ in $G_C$ are both equal to $\delta_G(C)$. We define a single ordering $\oset^C$ of the edge set $\delta_G(C)$, that will serve both as the rotation  $\oset_{v_C}\in \hat \Sigma$, and as the rotation $\oset_{v^*_C}\in \Sigma_C$. The ordering $\oset^C$ is defined via the internal $C$-router $\qset(C)$, and the order in which the images of the paths of $\qset(C)$ enter the image of vertex $u(C)$. 
 On the one hand, letting  $\oset_{v_C}=\oset_{v^*_C}$ for every cluster $C\in \cset$ allows us to easily combine solutions $\phi(I')$ to instances $I'\in \iset$, in order to obtain a solution to instance $I$, whose cost is at most $O\left(\sum_{I'\in \iset'}\cro(\phi(I'))\right )$. On the other hand, defining $\oset^C$ via the set $\qset(C)$ of paths, for each cluster $C\in \cset$, allows us to bound $\sum_{I'\in \iset}\optcrors(I')$. 
 
 We now briefly describe how this latter bound is established, since it will motivate the remainder of the algorithm and clarify  the bottlenecks of this approach. We consider an optimal solution $\phi^*$ to instance $I$, and we use it in order to construct, for each instance $I'\in \iset$, a solution $\psi(I')$, such that $\sum_{I'\in \iset}\cro(\psi(I'))$ is relatively small compared to $\cro(\phi^*)+|E(G)|$.
  In order to construct  a solution $\psi(\hat I)$ to the global instance $\hat I$, we start with solution $\phi^*$ to instance $I$. We erase from this solution all edges and vertices that lie in the clusters of $\cset$. For each cluster $C\in \cset$, we let the image of the supernode $v_C$ coincide with the original image of the vertex $u(C)$ -- the center of the internal $C$-router $Q(C)$. In order to draw the edges that are incident to the supernode $v_C$ in $\hat G$ (that is, the edges of $\delta_G(C)$), we utilize the images of the paths of the internal $C$-router $\qset(C)$ in $\phi^*$, that connect, for each edge $e\in \delta_G(C)$, the original image of edge $e$ to the original image of vertex $u(C)$.
 
 Consider now some cluster $C\in \cset$. In order to construct a solution $\psi(I_C)$ to instance $I_C$, we start again with the solution $\phi^*$ to instance $I$. We erase from it all edges and vertices except for those lying in $C$. We let the image of the supernode $v^*_C$ be the original image of vertex $u'(C)$ -- the center of the external $C$-router $\qset'(C)$. In order to draw the edges that are incident to the supernode $v^*_C$ in $G_C$ (that is, the edges of $\delta_G(C)$), we utilize the images of the paths of the external $C$-router $\qset'(C)$, that connect, for each edge $e\in \delta_G(C)$, the original image of edge $e$ to the original image of vertex $u'(C)$.
 
 Observe that the only increase in $\sum_{I'\in \iset}\cro(\psi(I'))$, relatively to $\cro(\phi^*)$, is due to the crossings incurred by drawing the edges incident to the supernodes in $\set{v_C}_{C\in \cset}$ in instance $\hat I$, and for each subinstance $I_C$, drawing the edges incident to supernode $v^*_C$. All such edges are drawn along the images of the paths in $\bigcup_{C\in \cset}(\qset(C)\cup \qset'(C))$ in $\phi^*$. However, an edge may belong to a number of such paths. With careful accounting we can bound this number of new crossings as follows. Assume that, for every cluster $C\in \cset$, $\cong_G(\qset'(C))\leq \beta$. Assume further that, for each cluster $C\in \cset$, and for each edge $e\in E(C)$, $(\cong_G(\qset(C),e))^2\leq \beta$. Then $\sum_{I'\in \iset}\cro(\psi(I'))\leq O(\beta^2\cdot (\optcrors(I)+|E(G)|))$. Therefore, in order to ensure that the collection $\iset$ of subinstances of $I$ that we have obtained via the cluster disengagement procedure is an $\eta$-decomposition of $I$, we need to ensure that, for every cluster $C\in \cset$, $\cong_G(\qset'(C))\leq \beta$, and, for every edge $e\in E(C)$, $(\cong_G(\qset(C),e))^2\leq \beta$, for $\beta=O(\eta^{1/2})$. This requirement seems impossible to achieve. For example, if maximum vertex degree in graph $G$ is small (say a constant), then some edges incident to the center vertices $\set{u(C),u'(C)}_{C\in \cset}$ must incur very high congestion. In order to overcome this obstacle, we slightly weaken our requirements. Instead of providing, for every cluster $C\in \cset$, a single internal $C$-router $Q(C)$, and a single external $C$-router $Q'(C)$, it is sufficient for us to obtain, for each cluster $C\in \cset$, a \emph{distribution} $\dset(C)$ over the collection $\Lambda_G(C)$ of internal $C$-routers, such that, for every edge $e\in E(C)$, $\expect[\qset(C)\sim \dset(C)]{(\cong_G(\qset(C),e))^2}\leq \beta$, and a distribution $\dset'(C)$ over 
 the collection $\Lambda'_G(C)$ of external $C$-routers, such that for every edge $e$,   $\expect[\qset'(C)\sim \dset'(C)]{\cong_G(\qset'(C),e)}\leq \beta$.

To recap, in order to use the \BCD procedure described above to compute an $\eta$-decomposition of the input instance $I$ of \cnwrs into sufficiently small instances, it is now enough to design a procedure that, given an instance $I=(G,\Sigma)$ of \cnwrs, computes a collection $\cset$ of disjoint clusters of $G$, and, for every cluster $C\in \cset$, a distribution $\dset(C)$ over the collection $\Lambda_G(C)$ of internal $C$-routers, such that, for every edge $e\in E(C)$, $\expect[\qset(C)\sim \dset(C)]{(\cong_G(\qset(C),e))^2}\leq \beta$, together with a distribution $\dset'(C)$ over 
the collection $\Lambda'_G(C)$ of external $C$-routers, such that, for every edge $e$,  $\expect[\qset'(C)\sim \dset'(C)]{\cong_G(\qset'(C),e)}\leq \beta$, for $\beta=O(\sqrt\eta)$.
Additionally, we need to ensure that, for every cluster $C\in \cset$, $|E(C)|\leq |E(G)|/(2\mu)$, and that $|\Eout_G(\cset)|\leq |E(G)|/(2\mu)$. While computing a collection $\cset$ of clusters with the latter two properties seems possible (at least when the maximum vertex degree in $G$ is small), computing the distributions over the internal and the external routers for each cluster $C$ with the required properties seems quite challenging. As a first step towards this goal, we employ the standard notions of well-linkedness and bandwidth property of clusters as a proxy to constructing internal $C$-routers with the required properties. Before we turn to discuss these notions, we note that the \BCD procedure that we have just described can be easily generalized to the more general setting, where the set $\cset$ of clusters is laminar (instead of only containing disjoint clusters). 
This generalization will be useful for us later.

Assume that we are given a laminar family $\cset$ of clusters (that is, for every pair $C,C'\in \cset$ of clusters, either $C\subseteq C'$, or $C'\subseteq C$, or $C\cap C'=\emptyset$ holds), with $G\in \cset$. Assume further that we are given, for each cluster $C\in \cset$, a distribution $\dset(C)$ over the collection $\Lambda_G(C)$ of internal $C$-routers, in which, for every edge $e\in E(C)$, $\expect[\qset(C)\sim \dset(C)]{(\cong_G(\qset(C),e))^2}\leq \beta$, together with a distribution $\dset'(C)$ over 
the collection $\Lambda'_G(C)$ of external $C$-routers, where for every edge $e$,  $\expect[\qset'(C)\sim \dset'(C)]{\cong_G(\qset'(C),e)}\leq \beta$, for some parameter $\beta$.
The \BCD procedure, when applied to $\cset$, produces a collection $\iset=\set{I_C=(G_C,\Sigma_C)\mid C\in \cset}$ of instances. For every cluster $C\in \cset$, graph $G_C$ associated with instance $I_C$ is obtained from graph $G$, by first contracting all vertices of $V(G)\setminus V(C)$ into a supernode $v^*_C$, and then contracting, for each child-cluster $C'\in \cset$ of $C$, the vertices of $V(C')$ into a supernode $v_{C'}$. We define, for every cluster $C$, an ordering of the set $\delta_G(C)$ of edges via an internal $C$-router that is selected from the distribution $\dset(C)$, and we let the rotation $\oset_{v^*_C}$ in the rotation system $\Sigma_C$, and the rotation $\oset_{v_C}$ in the rotation system $\Sigma_{C'}$, where $C'$ is the parent-cluster of $C$, to be identical to this ordering. 
Using the same reasoning as in the case where $\cset$ is a set of disjoint clusters, we show that $\expect{\sum_{I'\in \iset}\optcrors(I')}\leq O\left( \beta^2\cdot \dep(\cset)\cdot(\optcrors(I)+|E(G)|)\right )$, where $\dep(\cset)$ is the depth of the laminar family $\cset$ of clusters. We then show that $\iset'$ is an $\eta'$-decomposition of instance $I$, where $\eta'=O(\beta^2\cdot \dep(\cset))$.

As noted already, one of the difficulties in exploiting the \BCD procedure in order to compute an $\eta$-decomposition of the input instance $\iset$  is the need to compute distributions over the sets of internal and the external $C$-routers for every cluster $C\in\cset$, with the required properties. We turn instead to the notions of well-linkedness and bandwidth properties of clusters. These notions are extensively studied, and there are many known algorithms for computing a collection $\cset$ of clusters that have bandwidth property in a graph. We will use this property as a proxy, that will eventually allow us to construct a distribution over the sets of internal $C$-routers for each cluster $C\in \cset$, with the required properties.

{\bf Well-Linkedness, Bandwidth Property, and Cluster Classification.}
We use the standard graph-theoretic notion of well-linkedness. Let $G$ be a graph,  let $T$ be a subset of the vertices of $G$, and let $0<\alpha<1$ be a parameter. We say that the set $T$ of vertices is \emph{$\alpha$-well-linked} in $G$ if for every partition $(A,B)$ of vertices of $G$ into two subsets, $|E_G(A,B)|\geq \alpha\cdot\min\set{|A\cap T|,|B\cap T|}$. 

We also use a closely related notion of \emph{bandwidth property} of clusters. Suppose we are given a graph $G$ and a cluster $C\subseteq G$. Intuitively, cluster $C$ has the $\alpha$-bandwidth property (for a parameter $0<\alpha<1$), if the edges of $\delta_G(C)$ are $\alpha$-well-linked in $C$. More formally, we consider the augmentation $C^+$ of cluster $C$, that is defined as follows. We start with the graph $G$, and subdivide every edge $e\in \delta_G(C)$ with a vertex $t_e$, denoting by $T=\set{t_e\mid e\in \delta_G(C)}$ this new set of vertices.  The augmentation $C^+$ of $C$ is the subgraph of the resulting graph induced by $V(C)\cup T$. We say that cluster $C$ has the $\alpha$-bandwidth property if set $T$ of vertices is $\alpha$-well-linked in $C^+$.

We note that, if a cluster $C$ has the $\alpha$-bandwidth property, then, using known techniques, we can efficiently construct a distribution $\dset$ over the set $\Lambda_G(C)$ of internal $C$-routers, such that, for every edge $e\in E(C)$, $\expect[\qset(C)\sim \dset(C)]{\cong(\qset(C),e)}\leq O(1/\alpha)$.
However, in order to use the \BCD procedure, we need a stronger property: namely, for every edge $e\in E(C)$, we require that $\expect[\qset(C)\sim \dset(C)]{(\cong(\qset(C),e))^2}\leq \beta$, for some parameter $\beta$. If we are given a distribution $\dset(C)$ over the set $\Lambda_G(C)$ of internal $C$-routers with this latter property, then we say that cluster $C$ is \emph{$\eta$-light} with respect to $\dset(C)$. Computing a distribution $\dset(C)$ for which cluster $C$ is $\eta$-light is a much more challenging task. We come close to achieving it in our Cluster Classification Theorem. Before we describe the theorem, we need one more definition. Let $C$ be a cluster of a graph $G$, and let $\eta'$ be some parameter. Assume that we are given some rotation system $\Sigma$ for graph $G$, and let $\Sigma^C$ be the rotation system for cluster $C$ that is induced by $\Sigma$. Let $I^C=(C,\Sigma^C)$ be the resulting instance of \cnwrs. We say that cluster $C$ is \emph{$\eta'$-bad} if $\optcrors(I^C)\geq |\delta_G(C)|^2/\eta'$. 

In the Cluster Classification Theorem, we provide an efficient algorithm, that, given an instance $I=(G,\Sigma)$ of \cnwrs with $|E(G)|=m$, and a cluster $C\subseteq G$ that has the $\alpha$-bandwidth property (where $\alpha=\Omega(1/\poly\log m)$), either correctly establishes that cluster $C$ is $\eta'$-bad, for $\eta'=2^{O((\log m)^{3/4}\log\log m)}$, or produces a distribution $\dset(C)$ over the set $\Lambda_G(C)$ of internal $C$-routers, such that cluster $C$ is $\beta$-light with respect to $\dset(C)$, for $\beta=2^{O(\sqrt{\log m}\cdot \log \log m)}$. 
In fact, the algorithm is randomized, and, with a small probability, it may erroneously classify cluster $C$ as being $\eta'$-bad, when this is not the case. This small technicality is immaterial to this high-level exposition, and we will ignore it here. 
The proof of the Cluster Classification Theorem is long and technically involved, and is partially responsible for the high approximation factor that we eventually obtain. It is our hope that a simpler and a cleaner proof of the theorem with better parameters will be discovered in the future. We believe that the theorem is a graph-theoretic result that is interesting in its own right. We now provide a high-level summary of the main challenges in its proof. 

At the heart of the proof is an algorithm that we called \algfindguiding. Suppose we are given an instance $I=(H,\Sigma)$ of \cnwrs, and a set $T$ of $k$ vertices of $H$ called terminals, that are $\alpha$-well-linked in $H$, for some parameter $0<\alpha<1$. Denote $C=H\setminus T$ and $|V(H)|=n$. The goal of the algorithm is to either establish that $\optcrors(H)+|E(H)|\geq k^2\poly(\alpha/\log n)$; or to compute a distribution $\dset(C)$ over internal $C$-routers, such that cluster $C$ is $\eta'=\poly(\log n/\alpha)$-light with respect to $\dset(C)$. 

Consider first a much simpler setting, where $H$ is a grid graph, and $T$ is the set of vertices on the first row of the grid. For this special case, the algorithm of \cite{Tasos-comm}  (see also Lemma D.10 in the full version of \cite{chuzhoy2011algorithm}) provides the construction of a distribution $\dset(C)$ over internal $C$-routers with the required properties. 
This result can be easily generalized to the case where $H$ is a bounded-degree planar graph,  since such a graph must contain a large grid minor.  If $H$ is a planar graph, but its maximum vertex degree is no longer bounded, we can still compute a grid-like structure in it, and apply the same arguments as in \cite{Tasos-comm} in order to compute the desired distribution $\dset(C)$. 
 The difficulty in our case is that the graph $H$ may be far from being planar, and, even though, from the Excluded Grid theorem of Robertson and Seymour \cite{robertson1986graph,robertson1994quickly},
it must contain a large grid-like structure, without having a drawing of $H$ in the plane with a small number of crossings, we do not know how to compute such a structure\footnote{We note that we need the grid-like structure to have dimensions $(k'\times k')$, where $k'$ is almost linear in $k$. Therefore, we cannot use the known bounds for the Excluded Minor Theorem (e.g. from \cite{chuzhoy2019towards}) for general graphs, and instead we need to use  an analogue of the stronger version of the theorem for planar graphs.}. We provide an algorithm that either establishes that $\optcrors(H)$ is large compared to $k^2$, or computes a grid-like structure in graph $H$, even if it is not a planar graph. Unfortunately, this algorithm only works in the setting where $|E(H)|$ is not too large comparable to $k$. Specifically, if we ensure that $|E(H)|\leq k\cdot \hat \eta$ for some parameter $\hat \eta$, then the algorithm either computes a distribution $\dset(C)$ over internal $C$-routers that is $\eta'$-light (with $\eta'=\poly(\log n/\alpha)$ as before), or it establishes that $\optcrors(H)+|E(H)|\geq k^2\poly(\alpha/(\hat \eta\log n))$.

Typically, this algorithm would be used in the following setting: we are given a cluster $C$ of a graph $G$, that has the $\alpha$-bandwidth property. We then let $H=C^+$ be the augmentation of $C$, and we let $T$ be the set of vertices of $C^+$ corresponding to the edges of $\delta_H(C)$. In order for this result to be meaningful, we need to ensure that $|E(C)|$ is not too large compared to $|\delta_H(C)|$. Unfortunately, we may need to apply the classification theorem to clusters $C$ for which $|E(C)|\gg |\delta_H(C)|$ holds. In order to overcome this difficulty, given such a cluster $C$, we construct a recursive decomposition of $C$ into smaller and smaller clusters. Let $\lset$ denote the resulting family of clusters, which is a laminar family of subgraphs of $C$. We ensure that every cluster $C'\in \lset$ has $\alpha=\Omega(1/\poly\log m)$-bandwidth property, and, additionally, if we let $\hat C'$ be the graph obtained from $C'$ by contracting every child-cluster of $C'$ into a supernode, then the number of edges in $\hat C'$ is comparable to $|\delta_H(C')|$. We consider the clusters of $\lset$ from smallest to largest. For each such cluster $C'$, we carefully apply Algorithm \algfindguiding to the corresponding contracted graph $\hat C'$, in order to either classify cluster $\hat C'$ as $\eta(C')$-bad, or to compute a distribution $\dset(C')$ over internal $C'$-routers, such that $C'$ is $\beta(C')$-light with respect to $\dset(C')$. Parameters $\eta(C')$ and $\beta(C')$ depend on the height of the cluster $C'$ in the decomposition tree that is associated with the laminar family $\lset$ of clusters. This recursive algorithm is eventually used to either establish that cluster $C$ is $\eta(C)$-bad, or to compute a distribution $\dset(C)$ over the set $\Lambda_G(C)$ of internal $C$-routers, such that cluster $C$ is $\beta(C)$-light with respect to $\dset(C)$. The final parameters $\eta(C)$ and $\beta(C)$ depend exponentially on the height of the decomposition tree associated with the laminar family $\lset$. This strong dependence on $\dep(\lset)$ is one of the reasons for the high approximation factor that our algorithm eventually achieves.

 \noindent{\bf Obstacles to Using \BCD.}
 Let us now revisit the \BCD routine. We start with an instance $I=(G,\Sigma)$ of \cnwrs, and denote $|E(G)|=m$. Throughout, we use a parameter $\eta=2^{O((\log m)^{3/4}\log\log m)}$, and $\beta=\eta^{1/8}$.
 Recall that the input to the procedure is a collection $\cset$ of disjoint clusters of $G$. For every cluster $C\in \cset$, we are also given a distribution $\dset'(C)$ over the set of external $C$-routers, such that,  for every edge $e$,  $\expect[\qset'(C)\sim \dset'(C)]{\cong_G(\qset'(C),e)}\leq \beta$, and a distribution $\dset(C)$ over the set of internal $C$-routers, such that cluster $C$ is $\beta$-light with respect to $\dset(C)$. 
 We are then guaranteed that the collection $\iset$ of subinstances of $I$ that is constructed via \BCD is an $\eta$-decomposition of $I$. We can slightly generalize this procedure to handle bad clusters as well.
 Specifically, suppose we are given a partition $(\cset^{\light},\cset^{\bad})$ of the clusters in $\cset$, and, for each cluster $C\in \cset^{\light}$, a distribution $\dset(C)$ over internal $C$-routers, such that cluster $C$ is $\beta$-light with respect to $\dset(C)$. Assume further that each cluster $C\in \cset^{\bad}$ is $\beta$-bad. Additionally,  assume that we are given, for every cluster $C\in \cset$,  a distribution $\dset'(C)$ over external $C$-routers, such that, for every edge $e$,  $\expect[\qset'(C)\sim \dset'(C)]{\cong_G(\qset'(C),e)}\leq \beta$, and that every cluster $C\in \cset$ has the $\alpha$-bandwidth property, for some $\alpha=\Omega(1/\poly\log m)$. We can then generalize the \BCD procedure to provide the same guarantees as before in this setting, to obtain an   $\eta$-decomposition of instance $I$.

 Assume now that we are given an instance $I=(G,\Sigma)$ of \cnwrs, with $|E(G)|=m$. For simplicity, assume for now that the maximum vertex degree in $G$ is quite small (it is sufficient, for example, that it is significantly smaller than $m$.) Using known techniques, we can compute a collection $\cset$ of disjoint clusters of $G$, such that, for every cluster $C\in \cset$, $|E(C)|\leq m/(2\mu)$; $|\Eout_G(\cset)|\leq m/(2\mu)$; and every cluster $C\in \cset$ has $\alpha$-bandwidth property. 
 If we could, additionally, compute, for each cluster $C\in \cset$,  a distribution $\dset'(C)$ over external $C$-routers, such that, 
for every edge $e$,  $\expect[\qset'(C)\sim \dset'(C)]{\cong_G(\qset'(C),e)}\leq \beta$, then we could use the Cluster Classification Theorem to partition the set $\cset$ of clusters into subsets $\cset^{\light}$ and $\cset^{\bad}$, and to compute, for every cluster $C\in \cset^{\light}$,  a distribution $\dset(C)$ over the set of its internal routers, such that  every cluster in $\cset^{\bad}$ is $\eta'$-bad, and every cluster $C\in \cset^{\light}$ is $\eta'$-light with respect to $\dset(C)$, for some parameter $\eta'$. We could then apply the \BCD procedure in order to compute the desired $\eta$-decomposition of the input instance $I$. 
 Unfortunately, we currently do not  have an algorithm that computes both the collection $\cset$ of clusters of $G$ with the above properties, and the required distributions over the external $C$-routers for each such cluster $C$. In order to overcome this difficulty, we design \ACD procedure, that generalizes \BCD, and no longer requires the distribution over external $C$-routers for each cluster $C\in \cset$.

 \noindent{\bf \ACD.}
 The input to the \ACD procedure is an instance $I=(G,\Sigma)$ of \cnwrs, and a set $\cset$ of disjoint clusters of $G$, that we refer to as \emph{basic clusters}. Let $m=|E(G)|$, and $\eta=2^{O((\log m)^{3/4}\log\log m)}$ as before. The output is an $\eta$-decomposition $\iset$ of $I$, such that every instance $I'=(G',\Sigma')\in \iset$ is a subinstance of $I$, and moreover, there is at most one basic cluster $C\in \cset$ with $E(C)\subseteq E(G')$, with all other edges of $G'$ lying in $\Eout_G(\cset)$. The algorithm for the \ACD and its analysis are significantly more involved than those of \BCD. We start with some intuition.
 
 Consider the contracted graph $H=G_{|\cset}$, and its Gomory-Hu tree $T$. This tree naturally defines a laminar family $\lset$ of clusters of $H$: for every vertex $v\in V(H)$, we add to $\lset$ the cluster $U_v$, that is the subgraph of $H$ induced by vertex set $V(T_v)$, where $T_v$ is the subtree of $T$ rooted at $v$. From the properties of Gomory-Hu trees, if $v'$ is the parent-vertex of vertex $v$ in $T$, there is an external $U_v$-router $\qset'(U_v)$ in graph $H$ with $\cong_H(\qset'(U_v))=1$. Laminar family $\lset$ of clusters of $H$ naturally defines a laminar family $\lset'$ of clusters of the original graph $G$, where for each cluster $U_v\in \lset$, set $\lset'$ contains a corresponding cluster $U'_v$, that is obtained from $U_v$, by un-contracting all supernodes that correspond to clusters of $\cset$. For each such cluster $U'_v\in \lset'$, we can use the external $U_v$-router $\qset'(U_v)$ in graph $H$ in order to construct  a distribution $\dset'(U'_v)$ over external $U'_v$-routers in graph $G$, where for every edge $e$, 
 $\expect[\qset'(U'_v)\sim \dset'(U'_v)]{\cong_G(\qset'(U'_v),e)}\leq O(1/\alpha)$. We can then apply the \BCD procedure to the laminar family $\lset'$ and the distributions $\set{\dset'(U'_v)}_{U'_v\in \lset'}$ in order to compute an $\eta^*$-decomposition $\iset$ of instance $I$, where every instance in $\iset$ is a subinstance of $I$. Recall that the parameter $\eta^*$ depends on the depth of the laminar family $\lset'$, which is equal to the depth of the laminar family $\lset$. Therefore, if $\dep(\lset)$ is not too large (for example, it is at most $2^{O((\log m)^{3/4}\log\log m)}$), then we will obtain the desired $\eta$-decomposition of $I$. But unfortunately we have no control over the depth of the laminar family $\lset$, and in particular the tools described so far do not work when the Gomory-Hu tree $T$ is a path.
 
 Roughly speaking, we would like to design a different disengagement procedure for the case where the tree $T$ is a path, and then reduce the general problem (by exploiting \BCD) to this special case. In fact we follow a similar plan. We define a special type of instances (that we call \emph{nice instances}), that resemble the case where the Gomory-Hu tree of the contracted graph $H=G_{|\cset}$ is a path. While the motivation behind the definition of nice instances is indeed this special case, the specifics of the definition are somewhat different, in that it is more general in some of its aspects, and more restrictive and well-structured in others. We provide an algorithm for Cluster Disengagment of nice instances, that ensures that, for each resulting subinstance $I'=(G',\Sigma')$, there is at most one cluster $C\in \cset$ with $C\subseteq G'$, and all other edges of $G'$ lie in $\Eout_G(\cset)$. We also provide another algorithm, that, given an instance $I=(G,\Sigma)$ of \cnwrs and a collection $\cset$ of disjoint basic clusters of graph $G$, computes a decomposition $\iset'$ of instance $I$, such that each resulting instance $I'=(G',\Sigma')\in \iset'$ is a subinstance of $I$ and a nice instance, with respect to the subset $\cset(I')\subseteq \cset$ of clusters, that contains every cluster $C\in \cset$ with $C\subseteq G'$.  
 Combining these two algorithms allows us to compute \ACD.
 
 \noindent{\bf Algorithm \algdecompose.}
 	Recall that Algorithm 
 	\algdecompose, given an instance $I=(G,\Sigma)$ of \cnwrs, needs to compute an $\eta$-decomposition $\iset$ of $I$, where  $\eta=2^{O((\log m)^{3/4}\log\log m)}$ and $m=|E(G)|$, such that, for each instance $I'=(G',\Sigma')\in \iset$, $|E(G')|\leq |E(G)|/\mu$. 
 We say that a vertex $v$ of $G$ is a \emph{high-degree vertex} if $|\delta_G(v)|\geq m/\poly(\mu)$ (here, $\mu=2^{O((\log m^*)^{7/8}\log\log m^*)}$, and $m^*$ is the number of edges in the original input instance $I^*$ of \cnwrs).
 
 Consider first the special case where no vertex of $G$ is a high-degree vertex. For this case, it is not hard to generalize known well-linked decomposition techniques to obtain a collection $\cset$ of disjoint clusters of $G$, such that each cluster $C\in \cset$ has $\alpha=\Omega(1/\poly\log m)$-bandwidth property, with $|E(C)|<O(m/\mu)$, and,  additionally, $|\Eout_G(\cset)|\leq O(m/\mu)$. We can now apply the \ACD procedure to the set $\cset$ of clusters, in order to obtain the desired $\eta$-decomposition of $I$. Recall that we are guaranteed that each resulting instance $I'=(G',\Sigma')\in \iset$ is a subinstance of $I$, and there is at most one cluster $C\in \cset$ with $C\subseteq G'$, with all other edges of $G'$ lying in $\Eout_G(\cset)$. This ensures that $|E(G')|\leq m/\mu$, as required.
 
 In general, however, it is possible that the input instance $I=(G,\Sigma)$ contains high-degree vertices. We then consider two cases. We say that instance $I$ is \emph{wide} if
 there is a vertex $v\in V(G)$, a partition $(E_1,E_2)$ of the edges of $\delta_G(v)$, such that the edges of $E_1$ appear consecutively in the rotation $\oset_v\in \Sigma$, and so do the edges of $E_2$, and a collection $\pset$  of at least $m/\poly(\mu)$ simple edge-disjoint cycles in $G$, such that every cycle $P\in \pset$ contains one edge of $E_1$ and one edge of $E_2$. An instance that is not wide is called \emph{narrow}. We provide separate algorithms for dealing with narrow and wide instances.

\begin{figure}[h]
 	\centering
 	\scalebox{0.13}{\includegraphics{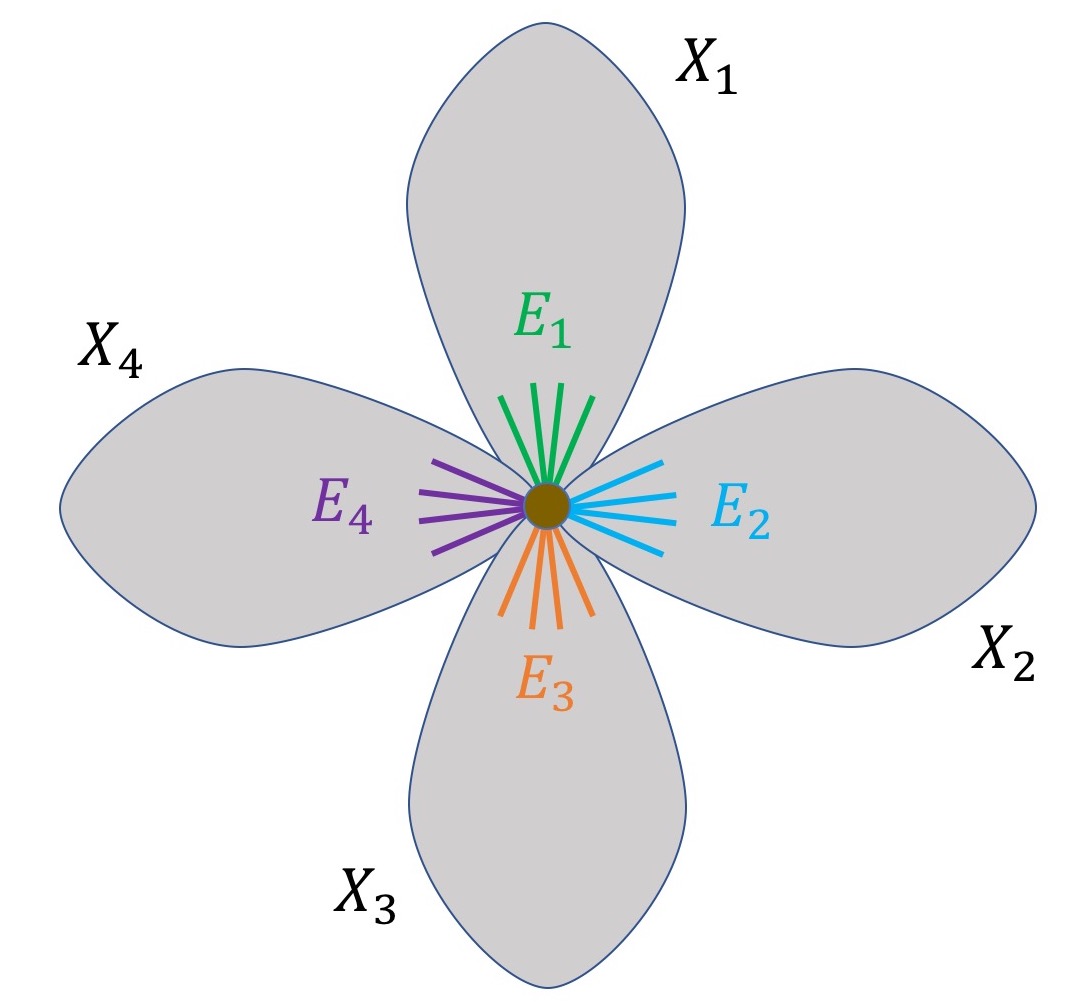}}
 	\caption{An illustration of a $4$-petal flower cluster.}\label{fig: flower_cluster intro}
 \end{figure}

 \noindent{\bf Narrow Instances.}
 The algorithm for decomposing narrow instances relies on and generalizes the algorithm for the special case where $G$ has no high-degree vertices. As a first step, we compute a collection $\cset$ of disjoint clusters of $G$, such that each cluster $C\in \cset$ has $\alpha=\Omega(1/\poly\log m)$-bandwidth property, and $|\Eout_G(\cset)|<O(E(G)/\mu)$. The set $\cset$ of clusters is partitioned into two subsets: set $\cset^{s}$ of small clusters, and set $\cset^{f}$ of \emph{flower clusters}. For each cluster $C\in \cset^s$, $|E(C)|<O(|E(G)|/\mu)$ holds. If $C$ is a cluster of $\cset^f$, then we no longer guarantee that $|E(C)|$ is small. Instead, we guarantee that cluster $C$ has a special structure. Specifically, $C$ must contain a single high-degree vertex $u(C)$, that we call the \emph{flower center}, and all other vertices of $C$ must be low-degree vertices. Additionally, there must be a set $\xset(C)=\set{X_1,\ldots,X_k}$ of subgraphs of $C$, that we call \emph{petals}, such that, for all $1\leq i<j\leq k$, $V(X_i)\cap V(X_j)=\set{u(C)}$. We also require that, for all $1\leq i\leq k$, there is a set $E_i$ of $\Theta(m/\poly(\mu))$ edges of $\delta_G(u(C))$ that are contiguous in the rotation $\oset_{u(C)}\in \Sigma$, and lie in $X_i$  (see \Cref{fig: flower_cluster intro}). Lastly, we require that, for all $1\leq i\leq k$, there is a set $\qset_i$ of edge-disjoint paths, connecting every edge of $\delta_G(X_i)\setminus\delta_G(u(C))$ to vertex $u(C)$, with all inner vertices on the paths lying in $X_i$.

We apply \ACD to the set $\cset$ of clusters, in order to compute an initial decomposition $\iset_1$ of the input instance $I$, such that every instance in $\iset_1$ is a subinstance of $I$. Unfortunately, it is  possible that, for some instances $I'=(G',\Sigma')\in \iset_1$, $|E(G')|>m/\mu$. For each such instance $I'$, there must be some flower cluster $C\in \cset^f$ that is contained in $G'$, and all other edges of $G'$ must lie in $\Eout_G(\cset)$.

We now consider each instance  $I'=(G',\Sigma')\in \iset_1$ with $|E(G')|>m/\mu$ one by one. Assume that $C\in \cset^f$ is the flower cluster that is contained in $G'$, and $\xset(C)=\set{X_1,\ldots,X_k}$ is the set of its petals. We further decompose instance $I'$ into a collection $\iset(C)$ of subinstances, that consists of a single global instance $\hat I(C)$, and $k$ additional instances $I_1(C),\ldots,I_k(C)$. We ensure that the graph $\hat G(C)$ associated with the global instance $\hat I(C)$ only contains edges of $\Eout_G(\cset)$, so $|E(\hat G(C))|<m/\mu$ holds. For all $1\leq j\leq k$, graph $G_j(C)$ associated with instance $I_j(C)\in \iset(C)$ contains the petal $X_j$, and all other edges of $G_j(C)$ lie in $\Eout_G(\cset)$. We note that unfortunately it is still possible that, for some $1\leq j\leq k$, graph $G_j(C)$ contains too many edges (this can only happen if $|E(X_j)|$ is large). However, our construction ensures that, for each such instance $I_j(C)$, no high-degree vertices lie in graph $G_j(C)$. We can then further decompose instance $I_j(C)$ into subinstances using the algorithm that we designed for the case where no vertex of the input graph is a high-degree vertex. After this final decomposition, we are guaranteed that each of the final subinstances of instance $I$ that we obtain contains fewer than $m/\mu$ edges, as required.
 
\noindent{\bf Wide Instances.}
Suppose we are given a wide instance $I=(G,\Sigma)$ of \cnwrs. In this case, we compute an $\eta$-decomposition $\iset$ of instance $I$, such that, for each resulting instance $I'=(G',\Sigma')\in \iset$, either $|E(G')|<m/\mu$ (in which case we say that $I'$ is a \emph{small instance}), or $I'$ is a narrow instance. We will then further decompose each resulting narrow instance using the algorithm described above.

In order to obtain the decomposition $\iset$ of $I$, we start with $\iset=\set{I}$. As long as set $\iset$ contains at least one wide instance $I'=(G',\Sigma')$ with $|E(G')|\geq m/\mu$, we perform a procedure that ``splits'' instance $I'$ into two smaller subinstances. We now turn to describe this procedure at a high level.

Let $I'=(G',\Sigma')\in \iset$ be a wide instance with $|E(G')|\geq m/\mu$.
Recall that from the definition of a wide instance, there is a vertex $v\in V(G')$, a partition $(E_1,E_2)$ of the edges of $\delta_{G'}(v)$, such that the edges of $E_1$ appear consecutively in the rotation $\oset_v\in \Sigma'$,  and a collection $\pset$ of at least $m/\poly(\mu)$ simple edge-disjoint cycles in $G'$, such that every cycle in $\pset$ contains one edge of $E_1$ and one edge of $E_2$. 
Consider the experiment, in which we choose a cycle $W\in \pset$ uniformly at random. Since $|\pset|$ is very large, with reasonably high probability, the edges of the cycle $W$ participate in relatively few crossings in the optimal solution to instance $I'$ of \cnwrs. We show that with high enough probability, there is a near-optimal solution to $I'$, in which cycle $W$ is drawn in the natural way. We use the cycle $W$ in order to partition instance $I'$ into two subinstances $I_1,I_2$ (intuitively, one subinstance corresponds to edges and vertices of $G'$ that are drawn ``inside'' the cycle $W$ in the near-optimal solution to $I'$, and the other subinstance contains all edges and vertices that are drawn ``outside'' $W$). Each of the resulting two instances contains the cycle $W$, and, in order to be able to combine the solutions to the two subinstances to obtain a solution to $I'$, we contract all vertices and edges of $W$, in each of the two instances, into a supernode. Let $I_1',I_2'$ denote these two resulting instances. The main difficulty in the analysis is to show that there is a solution to each resulting instance of \cnwrs, such that the sum of the costs of two solutions is close to $\optcrors(I')$. The difficulty arises from the fact that we do not know what the optimal solution to instance $I'$ looks like, and so our partition of $G'$ into two subgraphs that are  drawn on different sides of the cycle $W$ in that solution may be imprecise. Instead, we need to ``fix'' the solutions  to instances $I_1,I_2$  (that are induced by the optimal solution to $I'$) in order to move all edges and vertices of each subinstance to lie on one side of the cycle $W$. In fact we are  unable to do so directly. Instead, we show that we can compute a relatively small collection $E'$ of edges, such that, if we remove the edges of $E'$ from the graphs corresponding to instances $I_1,I_2$, then each of the resulting subinstances has the desired structure: namely, it can be drawn completely inside or completely outside the cycle $W$ with relatively few crossings compared to $\optcrors(I')$. After we solve the two resulting subinstances recursively, we combine the resulting solutions, and add the images of the edges of $E'$ back, in order to obtain a solution to instance $I'$.

 \subsection{Organization}
 We start with preliminaries in \Cref{sec: short_prelim}. We then provide, in \Cref{sec: high level}, the definitions of several main concepts that we use (such as wide and narrow instances), and state three main technical theorems that allow us to decompose wide and narrow instances. We then provide the proof of \Cref{thm: main_rotation_system} using these three theorems. In \Cref{sec:long prelim} we provide additional definitions, notation and summary of known results that we use, together with some easy extensions. This section can be thought of as an expanded version of preliminaries. 
 We then develop our main technical tools: \BCD in \Cref{sec: guiding paths orderings basic disengagement}, Cluster Classification Theorem  in \Cref{sec: routing within a cluster} (with parts of the proof delayed to \Cref{sec: guiding paths}), and \ACD in \Cref{sec: main disengagement}.
 In Sections \ref{sec: not well connected}, \ref{sec: many paths} and \ref{sec: computing the decomposition} we provide the proofs of the three main theorems. Sections \ref{sec: not well connected} and \ref{sec: many paths}  deal with decomposing wide instances, and  \Cref{sec: many paths} provides an algorithm for decomposing a narrow instance.

\section{Preliminaries}
\label{sec: short_prelim}

By default, all logarithms in this paper are to the base of $2$. All graphs are undirected and finite. Graphs may contain parallel edges but they may not contain self loops.
Graphs without parallel edges are explicitly referred to as simple graphs.

\subsection{Graph-Theoretic Notation}

We follow standard graph-theoretic notation. Let $G=(V,E)$ be a graph.
For a vertex $v\in V$, we denote by $\delta_G(v)$ the set of all  edges of $G$ that are incident to $v$, and we denote $\deg_G(v)=|\delta_G(v)|$.
For two disjoint subsets $A,B$ of vertices of $G$, we denote by $E_G(A,B)$ the set of all edges with one endpoint in $A$ and the other in $B$.
For a subset $S\subseteq V$ of vertices, we denote by $G[S]$ the subgraph of $G$ induced by $S$, by $E_G(S)$ the set of all edges with both endpoints in $S$, and by $\delta_G(S)$ the set of all edges  with exactly one endpoint in $S$.
Abusing the notation, for a subgraph $C$ of $G$, we use  $\delta_G(C)$ to denote $\delta_G(V(C))$.

\begin{definition}[Congestion]
	Let $G$ be a graph, let $e$ be an edge of $G$, and let $\qset$  be a set of paths in $G$. The \emph{congestion that the set $\qset$ of paths causes on edge $e$}, denoted by $\cong_G(\qset,e)$, is the number of paths in $\qset$ that contain $e$. The \emph{total congestion of $\qset$ in $G$} is $\cong_G(\qset)=\max_{e\in E(G)}\set{\cong_G(\qset, e)}$.  
\end{definition}

\subsection{Curves in General Position, Graph Drawings, Faces, and Crossings}

Let $\gamma$ be an open curve in the plane, and let $P$ be a set of points in the plane. We say that $\gamma$ is \emph{internally disjoint} from $P$ if no inner point of $\gamma$ lies in $P$. In other words, $P\cap \gamma$ may only contain the endpoints of $\gamma$.
Given a set $\Gamma$ of open curves in the plane, we say that the curves in $\Gamma$ are \emph{internally disjoint} if, for every pair $\gamma,\gamma'\in \Gamma$ of distinct curves, every point $p\in \gamma\cap\gamma'$ is an endpoint of both curves.
We use the following definition of curves in general position.

\begin{definition}[Curves in general position]
	Let $\Gamma$ be a finite set of open curves in the plane.
	We say that the curves of $\Gamma$ are \emph{in general position}, if the following conditions hold:
	
	\begin{itemize}
		\item for every pair $\gamma,\gamma'\in \Gamma$ of distinct curves, there is a finite number of points $p$ with $p\in \gamma\cap  \gamma'$; 
		
		\item for every pair $\gamma,\gamma'\in \Gamma$ of distinct curves, an endpoint of $\gamma$ may not serve as an inner point of $\gamma'$ or of $\gamma$; and
		\item for every triple $\gamma,\gamma',\gamma''\in \Gamma$ of distinct curves, if some point $p$ lies on all three curves, then it must be an endpoint of each of these three curves.
	\end{itemize}
\end{definition}

Let $\Gamma$ be a set of curves in general position, 
and let $\gamma,\gamma'\in \Gamma$ be a pair of curves. Let $p$ be any point that lies on both $\gamma$ and $\gamma'$, but is not an endpoint of either curve. We then say that point $p$ is
a \emph{crossing} between $\gamma$ and $\gamma'$, or that curves $\gamma$ and $\gamma'$ \emph{cross} at point $p$.
We are now ready to formally define graph drawings.

\begin{definition}[Graph Drawings]
	A \emph{drawing}  $\phi$ of a graph $G$ in the plane is a map $\phi$,  that maps every vertex $v$ of $G$ to a point $\phi(v)$ in the plane (called the \emph{image of $v$}), and every edge $e=(u,v)$ of $G$ to a simple curve $\phi(e)$ in the plane whose endpoints are $\phi(u)$ and $\phi(v)$  (called the \emph{image of $e$}), such that all points in set $\set{\phi(v)\mid v\in V(G)}$ are distinct, and the set $\set{\phi(e)\mid e\in E(G)}$ of curves is in general position.  Additionally, for every vertex $v\in V(G)$ and edge $e\in E(G)$, $\phi(v)\in \phi(e)$ only if $v$ is an endpoint of $e$.
\end{definition}

Assume now that we are given some drawing $\phi$ of graph $G$ in the plane, and assume that for some pair $e,e'$ of edges, their images $\phi(e),\phi(e')$ cross at point $p$. Then we say that $(e,e')_p$ is a \emph{crossing} in the drawing $\phi$ (we may sometimes omit the subscript $p$ if the images of the two edges only cross at one point).	
We also say that $p$ is a \emph{crossing point} of drawing $\phi$.
We denote by $\cro(\phi)$ the total number of crossings in the drawing $\phi$. 

Note that a drawing of a graph $G$ in the plane naturally defines a drawing of $G$ on the sphere and vice versa; we use both types of drawings. 

For convenience, given a drawing $\phi$ of a graph $G$, we sometimes will not distinguish between the edges of $G$ and their images. For example, we may say that edges $e,e'$ cross in drawing $\phi$ to indicate that their images cross. Similarly, we may not distinguish between vertices and their images. For example, we may talk about the order in which edges of $\delta_G(v)$ enter vertex $v$ in drawing $\phi$, to mean the order in which the images of the edges of $\delta_G(v)$ enter the image of $v$. We denote by $\phi(G)=\left (\bigcup_{e\in E(G)}\phi(e)\right )\cup \set{\phi(v)\mid v\in V(G)}$. 

\paragraph{Images of Paths.}
Assume that we are given a graph $G$, its drawing $\phi$, and a path $P$ in $G$. The \emph{image of path $P$ in $\phi$}, denoted by $\phi(P)$, is the curve that is obtained by concatenating the images of all edges $e\in E(P)$. Equivalently, $\phi(P)=\bigcup_{e\in E(P)}\phi(e)$. If $P=\set{v}$ for some vertex $v$, then $\phi(P)=\phi(v)$.

\paragraph{Planar Graphs and Planar Drawings.}
A graph $G$ is \emph{planar} if there is a drawing of $G$ in the plane with no crossings. A drawing $\phi$ of a graph $G$ in the plane with $\cro(\phi)=0$ is called a \emph{planar drawing} of $G$. We use the following result by Hopcroft and Tarjan.

\begin{theorem}[\cite{hopcroft1974efficient}]\label{thm: testing planarity}
There is an algorithm, that, given a graph $G$, correctly establishes whether $G$ is planar, and if so, computes a planar drawing of $G$. The running time of the algorithm in $O(|V(G)|)$.
\end{theorem}



\paragraph{Faces of a Drawing.}
Suppose we are given a graph $G$ and a drawing $\phi$ of $G$ in the plane or on the sphere. The set of faces of $\phi$ is the set of all connected regions of $\mathbb{R}^2\setminus \phi(G)$. If $G$ is drawn in the plane, then we designate a single face of $\phi$ as the ``outer'', or the ``infinite'' face.

\paragraph{Identical Drawings and Orientations.}
Assume that we are given some planar drawing $\phi$ of a graph $G$. 
We can associate, with every face $F$ of this drawing, a subgraph $\partial(F)$ of $G$, containing all vertices and edges of $G$ whose images are contained in the boundary of $F$. 
 Drawing $\phi$ of $G$ can be uniquely defined by the list $\fset$ of all its faces, and, for every face $F\in \fset$, the corresponding subgraph $\partial(F)$ of $G$. In particular, if $\phi,\phi'$ are two planar drawings of the graph $G$, and there is an one-to-one mapping between the set $\fset$ of the faces of $\phi$ and the set $\fset'$ of the faces of $\phi'$, and, for every face $F$, the graph $\partial(F)$ is identical in both drawings, then we say that drawings $\phi$ and $\phi'$ are \emph{identical}.

Assume now that we are given a (possibly non-planar) drawing $\phi$ of a graph $G$. Let $G'$ be the graph obtained from $G$ by placing a vertex on every crossing of $\phi$. We then obtain a planar drawing $\psi$ of the resulting graph $G'$, where every vertex $v\in V(G')\setminus V(G)$ corresponds to a unique crossing point of $\phi$. For every edge $e\in E(G)$, let $L(e)$ be the list of all vertices of $G'$ that correspond to crossings in which edge $e$ participates in $\phi$, ordered in the order in which these crossings appear on the image of edge $e$ in $\phi$, as we traverse it from one endpoint to another. Graph $G'$, its planar drawing $\psi$, and the lists $\set{L(e)}_{e\in E(G)}$ uniquely define the drawing $\phi$ of $G$. In other words, if $\phi,\phi'$ are two drawings of the graph $G$, for which (i) the corresponding graphs $G'$ are the same (up to renaming the vertices of $V(G')\setminus V(G)$); (ii) the induced planar drawings $\psi$ of $G'$ are identical; and (iii) the vertex lists $\set{L(e)}_{e\in E(G)}$ are identical, then $\phi$ and $\phi'$ are \emph{identical drawings} of $G$.

Assume now that $\phi$ is a drawing of a graph $G$ in the plane, and let $\phi'$ be the drawing of $G$ that is the mirror image of $\phi$. We say that $\phi$ and $\phi'$ are \emph{identical} drawings of $G$, and that their \emph{orientations} are \emph{different}, or \emph{opposite}. We sometime say that $\phi'$ is obtained by \emph{flipping} the drawing $\phi$. 

We say that a graph $G$ is $3$-connected, if for every pair $u,v\in V(G)$ of its vertices, $G\setminus\set{u,v}$ is a connected graph. We use the following well known result.

\begin{theorem}[\cite{whitney1992congruent}]
	Every $3$-connected planar graph has a unique planar drawing.
\end{theorem}

\subsection{Grids and Their Standard Drawings}


The $(r\times r)$-grid is a graph whose vertex set is $\set{v_{i,j}\mid 1\le i,j\le r}$, and edge set is the union of the set $\set{(v_{i,j},v_{i,{j+1}})\mid 1\le i\le r, 1\le j< r}$ of  \emph{horizontal edges},  and the set $\set{(v_{i,j},v_{i+1,{j}})\mid 1\le i< r, 1\le j\le r}$ of \emph{vertical edges}.
For $1\le i\le r$, the \emph{$i$th row} of the grid is the subgraph of the grid graph induced by vertex set $\set{v_{i,j}\mid 1\le j\le r}$. Similarly, for $1\leq j\leq r$, the \emph{$j$th column} of the grid is the subgraph of the grid graph induced by vertex set $\set{v_{i,j}\mid 1\le i\le r}$.
Given an $(r\times r)$-grid, we refer to vertices $v_{1,1},v_{1,r},v_{r,1}$, and $v_{r,r}$ as the \emph{corners} of the grid. 
We also refer to the graph that is obtained from the union of row $1$, row $r$, column $1$, and column $r$, as the \emph{boundary} of the grid.

\begin{figure}[h]
	\centering
	\includegraphics[scale=0.1]{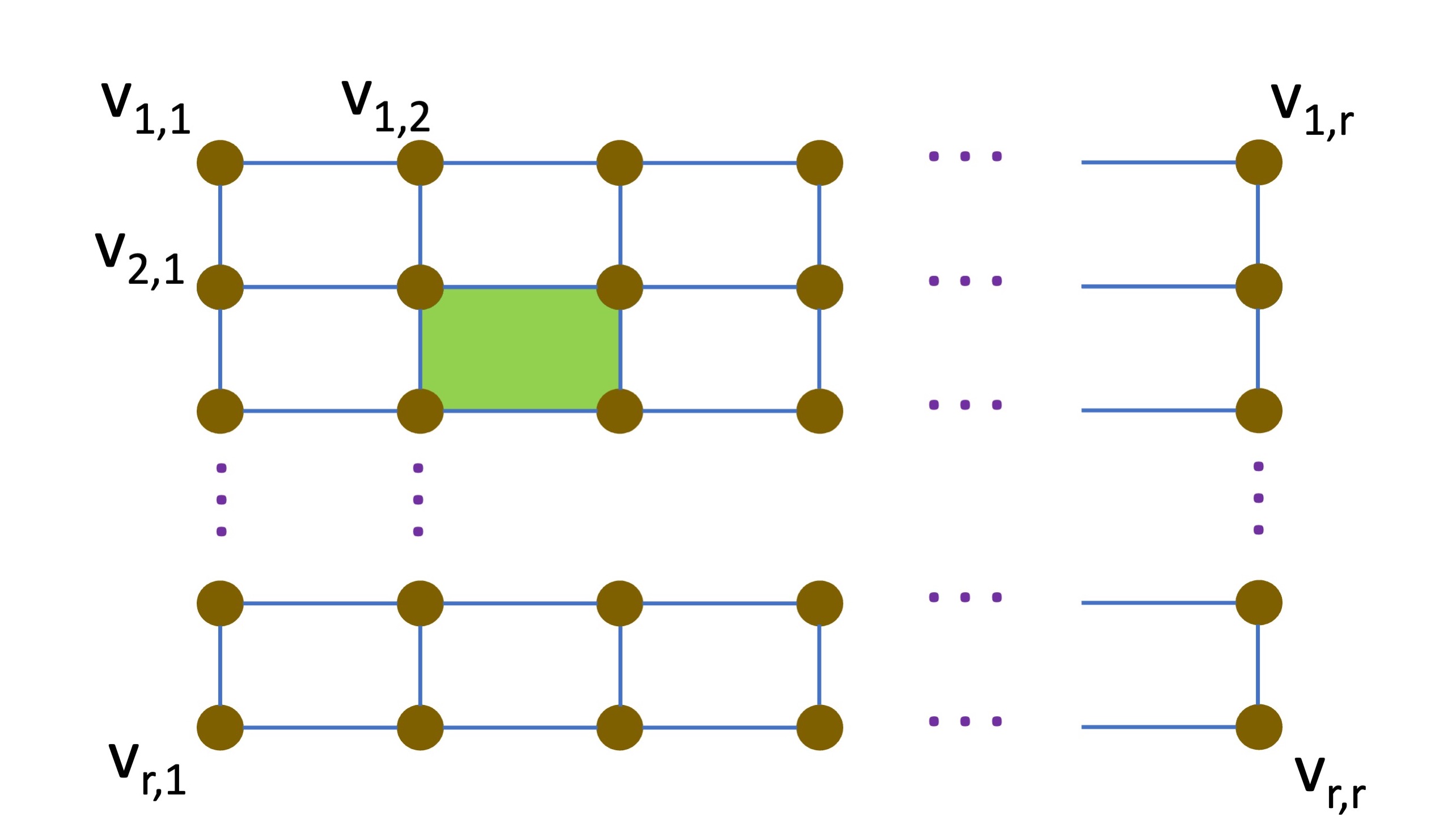}
	\caption{The standard drawing of the $(r\times r)$-grid with $\cell_{2,2}$ shown in green.
	}\label{fig: grid}
\end{figure}

It is not hard to see that the $(r\times r)$-grid has a unique planar drawing (this is since the $(1\!\times \!1)$-grid and the $(2\!\times \!2)$-grid have unique planar drawings, and for all $r\ge 3$, if we suppress the corner vertices of the grid, we obtain a planar $3$-connected graph, that has a unique planar drawing). We refer to this unique planar drawing of the grid as its \emph{standard drawing}
(see \Cref{fig: grid}).
For all $1\le i,j\le r-1$, we let $\cell_{i,j}$ be the face of the standard drawing, that contains the images of the vertices $v_{i,j},v_{i,j+1},v_{i+1,j},v_{i+1,j+1}$ on its boundary.

\subsection{Circular Orderings, Orientations, and Rotation Systems}
Suppose we are given a collection $U=\set{u_1,\ldots,u_r}$ of elements. Let $D$ be any disc in the plane.
Assume further that we are given, for every element $u_i\in U$, a point $p_i$ on the boundary of $D$, so that all resulting points in $\set{p_1,\ldots,p_r}$ are distinct. As we traverse the boundary of the disc $D$ in the clock-wise direction, the order in which we encounter the points $p_1,\ldots,p_r$ defines a \emph{circular ordering $\oset$ of the elements of $U$}. If we traverse the boundary of the disc $D$ in the counter-clock-wise direction, we obtain a circular ordering $\oset'$ of the elements of $U$, which is the mirror image of the ordering $\oset$. We say that the orderings $\oset$ and $\oset'$ are \emph{identical} but their \emph{orientations} are different, or opposite: $\oset$ has a negative and $\oset'$ has a positive orientation. Whenever we refer to an ordering $\oset$ of elements, we view it as \emph{unoriented} (that is, the orientation can be chosen arbitrarily). When the orientation of the ordering is fixed, we call it an \emph{oriented ordering}, and denote it by $(\oset,b)$, where $\oset$ is the associated (unoriented) ordering of elements of $U$, and $b\in \set{-1,1}$ is the orientation, with $b=-1$ indicating a negative (that is, clock-wise), orientation.

We will also consider graph drawings on the sphere. 
In this case, when we say we traverse the boundary of a disc $D$ in the clock-wise  direction, we mean that we traverse the boundary of $D$ so that the interior of $D$ lies to our right. Similarly, we traverse the boundary of $D$ in the counter-clock-wise direction, if the interior of $D$ lies to our left.
Circular orderings and orientations are then defined similarly.

Given a graph $G$ and a vertex $v\in V(G)$, a circular ordering $\oset_v$ of the edges of $\delta_G(v)$ is called a \emph{rotation}. A collection of circular orderings $\oset_v$ for all vertices $v\in V(G)$ is called a \emph{rotation system} for graph $G$.

\subsection{Tiny $v$-Discs and Drawings that Obey Rotations}

Given a graph $G$, its drawing $\phi$, and a vertex $v\in V(G)$, we will sometimes utilize the notion of a \emph{tiny $v$-disc}, that we define next.

\begin{definition}[Tiny $v$-Disc]\label{def: tiny v-disc}
	Let $G$ be a graph and let $\phi$ be a drawing of $G$ on the sphere or in the plane. For each vertex $v\in V(G)$, we denote by $D_{\phi}(v)$ a very small disc containing the image of $v$ in its interior, and we refer to $D_{\phi}(v)$ as \emph{tiny $v$-disc}. Disc $D_{\phi}(v)$ must be small enough to ensure that, for every edge $e\in \delta_G(v)$, the image $\phi(e)$ of $e$ intersects the boundary of $D_{\phi}(v)$ at a single point, and $\phi(e)\cap D_{\phi}(v)$ is a contiguous curve. Additionally, we require that for every vertex $u\in V(G)\setminus\set{v}$, $\phi(u)\not\in D_{\phi}(v)$; for every edge $e'\in E(G)\setminus\delta_G(v)$, $\phi(e')\cap D_{\phi}(v)=\emptyset$; and that no crossing point of drawing $\phi$ is contained in $D_{\phi}(v)$. Lastly, we require that all discs in $\set{D_{\phi}(v)\mid v\in V(G)}$ are mutually disjoint.
\end{definition}

Consider now a graph $G$, a vertex $v\in V(G)$, and a drawing $\phi$ of $G$. Consider the tiny $v$-disc $D=D_{\phi}(v)$. For every edge $e\in \delta_G(v)$, let $p_e$ be the (unique) intersection of the image $\phi(e)$ of $e$ and the boundary of the disc $D$. Let $\oset$ be the (unoriented) circular ordering in which the points of $\set{p_e}_{e\in \delta_G(v)}$ appear on the boundary of $D$. Then $\oset$ naturally defines a circular ordering $\oset^*_v$ of the edges of $\delta_G(v)$: ordering $\oset^*_v$ is obtained from $\oset$ by replacing, for each edge $e\in \delta_G(v)$, point $p_e$ with the edge $e$. We say that \emph{the images of the edges of $\delta_G(v)$ enter the image of $v$ in the order $\oset^*_v$} in the drawing $\phi$. For brevity, we may sometimes say that the edges of $\delta_G(v)$ enter $v$ in the order $\oset^*_v$ in $\phi$. While we view the ordering $\oset^*_v$ as unoriented, drawing $\phi$ also defines an orientation for this ordering. If the points in set $\set{p_e\mid e\in \delta_G(v)}$ are encountered in the order $\oset^*_v$ when traversing the boundary of $D$ in the counter-clock-wise direction, then the orientation is $1$, and otherwise it is $-1$.

Assume now that we are given a graph $G$ and a rotation system $\Sigma$ for $G$. Let $\phi$ be a drawing of $G$. Consider any vertex $v\in V(G)$, and its rotation $\oset_v\in \Sigma$. We say that the drawing $\phi$  \emph{obeys the rotation $\oset_v\in \Sigma$}, if the order in which the edges of $\delta_G(v)$ enter $v$ in $\phi$ is precisely $\oset_v$ (note that both orderings are unoriented). 
We say that the \emph{orientation of $v$ is $-1$}, or \emph{negative}, in the drawing $\phi$ if the orientation of the ordering $\oset_v$ of the edges of $\delta_G(v)$ as they enter $v$ is $-1$, and otherwise, the orientation of $v$ in $\phi$ is $1$, or positive. 
We say that drawing $\phi$ of $G$ \emph{obeys the rotation system $\Sigma$}, if it obeys the rotation $\oset_v\in \Sigma$ for every vertex $v\in V(G)$.

Assume now that we are given a set $\Gamma$ of curves in general position, where each curve $\gamma\in \Gamma$ is an open curve. Let $p$ be any point that serves as an endpoint of at least one curve in $\Gamma$, and let $\Gamma'\subseteq \Gamma$ be the set of curves for which $p$ serves as an endpoint. We then define a \emph{tiny $p$-disc} $D(p)$ to be a small disc that contains the point $p$ in its interior; does not contain any other point that serves as an endpoint of a curve in $\Gamma$; and does not contain any crossing point of curves in $\Gamma$. Additionally, we ensure that, for every curve $\gamma\in \Gamma$, if $\gamma\in \Gamma'$, then $\gamma\cap D(p)$ is a simple curve, and otherwise $\gamma\cap D(p)=\emptyset$. For every curve $\gamma\in \Gamma'$, let $q(\gamma)$ be the unique point of $\gamma$ lying on the boundary of the disc $D(p)$. Note that all points in $\set{q(\gamma)\mid \gamma\in \Gamma'}$ are distinct. Let $\oset$ be the circular order in which these points are encountered when we traverse the boundary of $D(p)$. As before, this ordering naturally defines a circular ordering $\oset'$ of the curves in $\Gamma'$. We then say that the curves of $\Gamma'$ \emph{enter the point $p$ in the order $\oset'$}.

\subsection{Problem Definitions and Trivial Algorithms}
\label{subsec: prelim problem definitions}

In the \mcn~problem, the input is an $n$-vertex graph $G$, and the goal is to compute a drawing of $G$ in the plane with minimum number of crossings.
The value of the optimal solution, also called the \emph{crossing number} of $G$, is denoted by $\optcro(G)$.

We also consider a closely related problem called Minimum Crossing Number with Rotation System (\CNwRS). In this problem, the input is a graph $G$, and a rotation system $\Sigma$ for $G$. 
Given an instance $I=(G,\Sigma)$ of the \CNwRS problem, we say that a drawing $\phi$ of $G$ is a \emph{feasible solution} for $I$ if $\phi$ obeys the rotation system $\Sigma$. The \emph{cost} of the solution is the number of crossings in $\phi$. The goal in the \CNwRS problem is to compute a feasible solution to the given input instance $I$ of smallest possible cost. We denote the cost of the optimal solution of the \CNwRS instance $I$ by $\optcrors(I)$.

We use the following two simple theorems about the \CNwRS problem, whose proofs are deferred to 
Appendix~\ref{apd: Proof of crwrs_planar} and Appendix~\ref{apd: Proof of crwrs_uncrossing}, respectively.
\begin{theorem}
	\label{thm: crwrs_planar}
	There is an efficient algorithm, that, given an instance $I=(G,\Sigma)$ of \CNwRS, correctly determines whether $\optcrors(I)=0$, and, if so, computes a feasible solution to instance $I$ of cost $0$.
\end{theorem}

\begin{theorem}
	\label{thm: crwrs_uncrossing}
	There is an efficient algorithm, that given an instance $I=(G,\Sigma)$ of \CNwRS, computes a feasible solution to $I$, of cost at most $|E(G)|^2$.
\end{theorem}

We refer to the solution computed by the algorithm from Theorem~\ref{thm: crwrs_uncrossing} as a \emph{trivial solution}.
We will also use the following lemma from \cite{chuzhoy2020towards}, that allows us to insert edges into a partial solution to \cnwrs problem instance.

\begin{lemma}[Lemma 9.2 of \cite{chuzhoy2020towards}]
	\label{lem: edge insertion}
	There is an efficient algorithm, that, given an instance $I=(G,\Sigma)$ of the \CNwRS problem, a subset $E'\subseteq E(G)$ of edges of $G$, and a drawing $\phi$ of graph $G\setminus E'$ that obeys $\Sigma$, computes a solution $\phi'$ to instance $I$, with $\cro(\phi')\le \cro(\phi)+|E'|\cdot |E(G)|$.
\end{lemma}

\subsection{A $\nu$-Decomposition of an Instance}
\label{subsec: subinstances}

A central tool that we use in our divide-and-conquer algorithm is a $\nu$-decomposition of instances. 

\begin{definition}[$\nu$-Decomposition of Instances]
	Let $I=(G,\Sigma)$ be an instance of \cnwrs  with $|E(G)|=m$, and let $\nu\geq 1$ be a parameter. We say that a collection $\iset$ of instances of \cnwrs is a \emph{$\nu$-decomposition of $I$}, if the following hold:
	
	\begin{properties}{D}
		\item  $\sum_{I'=(G',\Sigma')\in \iset}|E(G')|\leq m\cdot (\log m)^{O(1)}$;\label{prop: few edges}
		\item $\sum_{I'\in \iset}\optcrors(I')\le \left (\optcrors(I)+m\right )\cdot \nu$; and \label{prop: small solution cost}
		
		\item 	there is an efficient algorithm $\alg(\iset)$, that, given, a feasible solution $\phi(I')$  to every instance $I'\in \iset$, computes a feasible solution $\phi$ to instance $I$, of cost $\cro(\phi)\leq O\left (\sum_{I'\in \iset}\cro(\phi(I'))\right )$. \label{prop: alg to put together}
	\end{properties}

	We say that a randomized algorithm $\alg$ is a \emph{$\nu$-decomposition algorithm for a family $\iset^*$ of instances of \cnwrs} if $\alg$ is an efficient algorithm, that, given an instance $I=(G,\Sigma)\in \iset^*$, produces a collection $\iset$ of instances that has properties \ref{prop: few edges} and \ref{prop: alg to put together}, and ensures the following additional property (that replaces Property \ref{prop: small solution cost}):

	\begin{properties}[1]{D'}
		\item $\expect{\sum_{I'\in \iset}\optcrors(I')}\le \left (\optcrors(I)+|E(G)|\right )\cdot \nu$.\label{prop: modified expectation}
	\end{properties}
\end{definition}

In the following claim,  whose proof appears in \Cref{apd: Proof of compose algs}, we show that algorithms for computing $\nu$-decompositions can be naturally composed together.

\begin{claim}\label{claim: compose algs}
	Let $\alg_1$ be a randomized $\nu'$-decomposition algorithm for some family $\iset^*$ of instances of \cnwrs. Assume that, given an instance $I\in \iset^*$, algorithm $\alg_1$ produces a collection $\iset'$ of instances, all of which belong to some family $\iset^{**}$ of instances of \cnwrs. Let $\alg_2$ be a randomized $\nu''$-decomposition algorithm for family $\iset^{**}$ of instances of \cnwrs. Lastly, let $\alg$ be a randomized algorithm, that, given an instance $I\in \iset^*$ of \cnwrs, applies Algorithm $\alg_1$ to $I$, to obtain a collection $\iset'$ of instances, and then, for every instance $I'\in \iset'$, applies Algorithm $\alg_2$ to $I'$, obtaining a collection $\iset''(I')$ of instances. The output of algorithm $\alg$ is the collection $\iset=\bigcup_{I'\in \iset'}\iset''(I')$ of instances of \cnwrs. Then $\alg$ is a randomized $\nu$-decomposition algorithm for family $\iset^*$ of instances of $\cnwrs$, for $\nu=\nu''\cdot\max\set{2\nu',(\log m)^{O(1)}}$, where $m$ is the number of edges in instance $I$.
\end{claim}

\subsection{Subinstances}
We use the following definition of subinstances.

\begin{definition}[Subinstances]\label{def: subinstance}
	Let $I=(G,\Sigma)$ and $I'=(G',\Sigma')$ be two instances of \cnwrs. We say that instance $I'$ is a \emph{subinstance} of instance $I$, if there is a subgraph $\tilde G\subseteq G$, and a collection $S_1,\ldots,S_r$ of mutually disjoint subsets of vertices of $\tilde G$, such that graph $G'$ can be obtained from $\tilde G$ by contracting, for all $1\leq i\leq r$, every vertex set $S_i$ into a supernode $u_i$;  we keep parallel edges but remove self-loops\footnote{Note that this definition is similar to the definition of a minor, except that we do not require that the induced subgraphs $G[S_i]$ of $G$ are connected.}. We do not distinguish between the edges incident to the supernodes in graph $G'$ and their counterparts in graph $G$.  For every vertex $v\in V(G')\cap V(G)$, its rotation  $\oset'_v$ in $\Sigma'$ must be consistent with the rotation $\oset_v\in \Sigma$, while for every supernode $u_i$, its rotation $\oset'_{u_i}$ in $\Sigma'$ can be defined arbitrarily.
\end{definition}	

Observe that, if instance $I'=(G',\Sigma')$ is a subinstance of $I=(G,\Sigma)$, then $|E(G')|\le |E(G)|$.
Also notice that the subinstance relation is transitive: if instance $I_1$ is a subinstance of instance $I_0$, and instance $I_2$ is a subinstance of $I_1$, then $I_2$ is a subinstance of $I_0$.

\section{An Algorithm for \cnwrs -- Proof of \Cref{thm: main_rotation_system}}
\label{sec: high level}


In this section we provide the proof of Theorem \ref{thm: main_rotation_system}, with some of the details deferred to subsequent sections.
Throughout the paper, we denote by $I^*=(G^*,\Sigma^*)$  the input instance of the \CNwRS problem, and we denote $m^*=|E(G^*)|$. We also use the following parameter that is central to our algorithm: ${\mu=2^{c^*(\log m^*)^{7/8}\log\log m^*}}$, 
where $c^*$ is a large enough constant. 

As mentioned already, our algorithm for solving the \CNwRS problem is recursive, and, over the course of the recursion, we will consider various other instances $I$ of \cnwrs. Throughout the algorithm, parameters $m^*$ and $\mu$ remain unchanged, and are defined with respect to the original input instance $I^*$. 
The main technical ingredient of the proof is the following theorem.

\begin{theorem}
	\label{thm: main}
	There is a constant $c''$, and an efficient randomized algorithm, that, given an instance $I=(G,\Sigma)$ of \cnwrs with $m=|E(G)|$, such that $\mu^{c''}\leq m\leq m^*$, either returns FAIL, or computes a collection $\iset$ of instances of \cnwrs with the following properties:
	
\begin{itemize}
\item for every instance $I'=(G',\Sigma')\in \iset$, $|E(G')|\leq m/\mu$;

\item $\sum_{I'=(G',\Sigma')\in \iset}|E(G')|\leq  m\cdot (\log m)^{O(1)}$;

\item there is an efficient algorithm called \algcombine, that, given a solution $\phi(I')$ to every instance $I'\in \iset$, computes a solution $\phi$ to instance $I$; and

\item if $\optcrors(I)\leq |E(G)|^2/\mu^{c''}$, then with probability at least $15/16$, all of the following hold:

\begin{itemize}
	\item the algorithm does not return FAIL;
	
\item $\iset\neq \emptyset$;
	
\item 	$\sum_{I'\in \iset}\optcrors(I')\leq (\optcrors(I)+m)\cdot 2^{O((\log m)^{3/4}\log\log m)}$; and

\item if algorithm \algcombine is given as input a solution $\phi(I')$ to every instance $I'\in \iset$, then the resulting solution $\phi$ to instance $I$ that it computes has cost at most: $$O\bigg(\sum_{I'\in \iset}\cro(\phi(I'))\bigg) +(\optcrors(I)+m)\cdot\mu^{O(1)}.$$
\end{itemize} 
\end{itemize}
\end{theorem}

The remainder of this paper is dedicated to the proof of \Cref{thm: main}. In the following subsection, we complete the proof of \Cref{thm: main_rotation_system} using \Cref{thm: main}.

\subsection{Proof of \Cref{thm: main_rotation_system}}

Throughout the proof, we assume that $m^*$ is larger than a sufficiently large constant, since otherwise we can return a trivial solution to instance $I^*$, from \Cref{thm: crwrs_uncrossing}.

We let $c_g>100$ be a large enough constant, so that, for example, when the algorithm from \Cref{thm: main} is applied to an instance  $I=(G,\Sigma)$ with $m=|E(G)|$, such that $\mu^{c''}\leq m\leq m^*$ holds, it is guaranteed to return a family $\iset$ of instances of \cnwrs, with $\sum_{I'=(G',\Sigma')\in \iset}|E(G')|\leq  m\cdot (\log m)^{c_g}$. 
We say that the algorithm from \Cref{thm: main} is \emph{successful} if all of the following hold: 

\begin{itemize}
	\item the algorithm does not return FAIL;
	\item if $\iset$ is the collection of instances returned by the algorithm, then $\iset\neq \emptyset$;
	\item $\sum_{I'\in \iset}\optcrors(I')\leq (\optcrors(I)+m)\cdot 2^{c_g((\log m)^{3/4}\log\log m)}$; and
	\item if algorithm \algcombine is given a solution $\phi(I')$ to every instance $I'\in \iset$, then it computes a solution $\phi$ to instance $I$, of cost at most $c_g\cdot (\sum_{I'\in \iset}\cro(\phi(I')) +(\optcrors(I)+m)\cdot\mu^{c_g}$.
\end{itemize}

By letting $c_g$ be a large enough constant,  \Cref{thm: main} guarantees that, if  $\optcrors(I)\leq |E(G)|^2/\mu^{c''}$, then with probability at least $15/16$ the algorithm is successful.
We assume that the parameter $c^*$ in the definition of $\mu$ is sufficiently large, so that, e.g., $c^*>2c_g$.

We use a simple recursive algorithm called \algrec, that appears in Figure \ref{fig: algrec}.

\program{\algrec}{fig: algrec}{ 
	\noindent{\bf Input:} an instance $I=(G,\Sigma)$  of the \CNwRS problem, with $|E(G)|\leq m^*$.
	
	\noindent{\bf Output:} a feasible solution to instance $I$.
	
	\begin{enumerate}
		\item Use the algorithm from \Cref{thm: crwrs_planar} to determine whether $\optcrors(I)=0$. If so, use the algorithm from \Cref{thm: crwrs_planar} to compute a solution to $I$ of cost $0$. Return this solution, and terminate the algorithm.
		
		\item Use the algorithm from Theorem~\ref{thm: crwrs_uncrossing} to compute a trivial solution $\phi'$ to instance $I$.
		
		\item If $|E(G)|\leq \mu^{c''}$, return the trivial solution $\phi'$ and terminate the algorithm.
		
		\item For $1\leq j\leq \ceil{\log m^*}$:
		\begin{enumerate}
			\item Apply the algorithm from \Cref{thm: main} to instance $I$. 
			
			\item If the algorithm returns FAIL, let $\phi_j=\phi'$ be the trivial solution to instance $I$, and set $\iset_j(I)=\emptyset$.
			
			\item Otherwise: 
			\begin{enumerate}	
				\item Let $\iset_j(I)$ be the collection of instances computed by the algorithm.
				
				\item For every instance $I'\in \iset_j(I)$, apply Algorithm \algrec to instance $I'$, to obtain a solution $\phi(I')$ to this instance.
				
				\item Apply Algorithm \algcombine from \Cref{thm: main} to solutions $\set{\phi(I')}_{I'\in \iset_j(I)}$, to obtain a solution $\phi_j$ to instance $I$.
			\end{enumerate}
		\end{enumerate} 
		Return a solution to instance $I$ from among $\set{\phi',\phi_1,\ldots,\phi_{\ceil{\log m^*}}}$ that has fewest crossings.
	\end{enumerate}
}

In order to analyze the algorithm, it is convenient to associate a \emph{partitioning tree} $T$ with it, whose vertices correspond to all instances of \cnwrs considered over the course of the algorithm. Let $L=\ceil{\log m^*}$. We start with the tree $T$ containing a single root vertex $v(I^*)$, representing the input instance $I^*$. Consider now some vertex $v(I)$ of the tree, representing some instance $I=(G,\Sigma)$.
When Algorithm \algrec was applied to instance $I$, if it did not terminate after the first three steps, it constructed $L$ collections $\iset_1(I),\ldots,\iset_L(I)$ of instances (some of which may be empty, in case the algorithm from \Cref{thm: main} returned FAIL in the corresponding iteration). For each such instance $I'\in \bigcup_{j=1}^L\iset_j(I)$, we add a vertex $v(I')$ representing instance $I'$ to $T$, that becomes a child vertex of $v(I)$. This concludes the description of the partitioning tree $T$.

We denote by $\iset^*=\set{I\mid v(I)\in V(T)}$ the set of all instances of \cnwrs, whose corresponding vertex appears in the tree $T$. For each such instance $I\in \iset^*$, its  \emph{recursive level} is the distance from vertex $v(I)$ to the root vertex $v(I^*)$ in the tree $T$ (so the recursive level of $v(I^*)$ is $0$).
For $j\geq 0$, we denote by $\hat \iset_j\subseteq \iset^*$ the set of all instances $I\in \iset^*$, whose recursive level is $j$.
Lastly, the \emph{depth} of the tree $T$, denoted by $\dep(T)$, is the largest recursive level of any instance in $\iset^*$.
In order to analyze the algorithm, we start with the following two simple observations.

\begin{observation}\label{obs: few recursive levels}
	$\dep(T)\leq \frac{(\log m^*)^{1/8}}{c^*\log\log m^*}$.
\end{observation}
\begin{proof}
	Consider any non-root vertex $v(I)$ in the tree $T$, and let $v(I')$ be the parent-vertex of $v(I)$. Denote $I=(G,\Sigma)$ and $I'=(G',\Sigma')$.
	From the construction of tree $T$, instance $I$ belongs to some collection of instances obtained by applying the algorithm from \Cref{thm: main} to instance $I'$. Therefore, from \Cref{thm: main}, $|E(G)|\leq |E(G')|/\mu$ must hold. Therefore, for all $j\geq 0$, for every instance $I=(G,\Sigma)\in \hat \iset_j$, $|E(G)|\leq m^*/\mu^j$. Since $\mu=2^{c^*(\log m^*)^{7/8}\log\log m^*}$, we get that $\dep(T)\leq \frac{(\log m^*)^{1/8}}{c^*\log\log m^*}$.
\end{proof}

\begin{observation}\label{obs: num of edges}
	$\sum_{I=(G,\Sigma)\in \iset^*}|E(G)|\le m^*\cdot 2^{(\log m^*)^{1/8}}$.
\end{observation}
\begin{proof}
	Consider any non-leaf vertex $v(I)$ of the tree $T$, and denote $I=(G,\Sigma)$. 
	Recall that, when Algorithm \algrec is applied to  instance $I$, it uses the algorithm from \Cref{thm: main} to compute $L$ collections $\iset_1(I),\ldots,\iset_L(I)$ of instances, such that, if we denote $|E(G)|=m$, then, for all $1\leq j\leq L$:
	$$\sum_{I'=(G',\Sigma')\in \iset_j(I)}|E(G')|\leq m\cdot (\log m)^{c_g}\leq m\cdot (\log m^*)^{c_g}$$
	(since $m\leq m^*$ must hold). Since $L\leq 2\log m^*$, and $m^*$ is sufficiently large, we get that:
	$$\sum_{j=1}^L\sum_{I'=(G',\Sigma')\in \iset_j(I)}|E(G')|\leq m\cdot (\log m^*)^{c_g+2}.$$
	For all $j\geq 0$, we denote by $N_j$ the total number of edges in all instances in set $\hat \iset_j$, $N_j=\sum_{I=(G,\Sigma)\in \hat \iset_j}|E(G)|$. Clearly, $N_0=m^*$, and, from the above discussion, for all $j>0$, $N_j\leq N_{j-1}\cdot (\log m^*)^{c_g+2}$. 
	
	Since $\dep(T)\leq \frac{(\log m^*)^{1/8}}{c^*\log\log m^*}$,
	we conclude that:
	\[
	\sum_{I=(G,\Sigma)\in \iset^*}|E(G)|\leq m^*\cdot (\log m^*)^{2c_g\cdot  (\log m^*)^{1/8}/(c^*\log\log m^*)}\leq m^*\cdot 2^{(\log m^*)^{1/8}}, 
	\]	
	since $c_g\leq c^*/2$.
\end{proof}

We use the following corollary, that follows  immediately from \Cref{obs: num of edges}.

\begin{corollary}\label{cor: num of instances}
	The number of instances $I=(G,\Sigma)\in \iset^*$ with $|E(G)|\geq \mu^{c''}$ is at most $m^*$.
\end{corollary}

We say that an instance $I\in \iset^*$ is a \emph{leaf instance}, if  vertex $v(I)$ is a leaf vertex of the tree $T$, and we say that it is a non-leaf instance otherwise.
Consider now a non-leaf instance $I=(G,\Sigma)\in \iset^*$. We say that a bad event $\event(I)$ happens, if  $0<\optcrors(I)\leq |E(G)|^2/\mu^{c''}$, and, for all $1\leq j\leq L$, the $j$th application of the algorithm  from \Cref{thm: main} to instance $I$ was unsuccessful.
Clearly, from \Cref{thm: main},  $\prob{\event(I)}\le (1/16)^L\leq 1/(m^*)^4$.
Let $\event$ be the bad event that event $\event(I)$ happened for any  instance $I\in \iset^*$. From the Union Bound and \Cref{cor: num of instances}, we get that $\prob{\event}\leq 1/(m^*)^2$.
We use the following immediate observation.

\begin{observation}\label{obs: leaf}
	If Event $\event$ does not happen, then for every leaf vertex $v(I)$ of $T$ with $I=(G,\Sigma)$, either $|E(G)|\leq \mu^{c''}$; or $\optcrors(I)=0$; or $\optcrors(I)> |E(G)|^2/\mu^{c''}$.
\end{observation}

We use the following lemma to complete the proof of \Cref{thm: main_rotation_system}.

\begin{lemma}\label{lem: solution cost}
	If Event $\event$ does not happen, then  Algorithm $\algrec$ computes a solution for instance $I^*=(G^*,\Sigma^*)$ of cost at most $2^{O((\log m^*)^{7/8}\log\log m^*)}\cdot \left(\optcrors(I^*)+|E(G^*)|\right)$.
\end{lemma}
\begin{proof}
	Consider a non-leaf instance $I=(G,\Sigma)$, and let $\iset_1(I),\ldots,\iset_L(I)$ be families of instances of \cnwrs that  Algorithm \algrec computed, when applied to instance $I$. Recall that, for each $1\le j\le L$ with $\iset_j(I)\neq \emptyset$, the algorithm computes a solution $\phi_j$ to instance $I$, by first solving each of the instances in $\iset_j(I)$ recursively, and then combining the resulting solutions using Algorithm \algcombine. Eventually, the algorithm returns the best solution of $\set{\phi',\phi_1,\ldots,\phi_L}$, where $\phi'$ is the trivial solution, whose cost is at most  $|E(G)|^2$. We fix an arbitrary index $1\leq j\leq L$, such that the $j$th application of the algorithm  from \Cref{thm: main} to instance $I$ was successful. Note that the cost of the solution to instance $I$ that the algorithm returns is at most $\cro(\phi_j)$. We then \emph{mark} the vertices of $\set{v(I')\mid I'\in \iset_j(I)}$ in the tree $T$. We also mark the root vertex of the tree. 
	
	Let $T^*$ be the subgraph of $T$ induced by all marked vertices. It is easy to verify that $T^*$ is a tree, and moreover, if $\event$ did not happen, every leaf vertex of $T^*$ is also a leaf vertex of $T$. For a vertex $v(I)\in V(T^*)$, we denote by $h(I)$ the length of the longest path in tree $T^*$, connecting vertex $v(I)$ to any of its descendants in the tree. 
	We use the following claim, whose proof is straightforward conceptually but somewhat technical; we defer the proof to \Cref{Appx: inductive bound proof}.

\begin{claim}\label{claim: bound by level}
		Assume that Event $\event$ did not happen. Then there is a fixed constant $\tilde c\geq \max\set {c'',c_g,c^*}$, such that, for every vertex $v(I)\in V(T^*)$, whose corresponding instance is denoted by $I=(G,\Sigma)$, the cost of the solution that the algorithm computes for  $I$ is at most:
		$$2^{\tilde c\cdot h(I)\cdot (\log m^*)^{3/4}\log\log m^*}\cdot \mu^{c''\cdot c_g}\cdot \optcrors(I)+(\log m^*)^{4c_g h(I)}\mu^{2c''\cdot \tilde c}\cdot|E(G)|.$$
	\end{claim}

	We are now ready to complete the proof of \Cref{lem: solution cost}. Recall that $h(I^*)=\dep(T^*)\leq \dep(T)\leq  \frac{(\log m^*)^{1/8}}{c^*\log\log m^*}$  from \Cref{obs: few recursive levels}. Therefore, from \Cref{claim: bound by level}, the cost of the solution that the algorithm computes for instance $I^*$ is bounded by:
	\[
	\begin{split}
	& 2^{O(\dep(T))\cdot (\log m^*)^{3/4}\log\log m^*}\cdot \mu^{O(1)}\cdot \optcrors(I^*)+(\log m^*)^{O(\dep(T))}\cdot\mu^{O(1)}\cdot m^*\\
	 & \quad \quad \quad \quad \quad \quad \leq2^{O((\log m^*)^{7/8})}\cdot \mu^{O(1)}\cdot \optcrors(I^*)+(\log m^*)^{O((\log m^*)^{1/8}/\log\log m^*) }\cdot\mu^{O(1)}\cdot m^*\\
	 &\quad \quad \quad \quad \quad \quad \leq 2^{O((\log m^*)^{7/8}\log\log m^*)}\cdot \left(\optcrors(I^*)+|E(G^*)|\right),
	\end{split}
	\]
	since $\mu=2^{O((\log m^*)^{7/8}\log\log m^*)}$.
\end{proof}

In order to complete the proof of \Cref{thm: main_rotation_system}, it is now enough to prove Theorem~\ref{thm: main}. The remainder of the paper is dedicated to the proof of \Cref{thm: main}.

\subsection{Proof of Theorem~\ref{thm: main} -- Main Definitions and Theorems} 

We classify instances of \cnwrs into \emph{wide} and \emph{narrow}. Wide instances are, in turn, classified into \emph{well-connected} and not well-connected instances. We then provide different algorithms for decomposing instances of each of the resulting three kinds. 
We use the following notion of a high-degree vertex.

\begin{definition}[High-degree vertex]
Let $G$ be any graph.	A vertex $v\in V(G)$ is a \emph{high-degree} vertex, if $\deg_G(v)\geq |E(G)|/\mu^4$. 
\end{definition}

We are now ready to define wide and narrow instances.

\begin{definition}[Wide and Narrow Instances]
Let $I=(G,\Sigma)$ be an instance of \cnwrs with $|E(G)|=m$. We say that $I$ is a \emph{wide} instance, if there is a high-degree vertex $v\in V(G)$, a partition $(E_1,E_2)$ of the edges of $\delta_G(v)$, such that the edges of $E_1$ appear consecutively in the rotation $\oset_v\in \Sigma$, and so do the edges of $E_2$, and there is a collection $\pset$  of at least $\floor{m/\mu^{\interestconst}}$ simple edge-disjoint cycles in $G$, such that every cycle $P\in \pset$ contains one edge of $E_1$ and one edge of $E_2$.
An instance that is not wide is called \emph{narrow}.
\end{definition}

Note that there is an efficient algorithm to check whether a given instance $I$ of \cnwrs is wide, and, if so, to  compute the corresponding set $\pset$ of cycles, via standard algorithms for maximum flow. (For every vertex $v\in V(G)$, we  try all possible partitions $(E_1,E_2)$ of $\delta_G(v)$ with the required properties, as the number of such partitions is bounded by $|\delta_G(v)|^2$.)
We will use the following simple observation regarding narrow instances.

\begin{observation}\label{obs: narrow prop 2}
	If an instance $I=(G,\Sigma)$ of \cnwrs is narrow, then for every pair $u,v$ of distinct high-degree vertices of $G$, and any set $\pset$ of edge-disjoint paths connecting $u$ to $v$  in $G$,  $|\pset|\leq 2\ceil{|E(G)|/\mu^{\interestconst}}$ must hold.
\end{observation}

\begin{proof}
	Assume for contradiction that $I=(G,\Sigma)$ is a narrow instance of \cnwrs, with $|E(G)|=m$, and that there are two high-degree vertices $u,v$ of $G$, and a set $\pset$ of more than $2\ceil{m/\mu^{\interestconst}}$ edge-disjoint paths in $G$ connecting $u$ to $v$.  We denote $|\pset|=k$.
	Let $E'\subseteq \delta_G(v)$ be the set of all edges $e\in \delta_G(v)$, such that $e$ is the first edge on some path in $\pset$. We denote $E'=\set{e_1,\ldots,e_k}$, where the edges are indexed according to their ordering in the rotation $\oset_v\in \Sigma$. We also denote $\pset=\set{P(e_i)\mid 1\leq i\leq k}$, where path $P(e_i)$ contains the edge $e_i$ as its first edge. We can then compute a partition $(E_1,E_2)$ of $\delta_G(v)$, such that the edges of $E_1$ appear consecutively in the rotation $\oset_v\in \Sigma$, and so do the edges of $E_2$. Additionally, we can ensure that $e_1,\ldots,e_{\ceil{k/2}}\in E_1$, while the remaining edges of $E'$ lie in $E_2$. For each $1\leq i\leq \ceil{m/\mu^{\interestconst}}$, we let $Q_i$ be the cycle obtained by concatenating the paths $P(e_i)$ and $P(e_{k-i+1})$. We turn $Q_i$ into a simple cycle, by removing from it all cycles that are disjoint from vertex $v$. It is then immediate to verify that the set $\set{Q_i\mid 1\leq i\leq \ceil{m/\mu^{\interestconst}}}$ of cycles has all the required properties to establish that instance $I$ is wide, a contradiction.
\end{proof}

Next, we define well-connected wide instances.

\begin{definition}[Well-Connected Wide Instances]
	Let $I=(G,\Sigma)$ be a wide instance of \cnwrs with $|E(G)|=m$. We say that it is a \emph{well-connected} instance iff for every pair $u,v$ of distinct vertices of $G$ with $\deg_G(v),\deg_G(u)\geq m/\mu^5$, there is a collection of at least $\frac{8m}{\mu^{\interestconst}}$ edge-disjoint paths connecting $u$ to $v$ in $G$.
\end{definition}

The proof of \Cref{thm: main} relies on the following three theorems. The first theorem deals with wide instances that are not necessarily well-connected. Its proof is deferred to \Cref{sec: not well connected}.

\begin{theorem}\label{thm: not well connected}
	There is an efficient randomized algorithm,  whose input is a wide instance  $I=(G,\Sigma)$ of \cnwrs, with $m=|E(G)|$, such that $ \mu^{20}\leq m\leq m^*$ holds. The algorithm computes  a $\nu$-decomposition $\iset$ of $I$, for $\nu= 2^{O((\log m)^{3/4}\log\log m)}$,   such that every instance $I'=(G',\Sigma')\in \iset$ is a subinstance of $I$, and one of the following holds for it:
	
	\begin{itemize}
		\item either $|E(G')|\le m/\mu$;
		\item or $I'$ is a narrow instance;
		\item or $I'$ is a wide and well-connected instance.
	\end{itemize} 
\end{theorem}

The second theorem deals with wide well-connected instances. Its proof appears in \Cref{sec: many paths}.

\begin{theorem}\label{lem: many paths}
	There is an efficient randomized algorithm, whose input is a wide and well-connected instance  $I=(G,\Sigma)$ of \cnwrs, with $m=|E(G)|$, such that $\mu^{c'}\leq m\leq m^*$ holds, for some large enough constant $c'$. The algorithm either returns FAIL, or computes a non-empty collection $\iset$ of instances of \cnwrs, such that the following hold:
	
	\begin{itemize}
		\item  $\sum_{I'=(G',\Sigma')\in \iset}|E(G')|\le 2|E(G)|$;
		
		\item for every instance $I'=(G',\Sigma')\in \iset$, either $|E(G')|\le m/\mu$, or instance $I'$ narrow;
		
		\item 
		there is an efficient algorithm called $\algcombine'$, that, given a solution $\phi(I')$ to every instance $I'\in \iset$, computes a solution $\phi$ to instance $I$; and
		
		\item  if $\optcrors(I)\leq m^2/\mu^{c'}$ then, with probability at least $1-1/\mu^2$,  all of the following hold: 
		
		\begin{itemize}
			\item the algorithm does not return FAIL;
			\item  $\sum_{I'\in \iset}\optcrors(I')\leq \optcrors(I)\cdot  (\log m)^{O(1)}$; and
			\item if algorithm $\algcombine'$ is given as input a solution $\phi(I')$ to every instance $I'\in \iset$, then the resulting solution $\phi$ to instance $I$ that it computes has cost at most: $\cro(\phi)\leq  \sum_{I'\in \iset}\cro(\phi(I')) + \optcrors(I)\cdot\mu^{O(1)}$.
		\end{itemize}
	\end{itemize}
\end{theorem}

The third theorem deals with narrow instances, and its proof appears in \Cref{sec: computing the decomposition}.

\begin{theorem}\label{lem: not many paths}
	There is an efficient randomized algorithm,  whose input is a narrow instance  $I=(G,\Sigma)$ of \cnwrs, with $m=|E(G)|$, such that $ \mu^{50}\leq |E(G)|\leq 2m^*$. The algorithm either  returns FAIL, or computes  a $\nu$-decomposition $\iset$ of $I$, for $\nu= 2^{O((\log m)^{3/4}\log\log m)}$,   such that, for every instance $I'=(G',\Sigma')\in \iset$, $|E(G')|\le m/(2\mu)$. Moreover, if $\optcrors(I)<m^2/\mu^{21}$, then the probability that the algorithm returns FAIL is at most $O(1/\mu^2)$.
\end{theorem}

The majority of the remainder of this paper is dedicated to the proofs of the above three theorems. Before we provide these proofs, we develop  central technical tools that they use, in Sections \ref{sec: guiding paths orderings basic disengagement} -- \ref{sec: main disengagement}.
In the remainder of this section, we complete the proof of Theorem \ref{thm: main} using Theorems \ref{thm: not well connected}, \ref{lem: many paths}, and \ref{lem: not many paths}.

Recall that we are given an instance $I=(G,\Sigma)$ of \cnwrs, with $\mu^{c''}\leq |E(G)|\leq m^*$, for some large enough constant $c''$. We assume that $c''>100c'$, where $c'$ is the constant in \Cref{lem: many paths}. We use another large constant $c'_g$, and we assume that $c^*> c'_g >c''$, where $c^*$ is the constant in the definition of the parameter $\mu$. Throughout, we denote $m=|E(G)|$. We compute the desired collection $\iset^*$ of instances in three steps. 

\subsubsection*{Step 1}

Assume first that the input instance $I$ is a wide instance. 
We apply the algorithm from \Cref{thm: not well connected} to $I$.
Let $\hat \iset$ be the resulting collection of instances. We partition the set $\hat\iset$ of instances into three subsets. The first set, denoted by $\hat \iset_{\textsf {small}}$, contains all instances $I'=(G',\Sigma')\in \hat \iset$ with $|E(G')|\le m/\mu$. The second set, denoted by $\hat \iset^{(n)}_{\textsf {large}}$, contains all narrow instances in $\hat \iset\setminus \hat \iset_{\textsf {small}}$. The third set, denoted by $\hat \iset^{(w)}_{\textsf {large}}$, contains all remaining instances of $\hat \iset$. From  \Cref{thm: not well connected}, every instance in $\hat \iset^{(w)}_{\textsf {large}}$ is wide and well-connected. Since every instance  $I'=(G',\Sigma')\in \hat \iset$ is a subinstance of $I$, $|E(G')|\leq |E(G)|\leq m^*$ must hold.
Recall that, from \Cref{thm: not well connected}, $\hat \iset$ is a $\nu_1$-decomposition for $I$, for 
$\nu_1= 2^{O((\log m)^{3/4}\log\log m)}$. Therefore:

\begin{equation}
\sum_{I'=(G',\Sigma')\in \hat \iset}|E(G')|\le m\cdot (\log m)^{c'_g},\label{eq: num of edges step 1}
\end{equation}

and 
$$\expect{\sum_{I'\in \hat\iset}\optcrors(I')}\le \left (\optcrors(I)+m\right )\cdot \nu_1.$$

\paragraph{Bad Event $\event_1$.}
We say that a bad event $\event_1$ happens if $\sum_{I'\in \hat\iset}\optcrors(I')> 100\cdot\left (\optcrors(I)+m\right )\cdot \nu_1$. From the Markov Bound, $\prob{\event_1}\le 1/100$. 
Note that, if event $\event_1$ did not happen, then for each instance $I'\in \hat\iset$, $\optcrors(I')\le 100\cdot\left (\optcrors(I)+m\right )\cdot \nu_1$.
We need the following simple observation.

\begin{observation}\label{obs: optbound for narrow 1}
	Assume that $\optcrors(I)\leq m^2/\mu^{c''}$, and that Event $\event_1$ did not happen. Then for every instance  $I'=(G', \Sigma')\in \hat \iset^{(n)}_{\textsf {large}}\cup \hat \iset^{(w)}_{\textsf {large}}$, $\optcrors(I')\leq |E(G')|^2/\mu^{c'}$.
\end{observation}

\begin{proof}
	If $\optcrors(I)\le m^2/\mu^{c''}$, and Event $\event_1$ did not happen, then for every instance $I'\in \hat \iset^{(n)}_{\textsf {large}}\cup \hat \iset^{(w)}_{\textsf {large}}$: 
	$$\optcrors(I')\le 100\cdot\left (\optcrors(I)+m\right )\cdot \nu_1\le \mu\cdot \left (\optcrors(I)+m\right )\le \mu\cdot \left (\frac{m^2}{\mu^{c''}}+m\right )\le \frac{m^2}{\mu^{c'}}$$
	(since $c''>100c'$ is a large enough constant and $m\ge \mu^{c''}$).
\end{proof}

Assume now that instance $I$ is a narrow instance. Then we simply set $\hat\iset=\hat \iset^{(n)}_{\textsf {large}}=\set{I}$ and $\hat \iset_{\textsf {small}}=\hat \iset^{(w)}_{\textsf {large}}=\emptyset$.
This completes the description of the first step.

\subsubsection*{Step 2}

In the second step, we apply the algorithm from \Cref{lem: many paths} to every instance $I'\in\hat \iset^{(w)}_{\textsf {large}}$. 
If the algorithm returns FAIL, then we terminate our algorithm and return FAIL as well. Assume now that the algorithm from \Cref{lem: many paths}, when applied to instance $I'$, did not return FAIL. We let $\tilde\iset(I')$ be the collection of instances that the algorithm computes.
Recall that we are guaranteed that, for each instance $\tilde I=(\tilde G,\tilde \Sigma)\in \tilde\iset(I')$, either $\tilde I$ is a narrow instance, or $|E(\tilde G)|\le \frac{|E(G')|}{\mu}\le \frac m {\mu}$ (we have used the fact that $|E(G')|\le m$, since $I'=(G',\Sigma')$ is a subinstance of $I$). 
Additionally, we are guaranteed that: 

\begin{equation}
\sum_{\tilde I=(\tilde G, \tilde\Sigma)\in \tilde\iset(I')}|E(\tilde G)|\le 2|E(G')|.\label{ineq: edges step 2}
\end{equation}

In particular, for every instance  $\tilde I=(\tilde G, \tilde\Sigma)\in \tilde\iset(I')$, $|E(\tilde G)|\leq 2|E(G')|\leq 2m\leq 2m^*$.

We say that the application of the algorithm from \Cref{lem: many paths} to an instance $I'=(G', \Sigma')\in \hat \iset^{(w)}_{\textsf {large}}$ is \emph{successful}, if (i) the algorithm does not return FAIL; (ii)
$\sum_{\tilde I\in \tilde\iset(I')}\optcrors(\tilde I)\leq   \optcrors(I')\cdot(\log m)^{c'_g}$; and (iii) there is an efficient algorithm $\algcombine'$, that, given  a solution $\phi(\tilde I)$ to every instance $\tilde I\in \tilde \iset(I')$, computes a solution $\phi(I')$ to instance $I'$ with  
$\cro(\phi(I'))\leq  \sum_{\tilde I \in \tilde \iset(I')}\cro(\phi(\tilde I)) + \optcrors(I')\cdot\mu^{c'_g}$.

\paragraph{Bad Event $\event_2$.}
For an instance $I'=(G', \Sigma')\in \hat \iset^{(w)}_{\textsf {large}}$, we say that a bad event $\event_2(I')$ happens if the algorithm from \Cref{lem: many paths}, when applied to instance $I'$, was not successful. From  \Cref{lem: many paths} and \Cref{obs: optbound for narrow 1}, if $\optcrors(I)\leq m^2/\mu^{c''}$, then $\prob{\event_2(I')\mid\neg\event_1}\leq 1/\mu^2$ (since we can assume that $c'_g$ is a large enough constant).  

We let $\event_2$ be the bad event that at least of the events in $\set{\event_2(I')\mid I'\in \hat \iset^{(w)}_{\textsf {large}}}$ happened. 
 
Recall that, from the definition of the set $\iset^{(w)}_{\textsf {large}}$ of instances, for every instance $I'=(G',\Sigma')\in \hat \iset^{(w)}_{\textsf {large}}$, $|E(G')|\ge \frac{m}{\mu}$ holds. On the other hand, from Equation \ref{eq: num of edges step 1},

$$ \sum_{I'=(G',\Sigma')\in \hat \iset^{(w)}_{\textsf {large}}}|E(G')|\le \sum_{I'=(G',\Sigma')\in \hat \iset}|E(G')|\le  m\cdot (\log m)^{c'_g}.$$

Therefore,  $|\hat \iset^{(w)}_{\textsf {large}}|\le \mu\cdot (\log m)^{c'_g}$. From the Union Bound,  if   $\optcrors(I)\leq \frac{m^2}{\mu^{c''}}$, then $\prob{\event_2\mid\neg\event_1}\le \frac{\mu\cdot (\log m)^{c'_g}}{\mu^2}\le \frac 1 {100}$.

Let $\tilde \iset=\bigcup_{I'\in \hat \iset^{(w)}_{\textsf {large}}}\tilde\iset(I')$.
Note that, from Inequalities \ref{eq: num of edges step 1} and \ref{ineq: edges step 2}, we get that:

\begin{equation}\label{ineq: total edges step 2}
\sum_{\tilde I=(\tilde G,\tilde \Sigma)\in \tilde\iset}|E(\tilde G)|\leq 2m\cdot (\log m)^{c'_g}.
\end{equation}

We partition the instances in set $\tilde \iset$ into two subsets: set $\tilde{\iset}_{\textsf {small}}$, containing all instances $\tilde I=(\tilde G,\tilde \Sigma)$ in $\tilde\iset$ with $|E(\tilde G)|\le m/\mu$, and set $\tilde{\iset}^{(n)}_{\textsf {large}}$, containing all remaining instances. From \Cref{lem: many paths}, every instance $\tilde I\in \tilde{\iset}^{(n)}_{\textsf {large}}$ is narrow. This completes the description of the second step.

\subsection*{Step 3}

We focus on four sets of instances that we have constructed so far: $\hat{\iset}_{\textsf {small}}, \hat{\iset}^{(n)}_{\textsf {large}}, \tilde{\iset}_{\textsf {small}}, \tilde{\iset}^{(n)}_{\textsf {large}}$. Recall that, if instance $I'=(G',\Sigma')$ belongs to set $\hat{\iset}_{\textsf {small}}\cup \tilde{\iset}_{\textsf {small}}$, then $|E(G')|\leq m/\mu$. If instance $I'=(G',\Sigma')$ belongs to set $\hat{\iset}^{(n)}_{\textsf {large}},\cup \tilde{\iset}^{(n)}_{\textsf {large}}$, then $m/\mu< |E(G')|\leq 2m$, and instance $I'$ is narrow.
We use the following simple observation.

\begin{observation}\label{obs: optbound for narrow}
If $\optcrors(I)\leq m^2/\mu^{c''}$, and neither of the events $\event_1,\event_2$ happened, then  for every instance $I'=(G',\Sigma')\in \hat{\iset}^{(n)}_{\textsf {large}}\cup \tilde{\iset}^{(n)}_{\textsf {large}}$, $\optcrors(I')< |E(G')|^2/\mu^{21}$.
\end{observation}
\begin{proof}
From \Cref{obs: optbound for narrow 1}, if $\optcrors(I)\leq m^2/\mu^{c''}$, and the bad event $\event_1$ did not happen, then for every instance  $I'=(G', \Sigma')\in \hat \iset^{(n)}_{\textsf {large}}\cup \hat \iset^{(w)}_{\textsf {large}}$, $\optcrors(I')\leq |E(G')|^2/\mu^{c'}$.

Consider now some instance $I'=(G',\Sigma')\in \hat{\iset}^{(w)}_{\textsf {large}}$. If, additionally, event $\event_2$ did not happen, then:  

$$\sum_{\tilde I\in \tilde\iset(I')}\optcrors(\tilde I)\le \optcrors(I')\cdot (\log m)^{c'_g}.$$

Therefore, for every instance $\tilde I\in \tilde\iset(I')$:

$$\optcrors(\tilde I)\le \optcrors(I')\cdot (\log m)^{c'_g}\le \frac{|E(G')|^2}{\mu^{c'}}\cdot (\log m)^{c'_g}\le \frac{m^2}{\mu^{c'-1}}.$$

We conclude that for every instance $\tilde I=(\tilde G,\tilde \Sigma)\in \tilde{\iset}^{(n)}_{\textsf {large}}$, $\optcrors(\tilde I)\leq  \frac{m^2}{\mu^{c'-1}}$. Since, from the definition of the set $ \tilde{\iset}^{(n)}_{\textsf {large}}$ of instances,  $|E(\tilde G)|\geq \frac m{\mu}$, we get that:
$$\optcrors(\tilde I)\leq \frac{m^2}{\mu^{c'-1}}< \frac{|E(\tilde G)|^2}{\mu^{21}},$$
assuming that $c'$ is a large enough constant.
\end{proof}

Next, we process every instance $I'\in \hat{\iset}^{(n)}_{\textsf {large}}\cup \tilde{\iset}^{(n)}_{\textsf {large}}$  one by one. Notice that for each such instance $I'=(G',\Sigma')$, $|E(G')|\geq m/\mu\geq \mu^{50}$ must hold, since $m\geq \mu^{c''}$. Additionally, as observed already, $|E(G')|\leq 2m\leq 2m^*$. When instance $I'=(G',\Sigma')$ is processed, we apply the algorithm from \Cref{lem: not many paths} to it. 
If the algorithm returns FAIL, then we terminate the algorithm and return FAIL as well. Otherwise, we obtain a collection $\overline\iset(I')$ of instances of \cnwrs. From \Cref{lem: not many paths}, for every instance $I''=(G'',\Sigma'')\in \overline\iset(I')$, $|E(G'')|\leq \frac{|E(G')|}{2\mu} \leq \frac m {\mu}$. Moreover, from the definition of a $\nu$-decomposition of an instance, and from the fact that $|E(G')|\leq 2m$, we get that: 

\begin{equation}\label{eq: few edges 3}
\sum_{I''=(G'',\Sigma'')\in \overline \iset(I')}|E(G'')|\leq |E(G')|\cdot (\log m)^{c'_g}.
\end{equation}

\paragraph{Bad Events $\event_3$ and $\event$.}
For an instance $I'=(G',\Sigma')\in \hat{\iset}^{(n)}_{\textsf {large}}\cup \tilde{\iset}^{(n)}_{\textsf {large}}$, we say that the bad event $\event_3(I')$ happens if the algorithm from \Cref{lem: not many paths}, when applied to instance $I'$, returns FAIL. From \Cref{lem: not many paths}, if 
$\optcrors(I')<|E(G')|^2/\mu^{21}$, then the probability that the algorithm returns FAIL is at most $O(1/\mu^2)$. Therefore, from \Cref{obs: optbound for narrow}, if  $\optcrors(I)\leq m^2/\mu^{c''}$, then $\prob{\event_3(I')\mid \neg\event_1\band\neg\event_2}\leq O(1/\mu^2)$. We let $\event_3$ to be the bad event that $\event_3(I')$ happened for any instance $I'\in \hat{\iset}^{(n)}_{\textsf {large}}\cup \tilde{\iset}^{(n)}_{\textsf {large}}$.  Recall that, for every instance $I'=(G',\Sigma')\in \hat{\iset}^{(n)}_{\textsf {large}}\cup \tilde{\iset}^{(n)}_{\textsf {large}}$, $|E(G')|\geq \frac m{\mu}$.
On the other hand, from Inequality \ref{eq: num of edges step 1}, 
$\sum_{I'=(G',\Sigma')\in \hat \iset^{(n)}_{\textsf {large}}}|E(G')|\le m\cdot (\log m)^{c'_g}$,
and from Inequality \ref{ineq: total edges step 2}, 
$\sum_{I'=(G',\Sigma')\in \tilde{\iset}^{(n)}_{\textsf {large}}}|E(G')|\leq 2m\cdot (\log m)^{c'_g}$.
Therefore, $|\hat{\iset}^{(n)}_{\textsf {large}}\cup \tilde{\iset}^{(n)}_{\textsf {large}}|\leq 3\mu\cdot (\log m)^{c'_g}$. From the Union Bound, assuming that the constant $c^*$ in the definition of the parameter $\mu$ is large enough, if $\optcrors(I)\leq m^2/\mu^{c''}$, then $\prob{\event_3\mid \neg\event_1\band\neg\event_2}\leq O\left ( \frac{\mu\cdot(\log m)^{c'_g}}{\mu^2}\right )\leq \frac 1 {100}$. 
Lastly, we define bad event $\event$ to be the event that at least one of the events $\event_1,\event_2,\event_3$ happened. 
Note that $\prob{\event}\leq \prob{\event_1}+\prob{\event_2\mid \neg\event_1}+\prob{\event_3\mid\neg \event_1\band\neg\event_2}$. Therefore, altogether, if  $\optcrors(I)\leq m^2/\mu^{c''}$, then
 $\prob[]{\event}\le \frac{3}{100}\leq \frac 1{30}$. Note that, if bad event $\event$ does not happen, then the algorithm does not return FAIL. 
 
 If the third step of the algorithm did not terminate with a FAIL, we let $\overline\iset_{\textsf{small}}= \bigcup_{I'\in \hat{\iset}^{(n)}_{\textsf {large}}\cup  \tilde{\iset}^{(n)}_{\textsf {large}}} \overline\iset(I')$.
By combining Equations  \ref{eq: num of edges step 1},  \ref{ineq: total edges step 2} and \ref{eq: few edges 3}, we get that:

\begin{equation}\label{eq: few edges 4}
\sum_{I''=(G'',\Sigma'')\in \overline \iset_{\textsf{small}}}|E(G'')|\leq 3m\cdot (\log m)^{2c'_g}.
\end{equation}

  The output of the algorithm is the collection 
$\iset^*=\hat{\iset}_{\textsf {small}}\cup\tilde{\iset}_{\textsf {small}}
\cup  \overline\iset_{\textsf{small}}$ of instances of \cnwrs. From the above discussion, for every instance $I''=(G'',\Sigma'')\in \iset^*$, $|E(G'')|\leq m/\mu$. As discussed already, if bad event $\event$ does not happen, then the algorithm does not return FAIL.

From now on we assume that the algorithm did not return FAIL.
From Inequalities \ref{eq: num of edges step 1},  \ref{ineq: total edges step 2} and \ref{eq: few edges 4}, we get that:

$$\sum_{I''=(G'',\Sigma'')\in \iset^*}|E(G'')|\leq 6m\cdot(\log m)^{2c'_g}.$$

Next, we provide Algorithm \algcombine in the following claim, whose proof is conceptually straightforward but somewhat technical, and is deferred to Section \ref{Appx: Proof of combine drawings} of Appendix.

\begin{claim}\label{claim: combine drawings}
	There is an efficient algorithm, called \algcombine, that, given a solution $\phi(I'')$ to every instance $I''\in \iset^*$, computes a solution $\phi(I)$ to instance $I$.
	Moreover, if $\optcrors(I)\le m^2/\mu^{c''}$, and event $\event$ did not happen, then $\cro(\phi(I))\le O(\sum_{I''\in \iset^*}\cro(\phi(I''))) +(\optcrors(I)+m)\cdot\mu^{O(1)}$.
\end{claim}

The following observation, whose proof is deferred to Section \ref{Appx: Proof of bound sum of opts} of Appendix,  will complete the proof of \Cref{thm: main}.

\begin{observation}\label{obs: bound sum of opts}
	If $\optcrors(I)\leq |E(G)|^2/\mu^{c'}$ and bad event $\event$ did not happen, then for some constant $c$, with probability at least $99/100$: $$\sum_{I''\in \iset^*}\optcrors(I'')\leq (\optcrors(I)+m)\cdot2^{c(\log m)^{3/4}\log\log m}.$$
\end{observation}

Let $\event'$ be the bad event that $\sum_{I''\in \iset^*}\optcrors(I'')> (\optcrors(I)+m)\cdot2^{c(\log m)^{3/4}\log\log m}$. Clearly, if  $\optcrors(I)\leq m^2/\mu^{c''}$, then the probability that either of the events $\event$ or $\event'$ happens is at most $\prob{\event}+\prob{\event'\mid\neg\event}\leq 1/16$. Therefore, we conclude that, if  $\optcrors(I)\leq m^2/\mu^{c''}$, then with probability at least $15/16$, all of the following hold:  (i) the algorithm does not return FAIL; (ii) $\iset^*\neq \emptyset$; (iii) 	$\sum_{I''\in \iset^*}\optcrors(I'')\leq (\optcrors(I)+m)\cdot 2^{O((\log m)^{3/4}\log\log m)}$; and (iv) if algorithm \algcombine is given as input a solution $\phi(I'')$ to every instance $I''\in \iset^*$, then the resulting solution $\phi$ to instance $I$ that it computes has cost at most: $O\bigg(\sum_{I''\in \iset^*}\cro(\phi(I''))\bigg) +(\optcrors(I)+m)\cdot\mu^{O(1)}$. This concludes
the proof of Theorem \ref{thm: main} from Theorems \ref{thm: not well connected}, \ref{lem: many paths}, and \ref{lem: not many paths}.

\section{Definitions, Notation, Known Results, and their Easy Extensions}
\label{sec:long prelim}

In this section we provide additional definitions and notation, together with known results and their easy extensions that we use throughout the paper.


\subsection{Clusters, Paths, Flows, and Routers}
\label{subsec: clusters and paths}

\subsubsection{Clusters and Augmentations of Clusters}
Let $G$ be a graph. A \emph{cluster} of $G$ is a vertex-induced connected subgraph of $G$.  For a set $\cset$ of mutually disjoint clusters of $G$, we denote by $\inE_G(\cset)$ the set of all edges $e=(u,v)$ of $G$, with endpoints $u$ and $v$ lying in distinct clusters of $\cset$.
We sometimes omit the subscript $G$ when clear from the context. 

Next, we define the notion of augmentation of a cluster.

\begin{definition}[Augmentation of Clusters]
	\label{def: Graph C^+}
	Let $C$ be a cluster of a graph $G$. The \emph{augmentation} of cluster $C$, denoted by $C^+$, is a graph that is obtained from $G$ as follows. First, we subdivide every edge $e\in \delta_G(C)$ with a vertex $t_e$, and let $T(C)=\set{t_e\mid e\in \delta_G(C)}$ be the resulting set of newly added vertices. We then let $C^+$ be the subgraph of the resulting graph induced by the set $V(C)\cup T(C)$ of vertices.
\end{definition}

\subsubsection{Paths and Flows}
\label{sec: routing paths}

As mentioned already, all graphs that we consider in this paper are undirected. However, sometimes it will be convenient for us to assign direction to paths in such graphs. We do so by designating one endpoint of the path as its first endpoint, and another endpoint as its last endpoint. We will then view the path as being directed from its first endpoint towards its last endpoint. We will sometimes refer to a path with an assigned direction as a directed path, even though the underlying graph is an undirected graph. 

Let $G$ be a graph, and let $\pset$ be a collection of paths in $G$. We say that the paths of $\pset$ are \emph{edge-disjoint} if every edge of $G$ belongs to at most one path of $\pset$. We say that the paths in $\pset$ are \emph{vertex-disjoint} if every vertex of $G$ belongs to at most one path of $\pset$. We say that the paths in $\pset$ are \emph{internally disjoint} if every vertex $v\in V(G)$ that serves as an inner vertex of some path in $\pset$ only belongs to one path of $\pset$. Given a subset $S$ of vertices of $G$, we say that the paths in $\pset$ are \emph{internally disjoint from $S$} if no vertex of $S$ serves as an inner vertex of any path in $\pset$. Abusing the notation, for a subgraph $C$ of $G$, we sometimes say that a set $\pset$ of paths is internally disjoint from $C$ to indicate that it is internally disjoint from $V(C)$.

\paragraph{Flows.}
Let $G$ be a graph, and let $\pset$ be a collection of directed paths in graph $G$. A \emph{flow} over the set $\pset$ of paths is an assignment of non-negative values $f(P)\geq 0$, called \emph{flow-values}, to every path $P\in \pset$. We sometimes refer to paths in $\pset$ as \emph{flow-paths for flow $f$}. For each edge $e\in E(G)$, let $\pset(e)\subseteq \pset$ be the set of all paths whose first edge is $e$, and let $\pset'(e)\subseteq \pset$ be the set of all paths whose last edge is $e$. We say that edge $e$ \emph{sends $z$ flow units} in $f$ if $\sum_{P\in \pset(e)}f(e)=z$, and we say that edge $e$ \emph{receives $z$ flow units} in $f$ if $\sum_{P\in \pset'(e)}f(P)=z$. Similarly, for a vertex $v\in V(G)$, we say that $v$ sends $z$ flow units in $f$ if the sum of flow-values of all paths $P\in \pset$ that originate at $v$ is $z$. We say that $v$ receives $z$ flow units in $f$ if the sum of the flow-values of all paths $P\in \pset$ that terminate at $v$ is $z$. The \emph{congestion} that flow $f$ causes on an edge $e$ is $\sum_{\stackrel{P\in \pset:}{e\in E(P)}}f(P)$, and the \emph{total congestion} of the flow $f$ is the maximum congestion that it causes on any edge $e\in E(G)$.

An \emph{$s$-$t$ flow network} consists of a graph $G$, non-negative capacities $c(e)\geq 0$ for each edge $e\in E(G)$, and two special vertices: source $s$ and destination $t$. Let $\pset$ be the set of all paths in graph $G$ originating at $s$ and terminating at $t$. An $s$-$t$ flow in $G$ is a flow $f$ that is defined over the set $\pset$ of paths, such that for every edge $e\in E(G)$, the congestion that $f$ causes on edge $e$ is at most $c(e)$. The \emph{value} of the flow is $\sum_{P\in \pset}f(P)$. Maximum $s$-$t$ flow is an $s$-$t$ flow of largest possible value. We say that a flow $f$ is \emph{integral} if, for every path $P$, value $f(P)$ is an integer. It is a well known fact (called \emph{integrality of flow}) that, if all edge capacities in a flow network are integral, then there is a maximum $s$-$t$ flow that is integral, and such a flow can be found efficiently. In case where the capacity of every edge is unit, such a flow defines a maximum-cardinality collection of edge-disjoint $s$-$t$ paths.

\paragraph{Congestion Reduction.}
We repeatedly use the following simple claim, whose proof follows from integrality of flow, and appears in \Cref{apd: Proof of remove congestion}.

\begin{claim}\label{claim: remove congestion}
	Let $G$ be a graph and let $\pset$ be a set of directed paths in $G$. For each vertex $v\in V(G)$, let $n_S(v)$ and $n_T(v)$ denote the numbers of paths in $\pset$ originating and terminating at $v$, respectively. Then there is a set $\pset'$ of at least $|\pset|/\cong_G(\pset)$ 
	edge-disjoint directed paths in $G$, such that, for every vertex $v$, at most $n_S(v)$ paths of $\pset'$ originate at $v$, and at most $n_T(v)$ paths of $\pset'$ terminate at $v$. Moreover, there is an efficient algorithm, that, given $G$ and $\pset$, computes a set $\pset'$ of paths with these properties.
\end{claim}

\subsubsection{Routing Paths, Internal Routers and External Routers}
\label{subsubsection: routing paths}

\paragraph{Routing Paths.} Suppose we are given a graph $G$, two sets $S,T\subseteq V(G)$ of its vertices, and a set $\qset$ of paths. We say that $\qset$ is
a \emph{routing of vertices of $S$ to vertices of $T$}, or that $\qset$ \emph{routes vertices of $S$ to vertices of $T$} if $\qset=\set{Q_v\mid v\in S}$, and, for every vertex $v\in S$, path $Q_v$ originates at $v$ and terminates at a vertex of $T$. 
If, additionally, for every vertex $t\in T$, exactly one path in $\qset$ terminates at $t$, then we say that $\qset$ is a \emph{one-to-one routing} of vertices of $S$ to vertices of $T$.

Similarly, given two sets $E_1,E_2$ of edges of $G$, we say that a set $\qset=\set{Q_e\mid e\in E_1}$ of paths is a \emph{routing of edges of $E_1$ to edges of $E_2$}, or that $\qset$ \emph{routes edges of $E_1$ to edges of $E_2$}, if, for every edge $e\in E_1$, path $Q_e$ has $e$ as its first edge, and some edge of $E_2$ as its last edge. If, additionally, every edge of $E_2$ serves as the last edge of exactly one path in $\qset$, then we say that $\qset$ is a \emph{one-to-one routing of edges of $E_1$ to edges of $E_2$}.

Next, we define the notions of internal and external routers for clusters, which are central notions that are used throughout our algorithms.

\begin{definition}[Internal and External Routers for Clusters]\label{def: routers}
	Let $G$ be a graph, let $C$ be a cluster of $G$, and let $\qset(C)$ be a set of paths in $G$. We say that $\qset(C)$ is an \emph{internal router} for $C$, or an \emph{internal $C$-router}, if there is some vertex $u\in V(C)$, such that $\qset(C)=\set{Q_e\mid e\in \delta_G(C)}$, and, for each edge $e\in \delta_G(C)$, path $Q_e$ has $e$ as its first edge, $u$ as its last vertex, and all edges of $E(Q_e)\setminus\set{e}$ lie in $C$. We refer to vertex $u$ as the \emph{center of the router}.
	Similarly, we say that a set $\qset'(C)$ of paths in $G$ is an \emph{external router} for $C$, or an \emph{external $C$-router}, if there is some vertex $u'\in V(G)\setminus V(C)$, such that $\qset'(C)=\set{Q'_e\mid e\in \delta_G(C)}$, and, for each edge $e\in \delta_G(C)$, path $Q_e$ has $e$ as its first edge, $u'$ as its last vertex, is internally disjoint from $C$.  We refer to $u'$ as the \emph{center of the router}.
	We denote by $\Lambda_G(C)$ the set of all internal $C$-routers, and by $\Lambda'_G(C)$ the set of all external $C$-routers in $G$. We may omit the subscript $G$ when clear from the context.
\end{definition}

Throughout the paper, we will be working with distributions over the set $\Lambda_G(C)$ of internal $C$-routers and distributions over the set $\Lambda'_G(C)$ of external $C$-routers for various clusters $C$ of a given graph $G$. We say that a distribution $\dset$ over a set $U$ of elements is given \emph{explicitly}, if we are given a list $U'\subseteq U$ of elements, whose probability in $\dset$ is non-zero, together with their associated probability values. We say that distribution $\dset$ is given \emph{implicitly} if we are given an efficient randomized algorithm that draws an element from $U$ according to the distribution. When the distribution $\dset$ is over a set of routers in a graph $G$, the running time of the algorithm should be bounded by $\poly(|E(G)|)$.


\subsubsection{Non-Transversal Paths and Path Splicing}
\label{subsec: non-transversal paths and splicing}

We start by defining the notions of transversal and non-transversal intersections of paths and cycles, which we then use to define non-transversal paths.

\begin{figure}[h]
	\centering
	\subfigure[The intersection of paths $P_1$ (red) and  $P_2$ (purple) is transversal at $v$.]{\scalebox{0.14}{\includegraphics{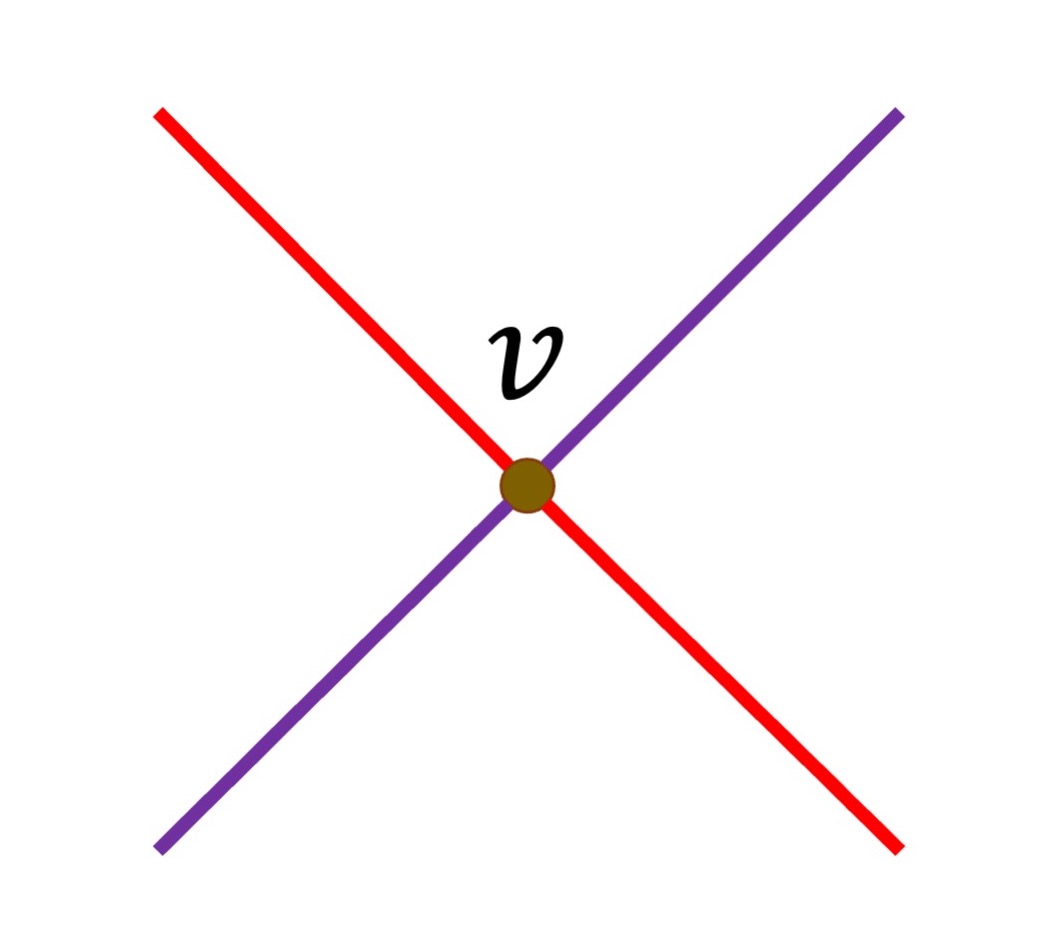}}}
	\hspace{0.2cm}
	\subfigure[The intersection of path $P_1$ (red) and path $P_2$ (purple) is non-transversal at $v$.]{
		\scalebox{0.14}{\includegraphics{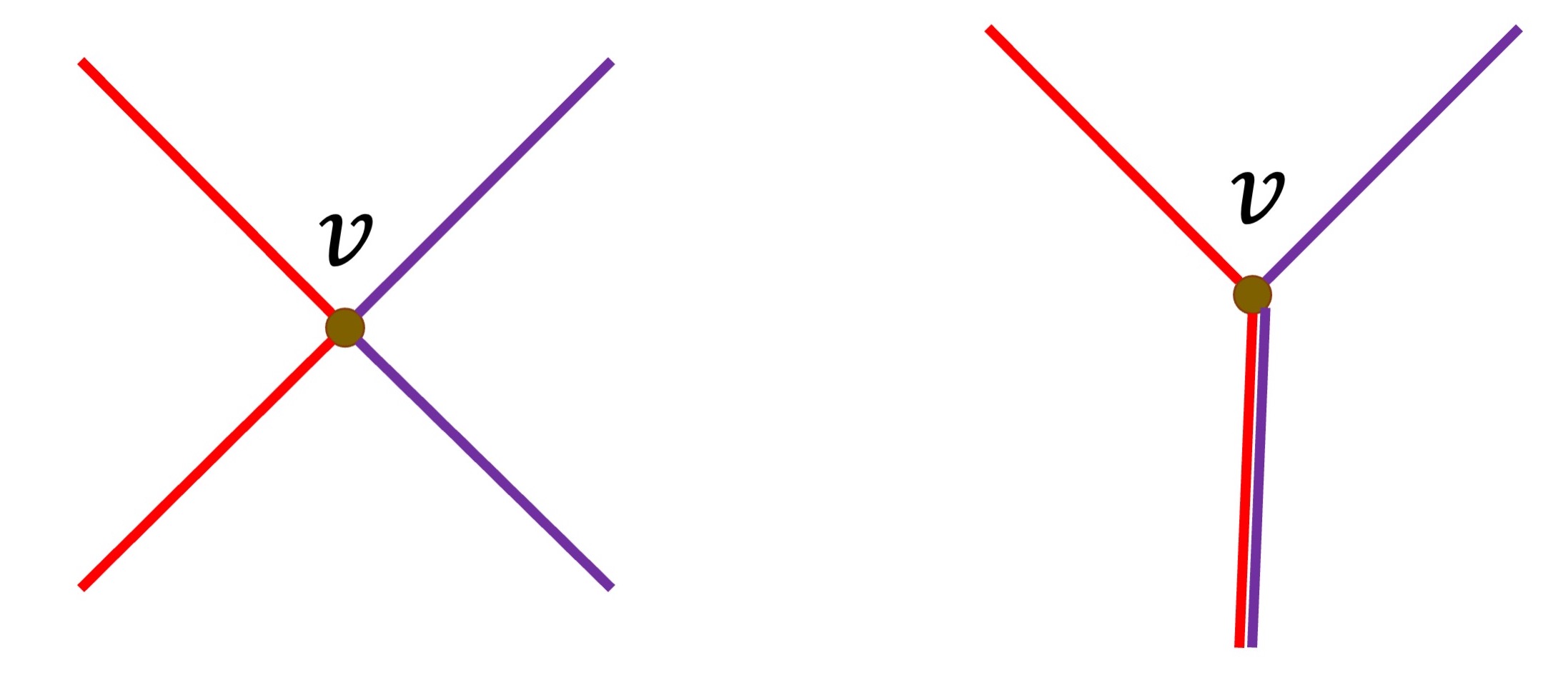}}}
	\caption{Transversal and non-transversal intersections of paths. 
	}\label{fig:non_trans}
\end{figure}

\begin{definition}[Non-transversal Intersection of Paths and Cycles]
	Let $I=(G,\Sigma)$ be an intance of \cnwrs,	let $P_1,P_2$ be two simple paths in $G$, and let $u$ be a vertex in $V(P_1)\cap V(P_2)$. Denote by $E_1$ the set of (one or two) edges of $P_1$ that are incident to $u$, and similarly denote by $E_2$ the set of (one or two) edges of $P_2$ that are incident to $u$. We say that the intersection of the paths $P_1,P_2$ at vertex $u$ is \emph{non-transversal with respect to $\Sigma$} if one of the following holds:
	\begin{itemize}
		\item either the set $E_1\cup E_2$ contains fewer than $4$ distinct edges; or
		\item $E_1=\set{e_1,e_1'}$ and $E_2=\set{e_2,e_2'}$, all edges in set $\set{e_1,e_1',e_2,e_2'}$ are distinct,
		and they appear in the ordering $\oset_u\in \Sigma$ in one of the following circular orders:  $(e_1,e_1',e_2,e_2')$, or $(e_1,e_1',e_2',e_2)$ (recall that the orderings are unoriented, so the reversals of the above two orderings are also included in this definition).
	\end{itemize}
	Otherwise, we say that the intersection of the paths $P_1,P_2$ at vertex $u$ is \emph{transversal} (see \Cref{fig:non_trans}). 
	If $R_1,R_2$ are simple cycles in $G$, and $u$ is a vertex in $V(R_1)\cap V(R_2)$, then we classify the intersection of $R_1$ and $R_2$ and $u$ as transversal or non-transversal with respect to $\Sigma$ similarly.
\end{definition}

\begin{definition}[Non-transversal Set of Paths]\label{def: non-transversal paths}
	Let $I=(G,\Sigma)$ be an intance of \cnwrs,	and let $\pset$ be a collection of simple paths in $G$. We say that the set $\pset$ of paths is \emph{non-transversal with respect to $\Sigma$} if, for every pair $P_1,P_2\in \pset$ of paths, for every vertex $u\in V(P_1)\cap V(P_2)$, the intersection of $P_1$ and $P_2$ at $u$ is non-transversal with respect to $\Sigma$.
\end{definition}

Assume now that we are given some instance $I=(G,\Sigma)$ of \cnwrs, and a collection $\qset$ of simple paths in $G$. 
We let $\Pi^T(\qset)$ denote the set of all triples $(Q,Q',v)$, such that $Q,Q'\in \qset$, $v$ is an inner vertex of  both $Q$ and $Q'$, and the intersection of $Q$ and $Q'$ at $v$ is transversal with respect to $\Sigma$. 

We need to design a subroutine, that, given a set $\qset$ of simple directed paths in a graph $G$, transforms it into a set  $\qset'$ of paths that is  non-transversal with respect to the given rotation system $\Sigma$ for $G$. We need to ensure that the multisets containing the first vertex of every path in $\qset$ and in $\qset'$, respectively, remain unchanged, and the same holds for multisets containing the last vertex of every path in both path sets. We also need to ensure that for each edge $e\in E(G)$, $\cong_G(\qset')\leq \cong_G(\qset)$.
Below we provide a procedure for performing such a transformation. The procedure uses a simple subroutine that we call \emph{path splicing} and describe next.

\begin{figure}[h]
	\centering
	\subfigure[Before: Path $P$ is shown in red and path $P'$ is shown in purple.]{\scalebox{0.12}{\includegraphics{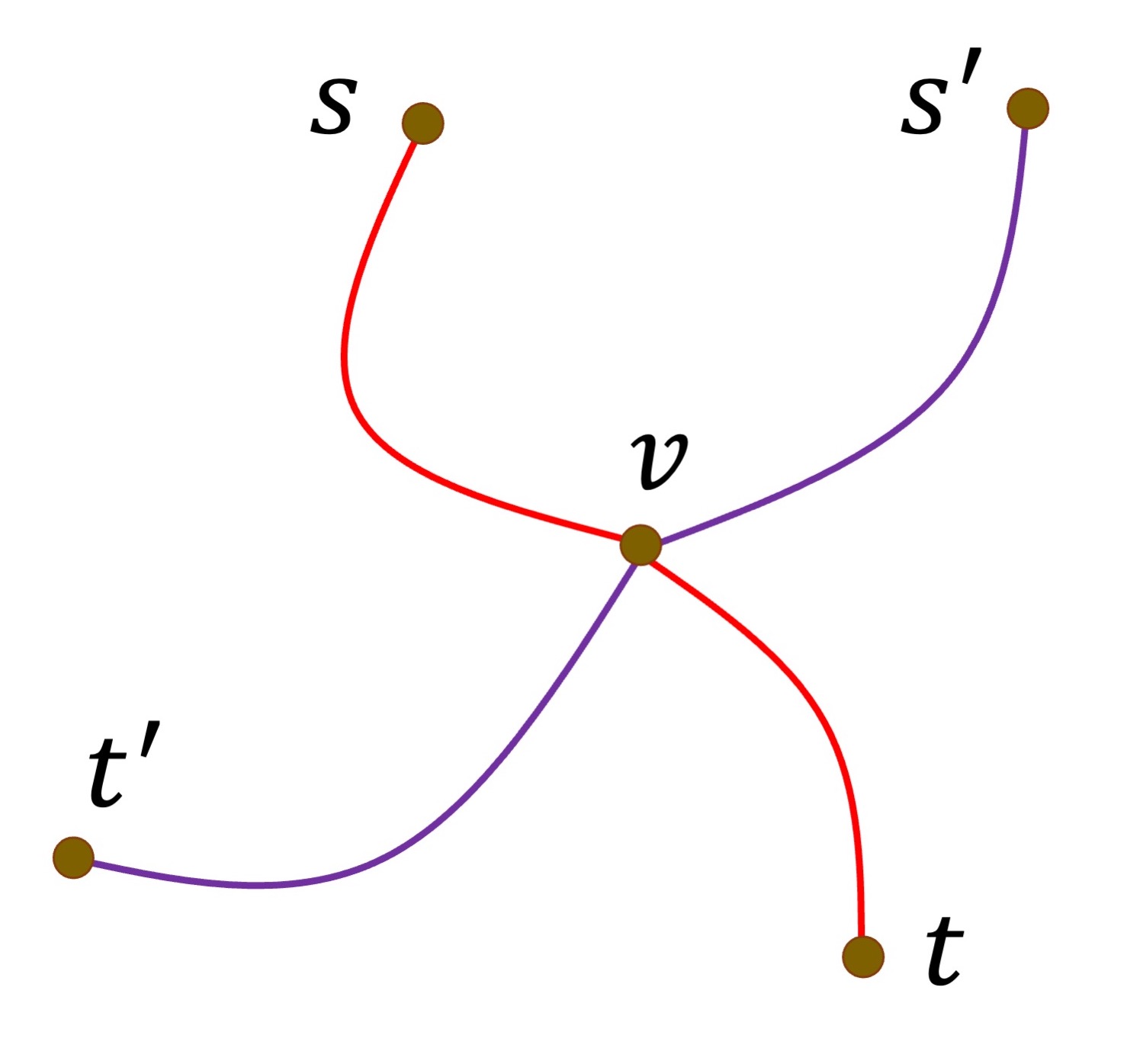}}}
	\hspace{0.7cm}
	\subfigure[After: Path $\tilde P$ is shown in red and path $\tilde P'$ is shown in purple.]{
		\scalebox{0.12}{\includegraphics{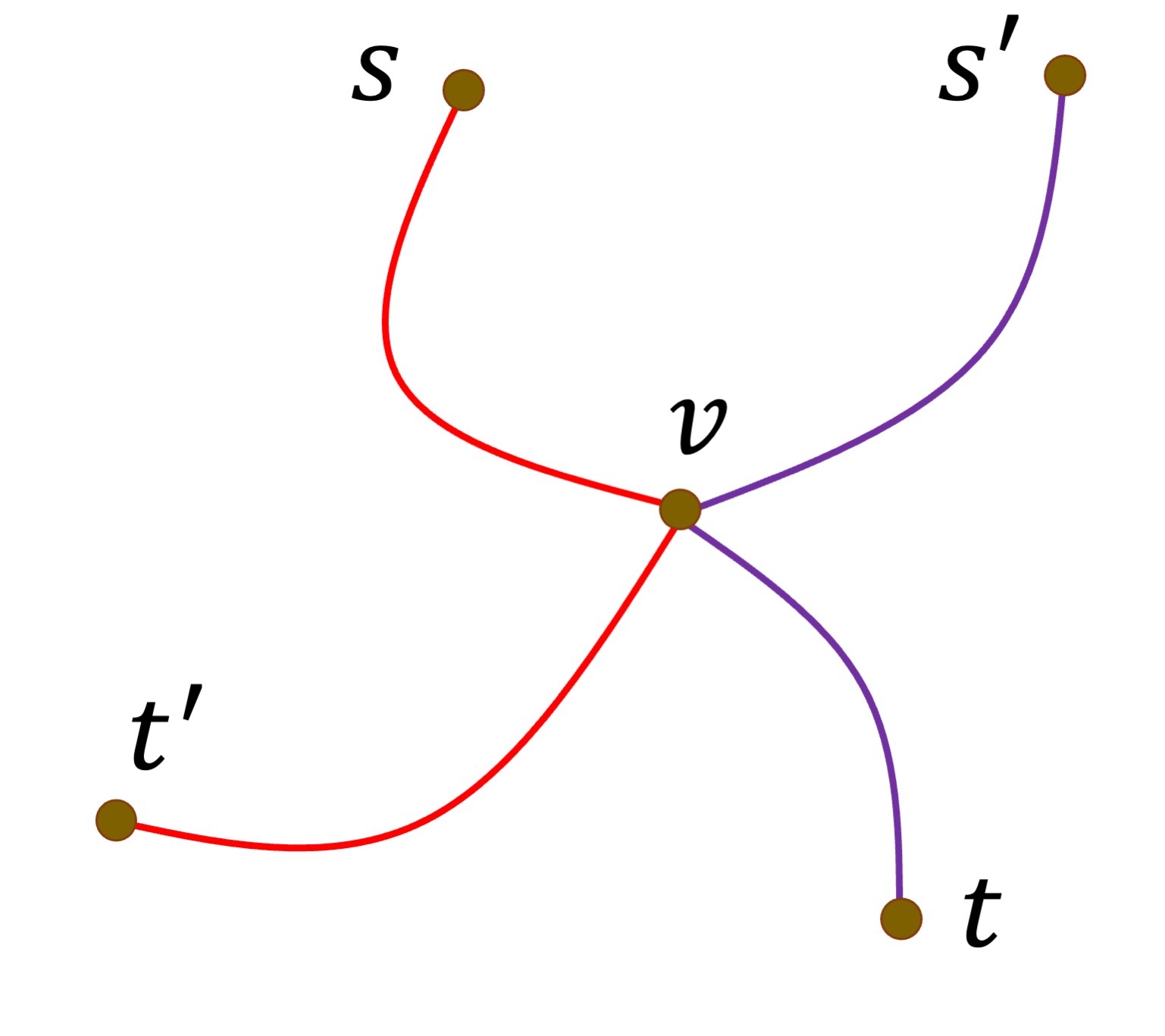}}}
	\caption{An illustration of path splicing at vertex $v$.}\label{fig: path_splicing}
\end{figure}

\paragraph{Path Splicing.}
Suppose we are given an instance $I=(G,\Sigma)$ of \cnwrs, two simple paths $P,P'$ in $G$, and a vertex $v$, that serves as an inner vertex of both $P$ and $P'$, such that the intersection of $P$ and $P'$ at vertex $v$ is transversal with respect to $\Sigma$. We assume that each of the paths $P,P'$ is assigned a direction, and we denote by $s$ and $t$ the first and the last endpoints of $P$, respectively, and by $s'$ and $t'$ the first and the last endpoints of $P'$, respectively. The \emph{splicing} of $P$ and $P'$ at vertex $v$ produces two new paths: path $\tilde P$, that is a concatenation of the subpath of $P$ from $s$ to $v$, and the subpath of $P'$ from $v$ to $t'$; and path $\tilde P'$, that is a concatenation of the  subpath of $P'$ from $s'$ to $v$, and the subpath of $P$ from $v$ to $t$. 
See \Cref{fig: path_splicing} for an illustration.

For a set $\pset$ of directed paths in a graph $G$, we denote by $S(\pset)$ and $T(\pset)$ the multisets containing the first vertex on every path in $\pset$, and the last vertex on every path in $\pset$, respectively.
We use the following simple observation regarding the splicing procedure, whose proof is deferred to Section \ref{apd: Proof of splicing} of Appendix.

\begin{observation}\label{obs: splicing}
	Let $I=(G,\Sigma)$ be an instance of $\cnwrs$, let $\pset$ be a set of simple directed paths in $G$, 
	and let $(P,P',v)$ be a triple in $\Pi^T(\pset)$.  Let $\tilde P,\tilde P'$ be the pair of paths obtained by splicing 	$P$ and $P'$ at $v$, and let $\pset'=\big(\pset\setminus\set{P,P'}\big)\cup \set{\tilde P,\tilde P'}$.
	Then $S(\pset')=S(\pset)$ and $T(\pset')=T(\pset)$. Additionally,	either (i) at least one of the paths $\tilde P,\tilde P'$ is a non-simple path; or (ii) $|\Pi^T(\pset')|<|\Pi^T(\pset)|$.
\end{observation}

Using \Cref{obs: splicing}, we can prove the following lemma that allows us to transform an arbitrary set $\rset$ of paths into a set $\rset'$ of non-transversal paths, while preserving the multisets containing the first endpoint and the last endpoint of every path, and without increasing the congestion on any edge. 
The proof of the lemma below is similar to the proof of Lemma 9.5 in~\cite{chuzhoy2020towards}, and is provided in Appendix~\ref{apd: Proof of non_interfering_paths} for completeness.

\begin{lemma}
	\label{lem: non_interfering_paths}
	There is an efficient algorithm, that, given an instance $(G, \Sigma)$ of \CNwRS and a set $\rset$ of directed paths in $G$, computes another set $\rset'$ of simple directed paths in $G$, such that $S(\rset')=S(\rset)$, $T(\rset')=T(\rset)$, and the paths in $\rset'$ are non-transversal with respect to $\Sigma$. Moreover, for every edge $e\in E(G)$, $\cong_G(\rset',e)\leq \cong_G(\rset,e)$.
\end{lemma}

\subsection{Cuts, Well-Linkedness, and Related Notions}
\label{subsec: cuts, wl}

\subsubsection{Minimum Cuts}
A \emph{cut} in a graph $G$ is a bipartition $(A,B)$ of its vertices into non-empty subsets. The \emph{value} of the cut is $|E(A,B)|$. 
 We sometimes consider cuts in edge-capacitated graphs. Given a graph $G$ with capacities $c(e)\geq 0$ on edges $e\in E(G)$ and a cut $(A,B)$ in $G$, the value of the cut  is $\sum_{e\in E_G(A,B)}c(e)$.
When edge capacities are not specified, we assume that they are unit.

Given two disjoint subsets $S,T$ of vertices of $G$, an \emph{$S$-$T$} cut, or a \emph{cut separating $S$ from $T$} is a cut $(A,B)$ with $S\subseteq A$, $T\subseteq B$. 
A \emph{minimum $S$-$T$ cut} is an $S$-$T$ cut $(A,B)$ of minimum value. When $S=\set{s}$ and $T=\set{t}$, we refer to $S$-$T$ cuts as $s$-$t$ cuts.
We will use the following lemma, whose proof is provided in Section~\ref{apd: Proof of multiway cut with paths sets} of Appendix.
\begin{lemma}\label{lem: multiway cut with paths sets}
	There is an efficient algorithm, that, given a graph $G$ and a collection $S=\set{s_1,\ldots,s_k}$ of its vertices, computes, for all $1\leq i\leq k$, a set $A_i$ of vertices of $G$, and a collection $\qset_i$ of paths in $G$, such that the following hold:
	
	\begin{itemize}
		\item for all $1\leq i\leq k$, $S\cap A_i=\set{s_i}$, and moreover, $(A_i,V(G)\setminus A_i)$ is a minimum cut separating $s_i$ from the vertices of $S\setminus\set{s_i}$ in $G$;
		\item for all $1\leq i<i'\leq k$, $A_i\cap A_{i'}= \emptyset$; and
		\item for all $1\leq i\leq k$, $\qset_i=\set{Q_i(e)\mid e\in \delta_G(A_i)}$, where for each $e\in \delta_G(A_i)$, path $Q_i(e)$ has $e$ as its first edge, $s_i$ as its last vertex, and all internal vertices of $Q_i(e)$ lie in $A_i$. Moreover, the paths in set $\qset_i$ are edge-disjoint.
	\end{itemize}
\end{lemma}

\subsubsection{Gomory-Hu Trees}
\label{subsec: GH tree}

Gomory-Hu tree is a convenient structure that represents all  minimum $s$-$t$ cuts in a given graph $G$. We summarize its properties in the following theorem.
\begin{theorem}[\cite{gomory1961multi}]\label{thm: GH tree properties}
	There is an efficient algorithm, that, given a graph $G=(V,E)$ with capacities $c(e)\geq 0$ on its edges $e\in E$, computes a tree $\tau=(V,E')$ with capacities $c'(e)\geq 0$ on its edges $e\in E'$, such that the following hold:
	
	\begin{itemize}
		\item for every pair $s,t$ of distinct vertices of $V$, the value of the minimum $s$-$t$ cut in $G$ is equal to $\min_{e\in E(P_{s,t})}\set{c'(e)}$, where $P_{s,t}$ is the unique path connecting $s$ to $t$ in $\tau$; and
		\item for every pair $s,t$ of distinct vertices of $V$, if $(A,B)$ is a minimum $s$-$t$ cut in graph $G$, then $(A,B)$ is a minimum $s$-$t$ cut in graph $\tau$, and vice versa.
	\end{itemize}
\end{theorem}

We obtain the following immediate corollary of \Cref{thm: GH tree properties}.

\begin{corollary}
	\label{cor: G-H tree_edge_cut}
	Let $G$ be an edge-capacitated graph, and let $\tau$ be a Gomory-Hu tree of graph $G$. Then for every edge $e=(u,u')\in E(\tau)$, if we denote by $U,U'$ the vertex sets of the two connected components of $\tau\setminus\set{e}$, with $u\in U$, then $(U,U')$ is a minimum $u$-$u'$ cut in graph $G$.
\end{corollary}

\subsubsection{Balanced Cut and Sparsest Cut}

Suppose we are given a graph $G=(V,E)$, and a subset $T\sse V$ of its vertices. We say that a cut $(X,Y)$ in $G$ is a valid $T$-cut iff $X\cap T,Y\cap T\neq \emptyset$. The \emph{sparsity} of a valid $T$-cut $(X,Y)$ with respect to $T$ is $\frac{|E(X,Y)|}{\min\set{|X\cap T|, |Y\cap T|}}$. 
In the Sparsest Cut problem, given a graph $G$ and a subset $T$ of its vertices, the goal is to compute a valid $T$-cut of minimum sparsity. Arora, Rao and Vazirani~\cite{ARV} designed an $O(\sqrt {\log n})$-approximation algorithm for the sparsest cut problem\footnote{The algorithm was originally designed for simple graphs, but it can be easily generalized to graphs with parallel edges by exploiting edge capacities.}, where $n=|V(G)|$. 
We denote this algorithm by \algsc, and its approximation factor by $\alphasc(n)=O(\sqrt{\log n})$.

We say that a cut $(A,B)$ in a graph $G$ is \emph{$\eta$-edge-balanced},  or just \emph{$\eta$-balanced}, for a parameter $0<\eta<1$, if $|E(A)|,|E(B)|\leq  \eta\cdot|E(G)|$. We say that a cut $(A,B)$ is a \emph{minimum $\eta$-balanced cut in $G$} if $(A,B)$ is an $\eta$-balanced cut of minimum value $|E(A,B)|$.
We will use the following theorem that follows from the work of \cite{ARV}.
The proof is provided in Section~\ref{apd: Proof of approx_balanced_cut} of Appendix.

\begin{theorem}
	\label{cor: approx_balanced_cut}
	For every constant $1/2<\hat \eta <1$, there is another constant $\hat \eta<\hat \eta'<1$ and an efficient algorithm, that, given a connected (not necessarily simple) graph $G$ with $m$ edges, computes a $\hat \eta'$-balanced cut $(A,B)$ in $G$, whose value is at most $O(\alphasc(m))$ times the value of the minimum $\hat \eta$-balanced cut in $G$.
\end{theorem}

The following lemma is a simple consequence of the Planar Separator Theorem of  Lipton and Tarjan \cite{lipton1979separator}.
A version of the lemma for vertex-balanced cuts was proved in~\cite{pach1996applications}. 
For completeness, we provide the proof of the lemma in Section~\ref{apd: Proof of min_bal_cut} of Appendix. 

\begin{lemma}
	\label{lem:min_bal_cut}
	Let $G$ be a connected (not necessarily simple) graph with $m$ edges and maximum vertex degree $\Delta\le m/2^{40}$. If $\optcro(G)\le m^2/2^{40}$, then the value of a minimum $(3/4)$-edge-balanced cut in $G$ is at most $O(\sqrt{\optcro(G)+\Delta\cdot m})$.
\end{lemma}

\subsubsection{Well-Linkedness, Bandwidth Property, and Routing Well-Linked Vertex Sets} 

The notion of well-linkedness plays a central role in graph theory and graph algorithms (see e.g. \cite{racke2002minimizing,chekuri2004edge,andrews2010approximation,chuzhoy2012polylogarithmic,chuzhoy2012routing,chekuri2016polynomial,chuzhoy2016improved,chuzhoy2019towards}). 
We use the following standard definitions, which are equivalent to
those used in much of previous work. 

\begin{definition}[Well-Linkedness]
We say that a set $T$ of vertices in a graph $G$ is \emph{$\alpha$-well-linked}, for a parameter $0<\alpha<1$, if the sparsity of every valid $T$-cut in graph $G$ is at least $\alpha$.  Equivalently, for every partition $(A,B)$ of $V(G)$ with $A\cap T,B\cap T\neq \emptyset$, $|E_G(A,B)|\geq \alpha\cdot \min\set{|A\cap T|,|B\cap T|}$ must hold.
\end{definition}

The next simple observation, that has been used extensively in previous work, shows that the set of vertices lying on the first row of the $(r\times r)$-grid is $1$-well-linked. For completeness, we provide its proof in Section \ref{apd: Proof grid 1st row well-linked} of Appendix.

\begin{observation}
	\label{obs: grid 1st row well-linked}
	Let $r\geq 1$ be an integer, and let $H$ be the $(r\times r)$-grid graph. Let $S$ be the set of vertices lying on the first row of the grid. Then vertex set $S$ is $1$-well-linked in $H$.
\end{observation}

Next, we define the notion of bandwidth property, that was also used extensively in graph algorithms.

\begin{definition}[$\alpha$-Bandwidth Property]
We say that a cluster $C$ of a graph $G$ has the \emph{$\alpha$-bandwidth property} in $G$, for some parameter $0<\alpha<1$, if, for every partition $(A,B)$ of vertices of $C$, 
$|E_G(A,B)|\ge \alpha\cdot\min\set{|\delta_G(A)\cap \delta_G(C)|, |\delta_G(B)\cap \delta_G(C)|}$.
\end{definition}

The following immediate observation provides an equivalent definition  of the bandwidth property that is helpful to keep in mind. Recall that, for a cluster $C$ of a graph $G$, its augmentation  $C^+$ is a graph that is obtained from graph $G$ as follows. We subdivide every edge $e\in \delta_G(C)$ with a vertex $t_e$, and let $T(C)=\set{t_e\mid e\in \delta_G(C)}$ be the resulting set of newly added vertices. We then let $C^+$ be the subgraph of the resulting graph induced by vertex set $V(C)\cup T(C)$.

\begin{observation}\label{obs: wl-bw}
	Let $G$ be a graph, let $C\subseteq G$ a cluster of $G$, and let $0<\alpha<1$ be a parameter. Cluster $C$ has the $\alpha$-bandwidth property iff the set $T(C)=\set{t_e\mid e\in \delta_G(C)}$ of vertices is $\alpha$-well-linked in graph $C^+$, which is the augmentation of cluster $C$ in $G$. 
\end{observation}

One useful property of well-linked sets of vertices is that routing is easy between vertices of such sets. We summarize this property, that has been used extensively in past work, in the following theorem, and we provide its proof in Appendix~\ref{apd: Proof of bandwidth_means_boundary_well_linked} for completeness. The theorem uses the notion of one-to-one routing that was defined in \Cref{subsubsection: routing paths}.

\begin{theorem}
\label{thm: bandwidth_means_boundary_well_linked}
There is an efficient algorithm, that, given a graph $G$, a set $T$ of vertices of $G$ that is $\alpha$-well-linked, and a pair $T_1,T_2$ of disjoint equal-cardinality subsets of $T$, computes a one-to-one routing $\qset$ of vertices of $T_1$ to vertices of $T_2$, with $\cong_G(\qset)\leq \ceil{1/\alpha}$.
\end{theorem}

The next corollary follows immediately from \Cref{obs: wl-bw} and \Cref{thm: bandwidth_means_boundary_well_linked}.
\begin{corollary}
\label{cor: bandwidth_means_boundary_well_linked}
There is an efficient algorithm, that, given a graph $G$, a cluster $S$ of $G$ that has the $\alpha$-bandwidth property for some $0<\alpha<1$, and a pair $E_1,E_2$ of disjoint equal-cardinality subsets of the edge set $\delta_G(S)$, computes a one-to-one routing $\qset$ of edges of $E_1$ to edges of $E_2$, with $\cong_G(\qset)\leq \ceil{1/\alpha}$, such that, for every path $Q\in\qset$, all inner vertices of $Q$ lie in $S$.
\end{corollary}

\subsubsection{Basic Well-Linked Decomposition}
Typically, in a well-linked decomposition, we are given a graph $G$ together with a cluster $S$ of $G$, and our goal is to compute a partition of $S$ into clusters, each of which has the $\alpha$-bandwidth property in graph $G$, for some given parameter $0<\alpha<1$. Algorithms for computing well-linked decompositions were used extensively in prior work on graph-based problems (see e.g. \cite{racke2002minimizing,chekuri2004edge,andrews2010approximation,chuzhoy2012polylogarithmic,chuzhoy2012routing,chekuri2016polynomial,chuzhoy2016improved,chuzhoy2019towards}). We use a variation of this technique, that, in addition to ensuring that each cluster $R$ in the decomposition has the $\alpha$-bandwidth property, provides a collection  $\pset(R)$ of paths routing the edges of $\delta_G(R)$ to edges of $\delta_G(S)$, such that the paths in $\pset(R)$ are internally disjoint from $R$ and cause low congestion. The proof uses standard techniques and is deferred to Section~\ref{apd: Proof of well_linked_decomposition} of Appendix.

\begin{theorem}
	\label{thm:well_linked_decomposition}
	There is an efficient algorithm, whose input is a graph $G$, a connected cluster $S$ of $G$, and parameters $m$ and $\alpha$, for which $|E(G)|\leq m$ and $0<\alpha< \min\set{\frac 1 {64\alphasc(m)\cdot \log m},\frac 1 {48\log^2 m}}$ hold. 
	The algorithm computes a collection $\rset$ of vertex-disjoint clusters of $S$, such that:
	\begin{itemize}
		\item $\bigcup_{R\in \rset}V(R)=V(S)$;
		\item for every cluster $R\in\rset$, $|\delta_G(R)|\le |\delta_G(S)|$;
		\item every cluster $R\in\rset$ has the $\alpha$-bandwidth property in graph $G$; and 
		\item $\sum_{R\in \rset}|\delta_G(R)|\le |\delta_G(S)|\cdot\left(1+O(\alpha\cdot \log^{1.5} m)\right)$. 
	\end{itemize}

Additionally, the algorithm computes, 
 for every cluster $R\in \rset$, a set $\pset(R)=\set{P(e)\mid e\in \delta_G(R)}$ of paths in graph $G$ with $\cong_G(\pset(R))\leq 100$, such that, for every edge $e\in \delta_G(R)$, path $P(e)$ has $e$ as its first edge and some edge of $\delta_G(S)$ as its last edge, and all inner vertices of $P(e)$ lie in $V(S)\setminus V(R)$. 
\end{theorem}

We note that, while the above theorem requires that cluster $S$ is connected, it can also be used when this is not the case, by simply applying the algorithm to every connected component of $S$ and then taking the union of all resulting sets of clusters; all properties that the theorem guarantees will continue to hold.

\subsubsection{Layered Well-Linked Decomposition}
\label{subsec: layered wld}

To the best of our knowledge, layered well-linked decomposition was first introduced by Andrews \cite{andrews2010approximation}. It is similar to the basic well-linked decomposition, except that it has some additional useful properties.  We start by defining a layered well-linked decomposition formally. Our definition is very similar to that of \cite{andrews2010approximation}, except that we require some additional properties.

\begin{figure}[h]
	\centering
	\includegraphics[scale=0.08]{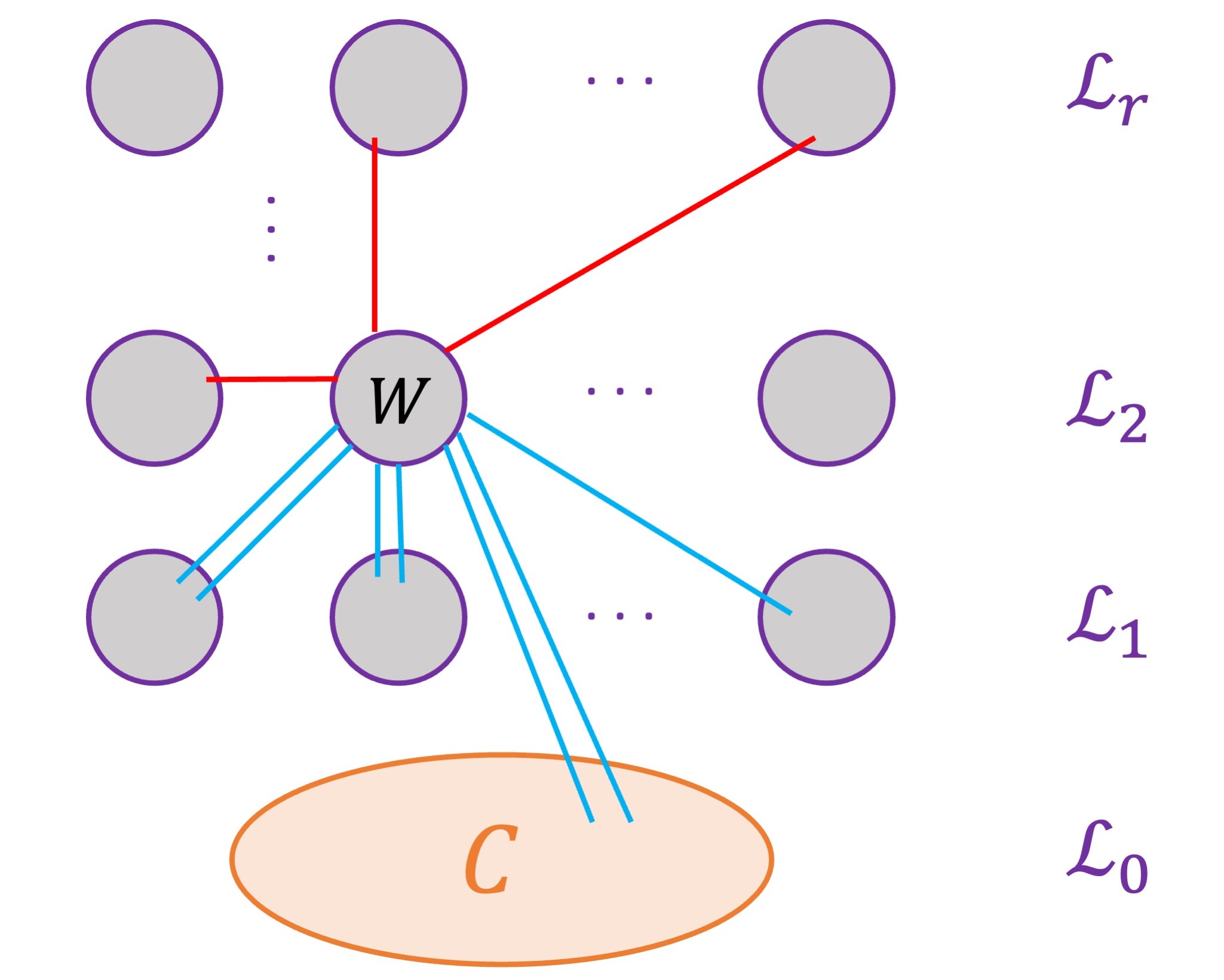}
	\caption{An illustration of a layered well-linked decomposition of $H$ with respect to $C$. For cluster $W\in \lset_2$, the edges of $\delta^{\up}(W)$ are shown in red, and the edges of $\delta^{\down}(W)$ are shown in blue.}\label{fig: LWLD}
\end{figure}

Let $H$ be a graph with $|E(H)|=m$ and $C\subseteq H$ a cluster of $H$. Let $\wset$ be a collection of disjoint clusters of $H\setminus C$ with $\bigcup_{W\in \wset}V(W)=V(H\setminus C)$, and let $(\lset_1,\lset_2,\ldots,\lset_r)$ be a partition of $\wset$ into subsets that we call \emph{layers}. We denote $\lset_0=\set{C}$, and, for all $1\leq i\leq r$, for every cluster $W\in \lset_i$, we partition the set $\delta_H(W)$ of edges into two subsets: set $\delta^{\down}(W)$ containing all edges $(u,v)$ with $u\in V(W)$ and $v$ lying in a cluster of $\lset_0\cup\cdots\cup\lset_{i-1}$, and set $\delta^{\up}(W)$ containing all remaining edges of $\delta(W)$, namely: all edges $(u,v)$ with $u\in V(W)$ and $v$ lying in a cluster of $\lset_i\cup\cdots\cup\lset_{r}$ (see \Cref{fig: LWLD}). We say that the collection $\wset$ of clusters, together with its partition $(\lset_1,\lset_2,\ldots,\lset_r)$ into layers is a \emph{valid layered $\alpha$-well-linked decomposition of $H$ with respect to $C$}, for some parameter $0<\alpha<1$, iff the following conditions hold:

\begin{properties}{L}
	\item  For every pair $W,W'$ of distinct clusters in $\wset$, $V(W)\cap V(W')=\emptyset$, and $\bigcup_{W\in \wset}V(W)=V(H)\setminus V(C)$; \label{condition: layered decomposition is partition}
	\item each cluster $W\in \wset$ has the $\alpha$-bandwidth property in $H$; \label{condition: layered well linked}
	\item for every cluster $W\in \wset$, $|\delta_H(W)|\leq |\delta_H(C)|$, and $|E_H(W)|\geq |\delta_H(W)|/(64\log m)$;  \label{condition: layered decomp each cluster prop}
	\item for every cluster $W\in \wset$, $|\delta^{\up}(W)|<|\delta^{\down}(W)|/\log m$; \label{condition: layered decomp edge ratio} 
	\item $\sum_{W\in \wset}|\delta_H(W)|\leq 4|\delta_H(C)|$; \label{condition: layered decomposition few edges}         
	and
	\item for every cluster $W\in \wset$, there is a collection $\pset(W)=\set{P(e)\mid e\in \delta_H(W)}$ of paths in $H$, that cause congestion at most $200/\alpha$, and for all $e\in \delta_H(W)$, path $P(e)$ contains $e$ as its first edge, some edge $e'\in \delta_H(C)$ as its last edge, and all inner vertices of $P(e)$ are disjoint from $W$. \label{condition: layered decomposition routing}
\end{properties}

Recall that, given a graph $H$ and two sets $E',E''$ of its edges, we say that a set $\pset$ of paths in $H$ routes edges of $E'$ to edges of $E''$ if $\pset=\set{P(e)\mid e\in E'}$, and, for each edge $e\in E'$, path $P(e)$ has $e$ as its first edge and some edge of $E''$ as its last edge. Given a cluster $W$ of $H$, we say that the set $\pset$ of paths \emph{avoids} $W$ if, for every path $P\in \pset$, no inner vertex of $P$ lies in $W$. Therefore, Condition \ref{condition: layered decomposition routing} equivalently requires that for every cluster $W\in \wset$, there is a collection $\pset(W)$ of paths in $H$
routing the edges of $\delta_H(W)$ to the edges of $\delta_H(C)$, such that the paths in $\pset(W)$ avoid $W$. This property is the main difference between our definition of a layered well-linked decomposition and that of \cite{andrews2010approximation}, which did not require this property.

The following theorem allows us to compute a layered well-linked decomposition in any graph. Its proof is practically identical to the algorithm of \cite{andrews2010approximation}. The main difference is that we need to prove that the resulting decomposition has property \ref{condition: layered decomposition routing}.
The proof of the theorem is deferred to Section \ref{sec: layered well linked} of Appendix.

\begin{theorem}\label{thm: layered well linked decomposition}
	There is a large enough constant $c$, and an efficient algorithm, that given a connected graph $H$ with $|E(H)|=m\geq c$ and a cluster $C$ of $H$, 
	computes a valid layered $\alpha$-well-linked decomposition $(\wset, (\lset_1,\ldots,\lset_r))$  of $H$ with respect to $C$, for $\alpha=\frac{1}{c\log^{2.5}m}$. The number of layers in the decomposition is $r\leq \log m$.
\end{theorem}

\subsection{Expanders, Graph Embeddings, and Routing Well-Linked Sets}

We will use the notion of expanders, that we define next. 

\begin{definition}[Expanders]
	We say that a graph $W$ is an $\alpha$-expander, for some  $0<\alpha<1$, if, for every partition $(A,B)$ of $V(W)$ into non-empty subsets, $|E_W(A,B)|\geq \alpha\cdot\min\set{|A|,|B|}$; equivalently, the set $V(W)$ of vertices is $\alpha$-well-linked in $W$.
\end{definition}	

We will also use a standard notion of graph embeddings.

\begin{definition}[Embedding of Graphs]
	Let $H$, $G$ be a pair of graphs with $V(H)\subseteq V(G)$. An \emph{embedding} of $H$ into $G$ is a collection $\pset=\set{P(e)\mid e\in E(H)}$ of paths in graph $G$, where for each edge $e=(u,v)\in E(H)$, path $P(e)$ has endpoints $u$ and $v$. The \emph{congestion of the embedding} is $\cong_{G}(\pset)$.
\end{definition}

The following well known claim shows a connection between well-linked sets of vertices and embeddings of expanders. The proof is standard and deferred to Section~\ref{apd: Proof of embed expander} of Appendix. 

\begin{claim}\label{claim: embed expander}
	There is a universal constant $\cCMG$, and an efficient randomized algorithm that, given a graph $G$ together with a subset $T$ of its vertices of cardinality $k$, such that $T$ is $\alpha$-well-linked in $G$, for some $0<\alpha<1$, constructs another graph $W$ with $V(W)=T$ and maximum vertex degree at most $\cCMG\log^2k$, together with an embedding $\pset$ of $W$ into $G$ with congestion at most $\frac{\cCMG \log^2k}{\alpha}$, such that with high probability graph $W$ is an $(1/4)$-expander.
\end{claim}

We show in the following observation that, if $W$ is the outcome expander of the algorithm from \Cref{claim: embed expander}, then it has a high crossing number. The proof is provided in Section~\ref{apd: Proof of cr of exp} of Appendix.

\begin{observation}\label{obs: cr of exp}
	There is some constant $c$, such that, if $W$ is an $(1/4)$-expander, with $|V(W)|=k>c$ and maximum vertex degree $O(\log^2k)$, then  $\optcro(W)\geq k^2/(c\log^8k)$.
\end{observation}

We obtain the following useful corollary of  \Cref{claim: embed expander}, that allows us to route specific pairs of vertices of a well-linked vertex set $T$. We provide its proof in Appendix~\ref{apd: Proof of routing well linked vertex set}.

\begin{corollary}\label{cor: routing well linked vertex set}
	There is an efficient randomized algorithm that, given a graph $G$, a subset $T$ of its vertices of cardinality $k$, that is $\alpha$-well-linked in $G$, for some $0<\alpha<1$, together with a partial matching $M$ over the vertices of $T$, computes a set $\rset(M)=\set{R(u,v)\mid (u,v)\in M}$ of paths in graph $G$, such that for every pair $(u,v)\in M$ of vertices, path $R(u,v)$ connects $u$ to $v$. Moreover, with high probability, the congestion caused by the paths in $\rset(M)$ in $G$ is $O((\log^4 k)/\alpha)$.
\end{corollary}

Let $K_z$ be a complete graph, whose vertex set has cardinality $z$. We obtain the following immediate corollary of \Cref{cor: routing well linked vertex set}, whose proof appears in Section~\ref{apd: Proof of embed complete graph} of Appendix.

\begin{corollary}\label{cor: embed complete graph}
	There is an efficient randomized algorithm that, given a graph $G$ and a subset $T$ of its vertices of cardinality $z$, such that $T$ is $\alpha$-well-linked in $G$, for some $0<\alpha<1$, computes an embedding $\tilde \pset$ of the complete graph $K_z$ with $V(K_z)=T$ into $G$, such that, with high probability, the congestion of the embedding is $O((z\log^4z)/\alpha)$.
\end{corollary}


\subsubsection{Constructing Internal  Routers}

We now provide an efficient algorithm, that, given a graph $G$ and a cluster $C$ of $G$ that has the $\alpha$-bandwidth property, constructs a distribution $\dset(C)$ over the internal $C$-routers, such that the expected congestion on every edge of $C$ is small. We start with the following lemma, that provides a similar result for a graph $G$ and a set $T$ of vertices of $G$ that is well-linked.


\begin{lemma}\label{lem: simple guiding paths}
	There is an efficient randomized algorithm, whose input is a graph $G$ and set $T$ of its vertices called terminals, such that $|T|=z$, and $T$ is $\alpha$-well-linked in $G$, for some $0<\alpha<1$. The algorithm computes, for every terminal $t\in T$, a set $\qset_t=\set{Q_t(t')\mid t'\in T\setminus\set{t}}$ of paths, where, for all $t'\in T\setminus\set{t}$, path $Q_t(t')$ connects $t'$ to $t$. Moreover, if we select a vertex $t\in T$ uniformly at random, then, for every edge $e\in E(G)$, $\expect{\cong(\qset_t,e)}\leq  O(\log^4z/\alpha)$. 
\end{lemma}

\begin{proof}
	We use the algorithm from \Cref{cor: embed complete graph}, in order to compute an embedding $\tilde\pset$ the complete graph $K_z$ with $V(K_z)=T$ into $G$. Recall that the algorithm ensures that, with high probability, the congestion of the embedding is at most $(cz\log^4z)/\alpha$, for some constant $c$. If the congestion caused by the paths in $\tilde \pset$ is greater than this bound, then we 
	say that the algorithm from  \Cref{cor: embed complete graph} failed. 
	We repeat the algorithm from \Cref{cor: embed complete graph} $O(\log |E(G)|)$ times. Let $\event_1$ be the event that the algorithm failed in each of these applications. Then $\prob{\event_1}\leq 1/\poly(z)$. In this case, for every terminal $t\in T$, we return a set $\qset_t=\set{Q_t(t')\mid t'\in T\setminus\set{t}}$ of paths, where for every terminal $t'\in T\setminus\set{t}$, $Q_t(t')$ is an arbitrary path connecting $t$ to $t'$ in $G$. Clearly, for all $t\in T$, for every edge $e\in E(G)$, $\cong_G(\qset_t,e)\leq z$.
	
	We assume from now on that, in some application of the algorithm from \Cref{cor: embed complete graph}, it returned a set $\tilde \pset$ of paths with  $\cong_G(\tilde \pset)\leq O((z\log^4z)/\alpha)$.
	
	We now fix a terminal $t\in T$, and define the corresponding set $\qset_t=\set{Q_t(t')\mid t'\in T\setminus\set{t}}$ of paths. For every terminal $ t'\in T\setminus\set{t}$, we let $Q_t(t')$ be the unique path in set $\tilde \pset$ that serves as the embedding of the edge $(t,t')\in E(K_z)$. Clearly, path $Q_t(t')$ connects $t'$ to $t$ as required.
	
	Consider now an edge $e$, and let $\eta_e=\cong_G(\tilde \pset,e)\leq  O((z\log^4z)/\alpha)$. Since every path of $\tilde \pset$ may lie in at most two path sets of $\set{\qset_t}_{t\in T}$, we get that $\sum_{t\in T}\cong_G(\qset_t,e)\leq 2\eta_e $. Therefore, if Event $\event_1$ did not happen, and a terminal $t\in T$ is selected uniformly at random, then
	$\expect{\cong(\qset_t,e)}\leq  2\eta_e/z\leq O(\log^4z/\alpha)$. Overall, for every edge $e\in E(G)$, $\expect{\cong(\qset_t,e)}\leq \expect{\cong(\qset_t,e)\mid \neg\event_1}+\expect{\cong(\qset_t,e)\mid \event_1}\cdot \prob{\event_1}\leq  O(\log^4z/\alpha)+O(1/z)\leq O(\log^4z/\alpha)$.
\end{proof}

The following corollary allows us to compute a distribution over internal $C$-routers for a cluster $C$ of a graph $G$, such that the expected congestion on every edge of $C$ is small. The corollary follows immediately by applying the algorithm from \Cref{lem: simple guiding paths} to the augmentation $C^+$ of the cluster $C$ in graph $G$. The proof of the corollary is omitted.

\begin{corollary}\label{cor: simple guiding paths}
	There is an efficient randomized algorithm, whose input is a graph $G$ and a cluster $C$ of $G$ that has the $\alpha$-bandwidth property for some $0<\alpha<1$. The algorithm  returns (explicitly) a distribution $\dset$ over the set $\Lambda(C)$ of internal $C$-routers, such that, for every edge $e\in E(C)$, $\expect[\qset\sim \dset]{\cong(\qset,e)}\leq  O((\log |\delta_G(C)|)^4/\alpha)$. 
\end{corollary}


\subsection{Curves in the Plane or on a Sphere}

\subsubsection{Reordering Curves}

Assume that we are given two oriented orderings $(\oset,b), (\oset',b')$ on a set $U=\set{u_1,\ldots,u_r}$ of elements. Assume for simplicity that $b=b'=1$ (otherwise the corresponding ordering can be flipped).  Consider a disc $D$, with a collection $\set{p_1,\ldots,p_r}$ of distinct points appearing on the boundary of $D$ (we will view each point $p_i$ as representing element $u_i$ of $U$), such that the order in which these points are encountered, as we traverse the boundary of $D$ in the counter-clock-wise direction, is precisely $\oset$. Let $D'\subseteq D$ be another disc that is contained in $D$, whose boundary is disjont from the boundary of $D$. Assume that a collection $\set{p'_1,\ldots,p'_r}$ of points appear on the boundary of $D'$, and that the order in which these points are encountered as we traverse the boundary of $D'$ in the counter-clock-wise direction is precisely $\oset'$. As before, for each $1\leq i\leq r$, we view point $p'_i$ as representing element $u_i\in U$. 
We now define reordering curves between the oriented orderings  $(\oset,b)$ and $(\oset',b')$, which are then used in order to define the distance between the two orderings.

\begin{definition}[Reordering curves]
We say that a collection $\Gamma=\set{\gamma_1,\ldots,\gamma_r}$ of curves is a \emph{set of reordering curves for oriented orderings $(\oset,b)$ and $(\oset',b')$} iff (i) the  curves in $\Gamma$ are in general position; (ii) each curve $\gamma_i\in \Gamma$ is simple and its interior is contained in $D\setminus D'$; and (iii) for all $1\leq i\leq r$, curve $\gamma_i$ has $p_i,p'_i$ as its endpoints. The \emph{cost} of the collection $\Gamma$ is the total number of crossings between its curves.
\end{definition}

\begin{definition}[Distance between orderings]
	Let $(\oset,b)$ and $(\oset',b')$ be two oriented orderings on a set $U$ of elements. The \emph{distance} between the two oriented orderings, denoted by $\dist((\oset,b),(\oset',b'))$, is the smallest cost of any collection $\Gamma$ of reordering curves for  $(\oset,b)$ and $(\oset,b')$. For two unoriented orderings $\oset,\oset'$ on $U$, we define $\dist(\oset,\oset')=\min_{b,b'\in \set{-1,1}}\set{\dist((\oset,b),(\oset',b'))}$.
\end{definition}

The following lemma, that follows from Section 4 of~\cite{pelsmajer2009odd} and Section 5.2 of~\cite{pelsmajer2011crossing}, provides an efficient algorithm to compute a collection of reordering curves of near-optimal cost for a given pair of oriented orderings. The proof is deferred to Section \ref{subsec: compute reordering} of Appendix. 

\begin{lemma}\label{lem: find reordering}
	There is an efficient algorithm, that, given a pair  $(\oset,b)$, $(\oset',b')$ of oriented orderings on a set $U$ of elements, computes a collection $\Gamma$ of reordering curves for $(\oset,b)$ and $(\oset',b')$, of cost at most $2\cdot\dist((\oset,b),(\oset',b'))$.
\end{lemma}

We will use the following simple corollary of the lemma, whose proof is provided Appendix~\ref{apd: Proof of find reordering}.

\begin{corollary}
	\label{lem: ordering modification}
	There is an efficient algorithm, whose input is a graph $G$, a drawing $\phi$ of $G$ in the plane, a vertex $v\in V(G)$, and a circular ordering $\oset_v$ of the edges of $\delta_G(v)$. Let $\oset'_v$ be the circular order in which the edges of $\delta_G(v)$ enter the image of $v$ in $\phi$, and let $D=D_{\phi}(v)$ be a tiny $v$-disc. The algorithm produces a new drawing $\phi'$ of $G$, with $\cro(\phi')\leq \cro(\phi)+2\cdot\dist(\oset_v,\oset'_v)$, such that the following hold:
	
	\begin{itemize}
		\item the images of the edges of $\delta_G(v)$ enter the image of $v$ in the order $\oset_v$ in $\phi'$; and
		
		\item the drawings $\phi$ and $\phi'$ are identical, except that, for each edge $e\in \delta_G(v)$, the segment of the image of $e$ lying inside the disc $D$ may be different in the two drawings.
	\end{itemize}
\end{corollary}

\subsubsection{Type-1 Uncrossing of Curves}
\label{subsec: uncrossing type 1}

In this subsection we consider a set $\Gamma$ of curves in the plane (or on a sphere) that are in general position, and provide a simple operation, called \emph{type-$1$ uncrossing}, whose goal is to ``simplify'' this collection of curves by eliminating some of the crossings between them. Specifically, we modify the curves in $\Gamma$, without changing their endpoints, to ensure that every pair of curves cross at most once. We now describe the type-1 uncrossing operation formally.

Let $\Gamma$ be a set of simple curves in the plane that are in general position. 
For a pair $\Gamma_1,\Gamma_2$ of disjoint subsets of $\Gamma$, we denote by $\chi(\Gamma_1,\Gamma_2)$ the total number of crossings between the curves in $\Gamma_1$ and the curves in $\Gamma_2$. In other words, $\chi(\Gamma_1,\Gamma_2)$ is the number of points $p$, such that $p$ lies on a curve in $\Gamma_1$ and on a curve in $\Gamma_2$, and $p$ is not an endpoint of these curves. If $\Gamma_1=\set{\gamma}$, then we use the shorthand $\chi(\gamma,\Gamma_2)$ instead of $\chi(\set{\gamma},\Gamma_2)$. 

\begin{figure}[h]
	\centering
	\subfigure[Before: Curves $\gamma$ and $\gamma'$ cross twice,  at points $p$ and $q$. The crossing points of both curves with the third curve are circled as well.]{\scalebox{0.12}{\includegraphics{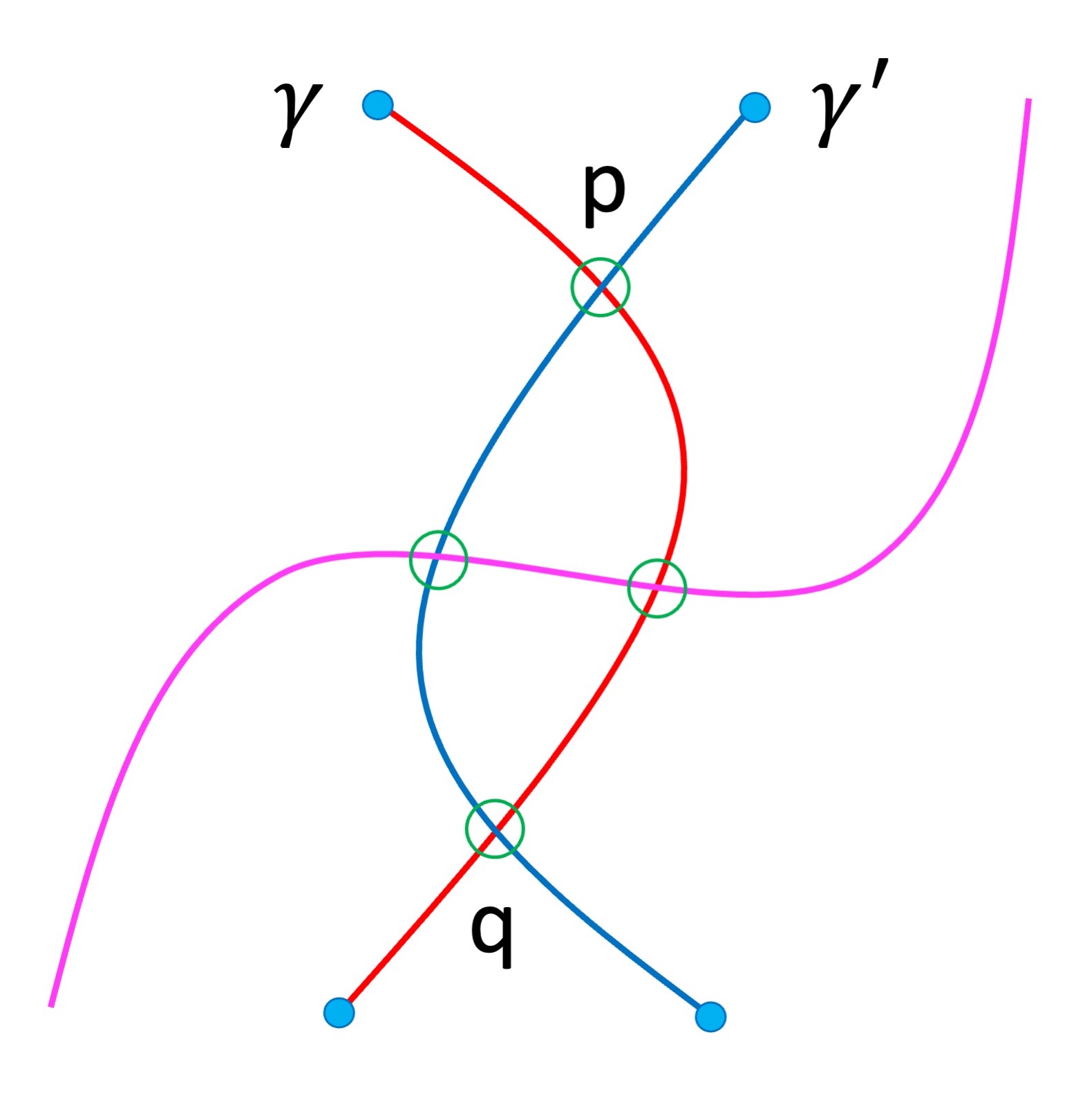}}}
	\hspace{0.8cm}
	\subfigure[After: Each of the new curves $\gamma$ and $\gamma'$ has same endpoints as before. The two curves no longer cross each other, and the pink curve still participates in two crossings with $\gamma$ and $\gamma'$.]{
		\scalebox{0.12}{\includegraphics{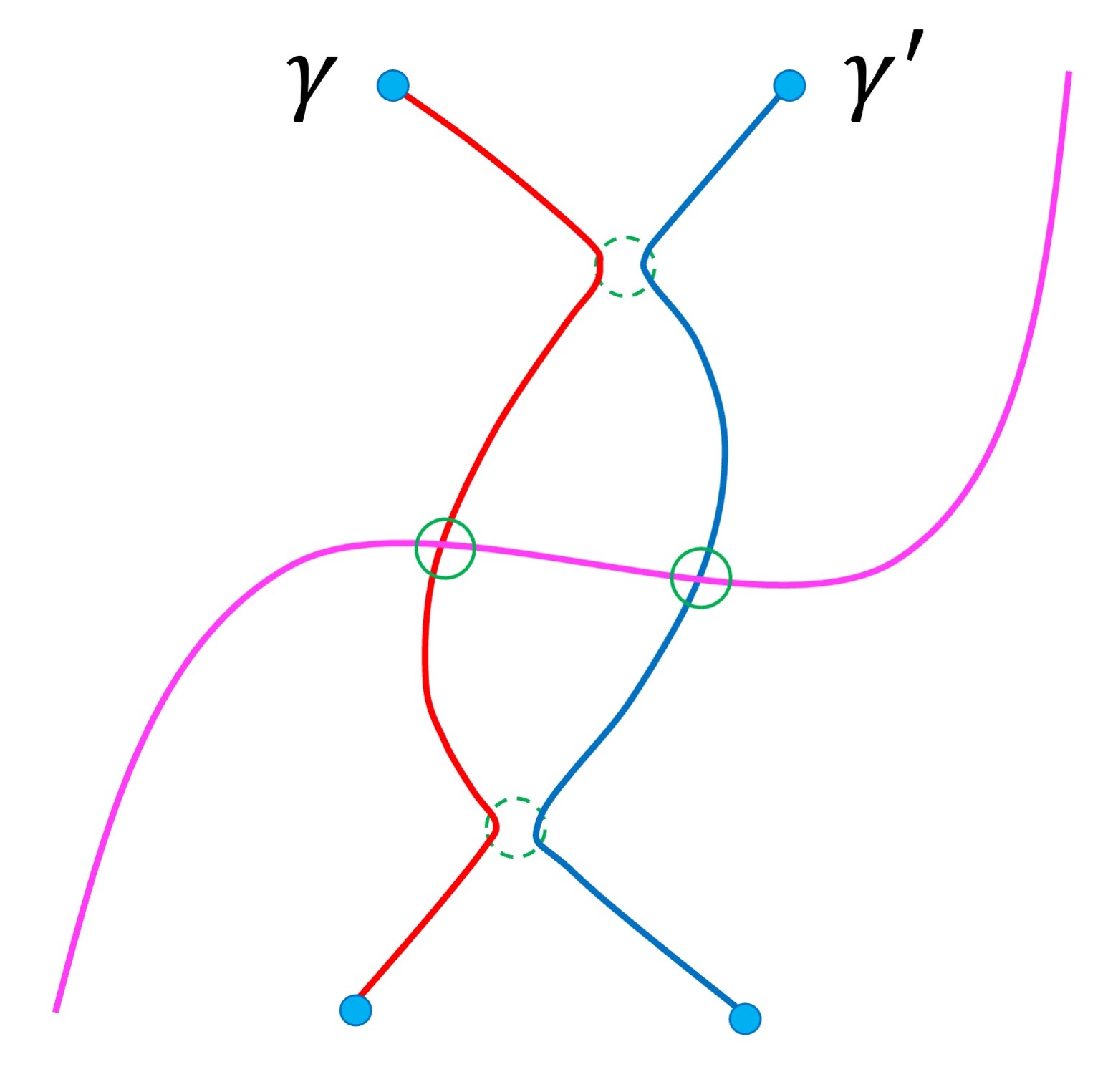}}}
	\caption{Type-1 uncrossing operation.}\label{fig:type_1_uncrossing}
\end{figure}

The type-1 uncrossing operation iteratively considers pairs  $\gamma,\gamma'\in \Gamma$ of distinct curves that cross more than once, and then locally modifies them, as shown in Figure~\ref{fig:type_1_uncrossing}, to eliminate two crossings. This operation ensures that no new crossings are created, and preserves the endpoints of both curves. The following theorem summarizes this operation. The proof of the theorem is standard and is deferred to Section \ref{apd: type-1 uncrossing} of Appendix for completeness.

\begin{theorem}[Type-1 Uncrossing]
	\label{thm: type-1 uncrossing}
	There is an algorithm, that, given a set $\Gamma$ of simple curves in general position, that are partitioned into two disjoint subsets $\Gamma_1,\Gamma_2$, computes, for each curve $\gamma\in \Gamma_1$, a simple curve $\gamma'$ that has same endpoints as $\gamma$, such that, if we denote by $\Gamma_1'=\set{\gamma'\mid \gamma\in \Gamma_1}$, then the following hold:
	\begin{itemize}
		\item the curves  in set $\Gamma'_1\cup \Gamma_2$ are in general position;
		\item every pair of distinct curves in $\Gamma_1'$ cross  at most once;
		\item for every curve $\gamma\in \Gamma_2$, $\chi(\gamma,\Gamma'_1)\le \chi(\gamma,\Gamma_1)$; and
		\item the total number of crossings between the curves of $\Gamma_1'\cup \Gamma_2$ is bounded by the total number of crossings between the curves of $\Gamma$.
	\end{itemize}
The running time of the algorithm is bounded by $\poly(n\cdot N)$, where $n$ is the number of bits in the representation of the set $\Gamma$ of curves, and $N$ is the number of crossing points between the curves of $\Gamma$.
\end{theorem}

\subsubsection{Curves in a Disc and Nudging of Curves}
\label{sec: curves in a disc}

Suppose we are given a disc $D$, and a collection $\set{s_1,t_1,\ldots,s_k,t_k}$ of distinct points on its boundary. For all $1\leq i< j\leq k$, we say that the two pairs  $(s_i,t_i),(s_j,t_j)$ of points \emph{cross} iff the unoriented circular ordering of the points $s_i,s_j,t_i,t_j$ on the boundary of $D$ is $(s_i,s_j,t_i,t_j)$.
We use the following simple claim, whose proof is deferred to \Cref{apd: Proof of curves in a disc}.

\begin{claim}\label{claim: curves in a disc}
	There is an efficient algorithm that,  given a disc $D$, and a collection $\set{s_1,t_1,\ldots,s_k,t_k}$ of distinct points on the boundary of  $D$, computes a collection $\Gamma=\set{\gamma_1,\ldots,\gamma_k}$ of curves, such that, for all $1\leq i\leq k$, curve $\gamma_i$ has $s_i$ and $t_i$ as its endpoints, and its interior is contained in the interior of $D$. Moreover, for every pair $1\leq i<j\leq k$ of indices, if the two pairs  $(s_i,t_i),(s_j,t_j)$ of points cross then curves $\gamma_i,\gamma_j$ intersect at exactly one point; otherwise, curves $\gamma_i,\gamma_j$ do not intersect. Lastly, every point in the interior of $D$ may be contained in at most two curves of $\Gamma$.
\end{claim}

\paragraph{Nudging Procedure.}
In a nudging procedure, we are given an instance $I=(G,\Sigma)$ of \cnwrs, a subset $U$ of vertices of $G$, and a collection $\pset$ of edge-disjoint paths, such that, for every path $P\in \pset$, all inner vertices of $P$ lie in $U$, and the endpoints of $P$ do not lie in $U$. Additionally, we are given some solution $\phi$ to instance $I$. For every path $P\in \pset$, we denote by $\gamma(P)$ the image of path $P$ in $\phi$, that is, $\gamma(P)$ is the concatenation of the images of all edges of $P$. Notice that the resulting collection $\Gamma=\set{\gamma(P)\mid P\in \pset}$ may not be in general position. This is since some vertices $u\in U$ may lie on more than $2$ paths in $\pset$, and in such a case more than $2$ curves in $\Gamma$ contain the point $\phi(u)$. The purpose of the nudging procedure is to slightly modify the curves in $\Gamma$ in the viccinity of the images of such vertices to ensure that the resulting collection of curves $\Gamma'=\set{\gamma'(P)\mid P\in \pset}$ is in general position, while introducing relatively few crossings. Additionally, the procedure ensures that, for every path $P\in \pset$, the endpoints of the new curve $\gamma'(P)$ are identical to those of the original curve $\gamma(P)$. 

We start by letting, for every path $P$, curve $\gamma'(P)$ be the original curve $\gamma(P)$, and we set $\Gamma'=\set{\gamma'(P)\mid P\in \pset}$. We then process every vertex $u\in U$ one by one. Consider an iteration when any such vertex $u$ is processed. Let $\pset(u)\subseteq \pset$ be a set of all paths $P\in \pset$ with $u\in V(P)$.  We denote $\pset(u)=\set{P_1,\ldots,P_k}$. Consider the tiny $u$-disc $D(u)=D_{\phi}(u)$ in the drawing $\phi$ of graph $G$. For all $1\leq i\le k$, we let $s_i,t_i$ be the two points at which curve $\gamma'(P_i)$ intersects the boundary of the disc $D(u)$. Note that all points $s_1,t_1,\ldots,s_k,t_k$ must be distinct, as the paths in $\pset$ are edge-disjoint. We use the algorithm from \Cref{claim: curves in a disc} in order to construct a collection $\set{\sigma_1,\ldots,\sigma_k}$ of curves, such that, for all $1\leq i\leq k$, curve $\sigma_i$ has $s_i$ and $t_i$ as its endpoints, and is completely contained in $D(u)$. Recall that the claim ensures that, for every pair $1\leq i<j\leq k$ of indices, if the two pairs  $(s_i,t_i),(s_j,t_j)$ of points cross, then curves $\sigma_i,\sigma_j$ intersect at exactly one point; otherwise, curves $\sigma_i,\sigma_j$ do not intersect. The former may only happen if paths $P_i,P_j$ have a transversal intersection at vertex $u$.
For all $1\leq i\leq k$, we modify the curve $\gamma'(P_i)$ as follows: we replace the segment of the curve between points $s_i,t_i$ with the curve $\sigma_i$. 
Once every vertex of $U$ is processed, we obtain the final collection $\Gamma'=\set{\gamma'(P)\mid P\in \pset}$ of curves. From the above discussion, we get the following observation.

\begin{observation}\label{obs: nudging summary}
	The set $\Gamma'=\set{\gamma'(P)\mid P\in \pset}$ of curves is in general position, and, for every path $P\in \pset$, the endpoints of curve $\gamma'(P)$ are identical to the endpoints of curve $\gamma(P)$. Moreover, if $\chi$ denotes the set of all crossings $(e,e')_p$ in $\phi$, where $e$ and $e'$ are edges of $\bigcup_{P\in \pset}E(P)$, then the number of crossings between the curves of $\Gamma'$ is bounded by $|\chi|+|\Pi^T(\pset)|$. Lastly, if the paths in $\pset$ are non-transversal with respect to $\Sigma$, then for every path $P\in \pset$, the number of crossings between $\gamma'(P)$ and $\Gamma'\setminus\set{\gamma'(P)}$ is bounded by the number of crossings $(e,e')_p$ in $\phi$ where exactly one of the edges $e,e'$ belongs to $P$.
\end{observation}

\subsubsection{Type-2 Uncrossing of Curves}
\label{subsec: type-2 uncrossing}

In this subsection we provide another subroutine, called \emph{type-2 uncrossing of curves}, that allows us to simplify a given set $\Gamma$ of curves by removing some of the crossings between them. Unlike the type-1 uncrossing operation, we no longer preserve the endpoints of every curve, but we ensure that the multisets containing the endpoints of the curves are preserved under this operation.

 It will sometimes be useful for us to assign a \emph{direction} to a curve $\gamma$, by designating one of its endpoints, that we denote by $s(\gamma)$, as its first endpoint,  and the other endpoint, denoted by $t(\gamma)$, as its last endpoint. If $\Gamma$ is a collection of curves, and each curve in $\Gamma$ is assigned a direction, then we say that $\Gamma$ is a collection of \emph{directed curves}. In such a case, we let $S(\Gamma)$ be the multiset of points containing the first endpoint of every curve in $\Gamma$, and we let $T(\Gamma)$ be the multiset of points containing the last endpoint of every curve in $\Gamma$.

For the type-2 uncrossing operation, we will consider curves that arise from some drawing $\phi$ of a graph $G$. We first need to define curves that are aligned with a graph drawing. For intuition, consider first some planar graph $G$, and its planar drawing $\phi$. In this case, curve $\gamma$ is aligned with the drawing $\phi$ of $G$, if there is some path $P$ in $G$, such that $\gamma$ can be obtained by first concatenating the images of all edges if $P$, and then possibly modifying the resulting curve within tiny discs $D_{\phi}(v)$ for vertices $v\in V(P)$ (typically via a nudging operation). If $\phi$ is a non-planar drawing of some graph $G$, then the definition of a curve $\gamma$ being aligned with the drawing is similar, but now we allow the curve $\gamma$ to ``switch'' from the image of one edge to another, at a crossing point between the two edges. Therefore, we can define a sequence  $e_1,e_2,\ldots,e_{r-1}$ of edges of $G$, such that the curve ``follows'' segments of these edges. The curve $\gamma$ itself can then be partitioned into segments $\sigma_1,\sigma_1',\sigma_2,\sigma_2',\ldots,\sigma'_{r-1},\sigma_r$, where for all $1\leq i\leq r-1$, $\sigma_i'$ is a contiguous segment of the image of edge $e_i$. For a pair $\sigma'_i,\sigma'_{i+1}$ of such segments, either the last endpoint of $\sigma'_i$ and the first endpoint of $\sigma_{i+1}'$ are identical (and it is a crossing point between the images of $e_i$ and $e_{i+1}$); or segment $\sigma_{i+1}$ is contained in disc $D_{\phi}(v_{i+1})$, where $v_{i+1}$ is a common endpoint of $e_i$ and $e_{i+1}$. With this intuition in mind, we now define the notion of alignment of a curve with a drawing of a graph.

\begin{definition}[Curve aligned with a drawing of a graph]
	Let $G$ be a graph and $\phi$ a drawing of $G$ in the plane.  We say that a curve $\gamma$ is \emph{aligned} with the drawing $\phi$ of $G$ if there is a sequence $(e_1,e_2,\ldots,e_{r-1})$ of edges of $G$, and a partition $(\sigma_1,\sigma_1',\sigma_2,\sigma_2',\ldots,\sigma'_{r-1},\sigma_r)$ of $\gamma$ into consecutive segments, such that, if we denote, for all $1\leq i< r$, $e_i=(v_i,v_{i+1})$, then the following hold:
	
	\begin{itemize}
		\item for all $1\leq i\leq r-1$, $\sigma'_i$ is a contiguous segment of non-zero length of $\phi(e_i)$, and it is disjoint from all discs in $\set{D_{\phi}(u)}_{u\in V(G)}$, except that its first endpoint may lie on the boundary of $D_{\phi}(v_i)$, and its last endpoint may lie on the boundary of $D_{\phi}(v_{i+1})$;
		
		\item for all $1\leq i\leq r$, segment $\sigma_i$ is either contained in disc $D_{\phi}(v_i)$, or it is contained in a tiny $p$-disc $D(p)$, where $p$ is a crossing point of $\phi(e_{i-1})$ and $\phi(e_{i})$; 
		
		\item $\sigma_1=\phi(e_1)\cap D_{\phi}(v_1)$;  and 
		
		\item $\sigma_r=\phi(e_{r-1})\cap D_{\phi}(v_r)$.
	\end{itemize}
\end{definition}

In order to perform a type-2 uncrossing operation, we consider a graph $G$, a drawing $\phi$ of $G$, and a set $\qset$ of simple directed paths in $G$. 
We assume that no vertex of $G$ may serve simultaneously as an endpoint of a path of $\qset$ and an inner vertex of some other path of $\qset$.
We can then define a set $\Gamma=\set{\gamma(Q)\mid Q\in \qset}$ of curves, where, for every path $Q\in \qset$, curve $\gamma(Q)$ is obtained by concatenating the images of the edges of $Q$. Note however that the curves in the resulting set $\Gamma$ are not necessarily in general position. Type-2 uncrossing allows us to fix this, and moreover to eliminate all crossings between the resulting set $\Gamma'$ of curves. Unlike type-1 uncrossing, we only guarantee that the multisets containing the first and last endpoints of the curves in $\Gamma'$ remain identical to those corresponding to $\Gamma$, but we no longer guarantee that they are matched in the same way to each other. For technical reasons, we need to consider two different settings for the type-2 uncrossing: one where the paths in set $\qset$ are edge-disjoint, in which case we can provide somewhat stronger guarantees, and another where this is not the case. These two settings for type-2 uncrossing are provided in the following theorem and its corollary, whose proofs are simple and are deferred to Sections \ref{apd: new type 2 uncrossing} and \ref{apd: cor new type 2 uncrossing} of Appendix, respectively. We start with the setting where the paths in set $\qset$ are edge-disjoint.

\begin{theorem}
	\label{thm: new type 2 uncrossing}
	There is an efficient algorithm, whose input consists of a graph $G$, a drawing $\phi$ of $G$ on the sphere, and a collection $\qset$ of edge-disjoint paths in $G$, such that no vertex of $G$ may simultaneously serve as an endpoint of some path in $\qset$ and an inner vertex of some path in $\qset$. Additionally, for each path $Q\in \qset$, one of its endpoints is designated as its first endpoint and is denoted by $s(Q)$, and the other endpoint is designated as its last endpoint and denoted by $t(Q)$. 
	The algorithm computes a set $\Gamma=\set{\gamma(Q)\mid Q\in \qset}$ of directed simple curves on the sphere with the following properties:
	\begin{itemize}
		
		\item every curve $\gamma(Q)\in \Gamma$ is aligned with the drawing of the graph $\bigcup_{Q'\in \qset}Q'$ induced by $\phi$;
		
		\item for each path $Q\in \qset$, $s(\gamma(Q))=\phi(s(Q))$; moreover, if $e_1(Q)$ is the first edge of $Q$, then curve $\gamma(Q)$ contains the segment $\phi(e_1(Q))\cap D_{\phi}(s(Q))$;

		\item the multiset $T(\Gamma)$, containing the last endpoint of every curve in $\Gamma$, is precisely the multiset $\set{\phi(t(Q))\mid Q\in \qset}$, containing the image of the last vertex on every path of $\qset$ in $\phi$; and
		\item the curves in $\Gamma$ do not cross each other.
	\end{itemize}

\end{theorem}	

We emphasize that the curves in $\Gamma$ may match the mutisets $\set{\phi(s(Q))\mid Q\in \qset}$ and $\set{\phi(t(Q))\mid Q\in \qset}$ differently from the paths in $\qset$.

We will sometimes use \Cref{thm: new type 2 uncrossing} in a setting where we are additionally given a subgraph $C\subseteq G$, and the paths of $\qset$ are internally disjoint from $C$. In such a case, from the definition of aligned curves, and from the fact that the curves of $\Gamma$ do not cross each other, for every edge $e\in E(C)$, the number of crossings between $\phi(e)$ and the curves in $\Gamma$ is bounded by the number of crossings between $\phi(e)$ and the curves of $\set{\phi(e')\mid e'\in \bigcup_{Q\in \qset}E(Q)}$. 

We use the following corollary of  \Cref{thm: new type 2 uncrossing}, that deals with the setting where paths in set $\qset$ may share edges. The proof is deferred to Section \ref{apd: cor new type 2 uncrossing} of Appendix.

\begin{corollary}
	\label{cor: new type 2 uncrossing}
There is an efficient algorithm, whose input consists of a graph $G$, a drawing $\phi$ of $G$ on the sphere,  a subgraph $C$ of $G$, and a collection $\qset$ of paths in $G$, that are internally disjoint from $C$, such that no vertex of $G$ may simultaneously serve as an endpoint of some path in $\qset$ and an inner vertex of some path in $\qset$. Additionally, for each path $Q\in \qset$, one of its endpoints is designated as its first endpoint and is denoted by $s(Q)$, and the other endpoint is designated as its last endpoint and is denoted by $t(Q)$. 
The algorithm computes a set $\Gamma=\set{\gamma(Q)\mid Q\in \qset}$ of directed simple curves on the sphere with the following properties:
\begin{itemize}

	\item for every path $Q\in \qset$, $s(\gamma(Q))=\phi(s(Q))$;
	
	\item the multiset $T(\Gamma)$, containing the last endpoint of every curve in $\Gamma$, is precisely the multiset $\set{\phi(t(Q))\mid Q\in \qset}$, containing the image of the last vertex on every path of $\qset$; 
	\item the curves in $\Gamma$ do not cross each other; and
	
	\item for each edge $e\in E(C)$, the number of crossings between $\phi(e)$ and the curves in $\Gamma$ is bounded by 
	$\sum_{e'\in E(G)\setminus E(C)}\chi(e,e')\cdot \cong_G(\qset,e')$, where $\chi(e,e')$ is the number of crossings between $\phi(e)$ and $\phi(e')$.
\end{itemize}
\end{corollary}


\subsection{Contracted Graphs} 
Let $G$ be a graph and let $\cset$ be a collection of disjoint clusters of $G$. We define the \emph{contracted graph} $G_{|\cset}$ to be the graph obtained from $G$ by contracting each cluster $C\in \cset$ into a supernode $v_C$; we remove self-loops but keep parallel edges. Note that every edge of graph $G_{|\cset}$ corresponds to some edge of graph $G$. We do not distinguish between such edges, so $E(G_{|\cset})\subseteq E(G)$. We refer to vertices of $G_{|\cset}$ that are not supernodes as \emph{regular} vertices.
In the following claim, we derive well-linkedness properties of a set $T$ of vertices in a graph $G$ from well-linkedness of $T$ in a contracted graph $G_{|\cset}$ and bandwidth properties of the clusters of $\cset$ in $G$. The proof is deferred to Section \ref{apd: Proof of contracted_graph_well_linkedness} of Appendix.

\begin{claim}
	\label{clm: contracted_graph_well_linkedness}
	Let $G=(V,E)$ be a graph, $T\subseteq V$ a subset of its vertices, and $\cset$ a collection of disjoint clusters of $G$, such that $T\cap (\bigcup_{C\in \cset}V(C))=\emptyset$. Assume that each cluster $C\in \cset$ has the $\alpha_1$-bandwidth property in $G$, and that the set $T$ of vertices is $\alpha_2$-well-linked in the contracted graph $G_{|\cset}$, for some parameters $0<\alpha_1,\alpha_2<1$. Then $T$ is $(\alpha_1\cdot \alpha_2)$-well-linked in $G$. 
\end{claim}

The following corollary of \Cref{clm: contracted_graph_well_linkedness} essentially replaces the well-linkedness property of the set $T$ of vertices with the equivalent bandwidth property of a cluster of a given graph $G$.
The proof is deferred to Section \ref{apd: Proof of cor contracted_graph_well_linkedness} of Appendix.

\begin{corollary}
	\label{cor: contracted_graph_well_linkedness}
	Let $G$ be a graph, and let $R$ be a cluster of $G$. Let $\cset$ be a collection of disjoint clusters of $R$, such that every cluster $C\in \cset$ has the $\alpha_1$-bandwidth property in graph $G$, for some parameter $0<\alpha_1<1$.  Denote $\hat R=R_{\mid\cset}$ and $\hat G=G_{\mid\cset}$, and assume further that $\hat R$ has the $\alpha_2$-bandwidth property in graph $\hat G$, for some $0<\alpha_2<1$. Then cluster $R$ has the $(\alpha_1\cdot\alpha_2)$-bandwidth property in graph $G$. 
\end{corollary}

 The following simple claim allows us to transform a routing in a contracted graph $G_{|\cset}$ into a routing in the original graph $G$. The proof appears in Section \ref{apx: contracted graph routing} of Appendix.
 
\begin{claim}
	\label{claim: routing in contracted graph}
	There is an efficient algorithm, that takes as input a graph  $G$, a set $\cset$ of disjoint clusters of $G$, such that each cluster $C\in \cset$ has the $\alpha$-bandwidth property in $G$ for some $0<\alpha<1$, and a collection $\pset$ of edge-disjoint paths in the contracted graph $G_{|\cset}$, routing some set $T\subseteq V(G)\cap V(G_{|\cset})$ of vertices to some vertex $x\in V(G)\cap V(G_{|\cset})$. The algorithm produces a collection $\pset'$ of paths in graph $G$, routing the vertices of $T$ to vertex $x$, such that, for each edge $e\in E(G)\setminus \left( \bigcup_{C\in \cset}E(C)\right )$, $\cong_G(\pset',e)\le 1$, and for each edge $e\in \bigcup_{C\in \cset}E(C)$, $\cong_{G}(\pset',e)\leq \ceil{1/\alpha}$. Additionally, the algorithm produces another set $\pset''$ of edge-disjoint paths in graph $G$, of cardinality at least $\alpha \cdot |T|/2$, routing a subset $T'\subseteq T$ of vertices to $x$. 
\end{claim}

The following claim  allows us to bound the crossing number of a contracted graph. The proof is provided in Section \ref{apd: Proof of crossings in contr graph} of the Appendix.

\begin{claim}\label{lem: crossings in contr graph}
	Let $I=(G,\Sigma)$ be an instance of the \cnwrs problem, and let $\cset$ be a collection of disjoint clusters of $G$, such that each cluster in $\cset$ has the $\alpha$-bandwidth property, for some $0<\alpha<1$. Then there is a drawing $\phi$ of the contracted graph $G_{|\cset}$, with $\cro(\phi)\leq O(\optcrors(I)\cdot \log^8m/\alpha^2)$, where $m=|E(G)|$. Moreover, for every regular vertex $x\in V(G_{|\cset})\cap V(G)$, the ordering of the edges of $\delta_G(x)$ as they enter $x$ in $\phi$ is consistent with the rotation $\oset_x\in \Sigma$.
\end{claim}



\section{First Set of Tools: Light Clusters, Bad Clusters, Path-Guided Orderings, and Basic Cluster Disengagement}
\label{sec: guiding paths orderings basic disengagement}

The main goal of this section is to define and analyze the \BCD procedure. Along the way we will define several other central tools that we use throughout the paper, such as light clusters, bad clusters, and path-guided orderings. 

We start by defining and analyzing laminar family-based disengagement procedure, which will serve as the basis of the basic disengagement procedure.


\subsection{Laminar Family-Based Disengagment}

\label{subsec: laminar-based decomposition}

We start by defining a laminar family of clusters and its associated partitioning tree.

\subsubsection{Laminar Family of Clusters and Partitioning Tree}
\label{subsubsec: laminar}

\begin{definition}[Laminar family of clusters]
Let $G$ be a graph, and let $\lset$ be a family of clusters of $G$. We say that $\lset$ is a \emph{laminar family}, if $G\in \lset$, and additionally, for all $S,S'\in \lset$, either $S\cap S'=\emptyset$, or $S\subseteq S'$, or $S'\subseteq S$ holds.
\end{definition}

Given a laminar family $\lset$ of clusters of $G$, we associate a \emph{paritioning tree} $\tau(\lset)$ with it, that is defined as follows. The vertex set of the tree is $\set{v(S)\mid S\in \lset}$; for every cluster $S\in \lset$, we view vertex $v(S)$ as representing the cluster $S$. The root of the tree is $v(G)$ -- the vertex associated with the graph $G$ itself. 
In order to define the edge set, consider a pair $S,S'\in \lset$ of clusters. If $S\subsetneq S'$, and there is no other cluster $S''\in \lset$ with $S\subsetneq S''\subsetneq S'$, then we add an edge $(v(S),v(S'))$ to the tree $\tau(\lset)$; vertex $v(S)$ becomes  a child vertex of $v(S')$ in the tree. We also say that cluster $S$ is a \emph{child cluster} of cluster $S'$, and cluster $S'$ is the \emph{parent cluster} of $S$. 
Similarly, we define an ancestor-descendant relation between clusters in a natural way: cluster $S\in \lset$ is a descendant-cluster of a cluster $S'\in \lset$ if vertex $v(S')$ lies on the unique path connecting $v(S)$ to $v(G)$ in the tree $\tau(\lset)$. If $S$ is a descendant-cluster of $S'$, then $S'$ is an ancestor-cluster of $S$. Notice that every cluster is its own ancestor and its own descendant.

The \emph{depth} of the laminar family $\lset$ of clusters, denoted by $\dep(\lset)$, is the length of the longest root-to-leaf path in tree $\tau(\lset)$. We also say that cluster $S$ \emph{lies at level $i$ of the laminar family $\lset$} iff the distance from $v(S)$ to the root of the tree $\tau(\lset)$ is exactly $i$.


\subsubsection{Definition of Laminar Family-Based Disengagement}
\label{subsubsec: laminar based disengagment def}

The input to the Laminar Family-Based Disengegement is an instance $I=(G,\Sigma)$ of \cnwrs, a laminar family $\lset$ of clusters of $G$, and, for every cluster $C\in \lset$, a circular ordering $\oset(C)$ of the edges of $\delta_G(C)$ (for $C=G$, $\delta_G(C)=\emptyset$ and the ordering $\oset(C)$ is trivial). The output of the procedure is a collection $\iset=\set{I_C=(G_C,\Sigma_C)\mid C\in \lset}$ of subinstances of $I$, that are defined as follows.

\begin{figure}[h]
	\centering
	\subfigure[Layout of graph $G$ with respect to $C$.]{\scalebox{0.13}{\includegraphics{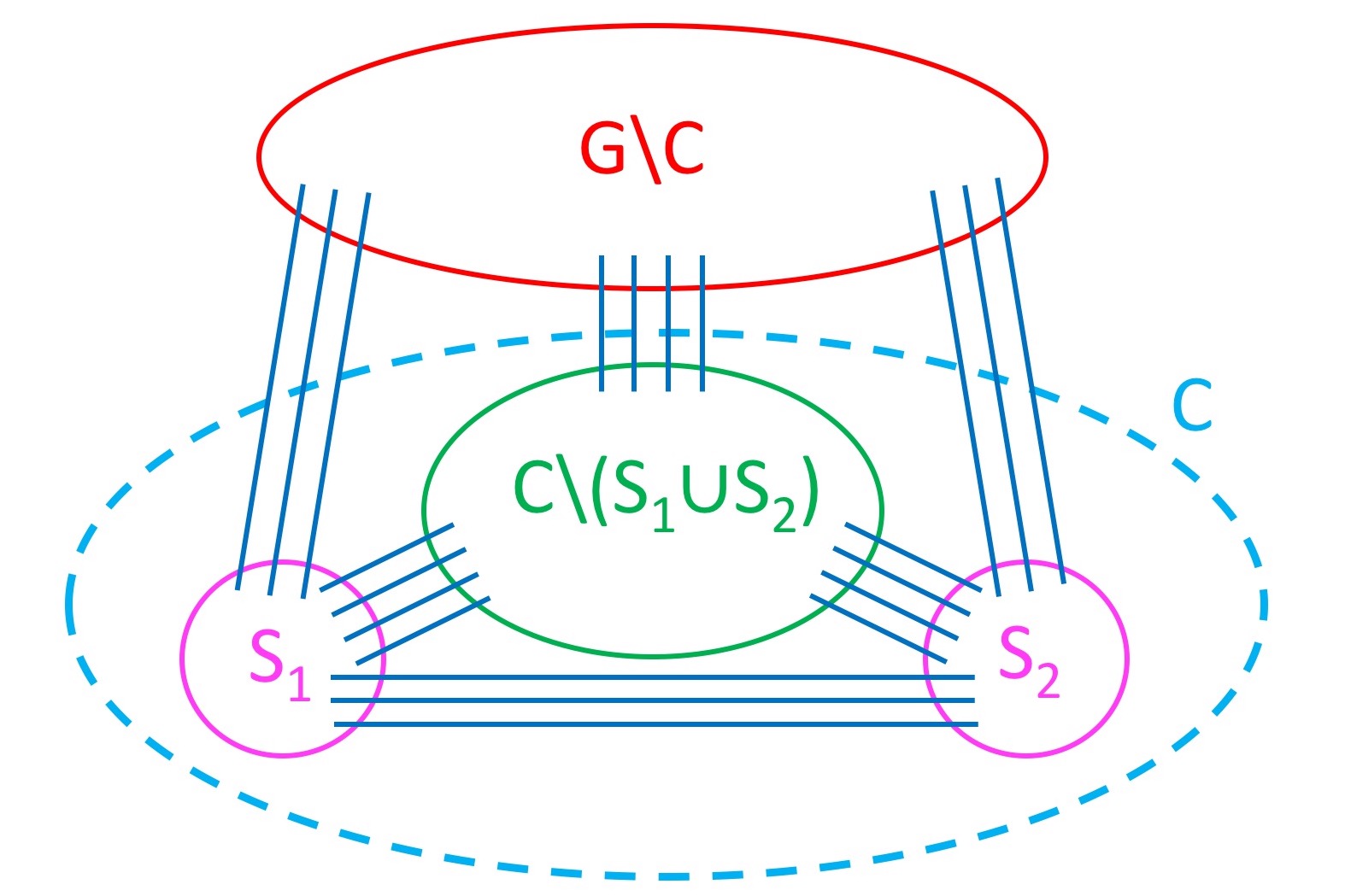}}\label{fig: graph G}}
	\hspace{0cm}
	\subfigure[Graph $G_C$.]{
		\scalebox{0.14}{\includegraphics{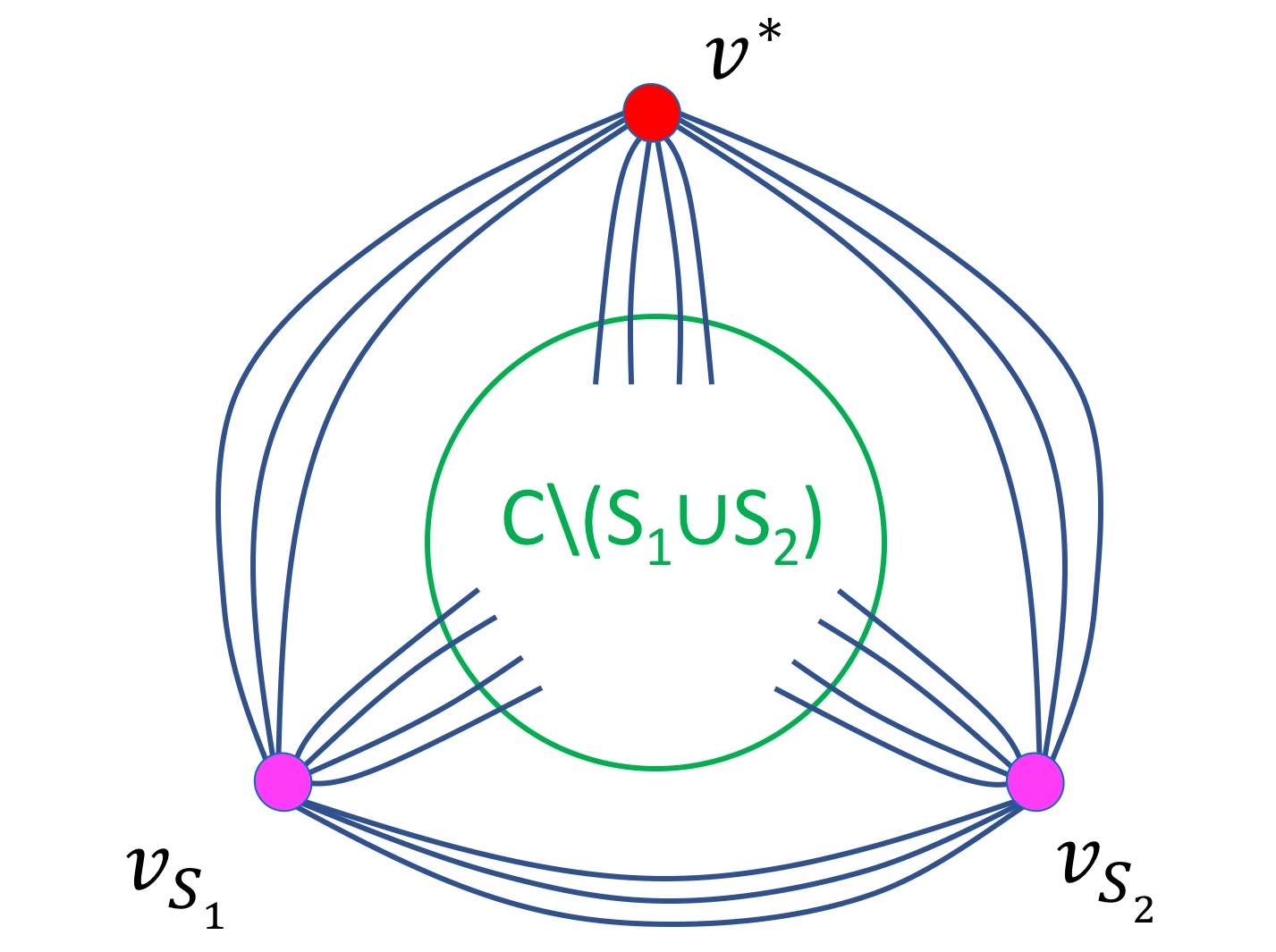}}\label{fig: C and disc}}
	\caption{Construction of graph $G_C$, where $C\in \lset$ is a cluster with two child-clusters $S_1,S_2$.\label{fig: disengaged_instance}}
\end{figure}

Consider a cluster $C\in \lset$, and denote by $\wset(C)\subseteq \lset$ the set of all child-clusters of $C$. In order to construct the graph $G_C$, we start with $G_C=G$. For every cluster $C'\in \wset(C)$, we contract the vertices of $C'$ into a supernode $v_{C'}$. Additionally, if $C\neq G$, then we contract all vertices of $V(G)\setminus V(C)$ into a supernode $v^*$.  This completes the definition of the graph $G_C$ (see \Cref{fig: disengaged_instance}). We now define the rotation system $\Sigma_C$ for $G_C$.
If $C\neq G$, then the set of edges incident to $v^*$  in $G_C$  is exactly $\delta_G(C)$. We set the rotation $\oset_{v^*}\in \Sigma_C$ to be $\oset(C)$. For every cluster $C'\in \wset(C)$, the set of edges incident to $v_{C'}$ in $G_C$ is $\delta_G(C')$. We set the rotation $\oset_{v_{C'}}\in \Sigma_C$ to be $\oset(C')$. For every regular vertex $x\in V(G_C)\cap V(G)$, $\delta_{G_C}(v)=\delta_G(v)$ holds, and its rotation $\oset_v\in \Sigma_C$ remains the same as in $\Sigma$.

We refer to the resulting collection $\iset$ of clusters as \emph{disengagement of instance $I$ via the laminar family $\lset$ and the collection $\set{\oset(C)}_{C\in \lset}$ of orderings}.

\subsubsection{Analysis}

We start by showing that the total number of edges in all instances that we obtain via the laminar family-based disengagement procedure is small compared to $|E(G)|$.

\begin{lemma}\label{lem: number of edges in all disengaged instances}
	Let $I=(G,\Sigma)$ be an instance of \cnwrs, let $\lset$ be a laminar family of clusters of $G$, and let $\set{\oset(C)}_{C\in \lset}$ be a collection of orderings of the edges of $\delta_G(C)$, for every cluster $C\in \lset$. Consider the collection $\iset=\set{I_C=(G_C,\Sigma_C)\mid C\in \lset}$ of subinstances of $I$ obtained by applying the laminar family-based decomposition to instance $I$ via the laminar family $\lset$ and the orderings in  $\set{\oset(C)}_{C\in \lset}$. Then
	$\sum_{C\in \lset}|E(G_C)|\leq O(\dep(\lset)\cdot|E(G)|)$. 
\end{lemma}
\begin{proof}
	Fix an integer $1\le i\le \dep(\lset)$ and denote by $\lset_i\subseteq \lset$ the set of all clusters of $\lset$ that lie at level $i$ of the partitioning tree. 
	From the definition of the laminar family $\lset$ and the partitioning tree, all clusters in set $\lset_i$ are mutually disjoint. 
	Consider now some cluster $C\in \lset_i$, and its corresponding graph $G_C$. Note that every edge of $G_C$ corresponds to some distinct edge of cluster $C$, except for the edges incident to the supernode $v^*$. However, the number of edges incident to $v^*$ is at most $|\delta_G(C)|$. Therefore, overall, $|E(G_C)|\leq |E_G(C)|+|\delta_G(C)|$. Since all clusters in $\lset_i$ are mutually disjoint, we get that:
\[\sum_{C\in \lset_i}|E(G_C)|\leq \sum_{C\in \lset_i}(|E_G(C)|+|\delta_G(C)|)\leq  O(|E(G)|).\]
Summing over all indices $1\le i\le \dep(\lset)$, we get that $\sum_{C\in \lset}|E(G_C)|=O(\dep(\lset)\cdot|E(G)|)$.
\end{proof}

Next, we show that solutions to the instances in $\iset$ can be efficiently combined in order to obtain a solution to instance $I$ of relatively low cost. The proof is conceptually simple but somewhat technical, and is deferred to Section \ref{appx: proof of basic disengagement combining solutions} of Appendix.

\begin{lemma}\label{lem: basic disengagement combining solutions}
	There is an efficient algorithm, that takes as input an instance $I=(G,\Sigma)$ of \cnwrs, a laminar family $\lset$ of clusters of $G$, a collection $\set{\oset(C)}_{C\in \lset}$ containing an ordering of the edges of $\delta_G(C)$ for every cluster $C\in \lset$, and, for every cluster $C\in \lset$, a solution $\phi(I_C)$ to the instance $I_C\in \iset$ of \cnwrs, where $\iset=\set{I_C\mid C\in \lset}$ is the collection of subinstances of $I$ obtained via laminar family-based disengagement of $I$ via $(\lset,\set{\oset(C)}_{C\in \lset})$. The output of the algorithm is a solution to instance $I$ with cost at most  $\sum_{I_C\in \iset}\cro(\phi(I_C))$.
\end{lemma}

So far we have shown that, if $\iset$ is the collection of instances that is constructed via a laminar family-based disengagement of instance $I$, then the total number of edges in all resulting instances is at most $O(|E(G)|\cdot \dep(\lset))$, and that there is an efficient algorithm for combining the solutions to the resulting instances, in order to obtain a low-cost solution to the original instance $I$. Another highly desirable property of the family $\lset$ of clusters would be for $\sum_{I'\in \iset}\optcrors(I')$ to be small compared to $\optcrors(I)+|E(G)|$. Unfortunately, we cannot show that this property holds, except for some special cases. The \BCD procedure essentially considers one such special case, in which $\sum_{I'\in \iset}\optcrors(I')$ can be appropriately bounded. In order to define this procedure formally, we first need to define several central notions that are used throughout our algorithm, namely: light clusters, bad clusters, and path-guided orderings.

\subsection{Light Clusters, Bad Clusters, and Path-Guided Orderings}
\label{subsec: guiding paths rotations}

\begin{definition}[Light Cluster]
	\label{def: light cluster}
	Let $I=(G,\Sigma)$ be an instance of \emph{\CNwRS}, and let $C$ be a cluster of $G$. Assume further that we are given a distribution $\dset(C)$ over the set $\Lambda(C)$ of internal $C$-routers. We say that cluster $C$ is \emph{$\beta$-light with respect to $\dset(C)$} if, for every edge $e\in E(C)$:
	$$\expect[\qset\sim\dset(C)]{(\cong_G(\qset,e))^2}\leq \beta.$$
\end{definition}	

\begin{definition}[Bad Cluster]\label{def: bad cluster}
	Let $I=(G,\Sigma)$ be an instance of \emph{\CNwRS}, let $C$ be a cluster of $G$, and let $\Sigma(C)$ be the rotation system for $C$ induced by $\Sigma$. We say that $C$ is a \emph{$\beta$-bad} cluster, if: 
	\[\optcrors(C,\Sigma(C))+|E(C)|\geq \frac{|\delta_G(C)|^2}{\beta}. \]
\end{definition}


\paragraph{Path-Guided Orderings.}
Let $I=(G,\Sigma)$ be an instance of \CNwRS, and let $C$ be a cluster of $G$. Consider an internal $C$-router $\qset=\set{Q(e)\mid e\in \delta_G(C)}$. Recall that there is some vertex $u\in V(C)$ (the center of the router), such that, for all $e\in \delta_G(C)$, path $Q(e)$ has edge $e$ as its first edge, vertex $u$ as its last vertex, and all inner vertices of $Q(e)$ lie in $C$.

We will use the internal $C$-router $\qset$, and the rotation system $\Sigma$ for $G$, in order to define a circular ordering $\oset$ of the edges of $\delta_G(C)$. We refer to the ordering $\oset$ as \emph{an ordering guided by $\qset$ and $\Sigma$}. Ordering $\oset$ of the edges of $\delta_G(C)$ is constructed as follows. 
Denote $\delta_G(u)=\set{a_1,\ldots,a_r}$, where the edges are indexed according to their circular ordering $\oset_u\in \Sigma$. For all $1\leq i\leq r$, let $\qset_i\subseteq \qset$ be the set of paths whose last edge is $a_i$. We first define an ordering $\hat \oset$ of the paths in $\qset$, where the paths in sets $\qset_1,\ldots,\qset_r$ appear in the natural order of their indices, and for all $1\leq i\leq r$, the ordering of the paths in $\qset_i$ is arbitrary. Ordering $\hat\oset$ of the paths in $\qset$ naturally defines the ordering $\oset$ of the edges of $\delta_G(C)$: we obtain the ordering $\oset$ from $\hat \oset$ by replacing, for every path $Q(e)\in \qset$, the path $Q(e)$ in $\hat \oset$ with the edge $e$ (the first edge of $Q(e)$). We refer to $\oset$ as the ordering of the edges of $\delta_G(C)$ that is guided by $\qset$ and $\Sigma$, and we denote  it by $\oset^{\guided}(\qset,\Sigma)$. A convenient way to think of the ordering $\oset^{\guided}(\qset,\Sigma)$ of the edges of $\delta_G(C)$ is that this order is determined by the order in which the paths of $\qset$ enter the vertex $u$, which in turn is determined by the rotation $\oset_u\in \Sigma$ (as the last edge on each path in $\qset$ lies in $\delta_G(u)$).

\subsection{Basic Cluster Disengagement}
\label{subsec: basic disengagement}

The input to the \BCD procedure consists of an instance $I=(G,\Sigma)$ of \cnwrs and a laminar family $\lset$ of clusters of $G$. Recall that, by the definition, $G\in \lset$ must hold. We further assume that we are given a partition $(\lset^{\gd},\lset^{\bad})$ of the clusters of $\lset\setminus \set{G}$, and, for every cluster $C\in \lset^{\gd}$, a distribution $\dset(C)$ over its internal $C$-routers (that may be given implicitly). The output of the procedure is a collection $\iset=\set{I_C\mid C\in \lset}$ of subinstances of $I$. In order to define the instances of $\iset$, we will define, for every cluster $C\in \lset$, an ordering $\oset(C)$ of the edges of $\delta_G(C)$. Family $\iset$ of subinstances of $I$ is then constructed by disengaging instance $I$ via the laminar family $\lset$ and the collection $\set{\oset(C)}_{C\in \lset}$ of orderings.

In order to describe the algorithm for computing the collection $\iset$ of subinstances of $I$, it is now enough to describe an algorithm that computes, for every cluster $C\in \lset$, an ordering $\oset(C)$ of the edges of $\delta_G(C)$.
Consider any such cluster $C\in \lset$. If $C=G$, then $\delta_G(C)=\emptyset$, and the ordering $\oset(C)$ is trivial. If $C\in \lset^{\bad}$, then we let $\oset(C)$ be an arbitrary ordering of the edges of $\delta_G(C)$. 
Lastly, consider a cluster $C\in \lset^{\gd}$. We select an internal router $\qset(C)\in \Lambda_G(C)$ from the given distribution $\dset(C)$ at random. 
We view the paths in $\qset(C)$ as being directed towards the center vertex $u(C)$ of the router. We use the algorithm from \Cref{lem: non_interfering_paths} to compute a non-transversal path set $\tilde \qset(C)$, routing all edges of $\delta_G(C)$ to vertex $u(C)$, so $\tilde \qset(C)$ is also an internal $C$-router. The algorithm ensures that, for every edge $e\in E(G)$,  $\cong_G(\tilde \qset(C),e)\leq \cong_G(\qset(C),e)$.
We then let the ordering $\oset(C)$ of the edges of $\delta_G(C)$ be an ordering guided by the set $\tilde \qset(C)$ of paths in graph $G$, and the rotation system $\Sigma$, so $\oset(C)=\oset^{\guided}(\tilde \qset(C),\Sigma)$.

This completes the description of the algorithm for selecting an ordering $\oset(C)$ of the edges in $\delta_G(C)$ for each cluster $C\in \lset$; note that this algorithm is randomized. This also completes the description of the algorithm for performing \BCD, that we refer to as \algbasicdisengagement in the remainder of the paper. Since this algorithm essentially performs disengagement via a laminar family of clusters of $G$, \Cref{lem: number of edges in all disengaged instances} and \Cref{lem: basic disengagement combining solutions} continue to hold for the resulting collection $\iset$ of instances. But we can now show that, under some conditions, we can bound the expected value of $\sum_{I'\in \iset}\optcrors(I')$. 

When using the algorithm \algbasicdisengagement for performing \BCD of an instance $I$ of \cnwrs via a laminar family $\lset$, we will typically require that the following properties hold, for some parameter $\beta$:

\begin{properties}{P}
	\item every cluster $C\in \lset^{\bad}$ is $\beta$-bad, and has the $\alpha_0$-bandwidth property in $G$, for some $\alpha_0\geq \Omega(1/\log^{12}m)$; \label{prop: for bad clusters}
	\item every cluster $C\in \lset^{\light}$ is $\beta$-light with respect to the given distribution $\dset(C)$ over the set $\Lambda(C)$ of its internal routers; and \label{prop: for light clusters}
	\item for every cluster $C\in \lset$, there is a distribution $\dset'(C)$ over the set $\Lambda'(C)$ of external $C$-routers, such that for every edge $e\in E(G\setminus C)$, $\expect[\qset'(C)\sim\dset'(C)]{\cong_G(\qset'(C),e)}\leq \beta$.\label{prop: external routers}
\end{properties}

Observe that the algorithm for computing the family $\iset$ of clusters is randomized.
We show in the following lemma that, if all the above conditions hold, then the expected value of $\sum_{I'\in \iset}\optcrors(I')$ is suitably bounded. The proof is somewhat technical, and  is deferred to Section \ref{subsec: appx basic diseng opt bounds} of Appendix.

\begin{lemma}\label{lem: disengagement final cost}
	Let $I=(G,\Sigma)$ be an instance of the  \cnwrs problem, $\lset$ a laminar family of clusters of $G$,  $(\lset^{\gd},\lset^{\bad})$ a partition of cluster set $\lset\setminus \set{G}$, and, for every cluster $C\in \lset^{\gd}$, $\dset(C)$ a distribution over internal $C$-routers. Let $\iset$ be the collection of subinstances of $I$ obtained by applying Algorithm \algbasicdisengagement to instance $I$, with laminar family $\lset$, cluster sets $\lset^{\gd},\lset^{\bad}$, and distributions $\set{\dset(C)}_{C\in \lset^{\gd}}$. Assume further that Properties \ref{prop: for bad clusters} -- \ref{prop: external routers} hold for some parameter $\beta\geq c(\log |E(G)|)^{18}$,  where $c$ is a large enough constant. Then:
	\[\expect{\sum_{I'\in \iset}\optcrors(I')}\leq O(\dep(\lset)\cdot\beta^2\cdot (\optcrors(I)+|E(G)|)).\]
\end{lemma}

We note that the distributions $\set{\dset'(C)}_{C\in \lset}$ over external routers of the clusters play no role in constructing the collection $\iset$ of subinstances of $I$, but they are essential in order to ensure that the expectation of $\sum_{I'\in \iset}\opt(I')$ is suitably bounded. Since this property is essential to us, we will only use \BCD when such distributions are given. Therefore, abusing the notation, we refer to the family $\iset$ of subinstances of $I$ computed via \BCD of instance $I$ as described above, as a \BCD of $I$ via the tuple $(\lset,\lset^{\bad},\lset^{\gd}, \set{\dset'(C)}_{C\in \lset},\set{\dset(C)}_{C\in \lset^{\gd}})$.

\section{Second Main Tool: Cluster Classification}
\label{sec: routing within a cluster}

In this section we introduce our second main tool, the algorithm \algclassifycluster, that is summarized in the following theorem. 

\begin{theorem}\label{thm:algclassifycluster}
There is a randomized algorithm, that, given an instance $I=(G,\Sigma)$ of \cnwrs problem with $|E(G)|=m$, a cluster $J\subseteq G$ that has the $\alpha_0$-bandwidth property in $G$, for $\alpha_0=1/\log^{50}m$, and a parameter $0<p<1$, either returns FAIL, or computes a distribution $\dset(J)$ over the set $\Lambda(J)$ of internal $J$-routers, such that cluster $J$ is $\beta^*$-light with respect to $\dset(J)$, where $\beta^*=2^{O(\sqrt{\log m}\cdot \log\log m)}$. Moreover, if cluster $J$ is not $\eta^*$-bad, for $\eta^*=2^{O((\log m)^{3/4}\log\log m)}$, then the probability that the algorithm returns FAIL is at most $p$. The running time of the algorithm is $\poly(m\cdot \log(1/p))$.
\end{theorem}

We will sometimes say that the algorithm \algclassifycluster \emph{errs} if it returns FAIL and yet cluster $J$ is not $\eta^*$-bad. Clearly, the probability that the algorithm errs is at most $p$. We note that the distribution $\dset(J)$ over the set $\Lambda(J)$ of internal $J$-routers that the algorithm computes may be returned by the algorithm implicitly, by providing another efficient algorithm to draw a router from the distribution.

In order to prove \Cref{thm:algclassifycluster}, it is sufficient to prove the following theorem.

\begin{theorem}\label{thm:algclassifycluster easier}
	There is an efficient randomized algorithm, that, given an instance $I=(G,\Sigma)$ of \cnwrs with $|E(G)|=m$, a 
	cluster $J\subseteq G$ that has the $\alpha_0$-bandwidth property in $G$, for $\alpha_0=\Omega(1/\log^{50}m)$, either returns FAIL, or computes a distribution $\dset(J)$ over the set of internal $J$-routers, such that cluster $J$ is $\beta^*$-light with respect to $\dset(J)$, where $\beta^*=2^{O(\sqrt{\log m}\cdot \log\log m)}$. Moreover, if cluster $J$ is not $\eta^*$-bad, for $\eta^*=2^{O((\log m)^{3/4}\log\log m)}$, then the probability that the algorithm returns FAIL is at most $1/2$.
\end{theorem}

Indeed, given a graph $G$, a cluster $J$ of $G$ and a parameter $0<p<1$, as in the statement of  \Cref{thm:algclassifycluster}, we simply run the algorithm from \Cref{thm:algclassifycluster easier} $\ceil{\log(1/p)}$ times on the input instance $(G,\Sigma)$ and cluster $J$ of $G$. If the algorithm returns FAIL in every of these iterations, then we also return FAIL. Otherwise, in at least one of the iterations, the algorithm from \Cref{thm:algclassifycluster easier} returns a distribution $\dset(J)$ over the set $\Lambda(J)$ of internal $J$-routers, such that cluster $J$ is $\beta^*$-light with respect to $\dset(J)$. We then return the distribution $\dset(J)$ as the algorithm's outcome. It is immediate to verify that the probability that the algorithm errs is at most $p$, and that its running time is $\poly(m\cdot \log(1/p))$, as required.

In the remainder of this section, we focus on the proof of \Cref{thm:algclassifycluster easier}.
It will be convenient for us to consider the augmentation $J^+$ of cluster $J$. Recall that this is the graph that is obtained from $G$ by subdiving every edge $e\in \delta_G(J)$ with a vertex $t_e$, letting $T=\set{t_e\mid e\in \delta_G(J)}$ be the set of the new vertices, and then letting $J^+$ be the subgraph of the resulting graph induced by $T\cup V(J)$. We refer to vertices of $T$ as \emph{terminals}, and we denote $|T|=k$. Recall that, from the $\alpha_0$-bandwidth property of cluster $J$, the set $T$ of terminals is  $\alpha_0$-well-linked in $J^+$. Since the degree of every terminal in $J^+$ is $1$, the rotation system $\Sigma$ for graph $G$ naturally defines a unique rotation system $\Sigma(J^+)$ for $J^+$. Moreover, cluster $J$ is $\eta^*$-bad iff $\optcrors(J^+,\Sigma(J^+))+|E(J^+\setminus T)|\geq k^2/\eta^*$.

Let $\Lambda(J^+,T)$ denote the collection of all sets $\qset$ of paths, such that paths in $\qset$ route all vertices of $T$ to some vertex $x\in  V(J^+)\setminus T$,  in graph $J^+$. 
We sometimes also call $\qset$ a router, and refer to $x$ as the \emph{center vertex} of the router.
Notice that, if we are given a distribution $\dset$ over sets of paths in $\Lambda(J^+,T)$, such that, for every edge $e\in E(J^+)$, $\expect[\qset\sim \dset]{(\cong_{J^+}(\qset,e))^2}\leq \beta^*$, then we can immediately obtain a distribution $\dset(J)$ over the set $\Lambda(J)$ of internal $J$-routers, such that cluster $J$ is $\beta^*$-light with respect to $\dset(J)$. 

From now on we will only focus on graph $J^+$ and the corresponding rotation system $\Sigma(J^+)$, so it will be convenient for us to denote graph $J^+$ by $G$ and $\Sigma(J^+)$ by $\Sigma$. We denote by $I=(G,\Sigma)$ the resulting instance of \cnwrs.
From now on our goal is to design a randomized algorithm, that either computes a distribution $\dset$ over the set $\Lambda(G,T)$ of internal routers, such that, for every edge $e\in E(G)$, $\expect[\qset\sim \dset]{(\cong_{G}(\qset,e))^2}\leq \beta^*$, or returns FAIL. We need to ensure that, if $\optcrors(I)+|E(G\setminus T)|< k^2/\eta^*$, then the probability that the algorithm returns FAIL is at most $1/2$.

We now provide some intuition. 
We first show below an algorithm called \algfindguiding\xspace that ``almost'' provides the required guarantees. 
Specifically, if we are guaranteed that $|E(G)|\leq k\cdot \eta$ for some small parameter $\eta$, then the algorithm either computes a distribution $\dset$ over sets of paths in $\Lambda(G,T)$, such that, for every edge $e\in E(G)$, $\expect[\qset\sim \dset]{(\cong_G(\qset,e))^2}\leq \poly(\log m)/\poly(\alpha_0)$, or it returns FAIL, with the guarantee that, if $\optcrors(G,\Sigma)+|E(G\setminus T)|<\frac{k^2\poly(\alpha_0)}{\poly(\eta\log m)}$, then the probability that the algorithm returns FAIL is at most $1/2$. We could use this theorem directly if  $|E(G)|\leq k\cdot \eta$ holds for some $\eta\leq (\eta^*)^{\eps}$ where $\eps$ is a constant, but unfortunately this is not guaranteed in the statement of \Cref{thm:algclassifycluster}, and $|E(G)|$ may be arbitrarily large compared to $k$. In order to overcome this difficulty, we use another algorithm, that, given a graph $G$ and a set $T$ of its terminals as above, computes a collection $\cset$ of disjoint clusters of $G\setminus T$, such that, for each cluster $C\in \cset$, either (i) there is an internal $C$-router $\qset(C)\in \Lambda(C)$ such that the paths in $\qset(C)$ are edge-disjoint; or (ii) $C$ is $\eta$-bad for some parameter $\eta\ll \eta^*$;  or (iii) $|E(C)|\leq |\delta_G(C)|\cdot O(\poly(\eta\log m))$. In the latter case, we say that $C$ is a \emph{concise} cluster. 
We then apply the algorithm \algfindguiding to each concise cluster. As a result, for each such concise cluster $C\in \cset$, we will either establish, with high probability, that it is a $\eta'$-bad cluster, for some parameter $\eta'$, or we will compute a distribution $\dset(C)$ over the set $\Lambda(C)$ of internal $C$-routers, such that cluster $C$ is $\beta'$-light with respect to $\dset(C)$, for some parameter $\beta'$. The algorithm for computing the collection $\cset$ of clusters of $G$ also guarantees that each cluster $C\in \cset$ has the $\alpha'$-bandwidth property, for $\alpha'=\Omega(1/\log^{1.5}m)$, and that the corresponding contracted graph $G_{|\cset}$ contains significantly fewer edges: $|E(G_{|\cset})|\leq |E(G)|/\eta$. Intuitively, we would then like to continue with the contracted graph $G_{|\cset}$, applying exactly the same algorithm to this graph. We could continue this process, obtaining a clustering $\cset'$ of this new contracted graph, and so on, until we reach a final contracted graph $\hat G$, with $|E(\hat G)|\leq O(k\eta)$. At this point we can apply the algorithm \algfindguiding to graph $\hat G$ directly, and as a result, we either obtain the desired distribution $\dset$ over path sets in $\Lambda(G,T)$, or establish, with high probability, that $\optcrors(G,\Sigma)+|E(G\setminus T)|$ is sufficiently high. A problem with this approach is that the algorithm \algfindguiding requires a rotation system $\Sigma'$ for its input graph $H$. Recall that the algorithm guarantees that, if $\optcrors(H,\Sigma')+|E(H)|$ is sufficiently low, then it only returns FAIL with probability at most $1/2$. The difficulty is that, if $H$ is the contracted graph $G_{|\cset}$, then it is not immediately clear how to define the rotation system $\Sigma'$ for $H$, such that $\optcrors(H,\Sigma')$ is not much higher than $\optcrors(G,\Sigma)$. 

In order to overcome this difficulty, we design the algorithm \algfindguiding for a more general setting. In this setting, the input is a graph $H$, a rotation system $\Sigma'$ for $H$, and a set $T'$ of terminals of $H$, such that the terminals of $T'$ are $\alpha$-well-linked in $H$. Additionally, we are given some collection $\cset'$ of disjoint $\eta'$-bad clusters in $H$. We also require  that $|E(H_{|\cset'})|\leq |T'|\eta'$, for some parameter $\eta'$. 
The algorithm either returns FAIL, or computes a distribution $\dset'$ over the set $\Lambda(H,T')$ of routers, such that, for every edge $e\in E(H)$, 
 $\expect[\qset\sim \dset']{(\cong(\qset,e))^2}$ is sufficiently low. We are also guaranteed that, if $|\optcrors(H,T')|+|E(H\setminus T')|$ is sufficiently small compared to $|T'|^2$, then the probability that the algorithm  returns FAIL is at most $1/2$.
This stronger version of algorithm \algfindguiding will allow us to carry out the algorithm outlined above. 

We now provide formal descriptions of the two main tools that our algorithm uses. The first tool allows us to compute a decomposition of an input graph $G$ into a collection of clusters, each of which is either light, bad, or concise. 
The proof of the theorem uses rather standard techniques and is deferred to Section \ref{sec: appx-decomposition-good-bad-other} of Appendix. 

\begin{theorem}\label{thm: basic decomposition of a graph}
	There is an efficient algorithm, that we refer to as \alginitpartition, whose input consists of a connected $m$-edge graph $G$, a set $T\subseteq V(G)$ of $k$ vertices called \emph{terminals}, such that each vertex of $T$ has degree $1$ in $G$, and a parameter  $\eta>\log m$, such that  $k\leq \frac{m}{16\eta\log m}$.
The algorithm computes a collection $\cset$ of vertex-disjoint clusters of $G\setminus T$, a partition $(\cset_1,\cset_2,\cset_3)$ of $\cset$ into three subsets, and, for every cluster $C\in \cset_3$, an internal $C$-router $\qset(C)\in \Lambda(C)$, where the paths of $\qset(C)$ are edge-disjoint, such that the following additional properties hold:

\begin{itemize}
	\item every cluster $C\in \cset$ has the $\alpha'$-bandwidth property, where $\alpha'=\frac{1}{16\alphasc(m)\cdot \log m}=\Omega\left(\frac{1}{\log^{1.5}m}\right )$;
	
	\item for every cluster $C\in \cset_1$, $|E(C)|\leq O(\eta^4\log^8m)\cdot |\delta_G(C)|$; 
	
	\item for every cluster $C\in \cset_2$, $\optcro(C)\geq \Omega(|E(C)|^2/(\eta^2\poly\log m))$, and $|E(C)|\ge \Omega(\eta^4 |\delta_G(C)|\log^8m)$;
	\item $\bigcup_{C\in \cset}V(C)=V(G)\setminus T$; and
	\item $|E(G_{|\cset})|\leq |E(G)|/\eta$.
\end{itemize}
\end{theorem}

Note that, from the theorem statement, for every cluster $C\in \cset_2$,  $\optcro(C)\geq \Omega(|\delta_G(C)|^2\eta^6/\poly\log m)$. We will informally refer to clusters in $\cset_1$ as concise clusters. 

Let $H$ be a graph and let $T$ be a set of vertices of $H$ called terminals. We say that a set $\qset=\set{Q(t)\mid t\in T}$ of paths in graph $H$ is a \emph{router for $H$ and $T$} if there is a vertex $x\in V(H)$, such that, for every terminal $t\in T$, path $Q(t)$ originates at $t$ and temrinates at $x$. We denote by $\Lambda(H,T)$ the set of all routers for $H$ and $T$.
Our second tool is algorithm \algfindguiding, summarized in the following theorem.


\begin{theorem}\label{thm: find guiding paths}
	There are universal constants $c_0$ and $c^*$, and an efficient randomized algorithm, called \algfindguiding, that receives as input an instance $I=(H,\Sigma)$ of \cnwrs, where $|E(H)|=m$, a set $T\subseteq V(H)$ of $k$ vertices of $H$ called terminals, and a collection $\cset$ of disjoint clusters of $H\setminus T$. Additionally, the algorithm receives as input parameters $0\leq \alpha,\alpha'\leq 1$ and $\eta,\eta'\geq 1$, such that the following conditions hold:
	\begin{itemize}
		\item $\eta\geq \frac{c^*\log^{46}m}{\alpha^{10}(\alpha')^2}$ and   $\eta'\geq \eta^{13}$;
		\item $k\geq |E(H_{|\cset})|/\eta$; 
		\item every terminal $t\in T$ has degree $1$ in $H$;
		\item the set $T$ of terminals is $\alpha$-well-linked in the contracted graph $H_{|\cset}$; and
		\item every cluster $C\in \cset$ is $\eta'$-bad and  has the $\alpha'$-bandwidth property in $H$.
	\end{itemize}
	The algorithm either returns FAIL or (explicitly) returns a distribution $\dset$ over the routers in $\Lambda(H,T)$, such that,
	for every edge $e\in E(H)$, 
	$\expect[\qset\sim \dset]{(\cong(\qset,e))^2}\leq  O\left (\frac{\log^{32}m}{\alpha^{12}(\alpha')^8}\right )$. Moreover, if $\optcrors(I)+|E(H\setminus T)|\leq \frac{(k\alpha^4 \alpha')^2}{c_0\eta'\log^{50}m}$, then the probability that the algorithm returns FAIL is at most $1/2$. 
\end{theorem}

The proof of \Cref{thm: find guiding paths} is quite technical, and is deferred to Section \ref{sec: guiding paths}. We note that the algorithm returns the distribution $\dset$ explicitly, that is, it lists all routers $\qset \in \Lambda(H,T)$ that have a non-zero probability, together with their probability values in $\dset$.
In the remainder of this section, we complete the proof of \Cref{thm:algclassifycluster}, using the algorithms \alginitpartition and \algfindguiding.

Note that we can assume, throughout the proof, that $m$ is sufficiently large (larger than some large enough constant). Otherwise, since the vertices of $T$ are $\alpha$-well-linked in $G$, we get that $k\leq m\leq O(1)$. We can then let $\dset$ be a distribution that gives a probability $1$ to an internal router $\qset$ with target vertex $u$, where $u$ is an arbitrary vertex of $V(G)\setminus T$, and $\qset$ is an arbitrary collection of simple paths routing the vertices of $T$ to $u$. Clearly, for every edge $e\in E(G)$, $(\cong_G(\qset,e))^2\leq k^2\leq O(1)$.

\subsection{Main Parameters}

We now introduce some parameters that our algorithm uses. The main parameter is $\eta=2^{(\log m)^{3/4}}$. Our algorithm will consist of  $(\ell-1)$ phases, for $\ell\leq O\left(\frac{\log m}{\log \eta}\right)=O((\log m)^{1/4})$. 
At the beginning of the $i$-th phase, we will be given a collection $\cset_i$ of disjoint clusters of $G\setminus T$.

\paragraph{Parameters for bandwidth property.}
We will use the following parameters for bandwidth property of the clusters. Recall that $\alpha_0=\frac{1}{(\log m)^{50}}$ is the parameter from the statement of \Cref{thm:algclassifycluster}. For $1\leq i\leq \ell$, $\alpha_i=(\alpha_0)^i=\frac{1}{(\log m)^{50\cdot i}}$. We will ensure that for all $1\leq i\leq \ell$, every cluster in $\cset_i$ has the $\alpha_{i}$-bandwidth property. Note that $\alpha_{\ell}=1/(\log m)^{50\cdot\ell}=1/2^{O((\log m)^{1/4}\log\log m)}$.

\paragraph{Parameters for light clusters.}
We set $\beta_0=1$, and for $1\leq i\leq \ell$, $\beta_i=\frac{(\log m)^{56}}{(\alpha_0)^{12}\cdot (\alpha_{i})^8}\cdot \beta_{i-1}$. It is easy to verify that $\beta_{\ell}\leq (\log m)^{O(\ell^2)}\leq 2^{O((\log m)^{1/2}\log\log m)}\leq \beta^*$.

\paragraph{Parameters for bad clusters.}
We define the following parameters: $\eta_0=\eta^4\cdot \log^9m=2^{O ((\log m)^{3/4} )}$; $\eta_1=\max\set{(2\eta_0)^{13},\eta_0\cdot (\log m)^{80}}=2^{O ((\log m)^{3/4} )}$, and for each $1\leq i\leq \ell$, 
$\eta_i=\eta_{i-1}\cdot \max\set{(\log m)^{80+50\cdot i},\beta^3_{i-1}}$. 
Clearly, 
$\eta_i\leq (\beta_{\ell})^3\cdot\eta_{i-1}$, and
$\eta_{\ell}\leq \eta_1\cdot (\beta_{\ell})^{3\ell}\leq 2^{O ((\log m)^{3/4}\log\log m )}\leq \eta^*$.

\subsection{Algorithm Execution}
As already mentioned, the algorithm consists of $(\ell-1)$ phases, where $\ell= O((\log m)^{1/4})$. At the beginning of the $i$th phase, we will be given a collection $\cset_{i}$ of disjoint clusters of $G\setminus T$, which is partitioned into two subsets: $\cset_{i}^{\gd}$ and $\cset_{i}^{\bad}$. For each cluster $C\in \cset_{i}^{\gd}$, we will also be given a distribution $\dset(C)$ over the set $\Lambda(C)$ of internal $C$-routers. 
For all $1\leq i\leq \ell$, we will also define a bad event $\event_i$, and we will ensure that it happens with probability at most $i/m^{10}$. 
We will ensure that the following properties hold for all $1\leq i\leq \ell$:

\begin{properties}{R}
	\item each cluster $C\in \cset_{i}$ has the $\alpha_{i}$-bandwidth property; \label{prop 1 of clusters bw}
	\item the number of edges in the contracted graph $G_{|\cset_{i}}$ is at most $m/\eta^{i-1}$; \label{prop 2 of clusters small contracted graph}
	
	\item each cluster $C\in \cset_{i}^{\gd}$ is $\beta_{i}$-light with respect to the distribution $\dset(C)$; \label{prop 3 of clusters good} and
	
	\item if the bad event $\event_i$ does not happen, then each cluster $C\in \cset_{i}^{\bad}$ is $\eta_{i}$-bad.\label{prop 4 last of clusters bad}
\end{properties}


The input to the first phase, $\cset_1=\emptyset$. Clearly, all  properties \ref{prop 1 of clusters bw}--\ref{prop 4 last of clusters bad} hold for this set of clusters. 
We now describe the execution of the $i$th phase.
We assume that we are given as input a collection $\cset_{i}$ of disjoint clusters of $G\setminus T$, which is partitioned into two subsets, $\cset_{i}^{\gd}$ and $\cset_{i}^{\bad}$. We are also given, for each cluster $C\in \cset_{i}^{\gd}$, a distribution $\dset(C)$ over the set $\Lambda(C)$ of internal $C$-routers, and we are guaranteed that Properties \ref{prop 1 of clusters bw}--\ref{prop 4 last of clusters bad} hold.

We consider the contracted graph $G'=G_{|\cset_{i}}$.  The execution of the $i$th phase consists of two steps: in the first step, we apply the algorithm \alginitpartition to the contracted graph $G'$, obtaining a collection $\cset$ of clusters of $G'$, which we then convert into clusters of $G$. The set $\cset$ of clusters is partitioned into three subsets.  Informally, the clusters in the first subset are concise, the clusters in the second subset are $\eta_{i+1}$-bad if event $\event_{i}$ did not happen, and the clusters in the third set are $\beta_{i+1}$-light with respect to a distribution over the internal routers that we construct. In the second step we further process each concise cluster, using the algorithm \algfindguiding, in order to determine whether it is a $\beta_{i+1}$-light or an $\eta_{i+1}$-bad cluster, and in the former case, to compute the corresponding distribution $\dset(C)$ over the set $\Lambda(C)$ of internal $C$-routers.

\subsection{Step 1: Partition}
We assume first that $|E(G')|>16\eta k\log m $, where $\eta=2^{(\log m)^{3/4}}$ is the parameter that we have defined above. 
If the inequality does not hold, then the current phase is the last phase of the algorithm, and we show how to execute this phase at the end of this subsection.
 
We apply the algorithm \alginitpartition from \Cref{thm: basic decomposition of a graph} to graph $G'$, the set $T$ of terminals, and the parameter $\eta$ that we have defined. Notice that, since $\eta=2^{(\log m)^{3/4}}$, and since we have assumed that $m$ is greater than some large enough constant, $\eta>\log m\geq \log (|E(G')|)$ must hold. 
We now consider the output of the algorithm, that consists of a collection $\cset$ of disjoint clusters of $G'\setminus T$, a partition $(\cset_1',\cset_2',\cset_3')$ of $\cset$ into three subsets, and, for every cluster $C\in \cset_3'$, a vertex $u(C)\in V(C)$, and an internal $C$-router $\qset(C)$, consisting of edge-disjoint paths routing the edges of $\delta_G(C)$ to $u(C)$. Recall that we are also guaranteed that every cluster $C\in \cset$ has the $\alpha'$-bandwidth property, where $\alpha'=\frac{1}{16\alphasc(m)\cdot \log m}\geq \alpha_0$.

 Consider any cluster $C\in \cset$. Recall that $C$ is a cluster of the contracted graph $G'$, and it has the $\alpha_0$-bandwidth property. Let $\wset(C)$ be the set of all clusters $W\in \cset_{i}$, whose corresponding supernode $v_W\in V(C)$. Recall that every cluster of $\cset_{i}$ has the $\alpha_{i}$-bandwidth property from Property \ref{prop 1 of clusters bw}. Let $U_C$ be the set of vertices of $G$, that contains every regular (non-supernode) vertex of $C$, and every vertex lying in clusters of $\wset(C)$. In other words, $U_C=(V(G)\cap V(C))\cup \left(\bigcup_{W\in \wset(C)}V(W)\right )$. We then let $\tilde C=G[U_C]$. 
Since cluster $C$ has the $\alpha_0$-bandwidth property, and every cluster in $\wset(C)$ has the $\alpha_{i}$-bandwidth property, from \Cref{clm: contracted_graph_well_linkedness} and \Cref{obs: wl-bw}, cluster $\tilde C$ has the $\alpha_{i}\cdot \alpha_0=\alpha_{i+1}$-bandwidth property. We let $\cset_{i+1}=\set{\tilde C\mid C\in \cset}$.
Notice that we have just established Property \ref{prop 1 of clusters bw} for clusters in $\cset_{i+1}$. It is immediate to verify that $G_{|\cset_{i+1}}=G'_{|\cset}$. Since \Cref{thm: basic decomposition of a graph} guarantees that $|E(G'_{|\cset})|\leq |E(G')|/\eta$, and, from Property \ref{prop 2 of clusters small contracted graph}, $|E(G')|=|E(G_{|\cset_{i}})|\leq m/\eta^{i-1}$, we get that $|E(G_{|\cset_{i+1}})|\leq |E(G')|/\eta\leq m/\eta^i$, establishing Property \ref{prop 2 of clusters small contracted graph} for the set $\cset_{i+1}$ of clusters.

We now construct the partition $(\cset_{i+1}^{\bad}, \cset_{i+1}^{\gd})$ of the set $\cset_{i+1}$ of clusters.
We start by letting $\cset_{i+1}^{\bad}=\set{\tilde C\mid C\in \cset_2'}$ and $\cset_{i+1}^{\gd}=\emptyset$. We then consider every cluster $C\in \cset_3'$ one by one. Recall that for each such cluster $C$,  the algorithm from \Cref{thm: basic decomposition of a graph} provides an internal router $\qset(C)$, routing the edges of $\delta_{G'}(C)$ to $u(C)$, such that the paths in $\qset(C)$ are edge-disjoint. If vertex $u(C)$ is a supernode, whose corresponding cluster $W\in \cset_{i}$ lies in set $\cset_{i}^{\bad}$, then we  add $\tilde C$ to $\cset_{i+1}^{\bad}$; otherwise, we add $\tilde C$ to $\cset_{i+1}^{\gd}$.
%
Lastly, we set $\cset_{i+1}^{\con}=\set{\tilde C\mid C\in \cset_1'}$, and we refer to clusters in $\cset_{i+1}^{\con}$ as \emph{concise clusters}. In Step 2, we will further process clusters in $\cset_{i+1}^{\con}$, and we will eventually add each such cluster to either $\cset_{i+1}^{\bad}$ or to $\cset_{i+1}^{\gd}$. Before we do so, we establish Property \ref{prop 4 last of clusters bad} for clusters that are currently in $\cset_{i+1}^{\bad}$, and we define a distribution $\dset(\tilde C)$ over the set $\Lambda_G(\tilde C)$ of internal $\tilde C$-routers for every cluster $\tilde C$ that is currently in $\cset_{i+1}^{\gd}$, such that $\tilde C$ is $\beta_{i+1}$-light with respect to $\dset(\tilde C)$ (that is, we establish Property \ref{prop 3 of clusters good} for clusters that are currently in $\cset_{i+1}^{\gd}$).

\paragraph{Bad Clusters.}
Recall that  \Cref{thm: basic decomposition of a graph} guaranteed that, for every cluster $C\in \cset_2'$, $\optcro(C)\geq \Omega(|E(C)|^2/(\eta^2\poly\log m))$, and $|E(C)|> \Omega(\eta^4 |\delta_G(C)|\log^8m)$. Therefore:
\[ \optcro(C)\geq \Omega\left (\frac{|E(C)|^2}{\eta^2\poly\log m}\right )\geq \Omega\left (\frac{|\delta_G(C)|^2\eta^6}{\poly\log m}\right )\geq |\delta_G(C)|^2.\] 
%
From the definition of cluster $\tilde C$, graph $C$ is a contracted graph of $\tilde C$ with respect to clusters in $\wset(C)$, that is, $C=\tilde C_{|\wset(C)}$. As each cluster in $\wset(C)\subseteq \cset_i$ has the $\alpha_{i}$-bandwidth property (from Property \ref{prop 1 of clusters bw}), from \Cref{lem: crossings in contr graph}, there is a drawing of $C$ containing at most 
$O(\optcrors(\tilde C, \Sigma_{\tilde C})\cdot \log^8m/(\alpha_{i})^2)$ crossings, where $\Sigma_{\tilde C}$ is the rotation system for $\tilde C$ induced by $\Sigma$. Since we have established that $\optcro(C)\geq |\delta_G(C)|^2$, we get that:
\[\optcrors(\tilde C, \Sigma_{\tilde C})\geq  \Omega \left (\frac{|\delta_G(C)|^2\cdot (\alpha_{i})^2}{\log^8m} \right )\geq \frac{|\delta_G(C)|^2}{\eta_{i+1}}, \]
since, by the definition, $\eta_{i+1}> \eta \geq 2^{(\log m)^{3/4}}$, while 
$\alpha_{i}\geq \alpha_{\ell}\geq 1/2^{O((\log m)^{1/4}\log\log m)}$, and since we have assume that $m$ is large enough. We conclude that every cluster in
$\set{\tilde C\mid C\in \cset_2'}$ is $\eta_{i+1}$-bad.

Consider now some
cluster $C\in \cset_3'$, such that vertex $u(C)$ that serves as the center of the router $\qset(C)$ provided by  the algorithm from \Cref{thm: basic decomposition of a graph} is a supernode, whose corresponding cluster $W\in \cset_{i}^{\bad}$. From Property \ref{prop 4 last of clusters bad}, if Event $\event_{i}$ did not happen, cluster $W$ is an $\eta_{i}$-bad cluster, that is, $\optcrors(W,\Sigma_W)+|E(W)|\geq \frac{|\delta_G(W)|^2}{\eta_{i}}$, where $\Sigma_W$ is the rotation system for $W$ induced by $\Sigma$. Since $W\subseteq \tilde C$, we get that $\optcrors(\tilde C,\Sigma_{\tilde C})+|E(\tilde C)|\geq \frac{|\delta_G(W)|^2}{\eta_{i}}$. Lastly, since there is a set $\qset(C)$ of edge-disjoint paths routing the edges of $\delta_{G'}(C)$ to vertex $u(C)$ inside $C$, we conclude that $|\delta_G(\tilde C)|\leq |\delta_G(W)|$. Altogether, from the fact that $\eta_{i+1}>\eta_{i}$, we get that
$\optcrors(\tilde C,\Sigma_{\tilde C})+|E(\tilde C)|\geq \frac{|\delta_G(\tilde C)|^2}{\eta_{i+1}}$.

We conclude that, if event $\event_{i}$ did not happen, then every cluster that we have added to set $\cset_{i+1}^{\bad}$ so far is an $\eta_{i+1}$-bad cluster.

\paragraph{Light Clusters.}
Consider now some cluster $\tilde C\in \cset_{i+1}^{\gd}$, and let $C\in \cset$ be its corresponding cluster in graph $G'$. Recall that the algorithm from \Cref{thm: basic decomposition of a graph} provides a collection $\qset(C)$ of edge-disjoint paths routing the edges of $\delta_{G'}(C)$ to $u(C)$, such that, for every path in $\qset(C)$, all inner vertices of the path lie in $C$. We will now define a distribution $\dset(\tilde C)$ over the set $\Lambda_G(\tilde C)$ of internal $\tilde C$-routers, so that $\tilde C$ is $\beta_{i+1}$-light with respect to $\dset(\tilde C)$.

Assume first that vertex $u(C)$ is a regular vertex in cluster $C$, that is, it is not a supernode. Since every cluster $W\in \wset(C)$ has the $\alpha_{i}$-bandwidth property, we can use the algorithm from 
\Cref{claim: routing in contracted graph} to compute a collection $\qset(\tilde C)$ of paths, routing the edges of $\delta_G(\tilde C)$ to $u(C)$, such that, for every path of $\qset(\tilde C)$, all its inner vertices lie in  $\tilde C$, and the largest congestion on an edge of $\tilde C$ is bounded by $\ceil{1/\alpha_{i}}$. 
 The resulting distribution $\dset(\tilde C)$ then consists of a single internal $\tilde C$-router $\qset(\tilde C)$, that is chosen with probability $1$. 
Clearly, for every edge $e\in E(\tilde C)$:
$$\expect[\qset(\tilde C) \sim \dset(\tilde C)]{(\cong_G(\qset(\tilde C),e))^2}\leq \ceil{1/\alpha_{i}}^2\leq \beta_{i+1},$$
since by definition $\beta_{i+1}=\frac{(\log m)^{56}}{(\alpha_0)^{12}\cdot (\alpha_{i+1})^8}\cdot \beta_{i}$.

Assume now that $u(C)$ is a supernode, corresponding to some cluster $W^*\in \wset(C)$, such that $W^*\in \cset_{i}^{\gd}$. Let $\tilde C'$ be the cluster obtained from $\tilde C$, after we contract cluster $W^*$ into a supernode $v_{W^*}=u(C)$. Using the same reasoning as in the previous case, we can compute a set $\qset(\tilde C')$ of paths in graph $\tilde C'$, routing the edges of $\delta_G(\tilde C)$ to $u(C)$, such that, for every path in $\qset(\tilde C')$, every inner vertex on the path lies in $\tilde C'$, and the largest congestion on an edge of $\tilde C'$ is bounded by $\ceil{1/\alpha_{i}}$. Moreover, the algorithm from \Cref{claim: routing in contracted graph} guarantees that every edge in $\delta_{\tilde C'}(u(C))$ belongs to at most one path in $\qset(\tilde C')$ (and it is the last edge on that path). 

Recall that cluster $W^*$ is $\beta_{i}$-light with respect to the distribution $\dset(W^*)$ over the set $\Lambda_G(W^*)$ of internal $W^*$-routers.
We choose an internal $W^*$-router $\qset(W^*)\in \Lambda(W^*)$ from the distribution $\dset(W^*)$, routing the edges of $\delta_G(W^*)$ to a vertex $u(W^*)$ of $W^*$. We now consider every path $Q\in \qset(\tilde C')$ one by one. Let $e\in \delta_G(\tilde C)$ be the first edge on $Q$, and let $e'\in \delta_G(W^*)$ be the last edge on $Q$. Let $Q^*$ be the unique path in $\qset(W^*)$ whose first edge is $e'$, and let $Q'$ be obtained by first deleting the edge $e'$ from $Q$, and then concatenating the resulting path with path $Q^*$. Notice that path $Q'$ connects the edge $e$ to the vertex $u(W^*)$, in graph $\tilde C\cup \delta_G(\tilde C)$. We then set $\qset(\tilde C)=\set{Q'\mid Q\in \qset(\tilde C')}$, so that $\qset(\tilde C)$ is an internal $\tilde C$-router in $\Lambda_G(\tilde C)$. 
This finishes the definition of the distribution $\dset(\tilde C)$ over the set $\Lambda(\tilde C)$ of internal $\tilde C$-routers. Note that the distribution is given implicitly, that is, we provide an efficient algorithm to draw a router from the distribution.

Consider now some edge $e\in E(\tilde C)$. If $e\not\in E(W^*)$, then with probability $1$ (over the choices of internal $\tilde C$-routers $\qset(\tilde C)$ from $\dset(\tilde C)$), $\cong_{G}(\qset(\tilde C),e)\le \ceil{1/\alpha_{i}}^2$, and so $(\cong_{G}(\qset(\tilde C),e))^2\leq \beta_{i+1}$ as argued before. If $e\in E(W^*)$, then $\cong_G(\qset(\tilde C),e)=\cong_{G}(\qset(W^*),e)$, and so:
$$\expect[\qset(\tilde C) \sim \dset(\tilde C)]{(\cong_G(\qset(\tilde C),e)^2}=\expect[\qset(W^*) \sim \dset(W^*)]{(\cong_G(\qset(W^*),e)^2}
\leq \beta_{i}\leq \beta_{i+1}.$$
We conclude that cluster $\tilde C$ is $\beta_{i+1}$-light with respect to the distribution $\dset(\tilde C)$ over the set $\Lambda_G(\tilde C)$ of intenral $\tilde C$-routers that we have computed.

Recall that so far we assumed that $|E(G')|>16\eta k\log m$ held, where $G'=G_{|\cset_i}$. Assume now that $|E(G')|\leq 16\eta k\log m $. In this case, we let $\cset_{i+1}^{\bad}=\cset_{i+1}^{\gd}=\emptyset$ and we let $\cset_{i+1}^{\con}$ contain a single cluster, $\tilde C=G\setminus T$. We also let $\cset_1'$ contain a single cluster, $C=G'\setminus T$, and we set $\cset_2'=\cset_3'=\emptyset$. We denote $\wset(C)=\cset_i$. The current phase will become the final phase of the algorithm.

\subsection{Step 2: Concise Clusters}

Observe that every cluster $C\in \cset_1'$ has the $\alpha_0$-bandwidth property in $G'$, and $|E(C)|\leq \eta_0\cdot |\delta_{G'}(C)|$ holds. Indeed, if $|E(G')|\leq 16\eta k\log m $ holds, then set $\cset_1'$ contains a single cluster $C=G'\setminus T$, and $|E(C)|\leq 16\eta k\log m\leq \eta_0|\delta_{G'}(C)|$ holds. 
Since the set $T$ of terminals is $\alpha_0$-well-linked in $G$ (from the statement of \Cref{thm:algclassifycluster}), cluster $\tilde C=G\setminus T$ has the $\alpha_0$-bandwidth property in $G$, and cluster $C$ has the $\alpha_0$-bandwidth property in $G'$.

Otherwise,  \Cref{thm: basic decomposition of a graph} guarantees that every cluster $C\in \cset_1'$ has the $\alpha_0$-bandwidth property, and moreover,
$|E(C)|\leq O(\eta^4\log^8m)\cdot |\delta_{G'}(C)|\leq \eta_0\cdot |\delta_G(C)|$ 
(since $\eta_0=\eta^4\cdot \log^9m$, and we have assumed that $m$ is large enough). 
Recall that for every cluster $C\in \cset_1'$, we have defined a collection $\wset(C)\subseteq \cset_{i}$ of clusters, such that for each cluster $W\in \wset(C)$, its corresponding supernode $v_W$ lies in $C$. We have also defined a cluster $\tilde C\in \cset^{\con}_{i+1}$, that is a subgraph of $G$ correpsonding to $C$. In other words, we can think of $\tilde C$ as being obtained from $C$ by un-contracting every cluster $W\in \wset(C)$.

We would now like to apply the algorithm \algfindguiding to each such cluster $ C\in \cset_1'$, in order to classify the corresponding cluster $\tilde C$ of $G$ as either an $\eta_{i+1}$-bad or a $\beta_{i+1}$-light cluster. Notice however that the algorithm requires that we define a rotation system for $C$, and, if the algorithm classifies $C$ as an $\eta_{i+1}$-bad cluster (by returning ``FAIL''), then we are only guaranteed that it is likely that the value of the optimal solution of the resulting instance is high. Therefore, ideally we would like to define a rotation system $\hat \Sigma(C)$ for cluster $C$ of $G'$, such that $\optcrors(C,\hat \Sigma(C))$ is not much higher than $\optcrors(\tilde C,\Sigma_{\tilde C})$. 
Unfortunately, it is not immediately clear how to define such a rotation system, mainly because it is unclear how to define the orderings on edges incident to supernodes. In order to overcome this difficulty, we consider a different graph, that can be thought of as an intermediate graph between $C$ and $\tilde C$. This new graph, that we denote by $H(\tilde C)$, is obtained as follows.
We start
 from graph $\tilde C^+$.
Recall that $\tilde C^+$ is obtained from graph $G$ by first subdividing every edge $e\in \delta_G(\tilde C)$ with a vertex $t_e$, and then letting $\tilde C^+$ be the subgraph of the resulting graph  induced by $V(\tilde C)\cup \set{t_e\mid e\in \delta_G(\tilde C)}$. We denote $T'=\set{t_e\mid e\in \delta_G(\tilde C)}$. We partition the set $\wset(C)$ of clusters into two subsets: set $\wset^{\gd}(C)=\wset(C)\cap \cset_{i}^{\gd}$ and $\wset^{\bad}(C)=\wset(C)\cap \cset_{i}^{\bad}$.
Graph $H(\tilde C)$ is then obtained from graph $\tilde C^+$, by contracting every cluster $W\in \wset^{\gd}(C)$ into a supernode $v_W$. 
Additionally, we denote by $H'(\tilde C)$ the graph obtained from $\tilde C^+$ by contracting every cluster $W\in \wset(C)$ into a supernode $v_W$. 
Notice that graph $H'(\tilde C)$ is precisely the augmentation $C^+$ of the cluster $C$ in $G'$.

In the remainder of this step, we focus on one specific cluster $C\in \cset_1'$, so for convenience, we will denote $H(\tilde C)$ by $H$, $H'(\tilde C)$ by $H'$, $\wset(C)$ by $\wset$, and $\wset^{\gd}(C),\wset^{\bad}(C)$ by $\wset^{\gd}$ and $\wset^{\bad}$, respectively. From our construction, $H'=H_{|\wset^{\bad}}$.

Recall that we have already established that cluster $C$ has the $\alpha_0$-bandwidth property in $G'$. Therefore, the set $T'$ of vertices is $\alpha_0$-well-linked in graph $H'$.
Additionally, from Property \ref{prop 1 of clusters bw}, every cluster $W\in \wset^{\bad}$ has the $\alpha_{i}$-bandwidth property, and, if event $\event_{i}$ did not happen, each such cluster is $\eta_{i}$-bad. 
Recall also that we are guaranteed that $|E(C)|\leq \eta_0\cdot |\delta_{G'}(C)|=\eta_0\cdot |T'|$. Therefore, $|E(H')|=|E(C)|+|T'|\leq 2\eta_0|T'|$. 

Intuitively, we would now like to apply the algorithm \algfindguiding from \Cref{thm: find guiding paths} to graph $H$, the set $T'$ of terminals, and the correpsonding collection $\wset^{\bad}$ of clusters. In order to do so, we need to define a rotation system $\hat\Sigma$ for graph $H$. We do so using a randomized algorithm that exploits the distributions $\dset(W)$ over the set $\Lambda_G(W)$ of internal $W$-routers for clusters $W\in \wset^{\gd}$. We would like to use the algorithm \algfindguiding in order to decide whether to add cluster $\tilde C$ to the set $\cset_{i+1}^{\gd}$ of light clusters or to the set $\cset_{i+1}^{\bad}$ of bad clusters. Specifically, if the algorithm returns FAIL, we would like to add it to $\cset_{i+1}^{\bad}$, and otherwise we would like to add it to set $\cset_{i+1}^{\gd}$, together with the distribution $\dset(\tilde C)$ over the set $\Lambda_G(\tilde C)$ of internal $\tilde C$-routers that we can compute using the distribution over internal routers in $\Lambda(H,T')$ that the algorithm \algfindguiding computes. 
Notice however, that, even if $\optcrors(H,\hat \Sigma)$ is small, the algorithm may return FAIL with a constant probability. Additionally, the random choices that we make in defining the rotation system $\hat \Sigma$ for graph $H$ may also result in an instance whose solution value is too high (though this can only happen with relatively small probability). In order to ensure that our algorithm classifies cluster $\tilde C$ as a light or a bad cluster correctly with high probability, we will perform $m$ identical iterations (but in each iteration we construct the rotation system $\hat \Sigma$ for $H$ from scratch). We now describe a single iteration.

\paragraph{Execution of a single iteration.}
In order to perform a single iteration of the algorithm, we construct a rotation system $\hat \Sigma$ for graph $H$, as follows. Consider any vertex $v\in V(H)$. If $v\in T'$, then the degree of $v$ in $H$ is $1$, and the corresponding ordering of its incident edges is trivial. Assume now that $v\in V(H)\setminus T'$, and that $v$ is not a supernode. In this case, there is a one-to-one correpsondence between the edges in set $\delta_{H}(v)$ and the edges in set $\delta_G(v)$. We use the ordering $\oset_v\in \Sigma$ of the edges in $\delta_G(v)$ in order to define an ordering of the edges in $\delta_{H}(v)$, for the rotation system $\hat \Sigma$. Lastly, we assume that vertex $v\in V(H)\setminus T'$ is a supernode. In other words, $v=v_W$, where $W\in \wset^{\gd}$ is a light cluster. Recall that we are given a distribution $\dset(W)$ over the set $\Lambda_G(W)$ of internal $W$-routers, such that $W$ is $\beta_i$-light with respect to this distribution. We randomly select an internal $W$-router $\qset(W)\in \Lambda_G(W)$ from the distribution $\dset(W)$. Let $u(W)$ be the center vertex of $\qset(W)$, so that $\qset(W)$ is a set of paths routing the edges of $\delta_{G}(W)$ to vertex $u(W)$. Also recall that, from the definition of light clusters, for every edge $e\in E(W)$, $\expect[\qset(W)\sim\dset(W)]{(\cong_{G}(\qset(W),e))^2}\leq \beta_{i}$.

Observe that the edges of $\delta_H(v_W)$ are precisely the edges of $\delta_G(W)$. Next, we transform the set $\qset(W)$ of paths into a set of non-transversal paths, by applying the algorithm from \Cref{lem: non_interfering_paths} to the set $\qset(W)$ of paths. We denote the resulting set of paths by $\hat\qset(W)$; note that $\hat \qset(W)$ is an internal $W$-router.
Recall that we have defined an ordering of edges of $\delta_G(W)$ guided by the internal $W$-router $\hat\qset(W)$ (see \Cref{subsec: guiding paths rotations}). We let the ordering $\hat \oset_{v_W}\in \hat \Sigma$ be the ordering of the edges of $\delta_G(W)$ guided by the paths in $\hat \qset(W)$.
We need the following observation, whose proof follows arguments that are similar to those used in the proof of  \Cref{lem: disengagement final cost}, and is deferred to Section \ref{subsec: proof of obs opt is small} of Appendix.

\begin{observation}\label{obs: opt is small}
	$\expect{\optcrors(H, \hat \Sigma)}\leq O\left(\beta_{i}\cdot \left (\optcrors(\tilde C,\Sigma_{\tilde C})+|E(\tilde C)|\right )\right )$.
\end{observation}

A single iteration of our algorithm consists of computing a rotation system $\hat \Sigma$ for graph $H$ from scratch, and then applying the algorithm \algfindguiding from \Cref{thm: find guiding paths} to instance $I=(H,\hat \Sigma)$ of \cnwrs, 
with the set $T'$ of terminals, and the collection $\cset=\wset^{\bad}$ of clusters. We set the parameters  for the algorithm from \Cref{thm: find guiding paths} as follows: $\alpha=\alpha_0$, $\alpha'=\alpha_{i}$, $\eta=2\eta_0$, and $\eta'=\eta_{i}$. Recall that the set $T'$ of terminals is $\alpha_0$-well-linked in the contracted graph $H'=H_{|\wset^{\bad}}$, and $|T'|\geq |E(H'(\tilde C))|/(2\eta_0)$. Moreover, each cluster in $\wset^{\bad}(C)$ has the $\alpha_{i}$-bandwidth property, and, if event $\event_{i}$ did not happen, then each such cluster is $\eta_{i}$-bad (note that, if $i=1$ then $\wset^{\bad}(C)=\emptyset$).
Notice that $\eta'\geq \eta_1\geq (2\eta_0)^{13}\geq \eta^{13}$, from the definition of the parameter $\eta_i$.
It remains to verify that $\eta\geq \frac{c^*\log^{46}m}{\alpha^{10}(\alpha')^2}$, or, equivalently, that $\eta_0\geq \frac{c^*\log^{46}m}{2\alpha_0^{10}(\alpha_{i})^2}$.
 Recall that we set $\eta_0=\eta^4\cdot \log^9m=2^{O ((\log m)^{3/4} )}$, and we ensure that $\alpha_{i},\alpha_0\geq \alpha_{\ell}=1/(\log m)^{50\cdot\ell}=1/2^{O((\log m)^{1/4}\log\log m)}$. Since we assume that $m$ is large enough, the inequality clearly holds.
Therefore, all conditions of 
 \Cref{thm: find guiding paths}  hold, and we can apply the algorithm \algfindguiding to instance $I=(H,\hat \Sigma)$ of \cnwrs, 
 with the set $T'$ of terminals, the collection $\cset=\wset^{\bad}$ of clusters, and the parameters defined above.
 
 Recall that we perform $m$ such iterations. If, in every iteration, algorithm \algfindguiding returns FAIL, then we add cluster $\tilde C$ to set $\cset_{i+1}^{\bad}$.  We next show that, if Event $\event_{i}$ did not happen, 
 and $\tilde C$ is not an $\eta_{i+1}$-bad cluster in $G$, then the probability that $\tilde C$ is added to $\cset_{i+1}^{\bad}$ is small.
 
 \begin{claim}\label{claim: small probability of mistake for bad cluster}
 Let $\event_{i+1}(\tilde C)$ denote the bad event that $\tilde C$ is not an $\eta_{i+1}$-bad cluster, but our algorithm adds $\tilde C$ to  $\cset_{i+1}^{\bad}$. Then $\prob{\event_{i+1}(\tilde C)\mid \neg \event_i}\leq (3/4)^m$.
 \end{claim} 
 	
\begin{proof}
Consider a single iteration of the algorithm. Recall that, from \Cref{obs: opt is small},
$$\expect{\optcrors(H, \hat \Sigma)}\leq O\left(\beta_{i}^2\cdot \left (\optcrors(\tilde C,\Sigma_{\tilde C})+|E(\tilde C)|\right )\right ).$$ 
If cluster $\tilde C$ is not $\eta_{i+1}$-bad, then $|E(\tilde C)|+\optcrors(\tilde C,\Sigma_{\tilde C})<|T'|^2/\eta_{i+1}$. So for some constant $c'$: 
$$\expect{|E(H\setminus T')|+\optcrors(H(\tilde C), \hat \Sigma)}\leq c'\beta_{i}^2\cdot|T'|^2/\eta_{i+1}.$$
Let $\event'$ denote the event that $|E(H\setminus T')|+\optcrors(H, \hat \Sigma)>4c'\beta_{i}^2\cdot|T'|^2/\eta_{i+1}$. From Markov's bound, $\prob{\event'}\leq 1/4$.
Denote $k=|T'|$, and recall that we have set $\eta'=\eta_{i}$, $\alpha=\alpha_0$, and $\alpha'=\alpha_{i}$. 
Since $\eta_{i+1}\geq \eta_{i}\cdot \beta^3_{i}$, and 
$\beta_{i}=\frac{(\log m)^{56}}{(\alpha_0)^{12}\cdot (\alpha_{i})^8}\cdot \beta_{i-1}$,
we get that:
\[\frac{4c'\beta_{i}^2}{\eta_{i+1}}\leq \frac{4c'}{\beta_{i}\eta_{i}}\leq \frac{(\alpha_0)^{12}\cdot (\alpha_{i})^8}{c_0\eta_{i}\log^{50}m}.\]
We conclude that, if event $\event'$ did not happen, and $\tilde C$ is not an $\eta_{i+1}$-bad cluster, then $|E(H\setminus T')|+\optcrors(H, \hat \Sigma)\leq \frac{(k\alpha^4 \alpha')^2}{c_0\eta'\log^{50}m}$.
Let $\event''$ be the bad event that the algorithm  \algfindguiding returned FAIL. From \Cref{thm: find guiding paths}, if cluster $\tilde C$ is not $\eta_{i+1}$-bad, then $\prob{\event''\mid \neg \event'\band \neg\event_{i}}\leq 1/2$.

 Overall, assuming that the event $\event_{i}$ did not happen and cluster $\tilde C$ is not $\eta_i$-bad, then the algorithm \algfindguiding may only return FAIL if either $\event'$ or $\event''$ happen, which, from the above discussion, happens with probability at most $(3/4)$. Overall, since we repeat the above algorithm $m$ times, the probability that in every iteration the algorithm \algfindguiding returns FAIL is at most $(3/4)^m$.
\end{proof}

Assume now that, in any one of the iterations, the algorithm \algfindguiding did not return FAIL, and instead returned 
a distribution $\dset$ over the routers of $\Lambda(H,T')$, such that for every edge $e\in E(H)$, $\expect[\qset\sim\dset]{(\cong(\qset,e))^2}\leq  O\left (\frac{\log^{32}m}{\alpha_0^{12}\alpha_{i}^8}\right )$.
We now provide a distribution $\dset(\tilde C)$ over the set $\Lambda_G(\tilde C)$ of internal $\tilde C$-routers, such that $\tilde C$ is $\beta_{i+1}$-light with respect to $\dset(\tilde C)$. The distribution is provided implicitly: that is, we provide an efficient algorithm for drawing an internal $\tilde C$-router from the distribution.

In order to draw an internal $\tilde C$-router from distribution $\dset(\tilde C)$, we start by choosing a router $\qset\in \Lambda(H,T')$ from the distribution $\dset$. Let $x$ be the center vertex of $\qset$, so $x$ is a vertex of $V(H\setminus T')$, and $\qset$ is a collection of paths in $H$, routing all terminals in $T'$ to $x$. Equivalently, we can view $\qset$ as a collection of paths that route the edges of $\delta_G(\tilde C)$ to the vertex $x$, in the contracted graph $\tilde C_{|\wset^{\light}(C)}\cup \delta_G(\tilde C)$. Additionally, for every cluster $W\in \wset^{\light}(C)$, we select an internal $W$-router $\qset(W)\in \Lambda_G(W)$ from the distribution $\dset(W)$, and we denote by $u(W)\in V(W)$ its center vertex.

Assume first that $x$ is a regular vertex in graph $H$, that is, it is not a supernode representing a cluster of $\wset^{\light}(C)$. In this case, we set $u(\tilde C)=x$, and we will use $u(\tilde C)$ as the center vertex for internal router $\qset(\tilde C)\in \Lambda_G(\tilde C)$ that we construct. Otherwise, if $x=v_W$ for some cluster $W\in \wset^{\light}(C)$, then we set $u(\tilde C)=u(W)$, where $u(W)$ is the center vertex of the internal router $\qset(W)\in \Lambda_G(W)$ that we have selected for cluster $W$.

Next, we consider every path $Q\in \qset$ one by one. Let $Q$ be any such path, and assume that the first edge on $Q$ is $e\in \delta_G(\tilde C)$. We transform $Q$ into a path $Q'$ connecting $e$ to $u(\tilde C)$ in $G$, as follows. We consider supernodes $v_{W'}$ that lie on $Q$ one by one. For any such supernode $v_{W'}$ that is an inner vertex of $Q$, we let $e',e''$ be the two edges that appear immediately before and immediately after $v_{W'}$ on $Q$. Observe that $e',e''\in\delta_G(W')$. Therefore, there is a path $P(e')\in \qset(W')$ connecting $e'$ to $u(W')$, whose inner vertices lie in $W'$, and a path $P(e'')\in \qset(W')$ connecting $e''$ to $u(W')$, whose inner vertices lie in $W'$. By concatenating these two paths, we obtain a path $P^*(Q,W')$, whose first edge is $e'$, last edge is $e''$, and all remaining edges lie in $W'$. We then replace the segment of path $Q$ consisting of the edges $e',e''$ with the path $P^*(Q,W')$. Lastly, if $v_W$ is the last vertex on path $Q$ (in which case $x=v_W$), then we let $e'$ be the last edge on $Q$. Notice that $e'\in \delta_G(W)$ must hold. Then there must be a path $P(e')\in \qset(W)$, whose first edge is $e'$ and last vertex is $u(W)=u(\tilde C)$. We then replace the edge $e'$ on path $Q$ with the path $P(e')$. Let $Q'$ be the final path that is otbained from $Q$ after this transformation. Then $Q'$ is a path in graph $G$, whose first edge is $e$, last vertex is $u(\tilde C)$, and all inner edges and vertices are contained in $\tilde C$.

Lastly, we let $\qset(\tilde C)=\set{Q'\mid Q\in \qset}$ be the resulting router in $\Lambda_G(\tilde C)$.
This finishes the definition of the distribution $\dset(\tilde C)$ over the set $\Lambda(\tilde C)$ of internal $\tilde C$-routers. Notice that we do not provide the distribution explicitly, and instead we have described an algorithm that, given access to distribution $\dset$ computed by the algorithm \algfindguiding, and distributions $\set{\dset(W)}$ for clusters $W\in \wset^{\light}(C)$ (that may also be given implicitly), samples an internal $\tilde C$-router from the distribution $\dset(\tilde C)$.
We add cluster $\tilde C$ to set $\cset_{i+1}^{\gd}$, together with the distribution $\dset(\tilde C)$.
 It now remains to show that cluster $\tilde C$ is $\beta_{i+1}$-light with respect to the distribution $\dset(\tilde C)$, which we do in the following claim.
 
 \begin{claim}\label{claim: good cluster is good}
 	Cluster $\tilde C$ is $\beta_{i+1}$-light with respect to the distribution $\dset(\tilde C)$.
 \end{claim}

\begin{proof}
Consider some edge $e\in E(\tilde C)$. Assume first that edge $e$ does not lie in any cluster $W\in \wset^{\gd}(C)$. In this case:
\[\expect[\qset(\tilde C)\sim \dset(\tilde C)]{(\cong_G(\qset(\tilde C),e))^2}=\expect[\qset\sim \dset]{(\cong_{H}(\qset,e))^2}\leq   O\left (\frac{\log^{32}m}{\alpha_0^{12}(\alpha_{i})^8}\right ), \]
from \Cref{thm: find guiding paths}. Since $\beta_{i+1}=\frac{(\log m)^{56}}{(\alpha_0)^{12}\cdot (\alpha_{i+1})^8}\cdot \beta_{i}$, and $\alpha_{i+1}\leq \alpha_i$, we get that: 
$$\expect[\qset(\tilde C)\sim \dset(\tilde C)]{(\cong_G(\qset(\tilde C),e))^2}\leq \beta_{i+1}.$$
Next, we assume that $e$ lies in some cluster $W\in \wset^{\gd}(C)$. In order to analyze $\expect{(\cong_G(\qset(\tilde C),e))^2}$, we consider the following two-step process. In the first step, we select an internal $W$-router $\qset(W)\in \Lambda(W)$ from the distribution $\dset(W)$, and denote its center vertex by $u(W)$. Then, in the second step, we select a router $\qset\in \Lambda(H,T')$ from distribution $\dset$. Lastly, composing the paths in $\qset$ with the paths in $\qset(W)$, similarly to our construction of the final set $\qset(\tilde C)$ of paths, will establish the final congestion on edge $e$. 
	
Let $\qset(W)\in \Lambda(W)$ be the internal $W$-router that was chosen from distribution $\dset(W)$, and assume that the paths in $\qset(W)$ cause congestion $z$ on edge $e$. We denote $\qset(W)=\set{Q(e')\mid e'\in \delta_G(W)}$, where for every edge $e'\in \delta_G(W)$, path $Q(e')$ originates at edge $e'$ and terminates at vertex $u(W)$. Let $E'\subseteq \delta_G(W)$ be the set of edges $e'$ whose corresponding path $Q(e')$ contains the edge $e$, so  $|E'|=z$. Denoting $E'=\set{e_1,\ldots,e_z}$, and assuming that the set $\qset(W)$ of paths is fixed, we can now write:
\[
\begin{split}
\expect[\qset\sim \dset]{(\cong_{G}(\qset(\tilde C),e))^2}&=\expect[\qset\sim \dset]{\left(\sum_{i=1}^z\cong_{H}(\qset,e_i)\right )^2 }  \\
&\leq \expect[\qset\sim \dset]{\sum_{i=1}^z2z\cdot(\cong_{H}(\qset,e_i))^2 }\\
&\leq z^2\cdot O\left (\frac{\log^{32}m}{\alpha_0^{12}\alpha_i^8}\right ).
\end{split} \]
Recall that $z$ is the congestion caused by the set $\qset(W)$ of paths on edge $e$. Therefore, overall:
\[\begin{split}
	\expect[\qset(\tilde C) \sim \dset(\tilde C)]{(\cong_G(\qset(\tilde C),e))^2}
	&\leq \expect[\qset(W)\sim \dset(W)]{(\cong_G(\qset(W),e))^2}\cdot O\left (\frac{\log^{32}m}{\alpha_0^{12}\alpha_i^8}\right )\\
	&\leq \beta_{i}\cdot O\left (\frac{\log^{32}m}{\alpha_0^{12}\alpha_i^8}\right ) \leq \beta_{i+1}.
	\end{split}
 \]
(here we have used the fact that cluster $W$ is $\beta_{i}$-light with respect to distribution $\dset(W)$, that $\beta_{i+1}=\frac{(\log m)^{56}}{(\alpha_0)^{12}\cdot (\alpha_{i+1})^8}\cdot \beta_{i}$), and that $\alpha_{i+1}\leq \alpha_i$.
\end{proof}

We now summarize the second step of the algorithm. First, from \Cref{claim: good cluster is good}, every cluster $\tilde C$ that we have added to set $\cset_{i+1}^{\light}$ over the course of Step 2 is a $\beta_{i+1}$-light cluster with respect to the distribution $\dset(\tilde C)$ over the set $\Lambda_G(\tilde C)$ of internal $\tilde C$-routers that we have defined. Let $\event_{i+1}$ be the bad event that any cluster $\tilde C$ that was added to set $\cset_{i+1}^{\bad}$ over the course of the current phase is not $\eta_{i+1}$-bad. 
From the discussion above, event $\event_{i+1}$ may only happen if either event $\event_i$ happened, or there is some cluster $C\in \cset_1'$, for which event $\event_{i+1}(\tilde C)$ happened. From \Cref{claim: small probability of mistake for bad cluster}, and Property \ref{prop 4 last of clusters bad}, the probability of $\event_{i+1}$ is bounded by $\prob{\event_i}+m\cdot (3/4)^m\leq i/m^{10}+m\cdot (3/4)^m\leq (i+1)/m^{10}$, since we have assumed that $m$ is large enough.

Lastly, if the current phase is the last phase, that is, $|E(G')|\leq 16\eta k\log m$ holds, then set $\cset_1'$ contains a single cluster $C=G'\setminus T$.
If our algorithm added the corresponding cluster $\tilde C=G\setminus T$ to set $\cset_{i+1}^{\bad}$, then we return FAIL. Notice that, if $\optcrors(G,\Sigma)+|E(G)\setminus T|< k^2/\eta^*\leq k^2/\eta_{\ell}\leq k^2/\eta_{i+1}$, then the algorithm may only return FAIL if event $\event_{i+1}$ happened, with may only happen with probability at most $\frac{i+1}{m^{10}}<\frac 1 2$. Otherwise, our algorithm added cluster $\tilde C$ to set $\cset_{i+1}^{\light}$, and constructed a distribution $\dset(\tilde C)$ over the set $\Lambda(\tilde C)$ of internal $\tilde C$-routers, such that cluster $\tilde C$ is $\beta_{i+1}$-light with respect to $\dset(\tilde C)$. Notice that $\dset(\tilde C)$ can also be viewed as a distribution over the routers of $\Lambda(G,T)$, and we are guaranteed that, for every edge $e\in E(G)$, $\expect[\qset\sim \dset(\tilde C)]{(\cong_{G}(\qset,e))^2}\leq \beta_{\ell}\leq \beta^*$. From Property \ref{prop 2 of clusters small contracted graph}, after $(\ell-1)$ phases the algorithm terminates, for $\ell=O\left(\frac{\log m}{\log \eta}\right )$.

\section{Third Main Tool - Advanced Disengagement}
\label{sec: main disengagement}

The goal of this section is to prove the following theorem that allows us to perform disengagement in a more general setting than that from basic disengagement.


\begin{theorem}\label{thm: disengagement - main}
There is an efficient randomized algorithm, called \algadvanceddisengagement, whose input consists of an instance $I=(G,\Sigma)$ of \CNwRS, parameters $m$ and $\mu\geq 2^{c^*(\log m)^{7/8}\log\log m}$ for some large enough constant $c^*$, and a collection $\cset$ of disjoint clusters of $G$, for which the following hold:

\begin{itemize}
	\item $|V(G)|,|E(G)|\leq m$, and $m$ is greater than a sufficiently large constant;
	\item every cluster $C\in \cset$ has the $\alpha_0$-bandwidth property, for $\alpha_0=1/\log^3m$;
	\item $\bigcup_{C\in \cset}V(C)=V(G)$; and
	\item $\sum_{C\in \cset}|\delta_G(C)|\leq |E(G)|/\mu^{0.1}$. 
\end{itemize} 

The algorithm computes a $2^{O((\log m)^{3/4}\log\log m)}$-decomposition $\iset$ of instance $I$, such that every instance $I'\in \iset$ is a subinstance of $I$. Moreover, for each resulting instance $I'=(G',\Sigma')\in \iset$, there is at most one cluster $C\in \cset$ with $E(C)\subseteq E(G')$. If such a cluster exists, then $E(G')\subseteq E(C)\cup \Eout(\cset)$, and otherwise $E(G')\subseteq \Eout(\cset)$.
\end{theorem}

The remainder of this section is dedicated to the proof of \Cref{thm: disengagement - main}. 

Over the course of the proof, we will consider subinstances of the input instance $I$. 
Recall that an instance $I'=(G',\Sigma')$ of \cnwrs is a \emph{subinstance} of instance $I=(G,\Sigma)$  (see Definition \ref{def: subinstance}), if there is a subgraph $\tilde G\subseteq G$, and a collection $\rset$ of mutually disjoint subsets of vertices of $\tilde G$, such that graph $G'$ can be obtained from $\tilde G$ by contracting, for all $R\in \rset$, vertex set $R$ into a supernode $v_R$;  we keep parallel edges but remove self-loops. We do not distiguish between edges of $G'$ incident to supernodes and their corresponding edges in the original graph $G$. We call the non-supernode vertices of $G'$ \emph{regular vertices}.
We also require that, for every regular vertex $v\in V(G')\cap V(G)$, its rotation  $\oset'_v$ in $\Sigma'$ is the same as the rotation $\oset_v\in \Sigma$. For each supernode $v_R$, its rotation $\oset'_{v_R}$ can be defined arbitrarily.
We will consider special types of subinstances of a given instance, that we call \emph{canonical} subinstances.

\begin{definition}[Canonical Subinstances]
	Let $I'=(G',\Sigma')$ be an instance of \cnwrs, and let $\cset'$ be a collection of disjoint clusters $G'$. We say that instance $I''=(G'',\Sigma'')$ is a \emph{canonical} subinstance of $I'$ with respect to $\cset'$ if $I''$ is a subinstance of $I'$, and moreover, if $\tilde G\subseteq G'$, and $\rset$ is a collection of disjoint subsets of vertices of $\tilde G$, such that $G''$ is obtained from $\tilde G$ by contracting every vertex set $R\in \rset$ into a supernode $v_R$, then the following holds: 
		For every cluster $C\in \cset'$, either (i) there is some vertex set $R\in \rset$ with $V(C)\subseteq R$ (in which case we say that $C$ is \emph{contracted} in graph $G''$); or (ii) $C\subseteq \tilde G$, and for every vertex set $R\in \rset$, $R\cap V(C)=\emptyset$ (in which case we say that $C$ is \emph{not contracted} in $G''$); or (iii) $V(C)\cap V(\tilde G)=\emptyset$. 
	\end{definition}

We will consider canonical subinstances of instance $I$ with additional useful properties. We call such subinstances \emph{nice subinstances}. For each such subinstance, we will use a specific \emph{witness structure} to certify that it is indeed a nice subinstance. We will provide two theorems: the first theorem will be used to decompose the input instance $I$ into a collection of nice subinstances, and the second theorem will further decompose each resulting nice subinstance into a collection of subinstances that have the properties required by \Cref{thm: disengagement - main}. In the next subsection, we define the witness structure and the nice subinstances, and then provide the statements of the two  theorems that will allow us to complete the proof of \Cref{thm: disengagement - main}.

\subsection{Nice Witness Structure, Nice Subinstances, and Statements of Main Theorems}
\begin{figure}[h]
	\centering
	\includegraphics[scale=0.23]{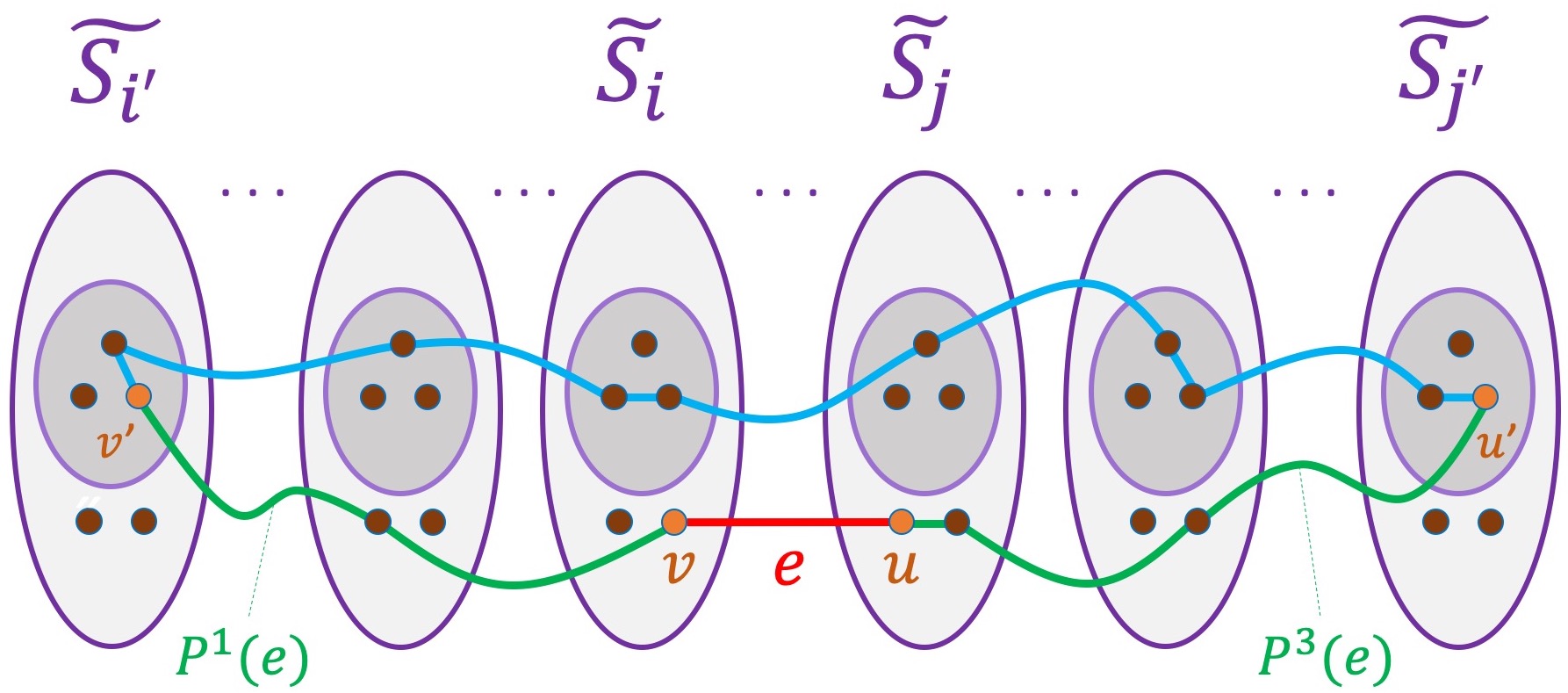}
	\caption{An illustration of a nice witness structure and a nice guiding path. An edge $e\in \hat E$ is shown in red. The prefix $P^1(e)$ and the suffix $P^3(e)$ of the nice guiding path $P(e)$ are shown in green, and the mid-part $P^2(e)$ is shown in blue.}\label{fig: witness_structure}
\end{figure}

Let $G'$ be a graph, let $m>0$ be an integer with $|E(G')|\leq m$, and let $\cset'$ be a collection of disjoint clusters of $G'$. A \emph{nice witness structure} for $G'$ with respect to $\cset'$ consists of the following three main ingredients (see \Cref{fig: witness_structure}):

\begin{enumerate}
	\item The first ingredient is a sequence $\tilde \sset=\set{\tilde S_1,\ldots,\tilde S_r}$ of disjoint vertex-induced subgraphs of $G'$, such that $\bigcup_{i=1}^rV(\tilde S_i)=V(G')$, and, for all $1\leq i\le r$, a cluster $\tilde S'_i\subseteq \tilde S_i$ that has the $\alpha^*=\Omega(1/\log^{12}m)$-bandwidth property in $G'$. We require that, for all $1\leq i\leq r$, there is at most one cluster $C \in\cset'$ with $C\subseteq \tilde S'_i$. Moreover, if such cluster $C$ exists then $E(\tilde S_i)\subseteq E(C)\cup E(G'_{|\cset'})$ must hold, and otherwise $E(\tilde S_i)\subseteq E(G'_{|\cset'})$. We also require that for each cluster $C\in \cset'$, there is an index $1\leq i\leq r$, such that $C\subseteq \tilde S_i'$. We refer to the sequence $\tilde \sset=\set{\tilde S_1,\ldots,\tilde S_r}$ as \emph{the backbone} of the nice witness structure, and to the clusters in $\tilde \sset'= \set{\tilde S'_1,\ldots,\tilde S'_r}$  as its \emph{verterbrae}. 
	
	\item The second ingredient is a partition of the edges of $E(G')$ into two disjoint subsets, $\tilde E'$ and $\tilde E''$. Set $\tilde E'$ contains all edges of $\bigcup_{i=1}^rE(\tilde S_i')$, and, additionally, for all $1\leq i<r$, it contains every edge $e=(u,v)$ with $u\in \tilde S_i'$, $v\in \tilde S_{i+1}'$. 
	Set $\tilde E''$ contains all remaining edges of $E(G')$.
	Additionally, we let $\hat E\subseteq \tilde E''$ be the set of all edges $(u,v)\in \tilde E''$, where $u$ and $v$ lie in different clusters of $\set{\tilde S_1,\ldots,\tilde S_r}$.
	
	\item The third ingerdient is a set $\hat \pset=\set{P(e)\mid e\in \hat E}$ of paths, that cause congestion at most $O(\log^{18} m)$ in $G'$ that we call \emph{nice guiding paths}. For each edge $e=(v,u)\in \hat E$, if we assume that $v\in \tilde S_i$, $u\in \tilde S_j$, and $i<j$, then path $P(e)$ connects vertex $v$ to vertex $u$, does not contain the edge $e$, and consists of three subpaths $P^1(e)$, $P^2(e)$ and $P^3(e)$, that have the following properties:
	
	\begin{itemize}
		\item There is an index $i'\leq i$, such that path $P^1(e)$ originates at vertex $v$ and terminates at some vertex $v'\in \tilde S'_{i'}$. Path $P^1(e)$ must be simple, and no vertex of $P^1(e)\setminus\set{v'}$ may lie in $\bigcup_{z=1}^rV(\tilde S_z')$.  Moreover, if we denote the sequence of vertices on path $P^1(e)$ by $v=v_0,v_1,\ldots,v_q=v'$, and, for all $0\leq z\leq q$, we assume that $v_z\in \tilde S_{i_z}$, then $i=i_0\geq i_1\geq\cdots\geq i_q=i'$. In other words, the path visits the sets $\tilde S_{a}$ in the non-increasing order of index $a$, possibly skipping over some of the indices.

		\item Similarly, there is an index $j'\geq j$, such that path $P^3(e)$ originates at vertex $u$ and terminates at some vertex $u'\in \tilde S'_{j'}$. Path $P^3(e)$ must be simple, and no vertex of $P^3(e)\setminus\set{u'}$ may lie in $\bigcup_{z=1}^rV(\tilde S_z')$.  Moreover, if we denote the sequence of vertices on path $P^3(e)$ by $u=u_0,u_1,\ldots,u_{q'}=u'$, and, for all $0\leq z'\leq q'$, we assume that $u_{z'}\in \tilde S_{j_{z'}}$, then $j=j_0\leq j_1\leq\cdots\leq j_{q'}=j'$. In other words, the path visits the sets $\tilde S_{a}$ in the non-decreasing order of index $a$, possibly skipping over some of the indices.
		
		\item Lastly, path $P^2(e)$ connects $v'$ to $u'$. It may only use edges of $\tilde E'$, and it can be partitioned into disjoint subpaths $Q_{i'}(e),Q_{i'+1}(e),\ldots,Q_{j'}(e)$, where for all $i'\leq x\leq j'$, $Q_x(e)\subseteq \tilde S'_x$, and $\bigcup_{i'\leq x\leq j'}V(Q_x(e))=V(P^2(e))$.
	\end{itemize}
\end{enumerate}

Note that, by definition, for every edge $e\in \hat E$, the paths $P^1(e)$ and $P^3(e)$ only use edges of $\tilde E''$, while path $P^2(e)$ only uses edges of $\tilde E'$. If $e=(v,u)\in \hat E$ is an edge for which $v\in \tilde S'_i$ holds, then $i'=i$ and $P^1(e)=\set{v}$ must hold. Similarly, if $u\in \tilde S'_j$, then $j'=j$ and $P^2(e)=\set{u}$.

Clearly, the edge sets $\tilde E',\tilde E'',\hat E$ in the nice witness structure are completely determined by the sequences $\tilde \sset=\set{\tilde S_1,\ldots,\tilde S_r}$ and $\tilde \sset'=\set{\tilde S'_1,\ldots,\tilde S'_r}$  of clusters. Therefore, the nice witness structure is completely determined by $\tilde \sset,\tilde\sset'$, and the set $\hat \pset=\set{P(e)\mid e\in \hat E}$ of nice guiding paths. We will use the shorthand $(\tilde \sset,\tilde \sset',\hat \pset)$ for a nice witness structure. For a path $P(e)\in \hat \pset$, we sometimes refer to $P^1(e),P^3(e)$ and $P^2(e)$ as the \emph{prefix}, the \emph{suffix}, and the \emph{mid-part} of path $P(e)$, respectively.
This completes the definition of a nice witness structure. Next, we define nice subinstances of instance $I$.

Consider a subinsance $I'=(G',\Sigma')$ of the input instance $I$, and assume that $I'$ is a canonical subinstance of $I$ with respect to the set $\cset$ of clusters. Recall that, from the definition of canonical subinstances, we are guaranteed that for every cluster $C\in \cset$, either $C\subseteq G'$, or $V(C)\cap V(G')=\emptyset$. We denote by $\cset(G')$ the set of all clusters $C\in \cset$ with $C\subseteq G'$.
Lastly, we say that a subinstance $I'=(G',\Sigma')$ of $I$ is a \emph{nice subinstance} of $I$ with respect to $\cset$, if it is a canonical subinstance with respect to $\cset$, and there is a nice witness structure for graph $G'$ with respect to the set $\cset'=\cset(G')$ of its clusters.
The remainder of the proof of \Cref{thm: disengagement - main} uses the following two theorems. The first theorem allows us to decompose a given instance $I$ into a collection of nice subinstances.

\begin{theorem}\label{thm: advanced disengagement get nice instances}
There is an efficient randomized algorithm, whose input consists of an instance $I=(G,\Sigma)$ of \CNwRS, parameters $m$ and $\mu\geq 2^{c^*(\log m)^{7/8}\log\log m}$ for some large enough constant $c^*$, and a collection $\cset$ of disjoint clusters of $G$, for which the following hold:

\begin{itemize}
	\item $|V(G)|,|E(G)|\leq m$, and $m$ is greater than a sufficiently large constant;
	\item every cluster $C\in \cset$ has the $\alpha_0$-bandwidth property, for $\alpha_0=1/\log^3m$;
	\item $\bigcup_{C\in \cset}V(C)=V(G)$; and
	\item $\sum_{C\in \cset}|\delta_G(C)|\leq |E(G)|/\mu^{0.1}$. 
\end{itemize} 

The algorithm either returns FAIL, or it computes a $2^{O((\log m)^{3/4}\log\log m)}$-decomposition $\iset_1$ of instance $I$, such that each resulting instance $I'=(G',\Sigma')\in \iset_1$ is a nice subinstance of $I$ with respect to $\cset$. In the latter case, the algorithm also computes, for each instance $(G',\Sigma')\in \iset_1$,  a nice witness structure for graph $G'$ with respect to the set $\cset(G')$ of clusters. The probability that the algorithm returns FAIL is at most $1/m^6$.
\end{theorem}

The second theorem allows us to further decompose nice subinstances of instance $I$ into subinstances that have the desired properties.

\begin{theorem}\label{thm: advanced disengagement - disengage nice instances}
	There is an efficient randomized algorithm, whose input consists of:
	
	\begin{itemize}
		\item an instance $I'=(G',\Sigma')$ of \cnwrs;
		\item a parameter $m$, such that $|E(G')|\leq m$;
		\item a collection $\cset'$ of disjont clusters of $G'$; and 
		\item a nice witness structure $(\tilde \sset,\tilde \sset',\hat\pset)$ for graph $G'$ with respect to the set $\cset'$ of clusters.
	\end{itemize} 

The algorithm either returns FAIL, or  computes a $2^{O((\log m)^{3/4}\log\log m)}$-decomposition $\iset_2(I')$ of instance $I'$, such that each resulting instance $I''=(G'',\Sigma'')\in \iset_2(I')$ is a subinstance of $I'$, and moreover,  there is at most one cluster $C\in \cset'$ with $E(C)\subseteq E(G'')$; if such a cluster exists then $E(G'')\subseteq E(C)\cup E(G'_{|\cset'})$ holds, and otherwise $E(G'')\subseteq E(G'_{|\cset'})$. The probability that the algorithm returns FAIL is $1/m^{6}$.
\end{theorem}

Note that \Cref{thm: disengagement - main} immediately follows from \Cref{thm: advanced disengagement get nice instances} and \Cref{thm: advanced disengagement - disengage nice instances}. Indeed, we start by applying the algorithm from  \Cref{thm: advanced disengagement get nice instances}  to the input instance $I$ and the collection $\cset$ of its clusters. Assume for now that the algorithm did not return FAIL. Then we obtain a collection $\iset_1$ of nice subinstances of $I$, and, for each instance $I'=(G',\Sigma')\in \iset_1$, a nice witness structure for $G'$ with respect to cluster set $\cset(G')$. 
From the definition of a nice subinstance, for every cluster $C\in \cset(G')$, $C\subseteq G'$, and for every cluster $C\in \cset\setminus\cset(G')$, $V(C)\cap V(G')=\emptyset$, so $E(G')\subseteq \left (\bigcup_{C\in \cset(G')}E(C)\right )\cup \Eout(\cset)$.

We then apply
the algorithm from \Cref{thm: advanced disengagement - disengage nice instances} to each such instance $I'=(G',\Sigma')\in \iset_1$ and the corresponding nice witness structure. Assume for now that this algorithm did not return FAIL. Then we obtain a collection $\iset_2(I')$ of subinstances of $I'$. We are guaranteed that, for each resulting instance $I''=(G'',\Sigma'')\in \iset_2(I')$,  there is at most one cluster $C\in \cset(G')$ with $E(C)\subseteq E(G'')$. If such a cluster exists, then $E(G'')\subseteq  E(C)\cup E(G'_{|\cset(G')})\subseteq  E(C)\cup \Eout(\cset)$ holds, and otherwise $E(G'')\subseteq E(G'_{|\cset(G')})\subseteq \Eout(\cset')$, since $E(G')\subseteq \left (\bigcup_{C'\in \cset(G')}E(C')\right )\cup \Eout(\cset)$.
If the algorithm from \Cref{thm: advanced disengagement get nice instances} did not return FAIL, and neither application of the algorithm from  \Cref{thm: advanced disengagement - disengage nice instances} returned FAIL, then we return the collection of instances $\iset=\bigcup_{I'\in \iset_1}\iset_2(I')$. From \Cref{claim: compose algs}, we obtain a randomized algorithm that computes a $2^{O((\log m)^{3/4}\log\log m)}$-decomposition  $\iset$ of the input instance $I$. 

It now remains to consider a case where the algorithm from \Cref{thm: advanced disengagement get nice instances} or any of the applications of the algorithm from  \Cref{thm: advanced disengagement - disengage nice instances} returned FAIL (which may only happen with probability at most $1/m^4$). In this case, we construct the collection $\iset$ of subinstances of $I$ directly, as follows. For every cluster $C\in \cset$, we let $\oset(C)$ be an arbitrary circular ordering of the edges of $\delta_G(C)$. Set $\iset$ will contain one global instance $\hat I=(\hat G,\hat \Sigma)$, and, for each cluster $C\in \cset$, a cluster-based instance $I_C=(G_C,\Sigma_C)$. Consider first a cluster $C\in \cset$. We let $G_C$ be the graph obtained from $G$ by contracting all vertices of $V(G)\setminus V(C)$ into a supernode $u_C$. We define the rotation system $\Sigma_C$ for graph $G_C$ as follows: for every vertex $v\in V(C)$, its rotation $\oset_v$ in $\Sigma_C$ remains the same as that in $\Sigma$. Observe that $\delta_{G_C}(u_C)=\delta_G(C)$. The rotation $\oset_{u_C}$ of vertex $u_C$ in $\Sigma_C$ is defined to be $\oset(C)$. This completes the definition of the cluster-based instance $I_C=(G_C,\Sigma_C)$. We now define the global instance $\hat I=(\hat G,\hat \Sigma)$. Graph $\hat G$ is obtained from graph $G$ by contracting, for every cluster $C\in \cset$, the set $V(C)$ of vertices into a supernode $u'_C$. Notice that the set of edges incident to $u'_C$ in $\hat G$ is precisely $\delta_G(C)$. We then define a rotation of $u'_C$ in $\hat \Sigma$ to be $\oset(C)$. This completes the definition of the global instance $\hat I$. Consider now the resulting collection $\iset$ of subinstances of $I$. It is immediate to verify that $\sum_{(G',\Sigma')\in \iset}|E(G')|\leq O(|E(G)|)$. 
Assume now that we are given, for each instance $I'\in \iset$, a feasible solution $\phi(I')$. We can combine these solutions together to obtain a solution $\phi$ to instance $I$, of cost at most 
$O\left (\sum_{I'\in \iset}\cro(\phi(I'))\right )$,  by employing an algorithm similar to that from \Cref{lem: basic disengagement combining solutions} (the algorithm that was used for basic disengagement). Lastly, from \Cref{thm: crwrs_uncrossing}, it is easy to verify that
$\sum_{I'\in \iset}\optcrors(I')\le O(m^2)$. Since the probability that the algorithm from \Cref{thm: advanced disengagement get nice instances}, or any of the applications of the algorithm from  \Cref{thm: advanced disengagement - disengage nice instances} return FAIL  is at most $1/m^4$, overall we have obtained a randomized algorithm that computes a $2^{O((\log m)^{3/4}\log\log m)}$-decomposition  $\iset$ of the input instance $I$ with required properties.

In order to complete the proof of \Cref{thm: disengagement - main}, it is now enough to prove \Cref{thm: advanced disengagement get nice instances} and \Cref{thm: advanced disengagement - disengage nice instances}, which we do in Sections \ref{sec: advanced disengagement - get nice instances} and \ref{subsec: disengagement with bad chain}, respectively.

\subsection{Decomposition into Nice Instances -- Proof of \Cref{thm: advanced disengagement get nice instances}}
\label{sec: advanced disengagement - get nice instances}

This subsection is dedicated to the proof of \Cref{thm: advanced disengagement get nice instances}. The main idea of the proof is to carefully construct a laminar family $\lset$ of clusters of $G$, whose depth is $2^{O((\log m)^{3/4}\log\log m)}$, and then apply Algorithm \algbasicdisengagement  from \Cref{subsec: basic disengagement}, to compute a $2^{O((\log m)^{3/4}\log\log m)}$-decomposition $\iset_1$ of instance $I$ via basic disengagement, using the laminar family $\lset$. The main challenge is to construct the laminar family $\lset$ in such a way that each resulting instance $I'=(G',\Sigma')\in \iset_1$ is a nice subinstance of $I$ with respect to $\cset$, and to compute a nice witness structure for each such graph $G'$.
We will construct the laminar family gradually, in the top-bottom fashion, using the notion of \emph{legal clustering}. 

In order to define the legal clustering, we consider a graph $G'$, together with a special vertex $v^*\in V(G)$. Intuitively, graph $G'$ represents some cluster $S\in \lset$ that we have constructed already, and it is a graph that is obtained from $G$ by contracting all vertices of $G\setminus S$ into the special vertex $v^*$. We will also consider the subset $\cset'\subseteq\cset$ of all clusters $C\in \cset$ with $C\subseteq S$. Intuitively, our goal is to construct a collection $\rset$ of disjoint clusters of $G'$, each of which must be a subgraph of $S$, that will then be added to $\lset$. Recall that, if $\lset$ is a laminar family of clusters of graph $G$, and $\iset_1$ is a collection of subinstances of $I$ obtained by decomposing $I$ via basic disengagement, then every cluster $S\in \lset$ has a subinstance $I(S)=(G(S),\Sigma(S))\in \iset_1$ associated with it. Graph  $G(S)$ is obtained from graph $G$ as follows. First, we contract the vertices of $V(G)\setminus V(S)$ into a supernode $v^*$, obtaining graph $G'$. Next, for every child-cluster $R\in \lset$ of $S$, we contract $R$ into a supernode $v_R$. Therefore, if $\rset$ is the set of child-clusters of $S$, then $G(S)=G'_{|\rset}$. Recall that we need to ensure that instance $I(S)$ is a nice instance. Given a graph $G'$, a special vertex $v^*$ in $G'$, and a collection $\cset'$ of disjoint basic clusters of $G'$, the notion of {legal clustering} of $G'$ with respect to $v^*$ and $\cset'$ is designed to ensure that every instance in our final decomposition $\iset_1$ of $I$, created via the process described above, is a nice instance.

Consider a graph $G'$ with a special vertex $v^*\in G'$.
We will consider clusters $R\subseteq G'\setminus \set{u^*}$. Recall we have defined a collection $\Lambda'_{G'}(R)$ of external routers for $R$, where each router $\qset'(R)\in \Lambda'_{G'}(R)$ is a collection of paths routing all edges of $\delta_{G'}(R)$ to a single vertex of $G'\setminus R$, such that all paths in $\qset'(R)$ are internally disjoint from $R$. We start by defining the notion of \emph{helpful clustering}, which will be used in the definition of legal clustering. We fix two parameters that will be used throughout this section: $\alpha_1=(\alpha_0)^2=1/\log^6m$ and $\beta=\log^{18}m$.

\begin{definition}[Helpful Clustering]
Let $G'$ be a graph with a special vertex $v^*\in V(G')$, and let $\cset'$ be a collection of disjoint vertex-induced subgraphs of $G'\setminus\set{v^*}$, that we call \emph{basic clusters}. Let $\rset$ be another collection of disjoint clusters of $G'$, and assume that for every cluster $R\in \rset$, we are given a distribution $\dset'(R)$ over the external routers in $\Lambda'_{G'}(R)$. We say that $(\rset,\set{\dset'(R)}_{R\in \rset})$ is a \emph{helpful clustering} of $G'$ with respect to $v^*$ and $\cset'$, iff the following conditions hold:

\begin{itemize}
	\item vertex $v^*$ does not belong to any of the clusters in $\rset$;
	\item for every basic cluster $C'\in \cset$, and for every cluster $R\in \rset$, either $C'\subseteq R$, or $V(C')\cap V(R)=\emptyset$; 
	\item every cluster $R\in \rset$ has the $\alpha_1$-bandwidth property in $G'$; and
	\item for every cluster $R\in \rset$, for every edge $e\in E(G')\setminus E(R)$, $\expect[\qset'(R)\sim\dset'(R)]{\cong_{G'}(\qset'(R),e)}\leq \beta$.
\end{itemize}
\end{definition}

Consider again a graph $G'$ with a special vertex $v^*\in V(G')$, and some cluster $R\subseteq G'$.
We say that an external $R$-router $\qset'(R)\in \Lambda'_{G'}(R)$ is \emph{careful} with respect to the special vertex $v^*$, if each edge of $\delta_{G'}(v^*)$ belongs to at most one path in $\qset'(R)$ (note that in general paths in $\qset'(R)$ may cause an arbitrarily large congestion in $G'$). We denote by $\Lambda''_{G'}(R)\subseteq \Lambda'_{G'}(R)$ the collection of all external $R$-routers $\qset'(R)$ that are careful with respect to $v^*$. We say that a distribution $\dset'(R)$ over the collection $\Lambda'_{G'}(R)$ of external $R$-routers is \emph{careful} with respect to $v^*$, if every router $\qset'(R)\in \Lambda'_{G'}(R)$ to which $\dset'(R)$ assigns a non-zero probability lies in $\Lambda''_{G'}(R)$. 

We will consider two different types of legal clustering. 
We start by defining the first, and the simpler type of legal clusterings.

\begin{definition}[Type-1 Legal Clustering]
	Let $G'$ be a graph with a special vertex $v^*\in V(G')$, and let $\cset'$ be a collection of disjoint vertex-induced subgraphs of $G'\setminus\set{v^*}$, that we call {basic clusters}. Let $\rset$ be another collection of disjoint clusters of $G'$, and assume that for every cluster $R\in \rset$, we are given a distribution $\dset'(R)$ over the external routers in $\Lambda'_{G'}(R)$. We say that $(\rset,\set{\dset'(R)}_{R\in \rset})$ is a \emph{type-1 legal clustering} of $G'$ with respect to $v^*$ and $\cset'$, if  the following conditions hold:
	
	\begin{itemize}
		\item  $(\rset,\set{\dset'(R)}_{R\in \rset})$ is a helpful clustering of $G'$ with respect to $v^*$ and $\cset'$;
		\item there is at most one cluster $C\in \cset'$, that is contained in $G'\setminus \left(\bigcup_{R\in \rset}R\right )$; 
  and
		\item for every cluster $R\in \rset$, distribution $\dset'(R)$ over  external routers is careful with respect to $v^*$.
	\end{itemize}
\end{definition}

While type-1 legal clustering would be ideal in order to construct the laminar family $\lset$ and to perform a basic disengagement of instance $I$ via $\lset$, we may not always succeed in computing a type-1 legal clustering of a given graph $G'$, and we may need to employ type-2 legal clustering, that is defined below, instead. Before we define the type-2 legal clustering formally, we provide some intuition. Type-2 legal clustering is defined somewhat similarly to type-1 legal clustering, except that we no longer require that, for every cluster $R\in \rset$, the distribution $\dset'(R)$ is careful with respect to $v^*$. We also no longer require that at most one cluster of $\cset'$ is contained in $G'\setminus \bigcup_{R\in \rset}R$. However, we require that, additionally, the decomposition provides a nice witness structure for the graph $G'_{|\rset}$ with respect to the set $\cset'(G'_{|\rset})$ of clusters (all clusers of $\cset'$ that are contained in graph $G'_{|\rset}$).
Unfortunately, the relaxation of the requirement that the distributions $\dset'(R)$ for clusters $R\in \rset$ is careful with respect to $v^*$ creates some major difficulties. For intuition, recall that we will construct the laminar family $\lset$ of clusters of $G$ gradually, in the top-bottom fashion. Assume that $S$ is some cluster of the current laminar family $\lset$, such that no cluster of $\lset$ is strictly contained in $S$. Let $G'$ be the graph obtained from $G$ by contracting all vertices of $V(G)\setminus V(S)$ into the special vertex $v^*$, and let $\cset'$ be the set of all clusters of $\cset$ that are contained in $S$. The idea of our algorithm is to compute a type-1 or a type-2 legal clustering $\rset$ in graph $G'$; assume that we compute a type-2 legal clustering. We then add the clusters of $\rset$ to the laminar family $\lset$, and continue to the next iteration. From the discussion so far, for each such cluster $R\in \rset$, the type-2 legal clustering provides a distribution $\dset'(R)$ over the  collection $\Lambda'_{G'}(R)$ of external routers in graph $G'$. However, in order to execute the basic disengagement via the laminar family $\lset$ (see \Cref{subsec: basic disengagement}), we need the distribution $\dset'(R)$ to be supported over the collection $\Lambda'_G(R)$ of external routers in graph $G$. In other words, the problem is that paths in sets $\qset'(R)\in \Lambda'_{G'}(R)$ that are assigned non-zero probability by $\dset'(R)$ may contain the special vertex $v^*$, which is not a vertex of $G$. Recall however that special vertex $v^*$ represents the cluster $G\setminus S$, and so edges incident to $v^*$ in $G'$ are precisely the edges of $\delta_G(S)$. Therefore, we could exploit the distribution $\dset'(S)$ over the external routers for cluster $S$ in $G$, in order to get rid of the special vertex $v^*$ on the paths of $\qset'(R)$, where $\qset'(R)\in \Lambda'_{G'}(R)$. In other words, by composing the distributions $\dset'(R)$ and $\dset'(S)$, we could obtain the desired distribution over the set $\Lambda'_{G}(R)$ of external routers for cluster $R$ in the original graph $G$. Unfortunately, this kind of recursive composition of distributions may lead to an explosion in  the congestion of the resulting sets of paths. Even if the depth of the laminar family $\lset $ is quite modest (say $O(\log m)$), we may obtain distributions $\dset''(R)$ over the routers in $\Lambda'_G(R)$, for which the maximum expected congestion on an edge of $G$ may be as large as $|\delta_G(R)|$, which is unacceptable. If we could ensure that the distributions $\dset'(R)$ obtained in type-2 legal clustering are careful with respect to $v^*$, then this accumulation of congestion could be avoided, but unfortunately we do not know how to ensure that. In order to overcome this difficulty, we will carefully alternate between type-1 and type-2 legal clusterings. Specifically, we will require that a type-2 legal clustering contains a single distinguished cluster $R^*$, whose corresponding distribution $\dset'(R^*)$ is careful with respect to $v^*$, and that $R^*$ contains a very large fraction of clusters of $\cset'$. We will also require that a type-1 legal clustering $\rset'$ of the graph associated with cluster $R^*$ is provided, and that for each cluster $R'\in \rset'$, the number of clusters of $\cset'$ contained in $R'$ is relatively small. This carefull alternation between type-1 and  type-2 legal clusterings will allow us to compute distributions $\dset'(S)$ over the routers of $\Lambda'_G(S)$ for each cluster $S\in \lset$ of the laminar family that we construct, such that the expected congestion on every edge of $G$ due to the router drawn from the distribution is not too large. We now formally define a type-2 legal clustering.

\begin{definition}[Type-2 Legal Clustering]
	Let $G'$ be a graph with a special vertex $v^*\in V(G')$, and let $\cset'$ be a collection of disjoint vertex-induced subgraphs of $G'\setminus\set{v^*}$, that we call {basic clusters}. 
	A \emph{type-2 legal clustering} of $G'$ with respect to $v^*$ and $\cset'$ consists of the following four ingredients:
	
	\begin{enumerate}
		
		\item  a helpful clustering $(\rset,\set{\dset'(R)}_{R\in \rset})$ of $G'$ with respect to $v^*$ and $\cset'$; 

		
		
		\item a nice witness structure for the graph $G'_{|\rset}$ with respect to the set $\cset''$ of clusters, where $\cset''$ contains every cluster $C\in \cset'$ with $C\subseteq G'\setminus\left(\bigcup_{R\in \rset}V(R)\right)$;
		
		\item a distinguished cluster $R^*\in \rset$, that contains at least $\floor{\left(1-1/2^{(\log m)^{3/4}}\right )|\cset'|}$ clusters of $\cset'$, such that the distribution $\dset'(R^*)$ is careful with respect to $v^*$; and
		
		\item a type-1 legal clustering $(\rset',\set{\dset'(R)}_{R\in \rset'})$ of 
		graph $G^*$, with respect to special vertex $v^{**}$, and cluster set $\cset^*$, where $G^*$ is the graph that is obtained from graph $G'$ by contracting all vertices of $G'\setminus R^*$ into the special vertex $v^{**}$, and $\cset^*$ contains all clusters $C\in \cset'$ with $C\subseteq R^*$. We also require that every cluster $R'\in \rset'$ contain at most $\floor{\left(1-1/2^{(\log m)^{3/4}}\right )|\cset'|}$ clusters of $\cset'$.
	\end{enumerate}
\end{definition}



The key ingerdient of the proof of \Cref{thm: advanced disengagement get nice instances} is the following theorem, that will allow us to gradually construct the desired laminar family $\lset$ of clusters.

\begin{theorem}\label{thm: construct one level of laminar family}
	There is an efficient randomized algorithm, whose input consists of:
	
	\begin{itemize}
		\item a graph $G'$, and a parameter $m$ that is greater than a sufficiently large constant, such that $|V(G')|,|E(G')|\leq m$;
		\item a special vertex $v^*\in V(G')$, such that the cluster $G'\setminus\set{v^*}$ has the $\alpha_1$-bandwidth property in $G'$; and
		
		\item a collection $\cset'$ of disjoint vertex-induced subgraphs of $G'\setminus \set{v^*}$ called \emph{basic clusters}, such that every cluster $C\in \cset'$ has the $\alpha_0$-bandwidth property, and $|\cset'|\geq 2$.
	\end{itemize}	
		
		The algorithm either returns FAIL, or computes a type-1 or a type-2 legal clustering of $G'$ with respect to $v^*$ and $\cset'$. The probability that the algorithm returns FAIL is $1/m^8$. Moreover, if the algorithm computes a type-1 legal clustering $(\rset,\set{\dset'(R)}_{R\in \rset})$, then every cluster $R\in \rset$ contains at most $\floor{\left(1-1/2^{(\log m)^{3/4}}\right )|\cset'|}$ clusters of $\cset'$.
	\end{theorem}

We prove  \Cref{thm: construct one level of laminar family} in the remainder of this subsection, after we complete
the proof of \Cref{thm: advanced disengagement get nice instances} using it.
We will construct a laminar family $\lset$ of clusters of graph $G$, in a top-down manner. For every cluster $R\in \lset$, we will define a distribution $\dset''(R)$ over external routers in $\Lambda'_G$, such that, for every edge $e\in E(G)\setminus E(R)$, $\expect[\qset'(R)\sim\dset''(R)]{\cong_{G}(\qset'(R),e)}\leq \beta^{O((\log m)^{3/4})}$. We will also define a partition $(\lset^{\light},\lset^{\light})$ of the clusters of $\lset$, and we will define, for each cluster $R\in \lset^{\light}$ a distribution $\dset(R)$ over the internal routers in $\Lambda_G(R)$, such that cluster $R$ is $\hat \beta$-light with respect to $\dset(R)$, for $\hat \beta=2^{O((\log m)^{3/4}\log\log m)}$.
 We will ensure that, with high probability, every cluser in $\lset^{\bad}$ is $\hat \beta$-bad. Once we complete consructing the laminar family $\lset$, we will apply algorithm \algbasicdisengagement from \Cref{subsec: basic disengagement} to the resulting tuple $(\lset,\lset^{\bad}, \lset^{\gd}, \set{\dset''(R)}_{R\in \lset},\set{\dset(R)}_{R\in \lset^{\gd}})$ to obtain the final collection $\iset_1$ of instances. We will also provide a nice witness structure for each such resulting instance. We now proceed to describe tha algorithm for constructing the laminar family $\lset$ of clusters.  

Initially, we start with the laminar family $\lset$ containing a single cluster -- graph $G$. Since $\Lambda'_G(G)=\emptyset$ (as $\delta_G(G)=\emptyset$), the distribution $\dset'(G)$ is defined in a trivial way (e.g. it selects $\emptyset$ with probability $1$). We also let $\lset^{\light}$ contain a single cluster -- the cluster $G$, whose distribution $\dset(G)$ over internal routers is defined in  a similar trivial way. Lastly, we set $\lset^{\bad}=\emptyset$.

We simultaneously consider the partitioning tree $\tau(\lset)$ associated with the laminar family $\lset$ (see \Cref{subsubsec: laminar} for a definition). Initially, tree $\tau(\lset)$ consists of a single vertex $v(G)$, associated with the cluster $G$. The algorithm then performs iterations, as long as there is some cluster $R\in \lset$, whose corresponding vertex $v(R)$ is a leaf vertex in the tree $\tau(\lset)$, and there are at least two clusters of $\cset$ that are contained in $R$. We will ensure that every cluster $R'\in \lset$ has the $\alpha_1$-bandwidth property in $G$. Notice that this trivially holds for the initial cluster $G$. 

We now describe an interation for processing  a  cluster $R\in \lset$. We assume that $v(R)$ is a leaf vertex in the current partitioning tree $\tau(\lset)$, and that there are at least two clusters of $\cset$ that are contained in $R$.

In order to process cluster $R$, we construct a graph $G'$, with a special vertex $v^*$, as follows. If $R=G$, then we let $G'$ be a graph that is obtained from $G$, by adding a new special vertex $v^*$ to it, that connects with an edge to an arbitrary fixed vertex $v_0\in V(G)$. Otherwise, if $R\subsetneq G$, then we let $G'$ be the graph that is obtained from $G$ by contracting all vertices of $V(G)\setminus V(R)$ into the special vertex $v^*$. Note that, since we are guaranteed that cluster $R$ has the $\alpha_1$-bandwidth property in $G$, cluster $G'\setminus\set{v^*}$ of $G'$ must have the $\alpha_1$-bandwidth property in $G'$ (in case where $R=G$ this property holds trivially, as $\delta_G'(G)$ contains a single edge). We let $\cset'\subseteq \cset$ be the set of all basic clusters $C\in \cset$ with $C\subseteq R$. We then apply the algorithm from 
\Cref{thm: construct one level of laminar family} to graph $G'$, special vertex $v^*$, and set $\cset'$ of clusters; parameter $m$ remains the same as in the input to \Cref{thm: advanced disengagement get nice instances}. If the algorithm returns FAIL, then we terminate our algorithm with a FAIL. Otherwise, consider the legal clustering that the algorithm produces (which may be a type-1 or a type-2 legal clustering), and let $\rset$ be the resulting set of clusters.

We add every cluster $R'\in \rset$ to the laminar family $\lset$, where it becomes a child cluster of cluster $R$. Recall that, from the properties of a helpful clustering, vertex $v^*$ may not lie in $R'$, so $R'\subseteq R$ must hold. Moreover, $R'$ must have the $\alpha_1$-bandwidth property in $G'$, and hence in $G$. 
Recall that we also obtain a distribution $\dset'(R')$ over external routers in $\Lambda'_{G'}(R')$, such that, for every edge $e\in E(G')\setminus E(R')$, $\expect[\qset'(R')\sim\dset'(R')]{\cong_{G'}(\qset'(R'),e)}\leq \beta$. Unfortunately, this distribution is not sufficiently good for us, since we need the distribution $\dset'(R')$ to be over the collection $\Lambda'_G(R')$ of external routers in graph $G$, and not in graph $G'$. We show how to modify this distribution later. 

Next, we process each cluster $R'\in \rset$ one by one. Consider  any such cluster $R'$. 
We apply Algorithm \algclassifycluster from \Cref{thm:algclassifycluster} to instance $I=(G,\Sigma)$ of \cnwrs, and cluster $R'$, that has the $\alpha_1$-bandwidth property in $G$, together with parameter $p=1/m^{10}$. Recall that the running time of the algorithm is $O(\poly(m\log m))$. If the algorithm returns FAIL, then we add $R'$ to the  set $\lset^{\bad}$ of clusters. Otherwise, the algorithm computes a distribution $\dset(R')$ over  internal routers in $\Lambda_G(R')$, such that cluster $R'$ is $\beta^*$-light with respect to $\dset(R')$, where $\beta^*=2^{O(\sqrt{\log m}\cdot \log\log m)}$. We then add cluster $R'$ to set $\lset^{\light}$.
Recall that, if cluster $R'$ is not $\eta^*$-bad, for $\eta^*=2^{O((\log m)^{3/4}\log\log m)}$, then the probability that the algorithm returns FAIL (that is, the algorithm \emph{errs}), is at most $1/m^{10}$. If the clustering $\rset$ is a type-1 legal clustering, then we also \emph{mark} the vertex $v(R')$ in the decomposition tree $\tau(\lset)$, to indicate that the distribution $\dset'(R')$ is careful with respect to $v^*$. Otherwise, $\rset'$ is a type-2 legal clustering, and we only mark vertex $v(R')$ if $R'=R^*$, where $R^*$ is the distinguished cluster. Recall that in this case, the distribution $\dset'(R^*)$ over external routers of $R^*$ is also careful with respect to $v^*$.

If the algorithm from \Cref{thm: construct one level of laminar family} returned a type-2 legal clustering, then we also consider the type-1 legal clustering $\rset'$ of $R^*$, that is given as part of the type-2 legal clustering of $R$. We process every cluster $R''\in \rset'$ one by one. When cluster $R''$ is processed, we add it to the laminar family $\lset$ and we add vertex $v(R'')$ to the partitioning tree $\tau(\lset)$ as a child of vertex $v(R^*)$; we also mark  vertex $v(R'')$ in the tree, to indicate that the distribution $\dset'(R'')$ over the external routers of $R''$ is careful with respect to $v^{**}$. As before, we apply the  Algorithm \algclassifycluster from \Cref{thm:algclassifycluster} to instance $I=(G,\Sigma)$ of \cnwrs, and cluster $R''$, that has the $\alpha_1$-bandwidth property in $G$, together with parameter $p=1/m^{10}$, As before, if the algorithm returns FAIL, then we add $R''$ to $\lset^{\bad}$, and otherwise we add it to $\lset^{\light}$, together with the  distribution $\dset(R'')$ over  internal routers in $\Lambda_G(R'')$, such that cluster $R''$ is $\beta^*$-light with respect to $\dset(R'')$. This completes the description of the algorithm for constructing the laminar family $\lset$ of clusters. We now establish some of its useful properties.
The following claim, whose proof appears in \Cref{subsec: bounding tree height} will be used to bound the height of the tree $\tau$, and the number of marked vertices on any root-to-leaf path.

\begin{claim}\label{claim: path length in decomposition tree}
	Consider any root-to-leaf path $P$ in the decomposition tree $\tau(\lset)$. Then $P$ contains at most $2^{O((\log m)^{3/4})}$ marked vertices, and at most $O(\log^{3/4} m)$ unmarked vertices. In particular, the depth of the tree $\tau(\lset)$ is at most $2^{O((\log m)^{3/4})}$.
\end{claim}

Next, we provide an algorithm for computing, for each cluster $R\in \lset$, the desired distribution $\dset''(R)$ over the external routers in $\Lambda'_G(R)$. The proof of the following claim is somewhat technical, and is deferred to \Cref{subsec: external routers}.

\begin{claim}\label{claim: compose distributions}
	There is an efficient algorithm that, given a cluster $R\in \lset$, computes a distribution  $\dset''(R)$ over the external $R$-routers in $\Lambda'_G(R)$, such that, for every edge $e\in E(G)\setminus E(R)$ $$\expect[\qset'(R)\sim\dset''(R)]{\cong_{G}(\qset'(R),e)}\leq \beta^{i+1},$$ 
	
	where $i$ is the total number of unmarked vertices on the unique path in tree $\tau(\lset)$, connecting $v(R)$ to the root of the tree.
\end{claim}


To summarize, if our algorithm did not return FAIL, we have now obtained a laminar family $\lset$ of clusters of graph $G$, with $G\in \lset$, so that the depth of the family $\lset$ is at most $2^{O((\log m)^{3/4})}$.
We have also computed a partition $(\lset^{\light},\lset^{\bad})$ of clusters in $\lset$, and, for each cluster $R\in \lset^{\light}$, a distribution $\dset(R)$ over  internal routers in $\Lambda_G(R)$, such that cluster $R$ is $\beta^*$-light with respect to $\dset(R)$, for $\beta^*=2^{O(\sqrt{\log m}\cdot \log\log m)}$. We are also guaranteed that, with probability at least $1-1/m^9$, every cluster $R\in \lset^{\bad}$ is $\eta^*$-bad, for $\eta^*=2^{O((\log m)^{3/4}\log\log m)}$. Lastly, we have computed, for every cluster $R\in \lset$, a distribution $\dset''(R)$ over external routers in $\Lambda'_G(R)$, such that, for every edge $e\in E(G)\setminus E(R)$, $\expect[\qset'(R)\sim\dset''(R)]{\cong_{G}(\qset'(R),e)}\leq \beta^{i+1}$, where $i$ is the total number of unmarked vertices on the unique path in tree $\tau(\lset)$, connecting $v(R)$ to the root of the tree. Since, from \Cref{claim: path length in decomposition tree}, the number of such vertices is bounded by $O(\log^{3/4} m)$, we get that, for every edge $e\in E(G)\setminus E(R)$, $\expect[\qset'(R)\sim\dset''(R)]{\cong_{G}(\qset'(R),e)}\leq \beta^{O((\log m)^{3/4})}\leq 2^{O((\log m)^{3/4}\log\log m)}$, since $\beta=\log^{18}m$.

We apply algorithm \algbasicdisengagement from \Cref{subsec: basic disengagement} to the resulting tuple $(\lset=\lset^{\bad}\cup \lset^{\gd}, \set{\dset''(R)}_{R\in \lset},\set{\dset(R)}_{R\in \lset^{\gd}})$, to obtain the final collection $\iset_1$ of subinstances of $I$.
Let $\hat \beta=2^{c(\log m)^{3/4}\log\log m}$ for some large enough constant $c$. We are then guaranteed that  every cluster $R\in \lset^{\light}$ is $\hat\beta$-light with respect to the distribution $\dset(R)$, and that, for every cluster $R\in \lset$ and every edge $e\in E(G)\setminus E(R)$, $\expect[\qset'(R)\sim\dset''(R)]{\cong_{G}(\qset'(R),e)}\leq \hat \beta$. We can set $c$ to be large enough so that $\eta^*\leq \hat \beta$ holds. We say that a bad event $\event$ happens if some cluster $R\in \lset^{\bad}$ is not $\hat \beta$-bad. From the above discussion, the probability of $\event$ happening is at most $1/m^9$. If Event $\event$ does not happen, then, from \Cref{lem: disengagement final cost}, 
$\expect{\sum_{I'\in \iset_1}\optcrors(I')}\leq O(\dep(\lset)\cdot\hat \beta^2\cdot (\optcrors(I)+|E(G)|))\leq  2^{O((\log m)^{3/4}\log\log m)}\cdot  (\optcrors(I)+|E(G)|)) $. 
If Event $\event$ happens (which happens with probability at most $1/m^9$), then clearly 
$\expect{\sum_{I'\in \iset_1}\optcrors(I')}\leq \sum_{I'=(G'',\Sigma'')\in \iset_1}|E(G'')|^2\leq m^3$.
 Therefore, overall, $\expect{\sum_{I'\in \iset_1}\optcrors(I')}\leq  2^{O((\log m)^{3/4}\log\log m)}\cdot  (\optcrors(I)+|E(G)|)) $
 
Additionally, from \Cref{lem: basic disengagement combining solutions}, 
	there is an efficient algorithm, that, given, for each instance $I'\in \iset$, a solution $\phi(I')$, computes a solution for instance $I$ of value at most $\sum_{I'\in \iset}\cro(\phi(I'))$.
In order to prove that the algorithm computes a valid $2^{O((\log m)^{3/4}\log\log m)}$-decomposition of instance $I$, it is now sufficient to prove that $\sum_{I'=(G'',\Sigma'')\in \iset_1}|E(G'')|\leq O(|E(G)|)$. We do so in the next claim, whose proof is deferred to \Cref{subsec:appx few edges}.

\begin{claim}\label{claim: few edges}
	 $\sum_{I'=(G'',\Sigma'')\in \iset_1}|E(G'')|\leq O(|E(G)|)$.
	\end{claim}

From the above discussion, if our algorithm did not return FAIL, it computed  a valid $2^{O((\log m)^{3/4}\log\log m)}$-decomposition of instance $I$. Consider now any resulting instance $\tilde I=(\tilde G,\tilde \Sigma)\in \iset_1$. From the definition of basic disengagment, this instance must correspond to some cluster $R\in \lset$. Assume first that $R\neq G$, and let $\rset$ be the set of all child clusters of $R$ in $\lset$ (if $R$ corresponds to a leaf vertex of $\tau(\lset)$, then $\rset=\emptyset$). Recall that graph $\tilde G$ is obtained from graph $G$ by first contracting all vertices of $G\setminus R$ into a supernode $v^*$; denote the resulting graph by $G'$. We then contract every cluster $R'\in \rset$ into a supernode, obtaining graph $\tilde G$. In other words, $\tilde G=G'_{|\rset}$.

 We now consider three cases. The first case is when $\rset$ is a type-1 legal clustering for cluster $R$. In this case, there is at most one cluster $C\in \cset$ that is contained in $G'\setminus \bigcup_{R'\in \rset}R'$, from the definition of type-1 legal clustering. Therefore, at most one cluster $C$ of $\cset$ is contained in $\tilde G$. We define the nice witness structure for graph $\tilde G$, with respect to the set $\cset'$ of basic clusters, where $\cset'=\set{C}$ if there is a cluster $C\in \cset$ that is contained in $\tilde G$, and $\cset'=\emptyset$ otherwise. We let $\tilde \sset=\set{\tilde S_1}$, where $\tilde S_1=\hat G$, and we let $\tilde \sset'=\set{\tilde S_1'}$, where $\tilde S_1'=C$ if $\cset'=\set{C}$, and $\tilde S_1'=\set{v}$, where $v$ is an arbitrary vertex of $\tilde G$ otherwise. Note that, under these definitions, $\hat E=\emptyset$, and so we can set $\hat \pset=\emptyset$. It is easy to verify that $(\tilde \sset,\tilde \sset',\hat \pset)$ is a nice witness structure for $\tilde G$.
 
 The second case is when cluster $R$ corresponds to a leaf vertex of tree $\tau(\lset)$. From our algorithm, this means that there is at most one cluster $C\in \cset$ with $C\subseteq R$. In this case, we define the nice witness structure for graph $\tilde G$ similarly to the first case.

Lastly, in the third case, when the algorithm from \Cref{thm: construct one level of laminar family} was applied to cluster $R$, it returned a type-2 legal clustering, with the corresponding cluster set $\rset$. In this case, the algorithm also must produce a nice witness structure for the graph $G'_{|\rset}=\tilde G$, with respect to the set $\cset''$ of clusters, that contains every cluster $C\in \cset$ with $C\subseteq G'\setminus\left(\bigcup_{R'\in \rset}V(R')\right)$. In other words, $\cset''=\cset(\tilde G)$. 

It remains to consider the case where $R=G$. As before, we let $\rset$ denote the set of all child-clusters of cluster $R$. Recall that in this case, graph $\tilde G$ is obtained from graph $G$ by contracting every cluster $R'\in \rset$ into a supernode. Graph $G'$ was obtained from graph $G$ by adding a special vertex $v^*$ that connects to some vertex $v_0\in V(G)$ with an edge. Therefore, graph $G'_{|\rset}$ is a graph that is obtained from $\tilde G$ by adding a special vertex $v^*$ to it and connecting it to some vertex of $\tilde G$. Recall that vertex $v^*$ may not lie in any cluster of $\rset$. We start by defining a nice structure for graph $G'_{|\rset}$, with respect to the collection $\cset''$ of clusters, that contains every cluster $C\in \cset$ with $C\subseteq G'_{|\rset}$ exactly as before (when we assumed that $R\neq G$). Since vertex $v^*$ has degree $1$, we can assume that no path in $\hat \pset$ contains $v^*$. Moreover, if $v^*\in \tilde S'_i$, for some $\tilde S'_i\in \tilde \sset'_i$, then cluster $\tilde S'_i\setminus\set{v^*}$ still has the $\alpha^*$-bandwidth property. By deleting vertex $v^*$ from the cluster of $\tilde \sset$ to which it belongs, and also from a cluster of $\tilde \sset'$ to which it belongs (if such a cluster exists), we obtain a nice witness structure for graph $\tilde G$, with respect to cluster set $\cset''$, as required.
The algorithm may return FAIL if any application of the algorithm from  \Cref{thm: construct one level of laminar family} returned FAIL. Since $|\lset|\leq m$, and the probability that a singel application of the algorithm from \Cref{thm: construct one level of laminar family} returns FAIL is at most $1/m^8$, overall, the probability that the algorithm returns FAIL is at most $1/m^6$.

In order to complete the proof of \Cref{thm: advanced disengagement get nice instances} it is now enough to prove \Cref{thm: construct one level of laminar family}, which we do next.
\subsubsection{Proof of \Cref{thm: construct one level of laminar family}}

\label{subsubsec-construct-one-level-of-laminar}

Throughout the proof, we will consider various graphs, sets of disjoint clusters in these graphs, and the corresponding contracted graphs. Let $H$ be any graph, and let $\rset$ be any set of disjoint vertex-induced subgraphs (clusters) of graph $H$. Let $\hat H=H_{|\rset}$ be the contracted graph corresponding to $H$ and $\rset$, that is obtained from $H$ by contracting every cluster $R\in \rset$ into a supernode $v_R$. Observe that every subset $\hat U\subseteq V(\hat H)$ of vertices of $\hat H$ naturally defines a vertex-induced subgraph of $H$, which is a subgraph of $H$ induced by vertex set $U=\left(\bigcup_{v_R\in \hat U}V(R)\right )\cup(V(H)\cap \hat U)$; in other words, $U$ contains all regular vertices of $\hat U$, and the vertices of every cluster $R\in \rset$ with $v_R\in \hat U$.
 We will refer to $H[U]$ as the \emph{subgraph of $H$ (or cluster of $H$) defined by the set $\hat U$ of vertices of $\hat H$}.
Similarly, if $S$ is a cluster of $\hat H$ induced by vertex set $\hat U$, we will refer to $H[U]$ as the cluster of $H$ defined by $S$.

Assume now that we are given any graph $H$, a special vertex $v^*$ in $H$, and a collection $\rset$ of disjoint clusters of $H$, such that vertex $v^*$ does not lie in any cluster of $\rset$, and every cluster $R\in \rset$ has $\alpha$-bandwidth property, for some parameter $0<\alpha<1$. As before, we denote $\hat H=H_{|\rset}$. Next, we consider a Gomory-Hu tree $\tau$ of the graph $\hat H$	
(see \Cref{subsec: GH tree} for a definition). We root the tree $\tau$ at the special vertex $v^*$. For every vertex $u\in V(\tau)$, we let $\tau_u$ be the subtree of $\tau$ rooted at $u$.

We will use the following useful observation multiple times. The proof is deferred to \Cref{appx: subtree to cluster}.

\begin{observation}\label{obs: subtree to cluster}
	Let $u\in V(\tau)\setminus \set{v^*}$ be any non-root vertex of the tree $\tau$, and let $S$ be the cluster of $H$ that is defined by the set $V(\tau_u)$ of vertices of $\hat H$. Then cluster $S$ has the $\alpha$-bandwidth property in $H$. Moreover, there is an efficient algorithm to compute a distribution $\dset'(S)$ over the external routers in $\Lambda'_{H}(S)$, such that distribution $\dset'(S)$ is careful with respect to $v^*$, and, for every edge $e\in E(H)\setminus E(S)$, $\expect[\qset'(S)\sim\dset'(S)]{\cong_{H}(\qset'(S),e)}\leq O(\log^4m/\alpha)$.
\end{observation}

For convenience, in the remainder of the proof, we denote graph $G'$ by $G$, and the set $\cset'$ of clusters by $\cset$.	
We start with the graph $G$ and the set $\cset$ of basic clusters, and we let $H=G_{|\cset}$ be the corresponding contracted graph. We consider the Gomory-Hu tree $\tau$ of the graph $H$. We root the tree $\tau$ at the special vertex $v^*$. For every vertex $u\in V(\tau)$, we let $\tau_u$ be the subtree of $\tau$ rooted at $u$, and we let the weight $w(u)$ be the number of supernodes (vertices corresponding to clusters in $\cset$) in the tree $\tau_u$. Let $u^{*}$ be the vertex of $\tau$ that is furthest from the root $v^*$, such that 
 $w(u^*)\geq \floor{\left(1-1/2^{(\log m)^{3/4}}\right )|\cset|}$. We now consider two cases.

The first case happens if $u^*=v^*$. In this case, we will compute a type-1 legal clustering of $G$. Let $u_1,\ldots,u_q$ denote all child vertices of $v^*$. For all $1\leq i\leq q$, let $R_i$ be the cluster of the graph $G$ defined by the vertex set $V(\tau_{u_i})$. Denote $\rset=\set{R_1,\ldots,R_q}$. Since every cluster $C\in \cset$ has the $\alpha_0$-bandwidth property, and $H=G_{|\cset}$, from \Cref{obs: subtree to cluster}, each cluster $R_i\in \rset$ has the $\alpha_0\geq\alpha_1$-bandwidth property. From the construction, vertex $v^*$ may not lie in any of the clusters of $\rset$, and, for each cluster $R\in \rset$, and for every basic cluster $C\in \cset$, either $C\subseteq R$ or $V(C)\cap V(R)=\emptyset$. We use the algorithm from \Cref{obs: subtree to cluster} to construct, for every cluster $R\in \rset$, a distribution $\dset'(R)$ over the external $S$-routers in $\Lambda'_{G}(S)$, such that the distribution is careful with respect to $v^*$, and, for every edge $e\in E(G)\setminus E(R)$, $\expect[\qset'(R)\sim\dset'(R)]{\cong_{G}(\qset'(R),e)}\leq O(\log^4m/\alpha_0)\leq \beta$. 

Note that $G\setminus\bigcup_{R\in \rset}R$ consists of only one vertex -- vertex $v^*$. Therefore, $\rset$ is a legal type-1 clustering of graph $G$. We terminate the algorithm, and return this clustering.

We assume from now on that $u^*\neq v^*$.  We will provide an algorithm for computing a type-2 legal clustering of $G$. Let $R^*$ be the subgraph of $G$ defined by vertex set $V(\tau_{u^*})$ of graph $H$.
As before, from 
 \Cref{obs: subtree to cluster}, cluster $R^*$ has the $\alpha_0\geq\alpha_1$-bandwidth property, it does not contain the verex $v^*$, and, for every basic cluster $C\in \cset$, either $C\subseteq R^*$ or $V(C)\cap V(R^*)=\emptyset$. 
 From the definition of vertex $u^*$, the total number of basic clusters of $\cset$ that are contained in $R^*$ is at least  $\floor{\left(1-1/2^{(\log m)^{3/4}}\right )|\cset|}$.
 We also use the algorithm from \Cref{obs: subtree to cluster} to construct a distribution $\dset'(R^*)$ over the external $R^*$-routers in $\Lambda'_{G}(S)$, such that the distribution is careful with respect to $v^*$, and, for every edge $e\in E(G)\setminus E(R^*)$, $\expect[\qset'(R^*)\sim\dset'(R^*)]{\cong_{G}(\qset'(R^*),e)}\leq O(\log^4m/\alpha_0)\leq \beta$. In the final type-2 legal clustering $\rset$ for graph $G$ that our algorithm will return, cluster $R^*$ will play the role of the distinguished cluster, and the distribution $\dset'(R^*)$ over the set of its external routers will remain unchanged. 
  Let $G^*$ be the graph associated with the cluster $R^*$: that is, graph $G^*$ is obtained from graph $G$ by contracting all vertices of $G\setminus R^*$ into a special vertex, that we denote by $v^{**}$. We also denote by $\cset^*\subseteq \cset$ the set of all basic clusters $C\in \cset$ with $C\subseteq R^*$. 
 We now construct a type-1 legal clustering $\rset'$ of $G^*$, which is required as part of definition of type-2 legal clustering of $G$. Denote the child vertices of vertex $u^*$ in the tree $\tau$ by $u_1,\ldots,u_q$. For all $1\leq i\leq q$, let $R_i$ be the cluster of the graph $G$ defined by the vertex set $V(\tau_{u_i})$. Denote $\rset'=\set{R_1,\ldots,R_q}$. Since every cluster $C\in \cset$ has the $\alpha_0$-bandwidth property, and $H=G_{|\cset}$, from \Cref{obs: subtree to cluster}, each cluster $R_i\in \rset$ has the $\alpha_0\geq\alpha_1$-bandwidth property. From the construction, vertex $v^{**}$ may not lie in any of the clusters of $\rset'$, and, for each cluster $R\in \rset'$, and for every basic cluster $C\in \cset^*$, either $C\subseteq R$ or $V(C)\cap V(R)=\emptyset$ holds. Consider now some cluster $R_i\in \rset'$, and denote $\delta_G(R_i)=E_i$.
 From the properties of the Gomory-Hu tree (see \Cref{thm: GH tree properties}), there is a collection $\qset'_i$ of edge-disjoint paths in graph $H$, routing the edges of $E_i$ to vertex $u^*$, that are internally disjoint from $V(\tau_{u_i})$. Let $H^*$ be the graph obtained from $H$, by contracting all vertices of $V(H)\setminus V(\tau_{u^*})$ into a supernode $\hat v^*$. A simple transformation of the paths in $\qset'_i$ shows that there is a collection $\qset''_i$ of edge-disjoint paths in graph $H^*$, routing the edges of $E_i$ to $u^*$. Observe that graph $H^*$ is precisely the contracted graph of $G^*$ with respect to the set $\cset^*$ of clusters, that is, $H^*=G^*_{|\cset^*}$, and recall that each cluster $C\in \cset^*$ has the $\alpha_0$-bandwidth property. 
 
If vertex $u^*$ is not a supernode, then we apply the algorithm from \Cref{claim: routing in contracted graph} to graph $H^*$, the set $\cset^*$ of clusters, and the set $\qset_i''$ of paths, to obtain a set $\qset^*_i$ of paths in graph $G^*$, routing the edges of $E_i$ to vertex $u^*$, such that every path in $\qset^*_i$ is internally disjoint from $R_i$. Moreover, for every edge $e\in \bigcup_{C\in \cset^*}E(C)$, the paths of $\qset^*_i$  cause congestion at most $ \ceil{1/\alpha_0}$, while for every edge $e\in E(G^*)\setminus \left(\bigcup_{C\in \cset^*}E(C)\right )$, the paths of $\qset^*_i$ cause congestion at most $1$. In particular, the set $\qset^*_i$ of paths is careful with respect to vertex $v^{**}$. We then define a distribution $\dset'(R_i)$ over the set $\Lambda'_{G^*}(R_i)$ of external $R_i$-routers to choose the  set $\qset^*_i$ of paths with probability $1$. 

Assume now that vertex $u^*$ is a supernode, and that it represents some cluster $C\in \cset^*$. We  apply the algorithm from \Cref{claim: routing in contracted graph} to graph $H^*$, the set $\cset^*\setminus\set{C}$ of clusters, and the set $\qset_i''$ of paths, to obtain a set $\qset^*_i$ of paths in graph $G^*$, routing the edges of $E_i$ to edges of $\delta_{G^*}(C)$, such that every path in $\qset^*_i$ is internally disjoint from $R_i$. As before, for every edge $e\in \bigcup_{C'\in \cset^*}E(C')$, the paths of $\qset^*_i$  cause congestion at most $ \ceil{1/\alpha_0}$, while for every edge $e\in E(G^*)\setminus \left(\bigcup_{C'\in \cset^*}E(C')\right )$, the paths of $\qset^*_i$ cause congestion at most $1$. As before, the set $\qset^*_i$ of paths is careful with respect to vertex $v^{**}$. 
We use the algorithm from \Cref{lem: simple guiding paths} to compute a distribution $\dset(C)$ over internal $C$-routers in $\Lambda_{G^*}(C)$, such that, for every edge $e\in E(C)$, $\expect[\qset(C)\sim \dset(C)]{\cong(\qset(C),e)}\leq \log^4m/\alpha_0$.
We now define a distribution $\dset'(R_i)$ over the set $\Lambda'_{G^*}(R_i)$ of external $R_i$-routers. In order to draw a router from the distribution, we first choose an internal $C$-router $\qset(C)$ from the distribution $\dset(C)$. Let $x$ be the vertex that serves the center of the router. For every edge $e\in E_i$, we let $\tilde Q(e)$ be the path obtained as follows. First, we let $Q^*(e)$ be the unique path of $\qset^*_i$ that originates from edge $e$. We let $e'$ be the last edge on path $\qset^*_i$, that must belong to $\delta_{G^*}(C)$. We then let $\tilde Q(e)$ be the path obtained by concatenating path $Q^*(e)$ with the unique path of $\qset(C)$ that originates at edge $e$. We let $\qset'(R_i)=\set{\tilde Q(e)\mid e\in E_i}$ be the resulting external $R_i$-router, that routes the edges of $E_i$ to $x$. Since every edge of $\delta_{G^*}(C)$ may lie on at most one path of $\qset^*_i$, it is immediate to verify that, for every edge $e\in E(C)$, $\cong(\qset'(R_i),e)\leq \cong(\qset(C),e)$, and so overall, for every edge $e'$, $\expect[\qset'(R_i)\sim \dset'(R_i)]{\cong_{G^*}(\qset'(R_i),e')}\leq \frac{\log^4m}{\alpha_0}\leq \beta$,
since $\alpha_0=1/\log^3m$ and $\beta=\log^{18}m$. As before, distribution $\dset'(R_i)$ is careful with respect to $v^{**}$.


 Lastly, observe that at most one cluster $C\in \cset$ may be contained in graph $R^*\setminus\bigcup_{R\in \rset'}R$ -- the cluster associated with vertex $u^*$, if $u^*$ is a supernode. 
 Therefore, $(\rset',\set{\dset'(R)}_{R\in \rset'})$ is a type-1 legal clustering of graph $G^*$, with cluster set $\cset^*$ and special vertex $v^{**}$.
Moreover, from the choice of vertex $u^*$, we are guaranteed that every cluster $R'\in \rset'$ contain at most $\floor{\left(1-1/2^{(\log m)^{3/4}}\right )|\cset'|}$ clusters of $\cset$.

The remainder of the algorithm is iterative. We start with a helpful clustering $(\rset=\set{R^*},\set{\dset'(R^*)})$ of $G$, and we view $R^*$ as the distinguished cluster of $\rset$. We then iterate. In every iteration, we either establish that the current helpful clustering $(\rset,\set{\dset'(R)}_{R\in \rset})$ is a  type-2 legal clustering, by computing a nice witness structure for graph $G_{|\rset}$, with respect to the set $\cset''$ of clusters, containing every cluster $C\in \cset$ with $C\subseteq G\setminus\left (\bigcup_{R\in \rset}V(R)\right )$; or we will compute another helpful clustering of $G$ that is ``better'' in some sense, and use it to replace the current helpful clustering $(\rset,\set{\dset'(R)}_{R\in \rset})$. We will ensure that the helpful clustering $\rset$ that the algorithm maintains always contains the cluster $R^*$ that we defined above, which will always remain the distinguished cluster of $\rset$. The distribution $\dset'(R^*)$ over the external $R^*$-routers in $\Lambda'_G(R^*)$, and the type-1 legal clustering  $(\rset',\set{\dset'(R)}_{R\in \rset'})$ of the graph $G^*$ associated with cluster $R^*$ will remain unchanged throughout the algorithm. We will use the following definition in order to compare different helpful clusterings of $G$.

\begin{definition}[Comparing clusterings]
	Let $\rset_1$, $\rset_2$ be two helpful clusterings of graph $G$, with respect to special vertex $v^*$ and set $\cset$ of basic clusters, such that $R^*\in \rset_1\cap \rset_2$. Denote by $\cset_1\subseteq \cset$ the set of all clusters $C\in \cset$ with $C\subseteq G\setminus\left(\bigcup_{R\in \rset_1}R\right )$, and define a subset $\cset_2\subseteq \cset$ of basic clusters for $\rset_2$ similarly. We say that clustering $\rset_2$ is \emph{better} than clustering $\rset_1$ if one of the following hold:
	
	\begin{itemize}
		\item either $|\cset_2|<|\cset_1|$; or
		\item $|\cset_1|=|\cset_2|$, and $|E(G_{|\rset_2})|<|E(G_{|\rset_1})|$.
	\end{itemize}
\end{definition}

The following lemma is key in the proof of \Cref{thm: construct one level of laminar family}.

\begin{lemma}\label{lemma: better clustering}
	There is an efficient randomized algorithm, that, given a helpful clustering $(\rset,\set{\dset'(R)}_{R\in \rset})$ of graph $G$ with respect to vertex $v^*$ and set $\cset$ of basic clusters, such that $R^*\in \rset$, either (i) establishes that $\rset$ is a type-2 legal clustering by providing a nice witness structure for graph $G_{|\rset}$, with respect to the set $\cset''$ of clusters, containing every cluster $C\in \cset$ with $C\subseteq G\setminus\left (\bigcup_{R\in \rset}V(R)\right )$, or (ii) computes another helpful clustering $(\tilde\rset,\set{\dset'(R)}_{R\in \tilde\rset})$ of graph $G$ with respect to $v^*$ and $\cset$ with $R^*\in \rset$, such that $\tilde \rset$ is a better clustering than $\rset$; or (iii) returns FAIL. The latter may only happen with probability at most $1/m^{10}$.
\end{lemma}

It is immediate to complete the proof of \Cref{thm: construct one level of laminar family}.
using \Cref{lemma: better clustering}. Our algorithm starts with the helpful custering $(\rset=\set{R^*},\set{\dset'(R^*)})$ of $G$, where $\dset'(R^*)$ is the distribution over external $R^*$-routers that we have computed above, and then iterates. In every iteration, we apply the algorithm from \Cref{lemma: better clustering} to the current helpful clustering $(\rset,\set{\dset'(R)}_{R\in \rset})$. If the algorithm  establishes that $\rset$ is a type-2 legal clustering by providing a nice witness structure for graph for graph $G_{|\rset}$, with respect to the set $\cset''$ of clusters, containing every cluster $C\in \cset$ with $C\subseteq G\setminus\left (\bigcup_{R\in \rset}V(R)\right )$, then we terminate the algorithm with the resulting type-2 legal clustering $(\rset,\set{\dset'(R)}_{R\in \rset})$; we view $R^*$ as the distinguished cluster of $\rset$, and the type-1 legal clustering $(\rset',\set{\dset'(R)}_{R\in \rset'})$ of the graph $G^*$ corresponding to $R^*$ remains unchanged. Otherwise, if the algorithm returns another helpful clustering $(\tilde\rset,\set{\dset'(R)}_{R\in \tilde\rset})$, then  we replace $(\rset,\set{\dset'(R)}_{R\in \rset})$ with $(\tilde\rset,\set{\dset'(R)}_{R\in \tilde\rset})$ and continue to the next iteration. Lastly, if the algorithm from 
\Cref{lemma: better clustering} returns FAIL, then we terminate the algorithm and return FAIL as well.
Let $\cset'$ be the set of all clusters $C\in\cset$ with $C\subseteq  G\setminus\left(\bigcup_{R\in \rset}V(R)\right )$, where $\rset$ is the current helpful clustering.
Since, in every iteration, either $|\cset'|$ decreases, or $|\cset'|$ remains the same but the number of edges in graph $G_{|\rset}$ decreases, the algorithm is guaranteed to terminate after at most $m^2$ iterations. 
Since the probability of the algorithm to return FAIL in each iteration is at most $1/m^{10}$, the total probability that the algorithm returns FAIL is at most $1/m^8$. If the algorithm does not return FAIL, then it returns a type-2 clustering of $G$ as required. In order to complete the proof of \Cref{thm: construct one level of laminar family}, it is now enough to prove \Cref{lemma: better clustering}, which we do next.

\subsubsection{Proof of \Cref{lemma: better clustering}}
\label{subsubsec: proof of improving clustering}

Recall that we are given a graph $G$ and a special vertex $v^*$ of $G$. We are also given a collection $\cset$ of disjoint vertex-induced subgraphs of $G\setminus\set{v^*}$ called basic clusters, such that every basic cluster $C\in \cset$ has the $\alpha_0$-bandwidth property. Lastly, we are given a helpful clustering $(\rset,\set{\dset'(R)}_{R\in \rset})$ of $G$ with respect to $v^*$ and $\cset$. Recall that vertex $v^*$ may not lie in any cluster of $\rset$, and every cluster $R\in \rset$ has the $\alpha_1$-bandwidth property. For every cluster $R\in \rset$, $\dset'(R)$ is a distribution over the external $R$-routers in $\Lambda'_G(R)$, and,  for every edge $e\in E(G)\setminus E(R)$, $\expect[\qset'(R)\sim\dset'(R)]{\cong_{G'}(\qset'(R),e)}\leq \beta$. 
Additionally, there is a distingiushed cluster $R^*\in \rset$, whose corresponding distribution $\dset'(R^*)$ is careful with respect to $v^*$, and $R^*$ contains at least $\floor{\left(1-1/2^{(\log m)^{3/4}}\right )|\cset|}$ clusters of $\cset$. 

\begin{figure}[h]
	\centering
	\subfigure[A schematic view of graph $G$. Clusters of $\rset$ are shown in red, clusters of $\cset$ are shown in blue, clusters of $\cset'$ are the blue clusters that are disjoint from the red clusters, and vertices of $V(G)\setminus \big( \bigcup_{C\in \cset}V(C)\big)$ are shown in black.]{\scalebox{0.093}{\includegraphics{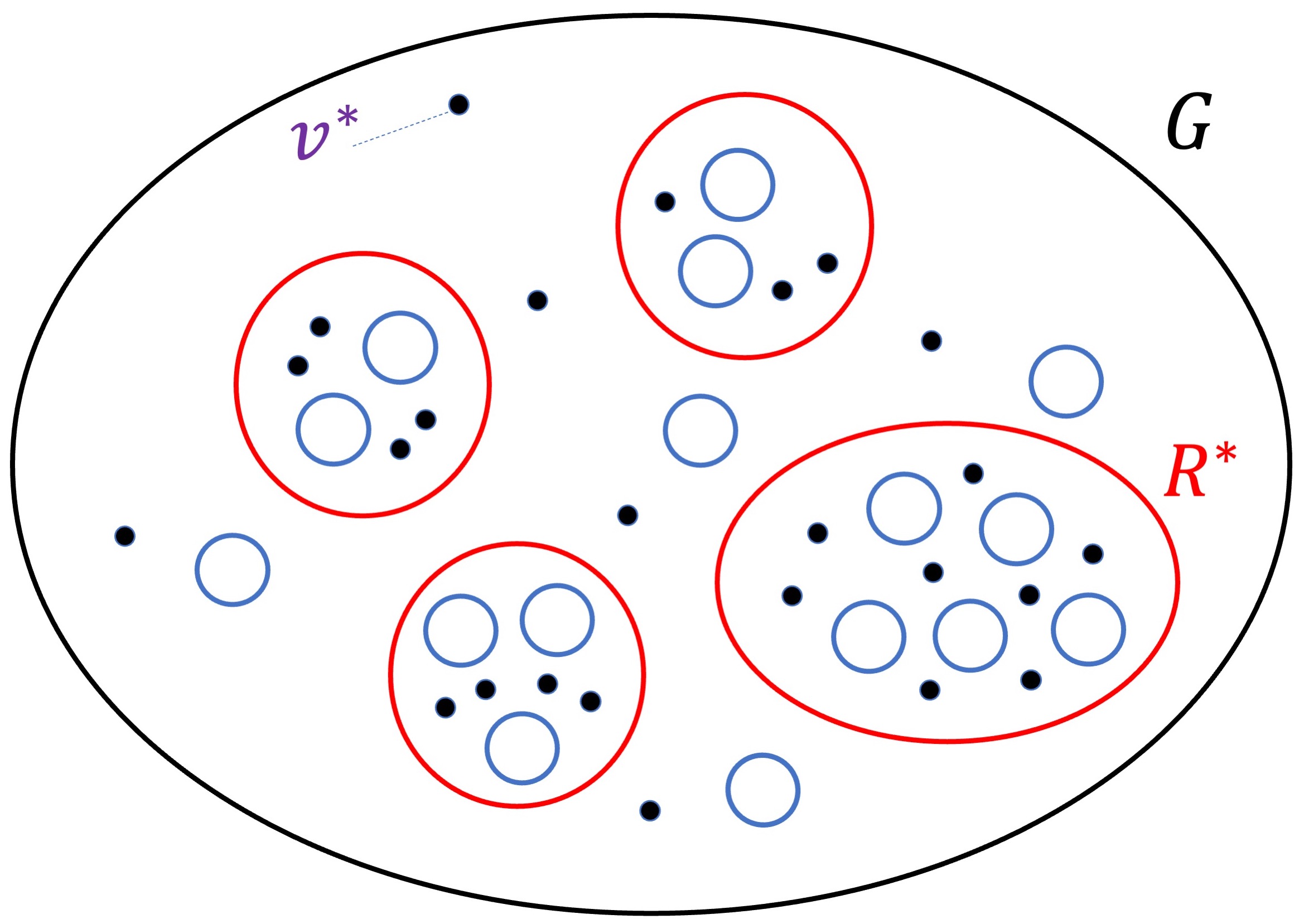}}\label{fig: NF1}}
	\hspace{1.2cm}
	\subfigure[A schematic view of graph $\hat H$. Regular vertices are shown in black, $C$-nodes are shown in blue and $R$-nodes are shown in red.]{\scalebox{0.093}{\includegraphics{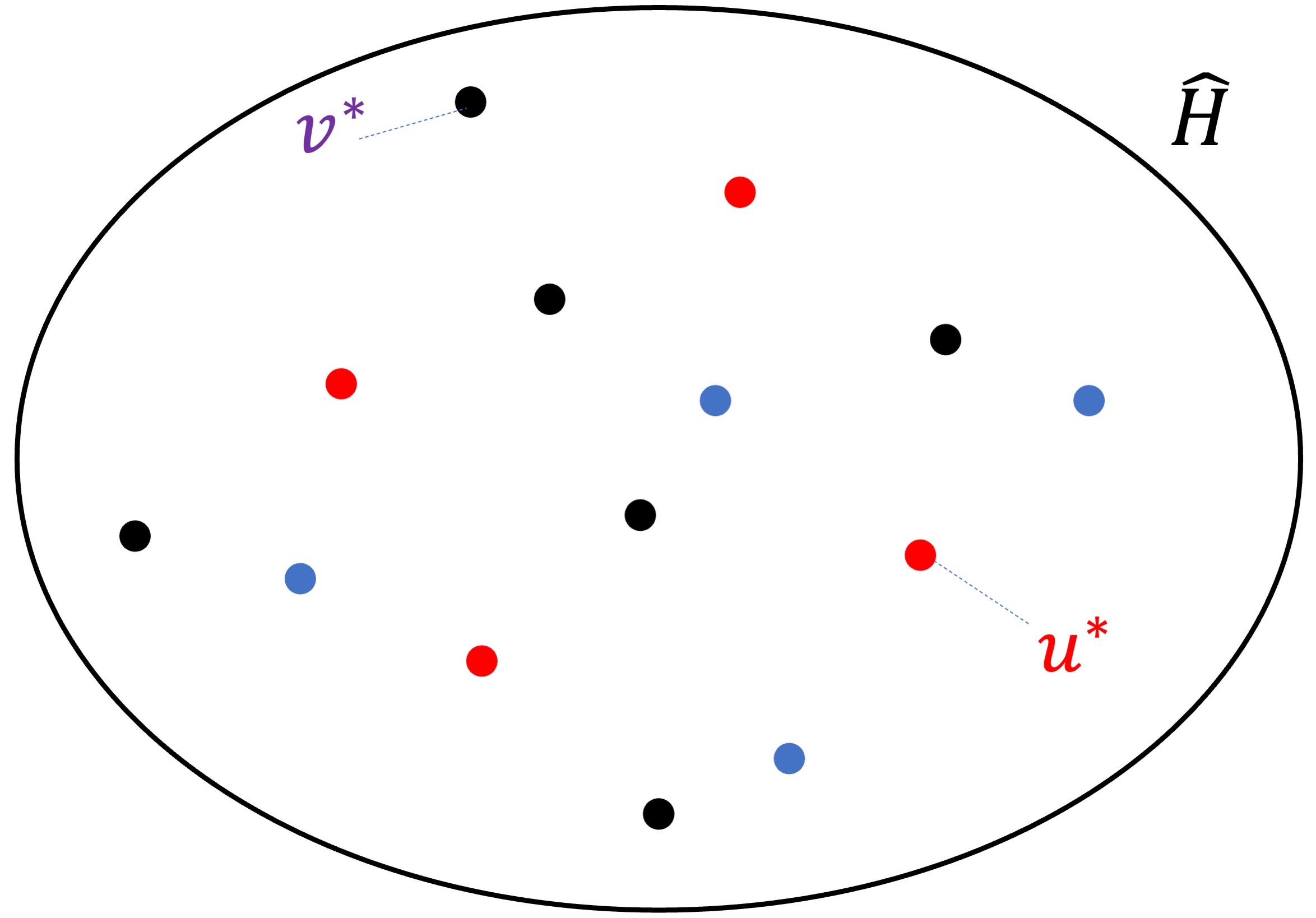}}\label{fig: NF2}}
	\caption{Graphs $G$ and $\hat H$.}
\end{figure}

We denote by $\cset'$ the set of all clusters $C\in \cset$, such that $C\subseteq  G\setminus\left(\bigcup_{R\in \rset}V(R)\right )$. Observe that $\rset\cup \cset'$ is a set of mutually disjoint clusters of graph $G$ (see \Cref{fig: NF1}).
It will be convenient for us to work with a slightly different contracted graph, that we denote by $\hat H=G_{|(\rset\cup\cset')}$. Note that every vertex $u\in V(\hat H)$ that is different from a special vertex $v^*$, is either a regular vertex (that is, it is a vertex of $G$), or a supernode corresponding to a cluster of $\cset'\cup \rset$. If supernode $u$ represents a cluster of $\cset'$, then we call it a $C$-node, and otherwise we call it an $R$-node  (see \Cref{fig: NF2}).
In order to prove \Cref{lemma: better clustering}, we will mostly work with graph $\hat H$. 
Note that $v^*\in V(\hat H)$. We denote by $u^*$ the $R$-node representing the distinguished cluster $R^*\in \rset$.
We will maintain a collection $\wset$ of clusters in graph $\hat H$, that we call $W$-clusters, and define next.

\begin{definition}[Valid set of $W$-clusters]
	A set $\wset$ of disjoint clusters of graph $\hat H$ is a \emph{valid set of $W$-clusters} if:
	
	\begin{itemize}
		\item for every cluster $W\in \wset$, every vertex of $W$ is an $R$-node or a regular vertex, and $W$ does not contain the special vertex $v^*$ or the 
		$R$-node $u^*$ representing cluster $R^*$; 
		
		\item for every cluster $W\in \wset$, $|E_{\hat H}(W)|\geq |\delta_{\hat H}(W)|/(64\log m)$; and
		\item every cluster $W\in \wset$ has the $\alpha'$-bandwidth property in graph $\hat H$, where $\alpha'=1/(c\log^{2.5}m)$, for some large enough constant $c$.
	\end{itemize}
\end{definition}

We will use the following lemma in order to prove \Cref{lemma: better clustering}.
\begin{lemma}\label{lemma: better clustering 2}
	There is an efficient randomized algorithm, that, given a valid $W$-clustering $\wset$ of graph $\hat H$, either (i) establishes that $(\rset,\set{\dset'(R)}_{R\in \rset})$ is a type-2 legal clustering of $G$, by providing a nice witness structure for graph $G_{|\rset}$, with respect to the set $\cset''$ of clusters, containing every cluster $C\in \cset$ with $C\subseteq G\setminus\left (\bigcup_{R\in \rset}V(R)\right )$; or (ii) computes another helpful clustering $(\tilde\rset,\set{\dset'(R)}_{R\in \tilde \rset})$ of graph $G$ with respect to vertex $v^*$ and set $\cset$ of basic clusters, such that   $R^*\in \rset$ and $\tilde \rset$ is a better clustering than $\rset$; or (iii) computes a new valid set $\wset'$ of $W$-clusters in the current graph $\hat H$, such that $|E(\hat H_{|\wset'})|<|E(\hat H_{|\wset})|$; or (iv) returns FAIL. The latter may only happen with probability at most $1/m^{11}$.
\end{lemma}

\Cref{lemma: better clustering} easily follows from \Cref{lemma: better clustering 2}. We start with $\wset=\emptyset$, which is a valid set of $W$-clusters for $\hat H$, and then iterate. In every iteration, we apply the algorithm from \Cref{lemma: better clustering 2} to the current valid set  $\wset$ of $W$-clusters. If the algorithm establishes that $\rset$ is a type-2 legal clustering of $G$, then we terminate the algorithm and return the correpsonding witness structure for graph $G_{|\rset}$. If the algorithm computes another helpful clustering $(\tilde\rset,\set{\dset'(R)}_{R\in \tilde \rset})$ of graph $G$ with respect to vertex $v^*$ and set $\cset$ of basic clusters with $R^*\in \rset$, such that $\tilde \rset$ is a better clustering than $\rset$, then we terminate the algorithm and return the clustering $(\tilde\rset,\set{\dset'(R)}_{R\in \tilde \rset})$. If the algorithm from \Cref{lemma: better clustering 2} returns FAIL, then we terminate and algorithm and return FAIL as well.  Otherwise, the algorithm from \Cref{lemma: better clustering 2} computes a valid set  $\wset'$ of $W$-clusers in the current graph $\hat H$, such that $|E(\hat H_{|\wset'})|<|E(\hat H_{|\wset})|$. We then replace $\wset$ with $\wset'$ and continue to the next iteration. Since the number of edges in graph $\hat H_{|\wset}$ decreases in every iteration, we are guaranteed that, after at most $m$ iterations the above algorithm terminates. 
Since the algorithm from \Cref{lemma: better clustering 2} only returns FAIL with probability at most $1/m^{11}$, the total probability that our algorithm returns FAIL is at most $1/m^{10}$. From now on we focus on the proof of \Cref{lemma: better clustering 2}.

\subsubsection{Proof of \Cref{lemma: better clustering 2}}
\label{subsec: getting nice structure better clustering}

\begin{figure}[h]
	\centering
	\subfigure[A schematic view of graph $G$. Clusters of $\wset'$ are shown in green. Clusters of $\rset$ are shown in red (with clusters of $\rset'$ shaded). Clusters of $\cset$ are shown in blue (with clusters of $\cset'$ shaded). Regular vertices lying outside of clusters of $\cset$ are shown in black. Note that, if there exist clusters $C\in \cset$ and $W'\in \wset'$ with $C\subseteq W'$, then there exists a cluster $R\in \rset$ with $C\subseteq R\subseteq W'$. Also, for every cluster $W'\in \wset'$, $R^*\not\subseteq W'$ and $v^*\notin W'$ hold.]{\scalebox{0.095}{\includegraphics{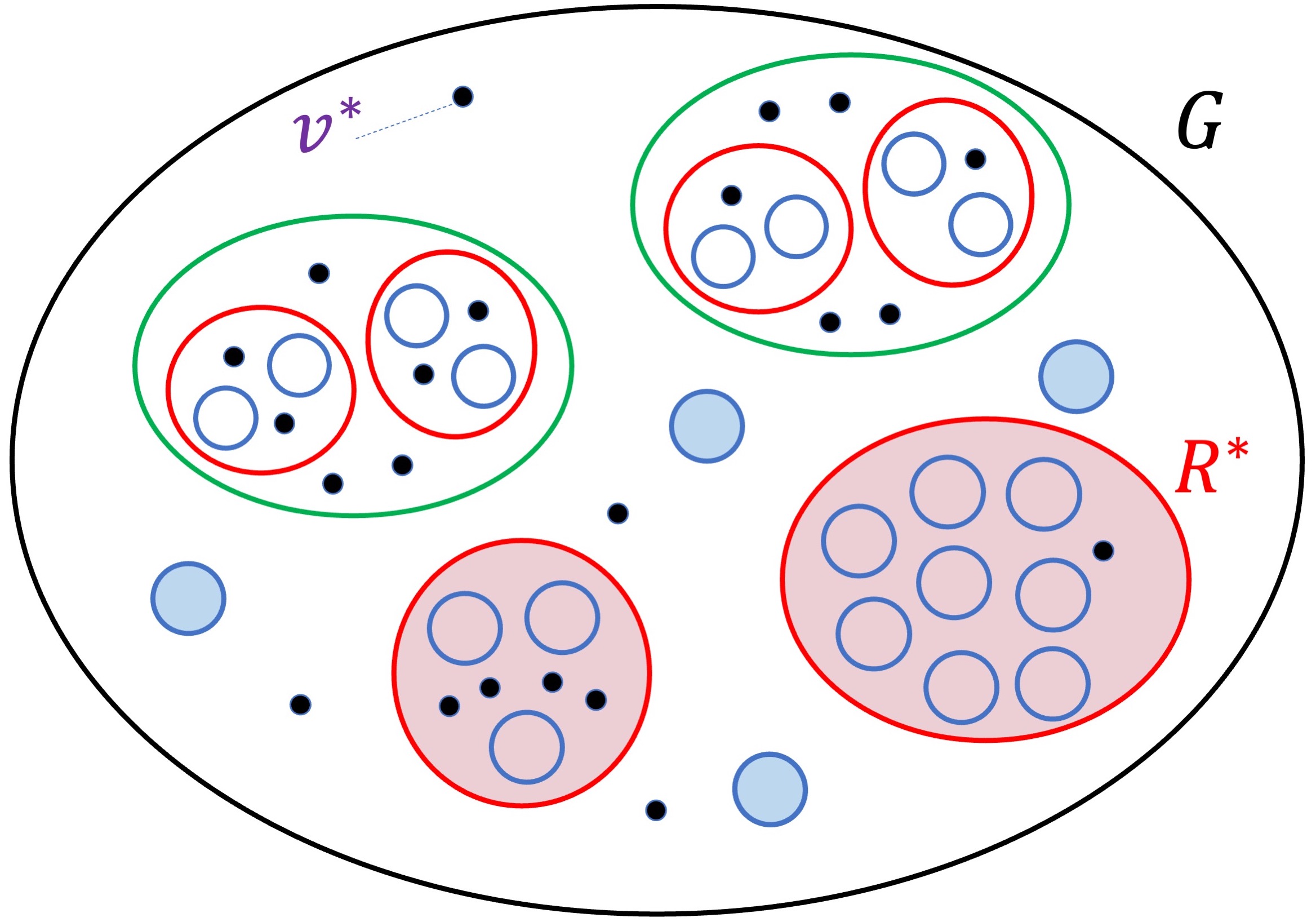}}\label{fig: NF3}}
	\hspace{1cm}
	\subfigure[A schematic view of graph $\hat H'$. Regular vertices are shown in black, $C$-nodes are shown in blue, $R$-nodes are shown in red, and $W$-nodes are shown in green.]{\scalebox{0.095}{\includegraphics{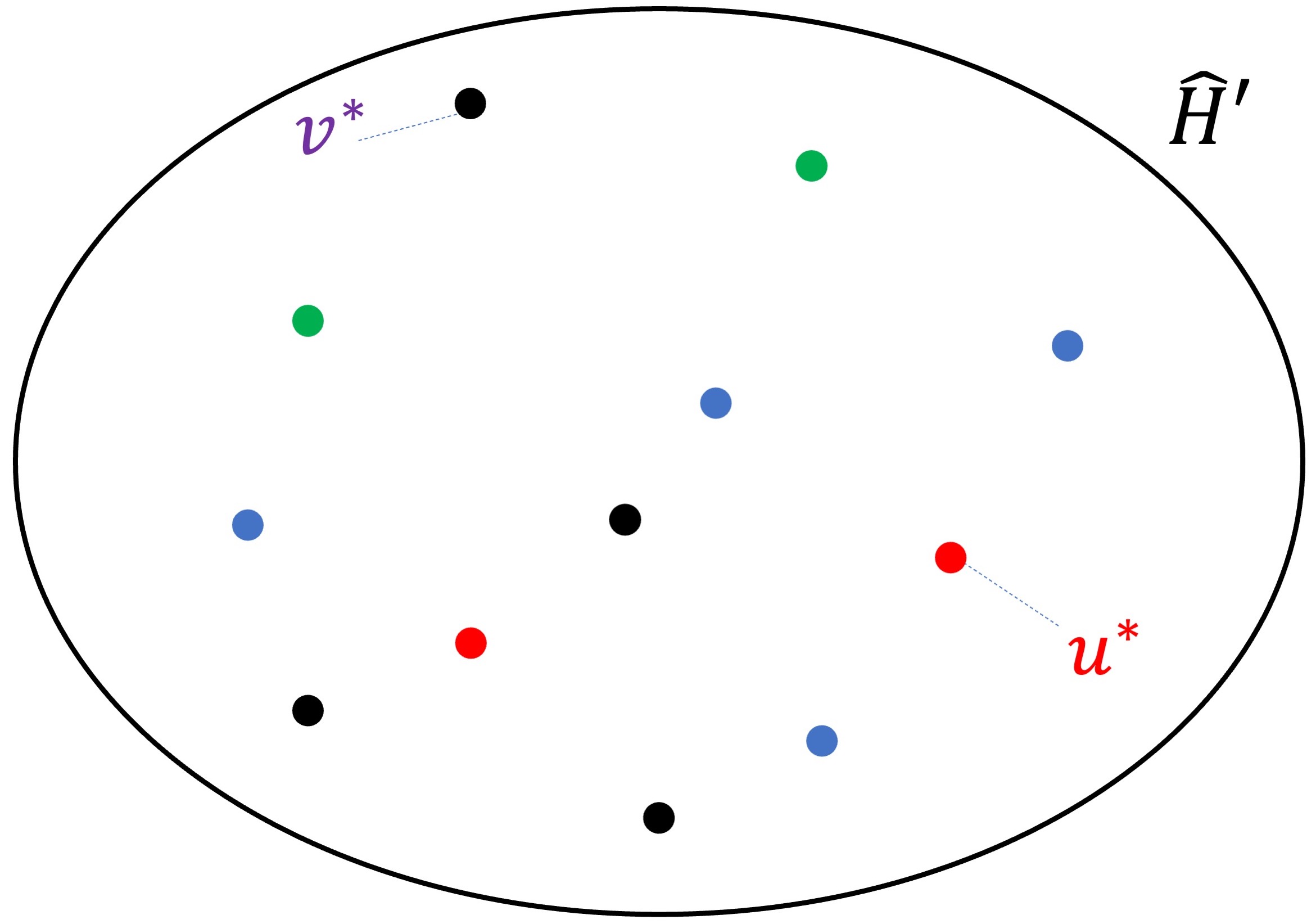}}\label{fig: NF4}}
	\caption{Graphs $G$ and $\hat H'$.}
\end{figure}

Observe that so far, we have constructed a 3-level hierarchical clustering of the graph $G$. The first level consists of the set $\cset$ of basic clusters of graph $G$. At the second level, there is a set $\rset$ of clusters of graph $G$. Recall that, for every basic cluster $C\in \cset$, either $C\subseteq G\setminus\left(\bigcup_{R\in \rset}V(R)\right )$, or 
there is some cluster $R\in \rset$, with $C\subseteq R$. As before, we denote by $\cset'\subseteq \cset$ the set of all basic clusters $C$ with $C\subseteq G\setminus\left(\bigcup_{R\in \rset}V(R)\right )$. We can use the valid set $\wset$ of $W$-clusters in graph $\hat H$, in order to construct another set $\wset'$ of clusters in the original graph $G$, as follows. Recall that every cluster $W\in \wset$ may only contain $R$-nodes or regular vertices of $\hat H$. For each such cluster $W$, let $\rset(W)\subseteq \rset$ be the set of all clusters $R\in \rset$ with $v_R\in V(W)$. We then let $W'$ be a subgraph of $G$ induced by vertex set $\left (\bigcup_{R\in \rset(W)}V(R)\right )\cup (V(G)\cap V(W))$. 
In other words, $V(W')$ contains all regular vertices of $W$, and all vertices lying in clusters of $\rset(W)$.
We will refer to $W'$ as the \emph{cluster of $G$ defined by $W$}. Finally, let $\wset'=\set{W'\mid W\in \wset}$. Note that every basic cluster $C\in \cset'$ must be disjoint from clusters of $\wset'$. We denote by $\rset'=\rset\setminus \left(\bigcup_{W\in \wset}\rset(W)\right )$. Note that each cluster $R\in \rset'$ is contained in  $G\setminus\left(\bigcup_{W'\in \wset'}V(W')\right )$, while for each cluster $R\in \rset\setminus \rset'$, there is some cluster $W'\in \wset'$ with $R\subseteq W'$. Therefore, $\cset'\cup \rset'\cup \wset'$ is a collection of disjoint clusters of graph $G$ (see \Cref{fig: NF3}).
Recall that we are guaranteed that every cluster $C\in \cset'$ has the $\alpha_0$-bandwidth property, where $\alpha_0=1/\log^3m$, and every cluster $R\in \rset'$ has the $\alpha_1$-bandwidth property, where $\alpha_1=1/\log^6m$. Lastly, every cluster $W\in \wset$ has the $\alpha'$-bandwidth property (for $\alpha'=1/(c\log^{2.5}m)$, where $c$ is a large enough constant) in graph $\hat H$.  From \Cref{cor: contracted_graph_well_linkedness}, every cluster $W'\in \wset'$ has the $\alpha_1\cdot \alpha'=1/(c\log^{8.5}m)$-bandwidth property in graph $G$.

In order to prove \Cref{lemma: better clustering}, it will be convenient for us to work with graph $\hat H'$, that is a contracted graph of $\hat H$, with respect to set $\wset$ of clusters, that is, $\hat H'=\hat H_{|\wset}$. Since graph $\hat H$ is itself a contracted graph of $G$ with respect to $\rset\cup \cset'$, it is easy to verify that $\hat H'=G_{|\cset'\cup\rset'\cup \wset'}$ (see \Cref{fig: NF4}).
The vertices of graph $\hat H'$ are partitioned into four types. The first type is regular vertices, which are also the vertices of the original graph $G$; note that the special vertex $v^*$ belongs to graph $\hat H$ as a regular vertex. The second type is supernodes corresponding to clusters of $\cset'$, that we refer to as $C$-nodes. The third type is supernodes corresponding to clusters of $\rset'$, that we call $R$-nodes, and it includes the vertex $u^*$, representing  the cluster $R^*$. The fourth type is supernodes corresponding to clusters of $\wset'$, that we call $W$-nodes. 
We denote the set of all regular vertices of $\hat H'$, excluding the special vertex $v^*$, by $U^*$. We denote the set of all $R$-nodes, excluding the vertex $u^*$, by $U^{R}$. 

The remainder of the proof of \Cref{lemma: better clustering} consists of three steps. In the first step, we perform some manipulations that will allow us to either compute a new valid set $\tilde \wset$ of $W$-clusters in the current graph $\hat H$, such that $|E(\hat H_{|\tilde \wset})|<|E(\hat H_{|\wset})|$, or to organize the vertices of $U^*\cup U^R$ into a nice layered structure. We also define a collection $\jset$ of clusters of the graph $\hat H'$ in this step.  In the second step, we will define another contracted graph $\cH$ with respect to the clustering $\jset$, and explore some of its properties. In particular, we define the notion of a ``simplifying cluster'' in $\cH$, and show an algorithm that, given a simplifying cluster in $\cH$, produces a helpful clustering $\tilde \rset$ of $G$ that is better than the current clustering $\rset$. Lastly, in the third setp, we either compute a nice witness structure in graph $G_{|\rset}$ as required, or compute a simplifying cluster in graph $\cH$, which in turn allows us to produce a helpful clustering $\tilde \rset$ in graph $G$ that is better than $\rset$.
We now describe these steps one by one.

\subsubsection*{Step 1: Layering the Vertices of $U^*\cup U^R$ and  Clustering $\jset$}

Consider the graph $\hat H'$, and let $C_0$ be the subgraph of $\hat H'$, induced by vertex set $V(\hat H')\setminus (U^*\cup U^R)$. We use the algorithm from \Cref{thm: layered well linked decomposition}, to compute a layered $\alpha'$-well-linked decomposition $(\sset,(\lset_1,\ldots,\lset_h))$ of $\hat H'$ with respect to $C_0$, where $\alpha'=\Theta(1/\log^{2.5}m)$ is the parameter that was used in the definition of a valid set of $W$-clusters, such that $h\leq \log m$. We say that a cluster $S\in \sset$ is a \emph{singleton cluster}, if it contains a single vertex of $\hat H'$. Assume first that $\sset$ contains a non-singleton cluster $S$. Recall that $S$ has the $\alpha'$-bandwidth property in $\hat H'$ (from Property \ref{condition: layered well linked} of layered well-linked decomposition; see \Cref{subsec: layered wld}), and it only contains verices of $U^*\cup U^R$. Therefore, $S$ is also a cluster of graph $\hat H$, and it has the $\alpha'$-bandwidth property in $\hat H$.
Moreover, from Property \ref{condition: layered decomp each cluster prop} of the layered well-linked decomposition, 
$|E_{\hat H}(S)|=|E_{\hat H'}(S)|\geq |\delta_{\hat H'}(S)|/(64\log m)=|\delta_{\hat H}(S)|/(64\log m)$ holds.
 Since $S$ is disjoint from clusters in set $\wset$, we get that $\tilde \wset=\wset\cup \set{S}$ is a valid set of $W$-clusters in graph $\hat H$, with $|E(\hat H_{|\tilde\wset})|<|E(\hat H_{|\wset})|$. We terminate the algorithm and return the new valid set $\tilde \wset$ of $W$-clusters.
Therfore, we will assume from now on that every cluster in set $\sset$ is a singleton cluster. The partition $(\lset_1,\ldots,\lset_h)$ of the clusters of $\sset$ into layers then immediately defines a partition $L_1,\ldots,L_h$ of vertices of $U^*\cup U^R$ into layers, where vertex $u$ lies in layer $U_i$ iff cluster $\set{u}\in \sset$ lies in $\lset_i$. For convenience, we denote by $L_0=V(\hat H')\setminus (U^*\cup U^R)$. For every vertex $u\in U^*\cup U^R$ that lies in some layer $L_i$, for $1\leq i\leq h$, we partition the edges of $\delta_{\hat H'}(u)$ into two subsets: set $\delta^{\down}(u)$ contains all edges $(u,u')$ with $u'\in L_0\cup L_1\cup \cdots\cup L_{i-1}$, and set $\delta^{\up}(u)$ contains all remaining edges of $\delta_{\hat H'}(u)$. Note that, from Property \ref{condition: layered decomp edge ratio}  of the layered well-linked decomposition, for every vertex $u\in U^*\cup U^R$, $|\delta^{\up}(u)|<|\delta^{\down}(u)|/\log m$. 

In the remainder of the proof of \Cref{lemma: better clustering}, we will attempt to construct a nice witness structure for graph $G_{|\rset}$, with respect to the set $\cset'$ of clusters, containing every cluster $C\in \cset$ with $C\subseteq G\setminus\left (\bigcup_{R\in \rset}V(R)\right )$. If we fail to do so, then we will compute another helpful clustering $\tilde \rset$ of graph $G$ with respect to vertex $v^*$ and set $\cset$ of basic clusters, such that $R^*\in \rset$ and $\tilde \rset$ is a better clustering than $\rset$.

In order to do so, we construct a collection $\jset$ of clusters in graph $\hat H'$. Every cluster $J\in \jset$ will contain exactly one vertex that is either a $C$-node or $W$-node, that we refer to as the \emph{center of the cluster}, and possibly a number of additional vertices from $U^*\cup U^R$. 
Initially, for every vertex $u$ of $\hat H'$ that is either a $C$-node or a $W$-node, we construct a cluster $J(u)\in \jset$, that only contains the vertex $u$ as its center node. We then iterate. As long as there exists a vertex $u'\in U^*\cup U^R$, such that at least $|\delta_{\hat H'}(u')|/128$ edges connect $u'$ to the vertices of some cluster $J\in \jset$, we add vertex $u'$, together with all edges connecting $u'$ to $V(J)$, to cluster $J$. We also delete $u'$ from vertex set $U^*$ or $U^R$ in which it lies. If $u'\in L_i$, for some $1\leq i\leq h$, then we delete $u'$ from $L_i$ and add it to $L_0$. 

Consider the set $\jset$ of clusters in graph $\hat H'$, that is obtained at the end of this procedure. It is immediate to verify that all clusters in $\jset$ are mutually disjoint; every cluster $J\in \jset$ contains a single center node that is a $C$-node or a $W$-node, and each remaining vertex of $J$ lies in $U^R\cup U^*$. We need the following observation, whose proof is deferred to Section \ref{subsec:J-clusters well-linked} of Appendix.

\begin{observation}\label{obs:J wl}
	Every cluster $J\in \jset$, has the $\Omega(1/\log m)$-bandwidth property in graph $\hat H'$.
\end{observation}

\subsubsection*{Step 2: New Contracted Graph and Simplifying Clusters}

We start by revisiting the current hierarchical (4-level) clustering of $G$ and defining a new contracted graph. Recall that our starting point is a graph $G$, with a special vertex $v^*$, and a collection $\cset$ of disjoint basic clusters in $G$, such that $v^*$ does not lie in any cluster of $\cset$. Recall that every cluster in $\cset$ has the $\alpha_0$-bandwidth property, where $\alpha_0=1/\log^3m$. This is the first-level clustering.

The second level of clustering is the helpful clustering $\rset$, which is also a collection of disjoint clusters, each of which has the $\alpha_1$-bandwidth property, where $\alpha_1=1/\log^6m$. Recall that $v^*$ may not lie in any cluster of $\rset$, and, for every cluster $C\in \cset$, either $C\subseteq G\setminus\left(\bigcup_{R\in \rset}R\right )$; or there is some cluster $R\in \rset$ with $C\subseteq R$. Recall that we have denoted by $\cset'\subseteq \cset$ the set of all clusters $C\in \cset$ with $C\subseteq G\setminus\left(\bigcup_{R\in \rset}R\right )$.  Recall also that we have defined a distinguished cluster $R^*\in \rset$.

The third level of clustering is a $W$-clustering $\wset$, that is defined with respect to the contracted graph $\hat H=G_{|\cset'\cup \rset}$. Recall that for every cluster $W\in \wset$, every vertex of $W$ is either a regular vertex or an $R$-node, and $W$ may not contain the special vertex $v^*$ or the $R$-node $u^*$ representing the distinguished cluster $R^*$. We have defined, for every cluster $W\in \wset$, the corresponding cluster $W'\subseteq G$, that is, intuitively, obtained from $W$ by un-contracting every cluster $R\in \rset$ with $v_{R}\in V(W)$. We have then set $\wset'=\set{W'\mid W\in \wset}$, and we have established that every cluster $W'\in \wset'$ has the $\Omega(1/\log^{8.5}m)$-bandwidth property in graph $G$. Observe that for every pair $C\in \cset'$, $W'\in \wset'$ of clusters, $C\cap W'=\emptyset$ must hold. For every pair $R\in \rset$, $W'\in \wset'$ of clusters, either $R\subseteq W'$, or $R\cap W'$ must hold. We denote by $\rset'$ the set of all custers $R\in \rset$ with $R\subseteq G\setminus\left(\bigcup_{W'\in \wset'}W'\right)$. Observe that $R^*\in \rset'$ must hold, and that $\cset'\cup \rset'\cup \wset'$ is a collection of disjoint clusters in graph $G$. Graph $\hat H'=\hat H_{|\wset}$ that we used in Step 1 is precisely the graph $G_{|\cset'\cup \rset'\cup \wset'}$.

The fourth and the last level of clustering is defined by the collection $\jset$ of clusters in graph $\hat H'$ that we have defined in Step 1. Recall that, for every cluster $J\in \jset$ (that is a subgraph of $\hat H'$), there is a unique center vertex, that is either a $C$-node or a $W$-node, and the remaining vertices of $J$ are regular vertices or $R$-nodes; however, $J$ may not contain the special vertex $v^*$ or the $R$-node $u^*$ representing the distinguished cluster $R^*$. Moreover, every $C$-node and every $W$-node is a center of some cluster in $J$.

\begin{figure}[h]
	\centering
	\includegraphics[scale=0.12]{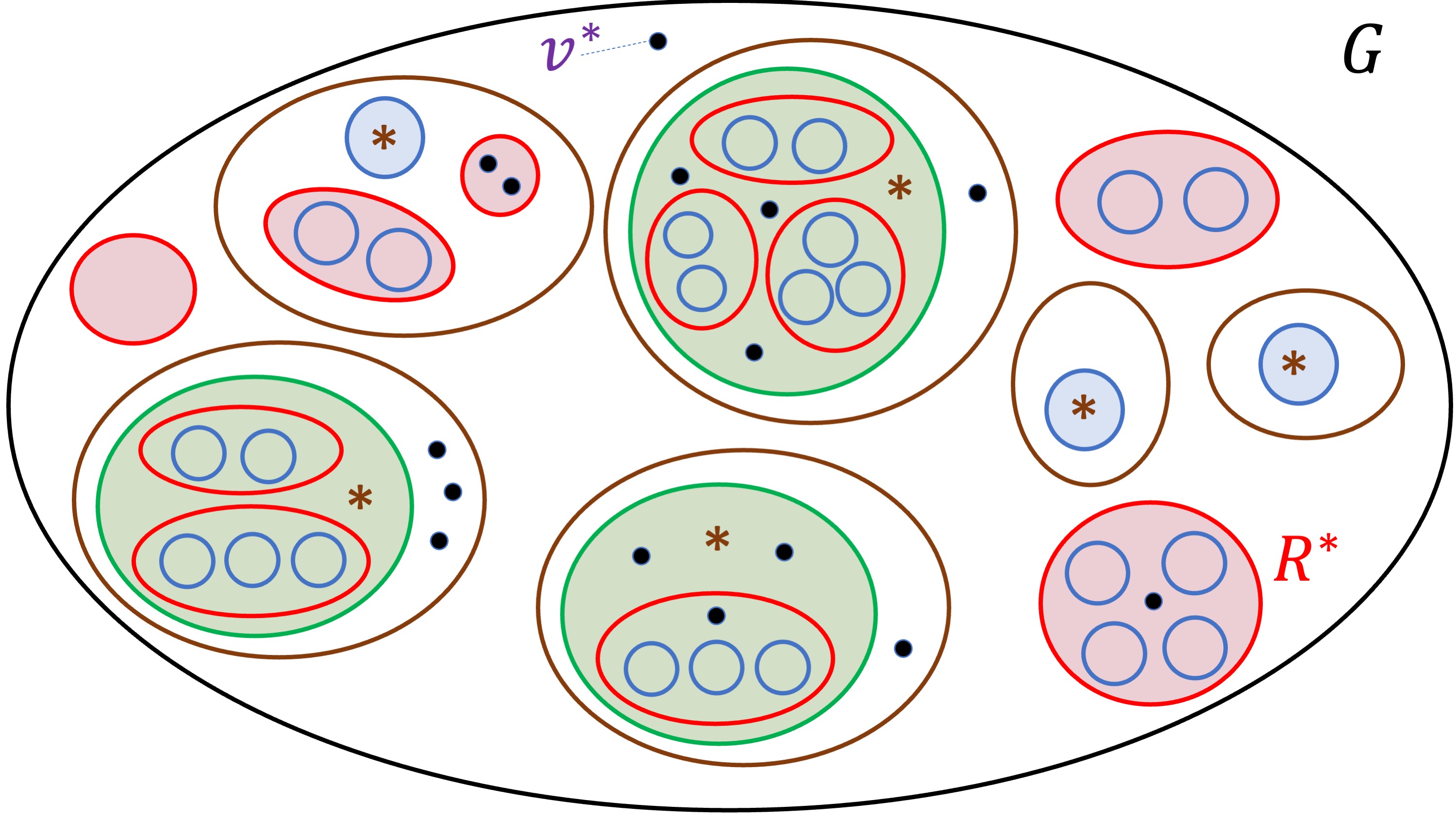}
	\caption{An illustration of a $\jset$-clustering. Clusters of $\cset$ are shown in blue (with clusters of $\cset'$ shaded).
		Clusters of $\rset$ are shown in red (with clusters of $\rset'$ shaded).
		Clusters of $\wset'$ are shown in green. Each cluster of $\wset'$ may contain clusters of $\rset$, but if a cluster $C\in \cset$ is contained in $W'$, then there exists $R\in \rset$ with $C\subseteq R\subseteq W'$.
		Vertices of $G$ that do not lie in clusters of $\cset$ are shown in black. 
		Clusters of $\jset'$ are shown in brown. Each cluster of $\jset'$ contains a cluster of $\wset'$ or $\cset'$ as its center cluster (indicated by $*$). Each cluster of $\wset'\cup \cset'$ is a center-cluster of some cluster of $\jset'$. In addition to the center-cluster, a cluster of $\jset'$ may contain clusters of $\rset$ and regular vertices. Some clusters of $\rset'$ and some regular vertices may not lie in any cluster of $\jset'$.
	}\label{fig: NF5}
\end{figure}

As before, we will define, for every cluster $J\in \jset$, a corresponding cluster $J'$ in graph $G$, in a natural way. We first define the vertex set $V(J')$, and then let $J'$ be the subgraph of $G$ induced by $V(J')$. First, we add to $V(J')$ every regular vertex that lies in $J$ -- each such vertex is a vertex of $G$. Next, for every $R$-node $v_{R}\in J$, we add all vertices of cluster $R$ to $V(J')$; observe that $R\in \rset'\setminus \set{R^*}$ must hold. Lastly, we consider the unique center vertex of $J$. If that vertex is a $C$-node, corresponding to a cluster $C\in \cset'$, then we add all vertices of $C$ to $V(J')$. Otherwise, the vertex is a $W$-node, representing some cluster $W'\in \wset'$. We then add to $V(J')$ all vertices of $V(W')$. Lastly, we set $J'=G[V(J')]$.
We denote by $\jset'=\set{J'\mid J\in \jset}$ the set of clusters in graph $G$ corresponding to the cluster set $\jset$ in $\hat H'$. Observe that for every cluster $C\in \cset'$, there is a unique cluster $J'(C)\in \jset'$ containing $C$; we call $C$ the \emph{center-cluster} of $J'(C)$. Similarly, for every cluster $W'\in \wset'$, there is a unique cluster $J'(W)\in \jset'$ containing $W'$; we similarly call $W'$ the \emph{center-cluster} of $J'(W)$. Lastly, for every pair of clusters $R\in \rset'$, $J'\in \jset'$, either $R\subseteq J'$ or $R\cap J'=\emptyset$ holds. We denote by $\rset''\subseteq \rset'$ the set of all custers $R\in \rset'$, with $R\subseteq G\setminus\left(\bigcup_{J'\in \jset}J'\right )$ (see \Cref{fig: NF5}). Note that $R^*\in \rset''$. 
Observe also that $\rset''\cup \jset'$ defines a collection of disjoint clusters in graph $G$. Note that the special vertex $v^*$ does not lie in any cluster of $\rset''\cup \jset'$, and that every cluster of $\cset$, $\rset$, and $\wset'$ is contained in exactly one cluster of $\rset''\cup \jset'$.

Since every cluster in $\cset$ has the $\alpha_0=1/\log^3m$-bandwidth property; every cluster $W\in \wset'$ has the  $\Omega(1/\log^{8.5}m)$-bandwidth property; and every cluster $R\in \rset$ has the $1/\log^6m$-bandwidth property in graph $G$, while, from \Cref{obs:J wl},
every cluster $J\in \jset$, has the $\Omega(1/\log m)$-bandwidth property in graph $\hat H'=G_{|\cset'\cup \rset'\cup \wset'}$, from \Cref{cor: contracted_graph_well_linkedness}, we get that every cluster $J'\in \jset'$ has the $\Omega(1/\log^{9.5}m)$-bandwidth property. This property will be useful for us later, so we summarize it in the following observation.

\begin{observation}\label{obs: J' clusters wl}
	 Every cluster $J'\in \jset'$ has the $\Omega(1/\log^{9.5}m)$-bandwidth property in graph $G$.
	\end{observation}


In the remainder of the proof of  \Cref{lemma: better clustering 2} we consider the contracted graph $\cH=G_{|\jset'\cup \rset''}$, which is exactly the contracted graph of $\hat H'$ with respect to cluster set $\jset$, that is, $\cH=\hat H'_{|\jset}$. 
The set of vertices of $\cH$ consists of three subsets: the set $V(G)\cap V(\cH)$ of regular vertices; the set $\set{v_{R}\mid R\in \rset''}$ of supernodes corresponding to clusters of $\rset''$ that we call $R$-nodes; and the set $\set{v_{J'}\mid J'\in \jset'}$ of supernodes corresponding to clusters of $\jset'$ that we call $J$-nodes.
For convenience, we denote by $\hat U^*$ the set of all regular vertices of $\cH$ excluding the special vertex $v^*$, and we denote by $\hat U^R$ the set of all $R$-nodes of $\cH$ excluding the node $u^*$ that represents the distinguished cluster $R^*$.
The algorithm from Step 1 ensures the following property of graph $\cH$:

\begin{properties}{H}
	\item for every vertex $u\in \hat U^*\cup \hat U^R$ and $J$-node $v_{J'}$, the number of edges connecting $u$ to $v_{J'}$ in $\cH$ is at most $|\delta_{\cH}(u)|/128$.
	\end{properties}

Indeed, if the above property does not hold for a vertex $u\in \hat U^*\cup \hat U^R$ and a $J$-node $v_{J'}$, then vertex $u$ should have been added to the cluster $J\in \jset$ that corresponds to cluster $J'\in \jset'$ by the algorithm that constructed the clusters in $\jset$.

Additionally, the algorithm from Step 1 defines  a partition $(L_1,L_2,\ldots,L_h)$ of the set $\hat U^*\cup \hat U^R$ of vertices, with $h\leq \log m$. Let $L_0$ be the set of vertices of $\cH$ containing all $J$-nodes and the special vertices $v^*,u^*$; equivalently, $L_0=V(\cH)\setminus (\hat U^*\cup \hat U^R)$. Recall that for all $1\leq i\leq h$, for every vertex $u\in L_i$, we have partitioned the edge set $\delta_{\cH}(u)$ into two subsets: set $\delta^{\down}(u)$ contains all edges connecting $u$ to vertices of $L_0\cup\cdots\cup L_{i-1}$, while  set $\delta^{\up}(u)$ contains all remaining edges of $\delta_{\cH}(u)$. Recall that we have also ensured that the following property holds:

\begin{properties}[1]{H}
	\item for every vertex $u\in \hat U^*\cup \hat U^R$, $|\delta^{\up}(u)|<|\delta^{\down}(u)|/\log m$. \label{prop: up and down}
\end{properties}

Next, we define the notion of a simplifying cluster in graph $\cH$. We will then show that, given a simplifying cluster in $\cH$, we can efficiently compute a helpful clustering $\tilde \rset$ of graph $G$ with respect to vertex $v^*$ and set $\cset$ of basic clusters, such that $R^*\in \tilde \rset$ and $\tilde \rset$ is a better clustering than $\rset$. 

\begin{definition}[Simplifying Cluster]
	Let $S$ be a vertex-induced subgraph of $\cH$. We say that $S$ is a \emph{simplifying cluster} if:
	
	\begin{itemize}
		\item vertices $v^*,u^*$ do not lie in $S$; 
		\item there is a set $\pset(S)$ of paths in graph $\cH$ (external $S$-router), routing the edges of $\delta_{\cH}(S)$ to a single vertex of $\cH\setminus S$, such that all paths in $\pset(S)$ are internally disjoint from $S$, and they cause congestion at most $\beta'=O(\log m)$;  and
		\item either $S$ contains at least one $J$-node, or $|E_{\cH}(S)|\geq |\delta_{\cH}(S)|/\log m$.
	\end{itemize}
\end{definition}

We will use the following simple observation.

\begin{observation}\label{obs: check simplifying}
	There is an efficient algorithm, that, given a cluster $S\subseteq \cH$, establishes whether $S$ is a simplifying cluster.
\end{observation}
\begin{proof}
	In order to establish whether $S$ is a simplifying cluster, we need to check whether $S$ contains a $J$-node, or $|E_{\cH}(S)|\geq |\delta_{\cH}(S)|/\log m$ holds, which can be done efficiently. Additionally, we need to check whether there is a set of paths in graph $\cH$, routing the edges of $\delta_{\cH}(S)$ to a single vertex of $\cH\setminus S$, such that the paths are internally disjoint from $S$ and cause congestion at most $\beta'$. The latter can be done efficiently by computing maximum flow between the vertices of $S$ and each vertex of $\cH\setminus S$ in turn.
\end{proof}

In the next claim we show that, if we are given a simplifying cluster $S$ in $\cH$, then 
we can efficiently compute a helpful clustering $\tilde \rset$ of graph $G$ with respect to $v^*$ and $\cset$, such that $R^*\in \tilde \rset$ and $\tilde \rset$ is a better clustering than $\rset$. The proof of the claim is somewhat technical and is deferred to \Cref{subsec: simplifying cluster is enough}

\begin{claim}\label{claim: simplifying cluster is enough}
	There is an efficient algorithm, that, given a simplifying cluster $S$ of $\cH$, computes a helpful clustering $(\tilde \rset,\set{\dset'(R)}_{R\in \tilde \rset})$ of graph $G$ with respect to the special vertex $v^*$ and the set $\cset$ of basic clusters, such that $R^*\in \tilde\rset$, and $\tilde \rset$ is a better clustering than $\rset$. 
\end{claim}

Let $\tau$ be the Gomory-Hu tree of the graph $\cH$. We root the tree at the special vertex $v^*$, and, for every vertex $u\in V(\tau)$, we denote by $\tau_u$ the subtree of $\tau$ rooted at vertex $u$. 

Assume first that there is some vertex $u\in V(\tau)$, such that the special $R$-node $u^*$ corresponding to the distinguished cluster $R^*$ does not lie in $\tau_u$, but some $J$-node $v_{J'}$ lies in $\tau_u$.
In this case, we let $S$ be a subgraph of $\cH$ that is induced by  $V(\tau_u)$. We claim that $S$ is a simplifying cluster. Indeed, from the construction, neither of $v^*,u^*$ may lie in $S$, and at least one $J$-node lies in $S$. Let $u'$ be the parent-vertex of $u$ in the tree $\tau$. Then from the properties of Gomory-Hu tree (see \Cref{cor: G-H tree_edge_cut}), $(V(S),V(\cH)\setminus V(S)$ is a minimum cut separating $u$ from $u'$ in $\cH$. From the max-flow / min-cut theorem, there is a collection $\pset$ of $|\delta_{\cH}(S)|$ edge-disjoint paths connecting $u$ to $u'$ in $\cH$. Clearly, each edge $e\in \delta_{\cH}(S)$ is contained in exactly one path of $\pset$, that we denote by $P(e)$. Let $P'(e)$ be the subpath of $P(e)$ that starts at edge $e$ and terminates at $u'$. Then $P'(e)$ must be internally disjoint from $S$. Therefore, $\pset(S)=\set{P'(e)\mid e\in \delta_{\cH}(S)}$ is a set of  edge-disjoint paths in graph $\cH$, routing the edges of $\delta_{\cH}(S)$ to vertex $u'\in \cH\setminus S$, and the paths in $\pset(S)$ are internally disjoint from $S$. We conclude that $S$ is a simplifying cluster. We can now use the algorithm from \Cref{claim: simplifying cluster is enough} to compute a helpful clustering $(\tilde \rset,\set{\dset'_R}_{R\in \tilde \rset})$ of graph $G$ with respect to the special vertex $v^*$ and the set $\cset$ of basic clusters, such that $R^*\in \tilde\rset$, and $\tilde \rset$ is a better clustering than $\rset$. 

Therefore, we assume from now on that, for every vertex $u\in V(\tau)$, if $u^*$ does not lie in $\tau_u$, then $\tau_u$ does not contain any $J$-node, and so $V(\tau_u)\subseteq \hat U^*\cup \hat U^R$. 
Let $P^*$ denote the path connecting $v^*$ to $u^*$ in the tree $\tau$. We denote the sequence of vertices on the path by $v^*=u_1,u_2,\ldots,u_r=u^*$. For all $1\leq i\leq r$, we define a cluster $S_i$ of $\cH$, associated with vertex $u_i$, as follows. We let $S_r$ be the subgraph of $\cH$ induced by the vertices of $\tau_{u_r}$. Consider now some index $1\leq i< r$. Let $\set{x_i^0,x_i^1,x_i^2,\ldots,x^{q_i}_i}$ be the set of all child-vertices of $u_i$ in the tree $\tau$, and assume that $x_i^0=u_{i+1}$. We then let $S_i$ be the subgraph of $\cH$ induced by the set $\set{u_i}\cup V(\tau_{x_i^1})\cup V(\tau_{x_i^2})\cup \cdots\cup V(\tau_{x^{q_i}_i})$ of vertices. In other words, we include in vertex set $V(S_i)$ the vertices lying in all subtrees of the children of $u_i$, except for the vertices lying in the subtree of $u_{i+1}$. From our assumption, for all $1\leq i\leq r$, the only vertex of $S_i$ that may be a $J$-node is the vertex $u_i$; all other vertices of $S_i$ are $R$-nodes or regular vertices (and it is also possible that $u_i$ is an $R$-node or a regular vertex).

For all $1\leq i\leq r$, we also define a subgraph $S'_i\subseteq S_i$, that is constructed as follows. We start by constructing the set $V(S'_i)$ of vertices. Initially, we let $V(S'_i)=\set{u_i}$. While there is any vertex $u\in S_i\setminus S'_i$, such that at the number of edges connecting $u$ to vertices of $V(S'_i)$ is at least $|\delta_{\cH}(u)|/128$, then we add $u$ to $V(S'_i)$. Once this algorithm terminates, we let $S'_i$ be the subgraph of $\cH$ induced by the set $V(S'_i)$ of vertices. 
Recall that we have established that, if $v$ is a vertex of $S_i\setminus S_i'$, for some $1\leq i\leq r$, then $v\in \hat U^*\cup \hat U^R$ must hold. The following observation easily follows from the construction of $J$-clusters.

\begin{observation}\label{obs: central path j-cluster}
	Consider any index $1<i<r$, for which $u_i$ is a $J$-node. Then $S'_i=\set{u_i}$.
\end{observation}

\begin{proof}
	Let $J\in \jset$ be the cluster of $\hat H'$ that node $u_i$ represents (recall that we can think of graph $\cH$ as a contracted graph of $\hat H'$ with respect to cluster set $\jset$). Assume for contradiction that $S'_i$ contains at least one vertex in addition to $u_i$, and let $v$ be the first vertex that was added to cluster $S'_i$. Then the number of edges connecting $v$ to $u_i$ is at least $|\delta_{\cH}(v)|/128$. But then $v$ is also a vertex of graph $\hat H'$, in which it serves as either an $R$-node distinct from $u^*$, or a regular vertex distinct from $v^*$. Moreover,  the number of edges connecting $v$ to vertices of $J$ is at least $|\delta_{\hat H'}(v)|/128$. Therefore, $v$ should have been added to cluster $J$ when it was constructed, a contradiction.
\end{proof}

 Additionally, we get the following observation, whose proof is identical to the proof of \Cref{obs:J wl} and is omitted here.

\begin{observation}\label{obs: S'i wl}
	For all $1\leq i\leq r$, cluster $S'_i$ has the $\Omega(1/\log m)$-bandwidth property in graph $\cH$.
\end{observation}

For all $1\leq i\leq r$, we employ the algorithm from \Cref{obs: check simplifying} in order to establish whether $S'_i$ is a simplifying cluster. Additionally, for all $1\leq i<r$, we use the algorithm from \Cref{obs: check simplifying} in order to establish whether the subgraph of $\cH$ induced by vertex set $V(S_i)\cup V(S_{i+1})$ is a simplifying cluster. If the algorithm from  \Cref{obs: check simplifying} establishes that any of the above clusters is a simplifying cluster,  then we can use the algorithm from \Cref{claim: simplifying cluster is enough} to compute a helpful clustering $(\tilde \rset,\set{\dset'(R)}_{R\in \tilde\rset}$ of graph $G$ with respect to the special vertex $v^*$ and the set $\cset$ of basic clusters, such that $R^*\in \tilde\rset$, and $\tilde \rset$ is a better clustering than $\rset$. 
Therefore, we assume from now on that, for all $1\leq i\leq r$, cluster $S'_i$ is not a simplifying cluster, and for all $1\leq i<r$, the subgraph of $\cH$ induced by $V(S_i)\cup V(S_{i+1})$ is not a simplifying cluster.  

We will show, in Step 3, an efficient algorithm that constructs a nice witness structure for graph $G_{|\rset}$, with respect to the set $\cset'$ of clusters, that contains every cluster $C\in \cset$ with $C\subseteq G\setminus\left (\bigcup_{R\in \rset}V(R)\right )$.

\subsubsection*{Step 3: Constructing the Nice Witness Structure}

The goal of this step is to construct a nice witness structure for graph $G_{|\rset}$, with respect to the set $\cset''$ of clusters, that contains every cluster $C\in \cset$ with $C\subseteq G\setminus\left (\bigcup_{R\in \rset}V(R)\right )$.

Intuitively, we will use the clusters $S_1,\ldots,S_r$ that we just defined in order to define the spine $\tilde \sset=\set{\tilde S_1,\ldots,\tilde S_r}$ of the nice witness structure in the natural way: cluster $\tilde S_i$ will be obtained from $S_i$ by first replacing every $R$-node and every $J$-node of $S_i$ with the corresponding cluster $R\in \rset''$ or $J'\in \jset'$, and then contracting the $R$-clusters back. Similarly, we will use the clusters $S'_1,\ldots,S'_r$ in order to define the verterbrae $\tilde S'_1,\ldots,\tilde S'_r$ of the nice witness structure.  

We partition the set $E(\cH)$ of edges into two disjoint subsets, $E'$ and $E''$, as follows. Set $E'$ contains all edges of $\bigcup_{i=1}^rE(S_i')$, and, additionally, for all $1\leq i<r$, it contains every edge $e=(u,v)$ with $u\in S_i'$, $v\in  S_{i+1}'$. Set $E''$ contains all remaining edges of $E(\cH)$. Additionally, we let $\hat E\subseteq E''$ be the set of all edges $(u,v)\in E''$, where $u$ and $v$ lie in different sets of $\set{ S_1,\ldots, S_r}$.

Next, we develop some tools that will allow us to define the set $\pset=\set{P(e)\mid e\in \hat E}$  of nice guiding paths for the nice witness structure that we construct.
Recall that in Step 1 of the algorithm, we have partitioned the set $U^*\cup U^R$ of vertices of graph $\hat H'$ into layers $L_1,\ldots,L_h$, where $h\leq \log m$. Recall that $U^*$ is the set of all regular vertices (excluding $v^*$), and $U^R$ is the set of all $R$-nodes (excluding $u^*$) of graph $\hat H'$. Recall that $\cH=\hat H'_{|\jset}$, that is, graph $\cH$ can be obtained from $\hat H'$ by contracting all clusters of $\jset$. Therefore, if we denote by $\hat U^*$ the set of all regular vertices of $\cH$ (excluding $v^*$), and by $\hat U^R$ the set of all $R$-nodes of $\cH$ (excluding $u^*$), then $\hat U^*\subseteq U^*$, and $\hat U^R\subseteq U^R$. Therefore, partition $(L_1,\ldots,L_h)$ of $U^*\cup U^R$ naturally defines a partition $(L'_1,\ldots,L_h')$ of $\hat U^R\cup \hat U^*$. Recall that, for all $1\leq i\leq r$, all vertices of $S_i\setminus S'_i$ lie in $\hat U^*\cup \hat U^R$. 
We denote by $L'_0=V(\cH)\setminus \left(\bigcup_{j=1}^hL'_j\right )$.
 As before, for all $1\leq j\leq r$, for every vertex $v\in L'_j$, we partition the set $\delta_{\cH}(v)$ of edges into two subsets: set $\delta^{\down}(v)$ containing all edges that connect $v$ to vertices of $L'_0\cup\cdots\cup L'_{j-1}$, and set $\delta^{\up}(v)$ containing all remaining edges incident to $v$. From Property \ref{prop: up and down} of graph $\cH$, for every vertex $v\in \hat U^*\cup \hat U^R$,  $|\delta^{\up}(v)|<|\delta^{\down}(v)|/\log m$.

For all $1\leq i\leq r$ and $1\leq j\leq h$, we denote by $U_{i,j}=L'_j\cap V(S_i)$ -- the set of all vertices of $S_i$ that lie in layer $L'_j$.

Consider some pair $1\leq i\leq r$, $1\leq j\leq h$ of indices, and some vertex $v\in U_{i,j}\setminus \set{u_i}$. We partition the edges of $\delta^{\down}(v)$ into four subsets, $\delta^{\down,\lef}(v),\delta^{\down,\rig}(v)$, $\delta^{\down,\straight'}(v)$, and $\delta^{\down,\straight''}(v)$, as follows. Let $e=(u,v)$ be an edge of $\delta^{\down}(v)$, and assume that $u\in U_{i',j'}$. Since $e\in \delta^{\down}(v)$, $j'<j$ must hold. If, additionally, $i'<i$ holds, then we add $e$ to $\delta^{\down,\lef}(v)$, Similarly, if $i'>i$, then we add $e$ to 
$\delta^{\down,\rig}(v)$.  If $i'=i$, and $u\in S'_i$, then $e$ is added to $\delta^{\down,\straight''}(v)$, and otherwise it is added to $\delta^{\down,\straight'(v)}$. We will use the following simple observation, whose proof appears in \Cref{subsec: left and right down-edges}.

\begin{observation}\label{obs: left and right down-edges}
	$S'_1=S_1$, and $S'_r=S_r$. Additionally,
	for every vertex $v\in V(\cH)\setminus\left(\bigcup_{i=1}^rS'_i\right )$: 
	
	\begin{itemize}
		\item $|\delta^{\down,\rig}(v)|+|\delta^{\down,\lef}(v)|+| \delta^{\down,\straight'}(v)|\geq 63|\delta(v)|/64$;
		\item $|\delta^{\down,\lef}(v)|\leq 2(|\delta^{\down,\rig}(v)|+|\delta^{\down,\straight'}(v)|)$; and
		\item $|\delta^{\down,\rig}(v)|\leq 2(|\delta^{\down,\lef}(v)|+|\delta^{\down,\straight'}(v)|)$.
	\end{itemize}
\end{observation}

We will also use the following simple observation, whose proof appears in \Cref{subsec: left and right mappings}

\begin{observation}\label{obs: left and right mappings}
	There is an efficient algorithm that defines, for every vertex  $v\in V(\cH)\setminus\left(\bigcup_{i=1}^rS'_i\right )$, two mappings: mapping $f^{\rig}(v)$, that maps every edge of $\delta^{\down,\straight''}(v)\cup\delta^{\up}(v)$ to a distinct edge of $\delta^{\down,\rig}(v)\cup \delta^{\down,\straight'}(v)$, and another mapping $f^{\lef}(v)$, that maps every edge of $\delta^{\down,\straight''}(v)\cup\delta^{\up}(v)$ to a distinct edge of $\delta^{\down,\lef}(v)\cup \delta^{\down,\straight'}(v)$.
\end{observation}

Next, we define the notion of a \emph{left-monotone} and a \emph{right-monotone} path.

\begin{definition}[Left-Monotone and Right-Monotone Paths]
	Let $P=(x_1,x_2,\ldots,x_q)$ be a path in graph $\cH$. For all $1\leq a\leq q$, assume that $x_a\in U_{i_a,j_a}$. We say that path $P$ is \emph{left-monotone} if either $q=1$ (that is, $P$ consists of a single vertex), or all of the following conditions holds:
	\begin{properties}{M}
		\item $j_1>j_2>\cdots>j_q$;\label{prop: monotone levels decrease}
		\item for all $1\leq a<q$, vertex $x_a\in S_{i_a}\setminus S'_{i_a}$, and $x_q\in S'_{i_q}$; \label{prop: monotone paths not in kernels} and
		\item $i_1\geq i_2\geq\cdots\geq i_q$, and $i_q<i_1$. \label{prop: monotone goes left}
	\end{properties}	
Similarly, we say $P$ is \emph{right-monotone} if either $q=1$, or properties \ref{prop: monotone levels decrease} and \ref{prop: monotone paths not in kernels} hold for it, together with the following property:
\begin{properties}[2]{M'}
	\item $i_1\leq i_2\leq\cdots\leq i_q$, and $i_q>i_1$
\end{properties}
\end{definition}

Observe that the vertices on a left-monotone path must appear in the decreasing order of their layers, and in the non-increasing order of the sets $S_i$ to which they belong. 
Similarly, vertices on a right-monotone path appear in the decreasing order of their layers, and in the non-decreasing order of the sets $S_i$ to which they belong.
The following lemma will allow us to construct prefix- and suffix-paths for each edge $e\in \hat E$, by constructing a left-monotone and a right-monotone path for each such edge in graph $\cH$; the proof is deferred to \Cref{subsubsec: monotone paths}.

\begin{lemma}\label{lem: prefix and suffix path}
	There is an efficient algorithm that constructs, for every edge $e=(u,v)\in \hat E$ two paths $P(e,u)$ and $P(e,v)$  in graph $\cH$, such that, if $u\in S_i$, $v\in S_{i'}$, and $i<i'$, then path $P(e,u)$ is left-monotone and path $P(e,v)$ is right-monotone. Moreover, the set $\set{P(e,v),P(e,u)\mid e=(u,v)\in \hat E}$ of paths causes congestion $O(\log m)$.
\end{lemma}

\begin{figure}[h]
	\centering
	\includegraphics[scale=0.12]{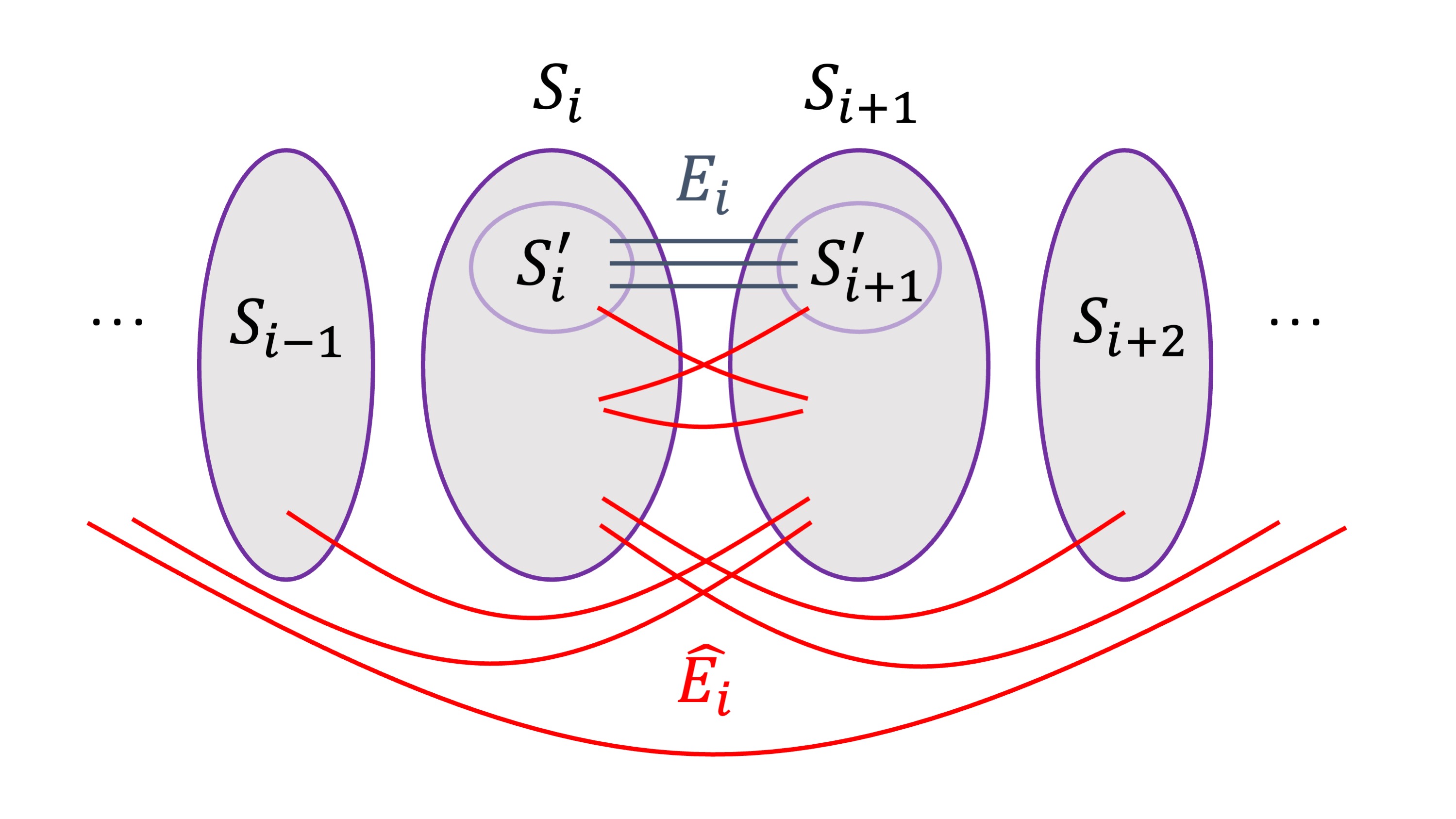}
	\caption{An illustration of edge sets $E_i$ and $\hat E_i$. 
	}\label{fig: NF8}
\end{figure}

Consider now some index $1\leq i< r$. We let $\hat E_i\subseteq \hat E$ contain all edges $e=(u,v)\in \hat E$, such that, if $u\in S_{i'}$, $v\in S_{i''}$, and $i'<i''$, then $i'\leq i$ and $i''\geq i+1$ must hold. We also denote by $E_i\subseteq E'$ the set of all edges $e=(u,v)$ with $u\in S'_i$ and $v\in S'_{i+1}$
(see \Cref{fig: NF8}).
Note that $E_i\cap \hat E_i=\emptyset$. 
The next lemma is crucial to the algorithm for constructing a nice witness structure in graph $G_{|\rset}$.

\begin{lemma}\label{lem: no bad indices}
	For all $1< i<r$, vertex $u_i$ is a $J$-node, and for all $1\leq i< r$, $|\hat E_i|\leq 1000\cdot |E_i|$.
\end{lemma}

The proof of \Cref{lem: no bad indices} is deferred to \Cref{subsubsec: no bad indices}.

From now on, we denote $H=G_{|\rset}$, and we denote by $\cset'\subseteq \cset$ the set of all basic clusters $C\in \cset$ with $C\subseteq G\setminus\left (\bigcup_{R\in \rset}V(R)\right )$; equivalently, $\cset'$ contains every cluster $C\in \cset$ that is contained in $H$. It now remains to construct a nice witness structure in graph $H$ with respect to the set $\cset'$ of clusters. We start by constructing the backbone and the vertebrae of the witness structure, and by defining the partition $(\tilde E',\tilde E'')$ of the edges of $H$. We then construct the prefix and the suffix of path $P(e)$ for each edge $e\in \hat E$. Lastly, we construct the mid-segment of each such path.

\paragraph{The Backbone and the Vertebrae of the Witness Structure.}
We use the clusters in set $\sset=\set{S_1,\ldots,S_r}$, and in set $\sset'=\set{S_1',\ldots,S'_r}$ in order to define the backbone and the vertebrae of the nice witness structure, respectively. Recall that every vertex of graph $\cH$ is either a regular vertex (that is, it lies in both $G$ and $H$); or it is an $R$-node $v_R$ representing some cluster $R\in \rset$ (in which case it lies in $H=G_{|\rset}$); or it is a $J$-node $v_{J'}$ for some cluster $J'\in \jset'$. As we have established, for all $1<i<r$, vertex $u_i$ is a $J$-node, and all other vertices of $\cH$ are either regular vertices of $R$-nodes.

Consider now any such $J$-node $v_{J'}$, and the corresponding cluster $J'\in \jset'$. Recall that for every cluster $R\in \rset$, either $R\subseteq J'$, or $R\cap J'=\emptyset$ holds. Denote by $\rset(J')\subseteq \rset$ the set of all clusters $R\in \rset$ with $R\subseteq J'$. Let $J''=J'_{|\rset(J')}$ be the graph obtained from $J'$ by contracting every cluster $R\in \rset(J')$ into a supernode. Then $J''\subseteq H$, and we will think of $J''$ as the cluster of $H$ that vertex $v_{J'}\in V(\cH)$ represents. We denote by $\jset''=\set{J''\mid J'\in \jset'}$ the resulting set of clusters in graph $H$. Note that equivalently we could define the graph $\cH$ as a graph that is obtained from $H$ by contracting every cluster in $\jset''$, that is, $\cH=H_{|\jset''}$. Recall that we have established, in \Cref{obs: J' clusters wl}, that
	every cluster $J'\in \jset'$ has the $\Omega(1/\log^{9.5}m)$-bandwidth property in graph $G$. It then immediately follows that  every cluster $J''\in \jset''$ has the $\Omega(1/\log^{9.5}m)$-bandwidth property in graph $H$.

We start by defining the sequence $\tilde \sset'=\set{\tilde S'_1,\ldots,\tilde S'_r}$ of the vertebrae of the nice witness structure. Consider an index $1\leq i\leq r$. If $i=1$, then $u_1=v^*$, and so every vertex of set $S'_1$ is a vertex of $H$. We then set $\tilde S'_1=S'_1$. If $i=r$, then $u_r=u^*$. As before, every vertex of $S'_r$ is then a vertex of $H$, and we set $\tilde S_r'=S'_r$. Lastly, assume that $1<i<r$. From \Cref{lem: no bad indices}, $u_i$ is a $J$-node, and, from \Cref{obs: central path j-cluster}, $S'_i=\set{u_i}$. Assume that $u_i=v_{J'}$, where $J'\in \jset'$. We then let $\tilde S'_i$ be the cluster $J''$ of $H$ that corresponds to vertex $v_{J'}$. This completes the definition of the sequence $\tilde \sset'=\set{\tilde S'_1,\ldots,\tilde S'_r}$ of the vertebrae of the nice witness structure. Consider
again some index $1\leq i\leq r$. If $i\in \set{1,r}$, then, from the construction, for every cluster $C\in \cset'$, $C\cap \tilde S'_i=\emptyset$. This is because every cluster of $\cset'$ must be contained in some cluster of $\jset'$. Otherwise, assume that $u_i=v_{J'}$, where $J'\in \jset'$. If the center-cluster of $J'$ is a basic cluster $C\in \cset'$, then $C\subseteq \tilde S_i'$, and for every other cluster $C'\in \cset'$, $C\cap \tilde S_i'=\emptyset$. Otherwise, the center-cluster of $J'$ is a cluster $W'\in \wset'$. In this case, no cluster of $\cset'$ may be contained in $\tilde S_i'$.

Consider again some index $1\leq i\leq r$.
Recall that we have established, in \Cref{obs: S'i wl}, that cluster $S'_i$ has the $\Omega(1/\log m)$-bandwidth property in graph $\cH$. We have also established above that every cluster in $J''\in \jset''$ has the $\Omega(1/\log^{9.5}m)$-bandwidth property. From \Cref{cor: contracted_graph_well_linkedness}, cluster $\tilde S'_i$ of $H$ has the $\Omega(1/\log^{10.5}m)\geq \alpha^*$-bandwidth property, since  $\alpha^*=\Omega(1/\log^{12}m)$. To conclude, we have shown that, for all $1\leq i\leq r$, cluster $\tilde S'_i$ of $H$ has the $\alpha^*$-bandwidth property. We have also shown that, for all $1\leq i\leq r$, there is at most one cluster $C \in\cset'$ with $C\subseteq \tilde S'_i$. It is easy to verify that, if such cluster $C$ exists, then $E(\tilde S_i')\subseteq E(C)\cup E(G_{|\cset'})$, and otherwise $E(\tilde S_i')\subseteq E(G_{|\cset'})$. This is because for every cluster $C\in \cset'$, and for all $1\leq i\leq r$, either $C\subseteq \tilde S'_i$, or $C\cap \tilde S'_i=\emptyset$ holds.

We now define the backbone $\tilde \sset=\set{\tilde S_1,\ldots,\tilde S_r}$ of the nice witness structure. Fix an index $1\leq i\leq r$. If $i\in \set{1,r}$, then, from \Cref{obs: left and right down-edges}, $S'_i=S_i$. We then set $\tilde S_i=\tilde S_i'$. Assume now that $1<i<r$. Recall that in this case, $S'_i=\set{u_i}$ holds, and $u_i$ is a $J$-node, from \Cref{lem: no bad indices}. Note that every vertex of $S_i\setminus\set{u_i}$ is either a regular vertex or an $R$-node, and so it must lie in graph $H$. 
We define the set $V(\tilde S_i)$ of vertices to contain all regular vertices and all $R$-nodes that lie in $S_i\setminus\set{u_i}$, and all vertices of $\tilde S_i'$. We then let $\tilde S_i$ be the subgraph of $H$ induced by the set $V(\tilde S_i)$ of vertices. In other words, we can think of cluster $\tilde S_i$ as being obtained from cluster $S_i$ of $\cH$, by un-contracting the $J$-node $u_i$ (into the corresponding cluster of $\jset''$). This completes the definition of the backbone of the nice witness structure. Since every vertex of $S_i\setminus S'_i$ is either a regular vertex or an $R$-node, either there is a single cluster $C \in\cset'$ with $C\subseteq \tilde S'_i$, in which case then $E(\tilde S_i)\subseteq E(C)\cup E(G_{|\cset'})$; or no such cluster exists, in which case $E(\tilde S_i)\subseteq E(G_{|\cset'})$.
Since vertex sets $V(S_1),\ldots,V(S_r)$ partition $V(\cH)$, it is easy to verify that vertex sets $V(\tilde S_1),\ldots,V(\tilde S_r)$ partition $V(H)$.

Recall that the second ingredient of the nice witness structure is a partition of the edges of $E(H)$ into two disjoint subsets, $\tilde E'$ and $\tilde E''$, that are defined as follows. Set $\tilde E'$ contains all edges of $\bigcup_{i=1}^rE(\tilde S_i')$, and, additionally, for all $1\leq i<r$, it contains every edge $e=(u,v)$ with $u\in \tilde S_i'$, $v\in \tilde S_{i+1}'$. 
Since, as observed already, $\cH=H_{|\jset''}$, it is easy to verify that $E'\subseteq \tilde E'$. Recall that we have denoted, for all $1\leq i<r$, by $E_i\subseteq E'$ the set of all edges $e=(u,v)$ of $\cH$ with $u\in S'_i$ and $v\in S'_{i+1}$. It is easy to verify that $E_i\subseteq E(H)$, and moreover, it is precisely the set of all edges $(u,v)$ in $H$ with $u\in \tilde S'_i$ and $v\in \tilde S'_{i+1}$. In particular, $E_i\subseteq \tilde E'$.
 The second edge set in the partition of $E(H)$ contains all remaining edges, $\tilde E''=E(H)\setminus \tilde E'$. From the fact that  $\cH=H_{|\jset''}$, and since, for all $1\leq i\leq r$, $S_i\setminus S'_i$ may only contain regular vertices or $R$-nodes, we get that $E''=\tilde E''$ holds. Lastly, we defined the set $\hat E\subseteq E''$ of all edges $e=(u,v)\in E''$ of graph $\cH$, where $u$ and $v$ lie in different clusters of $\set{ S_1,\ldots, S_r}$. It is immediate to verify that this is exactly the set of edges in graph $\cH$, containing all edges $e=(u,v)\in \tilde E''$ where $u$ and $v$ lie in different clusters of $\set{\tilde S_1,\ldots,\tilde S_r}$.
Recall that we have defined, for all $1\leq i< r$, the set $\hat E_i\subseteq \hat E$ of edges in graph $\cH$, that contains all edges $e=(u,v)\in \hat E$, such that, if $u\in S_{i'}$ and $v\in S_{i''}$ with $i'<i''$, then $i'\leq i$ and $i''\geq i+1$ hold. 
It is easy to verify that $\hat E_i$ is also precisely the set of all edges  $e=(u,v)\in \hat E$ in graph $H$, such that, if $u\in \tilde S_{i'}$ and $v\in \tilde S_{i''}$ with $i'<i''$, then $i'\leq i$ and $i''\geq i+1$ hold. As before,  $E_i\cap \hat E_i=\emptyset$.

In order to complete the construction of the nice witness structure, it now remains to define the paths in set $\hat \pset=\set{P(e)\mid e\in \hat E}$. Recall that each such path  $P(e)$ consists of three subpaths, prefix $P^1(e)$, suffix $P^3(e)$, and mid-segment $P^2(e)$.
We first construct the prefixes and the suffixes of the paths in $\hat \pset$, and then construct the mid-segment of each such path.

\paragraph{Prefixes and Suffixes of Paths in $\hat \pset$.}
 Consider an edge $e=(u,v)\in \hat E$ in graph $H$. Assume that $u\in \tilde S_i$, $v\in \tilde S_{i'}$, and $i<i'$. We now define vertices $u',v'$ of graph $\cH$ that correspond to $u$ and $v$.
If $u$ is also a vertex of cluster $S_i$ in $\cH$, then we set $u'=u$. Otherwise, $u_i$ must be a $J$-node corresponding to some cluster $J'\in \jset'$, with vertex $u$ lying in the corresponding cluster $J''\in \jset''$. In this case, we set $u'=u_i$. We define vertex $v'$ in graph $\cH$, that corresponds to vertex $v$ in graph $H$ similarly, so $v'\in S_{i'}$. Observe that $(u',v')$ is an edge of $\cH$, that lies in the edge set $\hat E$, and it corresponds to edge $e$ in $H$; we do not distinguish between the two edges. Consider now the left-monotone path $P(e,u')$ in graph $\cH$ given by \Cref{lem: prefix and suffix path}, and denote 
$P(e,u')=(u'=x_1,x_2,\ldots,x_q)$. For all $1\leq a\leq q$, assume that $x_a\in U_{i_a,j_a}$. Recall that, from the definition of the left-monotone path, $i_1\geq i_2\geq\cdots\geq i_q$, and, if $P(e,u')$ contains more than one vertex, then $i_q<i_1=i$ holds. Additionally, for all  $1\leq a<q$, vertex $x_a\not\in S'_{i_a}$, while $x_q\in S'_{i_q}$. In particular, every inner vertex on path $P(e,u')$
is an $R$-node or a regular vertex of $\cH$, and hence it lies in graph $H$. Clearly, every edge of path $P(e,u')$ is an edge of $H$ that lies in edge set $\tilde E''$. Therefore, path $P(e,u')$ is contained in graph $H$. We set the prefix $P^1(e)$ of the path $P(e)$ to be $P(e,u')$. We also denote by $i^{\lef}(e)=i_q$, and we denote by $e^{\lef}$ the last edge on path $P^1(e)$. Observe that $e^{\lef}\in \delta_{H}(\tilde S'_{i^{\lef}(e)})$. We define the suffix $P^3(e)$ using the right-monotone path $P(e,v)$ similarly. We denote by $e^{\rig}$ the last edge on that path, and by $i^{\rig}(e)$ the index $i^*$ such that the last vertex of path $P^3(e)$ belongs to $\tilde S'_{i^*}$. From the definition of monotone paths, if $u\not\in \tilde S_i'$, then $i^{\lef}(e)< i$, and, if $v\not\in \tilde S_{i'}'$, then $i^{\rig}(e)>i'$.
Lastly, we define the \emph{span} of edge $e$ to be $\spann(e)=\set{i^{\lef}(e),(i^{\lef}(e)+1),\ldots,(i^{\rig}(e)-1)}$. 
Recall that the congestion caused by the set $\set{P(e,v),P(e,u)\mid e=(u,v)\in \hat E}$ of paths in graph $\cH$ is $O(\log m)$, so the congestion caused by the set $\set{P^1(e),P^3(e)\mid e\in \hat E}$ of paths in graph $H$ is also $O(\log m)$.

\paragraph{Mid-Segments of Paths in $\hat \pset$.}
We now focus on constructing the mid-segment $P^2(e)$ of the nice guiding path $P(e)$ for every edge $e\in \hat E$. 
In order to do so, fix some index $1\leq i<r$, and let $\hat E'_i$ be the set of all edges $e\in \hat E$, such that $i\in \spann(e)$. Note that edge $e$ may only lie in $\hat E'_i$ if either (i) $e\in \hat E_i$; or (ii) some edge $e'\in \hat E_i$ belongs to path $P^1(e)$; or (iii) some edge $e''\in \hat E_i$ belongs to path $P^3(e)$. Since the paths in set $\set{P^1(e),P^3(e)\mid e\in \hat E}$ cause congestion at most $O(\log m)$ in graph $H$, from \Cref{lem: no bad indices}, $|\hat E_i'|\leq O(\log m)\cdot |E_i|$. Therefore, we can define an arbitrary mapping $f_i: \hat E_i'\rightarrow E_i$, such that, for every edge $e\in E_i$, at most $O(\log m)$ edges of $\hat E_i'$ are mapped to $e$.

In order to define the mid-segment of every path in $\set{P(e)\mid e\in \hat E}$, we proceed as follows. For all $1\leq i<r$, we will define a collection $M_i$ of pairs of edges in $\delta_{H}(\tilde S'_i)$, so that every edge of $\delta_{H}(\tilde S'_i)$ participates in at most $O(\log m)$ such pairs. We will later exploit the bandwidth property of cluster $\tilde S'_i$ in order to connect every pair of edges in $M_i$ with a path. We start with $M_i=\emptyset$ for all $1\leq i\leq r$, and then gradually add edge pairs to the sets $M_i$.

Consider again some edge $e\in \hat E$, and recall that $\spann(e)=\set{i^{\lef}(e),(i^{\lef}(e)+1),\ldots,(i^{\rig}(e)-1)}$. 
For convenience, denote $i^{\lef}(e)$ by $i'$ and $i^{\rig}(e)$ by $i''$.
Recall that the last edge on path $P^1(e)$, that we denoted by $e^{\lef}$, is an edge that is incident to cluster $\tilde S_{i'}$ in $H$. Let $e^{i'}$ be the edge of $E_{i'}$ to which edge $e$ is mapped by $f_{i'}$. We then add the pair $(e^{\lef},e^{i'})$ to $M_{i'}$.

Consider now any index $i'<i<i''-1$. Let $e^{i-1}\in E_{i-1}$ be the edge to which $e$ is mapped by $f_{i-1}$, and let $e^i\in E_i$ be the edge to which $e$ is mapped by $f_i$. We then add the pair $(e^{i-1},e^i)$ to $M_i$. Lastly, we add the edge pair $(e^{i''-1},e^{\rig})$ to $M_{i''}$. 

We will define a path $Q^{i'}(e)$ in graph $H$, whose first edge is $e^{\lef}$ and last edge is $e^{i'}$, such that all inner vertices of  $Q^{i'}(e)$ lie in $\tilde S_{i'}'$. Additionally, for all $i'<i<i''-1$, we will define a path $Q^i(e)$ in graph $H$, whose first edge is $e^{i-1}$ and last edge is $e^i$, such that all inner vertices of $Q^i(e)$ lie in $\tilde S_i'$. Laslty, we will define a path $Q^{i''}(e)$, whose first edge is $e^{i''-1}$, last edge is $e^{\rig}$, and all inner vertices lie in $\tilde S_{i''}'$. The final path $P^2(e)$ is then obtained by concatenating the paths $Q^{i'}(e),\ldots,Q^{i''}(e)$, and omitting the first and the last edge from the resulting path.

In order to define the paths of $\set{Q^i(e)\mid e\in \hat E; i^{\lef}(e)\leq i<i^{\rig}(e)}$, we consider the clusters $\tilde S_i'\in \tilde \sset'$ one by one. Consider any such cluster $\tilde S_i'$. Recall that we have defined a collection $M_i$ of pairs of edges from $\delta_H(\tilde S_i')$, such that every edge of $\delta_{H}(\tilde S_i')$ appears in at most $O(\log m)$ pairs. Using a standard greedy algorithm, we can compute $z=O(\log m)$ collections $M^1_i,\ldots,M^z_i$ of pairs of edges, such that $\bigcup_{j=1}^zM^j_i=M_i$, and, for all $1\leq j\leq z$, every edge of $\delta_{H}(\tilde S_i')$ participates in at most one pair of $M^j_i$. By applying the algorithm from \Cref{cor: routing well linked vertex set} to the augmented cluster $(\tilde S_i')^+$, we obtain, for each $1\leq j\leq z$, a collection $\qset^{j}_i=\set{\hat Q(e,e')\mid (e,e')\in M^j_i}$ of paths, where each path $Q(e,e')$ has $e$ as its first edge, $e'$ as its last edge, and all internal vertices of the path lie in $\tilde S_i'$. Moreover, since cluster $\tilde S_i'$ has $\alpha^*$-bandwidth property,
with high probability, the paths in $\qset^j_i$ cause edge-congestion at most $O(\log^4m/\alpha^*)\leq O(\log^{16}m)$, since $\alpha^*=\Omega(1/\log^{12}m)$. By letting $\qset_i=\bigcup_{j=1}^z\qset^j_i$, we obtain a collection $\qset_i=\set{\hat Q(e,e')\mid (e,e')\in M_i}$ of paths, where for every edge pair $(e,e')\in M_i$, path $Q(e,e')$ has $e$ as its first edge, $e'$ as its last edge, and every inner vertex on the path lies in $\tilde S_i'$. The total edge-congestion caused by paths in $\qset_i$ is then bounded by $O(\log^{17}m)$ with high probability.
This completes the definition of the nice routing paths $\hat\pset=\set{P(e) \mid e\in \hat E}$ in graph $H$. From the above discussion, the paths in $\hat \pset$ cause edge-congestion $O(\log^{18}m)$ with high probability. If the congestion caused by the paths in $\hat \pset$ is greater than $\Theta(\log^{18}m)$, we return FAIL.
Otherwise, we have established that $(\rset,\set{\dset'(R)}_{R\in \rset})$ is a type-2 legal clustering in $G$ with respect to $v^*$ and $\cset'$, by providing a nice witness structure in graph $H=G_{|\rset}$ with respect to set $\cset'$ of basic clusters. In order to complete the proof of \Cref{lemma: better clustering 2} and \Cref{thm: advanced disengagement get nice instances}, it now remains to prove \Cref{lem: no bad indices}, which we do next.
\subsubsection{Proof of \Cref{lem: no bad indices}}
\label{subsubsec: no bad indices}

\renewcommand{\tE}{\tilde E}

Throughout the proof, we will only consider the graph $\cH$, so we will omit subscript $\cH$ from various notations, such as, for example, $\delta_{\cH}(v)$ for vertices $v\in\cH$.
We start by considering the edges connecting different clusters in $\set{S_1,\ldots,S_r}$, and by establishing some useful relationships between them.

\subsubsection*{Edges Connecting Clusters in $\set{S_1,\ldots,S_r}$}

\begin{figure}[h]
	\centering
	\includegraphics[scale=0.12]{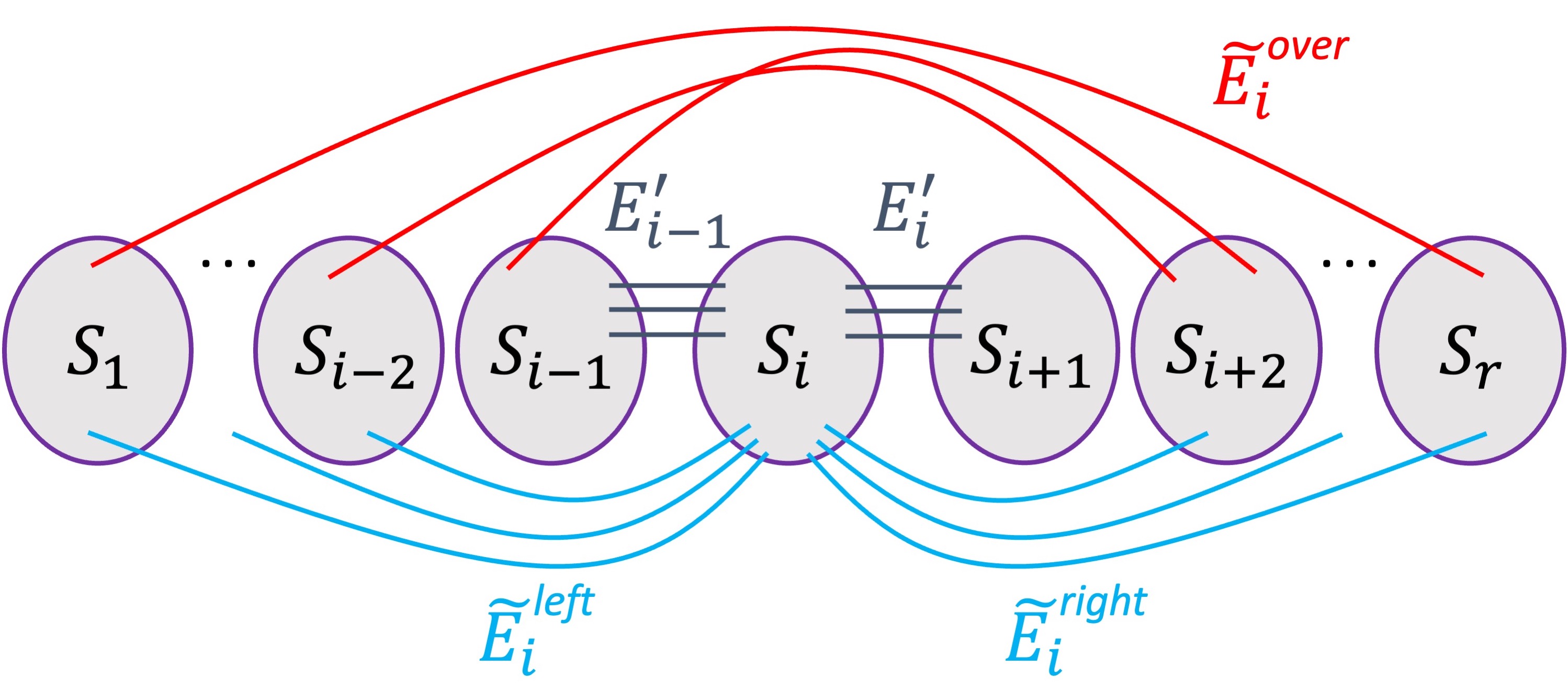}
	\caption{Edge sets $\tilde E^{\lef}_i, \tilde E^{\lef}_i$ and $\tilde E^{\through}_i$.}\label{fig: NF9}
\end{figure}

Fix an index $1\leq i< r$. We denote by $E'_i=E(S_i,S_{i+1})$, and by $\tilde E^{\through}_i$ the set of all edges $e=(u,v)$, such that, if $u\in S_{j},v\in S_{j'}$, then $j<i$ and $j'>i+1$ holds. For all $1<i\leq r$, we denote by $\tilde E_i^{\lef}$ the set of all edges $e=(u,v)$ with $u\in S_i$, such that, if $v\in S_j$, then $j<i-1$. Similarly, for all $1\leq i<r$, we denote by $\tilde E_i^{\rig}$ the set of all edges $e=(u,v)$ with $u\in S_i$, such that, if $v\in S_j$, then $j>i+1$ holds (see \Cref{fig: NF9}). Notice that $\delta(S_i)=E'_{i-1}\cup E'_i\cup \tilde E_i^{\lef}\cup \tilde E_i^{\rig}$. Notice also that, by the definition, if $i\in \set{1,2}$, then $E_i^{\lef}=\emptyset$; if $i\in \set{1,r-1,r}$, then $\tilde E_i^{\through}=\emptyset$, and, if $i\in \set{r-1,r}$, then $\tE_i^{\rig}=\emptyset$.
We prove the following observation that helps us relate the sizes of all these edge sets. The proof is deferred to \Cref{subsec: edges between Sis}.

\begin{observation}\label{obs: bad inded structure}
	For all $1< i<r$, the following hold:
	
	\begin{itemize}
\item 	$|\tE_i^{\through}|\leq 	|E_{i}'|$. 
\item $|\tE_{i+1}^{\lef}|\leq 
	|E_{i}'|+|\tE_{i}^{\rig}|$ 
	
	\item $|\tE_{i}^{\rig}|\leq 
	|E_{i}'|+|\tE_{i+1}^{\lef}|$.
	\end{itemize}
\end{observation}

\begin{figure}[h]
	\centering
	\includegraphics[scale=0.12]{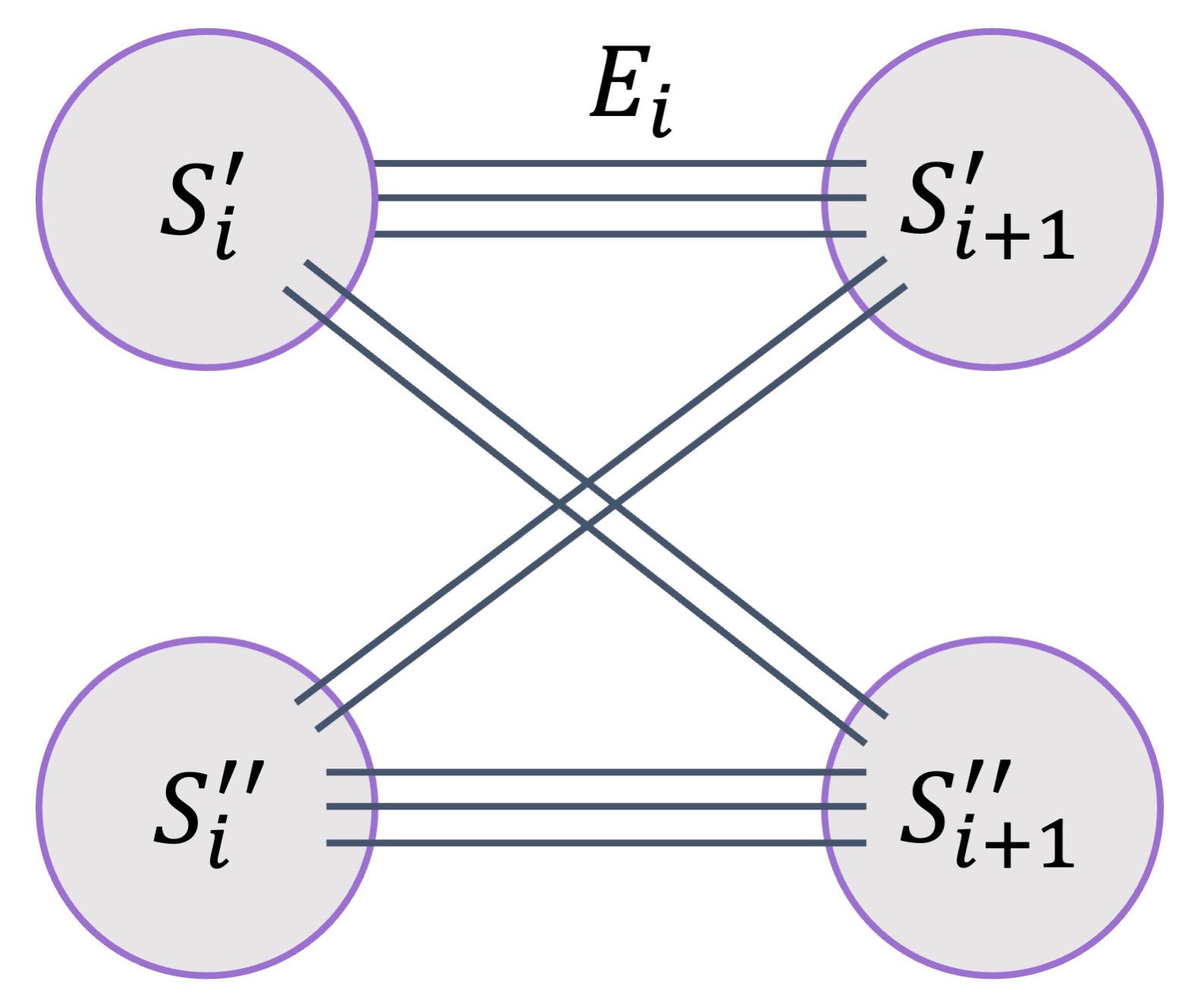}
	\caption{Edges in set $E'_i$ are shown in black.}\label{fig: NF12}
\end{figure}

For all $1\leq i\leq r$, we denote $S''_i=S_i\setminus S'_i$.
\Cref{obs: bad inded structure} allows us to bound the cardinality of the set 
$\tE_i^{\through}\subseteq \hat E_i$ of edges in terms of the cardinality of the set $E_i'$ of edges. Note that the set $E_i'$ of edges can be thought of as the union of four subsets: set $E_i$, and sets $E(S'_i,S''_{i+1})$, $E(S''_i,S'_{i+1})$, and $E(S''_i,S''_{i+1})$
(see \Cref{fig: NF12}).  
The latter three sets are all contained in $\hat E_i$. We will bound the cardinalities of these subsets in terms of $|E_i|$ in turn.
We start by considering edge sets incident to clusters of $\set{S''_i}_{1<i<r}$.

\subsubsection*{Edges Incident to Clusters of $\set{S''_i}_{1<i<r}$.}

\begin{figure}[h]
	\centering
	\includegraphics[scale=0.12]{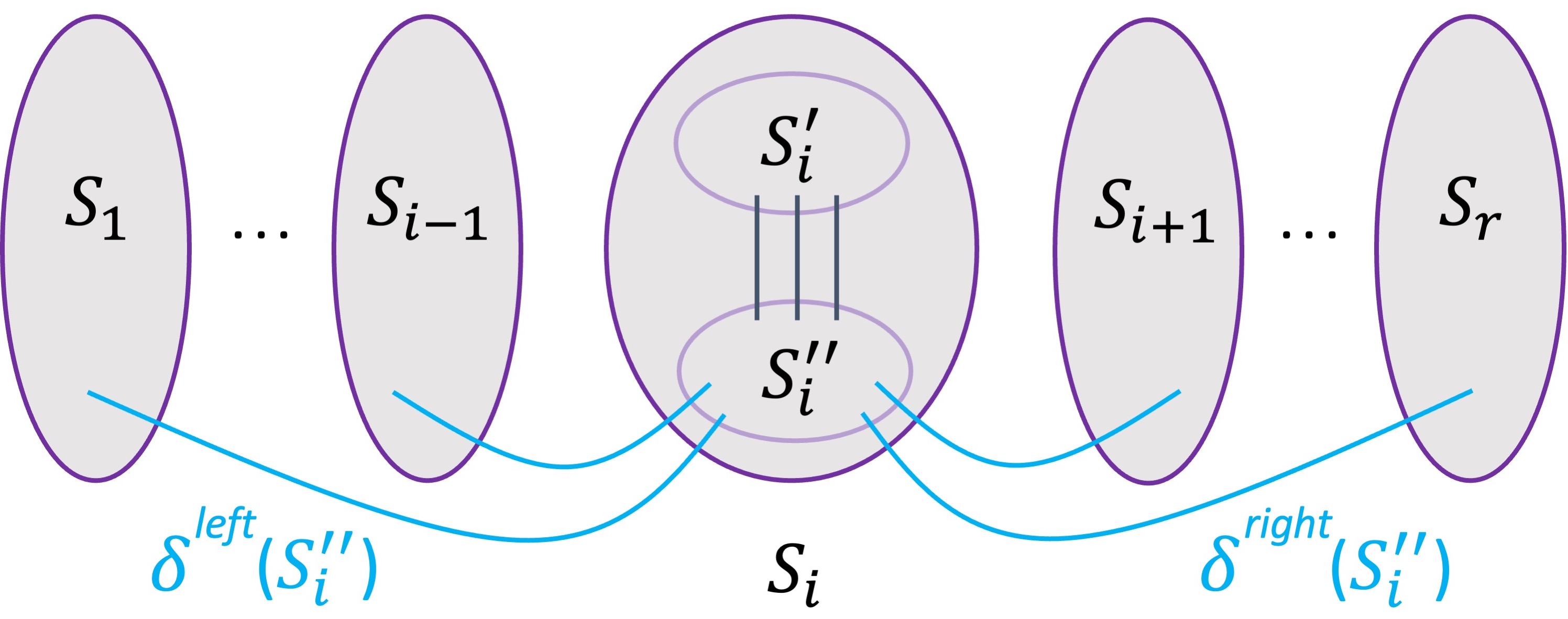}
	\caption{Edge sets $\delta^{\lef}(S''_i),\delta^{\rig}(S''_i)$ and $\delta^{\down}(S''_i)$ (shown in black).}\label{fig: NF13}
\end{figure}

Consider an index $1< i< r$ (recall that $S''_1=S''_r=\emptyset$ from \Cref{obs: left and right down-edges}). We partition the edges of $\delta(S''_i)$ into three subsets: set $\delta^{\down}(S''_i)=E(S_i',S''_i)$; set $\delta^{\lef}(S''_i)$ containing all edges $(u,v)$ with $u\in S''_i$ and $v\in V(S_1)\cup\cdots\cup V(S_{i-1})$; and set $\delta^{\rig}(S''_i)$ containing the remaining edges  (all edges $(u,v)$ with $u\in S''_i$ and $v\in V(S_{i+1})\cup\cdots\cup V(S_r)$) (see \Cref{fig: NF13}).

\newcommand{\leftedges}[1]{\delta^{\lef}(S''_{#1})}
\newcommand{\rightedges}[1]{\delta^{\rig}(S''_{#1})}
\newcommand{\downedges}[1]{\delta^{\down}(S''_{#1})}

We next show that for all $1<i<r$, $|\downedges{i}|\leq 0.1\min\set{|\rightedges{i}|,|\leftedges{i}|}$, and that the sizes of the edge sets $\rightedges{i},\leftedges{i}$ are close to each other, in the following two claims, whose proofs are deferred to \Cref{subsec: bound S' to S'' edges} and \Cref{subsec: bound left and right for S''}, respectively.

\begin{claim}\label{claim: bound S' to S'' edges}
	For all $1<i<r$, $|\downedges{i}|\leq 0.1\cdot \min\set{|\rightedges{i}|,|\leftedges{i}|}$ holds.
	Additionally, there is a set $\pset^{\lef}=\set{P^{\lef}(e)\mid e\in \downedges{i}}$ of edge-disjoint paths in $\cH$, where, for each edge  $ e\in \downedges{i}$, path $P^{\lef}(e)$ has $e$ as its first edge, some edge of $\leftedges{i}$ as its last edge, and all inner vertices of $P^{\lef}(e)$ are contained in $S''_i$. Similarly, there is a set $\pset^{\rig}=\set{P^{\rig}(e)\mid e\in \downedges{i}}$ of edge-disjoint paths in $\cH$, where, for each edge  $ e\in \downedges{i}$, path $P^{\rig}(e)$ has $e$ as its first edge, some edge of $\rightedges{i}$ as its last edge, and all inner vertices of $P^{\rig}(e)$ are contained in $S''_i$.
\end{claim}

\begin{claim}\label{claim: bound left and right for S''}
	For all $1<i<r$, $|\rightedges{i}|\leq 1.1|\leftedges{i}|$, and similarly, $|\leftedges{i}|\leq 1.1|\rightedges{i}|$.
\end{claim}

\newcommand{\leftCedges}[1]{\delta^{\lef}(S'_{#1})}
\newcommand{\rightCedges}[1]{\delta^{\rig}(S'_{#1})}

Next, we consider edges incident to the clusters of $\set{S'_1,\ldots,S'_r}$.

\subsubsection*{Edges Incident to the Clusters of $\set{S'_1,\ldots,S'_r}$}

\begin{figure}[h]
 	\centering
 	\includegraphics[scale=0.12]{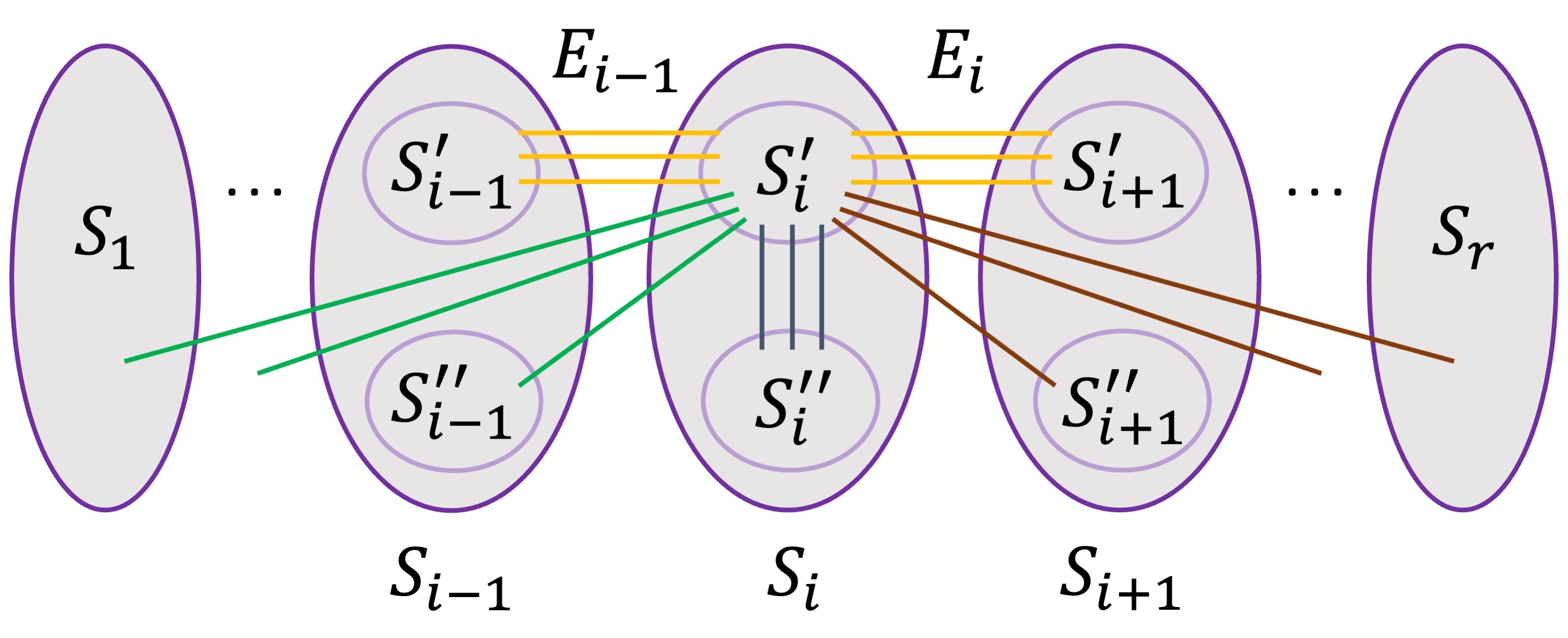}
 	\caption{Edges in set $\delta^{\lef}(S'_i)$ are shown in green, edges of $\delta^{\rig}(S'_i)$ are shown in brown, and edges of $\delta^{\down}(S''_i)$  in black.
 	}\label{fig: NF15}
 \end{figure}

Consider some index $1\leq i\leq r$, and consider the edges incident to the cluster $S_i'$ in graph $\cH$ (see \Cref{fig: NF15}).
Recall that we have denoted by $E_{i-1}=E(S_{i-1}',S_i')$, and by $E_i=E(S_i,S_{i+1})$. We have also denoted by $\downedges{i}=E(S_i',S_i'')$. The remaining edges that are incident to $S_i'$ can be partitioned into two subsets: set $\leftCedges{i}$, containing all edges $(u,v)$, with $u\in S'_i$ and $v\in (S_1\cup\cdots\cup S_{i-2})\cup S_{i-1}''$; and set 
 set $\rightCedges{i}$, containing all edges $(u,v)$, with $u\in S'_i$ and $v\in S_{i+1}''\cup (S_{i+2}\cup\cdots\cup S_{r})$. Next, for all $1<i<r$, we  bound the cardinality of edge set $\leftedges{i}$ in terms of the cardinality of $\leftCedges{i}$, and similarly we bound $|\rightedges{i}|$ in terms of $|\rightCedges{i}|$, in the following claim whose proof appears in \Cref{subsec: left edges for S' and S''}.

 \begin{claim}\label{claim left edges for S' and S''}
 	For all $1<i<r$: 
 	
 	\begin{itemize}
\item 	$|\rightedges{i}|\leq 1.3|E_{i}|+1.3|\rightCedges{i}|$; and 
\item $|\leftedges{i+1}|\leq 1.3|E_{i}|+1.3|\leftCedges{i+1}|$.
\end{itemize}
 \end{claim}

\subsection*{Accounting So Far}

\begin{figure}[h]
	\centering
	\includegraphics[scale=0.12]{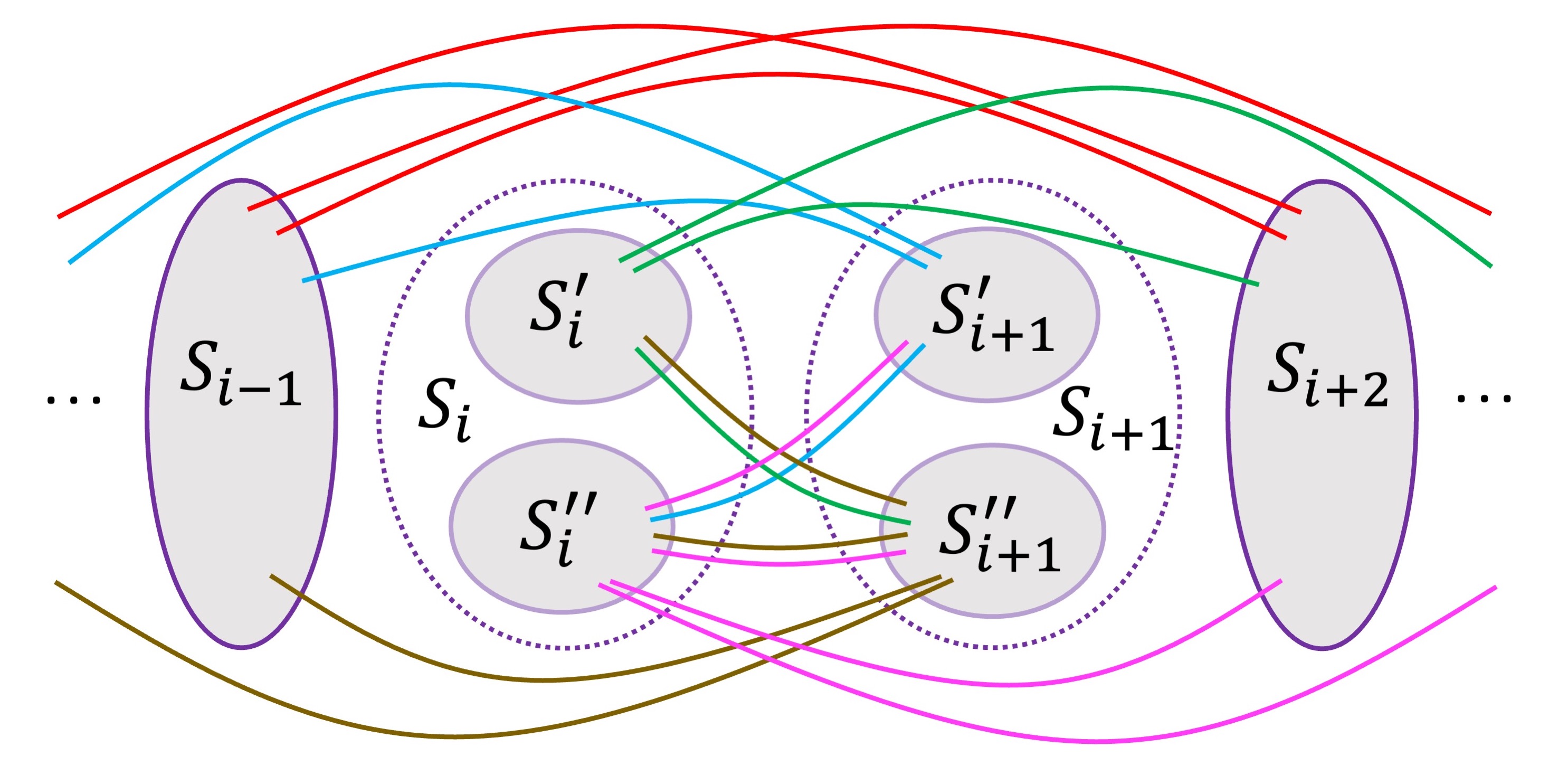}
	\caption{
The set $\hat E_i$ of edges, with the edges of $\tilde E^{\through}_i$  shown in red; the edges of $\delta^{\rig}(S_i'')$ and $\delta^{\lef}(S_{i+1}'')$ in pink and brown respectively; and the edges of $\delta^{\rig}(S_i')$ and  $\delta^{\lef}(S_{i+1}')$  in green and blue, respectively.
Note that the edges of $E(S''_i, S''_{i+1})$ belong to both $\delta^{\rig}(S_i'')$ and $\delta^{\lef}(S_i'')$. Also, the edges of $E(S'_i, S''_{i+1})$ belong to both $\delta^{\lef}(S_{i+1}'')$ and $\delta^{\rig}(S_i')$. Similarly, the edges of $E(S''_i, S'_{i+1})$ belong to both $\delta^{\rig}(S_i'')$ and $\delta^{\lef}(S_{i+1}')$). 
  }\label{fig: NF23}
\end{figure}

We now summarize what we have shown so far. Fix some index $1\leq i<r$. Recall that set $\hat E_i$ of edges contains every edge $e=(u,v)\in E(\cH)$, such that, if $u\in S_j$ and $v\in S_{j'}$, then $j\leq i$ and $j'\geq i+1$ holds, but it excludes the edges in the set $E_i=E(S_i',S_{i+1}')$. Therefore, $\hat E_i$ is the union of the following subsets (see \Cref{fig: NF23}): 

\begin{itemize}
	\item  edge set $\tE_i^{\through}$, connecting vertices of $V(S_1),\ldots,V(S_{i-1})$ to vertices of $V(S_{i+2}),\ldots,V(S_r)$ (see \Cref{fig: NF9}); 
	\item edges that lie in $\rightedges{i}\cup \leftedges{i+1}$ (see \Cref{fig: NF13}); 
	\item edges that lie in $\rightCedges{i}\cup \leftCedges{i+1}$ (see \Cref{fig: NF15}). 
\end{itemize}

In \Cref{obs: bad inded structure}, we have established that  	$|\tE_i^{\through}|\leq |E_{i}'|$, where $E_i'=E(S_i,S_{i+1})$. Notice that all edges of $E_i'$ are contained in $E_i\cup \rightedges{i}\cup \leftedges{i+1}$ (see \Cref{fig: NF12}), 
so we get that $|\tE_i^{\through}|\leq |E_i|+|\rightedges{i}|+|\leftedges{i+1}|$.
From \Cref{claim left edges for S' and S''}, $|\rightedges{i}|\leq 1.3|E_{i}|+1.3|\rightCedges{i}|$, and $|\leftedges{i+1}|\leq 1.3|E_{i}|+1.3|\leftCedges{i+1}|$.
Therefore, altogether, we have shown so far that:

\begin{equation}\label{eq: bound on hat E}
\begin{split}
|\hat E_i|&\leq |\tE_i^{\through}|+|\rightedges{i}|+|\leftedges{i+1}| +|\rightCedges{i}|+|\leftCedges{i+1}|\\
&\leq |E_i|+2|\rightedges{i}|+2|\leftedges{i+1}|+|\rightCedges{i}|+|\leftCedges{i+1}|\\
&\leq 7|E_i|+7|\rightCedges{i}|+7|\leftCedges{i+1}|.
\end{split} \end{equation}

Therefore, it now remains to bound $|\rightCedges{i}|$ and $|\leftCedges{i+1}|$ in terms of $|E_i|$. We start with the following claim that allows us to establish some useful connection between the cardinalities of the three edge sets. 
The proof appears in \Cref{subsec: left edges for S' only}

 \begin{claim}\label{claim left edges for S' only}
	For all $1\leq i<r$: $|\rightCedges{i}|\leq 2.5|E_{i}|+2.5|\leftCedges{i+1}|$, and $|\leftCedges{i+1}|\leq 2.5|E_{i}|+2.5|\rightCedges{i}|$.
\end{claim}

Next, we show that for all $1<i<r$, vertex $u_i$ must be a $J$-node.

\subsubsection*{Proving that $u_2,\ldots,u_{r-1}$ are $J$-nodes.}
We start with the following simple claim, whose proof appears in \Cref{subsec: non-J-node boundary size}.

\begin{claim}\label{claim: non-J-node-boundary size}
	Consider an index $1<i<r$, and assume that vertex $u_i$ is not a $J$-node. Then $|\bigcup_{v\in S'_i}\delta(v)|\leq \left(1+\frac{130}{\log m}\right )|\delta(u_i)|$. Moreover, if $u_i\in L'_j$, for some $1\leq j\leq h$, then every vertex of $S'_i\setminus\set{u_i}$ lies in $L'_{j+1}\cup\cdots\cup L'_h$.
\end{claim}

We are now ready to prove that, for all $1<i<r$, vertex $u_i$ must be a $J$-node.
\begin{lemma}\label{lem: each ui is a J-node}
For all $1<j<r$, vertex $u_i$ is a $J$-node.
\end{lemma}

\begin{proof}
	Assume for contradiction that the lemma is false. We fix an index $1<i^*<r$, such that $u_{i*}$ is not a $J$-node, and subject to this, $|\delta(u_{i^*})|$ is maximized, breaking ties arbitrarily.
	
	We first assume that there is some index $a$, such that at least $|\delta(u_{i^*})|/16$ edges connect $u_{i^*}$ to edges of $S''_a$. We show that in this case, $|\leftedges{a}|,|\rightedges{a}|$ are both large, and $u_a$ must be a $J$-node. (Note that it is impossible that $a\in \set{1,r}$, since $S''_1=S''_r=\emptyset$, as we have established in \Cref{obs: left and right down-edges}.)
The proof of the following claim is deferred to \Cref{subsec:many edges left right large}.

\begin{claim}\label{claim: many edges left right large}
	Suppose there is an index  $1\leq a\leq r$ (where possibly $a=i^*$), such that at least $|\delta(u_{i^*})|/16$ edges connect $u_{i^*}$ to vertices of $S''_a$. Then, $|\leftedges{a}|,|\rightedges{a}|\geq |\delta(u_{i^*})|\cdot\frac {\log m}{256}$, and moreover, $u_a$ is a $J$-node. 
\end{claim}

Consider again the vertex $u_{i^*}$, and assume that $u_{i^*}\in L'_j$, for some $1\leq j\leq h$. Recall that, from \Cref{claim: non-J-node-boundary size},  every vertex of $V(S'_{i^*})\setminus\set{u_{i^*}}$ lies in $L'_{j+1}\cup\cdots\cup L'_h$. 
Therefore, all edges connecting $u_{i^*}$ to vertices of $V(S'_{i^*})\setminus\set{u_{i^*}}$ 
lie in $\delta^{\up}(u_{i^*})$, and their number is bounded by $|\delta^{\up}(u_{i^*})|\leq |\delta(u_{i^*})|/\log m$. From \Cref{claim: many edges left right large}, the number of edges connecting $u_{i^*}$ to vertices of $S''_{i^*}$ must be bounded by $|\delta(u_{i^*})|/16$ (as $u_{i^*}$ is not a $J$-node). The remaining edges of $\delta(u_{i^*})$ must connect $u_{i^*}$ to vertices of $\bigcup_{a\neq i^*}V(S_a)$. Denote by $E^*$ the set of all edges connecting $u_{i^*}$ to vertices of $\bigcup_{a>i^*}V(S_a)$, and denote by $E^{**}$ the set of all edges connecting $u_{i^*}$ to vertices of $\bigcup_{a<i^*}V(S_a)$. From the above discusison, $|E^*\cup E^{**}|\geq 7|\delta(u_{i^*})|/8$, and so either $|E^*|\geq |\delta(u_{i^*})|/4$ or $|E^{**}|\geq |\delta(u_{i^*})|/4$ must hold. We assume w.l.o.g. that it is the former. Next we consider three cases. The first case is when neither $u_{i^*+1}$ or $u_{i^*+2}$ are $J$-nodes; the second case is when $u_{i^*+1}$ is a $J$-node; and the third case is when $u_{i^*+2}$ is a $J$-node. We show that neither of these cases is possible, by showing a simplifying cluster that should have been considered by our algorithm. For simplicity of notation, in the remainder of the proof, we denote $i^*$ by $i$.

\paragraph{Case 1: neither of $u_{i+1}$, $u_{i+2}$ is a $J$-node.}
Consider the set $E^*$ of edges; recall that these are all edges connecting $u_{i}$ to vertices of $\bigcup_{a>i}S_a$. We need the following observation:

\begin{observation}\label{obs: many through edges}
	At least $ |\delta(u_{i})|/16$ edges connect $u_{i}$ to vertices of $\bigcup_{a>i+2}V(S_a)$.
\end{observation}

(Note that in particular it follows from the observation that $r\geq i+3$ must hold).
\begin{proof}
	We partition the edges of $E^*$ into five subsets. The first subset, $E^*_1$, contains all edges of $E^*$ connecting $u_{i}$ to vertices of $S''_{i+1}$, and the second subset, $E^*_2$, contains all edges of $E^*$ connecting $u_{i}$ to vertices of $S''_{i+2}$. Notice that, from \Cref{claim: many edges left right large}, since we have assumed that neither of $u_{i+1}$, $u_{i+2}$ is a $J$-node, $|E^*_1|,|E^*_2|\leq |\delta(u_{i})|/16$. We let $E^*_3$ be the set of all edges of $E^*$ connecting $u_{i}$ to vertices of $\bigcup_{a>i+2}V(S_a)$. Lastly, we let $E^*_4$ and $E^*_5$ be the sets of all edges of $E^*$ connecting $u_{i}$ to vertices of $S'_{i+1}$ and of $S'_{i+2}$, respectively. Assume for contradiction that $|E^*_3|<|\delta(u_{i})|/16$. Then, since  $|E^*|\geq |\delta(u_{i})|/4$, either $|E^*_4|\geq  |\delta(u_{i})|/32$ or $|E^*_5|\geq  |\delta(u_{i})|/32$ must hold. We assume first that  $|E^*_4|\geq  |\delta(u_{i})|/32$. Since $|\delta^{\up}(u_{i})\leq |\delta(u_{i})|/\log m$, $|E^*_4\cap \delta^{\down}(u_{i})|\geq |\delta(u_{i})|/64$. Notice that, if $e=(u_{i},v)$ is an edge of $E^*_4\cap \delta^{\down}(u_{i})$, then $v\in S'_{i+1}$, and $e\in \delta^{\up}(v)$. Since, for every vertex $v$, $|\delta^{\up}(v)|\leq |\delta(v)|/\log m$, we get that:
	
	\[|\bigcup_{v\in S'_{i+1}}\delta(v)|\geq |E^*_4\cap \delta^{\down}(u_{i})|\cdot \log m\geq  \frac{|\delta(u_{i})|\cdot \log m}{32}.\]
	
	On the other hand, from \Cref{claim: non-J-node-boundary size}, $|\bigcup_{v\in S'_{i+1}}\delta(v)|\leq \left(1+\frac{130}{\log m}\right )|\delta(u_{i+1})|<2|\delta(u_{i+1})|$. Therefore, we get that $|\delta(u_{i+1})|>  \frac{|\delta(u_{i})|\cdot \log m}{64}>|\delta(u_{i})|$, contradicting the choice of index $i$. 
	
	In the case where  $|E^*_5|\geq  |\delta(u_{i})|/32$, the analysis is identical.
\end{proof}

Consider a cluster $S^*$, which is a subgraph of $\cH$ induced by $V(S_{i+1})\cup V(S_{i+2})$. In the following claim, whose proof is deferred to \Cref{subsec: simplifying cluster Case 1}, we prove that $S^*$ is a simplifying cluster, reaching a contradiction.

\begin{claim}\label{claim: simplifying cluster case 1}
	Cluster $S^*$ is a simplifying cluster.
\end{claim}

\paragraph{Case 2: $u_{i+1}$ is a $J$-node.}
Recall that in this case, from \Cref{obs: central path j-cluster}, $S'_{i+1}=\set{u_{i+1}}$. 
We show that cluster $S^*=\set{u_{i+1}}$ is a simplifying cluster in the following claim, whose proof is similar to but slightly more involved than the proof of \Cref{claim: simplifying cluster case 1}, and is deferred to \Cref{subsec: simplifying cluster Case 2}.

\begin{claim}\label{claim: simplifying cluster case 2}
	Cluster $S^*$ is a simplifying cluster.
\end{claim}

This is a constradiction, since our algorithm must have identified that $S^*=S'_{i+1}$ is a simplifying cluster.

\paragraph{Case 3: Neither Case 1 nor Case 2 happened.}

Since Cases 1 and 2 did not happen, vertex $u_{i+1}$ is not a $J$-node. 
We start with the following simple observation.

\begin{observation}\label{obs: few edges going right}
	The number of edges connecting $u_i$ to vertices of $S_{i+1}$ is at most $|\delta(u_i)|/8$.
\end{observation}
\begin{proof}
	Assume otherwise. 
	Since Case 3 happened, $u_{i+1}$ is not a $J$-node, and so, from \Cref{claim: many edges left right large}, at most $|\delta(u_i)|/16$ edges may connect $u_i$ to vertices of $S''_{i+1}$. Therefore, the number of edges connecting $u_i$ to $S'_{i+1}$ must be at least $|\delta(u_i)|/16$.
	But then at least $|\delta(u_i)|/32$ edges of $\delta^{\down}(u_i)$ connect $u_i$ to vertices of $S'_{i+1}$. For each such vertex $e'=(u,v)$, $e'\in \delta^{\up}(v)$ must hold. Since, for every vertex $v$, $\delta^{\up}(v)\leq |\delta(v)|/\log m$, we get that $|\bigcup_{v\in S'_{i+1}} \delta(v)|\geq \frac{|\delta(u_i)|\log m}{16}$ must hold. However, from \Cref{claim: non-J-node-boundary size}, $|\delta(u_{i+1})|\geq \frac{|\bigcup_{v\in S'_{i+1}}\delta(v)|}{2}\geq \frac{|\delta(u_i)|\log m}{32}>|\delta(u_i)|$, contradicting the choice of the index $i^*=i$. 
\end{proof}

Recall that we have assumed that $|E^*|\geq |\delta(u_i)|/4$, where $E^*$ is the set of all edges connecting $u_{i}$ to vertices of $\bigcup_{a>i}V(S_a)$. Since, from \Cref{obs: few edges going right}, at most $|\delta(u_i)|/8$ edges connect  $u_i$ to vertices of $S_{i+1}$, it must be the case that $i+2\leq r$, and at least $|\delta(u_i)|/8$ edges connect  $u_i$ to vertices of  $\bigcup_{a>i+1}V(S_a)$. Since we have assumed that Case 1 did not happen, vertex $u_{i+2}$ must be a $J$-node, and so, from \Cref{obs: central path j-cluster}, $S'_{i+2}=\set{u_{i+2}}$. 
We let $S^*=\set{u_{i+2}}$, and we show, in the next claim, that cluster $S^*$ is a simplifying cluster. The proof of the claim is deferred to
\Cref{subsec: simplifying cluster Case 3}.

\begin{claim}\label{claim: simplifying cluster case 3}
	Cluster $S^*$ is a simplifying cluster.
\end{claim}

This is a constradiction, since our algorithm must have identified that $S^*=S'_{i+2}$ is a simplifying cluster.
\end{proof}

In order to complete the proof of \Cref{lem: no bad indices}, it is enough to prove that for all $1\leq  i< r$, $|\hat E_i|\leq  1000|E_i|$. 
Assume for contradiction that there is some index $1\leq i< r$, for which  $|\hat E_i|>  1000|E_i|$ holds.
Recall that we have shown already, in \Cref{eq: bound on hat E}, that:

\[
|\hat E_i|\leq  7|E_i|+7|\rightCedges{i}|+7|\leftCedges{i+1}|.
 \]

If $|\hat E_i|>  1000|E_i|$, then either $|\rightCedges{i}|>64|E_i|$, or $|\leftCedges{i+1}|>64|E_i|$. We assume without loss of generality that it is the former; the other case is symmetric. Recall that, since $u_{i}$ is a $J$-node, $S'_{i}=\set{u_{i}}$. 
From the definition, set $\rightCedges{i}$ contains all edges $(u_i,v)$ with  $v\in S_{i+1}''\cup (S_{i+2}\cup\cdots\cup S_{r})$.
Note that,  from \Cref{obs: left and right down-edges},
 $S'_r=S_r$, and so $\rightCedges{r-1}=\emptyset$. Therefore, we can assume that $i<r-1$. 
We now prove that $S^*=S'_{i+1}=\set{u_{i+1}}$ is a simplifying cluster, in the following claim, whose proof is very similar to the analysis of Case 2 in the proof of \Cref{lem: each ui is a J-node}, and is deferred to \Cref{subsec: simplifying cluster last}. 

\begin{claim}\label{claim: simplifying cluster last case}
	Cluster $S^*$ is a simplifying cluster.
\end{claim}

We reach a contradiction, since our algorithm must have established that cluster $S^*=S'_{i+1}$ is a simplifying cluster.

\renewcommand{\tE}{\textbf{E}'}

\subsection{Disengagement of Nice Instances -- Proof of \Cref{thm: advanced disengagement - disengage nice instances}}
\label{subsec: disengagement with bad chain}

\newcommand{\midd}{\operatorname{mid}}

In this section we provide the proof of \Cref{thm: advanced disengagement - disengage nice instances}. 
Recall that we are given as input an instance $I'=(G',\Sigma')$ of the \cnwrs problem, that we will denote by $I=(G,\Sigma)$, in order to simplify the notation. Additionally, we are given a set $\cset'$ of disjoint clusters of $G'$; in order to simplify the notation, we will denote $\cset'$ by $\cset$. Lastly, we are given a nice witness structure $(\tilde\sset,\tilde\sset',\hat\pset)$ for graph $G$ with respect to the set $\cset$ of clusters, where $\tilde \sset=\set{\tilde S_1,\ldots, \tilde S_r}$ is the backbone of the witness structure, with vertex sets in $\set{V(\tilde S_i)}_{1\le i\le r}$ partitioning $V(G)$. For convenience, we will denote the set
$\tilde \sset'=\set{\tilde S'_1,\ldots, \tilde S'_r}$ of the vertebrae of the nice witness structure by $\sset=\set{S_1,\ldots,S_r}$. Recall that each cluster $S_i\in \sset$ has the $\alpha^*$-bandwidth property, for $\alpha^*=\Omega(1/\log^{12}m)$.
%
%

Recall that we are given a partition of the edges of $G$ into two subsets: set $\tilde E'$, containing all edges of $\bigcup_{1\leq i\leq r}E(S_i)$, and all edges of $\bigcup_{1\leq i<r}E(S_i,S_{i+1})$; and set $\tilde E''=E(G)\setminus \tilde E'$. 
Recall that set $\hat E\subseteq \tilde E''$ contains all edges $(v,u)\in \tilde E''$, where $v$ and $u$ lie in different clusters of $\tilde\sset$, and the set $\hat\pset$ of paths contains, for each edge $e\in \hat E$, a path $ P(e)$ that consists of three subpaths: $P^1(e), P^2(e)$, and $P^3(e)$, that are called the prefix, the mid-part and the suffix of $P(e)$, respectively.
We denote by $\hat \pset^1=\set{P^1(e)\mid e\in \hat E}$, $\hat\pset^2=\set{P^2(e)\mid e\in \hat E}$, and $\hat \pset^3=\set{P^3(e)\mid e\in \hat E}$, the sets of paths containing all prefixes, all mid-parts, and all suffixes of the paths in $\hat \pset$, respectively.
Throughout, we will use a parameter $\hat \eta=2^{O((\log m)^{3/4}\log\log m)}$.

In order to compute a decomposition of instance $I$ into subinstances, we need to define, for every edge $e\in \hat E$, a cycle $W(e)$, called an \emph{auxiliary cycle} that has some useful properties. As an intuition, we could obtain a cycle $W(e)$ by taking the union of the nice guiding path $P(e)\in \hat \pset$ with the edge $e$. The structure of the nice guiding paths ensures that the cycle $W(e)$ has a single contiguous segment $P^2(e)$ that visits a contiguous subset of the vertebrae in the order of their indices. The resulting set $\set{W(e)\mid e\in \hat E}$ of cycles is close to having the properties that we need, except that we would like to ensure that these cycles are non-transeversal (or close to being non-transversal) with respect to $\Sigma$. We discuss the construction of the family $\set{W(e)\mid e\in \hat E}$ of cycles with these properties below.
Next, we define a laminar family $\lset=\set{U_1,\ldots,U_r}$ of clusters of graph $G$, where for all $1\leq i\leq r$, $U_i$ is the subgraph of $G$ induced by vertex set $V(S_1)\cup\cdots\cup V(S_i)$. We define, for every vertebra $S_i\in \sset$, an internal $S_i$-router $\qset(S_i)$, and use these routers, together with the auxiliary cycles in $\set{W(e)\mid e\in \hat E}$ in order to define an internal $U_i$-router and an external $U_i$-router for every cluster $U_i\in \lset$. The final decomposition $\iset_2$ of instance $I$ into subinstances is simply a decomposition via the laminar family $\lset$ defined in \Cref{subsec: laminar-based decomposition}. Recall that for each cluster $U_z\in \iset_2$, there is a unique instance $I_z=(G_z,\Sigma_z)\in \iset_2$, where graph $G_z$ is obtained from $G$ by contracting all vertices of $S_1\cup\cdots \cup S_{z-1}$ into a special vertex $v_z^*$, and all vertices of $S_{z+1}\cup\cdots\cup S_r$ into a special vertex $v_z^{**}$ (for $z=1$, $G_1$ is obtained from $G$ by contracting all vertices of $S_2\cup\cdots \cup S_r$ into a special vertex $v_1^{**}$, and for $z=r$, graph $G_z$ is obtained from $G$ by contracting all vertices of $S_1\cup\cdots\cup S_{r-1}$ into a special vertex $v_r^*$).  For each $1\leq z<r$, we use the internal $U_z$-router that we computed, in order to define a circular ordering of the edges of $\delta_G(U_z)$, that will in turn be used in order to define the rotation systems $\set{\Sigma_z}_{z=1}^r$ associated with each subinstance. The techniques developed in \Cref{subsec: laminar-based decomposition} prove that there is an efficient algorithm that combines solutions to the resulting subinstances into a solution to instance $I$ that has a relatively low cost. However, since the depth of the laminar family $\lset$ may be quite high, we cannot use the tools from \Cref{sec: guiding paths orderings basic disengagement} in order to bound $\sum_{z=1}^r|E(G_z)|$ and $\sum_{z=1}^r\optcrors(I_z)$. Instead, we show a simple direct bound on $\sum_{z=1}^r|E(G_z)|$, and a more involved proof for bounding $\sum_{z=1}^r\optcrors(I_z)$. The latter proof exploits the internal and external $U_i$-routers that we construct, for $1\leq i\leq r$, which in turn are based on the auxiliary cycles $\set{W(e)\mid e\in \hat E}$, in order to show the existence of a low-cost solution to each instance $I_z\in \iset_2$. In order to ensure that the costs of these solutions is sufficiently low, it is crucial that the cycles in $\set{W(e)\mid e\in \hat E}$ are \emph{almost} non-transversal with respect to $\Sigma$: that is, for every pair $W(e),W(e')$ of cycles, there is at most one vertex $v$, such that $W(e)$ and $W(e')$ intersect transversally at $v$. In order to define the set $\wset=\set{W(e)\mid e\in \hat E}$ of cycles, we define two collections of paths: path set $\pset^{\out}=\set{P^{\out}(e)\mid e\in \hat E}$, which is obtained by modifying the paths of $\hat \pset^1\cup \hat \pset^3$, and path set  $\pset^{\inn}=\set{P^{\inn}(e)\mid e\in \hat E}$. For every edge $e\in \hat E$, the first and the last edges on paths $P^{\out}(e)$ and $P^{\inn}(e)$ are identical. Path $P^{\out}(e)$ contains the edge $e$, and all its edges lie in $\tilde E''$. All inner edges of path $P^{\inn}(e)$ lie in $\tilde E'$, and the path visits a consequtive subset of clusters of $\sset$ in their natural order.
The auxiliary cycle $W(e)$ is obtained by taking the union of the paths $P^{\out}(e)$ and $P^{\inn}(e)$.

The remainder of the proof of \Cref{thm: advanced disengagement - disengage nice instances} consists of four steps. In the first step, we construct the set $\pset^{\out}=\set{P^{\out}(e)\mid e\in \hat E}$ of paths. In the second step,  we construct the set $\pset^{\inn}=\set{P^{\inn}(e)\mid e\in \hat E}$ of paths and the collection $\wset=\set{W(e)\mid e\in \hat E}$ of auxiliary cycles. In the third step, we construct the laminar family $\lset=\set{U_1,\ldots,U_r}$ of clusters, and, for all $1\leq z\leq r$, an internal $U_z$-router $\qset(U_z)$ and an external $U_z$-router $\qset'(U_z)$. In the fourth and the final step, we compute the collection $\iset_2$ of subinstances of $I$ and analyze its properties. We now describe each of the steps in turn.



\subsubsection{Step 1. Constructing the Paths of $\pset^{\out}$}

In this step, we construct the set $\pset^{\out}=\set{P^{\out}(e)\mid e\in \hat E}$ of paths, by slightly modifying the prefixes and the suffixes of the paths in $\hat \pset$ to make them non-transversal. 

Throughout, we denote $V'=\bigcup_{S\in \sset}V(S)$ and $V''=V(G)\setminus V'$. 
Consider an edge $e=(u,v)\in \hat E$, and assume that $u\in S_i,v\in S_j$, and $i<j$. For convenience, we will call $u$ the \emph{left endpoint of edge $e$}, and $v$ the \emph{right endpoint of edge $e$}. 
We also define two sets of indices associated with edge $e$. The first set of indices is $\spann(e)=\set{i,i+1,\ldots,j-1}$. 
In order to define the second set of indices, assume that the last vertex on path $P^1(e)$ (vertex that lies in $V'$) belongs to cluster $S_{i'}$, while the last vertex on path $P^3(e)$ belongs to cluster $S_{j'}$. From the definition of nice guiding paths, $i'\leq i<j\leq j'$ must hold. We then let $\spann'(e)=\set{i',i'+1,\ldots,j'-1}$.

%
%
%
It will be convenient for us to define the notion of left-monotone and right-monotone paths. The definition is similar to the one used in \Cref{subsec: getting nice structure better clustering}, but not identical.

\begin{definition}
	Let $R$ be a (directed) path in graph $G$ that contains at least one edge, let $(v_1,\ldots,v_z)$ be the sequence of vertices appearing on $R$, and, for all $1\leq a\leq z$, assume that $v_a\in \tilde S_{i_a}$. We say that $R$ is a \emph{left-monotone path} if:
	
	\begin{itemize} 
		\item $v_z\in V'$;
		\item for $1<a<z$, $v_a\in V''$; and 
		\item $i_1\geq i_2\geq\cdots\geq i_z$. 
		\end{itemize}
	Similarly, we say that $R$ is a \emph{right-monotone path}, if:
	
		\begin{itemize} 
		\item $v_z\in V'$;
		\item for $1<a<z$, $v_a\in V''$; and 
		\item $i_1\leq i_2\leq \cdots\leq i_z$. 
	\end{itemize}
\end{definition}

For each edge $e\in \hat E$, we view the paths $P^1(e)\in \pset^1$, $P^3(e)\in \pset^3$ as directed paths that originate at an endpoint of edge $e$ and terminate at a vertex of $V'$ (notice that it is possible that one or even both endpoints of $e$ lie in $V'$).
For every vertex $v\in V'$, we denote by $n_1(v)$ the total number of paths in $\hat \pset^1$ that terminate at $v$, and by $n_3(v)$ the total number of paths in $\hat \pset^3$ that terminate at $v$. Let $\eta=O(\log^{18}m)$ be such that the set $\hat \pset$ of paths causes congestion at most $\eta$ in $G$.
We use the following claim in order to construct the paths of $\pset^{\out}$; the proof of the claim uses standard techniques and is deferred to \Cref{subsec: computing out-paths}.

\begin{claim}\label{claim: computing out-paths}
	There is an efficient algorithm to compute two sets $\pset^{\out,\lef}=\set{P^{\out,\lef}(e)\mid e\in \hat E}$ and $\pset^{\out,\rig}=\set{P^{\out,\rig}(e)\mid e\in \hat E}$ of simple paths in graph $G$, each of which causes congestion at most  $\eta$, such that the paths in set $\pset^{\out,\lef}$ are non-transversal with respect to $\Sigma$, and so are the paths in set $\pset^{\out,\rig}$. Additionally, for every edge $e\in \hat E$, path $P^{\out,\lef}(e)$ has $e$ as its first edge, and it is left-monotone, while path $P^{\out,\rig}(e)$  has $e$ as its first edge, and is right-monotone. Moreover, for every vertex $v\in V'$, exactly $n_1(v)$ paths of $\pset^{\out,\lef}$ terminate at $v$, and exactly $n_3(v)$ paths of  $\pset^{\out,\rig}$ terminate at $v$.
\end{claim}

Consider now some edge $e=(u,v)\in \hat E$, and assume that $u\in \tilde S_i$, $v\in \tilde S_j$, and $i<j$ holds. Consider the paths $P^{\out,\lef}(e),P^{\out,\rig}(e)$. Assume that the last vertex on path $P^{\out,\lef}$ is $u'$, and the last vertex on path $P^{\out,\rig}$ is $v'$.  We let $P^{\out}(e)$ be the path obtained by concatenating path $P^{\out,\lef}(e)$ with the reversed path $P^{\out,\rig}(e)$, after deleting the extra copy of edge $e$. We view path $P^{\out}(e)$ as being directed rom $u'$ to $v'$. Therefore, path $P^{\out}(e)$ originates at vertex $u'$ and termiantes at vertex $v'$, and it contains the edge $e$. All inner vertices on $P^{\out}(e)$ belong to $V''$. We will sometimes refer to $u'$ and to $v'$ as the first and the last endpoints of path $P^{\out}(e)$. We will also refer to the edge of $P^{\out}(e)$ that is incident to $u'$ as the \emph{first edge} of path $P^{\out}(e)$, and to the edge of $P^{\out}(e)$ that is incident to $v'$ as the \emph{last edge} of path $P^{\out}(e)$.
Assume that $u'\in V(S_{i''})$ and $v'\in V(S_{j''})$. We define another set of indices associated with edge $e$: $\spann''(e)=\set{i'',i''+1,\ldots,j''-1}$. Notice that $\spann(e)\subseteq \spann''(e)$ must hold by the definition of left-monotone and right-monotone paths. Lastly, we set $\pset^{\out}=\set{P^{\out}(e)\mid e\in \hat E}$.


For an index $1\leq i<r$, let $\hat E_i\subseteq \hat E$ be the set of all edges $e\in \hat E$, with $i\in \spann(e)$. We need the following simple claim.

\begin{claim}\label{claim: out-paths non-transversal}
	For every index $1\leq i<r$, the paths in set $\set{P^{\out}(e)\mid e\in \hat E_i}$ are non-transversal with respect to $\Sigma$. Moreover, for each edge $e=(u,v)\in \hat E_i$ whose left endpoint is $u$, every vertex of $P^{\out,\lef}(e)\setminus\set{v}$ lies in $\bigcup_{z\leq i}V(\tilde S_z)$, and every vertex of $P^{\out,\rig}(e)\setminus\set{u}$ lies in $\bigcup_{z> i}V(\tilde S_z)$.
\end{claim}
\begin{proof}
	The fact that, for every edge  $e\in \hat E_i$, every vertex of $P^{\out,\lef}(e)\setminus\set{v}$ lies in $\bigcup_{z\leq i}V(\tilde S_z)$, and every vertex of $P^{\out,\rig}(e)\setminus\set{u}$ lies in $\bigcup_{z> i}V(\tilde S_z)$ follows immediately from the definition of left-monotone and right-monotone paths and edge set $\hat E_i$.  Consider now any pair $e,e'\in \hat E_i$ of edges, and a vertex $v$ that is an inner vertex of both $P^{\out}(e)$ and $P^{\out}(e')$. Assume that $v\in V(\tilde S_j)$, for some index $1\leq j\leq r$. 
	From the definition of left-monotone and right-monotone paths, either $j\leq 
	i$ and $v$ is an inner vertex of both $P^{\out,\lef}(e)$ and $P^{\out,\lef}(e')$; or $j>i$, and $v$ is an inner vertex of both $P^{\out,\rig}(e)$ and $P^{\out,\rig}(e')$. In either case, \Cref{claim: computing out-paths} ensures that the intersection of $P^{\out}(e)$ and $P^{\out}(e')$ at $v$ is non-transversal.
\end{proof}

For every index $1\leq t<r$, we denote by $N_t$ the number of all edges 
$e\in \hat E$, with $t\in \spann'(e)$, and we denote by $N'_t$ be the number of all edges 
$e\in \hat E$, with $t\in \spann''(e)$. We also denote by $E_t\subseteq \tilde E'$ the set of all edges with one endpoint in $S_t$ and another in $S_{t+1}$. 
 We need the following claim, whose proof appears in \Cref{subsec: enough segments}.

\begin{claim}\label{claim: enough segments}
	For all $1\leq t<r$, $N'_t=N_t$, and $N_t\leq O(\log^{18}m)\cdot |E_t|$.
\end{claim}

\subsubsection{Step 2: Constructing the Paths of $\pset^{\inn}$ and the Auxiliary Cycles}

Consider some edge $e=(u,v)\in \hat E$, and assume that $u$ is the left endpoint of $e$. Assume that $\spann''(e)=\set{i'',i''+1,\ldots,j''-1}$, and denote by
 $\hat e_{i''-1}$ the first edge on path $P^{\out}(e)$, and by $\hat e_{j''}$ the last edge on path $P^{\out}(e)$.
In this step we will construct another path $P^{\inn}(e)$, whose first edge is $\hat e_{i''-1}$ and last edge is $\hat e_{j''}$. In order to do so, we will select, for  all $i''\leq z<j''$, some edge $\hat e_z\in E_z$, that we assign to the edge $e$, and we will compute a path $R_z(e)$ (that we call a \emph{segment}), whose first edge is $\hat e_{z-1}$, last edge is $\hat e_z$, and all inner vertices are contained in $S_z$. The final path $P^{\inn}(e)$ will be obtained by concatenating the segments $R_{i''}(e),\ldots,R_{j''}(e)$. Note that path $P^{\inn}(e)$ has  $\hat e_{i''-1}$ and $\hat e_{j''}$ as its first and last edges. By concatenating the paths $P^{\inn}(e)$ and $P^{\out}(e)$, we will then obtain the auxiliary cycle $W(e)$. Our goal in constructing the set $\pset^{\inn}=\set{P^{\inn}(e)\mid e\in \hat E}$ of paths is to ensure that these paths cause low congestion, and that these paths are \emph{mostly} non-transversal with respect to $\Sigma$. In fact, we will ensure that, for every pair $P,P'\in \pset^{\inn}$ of such paths, there is at most one vertex $v$, such that the intersection of $P$ and $P'$ at $v$ is transversal. Intuitively, the resulting auxiliary cycles in $\set{W(e)\mid e\in \hat E}$ will be exploited in order to show the existence of cheap solutions to the subinstances of the input instance $I$ that we compute. Each transversal intersection between a pair of such cycles may give rise to a crossing in these solutions, which motivates the requirement that the paths in  $\pset^{\inn}$ have few transversal intesrections is low.

Consider again an edge $e\in \hat E$, and assume that $\spann''(e)=\set{i'',i''+1,\ldots,j''-1}$. Recall that we have already defined edges $\hat e_{i''-1}\in \delta(S_{i''-1})$ and $\hat e_{j''}\in \delta(S_{j''-1})$. We  construct a collection $\tilde \rset(e)=\set{R_{i''}(e),\ldots, R_{j''}(e)}$ of paths, and define, for all $i''\leq z<j''$, edge $\hat e_z\in E_z$, such that, for all $i''\leq z\leq j''$, path $R_z(e)$ connects edge $\hat e_{z-1}$ to edge $\hat e_z$, and its inner vertices lie in $S_z$. In order to do so, we initially set $\tilde \rset(e)=\emptyset$ for every edge $e\in \hat E$. We then process indices $1\leq z\leq r$ one by one. When index $z$ is processed, we will define, for every edge $e\in \hat E$ with $z\in \spann''(e)$ or $z-1\in \spann''(e)$, the segment $R_z(e)$; if $z\in \spann''(e)$, we will also define the edge $\hat e_z\in E_z$, which is the last edge on path $R_z(e)$. 
We will ensure that every edge $e'\in E_z$ is assigned to at most $O(\log^{18}m)$ edges of $\hat E$.
We now describe an interation where index $1\leq z\leq r$ is processed.

\paragraph{Iteration Description.}
We fix an index $1\leq z\leq r$, and describe an iteration for processing index $z$. Let $A_z\subseteq \hat E$ be the set of all edges $e\in \hat E$, with $z\in \spann''(e)$. Note that for every edge $e\in A_z$, the corresponding edge $\hat e_{z-1}\in E_{z-1}$ is already fixed.  Let $A'_z\subseteq \hat E\setminus A_z$ be the set of all edges $e\in \hat E$, such that $z-1\in \spann''(e)$ but $z\not\in \spann''(e)$. Notice that, if $e\in A'_z$, then both edges $\hat e_{z-1},\hat e_z\in \delta_G(S_z)$ are already fixed (in this case, edge $\hat e_z$ is the last edge on path $P^{\out}(e)$).

Consider the augmentation $S_z^+$ of the cluster $S_z$, that we denote for convenience by $H$. Recall that, in order to obtain graph $H$, we start by subdividing every edge $e'\in \delta_G(S_z)$ by vertex $t_{e'}$, and then denote by $T=\set{t_{e'}\mid e'\in \delta_G(S_z)}$ the set of newly added vertices, that we call \emph{terminals}. We then let $H$ be the subgraph of the resulting graph induced by $V(S_z)\cup T$. 
Recall that, from the definition of nice witness structure, cluster $S_z$ has the $\alpha^*=\Omega(1/\log^{12}m)$-bandwidth property in $G$, and so, from \Cref{obs: wl-bw}, vertex set $T$ is $\alpha^*$-well-linked in $H$. 

We now define a collection $M$ of pairs of terminals, that we call \emph{demand pairs}, that are associated with the edges of $A'_z$. Consider an edge $e\in A'_z$, and recall that edges $\hat e_{z-1},\hat e_z\in \delta_G(S_z)$ are already defined. The demand pair associated with edge $e$ is $(x_e,y_e)$, where $x_e=t_{\hat e_{z-1}}$ (the terminal vertex associated with edge $\hat e_{z-1}$), and $y_e=t_{\hat e_z}$ (the terminal vertex associated with edge $\hat e_z$). We then set $M=\set{(x_e,y_e)\mid e\in A'_z}$.
Recall that the edges of $\pset^{\out}$ cause congestion at most $\eta=O(\log^{18}m)$, and every edge of $E_{z-1}$ is assigned to at most $\eta$ edges of $\hat E$. Therefore, a terminal $t_{e'}$ may participate in at most $\eta$ pairs in $M$. Using a standard greedy algorithm, we can now compute $2\eta$ sets of terminal pairs $M_1,\ldots,M_{2\eta}$, such that $M=\bigcup_{a=1}^{2\eta}M_a$, and, for all $1\leq a\leq 2\eta$, each terminal participates in at most one pair in $M_a$. For each $1\leq a\leq 2\eta$, we use the algorithm from \Cref{cor: routing well linked vertex set}, to compute a collection $\rset(M_a)=\set{R(x,y)\mid (x,y)\in M_a}$ of paths in graph $H$, where for every pair $(x,y)\in M_a$, path $R(x,y)$ connects $x$ to $y$. Since the vertices of $T$ are $\alpha^*$-well-linked in $G$, with high probability, the paths in $\rset(M_a)$ cause congestion $O(\log^{4}m/\alpha^*)=O(\log^{16}m)$. If the paths in $\rset(M_a)$ cause a higher congestion, then we terminate the algorithm and return FAIL.
Note that the set $\bigcup_{a=1}^{2\eta}\rset(M_a)$ of paths in graph $H$ naturally defines a set $\rset''=\set{R(e)\mid e\in A'_z}$ of paths in graph $G$, where, for every edge $e\in A'_z$, path $R(e)$ has $\hat e_{z-1}$ as its first edge and $\hat e_z$ as its last edge, while all inner vertices of $R(e)$ lie in $S_z$. From the above discussion, the paths in $\rset''$ cause congestion at most $\eta'=2\eta\cdot O(\log^{16}m)=O(\log^{34}m)$.

Next, we consider the set $A_z\subseteq E_{z-1}$ of edges. Let $X$ be a multiset of vertices of $T$ that contains, for every edge $e'\in A_z$, the corresponding vertex $t_{e'}$. Since every edge of $E_{z-1}$ may only be assigned to at most $\eta$ edges of $\hat E$, a vertex may appear in set $X$ at most $\eta$ times.
Recall that $|A_z|=N'_z$, and, from \Cref{claim: enough segments}, $N'_z\leq \eta\cdot |E_z|$. Therefore, we can define a multiset $Y$ that contains $|A|$ elements, each of which is a vertex from $\set{t_{e'}\mid e'\in E_z}$, such that at most $\eta$ copies of each such vertex $t_{e'}$ appear in set $Y$. We let $M'$ be an arbitrary matching between elements of $X$ and elements of $Y$. Using the same procedure as the one employed for the edges of $A'_z$, we construct a set $\rset'=\set{R(e)\mid e\in A_z}$ of paths in graph $G$, where, for every edge $e\in A_z$, path $R(e)$ has $\hat e_{z-1}$ as its first edge and some edge of $E_{z}$ as its last edge, while all inner vertices of $R(e)$ lie in $S_z$. Additionally,  the paths in $\rset''$ cause congestion at most $\eta'$, and every edge of $E_z$ appears on at most $\eta$ edges of $\rset$. (As before, if  the paths in set $\rset''$ cause a higher congestion, we terminate the algorithm and return FAIL.)
Note that we can assume without loss of generality that all paths in $\rset'\cup \rset''$ are simple paths. 

To summarize, we have now constructed two sets $\rset'=\set{R(e)\mid e\in A_z},\rset''=\set{R(e)\mid e\in A'_z}$ of paths, with the following properties:

\begin{properties}{I}
	\item Paths in each of the sets $\rset',\rset''$ cause congestion at most $\eta'=O(\log^{34}m)$; \label{inv: edge congestion}
	\item For every edge $e\in A_z$, path $R(e)\in \rset'$ has $\hat e_{z-1}$ as its first edge, some edge of $E_{z}$ as its last edge, and all inner vertices of $R(e)$ lie in $S_z$; \label{inv: first kind of paths}
\item Every edge of $E_z$ participates in at most $\eta$ paths of $\rset'$;
	\item For every edge $e\in A'_z$, path $R(e)\in \rset''$ has $\hat e_{z-1}$ as its first edge, $\hat e_z$ as its last edge,  and all inner vertices of $R(e)$ lie in $S_z$; and \label{inv: second kind of paths}
	\item All paths in set $\rset'\cup \rset''$ are simple. \label{inv: simple paths}
\end{properties}

Next, we will iteratively modify the paths in $\rset'\cup \rset''$, while ensuring that Properties \ref{inv: edge congestion}--\ref{inv: simple paths} hold at the end of each iteration. In every iteration, we will attempt to reduce the number of transversal intersections between the paths of $\rset'\cup \rset''$. In fact we will guarantee that, after each iteration, either $\sum_{R\in \rset'\cup \rset''}|E(R)|$ decreases, or $\sum_{R\in \rset'\cup \rset''}|E(R)|$ remains unchanged, and the number of triples in set $\Pi^{T}(\rset'\cup \rset'')$ strictly decreases (see definition immediately after \Cref{def: non-transversal paths})

We now describe a single iteration. Assume first that there are two paths $R(e),R(e')\in \rset'$, and some vertex $v$ that is an inner vertex of both $R(e)$ and $R(e')$, such that the intersection of $R(e)$ with $R(e')$ at $v$ is transversal. In this case, we splice paths $R(e)$ and $R(e')$ at vertex $v$ (see \Cref{subsec: non-transversal paths and splicing}), obtaining two new paths. The first path, that replaces $R(e)$ in $\rset'$, originates at edge $\hat e_{z-1}$ and terminates at the edge of $E_z$ that served as the last edge of $R(e')$. The second path, that replaces $R(e')$ in $\rset'$, originates at edge $e'_{z-1}$, and terminates at the edge of $E_z$ that served as the last edge of the original path $R(e)$. Both paths only contain vertices of $S_z$ as inner vertices.  From \Cref{obs: splicing}, either at least one of the two new paths $R(e), R(e')$ becomes a non-simple path; or both paths remain simple paths, but $|\Pi^T(\rset'\cup \rset'')|$ decreases. Notice that, in the latter case, $\sum_{R\in \rset'\cup \rset''}|E(R)|$ remains unchanged. If the former case happens, then we remove cycles from paths $R(e),R(e')$, until they become simple paths. In this case, $\sum_{R\in \rset'\cup \rset''}|E(R)|$ decreases. In either case, it is easy to verify that Invariants \ref{inv: edge congestion}--\ref{inv: simple paths} continue to hold. We then proceed to the next iteration.

Assume now that there is a path $R(e)\in \rset'\cup \rset''$ and another path $R(e')\in\rset''$, and two distinct vertices $v,v'$, both of which are inner vertices on both $R(e)$ and $R(e')$, such that $R(e)$ and $R(e')$ intersect transversally at both $v$ and $v'$. Let $Q$ be the subpath of $R(e)$ between $v$ and $v'$, and let $Q'$ be the subpath of $R(e')$ between $v$ and $v'$. We splice the paths $R(e)$ and $R(e')$ at both $v$ and $v'$. Equivalently, we modify path $R(e)$ by replacing its segment $Q$ with $Q'$, and we modify path $R(e')$ by replacing its segment $Q'$ with $Q$. Note that the first and the last edge on each path remains the same, and the congestion caused by the set $\rset'\cup \rset''$ of paths remains the same. If any of the resulting paths $R(e),R(e')$ becomes a non-simple path, then we delete cycles from it, until it becomes a simple path. In this case, $\sum_{R\in \rset'\cup \rset''}|E(R)|$ decreases. Otherwise, we can use \Cref{obs: splicing} to conclude that $|\Pi^T(\rset'\cup \rset'')$ has decreased. Indeed, it is easy to verify that, for any vertex $v''\in V(S_z)\setminus\set{v,v'}$, the number of triples $(R_1,R_2,v'')\in \Pi^T(\rset'\cup \rset'')$ did not grow. The number of triples $(R_1,R_2,v)\in \Pi^T(\rset'\cup \rset'')$, and the number of triples $(R_1,R_2,v')\in \Pi^T(\rset'\cup \rset'')$ have both decreased (as can be seen by applying \Cref{obs: splicing} to the set of paths that contains, for every path $R^*\in \rset'\cup \rset''$ with $v\in R^*$, a subpath of $R^*$ consisting of the two edges of $R^*$ incident to $v$, and doing the same for vertex $v'$). This completes the description of an iteration. It is easy to verify that Invariants \ref{inv: edge congestion}--\ref{inv: simple paths} continue to hold.

The algorithm for processing the index $z$ terminates when the path set $\rset'$ becomes non-traversal with respect to $\Sigma$, and, for every pair $R\in \rset'\cup\rset''$, $R'\in \rset''$ of paths, there is at most one vertex $v$ such that the intersection of $R$ and $R'$ at $v$ is transversal. We then denote $\rset_z=\rset'\cup \rset''$. For every edge $e\in A_z\cup A'_z$, we set $R_z(e)=R(e)$ (the unique path in $\rset_z$ that originates at edge $\hat e_{z-1}$). If $e\in A'_z$, then path $R(e)$ is guaranteed to terminate at edge $\hat e_z$, from Invariant \ref{inv: second kind of paths}. If $e\in A_z$, then we let $\hat e_z$ be the last edge on path $R_z(e)$. We then add path $R_z(e)$ to set $\tilde \rset(e)$. 

Once all indices $1\leq z\leq r$ are processed, we obtain, for every edge $e\in \hat E$, the desired set $\tilde \rset(e)$ of paths. If $\spann(e)=\set{i'',\ldots,j''-1}$, then $\tilde \rset(e)=\set{R_{i''}(e),\ldots,R_{j''}(e)}$. We then let $P^{\inn}(e)$ be the path obtained by concatenating the paths in $\tilde \rset(e)$. Recall that the first edge on $P^{\inn}(e)$ is $\hat e_{i''-1}$, which is the first edge of $P^{\out}(e)$, and similary, the last edge on $P^{\inn}(e)$ is $\hat e_{j''}$,  the last edge of $P^{\out}(e)$. We obtain the auxiliary cycle $W(e)$ by concatenating the paths $P^{\inn}(e)$ and $P^{\out}(e)$ (after deleting the extra copies of edges $\hat e_{i''-1},\hat e_{j''}$). It is immediate to verify that cycle $W(e)$ is a simple cycle. Lastly, we set $\pset^{\inn}=\set{P^{\inn}(e)\mid e\in \hat E}$ and $\wset=\set{W(e)\mid e\in \hat E}$. We refer to $\wset$ as the \emph{set of auxiliary cycles}. 
From the above discussion and \Cref{claim: computing out-paths},  we obtain the following immediate observation:

\begin{observation}\label{obs: bound congestion of cycles}
	Every edge $e\in \bigcup_{z=1}^rE(S_z)$ appears on at most  $\eta'=O(\log^{34}m)$ cycles of $\wset$. Every edge $e\in E(G)\setminus \left ( \bigcup_{z=1}^rE(S_z) \right )$ appears on at most $\eta=O(\log^{18}m)$ cycles of $\wset$.
\end{observation}

Recall that we denoted, for each index $1\leq z<r$, by $\hat E_z\subseteq \hat E$ the set of all edges $e\in \hat E$, with $z\in \spann(e)$.
We need the following observation.

\begin{observation}
	\label{obs: auxiliary cycles non-transversal at at most one}
	For every index $1\leq z<r$, for every pair $e, e'\in \hat E_z$ of distinct edges, there is at most one vertex $v\in V(W(e))\cap V(W(e'))$, such that the  intersection of cycles $W(e)$ and $W(e')$ is transversal at $v$. If such a vertex $v$ exists, then $v\in S_j$ for some index $z< j<r$, and either $j-1$ is the last index in both $\spann''(e),\spann''(e')$, or $j-1$ is the last index in one of these sets, while $j$ belongs to another.
\end{observation}
\begin{proof}
	Fix an index $1\leq z<r$ and a pair $e,e'\in \hat E_z$ of edges. Let $v$ be any vertex that lies on both $W(e)$ and $W(e')$. We consider two cases. The first case is when $v\in V''$, that is, there is some index $1\leq z'\leq r$, such that $v\in V(\tilde S_{z})\setminus V(S_{z})$. Since all inner vertices of paths $P^{\inn}(e),P^{\inn}(e')$ lie in $V'$, vertex $v$ must be an inner vertex of both $P^{\out}(e)$ and $P^{\out}(e')$. From \Cref{claim: out-paths non-transversal}, the intersection of $P^{\out}(e)$ and $P^{\out}(e')$ at vertex $v$ is non-transversal. Therefore, the intersection of $W(e)$ and $W(e')$ at $v$ is non-transversal. 
	
	The second case is when there is some index $1\leq z'\leq r$, such that $v\in V(S_{z'})$. Assume that the intersection of $W(e)$ and $W(e')$ at $v$ is transversal. From our construction of the path set  $\rset_{z'}$, it must be the case that at least one of the edges $e,e'$ lies in $A'_{z'}$. 
	Assume without loss of generality that this edge is $e$. Notice that, from the construction of set $A'_{z'}$, the last index of $\spann''(e)$ must be $z'-1$. Since $e'\in A_{z'}\cup A'_{z'}$, either $z'\in \spann''(e')$, or the last index in $\spann''(e')$ is $z'-1$. Therefore, if we denote by $j$ the last index of $\spann''(e')$, then $j\geq z'-1$ must hold. From our construction, $v$ is the only vertex of $S_{z'}$, such that the intersection of $P^{\inn}(e)$ and $P^{\inn}(e')$ at $v$ is transversal. Moreover, since $j\geq z'-1$, and $z'-1$ is the last index in $\spann''(e)$, for every index $z''\neq z'$, for every vertex $v'\in S_{z''}$ that lies on both $P^{\inn}(e)$ and $P^{\inn}(e')$, the intersection of the two paths at $v'$ must be non-transversal. 
	\end{proof}

\subsubsection{Step 3: Laminar Family $\lset$ of Clusters,  and Internal and External Routers for Clusters of $\lset$}

We define a laminar family $\lset=\set{U_1,\ldots,U_r}$ of clusters of $G$, that will be later used in order to compute a decomposition of the input instance $I$ into subinstances. 

For each $1\le i\le r$, we define cluster $U_i$ to be the subgraph of $G$ induced by $\bigcup_{1\le z\le i}V(\tilde S_z)$. For convenience, we also denote by $\notU_i$ the subgraph of $G$ induced by $\bigcup_{i< z\le r}V(\tilde S_z)$. We then define a laminar family $\lset=\set{U_1,\ldots,U_r}$ of clusters of $G$. Notice that $U_r=G$, and $U_1\subseteq U_2\subseteq\cdots\subseteq U_r$. 

We now fix an index $1\leq i\leq r$ and consider the set $\delta_G(U_i)=E(U_i,\notU_i)$ of edges. We can partition this edge set into two subsets: set $E_i=E(S_i,S_{i+1})$, and set $\hat E_i$, containing all remaining edges. Notice that $\hat E_i$ is precisely the set of all edges $e\in \hat E$ with $i\in \spann(e)$.

For all $1\leq i\leq r$, we will define an internal router $\qset(U_i)$, and an external router $\qset'(U_i)$ for cluster $U_i$. 
The internal routers $\qset(U_i)$ of the clusters $U_i\in \lset$ will be used in order to compute the decomposition of $I$ into subinstances. Both the internal and the external routers $\qset(U_i),\qset'(U_i)$ will be used in order to argue that the resulting instances have a relatively cheap solution. In order to define these routers, we first need to define, for all $1\leq i\leq r$, an internal router $\qset(S_i)$ for cluster $S_i\in \sset$. The algorithm for computing these routers is randomized, and is provided next.

\subsubsection*{Algorithm for Computing Internal Routers for the Vertebrae.}

Recall that we have defined a parameter $\hat \eta=2^{O((\log m)^{3/4}\log\log m)}$.
We also use  a new parameter
$\beta^*=2^{O(\sqrt{\log m}\cdot \log\log m)}$. 

We now provide a randomized algorithm that computes, 
for each cluster $S_i\in \sset$ an internal $S_i$-router $\qset(S_i)$.  Additionally, we compute a partition $(\sset^{\bad},\sset^{\light})$ of the clusters in $\sset$. Initially, we set $\sset^{\bad}=\sset^{\light}=\emptyset$.
Recall that, from the definition of the nice witness structure, every cluster $S_i\in \sset$ has the $\alpha^*$-bandwidth property, for $\alpha^*=\Omega(1/\log^{12}m)$.

For each $1\leq i\leq r$ in turn, we apply Algorithm \algclassifycluster from \Cref{thm:algclassifycluster} to instance $I=(G,\Sigma)$ of \CNwRS and cluster $J=S_i$, with parameter $p=1/m^{100}$.
If the algorithm returns a distribution $\dset(S_i)$ over internal $S_i$-routers in $\Lambda(S_i)$, such that $S_i$ is $\beta^*$-light with respect to $\dset(S_i)$, then we sample an internal $S_i$-router $\qset(S_i)$ from the distribution $\dset(S_i)$, and we let $u_i$ be the vertex of $S_i$ that serves as the common endpoint of all paths in $\qset(S_i)$. We refer to vertex $u_i$ as the \emph{center vertex of $S_i$}. We also add cluster $S_i$ to set $\sset^{\light}$ in this case.
Otherwise, Algorithm \algclassifycluster returns FAIL. In this case, we add cluster $S_i$ to $\sset^{\bad}$, and we apply the algorithm from \Cref{cor: simple guiding paths} to graph $G$ and cluster $S_i$. Let $\dset(S_i)$ be the distribution over the set $\Lambda_G(S_i)$ of internal $S_i$-routers that the algorithm returns. We then let $\qset(S_i)$ be an internal $S_i$-router sampled from the distribution $\dset(S_i)$. From \Cref{cor: simple guiding paths}, for each edge $e\in E(S_i)$, $\expect{\cong(\qset(S_i),e)}\leq  O(\log^4m/\alpha^*)=O(\log^{16}m)$.

\paragraph{Bad Event $\event$.} For an index $1\le i\le r$, we say that  bad event $\event_i$ happens, if $S_i$ is not a $\hat \eta$-bad cluster, but Algorithm \algclassifycluster returned  FAIL when applied to it.
From \Cref{thm:algclassifycluster}, $\Pr[\event_i]\le 1/m^{100}$. We also let $\event$  be the bad event that event $\event_i$ happens for any index  $1\le i\le r$. From the union bound, $\Pr[\event]\le 1/m^{99}$.

Consider again any index $1\leq i\leq r$. We will now slightly modify the paths in set $\qset(S_i)$, to ensure that they are non-traversal with respect to the rotation system $\Sigma$. In order to do so, we start by subdividing every edge $e\in \delta_G(S_i)$ with a vertex $t(e)$, and we let $X_i=\set{t(e)\mid e\in \delta_G(S_i)}$ be the resulting set of new vertices. We truncate the paths of $\qset(S_i)$, so that each such path originates at a distinct vertex of $X_i$ and terminates at vertex $u_i$. We then let $Y_i$ be the multiset of vertices containing the last vertex on every path of $\qset(S_i)$ (so $Y_i$ contains $|\qset(S_i)|$ copies of vertex $u_i$). We apply the algorithm from  \Cref{lem: non_interfering_paths} to the resulting instance of \CNwRS, set $\qset(S_i)$ of paths and vertex multisets $X_i,Y_i$, to obtain another set  $\tilde \qset(S_i)$ of paths that are non-transversal with respect to $\Sigma$.  Recall that for every edge $e\in E(G)$, $\cong_G(\tilde\qset(S_i),e)\leq \cong_G(\qset(S_i),e)$, so in particular all inner vertices of the paths in $\tilde \qset(S_i)$ lie in $S_i$. 
The path set $\tilde \qset(S_i)$ naturally defines a set of paths (internal $S_i$-router) in graph $G$,
that route the edges of $\delta_G(S_i)$ to vertex $u_i$, and  are non-transversal with respect to $\Sigma$.
For convenience of notation, we denote this internal $S_i$-router by $\qset(S_i)$ from now on. 
The following observation follows immediately from the above discussion, \Cref{thm:algclassifycluster} and the definition of $\beta^*$-light and $\hat \eta$-bad clusters  (see \Cref{def: light cluster} and \Cref{def: bad cluster} in \Cref{subsec: guiding paths rotations}), and from the fact that $\beta^*\leq \hat \eta$.


\begin{observation}
\label{obs: congestion square of internal routers}
For every cluster  $S_i\in \sset^{\light}$, for every edge $e\in E(S_i)$:
$\expect{\left (\cong_{G}(\qset(S_i),e)\right )^2}\le \hat \eta$.
%
%
Additionally, for every cluster $S_i\in \sset^{\bad}$, for every edge $e\in E(S_i)$:
$\expect{\cong_G(\qset(S_i),e)}\leq O(\log^{16}m)$. Moreover, if $\event$ did not happen, then every cluster $S_i\in \sset^{\bad}$ is $\hat \eta$-bad, that is: $$\optcrors(S_i,\Sigma(S_i))+|E(S_i)|\geq \frac{|\delta_G(S_i)|^2}{\hat \eta},$$ where $\Sigma(S_i)$ is the rotation system for graph $S_i$ induced by $\Sigma$. Lastly, $\prob{\event}\leq 1/m^{99}$.
\end{observation}

\subsubsection*{Internal and External Routers for Clusters of $\Lambda$}

Fix an index $1\leq i\leq r$. We now define an internal router $\qset(U_i)$ and an external router $\qset'(U_i)$ for cluster $U_i\in \Lambda$. We will ensure that all paths of $\qset(U_i)$ terminate at the center vertex $u_i$ of $S_i$, and all paths of $ \qset'(U_i)$ terminate at the center vertex $u_{i+1}$ of $S_{i+1}$.
In order to do so, we consider the edges $e\in \delta_G(U_i)$ one by one. For each such edge $e$, we define a path $Q(e)$, whose first edge is $e$, last vertex is $u_i$, and all inner vertices lie in $U_i$, and we define a path $Q'(e)$, whose first edge is $e$, last vertex is $u_{i+1}$, and all inner vertices lie in $U_{i+1}$. We will then set $\qset(U_i)=\set{Q(e)\mid e\in \delta(U_i)}$, and  $\qset'(U_i)=\set{Q'(e)\mid e\in \delta(U_i)}$.

We now fix an edge $e\in \delta_G(U_i)$, and define the two paths $Q(e),Q'(e)$. Recall that $\delta_G(U_i)=E_i\cup \hat E_i$. Assume first that $e\in E_i$. In this case, $e\in \delta_G(S_i)$ and $e\in \delta_G(S_{i+1})$ must hold. We let $Q(e)$ be the unique path of the internal $S_i$-router $\qset(S_i)$ whose first edge is $e$, and we let $Q'(e)$ be the unique path of the internal $S_{i+1}$-router $\qset(S_{i+1})$, whose first edge is $e$. As required, path $Q(e)$ connects $e$ to $u_{i}$ and only contains vertices of $U_i$ as inner vertices, while path $Q'(e)$ connects $e$ to $u_{i+1}$, and only contains vertices of $\notU_{i}$ as inner vertices.

\begin{figure}[h]
	\centering
	\includegraphics[scale=0.12]{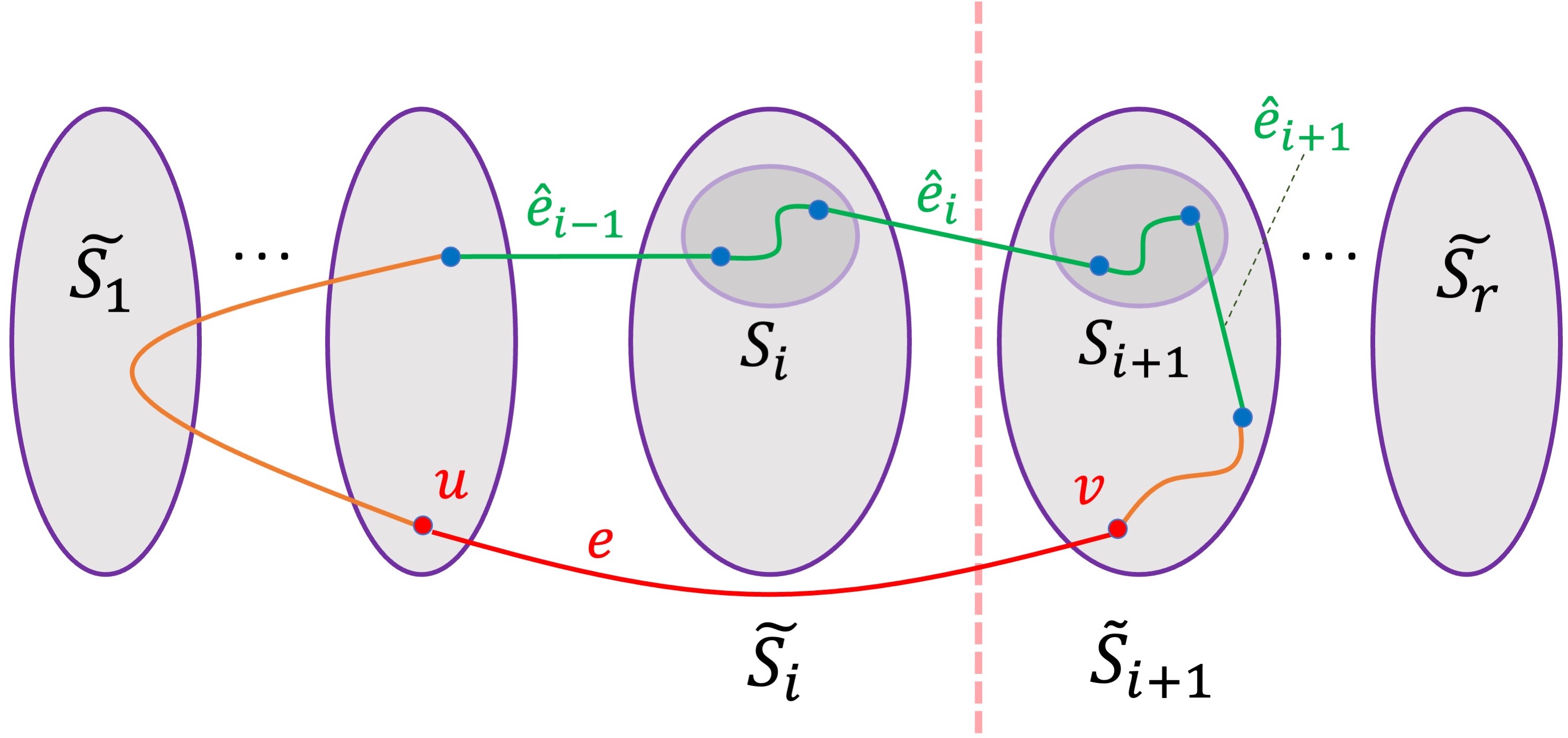}
	\caption{Construction of paths $Q(e)$ and $Q'(e)$ for an edge $e\in \hat E_i$. Path $\rho(e)$ is shown in green, and the cut $(U_i,V\setminus U_i)$ is shown in a pink dashed line. 
	}\label{fig: NN3a}
\end{figure}

\begin{figure}[h]
	\centering
	\includegraphics[scale=0.12]{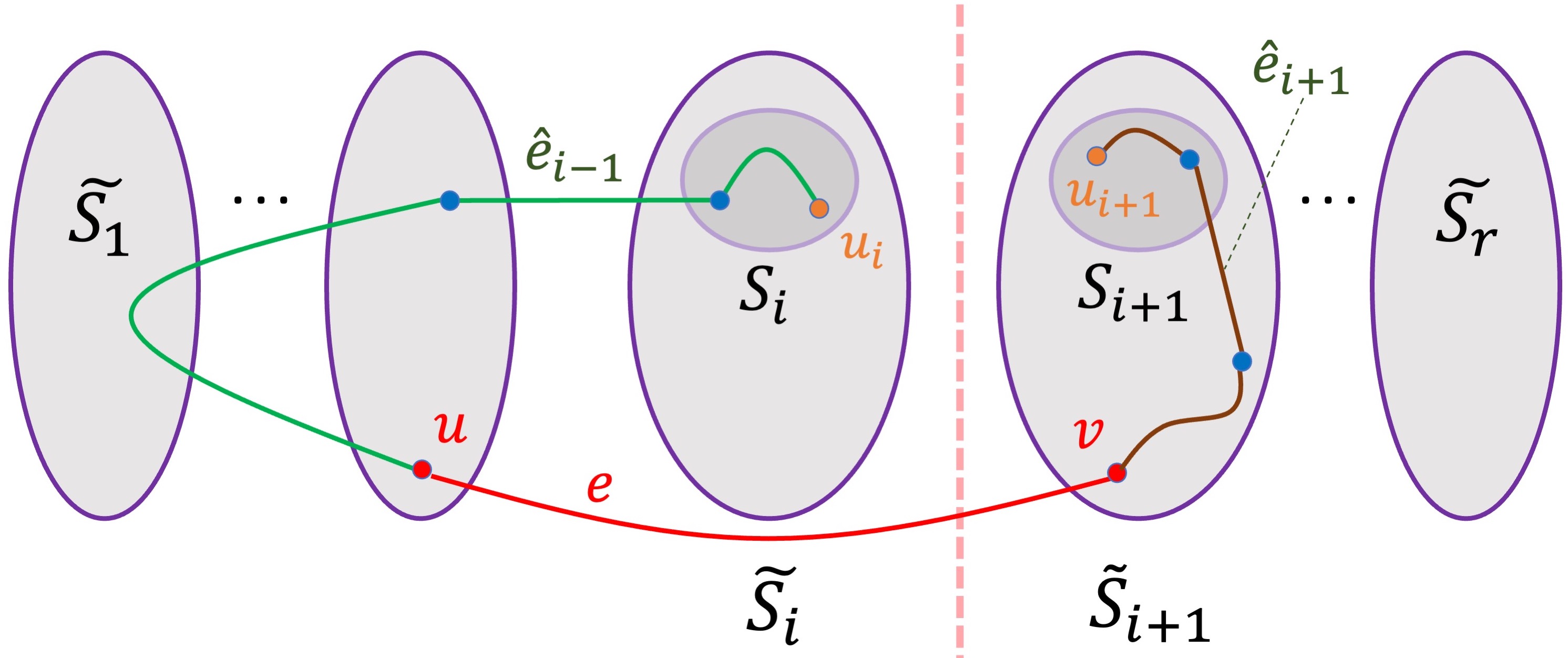}
	\caption{Paths $Q(e)$ is the union of edge $e$ and the green path; path $Q'(e)$ is the union of edge $e$ and the brown path.}\label{fig: NN3b}
\end{figure}

Assume now that $e\in \hat E_i$. We denote $e=(u,v)$, and we assume that $u$ is the left endpoint of $e$. 
	Since $e\in \hat E_i$, $i\in \spann(e)$, and, since $\spann(e)\subseteq \spann''(e)$, we get that $i\in \spann''(e)$. From the construction of the path $P^{\inn}(e)$, it must contain an edge $\hat e_i\in E_i$. Additionally, it must contain some edge $\hat e_{i-1}\in \delta_G(S_i)\setminus\set{\hat e_i}$ and some edge $\hat e_{i+1}\in \delta_G(S_{i+1})\setminus\set{\hat e_i}$ (if $e\in \delta_G(S_i)$, then $\hat e_{i-1}=e$, and if $e\in \delta_G(S_{i+1})$, then $\hat e_{i+1}=e$); see \Cref{fig: NN3a}.
	Let $\rho(e)\subseteq P^{\inn}(e)$ be the subpath of $P^{\inn}(e)$ that starts at edge $\hat e_{i-1}$ and terminates at edge $\hat e_{i+1}$. Consider now the graph that is obtained from the auxiliary cycle $W(e)$ by deleting the edge $e$, and all edges of $\rho(e)$ excluding $\hat e_{i-1}$ and $\hat e_{i+1}$. Once we delete all isolated vertices in the resulting graph, we obtain two contiguous paths. The first path, that we denote by $P$, originates at $u$ and has edge $\hat e_{i-1}\in \delta_G(S_i)$ as its last edge; all edges and vertices of $P$ lie in $U_i$ (if $e=\hat e_{i-1}$, then $P=\set{u}$). The second path, that we denote by $P'$, originates at $v$ and has $\hat e_{i+1}\in \delta_G(S_{i+1})$ as its last edge; all edges and vertices of $P'$ lie in $\notU_i$ (if $e=\hat e_{i+1}$, then $P'=\set{v}$). We let $Q(e)$ be the path obtained by concatenating the edge $e$ with the path $P$, and the unique path of the internal $S_i$-router $\qset(S_i)$ that originates at edge $\hat e_{i-1}$ (see \Cref{fig: NN3b}). Clearly, path $Q(e)$ has $e$ as its first edge, $u_i$ as its last vertex, and all its inner vertices lie in $U_i$. Similarly, we let $Q'(e)$ be the path obtained by concatenating the edge $e$ with the path $P'$, and the unique path of the internal $S_{i+1}$-router $\qset(S_{i+1})$ that originates at edge $\hat e_{i+1}$. Clearly, path $Q'(e)$ has $e$ as its first edge, $u_{i+1}$ as its last vertex, and all its inner vertices lie in $\notU_i$.
	
Once every edge of $\delta_G(U_i)$ is processed, we set  $\qset(U_i)=\set{Q(e)\mid e\in \delta(U_i)}$ and  $\qset'(U_i)=\set{Q'(e)\mid e\in \delta(U_i)}$. It is immediate to verify that $\qset(U_i)$ is an internal $U_i$-router, while $\qset'(U_i)$ is an external $U_i$-router. Note that the construction of both routers is randomized, and the only randomized component in the construction is the selection of the internal routers for the vertebrae.
We need the following simple observation.

\begin{observation}\label{obs: inner non-transversal}
	For all $1\leq i<r$, the set $\qset(U_i)$ of paths is non-transversal with respect to $\Sigma$. Additionally, for every pair $Q'(e),Q'(e')\in \qset'(U_i)$ of paths, there is at most one vertex $v$, such that  $Q'(e)$ and $Q'(e')$ have a transversal intersection at $v$. If such a vertex $v$ exists, then $v$ is the unique vertex, such that the auxiliary cycles $W(e),W(e')$ have a transversal intersection at $v$.
\end{observation}

\begin{proof}
We start by considering a pair of paths $Q(e),Q(e')\in \qset(U_i)$. Let $v$ be any vertex that serves as an inner vertex on both paths. We consider three cases. First, if $v\in V''$, then $e,e'\in \hat E_i$ must hold, and, from \Cref{obs: auxiliary cycles non-transversal at at most one}, the intersection of $Q(e)$ and $Q(e')$ at $v$ is non-transversal. The second case is when $v\in V(S_i)$. In this case, there must be two paths $\tilde Q_1,\tilde Q_2\in \qset(S_i)$, such that $\tilde Q_1\subseteq Q(e)$, $\tilde Q_2\subseteq Q(e')$, and $v$ is an inner vertex on both $\tilde Q_1$ and $\tilde Q_2$ (note that it is  possible that $\tilde Q_1=\tilde Q_2$). In this case, from the construction of the internal $S_i$-router $\qset(S_i)$, the intersection of $Q(e)$ and $Q(e')$ at $v$ is non-transversal. The third case is when $v\in V'\setminus V(S_{i})$. In this case, $e,e'\in \hat E_i$ must hold. Assume that $v\in V(S_z)$ for some index $z<i$. Clearly, $z$ may not be the last index of $\spann''(e)$ or of $\spann''(e')$. Therefore, from \Cref{obs: auxiliary cycles non-transversal at at most one}, the intersection of $Q(e)$ and of $Q(e')$ at $v$ is non-transversal. We conclude that  the set $\qset(U_i)$ of paths is non-transversal with respect to $\Sigma$.

Consider now some pair $Q'(e),Q'(e')\in \qset'(U_i)$ of paths. Let $v$ be any vertex that serves as an inner vertex on both paths. We again consider three cases. First, if $v\in V''$, then $e,e'\in \hat E_i$ must hold, and, from \Cref{obs: auxiliary cycles non-transversal at at most one}, the intersection of $Q'(e)$ and $Q'(e')$ at $v$ is non-transversal. The second case is when $v\in V(S_{i+1})$. In this case, there must be two paths $\tilde Q_1,\tilde Q_2\in \qset(S_{i+1})$, such that $\tilde Q_1\subseteq Q'(e)$, $\tilde Q_2\subseteq Q'(e')$, and $v$ is an inner vertex on both $\tilde Q_1$ and $\tilde Q_2$. In this case, from the construction of the internal $S_{i+1}$-router $\qset(S_{i+1})$, the intersection of $Q'(v)$ and $Q'(v')$ at $v$ is non-transversal. The third case is when $v\in V'\setminus V(S_i)$. In this case, $e,e'\in \hat E_i$ must hold, and $v$ also lies on cycles $W(e)$ and $W(e')$. Moreover, the intersection of $W(e)$ and $W(e')$ must be transversal at $v$.
From \Cref{obs: auxiliary cycles non-transversal at at most one}, there may be at most one vertex $v'$, such that the intersection of $W(e)$ and $W(e')$ at $v'$ is transversal. We conclude that there is at most one vertex $v\in V(Q'(e))\cap V(Q'(e'))$, such that the intersection of $Q'(e)$ and $Q'(e')$ at $v$ is transversal.  If such a vertex $v$ exists, then $v$ is the unique vertex, such that the auxiliary cycles $W(e),W(e')$ have a transversal intersection at $v$.
\end{proof}

From \Cref{obs: bound congestion of cycles}, and the construction of the internal and external $U_i$-routers, we obtain the following immediate observation.

\begin{observation}\label{obs: bound congestion of routers}
	For all $1\leq i< r$, an edge $e\in E(S_i)$ may appear on at most 
	$O(\log^{34}m)\cdot \cong_G(\qset(S_i),e)$ paths of $\qset(U_i)$, and an edge $e\in E(U_i)\setminus E(S_i)$ may appear on at most $O(\log^{34}m)$ paths of $\qset(U_i)$. Similarly, an edge $e\in E(S_{i+1})$ may appear on at most 	$O(\log^{34}m)\cdot \cong_G(\qset(S_{i+1}),e)$ paths of $\qset'(U_i)$, and an edge $e\in E(\notU_i)\setminus E(S_{i+1})$ may appear on at most $O(\log^{34}m)$ paths of $\qset'(U_i)$.
\end{observation}

We will also use the following simple corollary of the observation.

\begin{corollary}\label{cor: few edges crossing cuts}
	For all $1\leq z\leq r$, $|\delta_G(U_{z})|\leq |\delta_G(S_{z})|\cdot O(\log^{34}m)$, and  $|\delta_G(U_{z})|\leq |\delta_G(S_{z+1})|\cdot O(\log^{34}m)$.
\end{corollary}
\begin{proof}
Recall that we have defined a collection $\qset(U_{z})$ of paths, routing the edges of $\delta_G(U_z)$ to vertex $u_{z}$. Each such path must contain an edge of $\delta_G(S_{z})$. Moreover, from \Cref{obs: bound congestion of routers}, an edge of $\delta_G(S_{z})$ may lie on at most $O(\log^{34}m)$ paths of $\qset(U_{z})$. Therefore, $|\delta_G(U_{z})|\leq |\delta_G(S_{z})|\cdot O(\log^{34}m)$.

Similarly,  we have defined a collection $\qset'(U_{z})$ of paths, routing the edges of $\delta_G(U_z)$ to vertex $u_{z+1}$. Each such path must contain an edge of $\delta_G(S_{z+1})$. From \Cref{obs: bound congestion of routers}, an edge of $\delta_G(S_{z+1})$ may lie on at most $O(\log^{34}m)$ paths of $\qset'(U_{z})$. Therefore, $|\delta_G(U_{z})|\leq |\delta_G(S_{z+1})|\cdot O(\log^{34}m)$.
\end{proof}

\subsubsection{Step 4: Constructing the Collection of Subinstances}
\label{subsec: instances}

Consider again an index $1\leq i<r$, and the internal $U_i$-router $\qset(U_i)=\set{Q(e)\mid e\in \delta_G(U_i)}$. Recall that all paths in $\qset(U_i)$ terminate at vertex $u_i$.
Denote $\delta_G(u_i)=\set{e_1^i,\ldots,e_{|\delta(u_i)|}^i}$, where the edges are indexed according to their order in $\oset_{u_i}\in \Sigma$. For all $1\leq j\leq |\delta(u_i)|$, let $A_j^i\subseteq \delta_G(U_i)$ be the set of all edges $e\in \delta_G(U_i)$, such that the uniue path $Q(e)\in \qset(U_i)$ that has $e$ as its first edge contains edge $e^i_j$ as its last edge. We now define an ordering $\tilde \oset_i$ of the edges of $\delta_G(U_i)$: the edges in sets $A_1^i,A_2^i,\ldots,A_{|\delta(u_i)|}^i$ appear in the order of the indices of their sets, and, for each $1\leq j\leq |\delta(u_i)|$, the ordering of the edges in set $A_j^i$ is arbitrary. Notice that the resulting ordering $\tilde \oset_i$ of the edges of $\delta_G(U_i)$ is precisely the ordering $\oset^{\guided}(\qset(U_i),\Sigma)$, that is guided by the internal $U_i$-router $\qset(U_i)$ (see the definition in \Cref{subsec: guiding paths rotations}). For $i=r$, $\delta_G(U_i)=\emptyset$, and the ordering $\tilde \oset_r$  of the edges of $\delta_G(U_i)$ is the trivial one.

We let $\iset_2$ be a collection of subinstances of $I$ obtained by computing  a laminar family-based decomposition of $I$ (defined in \Cref{subsec: laminar-based decomposition}) via the laminar family $\lset=\set{U_i}_{1\leq i\leq r}$ of clusters, and the orderings $\tilde \oset_i$ of the edge sets $\delta(U_i)$ for all $1\leq i\leq r$. We denote $\iset_2=\set{I_1,\ldots,I_r}$, where, for $1\leq z\leq r$, instance $I_z=(G_z,\Sigma_z)$ is the instance associated with the cluster $U_z$. Recall that instance $I_z$ is defined as follows. Assume first that $1<z<r$. Then graph $G_z$ is obtained from graph $G$ by first contracting all vertices of $U_{z-1}$ into vertex $v^*_z$, and then contracting  all vertices of $U_{z+1}$ into vertex $v^{**}_z$. 
Notice that, equivalently, graph $G_z$ consists of the cluster $\tilde S_z\in \tilde \sset$, the two special vertices $v^*_z,v^{**}_z$, and possibly some additional edges that are incident to these two special vertices.

The rotation system $\Sigma_z$ is defined as follows. Observe first that, for every vertex $v\in V(G_z)\setminus\set{v^*_z,v^{**}_z}$, $\delta_{G_z}(v)=\delta_G(v)$. The ordering $\oset_v\in \Sigma_z$ of the edges incident to $v$ in $\Sigma_z$ remains the same as in $\Sigma$. Notice that $\delta_{G_z}(v^*_z)=\delta_G(U_{z-1})$. We let the ordering $\oset_{v^*_z}\in \Sigma_z$ of the edges of $\delta_{G_z}(v^*_z)$ be the ordering $\tilde \oset_{z-1}$ that we defined above. Lastly, observe that $\delta_{G_z}(v^{**}_z)=\delta_G(U_{z})$. We let the ordering $\oset_{v^{**}_z}\in \Sigma_z$ of the edges of $\delta_{G_z}(v^{**}_z)$ be the ordering $\tilde \oset_{z}$ that we defined above.
For $z=1$, instance $I_1=(G_1,\Sigma_1)$ is defined similarly, except that the instance does not contain vertex $v^*_1$. Instance $I_r=(G_r,\Sigma_r)$ is also defined similarly, except that it does not contain vertex $v^{**}_r$.

We now verify that the resulting collection $\iset_2$ of instances has all required properties. Fix some index $1\leq z\leq r$. Recall that, from the definition of a nice witness structure, there is at most one cluster $C\in \cset$ with $C\subseteq \tilde S_z$.  Recall also that, for each cluster $C\in \cset$, there is exactly one cluster $S_i\in \sset$ that contains $C$. If some cluster $C\in \cset$ is contained in $\tilde S_z$, then  $E(\tilde S_z)\subseteq E(C)\cup E(G_{|\cset})$ must hold, and so $E(G_z)\subseteq E(C)\cup E(G_{|\cset})$ as well. Otherwise, $E(\tilde S_z)\subseteq E(G_{|\cset})$, and so $E(G_z)\subseteq E(G_{|\cset})$.

From \Cref{lem: basic disengagement combining solutions},
	there is an efficient algorithm, that, given, for each instance $I_z\in \iset$, a solution $\phi_z$, computes a solution for instance $I$ of value at most $\sum_{t=1}^r\cro(\phi_z)$.
Next, we bound the total number of edges in all resulting instances.

\begin{observation}
\label{obs: disengaged instance size}
$\sum_{1\le z\le r}|E(G_z)|\le O(|E(G)|\cdot \log^{34}m)$.
\end{observation}
\begin{proof}
	Fix an index $1\leq z\leq r$. From our construction, $|E(G_z)|\leq |E(\tilde S_z)|+|\delta_G(U_{z-1})|+|\delta_G(U_z)|$. 
	From \Cref{cor: few edges crossing cuts}, $|\delta_G(U_{z-1})|\leq |\delta_G(S_{z-1})|\cdot O(\log^{34}m)$, and $|\delta_G(U_z)|\leq |\delta_G(S_{z+1})|\cdot O(\log^{34}m)$.
Therefore, $|E(G_z)|\leq |E(\tilde S_z)|+(|\delta_G(S_{z-1})|+|\delta_G(S_{z+1})|)\cdot O(\log^{34}m)$. Summing up over all indices $z$, we get that:

	\[\sum_{z=1}^r|E(G_z)|\leq \sum_{z=1}^r|E(\tilde S_z)|+ O(\log^{34}m)\cdot \sum_{z=1}^r|\delta_G(S_z)|\leq O(|E(G)|\cdot \log^{34}m). \]
\end{proof}

Recall that we have used a randomized algorithm to compute, for all $1\leq i\leq r$, an internal $U_i$-router $\qset(U_i)$ and an external $U_i$-router $\qset'(U_i)$. 
In order to complete the proof of \Cref{thm: advanced disengagement - disengage nice instances}, it is now enough to show that the expected total optimal solution costs of all instances in $\iset_2$ (over the random choices performed by the algorithm that computed the internal and the external $U_i$-routers) is bounded by $2^{O((\log m)^{3/4}\log\log m)}\cdot (\optcrors(I)+|E(G)|)$. The following claim, whose proof appears in \Cref{subsec: bound opt costs} will finish the proof of \Cref{thm: advanced disengagement - disengage nice instances}.

\begin{claim}
\label{claim: existence of good solutions special}
$\expect{\sum_{z=1}^r\optcrors(I_z)}\leq  2^{O((\log m)^{3/4}\log\log m)}\cdot (\optcrors(I)+|E(G)|)$.
\end{claim}

\subsection{Proof of \Cref{claim: existence of good solutions special}}
\label{subsec: bound opt costs}

Notice that  graphs $G_1,G_r$ have a somewhat different structure than graphs of  $\set{G_z}_{1<z<r}$: specifically, graph $G_1$ does not contain vertex $v^*_1$, and graph $G_r$ does not contain vertex $v^{**}_r$. It would be convenient for us to modify these two graphs so that we can treat all resulting graphs uniformly. In order to do so, we add a new dummy vertex $v^*_1$ to graph $G_1$, and connect it with an edge to an arbitrary vertex $v_1\in S_1$. We modify the rotation $\oset_{v_1}\in \Sigma_1$ to include the new edge $(v_1,v^*_1)$ at an arbitrary position in the rotation. Notice that any solution to the resulting new instance $I_1=(G_1,\Sigma_1)$ of \cnwrs immediately provides a solution to the original instance $I_1=(G_1,\Sigma_1)$, of the same cost. We similarly modify graph $G_r$, by adding a new dummy vertex $v^{**}_r$, which is connected with an edge to an arbitrary vertex $v_r\in S_r$. We modify the rotation $\oset_{v_r}\in \Sigma_r$ as before. In order to be consistent, we also add the vertices $v^*_1,v^{**}_r$, and edges $(v^*_1,v_1)$ and $(v^{**}_r,v_r)$ to  the original graph $G$, and modify the rotations $\oset_{v_1},\oset_{v_r}\in \oset$, so that they remain consistent with the rotations  $\oset_{v_1}\in \Sigma_1$ and  $\oset_{v_r}\in \Sigma_r$, respectively. Notice that this modification does not increase $\optcrors(I)$.
We also modify the nice witness structure, by adding two new clusters $\tilde S_0=\set{v^*_1}$ and $\tilde S_{r+1}=\set{v^{**}_r}$ to $\tilde \sset$, and their subclusters $S_0=\set{v^*_1}$ and $ S_{r+1}=\set{v^{**}_{r}}$. We note that all the above modifications are only performed for ease of exposition and are not strictly necessary.

Let $\phi^*$ be an optimal solution to instance $I$ of \cnwrs. We can assume that no pair of edges cross twice, and that the image of each edge does not cross itself in $\phi^*$. 
We denote by $\chi^*$ the set of all unordered pairs $(e,e')$ of edges of $G$, such that the images of $e$ and $e'$ cross in $\phi^*$.
For all $1\leq z\leq r$, we denote by $\chi^*_z\subseteq \chi^*$ the set of all unordered pairs $(e,e')$ of edges of $G$ whose images cross, such that either $e\in E(\tilde S_z)\cup \delta_G(\tilde S_z)$, or $e'\in E(\tilde S_z)\cup \delta_G(\tilde S_z)$, or both. We will use the drawing $\phi^*$ in order to construct, for each $1\leq z\leq r$, a solution $\phi_z$ to instance $I_z=(G_z,\Sigma_z)$

For each $1\leq z\leq r$, we construct a solution $\phi_z$ to instance $I_z$, and then argue that the total expected costs of all these solutions is relatively small. We now fix an index $1\leq z\leq r$, and focus on constructing solution $\phi_z$ to instance $I_z$.
The construction of the solution consists of four steps. In the first step, we construct an auxiliary graph $H_z$ and its drawing $\psi_z$. This graph and its drawing are used in the second step, in order to construct an initial drawing $\phi'_z$ of graph $G_z$. The number of crossings in drawing $\phi'_z$ may be quite large, and we modify the drawing in order to lower the number of crossings in the third step. The resulting drawing, $\phi''_z$, will have a sufficiently low expected number of crossings, but unfortunately it may not obey the rotation $\oset_{v^{**}_z}\in \Sigma_z$. In the fourth and the last step, we modify this drawing in order to obtain a feasible solution $\phi_z$ to instance $I_z$ of \cnwrs, while only slightly increasing the number of crossings. 
We now fix an index $1\leq z\leq r$, and describe a construction of a solution $\phi_z$ for instance $I_z$ of \cnwrs step by step.

\subsubsection{Step 1: Computing Auxiliary Graph $H_z$ and Its Drawing}

Recall that graph $G_z$ is obtained from graph $G$ by contracting all vertices of $\bigcup_{1\leq i<z}V(\tilde S_i)$ into the special vertex $v^*_z$, and then contracting all vertices of $\bigcup_{z<i\leq r}$ into the special vertex $v^{**}_z$. Clearly, every edge of $G_z$ corresponds to some edge of $G$, and we do not distinguish between these edges. In order to simplify the notation, when the index $z$ is fixed, we denote vertices $v^*_z$ and $v^{**}_z$ by $v^*$ and $v^{**}$, respectively. Notice that $\delta_{G_z}(v^*)=\delta_G(U_{z-1})=E_{z-1}\cup \hat E_{z-1}$, while $\delta_{G_z}(v^{**})=\delta_G(U_z)=E_{z}\cup \hat E_z$. In order to obtain the  drawing $\phi_z$ of graph $G_z$, we will exploit the internal $U_{z-1}$-router $\qset(U_{z-1})$, that routes the edges of $\delta_G(U_{z-1})$ to vertex $u_{z-1}$, and the external $U_z$-router $\qset'(U_z)$, that routes the edges of $\delta_G(U_z)$ to vertex $u_{z+1}$. For each edge $e\in \delta_G(U_{z-1})$, we denote by $Q(e)$ the unique path of $\qset(U_{z-1})$ whose first edge is $e$, and for each edge $e\in \delta_G(U_z)$, we denote by $Q'(e)$ the unique path of $\qset'(U_z)$ whose first edge is $e$.


Denote $\qset^*_z=\qset(U_{z-1})\cup \qset(U_z)$. For every edge $e\in E(G)$, we define a value $N_z(e)$, as follows. If $e\in E(G) \setminus E(G_z)$, then $N_z(e)=\cong_G(\qset^*_z,e)$ -- the number of paths in $\qset^*_z$ that contain the edge $e$. For each edge $e\in \delta_G(U_{z-1})\cup \delta_G(U_z)\cup E(\tilde S_z)$, we set $N_z(e)=1$.
The will use the following observation.
\begin{observation}\label{obs: bound on num of copies}
	Let $e$ be an edge of $ E(G) \setminus E(G_z)$. If $e\not\in E(S_{z-1})\cup E(S_{z+1})$, then $N_z(e)\leq O(\log^{34}m)$; otherwise, $\expect{N_z(e)}\leq \hat \eta$. Moreover, if $e\in E(S_{z-1})$ and $S_{z-1}\in \sset^{\light}$, then $\expect{(N_z(e))^2}\leq \hat \eta$. Similarly, if $e\in E(S_{z+1})$ and $S_{z+1}\in \sset^{\light}$, then $\expect{(N_z(e))^2}\leq \hat \eta^2$. (All expectations here are over the selections of the internal routers $\qset(S_{z-1})$ and $\qset(S_{z+1})$).
\end{observation}

\begin{proof}
	Consider an edge $e\in E(G)\setminus E(G_z)$. Notice that either $e\in E(U_{z-1})$, or $e\in E(\notU_z)$ must hold. We assume that it is the former; the other case is symmetric. In this case, $N_z(e)=\cong_G(\qset(U_{z-1}),e)$, and, from \Cref{obs: bound congestion of routers}, $N_z(e)\leq O(\log^{34}m)$.
		
		Assume now that $e\in E(U_{z-1})$. 
		From \Cref{obs: bound congestion of routers}, edge $e$ may appear on at most 
		$\cong_G(\qset(S_{z-1}),e)\cdot O(\log^{34}m)$ paths of $\qset(U_{z-1})$, that is, $N_z(e)\leq \cong_G(\qset(S_{z-1}),e)\cdot O(\log^{34}m)$.
		From \Cref{obs: congestion square of internal routers}, for $1\leq i\leq r$, if $S_i\in \sset^{\light}$, then
		$\expect{\left (\cong_{G}(\qset(S_i),e)\right )^2}\le \hat \eta$, while, if  $S_i\in \sset^{\bad}$, then
		$\expect{\cong(\qset(S_i),e)}\leq O(\log^{16}m)\leq \hat \eta$. The observation now follows immediately.
\end{proof}

We now construct an auxiliary graph $H_z$, and its drawing $\psi_z$. In order to do so, we start with $H_z=G$, and $\psi_z=\phi^*$. We call the edges of $E(\tilde S_z)\cup \delta_G(\tilde S_z)$ \emph{primary edges}, and the remaining edges of $G$ \emph{secondary edges}. We now  process every secondary edge $e$ one by one. If edge $e$ does not participate in any path of $\qset^*_z$ (that is, $N_z(e)=0$), then we delete $e$ from $H_z$ and we delete its image from $\psi_z$. Otherwise, we replace $e$ with  a set $J(e)$ of $N_z(e)$ parallel copies of $e$ in graph $H_z$, and we replace the image of $e$ in $\psi_z$ with images of these copies, that follow the original image of $e$ in parallel to it, without crossing each other. For convenience, for each edge $e\in E(\tilde S_z)\cup \delta_G(U_{z-1})\cup \delta_G(U_z)$, we define $J(e)=\set{e}$, and we think of the graph $H_z$ as having a single copy of the edge $e$ (the edge $e$ itself).
This completes the definition of the graph $H_z$ and its drawing $\psi_z$. 

For every edge $e\in E(G)\setminus E(G_z)$, we can now assign, to every path of $\qset^*_z$ containing $e$, a distinct copy of this edge from $J(e)$. If edge $e\not\in \delta_G(u_{z-1})$, we assign each copy of $e$ in $J(e)$ to a distinct path of $\qset^*_z$ containing $e$ arbitrarily. If edge $e\in \delta_G(u_{z-1})$, then we perform the assignment more carefully. Intuitively, this assignment is performed in a way that is consistent with the ordering $\tilde \oset_{z-1}$ of the edges of $\delta_G(U_{z-1})$ that we have defined, and the ordering of the paths of set $\qset(U_{z-1})=\set{Q(e)\mid e\in \delta_G(U_{z-1})}$  that it induces.

\paragraph{Assigning the copies of edges of $\delta_G(u_{z-1})$ to paths.}
Consider the set $\delta_G(U_{z-1})$ of edges. Recall that we have defined an ordering $\tilde \oset_{z-1}$ of the edges of $\delta_G(U_{z-1})$, which is precisely the ordering $\oset^{\guided}(\qset(U_{z-1}),\Sigma)$, that is guided by the internal $U_{z-1}$-router $\qset(U_{z-1})$. Denote $\delta_G(U_{z-1})=\set{\hat a_1,\hat a_2,\ldots,\hat a_q}$, where the edges are indexed according to the ordering $\tilde \oset_{z-1}$. 

 Recall the procedure that we used in order to define the ordering $\tilde \oset_{z-1}$ of the edges of $\delta_G(U_{z-1})$ (for convenience we omit the superscript $z-1$): we have denoted $\delta_G(u_{z-1})=\set{e_1,\ldots,e_{|\delta(u_{z-1})|}}$, where the edges are indexed according to their order in the rotation $\oset_{u_{z-1}}\in \Sigma$. For all $1\leq j\leq |\delta(u_{z-1})|$, we denoted by $A_j\subseteq \delta_G(U_{z-1})$ the set of all edges $e'\in \delta_G(U_{z-1})$, such that the unique path $Q(e')\in \qset(U_{z-1})$ originating at edge $e'$ terminates at edge $e_j$.  

We have defined the ordering $\tilde \oset_{z-1}=(\hat a_1,\hat a_2,\ldots,\hat a_q)  $ of the edges of $\delta_G(U_{z-1})$ as follows: the edges that lie in sets $A_1,A_2,\ldots,A_{|\delta(u_{z-1})|}$ appear in the order of the indices of their sets, and, for each $1\leq j\leq |\delta(u_{z-1})|$, the ordering of the edges within each set $A_j$ is arbitrary; 
denote this latter ordering by $\hat \oset_j=\set{a^{j}_1,a^{j}_2,\ldots,a^{j}_{q_j}}$. 
The current drawing $\psi_z$ of graph $H_z$ naturally defines a circular ordering of the edges of $\delta_H(u_{z-1})$, which is precisely the order in which the images these edges enter the image of $u_{z-1}$. In this circular ordering, the edges of each set $J(e_1),J(e_2),\ldots,J(e_{|\delta_G(u_{z-1})|})$ appear consecutively, in the order of the indices of their sets. For each index $1\leq j\leq  |\delta_G(u_{z-1})|$, the above circular ordering defines an ordering $\hat \oset'_j$ of the edges of $J(e_j)$. 

Consider now some edge $e_j\in \delta_G(u_{z-1})$, and assume that $|J(e_j)|=q_j$. On the one hand, we have defined the  ordering $\hat \oset'_j$ of the edges of $J(e_j)$ -- the order in which the images of these edges in $\psi_z$ enter the image of $u_{z-1}$. On the other hand, we have defined an ordering   $\hat  \oset_j=\set{a^{j}_1,a^{j}_2,\ldots,a^{j}_{q_j}}$ of the edges of $A_j$ -- that is, the edges $e'\in \delta_G(u_{z-1})$, whose corresponding path $Q(e')$ contains edge $e_j$. For all $1\leq h\leq q_j$, we then assign the $h$th edge of $J(e_j)$  in the ordering $\hat \oset'_j$ to path $Q(a^{j}_{h})$. This completes the assignment of edges of $H_z$ that are incident to vertex $u_{z-1}$ to the paths of $\qset(U_{z-1})$. 

For every edge $\hat a_i\in \delta_G(U_{z-1})$, we can now obtain a path $\hat Q(\hat a_i)$ in graph $H_z$, that originates at edge $\hat a_i$ and terminates at vertex $u_{z-1}$, with all inner vertices of $\hat Q(a_i)$ lying in $V(U_{z-1})$, by starting from the path $Q(\hat a_i)\in \qset(U_{z-1})$, and replacing every edge $e'\in E(G)\setminus E(G_z)$ with the copy of $e'$ that is assigned to path $Q(\hat a_i)$. Denote the resulting set of paths in graph $H_z$ by $\hat \qset_z=\set{\hat Q(\hat a_i)\mid \hat a_i\in \delta_G(U_{z-1})}$. For each edge $\hat a_i\in \delta_G(U_{z-1})$, denote by $\hat a'_i$ the last edge on path $\hat Q(\hat a_i)$. Then the paths of $\hat \qset_z$ are mutually edge-disjoint, and they route the edges of $\delta_{G}(U_{z-1})$ to vertex $u_{i-1}$ in $H_z$.
All inner vertices on the paths of $\hat \qset_z$ lie in $V(U_{z-1})$.
 Moreover, the images of edges $\hat a'_1,\ldots,\hat a'_q$ enter the image of $u_{i-1}$ in the drawing $\psi_z$ of $H_z$ in the circular order of their indices (and recall that edges $\hat a_1,\ldots,\hat a_q$ are indexed in the order of their appearance in $\tilde \oset_{z-1}$).

Similarly, for every edge $a\in \delta_G(U_z)$, we can now obtain a path $\hat Q'(a)$ in graph $H_z$, that originates at edge $a$ and terminates at vertex $u_{z+1}$, with all inner vertices of $\hat Q'(a)$ lying in $V(\notU_{z})$, by starting from the path $Q'( a)\in \qset'(U_{z})$, and replacing every edge $e'\in E(G)\setminus E(G_z)$ with the copy of $e'$ that is assigned to path $Q(a)$. Denote the resulting set of paths in graph $H_z$ by $\hat \qset'_z=\set{\hat Q'(a)\mid a\in \delta_G(U_{z})}$. 
 Then the paths of $\hat \qset'_z$ are mutually edge-disjoint, and they route the edges of $\delta_{G}(U_{z})$ to vertex $u_{i+1}$. All inner vertices on paths of  $\hat \qset'_z$ lie in $V(\notU_z)$. 
 This completes the construction of graph $H_z$ and its drawing $\psi_z$. 
 We now analyze the number of crossings in this graph.

\paragraph{Bounding the Number of Crossings in $\psi_z$.}

\begin{figure}[h]
	\centering
	\includegraphics[scale=0.12]{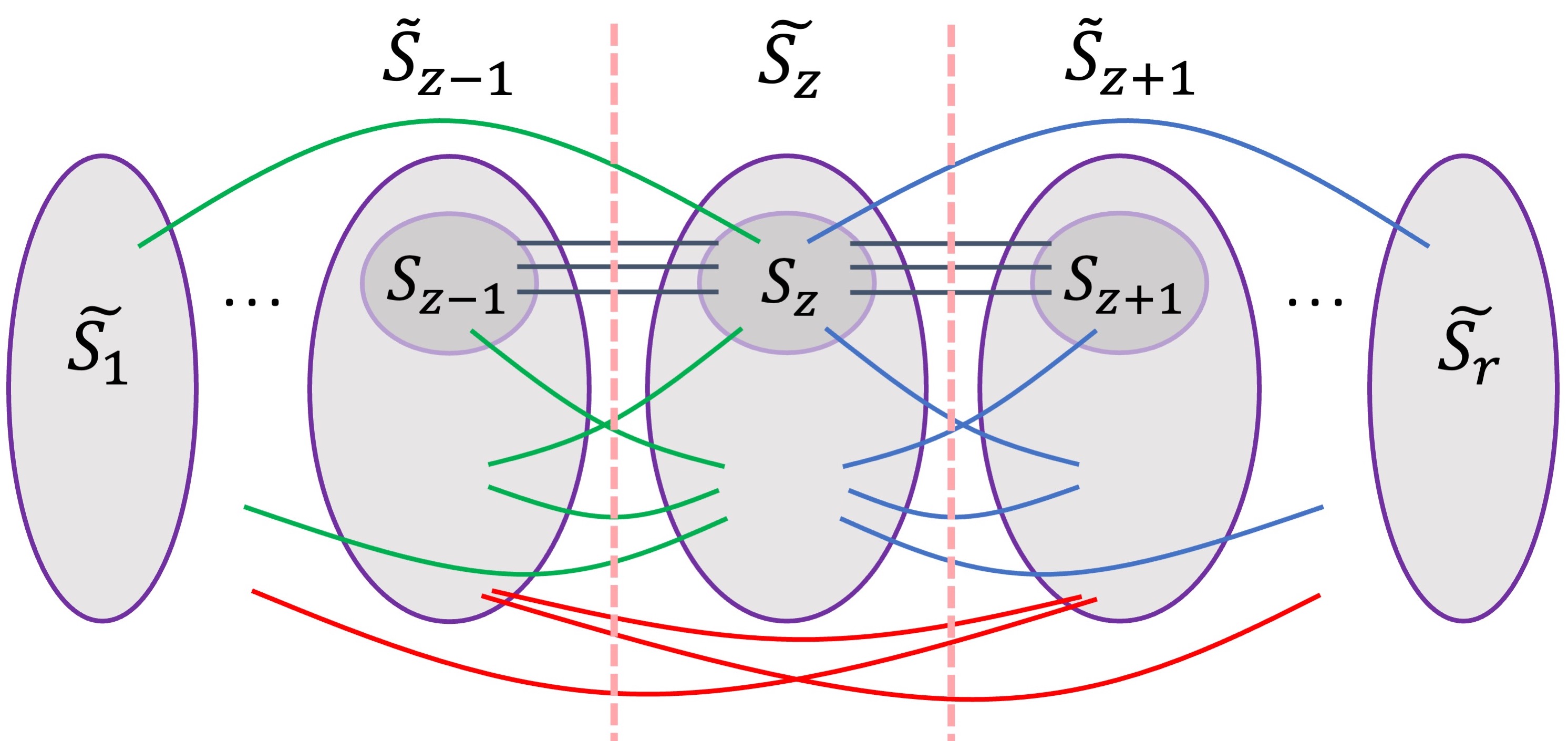}
	\caption{Set $E^{\through}_z$ of edges is shown in red, set $E^{\rig}_z$ in blue, and set $E^{\lef}_z$ in green. The left pink dashed line shows the cut $(U_{z-1}, V\setminus U_{z-1})$, and the right pink dashed line shows the cut $(U_{z}, V\setminus U_{z})$.}\label{fig: NN1}
\end{figure}

Recall that $\delta_G(U_{z-1})=E_{z-1}\cup \hat E_{z-1}$, while $\delta_G(U_{z})=E_{z}\cup \hat E_{z}$.
We denote $E_z^{\through}=\hat E_{z-1}\cap \hat E_z$; note that every edge $e\in E_z^{\through}$ has one endpoint in $\bigcup_{1\leq i<z}V(\tilde S_i)$, and another endpoint in  $\bigcup_{z< i\leq r}V(\tilde S_i)$ (see \Cref{fig: NN1}).
We also denote by $E_z^{\lef}=\hat E_{z-1}\setminus E_z^{\through}$, and by $E_z^{\rig}=\hat E_z\setminus E_z^{\through}$. Notice that every edge $e\in E_z^{\lef}$ has one endpoint in $\bigcup_{1\leq i<z}V(\tilde S_i)$, and another endpoint in  $V(\tilde S_z)$, while every edge $e\in E_z^{\rig}$ has one endpoint in $V(\tilde S_z)$ and another endpoint in $\bigcup_{z< i\leq r}V(\tilde S_i)$ (see \Cref{fig: NN1}). 
From the above definitions, $\delta_{G_z}(v^*)=\delta_G(U_{z-1})=E_{z-1}\cup E_z^{\lef}\cup E_z^{\through}$, and $\delta_{G_z}(v^{**})=\delta_G(U_{z})=E_{z}\cup E_z^{\rig}\cup E_z^{\through}$.

We now reorganize the paths in $\hat \qset_z\cup \hat \qset'_z$ as follows. We let $\rset_1'=\set{\hat Q(e)\mid e\in E_{z-1}\cup E_z^{\lef}}$, $\rset_1''=\set{\hat Q'(e)\mid e\in E_{z}\cup E_z^{\rig}}$, and $\rset_1^{(z)}=\rset_1'\cup \rset_1''$. For each edge $e\in E_{z-1}\cup E_z^{\lef}\cup E_z^{\rig}\cup E_z$, we denote by $R(e)\in \rset^{(z)}_1$ the unique path that has edge $e$ as its first edge. For every edge $e\in E_z^{\through}$, we let $R(e)$ be the concatenation of the paths $\hat Q(e)\in \hat \qset_z$ and $\hat Q'(e)\in \hat \qset'_z$, so that path $R(e)$ is a simple path connecting vertices $u_{z-1}$ and $u_{z+1}$, and it contains the edge $e$. We then set $\rset^{(z)}_2=\set{R(e)\mid e\in E_z^{\through}}$.

For every secondary edge $e'$ in graph $G$, we denote by $N'_z(e')$ the number of paths in set $\rset_1^{(z)}$ that contain a copy of $e'$, and we denote by $N''_z(e')$ the number of paths in set $\rset_2^{(z)}$ that contain a copy of $e'$. Note that, equivalently, $N'_z(e')$ is the total number of paths in $\qset(U_{z-1})\cup \qset'(U_z)$ that originate at edges  of $E_{z-1}\cup E_z^{\lef}\cup E_z^{\rig}\cup E_z$ and contain $e'$, while $N''_z(e')$ is the total number of  paths in $\qset(U_{z-1})\cup \qset'(U_z)$ that originate at edges  of $E_{z}^{\through}$ and contain $e'$. For a primary edge $e'$, we set $N'_z(e')=1$ and $N''_z(e')=0$. Clearly, $N_z(e')=N'_z(e')+N''_z(e')$ holds for every edge $e'$. Intuitively, for each edge $e'$, the value $\sum_{z=1}^rN'_z(e')$ is relatively small, while the value $\sum_{z=1}^rN''_z(e')$ may be quite large. Indeed, 
recall that  the paths in set $\qset(U_{z-1})\cup \qset'(U_z)$ can be thought of as constructed by composing subpaths of cycles of $\set{W(e')\mid e'\in \hat E_{z-1}\cup \hat E_z}$ with the internal routers $\qset(S_{z-1})$ and $\qset(S_{z+1})$.
Consider an edge $e\in \hat E$, and the corresponding cycle $W(e)$. Assume that $\spann(e)=\set{i,\ldots,j-1}$. Then there is only one index $z$ for which $e\in E^{\lef}_z$ -- index $z=j$. Similarly, there is only one index $z$ for which $e\in E^{\rig}(z)$ -- index $z=i$. Therefore,  
cycle $W(e)$ contributes its subpath to set $\rset_1^{(z)}$ only for indices $z=i$ and $z=j$. On the other hand, cycle $W(e)$ may contribute a subpath to set $\rset_2^{(z)}$ for every index $i<z<j$. 
Because of this, we will try to bound the number of crossings in the final drawing $\phi_z$ that we construct for instance $I_z$ in terms of the values $\set{N'_z(e')}_{e'\in E(G)}$. For convenience, when the index $z$ is fixed, we omit the superscript $(z)$ in the notation $\rset_1^{(z)}$ and $\rset_2^{(z)}$.

Notice that, from our assumption about drawing $\phi^*$, no pair of edges in drawing $\psi_z$ of $H_z$ may cross more than once, and no edge has its image cross itself. Consider any crossing $(e_1,e_2)$ in drawing $\psi_z$. Assume that $e_1$ is a copy of edge $e_1'\in E(G)$, and that $e_2$ is a copy of edge $e_2'\in E(G)$. Then the images of edges $e_1',e_2'$ must cross in $\phi^*$, and we say that 
crossing $(e'_1,e_2')$ in $\phi^*$ is \emph{responsible} for crossing $(e_1,e_2)$ in $\psi_z$.


We classify the crossings in drawing $\psi_z$ into several types, and we bound the number of crossings of each of these types separately.
Consider now a crossing $(e_1,e_2)$ in drawing $\psi_z$ of graph $H_z$. Let $e_1',e_2'$ be the edges of $G$, such that $e_1\in J(e_1')$ and $e_2\in J(e_2')$, so crossing $(e_1',e_2')$ of $\phi^*$ is responsible for crossing $(e_1,e_2)$.

\paragraph{Type-1 Crossings.}
We say that crossing $(e_1,e_2)$ in $\psi_z$ is a \emph{type-1 crossing} if at least one of the two edges $e_1',e_2'$ lies in $E(\tilde S_z)\cup \delta_G(\tilde S_z)$. We assume w.l.o.g. that $e_1'\in E(\tilde S_z)\cup \delta_G(\tilde S_z)$. Notice that crossing $(e_1',e_2')$ of $\phi^*$ may be responsible for at most $N_z(e_2')$ type-1 crossings in $\psi_z$. From \Cref{obs: bound on num of copies}, $\expect{N_z(e_2')}\leq \hat \eta$. We note that random variable $N_z(e_2')$ may only depend on the random selections of the internal routers $\qset(S_{z-1})$ and $\qset(S_{z+1})$, and in particular it is independent of the random selection of the internal router $\qset(S_{z})$.
 The expected number of crossings for which crossing  $(e_1',e_2')$  is responsible is then bounded by $\hat \eta$. From our definition, crossing $(e_1',e_2')$ must lie in $\chi^*_z$. Therefore, the total expected number of type-1 crossings is bounded by $|\chi^*_z|\cdot \hat\eta$.
 We note that the random variable corresponding to the total number of type-1 crossings only depends on the random selections of the internal routers $\qset(S_{z-1})$ and $\qset(S_{z+1})$, and it is independent of the random selection of the internal router $\qset(S_{z})$. We will use this fact later.

\paragraph{Type-2 Crossings.}
We say that a crossing $(e_1,e_2)$ in $\psi_z$ is a \emph{type-2 crossing} if it is not a type-1 crossing, and, additionally, one of the two edges (say $e_1$) lies on a path of $\rset'_1$, while the other edge (edge $e_2$) lies on a path of $\rset''_1$. Notice that, in this case, $e_1\in E(U_{z-1})$ and $e_2\in E(\notU_z)$ must hold. A crossing $(e_1',e_2')$ of $\psi^*$ may be responsible for at most $N_z'(e_1')\cdot N_z'(e_2')$ type-2 crossings of $\psi_z$. Moreover, the random variables $N_z'(e_1'), N_z'(e_2')$ are independent from each other. From \Cref{obs: bound on num of copies}, we can bound $\expect{N_z'(e_1')\cdot N_z'(e_2')}\leq \expect{N_z'(e_1')}\cdot \hat \eta$.
Therefore, we get that the total expected number of type-2 crossings is bounded by:

\[\sum_{(e_1',e_2')\in \chi^*} \left (\expect{N_z'(e_1')}+\expect{N_z'(e_2')}\right )\cdot \hat \eta.\]

\paragraph{Type-3 Crossings.}
We say that a crossing $(e_1,e_2)$ in $\psi_z$ is a \emph{type-3 crossing} if it is not a type-1 or a type-2 crossing, and, additionally, one of the two edges (say  $e_1$) lies on a path of $\rset'_1$, while the other edge (edge $e_2$) lies on a path of $\rset'_1\cup \rset_2$. 
We denote by $\tilde \chi_{z-1}$ the set of all crossings $(e_1',e_2')$ of $\phi^*$, where $e_1',e_2'\in E(S_{z-1})$.

Consider any crossing  $(e_1',e_2')$ of $\phi^*$ that does not lie in $\tilde \chi_{z-1}$.
This crossing may be responsible for at most $N_z'(e_1')\cdot N_z(e_2')+N_z(e_1')\cdot N'_z(e_2')$ type-3 crossings in $\psi_z$. Notice that random variables $N_z'(e_1')$,  $N_z(e_2')$ are independent from each other, as are random variables $N_z(e_1')$,  $N_z'(e_2')$. From \Cref{obs: bound on num of copies}, we can bound the expected number of type-3 crossings for which crossing 
$(e_1',e_2')$ of $\phi^*$ is responsible by $\expect{N_z'(e_1')}\cdot \hat \eta+\expect{N_z'(e_2')}\cdot \hat \eta$. 
Therefore, the total number of type-3 crossings, for which  crossings of $\chi^*\setminus \tilde \chi_{z-1}$ are responsible is bounded by:

\[\sum_{(e_1',e_2')\in \chi^*\setminus \tilde \chi_{z-1}} \left (\expect{N_z'(e_1')}+\expect{N_z'(e_2')}\right )\cdot \hat \eta.\]

A crossing $(e_1',e_2')\in \tilde \chi_{z-1}$ may be responsible for up to $N_z'(e_1')\cdot N_z(e_2')+N_z(e_1')\cdot N'_z(e_2')\leq 2N_z(e_1')\cdot N_z(e_2')$ type-3 crossings of $\psi_z$. But now the random variables 
$N_z(e_1'), N_z(e_2')$ are no longer indepdendent. We can, however, bound $N_z(e_1')\cdot N_z(e_2')\leq (N_z(e_1'))^2+(N_z(e_2'))^2$. 
From \Cref{obs: bound congestion of routers}, for an edge $e\in E(S_{z-1})$, $N_z(e)\leq O(\log^{34}m)\cdot \cong_G(\qset(S_{z-1}),e)$. 
Therefore, the total expected number of type-3 crossings is at most:

\[
\begin{split}
&\hat \eta\cdot \sum_{(e_1',e_2')\in \chi^*\setminus \tilde \chi_{z-1}} \left (\expect{N_z'(e_1')}+\expect{N_z'(e_2')}\right )\\
&\quad\quad\quad\quad+ O(\log^{68}m)\cdot\sum_{(e_1',e_2')\in \tilde \chi_{z-1}}\left( \expect{ (\cong_G(\qset(S_{z-1}),e_1')^2}+\expect{(\cong_G(\qset(S_{z-1}),e_2')^2}\right ).
\end{split}\]

\paragraph{Type-4 Crossings.}
We say that a crossing $(e_1,e_2)$ in $\psi_z$ is a \emph{type-4 crossing} if it is not a crossing of one of the first three types, and, additionally, one of the two edges (say, edge $e_1$) lies on a path of $\rset''_1$, while the other edge (edge $e_2$) lies on a path of $\rset''_1\cup \rset_2$. 
We denote by $\tilde \chi_{z+1}$ the set of all crossings $(e_1',e_2')$ of $\phi^*$, where $e_1',e_2'\in E(S_{z+1})$.

Consider any crossing  $(e_1',e_2')$ of $\phi^*$ that does not lie in $\tilde \chi_{z+1}$.
As before, this crossing may be responsible for at most $N_z'(e_1')\cdot N_z(e_2')+N_z(e_1')\cdot N'_z(e_2')$ type-4 crossings in $\psi_z$. 
Using the same analysis as for type-3 crossings,  we can bound the expected number of type-4 crossings for which crossing 
$(e_1',e_2')$ of $\phi^*\setminus \tilde\chi_{z+1}$ is responsible by $\expect{N_z'(e_1')}\cdot \hat \eta+\expect{N_z'(e_2')}\cdot \hat \eta$. 
The total number of type-4 crossings, for which  crossings of $\chi^*\setminus \tilde \chi_{z+1}$ are responsible is bounded by:

\[\sum_{(e_1',e_2')\in \chi^*\setminus \tilde \chi_{z+1}} \left (\expect{N_z'(e_1')}+\expect{N_z'(e_2')}\right )\cdot \hat \eta.\]

As before, a crossing $(e_1',e_2')\in \tilde \chi_{z+1}$ may be responsible for up to $N_z'(e_1')\cdot N_z(e_2')+N_z(e_1')\cdot N'_z(e_2')\leq 2N_z(e_1')\cdot N_z(e_2')\leq 2(N_z(e_1'))^2+2( N_z(e_2'))^2$ type-4 crossings of $\psi_z$. Using the same reasoning as in type-3 crossings, the total expected number of type-4 crossings is bounded by:

\[
\begin{split}
&\hat \eta\cdot 
\sum_{(e_1',e_2')\in \chi^*\setminus \tilde \chi_{z+1}} \left (\expect{N_z'(e_1')}+\expect{N_z'(e_2')}\right )\\
&\quad\quad+ O(\log^{68}m)\cdot\sum_{(e_1',e_2')\in \tilde \chi_{z+1}}\left( \expect{ (\cong_G(\qset(S_{z-1}),e_1')^2}+\expect{(\cong_G(\qset(S_{z+1}),e_2')^2}\right ).
\end{split}\]

\paragraph{Type-5 Crossings}
All remaining crossings of $\psi_z$ are type-5 crossings. For each such crossing $(e_1,e_2)$, it must be the case that each of the edges $e_1,e_2$ belongs to a path of $\rset_2$. We do not bound the number of type-5 crossings, as we will eventually eliminate all such crossings.

\subsubsection{Step 2: Initial Drawing of $G_z$}
\label{subsubsec: step 2}

For every edge $e\in E_{z-1}\cup E_z^{\lef}$, we denote by $\Gamma(e)$ the curve corresponding to the image of path $R(e)\in \rset'_1$ in the drawing $\psi_z$ of $H_z$. Note that, if $v$ is an endpoint of the path $R(e)$ that lies in $V(\tilde S_z)$, then curve $\Gamma(e)$ connects the image of $v$ to the image of vertex $u_{z-1}$ in $\psi_z$. For every edge $e\in E_{z}\cup  E_z^{\rig}$, we denote by $\Gamma(e)$ the curve corresponding to the image of the path $R(e)\in \rset''_1$ in the drawing $\psi_z$ of $H_z$. Note that, if $v$ is an endpoint of $e$ that lies in $V(\tilde S_z)$, then curve $\Gamma(e)$ connects the image of $v$ to the image of vertex $u_{z+1}$ in $\psi_z$.
Lastly, for every edge $e\in E_z^{\through}$, we let $\Gamma(e)$ be the image of the path $R(e)\in \rset_2$ in $\psi_z$. Notice that curve $\Gamma(e)$ connects the image of $u_{z-1}$ to the image of $u_{z+1}$ in $\psi_z$. From the constructions of the paths in $\rset'_1\cup \rset_2$, if we denote $\delta_G(U_{z-1})=\set{\hat a_1,\ldots,\hat a_q}$, where the edges are indexed in the order of their apperance in the ordering $\tilde \oset_{z-1}$, the curves $\Gamma(\hat a_1),\ldots,\Gamma(\hat a_q)$ enter the image of vertex $u_{z-1}$ in this circular order.

In order to obtain the initial drawing $\phi'_z$ of $G_z$, we start with the drawing $\psi_z$ of graph $H_z$, and we delete from it the images of all vertices except those lying in $V(\tilde S_z)\cup \set{u_{z-1},u_{z+1}}$, and the images of all edges except those lying in $E(\tilde S_z)$; we view the image of vertex $u_{z-1}$ as the image of the special vertex $v^*$, and the image of vertex $u_{z+1}$ as the image of the secial vertex $v^{**}$. We then add to this drawing the curves in $\set{\Gamma(e)\mid e\in \delta_G(U_{z-1})\cup \delta_G(U_z)}$. Each such curve $\Gamma(e)$ becomes an image of the corresponding edge $e$. Notice that the edges of $\delta_G(U_{z-1})$ become incident to $v^*$ in graph $G_z$; the edges of $\delta_G(U_z)$ become incident to $v^{**}$, and the edges of $E_z^{\through}$ connect $v^*$ to $v^{**}$. From the above discussion, the circular order in which the images of the edges in $\delta_{G_z}(v^*)$ enter the image of $v^*$ in the current drawing is exactly the ordering $\tilde \oset_{z-1}$, which is precisely the ordering $\oset_{v^*}\in \Sigma_z$. However, the images of the edges of $\delta_{G_{z}}(v^{**})$ may not enter the image of vertex $v^{**}$ in the correct order. We will fix this in subsequent steps. There is one major problem with the current drawing of the graph $G_z$: it is possible that some point $p$ lies on a large number of curves in set  $\set{\Gamma(e)\mid e\in \delta_G(U_{z-1})\cup \delta_G(U_z)}$, and it is an inner point on each such curve. This may only happen if $p$ corresponds to an image of some vertex $v$, where $v\in V(U_{z-1})\setminus\set{u_{z-1}}$, or $v\in V(\overline U_{z})\setminus\set{u_{z+1}}$. We will now ``fix'' the images of the edges of $\delta_{G_z}(v^*)\cup \delta_{G_z}(v^{**})$ by slightly ``nudging'' them in the vicinity of each such vertex, to ensure that all resulting curves are in general position. We do so by performing a nudging operation (see \Cref{sec: curves in a disc}).

We process every vertex $v\in \left (V(U_{z-1})\cup V(\notU_z)\right)\setminus\set{u_{z-1},u_{z+1}}$ one by one. Consider an iteration when any such vertex $v$ is processed. Let $A(v)$ be the set of all edges $e\in \delta_G(U_{z-1})\cup \delta_G(U_z)$, such that curve $\Gamma(e)$ contains the image of vertex $v$ (in $\psi_z$). We denote $A(v)=\set{a_1,\ldots,a_k}$. Consider the tiny $v$-disc $D(v)=D_{\psi_z}(v)$ in the drawing $\psi_z$ of graph $H_z$. For all $1\leq i\le k$, we let $s_i,t_i$ be the two points at which curve $\Gamma(a_i)$ intersects the boundary of the disc $D(v)$. Note that all points $s_1,t_1,\ldots,s_k,t_k$ must be distinct. We use the algorithm from \Cref{claim: curves in a disc} in order to construct a collection $\set{\gamma_1,\ldots,\gamma_k}$ of curves, such that, for all $1\leq i\leq k$, curve $\gamma_i$ has $s_i$ and $t_i$ as its endpoints, and is completely contained in $D(v)$. Recall that the claim ensures that, for every pair $1\leq i<j\leq k$ of indices, if the two pairs  $(s_i,t_i),(s_j,t_j)$ of points cross, then curves $\gamma_i,\gamma_j$ intersect at exactly one point; otherwise, curves $\gamma_i,\gamma_j$ do not intersect.
For all $1\leq i\leq k$, we modify the curve $\Gamma(a_i)$ as follows: we replace the segment of the curve between points $s_i,t_i$ with the curve $\gamma_i$. 

Once every vertex  $v\in \left (V(U_{z-1})\cup V(\notU_z)\right)\setminus\set{u_{z-1},u_{z+1}}$ is processed in this way, the curves in set $\set{\Gamma(e)\mid e\in \delta_G(U_{z-1})\cup \delta_G(U_z)}$ are in general position, and we obtain a valid drawing of the graph $G_z$, that we denote by $\phi'_z$. The modification of the curves in $\set{\Gamma(e)\mid e\in \delta_G(U_{z-1})\cup \delta_G(U_z)}$ do not affect the endpoints of the curves, and so, for every vertex $x\in V(G_z)\setminus \set{v^{**}}$, the images of the edges of $\delta_{G_z}(x)$ enter the image of $x$ in the order consistent with the rotation $\oset_x\in \Sigma_z$. We now bound the number of crossings in drawing $\phi'_z$.

Consider some pair of edges $e,e'\in E(G_z)$ that cross at some point $p$ in the drawing $\phi'_z$. We say that this crossing is \emph{primary}  iff point $p$ does not belong to any of the discs in the set $$\set{D(v)\mid v\in \left (V(U_{z-1})\cup V(\notU_z)\right)\setminus\set{u_{z-1},u_{z+1}}};$$ otherwise we say that the crossing is \emph{secondary}.
	
	Notice that every primary crossing in $\phi'_z$ corresponds to a unique crossing in the drawing $\psi_z$ of the graph $H_z$. Recall that we have partitioned all such crossings into five types, and we have bounded the number of crossings of each of the first four types. This partition naturally defines a partition of all primary crossings in $\phi'_z$ into five types. Specifically, primary crossings of type 1 are all crossings $(e,e')$ where at least one of the edges $e,e'$ lies in $E(\tilde S_z)\cup \delta_G(\tilde S_z)$. Primary crossings of type 2 are primary crossings between curves $\Gamma(e),\Gamma(e')$, where $e\in E_{z-1}\cup E_z^{\lef}$, while $e'\in E_{z}\cup E_z^{\rig}$. Primary crossings of type $3$ are primary crossings between curves 
$\Gamma(e),\Gamma(e')$, where $e\in E_{z-1}\cup E_z^{\lef}$ and $e'\in E_{z-1}\cup E_z^{\lef}\cup E_z^{\through}$, while primary crossings  of type 4 are primary crossings between curves $\Gamma(e),\Gamma(e')$, where $e\in E_{z}\cup E_z^{\rig}$, while $e'\in E_{z}\cup E_z^{\rig}\cup E_z^{\through}$. Lastly, primary crossings of type 5 are primary crossings between curves $\Gamma(e),\Gamma(e')$, where $e,e'\in E_z^{\through}$.
The number of primary crossings of the first four types is bounded as before. 

We now consider secondary crossings of $\phi_z$. Notice that each such crossing must be between a pair of curves $\Gamma(e),\Gamma(e')$, where $e,e'\in \delta_G(U_{z-1})\cup \delta_G(U_z)$.

Consider a pair of curves $\Gamma(e),\Gamma(e')$, where $e,e'\in \delta_G(U_{z-1})\cup \delta_G(U_z)$, and some point $p$ at which the curves cross, such that the crossing is secondary. Let $v\in \left (V(U_{z-1})\cup V(\notU_z)\right)\setminus\set{u_{z-1},u_{z+1}}$ be the vertex such that $p$ lies in the interior of disc $D(v)$. Denote by $s,t$ the points on the boundary of $D$ that lie on $\Gamma(e)$, and define $s',t'$ similary for $\Gamma(e')$. From \Cref{claim: curves in a disc}, curves $\Gamma(e),\Gamma(e')$ may only cross inside the disc $D(v)$ if the pairs $(s,t),(s',t')$ of points on the boundary of $D$ cross. Consider now the paths $R(e)\in \rset_1\cup \rset_2$ and $R(e')\in \rset_1\cup \rset_2$. Denote by $e_1,e_2$ the two edges on path $R(e)$ that immediately precede and immediately follow vertex $v$, and define edges $e_1',e_2'$ similarly for path $R(e')$. We now consider three cases.

First, if $v\in V(S_{z-1})$, then there must be two paths $Q,Q'\in \qset(S_{z-1})$, such that $Q\subseteq R(e)$ and $Q'\subseteq R(e')$, where $Q,Q'$ both contain the vertex $v$. Since the paths of $\qset(S_{z-1})$ are non-transversal with respect to $\Sigma$, the only way for the two pairs $(s,t),(s',t')$ of points to cross is if the set $\set{e_1,e_2,e_1',e_2'}$ contains copies of fewer than four distinct edges of $\delta_G(v)$.  In other words, for some edge $e^*\in \delta_G(v)$, both $R(e)$ and $R(e')$ contain a copy of $e^*$. In this case, we say that edge $e^*$ is \emph{responsible} for this secondary crossing between $\Gamma(e)$ and $\Gamma(e')$. 
The second case is when $v\in V(S_{z+1})$. The analysis of this case is similar to the previous case: there must be an edge $e^*\in \delta_G(v)$ such that both $R(e)$ and $R(e')$ contain a copy of $e^*$. We say that $e^*$ is responsible for this crosisng.

We now consider the third case, when $v\not\in V(S_{z-1})\cup V(S_{z+1})$. We consider the paths $R(e),R(e')\in \rset_1\cup \rset_2$, and we define the edges $e_1,e_2,e_1',e_2'$ as before. Since $v\not\in  V(S_{z-1})\cup V(S_{z+1})$, 
it must be the case that $v\in W(e)\cap W(e')$. If  there is an edge $e^*\in \delta_G(v)$, such that both $R(e)$ and $R(e')$ contain a copy of $e^*$, then we designate $e^*$ to be responsible for this crossing as before. Otherwise, the edges in set $\set{e_1,e_2,e_1',e_2'}$ are copies of four distinct edges of $\delta_G(V)$. In this case, the cycles $W(e)$ and $W(e')$ must have a transversal intersection at vertex $v$, from \Cref{obs: inner non-transversal}, and $e,e'\in \hat E_z$ must hold. 
In this case, we say that the transversal intersection of $W(e)$ and $W(e')$ at $v$ is responsible for the crossing.

We now classify the secondary crossings into three types and bound the number of crossings of the first two types. We will eventually eliminate all crossings of the third type. 

\paragraph{Type-1 secondary crossing.}
Consider a secondary crossing between a pair $\Gamma(e),\Gamma(e')$ of curves, for $e,e'\in \delta_G(U_{z-1})\cup \delta_G(U_z)$, and a secondary crossing of the two curves at some point $p$. We say that the crossing is of type 1 if 
$e\in E_{z-1}\cup E_z^{\lef}$ and $e'\in E_{z-1}\cup E_z^{\lef}\cup E_z^{\through}$.

\paragraph{Type-2 secondary crossing.}
We say that a secondary crossing between a pair $\Gamma(e),\Gamma(e')$ of curves, for $e,e'\in \delta_G(U_{z-1})\cup \delta_G(U_z)$, is of type 2 if 
$e\in E_{z}\cup E_z^{\rig}$ and $e'\in E_z\cup E_z^{\rig}\cup E_z^{\through}$.

\paragraph{Type-3 secondary crossing.} All remaining secondary crossings are of tye 3. Consider any such crossing between a pair $\Gamma(e),\Gamma(e')$ of curves, for $e,e'\in \delta_G(U_{z-1})\cup \delta_G(U_z)$. Notice that it is impossible that one of the two edges $e,e'$ lies in $E_{z-1}\cup E_z^{\lef}$, while the other lies in $E_{z}\cup E_z^{\rig}$, since, in such a case, paths $R(e),R(e')$ cannot share any edges. Therefore, $e,e'\in E_z^{\through}$ must hold.

We now bound the expected number of type-1 secondary crossing. Consider any such crossing between a  pair  $\Gamma(e),\Gamma(e')$ of curves, and assume that the crossing point $p$ lies in disc $D(v)$, for some vertex $v$. From the definition of a type-1 crossing, $v\in V(U_{z-1})\setminus \set{u_{z-1}}$ must hold. 
In this case, it is impossible that a pair of auxiliary cycles $W(e),W(e')$ with $e,e'\in \hat E_z$ have a transversal intersection at vertex $v$, from \Cref{obs: auxiliary cycles non-transversal at at most one}. Therefore, some edge of $E_G(U_{z-1})$ must be responsible for this crossing. It is immediate to verify that every edge $e\in E_G(U_{z-1})$ may be responsible for at most $N'_z(e)\cdot N_z(e)$ type-1 secondary crossings. If $e\not\in E(S_{z-1})$, then, from \Cref{obs: bound on num of copies}, $N_z(e)\leq  O(\log^{34}m)$. If $e\in E(S_{z-1})$, then, from \Cref{obs: bound congestion of routers}, $N_z(e)\leq O(\log^{34}m)\cdot \cong_G(\qset(S_{z-1}),e)$. Therefore, the total expected  number of type-1 secondary crossings is bounded by:

\[O(\log^{34}m)\cdot \sum_{e\in E(U_{z-1})\setminus E(S_{z-1})}N'_z(e)  +O(\log^{68}m)\cdot \sum_{e\in E(S_{z-1})}\expect{\left (\cong _G(\qset(S_{z-1}),e)\right )^2}.\]

Next, we bound the expected number of type-2 secondary crossings. This time some of the crossigns are charged to individual edges (that is, some edge of $E(\notU_z)$ is responsible for the crossing), and some crossings are charged to transversal intersections of pairs of cycles $W(e'),W(e'')$, where  $e',e''\in \hat E_z$. The expected number of edges of the former type is bounded using the same reasoning as for type-1 crossings, and their expected number is at most:

 \[O(\log^{34}m)\cdot \sum_{e\in E(\notU_{z})\setminus E(S_{z+1})}N'_z(e)  +O(\log^{68}m)\cdot \sum_{e\in E(S_{z+1})}\expect{\left (\cong _G(\qset(S_z),e)\right )^2}.\]
 
Let $\Pi^T_z$ denote the set of triples $(e,e',v)$, where $e\in E_z^{\rig}$, $e'\in \hat E_z$, and $v$ is a vertex that lies on both $W(e)$ and $W(e')$, such that cycles $W(e)$ and $W(e')$ have a transversal intersection at $v$. Clearly, the number of type-2 secondary crossings that are charged to transversal intersections of pairs of cycles is bounded by $|\Pi^T_z|$. Overall, we get that the total expected number of type-2 secondary crossings is bounded by:

 \[O(\log^{34}m)\cdot \sum_{e\in E(\notU_{z})\setminus E(S_{z+1})}N'_z(e)  +O(\log^{68}m)\cdot \sum_{e\in E(S_{z+1})}\expect{\left (\cong _G(\qset(S_z),e)\right )^2}+|\Pi^T_z|.\]

This completes the analysis of the initial drawing $\phi'_z$ of graph $G_z$. Notice that we did not analyze the number of type-5 primary crossings and the number of type-3 secondary crossings. These are all crossings between the images of the edges of $E_z^{\through}$. Unfortunately, a crossing of the original drawing $\phi^*$ of $G$ may give rise to many crossings between edges of $E_z^{\through}$ in drawings $\phi'_z$ of graphs $G_z$, for $1\leq z\leq r$. In the next step, we will slightly modify the drawing $\phi'_z$ in order to eliminate all such crossings. Notice that our current bounds on the expected number of crossings in $\phi'_z$ contain terms like $\sum_{e\in E(S_{z-1})}\expect{\left (\cong_G(\qset(S_{z-1}),e)\right )^2}$. If cluster $S_{z-1}$ lies in set $\sset^{\light}$, then this expression can be bounded by $|E(S_{z-1})|\cdot  \hat \eta$. However, if $S_{z-1}\in \sset^{\bad}$ then this bound may no longer be valid. In such a case we will perform an additional uncrossing operation of the images of edges of $\delta_G(U_{z-1})$ in order to decrease this number of crossings. We also perform such an operation on the images of the edges of $\delta_G(U_z)$ if $S_{z+1}\in \sset^{\bad}$.

\subsubsection{Step 3: Modified Drawing of $G_z$}

In this step we modify the drawing $\phi'_z$ of $G_z$ to obtain a new modified drawing $\phi''_z$, by performing one or more uncrossing operations.
We first consider the cases where $S_{z-1}\in \sset^{\bad}$ or $S_{z+1}\in \sset^{\bad}$  hold, and perform some initial uncrossings to decrease the number of type-3 primary and type-1 secondary crossings (in case where $S_{z-1}\in \sset^{\bad}$), and the number type-4 primary  and type-2 secondary crossings (in case where $S_{z-1}\in \sset^{\bad}$). After that we perform one more uncrossing operation that will eliminate all type-5 primary and type-3 secondary crossings.

Recall that the expected number of type-3 primary crossings and type-1 secondary crossings (that is, all crossings between images of edge pairs $e,e'$ where $e\in E_{z-1}\cup E_z^{\lef}$ and $e'\in  E_{z-1}\cup E_z^{\lef}\cup E_z^{\through}$) is at most:

\[
\begin{split}&\hat \eta \cdot \sum_{(e_1',e_2')\in \chi^*\setminus \tilde \chi_{z-1}} \left (\expect{N_z'(e_1')}+\expect{N_z'(e_2')}\right )\\
&\quad +O(\log^{34}m)\cdot \sum_{e\in E(U_{z-1})\setminus E(S_{z-1})}N'_z(e)  \\
&\quad\quad+ O(\log^{68}m)\cdot\sum_{(e_1',e_2')\in \tilde \chi_{z-1}}\left( \expect{ (\cong_G(\qset(S_{z-1}),e_1')^2}+\expect{(\cong_G(\qset(S_{z-1}),e_2')^2}\right )\\
&\quad +O(\log^{68}m)\cdot \sum_{e\in E(S_{z-1})}\expect{\left (\cong _G(\qset(S_{z-1}),e)\right )^2}.
\end{split}
\]

From \Cref{obs: congestion square of internal routers}, if $S_{z-1}\in \sset^{\light}$, then for every edge $e\in E(S_{z-1})$, $\expect{(\cong_G(\qset(S_{z-1}),e)^2}\leq 
\hat \eta$.  Recalling that $\tilde \chi_{z-1}\subseteq \chi^*_{z-1}$, we get that, if $S_{z-1}\in \sset^{\light}$, then the expected number of type-3 primary and type-1 secondary crossings is bounded by:

\begin{equation}\label{eq: type 3 bound for light}
\begin{split}&\hat \eta\cdot \sum_{(e_1',e_2')\in \chi^*\setminus \tilde \chi_{z-1}} \left (\expect{N_z'(e_1')}+\expect{N_z'(e_2')}\right )\\
&\quad\quad\quad\quad +O(\log^{34}m)\cdot \sum_{e\in E(U_{z-1})\setminus E(S_{z-1})}N'_z(e)  \\
&\quad\quad \quad\quad+O(\hat \eta^2)\cdot (|\chi^*_{z-1}|+|E(S_{z-1}|).
\end{split}
\end{equation}

 Recall that the expected number of type-4 primary crossings and type-2 secondary crossings (that is, all crossings between images of edge pairs $e,e'$ where $e\in E_{z}\cup E_z^{\rig}$ and $e'\in E_z\cup E_z^{\rig}\cup E_z^{\through}$) is at most:

\[
\begin{split}
&\hat\eta \cdot 
\sum_{(e_1',e_2')\in \chi^*\setminus \tilde \chi_{z+1}} \left (\expect{N_z'(e_1')}+\expect{N_z'(e_2')}\right )\\
&\quad\quad\quad\quad
+O(\log^{34}m)\cdot \sum_{e\in E(\notU_{z})\setminus E(S_{z+1})}N'_z(e)  \\
&\quad\quad\quad\quad+ O(\log^{68}m)\cdot\sum_{(e_1',e_2')\in \tilde \chi_{z+1}}\left( \expect{ (\cong_G(\qset(S_{z+1}),e_1')^2}+\expect{(\cong_G(\qset(S_{z+1}),e_2')^2}\right )\\
&\quad\quad\quad\quad +O(\log^{68}m)\cdot \sum_{e\in E(S_{z+1})}\expect{\left (\cong _G(\qset(S_z),e)\right )^2}+|\Pi^T_z|
\end{split}\]

Here, $\Pi^T_z$ is the set of triples $(e,e',v)$, where $e\in E_z^{\rig}$, $e'\in \hat E_z$, and cycles $W(e)$ and $W(e')$ have a transversal intersection at vertex $v$.

From \Cref{obs: congestion square of internal routers}, if $S_{z+1}\in \sset^{\light}$, then for every edge $e\in E(S_{z+1})$, $\expect{(\cong_G(\qset(S_{z+1}),e)^2}\leq 
\hat \eta$.  Recalling that $\tilde \chi_{z+1}\subseteq \chi^*_{z+1}$, we get that, if $S_{z+1}\in \sset^{\light}$, then the expected number of type-4 primary and type-2 secondary crossings is bounded by:

\begin{equation}\label{eq: type 4 bound for light}
\begin{split}&\hat \eta\cdot \sum_{(e_1',e_2')\in \chi^*\setminus \tilde \chi_{z+1}} \left (\expect{N_z'(e_1')}+\expect{N_z'(e_2')}\right )\\
&\quad\quad\quad\quad +O(\log^{34}m)\cdot \sum_{e\in E(\notU_z)\setminus E(S_{z+1})}N'_z(e)  \\
&\quad\quad\quad\quad +O(\hat \eta^2)\cdot (|\chi^*_{z+1}|+|E(S_{z+1})|)+|\Pi^T_z|.
\end{split}
\end{equation}

We now consider four cases, depending on whether the clusters $S_{z-1},S_{z+1}$ lie in $\sset^{\light}$ or $\sset^{\bad}$.

\paragraph{Case 1: $S_{z-1}\in \sset^{\bad}$ and $S_{z+1}\in \sset^{\light}$.}

When $S_{z-1}\in \sset^{\bad}$, we no longer have a bound on $\expect{(\cong_G(\qset(S_{z-1}),e)^2}$ for edges $e\in E(S_{z-1})$, and so the bound from \Cref{eq: type 3 bound for light} on the number of type-3 primary and type-1 secondary crossings is no longer valid. Instead, we will perform a type-1 uncrossing of the images of the edges of $E_{z-1}\cup E_z^{\lef}\cup E_z^{\through}$. Let $\Gamma_1$ denote the set of curves reprsenting the images of these edges in $\phi_z'$. Let $\Gamma_2$ denote the set of curves reprsenting the images of all remaining edges of $G_z$ in $\phi_z'$. We apply the algorithm from \Cref{thm: type-1 uncrossing} to compute a new collection $\Gamma_1'$ of curves, where, for each edge $e\in E_{z-1}\cup E_z^{\lef}\cup E_z^{\through}$, there is a curve $\gamma(e)\in \Gamma_1'$ connecting the images of the endpoints of $e$. Intuitively, the algortihm for type-1 uncrossing proceeds in iterations, as long as there is a pair $\gamma_1,\gamma_2\in \Gamma_1$ of curves that cross at least twice. Assume that $p$ and $q$ are two points lying on both $\gamma_1$ and $\gamma_2$. The algorithm then uncrosses the two curves, by ``swapping'' the segments of these curves that connect $p$ and $q$ (see  Figure~\ref{fig: type_1_uncross_proof} in \Cref{apd: type-1 uncrossing} for an illustration.)
At the end of this procedure, every pair of curves in $\Gamma_1'$ may cross each other at most once. For each curve $\gamma\in \Gamma_2$, the number of crossings between $\gamma$ and the curves in $\Gamma_1'$ is no higher than the number of crossings between $\gamma$ and the curves in $\Gamma_1$. We modify the images of the edges in $E_{z-1}\cup E_z^{\lef}\cup E_z^{\through}$, so that for each such edge $e$, its new image is the curve $\gamma(e)\in \Gamma_1'$. Observe that the total number of primary crossings of types 1,2, and 4 does not increase, and neither does the number of secondary crossings of type 2. We note however that a primary crossing of type 2 (a crossing between images of edges $e,e'$ where $e\in E_{z-1}\cup E_z^{\lef}$ and $e'\in E_z\cup E_z^{\rig}$) may become a primary crossing of type 4 (a crossing between images of edges $e,e'$ where $e'\in E_z\cup E_z^{\rig}$ and $e\in E_z^{\through}$), and vice versa.
 The total number of type-3 primary crossings and of type-1 secondary crossings is now bounded by:
$\left (|E_{z-1}|+ |E_z^{\lef}|+|E_z^{\through}|\right )^2\leq |\delta_G(U_{z-1})|^2$. 

From \Cref{cor: few edges crossing cuts},
 $|\delta_G(U_{z-1})|\leq |\delta_G(S_{z-1})|\cdot O(\log^{34}m)$.
 Recall that, if $S_{z-1}\in \sset^{\bad}$, and the bad event $\event$ does not happen, then, from \Cref{obs: congestion square of internal routers}, 
$\optcrors(S_{z-1},\Sigma(S_{z-1}))+|E(S_{z-1})|\geq \frac{|\delta_G(S_{z-1})|^2}{\hat \eta}$,
 where $\Sigma(S_{z-1})$ is the rotation system for graph $S_{z-1}$ induced by $\Sigma$. Therefore, 
$|\delta_G(S_{z-1})|^2\leq\hat \eta \cdot  \left (|\chi^*_{z-1}|+|E(S_{z-1})|\right )$ must hold.

Overall, if $S_{z-1}\in \sset^{\bad}$, and event $\event$ did not happen, then the total number of type-3 primary crossings and of type-1 secondary crossings is now bounded by:

\[|\delta_G(U_{z-1})|^2\leq \hat \eta^2\cdot |\delta_G(S_{z-1})|^2\leq \hat \eta^2\cdot  \left (|\chi^*_{z-1}|+|E(S_{z-1})|\right ).\]

Combining this with the bound from Equation \ref{eq: type 3 bound for light}, we get that, regardless of whether Case 1 happened or not, if event $\event$ did not happen, then, after the current modification, the total expected number of type-3 primary crossings and of type-1 secondary crossings is bounded by:

\begin{equation}\label{eq: type 3 final}
\begin{split}&\hat \eta\cdot \sum_{(e_1',e_2')\in \chi^*\setminus \tilde \chi_{z-1}} \left (\expect{N_z'(e_1')}+\expect{N_z'(e_2')}\right )\\
	&\quad\quad \quad\quad+O(\log^{34}m)\cdot \sum_{e\in E(U_{z-1})\setminus E(S_{z-1})}N'_z(e)  \\
	&\quad\quad\quad\quad+ \hat \eta^2 \cdot (|\chi^*_{z-1}|+|E(S_{z-1}|).
\end{split}
\end{equation}

\paragraph{Case 2: $S_{z-1}\in \sset^{\light}$ and $S_{z+1}\in \sset^{\bad}$.}

We now consider the case where $S_{z-1}\in \sset^{\light}$ and $S_{z+1}\in \sset^{\bad}$. The modification that we perform is almost identical to that performed in the case where $S_{z-1}\in \sset^{\bad}$, except that now we uncross the images of the edges in $\delta_G(U_z)$.

As before, when $S_{z+1}\in \sset^{\bad}$, we no longer have a bound on $\expect{(\cong_G(\qset(S_{z+1}),e)^2}$ for edges $e\in E(S_{z+1})$,  and so the bound from \Cref{eq: type 4 bound for light} on the number of type-4 primary and type-2 secondary crossings is no longer valid. Instead, we will perform a type-1 uncrossing of the images of the edges of $E_{z}\cup E_z^{\rig}\cup E_z^{\through}$, similarly to the first case. Let $\Gamma_1$ denote the set of curves reprsenting the images of these edges in the current drawing $\phi_z'$. Let $\Gamma_2$ denote the set of curves reprsenting the images of all remaining edges of $G_z$ in $\phi_z'$. We apply the algorithm from \Cref{thm: type-1 uncrossing} to compute a new collection $\Gamma_1'$ of curves, where, for each edge $e\in E_{z}\cup E_z^{\rig}\cup E_z^{\through}$, there is a curve $\gamma(e)\in \Gamma_1'$ connecting the images of the endpoints of $e$. 
Recall that every pair of curves in $\Gamma_1'$ may cross at most once, and, for each curve $\gamma\in \Gamma_2$, the number of crossings between $\gamma$ and the curves in $\Gamma_1'$ is no higher than the number of crossings between $\gamma$ and the curves in $\Gamma_1$. We modify the images of the edges in $E_{z}\cup E_z^{\rig}\cup E_z^{\through}$, so that for each such edge $e$, its new image is the curve $\gamma(e)\in \Gamma_1'$. As before, the total number of primary crossings of types 1,2, and 3 does not increase, and neither does the number of secondary crossings of type 1. As before, a primary crossing of type 2 (a crossing between images of edges $e,e'$ where $e\in E_{z-1}\cup E_z^{\lef}$ and $e'\in E_{z}\cup E_z^{\rig}$) may become a primary crossing of type 3 (a crossing between images of edges $e,e'$ where $e'\in E_{z-1}\cup E_z^{\lef}$ and $e\in E_z^{\through}$), and vice versa.
The total number of type-4 primary crossings and of type-2 secondary crossings is now bounded by:
$\left (|E_{z}|+ |E_z^{\rig}|+|E_z^{\through}|\right )^2\leq |\delta_G(U_{z})|^2$. 
Using the same arguments as in the first case, and the second statement from 
\Cref{cor: few edges crossing cuts}, we conclude that $|\delta_G(U_{z})|\leq |\delta_G(S_{z+1})|\cdot O(\log^{34}m)$. As before, if $S_{z+1}\in \sset^{\bad}$, and the bad event $\event$ does not happen, then, from \Cref{obs: congestion square of internal routers}, 
$\optcrors(S_{z+1},\Sigma(S_{z+1}))+|E(S_{z+1})|\geq \frac{|\delta_G(S_{z-1})|^2}{\hat\eta}$,
where $\Sigma(S_{z+1})$ is the rotation system for graph $S_{z+1}$ induced by $\Sigma$. As before, we get that
$|\delta_G(S_{z+1})|^2\leq \hat \eta\cdot  \left (|\chi^*_{z+1}|+|E(S_{z+1})|\right )$.

Overall, if $S_{z+1}\in \sset^{\bad}$, and event $\event$ did not happen, then the total number of type-4 primary type-2 secondary crossings is now bounded by:

\[|\delta_G(U_{z+1})|^2\leq O(\log^{68}m)\cdot |\delta_G(S_{z+1})|^2\leq \hat\eta^2\cdot  \left (|\chi^*_{z+1}|+|E(S_{z+1})|\right ).\]

Combining this with the bound from Equation \ref{eq: type 4 bound for light}, we get that, regardless of whether Case 2 happened or not, if event $\event$ did not happen, then, after the current modification, the total expected number of type-4 primary crossings and of type-2 secondary crossings is bounded by:

\begin{equation}\label{eq: type 4 final}
\begin{split}&\hat \eta \cdot \sum_{(e_1',e_2')\in \chi^*\setminus \tilde \chi_{z+1}} \left (\expect{N_z'(e_1')}+\expect{N_z'(e_2')}\right )\\
&\quad\quad\quad\quad +O(\log^{34}m)\cdot \sum_{e\in E(\notU_z)\setminus E(S_{z+1})}N'_z(e)  \\
&\quad\quad\quad\quad +\hat \eta^2\cdot (|\chi^*_{z+1}|+|E(S_{z+1})|)+|\Pi^T_z|.
\end{split}
\end{equation}

\paragraph{Case 3: $S_{z-1},S_{z+1}\in \sset^{\light}$, and accounting so far.}

If $S_{z-1},S_{z+1}\in \sset^{\light}$, then we do not perform any modifications for now. We now bound the total number of crossings in the current drawing $\phi'_z$ of graph $G_z$ for cases 1--3, excluding the crossings between pairs of edges in $E_z^{\through}$. If Case 3 happened, then the number of crossings did not increase in this step. If Case 1 happened, then the total number of primary crossings of types 1,2 and 4, and secondary crossings of type 2 did not change, and the number of primary crossings of type 3 and secondary crossings of type 1 is bounded by \Cref{eq: type 3 final}. Similarly, If Case 2 happened, then the total number of primary crossings of types 1,2 and 3, and secondary crossings of type 1 did not change, and the number of primary crossings of type 4 and secondary crossings of type 2 is bounded by \Cref{eq: type 4 final}.
 Therefore, if any of the cases 1--3 happened, and event $\event$ did not happen, then the total expected number of crossings in the current drawing $\phi'_z$ of graph $G_z$, excluding  the crossings between pairs of edges in $E_z^{\through}$, is at most:

\begin{equation}\label{eq: bound all crossings}
\begin{split}
&\hat \eta^2\left ( |\chi^*_{z-1}|+|\chi^*_{z}| +|\chi^*_{z+1}|+|E(S_{z-1})|+|E(S_{z+1})|\right )\\
&\quad\quad\quad\quad+ \hat \eta \sum_{(e,e')\in \chi^*} \left (\expect{N_z'(e_1')}+\expect{N_z'(e_2')}\right )  \\
&\quad\quad\quad\quad+ \hat \eta\cdot \sum_{e\in E(U_{z-1})\setminus E(S_{z-1})}N'_z(e)+ \hat \eta\cdot \sum_{e\in E(\notU_{z})\setminus E(S_{z+1})}N'_z(e)+|\Pi^T_z|.
\end{split}
\end{equation}

\paragraph{Case 4: $S_{z-1},S_{z+1}\in \sset^{\bad}$.}
In this case, we will perform a type-1 uncrossing of the images of the edges of $E_{z-1}\cup E_z^{\lef}\cup E_z^{\through}\cup E_z^{\rig}\cup E_z=\delta_G(U_{z-1})\cup \delta_G(U_z)$. Let $\Gamma_1$ denote the set curves reprsenting the images of these edges in $\phi_z'$. Let $\Gamma_2$ denote the set of curves reprsenting the images of all remaining edges of $G_z$ in $\phi_z'$. We apply the algorithm from \Cref{thm: type-1 uncrossing} to compute a new collection $\Gamma_1'$ of curves, where, for each edge $e\in \delta_G(U_{z-1})\cup \delta_G(U_z)$, there is a curve $\gamma(e)\in \Gamma_1'$ connecting the images of the endpoints of $e$. 
We are guaranteed that every pair of curves in $\Gamma_1'$ may cross each other at most once, and, for each curve $\gamma\in \Gamma_2$, the number of crossings between $\gamma$ and the curves in $\Gamma_1'$ is no greater than the number of crossings between $\gamma$ and the curves in $\Gamma_1$. We modify the images of the edges in $\delta_G(U_{z-1})\cup \delta_G(U_z)$, so that for each such edge $e$, its new image is the curve $\gamma(e)\in \Gamma_1'$. 
Note that the total number of type-1 primary crossings does not change. The total number of all other crossings is bounded by $(|\delta_G(U_{z-1})|+| \delta_G(U_z)|)^2\leq O(|\delta_G(U_{z-1})|^2)+O(|\delta_G(U_{z})|^2)$. Using the same reasoning as in Cases 1 and 2, if event $\event$ did not happen, 
then:

$$|\delta_G(U_{z-1})|^2\leq \hat \eta^2\cdot  \left (|\chi^*_{z-1}|+|E(S_{z-1})|\right ),$$

and 

$$|\delta_G(U_{z})|^2\leq \hat \eta^2\cdot  \left (|\chi^*_{z+1}|+|E(S_{z+1})|\right ),$$

Therefore, if event $\event$ did not happen, the total expected number of crossings in the current drawing  is bounded by:

\begin{equation}\label{eq: case 4 final}
\hat \eta^2\cdot \left(|\chi^*_{z+1}|+|\chi^*_z|+ |\chi^*_{z+1}|+|E(S_{z-1})|+|E(S_{z+1})|\right).
\end{equation}

\paragraph{Uncrossing the Edges of $E_z^{\through}$.}

So far we have constructed a drawing $\phi'_z$ of graph $G_z$ and bounded the expected number of crossings in $\phi'_z$, excluding the crossings between the images of the edges in $E_z^{\through}$. In this step, we eliminate all crossings of the latter type, by performing a type-2 uncrossing of the images of the edges in $E_z^{\through}$.
Specifically, we let $\tilde \qset$ be the set of paths  in graph $G_z$ that contains, for each edge $e\in E_z^{\through}$, a path $\tilde Q(e)$, that consists of the edge $e$ only. Recall that each edge $e\in E_z^{\through}$ connects the special vertices $v^*$, $v^{**}$ to each other. We view each such path $\tilde Q(e)$ as being directed from $v^*$ to $v^{**}$. We then apply the algorithm from \Cref{thm: new type 2 uncrossing} that performs a type-2 uncrossing on the images of the paths in $\tilde \qset$. Let $\Gamma$ be the resulting set of curves that it produces. Recall that, for every edge $e\in E_z^{\through}$, there must be a curve $\gamma(e)\in \Gamma$, that contains the segment of the image of edge $e$ that lies in the disc $D(v^*)$. We replace the current image of the edge $e$ with the curve $\gamma(e)$. Once the images of all edges $e\in E_z^{\through}$ are modified, we obtain the final modified drawing $\phi''_z$ of graph $G_z$. The algorithm from \Cref{thm: new type 2 uncrossing} ensures that the images of the edges in  $E_z^{\through}$ do not cross each other. Since the curves in $\Gamma$ are aligned with the graph that consists of the edges of $E_z^{\through}$, we are guaranteed that, for each edge $e\in E(G_z)\setminus E^{\through}_z$, the number of crossings in which edge $e$ participates does not increase. 
The algorithm from \Cref{thm: new type 2 uncrossing}, and the type-1 uncrossings that we performed in Cases 1 -- 3 ensure that the order in which the images of the edges of $\delta_{G_z}(v^*)$ enter the image of $v^*$ does not change, and remain consistent with the rotation $\oset_{v^*}\in \Sigma_z$. To summarize, we have obtained a drawing $\phi''_z$ of graph $G_z$, such that, for every vertex $v\in V(G_z)\setminus\set{v^{**}}$, the images of the edges of $\delta_{G_z}(v)$ enter the image of $v$ in the order consistent with the rotation $\oset_v\in\Sigma_z$, and the total expected number of crossings in $\phi''_z$ is bounded by:

\begin{equation}\label{eq: bound all crossings final}
\begin{split}
&\hat \eta^2\left ( |\chi^*_{z-1}|+|\chi^*_{z}| +|\chi^*_{z+1}|+|E(S_{z-1})|+|E(S_{z+1})|\right )\\
&\quad\quad\quad\quad+ \hat \eta\cdot \sum_{(e,e')\in \chi^*} \left (\expect{N_z'(e_1')}+\expect{N_z'(e_2')}\right )  \\
&\quad\quad\quad\quad+ \hat \eta\cdot \sum_{e\in E(U_{z-1})\setminus E(S_{z-1})}N'_z(e)+ \hat \eta\cdot \sum_{e\in E(\notU_{z})\setminus E(S_{z+1})}N'_z(e)+|\Pi^T_z|.
\end{split}
\end{equation}

In the next and the final step, we obtain the final solution $\phi_z$ to instance $I_z=(G_z,\Sigma_z)$ of \cnwrs, by modifying the current drawing $\phi''_z$ of graph $G_z$ inside the tiny $v^{**}$-disc $D(v^{**})$.

\subsubsection{Step 4: the Final Drawing of Graph $G_z$}

In this step we slightly modify the current drawing $\phi''_z$ of graph $G_z$ in order to obtain the final drawing $\phi_z$ of $G_z$, which is a valid solution to instance $I_z=(G_z,\Sigma_Z)$ of \cnwrs.

Consider the tiny $v^{**}$-disc $D=D_{\phi''_z}(v^{**})$. Denote $\delta_{G_z}(v^{**})=\set{e_1,\ldots,e_h}$, and, for all $1\leq i\leq h$, let $p_i$ be the point on the image of the edge $e_i$ in $\phi''_z$ that lies on the boundary of the disc $D$. We assume that the edges are indexed so that the points $p_1,\ldots,p_h$ are encountered in this order when traversing the boundary of $D$ in the clock-wise direction. We denote by $\oset$ this ordering of the edges $e_1,\ldots,e_h$. Let $\oset'$ be the ordering $\oset_{v^{**}}\in \Sigma_z$ of the edges of $\delta_{G_z}(v^{**})$.
We use the algorithm from \Cref{lem: ordering modification} to compute a collection $\Gamma=\set{\gamma(e_i)\mid 1\leq i\leq h}$ of curves, such that, for each edge $e_i$, curve $\gamma(e_i)$ only differs from the  image of the edge $e_i$ in the current drawing $\phi''_z$ of $G_z$ inside the disc $D$, and the curves of $\Gamma$ enter the image of $v^{**}$ in the order $\oset'$. We then replace, for each edge $e_i\in \delta_{G_z}(v^{**})$, the current image of the edge $e_i$ with the curve $\gamma(e_i)$. As the result, we obtain a valid solution $\phi_z$ to instance $I_z=(G_z,\Sigma_z)$ of \cnwrs, as the images of the edges in $\delta_{G_z}(v^{**})$ now enter the image of $v^{**}$ in the correct order. \Cref{lem: ordering modification} guarantees that the number of crossings between the curves in $\Gamma$ within the disc $D$ is bounded by $O(\dist(\oset,\oset'))$, and these are the only new crossings. Therefore, the number of crossings grows by at most $O(\dist(\oset,\oset'))$. In the next claim we bound $\dist(\oset,\oset')$.

\begin{claim}\label{claim: bound distance between rotations}
If event $\event$ did not happen, then the expectation of $\dist(\oset,\oset')$ is bounded by:

\[\begin{split}  &\hat \eta^{O(1)}\cdot \left (\sum_{e\in E(G)}\expect{N'_z(e)}+\sum_{(e,e')\in \chi^*}  \left (\expect{N'_z(e)}+\expect{N'_z(e')}\right )\right )\\
&\quad\quad\quad\quad+\hat \eta^{O(1)}\cdot \left (|\chi^*_{z-1}|+|\chi^*_z|+|\chi^*_{z+1}|+|E(S_{z-1})|+|E(\tilde S_z)|+|E(S_{z+1})|+| \delta_G(\tilde S_z)|+|\delta_G(S_{z-1})|\right )\\
&\quad\quad\quad\quad+\hat \eta \cdot \cro(\phi''_z)+|\Pi_z^T|.
\end{split}
\]

\end{claim}

We prove \Cref{claim: bound distance between rotations} below, after we complete the proof of  \Cref{claim: existence of good solutions special} using it.
For convenience, we denote by $E^*_z= E(S_{z+1})\cup E(S_{z-1})\cup E(\tilde S_z)\cup \delta_G(\tilde S_z)\cup   \delta_G(S_{z-1})$.
Combining the bound from \Cref{eq: bound all crossings final} with the bound from \Cref{claim: bound distance between rotations}, we get that, if Event $\event$ did not happen, then $\expect{\optcrors(I_z)}$ is bounded by:

\[
\begin{split} 
\hat \eta^{O(1)}\left ( |\chi^*_{z-1}|+|\chi^*_{z}| +|\chi^*_{z+1}|+|E^*_z|+ \sum_{(e,e')\in \chi^*} \left (\expect{N_z'(e_1')}+\expect{N_z'(e_2')}\right )  +\sum_{e\in E(G)}\expect{N'_z(e)}+|\Pi^T_z|\right ).
\end{split}
\]

Note that an edge $e\in E(G)$ may belong to at most $O(1)$ sets in $\set{E^*_z}_{z=1}^r$. Also, a crossing $(e,e')\in \chi^*$ may belong to at most two sets in $\set{\chi^*_z}_{z=1}^r$ Therefore, we get that:

\begin{equation}\label{eq: final bound}
\begin{split}
\expect{\sum_{z=1}^r\optcrors(I_z)}&\leq \hat \eta^{O(1)}(|E(G)|+|\chi^*|)\\
&+\hat \eta^{O(1)}\cdot \sum_{(e,e')\in \chi^*}\sum_{z=1}^r \left (\expect{N_z'(e_1')}+\expect{N_z'(e_2')}\right ) \\
&+ \hat \eta^{O(1)}\cdot\sum_{e\in E(G)}\sum_{z=1}^r\expect{N'_z(e)}  \\
&+ \hat \eta^{O(1)}\cdot\sum_{z=1}^r|\Pi^T_z|
\end{split}
\end{equation}

We use the following two observations, whose proofs appear in  \Cref{subsec:bound N' values} and \Cref{subsec: proof of obs bound Pi triples}, respectively, in order to complete the proof of \Cref{claim: existence of good solutions special}.

\begin{observation}\label{obs: bound N' values}
	For every edge $e\in E(G)$, $\sum_{z=1}^r\expect{N'_z(e)}\leq O(\hat \eta)$.
\end{observation}

\begin{observation}\label{obs: bound Pi triples}
	$\sum_{z=1}^r|\Pi^T_z|\leq  \left(|E(G)|+|\chi^*|\right )\cdot O(\log^{68}m)$.
\end{observation}


Combining \Cref{eq: final bound} with Observations \ref{obs: bound N' values}  and \ref{obs: bound Pi triples}, and recalling that $\hat \eta=2^{O((\log m)^{3/4}\log\log m)}$, we get that, if event $\event$ did not happen:

\[
\expect{\sum_{z=1}^r\optcrors(I_z)}\leq 2^{O((\log m)^{3/4}\log\log m)}\cdot (|E(G)|+|\chi^*|).
\]

Recall that $\prob{\event}\leq 1/m^{99}$, and, if $\event$ happens, $\sum_{z=1}^r\optcrors(I_z)\leq m^3$ must hold. Therefore, overall, $\expect{\sum_{z=1}^r\optcrors(I_z)}\leq 2^{O((\log m)^{3/4}\log\log m)}\cdot (|E(G)|+|\chi^*|)$.
In order to complete the proof of \Cref{claim: existence of good solutions special}, it is now enough to prove  \Cref{claim: bound distance between rotations}, which we do next.

\subsection{Proof of \Cref{claim: bound distance between rotations}}
\label{subsec: bound distance between rotations}
Assume first that $S_{z+1}\in \sset^{\bad}$. Clearly, $\dist(\oset,\oset')\leq |\delta_{G_z}(v^{**})|^2\leq |\delta_G(U_z)|^2$. 
From 
\Cref{cor: few edges crossing cuts}, 
$|\delta_G(U_{z})|\leq |\delta_G(S_{z+1})|\cdot O(\log^{34}m)$. If the bad event $\event$ does not happen, then, from \Cref{obs: congestion square of internal routers}, 
$\optcrors(S_{z+1},\Sigma(S_{z+1}))+|E(S_{z+1})|\geq \frac{|\delta_G(S_{z+1})|^2}{\hat \eta}$,
where $\Sigma(S_{z+1})$ is the rotation system for graph $S_{z-1}$ induced by $\Sigma$. Therefore: 

\[
\dist(\oset,\oset')\leq |\delta_G(U_z)|^2\leq O(\log^{68}m)\cdot |\delta_G(S_{z+1})|^2\leq \hat \eta^2\cdot  \left (|\chi^*_{z+1}|+|E(S_{z+1})|\right ).
\]

	Assume now that $S_{z}\in \sset^{\bad}$. As before, $\dist(\oset,\oset')\leq |\delta_G(U_z)|^2$. 
	From 
	\Cref{cor: few edges crossing cuts}, $|\delta_G(U_{z})|\leq |\delta_G(S_{z})|\cdot O(\log^{34}m)$. If the bad event $\event$ does not happen, then, from \Cref{obs: congestion square of internal routers}, 
	$\optcrors(S_{z},\Sigma(S_{z}))+|E(S_{z})|\geq \frac{|\delta_G(S_{z})|^2}{\hat \eta}$,
	where $\Sigma(S_{z})$ is the rotation system for graph $S_{z}$ induced by $\Sigma$. Therefore: 
	
	\[
	\dist(\oset,\oset') \leq |\delta_G(U_z)|^2\leq O(\log^{68}m)\cdot |\delta_G(S_{z})|^2\leq \eta^2\cdot \left (|\chi^*_{z}|+|E(S_{z})|\right ).
	\]
	
We assume from now on that $S_z,S_{z+1}\in \sset^{\light}$.
In order to complete the proof of \Cref{claim: bound distance between rotations},
we will define, for every edge $e\in \delta_G(U_z)$, a curve $\gamma(e)$, such that all curves in the resulting set $\Gamma^*=\set{\gamma(e)\mid e\in \delta_G(U_z)}$ are in general position; each one of the curves originates at the image of  vertex $v^{**}$ in the drawing $\phi''_z$ of $G_z$; and each one of the curves  terminates at the image of vertex $u_z$ in the drawing $\phi''_z$. We will ensure that the order in which the curves in set $\Gamma^*$ enter the image of $v^{**}$ is precisely the ordering $\oset$ of their corresponding edges, while the order in which they enter the image of $u_z$ is precisely the ordering $\oset'$ of their corresponding edges. We will then bound the number of crossings between the curves in $\Gamma^*$, thereby bounding  $\dist(\oset,\oset')$. 

In order to define the set $\Gamma^*$ of curves, we define, for every edge $e\in \delta_G(U_z)$, a path $\tilde R(e)$ in graph $G_z$, that connects vertex $v^{**}$ to vertex $u_{z}$, and originates at edge $e$. 
 For each edge $e\in \delta_G(U_z)$, the curve $\gamma(e)$ is then obtained by slightly altering the image of the path $\tilde R(e)$ in the drawing $\phi''_z$, in order to ensure that all resulting curves in $\Gamma^*$ are in general position.
We start by defining the set $\tilde \rset=\set{\tilde R(e)\mid e\in \delta_G(U_z)}$ of paths.
	
\paragraph{Set $\tilde \rset=\set{\tilde R(e)\mid e\in \delta_G(U_z)}$ of paths.}

\begin{figure}[h]
	\centering
	\includegraphics[scale=0.12]{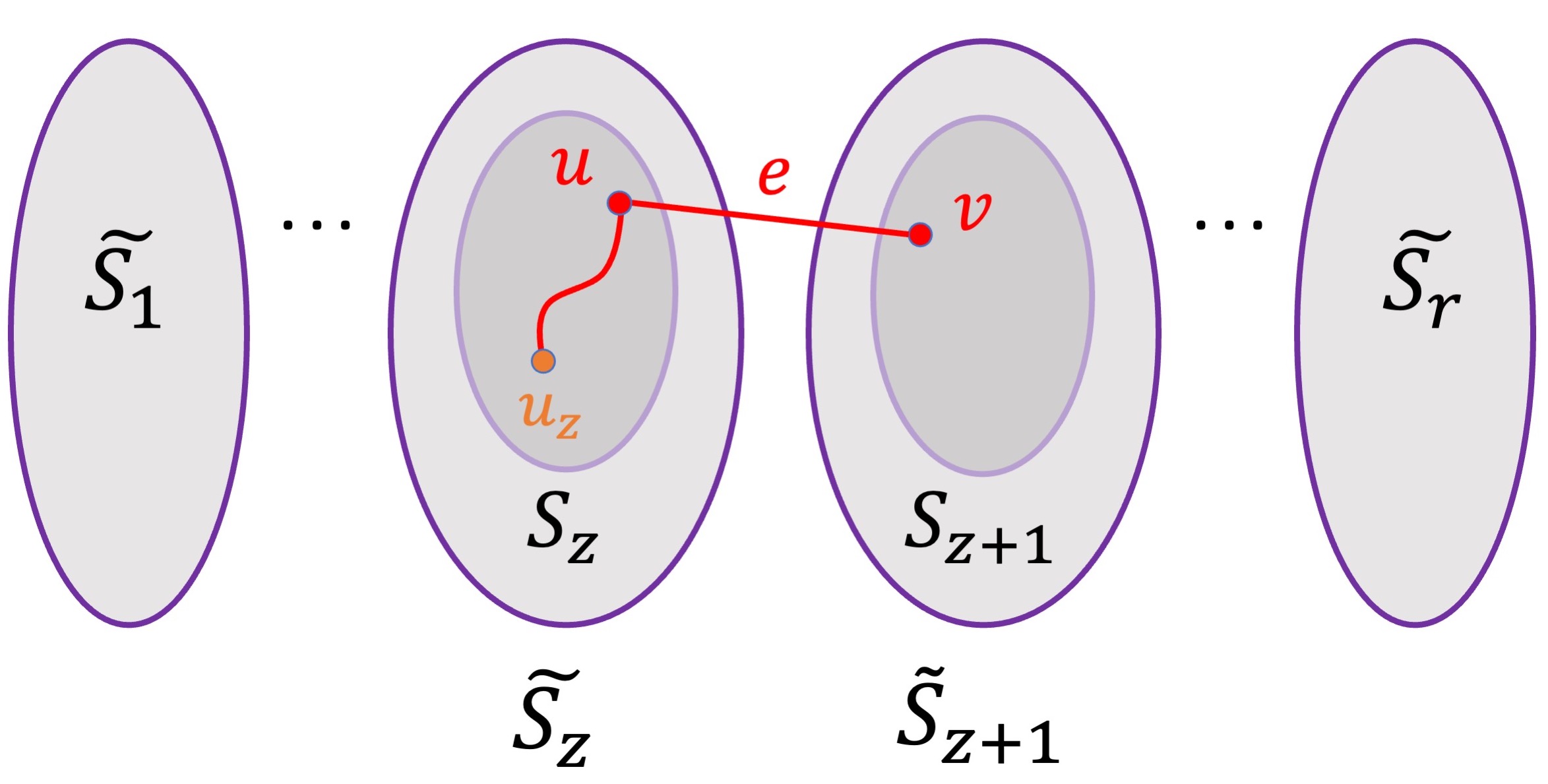}
	\caption{Definition of path $\tilde R(e)$ when $e\in E_z$.}\label{fig: NN4}
\end{figure}

Consider an edge $e\in \delta_G(U_z)$. Assume first that $e\in E_z$. Denote $e=(u,v)$, where $u\in S_z$ and $v\in S_{z+1}$ (see \Cref{fig: NN4}).
Note that edge $e$ belongs to graph $G_z$, where it connects vertex $u$ to vertex $v^{**}$. Let $\tilde Q(e)$ be the unique path of the internal $U_z$-router $\qset(U_z)$ that originates at edge $e$; recall that the path terminates at vertex $u_z$, and, from the construction of the path set $\qset(U_z)$, path $\tilde Q(e)$ is also the unique path of the internal $S_z$-router $\qset(S_z)$ that originates at edge $e$. Therefore, all internal vertices of path $\tilde Q(e)$ lie in $S_z$, and path $\tilde Q(e)$ is contained in graph $G_z$. We then let $\tilde R(e)$ be the path $\tilde Q(e)$ in graph $G_z$ (that now connects vertex $u_z$ to vertex $v^{**}$.)

\begin{figure}[h]
	\centering
	\includegraphics[scale=0.12]{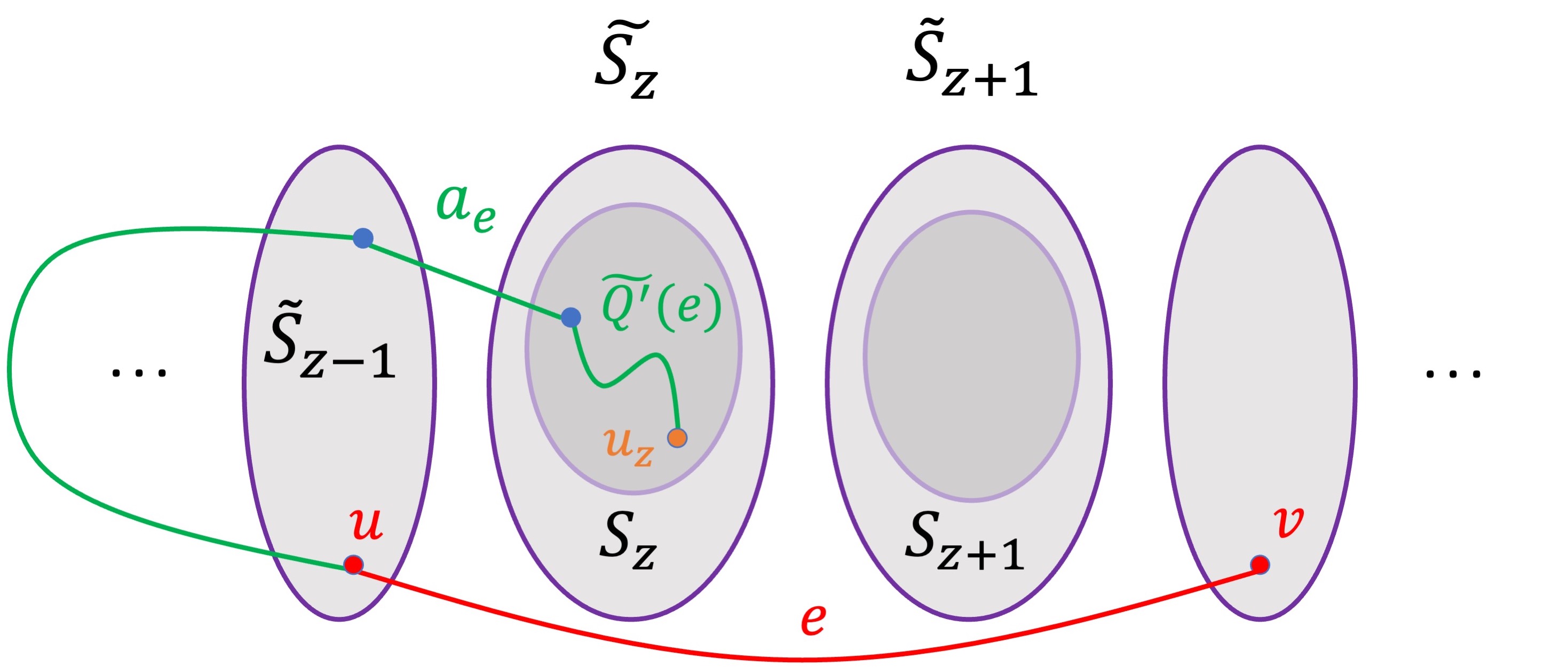}
	\caption{Construction of path $\tilde R(e)$ when $e\in E_z^{\through}$. Path $\tilde Q(e)$ is shown in green.}\label{fig: NN5}
\end{figure} 

Next, we consider an edge $e\in E^{\through}_z$. Assume that $e=(u,v)$, where $u\in \bigcup_{i<z}V(\tilde S_i)$ and $v\in \bigcup_{i>z}V(\tilde S_i)$ (see \Cref{fig: NN5}).
Consider the unique path $\tilde Q(e)\in \qset(U_z)$ that originates at edge $e$. Recall that the path terminates at vertex $u_z$, and it must contain some edge $a_e\in \delta_G(S_z)$ (in fact it may only contain one such edge). Note that, in graph $G_z$, all vertices of $U_{z-1}$ were contracted into vertex $v^*$, and so both edges $e$ and $a_e$ are incident to vertex $v^*$ in $G_z$. The subpath of the path $\tilde Q(e)$ from edge $a_e$ to vertex $u_z$ is precisely the unique path of the internal router $\qset(S_z)$ that originates at edge $a_e$, which we denote by $\tilde Q'(e)$. We then let $\tilde R(e)$ be the path obtained by appending the edge $e$ at the beginning of the path $\tilde Q'(e)$. Note that path $\tilde R(e)$ is contained in graph $G_z$, and it connects vertex $v^{**}$ to vertex $u_z$.
In fact, path $\tilde R(e)$ is a concatenation of edge $(v^*,v^{**})$ and path $\tilde Q'(e)$.

\begin{figure}[h]
	\centering
	\includegraphics[scale=0.12]{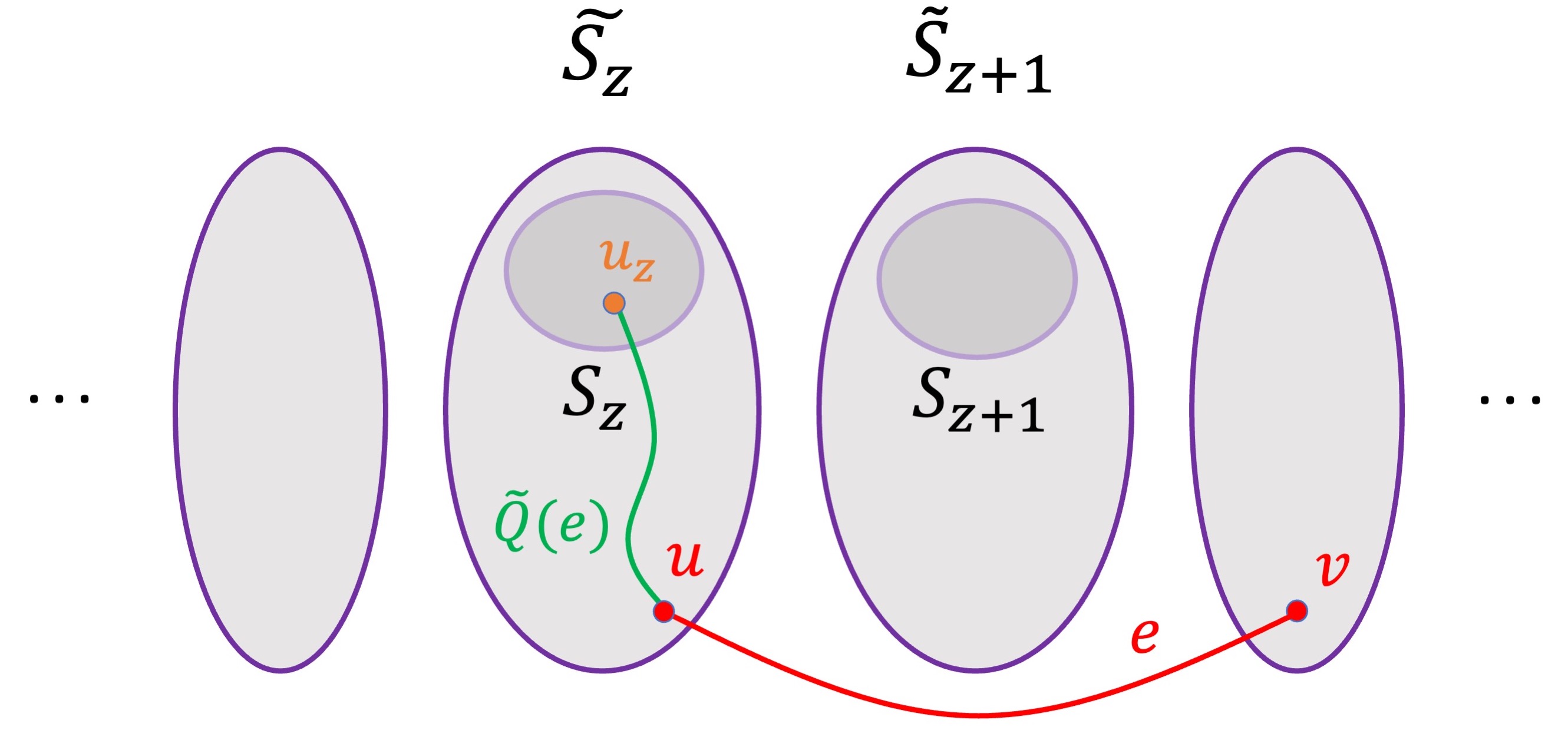}
	\caption{Construction of path $\tilde R(e)$ when $e\in E_z^{\rig}$ and $\tilde Q(e)\subseteq \tilde S_z$. Path $\tilde Q(e)$ is shown in green.}\label{fig: NN6a}
\end{figure}

\begin{figure}[h]
	\centering
	\includegraphics[scale=0.12]{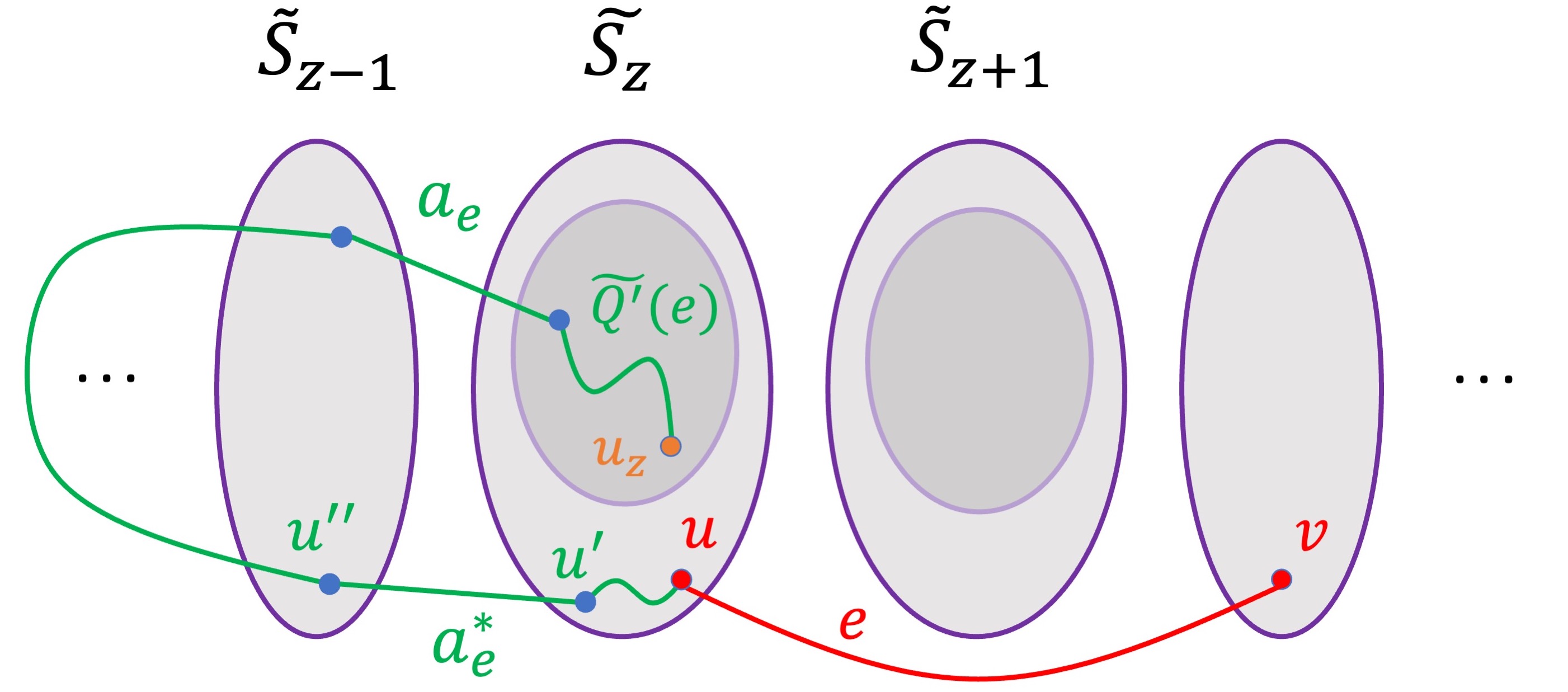}
	\caption{Construction of path $\tilde R(e)$ when $e\in E_z^{\rig}$ and $\tilde Q(e)$ is not contained in $\tilde S_z$. Path $\tilde Q(e)$ is shown in green.}\label{fig: NN6b}
\end{figure}

Lastly, we consider an edge $e\in E^{\rig}_z$. Assume that $e=(u,v)$, where $u\in V(\tilde S_z)$, and $v\in \bigcup_{i>z}V(\tilde S_i)$. Note that edge $e$ is also present in graph $G_z$, where it now connects vertex $u$ to vertex $v^{**}$. 
Consider the unique path $\tilde Q(e)\in \qset(U_z)$ that originates at edge $e$, and recall that this path terminates at vertex $u_z$.
We now consider two cases. First, if path $\tilde Q(e)$ is contained in cluster $\tilde S_z$, then it is contained in the current graph $G_z$, except that now it connects vertex $v^{**}$ to vertex $u_z$. We then set $\tilde R(e)=\tilde Q(e)$ (see \Cref{fig: NN6a}). We denote by $a_e$ the unique edge of $\tilde R(e)$ that lies in $\delta_G(S_z)$.
Otherwise, let $u''$ be the first vertex on path $\tilde Q(e)$ that does not belong to $\tilde S_z$, and let $u'$ be the vertex preceding $u''$ on the path (see \Cref{fig: NN6b}).
Denote $a^*_e=(u',u'')$. Note that edge $a^*_e$ also lies in graph $G_z$, where it connects vertex $u'$ to vertex $v^*$.  
Moreover, path $\tilde Q(e)$ must now contain some edge $a_e\in \delta_G(S_z)$. Since, in graph $G_z$, all vertices of $U_{z-1}$ were contracted into the vertex $v^*$, edge $a_e$ is now incident to vertex $v^*$. The subpath of the path $\tilde Q(e)$ from edge $a_e$ to vertex $u_z$, that we denote by $\tilde Q'(e)$, is precisely the unique path of the internal $S_z$-router $\qset(S_z)$ that originates at edge $a_e$. We then let $\tilde R(e)$ be the path obtained by concatenating the subpath of $\tilde Q(e)$ from edge $e$ to edge $a^*_e$ (that, in graph $G_z$, connects $v^{**}$ to $v^*$),  and the path $\tilde Q'(e)$ (that originates at $v^*$ in $G_z$). Note that path $\tilde R(e)$ is contained in graph $G_z$, and it connects vertex $v^{**}$ to vertex $u_z$.
Since $\delta_G(U_z)=E_z\cup E_z^{\through}\cup E_z^{\rig}$, we have now constructed a path $\tilde R(e)$ for each edge $e\in \delta_G(U_z)$.

Consider the final set $\tilde \rset=\set{\tilde R(e)\mid e\in \delta_G(U_z)}$ of paths that we have defined in graph $G_z$. Notice that, for each edge $e\in \delta_G(U_z)$, there is some edge $a_e\in \delta_G(S_z)$ that lies both on the path $\tilde \qset(e)\in \qset(U_z)$, and on path $\tilde R(e)$ (in the case where $e\in E_z$, we set $a_e=e$; for the other two cases, we have defined the edge $a_e$ explicitly). Moreover, the unique path of the internal $S_z$-router $\qset(S_z)$ that originates at edge $a_e$ is a subpath of $\tilde R(e)$. Therefore, the
last edge on path $\tilde R(e)$ is identical to the last edge on the unique path  $\tilde Q(e)\in \qset(U_z)$ that originates at edge $e$. Recall that we have defined the ordering $\oset=\tilde \oset_z$  of the edges of $\delta_G(U_z)=\delta_{G_z}(v^{**})$ to be
$\oset^{\guided}(\qset(U_z),\Sigma)$ -- the ordering that is guided by the internal $U_z$-router $\qset(U_z)$ (see definition in \Cref{subsec: guiding paths rotations}). Since the rotation $\oset_{u_z}$ in $\Sigma$ and $\Sigma_z$ is identical, equivalently, $\oset'=\oset^{\guided}(\tilde\rset,\Sigma_z)$, that is, ordering $\oset'$ can be defined as an ordering that is guided by the set $\tilde \rset$ of paths in graph $G_z$, with respect to the rotation system $\Sigma_z$.

Notice that for every edge $e\in \delta_{G}(U_z)$, an edge $e'$ may lie on path $\tilde R(e)$ only if $e'$ lies on path $\tilde Q(e)\in \qset(U_z)$. 
From our construction, if $e'\in \delta_{G_z}(U_z)$, then $e'$ may lie on at most one path of $\tilde \rset$ -- the path $\tilde R(e')$. From \Cref{obs: bound congestion of routers}, an edge $e'\in E(G_z)\setminus E(S_z)$ may participate in at most $O(\log^{34}m)$ paths of $\tilde \rset$, and an edge $e'\in E(S_z)$ may participate in at most $O(\log^{34}m)\cdot \cong_G(\qset(S_z),e')$ paths of $\tilde \rset$.

Next, we construct an auxiliary graph $H'_z$, by replicating some edges of $G_z$ and deleting some other edges, similarly to our construction of graph $H_z$. We will use the paths of $\tilde \rset$ in order to define a collection of edge-disjoint paths $\tilde \rset'$ in the resulting graph $H'_z$, which will in turn be used in order to construct the collection $\Gamma^*$ of curves.

\paragraph{Graph $H'_z$ and its drawing $\psi'_z$}
For every edge $e\in E(G_z)$, we let $\tilde N(e)$ be the number of paths in $\tilde \rset$ that contain the edge $e$.
From the discussion so far, we obtain the following immediate observation.

\begin{observation}\label{obs: bound tilde N}
	For each edge $e\in E(S_z)$, $\tilde N(e)\leq O(\log^{34}m)\cdot \cong_G(\qset(S_z),e')$, and for each edge $e\in E(G_z)\setminus E(S_z)$, $\tilde N(e)\leq O(\log^{34}m)$.	
\end{observation}
 In order to construct the graph $H'_z$, we start with the set $V(H'_z)=V(G_z)$ of vertices. For every edge $e\in E(G_z)$ with $\tilde N(e)>0$, we add a collection $J'(e)$ of $\tilde N(e)$ parallel copies of the edge $e$ to graph $H'_z$. We also assign each copy of edge $e$ in set $J'(e)$ to a distinct path of $\tilde \rset$ that contains the edge $e$, arbitrarily. As in Step 1 of the algorithm for computing a drawing of graph $G_z$, we can now define a collection $\tilde \rset'=\set{\tilde R'(e)\mid e\in \delta_G(U_z)}$ of edge-disjoint paths in graph $H'_z$, as follows: for each edge $e\in \delta_G(U_z)$, path $\tilde R'(e)$ is obtained from path $\tilde R(e)$ by replacing each edge $e'\in \tilde R(e)$ with the copy of edge $e'$ that is assigned to path $\tilde R(e)$. 

Drawing $\phi''_z$ of graph $G_z$ naturally defines drawing $\psi'_z$ of graph $H'_z$: for each edge $e\in E(G_z)$ with $\tilde N(e)>0$, we draw all copies of $e$ to appear in parallel to the image of $e$, without crossing each other.  

As in Step 1 of the algorithm for constructing a drawing for graph $G_z$, we can assign the copies of the edges incident to vertex $u_z$ more carefully, to ensure that the images of the paths in $\tilde \rset'$ enter the image of $u_z$ according to the ordering $\tilde \oset_z=\oset'$. In other words, if we denote $\delta_G(U_z)=\set{\tilde a'_1,\tilde a'_2,\ldots\tilde a'_h}$, and the edges are indexed in the order of their appearance in the ordering $\tilde \oset_z=\oset'$, and if, for all $1\leq i\leq h$, we denote by $\tilde a''_i$ the last edge on the path $\tilde R'(\tilde a'_i)$, then the images of the edges $\tilde a''_1,\ldots,\tilde a''_h$ enter the image of vertex $u_z$ in the natural order of their indices. Note that the images of edges $\tilde a_1',\ldots,\tilde a_h'$ enter the image of $v^{**}$ in the ordering $\oset$.

We now bound the number of crossings in the drawing $\psi'_z$ of graph $H'_z$. Consider any crossing in $\psi'_z$ between a pair of edges $e_1',e_2'$. Assume that $e_1'$ is a copy of edge $e_1\in E(G_z)$, and $e_2'$ is a copy of edge $e_2\in E(G_z)$. Clearly, the images of the edges $e_1$ and $e_2$ must cross in drawing $\phi_z''$, and we say that this crossing is responsible for the crossing $(e_1',e_2')$. It is easy to see that a crossing $(e_1,e_2)$ in drawing $\phi_z$ may be responsible for at most $\tilde N(e_1)\cdot \tilde N(e_2)$ crossings in $\psi'_z$. If neither of the edges $e_1,e_2$ lie in $E(S_z)$, then, from \Cref{obs: bound tilde N}, $\tilde N(e_1),\tilde N(e_2)\leq O(\log^{34}m)$. Therefore, the total number of crossings of $\psi_{z}'$, for which crossings $(e_1,e_2)$ of $\phi''_z$ with $e_1,e_2\not\in E(S_z)$ are responsible is at most:
$O(\log^{68}m)\cdot \cro(\phi''_z)$.
 
If exactly one of the two edges (say $e_1$) lies in $E(S_z)$, then, from \Cref{obs: bound tilde N}, $\tilde N(e_1)\leq O(\log^{34}m)\cdot \cong_G(\qset(S_z),e_1)$, while $\tilde N(e_2)\leq  O(\log^{34}m)$. Moreover, crossing $(e_1,e_2)$ must be a type-1 primary crossing of $\phi''_z$. Recall that the expected number of type-1 crossings in $\phi''_z$ is bounded by $|\chi^*_z|\cdot \hat\eta$.
Recall also that the random variable corresponding to the total number of type-1 primary crossings only depends on the random choices of the internal routers $\qset(S_{z-1})$ and $\qset(S_{z+1})$, and it is independent of the random choice of the internal router $\qset(S_{z})$. For each edge $e\in E(S_z)$, $\expect{\cong_G(\qset(S_z),e)}\leq \hat \eta$ (from \Cref{obs: congestion square of internal routers} and our assumption that $S_z\in \sset^{\light}$), and random variable $\cong_G(\qset(S_z),e)$ only depends on the selection of the internal $S_z$-router $\qset(S_z)$. Since the random variable representing the number of type-1 primary crossings of $\phi''_z$ is independent from the random variables $\set{\cong_G(\qset(S_z),e)}_{e\in E(S_z)}$, we get that the total expected number of crossings of $\psi_z''$, for which crossings $(e_1,e_2)$ of $\phi''_z$, with exactly one of $e_1,e_2$ lying in $E(S_z)$ are responsible, is at most: $|\chi^*_z|\cdot \hat \eta^{O(1)}$.

Lastly, assume that both edges $e_1,e_2\in E(S_z)$. Then, from \Cref{obs: bound tilde N}, $\tilde N(e_1)\leq O(\log^{34}m)\cdot \cong_G(\qset(S_z),e_1)$ and $\tilde N(e_2)\leq O(\log^{34}m)\cdot \cong_G(\qset(S_z),e_2)$. The number of crossigns in drawing $\psi_z'$ for which crossing $(e_1,e_2)$ is responsible is then bounded by:

\[\begin{split}& O(\log^{68}m)\cdot \cong_G(\qset(S_z),e_1)\cdot \cong_G(\qset(S_z),e_2)\\&\quad\quad\quad\quad\leq  O(\log^{68}m)\left(    (\cong_G(\qset(S_z),e_1))^2+ (\cong_G(\qset(S_z),e_2))^2   \right ) .
\end{split} \]

From \Cref{obs: congestion square of internal routers}, and since we have assumed that $S_z\in \sset^{\light}$, for every edge $e\in E(S_z)$, $\expect{\left (\cong_{G}(\qset(S_z),e)\right )^2}\le \hat \eta$. As before, random variable $\left (\cong_{G}(\qset(S_z),e)\right )^2$ only depends on the random selection of the internal $S_z$-router $\qset(S_z)$.
Clearly, the expected number of crossings of $\psi'_z$ for which crossing $(e_1,e_2)$ is responsible is at most  $O(\hat \eta^2)$.
Note also that crossing $(e_1,e_2)$ must a type-1 primary crossing of $\phi''_z$, and the expected number of such crossings is bounded by  $|\chi^*_z|\cdot \hat \eta$. As before, the random variable corresponding to the number of type-1 primary crossings of $\phi''_z$ does not depend on the selection of  the internal $S_z$-router $\qset(S_z)$. Therefore, the total expected number of crossings of $\psi_z''$, for which crossings $(e_1,e_2)$ of $\phi''_z$, with  $e_1,e_2\in E(S_z)$ are responsible is at most: $|\chi^*_z|\cdot \hat \eta^{O(1)}$.

Overall, the total expected number of crossings in drawing $\psi'_z$ of graph $H'_z$ is bounded by:

\[ \cro(\phi''_z)\cdot O(\log^{68}m)+|\chi^*_z|\cdot \hat \eta^{O(1)}.\]

\paragraph{Constructing the set $\Gamma^*$ of curves.}
For every edge $e\in \delta_G(U_z)$, we initially let $\gamma(e)$ be the image of the path $\tilde R'(e)$ in the drawing $\psi'_z$ of graph $H'_z$. From our construction, the curves in set $\Gamma^*=\set{\gamma(e)\mid e\in \delta_G(U_z)}$ all originate at the image of vertex $v^{**}$, and terminate at the image of vertex $u_z$ in $\psi'_z$. Moreover, from our construction, the order in which the curves of $\Gamma^*$ enter the image of $v^{**}$ is according to the ordering $\oset$ of the edges of $\delta_G(U_z)$, while the order in which the curves of $\Gamma^*$ enter the image of $u_z$ is according to the ordering $\oset'$ of the edges of $\delta_G(U_z)$. However, the curves of $\Gamma^*$ are not in general position, as a point $p$ may serve as an inner point on more than $2$ such curves; this, however, may only happen if $p$ is an image of some vertex $v\in V(G_z)\setminus\set{u_z,v^{**}}$. 
We will now ``nudge'' the curves in the vicinity of each such vertex to ensure that the resulting set of curves is in general position. The nudging procedure is identical to that we have employed in Step 2 (see \Cref{subsubsec: step 2}).

We process every vertex $v\in V(G_z)\setminus\set{u_z,v^{**}}$ one by one. Consider an iteration when any such vertex $v$ is processed. Let $A(v)\subseteq \delta_G(U_{z})$ be the set of all edges $e\in \delta_G(U_{z})$, such that curve $\gamma(e)$ contains the image of vertex $v$ (in $\psi'_z$). We denote $A(v)=\set{a_1,\ldots,a_k}$. Consider the tiny $v$-disc $D(v)=D_{\psi'_z}(v)$ in the drawing $\psi'_z$ of graph $H'_z$. For all $1\leq i\le k$, we let $s_i,t_i$ be the two points at which curve $\gamma(a_i)$ intersects the boundary of the disc $D(v)$. Note that all points $s_1,t_1,\ldots,s_k,t_k$ must be distinct. We use the algorithm from \Cref{claim: curves in a disc} in order to construct a collection $\set{\gamma'_1,\ldots,\gamma'_k}$ of curves, such that, for all $1\leq i\leq k$, curve $\gamma'_i$ has $s_i$ and $t_i$ as its endpoints, and is completely contained in $D(v)$. Recall that the claim ensures that, for every pair $1\leq i<j\leq k$ of indices, if the two pairs  $(s_i,t_i),(s_j,t_j)$ of points cross, then curves $\gamma_i,\gamma_j$ intersect at exactly one point; otherwise, curves $\gamma_i,\gamma_j$ do not intersect.
For all $1\leq i\leq k$, we modify the curve $\Gamma(a_i)$ as follows: we replace the segment of the curve between points $s_i,t_i$ with the curve $\gamma_i$. 

Once every vertex  $v\in V(G_z)\setminus\set{u_z,v^{**}}$ is processed, we obtain the final collection $\Gamma^*=\set{\gamma(e)\mid e\in \delta_G(U_z)}$ of curves, which are now in general position. The order in which these curves enter the images of vertices $u_{z}$ and $v^{**}$ did not change, but we may have added some new crossings over the course of this modification of the curves in $\Gamma^*$. For convenience, we say that a crossing between a pair of curves in $\Gamma^*$ is \emph{primary} if this crossing existed before this last modification, and otherwise it is called \emph{secondary}. For each point $p$ corresponding to a secondary crossing, there must be a vertex $v$, with $p\in D(v)$. 
	The expected number of all primary crossings remains unchanged, and is bounded by $\cro(\phi''_z)\cdot O(\log^{68}m)+|\chi^*_z|\cdot \hat \eta^{O(1)}$. We now bound the expected number of all secondary crossings.
	
	Consider a pair $e_1,e_2\in  \delta_G(U_z)$ of distinct edges, and assume that there is some vertex $v\in V(G_z)\setminus\set{u_z,v^{**}}$, such that the curves $\gamma(e_1),\gamma(e_2)$ cross at some point $p\in D(v)$. Using the same arguments as before, this may only happen in one of two cases: either (i) some edge $e\in \delta_{G_z}(v)$ lies on both $\tilde R(e_1)$ and $\tilde R(e_2)$; or (ii) paths $\tilde R(e_1),\tilde R(e_2)$ have a transversal intersection at vertex $v$. In the former case, we say that the crossing is a type-1 secondary crossing, and that edge $e$ is responsible for it, while in the second case we say that the crossing is a type-2 secondary crossing, and that the transversal intersection of paths  $\tilde R(e_1),\tilde R(e_2)$ at vertex $v$ is responsible for it. 
	
	Clearly, for every edge $e\in E(G_z)$, the total number of type-1 secondary intersections for which $e$ may be responsible is at most $(\tilde N(e))^2$. If $e\in \delta_{G_z}(v^{**})$, then $\tilde N(e)=1$, and $e$ may not be responsible for any type-1 secondary crossings. If $e\in E(G_z)\setminus (E(S_z)\cup \delta_G(v^{**}))$, then, from \Cref{obs: bound tilde N}, $\tilde N(e)\leq O(\log^{34}m)$. Otherwise, if $e\in E(S_z)$, then,  from \Cref{obs: bound tilde N}, $\tilde N(e)\leq O(\log^{34}m)\cdot \cong_G(\qset(S_{z}),e)$. 
	Moreover, since we have assumed that $S_z\in \sset^{\light}$, from \Cref{obs: congestion square of internal routers}, $\expect{\left (\cong_{G}(\qset(S_z),e)\right )^2}\le \hat \eta$.
	Overall, the total expected number of type-1 secondary crossings is bounded by: $\hat\eta\cdot |E(G_z)\setminus \delta_{G_z}(v^{**})|\leq \hat \eta \cdot (|E(\tilde S_z)|+|\delta_G(U_{z-1}|)$. Since, from \Cref{cor: few edges crossing cuts}, 
	$|\delta_G(U_{z-1})|\leq |\delta_G(S_{z})|\cdot O(\log^{34}m)$, and $\delta_G(S_z)\subseteq E(\tilde S_z)\cup \delta_G(\tilde S_z)$, we get that the total expected number of type-1 secondary crossings is at most:
$\hat \eta \cdot \left (|E(\tilde S_z)|+| \delta_G(\tilde S_z)|\right )$.	
	
We now turn to bound the number of type-2 secondary crossings. 
Consider any such crossing between a pair of curves $\gamma(e_1),\gamma(e_2)$, and assume that this crossing is charged to transversal intersection of the paths $\tilde R(e_1),\tilde R(e_2)$ at vertex $v$ with respect to $\Sigma_z$. We claim that in this case, $v=v^{*}$ must hold. Indeed, assume otherwise. As vertex $v^{**}$ may not serve as an inner vertex on any path of $\tilde \rset'$, it must be the case that $v\in V(\tilde S_z)$. If $v\in V(S_z)$, then there must be two paths $Q,Q'\in \qset(S_z)$, such that $Q\subseteq \tilde R(e_1)$ and $Q'\subseteq \tilde R(e_2)$. From the construction of the internal router $\qset(S_z)$, all paths in $\qset(S_z)$ are non-transversal with respect to $\Sigma$, so it is impossible that $Q$ and $Q'$ have a transversal intersection at $v$, and the same is true for paths $\tilde R(e_1)$ and $\tilde R(e_2)$. Otherwise, $v\in V(\tilde S_z)\setminus V(S_z)$. In this case, $v$ must be a vertex that lies on each of the two paths $\tilde Q(e_1),\tilde Q(e_2)$ of the internal router $\qset(U_z)$, and moreover, the two paths must have a transversal intersection at $v$. But that is impossible from \Cref{obs: inner non-transversal}. Therefore, $v=v^{*}$ must hold.

We denote by $\Pi$ the set of all pairs $(e_1,e_2)\in \delta_G(U_z)$ of edges, such that paths $\tilde R(e_1),\tilde R(e_2)$ have a transversal intersection at vertex $v^*$, with respect to $\Sigma_Z$. Clearly, the number of type-2 secondary crossings between the curves of $\Gamma^*$ is bounded by $|\Pi|$. 
We use the following claim, whose proof appears in \Cref{subsec: bound the Pi}, in order to bound $\expect{|\Pi|}$.

\begin{claim}\label{claim: bound on Pi}

\[\begin{split}
\expect{|\Pi|}&\leq \hat \eta^2\cdot \left(\sum_{e\in E(G)}\expect{N'_z(e)}+\sum_{(e,e')\in \chi^*} \left (\expect{N'_z(e)}+\expect{N'_z(e')} \right)\right )\\
&+\hat \eta^2\cdot \left (|E(S_{z-1})|+|E(\tilde S_z)|+|\delta_G(S_{z-1})|+|\chi^*_{z-1}|+|\chi^*_z|\right)+|\Pi_z^T|.
\end{split}\]

\end{claim}

 In order to complete the proof \Cref{claim: bound distance between rotations}, it now remains to bound the expected number of crossings between the curves of $\Gamma^*$.
Recall that the expected number of all primary crossings between the curves of $\Gamma^*$ is bounded by $\cro(\phi''_z)\cdot O(\log^{68}m)+|\chi^*_z|\cdot \hat \eta^{O(1)}$, while the expected number of type-1 secondary crossings is at most $\hat \eta \cdot \left (|E(\tilde S_z)|+| \delta_G(\tilde S_z)|\right )$. The expected number of type-2 secondary crossings is bounded by $\expect{|\Pi|}$. We conclude that, if $S_z,S_{z+1}\in \sset^{\light}$, then the expected number of all crossings between the curves in $\Gamma^*$ is bounded by:

\[\begin{split}  &\hat \eta^{O(1)}\cdot \left (\sum_{e\in E(G)}\expect{N'_z(e)}+\sum_{(e,e')\in \chi^*} \left (\expect{N'_z(e)}+\expect{N'_z(e')} \right)\right )\\
&\quad\quad\quad\quad+\eta^{O(1)}\cdot \left (|\chi^*_{z-1}|+|\chi^*_z|+|E(S_{z-1})|+|E(\tilde S_z)|+| \delta_G(\tilde S_z)|+|\delta_G(S_{z-1})|\right )\\
&\quad\quad\quad\quad+\hat \eta \cdot \cro(\phi''_z)+|\Pi_z^T|.
\end{split}
\]

In order to complete the proof of \Cref{claim: bound distance between rotations}, it now remains to prove \Cref{claim: bound on Pi}, which we do next.

\subsection{Proof of \Cref{claim: bound on Pi}}
	\label{subsec: bound the Pi}

Assume first that $S_{z-1}\in \sset^{\bad}$. Since we have assumed that Event $\event$ did not happen, from \Cref{obs: congestion square of internal routers}, $\optcrors(S_{z-1},\Sigma(S_{z-1}))+|E(S_{z-1})|\geq \frac{|\delta_G(S_{z-1})|^2}{\hat \eta}$, where $\Sigma(S_{z-1})$ is the rotation system for graph $S_{z-1}$ induced by $\Sigma$. We then get that $|\delta_G(S_{z-1})|^2\leq  \hat \eta\cdot (|\chi^*_{z-1}|+|E(S_{z-1})|)$.
On the other hand, if $(e_1,e_2)\in \Pi$, then paths $\tilde Q(e_1),\tilde Q(e_2)\in \qset(U_z)$ must each contain an edge of $\delta_G(S_{z-1})$. Since, from \Cref{obs: bound congestion of routers}, each edge of $\delta_G(S_{z-1})$ may appear on at most $O(\log^{34}m)$ paths of $\qset(U_{z-1})$, we get that $|\Pi|\leq O(\log^{68}m)\cdot |\delta_G(S_{z-1})|^2\leq  \hat \eta^2\cdot (|\chi^*_{z-1}|+|E(S_{z-1})|)$.
From now on we assume that  $S_{z-1}\in \sset^{\light}$.

Consider a pair of edges $(e_1,e_2)\in \Pi$. Note that both $\tilde R(e_1)$ and $\tilde R(e_2)$ must contain the vertex $v^*$, and $e_1,e_2\in \delta_G(U_z)$ must hold. Recall that $\delta_G(U_z)=E_z\cup E_z^{\rig}\cup E_z^{\through}$, and that, for each edge $e\in E_z$, path $\tilde R(e)$ may not contain vertex $v^*$. Therefore, $e_1,e_2\in E_z^{\rig}\cup E_z^{\through}$ must hold. 
Note that, for an edge $e\in E_z^{\rig}$, it is possible that path $\tilde R(e)$ does not contain the vertex $v^*$. For convenience, we denote by $E_z^{\rig'}$ the set of all edges $e\in E_z^{\rig}$ for which $v^*\in \tilde R(e)$.
We denote by $\Pi^1\subseteq \Pi$ the set of all pairs $(e_1,e_2)$ where at least one of the two edges $e_1,e_2$ lies in $E_z^{\rig'}$, and we denote by $\Pi^2=\Pi\setminus \Pi^1$. Clearly, for every pair $(e_1,e_2)\in \Pi^2$, $e_1,e_2\in E_z^{\through}$ must hold. 
We will now define, for each edge $e\in E_z^{\rig'}\cup E^{\through}_z$ three special edges $a^*_e,a_e$, and $\hat a_e$ associated with $e$, a new cycle $\hat W(e)$ in graph $G$, and some additional structures.

\begin{figure}[h]
	\centering
	\includegraphics[scale=0.12]{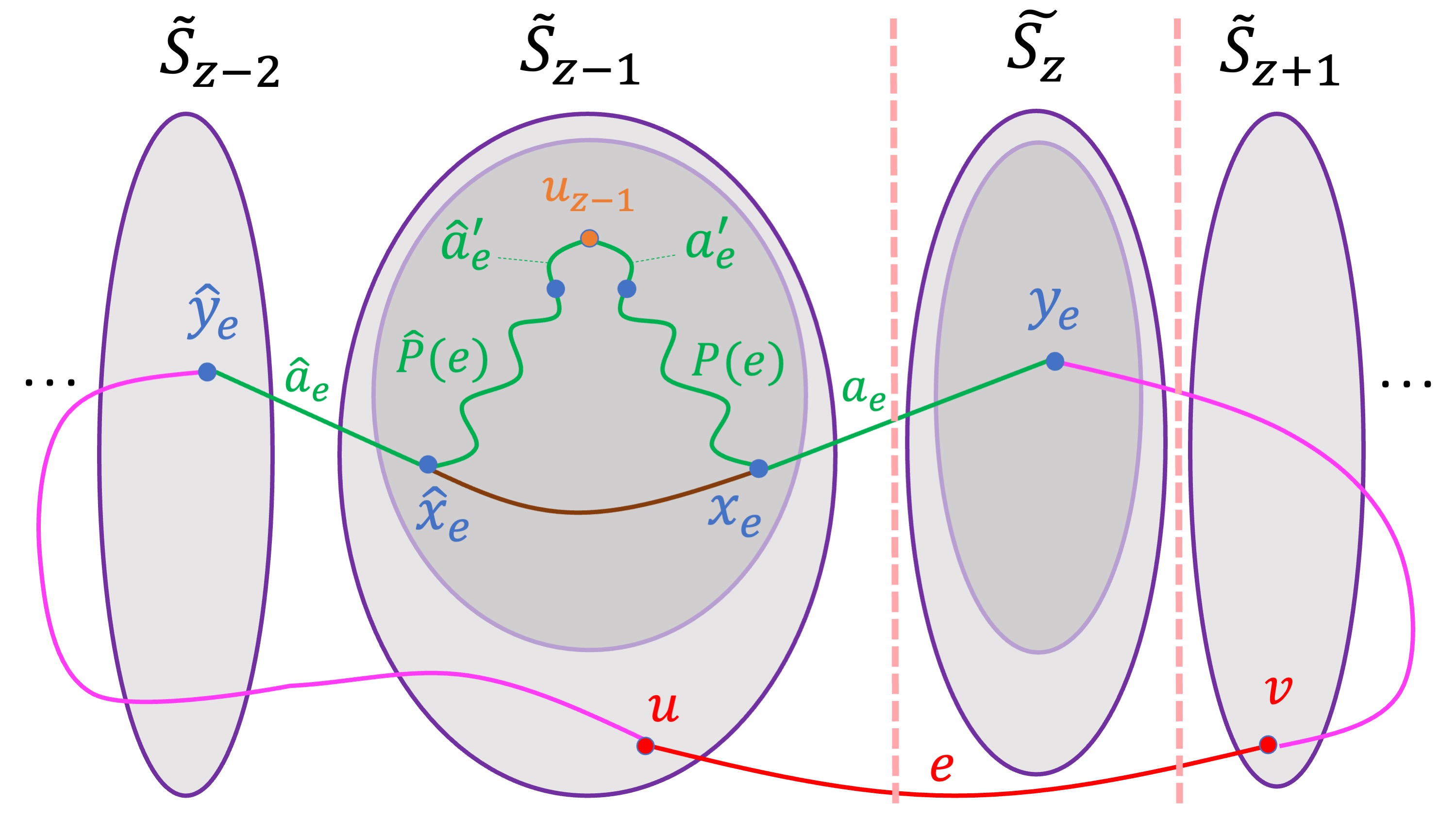}
	\caption{Definition of edges $a_e, \hat a_e$ and $a^*_e$ when edge $e\in E_z^{\through}$. Path $W'_e$ is the concatenation of the brown path and edges $a_e, \hat a_e$. Paths $\hat W^{\lef}(e)$ (connecting $e$ to $\hat y_e$) and $\hat W^{\rig}(e)$ (connecting $e$ to $y_e$) are shown in pink; both these paths also contain edge $e=a^*_e$.}\label{fig: NN7a}
\end{figure}

Consider first an edge $e\in E_z^{\through}$. Denote $e=(u,v)$, and assume that $u$ is the left endpoint of the edge. Then, from the definition of edge set $E_z^{\through}$, $u\in \bigcup_{i<z}V(\tilde S_i)$ and $v\in \bigcup_{i>z}V(\tilde S_i)$ must hold (see \Cref{fig: NN7a}).
In particular, $e\in \delta_G(U_{z-1})\cap \delta_G(U_z)$. We denote $a^*_e=e$. Clearly, $v^*\in \tilde R(e)$ must hold. We let $a_e$ be the edge immediately following vertex $v^*$ on path $\tilde R(e)$. Observe that $a_e$ is also an edge of graph $G$, where it must belong to edge set $E_{z-1}$ (see \Cref{fig: NN7a} and \Cref{fig: NN5}). 
We denote the endpoints of edge $a_e$ graph $G$ by $a_e=(x_e,y_e)$, with $x_e\in V(S_{z-1})$ and $y_e\in V(S_z)$.
Since edge $a_e$ lies on path $\tilde Q(e)$, that path visits the cluster $S_{z-1}$. We denote by $\hat x_e$ the first vertex on path $Q(e)$ that belongs to cluster $S_{z-1}$, by $\hat a_e$ the edge preceding vertex $\hat x_e$ on the path, and by $\hat y_e$ the other endpoint of edge $\hat a_e$ (see \Cref{fig: NN7a}). 
From the construction of the set $\qset(U_{z-1})$ of paths, and
the auxiliary cycles in $\wset$, edges $\hat a_e$ and $a_e$ must lie on the cycle $W(e)$. We denote by $W'(e)$ the subpath of the auxiliary cycle $W(e)$ that connects vertex $\hat y_e$ to vertex $y_e$, such that all inner vertices of $W'(e)$ lie in $S_{z-1}$.
We denote by $\hat W^{\lef}(e)$ the subpath of $W(e)$ from $\hat y_e$ to $v$, that is internally disjoint from $W'(e)$, and by $\hat W^{\rig}(e)$ the subpath of $W(e)$ from $u$ to $y_e$ that is internally disjoint from $W'(e)$.  Observe that edge $e$ lies on both $\hat W^{\rig}(e)$ and $\hat W^{\lef}(e)$.
Let $P(e)$ be the path of the internal $S_{z-1}$-router $\qset(S_{z-1})$ that originates at edge $a_e$, 
and let $\hat P(e)$ be the path of $\qset(S_{z-1})$ that originates at edge $\hat a_e$. 
We now define a new cycle $\hat W(e)$ associated with the edge $e$ to be the concatenation of the paths $\hat W^{\lef},\hat P(e),P(e)$, and $\hat W^{\rig}$ (after deleting the extra copy of the edge $e$, so that we obtain a cycle).

\begin{figure}[h]
	\centering
	\includegraphics[scale=0.12]{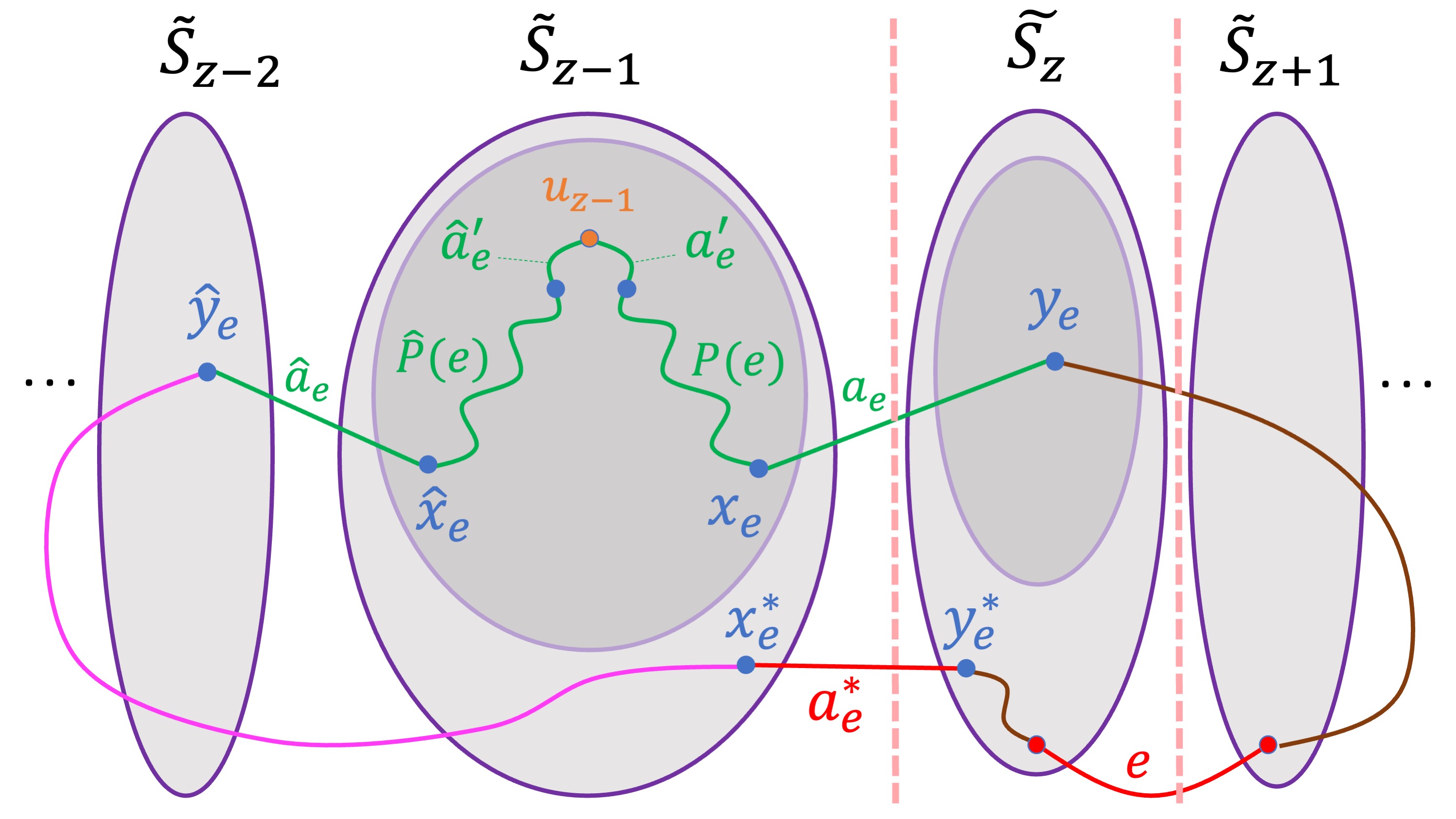}
	\caption{Definitions of edges $a_e,\hat a_e$ and $a^*_e$ when edge $e\in E_z^{\rig'}$. Path $\hat W^{\lef}(e)$ is the concatenation of the purple path and edge $a^*_e$. Path  $\hat W^{\rig}(e)$ is the concatenation of edges $a^*_e, e, a_e$ and the two brown paths.}\label{fig: NN7b}
\end{figure}

Next, we consider an edge  $e\in E_z^{\rig'}$. Denote $e=(u,v)$, and assume that $u$ is the left endpoint of the edge. Then, from the definition of edge set $E_z^{\rig}$, $u\in V(\tilde S_z)$ and $v\in \bigcup_{i>z}V(\tilde S_i)$ must hold (see \Cref{fig: NN7b}).
We let $a^*_e$ be the first edge on path $\tilde R(e)$ that is not contained in $E(\tilde S_z)$. Note that $a^*_e$ is also an edge of graph $G$, where it connects a vertex of $V(\tilde S_z)$, that we denote by $y^*_e$, to a vertex of $\bigcup_{i<z}V(\tilde S_i)$, that we denote by $x^*_e$. It is easy to see that edge $a^*_e$ must lie on the auxiliary cycle $W(e)$, and that it belongs to $\delta_G(U_{z-1})$, and more specifically to $E_z^{\lef}$.  We will say that edge $e$ \emph{owns} edge $a^*_e$, and that edge $a^*_e$ \emph{belongs} to edge $e$. Note that an edge of $E_z^{\lef}$ may belong to a number of edges of $E_z^{\rig'}$. 
Since edge $a^*_e$ lies on the auxiliary cycle $W(e)$, it must be the case that $z-1\in \spann''(e)$, and so cycle $W(e)$ must contain an edge of $E_{z-1}$, that we denote by $a_e$ (see \Cref{fig: NN7b}). 
As before, we denote the endpoints of edge $a_e$ in graph $G$ by $a_e=(x_e,y_e)$, with $x_e\in V(S_{z-1})$ and $y_e\in V(S_z)$. 
We also denote by $P(e)$ the unique path of the internal $S_{z-1}$-router $\qset(S_{z-1})$ that originates at edge $a_e$. We define the path $\hat W^{\rig}(e)$ to be the subpath of the auxiliary cycle $W(e)$, between vertices $x^*_e$ and $y_e$, that is disjoint from cluster $S_{z-1}$. Notice that this path contains both edges $a^*_e$ and $e$. Path $\hat P(e)$ and edge $\hat e_a$ are defined slightly differently. Recall again that $a^*_e\in \delta_G(U_{z-1})$. We consider the unique path $Q(a^*_e)$ of the internal $U_{z-1}$-router $\qset(U_{z-1})$ that originates at edge $a^*_e$. Recall that path $Q(a^*_e)$ terminates at vertex $u_{z-1}$, so it must contain some edge of $\delta_G(S_{z-1})$, and, from the definition of the internal router $\qset(U_{z-1})$, exactly one edge of $\delta_G(S_{z-1})$ lies on path $Q(a^*_e)$. We denote that edge by $\hat a_e=(\hat x_e,\hat y_e)$, where $\hat x_e$ is the endpoint of the edge that lies in $S_{z-1}$. We let $\hat W^{\lef}(e)$ be the subpath of $Q(a^*(e))$ from vertex $y^*_e$ to vertex $\hat y_e$. Observe that $a^*_e\in W^{\lef}(e)$; and that path $W^{\lef}(e)$ is a subpath of both the auxiliary cycle $W(a^*(e))$, and of path $W^{\out,\lef}(a^*(e))$. We denote by $\hat P(e)$ the unique path of the internal $S_{z-1}$-router $\qset(S_{z-1})$ that originates at edge $\hat a_e$. Lastly, we define a new cycle $\hat W(e)$ associated with edge $e$, to be the union of paths $\hat W^{\lef}(e),\hat P(e),P(e)$, and $\hat W^{\rig}(e)$, after we delete the extra copy of edge $a^*_e$ (see \Cref{fig: NN7b}). 

For consistency of notation, for an edge $e=(u,v)\in E_z^{\through}$, where $u$ is the left endpoint of $e$, we will also say that $e$ owns the edge $a^*_e=e$, and that edge $a^*_e$ belongs to $e$. We will also denote $x^*_e=u$ and $y^*_e=v$.

For every edge $e\in E^{\through}_z\cup E_z^{\rig'}$, we have now defined two paths $P(e),\hat P(e)\in \qset(S_{z-1})$. We denote by $a_e'$ the last edge on path $P(e)$, and by $\hat a_e'$ the last edge on path $\hat P(e)$; both these edges are incident to $u_{z-1}$ (see \Cref{fig: NN7a} and \Cref{fig: NN7b}). 
We now provide several observations that will be useful for us later.

\begin{observation}\label{obs: transversal pairs property}
	For every pair $(e_1,e_2)\in \Pi$, either edge set $\set{a'_{e_1},\hat a'_{e_1},a'_{e_2},\hat a'_{e_2}}$ contains fewer than four distinct edges, or edges $\hat a'_{e_1},\hat a'_{e_2},a'_{e_1},a'_{e_2}$ appear in this order in the rotation $\oset_{u_{z-1}}\in \Sigma$.
\end{observation}

\begin{proof}
	Since $(e_1,e_2)\in \Pi$, the paths $\tilde R(e_1),\tilde R(e_2)$ (that lie in graph $G_z$) have a transversal intersection at vertex $v^*$. Since vertex $v^*$ was obtained by contracting all vertices of $U_{z-1}$ in graph $G$, it is easy to verify that the edges of path $\tilde R(e_1)$ that immediately precede and follow vertex $v^*$ on the path are $a^*_{e_1}$ and $a_{e_1}$, respectively (see \Cref{fig: NN5} and \Cref{fig: NN6b}). 
	Similarly, 
	the edges of path $\tilde R(e_2)$ that immediately precede and follow vertex $v^*$ on the path are $a^*_{e_2}$ and $a_{e_2}$, respectively. 
	
	
Since the paths $\tilde R(e_1),\tilde R(e_2)$ have a transversal intersection at vertex $v^*$, edges $a^*_{e_1},a^*_{e_2},a_{e_1},a_{e_2}$ appear in this order in the circular ordering $\oset_{v^*}\in \Sigma_z$ (up to reversing the ordering). We now recall how the ordering $\oset_{v^*}\in \Sigma_z$ was constructed.

Recall that $\delta_{G_z}(v^*)=\delta_G(U_{z-1})$, and in particular, $a^*_{e_1},a_{e_1},a^*_{e_2},a_{e_2}\in \delta_G(U_{z-1})$. The ordering 
	$\oset_{v^*}\in \Sigma_z$ was defined to be identical to the ordering $\tilde \oset_{z-1}$ of the edges of $\delta_G(U_{z-1})$, which, in turn, is the ordering guided by the set $\qset(U_{z-1})$ of paths. 
	
	From our construction, it is immediate to verify that the last edge of the unique path in $\qset(U_{z-1})$ that originates at edge $a^*_{e_1}$ is $\hat a_{e_1}'$, and the last edge of the unique path in $\qset(U_{z-1})$ that originates at edge $a_{e_1}$ is $a_{e_1}'$ (see \Cref{fig: NN7a} and \Cref{fig: NN7b}). 
	Similarly, the last edge of the unique path in $\qset(U_{z-1})$ that originates at edge $a^*_{e_2}$ is $\hat a_{e_2}'$, and the last edge of the unique path in $\qset(U_{z-1})$ that originates at edge $a_{e_2}$ is $a_{e_2}'$. From the definition of the ordering $\tilde \oset_{z-1}$, it must be the case that either set $\set{a_{e_1},\hat a_{e_1},a_{e_2},\hat a_{e_2}}$ contains fewer than four distinct edges, or edges $\hat a'_{e_1},\hat a'_{e_2},a'_{e_1},a'_{e_2}$ appear in this order in the rotation $\oset_{u_{z-1}}\in \Sigma$.
\end{proof}

We will use the following simple observation in order to bound the congestion that is caused by the set $\hat \wset=\set{\hat W(e)\mid e\in E_z^{\through}\cup E_z^{\rig'}}$ of cycles.

\begin{observation}\label{obs: bound cong due to new cycles}
	Each edge $e\in E_z^{\lef}$ may belong to at most $O(\log^{34}m)$ edges of $E_z^{\rig'}$. Additionally, an edge $e\in E(G)\setminus E(S_{z-1})$ may lie on at most $O(\log^{68}m)$ cycles of $\hat \wset$, while an edge $e\in E(S_{z-1})$ may lie on at most $O(\log^{68}m)\cdot \cong_G(\qset(S_{z-1}),e)$ cycles of $\hat \wset$. Lastly, an edge $e\in E(G)\setminus \left ( E(\tilde S_z)\cup E(S_{z-1})\right )$ may lie on at most $O(\log^{34}m)\cdot N'_z(e)$ cycles of $\set{\hat W(e')\mid e'\in E_z^{\rig'}}$.
\end{observation}
\begin{proof}
	In order to prove the first assertion, consider any edge $e\in E_z^{\lef}$. From our construction, $e$ may belong to an edge $e'\in E_z^{\rig'}$ only if $e\in W(e')$. From \Cref{obs: bound congestion of cycles}, edge $e$ may lie on at most $O(\log^{34}m)$ cycles of $\wset$, and so $e$ may belong to at most $O(\log^{34}m)$ edges of $E_z^{\rig'}$.
	
	Consider now some edge  $e\in E(G)\setminus E(S_{z-1})$. Notice that, if $e$ lies on a cycle $\hat W(e')$ for some edge $e'\in E_z^{\rig'}\cup E_z^{\through}$, then either $e\in W(e')$, or $e\in W(a^*_{e'})$ must hold. Since, from  \Cref{obs: bound congestion of cycles}, edge $e$ may lie on at most $O(\log^{34}m)$ cycles of $\wset$, and, as we have shown, every edge $e''\in E_z^{\lef}$ may belong to at most $O(\log^{34}m)$ edges of $E_z^{\rig}$, we get that $e$ may lie on at most $O(\log^{68}m)$ cycles of $\hat \wset$.
	
	Consider now an edge $e\in  E(S_{z-1})$. Notice that, if $e$ lies on a cycle $\hat W(e')$ for some edge $e'\in E_z^{\rig'}\cup E_z^{\through}$, then either $e\in P(e')$, or $e\in \hat P(e')$ must hold. Consider some path $P\in \qset(S_{z-1})$ that contains the edge $e$, and let $a$ be the first edge on path $P$. Consider any edge $e'\in E_z^{\rig'}\cup E_z^{\through}$, for which $P=P(e')$ or $P=\hat P(e')$ holds. Then $a=a_{e'}$ or $a=\hat a_{e'}$ must hold, and in particular, edge $a$ must lie on $\hat W(e')$. As we have shown, every edge $e\in \delta_G(S_{z-1})$ may lie on at most $O(\log^{68}m)$ cycles of $\hat \wset$. Therefore, there are at most $O(\log^{68}m)$ edges $e'\in E_z^{\rig'}\cup E_z^{\through}$, for which $P=P(e')$ or $P=\hat P(e')$ holds. We conclude that $e$ 
	 may lie on at most $O(\log^{68}m)\cdot \cong_G(\qset(S_{z-1}),e)$ cycles of $\hat W$.
	 
	 It now remains to prove the last assertion. Consider an edge $e\in E(G)\setminus \left ( E(\tilde S_z)\cup E(S_{z-1})\right )$. Assume that $e\in \hat W(e')$ for some edge $e'\in E_z^{\rig'}$. From our construction, this may only happen if either $e$ lies on the unique path of $\qset(U_z)$ that originates at $e'$;  or $e$ lies on the  unique path of $\qset(U_{z-1})$ that originates at $a^*_{e'}$.
	 Recall that, in the latter case, $a^*_{e'}\in E_z^{\lef}$ must hold. Recall that  $N'_z(e)$ is the total number of paths in $\qset(U_{z-1})\cup \qset'(U_z)$ that originate at edges  of $E_{z-1}\cup E_z^{\lef}\cup E_z^{\rig}\cup E_z$ and contain $e$. 
	Since each edge $a^*\in E_z^{\lef}$ may belong to at most $O(\log^{34}m)$ edges of $E_z^{\rig'}$, we get that, overall, edge $e$ may lie on at most  $O(\log^{34}m)\cdot N'_z(e)$ cycles of $\set{\hat W(e')\mid e'\in E_z^{\rig'}}$.
\end{proof}

Recall that we have denoted by $\Pi^1\subseteq \Pi$ the set of all edge pairs $(e_1,e_2)\in \Pi$, where at least one of the two edges lies in $E_z^{\rig'}$. We will always assume w.l.o.g. that $e_1\in E_z^{\rig'}$ for each such pair. We bound the expected cardinalities of sets $\Pi^1$ and $\Pi^2$ separately in the following two claims.

\begin{claim}\label{claim: bound Pi1} 
	The expected cardinality of set $\Pi^1$ is at most:
	
	\[ \hat \eta^2\cdot \left(\sum_{e\in E(G)}\expect{N'_z(e)}+\sum_{(e,e')\in \chi^*}  \left (\expect{N'_z(e)}+\expect{N'_z(e')}\right )+|E(S_{z-1}|+|E(\tilde S_z)|+|\chi^*_{z-1}|+|\chi^*_z|\right )+ |\Pi^T_z|.\]
\end{claim}
\begin{proof}
  We denote by $\Pi^1_1\subseteq \Pi^1$ the set of all pairs $(e_1,e_2)\in \Pi^1$, for which cycles $\hat W(e_1),\hat W(e_2)$ share at least one edge. We let $e$ be any edge in $E(\hat W(e_1))\cap E(\hat W(e_2))$, and we say that $e$ is responsible for the pair $(e_1,e_2)$. 
Consider now any pair of edges $(e_1,e_2)\in \Pi^1\setminus \Pi^1_1$, and their two corresponding cycles $\hat W(e_1),\hat W(e_2)$. Note that the two cycles do not share edges, and, from \Cref{obs: transversal pairs property}, they have a transversal intersection at vertex $u_{z-1}$. Therefore, either there is a pair of edges $e_1'\in E(\hat W(e_1))$ and $e_2'\in E(\hat W(e_2))$ that cross in the drawing $\phi^*$ of $G$; or there is a vertex $v\neq u_{z-1}$, such that $\hat W(e_1),\hat W(e_2)$ have a transversal intersection at $v$. In the former case, we say that crossing $(e_1',e_2')$ is responsible for the edge pair $(e_1,e_2)$. In the latter case, we say that the transversal intersection of  $\hat W(e_1),\hat W(e_2)$ at $v$ is responsible for the pair $(e_1,e_2)$. We denote by $\Pi^1_2\subseteq \Pi^1\setminus \Pi^1_1$ the set of all pairs $(e_1,e_2)$, such that some crossing $(e_1',e_2')$ is responsible for $(e_1,e_2)$, and we denote by $\Pi^1_3=\Pi^1\setminus (\Pi^1_1\cup \Pi^1_2)$. We now bound the number of pairs in each one of the three sets one by one in the following three observations.

\begin{observation}\label{obs: bound first set of pairs}
	$\expect{|\Pi_1^1|}\leq \hat \eta^2\cdot \left (\sum_{e\in E(G)}\expect{N'_z(e)}+|E(S_{z-1}|+|E(\tilde S_z)|\right )$.
\end{observation}
\begin{proof}
	Consider an edge $e\in E(G)$. We will bound the expected number of  pairs $(e_1,e_2)\in \Pi^1_1$ of edges, for which edge $e$ is responsible. For each such pair, we assume w.l.o.g. that $e_1\in E_z^{\rig'}$. We distinguish between three cases.
	
	The first case is when $e\in E(G)\setminus \left ( E(\tilde S_z)\cup E(S_{z-1})\right )$.  From \Cref{obs: bound cong due to new cycles}, $e$ may lie on at most $O(\log^{34}m)\cdot N'_z(e)$ cycles of $\set{\hat W(e')\mid e'\in E_z^{\rig'}}$, and on at most $O(\log^{68}m)$ cycles of $\hat \wset$. Therefore,  such an edge may be responsible for at most $O(\log^{102}m)\cdot N'_z(e)\leq \hat \eta \cdot N'_z(e)$ edge pairs in $\Pi_1^1$.
	
	The second case is when $e\in E(\tilde S_z)$. In this case, from \Cref{obs: bound cong due to new cycles}, edge $e$ lies on at most $O(\log^{68}m)$ cycles of $\hat \wset$. Therefore, such an edge may be responsible for at most $O(\log^{136}m)\leq \hat \eta$ edge pairs in $\Pi_1^1$, and overall, the edges of $\tilde S_z$ may be responsible for at most $\hat \eta\cdot |E_G(\tilde S_z)|$ edge pairs in $\Pi^1_1$.
	
	The third and the last case is when $e\in E(S_{z-1})$. From \Cref{obs: bound cong due to new cycles}, such an edge may lie on at most $O(\log^{68}m)\cdot \cong_G(\qset(S_{z-1}),e)$ cycles of $\hat W$, and so it may be responsible for at most $O(\log^{138}m)\cdot \left(\cong_G(\qset(S_{z-1}),e)\right)^2\leq \hat \eta \cdot  \left(\cong_G(\qset(S_{z-1}),e)\right)^2$ edge pairs in $\Pi_1^1$. Since we have assumed that $S_{z-1}\in \sset^{\light}$, from  \Cref{obs: congestion square of internal routers}, $\expect{\left (\cong_{G}(\qset(S_{z-1}),e)\right )^2}\le \hat \eta$. Therefore, the expected number of edge pairs in $\Pi_1^1$ for which edge $e$  is responsible is at most $\hat \eta^2$, and the total expected number of edge pairs in $\Pi_1^1$ for which edges of $E(S_{z-1})$ are responsible is at most $\hat \eta^2\cdot |E(S_{z-1})|$. The bound now follows.
\end{proof}

\begin{observation}\label{obs: bound second set of pairs}
	$\expect{|\Pi_2^1|}\leq \hat \eta^2\cdot \left(\sum_{(e,e')\in \chi^*}   \left (\expect{N'_z(e)}+\expect{N'_z(e')}\right )+|\chi^*_{z-1}|+|\chi^*_z|\right )$.
\end{observation}

\begin{proof}
	Consider a crossing $(e,e')\in \phi^*$. We bound the number of pairs $(e_1,e_2)\in \Pi_2^1$ with $e_1\in E_z^{\rig'}$, for which the crossing $(e,e')$ is responsible. 
	Recall that, if crossing $(e,e')$ is responsible for a pair $(e_1,e_2)\in \Pi_2^1$, then $e\in \hat W(e_1)$ and $e'\in \hat W(e_2)$ must hold. 
	
	We first consider the case where neither of the edges $e,e'$ lie in $E(\tilde S_z)\cup E(S_{z-1})$. In this case, from \Cref{obs: bound cong due to new cycles}, edge $e'$ may lie on at most $O(\log^{68}m)$ cycles of $\hat \wset$, while edge $e$  may lie on at most $O(\log^{34}m)\cdot N'_z(e)$ cycles of $\set{\hat W(e')\mid e'\in E_z^{\rig'}}$. Therefore, crossing $(e,e')$ may be responsible for at most  $O(\log^{102}m)\cdot N'_z(e)\leq \hat \eta \cdot N'_z(e)$ edge pairs in $\Pi_2^1$. Note that, if either of the edges $e,e'$ lies in $E(\tilde S_z)\cup E(S_{z-1})$, then crossing $(e,e')$ belongs to $\chi^*_{z-1}\cup \chi^*_{z}$. We conclude overall, all crossings $(e,e')\in \chi^*\setminus (\chi^*_{z-1}\cup \chi^*_{z})$ may be responsible for at most $\sum_{(e,e')\in \chi^*}\hat \eta \cdot \left (N'_z(e)+N'_z(e')\right )$ pairs in $\Pi_2^1$.
	
	Next, we consider the case where at least one of the edges $e,e'$ lies in $E(\tilde S_z)\cup E(S_{z-1})$, so crossing $(e,e')$ belongs to $\chi^*_{z-1}\cup \chi^*_{z}$. We let $\hat N(e)$ be the random variable indicating the number of cycles in $\hat \wset$ containing edge $e$, and we define random variable $\hat N(e')$ for edge $e'$ similarly. Notice that random variables $\hat N(e),\hat N(e')$ may not be independent, if $e,e'\in E(S_{z-1})$. The total number of edge pairs in $\Pi^1_2$ for which crossing $(e,e')$ is responsible is bounded by $\hat N(e)\cdot \hat N(e')\leq (\hat N(e))^2+(\hat N(e'))^2$. From \Cref{obs: bound cong due to new cycles}, combined with \Cref{obs: congestion square of internal routers} and our assumption that $S_{z-1}\in \sset^{\light}$, we get that $\expect{(\hat N(e))^2}, \expect{(\hat N(e'))^2)}\leq \hat \eta^2$. We conclude that the expected number of pairs in $\Pi^1_2$ for which all crossings $(e,e')\in \chi^*_{z-1}\cup \chi^*_{z}$ are responsible is at most $\hat \eta^2\cdot (|\chi^*_{z-1}|+|\chi^*_z|)$.
\end{proof}

\begin{observation}\label{obs: bound third set of pairs}
	$|\Pi_3^1|\leq |\Pi^T_z|$.
\end{observation}
\begin{proof}
	Consider an edge pair $(e,e')\in \Pi^1_3$. Recall that there must be a vertex $v\neq u_{z-1}$, such that cycles $\hat W(e),\hat W(e')$ have a transversal intersection at $v$. Recall also that $e\in E_z^{\rig'}\subseteq E_z^{\rig}$. We claim that $v\in V(\notU_z)$ must hold, and moreover, auxiliary cycles $W(e),W(e')$ must have a transversal intersection at vertex $v$.
	
	Indeed, assume first that $v\in V(S_{z-1})\setminus\set {u_{z-1}}$. In this case, vertex $v$ must lie on $P(e)\cup \hat P(e)$, and on $P(e')\cup \hat P(e')$. Since paths $P(e),\hat P(e),P(e'),\hat P(e')$ all belong to the internal $S_{z-1}$-router $\qset(S_{z-1})$, they cannot have a transversal intersection at any vertex. 
	
	Assume now that $v\in V(U_{z-1})\setminus V(S_{z-1})$. In this case, by our construction, $v$ lies on the auxiliary cycle $W(a^*_e)$ of the edge $a^*_e\in E_z^{\lef}$ that belongs to $e$, and similarly, $v$ lies on the auxiliary cycle $W(a^*_{e'})$. 
Moreover, cycles $W(a^*_e),W(a^*_{e'})$ must have a transversal intersection at vertex $v$. 
	From \Cref{obs: auxiliary cycles non-transversal at at most one}, this is only possible if $v\in S_j$ for some index $1< j<r$, and either $j-1$ is the last index in both $\spann''(a^*_e),\spann''(a^*_{e'})$, or $j-1$ is the last index in one of these sets, while $j$ belongs to another. This is impossible, since $v\in V(U_{z-1})$, and $a^*_e,a^*_{e'}\in \delta_G(U_{z-1})$.

We conclude that vertex $v$ may not lie in $U_{z-1}$. But then, from the construction of the cycles $\hat W(e),\hat W(e')$, $v$ must lie on both the auxiliary cycles $W(e),W(e')$, 	and the two cycles must have a transversal intersection at $v$. From 
\Cref{obs: auxiliary cycles non-transversal at at most one}, this is only possible if $v\in S_j$ for some index $1< j<r$, and either $j-1$ is the last index in both $\spann''(e),\spann''(e')$, or $j-1$ is the last index in one of these sets, while $j$ belongs to another. Since $e,e'\in \delta_G(U_z)$, we conclude that $v\in V(\notU_z)$ must hold, and, from the above discusison, $W(e),W(e')$ must have a transversal intersection at vertex $v$. Recall that $\Pi^T_z$ is the set of triples $(\tilde e,\tilde e',\tilde v)$, where $\tilde e\in E_z^{\rig}$, $\tilde e'\in \hat E_z$, and  cycles $W(\tilde e)$ and $W(\tilde e')$ have a transversal intersection at $\tilde v$.  
We conclude that, if the transversal crossing of cycles  $\hat W(e),\hat W(e')$ at vertex $v$ is responsible for edge pair $(e,e')$, then triple $(e,e',v)$ must lie in $\Pi_z^T$, and so $|\Pi^1_3|\leq |\Pi^T_z|$.
\end{proof}

Combining the bounds from Observations \ref{obs: bound first set of pairs}--\ref{obs: bound third set of pairs}, we get that the expected cardinality of set $\Pi^1$ is at most:

\[ \hat \eta^2\cdot \left(\sum_{e\in E(G)}\expect{N'_z(e)}+\sum_{(e,e')\in \chi^*}  \left (\expect{N'_z(e)}+\expect{N'_z(e')}\right )+|E(S_{z-1}|+|E(\tilde S_z)|+|\chi^*_{z-1}|+|\chi^*_z|\right )+ |\Pi^T_z|.\]
\end{proof}

We use the following claim to bound the expected cardinality of $\Pi^2$.

\begin{claim}\label{claim: bound Pi2}
	$$\expect{|\Pi^2|}\leq \hat \eta^2\cdot \left(|E(S_{z-1})|+|\delta_G(S_{z-1})|+ |\chi^*_{z-1}|\right).$$
\end{claim}
\begin{proof}
Recall that set $\Pi^2$ contains all edge pairs $(e,e')\in \Pi$, with $e,e'\in E_z^{\through}$. Consider any edge $e\in E_z^{\through}$. Recall that we denoted by $W'(e)$ the subpath of the auxiliary cycle $W(e)$ between vertices $y_e$ and $\hat y_e$ that intersects cluster $S_{z-1}$. We denote by $W''(e)$ the subpath of $W(e)$ between vertices $y_e$ and $\hat y_e$ that does not share edges with $W'(e)$. Equivalently, $W''(e)$ is the concatenation of the paths $\hat W^{\lef}(e)$ and $\hat W^{\rig}(e)$ (after the extra copy of edge $e$ is deleted). We denote by $H(e)$ the 
 graph obtained from the union of the paths $W'(e),P(e)$ and $\hat P(e)$. 
The following observation, whose proof is deferred to \Cref{subsec: proof of bound transversal pairs} is central to the proof of \Cref{claim: bound Pi2}.

	\begin{observation}\label{obs: bound transversal pairs}
		Let $(e_1,e_2)$ be a pair of edges in $\Pi^2$. Then one of the following must happen: either (i) some edge lies in both $H(e_1)$ and $H(e_2)$; or (ii) there is a pair of edges $\tilde e_1\in H(e_1)$, $\tilde e_2\in H(e_2)\cup W''(e_2)$, whose images cross in the drawing $\phi^*$ of graph $G$; or (iii) 
		there is a pair of edges $\tilde e_1'\in H(e_1)\cup W''(e_1)$, $\tilde e_2'\in H(e_2)$,  whose images cross in the drawing $\phi^*$ of graph $G$.
	\end{observation}

As before, we partition the set $\Pi^2$ of edge pairs into two subsets. The first set, $\Pi^2_1$, containing all pairs $(e_1,e_2)\in \Pi^2$, such that there is an edge $e\in E(H_1)\cap E(H_2)$.  In this case, we say that edge $e$ is responsible for the pair $(e_1,e_2)$. Set $\Pi^2_2$ contains all remaining edge pairs $(e_1,e_2)\in \Pi^2$. From \Cref{obs: bound cong due to new cycles}, for each such pair $(e_1,e_2)\in \Pi^2_2$,  there must be a crossing in $\chi^*$ between a pair of edges
	$\tilde e_1$ and $\tilde e_2$, such that either (i) $\tilde e_1\in H(e_1)$ and $\tilde e_2\in H(e_2)\cup W''(e_2)$; or (ii) $\tilde e_1'\in H(e_1)\cup W''(e_1)$ and $\tilde e_2'\in H(e_2)$. For convenience, we will always assume that it is the former. Since $H(e_1)\subseteq S_{z-1}\cup \delta_G(S_{z-1})$, crossing $(e_1,e_2)$ must lie in $\chi^*_{z-1}$, and we say that this crossing is responsible for the pair $(e_1,e_2)\in \Pi$. 
	We bound the expected cardinalities of the sets $\Pi^2_1,\Pi^2_2$ separately, as before.	

In order to bound $\expect{|\Pi^2_1|}$, consider some edge $e\in E(S_{z-1})\cup \delta_G(S_{z-1})$. Note that edge $e$ may only lie in graph $H(e')$, for an edge $e'\in E_z^{\through}$, if $e\in W(e')$, or $e\in \hat W(e')$.  
	From \Cref{obs: bound congestion of cycles}, 
	edge $e$ may appear on at most $O(\log^{34}m)$ auxiliary cycles of $\wset$, and, from \Cref{obs: bound cong due to new cycles}, $e$ may lie on at most $O(\log^{68}m)\cdot \cong_G(\qset(S_{z-1}),e)$ cycles of $\hat W$. From   \Cref{obs: congestion square of internal routers}, since we have assumed that $S_{z-1}\in \sset^{\light}$, we get that  $\expect{\left (\cong_{G}(\qset(S_{z-1}),e)\right )^2}\le \hat \eta$. Therefore, the expected number of pairs in $\Pi^2_1$, for which edge $e$ is responsible is at most:
	
	\[O(\log^{136}m)\cdot \expect{(\cong_G(\qset(S_{z-1},e)))^2}\leq \hat \eta^2.  \]
	
	We conclude that $\expect{|\Pi^2_1|}\leq \hat \eta^2\cdot \left(|E(S_{z-1})|+|\delta_G(S_{z-1})|\right)$.

	In order to bound the expected cardinality of the set $\Pi_2^2$, consider some edge pair $(e_1,e_2)\in \Pi_2^2$, and the crossing $(\tilde e_1,\tilde e_2)$ that is responsible for it, where $\tilde e_1\in H(e_1)$ and $\tilde e_2\in H(e_2)\cup W''(e_2)$. Recall that $H(e_1)\subseteq  E(S_{z-1})\cup \delta_G(S_{z-1})$, and that crossing $(\tilde e_1,\tilde e_2)$ must lie in $\chi^*_{z-1}$. Consider now any crossing $(e,e')\in \chi^*_{z-1}$, and assume w.l.o.g. that $e\in  E(S_{z-1})\cup \delta_G(S_{z-1})$. 
	If $e'\in  E(S_{z-1})\cup \delta_G(S_{z-1})$ as well, then for every pair $(e_1,e_2)\in \Pi_2^2$ for which crossing $(e,e')$ is responsible, $e\in H(e_1)$ and $e'\in H(e_2)$ must hold. As observed above, the total number of edges $e_1\in  \delta_G(U_{z-1})$ with $e\in H(e_1)$ is $O(\log^{68}m)\cdot \cong_G(\qset(S_{z-1}),e)$, and similarly, the total number of edges $e_2\in  \delta_G(U_{z-1})$ with $e'\in H(e_2)$ is at most $O(\log^{68}m)\cdot \cong_G(\qset(S_{z-1}),e')$. Therefore, the total number of edge pairs $(e_1,e_2)\in \Pi_2^2$ for which crossing $(e,e')$ is responsible is bounded by:

	\[
	\begin{split}
	O(\log^{136}m)&\cdot \cong_G(\qset(S_{z-1}),e)\cdot \cong_G(\qset(S_{z-1}),e')\\ &\leq O(\log^{136}m)\cdot \left ( (\cong_G(\qset(S_{z-1}),e))^2+(\cong_G(\qset(S_{z-1}),e'))^2\right ).
	\end{split}
	\]

	As before, from \Cref{obs: congestion square of internal routers} and the assumption that $S_{z-1}\in \sset^{\light}$: $$\expect{\left (\cong_{G}(\qset(S_{z-1}),e)\right )^2}, \expect{\left (\cong_{G}(\qset(S_{z-1}),e')\right )^2}\le \hat \eta.$$
	Therefore, the expected number of edge pairs $(e_1,e_2)\in \Pi_2^2$ for which crossing $(e,e')$ is responsible is at most $\hat \eta^2$.
	
	Lastly, we consider a crossing $(e,e')\in \chi^*_{z-1}$ with $e\in  E(S_{z-1})\cup \delta_G(S_{z-1})$ and  $e'\not\in  E(S_{z-1})\cup \delta_G(S_{z-1})$. In this case, for every pair $(e_1,e_2)\in \Pi_2^2$ for which crossing $(e,e')$ is responsible, $e\in H(e_1)$, and $e'\in W''(e_2)\subseteq W(e_2)$ must hold. As before,  the total number of edges $e_1\in \delta_G(U_{z-1})$ with $e\in H(e_1)$ is at most $O(\log^{68}m)\cdot \cong_G(\qset(S_{z-1}),e)$, and, from \Cref{obs: bound congestion of cycles}, edge $e'$ appears on at most $O(\log^{34}m)$ cycles of $\wset$.
	Therefore, the total expected number of edge pairs $(e_1,e_2)\in \Pi_2^2$ for which crossing $(e,e')$ is responsible is bounded by:

	\[ O(\log^{102}m)\cdot \expect{\cong_G(\qset(S_{z-1}),e)}\leq \hat \eta^2, \]
	
	from \Cref{obs: congestion square of internal routers}. Overall, we get that $\expect{|\Pi_2^2|}\leq \eta^2\cdot |\chi^*_{z-1}|$, and:
	
	 $$\expect{|\Pi^2|}\leq \hat \eta^2\cdot \left(|E(S_{z-1})|+|\delta_G(S_{z-1})|+ |\chi^*_{z-1}|\right).$$
	\end{proof}

Combining the bounds from Claims \ref{claim: bound Pi1} and \ref{claim: bound Pi2}, we get that:

\[\begin{split}
\expect{|\Pi|}&\leq \hat \eta^2\cdot \left(\sum_{e\in E(G)}\expect{N'_z(e)}+\sum_{(e,e')\in \chi^*} \left (\expect{N'_z(e)}+\expect{N'_z(e')} \right)\right )\\
&+\hat \eta^2\cdot \left (|E(S_{z-1})|+|E(\tilde S_z)|+|\delta_G(S_{z-1})|+|\chi^*_{z-1}|+|\chi^*_z|\right)+|\Pi_z^T|,
\end{split}\]

completing the proof of \Cref{claim: bound on Pi}.

\section{Proof of \Cref{thm: not well connected}}
\label{sec: not well connected}

We assume that we are given a wide instance  $I=(G,\Sigma)$ of \cnwrs, with $m=|E(G)|$, such that $\mu^{20}\leq m\leq m^*$.
The high-level idea of the proof is to compute a collection $\cset$ of disjoint clusters in graph $G$ that have the $\alpha_0$-bandwidth property for $\alpha_0=1/\log^3m$ with $|E(G_{|\cset})|$ sufficiently small, and then to apply the algorithm for computing advanced disengagement from \Cref{thm: disengagement - main} to $\cset$, to obtain a $\nu$-decomposition of $I$ into subinstances. In order to ensure that all subinstances have the required properties, we need to ensure that, if $C$ is a cluster of $\cset$ with $|E(C)|> \frac{m}{\mu}$, then for any pair $u,v$ of disctinct vertices of $C$ whose degree in $C$ is at least $\frac m {\mu^6}$, there are at least $\frac{8m}{\mu^{50}}$ edge-disjoint paths connecting $u$ to $v$ in $C$.
We start with the following lemma that allows us to compute the desired collection $\cset$ of clusters.

\begin{lemma}\label{lem: initial clusters}
	There is an efficient algorithm, that, given  a wide instance  $I=(G,\Sigma)$ of \cnwrs, with $m=|E(G)|$, such that $\mu^{20}\leq m\leq m^*$, computes a collection $\cset$ of disjoint clusters of $G$, that have the following properties:
	
	\begin{itemize}
		\item $\bigcup_{C\in \cset}V(C)=V(G)$; 
		\item every cluster $C\in \cset$ has the $\alpha_0$-bandwidth property, for $\alpha_0=\frac 1 {\log^3m}$;
		\item for every cluster $C\in \cset$, for every pair $u,v$ of distinct vertices of $C$ with $\deg_G(v),\deg_G(u)\geq \frac{m}{\mu^6}$, there is a collection of at least $\frac{8m}{\mu^{50}}$ edge-disjoint paths in $C$ connecting $u$ to $v$; and
		\item $|\Eout(\cset)|\leq \frac{m}{\mu^{27}}$.
	\end{itemize}
\end{lemma}
\begin{proof}
	The proof of the lemma uses somewhat standard techniques and is similar, for example, to the proof of \Cref{thm:well_linked_decomposition}. We denote by $U$ the set of all vertices of $G$ whose degree is at least $\frac{m}{\mu^6}$. Clearly, $|U|\leq 2\mu^6$.
	Our algorithm maintains a collection $\rset$ of clusters of $G$. Throughout the algorithm, we ensure that the following invariants hold:

	\begin{properties}{I}
		\item all clusters in $\rset$ are mutually disjoint; and \label{inv1: disjointness2}
		\item  $\bigcup_{R\in \rset}V(R)=V(G)$.  \label{inv2: partition2}
	\end{properties}
	
	For a given collection $\rset$ of clusters with the above properties, we define a \emph{budget} $b(e)$ for every edge $e\in E(G)$, as follows. If both endpoints of $e$ lie in the same cluster of $\cset$, then we set the budget $b(e)=0$. Assume now that the endpoints of $e$ lie in different clusters $R,R'\in \rset$.
	We define $b_R(e)=1+8\alpha_0\cdot \alphasc(m)\cdot \log_{3/2}(|\delta_G(R)|)$, $b_{R'}(e)=1+8\alpha_0\cdot \alphasc(m)\cdot \log_{3/2}(|\delta_G(R')|)$, and $b(e)=b_R(e)+b_{R'}(e)$.  Notice that $b(e)\leq 3$ always holds.
	
	For every cluster $R\in \rset$, we denote $U_R=U\cap V(R)$. For every vertex $u\in U$, we define a budget $b(u)$ of $u$ as follows. Assume that $u\in V(R)$ for some cluster $R\in \rset$. Then $b(u)=4|U_R|\cdot \frac{m}{\mu^{40}}$. 
	We denote by $B=\sum_{e\in E(G)}b(e)+\sum_{u\in U}b(u)$ the \emph{total budget in the system}. Clearly, throughout the algorithm, $B\geq \sum_{R\in \rset}|\delta_G(R)|$ holds. 
	
	At the beginning of the algorithm, we set $\rset=\set{G}$. Clearly, both invariants hold for $\rset$. Moreover, the budget of every vertex $u\in U$ is at most $4|U|\cdot \frac{m}{\mu^{40}}$, so the total budget of all vertices in $U$ is at most $4|U|^2\cdot \frac{m}{\mu^{40}}\leq \frac{16m}{\mu^{28}}<\frac{m}{\mu^{27}}$ (since $|U|\leq 2\mu^6$), while the budget of every edge of $G$ is $0$. Therefore, at the beginning of the algorithm, $B\leq \frac{m}{\mu^{27}}$ holds. We will ensure that, throughout the algorithm, the total budget $B$ never increases. Since $B\ge \sum_{R\in \rset}|\delta_G(R)|$ always holds, this will ensure that, at the end of the algorithm, 
	$\sum_{R\in \rset}|\delta_G(R)|\leq \frac{m}{\mu^{27}}$ will hold.

	Throughout the algorithm, we maintain a partition of the set $\rset$ of clusters into two subsets: set $\rset^A$ of \emph{active} clusters, and set $\rset^I$ of \emph{inactive} clusters. We will ensure that the following additional invariant holds:
	
	\begin{properties}[2]{I}
		\item every cluster $R\in \rset^I$ has the $\alpha_0$-bandwidth property; and \label{inv: last - bw2}
		\item for every cluster $R\in \rset^I$, for every pair $u,v$ of distinct vertices of $U_R$, there is a collection of at least $\frac{8m}{\mu^{50}}$ edge-disjoint paths in $R$ connecting $u$ to $v$.
	\end{properties}
	
 At the beginning of the algorithm, we set $\rset^A=\rset=\set{G}$ and $\rset^I=\emptyset$. Clearly, all invariants hold then. We then proceed in iterations, as long as $\rset^A\neq \emptyset$.
	
In order to execute an iteration, we select an arbitrary cluster $R\in \rset^A$ to process. We will  either establish that $R$ has the $\alpha_0$-bandwidth property in graph $G$, and that for every pair $u,v\in U_R$ of distinct vertices there is a collection of at least  $\frac{8m}{\mu^{50}}$ edge-disjoint paths in $R$ connecting $u$ to $v$ (in which case $R$ is moved from $\rset^A$ to $\rset^I$); or we will modify the set $\rset$ of clusters in a way that ensures that the total budget decreases by at least $1/m$. An iteration that processes a cluster $R\in \rset^A$ consists of two steps. The purpose of the first step is to either establish the $\alpha_0$-bandwidth property of cluster $R$, or to replace it with a collection of smaller clusters in $\rset^A$. The purpose of the second step is to either establish that, for every pair $u,v\in U_R$ of distinct vertices there is a collection of at least  $\frac{8m}{\mu^{50}}$ edge-disjoint paths in $R$ connecting $u$ to $v$, or to modify the set $\rset$ of clusters in a way that decreases the total budget by at least $1/m$. We now describe each of the two steps in turn.

\paragraph{Step 1: Ensuring the Bandwidth Property.}
	
Let $R^+$ be the augmentation of the cluster $R$ in graph $G$. Recall that $R^+$ is a graph that is obtained from $G$ through the following process. First, we subdivide every edge $e\in \delta_G(R)$ with a vertex $t_e$, and we let $T=\set{t_e\mid e\in \delta_G(R)}$ be the resulting set of vertices. We then let $R^+$ be the subgraph of the resulting graph induced by vertex set $V(R)\cup T$. We apply  Algorithm  \algsc  for computing approximate sparsest cut to graph $R^+$, with the set $T$ of vertices, to obtain a $\alphasc(m)$-approximate sparsest cut $(X,Y)$ in graph $R^+$ with respect to vertex set $T$. 
	We now consider two cases. The first case happens if $|E_R(X,Y)|\geq \alpha_0\cdot \alphasc(m)\cdot \min\set{|X\cap T|,|Y\cap T|}$. In this case, we are guaranteed that the minimum sparsity of any $T$-cut in graph $R^+$ is at least $\alpha_0$, or equivalently, set $T$ of vertices is $\alpha_0$-well-linked in $R^+$. From \Cref{obs: wl-bw}, cluster $R$ has the $\alpha_0$-bandwidth property in graph $G$. In this case, we proceed to the second step of the algorithm.
	
	Assume now that $|E_R(X,Y)|< \alpha_0\cdot \alphasc(m)\cdot \min\set{|X\cap T|,|Y\cap T|}$. 
	Since $\alpha_0=1/\log^3m$, and $m$ is larger than a large enough constant (because $m\geq \mu^{20}$), and since $\alphasc(m)=O(\sqrt{\log m})$,  we get that the sparsity of the cut $(X,Y)$ is less than $1$.  Consider now any vertex $t\in T$, and let $v$ be the unique neighbor  of $t$ in $R^+$. We can assume w.l.o.g. that either $t,v$ both lie in $X$, or they both lie in $Y$. Indeed, if $t\in X$ and $v\in Y$, then moving vertex $t$ from $X$ to $Y$ does not increase the sparsity of the cut $(X,Y)$. This is because, for any two real numbers $1\leq a<b$, $\frac{a-1}{b-1}\leq \frac a b$. Similarly, if $t\in Y$ and $v\in X$, then moving $t$ from $Y$ to $X$ does not increase the sparsity of the cut $(X,Y)$. Therefore, we assume from now on, that for every vertex $t\in T$, if $v$ is the unique neighbor of $t$ in $R^+$, then either both $v,t\in X$, or both $v,t\in Y$.
	
	Consider now the partition $(X',Y')$ of $V(R)$, where $X'=X\setminus T$ and $Y'=Y\setminus T$. It is easy to verify that $|\delta_G(R)\cap \delta_G(X')|=|X\cap T|$, and $|\delta_G(R)\cap \delta_G(Y')|=|Y\cap T|$. Let $E'=E_G(X',Y')$, and assume w.l.o.g. that $|\delta_G(R)\cap \delta_G(X')|\leq |\delta_G(R)\cap \delta_G(Y')|$. Then $|E'|< \alpha_0\cdot \alphasc(m)\cdot |\delta_G(R)\cap \delta_G(X')|$ must hold. We remove cluster $R$ from sets $\rset$ and $\rset^A$, and we add instead every connected component of graphs $G[X']$ and $G[Y']$ to both sets. It is immediate to verify that $\rset$ remains a collection of disjoint clusters of $G$, and that $\bigcup_{R'\in \rset}V(R')=V(G)$. Therefore, all invariants continue to hold. We now show that the total budget $B$ decreases by at least $1/m$ as the result of this operation.

	Note that the only edges whose budgets may change as the result of this operation are edges of $\delta_G(R)\cup E'$. Observe that, for each edge $e\in \delta_G(R)\cap \delta_G(Y')$, its budget $b(e)$ may not increase. Since we have assumed that $|\delta_G(R)\cap \delta_G(X')|\leq |\delta_G(R)\cap \delta_G(Y')|$, and since $|E'|<|\delta_G(R)|/8$, we get that $|\delta_G(X')|\leq 2|\delta_G(R)|/3$. Therefore, for every edge $e\in \delta_G(X')\cap \delta_G(R)$, its budget $b(e)$ decreases by at least $8\alpha_0 \cdot \alphasc(m)\cdot\log_{3/2}(|\delta_G(R)|)-8\alpha_0 \cdot \alphasc(m)\cdot\log_{3/2}(|\delta_G(X')|)$. Since $|\delta_G(X')|\leq 2|\delta_G(R)|/3$, we get that $ \log_{3/2}(|\delta_G(R)|)\geq \log_{3/2}(3|\delta_G(X')|/2)\geq 1+\log_{3/2}(|\delta_G(X')|$. We conclude that the budget $b(e)$ of each edge $e\in \delta_G(X')\cap \delta_G(R)$ decreases by at least $8\alpha_0\cdot \alphasc(m)$.
	On the other hand, the budget of every edge $e\in E'$ increases by at most $3$. Since $|E'|\leq \alpha_0\cdot \alphasc(m)\cdot |\delta_G(R)\cap \delta_G(X')|$, we get that the decrease in the budget $B$ is at least:
	\[
	\begin{split}
	&8\alpha_0\cdot \alphasc(m)\cdot |\delta_G(X')\cap \delta_G(R)|-3|E'|\\&\hspace{3cm}\geq 8\alpha_0\cdot \alphasc(m)\cdot |\delta_G(X')\cap \delta_G(R)|- 3\alpha_0\cdot \alphasc(m)\cdot |\delta_G(R)\cap \delta_G(X')| 
	\\&\hspace{3cm}\geq 5 \alpha_0\cdot \alphasc(m)\cdot |\delta_G(R)\cap \delta_G(X')|\\&\hspace{3cm}>1/m,\end{split}\]
	since $\alpha_0\geq 1/m$.
		Therefore, the total budget of all edges decreases by at least $1/m$. Since the clusters only become smaller, it is easy to verify that the budgets of the vertices of $U$ do not increase.
	To conclude, if $|E_R(X,Y)|< \alpha_0\cdot \alphasc(m)\cdot \min\set{|X\cap T|,|Y\cap T|}$, then we have modified the set $\rset$ of clusters, so that all invariants continue to hold, and the total budget $B$ decreases by at least $1/m$. In this case, we terminate the current iteration.
	
	From now on we assume that $|E(X,Y)|> \alpha_0\cdot \alphasc(m)\cdot \min\set{|X\cap T|,|Y\cap T|}$, which, as observed already, implies that cluster $R$ has the $\alpha_0$-bandwidth property. We now proceed to describe the second step of the algorithm.

\paragraph{Step 2: Ensuring Connectivity of Vertices of $U$.}
If, for every pair $u,v\in U_R$ of distinct vertices, there is a collection of at least $\frac{8m}{\mu^{50}}$ edge-disjoint paths in $R$ connecting $u$ to $v$, then we move cluster $R$ from $\rset^A$ to $\rset^I$ and terminate the current iteration. It is easy to verify that all invariants continue to hold.

Assume now that there is a pair $u,v\in U_R$ of distinct vertices, such that the largest collection of edge-disjoint paths in graph $R$ connecting $u$ to $v$ contains fewer than $\frac{8m}{\mu^{50}}$ paths. From the max-flow /  min-cut theorem, there is a cut $(X,Y)$ of $R$, with $u\in X$, $v\in Y$, and $|E_R(X,Y)|<\frac{8m}{\mu^{50}}$. We assume w.l.o.g. that $|X\cap U_R|\leq |Y\cap U_R|$. We delete cluster $R$ from $\rset$ and from $\rset^A$, and we add instead every connected component of $G[X]$ and $G[Y]$ to both sets. We now show that the total budget decreases by at least $1/m$ as the result of this procedure.

Notice that for every vertex $x\in U$, the budget of $x$ did not increase. Moreover, if $x$ is a vertex of $X\cap U_R$, then its original budget was $4|U_R|\cdot \frac{m}{\mu^{40}}$, and its new budget is $4|U\cap X|\cdot \frac{m}{\mu^{40}}\leq 2|U_R|\cdot \frac{m}{\mu^{40}}$. Since $U\cap X\neq\emptyset$, we get that $\sum_{x\in U}b(x)$ decreased by at least $\frac{2m}{\mu^{40}}$.

Next, we consider the changes to the budgets of the edges. First, every edge in set $E'=E_R(X,Y)$ had budget $0$ at the beginning of the iteration, and has budget at most $3$ at the end of the iteration. Since $|E_R(X,Y)|\leq \frac{8m}{\mu^{50}}$, the increase in the budget of the edges of $E'$ is bounded by $\frac{24m}{\mu^{50}}$.

We now consider two cases. The first case happens if $|\delta_G(R)|\leq \frac{m}{3\mu^{40}}$. In this case, the increase in the budget of every edge $e\in \delta_G(R)$ is bounded by $3$ (since edge budgets may not exceed $3$), and so the total increase in the budgets of edges $e\in \delta_G(R)$ is bounded by $3|\delta_G(R)|\leq \frac{m}{\mu^{40}}$. The total increase in all edge budgets is then bounded by $\frac{m}{\mu^{40}}+\frac{24m}{\mu^{50}}$, and, since the total budgets of all vertices in $U$ decreases by at least  $\frac{2m}{\mu^{40}}$, we get that the total budget $B$ decreases by at least $\frac{m}{4\mu^{40}}\leq \frac{1}{m}$.

Lastly, we assume that $|\delta_G(R)|>\frac{m}{3\mu^{40}}$.
Consider some edge $e\in \delta_G(R)$. Since $|\delta_G(X)|,|\delta_G(Y)|\leq |\delta_G(R)|+|E'|$, the increase in the budget of $e$ is bounded by: 

\[8\alpha_0\cdot \alphasc(m)\cdot (\log_{3/2}(|\delta_G(R)|+|E'|)-\log_{3/2}(|\delta_G(R)|))\leq 8\alpha_0\cdot \alphasc(m)\cdot \log_{3/2}\left (1+\frac{|E'|}{|\delta_G(R)|}\right ).\]

 Since we have assumed that $|\delta_G(R)|>\frac{m}{3\mu^{40}}$, while $|E'|\leq \frac{3m}{\mu^{50}}$, we get that $\frac{|E'|}{|\delta_G(R)|}<1/2$.
Since for all $\eps\in (0,1/2)$, $\ln(1+\eps)\leq \eps$, we get that the increase in the budget of $e$ is bounded by $8\alpha_0\cdot \alphasc(m)\cdot \frac{|E'|}{|\delta_G(R)|\cdot \ln(3/2)}\leq 24\alpha_0\cdot \alphasc(m)\cdot \frac{|E'|}{|\delta_G(R)|}$. The increase in the budget of all edges of $\delta_G(R)$ is then bounded by $24\alpha_0\cdot \alphasc(m)\cdot|E'|\leq \frac{576m\alpha_0\alphasc(m)}{\mu^{50}}\leq \frac{m}{\mu^{49}}$. Since the budget of all edges in $E'$ increases by at most $\frac{24m}{\mu^{50}}$, and the budget of the vertices of $U$ decreases by at least $\frac{2m}{\mu^{40}}$, the total budget in the system decreases by at least $\frac{m}{\mu^{40}}\geq \frac 1 m$.

Since the initial budget $B$ is bounded by $\frac{m}{\mu^{27}}$, and in every iteration, either a new cluster is added to set $\rset^I$, or the budget $B$ decreases by at least $1/m$, the number of iterations is bounded by $\poly(m)$, so the algorithm is efficient. Once the algorithm terminates, $\rset^I=\rset$ holds. We then return the set $\cset=\rset$ of clusters as the outcome of the algorithm. From our invarinats, we are guaranteed that $\bigcup_{C\in \cset}V(C)=V(G)$,  every cluster $C\in \cset$ has the $\alpha_0$-bandwidth property, and for every cluster $C\in \cset$, for every pair $u,v$ of distinct vertices of $C$ with $\deg_G(v),\deg_G(u)\geq \frac{m}{\mu^6}$, there is a collection of at least $\frac{8m}{\mu^{50}}$ edge-disjoint paths in $C$ connecting $u$ to $v$. Since the total budget $B$ remains bounded by  $\frac{m}{\mu^{27}}$, and
 $\sum_{C\in \cset}|\delta_G(C)|\leq B$, we get that $\sum_{C\in \cset}|\delta_G(C)|\leq \frac{m}{\mu^{27}}$ holds.
\end{proof}

We are now ready to complete the proof of \Cref{thm: not well connected}. We start by applying the algorithm from \Cref{lem: initial clusters} to instance $I$, to obtain a collection $\cset$ of clusters. We then apply Algorithm \algadvanceddisengagement from \Cref{thm: disengagement - main} to instance $I=(G,\Sigma)$, cluster set $\cset$, parameter $\mu$ that remains unchanged, and parameter $m=|E(G)|$. Let 	$\iset$ be the $2^{O((\log m)^{3/4}\log\log m)}$-decomposition of instance $I$ that the algorithm returns. Recall that every instance $I'\in \iset$ is a subinstance of $I$.
Consider any instance $I'=(G',\Sigma')\in \iset$, and assume that $|E(G')|>m/\mu$, and that $I'$ is a wide instance. It is enough to prove that instance $I'$ is well-connected.
Indeed, the algorithm from \Cref{thm: disengagement - main}  ensures that there is at most one cluster $C\in \cset$ with $E(C)\subseteq E(G')$. 
If no such cluster exists, then $E(G')\subseteq \Eout(\cset)$. Since $|\Eout(\cset)|\leq \frac{m}{\mu^{27}}$, $|E(G')|\leq \frac{m}{\mu^{27}}$ must hold in this case, contradicting our assumption that $|E(G')|>m/\mu$. Therefore, there must be a cluser $C\in \cset$ with $C\subseteq G'$. The algorithm from \Cref{thm: disengagement - main} then guarantees that  $E(G')\subseteq E(C)\cup \Eout(\cset)$. 
Consider some vertex $v\in V(G')$ with $\deg_{G'}(v)\geq \frac{|E(G')|}{\mu^5}$. Since $|E(G')|\geq \frac{m}{\mu}$, $\deg_{G'}(v)\geq\frac{|E(G')|}{\mu^5}\geq \frac{m}{\mu^6}$ must hold. In particular, since $I'$ is a subinstance of $I$, and since $|\Eout(\cset)|\leq  \frac{m}{\mu^{27}}$, vertex $v$ must lie in cluster $C$ (as otherwise all edges incident to $v$ in $G'$ belong to $\Eout(\cset)$), and so $\deg_{G}(v)\geq \frac{m}{\mu^6}$ must hold as well. The algorithm from \Cref{lem: initial clusters} ensures that, for every pair $u,v$ of vertices of $C$ with $\deg_G(u),\deg_G(v)\geq \frac{m}{\mu^6}$, there is a collection $\pset$ of at least $\frac{8m}{\mu^{50}}\geq\frac{8|E(G')|}{\mu^{50}}$ edge-disjoint paths in $C$ connecting $u$ to $v$. Since $C\subseteq G'$, every path in $\pset$ is also contained in $G'$. We conclude that instance $I'$ must be well-connected.

\section{An Algorithm for Wide and Well-Connected Instances -- Proof of \Cref{lem: many paths}}
\label{sec: many paths}

\newcommand{\cI}{\check I}
\newcommand{\cG}{\check G}
\newcommand{\cSigma}{\check \Sigma}
\newcommand{\cm}{\check m}
\newcommand{\npaths}{m/\mu^{50}}
\newcommand{\thec}{50}
\renewcommand{\tpsi}{\tilde \psi}
\renewcommand{\tG}{\tilde G}
\newcommand{\tI}{\tilde I}

This section is dedicated to the proof of  \Cref{lem: many paths}. 
Recall we are given a wide and a well-connected instance $I=(G,\Sigma)$ of the \CNwRS problem.
For convenience, throughout this section, we refer to instance $I$ as $\check I^*=(\check G^*,\check \Sigma^*)$ and denote $\check m=|E(\check G^*)|$. 
We let $\cG$ be the graph that is obtained from graph $\cG^*$ by subdividing every edge of $\cG^*$ with a vertex. Since every vertex in $V(\cG)\setminus V(\cG^*)$ has degree $2$, we can extend the rotation system $\cSigma^*$ for graph $\cG^*$ to a rotation system $\cSigma$ for graph $\cG$, in a natural way. We denote by $\cI=(\cG,\cSigma)$ the resulting instance of  \cnwrs. We sometimes refer to $\cI$ as the \emph{subdivided instance corresponding to $\cI^*$}. Note that $|E(\cG)|=2\cm$ and $\optcrors(\cI)=\optcrors(\cI^*)$. Throughout this section, we will mostly be working with instance $\cI$. 
We will use notation $I$ and $m$ when discussing various subinstances of $\check I$. Recall that $\cm\geq \mu^{c'}$ must hold, where $c'$ is a large enough constant.
Given a subgraph $G\subseteq \cG$, we let $\Sigma$ be the rotation system for $G$ induced by $\cSigma$, and we will refer to instance $I=(G,\Sigma)$ as the \emph{subinstance of $\cI$ defined by graph $G$}.

We start with intuition. Fix an optimal solution $\phi^*$ to instance $\cI$, where $\phi^*$ is a drawing of graph $\cG$ on the sphere. In order to simplify the exposition, assume that $\cro(\phi^*)= \optcrors(I)\leq \cm^2/\mu^{c'}$, where $c'$ is a large enough constant. 
Since instance $\cI^*$ is  wide, there is a vertex $v\in V(\cG)$, a partition $(E_1,E_2)$ of the edges of $\delta_{\cG}(v)$, with the edges of $E_1$ appearing consecutively in the rotation $\oset_v\in \Sigma$, and a collection $\pset$ of at least $\cm/\mu^{50}$ edge-disjoint cycles in graph $G$, where every cycle $P\in \pset$ contains an edge of $E_1$ and an edge of $E_2$.
Informally, we say that a vertex $u$ of $\cG$ is \emph{heavy} if it lies on a large number of cycles of $\pset$, otherwise we say it is \emph{light}.

 Let $P^*\in \pset$ be a cycle that we select uniformly at random. Since  $|\pset|\geq \cm/\mu^{50}$, the expected number of crossings in which the edges of $E(P^*)$ participate in $\phi^*$ is relatively small -- at most $\optcrors(I)\cdot \mu^{50}/\cm$. We can use this fact in order to show that, with a high enough probability, there is a near-optimal solution $\phi'$ to instance $\cI$, in which the images of the edges of $E(P^*)$ do not cross each other, and they participate in relatively few crossings. 
 Let $E'$ denote the set of all edges $e\in E(\cG)$, such that $e$ is incident to a light vertex $u\in V(P^*)$, and $e\not\in E(P^*)$. From the definition of light vertices and the fact that $|\pset|$ is large, we can show that with high enough probability, $|E'|$ is quite small. 
 Additionally, using the cycles in $\pset$, we can compute, for every heavy vertex $u\in V(P^*)$, an orientation $b_u\in \set{-1,1}$. We show that with high probability, this orientation is consistent with the solution $\phi'$ to instance $\cI$. In other words, if $b_u=-1$, then the images of the edges of $\delta_G(u)$ enter the image of $u$ in $\phi'$ according to the ordering $\oset_u\in \Sigma$ in clock-wise direction, and otherwise the direction is counter-clock-wise.
 
 Note that the image of the cycle $P^*$ in the near-optimal solution $\phi'$ to instance $\cI$ partitions the sphere into two internally disjoint discs $D$ and $D'$. Let $E''$ be the set of edges of $\cG$ whose images cross the images of the edges of $P^*$. From our construction, with reasonably high probability, $|E''|$ is relatively small. We can view the image of cycle $P^*$ in $\phi'$  as splitting graph $G\setminus (E'\cup E'')$ into two subgraphs, $G_1$ and $G_2$, where $G_1$ contains all edges and vertices that are drawn inside $D$, while $G_2$ contains all edges and vertices that are drawn inside $D'$.
 The only vertices and edges that the two graphs share are the vertices and edges of $P^*$. We can further ensure that each of the subinstances contains a significant number of edges, so $|E(G_1)|,|E(G_2)|$ are both significantly smaller than $\cm$. 
 
 If we could compute the two graphs $G_1,G_2$ efficiently, then we could construct two subinstances $I_1=(G'_1,\Sigma'_1),I_2=(G'_2,\Sigma'_2)$ of instance $\cI$, where graph $G_1'$ is obtained from $G_1$ by contracting the vertices of $V(P^*)$ into a supernode $v^*$, and letting the rotation of this supernode in $\Sigma'_1$ be determined by the rotations of the vertices of $V(P^*)$ in instance $\cI$, the orientations of the heavy vertices of $P^*$ that we have computed, and the order of the vertices of $P^*$ on the cycle; instance $I_2$ is computed similarly from $G_2$. Given solutions $\phi_1,\phi_2$ to the instances $I_1,I_2$, respectively, we can combine them to obtain a solution $\phi$ to instance $\cI$, as follows. First, we un-contract the supernode $v^*$ in each of the two instances to obtain a cycle $P^*$, and then we ``glue'' the two drawings together via this cycle. Next, we insert the edges of $E'\cup E''$ into the resulting drawing, obtaining a drawing $\phi$ of $G$. Since $|E'|,|E''|$ are relatively small, so is the increase in the number of crossings relatively to $\cro(\phi_1)+\cro(\phi_2)$. We could then apply the same decomposition process recursively to instances $I_1$ and $I_2$ (if these instances are wide). We refer to $I_1$ and $I_2$ as the \emph{contracted instances} corresponding to subgraphs $G_1$ and $G_2$ of $G$, respectively.
 
 The main difficulty with this approach is that we do not know the optimal solution $\phi^*$ to instance $\cI$, or the near-optimal solution $\phi'$, and so we cannot compute the two graphs $G_1,G_2$ with the required properties. We also do not know the set $E''$ of edges whose images cross the images of the edges of $E(P^*)$ in $\phi'$. Instead, we compute a relatively small edge set $E^*$, and two subgraphs $\tilde G_1,\tilde G_2$ of $G\setminus (E'\cup E^*)$, for which the following hold. First, $E(\tilde G_1)\cup E(\tilde G_2)\cup E'\cup E^*=E(G)$. Second, the only vertices and edges that these two graphs share are the vertices and edges of $P^*$. Third, the number of edges in each of the graphs $\tilde G_1$, $\tilde G_2$ is significantly smaller than $|E(G)|$. Lastly, there is a near-optimal drawing $\phi'$ of $G\setminus (E'\cup E^*)$, in which the edges of $P^*$ do not cross each other. Moreover, if we denote by $D$ and $D'$ the two discs of the sphere whose boundaries are the image of $P^*$  in $\phi'$, then, by slightly modifying the drawing, we can ensure that all vertices of $\tilde G_1$ are drawn inside $D$, all vertices of $\tilde G_2$ are drawn inside $D'$, and the number of crossings in which the edges of $P^*$ participate is quite small. Unfortunately, we can no longer guarantee that every edge of $\tilde G_1$ is drawn inside $D$ and every edge of $\tilde G_2$ is drawn inside $D'$.  We can then define the contracted instances $I_1$ and $I_2$ associated with the graphs $\tilde G_1$ and $\tilde G_2$ exactly as before. As before, given solutions to instances $I_1$ and $I_2$, we can efficiently combine them to obtain a solution to instance $\cI$. But unfortunately we can no longer claim that $\optcrors(I_1)+\optcrors(I_2)$ is small. This is since, in the near-optimal drawing of $G\setminus (E'\cup E^*)$, some edges of $\tilde G_1$ may be partially drawn inside the disc $D'$, and it is not clear how to ``move'' them to the interior of $D$ without significantly increasing the number of crossings. The same is true for edges of $\tilde G_2$ that may be partially drawn inside the disc $D$. This is the main difficulty in designing our algorithm using the framework outlined above.
 
 Our algorithm consists of two phases. In the first phase, we follow the above framework to construct an initial collection $\iset$ of subinstances of $\cI$, that  have all required properties, except that we will not be able to ensure that $\sum_{I\in \iset}\optcrors(I)$ is suitably bounded, even if $\optcrors(\cI)$ is small. In the second phase, we will try to ``repair'' each one of the instances $I=(G,\Sigma)\in \iset$, by removing a small subset of edges from $G$. We will show that, after the removal of these edges, the expectation of $\sum_{I\in \iset}\optcrors(I)$ is suitably bounded. In both phases, we rely on the same algorithm outlined above, that gradually decomposes an input instance $I=(G,\Sigma)$ into smaller and smaller subinstances. 
 The algorithms for both Phase 1 and Phase 2 follow this high-level framework, though the specifics are somewhat different.  
We start with the main definitions that we use throughout this section.

\subsection{Main Definitions}
\label{subsec: main defs for interesting}

Throughout, given a graph $H$ and a drawing $\phi$ of $H$ on the sphere or in the plane, we denote by $\fset(\phi)$ the set of all faces in drawing $\phi$.

We use the notion of a \emph{subdivided graph}.

\begin{definition}[Subdivided graph]
	We say that a graph $G$ is a \emph{subdivided graph}, if $G$ does not contain parallel edges, and additionally, for every edge $e=(u,v)$, either $\deg_G(u)\leq 2$ or $\deg_G(v)\leq 2$ holds.
\end{definition}

Note that, if $G$ is any graph, and $G'$ is a graph obtained by suvdividing every edge of $G$, then $G'$ is a subdivided graph, and so is every subgraph of $G'$. In particular, graph $\cG$ associated with instance $\cI$ of \cnwrs is subdivided, and so is every subgraph of $\cG$.

\subsubsection{Cores and Core Structures}

The first central notion that we use is that of a core, and its associated core structure.

\begin{definition}[Core and Core Structure]\label{def: valid core 1}
Let $G$ be a subgraph of $\cG$, and let $I=(G,\Sigma)$ be the subinstance of $\cI$ defined by $G$. A \emph{core structure} for instance $I$ consists of the following:

\begin{itemize}
	\item a connected subgraph $J$ of $G$ called a \emph{core}, such that, for every edge $e\in E(J)$, graph $J\setminus \set{e}$ is connected (but we also allow $J$ to consist of a single vertex);
	\item an orientation $b_u\in \set{-1,1}$ for every vertex $u\in V(J)$; 
	\item a drawing $\rho_J$ of  $J$ on the sphere with no crossings, that is consistent with the rotation system $\Sigma$ and the orientations in $\set{b_u}_{u\in V(J)}$. In other words, for every vertex $u\in V(J)$, the images of the edges in $\delta_J(u)$ enter the image of $u$ in $\rho_J$ according to their order in the rotation $\oset_u\in \Sigma$ and orientation $b_u$ (so, e.g. if $b_u=1$ then the orientation is counter-clock-wise); and
	\item a distinguished face $F^*(\rho_J)\in \fset(\rho_J)$, such  that the image of every vertex $u\in V(J)$, and the image of every edge $e\in E(J)$  is contained in the boundary of face $F^*(\rho_J)$ in drawing $\rho_J$.
\end{itemize}
\end{definition}

We denote a core structure by $\jset=(J,\set{b_u}_{u\in V(J)},\rho_J, F^*(\rho_J))$, and we refer to graph $J$ as the \emph{core associated with $\jset$}. We denote $\fset^{\forbidden}(\rho_J)=\fset(\rho_J)\setminus\set{F^*(\rho_J)}$, and we refer to the faces in $\fset^{\forbidden}(\rho_J)$ as the \emph{forbidden faces} of the drawing $\rho_J$. 
	
\begin{figure}[h]
	\centering
	\subfigure[A core $J$ and its drawing $\rho_J$, with the separator vertices of $J$ shown in green. The distinguished face $F^*(\rho_J)$ is the infinite face in this drawing.]{\scalebox{0.15}{\includegraphics{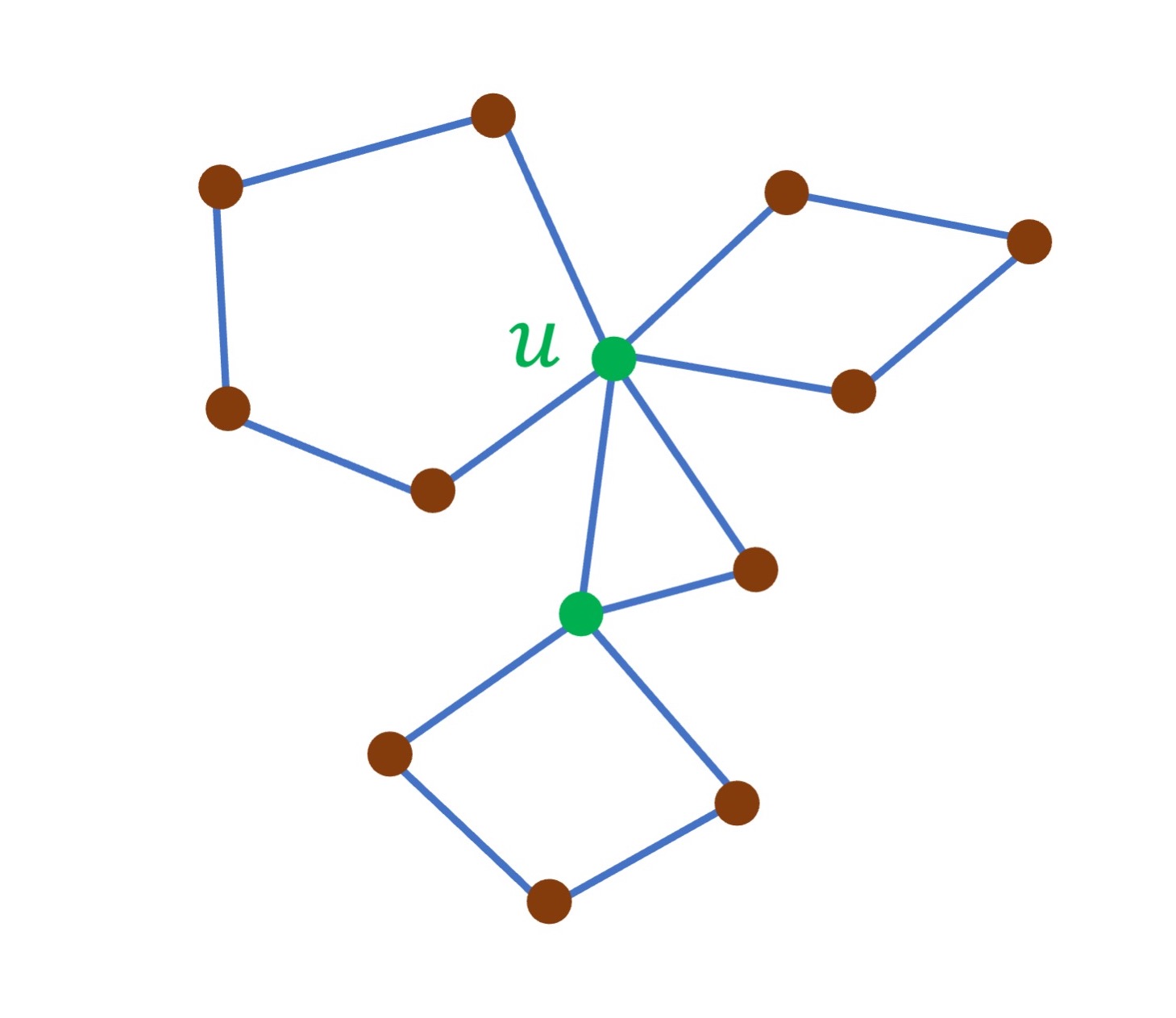}}\label{fig: core_disc_1}}
	\hspace{0.8cm}
	\subfigure[Disc $D(J)$ associated with core $J$, and its boundary (shown in pink). 
	]{\scalebox{0.15}{\includegraphics{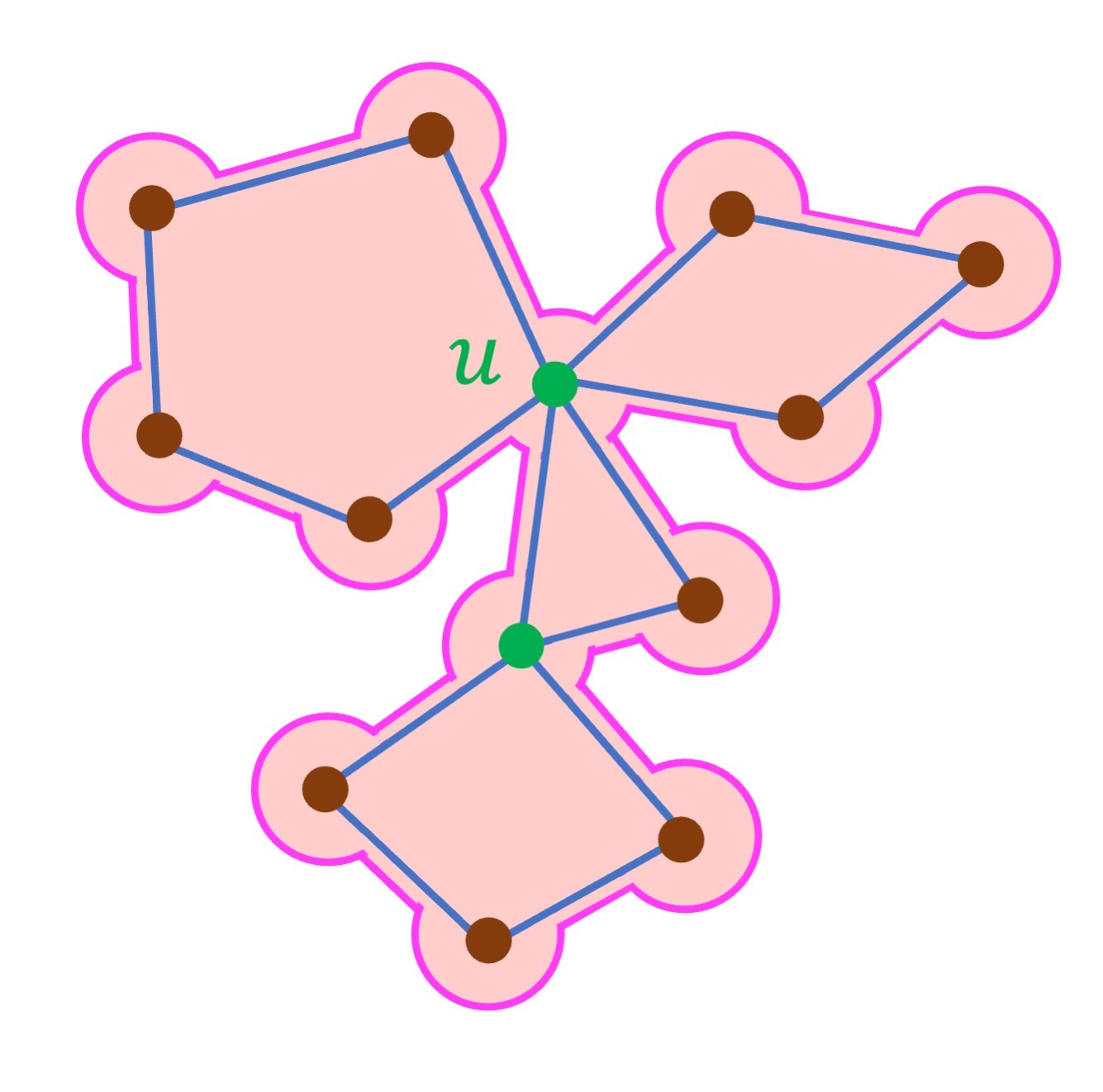}}\label{fig: core_disc_2}}
	\caption{An illustration of a core $J$ and its disc $D(J)$.}\label{fig: core_disc_12}
\end{figure}

The last two requirements in the above definition impose a certain structure on the core graph $J$. Specifically, there must be a collection $\wset$ of edge-disjoint cycles, with $\bigcup_{W\in \wset}E(W)=E(J)$, such that every pair $W,W'\in \wset$ of cycles share at most one vertex, which must be a separator vertex for $J$ (see \Cref{fig: core_disc_1} for an illustration).

Note that, since graph $G$ is subdivided, for every edge $e\in E(G)\setminus E(J)$, at most one endpoint of $e$ may lie in $J$. Indeed, assume that $e=(u,v)$. From the definition of subdivided graphs, either $\deg_G(u)\leq 2$ or $\deg_G(v)\leq 2$ holds. Assume w.l.o.g. that it is the former. If $u\in V(J)$, then graph $J$ contains at most one edge incident to $u$, that we denote by $e'$. But then $J\setminus\set{e'}$ is not a connected graph, contradicting the definition of a core.

Consider now a core structure $\jset=(J,\set{b_u}_{u\in V(J)},\rho_J,F^*(\rho_J))$, and specifically the drawing $\rho_J$ of the graph $J$ on the sphere. We define a disc $D(J)$, that contains the drawing of $J$ in its interior, such that the boundary of disc $D(J)$ is contained in face $F^*(\rho_J)$, and it is a simple closed curve that closely follows the boundary of $F^*(\rho_J)$ (see \Cref{fig: core_disc_2}).

\begin{figure}[h]
	\centering
	\includegraphics[scale=0.13]{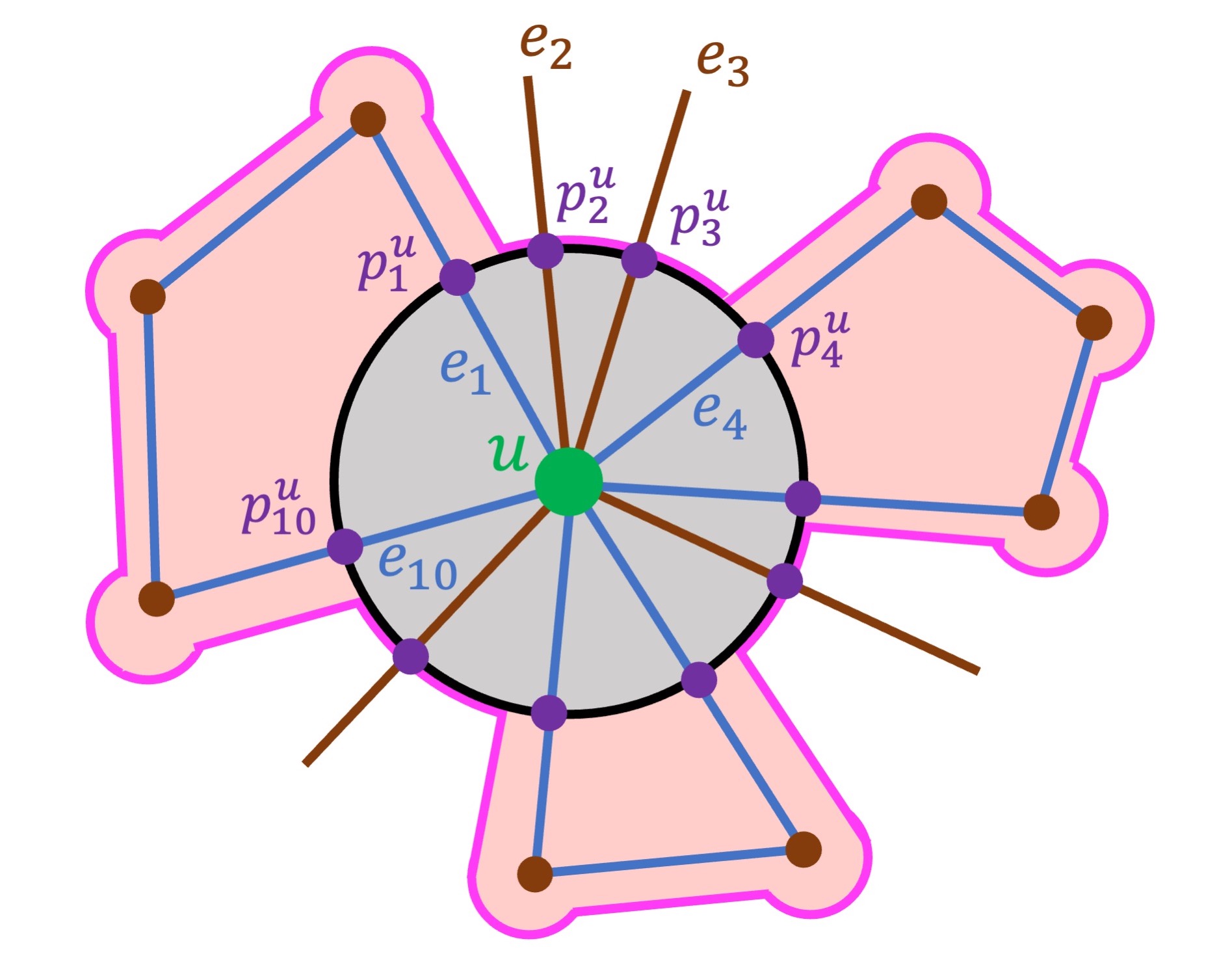}
	\caption{
	Illustration of disc $D(u)$ for a core that contains $u$. Disc $D(u)$ is shown in gray, and disc $D(J)$ is shown in pink. Note that both discs share portions of their boundaries that are contained in face $F^*(\rho_J)$ -- the infinite face in the current drawing. 
	Edges of $J$ incident to $u$ are shown in blue, and all other edges that are incident to $u$ are shown in brown.
}\label{fig: core_disc_3} 
\end{figure}

Consider some vertex $u\in V(J)$, and the tiny $u$-disc $D(u)=D_{\rho_J}(u)$ in the drawing $\rho_J$. Since vertex $u$ lies on the boundary of face $F^*(\rho_J)$, we can define disc $D(u)$ so that, for every maximal segment $\sigma$ on the boundary of $D(u)$ that is contained in $F^*(\rho_J)$, there is a contiguous curve $\sigma'\subseteq \sigma$ of non-zero length, that is contained in the boundary of the disc $D(J)$ (see \Cref{fig: core_disc_3}).

Denote $\delta_G(u)=\set{e^u_1,\ldots,e^u_{d_u}}$, where $d_u=\deg_G(u)$, so that the edges are indexed according to their ordering in the rotation $\oset_u\in \Sigma$. We can then define a collection $\set{p^u_1,\ldots,p^u_{d_u}}$ of distinct points on the boundary of the disc $D(u)$, such that the following hold:

\begin{itemize}
	\item points $p^u_1,\ldots,p^u_{d_u}$ are encountered in this order when traversing the boundary of $D(u)$ in the direction of the orientation $b_u$; 
	
	\item for every edge $e_i^u\in E(J)$, point $p_i^u$ is the unique point on the image of $e_i^u$ in $\rho_J$ lying on the boundary of $D(u)$; and
	
	\item if a point $p^u_i$ lies in the interior of face $F^*(\rho_J)$, then it lies on the boundary of the disc $D(J)$. 
\end{itemize}

For all $1\leq i\leq d_u$, we view point $p^u_i$ as representing the edge $e^u_i$.
There is one more property that we require from a core structure.

\begin{definition}[Valid Core Structure]\label{def: valid core 2}
	We say that a core structure $\jset=(J,\set{b_u}_{u\in V(J)},\rho_J,F^*(\rho_J))$ is \emph{valid} if, for every vertex $u\in V(J)$, for every edge $e_i^u\in \delta_G(u)\setminus E(J)$, the corresponding point $p_i^u$ lies in the interior of face $F^*(\rho_J)$ (and hence on the boundary of the disc $D(J)$).
\end{definition}

In the remainder of this section, whenever we use the term ``core structure", we assume that this core structure is valid. 

\paragraph{Ordering $\oset(J)$ of the Edges of $\delta_G(J)$.}
Consider a core structure $\jset=(J,\set{b_u}_{u\in V(J)},\rho_J,F^*(\rho_J))$, the drawing $\rho_J$ of $J$, and its corresponding disc $D(J)$. Recall that, for every vertex $u\in V(J)$ and every edge $e_i^u\in \delta_G(J)$, we have defined a point $p_i^u$ on the boundary of the disc $D(J)$ representing the edge $e_i^u$. Recall that each edge $e\in \delta_G(J)$ has exactly one endpoint in $J$. We define a circular oriented ordering $\oset(J)$ of the edges of $\delta_G(J)$ to be the circular order in which the  points $p_i^u$ corresponding to the edges of $\delta_G(J)$ are encountered, as we traverse the boundary of the disc $D(J)$ in the clock-wise direction.

\subsubsection{Drawings of Graphs}

Next, we define a valid drawing of a graph $G$ with respect to a core structure $\jset$.

\begin{definition}[A $\jset$-Valid Solution]\label{def: valid drawing}
Let $G$ be a subgraph of $\cG$, let $I=(G,\Sigma)$ be the subinstance of $\cI$ defined by $G$, and let $\jset=(J,\set{b_u}_{u\in V(J)},\rho_J,F^*(\rho_J))$ be a core structure for $I$. A solution $\phi$ of instance $I$ is \emph{$\jset$-valid} if we can define a disc $D'(J)$ that contains the images of all vertices and edges of the core $J$ in its interior, and the image of the core $J$ in $\phi$  is identical to $\rho_{J}$ (including the orientation), with disc $D'(J)$ in $\phi$ playing the role of the disc $D(J)$ in $\rho_{J}$. We sometimes refer to a $\jset$-valid solution to instance $I$ as a \emph{$\jset$-valid drawing of graph $G$}.
\end{definition}

Abusing the notation, we will not distinguish between disc $D(J)$ in $\rho_J$ and disc $D'(J)$ in $\phi$, denoting both discs by $D(J)$. 

Consider now some solution $\phi$ to instance $I$, that is $\jset$-valid, with respect to  some core structure $\jset$. The image of graph $J$ in $\phi$ partitions the sphere into regions, each of which corresponds to a unique face of $\fset(\rho_J)$. We do not distinguish between these regions and faces of $\fset(\rho_J)$, so we view $\fset(\rho_J)$ as a collection of regions in the drawing $\phi$ of $G$.

Note that the edges of $J$ may participate in crossings in $\phi$, but no two edges of $J$ may cross each other. 
Consider now a crossing  $(e,e')_p$ in drawing $\phi$. We say that it is a \emph{dirty} crossing if exactly one of the two edges $e,e'$ lies in $E(J)$. We denote by $\chi^{\dirty}(\phi)$ the set of all dirty crossings of drawing $\phi$. We say that an edge $e\in E(G)\setminus E(J)$ is \emph{dirty} in $\phi$ if it participates in some dirty crossing of $\phi$. 


Next, we define special types of $\jset$-valid drawings, called clean and semi-clean drawings.

\begin{definition}[Clean and Semi-Clean Drawings]\label{def: semiclean drawing}
Let $G$ be a subgraph of $\cG$, let $I=(G,\Sigma)$ be the subinstance of $\cI$ defined by $G$, let  $\jset=(J,\set{b_u}_{u\in V(J)},\rho_J,F^*(\rho_J))$ be a core structure for  $I$, and let $\phi$ be a solution to instance $I$.  We say that $\phi$ is a \emph{semi-clean solution to instance $I$}, or a \emph{semi-clean drawing of $G$}, with respect to $\jset$, if it is a $\jset$-valid drawing, and,  additionally, the image of every vertex of $V(G)\setminus V(J)$ lies outside of the disc $D(J)$ (so in particular it must lie in the interior of the region $F^*(\rho_J)\in \fset(\rho_J)$).

If, additionally, the image of every edge of $E(G)\setminus E(J)$ 
is entirely contained in region $F^*(\rho_J)$, then we say that $\phi$ is a \emph{clean} solution to $I$ with respect to $\jset$, or that it is a \emph{$\jset$-clean} solution.
\end{definition}

Notice that, from the definition, if $\phi$ is a clean solution to instance $I$ with respect to core structure $\jset$, then the edges of $J$ may not participate in any crossings in $\phi$.

\paragraph{Drawings of Subgraphs and Compatible Drawings.}
Notice that, if $G'$ is a subgraph of $G$ that contains $J$, then a core structure $\jset$ for instance $I=(G,\Sigma)$ remains a valid core structure for the subinstance $I'=(G',\Sigma')$ of $\cI$ defined by $G'$. Therefore,  $\jset$-valid drawings are well-defined for every subgraph $G'\subseteq G$.

Assume now that we are given a $\jset$-valid solution $\phi$ to instance $I$, and a subinstance $I'=(G',\Sigma')$ of $I$ that is defined as above. Intuitively, we will often obtain a $\jset$-valid solution $\phi'$ to instance $I'$ by slightly modifying the solution $\phi$ to instance $I$.  We will, however, restrict  the types of modifications that we allow. In particular we do  now allow adding any new images of edges (or their segments), or  new images of vertices to the forbidden regions in $\fset^{\forbidden}(\rho_J)$. We now define these restrictions formally.

\begin{definition}[Compatible Drawings.]
\label{def: compatible drawing}
Let $G$ be a subgraph of $\cG$, let $I=(G,\Sigma)$ be the subinstance of $\cI$ defined by $G$,  let $\jset=(J,\set{b_u}_{u\in V(J)},\rho_J, F^*(\rho_J))$ be a  core structure for $I$, and let $\phi$ be a $\jset$-valid solution to instance $I$. Let $G'$ be a subgraph of $G$ with $J\subseteq G'$, and let $I'=(G',\Sigma')$ be the subinstance of $\cI$ defined by $G'$. Finally, let $\phi'$ be a $\jset$-valid solution to instance $I'$. We say that drawing $\phi'$ of $G'$ is \emph{compatible} with drawing $\phi$ of $G$ with respect to $\jset$, if the following hold:

\begin{itemize}
	\item the image of the core $J$ and the correpsonding disc $D(J)$ in $\phi'$ are identical to those in $\phi$;
	
	\item if a point $p$ is an inner point of an image of an edge in $\phi'$, then it is an inner point of an image of an edge in $\phi$;
	
	\item if a point $p$ is a crossing point between a pair of edges in $\phi'$, then it is a crossing point between a pair of edges in $\phi$;
	
	\item if a point $p$ is an image of a vertex $v$ in $\phi'$, then either (i) point $p$ is an image of vertex $v$ in $\phi$; or (ii) vertex $v$ has degree $2$ in $G'$, and point $p$ is an inner point on an image of an edge in $\phi$;\; 

	\item if the image of a vertex $v\in V(G')$ lies outside the region $F^*(\rho_J)$ in $\phi'$, then $\phi'(v)=\phi(v)$; and

	\item if $\sigma$ is a maximal segment of an image of an edge $e\in E(G')$ in $\phi'$ that is internally disjoint from region $F^*(\rho_J)$, then $\sigma\subseteq \phi(e)$.

\end{itemize}
\end{definition}

Note that, if we obtain drawing $\phi'$ from drawing $\phi$, then the only changes that are allowed outside of region $F^*(\rho_J)$ is the deletion of images of vertices or (segments of) images of edges. In other words, $\big(\phi'(G')\setminus F^*(\rho_J)\big)\subseteq \big(\phi(G)\setminus F^*(\rho_J)\big)$.

\subsubsection{A $\jset$-Contracted Instance}

Suppose we are given a subgraph $G$ of $\cG$, together with a core structure $\jset=(J,\set{b_u}_{u\in V(J)},\rho_J, F^*(\rho_J))$ for the subinstance $I$ of $\cI$ defined by $G$. 
 We now define a \emph{$\jset$-contracted} subinstance $\hat I=(\hat G,\hat \Sigma)$ of instance $I$. Graph $\hat G$ is obtained from graph $G$, by contracting the vertices of the core $J$ into a supernode $v_{J}$.  In order to define the rotation system $\hat \Sigma$, for every vertex $u\in V(\hat G)\setminus\set{v_J}$, we let the rotation $\oset_u$ remain the same as in $\Sigma$, and for the supernode $v_{J}$, we set the corresponding rotation $\oset_{v_{J}}\in \Sigma'$ to be the ordering $\oset(J)$ of the edges of $\delta_{\hat G}(v_{J})= \delta_G(J)$ that we have defined above (recall that this is the order in which points $p_i^u$ appear on the boundary of disc $D(J)$, for all $u\in V(J)$ and $1\leq i\leq d_u$).

 Throughout our algorithm, we will consider subgraphs $G\subseteq \cG$. Each such subgraph will always be associated with a core structure 
$\jset=(J,\set{b_u}_{u\in V(J)},\rho_J, F^*(\rho_J))$ for the subinstance $I$ of $\cI$ defined by $G$.  The $\jset$-contracted subgraph of $I$ will always be denoted by $\hat I=(\hat G,\hat \Sigma)$. We denote by $\hat m(I)=|E(\hat G)|$ -- the number of edges in the $\jset$-contracted subinstance of $I$.
We need the following simple observation.
\begin{observation}\label{obs: clean solution to contracted}
	There is an efficient algorithm, whose input consists of a subgraph $G$ of $\cG$, a core structure $\jset=(J,\set{b_u}_{u\in V(J)},\rho_J, F^*(\rho_J))$ for the subinstance $I=(G,\Sigma)$ of $\cI$ defined by $G$, and a $\jset$-clean solution $\phi$ to instance $I$. The output of the algorithm is a solution $\hat \phi$ to the corresponding $\jset$-contracted  instance $\hat I=(\hat G,\hat \Sigma)$, with $\cro(\hat \phi)=\cro(\phi)$.
	\end{observation}

\begin{proof}
	Consider the solution $\phi$ to instance $I$. From the definition of a clean solution, there is a disc $D(J)$ that contains the drawing of $J$ in $\phi$, which is in turn identical to drawing $\rho_J$. For every vertex $u\in V(G)\setminus V(J)$, its image appears outside the disc $D(J)$ in $\phi$.  We are also guaranteed that, for every edge $e\in E(G)\setminus E(J)$, its drawing $\phi(e)$ is contained in region $F^*(\rho_J)$.
	
	By slightly manipulating the boundary of the disc $D(J)$, we can ensure that, for every edge $e\in E(G)\setminus E(J)$, if $e$ is not incident to any vertex of $J$, then $\phi(e)$ does not intersect disc $D(J)$, and otherwise, $\phi(e)\cap D(J)$ is a contiguous curve.

	For every edge $e\in \delta_G(J)$, denote by $p_e$ the unique intersection point between the boundary of $D(J)$ and the curve $\phi(e)$. 
	Since the drawing $\phi$ of $G$ is $\jset$-valid, the circular ordering of the points of $\set{p_e\mid e\in \delta_G(J)}$ on the boundary of disc $D(J)$ is precisely $\oset(J)$. For each edge $e\in \delta_G(J)$, we erase the segment of $\phi(e)$ that is contained in $D(J)$. We then contract the disc $D(J)$ into a single point, that becomes the image of the supernode $v_{J}$. We have now obtained a valid solution $\hat \phi$ to the $\jset$-contracted instance $\hat I$, with $\cro(\hat \phi)= \cro(\phi)$.
\end{proof}

We note that the converse of \Cref{obs: clean solution to contracted} is also true: given a solution $\hat \phi$ to the $\jset$-contracted instance $\hat I$, we can efficiently construct a clean solution $\phi$ to instance $I$, with $\cro(\phi)= \cro(\hat \phi)$, as follows. First, we expand the image of the supernode $v_{J}$ so it becomes a disc, that we denote by $D(J)$. We then plant the drawing $\rho_{J}$ of the core $J$ inside the disc. Note that the circular ordering in which the edges of $\delta_{\hat G}(v_{J})$ enter the boundary of the disc $D(J)$ from the outside is identical to the circular ordering in which the edges of $\delta_G(J)=\delta_{\hat G}(v_{J})$ enter the boundary of the disc $D(J)$ from the inside, and their orientations match. Therefore,  we can ``glue'' the corresponding curves to obtain, for each edge $e\in \delta_G(J)$, a valid drawing connecting the images of its endpoints.

\subsubsection{Core Enhancement and Promising Sets of Paths}
\label{subsubsec: sekelton enhancement}

Our main  subroutine, called \procsplit, starts with a subinstance $I=(G,\Sigma)$ of $\cI$ that is defined by a subgraph $G$ of $\cG$, and a core structure $\jset=(J,\set{b_u}_{u\in V(J)},\rho_J, F^*(\rho_J))$ for it. We ``enhance'' the corresponding core $J$ by adding either one cycle, or one path to it, that we refer to as \emph{core enhancenemt}. 
We also decompose instance $I=(G,\Sigma)$ into two subinstances, $I_1=(G_1,\Sigma_1)$ and $I_2=(G_2,\Sigma_2)$, where $G_1,G_2\subseteq G$. Using the enhancement for the core structure $\jset$, we then construct two new core structures: core structure $\jset_1$ for instance $I_1$, and core structure $\jset_2$ for instance $I_2$.

In order to avoid cumbersome notation, we will sometimes refer to simple cycles as paths. Given a simple cycle $W$, we will designate one of the vertices $v\in V(W)$ to be the ``endpoint'' of the cycle. When referring to two endpoints of $W$, we will think of both endpoints as being $v$.

We start by defining the notions of \emph{core enhancement} and  \emph{core enhancement structure}.

\begin{figure}[h]
	\centering
	\subfigure[When $P$ is a simple path.]{\scalebox{0.13}{\includegraphics{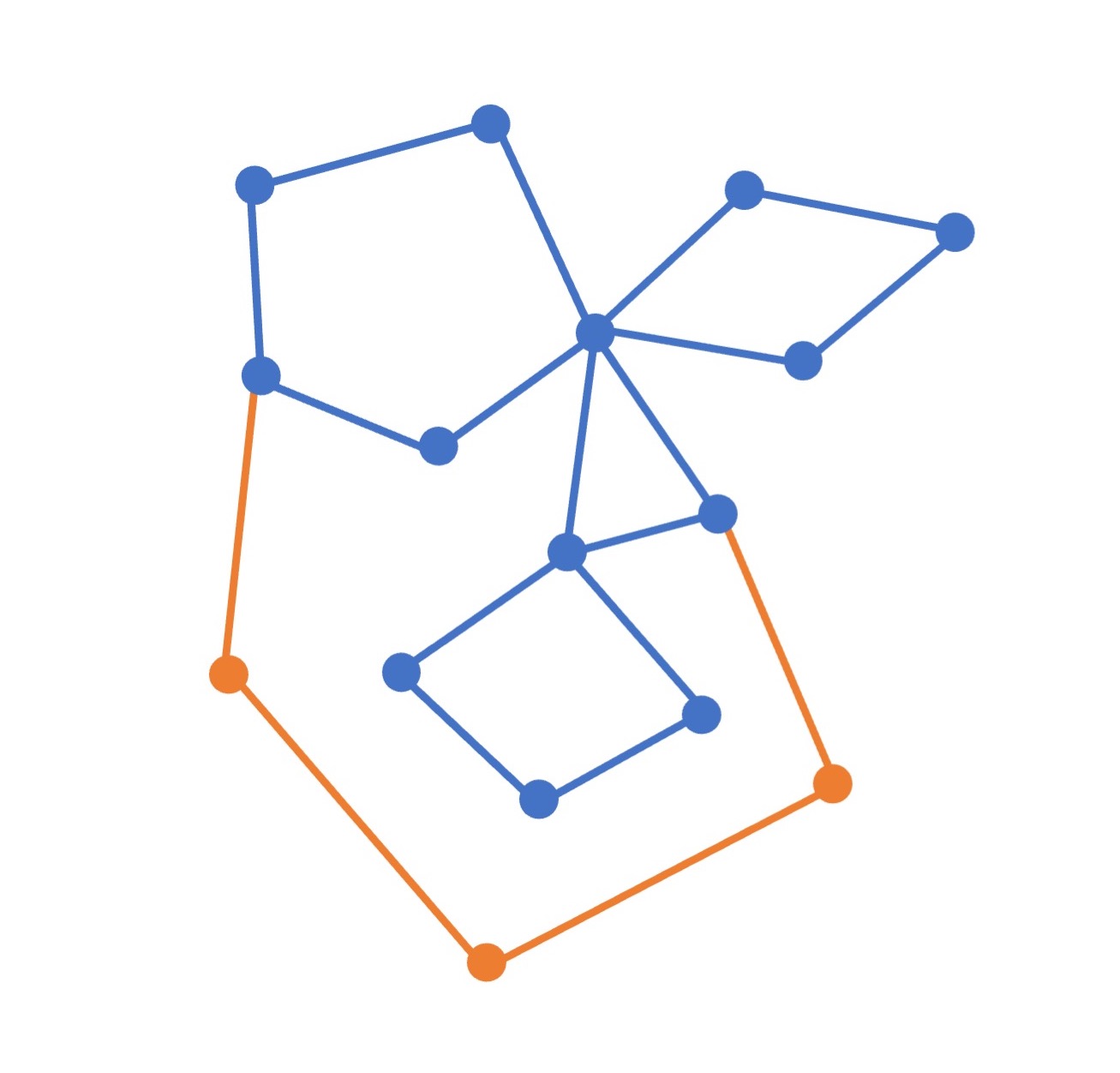}}\label{fig: enhance_1}}
	\hspace{0.08cm}
	\subfigure[When $P$ is a simple cycle.]{\scalebox{0.13}{\includegraphics{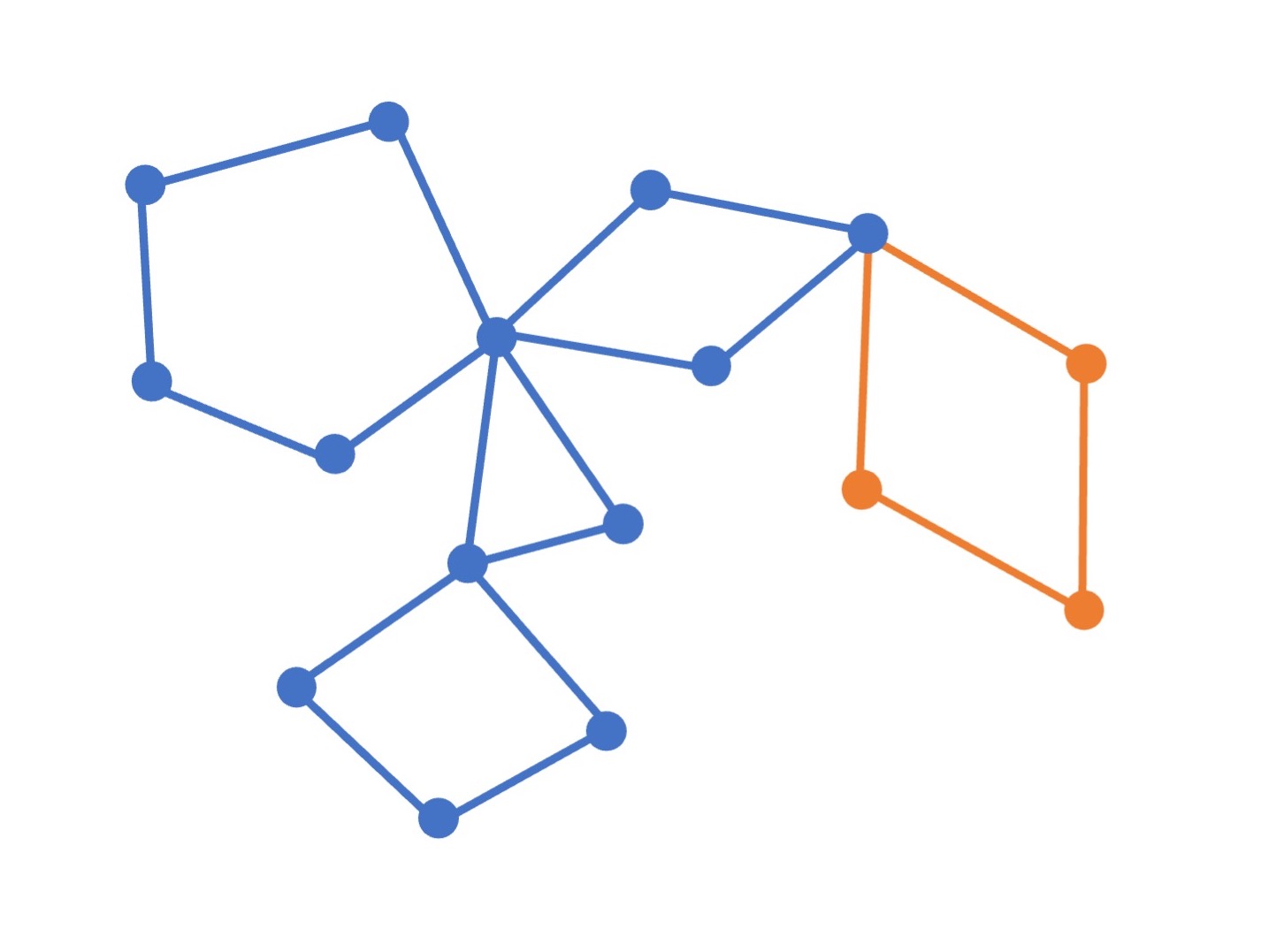}}\label{fig: enhance_2}}
	\caption{An illustration of an enhancement of a core structure $\jset$. The core $J$ is shown in blue, and the enhancement in orange. 
	}\label{fig: enhance}
\end{figure}

\begin{definition}[Core Enhancement]
	Given a subgraph $G$ of  $\cG$, and  a core structure \newline $\jset=(J,\set{b_u}_{u\in V(J)},\rho_J, F^*(\rho_J))$ for the subinstance $I=(G,\Sigma)$ of $\cI$ defined by $G$, an \emph{enhancement} of the core structure $\jset$ is a simple path $P\subseteq G$ (that may be a simple cycle), whose both endpoints belong to $J$, such that $P$ is internally disjoint from $J$ (see \Cref{fig: enhance}).
\end{definition}


For simplicity of notation, we will sometimes refer to an enhancement of a core structure $\jset$ as an enhancement of the corresponding core $J$, or as a $\jset$-enhancement. Next, we define a core enhancement structure.


\begin{definition}[Core Enhancement Structure]
	Given a subgraph $G$ of  $\cG$, and  a core structure $\jset=(J,\set{b_u}_{u\in V(J)},\rho_J, F^*(\rho_J))$ for the subinstance $I=(G,\Sigma)$ of $\cI$ defined by $G$,  a \emph{$\jset$-enhancement structure} consists of:

		\begin{itemize}
			\item a $\jset$-enhancement $P$;
			\item an orientation $b_u\in \set{-1,1}$ for every vertex $u\in V(P)\setminus V(J)$; and
			
			\item a drawing $\rho'$ of the graph $J'=J\cup P$ with no crossings, such that $\rho'$ is consistent with the rotation system $\Sigma$ and the orientations $b_u$ for all vertices $u\in V(J')$ (here, the orientations of vertices of $J$ are determined by $\jset$), and moreover, $\rho'$ is a clean drawing of $J'$ with respect to $\jset$.
		\end{itemize}
	\end{definition}

Intuitively, in the drawing $\rho'$ of graph $J'$, the drawing of graph $J$ should be identical to $\rho_J$, and the path $P$ should be drawn inside the region $F^*(\rho_J)$.
For convenience of notation, we denote a $\jset$-enhancement by
 $\aset=\left( P,\set{b_u}_{u\in V(J')},\rho'\right)$, where $J'=J\cup P$. We will always assume that, for every vertex $u\in V(J)$ its orientation $b_u$ in $\aset$ is identical to its orientation in $\jset$.

\paragraph{Promising Set of Paths.}
We now define promising sets of paths, that will be used in order to compute an enhancement of a given core structure.

\begin{definition}[Promising Set of Paths]
	Let $G$ be a subgraph of $\cG$,  let $I=(G,\Sigma)$ be the subinstance of $\cI$ defined by $G$, let $\jset=(J,\set{b_u}_{u\in V(J)},\rho_J, F^*(\rho_J))$ be a  core structure for $I$, and let $\pset$ be a collection of simple edge-disjoint paths in $G$, that are internally disjoint from $J$. We say that $\pset$ is a \emph{promising set of paths}, if  there is a partition $(E_1,E_2)$ of the edges of $\delta_G(J)$, such that the edges of $E_1$ appear consecutively in the ordering $\oset(J)$, and every path in $\pset$ has an edge of $E_1$ as its first edge, and an edge of $E_2$ as its last edge. \end{definition}

We note that some paths in a promising path set may be cycles.
We show an efficient algorithm to compute a large set of promising paths for an instance $I$ whose corresponding contracted instance is wide. The proof of the following claim is standard, and it is deferred to Section \ref{subsec: proof of finding potential augmentors} of Appendix.

\begin{claim}\label{claim: find potential augmentors}
There is an efficient algorithm that takes as input a subgraph $G\subseteq \cG$ and a core structure $\jset=(J,\set{b_u}_{u\in V(J)},\rho_J, F^*(\rho_J))$ for the subinstance $I=(G,\Sigma)$ of $\cI$ defined by graph $G$, such that the following properties hold:
	
	\begin{itemize}
		\item for every vertex $v\in V(G)$ with $\deg_G(v)\geq\frac{\hat m(I)}{\mu^4}$, there is a collection $\qset(v)$ of at least $\frac{2\hat m(I)}{\mu^{50}}$ edge-disjoint paths in $G$ connecting $v$ to the vertices of $J$; and
		\item  the $\jset$-contracted subinstance $\hat I$ of $I$ is wide.
	\end{itemize} 

The algorithm computes a  promising set of  paths for $I$ and $\jset$, of cardinality $\floor{\frac{\hat m(I)}{\mu^{50}}}$. 
\end{claim}

\subsubsection{Splitting a Core Structure and an Instance via an Enhancement Structure}

\paragraph{Splitting the Core Structure.}

Suppose we are given a subgraph $G$ of $\cG$, a core structure $\jset=(J,\set{b_u}_{u\in V(J)},\rho_J, F^*(\rho_J))$ for the subinstance $I=(G,\Sigma)$ of $\cI$ defined by $G$, and an enhancement structure   $\aset=\left (P,\set{b_u}_{u\in V(J')},\rho'\right )$ for $\jset$, where $J'=J\cup P$.
We now show an efficient algorithm that, given $\jset$ and $\aset$, splits the core structure $\jset$ into two new core structures, $\jset_1$ and $\jset_2$, using the enhancement structure $\aset$.
We refer to $(\jset_1,\jset_2)$ as a \emph{split} of the core structure $\jset$ via the enhancement structure $\aset$.

We let $\rho'$ be the drawing of the graph $J'$ on the sphere given by the enhancement structure $\aset$. Recall that there is a disc $D(J)$ that contains the image of $J$ in $\rho'$, whose drawing in $D(J)$ is identical to $\rho_J$. Additionally, all vertices and edges of $P$ must be drawn in region $F^*(\rho_J)$ of $\rho'$. Therefore, in the drawing $\rho'$ of $J'$, there are two faces incident to the image of the path $P$, that we denote by $F_1$ and $F_2$, respectively, and $F_1\cup F_2=F^*(\rho_J)$ holds. 
We let $J_1\subseteq J'$ be the graph containing all vertices and edges, whose images lie on the boundary of face $F_1$ in $\rho'$, and we define graph $J_2\subseteq J'$ similarly for face $F_2$. 

\begin{figure}[h]
	\centering
	\subfigure[Before the split. Core $J$ is shown in blue and face $F^*(\rho_J)$ is shown in gray.]{\scalebox{0.09}{\includegraphics{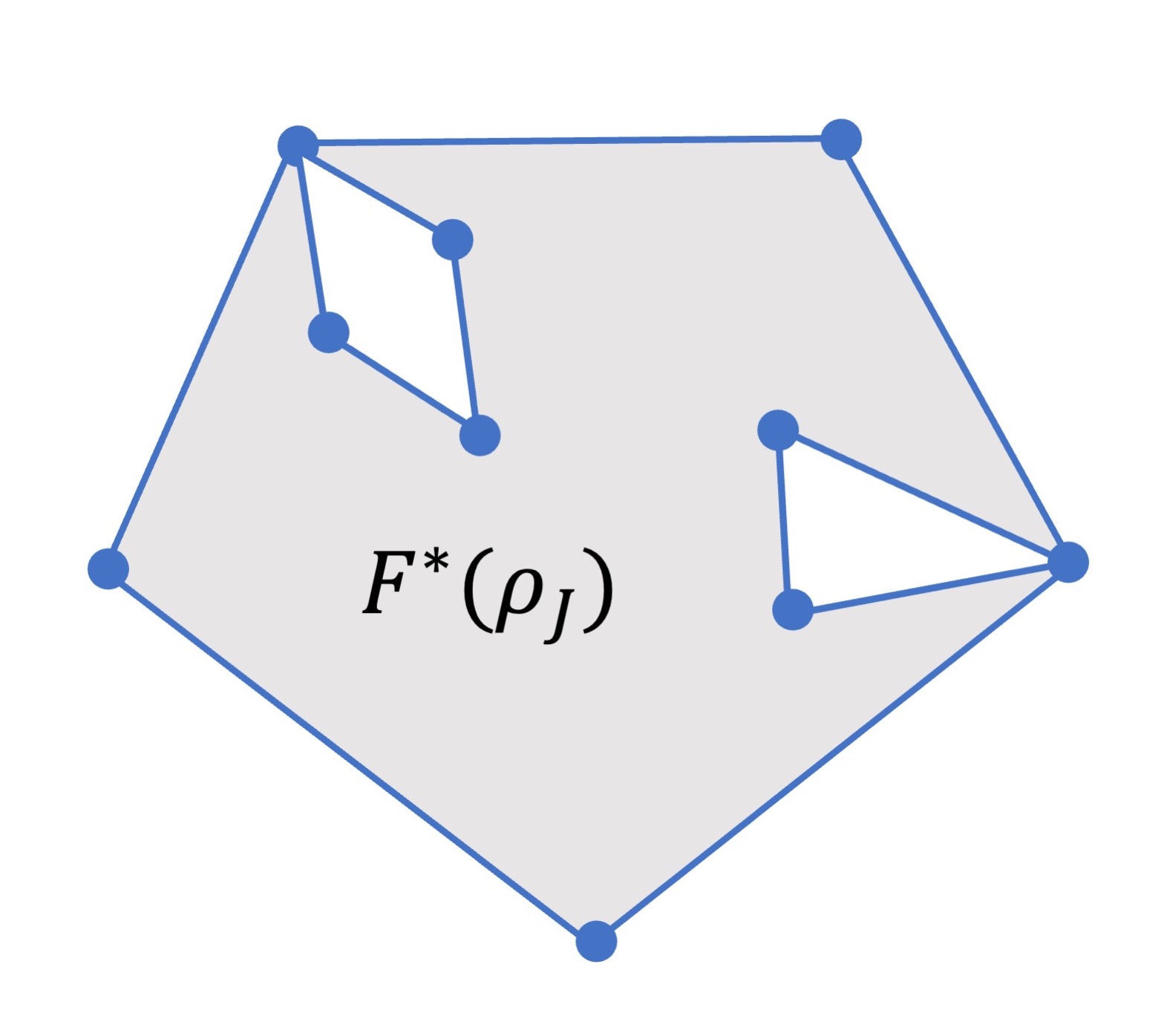}}}
	\hspace{0.08cm}
	\subfigure[Face $F^*(\rho_J)$ is split into faces $F_1$ and $F_2$ by the image of the enhancement path $P$ (shown in orange).]{\scalebox{0.09}{\includegraphics{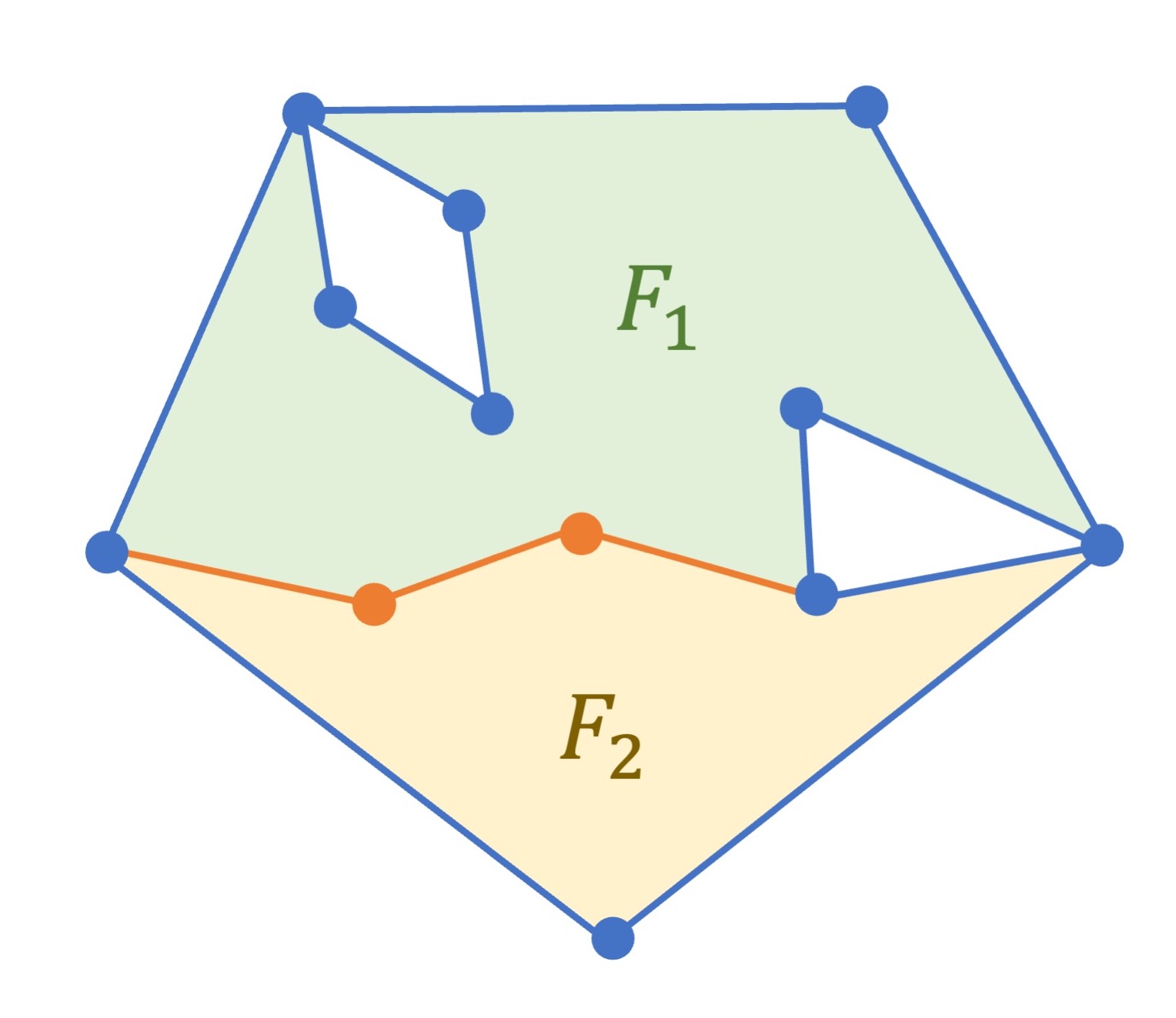}}}
	\hspace{0.08cm}
	\subfigure[New cores $J_1$  (top) and $J_2$ (bottom).]{\scalebox{0.09}{\includegraphics{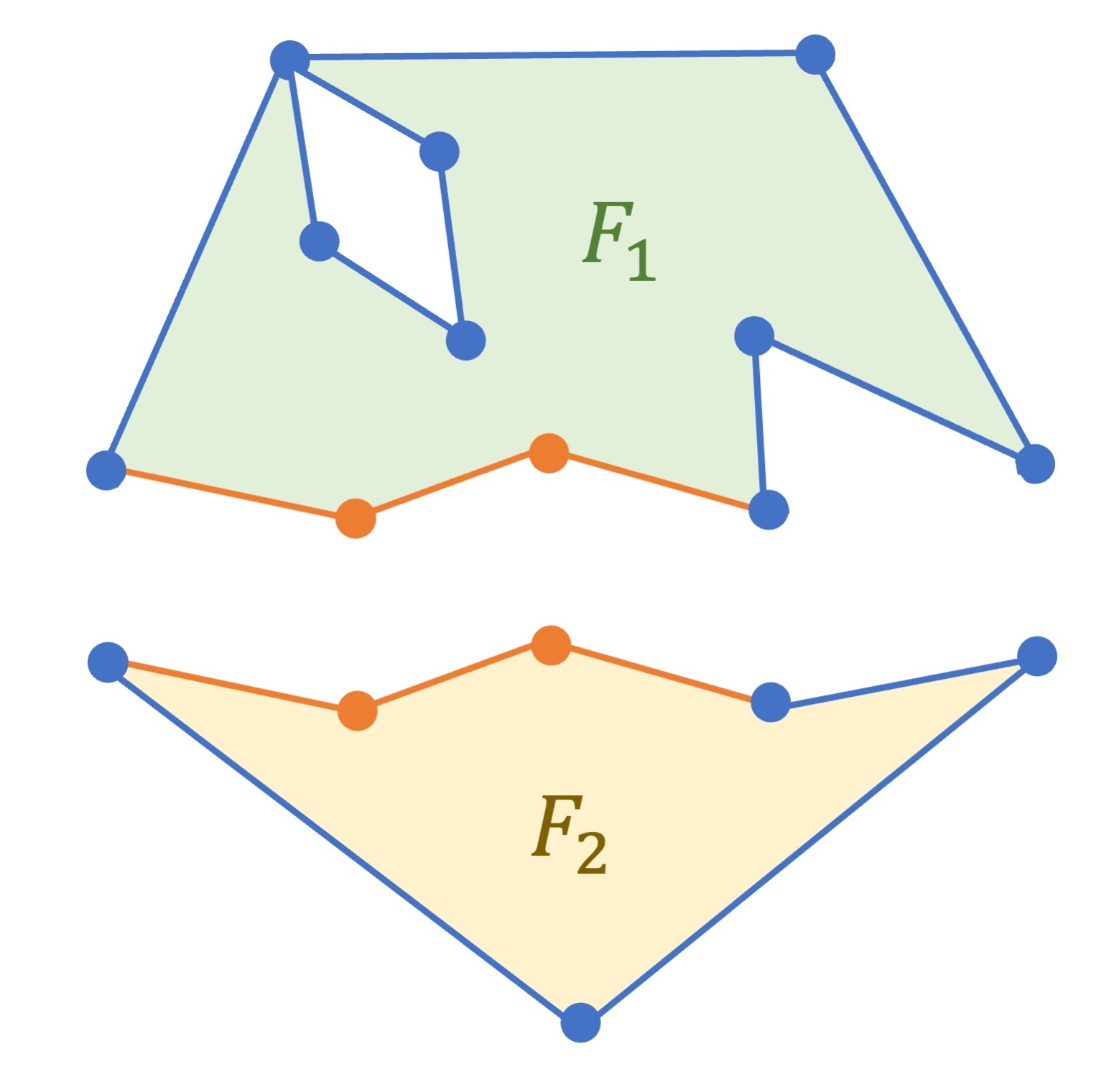}}}
	\caption{Splitting a core structure via an enhancement structure. 
	}\label{fig: type_2_enhance}
\end{figure}

We now define the core structure $\jset_1$, whose corresponding core graph is $J_1$. For every vertex $u\in V(J_1)$, its orientation $b_u$ is the same as in $\aset$. The drawing $\rho_{J_1}$ of $J_1$ is defined to be the drawing of $J_1$ induced by the drawing $\rho'$ of $J'$. Note that $F_1$ remains a face in the drawing $\rho_{J_1}$. We then let $F^*(\rho_{J_1})=F_1$.
The definition of the core structure $\jset_2$ is symmetric, except that we use core $J_2$ instead of $J_1$ and face $F_2$ instead of $F_1$ (see \Cref{fig: type_2_enhance} for an illustration). This completes the description of the algorithm for computing a split $(\jset_1,\jset_2)$ of the core structure $\jset$ via the enhancement structure $\aset$. 

Next, we define a split of an instance $I$ along a core enhancement structure $\aset$.

\begin{definition}[Splitting an Instance along an Enhancement Structure]\label{def: split}
Let $G$ be a subgraph of $\cG$, let $\jset=(J,\set{b_u}_{u\in V(J)},\rho_J, F^*(\rho_J))$ be a  core structure for the subinstance $I=(G,\Sigma)$ of $\cI$ defined by $G$, and let  $\aset=\left (P,\set{b_u}_{u\in V(J')},\rho'\right )$ be an enhancement structure for $\jset$, where $J'=J\cup P$. Let $(\jset_1,\jset_2)$ be the split of $\jset$ via the enhancement structure $\aset$, and denote by $J_1,J_2$ the cores of $\jset_1$ and $\jset_2$, respectively.
 A \emph{split} of instance $I$ along $\aset$ is a pair $I_1=(G_1,\Sigma_1), I_2=(G_2,\Sigma_2)$ of instances of \cnwrs, for which the following hold.

	\begin{itemize}		
		\item $V(G_1)\cup V(G_2)=V(G)$ and $E(G_1)\cup E(G_2)\subseteq E(G)$; 
		\item every vertex $v\in V(G_1)\cap V(G_2)$ belongs to $V(J_1)\cap V(J_2)$; 
		\item instance $I_1$ is the subinstance of $\cI$ defined by $G_1$, and instance $I_2$ is the subinstance of $\cI$ defined by $G_2$; and
		\item $\jset_1$ is a valid core structure for $I_1$, and $\jset_2$ is a valid core structure for $I_2$.
	\end{itemize}
\end{definition}

Notice that some edges of graph $G$ may not lie in $E(G_1)\cup E(G_2)$. We informally refer to such edges as \emph{deleted edges}, and we will sometimes denote the set of such deleted edges by $E^{\del}$. Typically we will ensure that $|E^{\del}|$ is quite small.

The following crucial observation shows that clean solutions to instances $I_1$ and $I_2$ can be combined to obtain a clean solution to instance $I$. The proof is deferred to Section \ref{subsec: combine solutions for split} of Appendix.

\begin{observation}\label{obs: combine solutions for split}
	There is an efficient algorithm, whose input consists of a subgraph $G$ of $\cG$,  a core structure $\jset$ for the subinstance $I=(G,\Sigma)$ of $\cI$ defined by $G$, a $\jset$-enhancement structure $\aset$, and  a split $(I_1,I_2)$ of $I$ along $\aset$, together with a clean solution $\phi_1$ to instance $I_1$ with respect to $\jset_1$,  and a clean solution $\phi_2$ to instance $I_{2}$ with respect to $\jset_2$, where $(\jset_1,\jset_2)$ is the split of $\jset$ along $\aset$. The algorithm computes a clean solution $\phi$ to instance $I$ with respect to $\jset$, with $\cro(\phi)\leq \cro(\phi_1)+\cro(\phi_2)+|E^{\del}|\cdot |E(G)|$, where $E^{\del}=E(G)\setminus (E(G_{1})\cup E(G_{2}))$.
\end{observation}

\subsubsection{Auxiliary Claim}

We will use the following simple auxiliary claim several times. The proof is  similar to the proof of Claim 9.9 in \cite{chuzhoy2020towards} and is deferred to Section \ref{subsec: curves orderings crossings} of Appendix.

\begin{claim}\label{claim: curves orderings crossings}
	Let $I=(G,\Sigma)$ be an instance of \cnwrs, and let $\pset=\set{P_1,\ldots,P_{4k+2}}$ be a collection of directed simple  edge-disjoint paths in $G$, that are non-transversal with respect to $\Sigma$. For all $1\leq i\leq 4k+2$, let $e_i$ be the first edge on path $P_i$. Assume that there are two distinct vertices $u,v\in V(G)$, such that all paths in $\pset$ originate at $u$ and terminate at $v$, and assume further that
	edges $e_1,\ldots,e_{4k+2}$ appear in this order in the rotation $\oset_u\in \Sigma$. Lastly, let $\phi$ be any solution to instance $I$, such that the number of crossings $(e,e')_p$ in $\phi$ with $e$ or $e'$ lying in $E(P_1)$ is at most $k$, and assume that the same is true for $E(P_{2k+1})$. Then $\phi$ does not contain a crossing between an edge of $P_1$ and an edge of $P_{2k+1}$.
\end{claim}

\subsection{Splitting a Subinstance: Procedure \procsplit}
\label{subsec: decomposition step}

In this subsection we describe the main subroutine that we use in our algorithm, called \procsplit. The goal of the subroutine is to split a given subinstance of the input instance $\cI$ into two. In order to simplify the statement of the main result of this subsection, we start by defining a valid input and a valid output of the subroutine. Recall that $\cI=(\cG,\cSigma)$ is the instance of $\cnwrs$ that was obtained by subdividing the instance that serves as input to \Cref{lem: many paths}. 

\paragraph{Valid Input for \procsplit.}
A valid input to Procedure \procsplit consists of a subgraph $G$ of $\cG$, a core structure $\jset=(J,\set{b_u}_{u\in V(J)},\rho_J, F^*(\rho_J))$ for the subinstance $I=(G,\Sigma)$ of $\cI$ defined by $G$, and a promising set $\pset$ of paths for $I$ and $\jset$ of cardinality $\floor{\frac{|E(G)|}{\mu^b}}$ for some constant $b$, such that
there exists a solution $\phi$ to instance $I$ that is $\jset$-valid, with $\cro(\phi)\leq \frac{|E(G)|^2}{\mu^{60b}}$ and $|\chi^{\dirty}(\phi)|\leq \frac{|E(G)|}{\mu^{60b}}$.

We emphasize that the solution $\phi$ to instance $I$ is not given as part of input and it is not known to the algorithm. 

\paragraph{Valid Output for \procsplit.}
A valid output for \procsplit consists of a
$\jset$-enhancement structure $\aset$, and
a split $(I_1=(G_1,\Sigma_1),I_2=(G_2,\Sigma_2))$ of $I$ along $\aset$. Let $P^*$ be the enhancement path of $\aset$, and let $(\jset_1,\jset_2)$ be the split of the core structure $\jset$ along $\aset$.
Denote $E^{\del}(I)=E(G)\setminus (E(G_1)\cup E(G_2))$. Let $G'=G\setminus E^{\del}(I)$, and let $I'=(G',\Sigma')$ be the subinstance of $\cI$ defined by graph $G'$. We require that the following properties hold:

\begin{properties}{P}
	\item $|E^{\del}(I)|\leq \frac{2\cro(\phi)\cdot \mu^{38b}}{m}+|\chi^{\dirty}(\phi)|$, where $m=|E(G)|$;\label{prop output deleted edges}
	\item  $|E(G_1)|,|E(G_2)|\leq |E(G)|-\frac{|E(G)|}{32\mu^{b}}$; and \label{prop: smaller graphs}
	\item there is a $\jset$-valid solution $\phi'$ for instance $I'$ that has the following properties:
	
	\begin{itemize}
		\item drawing $\phi'$ is compatible with $\phi$;
		\item the images of the edges of $E(J)\cup E(P^*)$ do not cross each other in $\phi'$;
		\item $\cro(\phi')\leq \cro(\phi)$;
		\item the number of crossings in which the edges of $P^*$ participate in $\phi'$ is at most $\frac{\cro(\phi)\cdot \mu^{12b}}{m}$;
		\item if we let $\phi_1$ be the solution to instance $I_1$ induced by $\phi'$, then drawing $\phi_1$ is $\jset_1$-valid, and similarly, if we let $\phi_2$ be the solution to instance $I_2$ induced by $\phi'$, then drawing $\phi_2$ is $\jset_2$-valid.
	\end{itemize} \label{prop output drawing}
\end{properties}

The main theorem of this subsection summarizes the properties of Procedure \procsplit.

\begin{theorem}\label{thm: procsplit}
	There is an efficient randomized algorithm, that, given a valid input  $(G,\jset,\pset)$ to procedure \procsplit, with probability at least $1-\frac{2^{20}}{\mu^{10b}}$ computes a valid output for the procedure.
\end{theorem}

We will refer to the algorithm from \Cref{thm: procsplit} as \procsplit. The remainder of this subsection is dedicated to the proof of \Cref{thm: procsplit}. Throughout the proof, we denote by $I=(G,\Sigma)$ the subinstance of $\cI$ defined by graph $G$, by  $\jset=(J,\set{b_u}_{u\in V(J)},\rho_J, F^*(\rho_J))$ the given core structure for $I$, and by $m=|E(G)|$. The algorithm consists of two steps. In the first step, we compute an enhancement $P^*$ of the core structure $\jset$ and analyze its properties. In the second step, we complete the construction of the enhancement structure $\aset$ and of the split $(I_1,I_2)$ of instance $I$ along $\aset$. We now describe each of the steps in turn. In order to simplify the exposition, throughout this subsection, we use ``enhancement'' and ``enhancement structure'' when we refer to an enhancement of $\jset$ and enhancement structure for $\jset$. For simplicity of notation, we also denote the drawing $\rho_J$ of the core $J$ by $\rho$, and the face $F^*(\rho_J)$ of this drawing by $F^*$.

\subsubsection{Step 1: Computing an Enhancement}
\label{subsub: step 1 of phase 1 of interesting}

We start by discarding from $\pset$ all paths that contain more than $32\mu^b$ edges. Since the paths in $\pset$ are edge-disjoint, the number of the discarded path is bounded by $\frac{m}{32\mu^b}$, and so $|\pset|\geq \frac{15m}{16\mu^b}$ continues to hold.

In order to compute an enhancement, we slightly modify the promising set $\pset$ of paths, to ensure that, for every pair $P,P'\in \pset$ of distinct paths, for every vertex $v\in (V(P)\cap V(P'))\setminus V(J)$, the intersection of $P$ and $P'$ at vertex $v$ is  non-transversal. 
Throughout, we denote by $(E_1,E_2)$ the partition of the edges of $\delta_G(J)$, with the edges of $E_1$ appearing consecutively in the ordering $\oset(J)$, suh that every path $P\in \pset$ has an edge of $E_1$ as its first edge and an edge of $E_2$ as its last edge.

In order to modify the set $\pset$ of paths, we  first subdivide every edge $e\in \delta_G(J)$ with a new vertex $t_e$, denoting $T_1=\set{t_e\mid e\in E_1}$ and $T_2=\set{t_e\mid e\in E_2}$. 
Let $H$ be a new graph obtained from $G$ after we delete the vertices and the edges of $J$ from it, contract all vertices of $T_1$ into a new vertex $s$, and contract all vertices of $T_2$ into a new vertex $t$.  We define a rotation system $\tilde \Sigma$ for graph $H$ in a natural way: For every vertex $v\in V(H)\setminus\set{s,t}$, its rotation $\oset_v$ in $\tilde \Sigma$ remains the same as in $\Sigma$. For vertex $s$ its rotation $\oset_s\in \tilde \Sigma$ is the circular ordering of the edges of $E_1$ induced by the ordering $\oset(J)$.
Similarly, rotation $\oset_t\in \tilde \Sigma$  is the circular ordering of the edges of $E_2$ induced by the ordering $\oset(J)$.

The set $\pset$ of paths in graph $G$ defines a set $\qset$ of $|\pset|$ edge-disjoint paths in $H$ that connect $s$ to $t$, and are internally disjoint from $s$ and $t$. We apply the algorithm from \Cref{lem: non_interfering_paths} to graph $H$ and the set $\qset$ of paths to obtain another set $\qset'$ of $|\pset|$ simple edge-disjoint paths in $H$, where every path connects  $s$ to $t$ and is internally disjoint from both $s$ and $t$ as before, but now the paths are non-transversal with respect to $\tilde \Sigma$. Lastly, path set $\qset'$ naturally defines a collection $\pset^*$ of $|\pset|$ simple edge-disjoint paths in graph $G$, where every path in $\pset^*$ contains an edge of $E_1$ as its first edge, and an edge of $E_2$ as its last edge, and is internally disjoint from $J$. Moreover,  for every vertex $u\in V(G)\setminus V(J)$, for every pair $P,P'\in \pset^*$ of paths that contain $u$, the intersection of $P$ and $P'$ at $u$ is non-transversal. Notice that $\pset^*$ remains a promising set of paths of cardinality $|\pset|$. We view the paths in $\pset^*$ as being directed from the edges of $E_1$ towards the edges of $E_2$. We denote by $k=|\pset^*|$, so $k=|\pset|\geq \frac{15m}{16\mu^b}$.

Let $E^*_1\subseteq E_1$  be the subset of edges that belong to the paths of $\pset^*$.
We denote $E^*_1=\set{e_1,\ldots,e_{k}}$, where the edges are indexed so that $e_1,\ldots,e_{k}$ appear in the order of their indices in the ordering $\oset(J)$. For all $1\leq j\leq k$, we denote by $P_j\in \pset^*$ the unique path originating at the edge $e_j$. We select an index $\floor{k/3}<j^*<\ceil{2k/3}$ uniformly at random, and we let $P^*=P_{j^*}$. Notice that $P^*$ is a valid enhancement for the core structure $\jset$, and it is either a simple path or a simple cycle. We say that path $P^*$ is \emph{chosen} from set $\pset^*$. 
Notice that the probability that a path of $\pset^*$ is chosen to be the enhancement path is at most $\frac{4}{k}\leq \frac{4\cdot 16\mu^b}{15m}\leq 
 \frac{16\mu^{b}}{m}$.

We will now define a number of bad events, and we will show that the probability that either of these events happens is low.

\paragraph{Good Paths and Bad Event $\event_1$.}

We need the following definition.

\begin{definition}[Good path]
We say that a path $P\in \pset^*$ is \emph{good} if the following hold:

\begin{itemize}
	\item the number of crossings in which the edges of $P$ participate in $\phi$ is at most $\frac{\cro(\phi)\cdot \mu^{12b}}{m}$; and
	
	\item there are no crossings in $\phi$ between edges of $P$ and edges of $J$.
\end{itemize} A path that is not good is called a \emph{bad path}.
\end{definition}

We now bound the number of bad paths in $\pset^*$.

\begin{observation}\label{obs: number of bad paths}
	The number of bad paths in $\pset^*$ is at most $\frac{4m}{\mu^{12b}}$.
\end{observation}

\begin{proof}
	Since the paths in $\pset^*$ are edge-disjoint, and every crossing involves two edges, the number of paths $P\in \pset^*$ such that there are more than $\frac{\cro(\phi)\mu^{12b}}{m}$ crossings in $\phi$ in which the edges of $P$ participate, is at most $\frac{2m}{\mu^{12b}}$. Additionally, we are guaranteed that $|\chi^{\dirty}(\phi)|\le \frac{m}{\mu^{60b}}$. Therefore, the number of paths $P\in \pset^*$, for which there is a crossing between an edge of $P$ and an edge of $J$, is bounded by  $\frac m {\mu^{60b}}$.
	Overall, the number of bad paths in $\pset^*$ is bounded by $\frac{2m}{\mu^{12b}}+\frac m {\mu^{60b}}\leq  \frac{4m}{\mu^{12b}}$.
\end{proof}

We say that bad event $\event_1$ happens if path $P^*$ is a bad path. Since the number of bad paths is bounded by $\frac{4m}{\mu^{12b}}$, and a path of $\pset^*$ is chosen to be the enhancement path with probability at most $\frac{16\mu^{b}}{m}$, we immediately get the following observation.

\begin{claim}\label{claim: event 1 prob2}
	$\prob{\event_1}\leq 64/\mu^{11b}$.
\end{claim}

\paragraph{Heavy and Light Vertices, and Bad Event $\event_2$.}

We use a parameter $h=\frac{512\cro(\phi)\cdot \mu^{26b}}{m}$. We say that a vertex $x\in V(G)$ is \emph{heavy} if at least $h$ paths of $\pset^*$ contain $x$; otherwise, we say that $x$ is \emph{light}. Recall that in order to define an enhancement structure using the enhancement $P^*$, we need to define an orientation for every inner vertex of $P^*$. 
Intuitively, we would like this orientation to be consistent with that in the drawing $\phi$ (which is not known to us). If $x$ is a heavy vertex lying on $P^*$, then computing such an orientation is not difficult, as we can exploit the paths of $\pset^*$ containing $x$ to do so. But if $x$ is a light vertex, we do not have enough information in order to determine its orientation in $\phi$. To get around this problem, we will simply delete all edges incident to the light vertices of $P^*$, except for the edges of $P^*\cup J$, and then let the orientation of each such vertex be arbitrary. We show below that with high probability, the number of edges that we delete is relatively small.

Specifically, we denote by $E'$ the set of all edges $e$, such that $e$ is incident to some light vertex $x\in V(P^*)$, and $e\not\in E(J)\cup E(P^*)$. We say that bad event $\event_2$ happens if $|E'|>\frac{\cro(\phi)\cdot \mu^{38b}}{m}$.
We bound the probability of Event $\event_2$ in the next simple claim.

\begin{claim}\label{claim: third bad event bound}
	$\prob{\event_2}\leq 2^{14}/\mu^{11b}$.
\end{claim}
\begin{proof}
	Consider some light vertex $x\in V(G)$. Since $x$ lies on fewer than $h= \frac{512\cro(\phi)\cdot \mu^{26b}}{m}$ paths of $\pset^*$, and each such path is chosen to the enhancement with probability at most $\frac{16\mu^{b}}{m}$, the probability that $x$ lies in $V(P^*)$ is at most $\frac{512\cro(\phi)\mu^{26b}}{m}\cdot \frac{16\mu^{b}}{m}\leq \frac{2^{13}\cro(\phi)\mu^{27b}}{m^2}$.
	
	Consider now some edge $e=(x,y)\in E(G)$. Edge $e$ may lie in $E'$ only if $x$ is a light vertex lying in $V(P^*)$, or the same is true for $y$. Therefore, the probability that $e\in E'$ is at most $\frac{2^{14}\cro(\phi)\mu^{27b}}{m^2}$, and $\expect{|E'|}\leq \frac{2^{14}\cro(\phi)\mu^{27b}}{m}$. From Markov's inequality, $\prob{|E'|>\frac{\cro(\phi)\cdot \mu^{38b}}{m}}\leq \frac{2^{14}}{\mu^{11b}}$.
\end{proof}

\paragraph{Unlucky Paths and Bad Event $\event_3$.}

If Event $\event_1$ does not happen, then we are guaranteed that, in drawing $\phi$, the edges lying on path $P^*$ do not cross the edges of $J$. However, it is still possible that there are crossings between edges that lie on path $P^*$ (that is, the image of path $P^*$ crosses itself). Intuitively, when the image of path $P^*$ crosses itself, then we obtain a loop. If we could show that all vertices lying on this loop are light vertices, then we can ``repair'' the drawing by straightening the loop. This is since we delete all edges incident to the light vertices lying on $P^*$, except for the edges of $E(P^*)\cup E(J)$. Unfortunately, it may happen that some of the vertices lying on these loops are heavy vertices. This may only happen in some limited circumstances, in which case we say that path $P^*$ is \emph{unlucky}. We now define the notion of unlucky paths, and show that the probability that $P^*$ is unlucky is small.

\begin{figure}[h]
	\centering
	\includegraphics[scale=0.13]{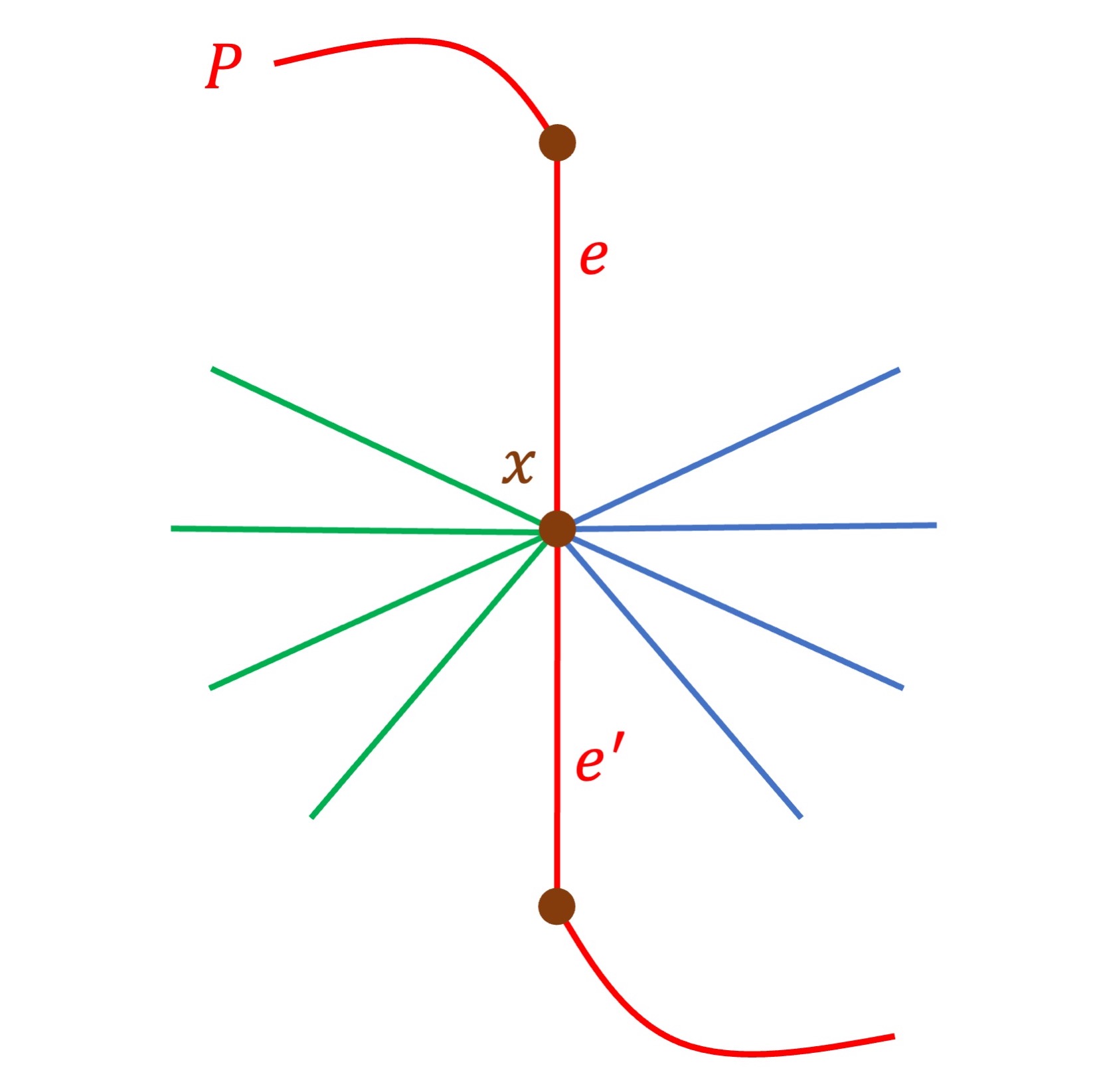}
	\caption{Definition of the sets $\hat E_1(x)$ and $\hat E_2(x)$ of edges. Path $P$ and its edges $e,e'$ are shown in red. 
Edges of $\delta(x)$ are depicted according to the circular order $\oset_x\in \Sigma$, and the set $\delta(x)\setminus \set{e,e'}$ is split into two subsets (green and blue).
Set $\hat E_1(x)$ contains every green edge that belongs to some good path of $\pset^*$, and set $\hat E_2(x)$ contains every blue edge that belongs to some good path of $\pset^*$.
}\label{fig: unlucky} 
\end{figure}

\begin{definition}[Unlucky Paths]
	Let $x\in V(G)\setminus V(J)$ be a vertex, and let $P\in \pset^*$ be a good path that contains $x$. Let $e,e'$ be the two edges of $P$ that are incident to $x$. Let  $\hat E_1(x)\subseteq \delta_G(x)$ be the set of edges $\hat e\in \delta_G(x)$, such that $\hat e$ lies between $e$ and $e'$ in the rotation $\oset_x\in \Sigma$ (in clock-wise orientation), and $\hat e$ lies on some good path of $\pset^*$. Let $\hat E_2(x)\subseteq \delta_G(x)$ be the set of edges $\hat e\in \delta_G(x)$, such that $\hat e$ lies between $e'$ and $e$ in the rotation $\oset_x\in \Sigma$ (in clock-wise orientation), and $\hat e$ lies on some good path of $\pset^*$ (see \Cref{fig: unlucky}).
	We say that path $P$ is \emph{unlucky with respect to vertex $x$} if either  $|\hat E_1(x)|<\frac{\cro(\phi)\mu^{13b}}{m}$ or $|\hat E_2(x)|< \frac{\cro(\phi)\mu^{13b}}{m}$ holds.
	We say that a path $P\in \pset^*$ is an \emph{unlucky path} if there is at least one heavy vertex $x\in V(G)\setminus V(J)$, such that $P$ is unlucky with respect to $x$.
\end{definition}

We will now show that the total number of good paths in $\pset^*$ that are unlucky  is small, and we will conclude that the probability that an unlucky path was chosen to be the enhancement path $P^*$ is also small. The proof of the following claim is somewhat technical and is defered to Section \ref{sec: bound unlucky paths} of Appendix.

\begin{claim}\label{claim: bound unlucky paths}
For every vertex $x\in V(G)\setminus V(J)$, the total number of good paths in $\pset^*$ that are unlucky with respect to $x$ is at most  $\frac{512\cro(\phi)\cdot \mu^{13b}}{m}$. 
\end{claim}

We say that bad event $\event_3$ happens if $P^*$ is an unlucky path. 

\begin{claim}\label{claim: third event bound}
	$\prob{\event_3}\leq 64/\mu^{10b}$.
\end{claim}

\begin{proof}
Recall that a heavy vertex must have degree at least $h$ in $G$. Therefore, the total number of heavy vertices is at most $\frac{2m}{h}$. From \Cref{claim: bound unlucky paths}, for every heavy vertex $x\in V(G)\setminus V(J)$, there are at most $\frac{512\cro(\phi)\cdot \mu^{13b}}{m}$ paths in $\pset^*$ that are good and unlucky for $x$. Since $h=\frac{512\cro(\phi)\cdot \mu^{26b}}{m}$, the total number of good paths in $\pset^*$ that are unlucky with respect to some heavy vertex is at most:

\[ \frac{2m}{h}\cdot \frac{512\cro(\phi)\cdot \mu^{13b}}{m}\leq \frac{2m}{ \mu^{13b}}.\]

Since the probability that a given path $P\in \pset^*$ is selected is at most $\frac{16\mu^b} m$, the probability that an unlucky path is selected is at most $\frac{32}{\mu^{12b}}$.	
\end{proof}

Recall that we have denoted by $E'$ the set of all edges $e$, such that $e$ is incident to some light vertex $x\in V(P^*)$, and $e\not\in E(J)\cup E(P^*)$.
 Denote $G'=G\setminus E'$, and let $\Sigma'$ be the rotation system for graph $G'$ induced by $\Sigma$. Denote $I'=(G',\Sigma')$, and  denote  $J'=J\cup P^*$.  Solution $\phi$ to instance $I$ naturally defines a soluton $\phi'$ to instance $I'$, that is compatible with $\phi$, with $\cro(\phi)\leq \cro(\phi')$. Moreover, if Event $\event_1$ did not happen, then there are no crossings in this drawing between the edges of $E(P^*)$ and the edges of $E(J)$. However, it is possible that this drawing contains crossings between pairs of edges in $E(P^*)$. In the next claim we show that, if events $\event_1$ and $\event_3$ did not happen, then drawing $\phi$ can be modified to obtain a solution $\phi'$ to instance $I'$ that is compatible with $\phi$,  in which the edges of $J'$ do not cross each other.
The proof of the following claim is deferred to Section \ref{sec:getting new drawing} of Appendix.

\begin{claim}\label{claim: new drawing}
	Assume that neither of the events $\event_1$ and $\event_3$ happened. Then there is a  solution $\phi'$ to instance $I'=(G',\Sigma')$ that is compatible with $\phi$, with $\cro(\phi')\leq \cro(\phi)$, such that the edges of $E(J')$ do not cross each other in $\phi'$. Moreover, if $(e,e')_p$ is a crossing in drawing $\phi'$, then there is a crossing between edges $e$ and $e'$ at point $p$ in drawing $\phi$.
\end{claim}

We emphasize that drawing $\phi'$ is derived from drawing $\phi$ and neither are known to our algorithm. From now on, we fix the solution $\phi'$ to instance $I'=(G',\Sigma')$ given by \Cref{claim: new drawing}.

\paragraph{Terrible Vertices and Bad Event $\event_4$.}

For a vertex $x\in V(G)$, let $N(x)$ denote the number of paths in $\pset^*$ containing $x$. We also denote by $N^{\bad}(x)$ the number of bad paths in $\pset^*$ containing $x$, and  by $N^{\good}(x)$ the number of good paths in $\pset^*$ containing $x$. Next, we define the notion of a terrible vertex.

\begin{definition}[Terrible Vertex]
	A vertex $x\in V(G)$ is \emph{terrible} if it is a heavy vertex, and $N^{\bad}(x)\geq N^{\good}(x)/64$.
\end{definition}

We say that a bad event $\event_4$ happens if any vertex of $P^*$ is a terrible vertex. We bound the probability of Event $\event_4$ in the following claim.

\begin{claim}\label{claim: no terrible vertices}
	$\prob{\event_4}\leq \frac{2^{18}}{\mu^{10b}}$.
\end{claim}

\begin{proof}
	Consider some terrible vertex $x\in V(G)$. Let $\pset'\subseteq \pset^*$ be the set of all bad paths in $\pset^*$ containing $x$, and let $\pset''\subseteq \pset^*$ be the set of all good paths in $\pset^*$ containing $x$. From the definition of a terrible vertex, $|\pset''|\leq 64|\pset'|$. Therefore, we can define a mapping $f_x:\pset''\rightarrow \pset'$, that maps every path in $\pset''$ to some path in $\pset'$, such that, for every path $P\in \pset'$, at most $64$ paths of $\pset''$ are mapped to $P$. If, for a pair $P''\in \pset''$, $P'\in \pset'$ of paths, $f_x(P'')=P'$, then we say that path $P'$ \emph{tags} path $P''$. Every bad path also tags itself.
	
	Notice that, if path $P\in \pset^*$ is a bad path, then the total number of  paths that it may tag is bounded by $64|V(P)|\leq 2^{12}\mu^b$ (as every path in $\pset^*$ contains at most $32\mu^b+2$ vertices). Every path of $\pset^*$ that contains a terrible vertex is now tagged. Since, from \Cref{obs: number of bad paths},
	the number of bad paths in $\pset^*$ is at most $\frac{4m}{\mu^{12b}}$, we conclude that the total number of  paths in $\pset^*$ that contain a terrible vertex is bounded by $\frac{2^{14}m}{\mu^{11b}}$.  Lastly, since the probability that a given path $P\in \pset^*$ is selected is at most $\frac{16\mu^b} m$, the probability that a path containing a terrible vertex is selected is bounded by:
	
	\[ \frac{2^{14}m}{\mu^{11b}}\cdot \frac{16\mu^b} m\leq \frac{2^{18}}{\mu^{10b}}. \]
\end{proof}

\paragraph{Bad Event $\event$.}
Let $\event$ be the bad event that either of the events $\event_1,\event_2,\event_3,\event_4$ happens. From the Union Bound and Claims \ref{claim: event 1 prob2}, \ref{claim: third bad event bound}, \ref{claim: third event bound} and \ref{claim: no terrible vertices}, $\prob{\event}\leq  \frac{2^{20}}{\mu^{10b}}$.
\subsubsection{Step 2: Computing the Enhancement Structure and the Split}
\label{subsub: step 2 of phase 1 of interesting}

In this step, we compute an orientation $b_u$ for every vertex $u\in V(P^*)\setminus V(J)$, that is  identical to the orientation of $u$ in drawing $\phi'$ (though drawing $\phi'$ itself is not known to the algorithm). We will then complete the construction of the enhancement structure $\aset$, and compute the split of instance $I$ along $\aset$. Throughout, we denote $J'=J\cup P^*$. Let $\rho'$ be the drawing of graph $J'$ that is induced by drawing $\phi'$ of $G'$. If Event $\event$ did not happen, then drawing $\rho'$ has no crossings, and the image of path $P^*$ is drawn in the region $F^*$. Let $\rho_{J'}$ be the unique drawing of graph $J'$ that has the following properties:

\begin{itemize}
	\item drawing $\rho_{J'}$ contains no crossings;
	\item drawing $\rho_{J'}$ obeys the rotation system $\Sigma$, and, for every vertex $u\in V(J)$, the orientation of $u$ in $\rho_{J'}$ is the orientation $b_u$ given by $\jset$;
	\item the drawing of graph $J$ induced by $\rho_{J'}$ is precisely $\rho_J$; and
	\item the image of path $P^*$ is contained in region $F^*$.
\end{itemize}

Note that there is a unique drawing $\rho_{J'}$ of $J'$ with the above properties, and it can be computed efficiently. Moreover, if event $\event$ did not happen, then $\rho_{J'}=\rho'$ must hold. The image of path $P^*$ partitions the region $F^*$ of $\rho_{J'}$ into two faces, that we denote by $F_1$ and $F_2$.  These two faces define regions in drawing $\phi'$ of $G'$, that we denote by $F_1$ and $F_2$, as well.

For every vertex $u\in V(J')$, we consider the tiny $u$-disc $D_{\phi'}(u)$.  For every edge $e\in \delta_{G'}(u)$, we denote by $\sigma(e)$ the segment of $\phi'(e)$ that is drawn inside the disc $D_{\phi'}(u)$. Let $\tilde E=\left (\bigcup_{u\in V(J')}\delta_{G'}(u)\right )\setminus E(J')$. Recall that, from the definiton of a valid core structure, and since the image of path $P^*$ is contained in region $F^*$ of $\phi'$, for every edge $e\in \tilde E$, segment $\sigma(e)$ must be contained in region  $F^*$. We partition edge set $\tilde E$ into a set $\tilde E^{\inn}$ of \emph{inner edges} and the set $\tilde E^{\out}$ of \emph{outer edges}, as follows.  Edge set $\tilde E^{\inn}$ contains all edges $e\in \tilde E$ with $\sigma(e)$ contained in region $F_1$ of $\phi'$, and $\tilde E^{\out}$ contains all remaining edges (so for every edge $e\in \tilde E^{\out}$, $\sigma(e)$ is contained in $F_2$). Let $e_1$ be the first edge of $E_1$ in the ordering $\oset(J)$. We will assume without loss of generality that $e_1\in \tilde E^{\inn}$. We now show an algorithm that correctly computes the orientation of every vertex $u\in V(P^*)\setminus V(J)$ in the drawing $\phi'$, and the partition $(\tilde E^{\inn},\tilde E^{\out})$ of the edges of $\tilde E$.

Before we describe the algorithm, we recall the definition of the oriented circular ordering $\oset(J)$ of the edges of $\delta_G(J)$.  In order to define the ordering, we considered the disc $D(J)$ in the drawing $\rho_{J}$ of $J$. In this drawing, the orientation of every vertex $u\in V(J)$ is the orientation $b_u$ given by the core structure  $\jset$. We have defined, for every edge $e\in \delta_G(J)$, a point $p(e)$ on the boundary of the disc $D(J)$, and we let $\oset(\thecore)$ be the circular ordering of the edges of $\delta_G(\thecore)$, in which the  points $p(e)$ corresponding to these edges  appear on the boundary of the disc $D(\thecore)$, as we traverse the boundary of the disc $D(\thecore)$ in the clock-wise direction. From the definition of a $\jset$-valid drawing, the drawing of the core $J$ induced by $\phi'$ is identical to $\rho_J$, including its orientation. Additionally, for every vertex $u\in V(J)$, the orientation of $u$ in $\phi'$ is the orientation $b_u$ given by $\jset$.

\paragraph{Computing Vertex Orientations and the Partition $(\tilde E^{\inn},\tilde E^{\out})$.}
Consider any vertex $u\in V(P^*)\setminus V(J)$. Let $\hat e(u)$, $\hat e'(u)$ be the two edges of $P^*$ that are incident to $u$, where we assume that $\hat e(u)$ appears before $\hat e'(u)$ on $P^*$ (we assume that $P^*$ is directed from an edge of $E_1$ to an edge of $E_2$). Edges $\hat e(u),\hat e'(u)$ partition the edge set $\delta_{G'}(u)\setminus\set{\hat e(u),\hat e'(u)}$ into two subsets, that we denote by $\hat E_1(u)$ and $\hat E_2(u)$, each of which appears consecutively in the rotation $\oset_u\in \Sigma'$. 
Note that either (i) $\hat E_1(u)\subseteq \tilde E^{\inn}$ and $\hat E_2(u)\subseteq \tilde E^{\out}$ holds, or (ii) $\hat E_2(u)\subseteq \tilde E^{\inn}$ and $\hat E_1(u)\subseteq \tilde E^{\out}$ holds.
While we do not know the orientation of vertex $u$ in $\phi'$, once we fix this orientation, we can efficiently determine which of the above two conditions holds. Therefore, we can assume w.l.o.g. that, if the orientation of $u$ in $\phi'$ is $1$ then $\hat E_1(u)\subseteq \tilde E^{\inn}$ and $\hat E_2(u)\subseteq \tilde E^{\out}$ hold  (as otherwise we can switch the names $\hat E_1(u)$ and $\hat E_2(u)$).

We now construct edge sets $\tilde E_1,\tilde E_2$, and fix an orientation $b_u$ for every vertex $u\in V(P^*)\setminus V(J)$. We then show that $\tilde E_1=\tilde E^{\inn}$, $\tilde E_2=\tilde E^{\out}$, and that the orientations of all vertices of $V(P^*)\setminus V(J)$ that we compute are consistent with the drawing $\phi'$.

Consider the drawing $\rho_{J'}$ of graph $J'$ that we have computed. Using this drawing, we can efficiently determine, for every edge $e\in \tilde E$ that is incident to a vertex of $J$, whether $e\in \tilde E^{\inn}$ or $e\in \tilde E^{\out}$ holds. In the former case, we add $e$ to $\tilde E_1$, and in the latter case we add it to $\tilde E_2$.
Notice that, for every path $P\in \pset^*$, the first and the last edges of $P$ are already added to either $\tilde E_1$ or $\tilde E_2$, and so far $\tilde E_1\subseteq \tilde E^{\inn}$ and $\tilde E_2\subseteq \tilde E^{\out}$ holds.

Next, we process every inner vertex $u$ on path $P^*$.
Consider any such vertex $u$. If $u$ is a light vertex, then there are exactly two edges that are incident to $u$ in $G'$ -- the edges of the path $P^*$. We can then set the orientation $b_u$ of $u$ to be arbitrary, and we can trivially assume that this orientaiton is identical to the orientation of $u$ in $\phi'$.

Assume now that $u$ is a heavy vertex. 
In order to establish the orientation of $u$, we let $\pset(u)$ contain all paths $P\in \pset^*\setminus\set{P^*}$ with $u\in P$.  We partition the set $\pset(u)$  of paths into four subsets: set $\pset_1(u)$ contains all paths $P$ whose first edge lies in $\tilde  E_1$, and the first edge of $P$ that is incident to $u$ lies in $\hat E_1(u)$. Set $\pset_2(u)$  contains all paths $P$ whose first edge lies in $\tilde  E_2$, and the first edge  of $P$  that is incident to $u$ lies in $\hat E_2(u)$. Similarly, set $\pset'_1(u)$ contains   all paths $P\in \pset(u)$ whose first edge lies in $\tilde  E_1$ and the first edge that is incident to $u$ lies in $\hat E_2(u)$, while set $\pset'_2(u)$ contains  all paths $P\in \pset'(u)$, whose first edge lies in $\tilde  E_2'$ and the first edge that is incident to $u$ lies in $\hat E_1(u)$. We let $w(u)=|\pset_1|+|\pset_2|$, and $w'(u)=|\pset'_1|+|\pset'_2|$.
If $w(u)\geq  w'(u)$, then we set $b_u=1$, add the edges of $\hat E_1(u)$ to $\tilde E_1$, and add the edges of $\hat E_2(u)$ to $\tilde E_2$.  Otherwise we set $b_u=-1$,  add the edges of $\hat E_1(u)$ to $\tilde E_2$, and add the edges of $\hat E_2(u)$ to $\tilde E_1$.

This completes the algorithm for computing the orientations of the inner vertices of $P^*$, and of the partition $(\tilde E_1,\tilde E_2)$ of the edge set $\tilde E$.
We use the following claim to show that both are computed correctly. 

\begin{claim}\label{claim: orientations and edge split is computed correctly Case2}
	Assume that Event $\event$ did not happen. Then for every vertex $u\in V(P^*)\setminus V(J)$, the orientation of $u$ in $\phi'$ is $b_u$.
\end{claim}

\begin{proof}
 It is now enough to show that, if $u\in V(P^*)\setminus V(J)$ is a heavy vertex, then the orientation of $u$ in $\phi'$ is $b_u$. We now consider any heavy vertex $u\in V(P^*)\setminus V(J)$.
	
	Recall that we denoted $N(u)=|\pset(u)|$, and we have denoted by $N^{\bad}(u)$ and $N^{\good}(u)$ the total number of bad and good paths in $\pset(u)$, respectively. Since we have assumed that bad event $\event_4$ did not happen, vertex $u$ is not a terrible vertex, that is, $N^{\bad}(u)< N^{\good}(u)/64$. Since $N(u)=N^{\bad}(u)+N^{\good}(u)$, we get that $N^{\bad}(u)<N(u)/65$.

	Assume first that the orientation of vertex $u$ in $\phi'$ is $1$, so $\hat E_1(u)\subseteq \tilde E^{\inn}$ and $\hat E_2(u)\subseteq \tilde E^{\out}$. 
	We claim that in this case $w(u)>w'(u)$ must hold, and so our algorithm sets $b_u=1$ correctly. Indeed, assume otherwise. 
	Then $w'(u)\geq N(u)/2$. Let $\qset$ denote the set of all good paths in $\pset'_1(u)\cup \pset'_2(u)$. Then $|\qset|\geq w'(u)-N^{\bad}(u)\geq N(u)/2-N(u)/65\geq N(u)/4\geq h/4$, since $u$ is a heavy vertex. 
	We now show that, for every path $Q\in \qset$, there must be a crossing between an edge of $Q$ and an edge of $P^*$ in $\phi'$.
	
	Indeed, consider any path $Q\in \qset$. Since $Q\in \pset'_1(u)\cup \pset'_2(u)$, either the first edge of $Q$ lies in $\tilde E^{\inn}$ and the last edge of $Q$ lies in $\tilde E^{\out}$, or the opposite is true. Therefore, the image of the path $Q$ must cross the boundary of the region $F_1$. Since path $Q$ is a good path, and it does not contain vertices of $J$ as inner vertices, no inner point of the image of $Q$ in $\phi'$ may belong to the image of $J$ in $\phi'$. Since, for every pair $P,P'\in \pset^*$ of paths, and for every vertex $v\in (V(P)\cap V(P'))\setminus V(J)$, the intersection of $P$ and $P'$ at $v$ is non-transversal, there must be a crossing between an edge of $Q$ and an edge of $P^*$ in $\phi'$. But then the edges of $P^*$ participate in at least $\frac h 4\geq \frac{128\cro(\phi)\cdot \mu^{26b}}{m}$ crossings in $\phi'$, and hence in $\phi$. However, since we have assumed that bad event $\event_1$ did not happen, $P^*$ is a good path, and so its edges may participate in at most $\frac{\cro(\phi)\cdot \mu^{12b}}{m}$ crossings in $\phi$, a contradiction. Therefore, when the orientation of $u$ in $\phi'$ is $1$, our algorithm correctly sets $b_u=1$. 
	
	In the case where the orientation of $u$ in $\phi'$ is $-1$, the analysis is symmetric. In this case, we consider the set  $\qset'\subseteq \pset_1(u)\cup \pset_2(u)$ containing all good paths. For each such path $P\in \qset'$,  the image of $P$ in $\phi$ must cross the image of $P^*$. If we assume that $w(u)\geq w'(u)$ in this case, then we reach a contradiction using the same argument as before. Therefore, $w(u)<w'(u)$ must hold, and our algorithm sets $b_u=1$ correctly.
\end{proof}

We have now obtained an enhancement structure 
$\aset=(P^*,\set{b_u}_{u\in V(J')},\rho_{J'})$. For every vertex $u\in V(J')$, if $u\in V(J)$, then its orientaiton $b_u$ remains the same as in $\jset$, and otherwise we let $b_u$ be the orientation that we have computed above. From the above discussion, if Event $\event$ did not happen, then for every vertex $u\in V(J')$ the orientation $b_u$ is identical to its orientation in $\phi'$, and $\rho_{J'}$ is the drawing of $J'$ induced by $\phi'$.
We denote by $(\jset_1,\jset_2)$ the split of $\jset$ via the enhancement structure $\aset$, where $\jset_1$ is the core structure associated with the face $F_1$. We denote $\jset_1=(J_1,\set{b_u}_{u\in V(J_1)},\rho_{J_1}, F^*(\rho_{J_1}))$, and $\jset_2=(J_2,\set{b_u}_{u\in V(J_2)},\rho_{J_2}, F^*(\rho_{J_2}))$, where $F^*(\rho_{J_1})=F_1$ and $F^*(\rho_{J_2})=F_2$.

\paragraph{Computing the Split.}

We now construct a split of instance $I$ along $\aset$. In order to do so, we construct a flow network $H$ as follows.  We start with $H=G'$, and then subdivide every edge $e\in \tilde E$ with a vertex $t_e$, denoting $T_1=\set{t_e\mid e\in \tilde E_1}$ and $T_2=\set{t_e\mid e\in \tilde E_2}$.  We delete all vertices of $J'$ and their adjacent edges from the resulting graph, contract all vertices of $T_1$ into a source vertex $s$, and contract all vertices of $T_2$ into a destination vertex $t$. We then compute a minimum $s$-$t$ cut $(A,B)$ in the resulting flow network $H$, and we denote by $E''=E_H(A,B)$.  We use the following claim, whose proof is provided in  Section \ref{subsec: small cut set in case 2} of Appendix,  in order to bound the cardinality of $E''$.

\begin{claim}\label{claim: cut set small case2}
	If Event $\event$ did not happen, then $|E''|\leq \frac{2\cro(\phi)\cdot \mu^{12b}}{m}+|\chi^{\dirty}(\phi)|$.
\end{claim}

Let $E^{\del}=E'\cup E''$. If bad event $\event$ did not happen, then $|E'|\leq \frac{\cro(\phi)\cdot \mu^{38b}}{m}$, and, from \Cref{claim: cut set small case2}, $|E''|\leq \frac{2\cro(\phi)\cdot \mu^{12b}}{m}+|\chi^{\dirty}(\phi)|$.
Therefore, overall, if bad event $\event$ did not happen, then $|E^{\del}|\leq \frac{2\cro(\phi)\cdot \mu^{38b}}{m}+|\chi^{\dirty}(\phi)|$.
Let $G''=G'\setminus E''=G\setminus E^{\del}$, let $\Sigma''$ be the rotation system for graph $G''$ induced by $\Sigma$, and let $I''=(G'',\Sigma'')$ be the resulting instance of \cnwrs. Solution $\phi'$ to instance $I'$ then naturally induces a solution to instance $I''$, that we denote by $\phi''$. From \Cref{claim: new drawing}, this solution is compatible with $\phi$, and $\cro(\phi'')\leq \cro(\phi)$. Moreover, if Event $\event$ did not happen, then  the number of crosings in which the edges of $P^*$ participate in $\phi''$ is at most $\frac{\cro(\phi)\cdot \mu^{12b}}{m}$, and the images of edges of $E(J)\cup E(P^*)$ do not cross each other in $\phi'$.

We are now ready to define a split $(I_1=(G_1,\Sigma_1),I_2=(G_2,\Sigma_2))$ of instance $I$ along the enhacement structure $\aset$. In order to do so, we define two sets $A',B'$ of vertices in graph $G'$, as follows. We start with $A'=A\setminus\set{s}$ and $B'=B\setminus\set{t}$, where $(A,B)$ is the cut that we have computed in graph $H$. We then add all vertices of the core $J_1$ to $A'$, and all vertices of the core $J_2$ to $B'$.
We let $G_1=G''[A']$ and $G_2=G''[B']$.  The rotation system $\Sigma_1$ for graph $G_1$ and the rotation system $\Sigma_2$ for graph $G_2$ are induced by $\Sigma$. 
Let $I_1=(G_1,\Sigma_1)$ and $I_2=(G_2,\Sigma_2)$ be the resulting two instances of \cnwrs. It is immediate to verify that $(I_1,I_2)$ is a valid split of instance $I$ along $\aset$. 
Note that $E(G_1)\cup E(G_2)=E(G'')$.

Let $\phi_1$ be the solution to instance $I_1$ induced by $\phi''$,
and let  $\phi_2$ be the solution to instance $I_2$ induced by $\phi'$. 
From our construction and \Cref{claim: orientations and edge split is computed correctly Case2}, if bad event $\event$ did not happen, 
then drawing $\phi_1$ of $G_1$ is $\jset_1$-valid, and drawing $\phi_2$ of $G_2$ is $\jset_2$-valid.

Lastly, we need the following observation, whose proof
appears in Section \ref{subsec: few edges in split Case 2} of Appendix.

\begin{observation}\label{obs: few edges in split graphs case2}
	If Event $\event$ did not happen, then $|E(G_1)|,|E(G_2)|\leq m-\frac{m}{32\mu^{b}}$.
\end{observation}

We conclude that, if bad event $\event$ did not happen, then our algorithm computes a valid output for Procedure \procsplit. Since $\prob{\event}\leq 2^{20}/\mu^{10b}$, this completes the proof of \Cref{thm: procsplit}.

\subsection{Phase 1 of the Algorithm}
\label{sec: phase 1 interesting}

In Phase 1, we compute a collection $\iset$ of subinstances of the subdivided instance $\cI$ that almost have all required properties, except that we will not be able to guarantee that the sum of the optimal solution costs of the resulting instances is suitably bounded. However, we will ensure that all resulting instances have a convenient structure, that will be utilized in Phase 2, in order to produce the final collection of subinstances of $\cI$. The algorithm that is used in Phase 1 is summarized in the following theorem.

\begin{theorem}\label{thm: phase 1}
	There is an efficient randomized algorithm, whose input consists of a wide and well-connected instance  $\cI^*=(\cG^*,\cSigma^*)$, with $\cm=|E(\cG^*)|\geq \mu^{c'}$, for some large enough constant $c'$. Let $\cI=(\cG,\cSigma)$ be the coresponding subdivided instance. The algorithm either returns FAIL, or computes a non-empty collection $\iset$ of subinstances of $\cI$, such that, for every instance $I=(G,\Sigma)\in \iset$, $G\subseteq \cG$, and $I$ is the subinstance of $\cI$ defined by $G$. Additionally, the algorithm computes, for every instance $I\in \iset$, a core structure $\jset(I)$ for $I$, such that, if we denote, for every instance $I\in \iset$, the $\jset(I)$-contracted subinstance of $I$  by $\hat I$, and let $\hat \iset=\set{\hat I\mid I\in \iset}$, then the following hold:
	
	\begin{itemize}
		\item for every instance $I\in \iset$, if the corresponding contracted instance $\hat I=(\hat G,\hat \Sigma)$  is a wide instance, then $|E(\hat G)|\le \cm/\mu$; 
		
		
		\item $\sum_{\hat I=(\hat G,\hat \Sigma)\in \hat \iset}|E(\hat G)|\le 2\cm$; and
		\item there is an efficient algorithm, called $\algcombine_1$, that, given a solution $\phi(\hat I)$ to every instance $\hat I\in \hat \iset$, computes a solution $\check\phi$ to instance $\cI$.
	\end{itemize}
	
	Moreover, if  $\optcrors(\cI)\leq \cm^2/\mu^{c'}$, then with probability at least $(1-1/\mu^{200})$, all of the following hold:
	
	\begin{enumerate}
		\item the algorithm does not return FAIL;\label{item: no fail}
		\item for every instance $I\in \iset$, there is a solution $\psi(I)$ to $I$, that is $\jset(I)$-valid, with 
		$\sum_{I\in \iset}\cro(\psi(I))\leq \optcrors(\cI)$, and $\sum_{I\in \iset}|\chi^{\dirty}(\psi(I))|\leq \frac{\optcrors(\cI)\cdot \mu^{900}}{\cm}$; and
		\item  if algorithm $\algcombine_1$ is given as input a solution $\phi(\hat I)$ to every instance $\hat I\in \hat \iset$, then the solution $\check\phi$ to instance $\cI$ that it computes has cost at most:  $\sum_{\hat I\in \hat \iset}\cro(\phi(\hat I)) + \optcrors(\cI)\cdot\mu^{8000}$.\label{item: combine}
	\end{enumerate}
\end{theorem}

If the algorithm from \Cref{thm: phase 1} returns a collection $\iset$ of subinstances of $\cI$ together with a core structure $\jset(I)$ for each such subinstance, such that properties (\ref{item: no fail}) -- (\ref{item: combine}) hold, then we say that the algorithm is successful, and otherwise we say that it is unsuccessful. Notice that, if $\optcrors(\cI)\leq \cm^2/\mu^{c'}$, then the probability that the algorithm is unsuccessful is bounded by $1/\mu^{200}$.

The remainder of this subsection is dedicated to the proof of \Cref{thm: phase 1}. The algorithm repeatedly applies Procedure $\procsplit$ to subinstances of the input instance $\cI$. Throughout, we set the constant $b$ used in Procedure $\procsplit$ to $b=52$.
We may not always be able to ensure that the input to Procedure $\procsplit$ is valid. We will always  ensure that the input consists of a graph $G\subseteq \cG$, a core structure $\jset=(J,\set{b_u}_{u\in V(J)},\rho_J, F^*(\rho_J))$ for the subinstance $I$ of $\cI$ defined by $G$, and a promising set $\pset$ of paths for $I$ and $\jset$ of cardinality $\floor{\frac{|E(G)|}{\mu^{52}}}$. 
But unfortunately we may not be able to ensure that there exists a solution $\phi$ to instance $I$ that is $\jset$-valid, with $\cro(\phi)\leq \frac{|E(G)|^2}{\mu^{60b}}$ and $|\chi^{\dirty}(\phi)|\leq \frac{|E(G)|}{\mu^{60b}}$, since drawing $\phi$ is not given explicitly as part of input. If the input to Procedure $\procsplit$ is not valid, then the procedure may fail during its execution. In this case, we will assume that the procedure returned FAIL (we will also say that the procedure fails). It is also possible that the output $(\aset,I_1,I_2)$ of the procedure is not a valid output. We can verify efficiently that $\aset$ is a valid enhancement structure for $\jset$, and that $(I_1,I_2)$ is a valid split of $I$ along $\aset$. We can also efficiently verify that Property \ref{prop: smaller graphs}  holds for the resulting output. If we establish that either of these properties does not hold, then we will also assume that the procedure returned FAIL, or that it failed. However, it is possible that all above properties hold for the procedure's output, but properties \ref{prop output deleted edges} or \ref{prop output drawing} do not. As we are unable to efficiently verify these latter two properties, we will say in such a case that the procedure did not fail, but that it was unsuccessful. If the input to procedure \procsplit is valid, it is still possible that, with small probability (up to $2^{10}/\mu^{520}$), its output is not valid. As before, if $\aset$ is not a valid enhancement structure for $\jset$, or $(I_1,I_2)$ is not a valid split of $I$ along $\aset$, or Property \ref{prop: smaller graphs} does not hold (which we can verify efficiently), we will say that the procedure returned FAIL, or that the procedure failed. Otherwise, if all these properties hold but the output of the procedure is not valid, we will say that the application of the procedure was unsuccessful. If the procedure returns a valid output, then we say that its application was successful.

As before, we denote $|E(\cG^*)|$ by $\cm$.
The algorithm for Phase 1 consists of a number of iterations. The input to iteration $j\geq 1$ consists of a collection $\iset_j$ of subinstances of  instance $\cI$, where for every instance $I=(G,\Sigma)\in \iset_j$, $G\subseteq \cG$, and $I$ is the subinstance of $\cI$ defined by $G$. Additionally, for  every instance $I\in \iset_j$, we are given a core structure $\jset(I)$ for $I$. We will ensure that, with high probability, the subinstances in $\iset_j$ satisfy the following properties:

\begin{properties}{A}
	\item for every instance $I=(G,\Sigma)\in \iset_j$,
	 either the $\jset(I)$-contracted subinstance $\hat I=(\hat G,\hat \Sigma)$ of $I$ is narrow, or 
	 $|E(\hat G)|\leq \max\set{\frac{\cm}{\mu},2\cm-(j-1)\cdot \frac{\cm}{32\mu^{53}}}$; \label{invariant: fewer edgesx}
	

	\item if we denote by $E^{\del}_j=E(\cG)\setminus \left(\bigcup_{I=(G,\Sigma)\in \iset_j}E(G)\right )$, then $|E^{\del}_j|\leq \frac{j\cdot \optcrors(\cI)\cdot \mu^{6000}}{\cm}$;\label{inv: few deleted edges}
	
	\item for every instance $I\in \iset_j$, there exists a solution $\psi(I)$ to instance $I$ that is $\jset(I)$-valid, such that $\sum_{I\in \iset_j}\cro(\psi(I)) \leq \optcrors(\cI)$, and $\sum_{I\in \iset_j}|\chi^{\dirty}(\psi(I))|\leq  \frac{j\cdot \mu^{800}\optcrors(\cI)}{\cm}$; \label{invariant: solutions to instances}

	\item for every instance $I=(G,\Sigma)\in \iset_j$, whose corresponding $\jset(I)$-contracted instance $\hat I=(\hat G,\hat \Sigma)$ is wide with  $|E(\hat G)|>\cm/\mu$, for every vertex $v\in V(G)$ with $\deg_G(v)\geq\frac{\cm}{\mu^5}$, there is a collection $\qset(v)$ of at least $\frac{8\cm}{\mu^{50}}-|E^{\del}_j|$ edge-disjoint paths in $G$ connecting $v$ to the vertices of the  core $J(I)$ associated with the core structure $\jset(I)$; \label{inv: route to core}
	
	\item if we denote, for every instance $I\in \iset_j$, by $\hat m(I)$ the number of edges in the corresponding $\jset(I)$-contracted instance $\hat I$, then $\sum_{I\in \iset_j}\hat m(I)\leq 2\cm$; \label{inv: disjoint edges} and
	
	\item there is an efficient algorithm, that, given, for every instance $I\in \iset_j$, a solution $\phi(I)$ that is clean with respect to $\jset(I)$, constructs a solution $\phi(\cI)$ to instance $\cI$, of cost at most $\sum_{I\in \iset_j}\cro(\phi(I))+ j\cdot \optcrors(\check I)\cdot \mu^{6000}$.\label{invariant: putting solutions together}
\end{properties}

Specifically, for all $j\geq 1$, the input to iteration $j$ consists of a collection $\iset_j$ of subinstances of $\cI$, where for every instance $I=(G,\Sigma)$, $G\subseteq \cG$, and $I$ is the subgraph of $\cI$ defined by $G$. Additionally, for every instance $I\in \iset_j$, we are given a core structure $\jset(I)$ for $I$. We denote, for each instance $I=(G,\Sigma)\in \iset_j$, the corresponding $\jset(I)$-contracted instance by $\hat I=(\hat G,\hat \Sigma)$, and we denote by $\hat m(I)=|E(\hat G)|$. We will guarantee that, if Properties \ref{invariant: fewer edgesx}--\ref{invariant: putting solutions together} hold for input $\iset_j$ to iteration $j$, then, with probability at least $1-\frac{1}{\mu^{400}}$, at the end of the iteration, we obtain a collection $\iset_{j+1}$ of subinstances of $\cI$, each of which is defined by a subgraph of $\cG$, and, for every instance $I\in \iset_{j+1}$, a core structure $\jset(I)$, for which Properties \ref{invariant: fewer edgesx}--\ref{invariant: putting solutions together} hold. With the remaining probability, the algorithm may either return FAIL, or produce an output for which some of the properties \ref{invariant: fewer edgesx}--\ref{invariant: putting solutions together} do not hold. In each of these two cases, we say that the iteration was unsuccessful. If the input $\iset_j$ to iteration $j$ does not have properties \ref{invariant: fewer edgesx}--\ref{invariant: putting solutions together}, then it is possible that the algorithm returns FAIL, or it returns output $\iset_{j+1}$ for which some of the invariants \ref{invariant: fewer edgesx}--\ref{invariant: putting solutions together} do not hold. In both of these cases, we say that the iteration was unsuccessful. If the iteration produces output $\iset_{j+1}$ for which properties \ref{invariant: fewer edgesx}--\ref{invariant: putting solutions together} hold, then we say that it was successful. For all $j\geq 1$, we denote by $\tilde \event_j$ the bad event that iteration $j$ was unsuccessful. 
The number of iterations in our algorithm is at most $z=128\ceil{\mu^{53}}$. 
Note that, from Invariant \ref{invariant: fewer edgesx}, for every instance $I=(G,\Sigma)\in \iset_z$, either the corresponding $\jset(I)$-contracted instance $\hat I=(\hat G,\hat \Sigma)$ is narrow, or $|E(\hat G)|\leq \cm/\mu$. We will ensure that, for all $1\leq j\leq z$, if  $\optcrors(\cI)\leq \cm^2/\mu^{c'}$, then $\prob{\tilde \event_j\mid \neg\tilde \event_1\band\cdots\band\neg\tilde \event_{j-1}}\leq \frac{1}{\mu^{400}}$. This will guarantee that, if  $\optcrors(\cI)\leq \cm^2/\mu^{c'}$, then with probability at least $1-1/\mu^{200}$, Properties   \ref{invariant: fewer edgesx}--\ref{invariant: putting solutions together} hold for $\iset_z$. 

The input to the first iteration is a set $\iset_1$ of instances, consisting of a single instance $\cI$. Since instance $\cI^*$ is wide,  there is at least one  vertex $v^*$ with $\deg_{\cG}(v^*)\geq \cm/\mu^4$. We define 
a core structure $\jset(\cI)=(J,\set{b_u}_{u\in V(J)},\rho_J, F^*(\rho_J))$ associated with instance $\cI$ as follows. The core $J$ consists of a single vertex $v^*$, and its orientation $b_v$ is set to be arbitrary (say $1$). Drawing $\rho_J$ is the unique trivial drawing of $J$, and face $F^*(\rho_J)$ is the unique face of this drawing.
It is easy to verify that Invariants \ref{invariant: fewer edgesx}--\ref{invariant: putting solutions together} hold for $\iset_1$. The only invariant that is not immediate is \ref{inv: route to core}. This invariant follows from the fact that the input instance $\cI^*$ is wide and well-connected. Therefore, for every vertex $u$ of $\cG$ with $\deg_{\cG}(u)\geq \cm/\mu^5$, $\deg_{\cG^*}(u)\geq \cm/\mu^5$ also holds, and there is a collection of at least $\frac{8\cm}{\mu^{50}}$ edge-disjoint paths connecting $u$ to $v^*$ in $\cG^*$ and hence in $\cG$.

We now describe the execution of iteration $j$. Consider an instance $I=(G,\Sigma)\in \iset_j$, and its corresponding core structure $\jset(I)$. We say that instance $I$ is \emph{inactive} if either  the $\jset(I)$-contracted subinstance $\hat I=(\hat G,\hat \Sigma)$ of $I$ is narrow, or $|E(\hat G)|\leq \cm/\mu$. Otherwise, we say that instance $I$ is \emph{active}. We denote by $\iset^A_j$ the set of all active instances in $\iset_j$, and by $\iset^I_j$ the set of all inactive instances. We start with the set $\iset_{j+1}$ containing every instance in $\iset^{I}_j$. We also maintain the set $E^{\del}_{j+1}$ of deleted edges, that is initialized to $E^{\del}_j$. We then process every active instance $I\in \iset^{A}_j$ one by one. We now describe the algorithm for processing one such instance $I=(G,\Sigma)$. 

\paragraph{Processing instance $I=(G,\Sigma)\in \iset^{A}_j$.}
Assume that Invariants \ref{invariant: fewer edgesx},  \ref{invariant: solutions to instances} and \ref{inv: route to core} hold for $\iset_j$. Denote $\jset(I)=(J,\set{b_u}_{u\in V(J)},\rho_J, F^*(\rho_J))$, $|E(G)|=m$, and $|E(\hat G)|=\hat m(I)$. Since instance $I$ is active, $\hat m(I)>\cm/\mu$. Since $G\subseteq \cG$, $m\leq 2\cm$. Therefore, $\frac{m}{2\mu}\leq  \hat m(I)\leq m$.

Consider any vertex  $v\in V(G)$ with $\deg_{G}(v)\geq\frac{\hat m(I)}{\mu^4}$.  Since $ \hat m(I)\geq \frac{\cm}{\mu}$, we get that $\deg_G(v)=\deg_{\hat G}(v)\geq \frac{\cm}{\mu^5}$. From Invariant \ref{inv: route to core}, 
 there is a collection $\qset(v)$ of at least $\frac{8\cm}{\mu^{50}}-|E^{\del}_j|\geq \frac{8\cm}{\mu^{50}}- \frac{j\cdot \optcrors(\cI)\cdot \mu^{6000}}{\cm}$ edge-disjoint paths in $G$ connecting $v$ to vertices of $J$. Since $j\leq z=128\ceil{\mu^{53}}$, if we assume that $\optcrors(\cI)\leq \cm^2/\mu^{c'}$ for a large enough constant $c'$, we get that $|\qset(v)|\geq \frac{4\cm}{\mu^{50}}\geq \frac{2\hat m(I)}{\mu^{50}}$.
We apply the algorithm from \Cref{claim: find potential augmentors}
to instance $I$ and core structure $\jset(I)$, to obtain a  promising set of  paths $\pset$, of cardinality $\floor{\frac{\hat m(I)}{\mu^{50}}}\geq \floor{\frac{m}{\mu^{52}}}$, since $\hat m(I)\geq \frac{m}{2\mu}$. 
We use the following claim.

\begin{claim}\label{claim: invariants give valid input}
	If $\optcrors(\cI)\leq \cm^2/\mu^{c'}$ for some large enough constant $c'$, and Invariants \ref{invariant: fewer edgesx}--\ref{invariant: putting solutions together} hold for $\iset_j$, then $(G,\jset(I),\pset)$ is a valid input to Procedure \procsplit.
\end{claim}
\begin{proof}	
From the invariants it is immediate to verify that $\jset(I)$ is a valid core structure for the subinstance $I$ of $\cI$ defined by $G$.

Consider the solution $\psi(I)$ to instance $I$, that is given by Invariant \ref{invariant: solutions to instances}. This solution is guaranteed to be $\jset(I)$-valid.  It is enough to verify that $\cro(\psi(I))\leq \frac{m^2}{\mu^{3120}}$ and $|\chi^{\dirty}(\psi(I))|\leq \frac{m}{\mu^{3120}}$.

Recall that Invariant \ref{invariant: solutions to instances} guarantees that
 $\sum_{I'\in \iset_j}\cro(\psi(I')) \leq \optcrors(\cI)$.
 In particular, $\cro(\psi(I))\leq \optcrors(\cI)\leq \frac{\cm^2}{\mu^{c'}}$ must hold for a large enough constant $c'$. Since $m\geq \hat m(I)\geq \frac{\cm}{\mu}$, we get that $\cro(\psi(I))\leq \frac{m^2}{\mu^{3120}}$ as required. Similarly, Invariant \ref{invariant: solutions to instances} guarantees that $\sum_{I'\in \iset_j}|\chi^{\dirty}(\psi(I'))|\leq  \frac{j\cdot \mu^{800}\optcrors(\cI)}{\cm}$. In particular, $|\chi^{\dirty}(\psi(I))|\leq  \frac{j\cdot \mu^{800}\optcrors(\cI)}{\cm}$. Since $\optcrors(\cI)\leq \frac{\cm^2}{\mu^{c'}}$ for a large enough constant $c'$, while $j\leq z=128\ceil{\mu^{53}}$, we get that $|\chi^{\dirty}(\psi(I))|\leq  \frac{\cm}{\mu^{c'-852}}$. Since 
$m\geq \hat m(I)\geq \frac{\cm}{\mu}$, we get that $|\chi^{\dirty}(\psi(I))|\leq \frac{m}{\mu^{3120}}$. 
\end{proof}

In order to process instance $I\in \iset_j^A$, we
  apply Procedure $\procsplit$ to input $(G,\jset(I),\pset)$. If the procedure returns FAIL, then we terminate the algorithm and return FAIL as well. In this case we say that the current iteration failed. Otherwise, the procedure returns a  $\jset(I)$-enhancement structure $\aset$, and
  a split $(I_1=(G_1,\Sigma_1),I_2=(G_2,\Sigma_2))$ of $I$ along $\aset$. Let $P^*$ be the enhancement path of $\aset$, and let $(\jset_1,\jset_2)$ be the split of the core structure $\jset$ along $\aset$.
  Denote $E^{\del}(I)=E(G)\setminus (E(G_1)\cup E(G_2))$. Let $G'=G\setminus E^{\del}(I)$, let $\Sigma'$ be the rotation system for $G'$ induced by $\Sigma$, and let $I'=(G',\Sigma')$ be the resulting instance of \cnwrs. 
  We add the edges of $E^{\del}(I)$ to set $E^{\del}_{j+1}$, and we add instances $I_1,I_2$ to the collection $\iset_{j+1}$ of instances, letting $\jset(I_1)=\jset_1$ and $\jset(I_2)=\jset_2$. From the definition of a split of an instance along an enhancement structure, $G_1,G_2\subseteq G$, $\jset_1$ is a valid core structure for $I_1$, and $\jset_2$ is a valid core structure for $I_2$.  This completes the description of the algorithm for processing an instance $I\in \iset_j^A$, and of the $j$th iteration. We now analyze its properties.

We say that iteration $j$ is \emph{good} if, for every instance $I\in \iset_j^A$, the algorithm from \Cref{claim: find potential augmentors}, when applied to instance $I$ and core structure $\jset(I)$ returned a  promising set of  paths $\pset$ of cardinality $\floor{\frac{\hat m(I)}{\mu^{50}}}$, and additionally,  the application of Procedure \procsplit to input $(G,\jset(I),\pset)$ was successful. We use the following claim to show that iteration $j$ is good with high probability.

\begin{claim}\label{claim: prob good iteration}
	If Invariants \ref{invariant: fewer edgesx}--\ref{invariant: putting solutions together} hold for $\iset_j$ and $\optcrors(\cI)\leq \cm^2/\mu^{c'}$,
	then the probability that iteration $j$ is good is at least $1-1/\mu^{498}$.
\end{claim}
\begin{proof}
From the discussion above, if Invariants \ref{invariant: fewer edgesx}--\ref{invariant: putting solutions together} hold for $\iset_j$, and $\optcrors(\cI)\leq \cm^2/\mu^{c'}$, then for every instance $I\in \iset^A_j$, the algorithm from \Cref{claim: find potential augmentors}, when applied to instance $I$ and  core structure $\jset(I)$ returns a  promising set of  paths $\pset(I)$ of cardinality $\floor{\frac{\hat m(I)}{\mu^{50}}}$. Additionally, from
\Cref{claim: invariants give valid input}, if Invariants \ref{invariant: fewer edgesx}--\ref{invariant: putting solutions together} hold for $\iset_j$, and $\optcrors(\cI)\leq \cm^2/\mu^{c'}$, then for every instance $I\in \iset_j^A$, $(G,\jset(I),\pset(I))$ is a valid input to Procedure \procsplit. In this case, from \Cref{thm: procsplit}, the probability that Procedure \procsplit is either unsuccessful or fails, when applied to $(I,\jset(I),\pset(I))$, is at most $2^{20}/\mu^{520}$.
Since, from Invariant \ref{inv: disjoint edges}, $\sum_{I\in \iset_j}\hat m(I)\leq 2\cm$, while for every active instance $I\in\iset_j^A$, $\hat m(I)\geq \cm/\mu$, we get that $|\iset^A_j|\leq 2\mu$. From the Union Bound, we conclude that, if Invariants \ref{invariant: fewer edgesx}--\ref{invariant: putting solutions together} hold, and $\optcrors(\cI)\leq \cm^2/\mu^{c'}$, then the probability that iteration $j$ is good is at least $1-1/\mu^{498}$. 
\end{proof}

Lastly, the next claim allows us to bound the probability of the bad event $\tilde \event_z$.

\begin{claim}\label{claim: success prob}
Assume that 
Invariants \ref{invariant: fewer edgesx}--\ref{invariant: putting solutions together} hold for $\iset_j$, $\optcrors(\cI)\leq \cm^2/\mu^{c'}$, and that iteration $j$ is good. Then the bad event $\tilde \event_j$ does not happen.
\end{claim}

\begin{proof}
	Throughout the proof, we assume that Invariants \ref{invariant: fewer edgesx}--\ref{invariant: putting solutions together} hold for $\iset_j$, $\optcrors(\cI)\leq \cm^2/\mu^{c'}$, and iteration $j$ is good. 
	From the definition of a split of an instance (see \Cref{def: split}), for every instance $I=(G,\Sigma)\in \iset_{j+1}$, $G\subseteq \cG$, and $I$ is the subinstance of $\cI$ defined by graph $G$.
	It is now enough to show that Invariants  \ref{invariant: fewer edgesx}--\ref{invariant: putting solutions together} continue to hold for the collection $\iset_{j+1}$ of instances.
	
	We first observe that, for each inactive instance $I\in \iset^I_{j}$, invariant \ref{invariant: fewer edgesx} continue to hold for $I$, and $\hat m(I)$ does not change.

	Consider now some active instance $I=(G,\Sigma)\in \iset^A_j$, and let $(I_1=(G_1,\Sigma_1),I_2=(G_2,\Sigma_2))$ be the split of $I$ that was computed by Procedure \procsplit. We also let $\aset$ be the core enhancement structure computed by the procedure, and we let $(\jset_1,\jset_2)$ be the split of the core structure $\jset(I)$ via $\aset$.
	Note that Property \ref{prop: smaller graphs} of a valid output for Procedure \procsplit ensures that
	 $|E(G_1)|,|E(G_2)|\leq |E(G)|-\frac{|E(G)|}{32\mu^{52}}$. Since $|E(G)|\geq \hat m(I)\geq \frac{\cm}\mu$, we get that $|E(G_1)|,|E(G_2)|\leq |E(G)|-\frac{\cm}{32\mu^{53}}\leq 2\cm-j\cdot \frac{\cm}{32\mu^{53}}$ (from the fact that Property \ref{invariant: fewer edgesx} holds for $\iset_j$). This establishes Property \ref{invariant: fewer edgesx}  for $\iset_{j+1}$.
	
	Recall that Property \ref{prop output deleted edges} of a valid output for Procedure $\procsplit$ ensures that 
	$|E^{\del}(I)|\leq \frac{2\cro(\psi(I))\cdot \mu^{2000}}{|E(G)|}+|\chi^{\dirty}(\psi(I))|$. Therefore, we get that:
	\[ 
	\begin{split}
	|E^{\del}_{j+1}|&\leq |E^{\del}_j|+\sum_{I=(G,\Sigma)\in \iset^A_j}\left (\frac{2\cro(\psi(I))\cdot \mu^{2000}}{|E(G)|}+|\chi^{\dirty}(\psi(I))|\right )\\
	&\leq \frac{j\cdot \optcrors(\cI)\cdot \mu^{6000}}{\cm}+\sum_{I\in \iset^A_j}\frac{\cro(\psi(I))\cdot \mu^{2002}}{\cm}+\sum_{I\in \iset^A_j}|\chi^{\dirty}(\psi(I))|.
	\end{split}
	\]
	(we have used the fact that Invariant \ref{inv: few deleted edges} holds for $\iset_j$, and that, for every instance $I=(G,\Sigma)\in \iset^A_j$, $|E(G)|\geq \hat m(I)\geq \frac{\cm}{\mu}$). Recall that, from Invariant \ref{invariant: solutions to instances}, $\sum_{I\in \iset_j}\cro(\psi(I)) \leq \optcrors(\cI)$, and $\sum_{I\in \iset_j}|\chi^{\dirty}(\psi(I))|\leq  \frac{j\cdot \mu^{800}\optcrors(\cI)}{\cm}\leq \frac{ \mu^{854}\optcrors(\cI)}{\cm}$ (since $j\leq z=128\ceil{\mu^{53}}$). Altogether, we get that $|E^{\del}_{j+1}|\leq \frac{(j+1)\cdot \optcrors(\cI)\cdot \mu^{6000}}{\cm}$, establising Invariant \ref{inv: few deleted edges} for $\iset_{j+1}$.
	
	Next, we establish Invariant \ref{invariant: solutions to instances} for $\iset_{j+1}$. For every instance $I=(G,\Sigma)\in \iset_j^I$, its solution $\psi(I)$ remains unchanged. Consider now some instance $I=(G,\Sigma)\in\iset_j^A$, and the two subinstances $(I_1,I_2)$ of $I$ that Procedure \procsplit produced. 
	Let $G'=G\setminus E^{\del}(I)$, let $\Sigma'$ be the rotation system for $G'$ induced by $\Sigma$, and let $I'=(G',\Sigma')$ be the resulting instance of \cnwrs. Consider the solution $\phi'$ for instance $I'$ that is guaranteed by Property \ref{prop output drawing} of valid output of Procedure \procsplit. Let $\phi_1$ be the solution to instance $I_1$ induced by $\phi'$, and let $\phi_2$ be the solution to instance $I_2$ induced by $\phi'$.
	Property \ref{prop output drawing} guarantees that $\phi_1$ is a $\jset_1$-valid solution to $I_1$, and $\phi_2$ is a $\jset_2$-valid solution to $I_2$. We implicitly set $\psi(I_1)=\phi_1$ and $\psi(I_2)=\phi_2$. 
From the definition of an instance split, (see \Cref{def: split}), the only edges that may be shared by graphs $G_1$ and $G_2$ are edges of $E(J)\cup E(P^*)$. Since no pair of edges in $E(J)\cup E(P^*)$ may cross each other in $\phi'$, we get that $\cro(\phi_1)+\cro(\phi_2)\leq \cro(\phi')$. Moreover, if $(e,e')_p\in \chi^{\dirty}(\phi_1)$, then either $(e,e')_p\in \chi^{\dirty}(\phi')$, or one of the edges $e,e'$ lies on $P^*$. Since the edges of $P^*$ participate in at most  $\frac{\cro(\psi(I))\cdot \mu^{624}}{|E(G)|}\leq \frac{\cro(\psi(I))\cdot \mu^{625}}{\cm}$ crossings (as $|E(G)|\geq \hat m(I)\geq \cm/\mu$), we get that $| \chi^{\dirty}(\phi_1)|+| \chi^{\dirty}(\phi_2)|\leq  |\chi^{\dirty}(\psi(I))|+\frac{\cro(\psi(I))\cdot \mu^{625}}{\cm}$.

Overall, we get that:
\[\sum_{I\in \iset_{j+1}}\cro(\psi(I)) \leq \sum_{I\in \iset_j}\cro(\psi(I)) \leq \optcrors(\cI);\]
and:
\[ \begin{split}
 \sum_{I\in \iset_{j+1}}|\chi^{\dirty}(\psi(I))|&\leq  \sum_{I\in \iset_{j}}|\chi^{\dirty}(\psi(I))|+\sum_{I\in \iset_j^A}\frac{\cro(\psi(I))\cdot \mu^{625}}{\cm}\\
 &\leq \frac{j\cdot \mu^{800}\optcrors(\cI)}{\cm}+\frac{\optcrors(\cI)\cdot \mu^{625}}{\cm}\\
 &\leq \frac{(j+1)\cdot \mu^{800}\optcrors(\cI)}{\cm},
\end{split} \]
establishing Invariant \ref{invariant: solutions to instances} for $\iset_{j+1}$.

Next, we establish Invariant \ref{inv: route to core}. Consider some instance $\tilde I=(\tilde G,\tilde \Sigma)\in \iset_{j+1}$ with $\hat m(\tilde I)>\cm/\mu$, whose corresponding $\jset(\tilde I)$-contracted graph is wide. If $\tilde I\in \iset_j$,  then $\tilde I$ is an inactive instance, and so Invariant  \ref{inv: route to core} holds for it. Otherwise, there is some instance $I\in \iset_j^A$, such that, if $(I_1,I_2)$ is the split of instance $I$ that we have computed, $\tilde I=I_1$ or $\tilde I=I_2$ holds. We assume w.l.o.g. that it is the former. We denote $I=(G,\Sigma)$, $I_1=(G_1,\Sigma_1)$, and we let $J$, $J_1$, and $J_2$ be the cores associated with the core structures $\jset(I)$, $\jset(I_1)$, and $\jset(I_2)$, respectively. 
Consider now any vertex $v\in V(G_1)$, whose degree in $G_1$ is at least $\frac{\cm}{\mu^5}$. Then $\deg_G(v)\geq \frac{\cm}{\mu^5}$ must hold as well. From  Invariant \ref{inv: route to core}, there is a collection $\qset(v)$ of at least $\frac{8\cm}{\mu^{50}}-|E^{\del}_j|$ edge-disjoint paths in $G$ connecting $v$ to the vertices of $J$. We assume w.l.o.g. that the paths in $\qset(v)$ are internally disjoint from $V(J)$. Let $\qset'(v)\subseteq \qset(v)$ be the set of paths that do not contain edges of $E^{\del}(I)$. Clearly, $|\qset'(v)|\geq \frac{8\cm}{\mu^{50}}-|E^{\del}_j|-|E^{\del}(I)|\geq \frac{8\cm}{\mu^{50}}-|E^{\del}_{j+1}|$. We direct the paths in $\qset'(v)$ from $v$ to the vertices of $V(J)$. Notice that every path $Q\in \qset'(v)$ is contained in graph $G'$. Consider now any such path $Q\in \qset'(v)$. If path $Q$ contains a vertex of $P^*$ as an inner vertex, then we truncate it so it connects $v$ to a vertex of $P^*$, and is internally disjoint from $V(J)\cup V(P^*)$.
We claim that the resulting path $Q$ must be contained in graph $G_1$. This is since, from the definition of a split of an instance, 
$V(G_1)\cup V(G_2)=V(G)$, and  every vertex $u\in V(G_1)\cap V(G_2)$ belongs to $V(J_1)\cap V(J_2)$, while $J_1,J_2\subseteq J\cup P^*$. Since $E(G')=E(G_1)\cup E(G_2)$, we get that every path $Q$ in the resulting set $\qset'(v)$ is contained in graph $G_1$, and it connects $v$ to a vertex of $J_1$. This establishes Invariant \ref{inv: route to core} for $\iset_{j+1}$.

Invariant \ref{inv: disjoint edges} follows from the fact that, for every instance $I\in \iset_j^A$, if $(I_1=(G_1,\Sigma_1),I_2=(G_2,\Sigma_2))$ is the split of instance $I$ that we have computed, then $E(G_1)\cap E(G_2)\subseteq E(J_1)\cap E(J_2)$ (since, from definition of a split, every vertex $u\in V(G_1)\cap V(G_2)$ belongs to $V(J_1)\cap V(J_2)$, and since a subdivided instance may not have parallel edges).

It now remains to establish Invariant \ref{invariant: putting solutions together}.
Assume we are given, for every instance $I'\in \iset_{j+1}$, a solution $\phi(I')$ that is clean with respect to $\jset(I')$.
Consider any active instance $I=(G,\Sigma)\in \iset_j^A$, and let $(I_1=(G_1,\Sigma_2), I_2=(G_2,\Sigma_2))$ be the split of $I$ that we have constructed. We apply the algorithm from \Cref{obs: combine solutions for split} in order to obtain  a solution $\phi(I)$ to instance $I$ that is clean with respect to $\jset(I)$, and $\cro(\phi(I))\leq \cro(\phi(I_1))+\cro(\phi(I_2))+|E^{\del}(I)|\cdot |E(G)|$. From Property \ref{prop output deleted edges} of a valid output for \procsplit,
$|E^{\del}(I)|\leq \frac{2\cro(\phi)\cdot \mu^{2000}}{|E(G)|}+|\chi^{\dirty}(\psi(I))|$.
 Overall, we have now obtained a solution $\phi(I)$ for every instance $I\in \iset_j$, that is clean with respect to $\jset(I)$, such that:
\[
\begin{split}
\sum_{I\in \iset_j}\cro(\phi(I))&\leq \sum_{I'\in \iset_{j+1}}\cro(\phi(I'))+\sum_{I=(G,\Sigma)\in \iset^A_j}|E(G)|\cdot \left ( \frac{\cro(\psi(I))\cdot \mu^{2000}}{|E(G)|}+|\chi^{\dirty}(\psi(I))| \right ) \\
&\leq \sum_{I'\in \iset_{j+1}}\cro(\phi(I'))+\sum_{I=(G,\Sigma)\in \iset^A_j}\cro(\psi(I))\cdot \mu^{2000} + 2\cm\cdot \sum_{I=(G,\Sigma)\in \iset^A_j} |\chi^{\dirty}(\psi(I))|.
\end{split}
\]

From Invariant \ref{invariant: solutions to instances}, $\sum_{I\in \iset_j}\cro(\psi(I)) \leq \optcrors(\cI)$, and $\sum_{I\in \iset_j}|\chi^{\dirty}(\psi(I))|\leq  \frac{j\cdot \mu^{800}\optcrors(\cI)}{\cm}$.
 Therefore, altogether: 
\[
\begin{split}
\sum_{I\in \iset_j}\cro(\phi(I))&\leq  \sum_{I'\in \iset_{j+1}}\cro(\phi(I'))+\optcrors(\cI)\cdot \mu^{2000} + 2z\cdot \mu^{800}\cdot \optcrors(\cI)\\
&\leq \sum_{I'\in \iset_{j+1}}\cro(\phi(I'))+2\optcrors(\cI)\cdot \mu^{2000}, 
\end{split}
\]
since $z=128\ceil{\mu^{53}}$.
We then apply the algorithm that is guaranteed by Invariant \ref{invariant: putting solutions together} to the collection $\iset_j$ of instances, to compute  a solution $\phi(\cI)$ to instance $\cI$, whose is at most:
\[
\begin{split}
\sum_{I\in \iset_j}\cro(\phi(I))+j\cdot \optcrors(\cI)\cdot \mu^{6000}&\leq \sum_{I'\in \iset_{j+1}}\cro(\phi(I'))+j\cdot \optcrors(\cI)\cdot \mu^{6000}+2\optcrors(\cI)\cdot \mu^{2000}\\
&\leq \sum_{I'\in \iset_{j+1}}\cro(\phi(I'))+(j+1)\cdot \optcrors(\cI)\cdot \mu^{6000}.
\end{split}\]
\end{proof}

From  Claims \ref{claim: prob good iteration} and \ref{claim: success prob}, for all $1\leq j\leq z$, if $\optcrors(\cI)\leq \cm^2/\mu^{c'}$, then $\prob{\tilde \event_j\mid \neg\tilde \event_1\band\cdots\band\neg\tilde \event_{j-1}}\leq 1/\mu^{498}$. Therefore,
$$\prob{\tilde \event_z}\leq \prob{\tilde \event_z\mid \neg\tilde \event_1\band\cdots\band\neg\tilde \event_{z-1}}+\prob{\tilde \event_{z-1}\mid \neg\tilde \event_1\band\cdots\band\neg\tilde \event_{z-2}}+\ldots+\prob{\tilde \event_1}\leq z/\mu^{498}\leq 1/\mu^{400},$$ 
since $z=128\ceil{\mu^{53}}$. 
If the algorithm did not return FAIL, then we return the set $\iset_z$ of subinstances of $\cI$, and, for every instance $I\in \iset_z$, the corresponding core structure $\iset(I)$. 
Assume that Event $\tilde \event_z$ did not happen.

From Invariant \ref{invariant: fewer edgesx}, we are guaranteed that, for every instance $I\in \iset$, if the corresponding contracted instance $\hat I=(\hat G,\hat \Sigma)$  is a wide instance, then $|E(\hat G)|\le \cm/\mu$. Invariant \ref{inv: disjoint edges} ensures that $\sum_{\hat I=(\hat G,\hat \Sigma)\in \hat \iset}|E(\hat G)|\le 2\cm$, and Invariant \ref{invariant: putting solutions together} provides an efficient algorithm for combining clean solutions $\phi(I)$ to instances in $I\in\iset_z$ to obtain a solution $\check \phi$ to instance $\check I$.
If Event $\tilde \event_z$ does not happen, then we are guaranteed that:
$$\cro(\check \phi)\leq \sum_{I\in \iset_z}\cro(\phi(I))+ z\cdot \optcrors(\check I)\cdot \mu^{6000}\leq \sum_{I\in \iset_z}\cro(\phi(I))+  \optcrors(\check I)\cdot \mu^{6054},$$
(since  $z=128\ceil{\mu^{53}}$). Since there is an efficient algorithm that, given, for every instance $I\in \iset_z$, a solution $\hat \phi(I)$ to the corresponding $\jset(I)$-contracted instance $\hat I$, computes a solution $\phi(I)$ to instance $I$ that is clean with respect to $\jset(I)$ with $\cro(\phi(I))\leq \cro(\hat \phi(I))$, we obtain the desired efficient algorithm for combining solutions to instances in $\hat\iset$ to obtain a solution to instance $\check I$.

Lastly, Invariant \ref{invariant: solutions to instances} ensures that,  if Event $\tilde \event_z$ did not happen, then,
 for every instance $I\in \iset_z$, there exists a solution $\psi(I)$ to $I$ that is $\jset(I)$-valid, such that $\sum_{I\in \iset_z}\cro(\psi(I))\leq \optcrors(\cI)$, and $\sum_{I\in \iset_z}|\chi^{\dirty}(\psi(I))|\leq 
 \frac{\optcrors(\cI)\cdot z\mu^{800}}{\cm}\le
 \frac{\optcrors(\cI)\cdot \mu^{900}}{\cm}$ (since  $z=128\ceil{\mu^{53}}$). 
 Notice also that, if Event $\tilde \event_z$ does not happen, then the algorithm does not return FAIL.
 This completes the proof of \Cref{thm: phase 1}.

\subsection{Phase 2 of the Algorithm}
\label{subsec: phase 2}

The goal of this phase is to ``repair'' each one of the instances $I\in \iset$ computed in the first phase in order to ensure that each such instance has a cheap solution that is clean with respect to the core structure $\jset(I)$. This, in turn, will ensure that the corresponding contracted graph has a cheap solution as well. In order to repair an instance $I=(G,\Sigma)\in \iset$, we will compute a collection $E^{\del}(I)$ of edges of $G$. We will ensure that no edge of the core $J(I)$ corresponding to the core structure $\jset(I)$ lies in $E^{\del}(I)$. We can then define a new instance $I'=(G',\Sigma')$, where $G'=G\setminus E^{\del}(I)$, and $\Sigma'$ is the rotation system for $G'$ that is induced by $\Sigma$. Note that $\jset(I)$ remains a valid core structure for $I'$. Our goal is to ensure that, on the one hand, $|E^{\del}(I)|$ is not too large, and, on the other hand, there is a solution $\psi(I')$ to instance $I'$ that is clean with respect to $\jset(I')$, and $\cro(\psi(I'))$ is not too large compared to $\cro(\psi(I))+|\chi^{\dirty}(\psi(I))|^2$, where $\psi(I)$ is the $\jset(I)$-valid solution for instance $I$ from \Cref{thm: phase 1}. 
We now state the main result of this subsection, summarizing the algorithm for Phase 2.

\begin{theorem}\label{thm: phase 2}
	There is an efficient randomized algorithm, whose input consists of a large enough constant $b$, a subinstance $I=(G,\Sigma)$ of $\cI$ with $|E(G)|=m$  and $G\subseteq \cG$, and a core structure $\jset(I)$ for $I$, whose corresponding core is denoted by $J(I)$.
	The algorithm computes a set $E^{\del}(I)\subseteq E(G)\setminus E(J(I))$ of edges, for which the following hold. 
	Let $G'=G\setminus E^{\del}(I)$, and let  $I'=(G',\Sigma')$ be the subinstance of $\cI$ defined by $G'$.
	The algorithm ensures that, if there is a solution $\psi(I)$ to instance $I$ that is $\jset(I)$-valid, with $\cro(\psi(I))\leq m^2/\mu^{240b}$, and  $|\chi^{\dirty}(\psi(I))|\leq m/\mu^{240b}$, then  with probability at least $1-1/\mu^{2b}$,  
	$|E^{\del}(I)|\leq \left (\frac{\cro(\psi(I))}{m}+|\chi^{\dirty}(\psi(I))|\right ) \cdot \mu^{O(b)}$, and  there is a solution $\psi(I')$ to instance $I'$ that is clean with respect to $\jset(I)$, with $\cro(\psi(I'))\leq \left (\cro(\psi(I))+|\chi^{\dirty}(\psi(I))|^2+\frac{|\chi^{\dirty}(\psi(I))|\cdot |E(G)|}{\mu^b}\right )\cdot (\log m)^{O(1)}$. 
\end{theorem}

If there is a solution $\psi(I)$ to instance $I$ that is $\jset(I)$-valid, with $\cro(\psi(I))\leq m^2/\mu^{240b}$, and  $|\chi^{\dirty}(\psi(I))|\leq m/\mu^{240b}$, then we let $\psi(I)$ be this solution, and we say that $\psi(I)$ is a good solution to instance $I$. Otherwise, we let $\psi(I)$ be any solution to instance $I$ that is $\jset(I)$-valid, and we say that $\psi(I)$ is a bad solution to instance $I$.
We say that an application of the algorithm from \Cref{thm: phase 2} is \emph{successful}, if  (i) $|E^{\del}(I)|\leq \left (\frac{\cro(\psi(I))}{m}+|\chi^{\dirty}(\psi(I))|\right ) \cdot \mu^{O(b)}$, and (ii)
there is a solution $\psi(I')$ to the resulting instance $I'=(G',\Sigma')$, that is  clean with respect to $\jset(I)$, with:  
$$\cro(\psi(I'))\leq \left (\cro(\psi(I))+|\chi^{\dirty}(\psi(I))|^2+\frac{|\chi^{\dirty}(\psi(I))|\cdot |E(G)|}{\mu^b}\right )\cdot (\log m)^{O(1)}.$$ 

From \Cref{thm: phase 2}, if there is a good solution $\psi(I)$ to instance $I$, then the algorithm is successful with probability at least $1-1/\mu^{2b}$.
We provide the proof of the theorem below, after we complete the proof of \Cref{lem: many paths} using it.

\subsubsection{Completing the Proof of \Cref{lem: many paths}}
\label{subsubsec: finish the proof}
Given an input instance $\cI^*=(\cG^*,\cSigma^*)$, we first apply the algorithm from \Cref{thm: phase 1} to this input. If the algorithm from \Cref{thm: phase 1} fails, then we terminate the algorithm and return FAIL as well.
 We denote by $\tilde \event_1'$ the bad event that the application of this algorithm is unsuccessful. Recall that, from \Cref{thm: phase 1}, if $\optcrors(\cI^*)\leq \cm^2/\mu^{c'}$, then  $\prob{\tilde \event_1'}\leq 1/\mu^{200}$, and, if bad event $\event_1'$ does not happen, then the algorithm does not fail.
We assume from now on that the algorithm from \Cref{thm: phase 1}  did not fail. Let $\iset$ be the collection of subinstances of $\cI$ computed by the algorithm from \Cref{thm: phase 1}. Recall that, if Event $\tilde \event'_1$ did not happen, then  for every instance $I\in \iset$, there is a solution $\psi(I)$ to $I$, that is $\jset(I)$-valid, such that: 
$\sum_{I\in \iset}\cro(\psi(I))\leq \optcrors(\cI)$, and $\sum_{I\in \iset}|\chi^{\dirty}(\psi(I))|\leq \frac{\optcrors(\cI)\cdot \mu^{900}}{\cm}$. We let $b$ be a large enough constant, so that $b\geq 4000$. We assume that the parameter $c'$ from the statement of \Cref{lem: many paths} is sufficiently large compared to $b$, for example, $c'\geq 400b$.

Recall that, for every instance $I=(G,\Sigma)\in \iset$, we have denoted by $\hat I=(\hat G,\hat \Sigma)$ the corresponding $\jset(I)$-contracted instance, and by $\hat m(I)=E(\hat G)$. We say that instance $I$ is \emph{small} if $\hat m(I)\leq \frac{\cm}{\mu^{1000}}$, and otherwise it is \emph{large}. We partition the set $\iset$ of instances into two subsets: set $\iset_1$ containing all small instances, and set $\iset_2$ containing all large instances. We let $\hat \iset_1=\set{\hat I\mid I\in \iset_1}$ contain the set of all contracted instances corresponding to the instances of $\iset_1$, and we define set $\hat \iset_2$ of instances corresponding to the instances of $\iset_2$ similarly.
We need the following obsevation.

\begin{observation}\label{obs: have invariants}
	 Assume that $\optcrors(\cI^*)\leq \cm^2/\mu^{c'}$, and that bad event $\tilde\event'$ did not happen. Then for every instance $I=(G,\Sigma)\in \iset_2$, there is a solution $\psi(I)$ to instance $I$ that is $\jset(I)$-valid, with $\cro(\psi(I))\leq m^2/\mu^{240b}$, and  $|\chi^{\dirty}(\psi(I))|\leq m/\mu^{240b}$, where $m=|E(G)|$.
	\end{observation}
\begin{proof}
 Assume that $\optcrors(\cI)=\optcrors(\cI^*)\leq \cm^2/\mu^{c'}$, and that bad event $\tilde \event'$ did not happen. Recall that the algorithm from \Cref{thm: phase 1} ensures that, for every instance $I\in \iset$, 
 there is a solution $\psi(I)$ to $I$, that is $\jset(I)$-valid, with 
 $\sum_{I\in \iset}\cro(\psi(I))\leq \optcrors(\cI)$, and $\sum_{I\in \iset}|\chi^{\dirty}(\psi(I))|\leq \frac{\optcrors(\cI)\cdot \mu^{900}}{\cm}$.
 
 Consider now some  instance $I=(G,\Sigma)\in \iset_2$, and denote $|E(G)|=m$.  From the above discussion, $\cro(\psi(I))\leq \optcrors(\cI)\leq \frac{\cm^2}{\mu^{c'}}$ must hold. Since, from definition of set $\iset_2$ of instances, $m\geq \hat m(I)>\frac{\cm}{\mu^{1000}}$ holds, we get that $\cro(\psi(I))\leq \frac{m^2}{\mu^{c'-2000}}\leq  \frac{m^2}{\mu^{240b}}$, since we can set $c'$ to be a large enough constant.
 
 Similarly, $|\chi^{\dirty}(\psi(I))|\leq \frac{\optcrors(\cI)\cdot \mu^{900}}{\cm}\leq \frac{\cm}{\mu^{c'-900}}$, since $\optcrors(\cI)\leq \cm^2/\mu^{c'}$. Using the fact that $m\geq \frac{\cm}{\mu^{1000}}$, we get that:
 
 \[ |\chi^{\dirty}(\psi(I))|\leq\frac{m}{\mu^{c'-1900}}\leq \frac{m}{\mu^{240b}}. \]
\end{proof}

We process every instance $I\in \iset_2$ one by one.
 For each such instance $I$, we apply the algorithm from \Cref{thm: phase 2} to instance $I$, core structure $\jset(I)$, and the constant parameter $b$ defined above. Let $\tilde \event'_2(I)$ be the bad event that this application of the algorithm was unsuccessful. 
 
 From \Cref{obs: have invariants} and \Cref{thm: phase 2}, if $\optcrors(\cI^*)\leq \cm^2/\mu^{c'}$, then $\prob{\tilde \event'_2(I)\mid \neg\tilde \event'_1}\leq 1/\mu^{2b}$. We denote by $I'=(G',\Sigma')$ the resulting instance of \cnwrs,  and by $\hat I'$ the corresponding $\jset(I)$-contracted instance. We then denote $\hat \iset_2'=\set{\hat I'\mid I\in \iset_2}$. The final output of our algorithm is the collection $\iset'=\hat\iset_1\cup\hat  \iset_2'$ of subinstances of $\cI$.

We now verify that the collection $\iset'$ of instances has all required properties. First, the algorithm from \Cref{thm: phase 1} ensures that, for every instance $I\in \iset$, if the corresponding contracted instance $\hat I=(\hat G,\hat \Sigma)$  is a wide instance, then $|E(\hat G)|\le \cm/\mu$. If instance $I$ lies in $\iset_2$, and $\hat I'=(\hat G',\hat\Sigma')$ is the $\jset(I)$-contracted instance corresponding to $I'$, then $|E(\hat G')|\leq |E(\hat G)|$, and, if $\hat I$ is not a wide instance, then neither is $\hat I'$. This is since graph $\hat G'$ can be obtained from graph $\hat G$ by deleting the edges of $E^{\del}(I)$ from it\footnote{This is slightly imprecise, since it is possible that $|E(\hat G')|<\hat m(I)$. Therefore, a vertex $v$ may be a high-degree vertex for $\hat G'$ but not for graph $\hat G$. It is therefore possible that $\hat I$ is narrow but $\hat I'$ is not, due to difference in the parameters  $|E(\hat G')|$ and $|E(\hat G)|$. However, we can easily fix this issue by adding $\hat m(I)-|E(\hat G')|$ new vertices to graph $\hat G'$, and connecting each of these vertices to a vertex whose degree in $\hat G'$ is smaller than in $\hat G$, so that for every vertex $v\in V(\hat G)$, $\deg_{\hat G'}(v)\leq \deg_{\hat G}(v)$ and $|E(\hat G')|=|E(\hat G)|$ holds. This ensures that, if $\hat I$ is a narrow instance, then so is $\hat I'$. Adding degree-1 vertices to an instance of \cnwrs does not increase its optimal solution value.}. Therefore, we are guaranteed that, for every instance $I'=(G',\Sigma')\in \iset'$, if $I'$ is a wide instance, then $|E(G')|\le \cm/\mu$.

The algorithm from \Cref{thm: phase 1} also guarantees that  $\sum_{I\in  \iset}\hm(I)\le 2\cm$. Since, for every instance $I\in \iset_2$, the corresponding instance $\hat I'=(\hat G',\Sigma')\in \hat \iset_2'$ has $|E(\hat G')|\leq \hat m(I)$, we get that $\sum_{I'=(G',\Sigma')\in \iset'}|E(G')|\le 2\cm$.

Let $\tilde \event_2'$ be the bad event that any of the events in $\set{\tilde \event'_2(I)\mid I\in \iset_2}$ happened. 
From the definition of the set $\iset_2$ of instances, for all $I\in \iset_2$, $\hat m(I)\geq \frac{\cm}{\mu^{1000}}$. Since, from \Cref{thm: phase 1}, $\sum_{I\in \iset}\hat m(I)\le 2\cm$, we get that $|\iset_2|\leq 2\mu^{1000}$.
Therefore, if $\optcrors(\cI^*)\leq \cm^2/\mu^{c'}$, then: 
$$\prob{\tilde \event'_2\mid \neg\tilde \event'_1}\leq \sum_{I\in \iset_2} \prob{\tilde \event'_2(I)\mid \neg\tilde \event'_1}\leq \frac{2\mu^{1000}}{\mu^{2b}}\leq \frac{1}{\mu^4},$$
since $b\geq 4000$.
Lastly, we let $\tilde \event'$ be the bad event that either of the events $\tilde \event'_1$ or $\tilde \event'_2$ happened. Then $\prob{\tilde \event'}\leq \prob{\tilde \event'_1}+\prob{\tilde \event'_2\mid \neg\tilde \event'_1}$. From the above discussion, if $\optcrors(\cI^*)\leq \cm^2/\mu^{c'}$, then $\prob{\tilde \event'}\leq \frac{1}{\mu^{200}}+\frac{1}{\mu^4}\leq \frac{1}{\mu^3}$.
We use the following two observations in order to complete the proof of \Cref{lem: many paths}.

\begin{observation}\label{obs: composing contracted solutions}
	Assume that $\optcrors(\cI^*)\leq \cm^2/\mu^{c'}$, and that event $\tilde \event'$ did not happen. Then there is an efficient algorithm, that, given a solution $\phi(I')$ to every instance $I'\in \iset'$, computes a solution $\check \phi$ to instance $\cI^*$, with $\cro(\check\phi)\leq  \sum_{I'\in \iset'}\cro(\phi(I')) + \optcrors(\check I^*)\cdot\mu^{O(1)}$.
	\end{observation}

\begin{proof}
We assume that $\optcrors(\cI^*)\leq \cm^2/\mu^{c'}$, that  Event $\tilde \event'$ did not happen, and that  we are given a solution $\phi(I')$ to every instance $I'\in \iset'$. We show an efficient algorithm to compute a solution $\check \phi$ to instance $\cI$. In order to do so, we consider every instance $I\in \iset_2$ one by one, and compute a solution $\phi(\hat I)$ to instance $\hat I$, from the solution $\phi(\hat I')$ to instance $\hat I'$. 

Consider now some instance $I=(G,\Sigma)\in \iset_2$. Let $\hat I=(\hat G,\hat \Sigma)$ be the corresponding  $\jset(I)$-contracted instance, and let $\hat I'=(\hat G',\hat \Sigma')$ be the $\jset(I)$-contracted instance corresponding to the instance $I'$. Note that $V(\hat G)=V(\hat G')$ and $E(\hat G')=E(\hat G)\setminus E^{\del}(I)$. We use the algorithm from \Cref{lem: edge insertion} in order to insert the edges of $E^{\del}(I)$ into the solution $\phi(\hat I')$ to instance $\hat I'$, obtaining a solution $\phi(\hat I)$ to instance $\hat I$, whose cost is at most $\cro(\phi(\hat I'))+|E^{\del}(I)|\cdot |E(\hat G)|$. 
Recall that, from \Cref{thm: phase 2}: 
$$|E^{\del}(I)|\leq \left (\frac{\cro(\psi(I))}{|E(G)|}+|\chi^{\dirty}(\psi(I))|\right ) \cdot \mu^{O(b)}.$$
Therefore:
\[
 \begin{split}
 \cro(\phi(\hat I))&\leq \cro(\phi(\hat I'))+\left (\cro(\psi(I))+|\chi^{\dirty}(\psi(I))|\cdot |E(\hat G)|\right ) \cdot \mu^{O(b)} \\ &\leq  \cro(\phi(\hat I'))+\left (\cro(\psi(I))+|\chi^{\dirty}(\psi(I))|\cdot \cm\right )\cdot \mu^{O(b)}.
 \end{split}
 \]
%
%
Lastly, using the algorithm from \Cref{thm: phase 1}, we obtain a solution 
$\check\phi$ to instance $\cI$, whose cost is bounded by:
\[
\begin{split}
\cro(\check \phi)&\leq  \sum_{\hat I\in \hat \iset}\cro(\phi(\hat I)) + \optcrors(\cI)\cdot\mu^{8000}\\
&\leq \sum_{I\in \iset_1}\cro(\phi(\hat I))+\sum_{I\in \iset_2}\left (\cro(\phi(\hat I'))+\cro(\psi(I))\cdot \mu^{O(b)}+|\chi^{\dirty}(\psi(I))|\cdot \cm\cdot \mu^{O(b)}\right )+\optcrors(\cI)\cdot\mu^{O(1)}\\
&=\sum_{I'\in  \iset'}\cro(\phi(I'))+\optcrors(\cI)\cdot\mu^{O(1)}+\sum_{I\in \iset_2}\cro(\psi(I))\cdot \mu^{O(1)}+\sum_{I\in \iset_2}|\chi^{\dirty}(\psi(I))|\cdot \cm\cdot \mu^{O(1)}.
\end{split}
\]
From \Cref{thm: phase 1}, if $\optcrors(\cI^*)\leq \cm^2/\mu^{c'}$, and Event $\tilde \event'$ did not happen, then 
$\sum_{I\in \iset}\cro(\psi(I))\leq \optcrors(\cI)$, and $\sum_{I\in \iset}|\chi^{\dirty}(\psi(I))|\leq \frac{\optcrors(\cI)\cdot \mu^{900}}{\cm}$.
Therefore, we get that $\cro(\check \phi)\leq \sum_{I'\in \hat \iset'}\cro(\phi(I'))+\optcrors(\cI)\cdot\mu^{O(1)}$. By suppressing the vertices that were used to subdivide the edges of graph $\cG^*$ to obtain graph $\cG$, we obtain a solution to the original instanc $\cI^*$ of the same cost.
\end{proof}

Lastly, the following observation will complete the proof of \Cref{lem: many paths}.

\begin{observation}\label{obs: cheap solutions to final instances}
	Assume that $\optcrors(I^*)\leq \cm^2/\mu^{c'}$, and that event $\tilde \event'$ did not happen. Then $\sum_{I'\in \iset'}\optcrors(I')\leq  \optcrors(\cI^*)\cdot (\log \cm)^{O(1)}$.
\end{observation}
\begin{proof}
	We bound $\sum_{I\in \iset_2}\optcrors(\hat I)$ and $\sum_{I\in \iset_1}\optcrors(\hat I')$ separately.
	
	From \Cref{thm: phase 2}, if Event $\tilde \event'$ did not happen, then, for every instance $I=(G,\Sigma)\in \iset_2$, 
	there is a solution $\psi(I')$ to the corresponding instance $I'$, that is clean with respect to $\jset(I)$, with $\cro(\psi(I'))\leq \left (\cro(\psi(I))+|\chi^{\dirty}(\psi(I))|^2+\frac{|\chi^{\dirty}(\psi(I))|\cdot |E(G)|}{\mu^b}\right )\cdot (\log \cm)^{O(1)}$. 
	From \Cref{obs: clean solution to contracted}, there is a solution to the corresponding contracted instance $\hat I'$, of cost at most $\cro(\psi(I'))$. Altogether, we get that:
	\[\begin{split}
	&\sum_{I\in \iset_2}\optcrors(\hat I')\leq \sum_{I\in \iset_2}\left(\cro(\psi(I))+|\chi^{\dirty}(\psi(I))|^2+\frac{|\chi^{\dirty}(\psi(I))|\cdot \cm}{\mu^b}\right )\cdot (\log \cm)^{O(1)}.\\
	&\leq \sum_{I\in \iset_2}\cro(\psi(I))\cdot (\log \cm)^{O(1)}+\left (\sum_{I\in \iset_2}|\chi^{\dirty}(\psi(I))|\right )^2\cdot (\log \cm)^{O(1)}+ \frac{\cm\cdot (\log \cm)^{O(1)}}{\mu^b}\cdot\left(\sum_{I\in \iset_2}|\chi^{\dirty}(\psi(I))|\right ).
	\end{split}
	\]
	From \Cref{thm: phase 1}, $\sum_{I\in \iset}\cro(\psi(I))\leq \optcrors(\cI)$ and $\sum_{I\in \iset}|\chi^{\dirty}(\psi(I))|\leq \frac{\optcrors(\cI)\cdot \mu^{900}}{\cm}$. Additionally, since we have assumed that $\optcrors(\cI^*)=\optcrors(\cI)\leq \cm^2/\mu^{c'}$ for a large enough constant $c'$, $\left (\sum_{I\in \iset_2}|\chi^{\dirty}(\psi(I))|\right )^2\leq \frac{(\optcrors(\cI))^2\cdot \mu^{1800}}{\cm^2}\leq \optcrors(\cI)$. Altogether, we get that:
	\[
	\begin{split}
	 \sum_{I\in \iset_2}\optcrors(\hat I')&\leq \optcrors(\cI)\cdot  (\log \cm)^{O(1)}+\frac{\optcrors(\cI)\cdot \mu^{900}\cdot  (\log \cm)^{O(1)}}{\mu^b}\\
	 &\leq \optcrors(\cI)\cdot  (\log \cm)^{O(1)}, \end{split}\]
	since $b\geq 4000$.
%

Next, we bound $\sum_{I\in \iset_1}\optcrors(\hat I)$. Consider some instance $I=(G,\Sigma)\in \iset_1$ and the solution $\psi(I)$ that is $\jset(I)$-valid. Let $J(I)$ be the core associated with the core structure $\jset(I)$. Let $E^{\dirty}(I)\subseteq E(G)\setminus E(J(I))$ be the set of all edges $e$, such that the image of $e$ in $\psi(I)$ crosses the image of some edge of $J(I)$.  Let $\hat I=(\hat G,\hat \Sigma)$ be the $\jset(I)$-contracted instance corresponding to instance $I$.

Denote $G'=G\setminus E^{\dirty}(I)$, and let $\Sigma'$ be the rotation system for graph $G'$ that is induced by $\Sigma$. Observe that $\jset(I)$ is a valid core structure for the resulting instance $I'=(G',\Sigma')$. Let $\hat I'=(\hat G',\hat \Sigma')$ be the $\jset(I)$-contracted instance associated with $I'$.

 Observe that we can easily modify the solution $\psi(I)$ to instance $I$ to obtain a solution $\psi(I')$ to instance $I'$ that is clean with respect to $\jset(I)$, with $\cro(\psi(I'))\leq \cro(\psi(I))$. Indeed, denote $\jset(I)=(J,\set{b_u}_{u\in V(J)},\rho_J, F^*)$.
 Let $\psi'(I')$ be the solution to instance $I'$ induced by $\psi(I)$.
 Since $G'=G\setminus E^{\dirty}(I)$,  for every connected component $C$ of $G'$, either the images of all edges and vertices of $C$ in $\psi'(I')$ are contained in the region $F^*$ of the drawing, or the images of all edges and vertices of $C$ in $\psi'(I')$ are disjoint from $F^*$ (note that, if $E(C)\cap \delta_G(J)\neq \emptyset$, then the image of $C$ must be contained in $F^*$, since the image of $C$ must intersect the interior of $F^*$, from the definition of a valid core structure (see \Cref{def: valid core 2})). If   the images of all edges and vertices of $C$ in $\psi'(I')$ are disjoint from $F^*$, then $C\cap J=\emptyset$ must hold, and so we can simply move the image of $C$ to lie in the interior of the region $F^*$ without changing the drawing of $C$ itself, and without introducing any new crossings. Once we move the image of each such connected component to lie inside region $F^*$, we obtain a solution $\psi(I')$ to instance $I'$ that is clean with respect to $\jset(I)$, and $\cro(\psi(I'))\leq \cro(\psi(I))$.

 From \Cref{obs: clean solution to contracted}, there is a solution $\psi(\hat I')$ to the contracted instance $\hat I'$ with $\cro(\psi(\hat I'))\leq \cro(\psi(I'))\leq \cro(\psi(I))$. 
We use the algorithm from \Cref{lem: edge insertion} in order to insert the edges of $E^{\dirty}(I)$ into the drawing $\psi(\hat I')$ to obtain a solution $\psi(\hat I)$ of instance $\hat I$, with the number of crossings bounded by $\cro(\psi(\hat I'))+|E^{\dirty}(I)|\cdot |E(\hat G)|\leq \cro(\psi(I))+|E^{\dirty}(I)|\cdot |E(\hat G)|$.
Since $I\in \iset_1$,  $|E(\hat G)|\leq \frac{\cm}{\mu^{1000}}$,  so $\cro(\psi(\hat I))
\leq \cro(\psi(I))+|\chi^{\dirty}(I)|\cdot  \frac{\cm}{\mu^{1000}}$. We then get that:
\[\sum_{I\in \iset_1} \optcrors(\hat I)\leq 
\sum_{I\in \iset_1}\cro(\psi(I))+\sum_{I\in \iset_1}|\chi^{\dirty}(I)|\cdot  \frac{\cm}{\mu^{1000}}.
\]
From \Cref{thm: phase 1}, $\sum_{I\in \iset}\cro(\psi(I))\leq \optcrors(\cI)$ and $\sum_{I\in \iset}|\chi^{\dirty}(\psi(I))|\leq \frac{\optcrors(\cI)\cdot \mu^{900}}{\cm}$.
Therefore: 
\[ \sum_{I\in \iset_1} \optcrors(\hat I)\leq  \optcrors(\cI)+\frac{\cm}{\mu^{1000}} \cdot  \frac{\optcrors(\cI)\cdot \mu^{900}}{\cm} \leq O(\optcrors(\cI)).\]
Overall, we get that $\sum_{I'\in \iset'}\optcrors(I')=\sum_{I\in \iset_1} \optcrors(\hat I)+\sum_{I\in \iset_2} \optcrors(\hat I')\leq \optcrors(\cI)\cdot (\log \cm)^{O(1)}=\optcrors(\cI^*)\cdot (\log \cm)^{O(1)}$.
\end{proof}

In the remainder of this section we focus on the proof of \Cref{thm: phase 2}. Throughout the proof, we denote the instance $I=(G,\Sigma)$ that serves as the input to the algorithm by $\cI'=(\cG',\cSigma')$, with $|E(\cG')|$ denoted by $\cm'$. We denote the core structure $\jset(I)$ by $\cjset=(\cJ,\set{b_u}_{u\in V(\cJ)},\rho_{\cJ}, F^*)$. We can assume that there is a $\cjset$-valid solution $\cpsi$ to instance $\cI'$ with  $\cro(\cpsi)\leq (\cm')^2/\mu^{240b}$, and  $|\chi^{\dirty}(\cpsi)|\leq \cm'/\mu^{240b}$, since otherwise we can set $E^{\del}(\cI')=E(\cG')\setminus E(\cJ)$, which trivially satisfies the requirements of the theorem. From now on we fix a  $\cjset$-valid solution $\cpsi$ to instance $\cI'$, with $\cro(\cpsi)\leq (\cm')^2/\mu^{240b}$ and  $|\chi^{\dirty}(\cpsi)|\leq \cm'/\mu^{240b}$. We emphasize that solution $\cpsi$ is not known to the algorithm.
\subsubsection{Proof of \Cref{thm: phase 2} -- Intuition}
\label{phase 2 intuition}

For simplicity of exposition, assume that the core $\cJ$ corresponding to the core structure $\cjset$ is a simple cycle.
Generally it is not difficult to modify the solution $\cpsi$ to instance $\cI'$ so that it becomes semi-clean, while only increasing the number of crossings by at most $|\chi^{\dirty}(\cpsi)|^2$. In order to do so, we let $E^{\dirty}$ be the set of all dirty edges -- that is, edges whose image in $\cpsi$ crosses the image of some edge of $\cJ$. Let $\cset$ be the set of all connected components of $\cG'\setminus E^{\dirty}$. It is easy to verify that for each component $C\in \cset$, either the images of all vertices and edges of $C$ in $\cpsi$ lie in the region $F^*$; or the images of all vertices and edges of $C$ in $\cpsi$ are disjoint from $F^*$. In the latter case, we move the image of $C$ to lie in the interior of the face $F^*$, without changing the image itself. We then need to modify the images of the edges in set $E^{\dirty}$, so that they connect the new images of their endpoints. This can be easily done while introducing at most $|\chi^{\dirty}(\cpsi)|^2$ new crossings. We do not provide the details here, since we do not use this algorithm eventually.

\begin{figure}[h]
	\centering
	\subfigure[Before: the image of $e$ (green) and its intersections $p,p'$ (red) with the image of $\cJ$ (blue). Region $F^*$ is shown in gray.]{\scalebox{0.16}{\includegraphics{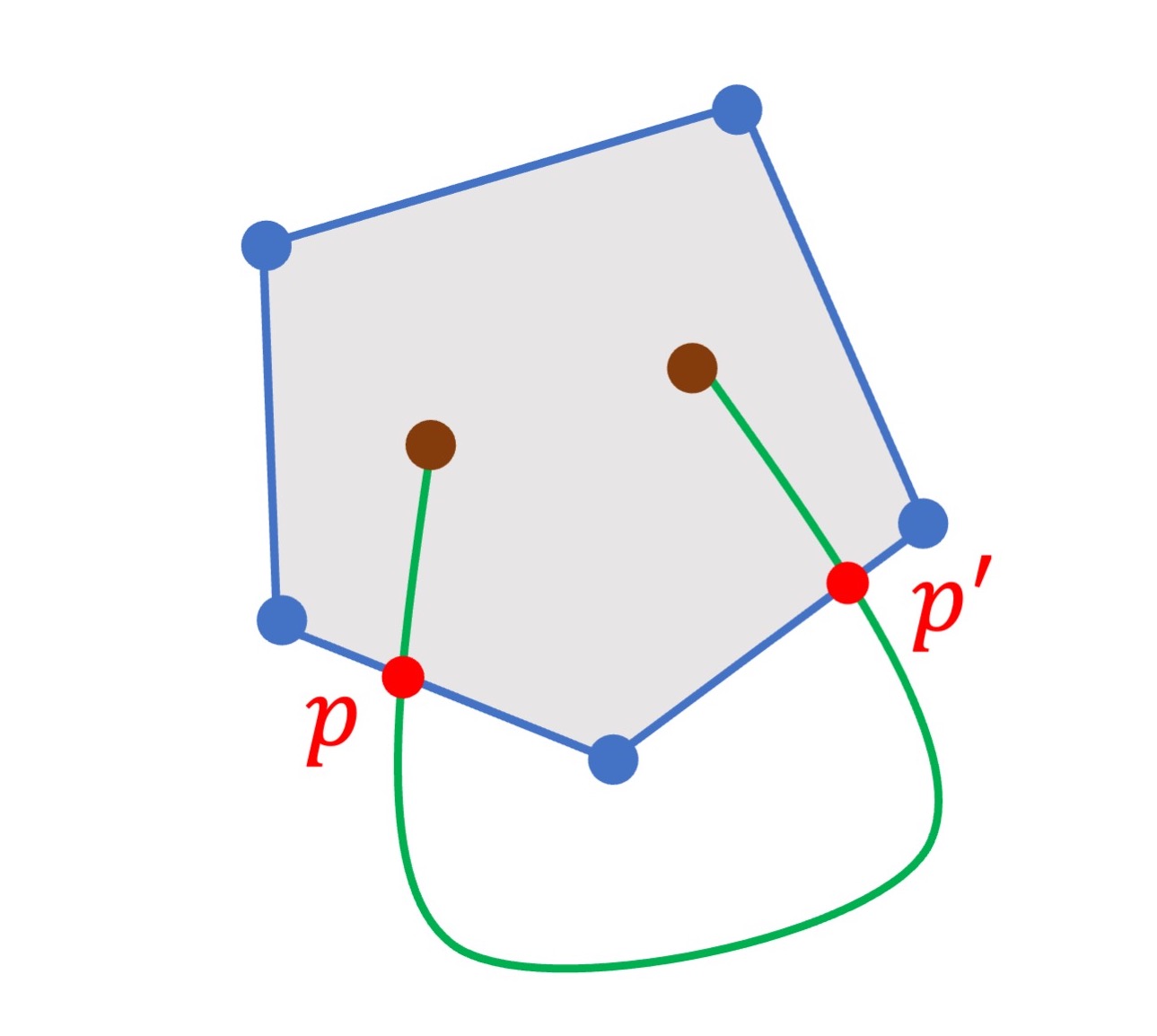}}\label{fig: fixdrawing_1}}
	\hspace{0.5cm}
	\subfigure[After: the new image of $e$ is shown in green.
	]{\scalebox{0.16}{\includegraphics{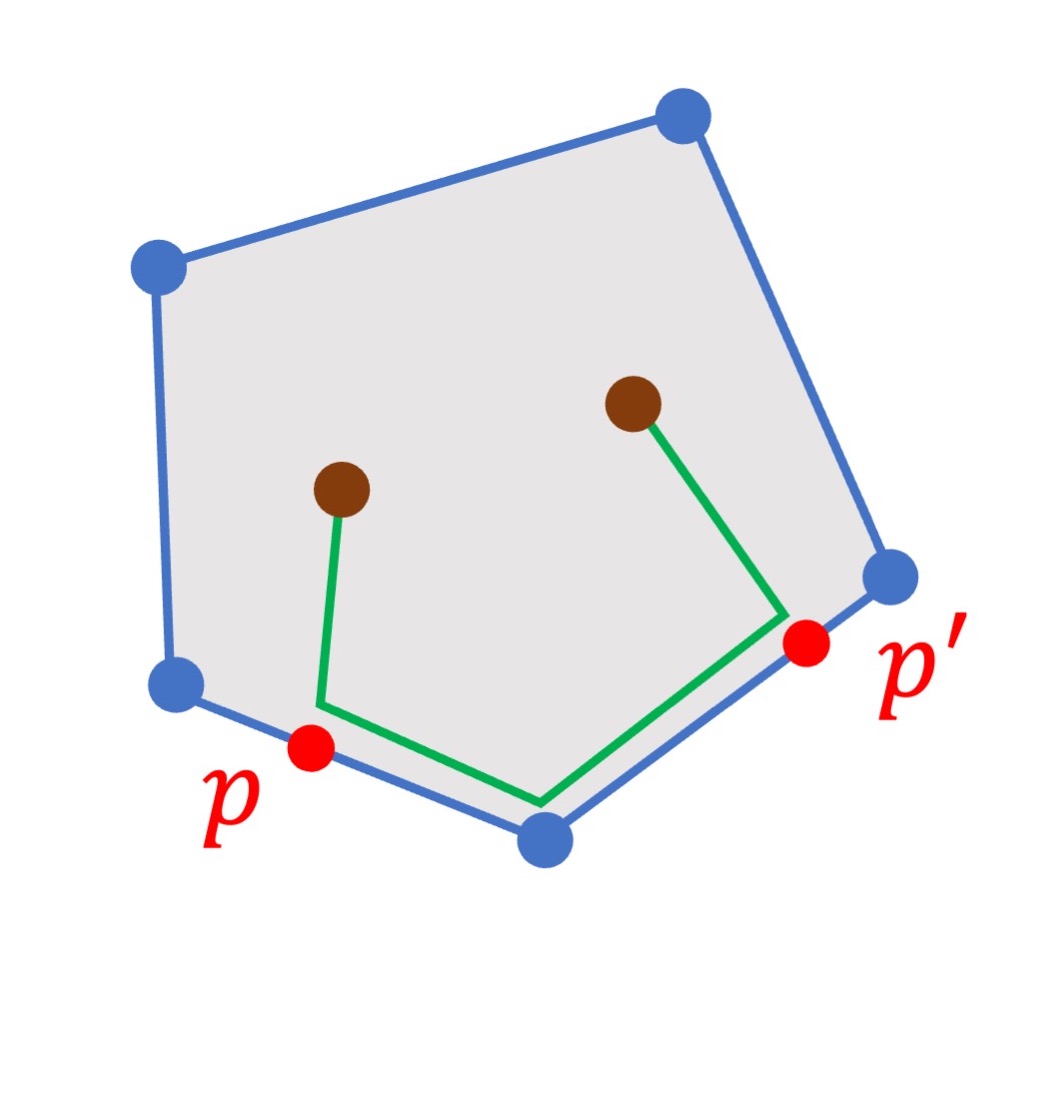}}\label{fig: fixdrawing_2}}
	\caption{Repairing the image of an edge $e\in E^{\dirty}$. 
	}\label{fig: fixdrawing}
\end{figure}

Let $\cpsi'$ denote this semi-clean solution to instance $\cI'$ with respect to $\cjset$.  Denote by $\gamma$ the image of the cycle $\cJ$ in $\cpsi$, which must be a simple closed curve. For convenience, we will now denote by $E^{\dirty}$ the set of all dirty edges of $\cG'$ -- edges whose image in $\cpsi'$ crosses the image of some edge of $\cJ$.
Consider now some dirty edge $e\in E^{\dirty}$. For simplicity of exposition, assume that $e$ is not incident to any vertex of $\cJ$.
Since $\cpsi'$ is a semi-clean drawing of $\cG'$ with respect to $\cjset$, the images of the endpoints of $e$ must lie in region $F^*$. Therefore, there must be at least two  points on $\cpsi'(e)$ that lie on $\gamma$. We assign the curve $\cpsi(e)$ an arbitrary direction, denote by $p$ the first  point on $\cpsi'(e)$ that lies on $\gamma$, and by $p'$ the last  point on $\cpsi'(e)$ that lies on $\gamma$. Points $p$ and $p'$ partition the curve $\gamma$ into two disjoint simple open curves, that we denote by $\gamma'$ and $\gamma''$, respectively. A simple way to ``repair'' the drawing of the edge $e$ so that it no longer crosses the edges of $\cJ$ would be to replace the segment of $\cpsi(e)$ between $p$ and $p'$ with a new segment $\sigma(e)$, that follows the curve $\gamma'$ closely, in the interior of region $F^*$ (see  \Cref{fig: fixdrawing}).

A problem with this approach is that this may greatly increase the number of crossings, as the segment $\sigma(e)$ may cross many edges in drawing $\cpsi'$. Intuitively, the requirements of \Cref{thm: phase 2} allow us to add up to  $\frac{\cm'(\log \cm')^{O(1)}}{\mu^b}$ new crossings to the drawing $\cpsi$ for each dirty edge whose image we modify, but unfortunately it is possible that, after the modification, $\sigma(e)$ crosses the images of many more edges.

Let $S$ denote the set of all vertices of $\cJ$ whose images lie on $\gamma'$, and let $T$ be defined similary for $\gamma''$. Consider the minimun cut $(X,Y)$ separating vertices of $S$ from vertices of $T$ in graph $\cG'\setminus E^{\dirty}$, so $S\subseteq X$ and $T\subseteq Y$. Assume first that $|E_{\cG'}(X,Y)|<\cm'/\mu^b$. In this case, we can rearrange the drawing $\cpsi'$, so that all vertices and edges of $\cG[X]\setminus E^{\dirty}$ are drawn very close to the segment $\gamma'$ (but in the interior of region $F^*$), and similarly all vertices and edges of $\cG[Y]\setminus E^{\dirty}$ are drawn very close to the segment $\gamma''$. We can then define a curve $\sigma(e)$ connecting points $p$ and $p'$, so that $\sigma(e)$ is contained in $F^*$, and it only crosses the images of the edges in $E_{\cG'}(X,Y)$. Therefore, we can ensure that $\sigma(e)$ participates in few crossings. We can then modify the image of edge $e$ to follow the segment $\sigma(e)$ as before, without increasing the number of crossings by too much.

Note that each dirty edge $e\in E^{\dirty}$ may define a different partition $(S,T)$ of the vertices of $\cJ$, and a different cut $(X,Y)$. However, if we can ensure that the number of edges crossing each such cut is sufficiently low, then we can still rearrange the drawing $\cpsi'$, and modify the drawings of all edges in $E^{\dirty}$, so that they become  contained in region $F^*$, while ensuring that the total number of crossings only increases moderately.

It is however possible that, for some edge $e\in E^{\dirty}$, and its corresponding partition $(S,T)$ of $V(\cJ)$, the minimum cut separating $S$ from $T$ in $\cG'$ contains more than $\cm'/\mu^b$ edges. In this case, there must be at least $\ceil{\cm'/\mu^b}$ edge-disjoint paths in $\cG'$ connecting vertices of $S$ to vertices of $T$. We can treat this set of paths as a promising set of paths, that can be used in order to define an enhancement $P$ of the core structure $\cjset$, using  Procedure \procsplit. We can also use the procedure in order to compute an enhancement structure $\aset$, and a split $(I_1=(G_1,\Sigma_1),I_2=(G_2,\Sigma_2))$ of instance $\cI'$ along $\aset$. Unlike the algorithm from Phase 1, we will not view the resulting two instances $I_1,I_2$ as separate instances. Instead, we will initialize the set $E^{\del}(\cI')$ of deleted edges to edge set $E(\cG')\setminus (E(G_1)\cup E(G_2))$. We then consider the graph $K=\cJ\cup P$, that we call a \emph{skeleton }, and fix a planar drawing $\rho_{K}$ of it (which is uniquely defined). 
From Property \ref{prop output drawing} of valid output to Procedure \procsplit, there is a drawing $\psi$ of graph $\cG'\setminus E^{\del}(\cI')$, that obeys the rotation system $\cSigma'$, such that drawing $\psi$ is $\cjset$-valid, and the edges of $\cJ\cup P$ do not cross each other in $\psi$, with $\cro(\psi)\leq \cro(\cpsi')$. If we consider the drawing $\rho_{\cJ}$ of the core ${\cJ}$, then the image of path $P$ partitions face $F^*$ into two new faces, that we denote by $F_1$ and $F_2$. Consider the split $(\jset_1,\jset_2)$ of the core structure $\cjset$ along $\aset$. The cores $J_1,J_2$ associated with the core structures $\jset_1$ and $\jset_2$, respectively, serve as the boundaries of the faces $F_1$, $F_2$, respectively, in the  drawing $\rho_K$ of graph $K$.

We note that the drawing of instance $I_1$ induced by $\cpsi$ is not necessarily semi-clean with respect to $\jset_1$, and the same is true regarding instance $I_2$ and core structure $\jset_2$. But we could modify $\cpsi$ to ensure this property, obtaining a new drawing $\psi'$ of graph $\cG'\setminus E^{\del}(\cI')$ (though this process would increase the number of crossings by factor $\poly\log \cm'$; we ignore this technicality for now). If we now consider some  edge $e$, whose image in $\psi'$ crosses the image of some edge in ${\cJ}$,  then edge $e$ must either lie in graph $G_1$, or in graph $G_2$. Assume w.l.o.g. that it is the former. 
In the modified  drawing $\psi'$ of $\cG'\setminus E^{\del}(\cI')$, both endpoints of edge $e$ are drawn inside the region $F_1$. The image of $e$ must then cross the boundary of region $F_1$ in at least two points (recall that the boundary of region $F_1$ in $\psi'$ is the image of the core $J_1$). We can again use these two points to define a partition $(S,T)$ of the vertices of $J_1$, and compute a minimum cut $(X,Y)$ separating $S$ from $T$ in $G_1$. As before, if the value of this minimum cut is small, then we can modify the current drawing $\psi'$ locally inside region $F_1$ and modify the image of the edge $e$, so that it is contained in $F_1$, and no longer crosses the edges of ${\cJ}$. If the value of this minimum cut is large, then we can again define a promising set of paths for instance $I_1$ and core structure $\jset_1$, and then invoke Procedure \procsplit in order to further split core structure $\jset_1$ and instance $I_1$, thereby  adding new edges to  set $E^{\del}(\cI')$.

At a high level, our algorithm can be thought of as maintaining a single \emph{skeleton} graph $K$ -- a  planar subgraph of $\cG'$ with ${\cJ}\subseteq K$, such that, for every edge $e\in E(K)$, graph $K\setminus\set{e}$ is connected. 
We also maintain a \emph{skeleton structure} $\kset$, that, in addition to the skeleton $K$, specifies the orientation $b_u$ of every vertex $u\in V(K)$, and a planar drawing $\rho_K$ of graph $K$ on the sphere. We require that, for every vertex $u\in V({\cJ})$, its orientations in $\kset$ and $\cjset$ are identical, and that drawing $\rho_K$ of $K$ is clean with respect to core structure $\cjset$, and is consistent with rotation system $\Sigma$ and orientations $b_u$ of the vertices $u\in V(K)$. Let $\fset(\rho_K)$ be the set of all faces of the drawing $\rho_K$. Since drawing $\rho_K$ is clean with respect to $\cjset$,  every forbidden face $F\in \fset^{\forbidden}(\rho_{\cJ})$ is also a face of $\fset(\rho_K)$. We denote by $\fset^{\forbidden}(\rho_K)=\fset^{\forbidden}(\rho_{\cJ})$ the set of all such faces of $\fset(\rho_K)$, that we refer to as \emph{forbidden faces} of drawing $\rho_K$.
For every face $F\in \fset(\rho_K)$, the set of vertices and edges of $K$ lying on its boundary define a core $J_F$. Using the skeleton structure $\kset$, we can define a core structure $\jset_F$ associated with the core $J_F$. 
We also maintain, for every face $F\in \fset(\rho_K)$, a subgraph $G_F$ of $\cG'$. We let $\Sigma_F$ be the rotation system for $G_F$ induced by $\Sigma$, and we let $\iset_F=(G_F,\Sigma_F)$ be the resulting instance of \cnwrs. We require that $\jset_F$ is a valid core structure for instance $I_F$, and, if face $F$ is a forbidden face, then $G_F=J_F$ holds. We will ensure that, for every pair $F,F'\in \fset(\rho_K)$ of distinct faces, every vertex and every edge of $G_F\cap G_{F'}$ belong to $J_F\cap J_{F'}$, and that $E^{\del}(\cI')=E(\cG')\setminus \left(\bigcup_{F\in \fset(\rho_K)}E(G_F)\right )$.

Consider now some face $F\in \fset(\rho_K)\setminus \fset^{{\forbidden}}(\rho_K)$. Intuitively (though somewhat imprecisely), we say that the corresponding instance $I_F$ is  \emph{acceptable} if, for every parition $(S,T)$ of the vertices of the core $J_F$, where the vertices of $S$ appear consecutively on the cycle $J_F$, there is a small cut in $G_F$ separating $S$ from $T$ (the definition is slightly more involved when $I_F$ is not a simple cycle). If instance $I_F$ is unacceptable, then we will use Procedure \procsplit in order to further augment the skeleton and partition instance $I_F$ into two subinstances. Once we reach a state where, for every face $F\in \fset(\rho_K)$ of the current skeleton $K$, the corresponding instance $I_F$ is acceptable, we terminate the algorithm. We show that the resulting drawing $\psi$ of graph $\cG'\setminus E^{\del}(\cI')$ can be modified so that all crossings with the edges of the original core $\cJ$ are eliminated, and the number of crossings only increases moderately.

\subsubsection{Proof of \Cref{thm: phase 2} -- Main Definitions and Notation}
\label{phase 2 defs}

In this subsection we define the main notions that we use in the proof of \Cref{thm: phase 2}, and also state the main lemmas from which the proof of \Cref{thm: phase 2} follows.

Suppose $G'$ is any subgraph of $\cG'$. Let $\Sigma'$ be the rotation system for $G'$ induced by $\cSigma'$, and let $I'=(G',\Sigma')$ be the resulting instance of \cnwrs.  For brevity, we will say that $I'=(G',\Sigma')$ is the \emph{subinstance of $\cI'$ defined by $G'$}. Recal that $\cG'\subseteq \cG$, so $I'$ is also the subinstance of $\cI$ defined by $G'$.

\paragraph{Planar Drawings and Face Boundaries.}
Suppose we are given a planar graph $H$, and a drawing $\phi$ of $H$ on the sphere with no crossings. As before, we denote by $\fset(\phi)$ the set of all faces of the drawing $\phi$. For a face $F\in \fset(\phi)$, we denote by $\partial_{\phi}(H,F)$ the subgraph of $H$ containing all vertices and edges whose images are contained in the boundary of the face $F$. We omit the subscript $\phi$ when clear of context. We sometimes say that the vertices and edges of $\partial_{\phi}(H,F)$ serve as the boundary of face $F$ in $\phi$.

\paragraph{Skeleton and Skeleton Structure.}

We now define the central notions that the proof of \Cref{thm: phase 2} uses, namely a skeleton and a skeleton structure, that can be thought of as extending the notions of a core and a core structure.

\begin{definition}[Skeleton]
	Let $K$ be a subgraph of $\cG'$. We say that $K$ is a \emph{skeleton graph}, if $\cJ'\subseteq K$, and, for every edge $e\in E(K)$, graph $K\setminus\set{e}$ is connected.
\end{definition}

While the definition of the skeleton $K$ is quite general, we will construct the skeleton using a specific procedure. At the beginning of the algorithm, we let $K=\cJ'$. In every iteration of the algorithm, we augment $K$ by adding to it a simple path (or a cycle) $P$, whose both endpoints belong to $K$, and all inner vertices are disjoint from $K$.
Next, we define a skeleton structure.

\begin{definition}[Skeleton Structure]
A skeleton structure $\kset$ consists of the following three ingredients:

\begin{itemize}
	\item a skeleton graph $K$;
	\item for every vertex $u\in V(K)$, an orientation $b_u\in \set{-1,1}$, such that, for every vertex $u\in V(\cJ')$, its orientation $b_u$ is identical to that given by the core structure $\cjset$; and
	\item a drawing $\rho_K$ of graph $K$ on the sphere with no crossings, such that $\rho_K$ obeys the rotation system $\cSigma'$, the orientation of every vertex $u\in V(K)$ in $\rho_K$ is $b_u$, and drawing $\rho_K$ is clean with respect to $\cjset'$.
\end{itemize}
\end{definition}

Consider now some skeleton structure $\kset=(K,\set{b_u}_{u\in V(K)},\rho_K)$. For brevity of notation, we denote by $\tfset(\kset)=\fset(\rho_K)$ the set of all faces in the drawing $\rho_K$. Since drawing $\rho_K$ is clean with respect to $\cjset'$, the image of every vertex and every edge of $K$ appears in the region $F^*=F^*(\rho_{\cJ'})$ of this drawing (including its boundary). Therefore, every forbidden face $F\in \fset^{\forbidden}(\rho_{\cJ'})$ is also a face of $\tfset(\kset)$. We denote by $\tfset^{\forbidden}(\kset)=\fset^{\forbidden}(\rho_{\cJ'})$ the set of all such faces, that we refer to as \emph{forbidden faces of drawing $\rho_K$}, and we denote by $\tfset'(\kset)=\tfset(\kset)\setminus\tfset^{\forbidden}(\kset)$ the set of all remaining faces.

Consider some face $F\in \tfset(\kset)$, and let $J_F=\partial_{\rho_K}(K,F)$ be the graph consisting of all vertices and edges of $K$ whose images appear on the boundary of face $F$ in drawing $\rho_K$. From the definition of a skeleton graph $K$, graph $J_F$ is a core. We can then define a core structure $\jset_F=(J_F,\set{b_u}_{u\in V(J_F)},\rho_{J_F}, F^*(\rho_{J_F}))$ as follows: for every vertex $u\in V(J_F)$, its orientation $b_u$ remains the same as in $\kset$. Drawing $\rho_{J_F}$ of graph $J_F$ is the drawing induced by $\rho_K$. Notice that face $F\in \tfset(\kset)$ remains a face in the drawing $\rho_{J_F}$. Face $F^*(\rho_{J_F})$ is then defined to be the face $F$. (We note that a core structure is generally defined for some specific graph $G'$, for which the properties specified in  \Cref{def: valid core 2} must hold. Therefore, for now we view the core structure $\jset_F$ as simply a tuple $(J_F,\set{b_u}_{u\in V(J_F)},\rho_{J_F}, F^*(\rho_{J_F}))$, and we will later define a subgraph $G_F$ of $\cG'$, for which $\jset_F$ will be a valid core structure.)

\paragraph{$\kset$-Valid Drawings.}

Next, we consider a skeleton structure $\kset=(K,\set{b_u}_{u\in V(K)},\rho_K)$ and a subgraph $G'\subseteq G$ with $K\subseteq G'$. We then define $\kset$-valid drawings of graph $G'$, in a natural way.

\begin{definition}[$\kset$-valid drawings]
	Let $\kset=(K,\set{b_u}_{u\in V(K)},\rho_K)$ be a skeleton structure, let  $G'$ be a subgraph of $\cG'$ with $K\subseteq G'$, and let $I'=(G',\Sigma')$ be the subinstance of $\cI'$ defined by $G'$. We say that a solution $\phi$ to instance $I'$ is $\kset$-valid, if the drawing of the skeleton $K$ induced by $\phi$ is identical to $\rho_K$, and the orientation of every vertex $u\in V(K)$ in drawing $\phi$ is identical to the orientation $b_u$ given by $\kset$. We also say that a $\kset$-valid solution $\phi$ to instance $I'$ is a \emph{$\kset$-valid drawing} of graph $G'$.
\end{definition}

Note that, if $\phi$ is a $\kset$-valid solution to instance $I'$, for any skeleton structure $\kset$, then it is also a $\cjset$-valid solution to $I'$.

Since we will be considering various core structures $\jset_F$ associated with faces $F\in \tfset(\kset)$ of a given skeleton structure $\kset$, the definition of dirty edges and dirty crossings will change depending on which core structure we consider. As the algorithm pogresses, the drawing $\psi$ that we maintain for the current graph $G'=\cG'\setminus E^{\del}(\cI')$ will evolve. We need to keep track of the crossings in which the edges of the original core $\cJ$ participate, and of the edges involved in these crossings.

Therefore, given a subgraph $G'\subseteq \cG'$, a skeleton structure $\kset=(K,\set{b_u}_{u\in V(K)},\rho_K)$, and a $\kset$-valid solution to subinstance $I'=(G',\Sigma')$ defined by graph $G'$, we denote by $\chi^*(\phi)$ the set of all crossings $(e,e')_p$ in $\phi$ where $e$ or $e'$ belong to the core $\cJ$. 

\paragraph{A $\kset$-Decomposition of $\cG'$ and Face-Based Instances.}

Over the course of the algorithm, we will maintain a skeleton structure $\kset$, and an associated decomposition of graph $\cG'$ into subgraphs. Intuitively, for every face $F\in \tfset(\kset)$, we will define a subgraph $G_F$ of $\cG'$ that we associate with face $F$. We now define a $\kset$-decomposition of $\cG'$.

\begin{definition}[A $\kset$-decomposition of $\cG'$]
Let $\kset=(K,\set{b_u}_{u\in V(K)},\rho_K)$ be a skeleton structure, and, for every face $F\in \tilde \fset(\kset)$, let $\jset_F=(J_F,\set{b_u}_{u\in V(J_F)},\rho_{J_F}, F^*(\rho_{J_F}))$ be the core structure that $\kset$ defines for face $F$. A $\kset$-decomposition of the input graph $\cG'$ is a collection $\gset=\set{G_{ F}\mid  F\in \tilde \fset(\kset)}$ of subgraphs of $\cG'$, for which the following hold: 

\begin{itemize}
	\item for every face $ F\in \tilde \fset(\kset)$, the core structure $\jset_F$ associated with face $F$ is a valid core structure for instance $I_F=(G_F,\Sigma_F)$ defined by graph $G_F$;
	
	\item for every forbidden face $F\in \tilde \fset^{\forbidden}(\kset)$, $G_F=J_F$;

	\item for every face $F\in \tilde \fset(\kset)$, $G_F\cap K=J_F$; and
	
	\item for every pair $F, F'\in \tilde\fset(\kset)$ of distinct faces, $V(G_{F})\cap V(G_{F'})\subseteq  V(J_F)\cap V(J_{F'})$, and $E(G_{F})\cap E(G_{F'})\subseteq  E(J_F)\cap E(J_{F'})$.
\end{itemize} 

We will sometimes refer to the subinstance $I_F=(G_F,\Sigma_F)$ of $\cI'$ defined by graph $G_F$ associated with a face $F\in \tilde \fset(\kset)$ as a \emph{face-based subinstance} associated with face $F$.
\end{definition}

In order to prove \Cref{thm: phase 2}, we will gradually construct a skeleton $K$ and its associated skeleton structure $\kset$, starting with $K=\cJ$. We will also maintain a  $\kset$-decomposition $\gset$ of the graph $\cG'$. We will denote by $E^{\del}=E(\cG')\setminus\left(\bigcup_{ F\in \tilde \fset(\kset)}E(G_{ F})\right )$, and we will view $E^{\del}$ as the set of \emph{deleted edges}, that will eventually be added to set $E^{\del}(\cI')$. We will ensure that $|E^{\del}|$ will remain small over the course of the algorithm. Consider the graph $G'=\cG'\setminus E^{\del}$ and its associated subinstance $I'=(G',\Sigma')$ of $\cI'$. We will ensure that throughout the algorithm, there is a solution $\phi$ to instance $I'$, that is compatible with the solution $\psi(\cI')$ to instance $\cI'$ (with respect to the core structure $\cjset'$), with $\cro(\phi)\leq \cro(\psi(\cI'))$ and $|\chi^{*}(\phi)|\leq  |\chi^{*}(\psi(\cI'))|$. The algorithm terminates once every instance $G_{ F}$ in the resulting decomposition $\gset$ becomes ``acceptable'' -- a notion that we define next. 

Given a skeleton structure $\kset=(K,\set{b_u}_{u\in V(K)},\rho_K)$, a $\kset$-decomposition $\gset=\set{G_F\mid F\in \tilde \fset(\kset)}$ of graph $\cG'$, and a face $F\in \tilde \fset(\kset)$, we denote by $\tilde E_F$ the set of all edges $e\in E(G_F)$, such that exactly one endpoint of $e$ lies in $J_F$; in other words, $\tilde E_F=\delta_{G_F}(J_F)$ (recall that, since $G_F$ is a subgraph of the subdivided graph $\cG$, no edge of $E(G_F)\setminus E(J_F)$ may have both its endpoints in the core $J_F$). Recall that, when we defined a core structure, we have also defined an ordering $\oset(J_F)$ of the edges of $\tilde E_F$. Intuitively, this is the order in which the edges of $\tilde E_F$ are encountered in any  $\jset_F$-valid solution to instance $I_F$, as we follow along the boudary of face $F^*(\rho_{J_F})$ inside the face, in the counter-clock-wise direction.

\begin{definition}[Acceptable Instances]\label{def: acceptable instance}
Let	$\kset=(K,\set{b_u}_{u\in V(K)},\rho_K)$ be skeleton structure, let $\gset=\set{G_F\mid F\in \tilde \fset(\kset)}$ be a $\kset$-decomposition of graph $\cG'$, and let $F\in \tilde \fset(\kset)$ be a face. Consider a graph $H_F$,  that is obtained from graph $G_{F}$ by first subdividing every edge $e\in \tilde E_F$ with a vertex $t_e$, and then deleting all vertices and edges of $J_{F}$ from it. We say that instance $G_{F}$ is \emph{acceptable} if, for every partition $(E_1,E_2)$ of the edges of $\tilde E_F$, such that the edges of $E_1$ appear consecutively in the ordering $\oset(J_F)$, there is a cut $(X,Y)$ in graph $H_{F}$ with vertex set $\set{t_e\mid e\in E_1}$ contained in $X$, vertex set $\set{t_e\mid e\in E_2}$ contained in $Y$, and $|E_{H_{F}}(X,Y)|\leq \cm'/\mu^{2b}$.
\end{definition}

The main ingredients in the proof of \Cref{thm: phase 2} are the following two lemmas.

\begin{lemma}\label{lem: compute phase 2 decomposition}
There is an efficient randomized algorithm, whose input consists of a large enough constant $b\geq 1$, a subinstance $\cI'=(\cG',\cSigma')$ of $\cI$ with $|E(\cG')|=\cm'$ and $\cG'\subseteq \cG$, and a core structure $\cjset$ for $\cI'$, whose corresponding core is denoted by $\cJ$.
The algorithm either returns FAIL, or computes a skeleton structure $\kset=(K,\set{b_u}_{u\in V(K)},\rho_K)$, and a $\kset$-decomposition $\gset$ of $\cG'$, such that, for every face $F\in \tilde \fset(\kset)$, the corresponding subinstance $I_F=(G_F,\Sigma_F)$ of $\cI'$ defined by the graph $G_F\in \gset$ is acceptable. 
Moreover, if there is a solution $\psi(\cI')$ to instance $\cI'$ that is $\cjset$-valid, with $\cro(\psi(\cI'))\leq (\cm')^2/\mu^{240b}$, and  $|\chi^{*}(\psi(\cI'))|\leq \cm'/\mu^{240b}$, then  with probability at least $1-1/\mu^{2b}$, the following hold:

\begin{itemize}
	\item the algorithm does not return FAIL;
	\item the cardinality of edge set $E^{\del}(\cI')=E(\cG')\setminus \left(\bigcup_{F\in \tilde \fset(\kset)}E(G_F)\right )$ is bounded by: $$\left (\frac{\cro(\psi(\cI'))}{\cm'}+|\chi^*(\psi(\cI'))|\right ) \cdot \mu^{O(b)};\mbox{ and }$$
	
	\item if we  denote $G'=\cG'\setminus E^{\del}(\cI')$, and let $I'=(G',\Sigma')$ be the subinstance of $\cI'$ defined by graph $G'$, then  there is a $\kset$-valid solution $\phi$ to instance $I'$, with $\cro(\phi)\leq \cro(\psi(\cI'))$,  $|\chi^*(\phi)|\leq |\chi^*(\psi(\cI'))|$, and the total number of crossings in which the edges of $E(K)\setminus E(\cJ)$ participate is at most $\frac{\cro(\psi(\cI'))\cdot  \mu^{50b}}{\cm'}$. 
\end{itemize}
\end{lemma}

\begin{lemma}\label{lem: computed decomposition is good enough}
Suppose we are given a subinstance $\cI'=(\cG',\cSigma')$ of $\cI$ with $\cG'\subseteq \cG$ and $|E(\cG')|=\cm'$, a core structure $\cjset$ for $\cI'$, a skeleton structure $\kset=(K,\set{b_u}_{u\in V(K)},\rho_K)$, and a $\kset$-decomposition $\gset$ of $\cG'$, such that, for every face $F\in \tilde \fset(\kset)$, the corresponding subinstance $I_F=(G_F,\Sigma_F)$ of $\cI'$ defined by the graph $G_F\in \gset$ is acceptable.  Let $E^{\del}(\cI')=E(\cG')\setminus \left(\bigcup_{F\in \tilde \fset(\kset)}E(G_F)\right )$, $G'=\cG'\setminus E^{\del}(\cI')$, and let $I'=(G',\Sigma')$ be the subinstance of $\cI'$ defined by graph $G'$. Assume further that there is a $\kset$-valid solution $\phi$ to instance $I'$, so that the total number of crossings in which the edges of $K$ participate is $N\leq \frac{\cm'}{\mu^{3b}}$ and $|\chi^{*}(\phi)|\leq \cm'/\mu^{240b}$. Then 
there is a solution $\psi(I')$ to instance $I'$ that is clean with respect to $\cjset$, with $\cro(\psi(I'))\leq \left (\cro(\phi)+N^2+|\chi^{*}(\phi)|^2+\frac{|\chi^{*}(\phi)|\cdot \cm'}{\mu^b}\right )\cdot (\log \cm')^{O(1)}$.
\end{lemma}


We provide the proofs of \Cref{lem: compute phase 2 decomposition} and \Cref{lem: computed decomposition is good enough} in Sections \ref{subsec: compute phase 2 decomposition} and \ref{subsec: computed decomposition is good enough}, respectively. The proof of  \Cref{thm: phase 2} easily follows from the above two lemmas. Indeed, consider the input instance 
$\cI'=(\cG',\cSigma')$ of \cnwrs that is a subinstance of $\cI$, with $\cG'\subseteq \cG$ and  $|E(\cG')|=\cm'$, and a core structure $\cjset$ for $\cI'$, whose corresponding core is denoted by $\cJ$. We apply the algorithm from \Cref{lem: compute phase 2 decomposition}  to this input. If the algorithm returns FAIL, then we set $E^{\del}(\cI')=E(\cG')\setminus E(\cJ)$, and return this edge set as the output. We say that the algorithm from \Cref{lem: compute phase 2 decomposition} \emph{fails} in this case. Otherwise, the algorithm from  \Cref{lem: compute phase 2 decomposition} computes a skeleton structure $\kset=(K,\set{b_u}_{u\in V(K)},\rho_K)$, and a $\kset$-decomposition $\gset$ of $\cG'$, such that, for every face $F\in \tilde \fset(\kset)$, the corresponding subinstance $I_F=(G_F,\Sigma_F)$ of $\cI'$ defined by the graph $G_F\in \gset$ is acceptable. 
Let $E^{\del}(\cI')=E(\cG')\setminus \left(\bigcup_{F\in \tilde \fset(\kset)}E(G_F)\right )$, let $G'=\cG'\setminus E^{\del}(\cI')$, and let $I'=(G',\Sigma')$ be the subinstance of $\cI'$ defined by graph $G'$.
We say that the algorithm from  \Cref{lem: compute phase 2 decomposition} is \emph{successful} if (i) it does not fail; (ii) $|E^{\del}(\cI')|\leq \left (\frac{\cro(\psi(\cI'))}{\cm'}+|\chi^{*}(\psi(\cI'))|\right ) \cdot \mu^{O(b)}$; and (iii)  there is a $\kset$-valid solution $\phi$ to instance $I'$, with $\cro(\phi)\leq \cro(\psi(\cI'))$,  $|\chi^*(\phi)|\leq |\chi^*(\psi(\cI'))|$, and the total number of crossings in which the edges of $E(K)\setminus E(\cJ)$ participate is at most $\frac{\cro(\psi(\cI'))\cdot  \mu^{50b}}{\cm'}$.  If the algorithm is not successful, then we say that it is \emph{unsuccessful}.  \Cref{thm: phase 2} guarantees that,  if there is a solution $\psi(\cI')$ to instance $\cI'$ that is $\cjset$-valid, with $\cro(\psi(\cI'))\leq (\cm')^2/\mu^{240b}$, and  $|\chi^{*}(\psi(\cI'))|\leq \cm'/\mu^{240b}$, then  the algorithm from \Cref{lem: compute phase 2 decomposition} is successful with probability at least $1-1/\mu^{2b}$.
If the algorithm from \Cref{lem: compute phase 2 decomposition}  does not fail, then we return edge set $E^{\del}(\cI')$ as the outcome of the algorithm.

In order to complete the proof of \Cref{thm: phase 2}, it is enough to show that,  if there is a solution $\psi(\cI')$ to instance $\cI'$ that is $\cjset$-valid, with $\cro(\psi(\cI'))\leq (\cm')^2/\mu^{240b}$ and $|\chi^{*}(\psi(\cI'))|\leq \cm'/\mu^{240b}$,  and the algorithm from \Cref{lem: compute phase 2 decomposition} is successful, then there is a solution $\psi(I')$ to instance $I'$ that is clean with respect to $\jset(I)$, with $\cro(\psi(I'))\leq \left (\cro(\psi(\cI'))+|\chi^{\dirty}(\psi(\cI'))|^2+\frac{|\chi^{\dirty}(\psi(\cI'))|\cdot \cm'}{\mu^b}\right )\cdot (\log \cm')^{O(1)}$.

Assume that  there is a solution $\psi(\cI')$ to instance $\cI'$ that is $\cjset$-valid, with $\cro(\psi(\cI'))\leq (\cm')^2/\mu^{240b}$ and $|\chi^{*}(\psi(\cI'))|\leq \cm'/\mu^{240b}$, and that the algorithm from  \Cref{lem: compute phase 2 decomposition} is successful. Then
there is a $\kset$-valid solution $\phi$ to instance $I'$, with $\cro(\phi)\leq \cro(\psi(\cI'))$,  $|\chi^*(\phi)|\leq |\chi^*(\psi(\cI'))|\leq \cm'/\mu^{240b}$, and the total number of crossings in which the edges of $K$ participate is at most $N=\frac{\cro(\psi(\cI'))\cdot  \mu^{50b}}{\cm'}+|\chi^{*}(\psi(\cI'))|\leq \frac{\cm'}{\mu^{3b}}$ (since $\cro(\psi(\cI'))\leq \frac{(\cm')^2}{\mu^{240b}}$ and $|\chi^{*}(\psi(\cI'))|\leq \frac{\cm'}{\mu^{240b}}$). From \Cref{lem: computed decomposition is good enough}, 
	there is a solution $\psi(I')$ to instance $I'$ that is clean with respect to $\cjset$, with $\cro(\psi(I'))\leq \left (\cro(\phi)+N^2+|\chi^{*}(\phi)|^2+\frac{|\chi^{*}(\phi)|\cdot \cm'}{\mu^b}\right )\cdot (\log \cm')^{O(1)}$.
Note that $\cro(\phi)\leq \cro(\psi(\cI'))$ and  $|\chi^*(\phi)|\leq |\chi^*(\psi(\cI'))|=|\chi^{\dirty}(\psi(\cI'))|$.
Additionally, 
$N\leq \frac{\cro(\psi(\cI'))\cdot  \mu^{50b}}{\cm'}+|\chi^{\dirty}(\psi(\cI'))|$, so
$N^2\leq \frac{2(\cro(\psi(\cI')))^2\cdot  \mu^{100b}}{(\cm')^2}+2|\chi^{\dirty}(\psi(\cI'))|^2\leq \cro(\psi(\cI'))$, since we have assumed that $\cro(\psi(\cI'))\leq (\cm')^2/\mu^{240b}$. Therefore, we get that: 
$$\cro(\psi(I'))\leq \left (\cro(\psi(\cI'))+|\chi^{\dirty}(\psi(\cI'))|^2+\frac{|\chi^{\dirty}(\psi(\cI'))|\cdot \cm'}{\mu^b}\right )\cdot (\log \cm')^{O(1)}$$ as required.

In order to complete the proof of \Cref{thm: phase 2}, it is now enough to prove 
\Cref{lem: compute phase 2 decomposition} and \Cref{lem: computed decomposition is good enough}, which we do next.  
 \subsubsection{Proof of \Cref{lem: compute phase 2 decomposition}}
 \label{subsec: compute phase 2 decomposition}

 The proof of \Cref{lem: compute phase 2 decomposition} is somewhat similar to the proof of \Cref{thm: phase 1}, in that we repeatedly apply Procedure \procsplit to obtain the desired decomposition, though the details are  different.
For simplicity of notation, we say that a solution $\psi$ to instance $\cI'$ is \emph{good}, if it is $\cjset$-valid, with $\cro(\psi)\leq (\cm')^2/\mu^{240b}$, and  $|\chi^{*}(\psi)|\leq \cm'/\mu^{240b}$. If such a solution exists, then we denote it by $\psi$ throughout the algorithm (though we note that the solution is not known to the algorithm). If such a solution does not exist, then we let $\psi$ be any $\cjset$-valid solution to instance $\cI'$.

Given a skeleton structure $\kset$, a $\kset$-decomposition $\gset$ of instance $\cI'$, and some face $F\in \tilde \fset(\kset)$, we denote by $m_F=|E(G_F)|$ and by $\hat m_F=|E(G_F)\setminus E(K)|$, where $G_F\in \gset$ is the graph associated with face $F$.
Our algorithm consists of at most $\ceil{32\mu^{6b}}$ stages. For $1\leq i\leq \ceil{32\mu^{6b}}$, the input to stage $i$ is a skeleton structure $\kset_i$, whose corresponding skeleton is denoted by $K_i$, and a $\kset_i$-decomposition $\gset_i$ of graph $\cG'$. We will ensure that, with high enough probability, the following properties hold.

\begin{properties}{A'}
	\item for every face $F\in \tilde \fset(\kset_i)$, either $m_F\leq \cm'\cdot \left (1-\frac{i-1}{32\mu^{6b}}\right )$, or the corresponding instance $I_F=(G_F,\Sigma_F)$ with $G_F\in \gset_i$ is acceptable; \label{invariant: fewer edges2}
	
	\item if we denote by $E^{\del}_i=E(\cG')\setminus\left(\bigcup_{G_F\in \gset_i}E(G_F)\right )$, then $|E^{\del}_i|\leq (i-1)\cdot \left( \frac{\cro(\psi)\mu^{100b}}{\cm'}+|\chi^{*}(\psi)|\cdot \mu^{3b}\right) $; and \label{invariant: few deleted edges}
	
	
	\item let $G_i=\cG'\setminus E^{\del}_i$, and let $\Sigma_i$ be the rotation system for $G_i$ induced by $\cSigma'$. Then there exists a solution $\phi_i$ to instance $(G_i,\Sigma_i)$, that is $\kset_i$-valid, and has the following additional properties:\label{invariant: drawing} 
	
	\begin{itemize}
		\item $\cro(\phi_i)\leq \cro(\psi)$;
		\item $|\chi^*(\phi_i)|\leq |\chi^{*}(\psi)|$; and
		\item if we denote by $N(\phi_i)$ the total number of crossings of $\phi_i$ in which the edges of $E(K_i)\setminus E(\cJ)$ participate, then $N(\phi_i)\leq \frac{\cro(\psi)\cdot i\cdot  \mu^{40b}}{\cm'}$.
	\end{itemize}
\end{properties}

We say that the execution of stage $i$ is \emph{successful} if the output of the stage satisfies the invariants \ref{invariant: fewer edges2}--\ref{invariant: drawing}. We denote by $\tilde \event_i'$ the bad event that the execution of stage $i$ is unsuccessful.

We start by constructing the input to the first iteration. The skeleton structure $\kset_1$ is defined by the core structure $\cjset=(\cJ,\set{b_u}_{u\in V(\cJ)},\rho_{\cJ}, F^*(\rho_{\cJ}))$ in a natural way: we let the skeleton $K_1$ be $\cJ$, the orientations $b_u$ for vertices $u\in V(\cJ)$ remain unchanged, and the drawing $\rho_{\cK_1}$ of skeleton $\cK_1$ is $\rho_{\cJ}$. Notice that the collection $\tilde \fset(\kset_1)$ of faces has a single non-forbidden face $F^*= F^*(\rho_{\cJ})$, and we let $G_{F^*}=\cG'$ be the graph associated with that face. For every other face $F\in \tilde \fset(\kset_1)$, skeleton structure $\kset_1$ defines a corresponding core structure $\jset_F$, whose core graph is denoted by $J_F$. We then let $G_F=J_F$ be the graph associated with face $F$. We let $\gset_1=\set{G_F\mid F\in \tilde \fset(\kset)}$ be the resulting $\kset_1$-decomposition of $\cG'$, so $E_1^{\del}=\emptyset$, and we let $\phi_1=\psi$ be the solution to instance $\cI'=(G_1,\Sigma_1)$ that we defined above. Note that $N(\phi_1)=0$.
 We have now obtained an input to Stage $1$ of the algorithm. If $\psi$ is a good solution to $\cI'$, then  Invariants  \ref{invariant: fewer edges2}--\ref{invariant: drawing} hold for this input.

\paragraph{Stage Execution.}
We now describe an execution of Stage $i$, for $1\leq i\leq \ceil{32\mu^{6b}}$. At the beginning of Stage $i$, we set $\kset_{i+1}=\kset_i$, denoting the corresponding skeleton by $K_{i+1}$, $\gset_{i+1}=\gset_i$, and $E^{\del}_{i+1}=E^{\del}_i$. Let $\fset'_i\subseteq \tilde \fset(\kset_i)$ be the collection of all faces $F$, for which the corresponding instance $I_F=(G_F,\Sigma_F)$ is not acceptable, where $G_F\in \gset_i$ is the graph associated with face $F$.
Denote $\fset'_i=\set{F_1,\ldots,F_q}$. Consider any face $F_j\in \fset'_i$. Since instance $I_{F_j}=(G_{F_j},\Sigma_{F_j})$ is not acceptable, from the definition of acceptable instances  (see \Cref{def: acceptable instance}), $|E(G_{F_j})|\geq \cm'/\mu^{2b}$ must hold. From the definition of a $\kset_i$-decomposition, for every pair $G_F,G_{F'}$ of graphs, if an edge $e$ lies in both graphs, then $e\in E(J_F)\cap E(J_{F'})$. Since an edge $e\in E(K_i)$ may lie on the boundary of at most two faces of the drawing $\rho_{K_i}$, we get that $\sum_{j=1}^q|E(G_{F_j})|\leq 2\cm'$. Therefore, $q\leq 2\mu^{2b}$ must hold. 

The algorithm for Stage $i$ consists of $q$ iterations. In iteration $j$, we apply Procedure \procsplit from \Cref{thm: procsplit}
to instance $I_{F_j}$ that is defined by graph $G_{F_j}$, and the corresponding core structure $\jset_{F_j}$. We will use parameter $b'=2b$ instead of parameter $b$ in Procedure \procsplit. The set $\pset_j$ of promising paths of cardinality $\floor{\frac{|E(G_{F_j})|}{\mu^{2b}}}$ is computed as follows. From the definition of acceptable instances (see \Cref{def: acceptable instance}), if instance $I_{F_j}$ is not acceptable, then there is a partition $(E_1,E_2)$ of the edges of $\tilde E_{F_j}$ (the edges of $G_{F_j}$ with exactly one endpoint in the core $J_{F_j}$), such that the edges of $E_1$ appear consecutively in the ordering $\oset(J_{F_j})$, and the minimum cut in the correpsonding graph $H_{F_j}$ separating separating these two sets of edges contains more than $\cm'/\mu^{2b}$ edges. From the Maximum Flow / Minimum Cut theorem, there is a collection $\pset_j$ of $\floor{\frac{\cm'}{\mu^{2b}}}\geq \floor{\frac{|E(G_{F_j})|}{\mu^{2b}}}$
edge-disjoint paths in graph $G_{F_j}$, where each path $P\in \pset_j$ has an edge of $E_1$ as its first edge, an edge of $E_2$ as its last edge, and is internally disjoint from $J_{F_j}$. Moreover, a set of paths with these properties can be computed efficiently. 

Let $\aset_j=\set{P_j,\set{b_u}_{u\in V(J_{F_j})\cup P_j},\rho'}$ be the  enhancement structure computed by Procedure \procsplit, and let $(I_1^j=(G_1^j,\Sigma_1^j), I_2^j=(G_2^j,\Sigma_2^j))$ be the split of instance $I_{F_j}$ along $\aset_j$ that the algorithm returns. We denote by $E^{\del}(I_{F_j})=E(G_j)\setminus (E(G^j_1)\cup E(G^j_2))$ the set of the deleted edges. We add the edges of $E^{\del}(I_{F_j})$ to $E^{\del}_{i+1}$, and we define a new skeleton $K'_{i+1}$ and skeleton structure $\kset'_{i+1}$ using the enhancement $\aset_j$ in a natural way: we  let $K'_{i+1}=K_{i+1}\cup P_j$. The orientations of vertices $u$ that belong to $K_{i+1}$  remain unchanged in $\kset'_{i+1}$, and the orientations of inner vertices on path $P_j$ are set to be identical to those given by $\aset_j$. Consider now drawing $\rho'$ of graph $J_{F_j}\cup P_j$. In this drawing, the image of the core $J_{F_j}$ is identical to that in the drawing $\rho_{K_{i+1}}$ of  skeleton $K_{i+1}$, and the image of path $P_j$ is drawn inside face $F_j$, partitioning it into two faces, $F_1^j$ and $F_2^j$. We obtain a drawing $\rho_{K_{i+1}'}$ of the new skeleton $K_{i+1}'$ by starting with the drawing $\rho_{K_{i+1}}$ of skeleton $K_{i+1}$, and then adding the drawing of path $P_j$ inside face $F_j$ exactly like in the drawing $\rho'$. This completes the definition of the new skeleton structure $\kset_{i+1}'$. Notice that $\tilde \fset (\kset'_{i+1})=(\tilde \fset(\kset_{i+1})\setminus\set{F_j})\cup \set{F_1^j\cup F_2^j}$. We modify the decomposition $\gset_{i+1}$ by replacing graph $G_{F_j}$ with the graphs $G_{F_1^{j}}$ (that becomes associated with face $F_1^j$) and  $G_{F_2^{j}}$ (that becomes associated with face $F_2^j$). We then replace skeleton structure $\kset_{i+1}$ with the new skeleton structure $\kset_{i+1}'$ and continue to the next iteration. We denote by $\event_j^i$ the bad event that the output of Procedure \procsplit computed in iteration $j$ is not a valid output.
This completes the definition of the algorithm for stage $i$. 

We start with the following observation.

\begin{observation}\label{obs: valid input to procsplit}
	Assume that there is a good solution $\psi$ to instance $\cI'$, that the Invariants \ref{invariant: fewer edges2}--\ref{invariant: drawing} hold at the beginning of stage $i$. 
	Then for all $1\leq j\leq q$, the input to Procedure \procsplit in iteration $j$ of stage $i$ is a valid input.
\end{observation}
\begin{proof}
	Since we have assumed that there is a good solution $\psi$ to instance $\cI'$, and that the Invariants \ref{invariant: fewer edges2}--\ref{invariant: drawing} hold at the beginning of stage $i$, 
	from Invariant \ref{invariant: drawing}, at the beginning of stage $i$, there exists a solution $\phi_i$ to instance $(G_i,\Sigma_i)$, that is $\kset_i$-valid, with $\cro(\phi_i)\leq \cro(\psi)$,
		 $|\chi^*(\phi_i)|\leq |\chi^{*}(\psi)|$, and
		 $N(\phi_i)\leq \frac{\cro(\psi)\cdot i\cdot  \mu^{40b}}{\cm'}$.

Consider now some index $1\leq j\leq q$. Let $\phi_{i,j}$ be the solution to instance $I_{F_j}$ defined by the graph $G_{F_j}\in \gset_i$ that is induced by $\phi_i$. 
This solution must be $\jset_{F_j}$-valid.
From the above discussion, 
$$\cro(\phi_{i,j})\leq \cro(\phi_i)\leq \cro(\psi)\leq \frac{(\cm')^2}{\mu^{240b}}\leq  \frac{|E(G_{F_j})|^2}{\mu^{120b}},$$ 
since $|E(G_{F_j})|\geq  \frac{\cm'}{\mu^{2b}}$. Next, we bound the number of dirty crossings of drawing $\phi_{i,j}$ with respect to the core structure $\jset_{F_j}$. The set of such dirty crossings  may include all crossings of $\chi^*(\phi_i)$ (whose number is bounded by $ |\chi^{*}(\psi)|\leq \frac{\cm'}{\mu^{240b}}\leq \frac{|E(G_{F_j})|}{\mu^{238b}}$), and the crossings in which the edges of $J_{F_j}\setminus \cJ'$ participate, whose number is bounded by:  
$$N(\phi_i)\leq \frac{\cro(\psi)\cdot i\cdot  \mu^{40b}}{\cm'}\leq \frac{i\cm}{\mu^{200b}}\leq \frac{|E(G_{F_j})|}{\mu^{150b}},$$
since  $\cro(\psi)\leq \frac{(\cm')^2}{\mu^{240b}}$, $i\leq \ceil{32\mu^{6b}}$ and $|E(G_{F_1})|\geq  \cm'/\mu^{2b}$. Therefore, the total number of dirty crossings in drawing $\phi_{i,j}$ with respect to the core structure $\jset_{F_j}$  is bounded by $\frac{|E(G_{F_1})|}{\mu^{120b}}$, as required. We conclude that there exists 
there exists a solution $\phi_{i,j}$ to instance $I_{G_{F_j}}$ that is $\jset_{F_j}$-valid, with $\cro(\phi_{i,j})\leq \frac{|E(G_{F_j})|^2}{\mu^{60b'}}$ and $|\chi^{\dirty}(\phi_{i,j})|\leq \frac{|E(G_{F_j})|}{\mu^{60b'}}$, and so the input to Procedure \procsplit in iteration $j$ of stage $i$ is valid.
\end{proof}

The following claim is central in the analysis of the algorithm.

\begin{claim}\label{claim: stage execution}
Assume that there is a good solution $\psi$ to instance $\cI'$, that Invariants \ref{invariant: fewer edges2}--\ref{invariant: drawing} hold at the beginning of stage $i$, and that neither of the bad events $\event_1^i,\ldots,\event_q^i$ happened over the course of the $i$th stage. Then Event $\tilde \event'_{i+1}$ does not happen either, and so Invariants  \ref{invariant: fewer edges2}--\ref{invariant: drawing}  hold at the end of stage $i+1$. 
\end{claim}

\begin{proof}
	Throughout the proof, we assume  that there is a good solution $\psi$ to instance $\cI'$, that Invariants \ref{invariant: fewer edges2}--\ref{invariant: drawing} hold at the beginning of stage $i$, and that neither of the bad events $\event_1^i,\ldots,\event_q^i$  happens.

	We start by establishing Invariant \ref{invariant: fewer edges2}. Consider some face $F\in \tilde \fset(\kset_{i+1})$. If $F\in \tilde \fset(\kset_i)$, then, since $F$ was not added to the set $\set{F_1,\ldots,F_q}$ of faces to be processed in stage $i$, the corresponding instance $I_F=(G_F,\Sigma_F)$ with $G_F\in \gset_i$ is acceptable. Since graph $G_F$ remains unchagnged in $\gset_{i+1}$, instance $I_F$ remains an acceptable instance. Assume now that $F\not\in \tilde \fset(\kset_i)$. Then there must be an index $1\leq j\leq q$, for which $F=F^j_1$ or $F=F^j_2$. In other words, face $F$ was created in iteration $j$ of Stage $i$.  Since Event $\event_j^i$ did not happen, the output produced by Procedure \procsplit in iteration $j$ is a valid output. From Property \ref{prop: smaller graphs},  $|E(G_{F^j_1})|, |E(G_{F^j_2})|\leq |E(G_{F})|-\frac{|E(G_{F_j})|}{32\mu^{b'}}\leq |E(G_{F_j})|-\frac{\cm'}{32\mu^{4b}}$, since $b'=2b$, and since $|E(G_{F_j})|\geq \cm/\mu^{2b}$ must hold, as instance $I_{F_j}$ is not acceptable. Since, from Invariant \ref{invariant: fewer edges2}, $|E(G_{F_j})|\leq  \cm'\cdot \left (1-\frac{i-1}{32\mu^{6b}}\right )$, we get that $|E(G_{F^j_1})|, |E(G_{F^j_2})|\leq  \cm'\cdot \left (1-\frac{i}{32\mu^{6b}}\right )$. Therefore, invariant \ref{invariant: fewer edges2} continues to hold.

Next, we establish Invariant \ref{invariant: few deleted edges}. Fix an index $1\leq j\leq q$. Using the arguments from the proof of \Cref{obs: valid input to procsplit}, 
there is a soluton $\phi_{i,j}$ to instance $I_{F_j}$ defined by the graph $G_{F_j}\in \gset_i$, that is $\jset_{F_j}$-valid, with 
$\cro(\phi_{i,j})\leq \cro(\psi)\leq  \frac{|E(G_{F_j})|^2}{\mu^{60b'}}$, and $|\chi^{\dirty}(\phi_{i,j})|\leq |\chi^{*}(\psi)|+N(\phi_i)\leq  \frac{|E(G_{F_j})|}{\mu^{60b'}}$.
From Property \ref{prop output deleted edges}:

\[\begin{split}
|E^{\del}(I_{F_j})|&\leq \frac{2\cro(\phi_{i,j})\cdot \mu^{38b'}}{|E(G_{F_j})|}+|\chi^{\dirty}(\phi_{i,j})|\\
&\leq \frac{2\cro(\psi)\cdot \mu^{78b}}{\cm'}+|\chi^{*}(\psi)|+N(\phi_i)\\
&\leq \frac{2\cro(\psi)\cdot \mu^{78b}}{\cm'}+|\chi^{*}(\psi)|  +\frac{\cro(\psi)\cdot i\cdot  \mu^{40b}}{\cm'}\\
&\leq \frac{2\cro(\psi)\cdot \mu^{78b}}{\cm'}+|\chi^{*}(\psi)|  +\frac{32\cro(\psi)\cdot \mu^{46b}}{\cm'}\\
&\leq \frac{\cro(\psi)\cdot \mu^{80b}}{\cm'}+|\chi^{*}(\psi)|.
\end{split}\]

(we have used the fact that $|E(G_{F_j})|\geq \cm/\mu^{2b}$, $b'=2b$, and $i\leq \ceil{32\mu^{6b}}$). Since $q\leq 2\mu^{2b}$, we get that $|E^{\del}_{i+1}|\leq |E^{\del}_i|+\frac{\cro(\psi)\cdot \mu^{100b}}{\cm'}+|\chi^{*}(\psi)|\cdot \mu^{3b}$, establishing invariant  \ref{invariant: few deleted edges}.

It now remains to establish Invariant \ref{invariant: drawing}. 
For convenience, we denote by $G^0=G_i$, and, for $1\leq j\leq q$, we denote by $G^j$ the graph obtained after the $j$th iteration of the $i$th stage, that is, $G^j=G^{j-1}\setminus E^{\del}(I_{F_j})$. We denote by $I^j$ the subinstance of $\cI'$ defined by graph $G^j$. We also denote by $K^0=K_i$ the initial skeleton at the beginning of phase $i$, and, for $1\leq j\leq q$, we denote by  $K^j$ the skeleton obtained at the end of iteration $j$, so $K^j=K^{j-1}\cup P_j$. We  denote by $\kset^j$ the skeleton structure associated with skeleton $K^j$, that can be obtained from $\kset^{j-1}$ and the enhancement structure $\aset_j$ as described above. Lastly, we will define, for all $0\leq j\leq q$,  a solution $\phi^j$ to instance $I^j$ that is $\kset^j$-valid. We will ensure that for all $0\leq j<j'\leq q$, the drawing of graph $G_{F_{j'}}$ in $\phi^j$ is identical to that in $\phi^0$.
Additionally, we will ensure that $\cro(\phi^j)\leq \cro(\psi)$ and $|\chi^*(\phi^j)|\leq |\chi^*(\psi)|$.

 Initially, we let $\phi^0=\phi_i$ be the solution for instance $I^0$ that is guaranteed to exist by Invariant \ref{invariant: drawing}. Recall that this solution is 
 $\kset^0$-valid; $\cro(\phi^0)\leq \cro(\psi)$;
 $|\chi^*(\phi^0)|\leq |\chi^{*}(\psi)|$; and the total number of crossings of $\phi^0$ in which the edges of $E(K^0)\setminus E(\cJ)$ participate is at most  $\frac{\cro(\psi)\cdot i\cdot  \mu^{40b}}{\cm'}$. For $0\leq j\leq q$, we denote by $N^j$ the total number of crossings in which the edges of $E(K^j)\setminus E(\cJ)$ participate in $\phi^j$.
 
 Consider now some index $1\leq j\leq q$, and assume that we are given a solution $\phi^{j-1}$ to instance $I^{j-1}$ that is $\kset^{j-1}$-valid. Recall that the drawing of graph $G_{F_j}$ induced by $\phi^{j-1}$ is identical to that in $\phi^0$. Therefore, the drawing of $G_{F_j}$ induced by $\phi^{j-1}$ is precisely $\phi_{i,j}$. From Property \ref{prop output drawing} of the valid output to procedure \procsplit, there is a $\jset_{F_j}$-valid solution $\phi'_j$ to the subinstance of $\cI'$ that is defined by graph $G'_{F_j}=G_{F_j}\setminus E^{\del}(I_{F_j})$, that is compatible with $\phi_{i,j}$, in which the edges of  $E(J_{F_j})\cup E(P_j)$ do not cross each other, and  the number of crosings in which the edges of $P_j$ participate is at most $\frac{\cro(\phi^{j-1})\cdot \mu^{12b'}}{|E(G_{F_j})|}\leq \frac{\cro(\psi)\cdot \mu^{26b}}{\cm'}$, since $\cro(\phi^{j-1})\leq \cro(\psi)$, $b'=2b$, and $|E(G_{F_j})|\geq \cm'/\mu^{2b}$.
Note that, from the definition of compatible drawings (see \Cref{def: compatible drawing}), the image of the core $J_{F_j}$ in $\phi'_j$ is identical to that in $\phi_{i,j}$. The only difference between drawing $\phi'_j$ and $\phi_{i,j}$ is that the images of the edges of $E^{\del}(I_{F_j})$ were deleted, and some additional local changes were made within the face $F_j$. We are guaranteed that,  if a point $p$ is an inner point on the image of some edge $\phi_j'$, then is an inner point on the image of some  edge  in $\phi_{i,j}$. 
Moreover, if $p$ is an image of some vertex $v$ in $\phi'_j$, then either (i) $p$ is the image of $v$ in $\phi_{i,j}$; or (ii) the degree of $p$ in $G_{F_j}\setminus E^{\del}(I_{F_j})$ is $2$, and $p$ is an inner point on the image of some edge in $\phi_{i,j}$.
In order to obtain drawing $\phi^{j}$ of graph $G^j$, we first delete, from drawing $\phi^{j-1}$, the images of all edges in $E^{\del}(I_{F_j})$. Next, we delete the image of the graph $G_{F_j}$ from the current drawing, and copy instead the image of the graph $G_{F_j}\setminus E^{\del}(I_{F_j})$ from drawing $\phi'_j$. Note that the two images of graph $G_{F_j}\setminus E^{\del}(I_{F_j})$ are identical except for some small local changes inside face $F_j$. While it is possible that edges and vertices of $G^j\setminus G_{F_j}$ are drawn inside face $F_j$ in $\phi^{j-1}$, it is easy to see that no new crossings between edges of $G_{F_j}$ and edges of $G^j\setminus G_{F_j}$ are introduced. Since $\cro(\phi'_j)\leq \cro(\phi_{i,j})$, we get that $\cro(\phi^j)\leq\cro(\phi^{j-1})\leq\cro(\psi)$. Since the only changes to the drawing outside face $F_j$ is the deletion of the segments of the images of some  edges, $|\chi^*(\phi^j)|\leq |\chi^*(\phi^{j-1})|\leq |\chi^*(\psi)|$. Since 
we are guaranteed from Property \ref{prop output drawing} that the number of crosings in which the edges of $P_j$ participate in $\phi'_j$ is at most $\frac{\cro(\psi)\cdot \mu^{26b}}{\cm'}$, we get that $N^j\leq N^{j-1}+\frac{\cro(\psi)\cdot \mu^{26b}}{\cm'}$.

We define the solution $\phi_{i+1}$ to instance $(G_{i+1},\Sigma_{i+1})$ to be $\phi^q$. From the above discussion, drawing $\phi^q$ is $\kset_{i+1}$-valid, $\cro(\phi_{i+1})\leq \cro(\psi)$, and 
 $|\chi^*(\phi_i)|\leq |\chi^{*}(\psi)|$. Since the number of iterations $q\leq  2\mu^{2b}$, and in every iteration we introduce at most $\frac{\cro(\psi)\cdot \mu^{26b}}{\cm'}$ new crossings with the edges of the new skeleton $K_{i+1}$, we get that  $N(\phi_{i+1})\leq N(\phi_i)+2\mu^{2b}\cdot \frac{\cro(\psi)\cdot \mu^{26b}}{\cm'}\leq \frac{\cro(\psi)\cdot i\cdot  \mu^{40b}}{\cm'}+ \frac{2\cro(\psi)\cdot \mu^{28b}}{\cm'}\leq \frac{\cro(\psi)\cdot (i+1)\cdot  \mu^{40b}}{\cm'}$.
\end{proof}

From \Cref{obs: valid input to procsplit} and \Cref{thm: procsplit}, for all $1\leq j\leq q$, $\prob{\event^i_j\mid \neg\tilde \event'_1\band\cdots\band\neg\event'_i}\leq \frac{2^{20}}{\mu^{10b'}}\leq \frac{2^{20}}{\mu^{20b}}$. Since $q\leq 2\mu^{2b}$, we get that $\prob{\event'_{i+1}\mid \neg\tilde \event'_1\band\cdots\band\neg\event'_i}\leq \frac{2^{21}}{\mu^{18b}}$.
Let $\tilde\event'$ be the bad event that either of the events $\tilde \event'_1,\ldots,\tilde \event'_z$ happened. Since  $z=\ceil{32\mu^{6b}}$, we get that $\prob{\tilde \event'_z}\geq \frac{z}{\mu^{20b}}\leq \frac{1}{\mu^{10b}}$.

We return the skeleton structure $\kset_z$, and the $\kset_z$-decomposition $\gset_z$ of $\cG'$. From Invariant \ref{invariant: fewer edges2}, for every face $F\in \tilde \fset(\kset_i)$, the corresponding instance $I_F=(G_F,\Sigma_F)$ with $G_F\in \gset_i$ must be acceptable (this is since the graph associated with an unacceptable instance must have at least $\cm'/\mu^{2b}$ edges). Assume now that there is a good solution $\psi$ to instance $\cI'$, and that bad event $\tilde \event'$ did not happen. Then we are guaranteed that the algorithm does not return FAIL, and, from Invariant \ref{invariant: few deleted edges},  $|E^{\del}_z|\leq z\cdot \left( \frac{\cro(\psi)\mu^{100b}}{\cm'}+|\chi^{*}(\psi)|\cdot \mu^{3b}\right) \leq \frac{\cro(\psi)\mu^{108b}}{\cm'}+|\chi^{\dirty}(\psi)|\cdot \mu^{10b}$, since $z=\ceil{32\mu^{6b}}$. Lastly, Invariant \ref{invariant: drawing}  guarantees that there is a solution $\phi$ to instance $(G_z,\Sigma_z)$, where $G_z=\cG'\setminus E^{\del}_z$, and $\Sigma_z$ is the rotation system for $G_z$ induced by $\cSigma'$, with the following properties. First, drawing $\phi$ is $\kset_z$-valid. Additionally,  $\cro(\phi)\leq \cro(\psi)$,  $|\chi^*(\phi)|\leq |\chi^{*}(\psi)|$, and the total number of crossings in which the edges of $E(K_z)\setminus E(\cJ)$ participate is at most $\frac{\cro(\psi)\cdot z\cdot 
	 \mu^{40b}}{\cm'}\leq\frac{\cro(\psi)\cdot  \mu^{47b}}{\cm'} $.
Since $\prob{\tilde \event'}\leq 1/\mu^{10b}$, this completes the proof of \Cref{lem: compute phase 2 decomposition}.

 \subsubsection{Proof of \Cref{lem: computed decomposition is good enough}}
 \label{subsec: computed decomposition is good enough}

We assume that we are given  
a subinstance $\cI'=(\cG',\cSigma')$ of $\cI$ with $|E(\cG')|=\cm'$, a core structure $\cjset$ for $\cI'$, whose corresponding core graph is denoted by $\cJ$, a skeleton structure $\kset=(K,\set{b_u}_{u\in V(K)},\rho_K)$, and a $\kset$-decomposition $\gset$ of $\cG'$, such that, for every face $F\in \tilde \fset(\kset)$, the corresponding subinstance $I_F=(G_F,\Sigma_F)$ of $\cI'$ defined by the graph $G_F\in \gset$ is acceptable.  Let $E^{\del}(\cI')=E(\cG')\setminus \left(\bigcup_{F\in \tilde \fset(\kset)}E(G_F)\right )$, $G=\cG'\setminus E^{\del}(\cI')$, and let $I=(G,\Sigma)$ be the subinstance of $\cI'$ defined by graph $G$. We also assume that there is a $\kset$-valid solution $\phi$ to instance $I$, so that the total number of crossings in which the edges of $K$ participate is at most $N\leq \frac{\cm'}{\mu^{3b}}$. Note that $\cjset$ remains a valid core structure, and $\kset$ remains a valid skeleton structure for instance $I$. From this point onward we will only work with instance $I$, and we will not need the initial subinstance $\cI'$ of instance $\cI$, or the set $E^{\del}(\cI')$ of edges, but we will use the parameter $\cm'=|E(\cG')|$. 
Our goal is to prove that 
there is a solution $\psi$ to instance $I$ that is clean with respect to $\cjset$, with $\cro(\psi)\leq \left (\cro(\phi)+N^2+|\chi^{*}(\phi)|^2+\frac{|\chi^{*}(\phi)|\cdot \cm'}{\mu^b}\right )\cdot (\log \cm')^{O(1)}$.

Since the statement of the lemma is existential, it is sufficient to show an algorithm that transforms the solution $\phi$ to instance $I$ into another solution $\psi$ that is clean with respect to $\cjset$, with the number of crossings bounded as above.
 In order to do so, we need to ``repair'' the drawing, so that for every edge $e\in E(G)$, the image of $e$ is disjoint from the interior of the forbidden faces in $\tfset^{\forbidden}(\kset)$. We do so in two steps. In the first step, we modify the drawing $\phi$ so that, for every non-forbidden face $F\in\tfset(\kset)\setminus \tfset^{\forbidden}(\kset)$, the images of all vertices of graph $G_F\in \gset$ associated with face $F$ lie in the region $F$ of the drawing. We say that an edge $e\in E(G)\setminus E(K)$ is \emph{bad} if the image of $e$ in the resulting drawing intersects the interior of at least one forbidden face in $\tfset^{\forbidden}(\kset)$. Notice that for each such bad edge $e$, there must be some face $F\in \tfset(\kset)\setminus \tfset^{\forbidden}(\kset)$ with $e\in G_F$, so the images of the endpoints of $e$ lie in region $F$ of the current drawing. In the second step, we further modify the drawing to obtain a $\kset$-clean solution to instance $I$. In order to do so, for every bad edge $e$, if $e\in E(G_F)$ for some face $F\in \tfset(\kset)\setminus \tfset^{\forbidden}(\kset)$, we ``move'' the image of $e$ so it is drawn completely inside face $F$. In order to ensure that the new image of $e$ crosses few edges, we may need to rearrange the current drawing inside the face $F$. This step exploits the fact that instance $I_F$ corresponding to face $F$ is acceptable.

 We now proceed to describe each of the steps in turn.
 For convenience, we denote $\tilde \fset'=\tilde \fset(\kset)\setminus \tilde \fset^{\forbidden}(\kset)$.

 \subsubsection*{Step 1: Moving the Vertices}
 The goal of the first step is to prove the following claim. 
 
 \begin{claim}\label{claim: semi-clean drawing}
 	There is a solution $\psi_1$ to instance $I$ that is $\kset$-valid, 
 	such that, for every face $F\in \tilde \fset'$, the images of all vertices of $G_F$ lie in region $F$ of the drawing. Additionally,  $\cro(\psi_1)\leq (\cro(\phi)+N^2)\cdot (\log \cm')^{O(1)}$, $|\chi^{*}(\psi_1)|\leq |\chi^{*}(\phi)| \cdot (\log \cm')^{O(1)}$, and    the total number of crossings in which the edges of $K$  participate in $\psi_1$ is at most $N\cdot (\log \cm')^{O(1)}$. 
 \end{claim}

\begin{proof}
	Consider some face $F\in \tilde \fset'$.
	Recall that we have defined a core structure $\jset_F$ associated with face $F$; we denote by $J_F$ the corresponding core graph, so $J_F\subseteq K$. We also denote by $I_F=(G_F,\Sigma_F)$ the instance associated with face $F$, where $G_F\in \gset$.
	
	Let $E'_F\subseteq E(G_F)$ be the set of edges $e=(x,y)\in E(G_F)$, such that either (i) the image of one of the vertices $x,y$ lies in region $F$  in $\phi$, and the image of the other vertex lies outside of $F$; or (ii) the images of both vertices lie outside region  $F$ in $\phi$, and the image of $e$ crosses the boundary of $F$.  Since $\jset_F$ is a valid core structure for instance $I_F$ (see \Cref{def: valid core 2}), for every edge $e\in E(G_F)$ that is incident to a vertex  $x\in V(J_F)$, a segment of $\phi(e)$ that contains $\phi(x)$ must be contained in region $F$ in $\phi$. Therefore, for every edge $e\in E'_F$, its image $\phi(e)$ intersects the interior of $F$, and it must cross the image of some edge of $J_F$. 
	Since the total number of crossings in which the edges of $K$ participate in $\phi$ is bounded by $N$, we get the following immediate observation:

\begin{observation}\label{obs: few boundary edges}
	$\sum_{F\in \tilde \fset'}|E'_F|\leq N$.
\end{observation}	
	
	Consider again some face $F\in \tilde \fset'$.
	 It will be convenient for us to subdivide every edge of $E'_F$ with one or two vertices, and to adjust the drawing $\phi$ of $G$ to include these new vertices, as follows. Consider an edge $e=(x,y)\in E'_F$. Assume first that for both $x$ and $y$, their images in $\phi$ lie outside the region $F$. In this case, we replace edge $e$ in both $G$ and $G_F$ with a path $(x,t_e,t'_e,y)$. Consider now the image $\phi(e)$ of edge $e$ in drawing $\phi$, and direct it from $\phi(x)$ to $\phi(y)$. Let $p$ be the first point on $\phi(e)$ that belongs to the boundary of face $F$, and let $p'$ be the last point on $\phi(e)$ lying on the boundary of $F$. We place the image of the new vertex $t_e$ on curve $\phi(e)$ immediately next to point $p$, in the interior of face $F$. Similarly, we place the image of the new vertex $t'_e$ on curve $\phi(e)$ immediately next to point $p'$, in the interior of face $F$. Assume now that the image of one of the vertices $x,y$ lies in $F$ (for example, vertex $y$), and the image of the other vertex (vertex $x$) lies outside of $F$. We direct $\phi(e)$ from $\phi(x)$ to $\phi(y)$, and we let $p$ be the first point on $\phi(e)$ that lies on the boundary of $F$. We replace edge $e$ with a path $(x,t_e,y)$ in graph $G$ and in graph $G_F$, and we place the image of vertex $t_e$ on $\phi(e)$, next to point $p$, in the interior of face $F$. For convenience, the graph that is obtained from $G$ after these modifications is still denoted by $G$, and for every face 
	$F\in \tilde \fset'$, the resulting graph associated with the face is still denoted by $G_F$. Since the newly added vertices all have degree $2$ in $G$, it is easy to extend the rotation system $\Sigma$ for graph $G$ to include these vertices, and we can similarly extend the rotation system $\Sigma_F$ for each graph $G_F\in \gset$.

	For a face $F\in \tilde \fset'$, let $H_F\subseteq G_F$ be the graph whose vertex set contains every vertex $x\in V(G_F)$ with $\phi(x)\not \in F$, and every edge $e\in E(G_F)$ whose image in $\phi$ is disjoint from $F$. Let $E''_F\subseteq E(G_F)$ be the set of all edges $e\in E(G_F)$ with one endpoint in $H_F$ and another in $G_F\setminus H_F$. Observe that each such edge $e\in E''_F$ was obtained by subdividing some edge of $E'_F$, so $|E''_F|\leq 2|E'_F|$. For every edge $e=(x,y)\in E''_F$ with $x\in V(H_F)$, the intersection of $\phi(e)$ with the region $F$ is a very short segment of $\phi(e)$ that is incident to $\phi(x)$ (recall that $\phi(x)$ lies inside $F$ very close to its boundary). We denote by $Z_F$ the set of all endpoints of edges $e\in E''_F$ whose image lies in region $F$. We also note that graph $H_F$ is a subgraph of $G_F$ induced by $V(H_F)$. Indeed, if $e=(x,y)$ is an edge of $G_F$ with $x,y\in V(H_F)$, then the image of edge $e$ must be entirely disjoint from region $F$ (otherwise it would have been subdivided).

We need a slight modification of the algorithm for computing a well-linked decomposition from \Cref{thm:well_linked_decomposition}, that is summarized in the following theorem. 
The proof is deferred to Section \ref{subsec: proof of wld cor} of Appendix.

\begin{theorem}\label{thm: wld all paths congestion}
	There is an efficient algorithm, whose input is a graph $G$, a  vertex-induced subgraph $S$ of $G$, and parameters $m$ and $\alpha$, for which $|E(G)|\leq m$ and $0<\alpha< \frac 1 {c\log^2 m}$ hold, for a large enough constant $c$. 
	The algorithm computes a collection $\rset$ of vertex-disjoint clusters of $S$, and, for every cluster $R\in \rset$, two sets $\pset_1(R)$, $\pset_2(R)$ of paths, such that the following hold:
	
	\begin{itemize}
		\item $\bigcup_{R\in \rset}V(R)=V(S)$;
		\item for every cluster $R\in\rset$, $|\delta_G(R)|\le |\delta_G(S)|$;
		\item every cluster $R\in\rset$ has the $\alpha$-bandwidth property in graph $G$;  
		\item $\sum_{R\in \rset}|\delta_G(R)|\le 4|\delta_G(S)|$;
		\item for every cluster $R\in \rset$, $\pset_1(R)=\set{P_1(e)\mid e\in \delta_G(R)}$, where for every edge $e\in \delta_G(R)$, path $P_1(e)$ has $e$ as its first edge and some edge of $\delta_G(S)$ as its last edge, and all inner vertices of $P_1(e)$ lie in $V(S)\setminus V(R)$. Additionally, $\cong_G(\pset_1(R))\leq 400/\alpha$; and
		\item for every cluster $R\in \rset$, there is a subset $\hat E_R\subseteq \delta_G(R)$ of at least $\floor{|\delta_G(R)|/64}$ edges, such that $\pset_2(R)=\set{P_2(e)\mid e\in \hat E_R}$, where for every edge $e\in \hat E_R$, path $P_2(e)$ has $e$ as its first edge and some edge of $\delta_G(S)$ as its last edge, and all inner vertices of $P_2(e)$ lie in $V(S)\setminus V(R)$. Moreover, $\cong_G\left(\bigcup_{R\in \rset}\pset_2(R)\right )\leq O\left (\frac{\log m}{\alpha}\right )$.
	\end{itemize}
\end{theorem}

We apply the algorithm from \Cref{thm: wld all paths congestion} to graph $G_F$, subgraph $S=H_F$ of $G_F$, parameter $\cm'$ and $\alpha=\frac{1}{\log^4\cm'}$, to compute a collection $\rset_F$ of vertex-disjoint clusters of $H_F$, such that $\bigcup_{R\in \rset_F}V(R)=V(H_F)$, $\sum_{R\in \rset_F}|\delta_{G_F}(R)|\le 4|E''_F|\leq 8|E'_F|$, and every cluster $R\in\rset_F$ has the $\alpha$-bandwidth property in graph $G_F$.
Additionally, the algorithm computes, 
for every cluster $R\in \rset_F$, a set $\pset_1(R)=\set{P_1(e)\mid e\in \delta_{G_F}(R)}$ of paths in graph $G_F$, with $\cong_{G_F}(\pset_1(R))\leq 400/\alpha\leq O(\log^4\cm')$, such that, for every edge $e\in \delta_{G_F}(R)$, path $P_1(e)$ has $e$ as its first edge and some edge of $E''_F$ as its last edge, and all inner vertices of $P_1(e)$ lie in $V(H_F)\setminus V(R)$. 
It also computes, for every cluster $R\in \rset_F$, 
 a subset $\hat E_R\subseteq \delta_G(R)$ of at least $\floor{|\delta_G(R)|/64}$ edges, and another set $\pset_2(R)=\set{P_2(e)\mid e\in \hat E_R}$ of paths, where for every edge $e\in \hat E_R$, path $P_2(e)$ has $e$ as its first edge and some edge of $E''_F$ as its last edge, such that all inner vertices of $P_2(e)$ lie in $V(H_F)\setminus V(R)$, and the total congestion caused by the paths in $\bigcup_{R\in \rset}\pset_2(R)$ is at most $ O\left (\frac{\log \cm'}{\alpha}\right )\leq O(\log^5\cm')$.
We let $E'''_F=\bigcup_{R\in \rset_F}\delta_{G_F}(R)$, so $E''_F\subseteq E'''_F$. 

It will be convenient for us to further slightly modify graph $G$, by subdividing some of its edges, as follows. Consider a face $F\in \tilde \fset'$ and an edge $e=(x,y)\in E'''_F$. If there are two distinct clusters $R,R'\in \rset_F$ with $x\in R$ and $y\in R'$, then we replace edge $e$ with a path $(x,t_{e}^R,t_{e}^{R'},y)$ in $G$, and we denote edge $\tilde e=(t_e^R,t_{e}^{R'})$. We also modify the current drawing $\phi$ by placing the images of the newly added vertices $t_e^R,t_e^{R'}$ on $\phi(e)$. Otherwise, there must be a cluster $R\in \rset_F$, such that one endpoint of $e$ (say $x$) lies in $R$, and the other endpoint (vertex $y$) lies in $Z_F$. In this case, we replace edge $e$ with a path $(x,t_e^R,y)$ in graph $G$, and we denote edge $\tilde e=(t_e^R,y)$.   We place the image of vertex $t_e^R$ on $\phi(e)$, outside the region $F$.

 Let $G'$ denote the final graph that is obtained from $G$ once every face $F\in \tilde \fset'$ and every edge $e\in E'''_F$ is processed. For each face $F\in \tilde \fset'$  we define a subgraph $G'_F\subseteq G'$ similarly, by subdividing the edges of $E'''_F$ as before, and we denote $\tilde E_F=\set{\tilde e\mid e\in E_F'''}$.  We similarly update graph $H_F$, to obtain a new graph $H'_F\subseteq G'_F$, as follows. First, for every edge $e\in E(H_F)$, whose endpoints lie in different clusters of $\rset_F$, we subdivide edge $e$ with two vertices as before. Additionally, for every edge $e=(x,y)\in E(G_F)$ with $x\in V(H_F)$ and $y\in Z_F$, we add the new edge $(x,t_e^R)$ that was obtained by subdividing $e$ to graph $H'_F$ (here, $R\in \rset_F$ is the cluster containing $x$).
 
For every cluster $R\in \rset_F$, the set $\pset_1(R)$ of paths naturally defines a set $\pset'_1(R)$ of paths in graph $G'_F$, where $\pset'_1(R)=\set{P'_1(e)\mid e\in \delta_{G'_F}(R)}$, with $\cong_{G'_F}(\pset_1'(R))\leq O(\log^4\cm')$, such that, for every edge $e\in \delta_{G'_F}(R)$, path $P_1'(e)$ has $e$ as its first edge, and it terminates at some vertex of $Z_F$. Furthermore, all inner vertices of $P_1'(e)$ lie in $V(H'_F)\setminus V(R)$.
Similarly, set $\pset_2(R)$ of paths naturally defines a set $\pset'_2(R)$ of paths in graph $G'_F$, where each path in $\pset'_2(R)$ starts at a distinct edge of $\delta_{G'_F}(R)$, terminates at some vertex of $Z_F$, and has all its inner vertices contained in $V(H'_F)\setminus V(R)$. As before, $|\pset'_2(R)|\geq \floor{ |\delta_{G'_F}(R)|/64}$, and all paths in $\bigcup_{R\in \rset_F}\pset'_2(R)$ cause congestion at most $O(\log^5\cm')$.

Notice that drawing $\phi$ of $G$ naturally defines a drawing $\phi'$ of graph $G'$. We denote $\tilde E=\bigcup_{F\in \tilde \fset'}\tilde E_F$. Observe that we have never subdivided the edges of the skeleton $K$, so $K\subseteq G'$ still holds. We can naturally extend the rotation system $\Sigma$ to graph $G'$, to obtain a rotation system $\Sigma'$, and we denote by $I'=(G',\Sigma')$ the resulting instance of \cnwrs.

Let $\phi''$ be any drawing of graph $G'$, and let $(e,e')_p$ be a crossing of $\phi''$. We say that crossing $(e,e')_p$ is \emph{uninteresting} if both $e,e'\in \tilde E$, and we say that this crossing is \emph{interesting} otherwise. We prove the following weaker analogue of \Cref{claim: semi-clean drawing}.

\begin{claim}\label{claim: semi-clean drawing2}
	There is a solution $\psi_2$ to instance $I'=(G',\Sigma')$ that is $\kset$-valid, 
	such that, for every face $F\in \tilde \fset'$, the images of all vertices of $G'_F$ lie in region $F$ of the drawing. Additionally, the number of interesting crossings in $\psi_2$ is bounded by $\cro(\phi)\cdot (\log \cm')^{O(1)}$; $|\chi^{*}(\psi_2))|\leq |\chi^{*}(\phi)| \cdot (\log \cm')^{O(1)}$;  and  the total number of crossings in which the edges of $K$  participate in $\psi_2$ is at most $N\cdot (\log \cm')^{O(1)}$. 
\end{claim}

The proof of \Cref{claim: semi-clean drawing} easily follows from \Cref{claim: semi-clean drawing2}. Indeed, consider the solution $\psi_2$ to instance $I'$. We apply type-1 uncrossing operation to the images of the edges in $\tilde E$ (see \Cref{subsec: uncrossing type 1} and \Cref{thm: type-1 uncrossing} for a formal description). The operation repeatedly selects pairs $e,e'\in \tilde E$ of edges that cross more than once, and then eliminates at least one of the crossings between these edges by a local uncrossing operation that ``swaps'' segments of images of these two edges without affecting the rest of the drawing. Therefore, if $\psi_3$ is the drawing of graph $G'$ obtained at the end of this procedure, then $\psi_3$ is a valid solution to instance $I'$ that remains $\kset$-valid. Since the edges of the core $\cJ$ may not belong to $\tilde E$,  $|\chi^{*}(\psi_3))|\leq |\chi^{*}(\psi_2))|\leq |\chi^{*}(\phi))| \cdot (\log \cm')^{O(1)}$. The number of interesting crossings in $\psi_3$ is bounded by the number of interesting crossings in $\psi_2$, which, in turn, is bounded by  $\cro(\phi)\cdot (\log \cm')^{O(1)}$. Since the edges of the skeleton $K$ may not lie in $\tilde E$,    the total number of crossings in which the edges of $K$  participate in $\psi_3$ remains  at most $N\cdot (\log \cm')^{O(1)}$. As before, for every face $F\in \tilde \fset'$, for every vertex $x\in V(G'_F)$, $\psi_3(x)\in F$. The number of uninteresting crossings in $\psi_3$ is now bounded by $|\tilde E|^2$, as every pair of edges in $\tilde E$ may now cross at most once. 
For every face $F\in \tilde F'$, $|\tilde E_F|=|E'''_F|=\sum_{R\in \rset_F}|\delta_{G_F}(R)|\leq 8|E'_F|$. From \Cref{obs: few boundary edges},
$\sum_{F\in \tilde \fset'}|E'_F|\leq N$. Therefore, $|\tilde E|\leq \sum_{F\in \tilde \fset'}|\tilde E_F|\leq 8N$. The number of uninteresting crossings in $\psi_3$ is then bounded by $|\tilde E|^2\leq 64N^2$. We conclude that the total number of crossings in $\psi_3$ is bounded by 
$(\cro(\phi)+N^2)\cdot (\log \cm')^{O(1)}$. Lastly, we can modify solution $\psi_3$ to instance $I'$ to obtain a solution $\psi_1$ to instance $I$ by suppressing the degree-$2$ vertices that we used to subdivide some of the edges of graph $G$. It is immediate to verify that this drawing has all required properties. In order to complete the proof of  \Cref{claim: semi-clean drawing}, it is now enough to prove \Cref{claim: semi-clean drawing2}, which we do next.

Consider a face $F\in \tilde \fset'$. We partition the edges of $\tilde E_F$ into two subsets: set $\tilde E'_F$ containing all edges $(x,y)$ with $x\in V(H'_F)$ and $y\in Z_F$, and set $\tilde E''_F$ containing all remaining edges. 
For a cluster  $R\in \rset_F$, we denote $R^+=R\cup \delta_{G'_F}(R)$ the augmentation of cluster $R$ with respect to graph $G'_F$. We also denote by $T_R=\set{t_e^R\mid e\in \delta_{G_F}(R)}$ the set of vertices that serve as endpoints of the edges of $\tilde E_F$ and lie in $R^+$.
Note that every edge $e\in \tilde E''_F$ connects a vertex of $R^+_1$ to a vertex of $R^+_2$ for some pair $R_1,R_2\in \rset_F$ of distinct clusters.

For each face $F\in \tilde \fset'$ and cluster $R\in \rset_F$, we denote by $\chi(R)$ the set of all crossings in the drawing $\phi'$ of $G'$ in which the edges of $R^+$ participate.

 We first use the drawing $\phi'$ of $G'$ to compute, for each cluster $R\in \rset_F$, a drawing $\psi_{R^+}$ of graph $R^+$ inside a disc $D(R)$, with the images of the vertices of $T_R$ lying on the boundary of the disc. We then select a location inside the region $F$, next to its boundary, into which we plant the disc $D(R)$ together with the drawing $\psi_{R^+}$ that is contained in it. Lastly, we modify the images of the edges of $\tilde E$ so that they connect the new images of their endpoints. All these modifications exploit the sets $\pset_1'(R)$ and $\pset_2'(R)$ of paths that we have defined for every cluster $R\in \rset_F$ and face $F\in \tilde \fset'$.
For a face $F\in \tilde \fset'$ and a cluster $R\in \rset_F$, we can now think of the paths in set $\pset_1'(R)$  as routing the vertices of $T_R$ to vertices of $Z_F$ in graph $G'_F$ (after we discard the first edge from each such path), and similarly we can think of paths in $\pset_2'(R)$ as routing a subset of at least $\floor{|T_R|/64}$ vertices of $T_R$ to vertices of $Z_F$ in graph $G'_F$. Recall that the paths in $\pset'_1(R)\cup \pset'_2(R)$ are internally disjoint from $V(R)$, and we can ensure that they are internally disjoint from $Z_F$. The paths in each set $\pset_1'(R)$ cause congestion at most $O(\log^4\cm')$, and the paths in $\bigcup_{R\in \rset(F)}\pset_2'(R)$ cause congestion at most $O(\log^5\cm')$ in graph $G'_F$. Recall also that our transformation of the graph $G$ and the initial drawing $\phi$ ensures that the image of every vertex $t\in Z_F$ in $\phi'$ appears in the interior of the region $F$, very close to its boundary. If $e$ is the unique edge of $\tilde E'$ incident to $t$, then only a small segment of $\phi'(e)$ that is incident to $\phi'(t)$ is contained in $F$, and that segment does not participate in any crossings.

\paragraph{Computing the Drawings $\psi_{R^+}$.}
Consider a face $F\in \tilde \fset'$ and a cluster $R\in \rset_F$. We view the drawing $\phi'$ of $G'$ as a drawing on the sphere.  Recall that, from the definition of graph $H_F$, for every edge $e\in E(H_F)$, the image of $e$ in $\phi'$ is disjoint from $F$. 
Consider the disc $D(J_F)$, that is associated with the core $J_F$. Recall that disc $D(J_F)$ is a disc that contains the image of the core $J_F$ in its interior, and the boundary of $D(J_F)$ closely follows the boundary of the region $F$ inside the region (see \Cref{fig: core_disc_2}). We can assume w.l.o.g. that the images of all vertices of $Z_F$ lie on the boundary of the disc $D(J_F)$. We let $D(R)$ be the disc that is the complement of disc $D(J_F)$, that is, the boundaries of both discs are identical, but their interiors are disjoint. Note that, from the definition of the graph $H_F$, the images of all vertices and edges of graph $R^+$ lie in disc $D(R)$ in drawing $\phi'$.

Consider a graph $\tilde H_R$, containing all edges and vertices of $R^+$, and all edges and vertices that lie on the paths of $\pset_1'(R)$. Let $\phi'(R)$ be the drawing of $\tilde H_R$ that is induced by the drawing $\phi'$ of $G'$.
We apply the algorithm from \Cref{cor: new type 2 uncrossing} to perform a type-2 uncrossing of the images of the paths in $\pset_1'(R)$. The input to the algorithm is graph $\tilde H_R$, its subgraph $C=R^+$, and a set $\qset=\pset_1'(R)$ of paths, together with the drawing $\phi'(R)$ of $\tilde H_R$. We direct all paths in $\pset'_1(R)$ away from the vertices of $T_R$.  The algorithm computes a collection $\Gamma=\set{\gamma(t)\mid t\in T_R}$ of curves, where, for every vertex $t\in T_R$, curve $\gamma(t)$ originates at the image of $t$ in $\phi'(R)$, and terminates at the image of some vertex of $Z_F$, which lies on the boundary of disc $D(R)$. 
For every pair $e,e'\in E(\tilde H_R)$ of distinct edges, let $N(e,e')$ denote the number of crossings in the drawing $\phi'(R)$ between edges $e$ and $e'$.
The algorithm from \Cref{cor: new type 2 uncrossing} also ensures that the curves in $\Gamma$ do not cross each other, and, for every edge $e\in E(R^+)$, the number of crossings between the image of $e$ in $\phi'(R)$ and the curves in $\Gamma$ is bounded by $\sum_{e'\in E(G_R)\setminus E(R^+)}N(e,e')\cdot \cong_{G_F}(\pset_1'(R),e')\leq (\log \cm')^{O(1)}\cdot \sum_{e'\in E(G_R)\setminus E(R^+)} N(e,e')$.

We are now ready to define the drawing $\psi_{R^+}$ of graph $R^+$. Recall that for every vertex $t\in T_R$, there is exactly one edge in $R^+$ that is incident to $t$, and we denote this edge by $e_t=(x_t,t)$, where $x_t\in V(R)$. The images of all vertices and edges of $R$ in $\psi_{R^+}$ remain the same as in $\phi'(R)$ (which are in turn identical to those in $\phi'$). For every vertex $t\in T_R$, the image of edge $e_t$ is obtained by concatenating the image of edge $e_t$ in $\phi'(R)$ and the curve $\gamma_t\in \Gamma$. The resulting curve connects the image of vertex $x_t$ in the current drawing, to some point $p$ on the boundary of disc $D(R)$. The image of vertex $t$ becomes that point $p$. We note that the image of $e_t$ is contained in disc $D(R)$. This completes the definition of the drawing $\psi_{R^+}$ of graph $R^+$. This drawing obeys the rotation system $\Sigma'$, is contained in disc $D(R)$, with the vertices of $T_R$ lying on the boundary of the disc. From the above discussion, the total number of crossings in this drawing is bounded by $(\log \cm')^{O(1)}\cdot |\chi(R)|$. For convenience, we define another disc $D'(R)$, that contains $D(R)$, so that the boundaries of both discs are disjoint.

\paragraph{Modifying the Drawing $\phi'$ of $G'$}
In this step we select, for every face $F\in \tilde \fset'$ and every cluster $R\in \rset_F$, a small disc  $D''(R)$ in the interior of the region $F$ in $\phi'$. We will then copy the contents of disc $D'(R)$ (including the drawing $\psi_{R^+}$) into the disc $D''(R)$, and extend the images of all edges of $\tilde E_F$ that are incident to the vertices of $T_R$, so that they terminate at the boundary of the disc $D''(R)$. We will then ``stitch'' the images of these edges inside the region $D''(R)\setminus D(R)$, so that the image of each edge terminates at the image of its endpoint.

Consider a face $F\in \tilde \fset'$, and a cluster $R\in \rset_F$. Since cluster $R$ has 
the $\alpha$-bandwidth property, for $\alpha=\frac{1}{\log^4\cm'}$, from \Cref{obs: wl-bw}, the set $T_R$ of vertices is $\alpha$-well-linked in $R^+$. We apply the algorithm from \Cref{lem: simple guiding paths} to compute, for every vertex $t\in T$, a set $\qset_t=\set{Q_t(t')\mid t'\in T_R\setminus\set{t}}$ of paths, where, for all $t'\in T_R\setminus\set{t}$, path $Q_t(t')$ connects $t'$ to $t$. Let $\hat T_R\subseteq T_R$ be the set containing all vertices $t\in T_R$, such that some path of $\pset'_2(R)$ originates from $t$. We then select a vertex $t_R\in \hat T_R$ uniformly at random, and we let $\qset_R=\set{Q_t(t')\mid t'\in T_R\setminus\set{t_R}}$ be a collection of path connecting every vertex in $T_R\setminus\set{t_R}$ to $t_R$.
We need the following observation.

\begin{observation}\label{obs: low congestion outer routing paths}
	For every edge $e\in E(R^+)$, $\expect{\cong(\qset_R,e)}\leq O(\log^8\cm')$.
\end{observation}

\begin{proof}
	Fix an edge $e\in E(R^+)$.
From \Cref{lem: simple guiding paths},  if we were to select a vertex $t\in T_R$ uniformly at random, then $\expect{\cong(\qset_t,e)}\leq  O\left (\frac{\log^4\cm'}{\alpha}\right )\leq O(\log^8\cm')$. Clearly, in the above process, a vertex $t\in T_R$ is selected with probability $1/|T_R|$. Our algorithm instead selects a vertex $t_R\in \hat T_R$ uniformly at random, so a vertex $t\in \hat T_R$ is selected with probability $\frac{1}{|\hat T_R|}\leq \frac{128}{|T_R|}$, since $|\hat T_R|\geq \floor{|T_R|}{64}$. Therefore, $\expect{\cong(\qset_R,e)}\leq 128\expect[t\sim T_R]{\cong(\qset_t,e)}\leq O(\log^8\cm')$.
\end{proof}

We construct another set $\qset'_R=\set{Q'(t')\mid t'\in T_R}$ of paths in graph $G'_F$, as follows. Consider the unique path $P_2'(t_R)\in \pset_2'(R)$ that originates at vertex $t_R$. We denote by $z_R\in Z_F$ the other endpoint of path $P_2'(t_R)$; recall that the image of $z_R$ lies in region $F$, very close to its boundary. For every vertex $t'\in T_R\setminus \set{t_R}$, we let $Q'(t')$ be the 
path obtained by concatenating path $Q(t')\in \qset_R$ with path $P'_2(t_R)$. For vertex $t_R$, we simply set $Q'(t_R)=P'_2(t_R)$. The resulting set $\qset'_R=\set{Q'(t')\mid t'\in T_R}$ of paths is contained in graph $G'_F$, and connects every vertex of $T_R$ to vertex $z_R\in Z_F$. Moreover, the only vertex of $G'_F\setminus H'_F$ that lies on the paths of $\qset'_R$ is vertex $z_R$. We assume w.l.o.g. that the paths in $\qset'_R$ are simple. 
For every face $F\in \tilde \fset'$, we denote $\qset(F)=\bigcup_{R\in \rset_F}\qset'_R$.
We need the following observation.

\begin{observation}\label{obs: bound expected congestion}
For every face $F\in \tilde \fset'$,	for every edge $e\in E(H'_F)\cup \tilde E'_F$, $\expect{\cong_{G'_F}(\qset(F),e)}\leq O(\log^8\cm')$.
\end{observation}
\begin{proof}
	Consider some edge $e\in  E(H'_F)\cup \tilde E'_F$. 
	If there is some cluster $R\in \rset_F$ with $e\in E(R)$, then denote $R_e=R$; otherwise we let $R_e$ be undefined.
	Recall that there are at most $O(\log^5\cm')$ paths in $\bigcup_{R\in \rset_F}\pset_2'(R)$ that contain the edge $e$. Let $S(e)$ be the collection of pairs $(R,t)$, where $R\in \rset_F\setminus\set{R_e}$, $t\in \hat T_R$, and the uniue path of $\pset_2'(R)$ that originates at $t$ contains the edge $e$. From the above discussion, $|S(e)|\leq O(\log^5\cm')$. Consider now a pair $(R,t)\in S(e)$. The probability that vertex $t$ is selected as vertex $t_R$ is bounded by $\frac{1}{|
		\hat T_R|}\leq \frac{128}{|T_R|}$, since $|\hat T_R|\geq \floor{\frac{|T_R|}{64}}$. If vertex $t$ is selected as $t_R$, then every path in set $\qset'_R$ may contain the edge $e$, and the number of such paths is $|T_R|$. Therefore, the expected number of paths in $\bigcup_{R\in \rset_F\setminus\set{R_e}}\qset'_R$ that contain edge $e$ is at most $O(\log^5\cm')$. If cluster $R_e$ is defined, then $\expect{\cong(\qset'_{R_e},e)}\leq \expect{\cong(\qset_{R_e},e)}\leq O(\log^8\cm')$. Therefore, overall, $\expect{\cong_{G'_F}(\qset(F),e)}\leq O(\log^8\cm')$.
\end{proof}

For every face $F\in \tilde \fset'$ and cluster $R\in \rset_F$, we let $D''(R)$ be a very small disc lying in the interior of region $F$ of $\phi'$, right next to the image of vertex $z_R$. Notice that it is possible that for several distinct clusters $R,R'\in \rset_F$, $z_R=z_{R'}$ holds. We ensure that all such discs in $\set{D''(R)}_{R\in \rset_F}$ are mutually disjoint. Eventually, for every component $R\in \rset_F$, we will plant the drawing $\psi_{R^+}$ inside disc $D''(R)$, so that the discs $D'(R)$ and $D''(R)$ will coincide. In order to modify the images of the edges of $\tilde E_F$, we  define, for every cluster $R\in \rset_F$, for every vertex $t\in T_R$, a curve $\gamma(t)$ connecting the image of $t$ in drawing $\phi'$ to the boundary of disc $D''(R)$. Consider now some edge $e\in \tilde E_F$. Assume first that $e\in \tilde E''_F$, and that $e=(t,t')$, with $t\in T_R$ and $t'\in T_{R'}$, for some clusters $R,R'\in \rset_F$. In order to define a new image of edge $e$, we start by concatenating the curve $\gamma(t)$ with the image of edge $e$ in $\phi'$, and curve $\gamma(t')$, thereby obtaining a curve connecting a point on the boundary of disc $D''(R)$ to a point on the boundary of disc $D''(R')$. We then extend the curve within $D''(R)\setminus D(R)$ so that it originates at the new image of vertex $t$, and we similarly extend the curve within $D''(R')\setminus D(R')$ so that it terminates at the new image of vertex $t'$. Notice that all crossings between the images of the edges of $\tilde E$ are uninteresting crossings, and \Cref{claim: semi-clean drawing2} allows us to introduce arbitrary number of such crossings. However, we need to ensure that the number of new crossings between the new images of the edges of $\tilde E$ and the remaining edges of $G'$ is small. In order to do so, we need to ensure that the curves in sets $\set{\gamma(t)\mid t\in T_R}$, for all $F\in \tilde \fset'$ and $R\in \rset_F$ have few crossings with the images of edges of $G'$.

We now proceed to define the curves $\gamma(t)$, wich is the main component in the remainder of the proof. Intuitively, these curves will follow the images of the paths in $\qset(F)$, for all $F\in \tilde \fset'$.
In order to do so, for every face $F\in \tilde \fset'$, for every edge $e\in E(H'_F)\cup \tilde E'_F$, let $N_e$ denote the number of paths in set $\qset(F)$ containing the edge $e$. 

Let $G''$ be a new graph that is obtained from $G'$ as follows. For every  face $F\in \tilde \fset'$, for every edge $e\in E(H'_F)\cup \tilde E'_F$,  we add a collection $J(e)$ of $N_e+1$ parallel copies of edge $e$ to the graph. We then let $\phi''$ be a drawing of graph $G''$ that is obtained from the drawing $\phi'$ of graph $G'$ in a natural way: for every face $F\in \tilde \fset'$ and edge $e\in E(H'_F)\cup \tilde E'_F$, we  add images of $N_e+1$ copies of edge $e$ in parallel to the original drawing of edge $e$, very close to it.

Consider a crossing $(e,e')_p$ in this new drawing $\phi''$. We say that the crossing is of  \emph{type 1}, if there is some face $F\in \tilde \fset'$, such that both $e$ and $e'$ are copies of edges that lie in $E(H'_F)\cup \tilde E'_F$. Otherwise, we say that the crossing is of \emph{type 2}. 
If a crossing $(e_1,e_2)_p$ in $\phi''$ is of type 2, then there is a crossing $(e_1',e_2')_{p'}$ in $\phi'$ in the vicinity of point $p$, such that either $e_1'=e_1$, or $e_1$ is a copy of edge $e_1'$, and similarly, either $e_2'=e_2$, or $e_2$ is a copy of edge $e_2'$. We say that crossing $(e_1',e_2')_{p'}$ in $\phi'$ is \emph{responsible} for crossing $(e_1,e_2)_p$ in $\phi''$. 
Since, 
from \Cref{obs: bound expected congestion}, for every face $F\in \tilde \fset'$ and edge $e\in E(H'_F)\cup \tilde E'_F$, $\expect{\cong_{G'_F}(\qset(F),e)}\leq O(\log^8\cm')$, and since the random choices made when computing sets $\qset(F)$ of paths for different faces $F\in \tilde \fset'$ are independent, the expected number of type-2 crossings in $\phi''$ for which a single crossing in $\phi'$ is responsible is bounded by  $O(\log^{16}\cm')$. Therefore, the expected number of type-2 crossings in $\phi''$ is at most $\cro(\phi')\cdot O(\log^{16}\cm')$.

Consider a face $F\in \tilde \fset'$. We can use the set $\qset(F)$ of paths in graph $G'$ in a natural way, in order to define a set $\tilde \qset(F)=\set{\tilde Q(t)\mid t\in \bigcup_{R\in \rset_F}T_R}$ of edge-disjoint paths in $G''$, where for every cluster $R\in \rset_F$, for every vertex $t\in T_R$, path $\tilde Q(t)$ connects the image of $t$ in $\phi'$ to vertex $z_R$. Since, for every edge $e\in E(H'_F)\cup \tilde E'_F$, we have added $N_e+1$ copies of edge $e$ to $G''$, for every edge $e\in \tilde E_F$, there is a copy $e^*\in J(e)$ of edge $e$ (that we call \emph{distinguished copy}), that does not belong to any path in $\tilde \qset(F)$.
For every component $R\in \rset_F$, for every vertex $t\in T_R$, we let $\gamma(t)$ be the image of path $\tilde Q(t)$ in $\phi''$. Notice that curve $\gamma(t)$ connects the image of $t$ in $\phi'$ to the image of vertex $z_R$. We slightly modify the final segment of $\gamma(t)$ so that it terminates at the boundary of disc $D''(R)$ (while ensuring that each such curve terminates at a different point on the boundary of the disc). We note that we are allowed to introduce arbitrary number of crossings between the curves in set $\set{\gamma(t)\mid t\in \bigcup_{R\in \rset_F}T_R}$, as all such crossings will become unimportant crossings in the final drawing of graph $G'$ that we construct. 

Consider now some edge $e\in \tilde E''_F$. Assume that $e=(t,t')$, with $t\in T_R$, $t'\in T_{R'}$, where $R,R'\in \rset_F$ are two distinct clusters. We define a curve $\gamma'(e)$ representing the edge $e$ by concatenating the curve $\gamma(t)$, the image of the distinguished copy $e^*$ of $e$ in $\phi''$; and curve $\gamma(t')$. Notice that the resulting curve $\gamma'(e)$ connects a point on the boundary of disc $D''(t)$ to a point on the boundary of disc $D''(t')$.

Next, we consider some edge $e\in \tilde E'_F$. Assume that $e=(t,z)$, with $z\in Z_F$, and $t\in T_R$, for some cluster $R\in \rset_F$. We let $\gamma'(e)$ be the curve obtained by concatenating the image  of the distinguished copy $e^*$ of $e$ in $\phi''$ with the curve $\gamma(t)$. Therefore, curve $\gamma'(e)$ connects the image of vertex $z$ to a point on the boundary of disc $D''(t)$.

Consider the resulting set $\Gamma_F=\set{\gamma'(e)\mid e\in \tilde E_F}$ of curves. Notice that these curves may not be in general position, since it is possible that for some vertex $v\in V(H'_F)$, the point $\phi''(v)$ lies on more than two such curves (for example, this can happen when $v$ lies on several paths in $\qset(F)$). In order to overcome this difficulty, for every vertex $v\in V(H'_F)$ with point $\phi''(v)$ lying on more than two  curves of $\Gamma_F$, we modify the  curves of $\Gamma_F$ containing point $\phi'(v)$ within the tiny $v$-disc $D_{\phi'}(v)$, for example, as described in the nudging procedure (see \Cref{sec: curves in a disc}). This may introduce new crossings between the curves in $\Gamma_F$, but since these crossings will eventually become unimportant crossings of drawing $\psi_2$, we can afford to introduce an arbitrary number of such crossings. 

We are now ready to define the solution $\psi_2$ to instance $I'$. Let $\tilde G=G'\setminus \bigcup_{F\in \tfset'}(H'_F\cup \tilde E'_F)$. The drawing of graph $\tilde G$ in $\psi_2$ remains the same as in $\phi'$. Consider now some face $F\in \tfset'$ and cluster $R\in \rset_F$. We plant drawing $\psi_{R^+}$ inside disc $D''(R)$, so that the boundaries of discs $D'(R)$ and $D''(R)$ coincide. Recall that, in drawing $\psi_{R^+}$, the image of every vertex $t\in T_R$ appears on the boundary of the disc $D(R)$, and the images of all other vertices, and of all edges, are disjoint from $D'(R)\setminus D(R)$. It now remains to add the images of the edges in $\tilde E$ to this drawing. Consider again a face $F\in \tfset'$, and an edge $e\in \tilde E$. Initially, we let the image of $e$ be the curve $\gamma'(e)\in \Gamma_F$. We now need to modify this curve slightly so it connects the images of the endpoints of edge $e$.  Assume first that $e\in \tilde E'_F$, and denote $e=(t,z)$, with $z\in Z_F$, and $t\in T_R$, for some cluster $R\in \rset_F$. Then curve $\gamma'(e)$ connects the image of $z$ to a point on the boundary of disc $D''(R)$. Recall that the image of $t$ appears on the boundary of disc $D(R)$. We extend curve $\gamma'(e)$ within the region $D''(R)\setminus D(R)$, so that it terminates at the image of vertex $t$. Assume now that $e\in \tilde E''_F$, and denote $e=(t,t')$, where $t\in T_R$, $t'\in T_{R'}$, for some distinct clusters $R,R'\in \rset_F$. Initially, we let the image of edge $e$ be the curve $\gamma'(e)$ that connects a point on the boundary of $D''(R)$ to a point on the boundary of  $D''(R')$. We extend the curve inside the regions $D''(R)\setminus D(R)$ and $D''(R')\setminus D(R')$, so that it connects the image of $t$ to the image of $t'$. The extensions to the curves in $\Gamma_F$ can be performed so that they remain in general position; we may introduce an arbitrary number of new crossings between these curves, but we will not introduce any crossings between these curves and the images of the edges of $\tilde G$. This completes the definition of the solution $\psi_2$ to instance $I'$. We now ensure that this drawing has the required properties. Observe first that every interesting crossing in $\psi_2$ is either between a pair of edges in $\tilde G$ (and so it must be a type-$2$ crossing in drawing $\phi''$ of $G''$); or it is a crossing between a pair of edges of graph $R^+$ for some cluster $R\in \bigcup_{F\in \tfset'}\rset_F$ (in which case it exists in drawing $\psi_{R^+}$); or it is a crossing between a curve in $\bigcup_{F\in \tfset'}\Gamma_F$ and an image of an edge of $\tilde G$ (in which case it corresponds to some type-2 crossing in drawing $\phi''$ of $G''$). Therefore, if we denote by $\chi_2(\phi'')$ the set of all type-2 crossings in drawing $\phi''$, then we get that the number of interesting crossings in $\psi_2$ is bounded by:
\[|\chi_2(\phi'')|+\sum_{F\in \tfset'}\sum_{R\in \rset_F}\cro(\psi_{R^+}). \]
Recall that for every cluster $R\in \bigcup_{F\in \tfset'}\rset_F$, $\cro(\psi_{R^+})\leq  |\chi(R)|$, where $\chi(R)$ the set of all crossings in the drawing $\phi'$ of $G'$ in which the edges of $R^+$ participate. Therefore, 
$\sum_{F\in \tfset'}\sum_{R\in \rset_F}\cro(\psi_{R^+})\leq 2\cro(\phi')\leq 2\cro(\phi)$.
Since, from the above discussion, the expected number of type-2 crossings in $\phi''$ is bounded by $c\cdot \cro(\phi')\cdot\log^{16}\cm'\leq c\cdot\cro(\phi)\cdot\log^{16}\cm'$ for some large enough constant $c$, we get that the expected number of type-2 crossings in $\phi''$ is at most $ 4c\cdot\cro(\phi)\cdot\log^{16}\cm'$. We say that a bad event $\event_1$ happens if the number of type-2 crossings in $\phi''$ is greater than $ 16c\cdot\cro(\phi)\cdot\log^{16}\cm'$. From Markov's inequality, $\prob{\event_1}\leq 1/4$.

Next, we bound $|\chi^*(\psi_2)|$ -- the number of crossings in which the edges of the core $\cJ$ participate. Notice that the edges of the core $\cJ$ may not lie in $\bigcup_{F\in \tfset'}(H'_F\cup \tilde E'_F)$. Consider any crossing $(e,e')_p\in \chi^*(\psi_2)$, and assume that $e\in E(\cJ)$. Then either $e'\in E(\tilde G)$ must hold (in which case crossing $(e,e')_p$ is a type-2 crossing in drawing $\phi''$ of $G''$), or $e'\in  \tilde E$ (in which case curve $\gamma'(e')$ crosses the image of $e$). Since curve $\gamma'(e')$ was constructed by following the images of one or two paths in $\bigcup_{F\in \tfset'}\tilde Q(F)$ in $\phi''$, and using the image of edge $e'$ in $\phi''$, there is a crossing $(e,e'')_p$ in drawing $\phi''$, such that the image of edge $e''$ has a non-zero length intersection with curve $\gamma'(e')$.

Therefore, $|\chi^*(\psi_2)|\leq |\chi^*(\phi'')|$. It now remains to bound the number of crossings in which the edges of $\cJ$ participate in $\phi''$. Consider any such crossing $(e,e'')_p$, with $e\in E(\cJ)$. Then either $e''\in E(\tilde G)$, and crossing $(e,e'')_p$ also exists in drawing $\phi'$ of $G'$; or $e''$ is a copy of some edge $e'\in \bigcup_{F\in \tfset'}(H_F\cup \tilde E_F)$, and crossing $(e,e')_p$ also exists in drawing $\phi'$ of $G'$.  In the former case, we say that crossing $(e,e'')_p$ in $\phi'$ is responsible for crossing $(e,e'')_p$ in $\phi''$, while in the latter case we say that crossing $(e,e')_p$ in $\phi'$ is responsible for crossing $(e,e'')_p$ in $\phi''$. Consider now some crossing $(e,e')_p$ in drawing $\phi'$. If $e'\in E(\tilde G)$, then this crossing may be responsible for at most one crossing in $\phi''$. Otherwise, there must be a face $F\in \tfset'$, with $e'\in E(H_F)\cup \tilde E_F$. In this case, crossing $(e,e')_p$ may be responsible for at most $N_{e'}+1$ crossings in $\phi''$. Since, from \Cref{obs: bound expected congestion}, for every face $F\in \tilde \fset'$,	for every edge $e'\in E(H'_F)\cup \tilde E'_F$, $\expect{N_{e'}+1}=\expect{\cong_{G'_F}(\qset(F),e')+1}\leq O(\log^8\cm')$, we get that $\expect{|\chi^*(\psi_2)|}\leq c|\chi^*(\phi')|\log^8\cm'\leq  c|\chi^*(\phi)|\log^8\cm'$ for some large enough constant $c$. We say that bad event $\event_2$ happens if $|\chi^*(\psi_2)|>4c|\chi^*(\phi)|\log^8\cm'$. As before, from Markov inequality, $\prob{\event_2}\leq 1/4$.

Lastly, we need to bound the number of crossings in which the edges of $E(K)$ participate in the drawing $\psi_2$. We denote by $N(\phi')$ and $N(\psi_2)$ the number of crossings in which the edges of $E(K)$ participate in drawings $\phi'$ and $\psi_2$, respectively. Recall that $N(\phi')$ is bounded by the number of crossings in which the edges of $K$ participate in $\phi$, which was denoted by $N$. From the definition, the edges of $E(K)$ may not lie in $\bigcup_{F\in \tfset'}(H'_F\cup \tilde E'_F)$. We can use the same argument that we used in bounding $|\chi^*(\psi_2)|$ to show that every crossing $(e,e'')_p$ in $\phi''$ in which $e\in E(K)$ can be mapped to some crossing $(e,e')_p$ in $\phi'$, such that the total number of crossings mapped to a single crossing $(e,e')_p$ is $1$ if $e'\in E(\tilde G)$ and $N_{e'}+1$ otherwise. Using the same reasoning as above, we get that $\expect{N(\phi'')}\leq  c\cdot N(\phi')\cdot \log^8\cm'\leq c\cdot N\cdot \log^8\cm'$, where $c$ is a large enough constant. We say that bad event $\event_3$ happens if $N(\psi_2)>4c\cdot N\cdot \log^8\cm'$. As before, from Markov inequality, $\prob{\event_3}\leq 1/4$.

Using the Union Bound, with probability at least $1/4$ neither of the bad events $\event_1,\event_2,\event_3$ happens. Therefore, there must exist  a solution $\psi_2$ to instance $I'=(G',\Sigma')$ that is $\kset$-valid, 
such that, for every face $F\in \tilde \fset'$, the images of all vertices of $G'_F$ lie in region $F$ of the drawing, the number of interesting crossings in $\psi_2$ is bounded by $4c\cdot\cro(\phi)\cdot \log^{16} \cm'$, $|\chi^{*}(\psi_2)|\leq 4c\cdot |\chi^{*}(\phi)| \cdot \log^8\cm'$,  and $N(\psi_2)\leq 4c\cdot N\cdot \log^8 \cm'$. 
This completes the proof of \Cref{claim: semi-clean drawing2} and of \Cref{claim: semi-clean drawing}.
\end{proof}

\subsection*{Step 2: Moving the Bad Edges}

In this step we consider the $\kset$-valid solution $\psi_1$ to instance $I$ that is guaranteed to exist from \Cref{claim: semi-clean drawing}. Recall that for every face $F\in \tilde \fset'$, the images of all vertices of $G_F$ lie in region $F$ of $\psi_1$. Additionally,  $\cro(\psi_1)\leq (\cro(\phi)+N^2)\cdot (\log \cm')^{O(1)}$, $|\chi^{*}(\psi_1))|\leq |\chi^{*}(\phi)| \cdot (\log \cm')^{O(1)}$, and    the total number of crossings in which the edges of $K$  participate in $\psi_1$ is at most $N\cdot (\log \cm')^{O(1)}$.

For every face $F\in \tfset'$, we define a disc $D(F)$, that is contained in region $F$ of $\psi_1$, so that the boundary of disc $D(F)$  closely follows the boundary of the region $F$. Equivalently, we can think of disc $D(F)$ as the complement of the disc $D(J_F)$ associated with the core $J_F$ (see e.g. \Cref{def: valid drawing}); the boundaries of both discs coincide, and their interiors are disjoint. Note that, if we traverse the boundary of the disc $D(F)$ in counter-clock direction, we encounter the images of the edges $\delta_{G_F}(J_F)$ in the oriented circular ordering $\oset(J_F)$. Notice however that images of additional edges of $G$ may cross the boundary of disc $D(F)$ -- edges whose images cross the image of some edge in $E(J_F)$. As before, we say that an edge of $G$ is \emph{bad} if its image in $\psi_1$ crosses the image of some edge of the original core $\cJ$. Note that for every bad edge $e=(u,v)$, there must be a face $F\in \tfset'$ with $e\in E(G_F)$. Intuitively, in this step, for each face $F\in \tfset'$, we will modify the drawing that is contained in disc $D(F)$ in $\psi_1$, and add the images of the bad edges $e\in E(G_F)$ to this drawing. In order to do so in a modular fashion, so that the resulting drawings for different faces $F$ can be ``glued'' together, we will define a new graph that is associated with the face $F$, that, intuitively, will contain all vertices and (segments of) edges that are drawn inside disc $D(F)$ in $\psi_1$. We will exclude all bad edges that do not lie in $G_F$, and include all bad edges that lie in $G_F$. We will also subdivide the image of every edge that crosses the boundary of disc $D(F)$ with a new vertex, whose image will be placed on the boundary of disc $D(F)$. We view these new vertices as ``anchors''. Intuitively, we will allow the image of the graph associated with face $F$ to be modified arbitrarily, as long as it is contained in disc $D(F)$, and as long as the images of the anchors remain unchanged. This will allow us to replace the part of the drawing of graph $G$ that is contained in $D(F)$ with a new drawing, to which the bad edges that lie in $G_F$ are added.

We start by constructing a new graph $\tilde G$, that is obtained from graph $G$, by subdividing some edges of $G$. Notice that, if graph $\tilde G$ is obtained from graph $G$ in this way, then the rotation system $\Sigma$ for $G$ naturally defines a rotation system $\tilde \Sigma$ for $\tG$. We then obtain an instance $\tI=(\tG,\tilde \Sigma)$ of \cnwrs problem that we call \emph{instance defined by graph $\tG$}.  We will also define a solution $\tilde \phi$ to instance $\tI$, that will be obtained from solution $\psi_1$ to instance $I$ in a natural way. Initially, we start with $\tilde G=G$ and $\tilde \phi=\psi_1$.
For every face $F\in \tfset'$, we also construct a set $A_F$ of \emph{anchor vertices}, whose image in $\tilde \phi$ appears on the boundary of the disc $D(F)$. Initially, we set $A_F=\emptyset$ for all $F\in \tfset'$.

Consider some face $F\in \tfset'$, an an edge $e=(x,y)$ that is incident to some vertex of the core $J_F$. Recall that exactly one endpoint of $e$ (say $x$) belongs to the core $J_F$. We subdivide edge $e$ with a new vertex $t_e$, so that edge $e$ is replaced with a path $(x,t_e,y)$. We also subdivide the image of edge $e$ in $\tilde \phi$ by placing the image of vertex $t_e$ on the first intersection point of $\tilde \phi(e)$ with the boundary of the disc $D(F)$, as we traverse $\tilde \phi(e)$ from $x$ to $y$. We add vertex $t_e$ to the set $A_F$ of anchor vertices. In our final graph $\tG$, we will delete the edge $(x,t_e)$, and we will view edge $(t_e,y)$ as representing the original edge $e$. 

Consider the current graph $\tG$, and its current drawing $\tilde \phi$. We say that an edge $e\in E(\tG)$ is \emph{good} if its image in $\tilde \phi$ does not cross the image of any edge in $K$. Since we have subdivided the edges incident to the vertices of $K$ that do not lie in $E(K)$, for every face $F\in \tfset'$, every edge of $\tG$ that is incident to a vertex of $J_F$ is a good edge. We say that an edge $e\in E(\tG)$ is \emph{bad} if its image in the current drawing $\tilde \phi$ crosses the image of some edge of the original core $\cJ$. Notice that the total number of bad edges is bounded by $|\chi^*(\tilde \phi)|\leq |\chi^*(\psi_1)|\leq |\chi^*(\phi)|\cdot (\log \cm')^{O(1)}$ from \Cref{claim: semi-clean drawing}. Notice also that for every bad edge $e\in E(\tG)$, there must be a face $F\in \tfset'$, such that the images of both endpoints of $e$ lie in the disc $D(F)$. Lastly, we say that an edge $e\in E(\tG)$ is a \emph{migrating edge} if it is neither good nor bad. In this case, the image of $e$ must cross the image of some edge in $E(K)\setminus E(\cJ)$. Next, we process all bad edges and all migrating edges.

Consider first a bad edge $e=(x,y)$, and assume that the images of both $x$ and $y$ lie inside disc $D(F)$ for some face $F\in \tfset'$. Then the image of edge $e$ in $\tilde \phi$ must intersect the boundary of the disc $D(F)$ in at least two points. We direct the image of $e$ in $\tilde \phi$ from $x$ to $y$, denote by $p(e)$ the first point on the boundary of disc $D(F)$ that lies on $\tilde \phi(e)$, and by $p'(e)$ the last point on the boundary of disc $D(F)$ that lies on $\tilde \phi(e)$. If $p(e)$ is the image of the vertex $x$, then we denote $t_e=x$ (in this case, $x$ is already added to the set $A_F$ of anchor vertices for face $F$). Otherwise, we subdivide the edge $e$ with a new vertex $t_e$, whose image is placed at point $p(e)$, and we add $t_e$ to the set $A_F$ of anchor vertices. Similarly, if $p'(e)$ is the image of the vertex $y$ in $\tilde \phi$, then we denote $t'_e=y$. Otherwise,  we subdivide the edge $e$ with a new vertex $t'_e$, whose image is placed at point $p'(e)$, and we add $t'_e$ to the set $A_F$ of anchor vertices. If edge $e$ has been subdivided twice, then we have replaced it with path $(x,t_e,t'_e,y)$. If $x\neq t_e$, then edge $(x,t_e)$ now becomes a good edge. This edge does not represent any edge in the original graph $G_F$, so we call it an \emph{extra edge}. Similarly, if $y\neq t'_e$, then edge $(y,t'_e)$ now becomes a good edge, and we also call it an extra edge. The edge $(t_e,t'_e)$ is a bad edge, and we view it as representing the bad edge $e$. Note that the images of both endpoints of this edge now appear on the boundary of disc $D(F)$. We say that face $F$ \emph{owns} this bad edge.

Lastly, we consider a migrating edge $e=(x,y)$. Note that there must be a face $F\in \tfset'$, such that the images of both $x$ and $y$ lie in disc $D(F)$ in $\tilde \phi$. We direct the image of the edge $e$ in $\tilde \phi$ from $x$ to $y$. Denote by $F_1,F_2,\ldots,F_r$ the sequence of the regions of $\tfset'$ that the image of the edge $e$ visits, in the order in which it visits them, so $F_1=F$ and $F_r=F$. 
For all $1<i<r$, we let $\sigma_i$ be the maximal segment on the image of $e$ that is contained in disc $D(F_i)$, and we denote by $p_i(e)$ and $p'_i(e)$ the first and the last endpoints of $\sigma_i$, respectively, that must lie on the boundary of disc $D(F_i)$. We also let $\sigma_1$ the segment of $\tilde \phi(e)$ from the image of $x$ to the first point that lies on the boundary of disc $D(F)$, denoting by $p'_1(e)$ the endpoint of $\sigma_1$ that is different from the image of $x$. Similarly, we let $\sigma_r$ be the segment of $\tilde \phi(e)$ from the last point that lies on the boundary of disc $D(F)$ to the image of $y$, denoting by $p_r(e)$ the endpoint of $\sigma_r$ that is different from the image of $y$. Note that $\tilde \phi(e)\setminus \left(\bigcup_{i=1}^r\sigma_i\right )$ is a collection of short segments, each of which lies outside of $\bigcup_{F\in \tfset'}D(F')$, and crosses some edge of $K$.
We subdivide edge $e$, replacing it with a path $(x,t'_1,t_2,t'_2,\ldots,t_r,y)$. For all $1<i\leq r$, we place the image of the new vertex $t_i$ at point $p_i(e)$, and for all $1\leq i<r$, we place the image of the new vertex $t'_i$ at point $p'_i(e)$. Denote $t_1=x$ and $t'_r=y$. For $1\leq i\leq r$, denote by $e_i$ the new edge $(t_i,t'_i)$, and for $1\leq i<r$, denote by $e'_i$ the new edge $(t'_i,t_{i+1})$. Notice that, for all $1\leq i\leq r$, the image of edge $e_i$ is precisely the segment $\sigma_i$, which is contained in disc $D(F_i)$. Both endpoints of the edge are drawn on the boundary of the disc $D(F_i)$ (except that, for $i=1$, the first endpoint of $\sigma_1$ is the image of $x$ that may lie in the interior of disc $D(F)$, and, for $i=r$, the last endpoint of $\sigma_r$ is the image of vertex $y$ that may lie in the interior of disc $D(F)$). For all $1\leq i\leq r$, edge $e_i$ now becomes a good edge, and we also call it an \emph{extra edge}. As discussed above, the images of edges $e'_1,\ldots,e'_{r-1}$ are short segments that lie outside of $\bigcup_{F'\in \tfset'}D(F')$ (except for their endpoints that lie on the boundaries of the discs), and each such segment crosses some edge of $K$. For all $1< i<r$, we add the new vertices $t_i$ and $t'_i$ to the set $A_{F_i}$ of anchor vertices for face $F_i$. Additionally, we add $t'_1$ and $t_r$ to $A_F$.

Consider the final graph $\tG$ obtained after all bad and migrating edges have been procesed, and the resulting solution $\tilde \phi$ to the instance $\tI$ defined by $\tG$. From the above discussion, the total number of extra edges in $\tG$ is bounded by 
$2N(\psi_1)+2|\chi^*(\psi_1)|$, where $N(\psi_1)$ is the number of crossings in which the edges of the skeleton $K$ participate in $\psi_1$. Recall that $N(\psi_1)\leq N\cdot (\log \cm')^{O(1)}\leq \frac{\cm'}{\mu^{3b}}\cdot (\log \cm')^{O(1)}$, and $|\chi^*(\psi_1)|\leq |\chi^{*}(\psi)| \cdot (\log \cm')^{O(1)}\leq \frac{\cm'}{ \mu^{240b}}\cdot (\log \cm')^{O(1)}$. Therefore, the total number of exta edges in 
graph $\tG$ is bounded by $\frac{\cm'}{\mu^{2b}}$.

For every face $F\in \tfset'$, we now define a subgraph $\tG_F$ of graph $\tG$ associated with face $F$. The set of vertices of $\tG_F$ contains all vertices whose images appear in disc $D(F)$ in the current drawing $\tilde \phi$. Notice that this includes all vertices in the original graph $G_F$ (except for vertices that belong to the core $J_F$), and, additionally, new vertices whose images were added to the boundary of $D(F)$, and  were added to the set $A_F$ of anchor vertices. Therefore, $V(\tG_F)=(V(G_F)\setminus V(J_F))\cup A_F$. The set of edges consists of all edges of $\tG$ whose image in $\tilde \phi$ is contained in the disc $D(F)$, and all bad edges that belong to face $F$ (that is, all bad edges whose both endpoints lie in $A_F$). Notice that, if $e$ is an edge of $\tG_F$, then either $e$ is an edge of $G_F$, or it was obtained by subdividing some edge of $G_F$, or $e$ was obtained by subdividing some migrating edge $e'$ that lied in some graph $G_{F'}$ for $F'\neq F$. In the latter case, the endpoints of $e'$ belong to the set $A_F$ of anchors, and edge $e'$ with its endpoints forms a separate connected component in graph $\tG_F$. 
We need the following observation.

\begin{observation}\label{obs: small boundary cuts}
	Let $F\in \tfset'$ be a face, and let $(S,T)$ be a partition of the set $A_F$ of the anchor vertices, so that the images of the vertices of $S$ in $\tilde \phi$ appear consecutively on the boundary of disc $D(F)$. Then there is a collection $E'$ of at most $4\cm'/\mu^{2b}$ edges in graph $\tG_F$, so that there is no path connecting a vertex of $S$ to a vertex of $T$ in $\tG_F\setminus E'$.
\end{observation}

\begin{proof}
	Assume otherwise. From the max-flow min-cut theorem, there is a collection $\pset$ of $\ceil{4\cm'/\mu^{2b}}$ edge-disjoint paths in graph $\tG_F$ connecting vertices of $S$ to vertices of $T$. Since there are at most $\cm'/\mu^{2b}$ extra edges in graph $\tG$, there is a subset $\pset'\subseteq \pset$ of at least $\ceil{2\cm'/\mu^{2b}}$ paths that do not contain extra edges. Therefore, every path in $\pset'$ only contains edges that were obtained by subdividing the edges of $G_F$. Observe that the anchor vertices representing the edges in $\delta_{G_F}(J_F)$ appear on the boundary of disc $D(F)$ in drawing $\tilde \phi$ according to the ordering $\oset(J_F)$. Therefore, partition $(S,T)$ of vertices of $A_F$ naturally induces a partition $(E_1,E_2)$ of the set $\delta_{G_F}(J_F)$ of edges, where the edges of $E_1$ appear consecutively in the ordering $\oset(J_F)$. But then the existence of the set $\pset'$ of paths contradicts the fact that $I_F$ is an acceptable instance (see \Cref{def: acceptable instance}).
\end{proof}

For a face $F\in \tfset'$, we denote by $\chi(F)$ the set of all crossings in the drawing $\tilde \phi$ in which the edges of $\tG_F$ participate, and we denote by $E^{\bad}(F)$ the set of all bad edges that face $F$ owns. 
The proof of \Cref{lem: computed decomposition is good enough} follows from the following claim.

\begin{claim}\label{claim: rearrange drawing in disc}
	For every face $F\in \tfset'$, there is a solution $\psi_F$ to instance $\tI_F$, with: 
	$$\cro(\psi_F)\leq \left(|\chi(F)|+|E^{\bad}(F)|^2+|E^{\bad}(F)|\cdot \frac{\cm'}{\mu^{2b}}\right )\cdot (\log \cm')^{O(1)},$$ such that the drawing of the graph $\tG_F$ is contained in disc $D(F)$, and, for every anchor vertex $t\in A_F$, the image of $t$ in $\psi_F$ is identical to its image in $\tilde \phi$.
\end{claim}

We provide the proof of \Cref{claim: rearrange drawing in disc} below, after we complete the  proof of \Cref{lem: computed decomposition is good enough} using it. We start from the solution $\tilde \phi$ to instance $\tI$. For every face $F\in \tfset'$, we  delete the contents of the disc $D(F)$, and replace them with the contents of disc $D(F)$ in drawing $\psi_F$ of graph $\tG_F$. Since the images of the anchor vertices of $A_F$ remain unchanged, once every face $F\in \tfset'$ is processed, we obtain a valid solution to instance $\tI$, that we denote by $\tilde \psi'$. Since, for every face $F\in \tfset'$, for every bad edge $e\in E^{\bad}(F)$, the image of $e$ now lies in disc $D(F)$, for every bad face $F'\in \tfset^{\forbidden}$, the image of every edge of $\tG$ in $\tilde \psi'$ is disjoint from the interior of $F'$.
Since graph $\tG$ was obtained from graph $G$ by subdividing some of its edges, solution $\tilde \psi'$ to instance $\tI$ naturally defines a solution $\psi$ to instance $I$, by suppressing the images of vertices that were used to subdivided edges. From the above discussion, the resulting solution $\psi$ to instance $I$ is clean with respect to  $\cjset$. Moreover:
\[
\begin{split}
\cro(\psi)&\leq \cro(\tilde \phi')\\
&\leq \cro(\tilde \phi)+\sum_{F\in \tfset'}\cro(\psi_F)\\
&\leq\cro(\tilde \phi)+\sum_{F\in \tfset'}\left(|\chi(F)|+|E^{\bad}(F)|^2+|E^{\bad}(F)|\cdot \frac{\cm'}{\mu^{2b}}\right )\cdot (\log \cm')^{O(1)}\\
&\leq \cro(\tilde \phi)\cdot (\log \cm')^{O(1)} +\sum_{F\in \tfset'}|E^{\bad}(F)|^2\cdot (\log \cm')^{O(1)}+|\chi^*(\psi_1)|\cdot \frac{\cm'\cdot (\log \cm')^{O(1)}}{\mu^{2b}}\\
&\leq (\cro(\phi)+N^2)\cdot (\log \cm')^{O(1)}+|\chi^*(\psi_1)|^2\cdot (\log \cm')^{O(1)}+  |\chi^{*}(\phi)|\cdot  \frac{\cm'\cdot (\log \cm')^{O(1)}}{\mu^{2b}}\\
&\leq \left(\cro(\phi)+N^2+|\chi^*(\phi)|^2+ \frac{\cm'\cdot|\chi^{*}(\phi)|}{\mu^{2b}}\right ) (\log \cm')^{O(1)}.
\end{split}
\]
In order to complete the proof of \Cref{lem: computed decomposition is good enough}, it is now enough to prove \Cref{claim: rearrange drawing in disc}, which we do next.

\subsubsection{Proof of \Cref{claim: rearrange drawing in disc}}

We fix a face $F\in \tfset'$. For convenience, we denote graph $\tG_F$ by $G$, and the corresponding instance $\tI_F=(\tG_F,\tilde \Sigma_F)$ of \cnwrs by $I=(G,\Sigma)$. We also denote the solution to instance $I$ induced by drawing $\tilde \phi$ of $\tilde G$ by $\phi$, so $|\chi(F)|\geq \cro(\phi)$. Recall that we are given a disc $D(F)$, that we denote by $D$, and a collection $A_F$ of anchor vertices (that we denote by $A$). The images of all vertices of $A$ in $\phi$ lie on the boundary of the disc $D$, and the images of all other vertices of $G$ lie in the interior of the disc. The set of edges of $G$ is partitioned into two subsets: set $E^{\bad}$ of bad edges, and set $E^{\good}$ of all remaining edges, that we refer to as good edges. The images of all good edges in $\phi$ are contained in disc $D$. For every bad edge $e\in E^{\bad}$, the endpoints of $e$ (which must be anchor vertices) lie on the boundary of disc $D$.  

Our goal is to show that there exists another solution $\psi$ to instance $I$, in which the images of the anchor vertices remain unchanged from $\phi$, the images of all vertices and edges of $G$ are contained in disc $D$, and $\cro(\psi)\leq \left (\cro(\phi)+|E^{\bad}|^2+|E^{\bad}|\cdot \frac{\cm'}{\mu^{2b}}\right )\cdot (\log \cm')^{O(1)}$. Recall that, from \Cref{obs: small boundary cuts}, for any partition $(S,T)$ of the vertices of $A$, so that the vertices of $S$ appear consecutively on the boundary of $D$ in $\phi$, there is a set $E'$ of at most $4\cm'/\mu^{2b}$ edges in $G$,  so that there is no path connecting a vertex of $S$ to a vertex of $T$ in $G\setminus E'$.

We let $A'\subseteq A$ be the set of vertices that serve as endpoints of the bad edges. Note that the edges of $E^{\bad}$ define a perfect matching between the vertices of $A'$. We then let $\Pi=\set{\phi(v)\mid v\in A'}$ be the set of points that serve as images of the vertices of $A'$. Let $r$ be the smallest integer, so that $|\Pi|\leq 2^r$. Clearly, $2^r\leq 4|E^{\bad}|\leq 4\cm'$, and $r\leq \log(4\cm')$. We add additional arbitrary points on the boundary of the disc $D$ to set $\Pi$ until $\Pi$ contains $2^r+1$ distinct points. We denote $\Pi=\set{p_0,p_1,\ldots,p_{2^r}}$, and we assume that the points appear on the boundary of disc $D$ in this order, as we traverse the boundary in counter-clock-wise direction. 

Next, we define a number of guiding curves, that we call \emph{corridors}. We will ensure that all these curves are disjoint, except for possibly sharing their endpoints. The curves are partitioned into $r$ levels.

The set $\Lambda_0$ of level-$0$ curves contains, for all $0\leq i<2^r$, a curve $\lambda_{0,i}$, that connects point $p_i$ to point $p_{i+1}$, and is contained in the interior of disc $D$ (except for its two endpoints that lie on the disc boundary). We ensure that all curves in set $\Lambda_0$ are disjoint from each other. Note that for all $0\leq i<2^r$, points $p_i$ and $p_{i+1}$ partition the boundary of the disc $D$ into two segments. Let $\sigma_{0,i}$ be the segment that is disjoint from point $p_{i+2}$. Then we can define a disc $D_{0,i}\subseteq D$ that corresponds to curve $\lambda_{0,i}$, whose boundary is the concatentation of curves $\sigma_{0,i}$ and $\lambda_{0,i}$.

For a level $0<j< r$,  we consider the points in $\set{p_{i\cdot 2^j}\mid 0\leq i\leq 2^{r-j}}$, and we connect every consecutive pair of such points with a curve. Specifically, the set $\Lambda_j$ of level-$j$ curves contains, for all $0\leq i<2^{r-j}$, a curve $\lambda_{j,i}$, that connects point $p_{i\cdot 2^j}$ to point $p_{(i+1)\cdot 2^j}$. We draw these curves so that they are internally disjoint from each other and from the curves in $\Lambda_0\cup\cdots\cup\Lambda_{j-1}$, and every curve is contained in the interior of the disc $D$ (except for its endpoints that lie on the disc's boundary).

\begin{figure}[h]
	\centering
	\includegraphics[scale=0.13]{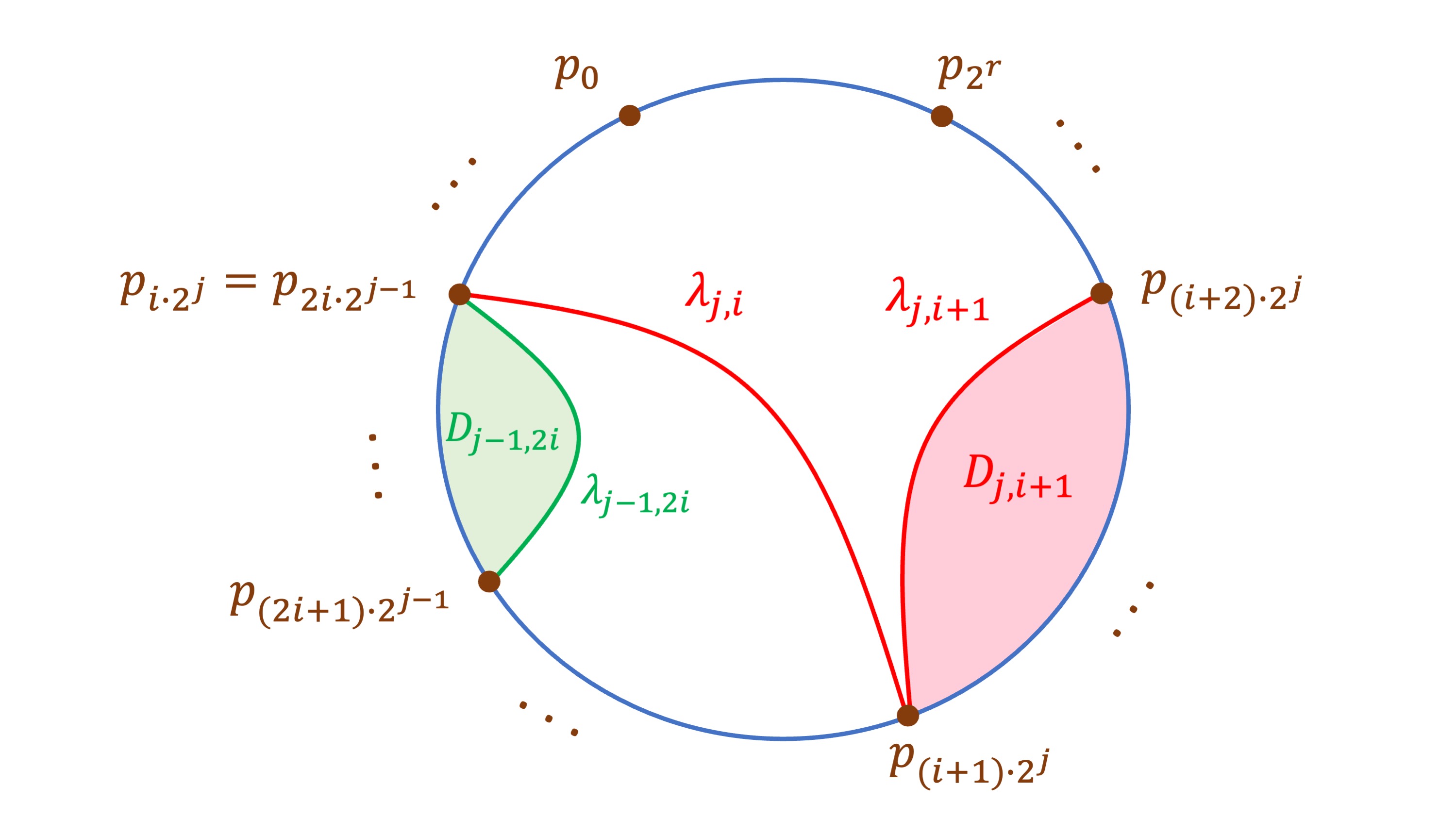}
	\caption{Level-$j$ curves $\lambda_{j,i}$ and $\lambda_{j,i+1}$, with  disc $D_{j,i+1}$ shown in red, and a level-$(j-1)$ curve $\lambda_{j-1,2i}$, with disc $D_{j-1,2_i}$ shown in green.
	}\label{fig: smalldiscs} 
\end{figure}

As before, for every index 
$0\leq i<2^{r-j}$, points $p_{i\cdot 2^j}$, $p_{(i+1)\cdot 2^j}$ partition 
the boundary of the disc $D$ into two segments. We denote by $\sigma_{j,i}$ the segment that does not contain the point $p_{(i+1)\cdot 2^j+1}$. We let $D_{j,i}$ be the disc that is contained in $D$, whose boundary is the concatenation of curves $\sigma_{j,i}$ and $\lambda_{j,i}$ (see \Cref{fig: smalldiscs}).

Lastly, the set $\Lambda_r$ of level-$r$ curves contains a single curve $\lambda_{r,0}$ that connects points $p_0$ and $p_{2^r}$. We ensure that this curve is internally disjoint from all curves in $\Lambda_0\cup\cdots\cup\Lambda_{r-1}$, and is contained in the interior of $D$, except for its endpoints that lie on the boundary of the disc. 
We define a disc $D_{r,0}$ associated with this curve exactly like before, so points $p_0,\ldots,p_r$ lie on the boundary of $D_{r,0}$. 

We denote by $\Lambda=\bigcup_{j=0}^r\Lambda_j$. Let $G'=G\setminus E^{\bad}$, and let $I'=(G',\Sigma')$ be the subinstance of $I$ that is defined by graph $G'$.
The crux of the proof of \Cref{claim: rearrange drawing in disc} is the following claim, that allows us to ``rearrange'' the image of graph $G'$, so that each guiding curve only crosses a small number of edges.

\begin{claim}\label{claim: avoid guiding curves}
	 There is a solution $\psi'$ to instance $I'$, with $\cro(\psi')\leq \cro(\phi)\cdot  (\log \cm')^{O(1)}$, so that the images of all vertices and edges of $G'$ lie in disc $D$, and the images of the anchor vertices in set $A$ remain the same as in $\phi$. Moreover, for every curve $\lambda\in \Lambda$, for every vertex $v\in V(G')$, the image of $v$ in $\psi'$ does not lie on an inner point of $\lambda$; for every edge $e\in E(G')$, the image of $e$ in $\psi'$ may intersect $\lambda$ in at most one point; and the total number of edges in $E(G')$ whose images intersect $\lambda$ is at most $4\cm'/\mu^{2b}$.
	\end{claim}

We provide the proof of \Cref{claim: avoid guiding curves} in Section \ref{subsec: proof of claim avoind guiding curves} of Appendix. Given two points $p_i,p_{i'}\in \Pi$, a \emph{tunnel} connecting $p_i$ to $p_{i'}$  is a sequence $L=(\lambda^1,\ldots,\lambda^z)$ of curves of $\Lambda$, such that the concatenation of the curves in $L$ is a simple curve connecting points $p_i$ and $p_{i'}$. The \emph{length} of the tunnel is $z$ -- the number of curves in the sequence. We also need the following simple observation, whose proof appears in Section \ref{subsec: appx tunnels} of Appendix.

\begin{observation}\label{obs: tunnels}
	For every pair $p_i,p_{i'}$ of distinct points of $\Pi$, there is a tunnel of length $O(\log \cm')$ connecting $p_i$ to $p_{i'}$.
\end{observation}

The proof of \Cref{claim: rearrange drawing in disc} easily follows from \Cref{claim: avoid guiding curves} and \Cref{obs: tunnels}. We start with the solution $\psi'$ to instance $I'$, that is given by \Cref{claim: avoid guiding curves}. Recall that  $\cro(\psi')\leq \cro(\phi)\cdot  (\log \cm')^{O(1)}$, the image of $G'$ is contained in disc $D$, and the images of the anchor vertices in set $A$ in $\psi'$ are identical to those in $\phi$. Next, we consider the bad edges one by one, and insert them into the drawing $\psi'$. Consider any such bad edge $e=(x,y)\in E^{\bad}$. Recall that there are two points $p_i,p_{i'}\in \Pi$, such that $\psi'(x)=\phi(x)=p_{i}$, and $\psi'(y)=\phi(x)=p_{i'}$. From our construction of graph $\tG$, vertices $x$ and $y$ each have degree $2$ in $G$. Let $L=(\lambda^1,\ldots,\lambda^z)$ be a tunnel connecting $p_i$ to $p_{i'}$, with $z\leq O(\log \cm')$, that is given by \Cref{obs: tunnels}. We let the image of the edge $e$ to be a simple curve that connect $p_i$ to $p_{i'}$, and closely follows the image of the curve $\gamma(L)$, obtained by concatenating all curves in $L$, next to this curve. From \Cref{claim: avoid guiding curves}, this new image of edge $e$ crosses the images of at most $O\left(\frac{\cm'\cdot \log \cm'}{\mu^{2b}}\right )$ edges of $G'$. We allow the images of the edges of $E^{\bad}$ to cross arbitrarily.

Once all bad edges are processed, we obtain a solution $\psi$ to instance $G$, where the images of all vertices and edges are contained in $D$, and the images of the anchor vertices in $A$ are identical to those in $\phi$. The number of crossings between pairs of edges in $E(G')$ is at most $ \cro(\phi)\cdot  (\log \cm')^{O(1)}$; the number of crossings between edges of $E^{\bad}$ and edges of $E(G')$ is at most 
$O\left(\frac{|E^{\bad}|\cdot \cm'\cdot \log \cm'}{\mu^{2b}}\right )$; and the number of crossings between the edges of $E^{\bad}$ may be arbitrary.
As our last step, we perform a type-1 uncrossing of the images of the bad edges (see \Cref{thm: type-1 uncrossing}). This procedure locally modifies the images of the bad edges by swapping segments between pairs of images of these edges (see \Cref{fig:type_1_uncrossing}). At the end of this procedure, we are guaranteed that every pair of edges in $E^{\bad}$ cross at most once, and the number of crossings between the new images of the edges in $E^{\bad}$ and the images of the edges in $E(G')$ does not grow. Therefore, the number of crossings in this final solution to instance $I$ is bounded by $\cro(\phi)\cdot  (\log \cm')^{O(1)}+O\left(\frac{|E^{\bad}|\cdot \cm'\cdot \log \cm'}{\mu^{2b}}\right )+|E^{\bad}|^2$, as required.

\section{An Algorithm for Narrow Instances -- Proof of \Cref{lem: not many paths}}
\label{sec: computing the decomposition}


We assume that we are given a narrow instance $I=(G,\Sigma)$ of the \CNwRS problem. Throughout this section, we denote $|E(G)|=m$.  We fix some optimal solution $\phi^*$ to instance $I$. 
We will gradually construct the desired family $\iset$ of instances, over the course of three phases. We will employ partitions of graphs  into clusters, that are defined as follows.

\begin{definition}[Partition into Clusters]
	Let $H$ be a graph, and let $\cset$ be a collection of subgraphs of $H$. We say that $\cset$ is a \emph{partition of $H$ into clusters}, if each subgraph $C\in \cset$ is a connected vertex-induced subgraph (cluster) of $H$, $\bigcup_{C\in \cset}V(C)=V(H)$, and for every pair $C,C'\in \cset$ of distinct subgraphs, $V(C)\cap V(C')=\emptyset$.
\end{definition}


Recall that a vertex $v\in V(G)$ is a high-degree vertex, if  $\deg_G(v)\geq m/\mu^4$.
It will be convenient for us to assume that, if $u$ is a neighbor vertex of a high-degree vertex, then the degree of $u$ is $2$, and that no vertex is a neighbor of two high-degree vertices. In order to achieve this, we simply subdivide every edge that is incident to a high-degree vertex with a single vertex; if, for an edge $e=(u,u')$, both its endpoints are high-degree vertices, then we subdivide this edge with two new vertices. 
Let $G'$ denote the resulting graph, and let $\Sigma'$ be the rotation system associated with $G'$, that is naturally defined from rotation system $\Sigma$ for $G$: for every vertex $v\in V(G)\cap V(G')$, the circular ordering of the edges of $\delta_G(v)=\delta_{G'}(v)$ remains the same, and for every vertex $v\in V(G')\setminus V(G)$, $|\delta_{G'}(v)|=2$, so its rotation is trivial. We denote by $I'=(G',\Sigma')$ the resulting instance of \cnwrs. Assume that we compute an $\eta$-decomposition $\iset'$ of instance $I'$.
It is easy to verify that $\iset'$ is also an $O(\eta)$-decomposition of instance $I$. This is since $|E(G')|\leq O(|E(G)|)$, $\optcrors(I')=\optcrors(I)$, and, 
if we are given, for every instance $\tilde I\in \iset$, a solution $\phi(\tilde I)$, then we can efficiently compute a solution $\phi(\tilde I')$ of the same cost to the corresponding instance $\tilde I'\in \iset'$. Lastly, a solution to instance $I'$ can be efficiently transformed into a solution to instance $I$ of the same cost. Therefore, from now on we will focus on decomposing instance $I'$ into a collection $\iset'$ of subinstances with required properties. To simplify the notation, we denote $G'$ by $G$, $\Sigma'$ by $\Sigma$, and $I'$ by $I$. Note that $|E(G)|\leq 3m$ now holds; every neighbor of a high-degree vertex in $G$ has degree at most $2$; and no vertex is a neighbor of two high-degree vertices. 

\begin{figure}[h]
	\centering
	\subfigure[A schematic view of the petals $X_1,X_2,X_3,X_4$.]{\scalebox{0.17}{\includegraphics{figs/flower_1.jpg}}}
	\hspace{30pt}
	\subfigure[Paths in set $\qset_2$ for petal $X_2$ are shown in pink.]{\scalebox{0.15}{\includegraphics{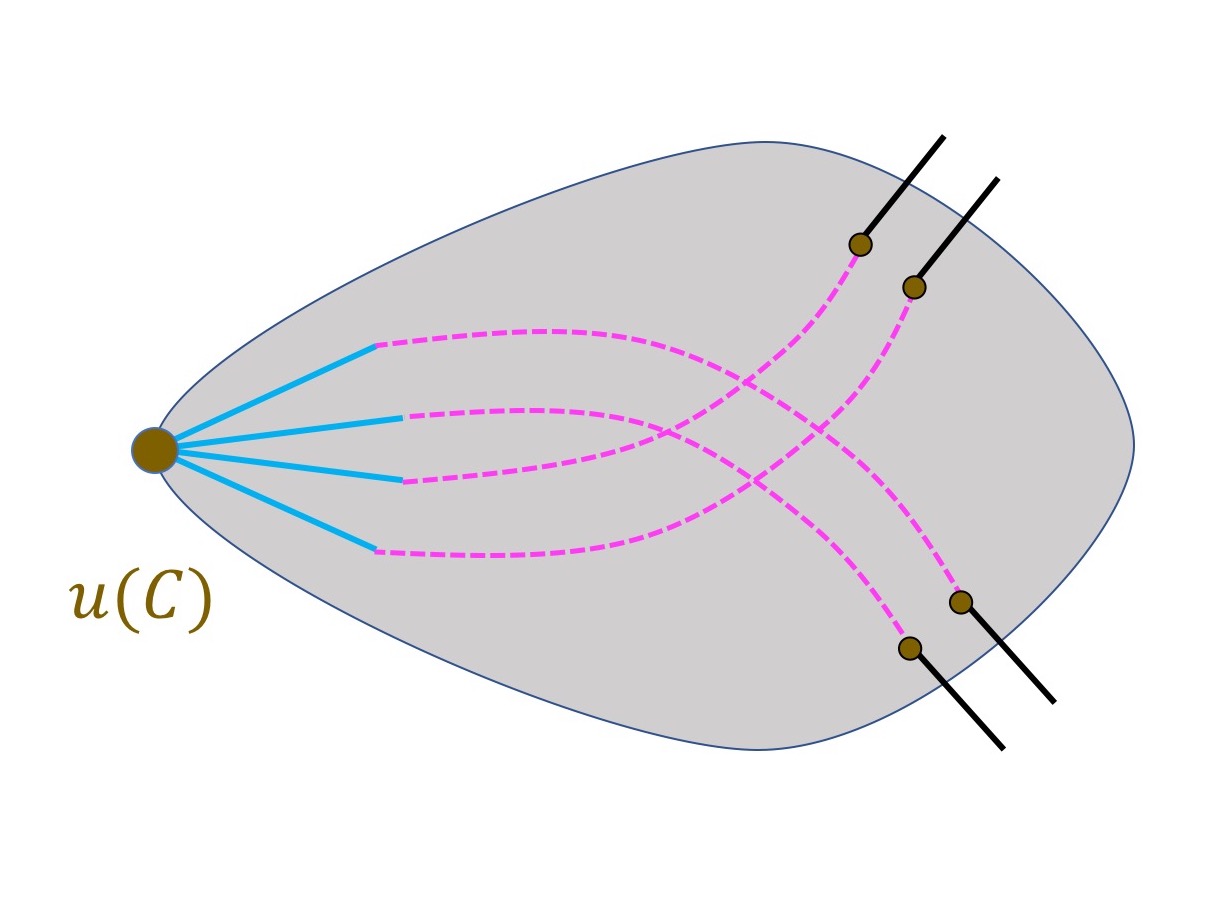}}}
	\caption{An illustration of a $4$-petal flower cluster.}\label{fig: flower_cluster}
\end{figure}

Intuitively, in order to prove \Cref{lem: not many paths}, it is enough to compute a partition $\cset$ of the input graph $G$ into clusters, such that, for every cluster $C\in \cset$,  $|E(C)|\leq m/\mu^2$ holds; we refer to such clusters as \emph{small} clusters. Additionally, we require that the total number of edges with endpoints in different clusters is small, namely, $|\inE(\cset)|\leq m/\mu^2$. 
Once such a collection $\cset$ of clusters is computed, we can use the algorithm from \Cref{thm: disengagement - main} in order to compute the desired decomposition $\iset$ of instance $I$. Unfortunately, we are unable to compute such a decomposition $\cset$ of graph $G$ into small clusters directly. The main obstacle to computing such a decomposition via standard techniques is that graph $G$ may contain high-degree vertices. In order to overcome this difficulty, we define a new type of clusters, called \emph{flower clusters}. Flower clusters will be used in order to isolate high-degree vertices in graph $G$. Eventually, we will compute a decomposition $\cset$ of $G$ into clusters, where each cluster in $\cset$ is either a small cluster or a flower cluster, and the number of edges whose endpoints lie in different clusters of $\cset$ is sufficiently small. We now formally define flower clusters (see \Cref{fig: flower_cluster} for an illustration). 


\begin{definition}[Flower Cluster]
	We say that a subgraph $C\subseteq G$ is a \emph{flower cluster} with a center vertex $u(C)\in V(C)$ and a set $\xset(C)=\set{X_1,\ldots X_k}$ of  petals, if the following hold:
	
	\begin{properties}{F}
		\item $C$ is a connected vertex-induced subgraph of $G$, and for all $1\leq i\leq k$, $X_i$ is a vertex-induced subgraph of $C$; \label{prop: flower cluster vertex induced petals too}

			\item $\bigcup_{i=1}^kV(X_i)=V(C)$, and for every pair $X_i,X_j\in \xset$ of clusters with $i\neq j$, $X_i\cap X_j=\set{u(C)}$; \label{prop: flower cluster petal intersection}

		\item the degree of $u(C)$ in $G$ is at least $m/\mu^4$, and all other vertices of $C$ have degrees below $m/\mu^4$ in $G$; \label{prop: flower center has large degree, everyone else no}
		
				\item $|\delta_G(C)|\leq 96m/\mu^{42}$, and the total number of edges $e=(u,v)$ with $u\in V(X_i)\setminus \set{u(C)}$ and $v\in V(X_j)\setminus \set{u(C)}$ for all $1\leq i<  j\leq k$ is at most $96m/\mu^{42}$; \label{prop: flower cluster small boundary size}

		\item there is a partition $E_1,\ldots,E_k$ of all edges of $\delta_C(u(C))$ into subsets, such that, for all $1\leq i\leq k$, edges in $E_i$ are consecutive in the ordering $\oset_{u(C)}\in \Sigma$, $|E_i|\leq m/(2\mu^4)$, and $E_i\subseteq E(X_i)$ (so in particular $\delta_C(u(C))=\delta_G(u(C))$; and \label{prop: flower cluster edges near center partition}

		\item for every cluster $X_i\in \xset$, there is a set $\qset_i$ of edge-disjoint paths, routing all edges of $\delta_G(X_i)\setminus \delta_G(u(C))$ to $u(C)$, such that all inner vertices on all paths of $\qset_i$ are contained in $X_i$. \label{prop: flower cluster routing}
	\end{properties}
\end{definition}

Notice that $\qset=\bigcup_{i=1}^k\qset_i$ is a set of edge-disjoint paths, routing all edges of $\delta_G(C)$ to vertex $u(C)$ inside $C$, and its existence certifies that a flower cluster must have $1$-bandwidth property.

The remainder of the proof of \Cref{lem: not many paths} consists of three phases. In the first phase, we compute a decomposition $\cset$ of the graph $G$ into clusters, such that the total number of edges with endpoints lying in different clusters is small, and every cluster in $\cset$ is either a small cluster or a flower cluster. We then use the algorithm from \Cref{thm: disengagement - main} in order to compute an initial collection $\iset_1$ of subinstances of instance $I$. This collection will have all required properties, except that for some instances $\tilde I=(\tilde G,\tilde \Sigma)\in \iset_1$, $|E(\tilde G)|\leq m/(2\mu)$ may not hold. We call such instances \emph{problematic}. Each such problematic instance consists of a single flower cluster $C\in \cset$, and possibly some additional edges that lie in $\Eout(\cset)$ (recall that $\Eout(\cset)$ is the set of all edges whose endpoints belong to different clusters of $\cset$). In the subsequent two phases, we consider each of the problematic instances in $\iset_1$ separately, and further decompose it into smaller subinstances. Specifically, suppose we are given some problematic instance  $\tilde I=(\tilde G,\tilde \Sigma)\in \iset_1$ with corresponding flower cluster $C\subseteq \tilde G$, whose center is vertex $u$ and the set of petals is $\xset$. Consider any petal $X\in \xset$. We say that petal $X$ is \emph{routable} in graph $\tilde G$ if there is a collection $\pset(X)=\set{P(e)\mid e\in \delta_{\tilde G}(X)\setminus \delta_{\tilde G}(u)}$ of paths in $\tilde G$ that cause congestion at most $4000$, such that for each edge $e\in \delta_{\tilde G}(X)\setminus \delta_{\tilde G}(u)$, path $P(e)$ has $e$ as its first edge and $u$ as its last vertex, and all inner vertices of $P(e)$ are disjoint from $X$. We show that, if every petal of a flower cluster $C$ is routable in $\tilde G$, then we can further decompose instance $\tilde I$ into a collection $\iset'(\tilde I)$ smaller subinstances, using reasoning similar to that in the basic disengagement procedure (see \Cref{subsec: basic disengagement}). It is still however possible that for some resulting instance $\tilde I'=(\tilde G',\tilde \Sigma')\in \iset'(\tilde I)$, $|E(\tilde G')|>m/(2\mu)$ holds. However, the disengagement procedure ensures that such an instance may not contain high-degree vertices. Therefore,  applying the algorithm from Phase 1 to each such instance $\tilde I'$ will yield a decomposition of the initial instance $I$ into subinstances with all required properties. One difficulty with the approach we have just outlined is that, 
if $\tilde I=(\tilde G,\tilde \Sigma)\in \iset_1$ is a problematic instance with corresponding flower cluster $C\subseteq \tilde G$, then we are not guaranteed that the petals of $C$ are routable in $\tilde G$. In order to overcome this difficulty, in Phase 2, we consider each such problematic instance $\tilde I\in \iset$ separately. By performing a layered well-linked decomposition and an additional disengagement step, we decompose each such instance into a collection of subinstances, such that at most one resulting subinstance contains the flower cluster, and we further modify this flower cluster to ensure that each of its petals is routable in the resulting graph. In the third phase, we perform further disengagement on instances containing flower clusters, that exploits the fact that now every petal of the flower cluster is routable. Some of the resulting instances may still contain too many edges, but, as we show, they may not contain high-degree vertices. We then perform one final disengagement on such instances in order to obtain the final decomposition. We now describe each of the three phases in turn.

\subsection{Phase 1: Flower Clusters, Small Clusters, and Initial Disengagement}
\label{subsec: phase 2 flower vs small cluster partition}

The first phase consists of three steps. In the first step, we carve flower clusters out of the graph $G$. In the second step, we decompose the remainder of the graph $G$ into small clusters. Lastly, we perform a disengagement step for all resulting clusters in the third step.

\subsubsection{Step 1: Carving out Flower Clusters}

Let $S=\set{s_1,\ldots,s_k}$ be the set of all vertices of $G$ that have degree at least $m/\mu^4$. Notice that, since we have assumed that no edge connects a pair of high-degree vertices, and $|E(G)|\leq 3m$, $k\leq 3\mu^4$ must hold. 
In this step, we use with the following lemma, in order to carve flower clusters out of graph $G$.

\begin{lemma}\label{lemma: carve out flower clusters}
	There is an efficient algorithm that computes a collection $\cset^f=\set{C^f_1,\ldots,C^f_k}$ of disjoint clusters of $G$, such that for all $1\leq i\leq k$, $C^f_i$ is a flower cluster with center $s_i$. The algorithm also computes, for each resulting flower cluster $C^f_i$, the corresponding set $\xset(C^f_i)$ of petals.
\end{lemma}
\begin{proof}	
We start with the following simple claim that allows us to compute an initial collection $\set{C_1,\ldots,C_k}$ of clusters with some useful properties. Eventually, for all $1\leq i\leq k$, we will define a flower cluster $C_i^f\subseteq C_i$.

\begin{claim}\label{claim: flowers initial}
	There is an efficient algorithm to compute $k$ disjoint clusters $C_1,\ldots, C_k$ of $G$, such that for all $1\leq i\leq k$, $s_i\in V(C_i)$, $\delta_G(s_i)\subseteq E(C_i)$, and $|\delta_G(C_i)|\leq 9m/\mu^{46}$. Additionally, the algorithm computes, for all $1\leq i\leq k$, a set $\qset_i$ of edge-disjoint paths routing the edges of $\delta_G(C_i)$ to $s_i$, such that all inner vertices of every path lie in $C_i$.
\end{claim}

\begin{proof}
We use the algorithm from \Cref{lem: multiway cut with paths sets} to compute, for all $1\leq i\leq k$, a set $A_i$ of vertices of $G$, such that $S\cap A_i=\set{s_i}$, and $(A_i,V(G)\setminus A_i)$ is a minimum cut separating $s_i$ from the vertices of $S\setminus\set{s_i}$ in $G$, and the vertex sets $A_1,\ldots,A_k$ are mutually disjoint. Recall that the algorithm also computes,
for all $1\leq i\leq k$, a set $\qset_i$ of edge-disjoint paths, routing the edges of $\delta_G(A_i)$ to vertex $s_i$, with all inner vertices on every path of $\qset_i$ lying in  $A_i$.

Recall that we have assumed that, for all $1\leq i\leq r$, for every edge $e=(s_i,v_e)\in \delta_G(s_i)$, the degree of vertex $v_e$ in $G$ is at most $2$, and $v_e$ is the neighbor of at most one vertex in $S$ -- vertex $s_i$. If such a vertex $v_e$ does not lie in $A_i$, then we move it to $A_i$ (if $v_e$ lies in some other vertex set $A_{i'}$, we remove it from that vertex set). Since the degree of $v_e$ in $G$ is $2$, this does not increase the cardinalities of edge sets $\delta_G(A_{i'})$ for any $1\leq i'\leq k$. Moreover, we can adjust the sets of paths $\set{\qset_i}_{i=1}^k$, such that, for all $1\leq i\leq k$, set $\qset_i$ remains a set of edge-disjoint paths routing the edges of $\delta_G(A_i)$ to $s_i$, with all inner vertices on every path lying in $A_i$.

For all $1\leq i\leq k$, we let $C_i=G[A_i]$. Note that, from the max-flow /  min-cut theorem, for each $i$, there is a collection $\pset_i$ of at least $|\delta_G(C_i)|$ edge-disjoint paths connecting $s_i$ to vertices of $S\setminus \set{s_i}$. Since $|S|=k\leq 3\mu^4$, there is some vertex $s_j\in S\setminus\set{s_i}$, such that at least $|\delta_G(C_i)|/k\geq |\delta_G(C_i)|/(3\mu^4)$ of the paths in $\pset_i$ connect $s_i$ to $s_j$. Since the original instance $I$ that served as input to \Cref{lem: not many paths} is narrow, from \Cref{obs: narrow prop 2}, the number of such paths must be bounded by $2\ceil{m/\mu^{50}}\le 3m/\mu^{50}$, and so for all $1\leq i\leq k$, $|\delta_G(C_i)|\leq 9m/\mu^{46}$.
\end{proof}

The following claim will complete the proof of \Cref{lemma: carve out flower clusters}.

\begin{claim}\label{claim: compute flowers}
	There is an efficient algorithm, that, for all $1\leq i\leq k$, computes a flower cluster $C^f_i\subseteq C_i$ with center $s_i$, together with a set $\xset_i$ of petals.
\end{claim}
\begin{proof}
	Fix an index $1\leq i\leq k$. For simplicity of notation, in the remainder of the proof, we denote $C_i$ by $C$, $s_i$ by $s$, and the set $\qset_i$ of paths by $\qset$. Recall that $|\delta_G(C)|\leq 9m/\mu^{46}$.

	Since $\deg_G(s)\geq m/\mu^4$, and $\delta_G(s)=\delta_C(s)$, we get that $\deg_C(s)\geq m/\mu^4$. Let $r=\floor{\deg_C(s)\cdot \frac{6\mu^4}{m}}$. Since $m/\mu^4 \leq \deg_C(s)\leq m$, we get that $6\leq r\leq 6\mu^4$. We compute a partition $E_1,\ldots,E_r$ of the edges of $\delta_C(s)$ into disjoint subsets, each of which contains at most $\floor{m/(2\mu^4)}$ edges that appear in the ordering $\oset_{s}\in \Sigma$ consecutively. From our choice of $r$, we get that $r\cdot\floor{m/(2\mu^4)}\geq \deg_G(s)$, so such a partition must exist.
	
	Consider now a graph $H$ that is defined as follows. We start with the graph $C\cup \delta_G(C)$. Let $R$ be the set of all endpoints of the edges of $\delta_G(C)$ that do not lie in $C$. We unify all  vertices of $R$ into a vertex $a_0$. Next, we subdivide every edge $e\in \delta_C(s)$ with a new vertex $v_e$, and delete vertex $s$ from the resulting graph. Finally, for all $1\leq j\leq r$, we unify the vertices in set $V_j=\set{v_e\mid e\in E_j}$ into a single vertex $a_j$, obtaining the graph $H$. 
	Denote $Z=\set{a_0,a_1,\ldots,a_r}$. Note that $\deg_H(a_0)=|\delta_G(C)|\leq 9m/\mu^{46}$, so for all $1\leq j\leq r$, there are at most $9m/\mu^{46}$ edge-disjoint paths connecting $a_0$ to $a_j$.
	Since the original instance $I$ that served as input to \Cref{lem: not many paths} is a narrow instance, from \Cref{obs: narrow prop 2}, for all $1\leq j<j'\leq r$, the maximum number of edge-disjoint paths connecting $a_j$ to $a_{j'}$ is at most $2m/\mu^{50}$. Therefore, for all $1\leq j\leq r$, there is a cut separating $a_{j}$ from all vertices of $Z\setminus a_j$, containing at most $9m/\mu^{46}+2rm/\mu^{50}\leq 16m/\mu^{46}$ edges.
	
	We apply the algorithm from  \Cref{lem: multiway cut with paths sets} to graph $H$ and vertex set $Z$. For all $1\leq j\leq r$, we denote by $A'_j$ the set of vertices that the algorithm returns for vertex $a_j$. Recall that, if $e=(s,x)$ is an edge of $E_j$, then vertex $x$ must have degree at most $2$ in graph $G$. We can then assume without loss of generality that $x$ lies in $A_j'$; if it lies in another set $A_{j'}'$, or it does not lie in any such set, we can simply move it to $A_j'$; this will not increase the values of the cuts $|E(A'_{j'},V(C)\setminus A'_{j'})|$ for any $j'$. We can also adjust the sets  $\set{\qset'_j}$ of paths that the algorithm from \Cref{lem: multiway cut with paths sets} returns for each $1\leq j\leq r$, so that the paths in $\qset'_j$ are edge-disjoint and route edges of $\delta_H(A'_j)$ to vertex $a_j$, with all inner vertices on every path lying in $A'_j$.
	
	For all $1\leq j\leq k$, let $X_j$ be the subgraph of $C$ induced by the vertex set $(A'_j\setminus\set{a_j})\cup \set{s}$. We then let $C^f$ be the subgraph of $G$ induced by vertex set $\bigcup_{j=1}^kV(X_j)$. We claim that $C^f$ is a flower cluster with center $s$ and set  $\xset=\set{X_1,\ldots,X_r}$ of petals.
	We now verify that $C^f$ and $\xset$ have all required properties. 
	
	First, it is immediate to verify that $C^f$ is a vertex-induced subgraph of $G$, and that for all $1\leq j\leq r$, $X_j$ is a vertex-induced subgraph of $C^f$ with $\bigcup_{j=1}^rV(X_j)=V(C^f)$, and, for all $1\leq j<j'\leq r$, $X_j\cap X_{j'}=\set{u}$.  Additionally, we can assume that $C$ is connected: if this is not the case, then connected components of $C$ that are disjoint from $u$ can be discarded. This establishes Properties \ref{prop: flower cluster vertex induced petals too} and \ref{prop: flower cluster petal intersection}. 
	
	Consider now some index $1\leq j\leq r$. Since we have assumed that, for every edge $e=(u,v)\in E_j$, $v\in A_j'$ holds, we get that $E_j\subseteq X_j$. Since $\delta_G(s)\subseteq \delta_C(s)$, and edge sets $E_1,\ldots,E_r$ partition $\delta_C(s)$, this establishes Property \ref{prop: flower cluster edges near center partition}. $\delta_G(s)\subseteq E(C^f)$. 

	Recall that for all $1\leq j\leq r$, $|\delta_C(X_j)\setminus \delta_C(s)|\le 16m/\mu^{46}$ (since $(A'_j,V(H)\setminus A'_j)$ is a minimum cut in $H$ separating $a_{j}$ from all vertices of $Z\setminus a_j$, whose value we have bounded above). Therefore, $|\delta_G(C^f)|\leq r\cdot 16m/\mu^{46}\leq (6\mu^4)\cdot (16m/\mu^{46})\leq 96m/\mu^{42}$, and moreover the total number of edges $e=(u,v)$ with $u,v$ lying in different sets $X_1,\ldots,X_r$ is also bounded by  $96m/\mu^{42}$. This  establishes Property \ref{prop: flower cluster small boundary size}.  Since no vertex of $S$ may lie in $C^f$ (as $C^f\subseteq C$), all vertices of $V(C)\setminus \set{s}$ have degrees at most $m/\mu^4$ in $G$. This establishes Property \ref{prop: flower center has large degree, everyone else no}.

	 Lastly, we need to establish Property \ref{prop: flower cluster routing}. Consider any petal $X_j\in \xset$. 
	 From the definition of graph $H$, and since vertex $a_0$ may not belong to  set $A'_j$, we get that $\delta_H(A'_j)=\delta_G(X_j)\setminus \delta_G(s)$. Recall that  the algorithm from  \Cref{lem: multiway cut with paths sets} provided a set $\qset'_j$ of edge-disjoint paths that route the edges of $\delta_H(A'_j)$ to vertex $a_j$, with all inner vertices on the paths lying in  $A'_j$. By replacing vertex $a_j$ with vertex $s$ on each such path,  we obtain a collection $\qset''_j$ of edge-disjoint paths in graph $C^f$, that route the edges of $\delta_G(X_j)$ to $s$, with all inner vertices on all paths lying in $X_j$. 
	 We conclude that $C^f$ is a valid flower cluster with center $s$ and set $\xset$ of petals.
%
\end{proof}
\end{proof}

\subsubsection{Step 2: Small Clusters}
Let $C_0$ be the cluster that is obtained from graph $G$ after we delete all vertices lying in $\bigcup_{C\in \cset^f}V(C)$ from it, that is, $C_0=G\setminus(\bigcup_{C\in \cset^f}V(C))$.
Note that, since $|\cset^f|\leq 3\mu^4$, and, for all $C\in \cset^f$, $|\delta_G(C)|\leq 96m/\mu^{42}$, we get that $|\bigcup_{C\in \cset^f}\delta_G(C)|\leq  (3\mu^4)\cdot (96m/\mu^{42})\leq 288m/\mu^{38}$, and so  $|\delta_G(C_0)|\leq 288m/\mu^{38}$.
 In this step, our goal is to further decompose cluster $C_0$ into a collection $\cset^s$ of clusters, that we refer to as \emph{small clusters}. We will require that each cluster $C\in \cset^s$ has the $\alpha_0$-bandwidth property, for an appropriately chosen parameter $\alpha_0$, and that $|E(C)|\leq |E(G)|/\mu^2$. Moreover, we will require that the total number of edges whose endpoints lie in different clusters of $\cset^f\cup \cset^s$ is relatively small. We show an algorithm that either computes such a decomposition of $C_0$, or establishes that $\optcrors(I)$ is sufficiently large, by utilizing the following lemma; the lemma will also be used later in this section in a slightly different setting, so it is stated for a more general setting than what is needed here.

\begin{lemma}\label{lem: decomposition into small clusters}
	There is an efficient algorithm, whose input consists of a graph $H$, a set  $T\subseteq V(H)$ of $k$ vertices called terminals (where possibly $T=\emptyset$), and parameters  $m,\tau\geq 0$, such that  $ |E(H)|\leq m$, 
	$k\leq m/(64\tau\log m)$; every vertex in $T$ has degree $1$ in $H$; and maximum vertex degree in $H$ is at most $\frac{m}{\check c\tau^3 \log^5 m}$, for a large enough constant $\check c$.
	The algorithm
	either correctly certifies that $\optcro(H)\geq  \Omega \left( \frac{m^2}{ \tau^4\log^5 m} \right )$, or computes a collection $\cset$ of disjoint clusters of $H\setminus T$ with the following properties:
	
	\begin{itemize}
		\item every cluster $C\in \cset$ has the $\alpha'$-bandwidth property, where $\alpha'=\frac{1}{16\alphasc(m)\cdot \log m}=\Omega\left(\frac{1}{\log^{1.5}m}\right )$;
		
		\item $\bigcup_{C\in \cset}V(C)=V(H)\setminus T$; 
		
		\item for every cluster $C\in \cset$, $|E(C)|\leq m/\tau$; and
		
		\item $|\bigcup_{C\in \cset}\delta_H(C)|\leq m/\tau$. 
		
	\end{itemize}
\end{lemma}

The proof of the lemma is very similar to the proof of \Cref{thm: basic decomposition of a graph} and is deferred to Section \ref{sec: appx-decomposition-small-clusters} of Appendix.

We consider the graph $C_0^+$, that is an augmentation of cluster $C_0$. Recall that $C_0^+$ is obtained from graph $G$, by first subdividing every edge $e\in \delta_G(C_0)$ with a vertex $t_e$, setting $T=\set{t_e\mid e\in \delta_G(C_0)}$, and letting $C_0^+$ be the subgraph of the resulting graph induced by $T\cup V(C_0)$. 
We apply the algorithm from \Cref{lem: decomposition into small clusters} to graph $H=C_0^+$, the set $T$ of terminals, and parameter $\tau=160\mu^{1.1}$.
Recall that, since $C_0$ contains no high-degree vertices, maximum vertex degree in $C_0$ is bounded by $\frac{m}{\mu^4}\leq \frac{m}{\check c\tau^3 \log^5 m}$. Recall also that $|T|=|\delta_G(C_0)|\leq 288m/\mu^{38}\leq m/(64\tau\log m)$.
If the algorithm from \Cref{lem: decomposition into small clusters} certifies that $\optcro(C_0^+)\geq  \Omega \left( \frac{m^2}{ \tau^4\log^5 m} \right )$, then we terminate the algorithm and  return FAIL. Notice that we are guaranteed that $\optcrors(I)\geq  \Omega \left( \frac{m^2}{ \mu^5} \right )$.

Therefore, we assume from now on that the algorithm from \Cref{lem: decomposition into small clusters} computed a collection $\cset^s$ of disjoint clusters, such that $\bigcup_{C\in \cset^s}V(C)=V(C_0)$, every cluster $C\in \cset^s$ has the $\alpha'$-bandwidth property in $G$, where $\alpha'=\frac{1}{16\alphasc(m)\cdot \log m}=\Omega\left(\frac{1}{\log^{1.5}m}\right )$, for every cluster $C\in \cset^s$, $|E(C)|\leq m/(160\mu^{1.1})$; and
 $|\bigcup_{C\in \cset^s}\delta_G(C)|\leq m/(160\mu^{1.1})$.
We refer to clusters in set $\cset^s$ as \emph{small clusters}.
Let $\cset=\cset^s\cup \cset^f$. Recall that $\Eout(\cset)$ is the set of all edges whose endpoints lie in different clusters of $\cset$. 
Since $|\bigcup_{C\in \cset^s}\delta_G(C)|\leq m/(160\mu^{1.1})$,
$|\cset^f|\leq 3\mu^4$, and, for all $C\in \cset^f$, $|\delta_G(C)|\leq 96m/\mu^{42}$, we get that:

\begin{equation}\label{eq: num of edges between clusters}
|\Eout(\cset)|\leq \frac m {160\mu^{1.1}}+(3\mu^4)\cdot \frac{96m}{\mu^{42}}\leq \frac m{80\mu^{1.1}}.
\end{equation}

\subsubsection{Step 3: Initial Disengagement}

In this step, we consider the set $\cset=\cset^s\cup \cset^f$ of clusters. Recall that all clusters in $\cset$ are disjoint and $\bigcup_{C\in \cset}V(C)=V(G)$. Moreover, every cluster in $\cset^s$ has the $\alpha'$-bandwidth property, for $\alpha'=\frac{1}{16\alphasc(m)\cdot \log m}$, while every cluster in $\cset^f$ has $1$-bandwidth property (which follows from the definition of flower clusters). Since $m\geq \mu^4$, we then get that every cluster in $\cset$ has the $\alpha_0$-bandwidth property, for $\alpha_0=1/\log^3m'$, where $m'=|E(G)|\leq 3m$.

 We apply the algorithm from \Cref{thm: disengagement - main} to instance $I=(G,\Sigma)$ of \CNwRS, with parameter $m'$ replacing $m$, and the set $\cset$ of clusters; parameter $\mu$ remains unchanged. Let $\iset_1$ be the resulting collection of instances that the algorithm computes, that is a $2^{O((\log m)^{3/4}\log\log m)}$-decomposition of instance $I$. 
 Recall that we are guaranteed that, for each instance $I'=(G',\Sigma')\in \iset_1$, $I'$ is a subinstance of $I$, and there is at most one cluster $C\in \cset$ with $E(C)\subseteq E(G')$, and all other edges of $G'$ lie in set $\Eout(\cset)$. 
We have therefore shown an efficient randomized algorithm that computes a $\nu_0$-decomposition $\iset_1$ of instance $I$, for $\nu_0= 2^{O((\log m)^{3/4}\log\log m)}$.

We partition instances in $\iset_1$ into two subsets, set $\iset'_1$ containing all instances $I'=(G',\Sigma')$ with $|E(G')|\leq m/(2\mu)$, and set $\iset''_1$ containing all remaining instances. We refer to instances in $\iset''_1$ as \emph{problematic instances}. Consider now any problematic instance $I'=(G',\Sigma')\in \iset''_1$. 
Recall that, from the guarantees of \Cref{thm: disengagement - main}, there is at most one cluster $C'\in \cset$ with $C'\subseteq G'$, and all edges of $G'$ lie in $E(C')\cup \Eout(\cset)$.
Since we are guaranteed that 
$$\bigg|\bigcup_{C\in \cset}\delta(C)\bigg|\leq \bigg|\bigcup_{C\in \cset^s}\delta(C)\bigg|+\bigg|\bigcup_{C\in \cset^f}\delta(C)\bigg|\leq m/(160\mu)+288m/\mu^{38}\leq m/(80\mu),$$ we get that $|E(G')\setminus E(C')|\leq m/(80\mu)$, and $|E(C')|\geq m/(2\mu)-m/(80\mu)$. In particular, cluster $C'$ may not be a small cluster, so it must be a flower cluster. We say that $C'$ is the flower cluster associated with the problematic instance $I'\in \iset''_1$. In the remaining phases, we will further decompose each problematic instance into subinstances, proving the following theorem.

\begin{theorem}\label{thm: decomposing problematic instances}
	There is an efficient randomized algorithm, that, given a problematic instance $I'=(G',\Sigma')\in \iset''_1$, either returns FAIL, or computes a $\nu_1$-decomposition $\tilde \iset(I')$ of $I'$, where $\nu_1=2^{O((\log m)^{3/4}\log\log m)}$, such that,  for each instance $\tilde I=(\tilde G,\tilde \Sigma)\in \tilde \iset(I')$, $|E(\tilde G)|\leq m/(2\mu)$. Moreover, if $\optcrors(I')< m^2/\left (\mu^{18}\cdot 2^{c'(\log m)^{3/4}\log\log m}\right )$ for some large enough constant $c'$, then the probability that the algorithm returns FAIL is at most $1/\mu^4$. (Here, $m=|E(G)|$, where $G$ is associated with the original instance $I$, that serves as input to \Cref{lem: not many paths}).
\end{theorem}

Before providing the proof of \Cref{thm: decomposing problematic instances}, we show that the proof of  \Cref{lem: not many paths} follows from it.
We apply the algorithm from  \Cref{thm: decomposing problematic instances}  to every problematic instance $I'\in \iset''_1$. Assume first that, for each such instance $I'$, the algortihm returns a $\nu_1$-decomposition $\tilde \iset(I')$ of $I'$, such that,  for each instance $\tilde I=(\tilde G,\tilde \Sigma)\in \tilde \iset(I')$, $|E(\tilde G)|\leq m/(2\mu)$. In this case, we return the collection 
$\iset=\iset'_1\cup \left(\bigcup_{I'\in \iset''_1}\tilde \iset(I')\right )$ of instances. From 
\Cref{claim: compose algs}, since $\nu_1\cdot\nu_2= 2^{O((\log m)^{3/4}\log\log m)}$, $\iset$ is indeed a $\nu$-decomposition of $I$, and, from the above discussion, we are guaranteed that, for every instance $\tilde I=(\tilde G,\tilde \Sigma)\in \iset$, $|E(\tilde G)|\leq m/(2\mu)$.
 
 If, for any problematic instance $I'\in \iset''_1$, the algorithm from  \Cref{thm: decomposing problematic instances} returned FAIL, then our algorithm returns FAIL as well.
 
 Recall that $\mu=2^{\check c(\log m^*)^{7/8}\log\log m^*}$, where $m^*\ge m$.
 Assume now that $\optcrors(I)<m^2/\mu^{21}$. 
 We will show that in this case, the probability that our algorithm returns FAIL is at most $O(1/\mu^2)$. Since we have assumed that $m>\mu^{50}$ (from the statement of \Cref{lem: not many paths}), 
 $|E(G)|<m^2/\mu^{20}$.
 
 Recall that $\iset_1$ is a $2^{O((\log m)^{3/4}\log\log m)}$-decomposition of instance $I$, and so:
 
  $$\expect{\sum_{I'\in \iset_1}\optcrors(I')}\leq 2^{O((\log m)^{3/4}\log\log m)}\cdot \left(\optcrors(I)+|E(G)|\right ).$$
  
  In particular, there is some constant $c$, such that  $\expect{\sum_{I'\in \iset_1''}\optcrors(I')}\leq 2^{c(\log m)^{3/4}\log\log m)}\cdot \left(\optcrors(I)+|E(G)|\right )$. Let $\event$ be the bad event that $\sum_{I'\in \iset_1''}\optcrors(I')> 8\mu^2\cdot 2^{c(\log m)^{3/4}\log\log m)}\cdot \left(\optcrors(I)+|E(G)|\right )$. From Markov's inequality, the probability that $\event$ happens is at most $1/(8\mu^2)$. 
  
  Assume now that 
  $\optcrors(I)<m^2/\mu^{21}$, and that the bad event $\event$ does not happen. Then for every problematic instance $I'\in \iset''_1$:
 \[  
 \begin{split}
 \optcrors(I')&\leq 8\mu^2\cdot  2^{c(\log m)^{3/4}\log\log m)}\cdot \left(\optcrors(I)+|E(G)|\right )\\
 &\leq 8\mu^2\cdot 2^{c(\log m)^{3/4}\log\log m)}\cdot \frac{2m^2}{\mu^{21}}\\
 &\leq  \frac{m^2}{\mu^{18}\cdot 2^{c'(\log m)^{3/4}\log\log m}},
 \end{split}
\]
where $c'$ is the constant from \Cref{thm: decomposing problematic instances}. Therefore, if  $\optcrors(I)<m^2/\mu^{21}$ and Event $\event$ does not happen, then, for every problematic instance $I'\in \iset''_1$, the probability that the algorithm from 
 \Cref{thm: decomposing problematic instances} returns FAIL is at most $1/\mu^4$. Since $\sum_{I'=(G',\Sigma')\in \iset_1}|E(G')|\leq |E(G)|\cdot (\log m)^{O(1)}$, and, for every problematic instance $I''=(G',\Sigma')\in \iset''_1$, $|E(G')|\geq m/(2\mu)$, we get that $|\iset''_1|\leq\mu\cdot (\log m)^{O(1)}$. Therefore, if $\optcrors(I)<m^2/\mu^{21}$ and $\event$ does not happen, then the probability that the algorithm from 
 \Cref{thm: decomposing problematic instances} returns FAIL for any problematic instance $I'\in \iset''_2$ is at most $O(1/\mu^2)$. Since the probability that event $\event$ happens is at most $1/(8\mu^2)$, we get that, if $\optcrors(I)<m^2/\mu^{21}$, then  the probability that our algorithm returns FAIL is at most $O(1/\mu^2)$.

In order to complete the proof of  \Cref{lem: not many paths}, it is now enough to prove  \Cref{thm: decomposing problematic instances}. From now on, we fix a single a problematic instance $I'=(G',\Sigma')\in \iset''_1$, and its corresponding flower cluster $C'\in\cset^f$, and provide an algorithm to compute a decomposition of $I'$ into subclusters with required properties.

\subsection{Phase 2: Layered Well-Linked Decomposition, Further Disengagement, and Fixing the Flower Cluster}

From now on we focus on the proof of \Cref{thm: decomposing problematic instances}. In order to simplify the notation, we denote the input problematic instance by $I=(G,\Sigma)$. Our goal is to design an efficient randomized algorithm that computes a collection $\iset'$ of instances of \cnwrs, such that, for each resulting instance $\tilde I=(\tilde G,\tilde \Sigma)\in \iset'$, $|E(\tilde G)|\leq m/\mu$,
and additionally,
$\sum_{\tilde I=(\tilde G,\tilde \Sigma)\in  \iset'}|E(\tilde G)|\leq O(|E(G)|)$, and $\expect{\sum_{\tilde I\in  \iset'}\optcrors(\tilde I)}\leq 2^{O((\log m)^{3/4}\log\log m)}\cdot \left(\optcrors(I)+|E(G)|\right )$. We also need to provide an efficient algorithm $\aset(I)$, that, given a solution $\phi(\tilde I)$ to each instance $\tilde I\in \iset'$, computes a solution $\phi$ to instance $I$, with $\cro(\phi)\leq O\left (\sum_{\tilde I\in  \iset'}\cro(\phi(\tilde I))\right )$.

We denote by $C$ the unique flower cluster of $\cset^f$ contained in $G$, by $u^*$ its center vertex, and by $\xset=\set{X_1,\ldots,X_k}$ its petals. This phase consists of two steps. In the first step, we compute a layered well-linked decomposition of the graph $G$ with respect to $C$, and perform disengagement of the resulting clusters. In the second step, we modify the flower cluster $C$ and its petals to ensure that every petal is routable in the resulting instance.

\subsubsection{Step 1: Layered Well-Linked Decomposition and Second Disengagement}

In this step, we apply the algorithm from \Cref{thm: layered well linked decomposition} to graph $G$ and cluster $C$, in order to compute 
 a valid layered $\alpha$-well-linked decomposition $(\wset, (\lset_1,\ldots,\lset_r))$ of $G$ with respect to $C$, for $\alpha=\frac{1}{c\log^{2.5}m}$, where $c$ is some large enough constant independent of $m$, and $r\leq \log m$. Note that, since we have assumed that $m$ is sufficiently large, every cluster $W\in \wset$ has the $\alpha_0$-bandwidth property, for $\alpha_0=1/\log^3m$. Recall that we are additionally guaranteed that   $\bigcup_{W\in \wset}V(W)=V(G)\setminus V(C)$, and that, for every cluster $W\in \wset$, $|\delta_G(W)|\leq |\delta_G(C)|$.
 Recall that, for every cluster $W\in \wset$ with $W\in \lset_i$, we have partitioned the set $\delta_G(W)$ of edges into two subsets: set $\delta^{\down}(W)$ connecting vertices of $W$ to vertices that lie in the clusters of $\set{C}\cup \lset_1\cup\cdots\cup \lset_{i-1}$; and set $\delta^{\up}(W)$ containing all remaining edges, and we are guaranteed that $|\delta^{\up}(W)|<|\delta^{\down}(W)|/\log m$. 
 
 Lastly, recall that, for every cluster $W\in \wset$, there is a collection $\pset(W)$ of paths in $G$, routing the edges of $\delta_G(W)$ to edges of $\delta_G(C)$, such that the paths in $\pset(W)$ avoid $W$, and cause congestion at most $200/\alpha$. Recall that, from  Properties \ref{prop: flower cluster petal intersection} and \ref{prop: flower cluster routing} of the flower cluster, there is a collection $\qset\subseteq \bigcup_{i=1}^k\qset_i$ of edge-disjoint paths, routing the edges of $\delta_G(C)$ to the vertex $u^*$, such that all inner vertices on every path lie in $C$. By concatenating the paths in $\pset(W)$ and the paths in $\qset$, we obtain a new collection $\qset'(W)$ of paths, that route the edges of $\delta_G(W)$ to vertex $u^*$, so that all inner vertices on every path lie outside $W$, and cause congestion at most $200/\alpha=O(\log^{2.5}m)$.

We partition the set $\wset$ of clusters into two subsets $\wset^{\light}$, and $\wset^{\bad}$, as follows. We apply the algorithm \algclassifycluster from \Cref{thm:algclassifycluster} to each cluster cluster $W\in \wset$ in turn, with parameter $p=1/(m^*)^4$. If the algorithm returns FAIL, then we add cluster $W$ to $\wset^{\bad}$. 
Recall that the probability that Algorithm \algclassifycluster errs, that is, it returns FAIL when $W$ is not $\eta^*$-bad, for $\eta^*=2^{O((\log m)^{3/4}\log\log m)}$, is at most $1/(m^*)^4$. Otherwise, the algorithm returns a distribution $\dset(W)$ over the set $\Lambda_G(W)$ of internal $W$-routers, such that cluster $W$ is $\beta^*$-light with respect to $\dset(W)$, where $\beta^*=2^{O(\sqrt{\log m}\cdot \log\log m)}$. We add $W$ to $\wset^{\light}$ in this case. This finishes the algorithm for partitioning the set $\wset$ of clusters into $\wset^{\light}$ and $\wset^{\bad}$. We let $\beta=\max\set{\beta^*,\eta^*}$, so that $\beta\leq 2^{O((\log m)^{3/4}\log\log m)}$.

We say that a bad event $\event_{\bad}$ happens if set $\wset^{\bad}$ contains a cluster that is not $\beta$-bad. From the above discussion, $\prob{\beta_{\bad}}\leq 1/(m^*)^3$, and every cluster $W\in \wset^{\light}$ is $\beta$-good with respect to the distribution $\dset(W)$ over the set $\Lambda_G(W)$ of internal $W$-routers.
Recall that we are also given, for every cluster $W\in \wset$, a set $\qset'(W)$ of paths routing the edges of $\delta_G(W)$ to vertex $u^*$ (external $W$-router), with congestion at most $O(\log^{2.5}m)\leq \beta$.

For every cluster $W\in \wset$, we define its distribution over the set $\Lambda'_G(W)$ of external $W$-routers to be the distribution that assign probability $1$ to the path set $\qset'(W)$.
We let $\lset$ be a laminar family of clusters of $G$, containing cluster $G$ and every cluster of $\wset$.
We now apply Algorithm \algbasicdisengagement to the instance $I=(G,\Sigma)$, using the laminar family $\lset$ of clusters, its partition $(\wset^{\bad},\wset^{\light}\cup \set{G})$, and distributions $\set{\dset(W)}_{W\in \wset^{\light}}$ and $\set{\dset'(W)}_{W\in \lset}$, that is given  in \Cref{subsec: basic disengagement}. 


We denote the resulting family of instances by $\iset_2(I)$. Recall that family $\iset_2(I)$ of instances contains a single global instance $\hat I$, and additionally, for every cluster $W\in \wset$, an instance $ I_W$.

If we denote the global instance by $\hat I=(\hat G,\hat \Sigma)$, then graph $\hat G$ is obtained from graph $G$ by contracting every cluster $W\in \wset$ into a supernode $v_W$. In particular, the flower cluster $C$ is contained in $\hat G$. For every cluster $W\in \wset$, if we denote corresponding instance by $I_W=(G_W,\Sigma_W)$, then graph $G_W$ is obtained from graph $G$ by contracting all vertices of $V(G)\setminus V(W)$ into a single vertex $v^*$.
In particular, no edge of cluster $C$ may lie in $G_W$, and so every edge of $G_W$ is an edge of set $\Eout(\cset)$, where $\cset$ is the set of clusters computed in Phase 1. Therefore, $|E(G_W)|\leq m/(2\mu)$.

Note that the depth of laminar family $\lset$ is $1$. From \Cref{lem: disengagement final cost}, if event $\event_{\bad}$ did not happen,

\[
\begin{split}
\expect{\sum_{I'\in \iset_2(I)}\optcrors(I')}& \leq  O\left (\beta^2\cdot \left(\optcrors(I)+|E(G)|\right )\right )\\
& \leq  2^{O((\log m)^{3/4}\log\log m)}\cdot \left (\optcrors(I)+|E(G)|\right ).
\end{split}
\]

If bad event $\event_{\bad}$ happened, then $\sum_{I'\in \iset_2(I)}\optcrors(I')\leq m^3$. Since $\prob{\event_{\bad}}\leq 1/(m^*)^3$, we get that overall:
\begin{equation}\label{eq: small total opt}
\begin{split}
\expect{\sum_{I'\in \iset_2(I)}\optcrors(I')}& 
 \leq  2^{O((\log m)^{3/4}\log\log m)}\cdot \left (\optcrors(I)+|E(G)|\right ).
\end{split}
\end{equation}
Additionally, from \Cref{lem: number of edges in all disengaged instances}, we get that $\sum_{I'=(G',\Sigma')\in \iset_2(I)}|E(G')|\leq O(|E(G)|)$.

Lastly, from \Cref{lem: basic disengagement combining solutions}, 
	there is an efficient algorithm, that, given, for each instance $I'\in \iset_2(I)$, a solution $\phi(I')$, computes a solution for instance $I$ of value at most $\sum_{I'\in \iset}\cro(\phi(I'))$.

Note that the set $\iset_2(I)$ of instances has all properties required in \Cref{thm: decomposing problematic instances}, with one exception: it is possible that, in the global instance $\hat I=(\hat G,\hat \Sigma)$, $|E(\hat G)|>m/(2\mu)$. 

In order to overcome this difficulty, we further decompose instance $\hat I$ into subinstances, proving the following lemma.

\begin{lemma}\label{lem: decomposing problematic instances by petal}
	There is an efficient randomized algorithm, that either returns FAIL, or computes a $\nu_2$-decomposition $\tilde \iset$ of instance $\hat I$, for $\nu_2=2^{O((\log m)^{3/4}\log\log m)}$, such that, for each instance $\tilde I=(\tilde G,\tilde \Sigma)\in \tilde \iset$, $|E(\tilde G)|\leq m/(2\mu)$. Moreover, if $\optcrors(\hat I)<\frac{ m^2}{c'' \mu^{13}}$ for some large enough constant $c''$, then the probability that the algorithm returns FAIL is  at most $1/(8\mu^4)$.
\end{lemma}

We prove the lemma below, after we complete the proof of 
\Cref{thm: decomposing problematic instances}
 using it. 
If the algorithm from \Cref{lem: decomposing problematic instances by petal} returns $\nu_2$-decomposition $\tilde \iset$ of instance $\hat I$, then we return the collection $\tilde \iset(I)=\tilde \iset\cup \set{I_W\mid W\in \wset}$ of instances, which is now guaranteed to be a $\nu_1$-decomposition of instance $I$, where $\nu_1=2^{O((\log m)^{3/4}\log\log m)}$. We are also guaranteed that,  for each instance $\tilde I=(\tilde G,\tilde \Sigma)\in \tilde \iset$, $|E(\tilde G)|\leq m/(2\mu)$. 

Assume now that
 $\optcrors(I)< m^2/\left (\mu^{18}\cdot 2^{c'(\log m)^{3/4}\log\log m}\right )$ for some large enough constant $c'$.

Recall that, from Equation \ref{eq: small total opt},
	$\expect{\sum_{I'\in \iset_2(I)}\optcrors(I')}\leq  2^{O((\log m)^{3/4}\log\log m)}\cdot \left (\optcrors(I)+|E(G)|\right )$, and in particular $\expect{\optcrors(\hat I)}\leq \nu^{*}\cdot \left (\optcrors(I)+|E(G)|\right )$, for some $\nu^{*}= 2^{O((\log m)^{3/4}\log\log m)}$. We say that a bad event $\event'$ happens if $\optcrors(\hat I)>8\mu^4\cdot \nu^{*}\cdot \left (\optcrors(I)+|E(G)|\right )$. From Markov's inequality, $\prob{\event'}\leq 1/(8\mu^4)$.
	
	By letting $c'$ be a large enough constant, we can assume that, if $\optcrors(I)< m^2/\left (\mu^{18}\cdot 2^{c'(\log m)^{3/4}\log\log m}\right )$, then  $\optcrors(I)<\frac{m^2}{16c''\mu^{18}\cdot \nu^{*}}$, where $c''$ is the constant from \Cref{lem: decomposing problematic instances by petal}. Since we have assumed (in the statement of \Cref{lem: not many paths}) that $m>\mu^{50}$ and since $\mu>\nu^*$, we get that
$|E(G)|<\frac{m^2}{16c''\mu^{18}\cdot \nu^{*}}$.
To conclude, if 	$\optcrors(I)< m^2/\left (\mu^{18}\cdot 2^{c'(\log m)^{3/4}\log\log m}\right )$, then $(\optcrors(I))+|E(G)|)<\frac{m^2}{8c''\mu^{18}\cdot \nu^{*}}$. If, additionally, Event $\event'$ did not happen, then $\optcrors(\hat I)<\frac{m^2}{c'' \mu^{11}}$. In this case, the algorithm from \Cref{lem: decomposing problematic instances by petal} may only return FAIL with probability at most $1/(8\mu^4)$. 
To conclude, if $\optcrors(I)< m^2/\left (\mu^{18}\cdot 2^{c'(\log m)^{3/4}\log\log m}\right )$, then our algorithm may return FAIL in only two cases: either (i) event $\event'$ happened (which happens with probability at most $1/(8\mu^4)$); or (ii) $\optcrors(\hat I)<\frac{c'' m^2}{ \mu^{11}}$, and yet the algorithm from \Cref{lem: decomposing problematic instances by petal} returns FAIL (which happens with probability at most  $1/(8\mu^3)$). Overall, if $\optcrors(I)<m^2/\left (\mu^{18}\cdot 2^{c'(\log m)^{3/4}\log\log m}\right )$, then the algorithm only returns FAIL with probability at most $1/(4\mu^4)$.

From now on we focus on the proof of \Cref{lem: decomposing problematic instances by petal}. Recall that we have denoted $\hat I=(\hat G,\hat \Sigma)$, and that graph $\hat G$ is obtained from $G$ by contracting every cluster $W\in \wset$ into a vertex $v_W$, that we refer to as a \emph{supernode}. We denote the resulting set of supernodes by $U=\set{v_W\mid W\in \wset}$. Recall that the flower cluster $C\subseteq\hat G$, and the edges of $E(\hat G)\setminus E(C)$ lie in $\Eout(\cset)$, where $\cset$ is the set of clusters computed in Phase 1. Therefore, $|E(\hat G)\setminus E(C)|\leq m/(160\mu)$.
Partition $(\lset_1,\ldots,\lset_r)$ of the set $\wset$ of clusters into $r\leq \log m$ layers naturally defines a partition $L_1,\ldots,L_r$ of the set $U$ of vertices into layers, where vertex $v_W$ lies in layer $L_i$ iff $W\in \lset_i$. For convenience, we denote $L_0=V(C)$. For all $1\leq i \leq r$, for every vertex $v\in L_i$, we partition the set $\delta(v)$ of its edges into two subsets: set $\delta^{\down}(v)$ connecting $v$ to vertices of $L_0\cup \cdots \cup L_{i-1}$ and set $\delta^{\up}(v)$ containing all remaining edges, that connect $v$ to vertices of $L_i\cup \cdots \cup L_r$. In the following step, we may move some vertices of $U$ from their current layer to layer $L_0$. The definition of the sets $\delta^{\down}(v'),\delta^{\up}(v')$ of edges is always with respect to the current partition of vertices of $\hat G$ into layers. Observe that Property \ref{condition: layered decomp edge ratio}  of layered well-linked decomposition ensures the following property:

\begin{properties}{P}
	\item  For every vertex $v\in U$,
$|\delta^{\up}(v)|<|\delta^{\down}(v)|/\log m$; \label{prop: layered decomp edge ratio for vertices}
\end{properties}

For convenience of notation, in the remainder of this proof we denote instance $\hat I=(\hat G,\hat \Sigma)$ by $I=(G,\Sigma)$. We use the parameter $m$ from before, so $|E(G)|\leq m$ holds.

\subsubsection{Step 2: Fixing Petals for Routability}

Recall that, as part of the definiton of the flower cluster $C$, we are given a collection $\xset=\set{X_1,\ldots,X_k}$ of petals of $C$. Consider now some petal $X_i\in \xset$. Let $\hat E_i=\delta_G(X_i)\setminus \delta_G(u^*)$, where $u^*$ is the center of the flower cluster $C$. 
We will use the following definition.

\begin{definition}
	Let $G$ be a graph, and let $C^f$ be a flower cluster in $G$, with center $u^*$ and a set $\xset=\set{X_1,\ldots,X_k}$ of petals. For $1\leq i\leq k$, we say that petal $X_i$ is \emph{routable} in $G$  if there is a collection $\qset'_i=\set{Q'(e)\mid e\in \hat E_i}$ of paths in $G$, where for each edge $e\in \hat E_i$, path $Q'(e)$ has $e$ as its first edge, terminates at vertex $u^*$, and its inner vertices are disjoint from $X_i$, such that the paths in $\qset'_i$ cause congestion at most $3000$.
\end{definition}

 As we show later, if every petal in $\xset$ is routable, then we can decompose the current instance $I$ into smaller instances, each of which will correspond to a distinct petal in $\xset$ (together with an additional ``global'' instance). Unfortunately, it is possible that some petals in $\xset$ are not routable. We overcome this difficulty by ``fixing'' the flower cluster $C$. We do so iteratively, while ensuring that Property \ref{prop: layered decomp edge ratio for vertices} continues to hold after each iteration. In every iteration, we select some vertex of $U$ to be added to some petal $X_i$ of $\xset$. The set 
$\hat E_i=\delta_G(X_i)\setminus \delta_G(u^*)$ of edges is always defined with respect to the current petal $X_i$. In addition to maintaining Property \ref{prop: layered decomp edge ratio for vertices}, we will maintain the following important property:

\begin{properties}[1]{P}
	\item  For every petal $X_i\in \xset$, there is a set $\qset_i=\set{Q(e)\mid e\in \hat E_i}$ of edge-disjoint paths, where for each edge $e\in \hat E_i$, path $Q(e)$ has $e$ as its first edge, vertex $u^*$ as its last vertex, and all inner vertices of $Q(e)$ lie in $X_i$.  \label{prop: routing inside clusters}
\end{properties}

We now describe the algorithm for fixing the petals of $C$. While there is some petal $X_i\in \xset$, and some vertex $v\in U$, such that at least $|\delta_G(v)|/2$ neighbors of $v$ in $G$ lie in $X_i$, we add $v$ to $X_i$, and remove it from $U$. 
In other words, we update $X_i$ to be the subgraph of $G$ induced by vertex set $V(X_i)\cup \set{v}$, and we update $C$ to be the subgraph of $G$ induced by vertex set $V(C)\cup\set{v}$. 
We also remove $v$ from its current layer $L_j$ and add it to $L_0$. It is immediate to verify that Property \ref{prop: layered decomp edge ratio for vertices} continues to hold after each iteration. We now show that the same is true for Property \ref{prop: routing inside clusters}.

Consider an iteration, when some vertex $v\in U$ was added to some petal $X_i\in \xset$. Partition the edges of $\delta_G(v)$ into two subsets: set $\delta'(v)$ connecting vertex $v$ to vertices of $X_i$, and set $\delta''(v)$ containing all remaining edges. From our definitions, at the beginning of the current iteration, $\delta'(v)\subseteq \hat E_i$ held. Therefore, set $\qset_i$ contained, for each edge $e\in \delta'(v)$, a path $Q(e)$, connecting $e$ to $u^*$, such that all inner vertices of $Q(e)$ belong to $X_i$. At the end of the current iteration, the edges of $\delta'(v)$ no longer lie in $\hat E_i$, and the edges of $\delta''(v)$ are added to $\hat E_i$ instead. Since $|\delta''(v)|\leq |\delta'(v)|$, we can define a mapping $M$, that maps every edge of $\delta''(v)$ to a distinct edge of $\delta'(v)$. We update the set $\qset_i$ of paths as follows: first, we remove from it all paths whose first edge lies in $\delta'(v)$. Next, for each edge $e\in \delta''(v)$, we add a new path $Q(e)$ to $\qset_i$, that is obtained by appending $e$ to the original path $Q(e')$, where $e'=M(e)$ is the edge of $\delta'(v)$ to which edge $e$ is mapped. Therefore, Property \ref{prop: routing inside clusters} continues to hold after each iteration. Lastly, we consider the cluster $C$ and its corresponding set $\xset$ of petals obtained at the end of the algorithm. 

We slightly modify Property \ref{prop: flower cluster small boundary size} of the flower cluster, and replace it with the following  property, that we refer to as Modified  Property \ref{prop: flower cluster small boundary size}: 

$$|\delta_G(C)|\leq 96m/\mu^{42} \mbox{ and\ \  }|\bigcup_{i=1}^k\delta(X_i)|\leq \frac{192m}{\mu^{42}}.$$ 

If Properties 
\ref{prop: flower cluster vertex induced petals too} --  \ref{prop: flower cluster routing} hold for a cluster $C'$, with Property \ref{prop: flower cluster small boundary size} replaced with its modified counterpart, then we say that $C'$ is a \emph{modified flower cluster}. We now prove that cluster $C$ is a valid modified cluster.

\begin{claim}\label{claim: remains flower cluster}
	Cluster $C$ is a valid modified flower cluster in the current graph $G$.
\end{claim}

\begin{proof}
	It is immediate to verify that throughout the algorithm, Property \ref{prop: flower cluster vertex induced petals too} continues to hold.
	Recall that, from Property \ref{prop: flower cluster small boundary size} in the definition of a flower cluster, at the beginning of the algorithm, $|\delta_G(C)|\leq 96m/\mu^{42}$ held. We claim that this property continues to hold throughout the algorithm. Indeed, when a vertex $v\in U$ is added to $C$, there is some petal $X_i\in \xset$, such that $|\delta'(v)|\geq |\delta''(v)|$, where $\delta'(v)$ contains all edges connecting $v$ to vertices of $X_i$, and $\delta''(v)$ contains all remaining edges of $\delta(v)$. Notice that edges of $\delta'(v)$ are removed from $\delta_G(C)$ at the end of the iteration, while only the edges of $\delta''(v)$ may be added to $\delta_G(C)$ at the end of the current iteration. Therefore, $|\delta_G(C)|$ does not increase, and modified Property \ref{prop: flower cluster small boundary size} continues to hold throughout the algorithm. From the above discussion, whenever a vertex $v$ is added to cluster $C$, $\deg_G(v)\leq 2|\delta_G(C)|\leq 192m/\mu^{42}$ (from Property \ref{prop: flower cluster small boundary size}). Therefore, Property \ref{prop: flower center has large degree, everyone else no} holds throughout the algorithm. 
It is immediate to verify that Properties  \ref{prop: flower cluster petal intersection} and \ref{prop: flower cluster edges near center partition} continue to hold throughout the algorithm, and we have already established Property \ref{prop: flower cluster routing} for the final cluster $C$.
\end{proof}

Lastly, we show that, once the algorithm terminates, every petal in $\xset$ is routable in $G$.

\begin{claim}\label{claim: routable petals}
	At the end of the algorithm, every petal of $\xset$ is routable in $G$.
\end{claim}

\begin{proof}
Consider some petal $X_i\in \xset$. Recall that we have defined the set $\hat E_i=\delta_G(X_i)\setminus \delta_G(u^*)$ of edges, where $u^*$ is the center of the flower cluster $C$. Recall that our goal is to show that  there is a collection $\qset'_i=\set{Q'(e)\mid e\in \hat E_i}$ of paths, where for each edge $e\in \hat E_i$, path $Q'(e)$ has $e$ as its first edge, terminates at vertex $u^*$, and is internally disjoint from $X_i$, such that the paths in $\qset'_i$ cause congestion at most $3000$. Let $\hat \qset_i$ be the set of all paths in graph $G$, where each path $Q\in \hat\qset_i$ contains some edge of $\hat E_i$ as its first edge, terminates at vertex $u^*$, and is internally disjoint from $X_i$. From the integrality of flow, it is enough to show that there exists a flow $\hat f_i$, defined over the set $\hat \qset_i$ of paths, in which every edge of $\hat E_i$ sends one flow unit, such that flow $\hat f_i$ causes congestion at most $3000$. From now on we focus on proving that such a flow indeed exists.

For the sake of the proof we will define layer $L_0$ slightly differently than before: we let $L_0=V(C)\setminus V(X_i\setminus\set{u^*})$.
For each index $0\leq j\leq r$, we let $S_j=L_0\cup L_1\cup\cdots\cup L_j$. We then let the set $E^*_j$ of edges contain all edges of $\delta_G(S_j)$, except for those insident to vertex $u^*$. Notice that in particular, since $S_r=V(G)\setminus (V(X_i)\setminus\set{u^*})$, edge set $E^*_r$ is precisely the edge set $\hat E_i=E_G(X_i)\setminus \delta_G(u^*)$.
For all $0\leq j\leq r$, we let $\pset^*_j$ be the set of all paths $P$, such that the first edge of $P$ lies in $E^*_j$, the last vertex of $P$ is $u^*$, and all inner vertices of $P$ lie in $S_j$. 
We prove the following claim.
	
	\begin{claim}\label{claim: layer by layer}
		For all $0\leq j\leq r$, there is a flow $f^*_j$ defined over the set $\pset^*_j$ of paths, in which every edge of $E^*_j$ sends one flow unit, such that the paths in $\pset^*_j$ cause congestion at most $\left (1+\frac 8 {\log m}\right )^j$.
	\end{claim}

Note that proof of \Cref{claim: layer by layer} will finish the proof of \Cref{claim: routable petals}. Indeed, 
as observed already, $E^*_r=\hat E_i$, and it is easy to verify that $\pset^*_r=\hat \qset_i$. In flow $f^*_r$, every edge of $\hat E_i$ sends one flow unit, as required, and the congestion of the flow is at most $\left (1+\frac 8 {\log m}\right )^r\leq 3000$, since $r\leq \log m$. Therefore, in order to complete the proof of  \Cref{claim: routable petals}, it is now enough to prove \Cref{claim: layer by layer}, which we do next.

\begin{proofof}{\Cref{claim: layer by layer}}
The proof is by induction on $j$. The base is when $j=0$. 
	Recall that $S_0=L_0=V(C)\setminus \left (V(X_i)\setminus u^*\right )$. The set $E^*_0$ of edges is then a subset of $\bigcup_{i'\neq i}\hat E_{i'}$. 
	Recall that, from the definition of the flower
	cluster, for all $1\leq i'\leq k$, there is a collection $\qset_{i'}$ of edge-disjoint paths routing the edges of $\hat E_{i'}$ to vertex $u^*$, with all inner vertices on every path contained in $X_{i'}$. 
	For each index $i\neq i'$, for each edge $e\in \hat E_{i'}$, let $Q(e)\in \qset_{i'}$ be the unique path whose first edge is $e$. Observe that $\bigcup_{i'\neq i}\qset_{i'}\subseteq \pset^*_0$. By sending one flow unit on each path in $\set{Q(e)\mid e\in \hat E_0}$, we obtain the desired flow $f^*_0$, defined over the set $\pset^*_0$ of paths, in which each edge of $E^*_0$ sends one flow unit. The congestion of the flow is $1$.

We now prove that the claim holds for an index $1\leq j\leq r$, provided that it holds for index $j-1$.

Consider some vertex $v\in L_j$. We partition the edges of $\delta^{\down}(v)$ into two subsets: set $\delta_1^{\down}(v)$ containing all edges connecting $v$ to vertices of $X_i$, and set $\delta_2^{\down}(v)$ containing all remaining edges of $\delta^{\down}(v)$. 
Notice that the edges of $\delta_2^{\down}(v)$ lie in $E^*_{j-1}$ but not in $E^*_j$, while edges of $\delta^{\up}(v)$ lie in $E^*_j$ but not in $E^*_{j-1}$. In fact, since $S_j=S_{j-1}\cup L_j$, $E^*_j\subseteq \left (E^*_{j-1}\setminus \left(\bigcup_{v\in L_j}\delta_2^{\down}(v)\right )\right )\cup  \left(\bigcup_{v\in L_j}\delta^{\up}(v)\right )$ (we use inclusion rather than equality since an edge of $\delta^{\up}(v)$ may connect $v$ to a vertex of $L_j$).

Consider again some vertex $v\in L_j$.
Recall that $|\delta^{\down}_1(v)|\leq |\deg_G(v)|/2$ (since the algorithm for fixing the flower cluster $C$ has terminated), while
$|\delta^{\up}(v)|\leq |\delta^{\down}(v)|/\log m\leq \deg_G(v)/\log m$. Therefore, $|\delta^{\down}_2(v)|\geq \left (\frac 1 2 -\frac 1 {\log m}\right )\deg_G(v)$, while $|E^*_j\cap \delta_G(v)|\subseteq |\delta^{\up}(v)|+|\delta^{\down}_1(v)|\leq \left (\frac 1 2 +\frac 1 {\log m}\right )\deg(v)$. Overall, we get that $|E_j^*\cap \delta_G(v)|\leq \left (1+\frac 8 {\log m}\right )|\delta_2^{\down}(v)|$.

Let $\rset_j(v)$ denote the collection of all paths that can be obtained by combining two edges: an edge of $E^*_j\cap \delta_G(v)$ and an edge of $\delta_2^{\down}(v)$; the paths are directed towards edges of $\delta_2^{\down}(v)$. Clearly, there is a flow $f'_v$, defined over the paths in $\rset_j(v)$, where every edge of $E^*_j\cap \delta_G(v)$  sends one flow unit, every edge of $\delta_2^{\down}(v)$ receives at most $\left (1+\frac 8 {\log m}\right )$ flow units, and the flow causes congestion at most $\left (1+\frac 8 {\log m}\right )$ (for example, we can obtain such a flow by spreading the flow originating at every edge of $E^*_j\cap \delta_G(v)$ evenly among the edges of $\delta_2^{\down}(v)$). 

Next, we define a new flow $f_j(v)$, in which every edge of $E^*_j\cap \delta_G(v)$ sends one flow unit via a subset of paths of $\pset^*_j$.
Consider any flow-path $R\in \rset_j(v)$, and assume that $R$ consists of two edges: $e\in  E^*_j\cap \delta_G(v)$ and $e'\in \delta_2^{\down}(v)$. Recall that edge $e'$ sends $1$ flow unit in flow $f^*_{j-1}$. For every path $P\in \pset^*_{j-1}$ whose first edge is $e'$, we consider a path $R^P$ obtained by appending the edge $e$ at the beginning of the path, so that path $R^P$ now starts with edge $e$, and terminates at vertex $u^*$ as before. Observe that path $R^P$ lies in path set $\pset^*_{j}$. Let $x=f_{j-1}^*(P)$ be the amount of flow sent via path $P$ in flow $f_{j-1}$, and let $x'=f'_v(R)$ be the amount of flow sent via path $R$ in flow $f'_v$. We then send $(x\cdot x')$ flow units via path $R^P$ in flow $f_j(v)$. Notice that, since every edge of $E^*_j\cap \delta_G(v)$ sends one flow unit in flow $f'_v$, and every edge of $\delta_2^{\down}(v)$ sends one flow unit in flow $f_{j-1}$, this ensures that every edge of $E^*_j\cap \delta_G(v)$ sends one flow unit in the new flow $f_j(v)$ that we just defined. Moreover, since every edge $e'\in \delta_2^{\down}(v)$ receives at most $\left (1+\frac 8 {\log m}\right )$ flow units in $f'_v$, for each flow-path $P\in \pset^*_{j-1}$ whose first edge is $e'$, 
the total amount of flow sent along path $P$ in the new flow $f_j(v)$ is at most $\left (1+\frac 8 {\log m}\right )$ times the amount of flow sent via path $P$ in $f^*_{j-1}$. In other words, we can think of flow $f_j(v)$ as obtained as follows: we start with flow $f^*_{j-1}$, and discard flow on all flow-paths except those whose first edge lies in $\delta_2^{\down}(v)$. Next, we scale the flow on each resulting flow-path by at most factor $\left (1+\frac 8 {\log m}\right )$. Lastly, we combine the resulting flow with flow $f'_v$.

We are now ready to define the final flow $f^*_j$. 
Recall that the set $E^*_j$ of edges can be obtained from edge set $E^*_{j-1}$ by first deleting the edges of $\bigcup_{v\in L_j}\delta_2^{\down}(v)$ from it, and then adding a subset of the edges of $\bigcup_{v\in L_j}\delta^{\up}(v)$ to it. For every edge $e\in E^*_j\cap E^*_{j-1}$, for every path $P\in \pset^*_{j-1}\cap \pset^*_{j}$, whose first edge is $e$, the flow $f^*_j(P)$ remains the same as the flow $f^*_{j-1}(P)$. This ensures that each edge of $E^*_j\cap E^*_{j-1}$ sends one flow unit in the new flow, as $\pset_{j-1}\subseteq \pset_j$. For every vertex $v\in L_j$, we use the flow $f_j(v)$ in order to send flow from the edges of $\delta^{\up}(v)\cap E^*_j$. Specifically, for each edge $e\in \delta^{\up}(v)\cap E^*_j$, for every path $P\in \pset^*_j$ whose first edge is $e$, we set the flow $f_j^*(P)$ to be equal to the flow sent via this path by $f_j(v)$. This ensures that every edge in $E^*_j\setminus E^*_{j-1}$ sends one flow unit in the new flow $f^*_j$. This finishes the description of the flow $f^*_j$. From the above discussion, every edge of $E^*_j$ sends one flow unit in $f^*_j$. It now remains to analyze the congestion of the flow. 

Observe that flow $f^*_j$ can be obtained as follows. We start with the flow $f^*_{j-1}$, and we scale the flow on some of the flow-paths by at most factor $\left (1+\frac{8}{\log m}\right)$ (this is since for every vertex $v\in L_j$, for each edge $e\in \delta_2^{\down}(v)$, edge $e$ receives at most $\left (1+\frac{8}{\log m}\right )$ flow units via flow $f'_v$, and this flow utilizes the flow that $e$ sends in $f^*_{j-1}$ in order to reach vertex $u^*$). Lastly, we combine the resulting flow with the flows $f'_v$ for all vertices $v\in L_j$. It is then easy to verify that the congestion caused by flow $f^*_j$ is bounded by the congestion caused by flow $f^*_{j-1}$ times $\left (1+\frac{8}{\log m}\right )$. Since, from the induction hypothesis, flow $f^*_{j-1}$ causes congestion at most 
$\left (1+\frac 8 {\log m}\right )^{j-1}$, flow $f^*_j$ causes congestion at most $\left (1+\frac 8 {\log m}\right )^j$.	
\end{proofof}

\end{proof}

\subsection{Phase 3: Petal-Based Disengagement and the Final Family of Instances}
\label{subsec: phase 3}

In this subsection we first compute a collection $\iset_3$ of subinstances of the current instance $I=(G,\Sigma)$. These subinstances will ``almost'' have all properties required in \Cref{lem: decomposing problematic instances by petal}, except that we will not be able to guarantee that for each resulting subinstance $\tilde I=(\tilde G,\tilde E)$, $|E(\tilde G)|\leq m/(2\mu)$. However, we will guarantee that each such resulting graph $\tilde G$ does not have vertices whose degree is at least $m/\mu^4$. This fact will be used in the second part of this subsection in order to further decompose each subinstance of $\iset_3$ into smaller subinstances. This will be done using an algorithm similar to that from Phase 1, except that now, since the instances we apply the algorithm to do not have high-degree vertices, we will not obtain any flower clusters, and so each subinstance obtained in this final phase will be sufficiently small. 

The intuition for the current phase is that we would like to define a set of clusters in the current graph $G$, using the set $\xset=\set{X_1,\ldots,X_k}$ of petals of the flower cluster, and then perform basic disengagement, described in \Cref{subsec: basic disengagement} on instance $I$ with the set $\xset$ of clusters. Unfortunately, the set $\xset$ of clusters is not laminar, as the clusters in $\xset$ all share vertex $u^*$. In order to overcome this obstacle, the algorithm in this phase consists of three steps. In the first step we ``split'' the vertex $u^*$, by creating new vertices $u_1,\ldots,u_k$, each of which is then added to a distinct cluster $X_i$. We show that the optimal solution value to the resulting split instance that we construct is bounded by $\optcrors(I)$, and that any solution to this new split instance can be efficiently transformed into a solution to the original instance, by only slightly increasing the solution cost. In the second step, we perform a basic disengagement of the new split instance using the modified set $\xset$ of clusters. We show that each of the resulting instances does not contain  vertices of degree at least $m/\mu^4$. We also bound the total solution cost and the total number of edges in the new instances. In the third and the final step we further decompose each resulting instance, exploiting the low degrees of its vertices. We now describe each of the steps in turn.

\subsubsection{Step 1: the Split Instance}
Recall that we are given an instance $I=(G,\Sigma)$ of the \cnwrs problem, and a (modified) flower cluster $C\subseteq G$ with center $u^*$, and a set $\xset=\set{X_1,\ldots,X_k}$ of petals, such that each petal is routable in $G$. 

\begin{figure}[h]
	\centering
	\subfigure[An original petal $X_i$. Paths of $\qset_i$ are shown in pink and paths of $\qset'_i$ are shown in orange.]{\scalebox{0.4}{\includegraphics[scale=0.35]{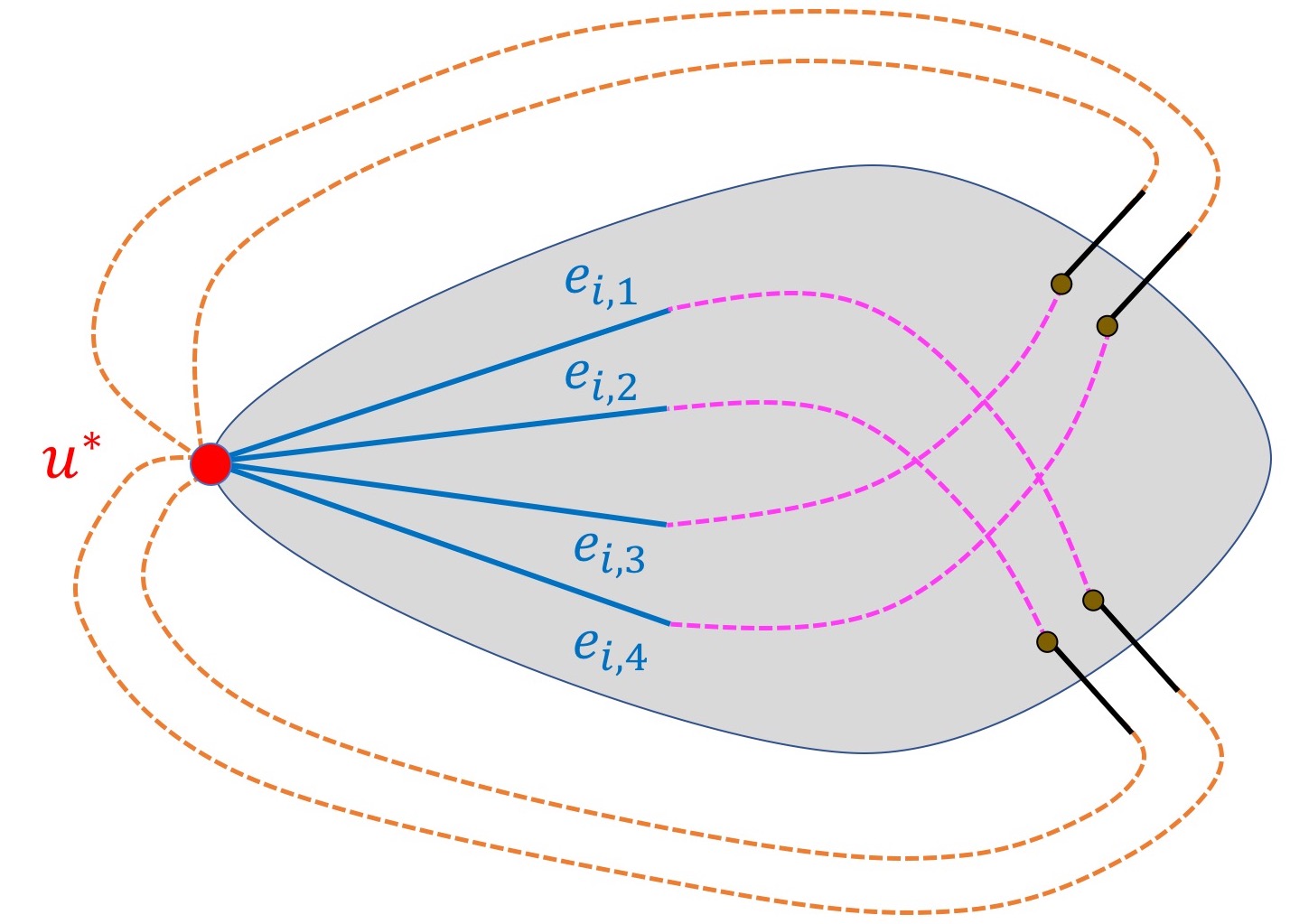}}\label{fig: original_petal}
	}
	\hspace{0.4cm}
	\subfigure[New cluster $X'_i$. 
	]{
		\scalebox{0.34}{\includegraphics[scale=0.5]{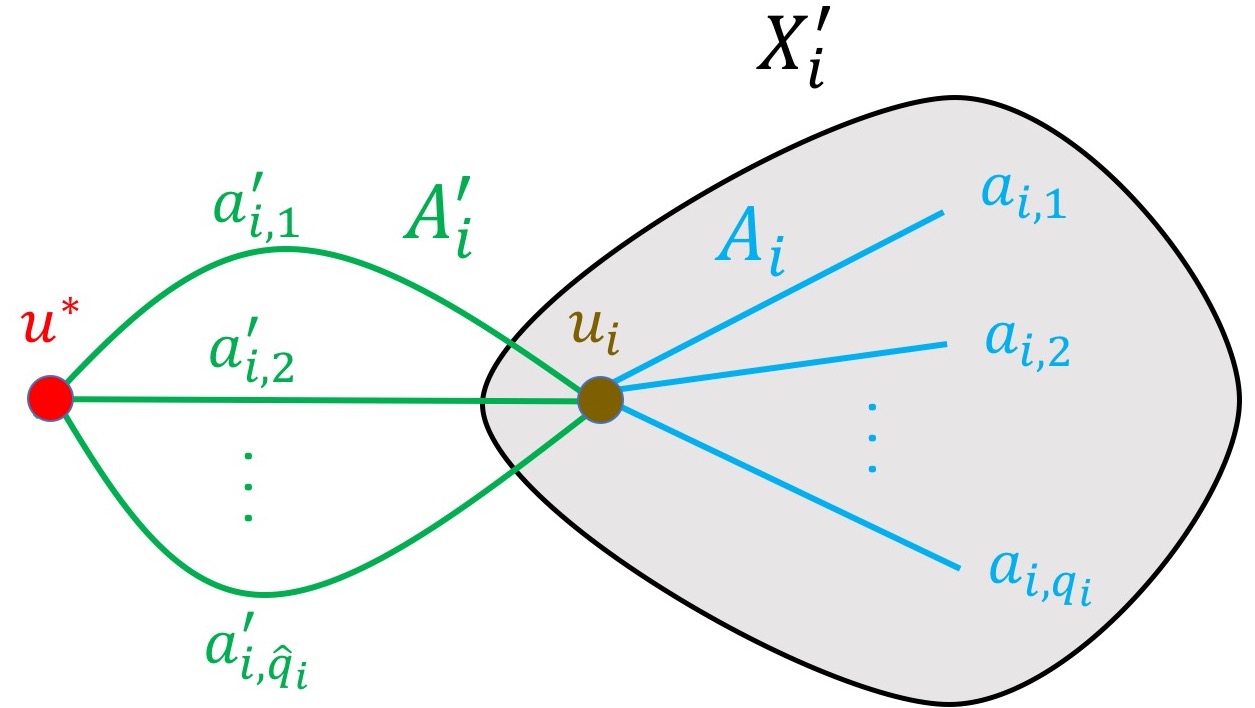}}\label{fig: new_petal}}
	\caption{Construction of a split instance $I'=(G',\Sigma')$.
	} 
\end{figure}

For all $1\leq i\leq r$, we let $E_i=E(X_i)\cap \delta_G(u^*)$, and we denote $E_i=(e_{i,1},\ldots,e_{i,q_i})$, where the edges are indexed according to their order in the ordering $\oset_{u^*}\in \Sigma$; in other words, the ordering of the set $\delta_G(u^*)$ of edges in $\Sigma$ is: $\oset_{u^*}=(e_{1,1},\ldots,e_{1,q_1},e_{2,1},\ldots,e_{2,q_2},\ldots,e_{k,1},\ldots,e_{k,q_k})$. For all $1\leq i\leq r$, we also let $\hat E_i=\delta_G(X_i)\setminus\delta_G(u^*)$, and we denote $|\hat E_i|=\hat q_i$; see \Cref{fig: original_petal} for an illustration.

Recall that, from Property \ref{prop: flower cluster routing} of a flower cluster, there is a set $\qset_i$ of edge-disjoint paths routing the edges of $\hat E_i$ to the edges of $E_i$, such that every inner vertex on every path lies in $X_i$, and, since petal $X_i$ is routable in $G$, there is a set $\qset'_i$ of paths in graph $G$, routing the edges of $\hat E_i$ to vertex $u^*$ such that the paths are internally disjoint from $X_i$ and cause congestion at most $3000$. 

In order to define the new split instance $I'=(G',\Sigma')$, we start with a graph $G'=G\setminus\delta_G(u^*)$. We then add $k$ new vertices $u_1,\ldots,u_k$ to $G'$. Next, we process each index $1\leq i\leq k$ one by one. When index $i$ is processed, 
we add a collection $A'_i=\set{a'_{i,1},\ldots,a'_{i, \hat q_i}}$ of $ \hat q_i$ parallel edges connecting $u^*$ to $u_i$ (recall that $\hat q_i=|\hat E_i|$). Additionally, for every edge $e_{i,j}=(u^*,x_{i,j})\in E_i$, we add a new edge $a_{i,j}=(u_i,x_{i,j})$ to graph $G'$; we view $a_{i,j}$ as a copy of edge $e_{i,j}$, and we will not distinguish between these edges. We denote $A_i=\set{a_{i,j}\mid 1\leq j\leq q_i}$. In order to complete the construction of graph $G'$, for every edge $e=(u,v)\not\in \delta_G(u^*)$ of the graph $G$ whose endpoints lie in different petals, we subdivide the edge $e$ with a new vertex $y_e$. 


For all $1\leq i\leq k$, we let $X'_i$ be the subgraph of $G'$ induced by $(V(X_i)\setminus \set{u^*})\cup \set{u_i}$. Notice that graph $X'_i$ is completely identical to graph $X_i$, except that vertex $u^*$ is replaced by vertex $u_i$. 
For all $1\leq i\leq k$, we denote by $\hat A_i=\delta_{G'}(X'_i)\setminus A'_i$, where $A'_i$ is the set of parallel edges connecting $u_i$ to $u^*$ (see \Cref{fig: new_petal}).  It is easy to see that there is a one-to-one correspondence between edges of $\hat A_i$ in graph $G'$ and edges of $\hat E_i$ in graph $G$.

In order to complete the definition of the split instance $I'$, we need to define its corresponding rotation system $\Sigma'$. It is easy to verify that, every vertex $v\in V(G')\setminus \set{u^*,u_1,\ldots,u_k}$ whose degree in $G'$ is greater than $2$, $\delta_{G'}(v)=\delta_G(v)$ holds (we do not distinguish here between edges whose endpoints lie in different petals of $G$ and their subdivided counterparts). For each such vertex, we set the ordering $\oset'_v\in \Sigma'$ of the edges of $\delta_{G'}(v)$ to be the same as the ordering $\oset_v\in \Sigma$ of the edges of $\delta_{G}(v)$. Note that $\delta_{G'}(u^*)=A'_1\cup\cdots\cup A'_k$. We set the ordering  $\oset'_{u^*} \in \Sigma'$ of the edges of $\delta_{G'}(u^*)$ to be $(a'_{1,1},\ldots,a'_{1, \hat q_1},a'_{2,1},\ldots,a'_{2, \hat q_2},\ldots,a'_{k,1},\ldots,a'_{k, \hat q_k} )$.  In other words, edges in sets $A'_1,\ldots,A'_k$ appear in this order of their sets, and within each set $A'_i$, the edges of $\set{a'_{i,j}}_{j=1}^{ \hat q_i}$ are ordered in the increasing order of index $j$.
Lastly, for all $1\leq i\leq k$, we define the ordering $\oset'_{u_i}\in \Sigma'$ of the edges of $\delta_{G'}(u_i)=A'_i\cup A_i$ to be: $(a'_{i,1},a'_{i,2},\ldots,a'_{i, \hat q_i}, a_{i, q_i}, a_{i, q_i-1},\ldots, a_{i,1})$.

\begin{figure}[h]
	\centering
	\subfigure[Schematic view of graph $G'$ when $C$ is a $4$-petal flower cluster.]{\scalebox{0.37}{\includegraphics[scale=0.4]{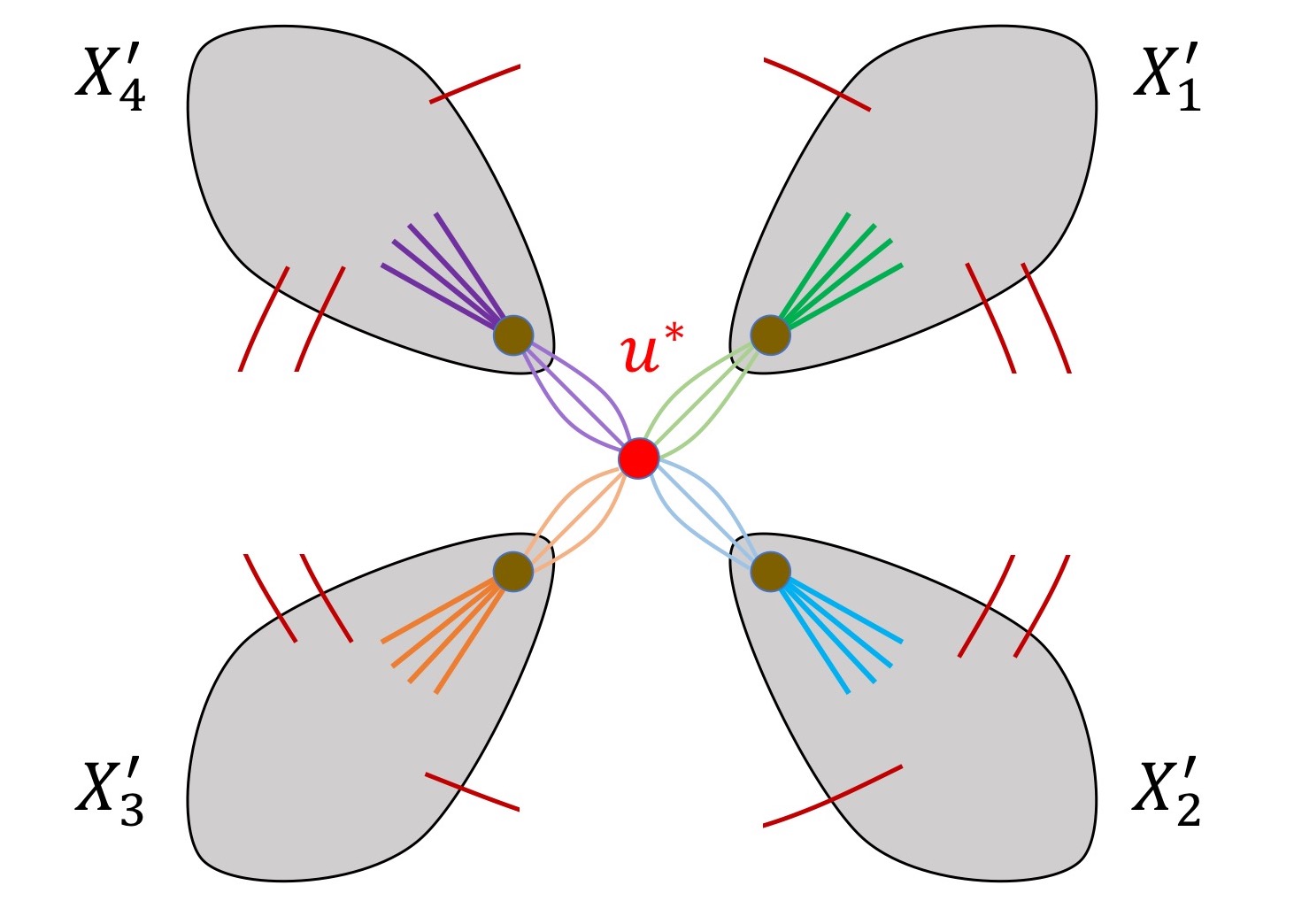}}
	}
	\hspace{0.1cm}
	\subfigure[The ordering $\oset'_{u^*}\in \Sigma'$ of edges of $\delta_{G'}(u^*)$.]{
		\scalebox{0.4}{\includegraphics[scale=0.444]{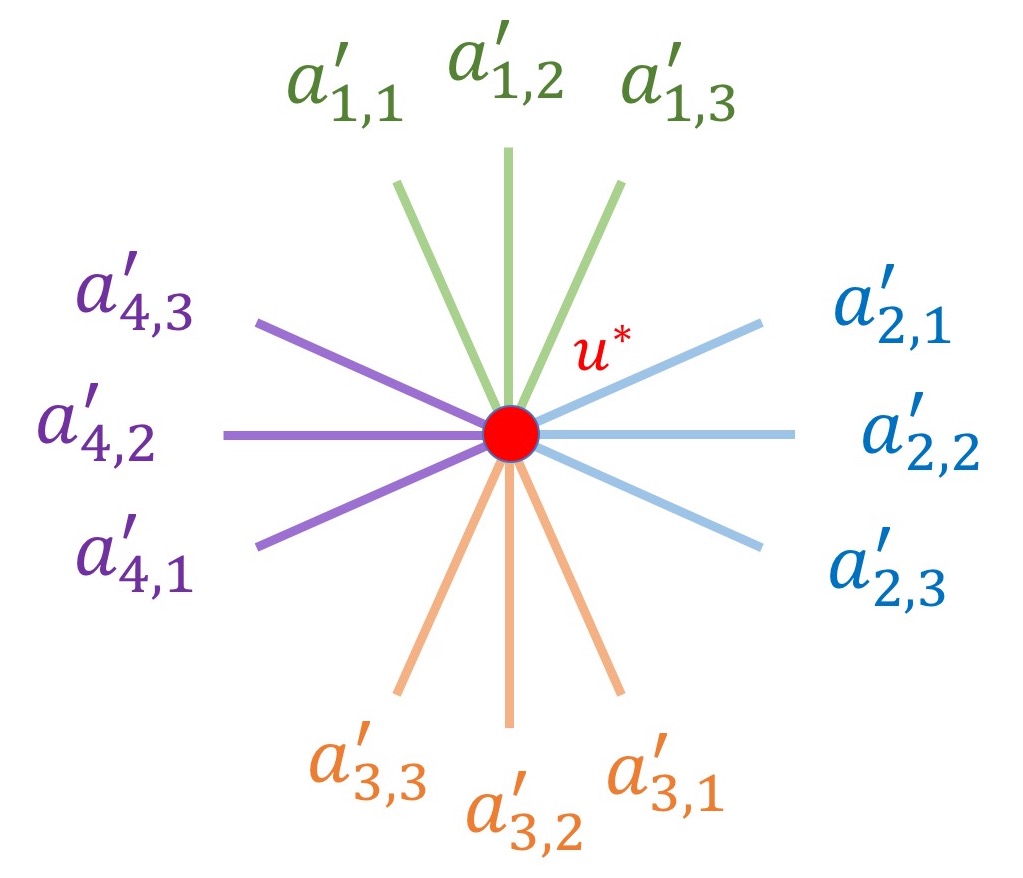}}\label{fig: rotation at u^*}}
	\caption{Split instance $I'=(G',\Sigma')$.
}\label{fig: split_flower}
\end{figure}

This completes the definition of the new split instance $I'=(G',\Sigma')$; 
see \Cref{fig: split_flower} for an illustration.
We now establish some of its properties. We start with the following easy observation:

\begin{observation}\label{obs: split instance cost}
	$|E(G')|\leq 4|E(G)|$, and $\optcrors(I')\leq \optcrors(I)$.
\end{observation}

\begin{figure}[h]
	\centering
	\subfigure[Before: the images of the original vertex $u^*$ and its incident edges in $\phi$.
	]{\scalebox{0.45}{\includegraphics[scale=0.53]{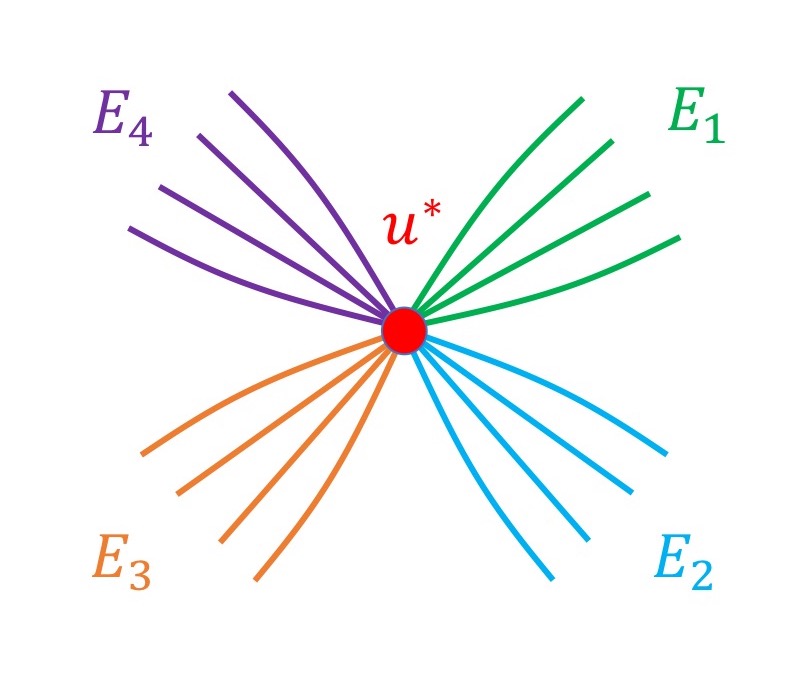}}
	}
	\hspace{2cm}
	\subfigure[After: the images of the new vertices $u, u_1,\ldots,u_k$ and their incident edges in $\phi'$. 
	]{
		\scalebox{0.45}{\includegraphics[scale=0.5]{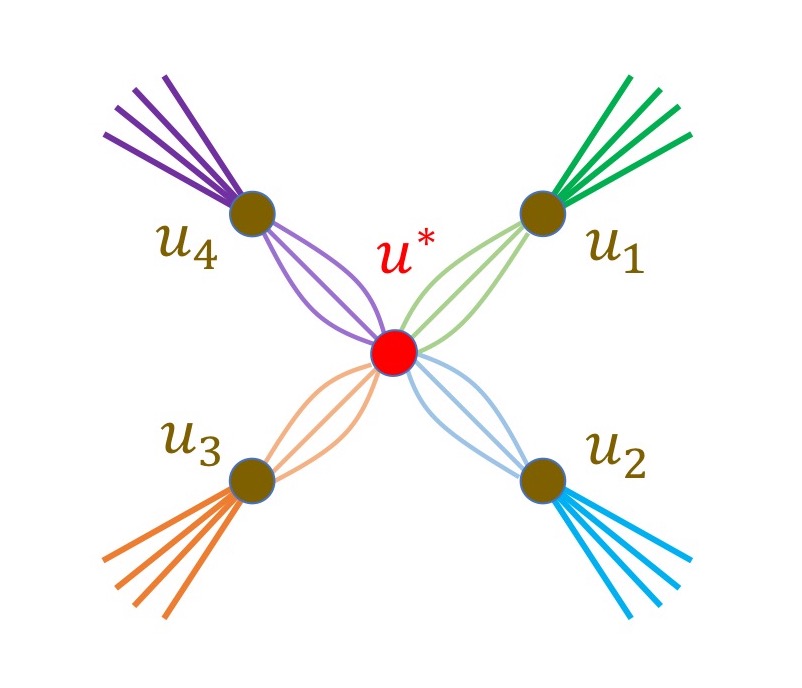}}}
	\caption{Transforming a solution for instance $I$ into a solution for instance $I'$.}\label{fig: split_drawing}
\end{figure}

The proof of \Cref{obs: split instance cost} is immediate. The first statement is immediate to see. For the second statement, given any solution $\phi$ to instance $I$, we can obtain a solution $\phi'$ to instance $I'$ by splitting the vertex $u^*$ to obtain images of vertices $u_1,\ldots,u_k$ and images of the edges in sets $A_1',\ldots,A_k'$ in a natural way  (see \Cref{fig: split_drawing}), and subdividing images of edges whose endpoints lie in different petals of $G$. 

The next lemma shows that a solution to instance $I'$ can be transformed into a solution to instance $I$ while only slightly increasing the solution cost.
The proof uses arguments similar to those used in basic and advanced disengagement, but is somewhat tedious, and is deferred to \Cref{apd: Proof of solution to split to solution to original}.

\begin{lemma}\label{lem: solution to split to solution to original}
	There is an efficient algorithm that, given a solution $\phi'$ to instance $I'$, computes a solution $\phi$ to instance $I$, with $\cro(\phi)\leq O(\cro(\phi'))$.
	\end{lemma}

\subsubsection{Step 2: Disengagement of the Petals}

In this step, we consider the split instance $I'=(G',\Sigma')$ that was constructed in Step 1 of the current phase, and we will apply Algorithm \algbasicdisengagement from \Cref{subsec: basic disengagement} to this instance, together with the family $\lset=\set{X'_1,\ldots,X'_k}$ of clusters in order to perform a basic disengagement of these clusers, with a parameter $\beta=c(\log m)^{18}$, where $c$ is a large enough constant. Note that the clusters of $\lset$ are disjoint, so $\lset$ is a laminar family of clusters. In order to be able to use \algbasicdisengagement, we need to define, for each cluster $X'_i$, a distribution $\dset'(X'_i)$ over the set $\Lambda'(X'_i)$ of external routers for $X'_i$.

Consider  some cluster $X'_i\in \lset$. Recall that petal $X_i$ is routable in $G$, and so there is a set $\qset'_i$ of paths in $G$, routing the edges of $\hat E_i=\delta_G(X_i)\setminus\delta_G(u^*)$ to vertex $u^*$, such that the paths in $\qset'_i$ cause congestion at most $3000$, and they are internally disjoint from $X_i$. By suitably subdividing the first edge of every path in $\qset'_i$, and by replacing the last edge $e_{i',j}$ on each such path by the corresponding edge $a_{i',j}$, we obtain a collection $\qset''_i$ of paths in graph $G'$, routing the edges of $\hat A_i$ to vertices of $\set{u_{i'}}_{i'\neq i}$, such that the paths in $\qset''_i$ are internally disjoint from $X'_i$, and cause congestion at most $3000$. Note that, for all $1\leq i'\leq k$ with $i'\neq i$, the number of paths terminating at vertex $u_{i'}$ is at most $3000|\hat A'_{i'}|\leq 3000\hat q_i\leq 3000|A'_{i'}|$. Therefore, by appending an edge of $A'_{i'}$ at the end of each such path, for all indices $i'\neq i$, we obtain a set $\pset'_i=\set{P(\hat a)\mid \hat a\in \hat A_i}$ of paths in graph $G'$, that cause congestion at most $3000$, such that for each edge $\hat a\in \hat A_i$, path $P(\hat a)$ has $\hat a$ as its first edge, terminates at vertex $u^*$, and is internally disjoint from $X'_i$. Lastly, for every edge $a'_{i,j}\in A'_i$, we define a path $P(a'_{i,j})$ consisting of only the edge $a'_{i,j}$ itself, and add that path to set $\pset'_i$. We have now obtained a set $\pset'_i$ of paths in graph $G'$, routing the edges of $\delta_{G'}(X'_i)$ to vertex $u^*$, such that the paths are internally disjoint from $X'_i$. Therefore, $\pset'_i\in \Lambda'(X'_i)$. We then let the distribution $\dset'(X'_i)$ choose the path set $\pset'_i$ with probability $1$.

We add each such cluster $X'_i$ to the set $\lset^{\light}$ of light clusters, and define, for each such cluster $X'_i$, a distribution  $\dset(X'_i)$ over the set $\Lambda(X'_i)$ of internal routers for $X'_i$, such that $X'_i$ is $\beta$-light with respect to $\dset(X'_i)$. 
In fact, the disctibution $\dset(X'_i)$ will select a single set $\pset_i\in \Lambda(X'_i)$ of paths with probability $1$. The set $\pset_i$ of paths is constructed as follows. From the properties of the flower cluster, there is a set $\qset_i$ of edge-disjoint paths in graph $G$, routing the edges of $\delta_G(X_i)\setminus \delta_G(u^*)$ to vertex $u^*$, such that every inner vertex on every path lies in  $X_i$. Since cluster $X'_i$ can be obtained from $X_i$ by replacing vertex $u_i$ with vertex $u^*$, we obtain a collection $\pset_i$ of edge-disjoint paths, routing the edges of $\hat A_i$ to vertex $u_i$, such that every inner vertex on every path lies in $X_i'$. For every edge $a'_{i,j}\in A'_i$, we add a path $P(a'_{i,j})$, consisting of the edge $a'_{i,j}$ only, to set $\pset_i$. We then obtain a set $\pset_i$ of edge-disjoint paths, routing the edges of $\delta_{G'}(X'_i)$ to vertex $u_i$ inside $X'_i$, that is, $\pset_i\in \Lambda(X_i)$.

Consider the set $\iset_3$ of subinstances of $I'$, that is obtained by performing a basic disengagement of instance $I'$ via the tuple $(\lset,\lset^{\bad}, \lset^{\gd}, \set{\dset'(X'_i)}_{i=1}^k,
\set{\dset(X'_i)}_{i=1}^k$) (here, we set $\lset^{\bad}=\emptyset$).

Recall that family $\iset_3$ of instances contains a single global instance $\hat I=(\hat G,\hat \Sigma)$, where graph $\hat G$ is obtained from graph $G'$ by contracting, for all $1\leq i\leq k$, the vertices of $X'_i$ into a supernode. 
Additionally, for every cluster $X'_i\in \lset$, we obtain an instance $I(X'_i)=(G_i,\Sigma_i)$, where graph $G_i$ is obtained from graph $G_i$, by contracting all vertices of $V(G')\setminus V(X'_i)$ into a supernode. 

We summarize the properties of the resulting family $\iset_3$ of instances in the following claim.

\begin{claim}\label{claim: properties of instances from petal-based disengagement}
	
	\begin{itemize}
		\item	$\sum_{\tilde I=(\tilde G,\tilde \Sigma)\in \iset_3}|E(\tilde G)|\leq O(|E(G)|)$;
		\item	$\expect{\sum_{\tilde I\in \iset_3}\optcrors(\tilde I)}\leq O
		\left ((\optcrors(I)+|E(G)|)\cdot \log^{36}m\right )$;

\item 	There is an efficient algorithm, that, given, for each instance $\tilde I\in \iset_3$, a solution $\phi(\tilde I)$, computes a solution to instance $I$ of cost at most $O\left (\sum_{\tilde I\in \iset_3}\cro(\phi(\tilde I))\right )$.
\end{itemize}\end{claim}

\begin{proof}
	For the first assertion, recall that, from \Cref{lem: number of edges in all disengaged instances} 	$\sum_{\tilde I=(\tilde G,\tilde \Sigma)\in \iset_3}|E(\tilde G)|\leq O(|E(G' )|)$. Since, from the construction of the split instance, $|E(G')|\leq O(|E(G)|)$, the  assertion follows.
	
	In order to prove the second assertion, recall that, from \Cref{lem: disengagement final cost}, 	$\expect{\sum_{\tilde I\in \iset_3}\optcrors(\tilde I)}\leq O\left (\beta^2\cdot (\optcrors(I')+|E(G')|)\right )$. Since, as discussed above, $\optcrors(I')\leq \optcrors(I)$, $|E(G')|\leq O(|E(G)|)$, and $\beta\leq O(\log^{18}m)$, the assertion follows.
	
In order to prove the last assertion, we use the algorithm from \Cref{lem: basic disengagement combining solutions}, that,  given, for each instance $\tilde I\in \iset_3$, a solution $\phi(\tilde I)$, computes a solution $\phi'$ for instance $I'$ of cost at most $\sum_{\tilde I\in \iset_3}\cro(\phi(\tilde I))$. We then use 
the algorithm from \Cref{lem: solution to split to solution to original} in order to compute a solution $\phi$ for instance $I$ of cost at most $O(\cro(\phi'))\leq O\left (\sum_{\tilde I\in \iset_3}\cro(\phi(\tilde I))\right )$.
\end{proof}

Consider now the global instance $\hat I=(\hat G,\hat \Sigma)$. Since graph $\hat G$ is obtained from $G'$ by contracting every cluster $X'_i$ into a supernode, for every edge $e\in E(\hat G)$, either $e$ is incident to $u^*$, or it corresponds to an edge of $\Eout(\cset)$, where $\cset$ is the initial collection of clusters that we computed in Phase 1. Recall that, from \Cref{eq: num of edges between clusters}, $
|\Eout(\cset)|\leq m/(80\mu)$. Recall that $\deg_{G'}(u^*)=\sum_{i=1}^k\hat q_i=\sum_{i=1}^k|\hat E_i|$. From Modified Property \ref{prop: flower cluster small boundary size} of the flower cluster, $\sum_{i=1}^k|\hat E_i|\leq 200m/\mu^{42}$. Therefore, overall, $|E(\hat G)|\leq |\Eout(\cset)|+\deg_{G'}(u^*)\leq m/(40\mu)$.

Next, we consider petal-based instances, and we prove that for each such instance, the maximum vertex degree is small.

\begin{claim}\label{claim: degree of petal instance}
	For all $1\leq i\leq k$, if $I(X'_i)=(G_i,\Sigma_i)$ is the instance of $\tilde \iset$ associated with cluster $X'_i$, then every vertex degree in graph $G_i$ is less than $m/\mu^4$.
\end{claim}

\begin{proof}
	Recall that graph $G_i$ is obtained from graph $G'$ by contracting all vertices of $V(G')\setminus V(X'_i)$ into a supernode, that we denote by $u'$. Recall that graph $X'_i$ is identical to the petal $X_i$, except that we replace vertex $u^*$ with vertex $u_i$. From the definition of a flower cluster, every vertex of $X_i$, except for vertex $u^*$, has degree less than $m/\mu^4$ in $G$. The degree of vertex $u_i$ in the new graph is bounded by $q_i+\hat q_i$. Here, $q_i=|A_i|=|E_i|\leq m/(2\mu^4)$ from Property \ref{prop: flower cluster edges near center partition} of the flower cluster, and $\hat q_i\leq 200m/\mu^{42}$ from Modified Property \ref{prop: flower cluster small boundary size} of the flower cluster. Therefore, the degree of $u_i$ in graph $G_i$ is less than $m/\mu^4$. It now remains to bound the degree of the supernode $u'$ in graph $G_i$. The edges incident to $u'$ are the edges of $A'_i\cup \hat A_i$, and their number is bounded by $2\hat q_i$, which, from the above discussion, is bounded by $400m/\mu^{42}$.
\end{proof}

\subsubsection{Step 3: Final Decomposition}

In this step, we consider each petal-based instance $I(X'_i)=(G_i,\Sigma_i)$ in which $|E(G_i)|>m/(2\mu)$. We further decompose each such instance into subinstances, by exploiting the fact that graph $G_i$ does not have high-degree vertices, using the following lemma.

\begin{lemma}\label{lem: last decomposition}
	There is an efficient randomized algorithm, that, given an instance $\tilde I=(\tilde G,\tilde \Sigma)$ of \cnwrs and parameters $m,\mu$, such that $m$ is greater than a large enough constant, $m/(2\mu)<|E(\tilde G)|\leq 3m$, $\mu\geq 2^{\Omega(\sqrt{\log m})}$, and maximum vertex degree in $\tilde G$ is less than $m/\mu^4$, either correctly establishes that $\optcrors(I)\geq  \Omega \left( \frac{m^2}{ \mu^{5.5}} \right )$, or
	computes a $\nu_3$-decomposition $\tilde \iset'$ of $\tilde I$, for $\nu_3=2^{O((\log m)^{3/4}\log\log m)}$, such that, for every instance $\tilde I'=(\tilde G',\tilde \Sigma')\in \tilde \iset'$, $|E(\tilde G')|\leq m/(2\mu)$.
\end{lemma}

We prove the lemma below, after we complete the proof of \Cref{lem: decomposing problematic instances by petal} using it.
Consider some index $1\leq i\leq k$. If $|E(G_i)|< m/(2\mu)$, then we let the set $\iset(X'_i)$ of subinstances of $I(X'_i)$ consist of a single instance -- instance $I(X'_i)$. Otherwise, 
we apply the algorithm from \Cref{lem: last decomposition} to  instance $I(X'_i)=(G_i,\Sigma_i)\in \iset_3$. If the algorithm from \Cref{lem: last decomposition} computes a $\nu_3$-decomposition $\tilde \iset$ of $I(X'_i)$, such that, for every instance $\tilde I'=(\tilde G',\tilde \Sigma')\in \tilde \iset$, $|E(\tilde G')|\leq m/(2\mu)$, then we set $\iset(X'_i)=\tilde \iset$. Otherwise, we terminate the algorithm and return FAIL.

If, every time \Cref{lem: last decomposition} is invoked, it returns a $\nu_3$-decomposition of the corresponding instance $I(X'_i)$, then we output a collection $\set{\hat  I}\cup \left (\bigcup_{i=1}^k\iset(X'_i)\right)$ of instances. From \Cref{claim: compose algs}, it is immediate to verify that this algorithm  produces a $\nu_2$-decomposition of instance $I$ for $\nu_2=O(\nu_3)$ (since the family $\iset_3$ of subinstances of $I$ computed in Step 2 of the current phase is an $O(\log^{36}m)$-decomposition of instance $I$, from \Cref{claim: properties of instances from petal-based disengagement}), and the graph associated with each instance has at most $m/(2\mu)$ edges.

Assume now that
$\optcrors(\hat I)<\frac{ m^2}{c'' \mu^{13}}$ for some large enough constant $c''$.
Recall that $|E(G)|\leq O(m)$, and, from the statement of \Cref{lem: not many paths}, $m\geq \mu^{50}$, so $|E(G)|<\frac{m^2}{\mu^{13}}$. Therefore, $(\optcrors(\hat I)+|E(G)|) <\frac{2m^2}{c'' \mu^{13}}$.

Recall that, from \Cref{claim: properties of instances from petal-based disengagement},	$\expect{\sum_{\tilde I\in \iset_3}\optcrors(\tilde I)}\leq O(\optcrors(I)+|E(G)|)$.
 We say that a bad event $\event''$ happens if $\sum_{i=1}^k\optcrors(I(X'_i))>c\mu^5(\optcrors(I)+|E(G)|)$ for some large enough constant $c$. From Markov's inequality, $\prob{\event''}\leq 1/(8\mu^4)$.

If $\optcrors(\hat I)<\frac{ m^2}{c'' \mu^{13}}$, and the bad event $\event''$ did not appen, then for all $1\leq i\leq k$, 
$\optcrors(I(X'_i))<c\mu^5(\optcrors(I)+|E(G)|)\leq \frac{2cm^2}{c''\mu^8}$. Note that our algorithm may only return FAIL if there is some index $1\leq i\leq k$, such that $|E(G_i)|\geq m/(2\mu)$, and  $\optcrors(I(X'_i))\geq  \Omega \left( \frac{|E(G_i)|^2}{ \mu^{5.5}} \right )\geq  \Omega \left(\frac{ m^2}{ \mu^{7.5}} \right )$. From the above discussion, and since we can choose $c''$ to be a large enough constant compared to $c$, if $\optcrors(\hat I)<\frac{ m^2}{c'' \mu^{13}}$, then the algorithm may only return FAIL if $\event''$ happens, which happens with probability at most $1/(8\mu^4)$.

In order to complete \Cref{lem: decomposing problematic instances by petal}, and \Cref{lem: not many paths}, it is now enough to prove \Cref{lem: last decomposition}.

\begin{proofof}{\Cref{lem: last decomposition}}
In order to simplify the notation, we denote instance $\tilde I=(\tilde G,\tilde \Sigma)$ by $I=(G,\Sigma)$. We will essentially repeat the algorithm from Phase 1, except that, since there are no high-degree vertices in $G$, we do not need to deal with flower cluster, and all instances that we will obtain in the final decomposition will be small.


We start by applying the algorithm from \Cref{lem: decomposition into small clusters} to graph $H=G$, with terminal set $T=\emptyset$, parameter $\tau=2\mu^{1.1}$, and the parameter $m$ replaced with $3m$. Recall that the maximum vertex degree in $G$ is less than $\frac{m}{\mu^4}< \frac{3m}{\check c\tau^3 \log^5 (3m)}$, as required. 

Assume first that the algorithm from \Cref{lem: decomposition into small clusters} establishes that that $\optcro(G)\geq  \Omega \left( \frac{m^2}{ \tau^4\log^5 m} \right )\geq  \Omega \left( \frac{m^2}{ \mu^{5.5}} \right )$. We then terminate the algorithm and report that $\optcrors(I)\geq  \Omega \left( \frac{m^2}{ \mu^{5.5}} \right )$.

Therefore, we assume from now on that the algorithm from \Cref{lem: decomposition into small clusters} computes a collection $\cset'$ of disjoint clusters of $G$, such that every cluster $C\in \cset'$ has the $\alpha'$-bandwidth property, where $\alpha'=\Omega\left(\frac{1}{\log^{1.5}m}\right )$.
Since $m\geq \mu^4$, we then get that every cluster in $\cset'$ has the $\alpha_0=1/\log^3m$-bandwidth property.
Additionally, we are guaranteed that, for each such cluster $C\in \cset'$, $|E(C)|\leq m/\tau\leq m/(4\mu)$,
 $\bigcup_{C\in \cset'}V(C)=V(G)$, and
 $|\bigcup_{C\in \cset'}\delta_G(C)|\leq m/\tau=m/(2\mu^{1.1})$. Notice that in particular, the number of edges of $G$ with endpoints in different clusters is $|\Eout(\cset')|\leq m/(2\mu^{1.1})$. Since we have assumed that $|E(G)|\geq m/(2\mu)$, we get that $\sum_{C\in \cset'}|\delta_G(C)|\leq |E(G)|/\mu^{0.1}$.

We apply the algorithm from \Cref{thm: disengagement - main} to instance $I=(G,\Sigma)$ of \CNwRS, and the set $\cset'$ of clusters. Let $\tilde \iset'$ be the resulting collection of subinstances of $I$ that the algorithm computes. Recall that  the algorithm guarantees that $\tilde \iset'$ is a $2^{O((\log m)^{3/4}\log\log m)}$-decomposition of $I$, and moreover, for each instance $\tilde I'=(\tilde G',\tilde \Sigma')\in \tilde \iset'$, there is at most one cluster $C\in \cset'$ with $E(C)\subseteq E(\tilde G')$, and all other edges of $\tilde G'$ lie in set $\Eout(\cset')$. Since $|\Eout(\cset')|\leq m/(2\mu^{1.1})$, and, for every cluster $C\in \cset'$, $|E(C)|\leq m/(4\mu)$, we are guaranteed that, for every instance $\tilde I'=(\tilde G',\tilde \Sigma')\in \tilde \iset'$, $|E(\tilde G')|\leq m/(2\mu)$. 
\end{proofof}

\section{Constructing Internal Routers - Proof of \Cref{thm: find guiding paths}}
\label{sec: guiding paths}


We will repeatedly use the following simple lemma, whose proof is provided in \Cref{apd: Proof of routing path extension}.
\begin{lemma}
	\label{lem: routing path extension}
	Let $G$ be a graph, let $T$ be a set of vertices that are $\alpha$-well-linked in $G$, for some $0<\alpha<1$, and let $T'$ be a subset of $T$. Suppose we are given a vertex $x\in V(G)$, and a set $\pset$ of paths in $G$, routing the vertices of $T'$ to $x$. Then there is a set $\pset'$ of paths routing the vertices of $T$ to $x$, such that, for every edge $e\in E(G)$,
	$\cong_G(\pset',e)\le \ceil{\frac{|T|}{|T'|}}(\cong_G(\pset,e)+\ceil{1/\alpha})$.
\end{lemma}

For conveninence, we denote the contracted graph $H_{|\cset}$ by $\hH$,  and we denote $|E(\hH)|=\hm$. From the statement of \Cref{thm: find guiding paths}, $k\geq \hm/\eta$. Observe that, from \Cref{clm: contracted_graph_well_linkedness}, the set $T$ of terminals is $(\alpha\alpha')$-well-linked in $H$. We will assume in the remainder of the proof that $\log m$ is greater than some large enough constant $c'_0$ (whose value we can set later). If this is not the case, then, $n$, and therefore $k$, is bounded by a constant $2^{c_0'}$. We can then use an arbitrary spanning tree $\tau$ of the graph $H$, 
rooted at an arbitrary vertex $y$, in order to define a set $\qset$ of paths routing all terminals of $T$ to $y$, where for each terminal $t\in T$, the corresponding path $Q_t\in \qset$ is the unique path connecting $t$ to $y$ in the tree $\tau$. Since $|T|$ is bounded by a constant, for every edge $e\in E(H)$, $\cong_H(\qset,e)\leq O(1)$. We then return a distribution $\dset$ consisting of a single set $\qset$ that has probability value $1$. Therefore, we assume from now on that $\log m>c_0'$ for some large enough constant $c_0'$ whose value we set later.

We start with some intuition. Assume first that graph $H$ contains a grid (or a grid minor) of size $(\Omega(k\alpha\alpha'/\poly\log m)\times \Omega(k\alpha \alpha'/\poly\log m))$, and a collection $\pset$ of paths connecting every terminal to a distinct vertex on the first row of the grid, such that the paths in $\pset$ cause a low edge-congestion. For this special case, the algorithm of \cite{Tasos-comm}  (see also the proof of Lemma D.10 in the full version of \cite{chuzhoy2011algorithm}) provides a distribution $\dset$ over routers $\qset\in  \Lambda(H,T)$ with the required properties. Moreover, if $H$ is  a bounded-degree planar graph, with a set $T$ of terminals that is $(\alpha\alpha')$-well-linked, 
then there is an efficient algorithm to compute such a grid minor, together with the required collection $\pset$ of paths. If $H$ is planar but no longer bounded-degree, we can still compute a grid-like structure in it, and apply the same arguments as in \cite{Tasos-comm} in order to compute the desired distribution $\dset$. The difficulty in our case is that the input graph $H$ may be far from being planar, and, even though, from the Excluded Grid theorem of Robertson and Seymour \cite{robertson1986graph}, 
it must contain a large grid-like structure, without having a drawing of $H$ in the plane with a small number of crossing, we do not know how to compute such a structure\footnote{We note that we need the grid-like structure to have dimensions $(k'\times k')$, where $k'$ is almost linear in $k$. Therefore, we cannot use the known bounds for the Excluded Minor Theorem (e.g. from \cite{chuzhoy2019towards}) for general graphs, and instead we need to use  an analogue of the stronger version of the theorem for planar graphs.}.

The proof of \Cref{thm: find guiding paths} consists of five steps. In the first step, we will  either establish that $\optcrors(I)$ is sufficiently large (so the algorithm can return FAIL), or compute a subgraph $\hH'\subseteq \hH$, and a partition $(X,Y)$ of $V(\hH')$, such that each of the clusters $\hH'[X],\hH'[Y]$ has the $\hat \alpha$-bandwidth property, for $\hat \alpha=\Omega(\alpha/\log^4m)$, together with a large collection of edge-disjoint paths routing the terminals to the edges of $E_{\hH'}(X,Y)$ in graph $\hH'$. Intuitively, we will view from this point onward the edges of $E_{\hH'}(X,Y)$ as a new set of terminals, that we denote by $\tilde T$ (more precisely, we subdivide each edge of $E_{\hH'}(X,Y)$ with a new vertex that becomes a new terminal). We show that it is sufficient to prove an analogue of \Cref{thm: find guiding paths} for this new set $\tT$ of terminals. The clusters $\hH'[X],\hH'[Y]$ of graph $\hH'$ naturally define a partition $(H_1,H_2)$ of the graph $H$ into two disjoint subgraphs. In the second step, we either establish that $\optcrors(I)$ is suffciently large 
(so the algorithm can return FAIL), or
compute some vertex $x$ of $H_1$, and a collection $\pset$ of paths in graph $H_1$, routing the terminals of $\tilde T$ to $x$, such that the paths in $\pset$ cause a relatively low edge-congestion.
We exploit this set $\pset$ of paths in order to define an ordering of the terminals in $\tilde T$, which is in turn exploited in the third step in order to compute a ``skeleton'' of the grid-like structure. We compute the grid-like structure itself in the fourth step. In the fifth and the final step, we generalize the arguments from \cite{Tasos-comm}  and  \cite{chuzhoy2011algorithm} in order to obtain the desired distribution $\dset$ over routers $\qset\in \Lambda(H,T)$, by exploiting this grid-like structure.

Before we proceed, we need to consider four simple special cases. In the first case,  $\sum_{C\in \cset}|\delta_H(C)|^2$ is large. In the second case, we can route a large subset of the terminals to a single vertex of $V(\hat H)\cap V(H)$ in the graph $\hat H$ via edge-disjont paths. The third case is when $\optcrors(H,\Sigma)=0$, and the fourth special case is when $k< \eta^6$.

\paragraph{Special  Case 1: $\sum_{C\in \cset}|\delta_H(C)|^2$ is large.}
We consider the case where $\sum_{C\in \cset}|\delta_H(C)|^2\geq \frac{(k \alpha^4 \alpha')^2}{c_0\log^{50}m}$, where $c_0$ is the constant from the statement of \Cref{thm: find guiding paths}. For every cluster $C\in \cset$, let $\Sigma_C$ be the rotation system for $C$ induced by $\Sigma$.
In this case, since we are guaranteed that every cluster $C\in \cset$ is $\eta'$-bad, that is,  $\optcrors(C,\Sigma_C)+|E(C)|\geq |\delta(C)|^2/\eta'$, we get that:
\[\optcrors(I)+|E(H\setminus T)|\geq \sum_{C\in \cset}\left (\optcrors(C,\Sigma_C)+|E(C)|\right )\geq \sum_{C\in \cset}\frac{|\delta_H(C)|^2}{\eta'}\geq \frac{(k \alpha^4 \alpha')^2}{c_0\eta'\log^{50}m}.\]
Therefore, if $\sum_{C\in \cset}|\delta_H(C)|^2\geq \frac{(k\alpha^4 \alpha')^2}{c_0\log^{50}m}$, the algorithm returns FAIL and terminates. We assume from now on that:

\begin{equation}\label{eq: boundaries squared sum bound}
\sum_{C\in \cset}|\delta_H(C)|^2<\frac{(k\alpha^4 \alpha')^2}{c_0\log^{50}m}.
\end{equation}

\paragraph{Special Case 2: Routing of terminals to a single vertex.}
The second special case happens if there exists a collection $\pset_0$ of at least $\frac{k\alpha^2}{1024\cCMG^3\log^6k}$ edge-disjoint paths in graph $\hH$ routing some subset $T_0\subseteq T$ of terminals to some vertex $x$ (here $\cCMG$ is the constant from \Cref{claim: embed expander}).
 
Note that, if Special Case 1 did not happen, and $c_0$ is a large enough constant, then $x$ may not be a supernode. Indeed, assume that $x=v_C$ for some cluster $C\in \cset$. Then:
\[|\delta_H(C)|^2\geq \Omega\left(\frac{(k\alpha^2)^2}{\log^{12}k}\right )\geq \frac{(k \alpha^4 \alpha')^2}{c_0\log^{50}m},\] 
which is, assuming that $c_0$ is a large enough constant, a contradiction. Therefore, we can assume that $x$ is not a supernode.
From \Cref{claim: routing in contracted graph}, since the clusters in $\cset$ have the $\alpha'$-bandwidth property, there is a collection $\pset'_0$ of paths in graph $H$, routing the vertices of $T_0$ to $x$, with edge-congestion at most $\ceil{1/\alpha'}\leq 2/\alpha'$. Since the set $T$ of terminals is $(\alpha\alpha')$-well-linked in graph $H$, from \Cref{lem: routing path extension}, there is a set $\qset$ of paths in graph $H$, routing the vertices of $T$ to $x$ with congestion at most:
\[\ceil{\frac{|T|}{|T_0|}}\left(\frac 2{\alpha'}+\ceil{\frac 1{\alpha\alpha'}}\right )\leq O\left (  \frac{\log^6k}{\alpha^3\alpha'} \right).\]

Note that a set $\qset$ of paths with the above properties can be computed efficiently via standard maximum flow algorithm.
We return a distribution $\dset$ consisting of a single router $\qset$ with probability value $1$, and terminate the algorithm. 
 Clearly, for every edge $e\in E(H)$, $\expect{(\cong(\qset,e))^2}\leq  O\left (\frac{\log^{32}m}{\alpha^{12}(\alpha')^8}\right )$.

\paragraph{Special Case 3: $\optcrors(I)=0$.}
Recall that we can efficiently check whether $\optcrors(I)=0$, using the algorithm from \Cref{thm: crwrs_planar}. Assume now that $\optcrors(H,\Sigma)=0$. 
We use the following theorem from \cite{chuzhoy2020towards}.

\begin{lemma}[Lemma E.2 in \cite{chuzhoy2020towards}]
\label{lem: find_guiding_in_planar}
There is an efficient algorithm, that, given a planar graph $H$ and a subset $T$ of $r$ vertices of $V(H)$ that are $\alpha$-well-linked in $H$ for some $0<\alpha<1$, computes a distribution $\dset$ over the routers in $\Lambda(H,T)$, such that the distribution has support size $O(r^2)$, and for each edge $e\in E(H)$,
\[\expect[(u^*,\qset)\sim \dset]{(\cong_H(\qset,e))^2}=O\bigg(\frac{\log r}{\alpha^4}\bigg).\]
\end{lemma}

Recall that the set $T$ of terminals is $\alpha$-well-linked in the contracted graph $H_{\mid \cset}$, and every cluster $C\in \cset$ has the $\alpha'$-bandwidth property.
From \Cref{clm: contracted_graph_well_linkedness}, the set $T$ of terminals is $(\alpha\alpha')$-well-linked in $H$.
We then apply the algorithm from \Cref{lem: find_guiding_in_planar} to graph $H$, terminal set $T$ and parameter $(\alpha\alpha')$. Let $\dset$ be the distribution over the set $\Lambda(H,T)$ of routers that we obtain. Then:
 
\[\expect[\qset\sim \dset]{(\cong_H(\qset,e))^2}=O\bigg(\frac{\log k}{(\alpha\alpha')^4}\bigg) \leq  O\left (\frac{\log^{32}m}{\alpha^{12}(\alpha')^8}\right ).\]

\paragraph{Special Case 4: $k< \eta^6$, but $\optcrors(I)>0$.}
Note that $\frac{(k\alpha^4 \alpha')^2}{c_0\eta'\log^{50}m}\leq \frac{\eta^{12}}{\eta'}<1$ in this case (as, from the statement of \Cref{thm: find guiding paths}, $\eta'>\eta^{13}$). 
Since we have assumed that $\optcrors(I)>0$, we get that $\optcrors(I)\geq 1>\frac{(k\alpha^4 \alpha')^2}{c_0\eta'\log^{50}m}$. We then simply return FAIL and terminate the algorithm.

In the remainder of the proof, we assume that neither of the four special cases happened. We now describe each step of the algorithm in detail.

\subsection{Step 1: Splitting the Contracted Graph}

In this step, we split the contracted graph $\hH$, using the algorithm  summarized in the following theorem.

\begin{theorem}\label{thm: splitting}
There is an efficient randomized algorithm that returns FAIL with probability at most $1/\poly(k)$, and, if it does not return FAIL, then it  computes a subgraph $\hH'\subseteq \hH$ and a partition $(X,Y)$ of $V(\hH')$ such that:
\begin{itemize}
\item clusters $\hH'[X]$ and $\hH'[Y]$ both have the $\halpha'$-bandwidth property in $\hH'$, for $\halpha'=\Omega(\alpha/\log^4m)$; and
\item there is a set $\rset$ of $\Omega(\alpha^3k/\log^8m)$ edge-disjoint paths in graph $\hH'$, routing a subset of terminals to edges of $E_{\hH'}(X,Y)$.
\end{itemize}
\end{theorem}

\begin{proof}
We start by applying the algorithm from \Cref{claim: embed expander} to graph $\hH$ and the set $T$ of terminals, to obtain a graph $W$ with $V(W)=T$ and maximum vertex degree at most $\cCMG\log^2k$, and an embedding $\hpset$ of $W$ into $\hH$ with congestion at most $(\cCMG\log^2k)/\alpha$. Let $\hat\event$ be the bad event that $W$ is not a $1/4$-expander. Then $\prob{\hat \event}\leq 1/\poly(k)$.
Define graph $\hH'$ as the union of all paths in $\hat{\pset}$.
We need the following observation.

\begin{observation}\label{obs: expansion and degree}
	If event $\hat \event$ did not happen, then the set $T$ of vertices is $\halpha$-well-linked in $\hH'$, for $\halpha=\frac{\alpha}{4\cCMG\log^2k}$, and the maximum vertex degree in $\hH'$ is at most $d=\frac{\alpha k}{512\cCMG\log^2k}$.
\end{observation}
\begin{proof}
	Assume that Event $\hat \event$ did not happen. We first prove that the set $T$ of terminals is $\halpha$-well-linked in $\hH$. Consider any paritition $(A,B)$ of vertices of $\hH'$, and denote $T_A=T\cap A$, $T_B=T\cap B$. Assume w.l.o.g. that $|T_A|\leq |T_B|$. Then it is sufficient to show that $|E_{\hH'}(A,B)|\geq \halpha\cdot |T_A|$.
	
	Consider the partition $(T_A,T_B)$ of the vertices of $W$, and denote $E'=E_{W}(T_A,T_B)$. Since $W$ is a $1/4$-expander, $|E'|\geq |T_A|/4$ must hold. Consider now the set $\hat \rset\subseteq \hpset$ of paths containing the embeddings $P(e)$ of every edge $e\in E'$. Each path $R\in \hat \rset$ connects a vertex of $T_A$ to a vertex of $T_B$, so it must contain an edge of $|E_{\hH}(A,B)|$. Since $|\hat\rset|\geq |T_A|/4$, and the paths in $\hpset$ cause edge-congestion at most $(\cCMG\log^2k)/\alpha$, we get that $|E_{\hH}(A,B)|\geq \alpha\cdot |T_A|/(4\cCMG\log^2k)\geq \halpha |T_A|$.
	
	Assume now that maximum vertex degree in $\hH'$ is greater than $d$, and let $x$ be a vertex whose degree is at least $d$. Let $\hat\qset\subseteq \hat\pset$ be the set of all paths containing the vertex $x$.
Consider any such path $Q\in \hat\qset$. The endpoints of this path are two distinct terminals $t,t'\in T$. We let $Q'\subseteq Q$ be the subpath of $Q$ between the terminal $t$ and the vertex $x$, and we let $\qset'=\set{Q'\mid Q\in \hat\qset}$.

Recall that every vertex in $W$ has degree at most $\cCMG\log^2k$, and so a terminal in $T$ may be an endpoint of at most $\cCMG\log^2k$ paths in $\hpset$. Therefore, there is a subset $\qset''\subseteq \qset'$ of at least $d/(2\cCMG\log^2k)$ paths in $\hH'$, each of which originates at a distinct terminal.
Since paths in $\qset''$ cause congestion at most $(\cCMG\log^2k)/\alpha$, from \Cref{claim: remove congestion}, there is a collection $\qset'''$ of edge-disjoint paths in graph $\hH'$, routing a subset of terminals to $x$ with:
$$|\qset'''|\geq |\qset''|\cdot \frac{\alpha}{\cCMG\log^2k}\geq \frac{d\alpha}{2\cCMG^2\log^4k}\geq\frac{\alpha^2k}{1024\cCMG^3\log^6k},$$
contradicting the fact that Special Case 2 did not happen.
\end{proof}

Next, we use the following lemma to compute the required sets $X$, $Y$ of vertices. The proof follows immediately from techniques that were introduced in \cite{chuzhoy2012routing} and then refined in \cite{chuzhoy2012polylogarithmic,chekuri2016polynomial,chuzhoy2016improved}. Unfortunately, all these proofs assumed that the input graph has a bounded maximum vertex degree, and additionally the proofs are somewhat more involved than the proof that we need here (this is because these proofs could only afford a $\poly\log k$ loss in the cardinality of the  set $\rset$ of paths relatively to $|T|$, while we can afford a $\poly\log m$ loss). Therefore, we provide a  proof of the lemma in Section \ref{sec: splitting} of the Appendix for completeness.

\begin{lemma}\label{lem: splitting}
	There is an efficient algorithm that, given as input an $m$-edge graph $G$, and a subset $T$ of $k$ vertices of $G$ called terminals, together with a parameter $0<\talpha<1$, such that the maximum vertex degree in $G$ is at most $\talpha k/64$, and every vertex of $T$ has degree $1$ in $G$, either returns FAIL, or computes a partition $(X,Y)$ of $V(G)$, such that:
	\begin{itemize}
		\item each of the clusters $G[X]$, $G[Y]$ has the $\talpha'$-bandwidth property, for $\talpha'=\Omega(\talpha/\log^2m)$; and
		\item there is a set $\rset$ of at least $\Omega(\talpha^3k/\log^2m)$ edge-disjoint paths in graph $G$, routing a subset of terminals to edges of $E_G(X,Y)$.
	\end{itemize}
Moreover, if the set $T$ of vertices is $\talpha$-well-linked in $G$, then the algorithm never returns FAIL.
\end{lemma}

We apply the algorithm from \Cref{lem: splitting} to graph $\hH'$, the set $T$ of terminals, and parameter $\talpha=\halpha=\frac{\alpha}{4\cCMG\log^2k}$. Recall that we are guaranteed that the maximum vertex degree in graph $\hat H'$ is at most $d=\frac{\alpha k}{512\cCMG\log^2k}\leq \frac{\talpha k}{64}$.
Note that the algorithm from  \Cref{lem: splitting} may only return FAIL if the set $T$ of terminals is not $\talpha$-well-linked in $\hH'$, which, from \Cref{obs: expansion and degree}, may only happen if event $\hat \event$ happened, which in turn may only happen with probability $1/\poly(k)$. If the algorithm from \Cref{lem: splitting} returned FAIL, then we terminate the algorithm and return FAIL as well. Therefore, we assume from now on that the algorithm from  \Cref{lem: splitting} did not return FAIL.
%
%
%
 Let $(X,Y)$ be the partition of $V(\hH')$ that the algorithm returns. We are then guaranteed that each of the clusters $\hH'[X],\hH'[Y]$ has the $\talpha'$-bandwidth property in $\tilde H$, where $\talpha'=\Omega(\talpha/\log^2m)=\Omega(\alpha/\log^4m)$. The algorithm also ensures that there is a collection $\rset$ of edge-disjoint paths in $\hH'$, routing a subset of the terminals to edges of $E_{\hH'}(X,Y)$, with $|\rset|\geq \Omega(\talpha^3k/\log^2m)\geq \Omega(\alpha^3k/\log^8m)$.
This completes the proof of \Cref{thm: splitting}.
\end{proof}

If the algorithm from \Cref{thm: splitting} returned FAIL (which may only happen with probability at most $1/\poly(k)$), then we terminate the algorithm and return FAIL as well. Therefore, we assume from now on that the algorithm from \Cref{thm: splitting} returned a subgraph $\hH'\subseteq \hH$ and a partition $(X,Y)$ of $V(\hH')$ such that each of the clusters $\hH'[X]$ and $\hH'[Y]$ has the $\halpha'$-bandwidth property, for $\halpha'=\Theta(\alpha/\log^4m)$, and
 there is a set $\rset$ of $\Omega(\alpha^3k/\log^8m)$ edge-disjoint paths in graph $\hH'$, routing a subset of terminals to edges of $E_{\hH'}(X,Y)$.
Notice that we can compute the path set $\rset$ with the above properties efficiently, using standard maximum flow algorithms. We assume w.l.o.g. that edges of $E_{\hH}(X,Y)$ do not serve as inner edges on paths in $\rset$.
Let $E'\subseteq E_{\hH'}(X,Y)$ be the subset of edges containing the last edge on every path in $\rset$, so, by reversing the direction of the paths in $\rset$, we can view the set $\rset$ of paths as routing the edges of $E'$ to the terminals.
In the remainder of this step, we will slightly modify the graphs $H$ and $\hat H$, and we will continue working with the modified graphs only in the following steps.

Let $\hH''\subseteq\hH'$ be the graph obtained from $\hH'$ by first deleting all edges of $E_{\hH'}(X,Y)\setminus E'$ from it, and then subdividing every edge $e\in E'$ with a vertex $t_e$. We denote $\tilde T=\set{t_e\mid e\in E'}$, and we refer to vertices of $\tilde T$ as \emph{pseudo-terminals}. Recall that $|\tilde T|=|\rset|=\Omega(\alpha^3k/\log^8m)$, and there is a set $\rset'$ of edge-disjoint paths in the resulting graph $\hH''$, routing the vertices of $\tilde T$ to the vertices of $T$.
We define $\hH_1=\hH''[X\cup \tilde T]$, the subgraph of $\hH''$ induced by the set $X\cup \tilde T$ of vertices, and we define $\hH_2=\hH''[Y\cup \tilde T]$ similarly. From the $\halpha'$-bandwidth property of the clusters $\hH'[X]$ and $\hH'[Y]$ in $\hat H'$, we are guaranteed that the vertices of $\tilde T$ are $ \halpha'$-well-linked in both $\hH_1$ and in $\hH_2$, where $\halpha'= \Theta(\alpha/\log^4m)$. Let $\cset'\subseteq \cset$ be the subset of all clusters $C$ whose corresponding supernode $v_C$ lies in graph $\hH''$.

For convenience, we also subdivide, in graph $H$, every edge $e\in E'$, with the vertex $t_e$, so graph $\hH''$ can be now viewed as a subgraph of the contracted graph $H_{|\cset}$.

Next, we let $H'\subseteq H$ be the subgraph of $H$ that corresponds to  graph $\hH''$. In other words, graph $H'$ is obtained from $\hH''$ by replacing every supernode $v_C$ with the corresponding cluster $C\in \cset'$. Equivalently, we can obtain graph $H'$ from $H$, by deleting every edge of $E(\hH)\setminus E(\hH'')$ and every regular (non-supernode) vertex of $V(\hH)\setminus V(\hH'')$. Additionally, for every cluster $C\in \cset\setminus \cset'$, we delete all edges and vertices of $C$ from $H'$. We also define a rotation system $\Sigma'$ for graph $H'$, which is naturally induced by $\Sigma$ (vertices $t_e\in \tilde T$ all have degree $2$, so their corresponding ordering $\oset_{t_e}$ of incident edges can be set arbitrarily). Let $I'=(H',\Sigma')$ be the resulting instance of \cnwrs.

We partition the set $\cset'$ of clusters into two subsets: set $\cset_X$ contains all clusters $C\in \cset'$ with $v_C\in X$, and set $\cset_Y$ contains all clusters $C\in \cset'$ with $v_C\in Y$. We can similarly define the graphs $H_1,H_2\subseteq H'$, that correspond to the contracted graphs $\hat H_1$ and $\hat H_2$, respectively: let $X'$ be the set of vertices of $H'$, containing every vertex $x\in V(H')$, such that either $x\in C$ for some cluster $C\in \cset_X$, or $x$ is a regular vertex of $\hat H''$ lying in $X$. Similarly, we let $Y'$ contain all vertices $y\in V(H)$, such that either $y\in C$ for some cluster $C\in \cset_Y$, or $y$ is a regular vertex of $\hat H''$ lying in $Y$. We then let $H_1=H'[X\cup \tilde T]$, and $H_2=H'[Y\cup \tilde T]$.

The following observation, summarizing  properties of instance $I'$, is immediate.
\begin{observation}\label{obs: properties of new graph}
Instance $I'=(H',\Sigma')$ of \cnwrs satisties the following properties:
	\begin{itemize}
		\item $\optcrors(I')\leq \optcrors (I)$;
		\item $\hat H''=H'_{|\cset'}$; 
		\item $|E(\hat H'')|\leq 2|E(\hat H)|\leq 2\eta k\leq O(|\tilde T|\eta \log^8m/\alpha^3)$; and
		\item graph $\hat H_1$ is a contracted graph of $H_1$ with respect to  $\cset_X$, and graph $\hat H_2$ is a contracted graph of $H_2$ with respect to $\cset_Y$. In other words, $\hat H_1=(H_1)_{|\cset_X}$, and $\hat H_2=(H_2)_{|\cset_Y}$.
	\end{itemize}
\end{observation}
For the third assertion we have used the fact that $k\geq |E(\hat H)|/\eta$ from the statement of \Cref{thm: find guiding paths}, and $|\tilde T|\geq \Omega(\alpha^3k/\log^8m)$.

Recall that $\Lambda(H',\tilde T)$ denotes the set of all routers in graph $H'$, with respect to the set $\tilde T$ of terminals. Each such router $\qset$ is a set of paths, routing the vertices of $\tilde T$ to some vertex of $H'$.
Intuitively, from now on we would like to work with instance $I'=(H',\Sigma')$ of \cnwrs, and the new set $\tilde T$ of terminals. To this end, we start by showing that,  in order to obtain the desired distribution $\dset$ over the routers of $\Lambda(H,T)$, it is now sufficient to compute a distribution $\dset'$ over the routers of  $\Lambda(H',\tilde T)$, such that for every edge $e\in E(H')$, $\expect[\qset'\sim \dset']{(\cong_{H'}(\qset,e))^2}$ is low.
\begin{observation}\label{obs: convert distributions}
	There is an efficient algorithm, that, given an explicit distribution $\dset'$ over the routers of $\Lambda(H',\tilde T)$,
	such that for every edge $e'\in E(H')$, $\expect[\qset'\sim\dset']{(\cong_{H'}(\qset',e'))^2}\leq \beta$ holds,
	 computes an explicit distribution $\dset$ over the routers of $\Lambda(H,T)$, such that for every edge $e\in E(H)$, $\expect[\qset\sim \dset]{(\cong_{H}(\qset,e))^2}\leq O\left(\frac{\beta \log^{16}m}{\alpha^8(\alpha')^4}\right )$.
\end{observation}
\begin{proof}
	Recall that $H'\subseteq H$. Consider some router $\qset'\in  \Lambda(H',\tilde T)$, whose probability value in distribution $\dset'$ is $p(\qset')>0$. We compute a router $\qset\in \Lambda(H,T)$ corresponding to $\qset'$, and we assign to $\qset$ the same probability value $p(\qset')$.

We now show an algorithm for computing a router $\qset\in \Lambda(H,T)$ from a router $\qset'\in \Lambda(H',\tilde T)$. We denote by $x'$ the vertex that serves as the center of the router $\qset'$.
	Recall that there is a set $\rset'$ of edge-disjoint paths in graph $\hH''$, routing the vertices of $\tilde T$ to the vertices of $T$, and moreover, a set of paths with these properties can be found efficiently via a standard maximum $s$-$t$ flow computation.
	Since $\hat H''=H'_{|\cset'}$, and every cluster in $\cset'$ has the $\alpha'$-bandwidth property in $H'$, from \Cref{claim: routing in contracted graph}, we can efficiently compute a set $\rset_0$ of edge-disjoint paths in graph $H'$, routing a subset $T_0\subseteq T$ of terminals to $\tT$, with $|\rset_0|\geq \alpha'\cdot |\rset'|/2=\alpha'\cdot |\tilde T|/2=\Omega(\alpha'\alpha^3k/\log^8m)$. By concatenating the paths in $\rset_0$ and the paths in $\qset'$, we obtain a collection $\rset_0'$ of paths in graph $H'$, routing the terminals of $T_0$ to vertex $x'$, such that for every edge $e\in E(H)$, $\cong_{H'}(\rset_0',e)\leq \cong_{H'}(\qset',e)+1$.
Since the set $T$ of terminals is $(\alpha\alpha')$-well-linked in graph $H$ (from \Cref{clm: contracted_graph_well_linkedness}), from \Cref{lem: routing path extension}, there exists a collection $\qset$ of paths in graph $H$, routing the terminals in $T$ to vertex $x'$, such that for every edge $e\in E(H)$:

\[
\cong_{H}(\qset,e)\leq  \ceil{\frac{|T|}{|T_0|}}\left(\cong_{H}(\rset'_0,e)+\ceil{\frac{1}{\alpha\alpha'}}\right )
\leq O\left (\frac{\log^8m}{\alpha'\alpha^3}\right )\cdot \left(\cong_{H'}(\qset',e)+\frac{2}{\alpha\alpha'}\right ).
\]

A set $\qset$ of paths with these properties can be computed efficiently via standard maximum $s$-$t$ flow algorithms.

For every router $\qset'\in \Lambda(H',\tT)$, whose probability value in distribution $\dset'$ is $p(\qset')>0$, we have computed a corresponding router $\qset\in \Lambda(H,T)$, and we have assigned to it the same probability value $p(\qset)=p(\qset')$. This completes the definition of the distribution $\dset$ over the routers in $\Lambda(H,T)$. From the above discussion,  for every edge $e\in E(H)$,
\[\expect[\qset\sim \dset]{(\cong_{H}(\qset,e))^2}\leq O\left(\frac{\log^{16}k}{\alpha^8(\alpha')^4}\right )\cdot \left (\expect[\qset'\sim \dset']{(\cong_{H'}(\qset',e))^2}+1\right) \leq O\left(\frac{\beta \log^{16}m}{\alpha^8(\alpha')^4}\right ). \]	
\end{proof}

The following immediate corollary is obtained by plugging in the bounds required by \Cref{thm: find guiding paths} into \Cref{obs: convert distributions}.

\begin{corollary}\label{cor: convert distribution fixed values}
	There is an efficient algorithm, that, given an explicit distribution $\dset'$ over the routers of $\Lambda(H',\tilde T)$,
	such that for every edge $e'\in E(H')$, $\expect[\qset'\sim \dset']{(\cong_{H'}(\qset',e'))^2}\leq  O\left (\frac{\log^{16}m}{(\alpha\alpha')^4}\right )$ holds, produces an explicit distribution $\dset$ over the routers of $\Lambda(H,T)$, such that, for every edge $e\in E(H)$: $$\expect[\qset\sim \dset]{(\cong_{H}(\qset,e))^2}\leq  O\left (\frac{\log^{32}m}{\alpha^{12}(\alpha')^8}\right ).$$
\end{corollary}

Denote $\tk=|\tilde T|$.
Recall that  $\tk\geq \Omega\left(\frac{\alpha^3k}{\log^8m}\right )$, so $\frac{(k\alpha^4 \alpha')^2}{c_0\log^{50}m}\leq O\left( \frac{(\tk\halpha'\alpha')^2}{c_0\log^{20}m}\right )$. We use a large enough constant $c_1$, whose value will be set later, and we set $c_0=c_1^2$. We can then assume that $\frac{(k\alpha^4 \alpha')^2}{c_0\log^{50}m}\leq  \frac{(\tk\halpha'\alpha')^2}{c_1\log^{20}m}$. In particular, from Equation \ref{eq: boundaries squared sum bound}, we get that $\sum_{C\in \cset}|\delta_{H'}(C)|^2<\frac{(k\alpha^4\alpha')^2}{c_0\log^{50}m}\leq   \frac{(\tk\halpha'\alpha')^2}{c_1\log^{20}m}$. Additionally, if 
$|E(H'\setminus\tilde  T)|+\optcrors(I')>\frac{(\tk\halpha'\alpha')^2}{c_1\eta'\log^{20}m}$, then $|E(H\setminus T)|+\optcrors(I)>\frac{(k\alpha^4 \alpha')^2}{c_0\eta'\log^{50}m}$.

In order to complete the proof of \Cref{thm: find guiding paths}, it is now enough to design a randomized algorithm, that either  returns FAIL, or computes a distribution $\dset'$ over the routers in $\Lambda(H',\tilde T)$, such that, for every edge $e\in E(H')$, $\expect[\qset'\sim \dset']{(\cong_{H'}(\qset',e))^2}\leq  O\left (\frac{\log^{16}m}{(\alpha\alpha')^4}\right )$. It is enough to ensure that,   if $|E(H'\setminus \tilde T)|+\optcrors(H',\Sigma')\leq \frac{(\tk\halpha'\alpha')^2}{c_1\eta'\log^{20}m}$, then the probability that the algorithm returns FAIL is at most $1/2$.

In the remainder of the proof we focus on the above goal. It would be convenient for us to simplify the notation, by denoting $H'$ by $H$, $\Sigma'$ by $\Sigma$, $I'$ by $I$,  $\hat H''$ by $\hat H$, and $\halpha'$ by $\tilde \alpha$. We also denote $\cset'$ by $\cset$. We now summarize all properties of the new graphs $H,\hat H$ that we have established so far, and in the remainder of the proof of \Cref{thm: find guiding paths} we will only work with these new graphs.

\paragraph{Summary of the Outcome of Step 1.}
We  assume from now on that we are given an instance $I=(H,\Sigma)$ of \cnwrs, a set $\tilde T$ of terminals in graph $H$, and a collection $\cset$ of disjoint subgraphs (clusters) of $H\setminus \tT$. We denote $|\tT|=\tk$. The corresponding contracted graph is denoted by $\hat H=H_{|\cset}$. 
We are also given a partition $(X,Y)$ of $V(H)\setminus \tilde T$ (note that for convenience of notation, $X$ and $Y$ are now subsets of vertices of $H$, and not of $\hat H$), and a parition $\cset_X,\cset_Y$ of $\cset$, such that each cluster $C\in \cset_X$ has $V(C)\subseteq X$, and each cluster $C\in \cset_Y$ has $V(C)\subseteq Y$. We denote $H_1=H[X\cup \tilde T]$ and $H_2=H[Y\cup \tilde T]$. We also denote by $\hat H_1=(H_1)_{|\cset_X}$ the contracted graph of $H_1$ with respect to $\cset_X$, and similarly by $\hat H_2=(H_2)_{|\cset_Y}$. 

We now summarize the properties of the graphs that we have defined and the relationships between the main parameters.
\begin{properties}{P}
	\item $\tk\geq \Omega(\alpha^3k/\log^8m)$;\label{prop after step 1: number of pseudoterminals}
	\item every cluster $C\in \cset$ has the $\alpha'$-bandwidth property in $H$; \label{prop after step 1: bandwidth property}
	\item $|E(\hat H)|\leq O\left(\frac{\tk\cdot \eta \log^8m}{\alpha^3}\right )$ (from \Cref{obs: properties of new graph}); \label{prop after step 1: few edges}
	\item every vertex of $\tilde T$ has degree $1$ in $H_1$, and vertex set $\tilde T$ is $\talpha$-well-linked in $\hat H_1$, for $\talpha=\Theta(\alpha/\log^4m)$;\label{prop after step 1: terminals in H1}
	\item similarly, every vertex of $\tilde T$ has degree $1$ in $H_2$, and vertex set $\tilde T$ is $\talpha$-well-linked in $\hat H_2$; and \label{prop after step 1: terminals in H2}
\item $\sum_{C\in \cset}|\delta_H(C)|^2<\frac{(\tk\talpha\alpha')^2}{c_1\log^{20}m}
$, where $c_1$ is some large enough constant, whose value we can set later. \label{prop after step 1: small squares of boundaries}
\end{properties}

Our goal is
to design an efficient randomized algorithm, that either returns FAIL, or computes a distribution $\dset$ over the routers in $\Lambda(H,\tilde T)$, such that,  for every edge $e\in E(H)$, $\expect[\qset\sim \dset]{(\cong_{H}(\qset,e))^2}\leq  O\left (\frac{\log^{16}m}{(\alpha\alpha')^4}\right )$.
It is enough to ensure that, if   
$\optcrors(I)+|E(H\setminus T)|<\frac{(\tilde k\tilde \alpha\alpha')^2}{c_1\eta'\log^{20}m}$,
then the probability that the algorithm returns FAIL is at most $1/2$.

\subsection{Step 2: Routing the Terminals to a Single Vertex, and an Expanded Graph}

In this step we start by considering the graph $\hat H_1$ and the set $\tT$ of terminals in it. Our goal is to compute a collection $\jset$ of paths in graph $\hat H_1$, routing all terminals of $\tT$ to a single regular  vertex, such that the paths in $\jset$ cause a relatively low congestion in graph $\hat H_1$. We show that, if such a collection of paths does not exist, then $\optcrors(I)$ is high. Intuitively, we will use the set $\jset$ of paths in order to define an ordering of the terminals in $\tT$, which will in turn be used in order to compute a grid-like structure in graph $H_2$. Once we compute the desired set $\jset$ of paths, we will replace the graph $H$ with its low-degree analogue $H^*$, that we refer to as the \emph{expanded graph}. The remaining steps in the proof of \Cref{thm: find guiding paths} will use this expanded graph only.


\subsubsection{Routing the Terminals to a Single Vertex in $\hat H_1$}
We process {\bf regular} vertices of $V(\hat H_1)$ (that is, vertices of $V(\hat H_1)\cap V(H_1)$) one by one. For each such vertex $x$, we compute a set $\jset(x)$ of paths in graph $\hat H_1$, with the following properties:

\begin{itemize}
	\item every path in $\jset(x)$ originates at a distinct vertex of $\tT$ and terminates at $x$;
	\item the paths in $\jset(x)$ are edge-disjoint; and
	\item $\jset(x)$ is a maximum-cardinality set of paths in $\hat H_1$ with the above two properties.
\end{itemize}

Note that such a set $\jset(x)$ of paths can be computed via a standard maximum $s$-$t$ flow computation.
Throughout, we use a parameter $\tk'=\tk\alpha^5/(c'\eta\log^{36}m)$, where $c'$ is a large enough constant whose value we set later. 

If, for every vertex $x\in V(H_1)$, $|\jset(x)|<\tk'$, then we reurn FAIL and terminate the algorithm.  In the following lemma, whose proof is deferred to Section \ref{sec: few paths high opt} of Appendix we show that, in this case, $\optcrors(I)\geq 
\Omega\left(\frac{(\tk\talpha \alpha')^2}{\eta'\log^{20}m}\right )$ must hold.
Note that, since we can set $c_1$ to be a large enough constant, we can ensure that $\optcrors(I)>\frac{(\tilde k\tilde \alpha\alpha')^2}{c_1\eta'\log^{20}m}$ holds in this case.
The value of the constant $c'$ that is used in the definition of the parameter $\tk'$ is set in the proof of the lemma.

\begin{lemma}\label{lem: high opt or lots of paths}
	If,  for every vertex $x\in V(\hat H_1)\cap V(H_1)$, $|\jset(x)|<\tk'$, then $\optcrors(I)\geq 
	\Omega\left(\frac{(\tk\talpha \alpha')^2}{\eta'\log^{20}m}\right )$.
\end{lemma}

From now on we assume that there is some vertex $x\in V(\hat H_1)\cap V(H_1)$, for which $|\jset(x)|\geq \tk'$.

\subsubsection{The Expanded Graph}

From now on we fix the vertex $x\in V(\hat H_1)\cap V(H_1)$, and we let $\jset=\jset(x)$ be a set of at least $\tk'$ edge-disjoint paths in graph $\hH_1$, routing a subset $\tT_0\subseteq \tT$ of terminals to vertex $x$. 

We are now ready to define the expanded graph $H^*$. We start with graph $H^*$ being empty, and then process every vertex $u\in V(H_2)\setminus \tilde T$ one by one. We now describe an iteration when a vertex $u\in V(H_2)\setminus \tilde T$ is processed. We denote by $d(u)$ the degree of the vertex $u$ in graph $H_2$.  Let $e_1(u),\ldots,e_{d(u)}(u)$ be the edges that are incident to $u$ in $H_2$, indexed according to their ordering in $ \oset_u\in  \Sigma$. We let $\Pi(u)$ be a $(d(u)\times d(u))$ grid, and we denote the vertices on the first row of this grid by $s_1(u),\ldots,s_{d(u)}(u)$ indexed in their natural left-to-right order. 
We add the vertices and the edges of the grid $\Pi(u)$ to graph $H^*$. We refer to the edges in the resulting grids $\Pi(u)$ as \emph{inner edges}.
Once every vertex $u\in V(H_2)\setminus \tT$ is processed, we add 
the vertices of $\tT$ to the graph $H^*$. Recall that every terminal $t\in \tT$ has degree $1$ in $H_2$. We denote the unique edge $e_t$ incident to $t$ by $e_1(t)$, and we denote $s_1(t)=t$.

Next, we add 
a collection of \emph{outer edges} to graph $H^*$, as follows. Consider any edge $e=(u,v)\in E(H_2)$. Assume that $e$ is the $i$th edge of $u$ and the $j$th edge of $v$, that is, $e=e_i(u)=e_j(v)$. Then we add an edge $e'=(s_i(u),s_j(v))$ to graph $H^*$, and we view this edge as the \emph{copy of the edge $e\in E(H_2)$}. We will not distinguish between the edge $e$ of $H_2$, and the edge $e'$ of $H^*$.

Our last step is to add vertex $x$ to graph $H^*$, that connects to every terminal $t\in \tT$ with an edge $(x,t)$, that is also viewed as an outer edge. The following lemma, whose proof is deferred to Section \ref{sec:ordering of terminals} of Appendix, allows us to compute an ordering $\tilde \oset$ of the terminals, such that the graph $H^*$ has a drawing $\phi$ with few crossings, in which the inner edges do not participate in any crossings, and the images of the edges incident to $x$ enter $x$ in order consistent with $\tilde \oset$.

\begin{lemma}\label{lem: find ordering of terminals}
	There is an efficient algorithm that computes an ordering $\tilde \oset$ of the terminals in $\tT$, such that there is a drawing $\phi$ of graph $H^*$ with at most $O\left(\optcrors(I)\cdot\frac{\eta^2\log^{74}m}{\alpha^{12}(\alpha')^4}\right ) +O \left (
	\frac{\tk \eta\log^{37}m}{\alpha^6(\alpha')^2}\right )$ crossings, in which all crossings are between pairs of outer edges. Moreover, if we denote $\tT=\set{t_1,\ldots,t_{\tk}}$, where the terminals are indexed according to the ordering $\tilde \oset$, and, for each $1\leq i\leq t_{\tk}$, denote by $e_i=(t_i,x)$ the edge of $H^*$ connecting $t_i$ to $x$, then the images of the edges $e_1,\ldots,e_{\tk}$ enter the image of $x$ in this circular order in the drawing $\phi$.
\end{lemma}

From now on we fix the ordering $\tilde \oset$ of the terminals in $\tilde T$ given by \Cref{lem: find ordering of terminals}, and the drawing $\phi$ of $H^*$ (which is not known to the algorithm).

It will be convenient for us to slightly modify the graph $H^*$ as follows. We denote 
the terminals by  $\tT=\set{t_1,\ldots,t_{\tk}}$, where the terminals are indexed according to the circular ordering $\tilde \oset$. Let $H'$ be a graph obtained from $H^*$, by first deleting the vertex $x$ from it, and then adding, for all $1\leq i<\tk$, an edge $e^*_i=(t_i,t_{i+1})$, and another edge $e^*_{\tk}=(t_{\tk},t_1)$. We denote this set of the newly added edges by $E^*$, and we view them as inner edges. Note that the edges of $E^*$ form a simple cycle, that we denote by $L^*$. We also denote $H''=H'\setminus E^*$.

We note that the drawing $\phi$ of $H^*$ can be easily extended to obtain a drawing $\phi'$ of  graph $H'$ in the plane, so that the inner edges of $H'$ do not participate in any crossings, and the image of the cycle $L^*$ (which must be a simple closed curve) is the boundary of the outer face.

In order to do so, we start with the drawing $\phi$ of $H^*$ on the sphere, and then consider the tiny $x$-disc $D=D_{\phi}(x)$, denoting its boundary by $\gamma^*$. 
For every terminal $t_i\in \tT$, we denote by $e_i$ the unique edge incident to $t_i$ in $H''$, and by $e'_i=(t_i,x)$. We also denote by $\gamma_i,\gamma'_i$ the images of the edges $e_i,e'_i$ in drawing $\phi$. Let $p_i$ be the unique point on the intersection of $\gamma'_i$ and $\gamma^*$. We move the image of terminal $t_i$ to point $p_i$. We then modify the image of the edge $e_i$, so that it becomes a concatenation of $\gamma_i$, and the portion of $\gamma'_i$ lying outside the interior of $D$. Lastly, we draw the edges of $E^*$ in a natural way, where edge $e^*_i$ is simply a segment of $\gamma^*$ between the images of $t_i$ and $t_{i+1}$, so that all resulting segments are mutually internally disjoint. Once we delete the vertex $x$ from this drawing, no part of the resulting drawing is contained in the interior of the disc $D$, and the image of the cycle $L^*$ is precisely $\eta$, so we can view the resuting drawing $\phi'$ of $H'$ as a drawing in the plane, with $D$ being its outer face. Note that this transformation does not increase the number of crossings.

The next observation follows by substituting parameters and bounds that we have already established. The proof is included in Section \ref{subsec: proof of obs on bounds on opt} of Appendix.

\begin{observation}\label{obs: bounds on opt}
Let $c_2$ be a  large enough constant. We can set the value of constant $c_1$ so that it is large enough, and,	if $\optcrors(I)<\frac{(\tilde k\tilde \alpha\alpha')^2}{c_1\eta'\log^{20}m}$, then  $\cro(\phi')\leq 
	\cro(\phi)\leq \frac{\tilde k^2}{c_2\eta^5}$.
\end{observation}

We will se the value of constant $c_2$ later, and the value of constant $c_1$ will then be set using \Cref{obs: bounds on opt}. 
It is now enough to ensure that, if $\cro(\phi')<
 \frac{\tilde k^2}{\eta^5}$, then the  probability that the algorithm returns FAIL is at most $1/2$.

Let $\Lambda'=\Lambda(H'',\tT)$ be the collection of all routers in graph $H''$ with respect to the set $\tT$ of terminals.
We need the following simple observation, whose proof is deferred to Section \ref{subsec: transform paths 2} of the Appendix.

\begin{observation}\label{obs: transform paths 2}
There is an efficient algorithm, that, given an explicit distribution $\dset$ over the routers of $\Lambda'$, such that for every {\bf outer} edge $e\in E(H'')$, $\expect[\qset\sim \dset]{(\cong_{H''}(\qset,e))^2}\leq \beta$, computes  an explicit distribution $\dset'$ over the routers in $\Lambda(H,\tT)$, where for every edge $e\in E(H)$, $\expect[\qset'\sim \dset']{(\cong_{H}(\qset',e))^2}\leq \beta$.
\end{observation}

\subsubsection{Summary of Step 2}
\label{step 2 summary}

In the remainder of the proof of \Cref{thm: find guiding paths} we will work with graph $H'$ only. Recall that graph $H'$ contains a set $E^*=\set{e^*_1,\ldots,e^*_{\tk}}$ of edges (that are considered to be inner edges), where for all $1\leq i\leq \tk$, $e^*_i=(t_i,t_{i+1})$ (we use indexing modulo $\tk$). The set $E^*$ of edges defines a cycle $L^*=(t_1,\ldots,t_{\tk})$ in graph $H'$. We also denoted $H''=H'\setminus E^*$. Recall that graph $H''$ is obtained from a subgraph $H_2\subseteq H$, by replacing every vertex $v\in V(H_2)\setminus \tT$ with a grid $\Pi(v)$. All edges lying in the resulting grids $\Pi(v)$, and the edges of $E^*$ are inner edges, while all other edges of $H'$ are outer edges. Each outer edge of $H'$ corresponds to some edge of graph $H_2$, and we do not distinguish between these edges. Note that in graph $H'$, all vertices have degrees at most $4$. We will also use the clustering $\cset_Y$ of graph $H_2$, and the fact that, from Property \ref{prop after step 1: small squares of boundaries}: 
\begin{equation}\label{eq: sum of squares}
\sum_{C\in \cset_Y}|\delta_H(C)|^2<\frac{(\tk\talpha\alpha')^2}{c_1\log^{20}m}. \end{equation}
We further partition the outer edges of graph $H''$ into two subsets: type-1 outer edges and type-2 outer edges. Consider any outer edge $e$ in graph $H''$, and let $e'=(u,v)$ be the corresponding edge in graph $H$. If $u$ and $v$ both lie in the same cluster $C\in \cset_Y$, then we say that $e$ is a \emph{type-2} outer edge, and otherwise it is a type-1 outer edge. 
Intuitively, for each type-1 outer edge, there is a corresponding edge in the contracted graph $\hat H=H_{|\cset}$. From Property \ref{prop after step 1: few edges}, we obtain the following observation.

\begin{observation}\label{obs: few outer edges}
	There is a universal constant $c$ (independent of $c_1$ and $c_2$), such that the total number of type-1 outer edges in $H''$ is bounded by ${c\tk\cdot \eta \log^8m/\alpha^3}$. 
\end{observation}

Recall that from Property \ref{prop after step 1: terminals in H1}, every vertex of $\tilde T$ has  $1$ in $H_2$, and vertex set $\tilde T$ is $\talpha$-well-linked in $\hat H_2$. Combining this with the $\alpha'$-bandwidth property of every cluster $C\in \cset_Y$ from Property \ref{prop after step 1: bandwidth property}, from \Cref{clm: contracted_graph_well_linkedness}, the set $\tT$ of terminals is $\talpha\cdot \alpha'$-well-linked in $H_2$. Lastly, using the fact that each graph in $\set{\Pi(v)\mid v\in V(H_2)}$ has the $1$-bandwidth property, from \Cref{clm: contracted_graph_well_linkedness}, we get the following observation.

\begin{observation}\label{obs: terminals well linked in H''}
	The set $\tT$ of terminals is $\alpha^*$-well-linked in $H''$, where ${\alpha^*=\talpha\cdot\alpha'=\Theta(\alpha\alpha'/\log^4m)}$. Moreover, each terminal in $\tT$ has degree $1$ in $H''$ and degree $3$ in $H'$.
\end{observation}
(we have used the fact that  
$\talpha=\Theta(\alpha/\log^4m)$ (see Property \ref{prop after step 1: terminals in H1})).


We will restrict our attention to special types of drawings of graph $H'$, called \emph{legal drawings}, that we define next.

\begin{definition}[Legal drawing of $H'$]
We say that a drawing $\phi^*$ of graph $H'$ in the plane is \emph{legal} if it has the following properties:

\begin{itemize}

	\item no inner edge of $H'$ participates in any crossing of $\phi^*$, and in particular the image of the cycle $L^*$ is a simple closed curve, denoted by $\gamma^*$; and
	\item $\gamma^*$ is the boundary of the outer face in the drawing.
\end{itemize}
\end{definition}

We let $\phi^*$ be a legal drawing of $H'$ with smallest number of crossings, and we denote by $\cro^*$ the number of crossings in $\phi^*$.
From \Cref{obs: bounds on opt}, if $\optcrors(I)<\frac{(\tilde k\tilde \alpha\alpha')^2}{c_1\eta'\log^{20}m}$, then  $\cro^*\leq 
 \frac{\tilde k^2}{c_2\eta^5}$.

Denote $\tT=\set{t_1,\ldots,t_{\tk}}$, where the terminals are indexed according their ordering in $\tilde \oset$. We partition the set $\tT$ of terminals into four subsets $T_1,\ldots,T_4$, where $T_1,T_2,T_3$ contain $\floor{\tk/4}$ consecutive terminals from $\tT$ each, and $T_4$ contains the remaining terminals, in a natural way using the ordering $\tilde \oset$, that is, $T_1=\set{t_1,\ldots,t_{\floor{\tk/4}}}$, $T_2=\set{t_{\floor{\tk/4}+1},\ldots,t_{2\floor{\tk/4}}}$, $T_3=\set{t_{2\floor{\tk/4}+1},\ldots,t_{3\floor{\tk/4}}}$, and $T_4=\set{t_{3\floor{\tk/4}+1},\ldots,t_{\tk}}$. Clearly, each of the four sets contains at least $\floor{\tk/4}$ terminals.

Recall that in a legal drawing $\phi$ of $H'$, the image of the cycle $L^*$ is a simple closed curve, that we denoted by $\gamma^*$. It will be convenient for us to view this curve $\gamma^*$ as the boundary of a rectangular area in the plane, that encloses the legal drawing of $H'$. We sometimes refer to this rectangular area as the \emph{bounding box} of the drawing, and denote it by $B^*$. We will think of the terminals in $T_1$ and $T_3$ as appearing on the left and on the right boundaries of $B^*$, respectively, and of the terminals in $T_2$ and $T_4$ as appearing on the top and the bottom boundaries of $B^*$, respectively. 

For all $1\leq i\leq 4$, we let $\tilde \oset_i$ be the ordering of the terminals in $T_i$ consistent with their ordering on the boundary of $B^*$ (where each ordering $\tilde\oset_i$ is no longer circular), so that the terminals in sets $T_1$ and in $T_3$ appear in the bottom-to-top order, and the terminals in $T_2$ and $T_4$ appear in their left-to-right order (so $\tilde \oset$ is obtained by concatenationg $\tilde\oset_1,\tilde\oset_2$, the reversed ordering $\tilde\oset_3$, and the reversed ordering $\tilde\oset_4$).

Recall that $\Lambda'=\Lambda(H'',\tT)$.
Our goal from now on is to design a randomized algorithm, that either computes a distribution $\dset$ over the routers of $\Lambda'$, such that for every outer edge $e\in E(H'')$, $\expect[\qset\sim\dset]{(\cong_{H'}(\qset,e))^2}\leq  O\left (\frac{\log^{16}m}{(\alpha\alpha')^4}\right )$, or returns FAIL.
It is enough to ensure that, if $\cro^*\leq 
\frac{\tilde k^2}{c_2\eta^5}$ for some large enough constant $c_2$, whose value we can set later, then the  probability that the algorithm returns FAIL is at most $1/4$.

\subsection{Step 3: Constructing a Grid Skeleton}

In this and the following step we will construct a grid-like structure in graph $H''$. Recall that the set $\tT$ of terminals is $\alpha^*$-well-linked in graph $H''$.  From \Cref{thm: bandwidth_means_boundary_well_linked} there is a set $\pset'$ of paths in $H''$, routing all terminals of $T_1$ to terminals of $T_3$, with edge-congestion at most $\ceil{1/\alpha^*}$, such that the routing is one-to-one. From \Cref{claim: remove congestion}, there is a collection $\pset''$ of at least $|T_1|/\ceil{1/\alpha^*}=\floor{\tk/4}/\ceil{1/\alpha^*}\geq \alpha^*\tk/8$ edge-disjoint paths in $H''$, routing some subset of terminals of $T_1$ to a subset of terminal of $T_3$, in graph $H''$. Moreover, since graph $H''$ has maximum vertex degree at most $4$, using arguments similar to those in the proof of \Cref{claim: remove congestion}, there is a collection $\pset$ of $\floor{\alpha^*\tk/32}$ {\bf node-disjoint} paths in graph $H''$, routing some subset $A\subseteq T_1$ of terminals, to some subset $A'\subseteq T_3$ of terminals. We can compute such a set $\pset$ of paths efficiently using standard maximum $s$-$t$ flow algorithms.

Using similar reasoning, we can compute a collection $\rset$ of $\floor{\alpha^*\tk/32}$ node-disjoint paths in graph $H''$, routing some subset $B\subseteq T_2$ of terminals, to some subset $B'\subseteq T_4$ of terminals.

Intuitively, after we discard a small subset of paths from each of the sets $\pset$ and $\rset$, the remaining paths will be used in order to construct a grid-like structure, where paths in $\pset$ will serve as horizontal paths of the grid, and paths in $\rset$ will serve as vertical paths. If the paths in the resulting sets do not form a grid-like structure, then we will terminate the algorithm with a FAIL. We will prove that, if $\cro^*\leq 
\frac{\tilde k^2}{c_2\eta^5}$ for a large enough constant $c_2$, then we will construct the grid-like structure successfully with probability at least $3/4$.

We denote $\pset_0=\pset$ and $\rset_0=\rset$. Recall that so far, ${|\pset_0|,|\rset_0|\geq \floor{\alpha^*\tk/32}}$.

Intuitively, if the dimensions of the grid-like structure that we construct are $(h\times h)$, then we need $h$ to be quite close to $\tk$, since this grid-like structure will be exploited in order to define the distribution $\dset$ over the routers of $\Lambda'$.
We will first construct a smaller grid-like structure, that we call a \emph{grid skeleton}. This grid skeleton will be associated with a grid $\Pi^*$ of smaller dimensions, that we sometimes call a \emph{supergrid}. We then extend this grid skeleton to construct a large enough grid-like structure.

We will use two additional parameters. The first parameter is:
\[\lambda=\frac{2^{24}c\cdot \eta \log^8m}{\alpha^*\alpha^3},\] 
where $c$ is the constant from \Cref{obs: few outer edges}. Notice that, since $\alpha^*=\Theta(\alpha\alpha'/\log^4m)$, we get that $\lambda=O\left( \frac{\eta \log^{12}m}{\alpha^4\alpha'}  \right )$. Moreover, since $\eta>\frac{c^* \log^{12}m}{\alpha^4\alpha'}$ for a large enough constant $c^*$ (from the statement of \Cref{thm: find guiding paths}), $\lambda<\eta^2$ holds. The supergrid that we construct will have dimensions $(\Theta(\lambda)\times \Theta(\lambda))$. The second parameter is:
\[\psi=\floor{\frac{\alpha^*\tk}{64\lambda}}=\floor{\frac{\alpha^3(\alpha^*)^2\tk}{2^{30}c\eta\log^8m}}.\]
Clearly, $|\rset_0|,|\pset_0|\geq \lambda\psi$.
Note that, since $\alpha^*=\Theta(\alpha\alpha'/\log^4m)$, $\psi\geq \Omega\left(\frac{\tk}{\eta}\cdot \frac{\alpha^5(\alpha')^2}{\log^{16}m}  \right)$. Since ${\eta>\frac{c^* \log^{16}m}{\alpha^5(\alpha')^2}}$ from the statement of \Cref{thm: find guiding paths}, we get that $\psi>\frac{16\tk}{\eta^2}$.  Every cell of the supergrid will be associated with a collection of $\Theta(\psi)$ horizontal paths and $\Theta(\psi)$ vertical paths, that will help us form the grid-like structure.

We discard paths from $\pset_0$ and from $\rset_0$ arbitrarily, until $|\pset_0|=|\rset_0|=\lambda\psi$ holds.

We denote by $A_0\subseteq T_1,A'_0\subseteq T_3$ the endpoints of the paths in $\pset_0$, and we denote by $B_0\subseteq T_4$, $B'_0\subseteq T_2$ the endpoints of the paths in $\rset_0$.

\subsubsection*{Grid Skeleton Construction}

We view the paths in $\pset_0$ as directed from vertices of $A_0$ to vertices of $A'_0$. Recall that $A_0\subseteq T_1$, so the ordering $\tilde\oset_1$ of the terminals in $T_1$ defines an ordering $\oset_{A_0}=\set{a_1,\ldots,a_{\lambda\psi}}$ of the terminals in $A_0$. This ordering in turn defines an ordering $\oset_{\pset_0}$ of the paths in $\pset_0$, as follows: if, for all $1\leq i\leq \lambda\psi$, $P_i\in \pset_0$ is the path originating from $a_i$, then $\oset_{\pset_0}=\set{P_1,\ldots,P_{\lambda\psi}}$.

Similarly, we view the paths in $\rset_0$ as directed from vertices of $B_0$ to vertices of $B_0'$. Ordering $\tilde\oset_4$ of terminals in $T_4$ defines an ordering $\oset_{B_0}=\set{b_1,\ldots,b_{\lambda\psi}}$ of the vertices in $B_0$, which in turn defines an ordering $\oset_{\rset_0}=\set{R_1,\ldots,R_{\lambda\psi}}$ of paths in $\rset_0$, where for all $i$, path $R_i$ originates at vertex $b_i$.

We partition the set $\pset_0$ of paths into  groups $\uset_1,\ldots,\uset_{\lambda}$ of cardinality $\psi$ each, using the ordering $\oset_{\pset_0}$, so for $1\leq i<\lambda$, set $\uset_i$ is the $i$th set of $\psi$ consecutive paths of $\pset_0$. 
Let $\lambda'= \floor{(\lambda-1)/2}$. 
For all $1\leq i\leq \lambda'$, we let $ P^*_i$ be a path that is chosen uniformly at random from set $\uset_{2i}$. Let $\pset^*=\set{ P^*_1,\ldots, P^*_{\lambda'}}$ be the resulting set of chosen paths. Intuitively, the path in $\pset^*$ will serve as the horizontal paths in the grid skeleton that we construct.
We then let $\pset_1\subseteq \pset_0$ be the set containing all paths in sets $\set{\uset_{2i-1}}_{i=1}^{\lambda'+1}$. 

We perform similar computation on the set $\rset_0$ of paths. First, we partition $\rset_0$ into  groups $\uset'_1,\ldots,\uset'_{\lambda}$ of cardinality $\psi$ each, using the ordering $\oset_{\rset_0}$, so for $1\leq i<\lambda$, set $\uset'_i$ is the $i$th set of $\psi$ consecutive paths of $\rset_0$. 
For all $1\leq i\leq \lambda'$, we let $ R^*_i$ be a path that is chosen uniformly at random from set $\uset'_{2i}$. Let $\rset^*=\set{ R^*_1,\ldots, R^*_{\lambda'}}$ be the resulting set of chosen paths. Intuitively, the path in $\rset^*$ will serve as the vertical paths in the grid skeleton that we construct.
We then let $\rset_1\subseteq \rset_0$ be the set containing all paths in sets $\set{\uset'_{2i-1}}_{i=1}^{\lambda'+1}$.

We let $\event_1$ be the bad event that there are two distinct paths $Q,Q'\in \rset^*\cup \pset^*$, and two distinct edges $e\in E(Q)$, $e'\in E(Q')$, such that the images of $e$ and $e'$ cross in the drawing $\phi^*$ of $H'$.

\begin{observation}\label{obs: first bad event}
	If $\cro^*<
	\frac{\tilde k^2}{c_2\eta^5}$, then $\prob{\event_1}\leq 1/64$.
\end{observation}

\begin{proof}
Consider any crossing $(e,e')$ in the drawing $\phi^*$. We say that crossing $(e,e')$ is \emph{selected} if there are two distinct paths $Q,Q'\in \rset^*\cup \pset^*$ with  $e\in E(Q)$, $e'\in E(Q')$. Notice that $e$ may belong to at most two paths in $\rset_0\cup \pset_0$ (one path in each set), and the same is true for $e'$. Each path of $\rset_0\cup\pset_0$ is chosen to $\rset^*\cup \pset^*$ with probability at most $1/\psi$. Therefore, the probability that a path containing $e$, and a path containing $e'$ are chosen to $\rset^*\cup \pset^*$ is at most $4/\psi^2$. Since $\event_1$ can only happen if at least one crossing is chosen, from the union bound, 
$\prob{\event_1}\leq 4\cro^*/\psi^2$. Since $\psi>\frac{16\tk}{\eta^2}$, if $\cro^*<
\frac{\tilde k^2}{c_2\eta^5}$, then:
\[\prob{\event_1}\leq \frac{4\cro^*}{\psi^2}\leq \frac{1}{64c_2\eta}\leq \frac{1}{64}.\]	
\end{proof}

We say that a path $Q\in \rset_0\cup \pset_0$ is \emph{heavy} iff there are at least $\frac{\psi}{64\lambda}$ crossings $(e,e')$ in $\phi^*$, such that at least one of the edges $e,e'$ lies on path $Q$. We say that a bad event $\event_2$ happens iff at least one path in $\rset^*\cup \pset^*$ is heavy.

\begin{observation}\label{obs: second bad event}
	If $\cro^*<
	\frac{\tilde k^2}{c_2\eta^5}$, then $\prob{\event_2}\leq 1/64$.
\end{observation}
\begin{proof}
	Note that every edge of $H''$ may lie on at most two paths of $\rset_0\cup \pset_0$, and every crossing $(e,e')$ involves two edges. Therefore, the total number of heavy paths in $\rset_0\cup \pset_0$ is bounded by $\frac{4\cro^*}{\psi/(64\lambda)}=\frac{2^8\lambda\cdot\cro^*}{\psi}$. Assuming that $\cro^*<
	\frac{\tilde k^2}{c_2\eta^5}$, and using the fact that $\psi=\Omega\left(\frac{\tk}{\eta}\cdot \frac{\alpha^5(\alpha')^2}{\log^{16}m}  \right)$ and $\lambda=O\left( \frac{\eta \log^{12}m}{\alpha^4\alpha'}  \right )$, we get that the total number of heavy paths in $\rset_0\cup \pset_0$ is bounded by:

	\[ \frac{2^8\lambda\cdot\cro^*}{\psi}\leq O\left( \frac{\tilde k^2}{c_2\eta^5}\cdot \frac{\eta \log^{12}m}{\alpha^4\alpha'}\cdot\frac{\eta}{\tk}\cdot \frac{\log^{16}m}{\alpha^5(\alpha')^2} \right ) 
\leq	 O\left( \frac{\tilde k\log^{28}m}{c_2\eta^3\alpha^9(\alpha')^3} \right ).
	\]
	
Note that each heavy path may be selected to $\rset^*\cup \pset^*$  with probability at most $1/\psi$. Therfore, using the union bound and the fact that $\psi=\Omega\left(\frac{\tk}{\eta}\cdot \frac{\alpha^5(\alpha')^2}{\log^{16}m}  \right)$, we get that:

\[\prob{\event_2}\leq 
O\left( \frac{\tilde k\log^{28}m}{\psi\cdot c_2\eta^3\alpha^9(\alpha')^3}\right )\leq
O\left( \frac{\log^{44}m}{c_2\eta^2\alpha^{14}(\alpha')^5}\right ).\]
	
Recall that, from the conditions of \Cref{thm: find guiding paths}, $\eta\geq c^*\log^{46}m/(\alpha^{10}(\alpha')^4)$, where $c^*$ is a sufficiently large constant. Therefore, if 
$\cro^*<
\frac{\tilde k^2}{c_2\eta^5}$, then $\prob{\event_2}\leq 1/64$.
\end{proof}

Let $\rset'\subseteq \rset_1$, $\pset'\subseteq \pset_1$ be the sets containing all paths $Q$, such that, in drawing $\phi^*$, the image of some edge of $Q$ crosses the image of some edge lying on the paths of $\rset^*\cup \pset^*$. Note that the drawing $\phi^*$ is not known to us, and so neither are the sets $\rset',\pset'$ of paths. We will also use the following observation:

\begin{observation}\label{obs: few bad paths}
	If $\event_2$ did not happen, then $|\rset'|,|\pset'|\leq \psi/32$.
\end{observation}
\begin{proof}
	Recall that $|\rset^*|+|\pset^*|\leq \lambda$. If bad event $\event_2$ did not happen, then for each path $Q\in \rset^*\cup \pset^*$, there are at most $\frac{\psi}{64\lambda}$ crossings in $\phi^*$, in which edges of $Q$ participate. Therefore, if event $\event_2$ did not happen, there are in total at most $\psi/64$ crossings $(e,e')$ in the drawing $\phi^*$, where at least one of the edges $e,e'$ lies on a path of $\pset^*\cup \qset^*$. Let $E'\subseteq E(H'')$ be the set of all edges $e$, such that there is an edge $e'$ lying on some path of $\pset^*\cup \qset^*$, and crossing $(e,e')$ is present in $\phi^*$. Then $|E'|\leq \psi/32$. Each path in $\rset'\cup \pset'$ must contain an edge of $E'$. As the paths in each of the sets $\rset',\pset'$ are disjoint, $|\rset'|,|\pset'|\leq \psi/32$ must hold.
\end{proof}

\paragraph{Summary of Step 3.}
In this step we have constructed a grid skeleton, that consists of two sets of paths: $\pset^*=\set{ P^*_1,\ldots, P^*_{\lambda'}}$, and $\rset^*=\set{ R^*_1,\ldots, R^*_{\lambda'}}$, where $\lambda'= \floor{(\lambda-1)/2}$. Recall that $\pset^*\subseteq \pset_0$, and the paths in $\pset^*$ are indexed according to their order in $\oset_{\pset_0}$. Recall that we have also defined the set  $\pset_1\subseteq \pset_0$ of paths, containing all paths in sets $\set{\uset_{2i-1}}_{i=1}^{\lambda'+1}$. It would be convinient for us to re-index the groups $\uset_i$ as follows: for  $0\leq i\leq \lambda'$, set $\uset_i=\uset_{2i+1}$. In other words, the paths of $\uset_0$ lie before path $P^*_1$  in the ordering $\oset_{\pset_0}$, the paths of $\uset_{\lambda'}$ lie after $P^*_{\lambda'}$ in this ordering, and, for $1\leq i<\lambda'$, the  paths of $\uset_i$ lie between paths $P^*_i$ and $P^*_{i+1}$.
Similarly, $\rset^*\subseteq \rset_0$, and the paths in $\rset^*$ are indexed according to their order in $\oset_{\rset_0}$. We have also defined the set  $\rset_1\subseteq \rset_0$ of paths, containing all paths in sets $\set{\uset'_{2i-1}}_{i=1}^{\lambda'+1}$. As before, we re-index them as follows: for $0\leq i\leq\lambda'$, we set $\uset'_i=\uset'_{2i+1}$. Therefore, the paths of $\uset'_0$ lie before path $R^*_1$  in the ordering $\oset_{\rset_0}$, the paths of $\uset'_{\lambda'}$ lie after $R^*_{\lambda'}$ in this ordering, and, for $1\leq i<\lambda'$, the  paths of $\uset'_i$ lie between paths $R^*_i$ and $R^*_{i+1}$.

From our definition, if $\event_1$ did not happen, then for every pair $Q,Q'\in \pset^*\cup \qset^*$ of distinct paths, their images in $\phi^*$ do not cross (but note that the image of a single path may cross itself). 

We have also defined a set $\pset'\subseteq \pset_1$ and a set $\rset'\subseteq \rset_1$ of paths, containing all paths $Q$ whose image crosses the image of some path in $\rset^*\cup \pset^*$ in drawing $\phi^*$. From \Cref{obs: few bad paths}, if Event $\event_2$ does not happen, then $|\pset'|,|\rset'|\leq \psi/32$.
Note that the sets $\pset',\rset'$ of paths are not known to the algorithm.

It will be convenient for us to consider the $((\lambda'+1)\times (\lambda'+1))$-grid $\Pi^*$. We view the columns of the grid as corresponding to the left boundary of the bounding box $B^*$, the paths in $\set{R^*_1,\ldots,R^*_{\lambda'}}$, and the right boundary of the bounding box $B^*$.
For convenience, we index the columns of the grid from $0$ to $\lambda'+1$, so the left boundary of the bounding box corresponds to column $0$, and, for $1\leq i\leq \lambda'$, path $P^*_i$ represents the $i$th column of the grid, with the right boundary of $B^*$ repersenting the last column. Similarly, we view the bottom boundary of $B^*$, the paths in  $\set{P^*_1,\ldots,P^*_{\lambda'}}$, and the top boundary of $B^*$ as representing the rows of the grid, in the bottom-to-top order. As before, we index the rows of the grid so that the botommost row has index $0$ and the topmost row has index $\lambda'+1$. Notice however that the union of the paths in $\pset^*\cup \rset^*$ does not necessarily form a proper grid graph, as it is possible that, for a pair $P\in \pset^*$, $R\in \rset^*$ of paths, $P\cap R$ is a collection of several disjoint paths.

We will now consider the drawing $\phi^*$ of $H''$, and we will use it to define vertical and horizontal strips corresponding to paths in $\pset^*$ and $\rset^*$, respectively. We will also associate, with each cell of the grid $\Pi^*$, some region of the plane. We assume in the following definitions that Event $\event_1$ did not happen.

\begin{figure}[h]
\centering
\includegraphics[scale=0.35]{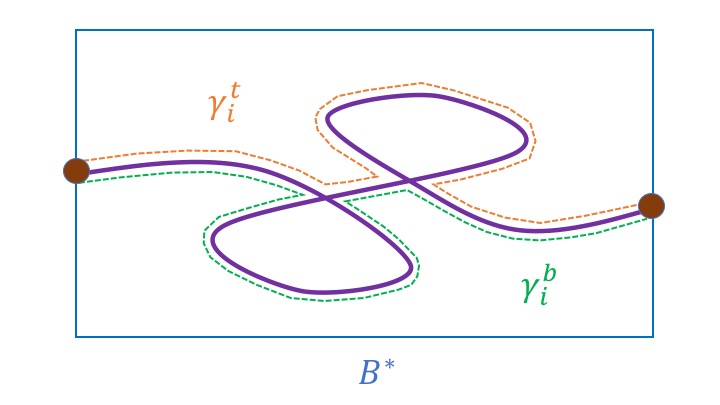}
\caption{An illustration of curves $\gamma^t_i$ and $\gamma^b_i$. The curve $\gamma_i$ is shown in purple.}\label{fig: top_bottom_curves}
\end{figure}

Consider first the image $\gamma_i$ of some path $P^*_i\in \pset^*$ in the drawing $\phi^*$.  Note that $\gamma_i$ is not necessarily a simple curve.
 We define two simple curves, $\gamma^t_i$ and $\gamma^b_i$, where $\gamma^t_i$ follows the image of $\gamma_i$ from the top, and $\gamma^b_i$ follows it from the bottom. In other words, we let $\gamma^b_i$ be a simple curve, whose every point lies on $\gamma_i$, that has the same endpoints as $\gamma_i$, such that the following holds: for every point $p\in \gamma_i$, either $p\in \gamma_i^b$, or $p$ lies above $\gamma_i^b$ in the bounding box $B^*$. We define the other curve, $\gamma^t_i$ symmetrically, so curve $\gamma_i$ is contained in the disc whose boundary is $\gamma_i^t\cup \gamma_i^b$ (see \Cref{fig: top_bottom_curves}). For convenience, we let $\gamma_0^t$ be the bottom boundary of the bounding box $B^*$, and $\gamma_{\lambda'+1}^b$ be the top boundary of the bounding box $B^*$.
 We now define, for all $0\leq i\leq \lambda'$, a region of the plane that we call the $i$th horizontal strip, and denote by $\hstrip_i$. This strip is simply the closed region of the bounding box between the curves $\gamma_i^t$ and $\gamma_{i+1}^b$. 

For every vertical path $R^*_i\in \rset^*$, we also define two curves, $\gamma^{\ell}_i$ and $\gamma^r_i$, that follow the image $\gamma_i$ of $R^*_i$ in $\phi^*$ on its left and on its right, respectively. We denote by $\gamma_0^r$ the left boundary of the bounding box $B^*$, and by $\gamma_{\lambda'+1}^{\ell}$ its right boundary. For all $0\leq i\leq \lambda'$, we define a vertical strip $\vstrip_i$ to be the closed region of the bounding box $B^*$ betwen $\gamma_i^r$ and $\gamma_{i+1}^{\ell}$.

The following observation is immediate from the fact that the paths in $\pset_0$ are node-disjoint, and so are the paths in $\rset_0$.

\begin{observation}\label{obs: must cross chosen paths}
	If $R\in \rset_1$ is a path whose image in $\phi^*$ intersects the interior of more than one vertical strip in $\set{\vstrip_0,\ldots,\vstrip_{\lambda'+1}}$, then $R\in \rset'$. Similarly, if $P\in \pset_1$ is a path whose image in $\phi^*$ intersects the interior of more than one horizontal strip in $\set{\hstrip_0,\ldots,\hstrip_{\lambda'+1}}$, then $P\in \pset'$. 
\end{observation}

Lastly, for all $0\leq i,j\leq \lambda'$, we let $\cellr_{i,j}=\hstrip_i\cap \vstrip_j$ be a closed region of the plane that we associate with cell $\cell_{i,j}$ of the grid $\Pi^*$.

\subsection{Step 4: Constructing a Grid-Like Structure}

In this step we further delete some paths from sets $\rset_1$ and $\pset_1$ to ensure that the resulting paths form a grid-like structure. This is done in three stages. In the first stage, we discard some paths to ensure that every remaining path in $\rset_1$ intersects the paths in $\pset^*$ ``in order'' (we formally define this notion later), and we process the paths in $\pset_1$ similarly. In the second stage, we associate, with every cell of the grid $\Pi^*$ a collection of horizontal paths and a collection of vertical paths. In the third stage, we ensure that, for every cell of the grid $\Pi^*$, there are many inersections between its corresponding horizontal and vertical paths.

Before we continue, we discard some paths of $\rset_1\cup \pset_1$ that must lie in $\rset'\cup \pset'$. Specifically, consider some path $P\in \pset_1$, and assume that it lies in group $\uset_i$, for some $0\leq i\leq \lambda'$. Let $(a,a')$ be the endpoints of path $P$, with $a\in T_1$ and $a'\in T_3$. Notice that from the definition, if $i>0$, then $a$ must lie, in the ordering $\tilde\oset_1$ of the terminals of $T_1$, after the endpoint of the path $P^*_i$ that belongs to $T_1$. Similarly, if $i<\lambda'$, then $a$ must lie before the endpoint of the path $P^*_{i+1}$ that belongs to $T_1$ in the same ordering. In particular, we are guaranteed that, in the drawing $\phi^*$, the image of $P$ must intersect the interior of the horizontal strip $\hstrip_i$. Consider now the endpoint $a'$ of $P$. If $i>0$, let $a'_i$ be the endpoint of path $P_i^*$ that lies in $T_3$, and if $i<\lambda'$, let $a'_{i+1}$ be the endpoint of path $P_{i'+1}$ that lies in $T_3$. Note that, if $a'$ lies before $a'_i$ in the ordering $\tilde\oset_3$ of $T_3$, or if $a'$ lies after $a'_{i+1}$ in the ordering $\tilde\oset_3$, then the image of $P$ has to intersect the interior of an additional horizontal strip, and, from \Cref{obs: must cross chosen paths}, path $P$ must lie in $\pset'$. We discard each such path from set $\pset_1$ (and from the corresponding set $U_i$). This ensures that, if $P\in U_i$, then its endpoint $a'$ must lie between $a'_i$ and $a'_{i+1}$ in $\tilde\oset_3$, if $1\leq i\leq \lambda'$; it must lie before $a'_{i+1}$ if $i=0$, and it must lie after $a'_i$ if $i=\lambda'$.
We process the paths in $\rset_1$ similarly, discarding paths as needed. Notice that so far all paths that we have discarded from $\pset_1\cup \rset_1$ lie in $\pset'\cup \rset'$.

\subsubsection{In-Order Intersection}

In this stage we discard some additional paths from $\pset_1\cup \rset_1$, to ensure that every remaining path in $\pset_1$ interesects the paths in $\rset^*$ in-order (notion that we define below); we do the same for paths in $\rset_1$. We will ensure that all paths discarded at this stage lie in $\pset'\cup \rset'$.

Since the definitions and the algorithms for the paths in $\pset_1$ and for the paths in $\rset_1$ are symmetric, we only describe the algorithm to process the paths in $\pset_1$ here.

Let $P\in \pset_1$ be any path, that we view as directed from its endpoint that lies in $T_1$ to its endpoint lying in $T_3$. Let $X(P)=\set{x_1,\ldots,x_r}$ denote all vertices of $P$ lying on paths in $R^*$, that is, $X(P)=V(P)\cap \left(\bigcup_{i=1}^{\lambda'}V(R^*_i)\right )$. We assume that the vertices of $X(P)$ are indexed in the order of their appearance on $P$. For each such vertex $x_j$, let $i_j$ be the index of the path $R^*_{i_j}\in \rset^*$ containing $x_j$.

\begin{definition}[In-order intersection]
	We say that path $P$ intersects the paths of $\rset^*$ in-order, if $r\geq \lambda'$, $i_1=1$, $i_r=\lambda'$, and, for $1\leq j<r$, $|i_j-i_{j+1}|\leq 1$.
\end{definition}

Notice that the definition requires that path $P$ intersects every path of $\rset^*$ at least once; the first path of $\rset^*$ that it intersects must be $R_1^*$, and the last path must be $R_{\lambda'}^*$, and for every consecutive pair $x_j,x_{j+1}$ of vertices in $X(P)$, either both vertices lie on the same path of $\rset^*$, or they lie on consecutive paths of $\rset^*$. Notice that path $P$ is still allowed to intersect a path of $\rset^*$ many times, and may go back and forth across all these paths several times.

\begin{observation}\label{obs: not in order intersection}
	Assume that Event $\event_1$ did not happen. Let $P\in \pset_1$ be a path that  intersect the paths of $\rset^*$ not in-order. Then $P\in \pset'$ must hold.
\end{observation}

\begin{proof}
	Assume first that $i_1\neq 1$, that is, vertex $x_1$ lies on some path $R^*_i$ with $i\neq 1$. Let $p$ be a point on the image of path $P$ in $\phi^*$ that is very close to its first endpoint, so $p$ lies in the interior of the vertical strip $\vstrip_1$, and let $p'$ be the image of the point $x_1$. Clearly, $p'$ does not lie in the interior or on the boundary of $\vstrip_1$, so the image of path $P$ must cross the right boundary of $\vstrip_1$, which means that the image of some edge of $P$ and the image of some edge of $R_1^*$ cross in $\phi^*$.
	
	The cases where $i_{r}\neq \lambda'$, or there is an index $1\leq j<r$ with $i_j-i_{j+1}> 1$ are treated similarly, as is the case when $r<\lambda'$.
\end{proof}

We discard from $\pset_1$ all paths $P$ that intersect the paths of $\rset^*$ not in-order. We denote by $\pset_2\subseteq \pset_1$ the set of remaining paths. We also update the groups $\uset_0,\ldots,\uset_{\lambda'}$ accordingly. Observe that so far all paths that we have discarded from $\pset_1$ lie in  $\pset'$. 
From \Cref{obs: few bad paths}, assuming that Events $\event_1$ and $\event_2$ did not happen, the number of paths that we have discarded so far from $\pset_1$ is at most $\psi/32$. In particular, for all $0\leq i\leq \lambda'$, $|\uset_i|\geq 31\psi/32$ still holds.

We perform the same transformation on set $\rset_1$ of paths, obtaining a new set $\rset_2$ of paths, each of which intersects the paths of $\pset^*$ in-order. We also update the groups $\uset'_0,\ldots,\uset'_{\lambda'}$. As before, for all $0\leq i\leq \lambda'$, $|\uset'_i|\geq 31\psi/32$ still holds.

\subsubsection{Definining Paths Associated with Grid Cells}

For every path $P\in \pset_2$, for all $1\leq i\leq \lambda'$, we denote by $v_i(P)$ the first vertex on path $P$ that belongs to the vertical path $R^*_i$; note that, from the definition of in-order intersection, such a vertex must exist. For all $1\leq i< \lambda'$, we define the $i$th segment of $P$, $\sigma_i(P)$, to be the subpath of $P$ between $v_i(P)$ and $v_{i+1}(P)$. We also let $\sigma_0(P)$ be the subpath of $P$ from its first vertex (which must be a terminal of $T_1$) to $v_1(P)$, and by $\sigma_{\lambda'}(P)$ the subpath of $P$ from $v_{\lambda'}(P)$ to the last vertex of $P$ (which must be a terminal of $T_3)$. Note that the sets of edges that lie on paths $\sigma_0(P),\ldots,\sigma_{\lambda'}(P)$ partition $E(P)$.

Similarly, for every path $R\in \rset_2$, for all $1\leq i\leq \lambda'$, we denote by $v_i(R)$ the first vertex on path $R$ that lies on the horizontal path $P^*_i$. For all $1\leq i< \lambda'$, we define the $i$th segment of $P$, $\sigma_i(R)$, to be the subpath of $R$ between $v_i(R)$ and $v_{i+1}(R)$. We also let $\sigma_0(R)$ be the subpath of $R$ from its first vertex (which must be a terminal of $T_4$) to $v_1(R)$, and by $\sigma_{\lambda'}(R)$ the subpath of $R$ from $v_{\lambda'}(R)$ to the last vertex of $R$ (which must be a terminal of $T_2$).

Consider now some cell $\cell_{i,j}$ of the grid $\Pi^*$, for some $0\leq i,j\leq \lambda'$. We define the set $\pset^{i,j}$ of horizontal paths, and the set $\rset^{i,j}$ of vertical paths associated with cell $\cell_{i,j}$, as follows. In order to define the set $\pset^{i,j}$ of horizontal paths, we consider the group $\uset_i\subseteq \pset_2$, and, for every path $P\in \uset_i$, we include its $j$th segment $\sigma_j(P)$ in $\pset^{i,j}$, so $\pset^{i,j}=\set{\sigma_j(P)\mid P\in \uset_i}$.
Similarly, we define $\rset^{i,j}=\set{\sigma_i(R)\mid R\in \uset'_j}$.
%
%
We need the following observation.
	
	\begin{observation}\label{obs: paths in cells don't cross}
		Let $P\in \uset_i,R\in \uset'_j$ be a pair of paths, for some $1< i,j< \lambda'$, and assume that their subpaths $\sigma_j(P)\subseteq P,\sigma_i(R)\subseteq R$ do not share any vertices. Then either $P\in \pset'$, or $R\in \rset'$, or the images of $\sigma_j(P)$ and $\sigma_i(R)$ cross in the drawing $\phi^*$.
	\end{observation}
\begin{proof}
	Assume that $P\not\in \pset'$ and $R\not\in \rset'$, that is, the images of the paths $P,R$ do not cross the images of the paths in $\pset^*\cup \rset^*$ in $\phi^*$. From the definition of set $\uset_i$, the image of $P$ intersects the interior of the horizontal strip $\hstrip_i$, and path $P$ does not share any vertices with the paths of $\pset^*$. Therefore, the image of $P$ must be contained in the strip $\hstrip_i$, and it is disjoint from its top and bottom boundaries $\gamma^t_i,\gamma^b_{i+1}$. Using similar reasoning, the image of $R$ is contained in the strip $\vstrip_j$, and it is disjoint from its left and right boundaries, $\gamma^{r}_j,\gamma^{\ell}_{j+1}$. Consider now the segment $\sigma_j(P)$ of $P$, whose endpoints lie on $R^*_j$ and $R^*_{j+1}$, respectively. Let $\sigma'_j(P)\subseteq \sigma_j(P)$ be the shortest subpath of $\sigma_j(P)$ whose first endpoint lies on $R^*_j$, and whose last endpoint lies on $R^*_{j+1}$; such a path must exist because we can let $\sigma'_j(P)=\sigma_j(P)$. From the definition of in-order intersection,  no inner vertex of $\sigma'_j$ may lie on any path of $\rset^*$. It is then easy to verify that the image of $\sigma'_j(P)$ in $\phi^*$ must be contained in $\cellr_{i,j}$, and it must split this region into two subregions: one whose top boundary contains a segment of $\gamma^b_{i+1}$, and one whose bottom boundary contains a segment of $\gamma^t_i$. 
	
	Using the same reasoning, we can select a segment $\sigma'_i(R)$, whose first endpoint lies on $P^*_i$, last endpoint lies on $P^*_{i+1}$, and all inner vertices are disjoint from the vertices lying on the paths in $\pset^*$.
	As before, the image of $\sigma'(R)$ must be contained in $\cellr_{i,j}$, but it connects a point on its top boundary to a point on its bottom boundary. Therefore, the image of $\sigma'_i(R)$ must cross the image of $\sigma'_j(P)$.
\end{proof}

\subsubsection{Completing the Construction of the Grid-Like Structure}

In order to complete the construction of the grid-like structure, we need to ensure that, for every pair $1< i,j< \lambda'$ of indices, there are many intersection between the sets $\pset^{i,j}$ and $\rset^{i,j}$ of paths. More specifically, we need to ensure that every path $\sigma\in \pset^{i,j}$ intersects many paths in $\rset^{i,j}$, and vice versa. This is needed in order to ensure well-linkedness properties: namely, that the collection of vertices containing the first and the last vertex on every path of $\pset^{i,j}$ is sufficiently well-linked in the graph obtained from the union of the paths in $\rset^{i,j}\cup \pset^{i,j}$. This property, in turn, will be exploited in order to construct the routers of $\Lambda'$ over which the distribution $\dset$ will be defined. This motivates the following definition.

\begin{definition}[Bad Paths]
	For a pair $0< i,j< \lambda'$ of indices, we say that a path $P\in \uset_i$ is \emph{bad for cell $\cell_{i,j}$} if there are at least $\psi/16$ paths in $\rset^{i,j}$ that are disjoint from $\sigma_j(P)$. Similarly, we say that a path $R\in \uset'_j$ is bad for cell $\cell_{i,j}$ if there are at least $\psi/16$ paths in $\pset^{i,j}$ that are disjoint from $\sigma_i(R)$.
	
	Consider now some index $0< i< \lambda'$. We say that a path $P\in \uset_i$ is \emph{bad} if it is bad for at least one cell in $\set{\cell_{i,j}\mid 0< j< \lambda'}$. Similarly, for an index  $0< j< \lambda'$, a path $R\in \uset'_j$ is bad if it is bad for at least one cell in $\set{\cell_{i,j}\mid 0< i< \lambda'}$.
\end{definition}

 The following observation bounds the number of bad paths in each group $\uset_i$ of horizontal paths, and in each group $\uset'_j$ of vertical paths.

\begin{observation}\label{obs: few bad paths in each group}
Assume that $\cro^*\leq \frac{\tilde k^2}{c_2\eta^5}$, and that neither of the events $\event_1,\event_2$ happenned. Then for all $0< i< \lambda'$, at most $\psi/16$ paths in $\uset_i$ are bad. Similarly, for all  $0< j<\lambda'$, at most $\psi/16$ paths in $\uset'_j$ are bad.
\end{observation}

\begin{proof}
Fix an index $0< i< \lambda'$, and the corresponding set $\uset_i\subseteq \pset_2$ of paths. We partition the set of all bad paths in $\uset_i$ into two subsets: set $\bset_1$ contains all bad paths lying in $\pset'$, and set $\bset_2$ contains all remaining bad paths. From \Cref{obs: few bad paths}, $|\bset_1|\leq \psi/32$.

We further partition the set $\bset_2$ of bad paths into subsets $\set{\bset_2^j\mid 0< j< \lambda'}$, where a path $P$ lies in $\bset_2^j$ if it is bad for cell $\cell_{i,j}$ (if path $P$ is bad for several cells, we add it to any of the corresponding sets). Consider now some index $0< j< \lambda'$, and some path $P\in \bset_2^j$. From the definition, there is a set $\Sigma'\subseteq \rset^{i,j}$ of at least $\psi/16$ paths that do not share any vertices with $P$. From \Cref{obs: few bad paths}, at most $\psi/32$ of these paths may lie in $\rset'$. Let $\Sigma''\subseteq \Sigma'$ be the collection of the remaining paths, whose cardinality is at least $\psi/32$. From \Cref{obs: paths in cells don't cross}, for every path $\sigma'\in \Sigma'$, the images of $\sigma_j(P)$, and of $\sigma'$ must cross. We let $\chi_j(P)$ denote the set of all crossings $(e,e')$, where $e\in \sigma_j(P)$, and $e'$ is an edge on a path of $\Sigma''$, so $|\chi_j(P)|\geq \psi/32$. We then let $\chi_{j}=\bigcup_{P\in \bset_2^j}\chi_j(P)$, so $|\chi_{j}|\geq |\bset_2^j|\cdot \psi/32$. Lastly, we let $\chi=\bigcup_{j=1}^{\lambda'-1}\chi_j$. Notice that set $\chi$ contains at least $|\bset_2|\cdot \psi/32$ distinct crossings in the drawing $\phi^*$. Assume for contradiction that $|\bset_2|>\psi/32$. Then:
\[\cro^*>\frac{\psi^2}{2^{10}}>\frac{\tk^2}{4\eta^4}>\frac{\tilde k^2}{c_2\eta^5},\]
since $\psi>\frac{16\tk}{\eta^2}$, a contradiction. 
Therefore, $|\bset_2|\leq \psi/32$, and overall there are at most $\psi/16$ bad paths in $\uset_i$.
The proof for path sets $\uset'_j\subseteq \rset_2$ is identical.
\end{proof}

For all $0< i< \lambda'$, we discard every bad path from $\uset_i$. If $|\uset_i| <\ceil{7\psi/8}$ for any $i$, then we terminate the algorithm and return FAIL. Notice that in this case, from \Cref{obs: few bad paths in each group}, if  $\cro^*< \frac{\tilde k^2}{c_2\eta^5}$, then at least one of the events $\event_1,\event_2$ must have happened, and the probability for this is at most $1/8$. Therefore, we assume that for all 
$0< i< \lambda'$, $|\uset_i|\geq \ceil{7\psi/8}$ holds. We discard additional arbitrary paths from $\uset_i$, until $|\uset_i|= \ceil{7\psi/8}$.
We then let $\pset_3=\bigcup_{i=1}^{\lambda'}\uset_i$ denote the resulting set of paths. 

Similarly, for all $0< j< \lambda'$, we discard every bad path from $\uset'_j$. If, as the result, $|\uset'_j|$ falls below  $\ceil{7\psi/8}$, we terminate the algorithm and return FAIL. Otherwise, we discard additional arbitrary paths as needed, so that $|\uset'_j|=\ceil{7\psi/8}$ holds. We also let $\rset_3=\bigcup_{j=1}^{\lambda'}\uset'_j$.

For all $0< i,j< \lambda'$, we also update the path sets $\pset^{i,j}$ and $\rset^{i,j}$ accordingly, discarding the paths that are no longer subpaths of paths in $\pset_3\cup \rset_3$. Since we are still guaranteed that $|\pset^{i,j}|,|\rset^{i,j}|= \ceil{7\psi/8}$, and since every path that is bad for cell $\cell_{i,j}$ was discarded, we are guaranteed that every path in $\pset^{i,j}$ intersects at least $\frac{7\psi}{8}-\frac{\psi}{16}=\frac{13\psi}{16}$ paths of $\rset^{i,j}$ and vice versa.
Since we use this fact later, we summarize it in the following observation.

\begin{observation}\label{obs: paths for cells}
	For all $0< i,j< \lambda'$, $|\pset^{i,j}|,|\rset^{i,j}|= \ceil{7\psi/8}$. Every path in 
$\pset^{i,j}$ intersects at least $\frac{13\psi}{16}$ paths of $\rset^{i,j}$ and vice versa.
\end{observation}

 This concludes the construction of the grid-like structure.

\subsection{Step 5: the Routing}

Recall that we have denoted by $\Lambda'=\Lambda(H'',\tT)$ the set of all  routers in graph $H''$ with respect to the set $\tT$ of terminals. In this final step we design an efficient algorithm to compute an explicit distribution $\dset$ over the routers of $\Lambda'$, such that for every outer edge $e\in E(H'')$, $\expect[\qset\sim\dset]{(\cong_{H''}(\qset,e))^2}\leq  O\left (\frac{\log^{16}m}{(\alpha\alpha')^4}\right )$.

Our algorithm closely follows the arguments of 
\cite{Tasos-comm}  (see also Lemma D.10 in the full version of \cite{chuzhoy2011algorithm}), who showed a similar result for a grid graph. In order to provide intuition, we first present their algorithm. Assume that we are given a $(q\times q)$ grid graph $G$ for some integer $q$, and let $T$ be the set of vertices lying on the first row of the grid, that we refer to as terminals. For convenience, assume that $q$ is an integral power of $2$. Our goal is to compute a distribution $\dset'$ over the routers of in $\Lambda(G,T)$. We need to ensure that, for every edge $e\in E(G)$, the expectation $\expect[\qset\sim\dset']{(\cong_G(\qset,e))^2}\leq O(\log q)$.

For every vertex $v$ in the top right quadrant of the grid, we will define a set $\qset(v)$ of paths in $G$, routing the terminals in $T$ to $v$. Our distribution $\dset$ then assigns, to each such router $\qset(v)$,  the same probability value $4/q^2$.

We now fix a vertex $v$ in the top right quadrant of the grid, and define the router $\qset(v)$. Let $r=\log(q/4)$. For $0\leq i\leq r$, let $S_i$ be a square subgrid of $G$, of size $(2^i\times 2^i)$, whose upper right corner has the same column-index as vertex $v$, and the same row-index as the bottom left corner of $S_{i-1}$ (we think of $S_0$ as a $(1\times 1)$-grid consisting only of vertex $v$). 
We refer to the subgrids $S_i$ of $G$ as \emph{squares}, and specifically to square $S_i$ as \emph{level-$i$ square}.
For all $0\leq i\leq r$, we denote by $T_i$ the set of vertices lying on the bottom boundary of square $S_i$. Using the well-linkedness of the grids, it is easy to show that for all $1\leq i\leq r$, there is a collection $\pset_i$  of paths in graph $S_i$, routing vertices of $T_i$ to vertices of $T_{i-1}$ with congestion at most $2$, such that every vertex  of $T_{i-1}$ serves as endpoint of at most two such paths. For $1\leq i\leq r$, let
Let $\pset'_i$ be a multipset obtained from set $\pset_i$ by creating $2^{r-i+1}$ copies of every path in $\pset_i$. 
Let $T_{r+1}\subseteq T$ be a set of $|T_r|$ vertices lying on the bottom boundary of the grid $G$, that contains, for every vertex $t\in T_r$, vertex $t'$ on the bottom boundary of the grid with the same column index as $t$. Let $P_t$ be the subpath of the corresponding column of $G$ connecting $t$ to $t'$, and denote $\pset'_{r+1}=\set{P_t\mid t\in T_r}$.

By concatenating the paths in $\pset_1',\ldots,\pset'_{r+1}$, we obtain a collection $\qset'(v)$ of paths in grid $G$, routing the terminals in $T_{r+1}$ to vertex $v$. 
Notice that for all $0\leq i\leq r$, for every edge $e$ lying in $S_i$, the congestion on 
edge $e$ due to paths in $\qset'(v)$ is at most $2^{r-i+2}$.
The key in analyzing the expectation $\expect[\qset\sim\dset']{(\cong_G(\qset,e))^2}$
is to notice that, for all $1\leq i\leq r$, square $S_i$ is a $(2^i\times 2^i)$-subgrid of $G$, whose upper right corner is chosen uniformly at random from a set of $q^2/4$ possible  points. The total number of subgrids of $G$ of size $(2^i\times 2^i)$ that contain $e$ is $2^{2i}$, so the probability that any of them is selected is bounded by $2^{2i+2}/q^2$. Therefore, for all $1\leq i\leq r$, with probability at most $2^{2i+2-2r}$, edge $e$ belongs to square $S_i$, and in this case, $\cong_G(\qset,e)\leq 2^{r-i+2}$. Therefore, we get that:
\[\expect[\qset\sim\dset']{(\cong_G(\qset,e))^2}
\leq \sum_{i=1}^{r}2^{2i+2-2r}\cdot 2^{2r-2i+4}\leq O(r)=O(\log q).
\]
Using the well-linkedness of the terminals in $T$, it is immediate to extend the set $\qset'(v)$ of paths to a set $\qset^*(v)$ routing all terminals in $T$ to $v$, while increasing the congestion on every edge of $G$ by at most an additive constant and a multiplicative constant factor. This provides the final distribution $\dset$ over the routers $\qset(v)\in \Lambda(G,T)$.

We will simulate a similar process on the grid $\Pi^*$, and its corresponding grid-like structure that we have constructed. Notice however that $\Pi^*$ is only a $(\lambda'\times\lambda')$-grid (where $\eta\leq \lambda'\leq \eta^2$), while the number of terminals that we need to route is much larger (comparable to $|\rset_3|$). Therefore, we will attempt to route all terminals to a single cell $\cell_{i,j}$ in the top right quadrant of the grid $\Pi^*$ (in other words, we will route them to vertices lying on paths  in $\pset^{i,j}\cup \rset^{i,j}$). This in itself is not sufficient, since we need to route them to a single vertex of $H''$. This means that we may need to perform some routing within the cell $\cell_{i,j}$, that is, within the graph obtained from the union of the paths in $\pset^{i,j}\cup \rset^{i,j}$. While generally such a routing (with low congestion on outer edges) may be difficult to compute, we will select a large collection of cells (called good cells) in the top right quadrant of the grid $\Pi^*$, for which such a routing is easy to obtain. We will then define, for each good cell, the corresponding set of paths routing the terminals to a single vertex $y^*\in V(H'')$. We do so by simulating the process described above: we define square subgrids $\set{S_i}$ of the grid $\Pi^*$, and we associate these subgrids with sets of horizontal and vertical paths (subpaths of some paths in $\pset_3\cup \rset_3$), so that the desired well-linkedness properties of graphs corresponding to each subgrid $S_i$ are achieved. Eventually, the distribution $\dset$ chooses one of the good cells uniformly at random, and uses the associated router $\qset\in \Lambda'$ in order to route the terminals to a single vertex of $H''$. The analysis of expected congestion squared on every outer edge of $H''$ is very similar to the one outlined above.

We start by defining the notion of good cells of the grid $\Pi^*$, and showing that a large enough number of such cells exist in the upper right quadrant of $\Pi^*$. We will then define square subgrids of $\Pi^*$ and associate sets of paths with each such subgrid to ensure the required well-linkedness properties. Lastly, we show how to construct the desired routing $\qset$ for each good cell.

\subsubsection{Good Cells}

Fix a pair of indices $0< i,j< \lambda'$, and consider the cell $\cell_{i,j}$ of the grid $\Pi^*$, and the two corresponding sets $\pset^{i,j}$, $\rset^{i,j}$ of paths.

\begin{definition}[Good cells]
A path $\sigma\in \pset^{i,j}$ is \emph{good} for cell $\cell_{i,j}$ if $\sigma$ contains no outer edges. 
We say that cell $\cell_{i,j}$ is \emph{good} if some path $\sigma \in \pset^{i,j}$ is good for $\cell_{i,j}$; otherwise we say it is \emph{bad}. 
\end{definition}

Assume that cell $\cell_{i,j}$ is good, and let $\sigma\in \pset^{i,j}$ be any horizontal path that is good for this cell. Since $\sigma$ contains no outer edges, there must be a vertex $y\in V(H)$, such that $V(\sigma)\subseteq V(\Pi(y))$. Recall that, from \Cref{obs: paths for cells}, $|\rset^{i,j}|=\ceil{7\psi/8}$, and that $\sigma$ intersects at least $13\psi/16$ paths of $\rset^{i,j}$. Let $\hat \rset^{i,j}\subseteq \rset^{i,j}$ be a set of $\ceil{13\psi/16}$ paths, each of which shares at least one vertex with $\sigma$. Note that each such path then must contain a vertex of $\Pi(y)$. We denote by $\portals^{i,j}$ the set of vertices that contains, for every path $\sigma'\in \hat\rset^{i,j}$, the first vertex of $\sigma'$ (by definition, each such vertex must lie on path $P^*_i$). For convenience, we denote vertex $y$  of $H$ by $y_{i,j}$.

Let $Z$ be the set of all pairs of indices $\floor{\lambda'/2}\leq i,j< \lambda'$, such that $\cell_{i,j}$ is good. Next, we show that $|Z|$ is sufficiently large. Our routing algorithm will then choose a pair $(i,j)$ of indices from $Z$ uniformly at random, and route the terminals to the vertices in set $\portals^{i,j}$, from where they will be routed to vertices of $\Pi(y_{i,j})$, and eventually to some specific vertex of $\Pi(y_{i,j})$.

\begin{claim}\label{claim: many good cells}
	$|Z|\geq (\lambda')^2/16$.
\end{claim}

\begin{proof}
	Let $\bset$ be a collection of all bad cells $\cell_{i,j}$ lying in the top right quadrant, that is, $\floor{\lambda'/2}\leq i,j< \lambda'$. It is enough to show that $|\bset|<(\lambda')^2/16$.
	
	Consider now some bad cell $\cell_{i,j}\in \bset$, and any path $Q\in \pset^{i,j}$. Since cell $\cell_{i,j}$ is bad, $Q$ must contain at least one outer edge. We say that $Q$ is a \emph{type-1} bad path for cell $\cell_{i,j}$ if it contains at least one type-1 outer edge (recall that a type-1 outer edge $e$ corresponds to some edge in graph $H$ that is {\bf not} contained in any cluster of $\cset$). Otherwise, every outer edge on path $Q$ is a type-2 outer edge, and in this case we say that $Q$ is a type-2 bad cluster for $\cell_{i,j}$. We say that cell $\cell_{i,j}$ is \emph{type-1 bad} if at least $\psi/32$ paths of $\pset^{i,j}$ are type-1 bad for this cell, and otherwise it is type-2 bad. We partition the set $\bset$ of bad cells into two subsets: set $\bset_1$ contains all type-1 bad cells, and set $\bset_2$ contains all type-2 bad cells. It is now enough to prove that $|\bset_1|,|\bset_2|<(\lambda')^2/32$, which we do in the following two observations.

	\begin{observation}\label{obs: few type-1 bad cells}
		$|\bset_1|< (\lambda')^2/32$.
	\end{observation}
\begin{proof}
	Assume for contradiction that $|\bset_1|\geq (\lambda')^2/32$.
	Consider a type-1 bad cell $\cell_{i,j}\in \bset_1$, and let $\qset^{i,j}\subseteq \pset^{i,j}$ be a set of $\ceil{\psi/32}$ paths that are type-1 bad paths for cell $\cell_{i,j}$. Each path in $\qset^{i,j}$ must contain at least one type-1 bad edge. Since the paths in $\qset^{i,j}$ are edge-disjoint, there is a set $E^{i,j}$ of at least $\psi/32$ type-1 outer edges of $H''$, lying on paths of $\qset^{i,j}$. Since every edge of $H''$ may lie on at most one path in $\pset$, the total number of outer edges in $H''$ must be at least: 
\[\frac{|\bset_1|\cdot \psi}{32}\geq \frac{(\lambda')^2\cdot \psi}{2^{10}}\geq  \frac{\lambda^2\cdot \psi}{2^{14}} ,\] 
as $\lambda'=\floor{(\lambda-1)/2}\geq \lambda/4$.
Recall that $\psi=\floor{\frac{\alpha^*\tk}{64\lambda}}$ and $\lambda=\frac{2^{24}c\cdot \eta \log^8m}{\alpha^*\alpha^3}$,  where $c$ is the constant from \Cref{obs: few outer edges}. Therefore, we get that the total number of outer edges in $H''$ is at least $\frac{2c\tk \eta \log^8m}{\alpha^3}$, contradicting \Cref{obs: few outer edges}.
\end{proof}

\begin{observation}\label{obs: few type-2 bad cells}
	$|\bset_2|< (\lambda')^2/32$.
\end{observation}	
\begin{proof}
		For a cluster $C\in \cset$, let $X(C)=\bigcup_{y\in V(C)}V(\Pi(y))$. Note that all terminals of $H$ lie outside of the clusters in $\cset$, and so $X(C)\cap \tT=\emptyset$. If a path $Q\in \pset_3\cup \rset_3$ contains a vertex of $X(C)$, then it must contain at least one edge of $\delta_H(C)$. As the paths in $\pset\cup \rset$ cause edge-congestion at most $2$, the total number of paths $Q\in \pset\cup \rset$ with a non-empty intersection with $X(C)$ is at most $2\delta_H(C)$.

	Let $\intpairs\subseteq \pset_3\times\rset_3$ be the collection of all pairs of paths $P\in \pset_3$, $R\in \rset_3$, such that $P$ and $R$ share at least one vertex. 
	For a cluster $C\in \cset$, let $\intpairs'_C\subseteq \intpairs$ denote the collection of all pairs $(P,R)\in \intpairs$ of paths, such that some vertex $v\in X(C)$ lies on both $P$ and $R$. Clearly, if $(P,R)\in \intpairs_C'$, then each of the paths $P$, $R$ must contain at least one edge of $\delta_H(C)$. Therefore, from the above discussion, $|\intpairs'_C|\leq 4|\delta_H(C)|^2$. Let $\intpairs'=\bigcup_{C\in \cset_Y}\intpairs'_C$. Then:
	\[ |\intpairs'|\leq \sum_{C\in \cset_Y}|\intpairs'_C|\leq  4\sum_{C\in \cset_Y}|\delta_H(C)|^2. \]
	From Equation \ref{eq: sum of squares} (see \Cref{step 2 summary}), $
	\sum_{C\in \cset_Y}|\delta_H(C)|^2<\frac{(\tk\talpha\alpha')^2}{c_1\log^{20}m}$, so we get that:
	\begin{equation}\label{eq: bounding num of intersection pairs}
	|\intpairs'|<\frac{4(\tk\talpha\alpha')^2}{c_1\log^{20}m},
	\end{equation} 
	where $c_1$ is an arbitrarily large constant.
	
	In the remainder of the proof, we assume for contradiction that $|\bset_2|\geq (\lambda')^2/32$, and we will show that $|\intpairs'|\geq \frac{4(\tk\talpha\alpha')^2}{c_1\log^{20}m}$ must hold, contradicting Equation \ref{eq: bounding num of intersection pairs}.

Consider a type-2 bad cell $\cell_{i,j}\in \bset_2$. Recall that every path in $\pset^{i,j}$ contains at least one outer edge, and at most $\psi/32$ such paths contain a type-1 bad edge. Since, from \Cref{obs: paths for cells}, $|\pset^{i,j}|= \ceil{7\psi/8}$, 
there is a collection $\Sigma\subseteq \pset^{i,j}$ of at least  $3\psi/4$ paths $P$, such that all edges on $P$ are either inner edges, or type-2 outer edges. Therefore, if $P\in \Sigma$ is any such path, then there is some cluster $C\in \cset$ with $V(P)\subseteq X(C)$. Recall that, from \Cref{obs: paths for cells}, each path $P\in \Sigma$ intersects at least $\frac{13\psi}{16}$ paths of $\rset^{i,j}$. Clearly, if a path $R\in \rset^{i,j}$ intersects a path $P\in \Sigma$, then $(P,R)\in \intpairs'$. Therefore, intersections between pairs of paths in $\pset^{i,j}\times \rset^{i,j}$ contribute at least $\frac{13\psi}{16}\cdot \frac{3\psi}{4}\geq \frac{\psi^2}{2}$ pairs to set $\intpairs'$.
Therfore, if we denote by $\intpairs'_{i,j}$ the collection of all pairs $(P,R)\in \intpairs'$, where a subpath $\sigma$ of $P$ lies in $\pset^{i,j}$,  and a subpath $\sigma'$ of $R$ lies in $\pset^{i,j}$, and $\sigma,\sigma'$ contain a vertex $v\in X(C)$, for some cluster $C\in \cset$, then, from the above discussion, $|\intpairs'_{i,j}|\geq \frac{\psi^2}{2}$. We claim that for every pair $(P,R)\in \intpairs'$ of paths, there is at most one pair of indices $0< i,j< \lambda'$, such that $(P,R)\in \intpairs_{i,j}'$. Indeed, assume that $P\in \uset_i$ and $R\in \uset'_j$. For a pair $0< i',j'< \lambda'$ of indices, $\pset^{i',j'}$ contains a subpath of $P$ iff $i'=i$, and $\rset^{i',j'}$ contains a subpath of $R$ iff $j'=j$. So the only pair $(i',j')$ of indices for which $(P,R) \in \intpairs'_{i',j'}$ may hold is $(i,j)$. Overall, we get that $|\intpairs'|\geq |\bset_2|\cdot \psi^2/2$. Assuming that $|\bset_2|\geq (\lambda')^2/32$, since $\lambda'=\floor{(\lambda-1)/2}\geq \lambda/4$, we get that $|\intpairs'|\geq \frac{\lambda^2\psi^2}{1024}$.

Recall that $\psi=\floor{\frac{\alpha^*\tk}{64\lambda}}$, and, from \Cref{obs: terminals well linked in H''},  $\alpha^*=\Theta(\talpha\alpha')$. 
We conclude that:
\[|\intpairs'|\geq \frac{(\alpha^*)^2\tk^2}{2^{22}}\geq \Omega\left( {(\talpha\alpha'\tk)^2} \right ).  \] 
Since we can choose $c_1$ to be a sufficiently large constant, this contradicts Equation \ref{eq: bounding num of intersection pairs}.
\end{proof}	
\end{proof}

\subsubsection{Square Subgrids and Corresponding Sets of Paths}

For integers $1\leq i,j< \lambda'$ and $\ell\leq \min\set{i,j}$, a \emph{square subgrid $S=S(i,j,\ell)$ of $\Pi^*$} (that we also refer to as a \emph{square}) is defined as the collection of cells $\cellset(S)=\set{\cell_{i',j'}\mid i-\ell+1\leq i'\leq i;\quad j-\ell+1\leq j'\leq j}$. Intuitively, $S(i,j,\ell)$ is a subgrid of $\Pi^*$ of size $(\ell\times \ell)$, whose top right corner is the cell $\cell_{i,j}$.

Given a square $S=S(i,j,\ell)$, we associate with it a collection $\pset(S)$ of horizontal paths, and $\rset(S)$ of vertical paths, as follows. Intuitively,
consider the graph  obtained by taking the union of all paths $\pset^{i',j'}$, where $\cell_{i',j'}\in \cellset(S)$.
This graph is a collection of disjoint paths, each of which is a subpath of a distinct path in $\bigcup_{i'=i-\ell+1}^i\uset_{i'}$; we let $\pset(S)$ be this set of paths.
Formally, for all $i-\ell+1\leq i'\leq i$, for every path $P\in \uset_{i'}$, we include in $\pset(S)$ the subpath of $P$ from the first vertex of $\sigma_{j-\ell+1}(P)$ to the last vertex of $\sigma_{j}(P)$.
Similarly, set $\rset(S)$ contains, for all $j-\ell+1\leq j'\leq j$, for every path $R\in \uset'_{j'}$, the subpath of $R$ from the first vertex of $\sigma'_{i-\ell+1}(R)$ to the last vertex of $\sigma'_{i}(R)$.
Notice that, from \Cref{obs: paths for cells}, $|\pset(S)|=|\rset(S)|= \ceil{7\psi/8}\cdot \ell$.

We denote by $\entryportals(S)$ the set of all vertices that serve as the first endpoint of the paths in $\pset(S)$, and by $\exitportals(S)$ the set of all vertices that serve as the last endpoint of the paths in $\pset(S)$. We denote by $G(S)$ the graph obtained by the union of the paths in $\pset(S)\cup \rset(S)$.

The following claim will be crucial for our algorithm for computing the routing paths for each good cell.

\begin{claim}\label{claim: routing in square}
Let $S=S(i,j,\ell)$ be a square of $\Pi^*$,	for some $1\leq i,j< \lambda'$ and $\ell\leq \min\set{i,j}$, and let $Y\subseteq \entryportals(S)$, $Y'\subseteq\exitportals(S)$ be two subsets of vertices of cardinality $z$ each, where $z\leq \psi \ell/2$. Then there is a collection $\qset$ of edge-disjoint paths in graph $G(S)$, which is a one-to-one routing from $Y$ to $Y'$.
\end{claim}
\begin{proof}
	Assume for contradiction that the claim is false. Then, from the maximum flow / minimum cut theorem, there is a collection $E'$ of at most $z-1$ edges in graph $G(S)$, such that $G(S)\setminus E'$ contains no path connecting a vertex of $Y$ to a vertex of $Y'$. Recall that each vertex of $Y$ is an endpoint of a distinct path in $\pset(S)$, and all paths in $\pset(S)$ are edge-disjoint. Since $|Y|=z$, while $|E'|\leq z-1$, there is some path $P\in \pset(S)$, whose endpoint $y$ belongs to $Y$, such that $P$ contains no edge of $E'$. Using the same arguments, there is some path $P'\in \pset(S)$, whose endpoint $y'$ belongs to $Y'$, that contains no edge of $E'$. Clearly, $P\neq P'$ must hold, as otherwise there is a path in $G(S)\setminus E'$ connecting $y$ to $y'$ -- the path $P$. It is now enough to show that there is some path $R\in \rset(S)$, that contains no edge of $E'$, but $R\cap P\neq \emptyset$ and $R\cap P'\neq\emptyset$ hold. Indeed, in this case, $P\cup R\cup P'\subseteq G(S)\setminus E'$, and so $y$ remains connected to $y'$ in $G(S)\setminus E'$.
	
	We now show that path $R$ with such properties must exist. Let $\tilde P\in \pset_3$ be the path with $P\subseteq \tilde P$, and assume that $\tilde P\in \uset_{i'}$ Similarly, let $\tilde P'\in \pset_3$ be the path with $P'\subseteq \tilde P'$, and assume that $\tilde P'\in \uset_{i''}$ (where possibly $i'=i''$). Consider some index $j-\ell+1\leq j'\leq \ell$. Recall
	$|\uset'_{j'}|=\ceil{7\psi/8}$, and, for every path $R\in \uset'_{j'}$, segment $\sigma'_{i'}(R)$, lies in $\rset^{i',j'}$, and segment $\sigma'_{i''}(R)$ lies in $\rset^{i'',j'}$. Moreover, from \Cref{obs: paths for cells}, path $\sigma_{j'}(\tilde P)$ must intersect at least $\frac{13\psi}{16}$ paths of $\set{\sigma'_{i'}(R)\mid \uset'_{j'}}$, and 
	similarly path $\sigma_{j'}(\tilde P')$ must intersect at least $\frac{13\psi}{16}$ paths of $\set{\sigma'_{i''}(R)\mid \uset'_{j'}}$.
	 Therefore, there is a subset $\uset''_{j'}\subseteq \uset'_{j'}$ of at least $\psi/2$ paths $R$, such that both $P$ and $P'$ intersect the subpath of $R$ that belongs to $\rset(S)$. Overall, there are at least $\ell \psi/2$ paths $R\in \rset(S)$ that intersect the subpaths of $P$ and of $P'$ that lie in $\pset(S)$. Since $z\leq \ell\psi/2$, at least one such path is disjoint from $E'$.
\end{proof}

\subsubsection{Routing the Terminals to Good Cells}

We fix some good cell $\cell_{i,j}$ in the top right quadrant of the grid, that is, $\floor{\lambda'/2}\leq i,j<\lambda'$. Recall that we have defined  a vertex $y_{i,j}\in V(H)$, and a collection $\hat \rset^{i,j}\subseteq \rset^{i,j}$ of $\psi'=\ceil{13\psi/16}$ paths, each of which contains a vertex of $\Pi(y_{i,j})$. We have also defined a set $\portals^{i,j}$ of vertices that contains, for every path $\sigma'\in \hat \rset^{i,j}$, the first vertex on $\sigma'$. Let $y^*_{i,j}$ be an arbitrary vertex of $\Pi(y_{i,j})$.

We define a set $\qset_{i,j}$ of paths in $H''$, routing the terminals of $\tT$ to vertex $y^*_{i,j}$, so $\qset_{i,j}\in \Lambda'$. In order to do so, we first define a set $\qset'_{i,j}$ of paths, routing a constant fraction of the terminals of $T_4$ to vertices of $\Pi(y_{i,j})$, and then extend this path set in order to obtain routing of all terminals to vertex $y^*_{i,j}$.

\paragraph{Routing to $\cell_{i,j}$.}
In order to define the routing, we let $z=\floor{\log(\lambda'/4)}$, and we define $z+1$ squares $S_0^{i,j},S_1^{i,j},\ldots,S_z^{i,j}$. In order to simplify the notation, we will omit the superscript $i,j$ for now.

Square $S_0$ is $S(i,j,1)$, so it consists of a single cell $\cell_{i,j}$. We denote by $\portals_0^{i,j}=\portals^{i,j}$ the set of $\psi'$ vertices that we have defined. We let $\qset_0$ be the set of $\psi'$ paths, containing, for every path $\sigma'\in \hat\rset^{i,j}$, a subpath of $\sigma'$ between a vertex of $\portals_0^{i,j}$ and a vertex of $\Pi(y_{i,j})$. Therefore, $\qset_0$ is a set of $\psi'$ edge-disjoint paths, routing vertices of $\portals_0^{i,j}$ to vertices of $\Pi(y_{i,j})$, and all paths of $\qset_0$ are contained in $\rset(S_0)$. We say that cell $\cell_{i,j}$ is the bottom right corner of square $S_0$.

Fix some index $1\leq r\leq z$, and 
assume that we have defined squares $S_0,\ldots,S_{r-1}$. We now   define square $S_r$. We let $S_r=(i_r,j,2^r)$, so the length of the side of the square is $2^r$, and the coordinates of the top right corner of $S_r$ are $(i_r,j)$; here, $j$ is the column index of the initial cell $\cell_{i,j}$, and $i_r$ is the cell immediately under the right bottom corner cell of $S_{r-1}$. In other words, if $S_{r-1}=(i_{r-1},j,2^{r-1})$, then $i_r=i_{r-1}+2^{r-1}$.
We assume that we have also defined a collection $\portals_{r-1}\subseteq \entryportals(S_{r-1})$, containing $2^{r-1}\cdot \psi'$ vertices. Note that the top boundary of square $S_r$ appears immediately under bottom boundary of square $S_{r-1}$, so $\entryportals(S_{r-1})\subseteq \exitportals(S_r)$, and in particular $\portals_{r-1}\subseteq \exitportals(S_r)$. We select an arbitrary subset $\portals_r\subseteq \entryportals(S_r)$ of $2^r\cdot \psi'$ vertices. By partitioning set $\portals_r$ into two equal-cardinality subsets $Y_1,Y_2$, and applying \Cref{claim: routing in square} to each of them separately, we obtain two collections $\qset_{r}^1,\qset_{r}^2$ of edge-disjoint paths in graph $G(S_r)$, routing vertex sets $Y_1$ and $Y_2$, respectively, to vertex set $\portals_{r-1}$, in a one-to-one routing. Therefore, there is a set $\qset_{r}$ of paths in  $G(S_r)$, routing vertex set $\portals_r$ to vertex set $\portals_{r-1}$ with edge-congestion at most $2$, such that every vertex in $\portals_{r-1}$ is an endpoint of at most two such paths. Moreover, we can compute such set $\qset_r$ of paths efficiently via standard maximum flow.

Lastly, consider the last square $S_z$. We define a subset $T^*\subseteq T_4$ of terminals, as follows. For every vertex $v\in \portals_z$, let $R_v\in \rset$ be the vertical path containing $v$, and let $t_v\in T_4$ be the terminal that serves as an endpoint of path $R_v$. We then let $T^*=\set{t_v\mid v\in \portals_z}$, and we let $\qset_{z+1}$ be a set of paths containing, for every vertex $v\in \portals_z$, the subpath of $R_v$ between $t_v$ and $v$. Therefore, set $\qset_{z+1}$ of paths routes terminals of $T^*$ to vertices of $\portals_z$, and the paths in $\qset_{z+1}$ are edge-disjoint. It is also easy to verify that the paths in $\qset_{z+1}$ do not contain any edges from graphs $G(S_0)\cup \cdots\cup G(S_z)$.
Note that, since $\psi=\floor{\frac{\alpha^*\tk}{64\lambda}}$ and  $\alpha^*=\Theta(\alpha\alpha'/\log^4m)$,
\[|T^*|=2^z\cdot \psi'= 2^{\floor{\log(\lambda'/4)}}\cdot \ceil{13\psi/16}=\Omega(\lambda'\cdot \psi)=\Omega(\lambda\psi)=\Omega(\alpha^*\tk)=\Omega(\tk\alpha\alpha'/\log^4m).
\]

To summarize, we have defined a collection $\set{S_0,\ldots,S_z}$ of squares in the grid $\Pi^*$, where for all $0\leq r\leq z$, square $S_r$ has dimensions $(2^r\times 2^r)$. The squares are aligned on the right, and are stacked on top of each other, with square $S_0$ containing a single cell, $\cell_{i,j}$. This guarantees that all corresponding graphs $G(S_r)$ are mutually disjoint, except that, for all $0\leq r<z$, $V(S_r)\cap V(S_{r+1})=\entryportals(S_r)$. We have defined, for all $0\leq r\leq z$, a set $\portals_r\subseteq \entryportals(S_r)$ of $2^r\cdot \psi'$ vertices, and a set $\qset_r$ of paths contained in $G(S_r)$, routing vertices of $\portals_r$ to vertices of $\portals_{r-1}$ with edge-congestion at most $2$, so that every vertex of $\portals_{r-1}$ serves as an endpoint of at most two such paths. Additionally, in graph $G(S_0)$, we have defined a set $\qset_0$ of $\psi'$ paths routing vertices of $\portals_0$ to vertices of $\Pi(y_{i,j})$, and an additional set $\qset_{z+1}$ of edge-disjoint paths routing terminals in $T^*$ to vertices of $\portals_z$ in a one-to-one routing, so that paths in $\qset_{z+1}$ do not contain edges of $G(S_0)\cup\cdots\cup G(S_z)$.

We are now ready to define a set $\qset'_{i,j}$ of paths, routing terminals of $T^*$ to vertices of $\Pi(y_{i,j})$. In order to do so, for all $0\leq r\leq z$, we let $\qset'_r$ be a multi-set of paths, contianing, for every path $\sigma'\in \qset_r$, $2^{z-r}$ copies of the path $\sigma'$. Therefore, paths in $\qset'_r$ cause edge-congestion $2^{z-r+1}$ in $G(S_r)$. Set $\qset'_{i,j}$ of paths is obtained by concatenating paths in sets $\qset_{z+1},\qset'_z,\ldots,\qset'_0$. It is easy to verify that paths in $\qset'_{i,j}$ route all terminals in $T^*$ to vertices of $\Pi(y_{i,j})$.

Recall that $|T^*|=\Omega(\tk\alpha\alpha'/\log^4m)$, and $|\tT|=\tk$. 
Moreover, from \Cref{obs: terminals well linked in H''},
	The set $\tT$ of terminals is $\alpha^*$-well-linked in $H''$, where $\alpha^*=\Theta(\alpha\alpha'/\log^4m)$.
From \Cref{lem: routing path extension}, 
there is a set $\qset_{i,j}$ of paths in graph $H''$, routing all vertices of $\tT$ to vertices of $\Pi(y_{i,j})$, such that, for every edge $e\in E(H'')$:
\[\cong_{H''}(\qset_{i,j},e)\le 
\ceil{\frac{\tk}{|T^*|}}\left (\cong_{H''}(\qset'_{i,j},e)+\ceil{1/\alpha^*}\right )\leq 
O\left (\frac{\log^4m}{\alpha\alpha'}\right )\cdot \left(\cong_{H''}(\qset'_{i,j},e)+\frac{\log^4m}{\alpha\alpha'}\right ).\]

\paragraph{Distribution $\dset$ and Analysis.}

The final distribution $\dset$ over the routers of $\Lambda'$ is defined as follows. For every pair $(i,j)$ of indices in $Z$, we extend the paths in set $\qset_{i,j}$ via the inner edges of $\Pi(y_{i,j})$ so that each such path terminates at vertex $y^*_{i,j}$, obtaining a router of $\Lambda'$. Each such resulting router $\qset_{i,j}$ is assigned the same distribution $1/|Z|$; recall that, from \Cref{claim: many good cells}, 
	$|Z|\geq (\lambda')^2/16$.

We now fix some outer edge $e\in E(H'')$, and analyze the expectation $\expect[\qset_{i,j}\sim \dset]{(\cong_{H''}(\qset_{i,j},e))^2}$.

Recall that there is at most one path $P\in \pset$ that contains $e$, and at most one path $R\in \rset$ containing $e$. Moreover, there is at most one pair $(i_1,j_1)$ of indices with edge $e$ lying on some path of $ \pset^{i_1,j_1}$, and at most one pair $(i_2,j_2)$ of indices with edge $e$ lying on a path of $\rset^{i_2,j_2}$.

We first focus on pair $(i_1,j_1)$ of indices, and the corresponding cell $\cell_{i_1,j_1}$. Fix some pair $(i,j)\in Z$ of indices, and $0\leq r\leq z$. If $\cell_{i_1,j_1}\in S_r^{i,j}$, then segment $\sigma_{j_1}(P)$ of $P$ may lie on at most $2^{z-r+1}$ paths in $\qset'_{i,j}$. Notice that there are at most $2^{2r+2}$ square subgrids $S$ of $\Pi^*$ of dimension $(2^r\times 2^r)$, that contain the cell $\cell_{i_1,j_1}$. For each such square $S$, there is exactly one pair $(i(S),j(S))$ of indices, for which $S_r^{i(S),j(S)}=S$. Since $|Z|\geq  (\lambda')^2/16$, the probability that an index $(i,j)\in Z$ is chosen for which $\cell_{i_1,j_1}$ lies in the square $S_r^{i,j}$ is at most $O(2^{2r+2}/(\lambda')^2)$. Recall that, if $\cell_{i_1,j_1}\in S_r^{i,j}$, then $\cong_{H''}(\qset'_{i,j})\leq 2^{z-r+1}\leq O(\lambda'/2^r)$. Moreover, if $\cell_{i_1,j_1}$ does not lie in any of the squares $S_0^{i,j},\ldots,S_z^{i,j}$, then $\cong_{H''}(\qset'_{i,j})\leq 1$. The analysis for cell $\cell_{i_2,j_2}$ is symmetric. Therefore, altogether (now taking into account both the cells $\cell_{i_1,j_1}$ and $\cell_{i_2,j_2}$), we get that:
\[\expect[(i,j)\in Z]{(\cong_{H''}(\qset'_{i,j},e))^2}\leq O(1)+\sum_{r=0}^zO\left (\frac{2^{2r+2}}{(\lambda')^2}\cdot\frac{(\lambda')^2}{2^{2r}}\right )=O(z)\leq O(\log m).
\] 
Lastly, since $\cong_{H''}(\qset_{i,j},e)\le 
O\left (\frac{\log^4m}{\alpha\alpha'}\right )\cdot \left(\cong_{H''}(\qset'_{i,j},e)+\frac{\log^4m}{\alpha\alpha'}\right )$, we get that:
\[\expect[\qset_{i,j}\sim\dset]{(\cong_{H''}(\qset_{i,j},e))^2}\leq O\left (\frac{\log^{16}m}{(\alpha\alpha')^4}\right ).\]

\newpage

\appendix

\section{Proof of \Cref{thm: main_result}}
\label{sec: proof of main theorem} 

In this section, we provide the proof of \Cref{thm: main_result} from  \Cref{thm: main_rotation_system} and \Cref{thm: MCN_to_rotation_system}.
Suppose we are given  a simple $n$-vertex  graph $G$ with maximum vertex degree $\Delta$. We  use the algorithm from \Cref{thm: MCN_to_rotation_system} in order to compute an instance $I=(G',\Sigma)$ of \CNwRS, with $m=|E(G')|\leq O\left(\optcro(G)\cdot \poly(\Delta\cdot\log n)\right)$, and   $\optcrors(I)\leq O\left(\optcro(G)\cdot \poly(\Delta\cdot\log n)\right )$. 
Notice that, since $G$ is a simple graph, $\optcro(G)\leq |E(G)|^2\leq n^4$, and  $\Delta\leq n$. Therefore, $m=|E(G')|\leq \poly(n)$.

We use the algorithm from 
\Cref{thm: main_rotation_system} to compute a solution to instance $I$ of \CNwRS, such that, w.h.p., the number of crossings in the solution is bounded by $2^{O((\log m)^{7/8}\log\log m)}\cdot \left(\optcrors(I)+m\right)$. Lastly, using the algorithm from \Cref{thm: MCN_to_rotation_system}, we efficiently compute a drawing of graph $G$, with the number of crossings bounded by:
\[\begin{split}
&\left (2^{O((\log m)^{7/8}\log\log m)}\cdot \left(\optcrors(I)+m\right) +\optcro(G)\right )\cdot \poly(\Delta\log n) \\
&\quad\quad\quad\quad\leq O\left (2^{O((\log n)^{7/8}\log\log n)}\cdot \poly(\Delta)\right )\cdot \optcro(G). 
\end{split}\]

\section{Proofs Omitted from \Cref{sec: short_prelim}}
\label{sec: apd_short_prelim}

\subsection{Proof of Theorem~\ref{thm: crwrs_planar}}

\label{apd: Proof of crwrs_planar}

For every vertex $u\in V(G)$, we denote $d_u=\deg_G(v)$, and we denote  $\delta_G(u)=\set{e_1(u),\ldots,e_{d_u}(u)}$, where the edges are indexed according to their order in the rotation $\oset_u\in \Sigma$. 

In order to prove the theorem, we construct a new graph $G'$, that is obtained from $G$ by replacing every vertex $u\in V(G)$ with the $(d_u\times d_u)$-grid.  We show that, if $\optcrors(I)=0$, then graph $G'$ is planar. We then provide an algorithm, that, given a planar drawing of $G'$, computes a solution to instance $I$ of \cnwrs whose cost is $0$.

We start by defining the graph $G'$. For every vertex $u\in V(G)$, we let $H_u$ be the $(d_u\times d_u)$ grid. We denote the vertices that appear on the first row of grid $H_u$ by $x_1(u),\ldots,x_{d_u}(u)$, in the natural order of their appearance. In order to construct graph $G'$, we start with the disjoint union of the graphs in $\set{H_u}_{u\in V(G)}$. We then consider every edge $e\in E(G)$ one by one. Let $e=(u,u')$ be any such edge, and assume that $e=e_i(u)=e_j(u')$ (that is, $e$ is the $i$th edge incident to $u$, and the $j$th edge incident to $u'$). We then add edge $e'=(x_i(u),x_j(u'))$ to graph $G'$, and we view edge $e'$ as \emph{representing} the edge $e\in E(G)$. This completes the construction of the graph $G'$. We call the edges of $G'$ that lie in set $\set{e'\mid e\in E(G)}$ \emph{primary edges}, and the remaining edges of $G'$ \emph{secondary edges}. Notice that, from our construction, a vertex of $G'$ may be incident to at most one primary edge.
We use the following two observations.

\begin{observation}\label{obs: G' planar}
 If $\optcrors(I)=0$, then graph $G'$ is planar.
\end{observation}

\begin{proof}
	Assume that $\optcrors(I)=0$, and let $\phi$ be a solution to instance $I$ of \cnwrs, with $\cro(\phi)=0$. We transform drawing $\phi$ to obtain a planar drawing $\psi$ of the graph $G'$.

	In order to do so, for every vertex $u\in V(G)$, we consider the tiny $u$-disc $D(u)=D_{\phi}(u)$. For every edge $e_i(u)\in \delta_G(u)$, we denote by $p_i(u)$ the unique point of the image of $e_i(u)$ in $\phi$ that lies on the boundary of the disc $D(u)$. Note that points $p_1(u),\ldots,p_{d_u}(u)$ must appear on the boundary of disc $D(u)$ in this circular order. If they are encountered in this order as we traverse the boundary of $D(u)$ in counter-clock-wise direction, then we say that vertex $u$ is positive; otherwise we say that it is negative.  We let $D'(u)$ be a disc that is contained in $D(u)$, such that the boundaries of $D(u)$ and $D'(u)$ are disjoint.
	
	Consider now a vertex $u\in V(G)$, and let $\psi_u$ be the standard drawing of the grid $H_u$. We let $\tilde D(u)$ be a disc in the drawing $\psi_u$, such that the image of the grid $H_u$ is contained in $\tilde D(u)$, and the images of vertices $x_1(u),\ldots,x_{d_u}(u)$, that we denote by $p'_1(u),\ldots,p'_{d_u}(u)$ lie on the boundary of $\tilde D(u)$, and are encountered in this order as we traverse the boundary of $\tilde D(u)$ in the counter-clock-wise direction.
	We also ensure that the only points of $\psi_u(H_u)$ that lie on the boundary of $\tilde D(u)$ are $p'_1(u),\ldots,p'_{d_u}(u)$.

	In order to define a planar drawing $\psi$ of graph $G'$, we process every vertex $u\in V(G)$ one by one. Consider any such vertex $u$. If $u$ is a positive vertex, then we plant the drawing $\psi_u$ of $H_u$ inside the disc $D'(u)$ that we defined before, so that the discs $\tilde D(u)$ and $D'(u)$ coincide. Observe that, in this case, points $p_1(u),\ldots,p_{d_u}(u)$ are encountered in this order on the boundary of $D(u)$ as we traverse it in counter-clock-wise direction; and similarly, points $p'_1(u),\ldots,p'_{d_u}(u)$ are encountered in this order on the boundary of $D'(u)$ as we traverse it in counter-clock-wise direction. For all $1\leq i\leq d_u$, we can then define a curve $\gamma_i(u)$ connecting points $p_i(u)$ and $p'_i(u)$, that is contained in $D(u)$, and is internally disjoint from $D'(u)$. Moreover, we can ensure that all curves in $\set{\gamma_i(u)\mid 1\leq i\leq d_u}$ are disjoint from each other and are internally disjoint from the boundary of $D_u$. 
If $u$ is a negative vertex, then we repeat the same process, except that we plant a mirror image of the drawing $\psi_u$ of $H_u$ inside the disc $D'(u)$. This allows us to define the set $\set{\gamma_i(u)\mid 1\leq i\leq d_u}$ of disjoint curves as before, where for $1\leq i\leq d_u$, curve $\gamma_i(u)$ connects points $p_i(u)$ and $p'_i(u)$, is contained in $D(u)$, and is internally disjoint from $D'(u)$.

	So far, for every vertex $u\in V(G)$, we have defined the images of the vertices and the edges of the grid $H_u$ in $\psi$. In order to complete the drawing $\psi$ of graph $G'$, we process the edges $e\in E(G)$ one by one. Consider any such edge $e=(u,u')$, and assume that $e=e_i(u)=e_{j}(u')$. Note that the image $\phi(e)$ of edge $e$ contains points $p_i(u)$ and $p_j(u')$. Let $\sigma(e)$ be the segment of $\phi(e)$ between these two points. Notice that, by construction, $\sigma(e)$ is internally disjoint from all discs in $\set{D(u'')}_{u''\in V(G)}$. Recall that graph $G'$ contains an edge $e'=(x_i(u),x_j(u'))$ representing edge $e$. We let the image of edge $e'$ in $\psi$ be the concatenation of curves $\gamma_i(u)$ (that connects the image of $x_i(u)$ to point $p_i(u)$; $\sigma(e)$ (connecting $p_i(u)$ to $p_j(u)$); and $\gamma_j(u')$ (connecting $p_j(u')$ to the image of $x_j(u')$). 
	This completes the definition of the drawing $\psi$ of $G'$. It is immediate to verify that it is a planar drawing.
\end{proof}

\begin{observation}
	\label{obs:grid_expansion_2}
	There is an efficient algorithm, that, given a planar drawing $\phi'$ of graph $G'$, computes a feasible solution $\phi$ to instance $I$ of \CNwRS with no crossings.
\end{observation}

\begin{proof}
	Consider the planar drawing $\phi'$ of graph $G'$ on the sphere. 
	Recall that for all $r\geq 1$, the $(r\times r)$-grid graph has a unique planar drawing. Therefore, for every vertex $u\in V(G)$, the drawing of grid $H_u$ that is induced by $\phi'$ is the standard drawing $\psi_u$ of the grid. Recall that the boundary of the grid $H_u$ is a simple cycle. Let $\gamma_u$ be the closed curve, that is obtained by taking the union of the images of all edges of the boundary of $H_u$. Notice that $\gamma_u$ must be a simple curve, and, moreover, for every pair $u',u''$ of distinct vertices of $G'$, $\gamma_{u'}\cap \gamma_{u''}=\emptyset$. For a vertex $u\in V(G)$, let $D'(u)$ be the disc whose boundary is $\gamma_u$, such that the drawing of $H_u$ in $\phi'$ is contained in $D'(u)$. We denote by $p^*(u)$ the image of vertex $v_{d_u,d_u}$ of the grid $H_u$, and, for $1\le i\leq r$, by $p_i(u)$ the image of vertex $x_i(u)$. Note that points $p_1(u),\ldots,p_{d_u}(u),p^*(u)$ appear in this circular order on the boundary of $D'(u)$. Notice also that it is possible that, for a pair $u'\neq u''$ of vertices of $G$, $D'(u')\subseteq D'(u'')$.
	
	Let $\Gamma$ denote the set of curves that contains, for every primary edge $e'$ of $G'$, its image $\phi'(e')$.
	We use the following claim.
	
	\begin{claim}\label{claim: construct curves along paths}
		There is an efficient algorithm that constructs, for each vertex $u\in V(G)$ and index $1\leq i\leq d_u$, a curve $\gamma_i(u)$ that is contained in $D'(u)$ and connects $p_i(u)$ to $p^*(u)$. Moreover, for every pair $\gamma,\gamma'$ of distinct curves in set $\Gamma\cup \set{\gamma_i(u)\mid u\in V(G); 1\leq i\leq d_u}$, every point $p\in \gamma\cap \gamma'$ must be an endpoint of both curves. 
	\end{claim}

We prove the claim below, after we complete the proof of \Cref{obs:grid_expansion_2} using it. We define a drawing $\phi$ of graph $G$ as follows. For every vertex $u\in V(G)$, the image $\phi(u)$ is defined to be $p^*(u)$. Consider now some edge $e=(u,u')\in E(G)$, and assume that $e=e_i(u)=e_j(u')$. We then let the image of $e$ in $\phi$ be the concatenation of three curves: (i) curve $\gamma_i(u)$, connecting $p^*(u)$ to $p_i(u)$; (ii) the image of edge $e'=(v_i(u),v_j(u'))\in E(G')$ in drawing $\phi'$, that connects $p_i(u)$ to $p_j(u')$; and (iii) curve $\gamma_j(u')$, connecting $p_j(u')$ to $p^*(u')$. Notice that the resulting curve connects $\phi(u)$ to $\phi(u')$, as required. This completes the definition of the drawing $\phi$ of $G$.

We now show that this is a legal drawing, and that the number of crossings in this drawing is $0$. Indeed, assume for contradition that there are two edges $e_1,e_2\in E(G)$, and that some point $p$ lies in $\phi(e_1)\cap \phi(e_2)$. Note that the endpoints of $\phi(e_2)$ may not be inner points of $\phi(e_1)$ and vice versa. Therefore, $p$ is an inner point on both $\phi(e_1)$ and $\phi(e_2)$. 
From our construction, there must be two curves $\gamma,\gamma'\in \Gamma\cup \set{\gamma_i(u)\mid u\in V(G); 1\leq i\leq d_u}$, with $\gamma\subseteq \phi(e_1)$ and $\gamma'\subseteq \phi(e_2)$ that contain $p$.
From \Cref{claim: construct curves along paths}, point $p$ must be an endpoint of both curves. Assume that $e_1=(u_1,u'_1)$, and that $e_1=e_i(u_1)=e_j(u'_1)$. Then, from our construction, $p=p_i(u_1)$ or $p=p_j(u_1')$ must hold. Similarly, assuming that  $e_2=(u_2,u'_2)$, and that $e_2=e_{i'}(u_2)=e_{j'}(u'_2)$, we get that $p=p_{i'}(u_2)$ or $p=p_{j'}(u_2')$ must hold. This may only happen if two distinct primary edges of $G'$ are incident to the same vertex of $G'$, which is impossible from our construction. 
We conclude that $\phi$ is a valid drawing of $G$ with $0$ crossings.

Next, we show that $\phi$ obeys the rotation system $\Sigma$. Consider some vertex $u\in V(G)$, and a tiny $u$-disc $D_{\phi}(u)$. For $1\leq i\leq d_u$, let $\tilde p_i(u)$ be the point on the boundary of $D_{\phi}(u)$ that lies on the image of edge $e_i(u)$ in $\phi$. In particular, point $\tilde p_i(u)$ belongs to the curve $\gamma_i(u)$, whose endpoints are $p_i(u),p^*(u)$. Since points $p_1(u),\ldots,p_{d_u}(u)$ appear on the boundary of disc $D'(u)$ in this circular order, and the curves $\gamma_1(u),\ldots,\gamma_{d_u}(u)$ are internally disjoint, points 
 $\tilde p_1(u),\ldots,\tilde p_{d_u}(u)$ must appear on the boundary of disc $ D_{\phi}(u)$ in this circular order. We conclude that drawing $\phi$ of $G$ obeys the rotation system $\Sigma$.

In order to complete the proof of \Cref{obs:grid_expansion_2}, it is now enough to prove \Cref{claim: construct curves along paths}, which we do next.

\begin{proofof}{\Cref{claim: construct curves along paths}} 
	Consider a vertex $u\in V(G)$.
	For convenience, for $1\leq i,j\leq d_u$, we denote by $v_{i,j}(u)$ the unique vertex of the grid $H_u$ lying in the intersection of its $i$th row and $j$th column.
	
	 Let $A(u)=\set{a_1(u),\ldots,a_{d_u-1}(u)}$ be the sequence of edges on the last row of the grid $H_u$. Recall that, for $1\leq i<d_u$, curve $\phi'(a_i(u))$ is contained in the boundary of the disc $D'(u)$. We
	denote $\sigma_i(u)=\phi'(a_i(u))$, and
	 draw another curve $\sigma'_i(u)$, whose endpoints are the same as those of $\sigma_i(u)$, such that $\sigma'_i(u)$ is contained in the interior of $D'(u)$; is internally disjoint from $\sigma_i(u)$ and the images of all edges of $G'$ in $\phi'$; and it is drawn in parallel to  $\sigma_i(u)$ right next to it. Next, we let $\hat D_i(u)$ be the disc, whose boundary is  the union of the curves $\sigma_i(u)$ and $\sigma_i'(u)$ 
	 (see \Cref{fig:extra_1}).
	 Lastly, we let $\hat D(u)\subseteq D'(u)$ be smallest disc, whose interior contains, for all $1\leq i\leq d_u-1$, the disc $\hat D_i(u)$, and, for all $1\leq i\leq d_u$, the intersection of the tiny $v_{d_u,i}(u)$-disc $D_{\phi'}(v_{d_u,i}(u))$ and the disc $D'(u)$ (see \Cref{fig:extra_2}).

\begin{figure}[h]
	\centering
	\subfigure[Curves in $\set{\sigma_i(u),\sigma'_i(u)}_i$, and the corresponding discs $\set{\hat D_i(u)}_i$. The boundary of disc $D'(u)$ is shown in black.]{\scalebox{0.1}{\includegraphics{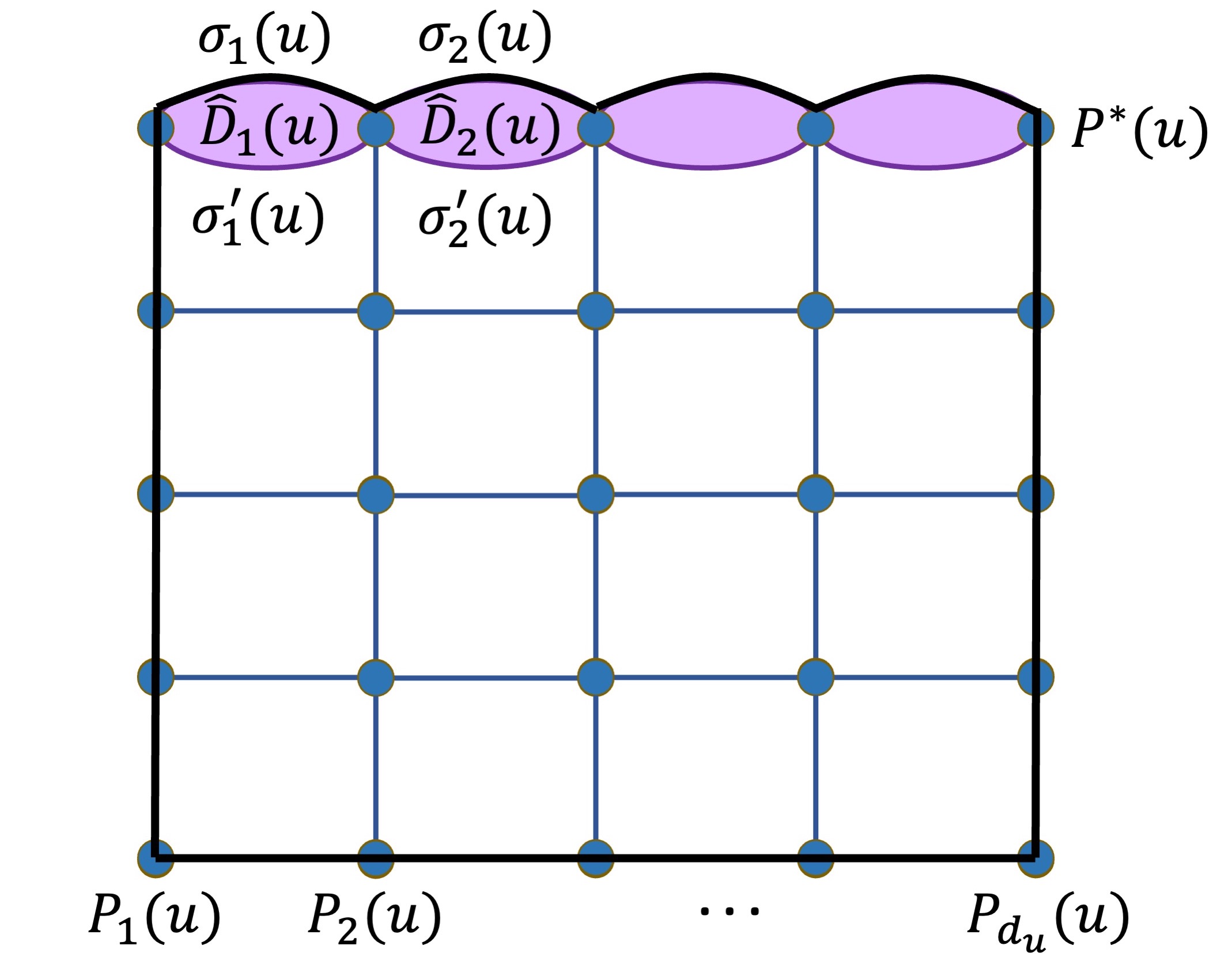}}\label{fig:extra_1}
	}
	\hspace{0.5cm}
	\subfigure[Disc $\hat D(u)$ is shown in red.]{\scalebox{0.1}{\includegraphics{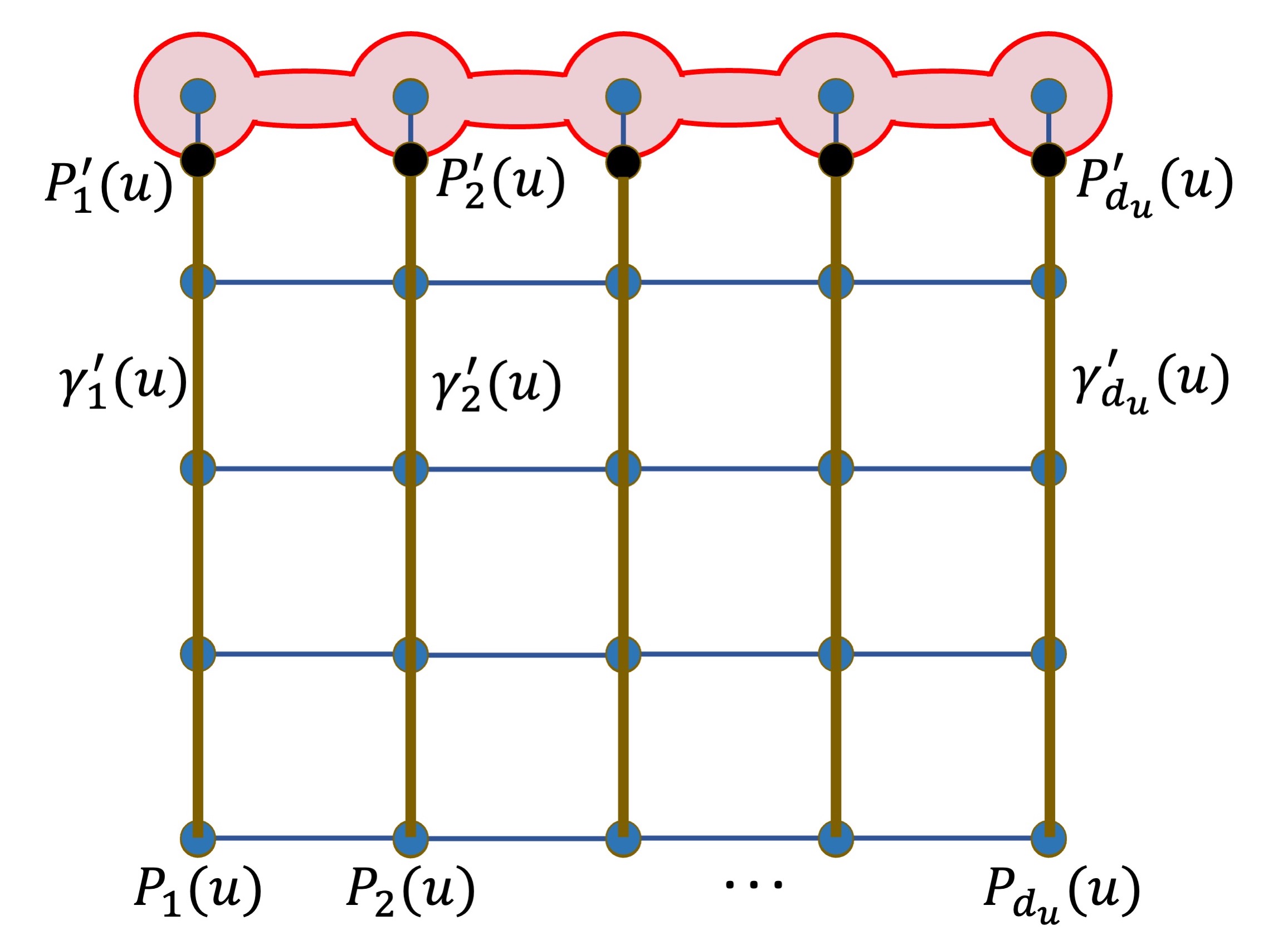}}\label{fig:extra_2}}
	\caption{Illustration of  curves in  $\set{\sigma_i(u),\sigma'_i(u)}_i$ and the corresponding discs.}
\end{figure}

	 From our construction, the only vertices of $G'$ whose images are contained in disc $\hat D(u)$ are the vertices lying in the last row of the grid $H_u$. The only edges of $G'$ that may have a non-empty intersection with disc $\hat D(u)$ are the edges of $H_u$ that are incident to the vertices of $H_u$ lying in the last row of the grid. 
	 
	 Consider now some index $1\leq i\leq d_u$. Let $\gamma'_i(u)$ be the curve obtained by concatenating the images of all edges that lie on the $i$th column of grid $H_u$. We truncate the curve $\gamma_i'(u)$, so that it terminates at a point on the boundary of disc $\hat D(u)$, and is internally disjoint from disc $\hat D(u)$. We denote by $p'_i(u)$ the point on the boundary of $\hat D(u)$ that lies on $\gamma_i'(u)$. Note that curve $\gamma_i'(u)$ connects points $p_i(u)$ and $p'_i(u)$; it is contained in disc $D'(u)$, and it is internally disjoint from disc $\hat D(u)$. It is easy to verify that, since drawing $\phi'$ of $G'$ is planar, curves $\gamma'_i(u),\ldots,\gamma'_{d_u}(u)$ are disjoint from each other. For all $1\leq i\leq d_u$, we then let $\gamma''_i(u)$ be any simple curve connecting $p_i'(u)$ to $p^*(u)$, that is contained in disc $\hat D(u)$; we construct the curves $\gamma''_1(u),\ldots,\gamma''_{d_u}(u)$ so that they are internally disjoint from each other. For all $1\leq i\leq d_u$, we then let $\gamma_i(u)$ be the curve obtained by concatenating $\gamma'_i(u)$ and $\gamma''_i(u)$. It is immediate to verify that curve $\gamma_i(u)$ is contained in $D'(u)$, and it connects $p_i(u)$ to $p^*(u)$. From our construction, it is easy to verify that,  for any pair $\gamma,\gamma'$ of distinct curves in set $\Gamma\cup \set{\gamma_i(u)\mid u\in V(G); 1\leq i\leq d_u}$, every point $p\in \gamma\cap \gamma'$ must be an endpoint of both curves. 	
\end{proofof}	
\end{proof}

We are now ready to complete the proof of Theorem~\ref{thm: crwrs_planar}. We  construct the graph $G'$ as described above, and use the algorithm from \Cref{thm: testing planarity} to test whether graph $G'$ is planar, and if so, to compute a planar drawing $\phi'$ of $G'$.
If $G'$ is not planar, then we correctly establish that $\optcrors(G,\Sigma)\ne 0$, from \Cref{obs: G' planar}.
If $G'$ is planar, then we apply the algorithm from  \Cref{obs:grid_expansion_2} to graph $G'$ and its planar drawing $\phi'$, to compute a valid solution $\phi$ to instance $I$ of \cnwrs with $\cro(\phi)=0$.

\subsection{Proof of Theorem~\ref{thm: crwrs_uncrossing}}

\label{apd: Proof of crwrs_uncrossing}

\begin{figure}[h]
	\centering
	\subfigure[Before: Curves $\phi(e)$ (red) and $\phi(e')$ (blue) cross at $p$ and $q$.]{\scalebox{0.12}{\includegraphics{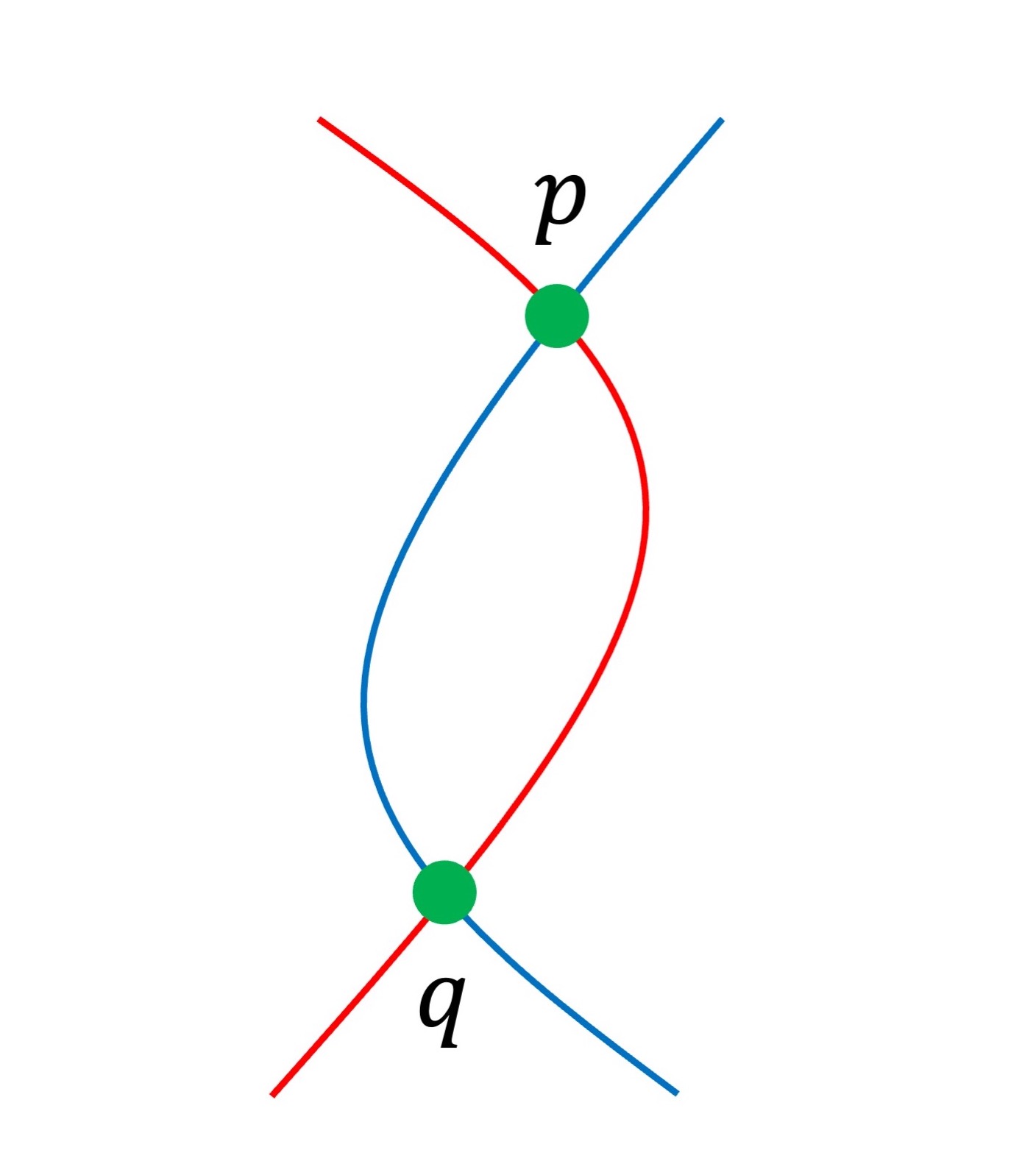}}
	}
	\hspace{1.5cm}
	\subfigure[After: The modified curves no longer cross at $p$ or at $q$.]{
		\scalebox{0.12}{\includegraphics{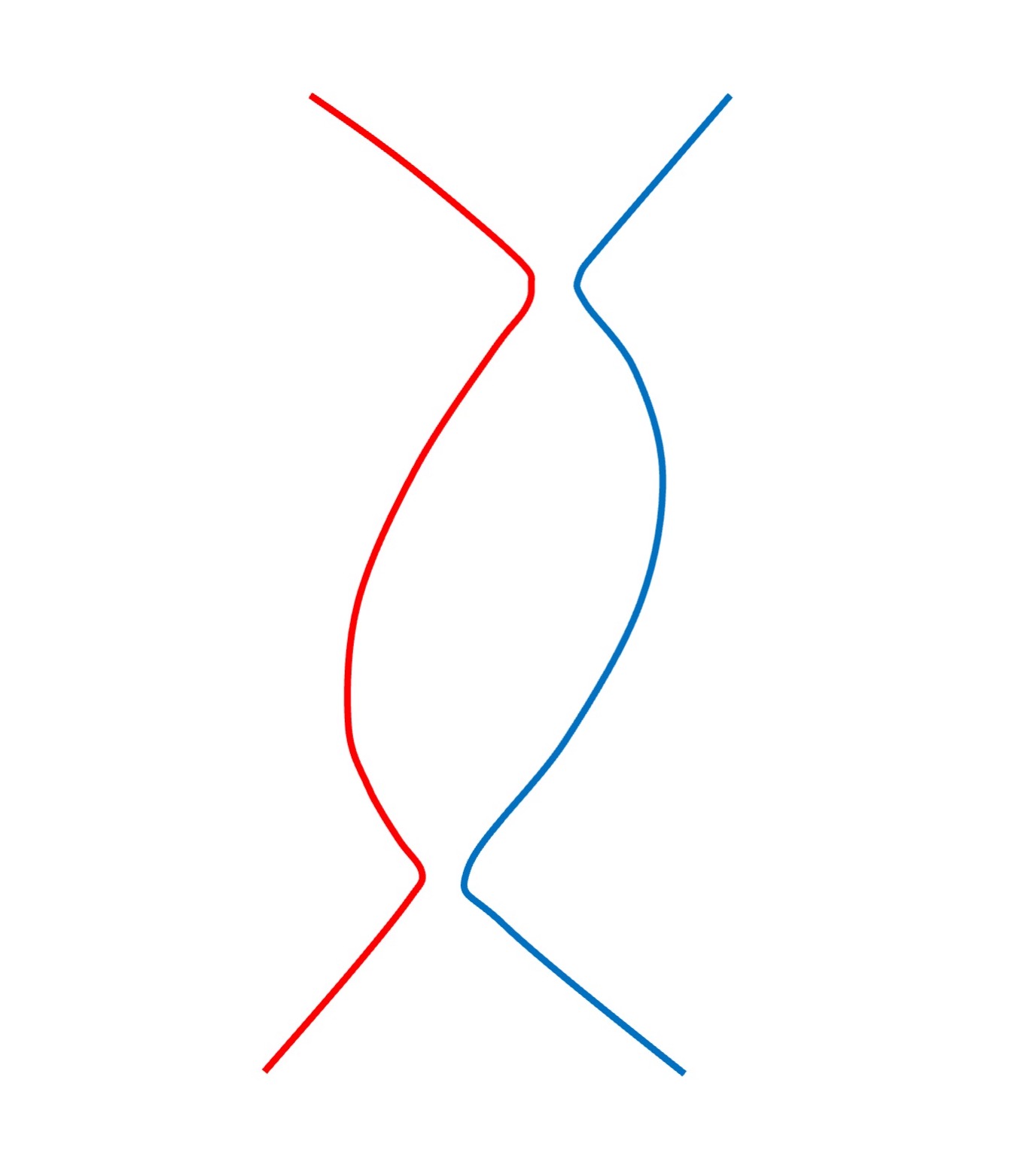}}}
	\caption{Uncrossing two curves.
	}\label{fig:uncross-curves}
\end{figure}

We start with any feasible solution $\phi$ to instance $I$, and then gradually modify it to ensure that every pair of edges in $G$ cross at most once. 
As long as there  is a pair $e,e'\in E(G)$ of distinct edges, whose images cross at least twice in ${\phi}$, we perform the following modification step. Let $p,q$ be two crossing points between curves $\phi(e),\phi(e')$, that appear consecutively on $\phi(e)$; in other words, the segment of $\phi(e)$ between $a$ and $b$ contains no other point that lies on $\phi(e')$. We ``uncross'' the images of edges $e$ and $e'$, as shown in Figure~\ref{fig:uncross-curves}.
(In Sections \ref{subsec: uncrossing type 1} and \ref{apd: type-1 uncrossing} we provide a more formal description of this uncrossing process, that we refer to as \emph{type-1 uncrossing}).
It is easy to see that, after this uncrossing step, the new drawing remains a feasible solution to instance $I$, and the number of crossings in the drawing decreases by at least $2$. 
We continue this process until every pair of edges of $G$ cross at most once in $\phi$. It is clear that the resulting drawing contains at most $|E(G)|^2$ crossings.

\subsection{Proof of \Cref{claim: compose algs}}
\label{apd: Proof of compose algs}

%
Denote $I=(G,\Sigma)$ and $m=|E(G)|$. Recall that from Property \ref{prop: few edges}, $\sum_{I'=(G',\Sigma')\in \iset'}|E(G')|\leq m\cdot (\log m)^{O(1)}$, so in particular, for every instance $I'=(G',\Sigma')\in \iset'$, $|E(G')|\leq m\cdot (\log m)^{O(1)}$.
From the same property, for every instance $I'=(G',\Sigma')\in \iset'$, $\sum_{I''=(G'',\Sigma'')\in \iset''(I')}|E(G'')|\leq |E(G')|\cdot (\log(|E(G')|))^{O(1)}\leq |E(G')|\cdot (\log m)^{O(1)}$. Therefore, altogether, we get that:
\[
\begin{split}
\sum_{I''=(G'',\Sigma'')\in \iset}|E(G'')|& = \sum_{I'\in \iset'}\sum_{I''=(G'',\Sigma'')\in \iset''(I')}|E(G'')|\\
& \leq \sum_{I'=(G',\Sigma')\in \iset'}|E(G')|\cdot (\log m)^{O(1)}\\
& \leq m\cdot (\log m)^{O(1)},\\
\end{split}
\]
establishing Property \ref{prop: few edges}.

Next, we establish Property \ref{prop: modified expectation}, using the same property of Algorithms $\alg_1$ and $\alg_2$:
%
%
%
\[\begin{split}
\expect{\sum_{I''\in \iset''}\optcrors(I'')}&=\sum_{I'\in \iset'}\expect{\sum_{I''\in \iset''(I')}\optcrors(I'')}\\
&\leq \sum_{I'=(G',\Sigma')\in \iset'}\expect{\bigg(\optcrors(I')+|E(G')|\bigg)\cdot \nu''}\\
&= \nu''\cdot\left(\sum_{I'=(G',\Sigma')\in \iset'}|E(G')|\right)+ \nu''\cdot\expect{\sum_{I'\in \iset'}\optcrors(I')}\\
&\leq O(\nu''\cdot m\cdot (\log m)^{O(1)})+\left (\optcrors(I)+m\right )\cdot (\nu'\nu'')\\
&\leq \left (\optcrors(I)+m\right )\cdot \nu''\cdot\max\set{2\nu',(\log m)^{O(1)}}\\
&\leq \left (\optcrors(I)+m\right )\cdot \nu.
\end{split}\]
	
Lastly, we establish Property \ref{prop: alg to put together} of Algorithm $\alg$, using the same property of algorithms $\alg_1$ and $\alg_2$.
Assume that we are  given, for every instance $I''\in \iset''$, a feasible solution $\phi(I'')$ to $I''$. 
We process instances $I'\in \iset'$ one by one. For each such instance, we apply Algorithm $\alg(\iset''(I'))$ that is given by Property \ref{prop: alg to put together} of the decomposition $\iset''(I')$ to solutions $\phi(I'')$ of instances $I''\in \iset''(I')$, to obtain  a solution $\phi(I')$ to instance $I'$ of cost at most  $O\left (\sum_{I''\in \iset''(I')}\cro(\phi(I''))\right )$. 
We then apply the algorithm $\alg(\iset')$, given by Property \ref{prop: alg to put together} of the decomposition $\iset'$ of $I$, to the resulting solutions $\phi(I')$ for instances $I'\in \iset'$, to obtain a solution $\phi(I)$ to instance $I$, whose cost is at most: 
$$O\left (\sum_{ I'\in \iset'}\cro(\phi(I'))\right )\le O\left (\sum_{ I''\in \iset}\cro(\phi(I''))\right ).$$


\section{Proofs Omitted from \Cref{sec: high level}}
\subsection{Proof of \Cref{claim: bound by level}}
\label{Appx: inductive bound proof}
The proof is by induction on $h(I)$. The base case is when $h(I)=0$, so $v(I)$ is a leaf vertex of $T^*$, and hence of $T$. Denote $I=(G,\Sigma)$.  From \Cref{obs: leaf}, either $|E(G)|\leq \mu^{c''}$; or $\optcrors(I)=0$; or $\optcrors(I)> |E(G)|^2/\mu^{c''}$.
If $\optcrors(I)=0$, then the algorithm returns a solution of cost $0$. Otherwise, $\optcrors(I)\geq 1$, and, if $|E(G)|\leq \mu^{c''}$, then the algorithm returns  the trivial solution, of cost at most $|E(G)|^2\leq |E(G)|\cdot \mu^{c''}$. Lastly, if $\optcrors(I)>|E(G)|^2/\mu^{c''}$, then, since the trivial solution $\phi'$ is considered by the algorithm, it returns a solution of cost at most $|E(G)|^2\leq \optcrors(I)\cdot \mu^{c''}$.

Assume now that  the claim holds for all vertices $v(I)$ of $T^*$ with $h(I)<q$,  for some $0<q\leq \dep(T)$. Consider any vertex $v(I)$ of the tree $T^*$ with $h(I)=q$. Let $v(I_1),\ldots,v(I_k)$ be the child vertices of $v(I)$ in the tree $T^*$. Denote $I=(G,\Sigma)$ and $|E(G)|=m$. Additionally, for all $1\leq r\leq k$, denote $I_r=(G_r,\Sigma_r)$ and $m_r=|E(G_r)|$.
 
Since instance $I$ is not a leaf instance of $T$, $|E(G)|\geq \mu^{c''}$ must hold. 
Since we have assumed that event $\event$ does not happen, either $\optcrors(I)> |E(G)|^2/\mu^{c''}$, or $\sum_{r=1}^k\optcrors(I_r)\leq (\optcrors(I)+m)\cdot 2^{c_g(\log m)^{3/4}\log\log m}$ must hold.
In the former case, the algorithm is guaranteed to return a solution to $I$ whose cost is at most $|E(G)|^2\leq \optcrors(I)\cdot \mu^{c''}$, since the trivial solution $\phi'$ is considered as one of the possible solutions. From now on we focus on latter case. For all $1\leq r\leq k$, let $\phi_r$ be the solution to instance $I_r$ that the algorithm computes recursively. From the induction hypothesis, for all $1\leq r\leq k$:
\[\cro(\phi_r)\leq  2^{\tilde c\cdot h(I_r)\cdot (\log m^*)^{3/4}\log\log m^*}\cdot \mu^{c''\cdot c_g}\cdot \optcrors(I_r)+(\log m^*)^{4c_g h(I_r)}\mu^{2c''\cdot \tilde c}\cdot m_r.\]
Notice that, for all $1\leq r\leq k$, $h(I_r)\leq q-1$. Moreover, from  \Cref{thm: main}, $\sum_{r=1}^km_r\leq m\cdot(\log m)^{c_g}\leq m\cdot(\log m^*)^{c_g}$. Lastly, as noted already, $\sum_{r=1}^k\optcrors(I_r)\leq (\optcrors(I)+m)\cdot 2^{c_g(\log m)^{3/4}\log\log m}$ must hold. 
Altogether, we get that:
\[
\begin{split}
\sum_{r=1}^k\cro(\phi_r)&\leq 2^{\tilde c\cdot (q-1)\cdot (\log m^*)^{3/4}\log\log m^*}\cdot \mu^{c''\cdot c_g}\cdot \sum_{r=1}^k\optcrors(I_r)+(\log m^*)^{4c_g (q-1)}\mu^{2c''\cdot \tilde c}\cdot \sum_{r=1}^km_r
\\
&\leq 2^{\tilde c\cdot (q-1)\cdot (\log m^*)^{3/4}\log\log m^*}\cdot \mu^{c''\cdot c_g}\cdot  (\optcrors(I)+m)\cdot 2^{c_g(\log m)^{3/4}\log\log m} \\
&\hspace{6cm}+(\log m^*)^{4c_g (q-1)}\mu^{2c''\cdot \tilde c}\cdot m\cdot(\log m^*)^{c_g}\\
&\leq 2^{\tilde c\cdot (q-0.5)\cdot (\log m^*)^{3/4}\log\log m^*}\cdot \mu^{c''\cdot c_g}\cdot \optcrors(I)+ (\log m^*)^{4c_gq-3c_g}\mu^{2c''\cdot \tilde c}\cdot m\\
&\hspace{6cm}+2^{\tilde c\cdot (q-0.5)\cdot (\log m^*)^{3/4}\log\log m^*} \cdot\mu^{c''\cdot c_g}\cdot m.
\end{split}\]
Since $q\leq \dep(T)\leq \frac{(\log m^*)^{1/8}}{c^*\log\log m^*}$  from \Cref{obs: few recursive levels},
the last term is bounded by:
\[ 2^{\tilde c\cdot (\log m^*)^{7/8}/c^*} \cdot\mu^{c''\cdot c_g}\cdot m  \leq \mu^{2c''\cdot c_g}\cdot m\]

(since $\mu =2^{c^*(\log m^*)^{7/8}\log\log m^*}$). Therefore, we get that:
\[\sum_{r=1}^k\cro(\phi_r)\leq  2^{\tilde c\cdot (q-0.5)\cdot (\log m^*)^{3/4}\log\log m^*}\cdot \mu^{c''\cdot c_g}\cdot \optcrors(I)+ (\log m^*)^{4c_gq-2c_g}\mu^{2c''\cdot \tilde c}\cdot m. \]

The solution that our algorithm returns for instance $I$ is obtained by applying Algorithm \algcombine from \Cref{thm: main} to solutions $\phi_1,\ldots,\phi_k$ to instances $I_1,\ldots,I_k$ (or some other solution the algorithm considers, if its cost is smaller). Since event $\event$ does not happen, the cost of the resulting solution is bounded by:
\[
\begin{split}
&c_g\cdot\bigg(\sum_{r=1}^k\cro(\phi_r)\bigg) +(\optcrors(I)+m)\cdot\mu^{c_g}\\
&\hspace{1cm}\leq c_g\cdot 2^{\tilde c\cdot (q-0.5)\cdot (\log m^*)^{3/4}\log\log m^*}\cdot \mu^{c''\cdot c_g}\cdot \optcrors(I)+\mu^{c_g}\cdot \optcrors(I)\\
&\hspace{3cm}+ c_g\cdot(\log m^*)^{4c_gq-2c_g}\mu^{2c''\cdot \tilde c}\cdot m+\mu^{c_g}\cdot m\\
&\hspace{1cm}\leq 2^{\tilde c\cdot q\cdot (\log m^*)^{3/4}\log\log m^*}\cdot \mu^{c''\cdot c_g}\cdot \optcrors(I)+(\log m^*)^{4c_gq}\mu^{2c''\cdot \tilde c}\cdot m,
\end{split}
\]
as required.

\subsection{Proof of \Cref{claim: combine drawings}}
\label{Appx: Proof of combine drawings}

We construct the solution $\phi(I)$ to instance $I$ in three steps.
In the first step, we compute a solution $\phi(I')$ to every instance $I'\in \hat{\iset}^{(n)}_{\textsf {large}}\cup \tilde{\iset}^{(n)}_{\textsf {large}}$, as follows. Consider any instance $I'=(G',\Sigma')\in \hat{\iset}^{(n)}_{\textsf {large}}\cup \tilde{\iset}^{(n)}_{\textsf {large}}$. Recall that we are given a solution $\phi(I'')$  to every instance $I''\in \overline\iset(I')$. Recall also that $\overline\iset(I')$ is a $\nu$-decomposition  of instance $I'$. We apply the efficient algorithm $\alg(\overline\iset(I'))$ from the definition of $\nu$-decomposition to the drawings in set  $\set{\phi(I'')}_{I''\in \overline\iset(I')}$, to compute a feasible solution $\phi(I')$ to instance $I'$, of cost $\cro(\phi(I'))\leq O\left (\sum_{I''\in \overline\iset(I')}\cro(\phi(I''))\right )$.
Overall, we get that:

\begin{equation}\label{eq: solutoins to narrow instances}
\sum_{I'\in \hat{\iset}^{(n)}_{\textsf {large}}\cup \tilde{\iset}^{(n)}_{\textsf {large}}}\cro(\phi(I'))\leq \sum_{I''\in \iset^*}O(\cro(\phi(I''))).
\end{equation}

We have now obtained a solution $\phi(I')$ to every instance $I'\in \big(\hat{\iset}_{\textsf {small}}\cup \hat{\iset}^{(n)}_{\textsf {large}}\cup \tilde{\iset}_{\textsf {small}}\cup \tilde{\iset}^{(n)}_{\textsf {large}}\big)$.

In the second step, we compute a solution $\phi(I')$ to every instance $I'\in \hat{\iset}^{(w)}_{\textsf {large}}$. 
Consider any such instance $I'\in \hat{\iset}^{(w)}_{\textsf {large}}$.
Recall that we applied the algorithm from  \Cref{lem: many paths} to instance $I'$, to obtain a collection $\tilde \iset(I')$ of instances of \cnwrs. Every instance in the resulting collection belongs to $\tilde \iset_{\textsf{small}}$ or to $ \tilde{\iset}^{(n)}_{\textsf {large}}$.
We use Algorithm $\algcombine'$, that is guaranteed from \Cref{lem: many paths}, to compute a solution $\phi(I')$ to instance $I'$.
Since we have assumed that event $\event_2$ did not happen, the cost of the solution is bounded by:
$\cro(\phi(I')) \leq \sum_{\tilde I=(\tilde G, \tilde \Sigma)\in \tilde \iset(I')}\cro(\phi(\tilde I)) + \optcrors(I')\cdot\mu^{c'_g}$.
Overall, we get that:

\begin{equation}\label{eq: solutions to wide not connected}
\sum_{I'\in \hat{\iset}^{(w)}_{\textsf {large}}}\cro(\phi(I'))\leq \sum_{I''\in \iset^*}O(\cro(\phi(I''))+\sum_{I'\in \hat{\iset}^{(w)}_{\textsf {large}}}\optcrors(I')\cdot\mu^{c'_g}.
\end{equation}

We now describe the third step. We have so far obtained a solution $\phi(I')$ to every instance $I'\in \big(\hat{\iset}_{\textsf {small}}\cup \hat{\iset}^{(n)}_{\textsf {large}}\cup 
\hat{\iset}^{(w)}_{\textsf {large}}\big)$, that is, a solution to every instance in $\hat \iset$.
Recall that $\hat \iset$ is a $\nu_1$-decomposition of the input instance $I$. Since we have assumed that Event $\event_1$ did not happen, $\sum_{I'\in \hat\iset}\optcrors(I')\leq  100\cdot\left (\optcrors(I)+m\right )\cdot \nu_1$. By combining Inequalities \ref{eq: solutoins to narrow instances} and \ref{eq: solutions to wide not connected}, we get that:

\begin{equation}\label{eq: solution for step 1}
\begin{split}
\sum_{I'\in \hat \iset}\cro(\phi(I'))&\leq \sum_{I''\in \iset^*}O(\cro(\phi(I''))+\sum_{I'\in \hat{\iset}^{(w)}_{\textsf {large}}}\optcrors(I')\cdot\mu^{c'_g}\\
&\leq \sum_{I''\in \iset^*}O(\cro(\phi(I''))+100\cdot\left (\optcrors(I)+m\right )\cdot \nu_1\cdot \mu^{c'_g}\\
&\leq \sum_{I''\in \iset^*}O(\cro(\phi(I''))+\left (\optcrors(I)+m\right )\cdot \mu^{O(1)},
\end{split}
\end{equation}

since $\nu_1= 2^{O((\log m)^{3/4}\log\log m)}$ and $\mu\gg \nu_1$.

Lastly, we apply the efficient algorithm $\alg(\hat \iset)$ that is guaranteed by the definition of $\nu_1$-decomposition to the solutions $\set{\phi(I')}_{I'\in \hat \iset}$, to obtain a feasible solution $\phi(I)$ to instance $I$. 
The cost of the solution is bounded by $\sum_{I'\in \hat \iset}O(\cro(\phi(I')))\leq 
\sum_{I''\in \iset^*}O(\cro(\phi(I''))+\left (\optcrors(I)+m\right )\cdot \mu^{O(1)}$, as required.

\subsection{Proof of \Cref{obs: bound sum of opts}}
\label{Appx: Proof of bound sum of opts}

Throughout the proof, we assume that $\optcrors(I)\leq |E(G)|^2/\mu^{c'}$ and bad event $\event$ did not happen.

Since event $\event_1$ does not happen:

\begin{equation}\label{eq: bound on opts step 1}
\sum_{I'\in \hat\iset}\optcrors(I')\leq  100\nu_1\cdot (\optcrors(I)+m).
\end{equation}

Recall that $\hat \iset=\hat{\iset}_{\textsf {small}}\cup \hat{\iset}^{(n)}_{\textsf {large}}\cup 
\hat{\iset}^{(w)}_{\textsf {large}}$. In Step 2 of the algorithm, applied the algorithm from \Cref{lem: many paths} to every instance $I'=(G',\Sigma')\in \hat{\iset}^{(w)}_{\textsf {large}}$, to compute a collection $\tilde \iset(I')$ of instances of \cnwrs. Consider now any such instance $I'\in \hat{\iset}^{(w)}_{\textsf {large}}$. Since we have assumed that Event $\event_2$ did not happen:

\[
\sum_{\tilde I\in \tilde \iset(I')}\optcrors(\tilde I)\le \optcrors(I')\cdot (\log |E(G')|)^{c'_g}\leq \optcrors(I')\cdot (\log m)^{c'_g}.
\]

Recall that we have defined $\tilde \iset=\bigcup_{I'\in \hat{\iset}^{(w)}_{\textsf {large}}}\tilde \iset(I')$. Combining the above inequality with Equation \ref{eq: bound on opts step 1}, and recalling that $\nu_1=2^{O((\log m)^{3/4}\log\log m)}$, we get that:

\begin{equation}\label{eq: bound opts second}
\sum_{\tilde I\in \tilde \iset}\optcrors(\tilde I)\le \sum_{I'\in \hat{\iset}^{(w)}_{\textsf {large}}}\optcrors(I')\cdot (\log m)^{c'_g}\leq (\optcrors(I)+m)\cdot 2^{O((\log m)^{3/4}\log\log m)}.
\end{equation}

Consider now an instance $I'=(G',\Sigma')\in \hat{\iset}^{(n)}_{\textsf {large}}\cup \tilde{\iset}^{(n)}_{\textsf {large}}$. Since we have assumed that event $\event_3$ does not happen, from the definition of a $\nu$-decomposition:

\[\expect{\sum_{I''\in \overline\iset(I')}\optcrors(I'')}\le \left (\optcrors(I')+|E(G')|\right )\cdot \nu. \]

Recall that $\overline\iset_{\textsf{small}}=\bigcup_{I'\in \hat{\iset}^{(n)}_{\textsf {large}}\cup \tilde{\iset}^{(n)}_{\textsf {large}}}\overline\iset(I')$. Recall also that, from  Inequality \ref{eq: num of edges step 1}, 
$\sum_{I'=(G',\Sigma')\in \hat \iset^{(n)}_{\textsf {large}}}|E(G')|\le m\cdot (\log m)^{c'_g}$,
and from Inequality \ref{ineq: total edges step 2}, 
$\sum_{I'=(G',\Sigma')\in \tilde{\iset}^{(n)}_{\textsf {large}}}|E(G')|\leq 2m\cdot (\log m)^{c'_g}$.  Altogether, we get that:

\begin{equation}\label{eq: bound ops last step}
\begin{split}
\expect{\sum_{I''\in \overline\iset_{\textsf{small}}}\optcrors(I'')}&\leq \sum_{I'=(G',\Sigma')\in \hat \iset^{(n)}_{\textsf {large}}\cup \tilde \iset^{(n)}_{\textsf {large}}} \left (\optcrors(I')+|E(G')|\right )\cdot \nu\\
&\leq \sum_{I'\in \hat \iset^{(n)}_{\textsf {large}}\cup \tilde \iset^{(n)}_{\textsf {large}}} \optcrors(I')\cdot \nu +4m\cdot (\log m)^{c'_g}\cdot \nu\\
&\leq (\optcrors(I)+m)\cdot 2^{O((\log m)^{3/4}\log\log m)}
\end{split}
\end{equation}

(we have used Equations \ref{eq: bound on opts step 1} and \ref{eq: bound opts second} in order to bound $\sum_{I'\in \hat \iset^{(n)}_{\textsf {large}}}\optcrors(I')$ and $\sum_{I'\in \hat \iset^{(n)}_{\textsf {large}}}\optcrors(I')$, respectively, and the fact that $\nu_1,\nu\leq  2^{O((\log m)^{3/4}\log\log m)}$).

Finally, by combining Equations \ref{eq: bound on opts step 1}, \ref{eq: bound opts second} and \ref{eq: bound ops last step}, we get that:

\[
\begin{split}
\expect{\sum_{I''\in \iset^*}\optcrors(I'')}&\leq \expect{\sum_{I''\in \hat \iset_{\textsf {small}}}\optcrors(I'')+ \sum_{I''\in \tilde \iset_{\textsf {small}}}\optcrors(I'')+\sum_{I''\in \overline \iset_{\textsf {small}}}\optcrors(I'')}\\
&\leq (\optcrors(I)+m)\cdot 2^{O((\log m)^{3/4}\log\log m)}.
\end{split}
\]

We denote this expectation by $\eta'$. 
Let $\hat \event$ be the bad event that 
$\sum_{I''\in \iset^*}\optcrors(I'')>100\eta'$. From Markov inequality, $\prob{\hat \event\mid \neg\event}<1/100$.

\section{Proofs Omitted from Section~\ref{sec:long prelim}}
\label{sec: apd_prelim}


\subsection{Proof of \Cref{claim: remove congestion}}
\label{apd: Proof of remove congestion}

Denote $k=|\pset|$ and $\rho =\cong_G(\pset)$.
We define an undirected $s$-$t$ flow network $H$, as follows. We start with the graph $G$, and set the capacity of every edge in $G$ to be $1$. We then add a source vertex $s$, that connects to every vertex $v\in V(G)$ with an edge of capacity $n_S(v)$, and a destination vertex $t$, that connects to every vertex $v\in V(G)$ with  an edge of capacity $n_T(v)$. Notice that, by sending $1/\rho$ flow units on every path $P\in \pset$, we obtain an $s$-$t$ flow of value $k/\rho$ in this network. From the integrality of flow, since all edge capacities in $H$ are integral, there is an integral $s$-$t$ flow in $H$, of value at least $k/\rho$. This integral flow  defines the desired collection $\pset'$ of  at least $k/\rho$ edge-disjoint paths in graph $G$. We can use standard algorithms for computing maximum $s$-$t$ flow in order to obtain the set $\pset'$ of paths with these properties.

\subsection{Proof of \Cref{obs: splicing}}
\label{apd: Proof of splicing}

We first show that $S(\pset')=S(\pset)$ and $T(\pset')=T(\pset)$.
We denote by $s$ and $t$ the first and the last endpoints of $P$, respectively, and by $s'$ and $t'$ the first and the last endpoints of $P'$, respectively.
From the construction, the first endpoint of $\tilde P$ is $s$, the last endpoint of $\tilde P$ is $t'$, the first endpoint of $\tilde P'$ is $s'$, and the last endpoint of $\tilde P'$ is $t$. It is then immediate to verify that  $S(\pset')=S(\pset)$ and $T(\pset')=T(\pset)$.

We now prove the second assertion. In order to do so, we assume that both $\tilde P,\tilde P'$ are simple paths, and we will show that $|\Pi^T(\pset')|<|\Pi^T(\pset)|$.

For every vertex $u\in V(G)$, let $N_1(u)$ be the number of triples of $\Pi^T(\pset)$ in which $u$ participates, and let $N_2(u)$ be the number of triples of $\Pi^T(\pset')$ in which $u$ participates. It is enough to show that, for every vertex $u\in V(G)\setminus\set{v}$, $N_2(u)\leq N_1(u)$, and that $N_2(v)<N_1(v)$.

Consider some vertex $u\in V(G)\setminus\set{v}$. We will assign, to every triple $(Q,Q',u)\in \Pi^T(\pset')$, a unique triple in $\Pi^T(\pset)$ that is responsible to it, and we will ensure that every triple in $\Pi^T(\pset)$ is responsible for at most one such triple. 

Consider some triple $(Q,Q',u)\in \Pi^T(\pset')$. If neither of the two paths $Q,Q'$ lies in $\set{\tilde P,\tilde P'}$, then triple $(Q,Q',u)$ lies in $\Pi^T(\pset)$ as well, and we make $(Q,Q',u)$ responsible for itself. If $Q=\tilde P$ and $Q'=\tilde P'$ (or the other way around), then  either $Q$ is a subpath of $P$ and $Q'$ is a subpath of $P'$, or the other way around (we use the fact that paths $P,P'$ are simple, so $Q,Q'$ may not be subpaths of the same path). In either case, it is easy to see that triple $(P,P',u)$ lied in  $\Pi^T(\pset)$. We make the triple $(P,P',u)$ responsible for triple $(Q,Q,u)$. The last case is when exactly one of the paths $Q,Q'$ is in $\set{\tilde P,\tilde P'}$. We assume w.l.o.g. that $Q=\tilde P$, and $Q'\not\in\set{\tilde P, \tilde P'}$. If $u$ lies on path $P$ between its first vertex and $v$, then triple $(P,Q',u)$ lies in $\Pi^T(\pset)$, and we make it responsible for $(Q,Q',u)$. Otherwise, triple $(P',Q',u)$ lies in $\Pi^T(\pset)$, and we make it responsible for $(Q,Q',u)$. 

It is easy to see that every triple $(\hat Q,\hat Q',u)\in \Pi^T(\pset)$ is responsible for at most one triple in $\Pi^T(\pset')$. Indeed, if neither of $\hat Q,\hat Q'$ lies in $\set{P,P'}$, then triple $(\hat Q,\hat Q',u)$ may only be responsible for itself. If both $\hat Q,\hat Q'\in \set{P,P'}$, then triple $(P,P',u)$ may only be responsible for triple $(\tilde P,\tilde P',u)$. If exactly one of $\hat Q,\hat Q'$ lies in $\set{P,P'}$, for example, $\hat Q=P$, then two cases are possible: if vertex $u$ lies between the first endpoint of $P$ and $v$, then triple $(P,Q',u)$ may only be responsible for triple $(\tilde P,Q',u)$, and otherwise it may only be responsible for triple $(\tilde P',Q',u)$. We conclude that $N_2(u)\leq N_1(u)$.

Consider now the case where $u=v$, and consider some triple $(Q,Q',v)\in \Pi^T(\pset')$. If neither of the two paths $Q,Q'$ lies in $\set{\tilde P,\tilde P'}$, then triple $(Q,Q',v)$ lies in $\Pi^T(\pset)$, and we make $(Q,Q',v)$ responsible for itself. Note that, in case where $u=v$, it is impossible that the triple $(\tilde P,\tilde P',v)$ lies in $\Pi^T(\pset')$. Therefore, it remains to consider the triples $(Q,Q',v)$, where exactly one of the paths $Q,Q'$ lies in $\set{\tilde P,\tilde P'}$.
We call such triples \emph{problematic triples}, and we assume w.l.o.g. that in each such triple, $Q\not \in \set{\tilde P,\tilde P'}$. If path $Q$ participates in a problematic triple, then we say that path $Q$ is a \emph{problematic path}.

\begin{figure}[h]
	\centering
	\scalebox{0.8}{\includegraphics[scale=0.1]{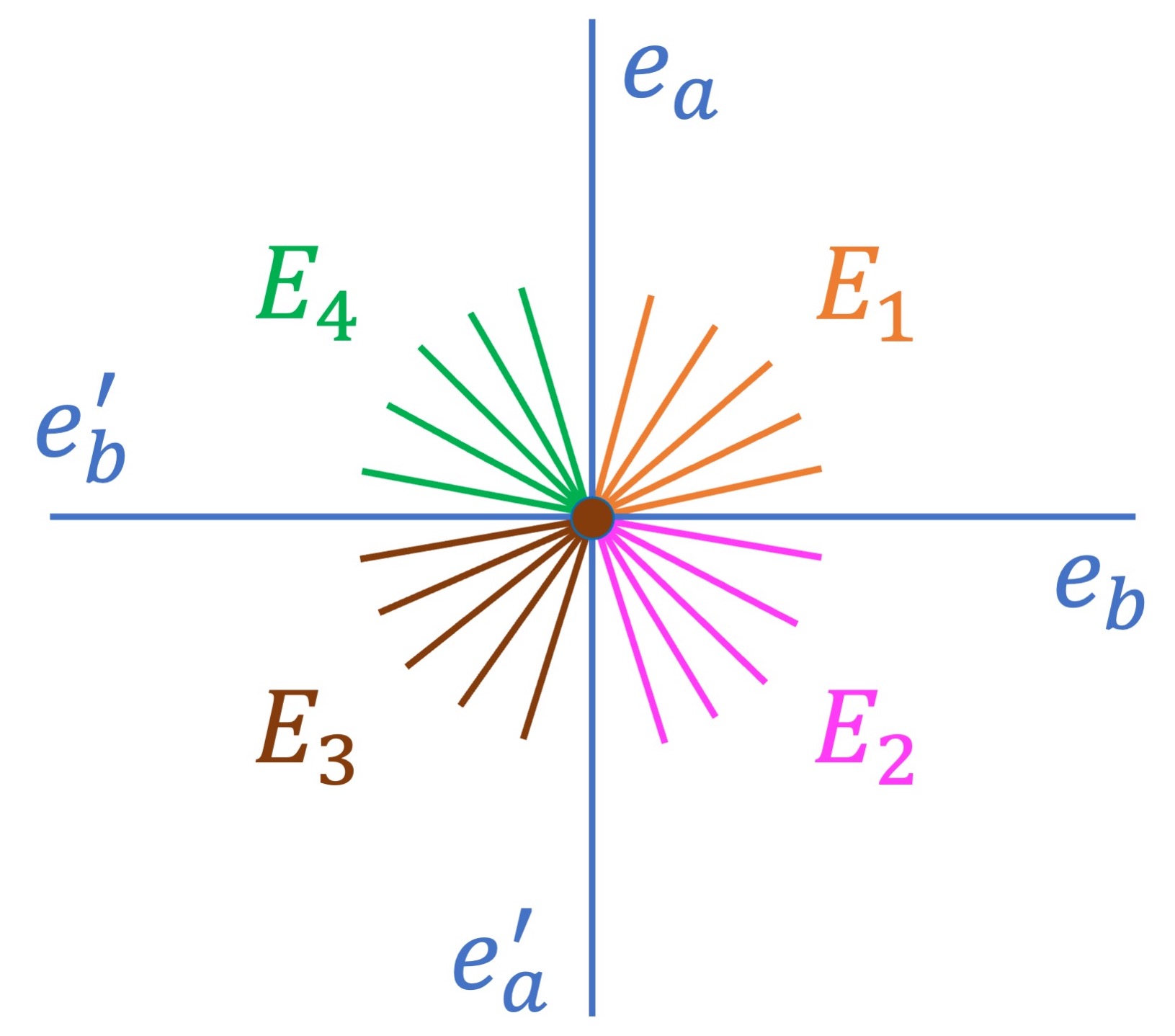}}
	\caption{A schematic view of edges $e_a$, $e_b$, $e'_a$, $e'_b$ and edge sets $\set{E_i}_{1\le i\le 4}$.}\label{fig: splicing_1}
\end{figure}

We denote by $e_a,e'_{a}$ the two edges on path $P$ that are incident to vertex $v$, and we assume that $e_a$ appears before $e'_{a}$ on $P$. We denote by $e_b,e'_{b}$ the two edges on path $P'$ that are incident to $v$, and we assume that $e_b$ appears before $e'_{b}$ on $P'$. Recall that path $\tilde P$ contains edges $e_a$ and $e'_{b}$, while path $\tilde P'$ contains edges $e_{b}$ and $e'_{a}$. Recall that edges $e_a,e_b,e'_{a},e'_{b}$ must appear in this circular order in $\oset_v\in \Sigma$, since paths $P$ and $P'$ are transversal (recall that the ordering is unoriented). We use the edges of $\set{e_a,e_b,e'_{a},e'_{b}}$ to partition the edge set $\delta_G(v)\setminus \set{e_a,e_b,e'_{a},e'_{b}}$ into four subsets: set $E_1$ of edges appearing between $e_a$ and $e_b$ in $\oset_v$; set $E_2$ of edges appearing between $e_b$ and $e'_a$;  set $E_3$ of edges appearing between $e'_a$ and $e'_b$, and set $E_4$ of all remaining edges, that must appear between $e'_b$ and $e_a$ (see \Cref{fig: splicing_1}). 

Consider now some problematic path $Q$, and denote by $e(Q),e'(Q)$ the two edges that lie on $Q$ and are incident to $v$.  Note that $e(Q),e'(Q)$ must lie in different sets of $\set{E_1,\ldots,E_4}$. Since path $\tilde P$ contains edges $e_a,e'_b$, while path $\tilde P'$ contains edges $e'_a,e_b$, in order for path $Q$ to be problematic, at least one of the two edges $e(Q),e'(Q)$ must lie in one of the sets $E_2,E_4$. We assume w.l.o.g. that $e(Q)\in E_2$. If $e'(Q)\in E_4$, then both $(Q,\tilde P,v)$ and $(Q,\tilde P',v)$ are problematic pairs. But in this case, both $(Q,P,v)$ and $(Q,P',v)$ lied in $\Pi^T(\pset)$. We make triple $(Q,P,v)$ responsible for $(Q,\tilde P,v)$, and we make triple $(Q,P',v)$ responsible for $(Q,\tilde P',v)$. Otherwise, $e'_Q\in E_1$ or $e'_Q\in E_3$ must hold. In the either case, the only problematic triple involving path $Q$ is $(Q,\tilde P',v)$. In the former case, $(Q,P',v)\in \Pi^T(\pset)$, and we make this triple responsible for $(Q,\tilde P',v)$, while in the latter case, $(Q,P,v)\in \Pi^T(\pset)$, and we make this triple responsible for $(Q,\tilde P',v)$. 
So far, we have assigned, to every triple $(Q,Q',v)\in \Pi^T(\pset')$, a distinct triple $(\hat Q,\hat Q',v)\in \Pi^T(\pset)$ that is responsible for it. Note that triple $(P,P',v)$ is not responsible for any triple $(Q,Q',v)\in \Pi^T(\pset')$, so $N_2(v)<N_1(v)$. We conclude that $|\Pi^T(\pset')|<|\Pi^T(\pset)|$.

\subsection{Proof of Lemma~\ref{lem: non_interfering_paths}}
\label{apd: Proof of non_interfering_paths}


We first preprocess the set $\rset$ of paths by removing cycles from the paths, to obtain a collection $\rset$ of simple paths. 
The algorithm is iterative. Throughout the algorithm, we maintain a set $\hat \rset$ of paths in $G$, that is initialized to be $\rset$. The algorithm proceeds in iterations, as long as $\Pi^T(\hat\rset)\neq \emptyset$. An iteration is executed as follows.
Let $(P,P',v)$ be any triple in $\Pi^T(\hat\rset)$. We perform path splicing of $P$ and $P'$ at vertex $v$, obtaining two new paths $\tilde P$ and $\tilde P'$. We then remove cycles from $\tilde P$ and $\tilde P'$, to obtain two simple paths, which are then added to $\hat \rset$, replacing the paths $P$ and $P'$. Note that, from \Cref{obs: splicing}, multisets $S(\hat \rset),T(\hat \rset)$ remain unchanged after the execution of the iteration. It is also easy to verify, from the definition of the splicing procedure, that, for every edge $e\in E(G)$, $\cong_G(\hat \rset,e)$ may not increase after the iteration execution. Moreover, if the paths $\tilde P,\tilde P'$ obtained after the splicing procedure are simple, then $|\Pi^T(\rset)|$ is guaranteed to decrease after the current iteration, while otherwise, $\sum_{R\in \hat\rset}|E(R)|$ must decrease. We conclude that, after every iteration of the algorithm, either $\sum_{R\in \hat\rset}|E(R)|$ decreases, or $\sum_{R\in \hat\rset}|E(R)|$ remains unchanged and $|\Pi^T(\hat \rset)|$ decreases. Since $|\Pi^T(\hat \rset)|\leq |\hat \rset|^2\cdot |V(G)|$, the number of iterations in the algorithm is bounded by $|\hat \rset|^2\cdot |V(G)|\cdot |E(G)|$, and so the algorithm is efficient. The output $\rset'$ of the algorithm  is the set $\hat \rset$ of paths that is obtained when the algorithm terminates. From the above discussion, we get that $S(\rset')=S(\rset)$ and $T( \rset')=T(\rset)$. Once the algorithm terminates, the paths in set $\rset'=\hat \rset$ are non-transversal with respect to $\Sigma$. Lastly, from the above discussion, for every edge $e\in E(G)$, $\cong_G(\hat \rset,e)$ may not increase over the course of the algorithm, and so $\cong_G(\rset',e)\le \cong_G(\rset,e)$ must hold.

\subsection{Proof of Lemma~\ref{lem: multiway cut with paths sets}}

\label{apd: Proof of multiway cut with paths sets}

We use the following claim.
\begin{claim}
\label{lem: cut_uncrossing}
Let $G$ be a graph, $S$ a subset of vertices in $G$, and $x,y\in S$ two distinct vertices. Assume that $(A,B)$ is a minimum cut separating $x$ from $S\setminus\set{x}$ with $x\in A$, and $(A',B')$ is a minimum cut separating $y$ from $S\setminus \set{y}$ with $y\in A'$. Consider another cut $(\hat A,\hat B)$, where $\hat A=A\setminus A'$, and $\hat B=V(G)\setminus \hat A$. Then $(\hat A,\hat B)$ is a minimum cut separating $x$ from $S\setminus \set{x}$ in $G$.
\end{claim}

\begin{proof}
	Since $(A',B')$ is a cut separating $y$ from $S\setminus\set{y}$ with $y\in A'$, we get that $A'\cap S=\set{y}$. Similarly, $A\cap S=\set{x}$. Therefore, $\hat A\cap S=\set{x}$, and so $(\hat A,\hat B)$ is indeed a cut separating $x$ from $S\setminus\set{x}$. It now remains to show that $|E(\hat A,\hat B)|\leq |E(A,B)|$.
	Denote $\hat A'=A'\setminus A$, and $\hat B'=V(G)\setminus \hat A'$. Using the same argument as above, $(\hat A',\hat B')$ is a cut separating $y$ from $S\setminus \set{y}$.

	From submodularity of cuts, for any pair $X,Y$ of vertex subsets in a graph $G$, $|\delta_G(X)|+|\delta_G(Y)|\geq |\delta_G(X\setminus Y)|+|\delta_G(Y\setminus X)|$.
	Therefore:
$$|\delta_G(A)|+|\delta_G(A')|\geq |\delta_G(A\setminus A')|+ |\delta_G(A'\setminus A)|=|\delta_G(\hat A)|+|\delta_G(\hat A')|.$$
	 Notice however that $(A,B)$ is a minimum cut separating $x$ from $S\setminus \set{x}$, so $|\delta_G(A)|=|E(A,B)|\leq |E(\hat A,\hat B)|=|\delta_G(\hat A)|$. Similarly, since $(A',B')$ is a minimum cut separating $y$ from $S\setminus \set{y}$, we get that $|\delta_G(A')|=|E(A',B')|\leq |E(\hat A',\hat B')|=|\delta_G(\hat A')|$. We conclude that $|\delta_G(A)|+|\delta_G(A')|=|\delta_G(\hat A)|+|\delta_G(\hat A')|$ must hold. If we assume for contradiction that $(\hat A,\hat B)$ is not a minimum cut separating $x$ from $S\setminus\set{x}$, then $|\delta_G(A)|<|\delta_G(\hat A)|$ must hold, and so $|\delta_G(A')|>|\delta_G(\hat A')|$, a contradiction to the minimality of the cut $(A',B')$. We conclude that $(\hat A,\hat B)$ is a minimum cut  separating $x$ from $S\setminus\set{x}$.
\end{proof}	 

We now complete the proof of Lemma~\ref{lem: multiway cut with paths sets} using Claim~\ref{lem: cut_uncrossing}.
Recall that we are given a set $S=\set{s_1,\ldots,s_k}$ of vertices of graph $G$. We first compute, for all $1\le i\le k$, a minimum cut  separating $\set{s_i}$ from $S\setminus \set{s_i}$ in $G$, that we denote by $(U_i,\overline{U_i})$, with $s_i\in U_i$. For each $1\leq i\leq k$, we then  let $A_i=U_i\setminus (\bigcup_{1\le j\le k, j\ne i}U_j)$. Clearly, for all $1\leq i<j\leq k$, $A_i\cap A_j=\emptyset$.

Consider now some index $1\leq i\leq k$. We claim that $(A_i,V(G)\setminus A_i)$ is a minimum cut separating $s_i$ from $S\setminus \set{s_i}$ in graph $G$. For convenience, assume that $i=k$ (the other cases are symmetric). For all $1\leq j<k$, let $Z_j=U_k\setminus (U_1\cup U_2\cup\cdots\cup U_j)$, so that $Z_{k-1}=A_k$. Set $Z_0=U_k$. By applying 
\Cref{lem: cut_uncrossing} to each of the sets $Z_0,\ldots,Z_{k-1}$ in turn, and using the fact that, for all $1\leq j\leq k-1$, $Z_j=Z_{j-1}\setminus U_j$,
we get that, for all $0\leq j\leq k-1$, $(Z_j,V(G)\setminus Z_j)$ is a minimum cut separating $s_k$ from $S\setminus\set{s_k}$ in $G$.

It remains to compute, for each $1\le i\le k$, a set $\qset_i$ of paths routing the edges of $\delta(A_i)$ to $s_i$. Fix an index $1\leq i\leq k$. We construct a flow network as follows. Let $H_i$ be the graph obtained from $G[A_i]\cup \delta_G(A_i)$, by contracting all vertices that do not belong to $A_i$ into a single vertex, that we denote by $t_i$. We set the capacity of every edge in $H_i$ to be $1$, and compute a maximum $s_i$-$t_i$ flow in the resulting network. From the max-flow / min-cut theorem, the value of the resulting flow must be $|\delta_G(A_j)|$, and from the integrality of flow we can ensure that the resulting flow is integral. We can then use this flow to obtain a set $\qset_i=\set{Q_i(e)\mid e\in \delta_G(e)}$ of edge-disjoint paths, where, for all $e\in \delta_G(e)$, path $Q_i(e)$ has $e$ as its first edge, $s_i$ as its last vertex, and all inner vertices of $Q_i(e)$ are contained in $A_i$.


\subsection{Proof of \Cref{cor: approx_balanced_cut}}

\label{apd: Proof of approx_balanced_cut}

Let $0< \eta <1$ be some parameter. 
In order to avoid confusion, throughout this proof, we will refer to $\eta$-balanced cuts as $\eta$-edge-balanced cuts. We now define the notion of $\eta$-vertex-balanced cuts, that will be used in this proof.
We say that a cut $(A,B)$ in a graph $G$ is  \emph{$\eta$-vertex-balanced} if $|A|,|B|\leq \eta\cdot |V(G)|$. We say that a cut $(A,B)$ is a \emph{minimum $\eta$-vertex-balanced cut in $G$} if $(A,B)$ is an $\eta$-vertex-balanced cut of minimum value $|E(A,B)|$. We need the following theorem. 

\begin{theorem}[Corollary 2 in~\cite{ARV}]
	\label{thm: ARV}
	For every constant $1/2<\eta<1$, there is another constant $\eta<\eta'<1$, and an efficient algorithm, that, given any \textbf{simple} connected graph $G$ with $n$ vertices, computes an $\eta'$-vertex-balanced cut $(A,B)$ in $G$, whose value $|E(A,B)|$ is at most $\alphasc(n)$ times the value of a minimum $\eta$-vertex-balanced cut of $G$.
\end{theorem}

We now turn to prove \Cref{cor: approx_balanced_cut}.
Let $G$ be the input graph, with $|E(G)|=m$.
For every vertex $v\in V(G)$, we denote by $d_v=\deg_G(v)$ the degree of $v$ in $G$. For each such vertex $v\in V(G)$, we denote $\delta_G(v)=\set{e_1(v),\ldots,e_{d_v}(v)}$, where the edges are indexed arbitrarily, and we let $K_v$ be a complete graph on $d_v$ vertices. We denote $V(K_v)=\set{x_1(v),\ldots,x_{d_v}(v)}$.

We construct a new graph $H$ as follows. First, we let $H$ be a disjoint union of graphs $K_v$, for all $v\in V(G)$. We call all edges in $\bigcup_{v\in V(G)}E(K_v)$ \emph{internal edges}. Next, we consider the edges of the graph $G$ one by one. Consider any such edge $e=(v,v')$, and assume that $e=e_i(v)=e_j(v')$. In other words, $e$ is the $i$th edge incident to $v$ and the $j$th edge incident to $v'$. We add the edge $e'=(x_i(v),x_j(v'))$ to graph $H$, and we view this edge as the \emph{copy of the edge $e$}. We call the resulting set $\set{e'\mid e\in E(H)}$ of edges \emph{external edges of $H$}. This completes the definition of the graph $H$. Note that $|V(H)|=\sum_{v\in V(G)}d_v=2m$, and every vertex of $H$ is incident to exactly one external edge.

Consider now any cut $(A',B')$ in graph $H$. We say that cut $(A',B')$ is \emph{canonical} if, for every vertex $v\in V(G)$, either $V(K_v)\subseteq A'$, or $V(K_v)\subseteq B'$.

Let $(X,Y)$ be a minimum $\hat \eta$-edge-balanced cut in graph $G$, and let $\rho=|E_G(X,Y)|$ denote its value. We start with the following observation.

\begin{observation}\label{small balanced cut in G}
	There is an $\eta_1$-vertex-balanced cut in graph $H$ of value at most $\rho$, for $\eta_1=\frac{1+\hat \eta}{2}$.
\end{observation}

\begin{proof}
	We construct a cut $(X',Y')$ in graph $H$ using the cut $(X,Y)$ in $G$, as follows. We start with $X',Y'=\emptyset$. For every vertex $v\in V(H)$, if $v\in X$, then we add all vertices of $K_v$ to $X'$, and otherwise we add them to $Y'$. It is immediate to verify that the value of the resulting cut $(X',Y')$ is $|E_H(X',Y')|=|E_G(X,Y)|=\rho$.
	
	We now show that cut $(X',Y')$ is $\eta_1$-vertex-balanced. In order to do so, it is enough to show that $|X'|,|Y'|\leq \eta_1\cdot |V(H)|$. We show that $|X'|\leq  \eta_1\cdot |V(H)|$. The proof for $Y'$ is symmetric.
	Indeed:
	\[|X'|=\sum_{v\in X}|V(K_v)|=\sum_{v\in X}d_v=2|E_G(X)|+|E_G(X,Y)|\leq |E_G(X)|+m\leq \hat \eta m+m\leq \frac{1+\hat \eta}{2}\cdot |V(H)|,
	\]
	which is bounded by $\eta_1|V(H)|$ (we have used the fact that $|V(H)|=2m$).
\end{proof}

We can now use the algorithm from \Cref{thm: ARV} to compute an $\eta_2$-vertex-balanced cut $(X',Y')$ in graph $H$, whose value is at most $\rho'=\alphasc(2m)\cdot \rho$. Here, $\eta_1<\eta_2<1$ is some constant. Note that, if cut $(X',Y')$ were canonical, we could immediately obtain a corresponding cut $(A,B)$ in graph $G$, whose value is at most $\rho'$, with the guarantee that $(A,B)$ is a $\hat \eta'$-edge-balanced cut, for some constant $\hat \eta'$. We use the following observation in order to convert the cut into a canonical one.

\begin{observation}\label{obs: make canonical}
	There is an efficient algorithm, that, given an $\eta'$-vertex-balanced cut $(X',Y')$ in graph $H$ of value $\rho'$, for some $0<\eta'<1$, computes a canonical $\eta^*$-vertex-balanced cut $(X^*,Y^*)$ in graph $H$ of value $\rho^*\leq O(\rho')$, for $\eta^*=\max\set{\frac{1+\eta'}2,0.95}$.
\end{observation}

\begin{proof}
	For every vertex $v\in V(G)$, we denote $X_v=X'\cap V(K_v)$ and $Y_v=Y'\cap V(K_v)$. Notice that graph $K_v$ contributes $|X_v|\cdot |Y_v|$ edges to the cut $(X',Y')$. We say that vertex $v$ is \emph{indecisive} if $|X_v|,|Y_v|\geq \frac{1-\eta'}{2}\cdot d_v$, and we say that it is \emph{decisive} otherwise.
	
	We modify the cut $(X',Y')$ in two steps. In the first step, we construct a new cut $(X'',Y'')$ in graph $H$ as follows. We start from $(X'',Y'')=(X',Y')$. We then consider every decisive vertex $v\in V(G)$ one by one. Consider any such vertex $v$, and recall that either $|X_v|<\frac{1-\eta'}{2}\cdot d_v$ holds, or $|Y_v|<\frac{1-\eta'}{2}\cdot d_v$. In the former case, we move the vertices of $X_v$ to $Y''$, while in the latter case we move the vertices of $Y_v$ to $X''$. Notice that $|X_v|\cdot |Y_v|$ edges of $K_v$ lie in the cut $(X',Y')$. At the end of the current iteration, no internal edges of $K_v$ contribute to the cut $(X'',Y'')$, but we may have added new external edges to the cut: if vertices of $X_v$ were moved to $Y''$, then we may have added up to $|X_v|$ such new edges (edges incident to vertices of $X_v$), and otherwise we may have added up to $|Y_v|$ such new edges. In either case, it is easy to see that $|E(X'',Y'')|$ may not grow as the result of the current iteration.
	The first step terminates once every decisive vertex of $G$ is processed. Notice that the total number of new vertices that we may have added to set $X''$ over the course of this step is at most:
	\[ \frac{1-\eta'}{2}\cdot\sum_{v\in V(G)}d_v\leq (1-\eta') m.\] 
	Since we are guaranteed that $|X'|\leq \eta'\cdot (2m)$, we get that, at the end of the current step, $|X''|\leq \eta'\cdot (2m)+  (1-\eta') m\leq \frac{1+\eta'}{2}\cdot (2m)$ holds. Similarly, $|Y''|\leq  \frac{1+\eta'}{2}\cdot (2m)$. We conclude that $(X'',Y'')$ is an $\eta''$-vertex-balanced cut in $H$, of value at most $\rho'$, where $\eta''=\frac{1+\eta'}{2}$.

	In the second step, we construct the final cut $(X^*,Y^*)$ in $H$ by taking care of indecisive vertices. Assume first that there is some indecisive vertex $v\in V(H)$ with $d_v\geq m/10$. Notice that, in this case, the number of edges that graph $K_v$ contributes to cut $(X',Y')$ is at least $|X_v|\cdot |Y_v|\geq \frac{1-\eta'}{4}\cdot (d_v)^2\geq  \frac{1-\eta'}{400}\cdot m^2$. Therefore, $\rho'>\frac{1-\eta'}{400}\cdot m^2$ must hold. Consider now a new cut $(X^*,Y^*)$ in graph $H$, where $X^*=V(K_v)$ and $Y^*=V(H)\setminus X^*$. Notice that $|X^*|=d_v\leq m$ and $|Y^*|\leq 2m-|X^*|\leq 2m\cdot 0.95$ holds. Therefore, cut $(X^*,Y^*)$ is $0.95$-vertex-balanced. Additionally, $|E_H(X^*,Y^*)|\leq d_v\leq m\leq O(\rho')$. We then return the cut $(X^*,Y^*)$ as the outcome of the algorithm. We assume from now on that for every indecisive vertex $v\in V(H)$, $d_v< m/10$.
	
	We start with $(X^*,Y^*)=(X'',Y'')$, and then process every indecisive vertex $v\in V(G)$ one by one. Consider an iteration when vertex $v$ is processed. Recall that graph $K_v$ contributes at least $|X_v|\cdot |Y_v|\geq \max\set{|X_v|,|Y_v|}$ edges to the cut $(X^*,Y^*)$. If $|X^*|<|Y^*|$, then we move the vertices of $Y_v$ from $Y^*$ to $X^*$. Notice that, after this transformation, the inner edges of $K_v$ no longer contribute to the cut, and at most $|Y_v|$ new outer edges are added to the cut. Therefore, the value of the cut does not increase. Otherwise, $|X^*|\geq |Y^*|$, and we move the vertices of $X_v$ from $X^*$ to $Y^*$. Using the same argument as before, the value of the cut does not increase. This completes the description of an iteration. Consider the cut $(X^*,Y^*)$ that is obtained at the end of the algorithm, after all indecisive vertices are processed. 
	
	We now show that $|X^*|,|Y^*|\leq \eta^*\cdot (2m)$. We prove this for $X^*$, and the proof for $Y^*$ is symmetric. We consider two cases. First, if no new vertices were added to $X^*$ over the course of the second step, then $|X^*|\leq |X''|\leq \frac{1+\eta'}{2}\cdot (2m)$. Assume now that some vertices were added to $X^*$, and let $v$ be the last indecisive vertex of $G$, for which the vertices of $K_v$ were added to $X^*$. Then before vertex $v$ was processed, $|X^*|\leq |Y^*|$ held. Since $d_v\leq m/10$ from our assumption, at the end of the iteration when $v$ was processed, $|X^*|\leq 1.1m$ held. Since no new vertices were added to $X^*$ in subsequent iterations, we get that $|X^*|\leq 1.1m\leq \eta^*\cdot (2m)$ holds at the end of the algorithm. We conclude that $(X^*,Y^*)$ is an $\eta^*$-balanced cut, of value at most $O(\rho')$. 
\end{proof}

By applying the algorithm from \Cref{obs: make canonical} to the $\eta'$-vertex-balanced cut $(X',Y')$ in graph $H$, we obtain a canonical $\eta^*$-vertex-balanced cut $(X^*,Y^*)$ in graph $H$, with $\eta^*=\max\set{\frac{1+\eta'}2,0.95}$, whose value is $\rho^*\leq O(\rho')=O(\alphasc(m))\cdot \rho$. We use this cut in order to construct a cut $(A,B)$ in $G$ as follows: for every vertex $v\in V(G)$, if $V(K_v)\subseteq X^*$, then vertex $v$ is added to $A$, and otherwise it is added to $B$. Notice that $|E_G(A)|\leq \sum_{v\in A}d_v/2\leq |X^*|/2\leq \eta^*\cdot m$. Similarly, $|E_G(B)|\leq \eta^*\cdot m$. Therefore, cut $(A,B)$ is $\eta^*$-edge-balanced. Additionally, $|E_G(A,B)|\leq |E_H(A^*,B^*)|\leq  O(\alphasc(m))\cdot \rho$.


\subsection{Proof of Theorem~\ref{lem:min_bal_cut}}

\label{apd: Proof of min_bal_cut}

We use the following theorem from~\cite{lipton1979separator}.

\begin{theorem}[Theorem 4 from~\cite{lipton1979separator}]
	\label{thm: weighted_planar_separator}
	Let $G$ be any  \textbf{simple} $n$-vertex planar graph with  weights $w_v\geq 0$  on its vertices $v\in V(G)$, such that $\sum_{v\in V(G)}w_v\leq 1$. Then there is a partition $(A,B,C)$ of $V(G)$, such that no edge connects a vertex of $A$ to a vertex in $B$; $\sum_{v\in A}w_v,\sum_{v\in B}w_v\leq 2/3$; and $|C|\leq \sqrt{8n}$.
\end{theorem}

In order to prove \Cref{lem:min_bal_cut}, we define a new simple planar graph $G'$ that is obtained by modifying graph $G$, using its optimal drawing. We then apply \Cref{thm: weighted_planar_separator} to graph $G'$, and transform the resulting partition $(A,B,C)$ of $V(G')$ into a  $(3/4)$-edge-balanced cut of graph $G$, whose value is at most $O(\sqrt{\optcro(G)+\Delta\cdot m})$.

In order to define graph $G'$, we first define an intermediate graph $G_1$. Consider the input graph $G$ and its optimal drawing $\phi$ in the plane. For every vertex $v\in V(G)$, we denote $d_v=\deg_G(v)$, and we denote $\delta_G(v)=\set{e_1(v),\ldots,e_{d_v}(v)}$, where the edges are indexed according to the order in which their images enter the image of $v$ in $\phi$, in the counter-clock-wise direction. We let $H_v$ be the $(d_v\times d_v)$-grid, and we denote the set of the vertices on the first row of the grid by $X(v)=\set{x_1(v),\ldots,x_{d_v}(v)}$, where the vertices are indexed in their natural order. In order to define graph $G_1$, we start with the disjoint union of all grids in $\set{H_v}_{v\in V(G)}$. We refer to the edges that lie in these grids as \emph{internal edges}. Next, we process every edge $e\in E(G)$ one by one. Consider any such edge $e=(v,v')$, and assume that $e=e_i(v)=e_j(v')$, that is, $e$ is the $i$th edge incident to $v$ and the $j$th edge incident to $v'$. We then add an edge $e'=(x_i(v),x_j(v'))$ to graph $G_1$. We think of edge $e'$ as the \emph{copy} of the edge $e$ in $G_1$. The edges in set $\set{e'\mid e\in E(G)}$ are called \emph{external edges} of graph $G_1$. We say that a cut $(A,B)$ in graph $G_1$ is \emph{canonical} if, for every vertex $v\in V(G)$, either $V(H_v)\subseteq A$, or $V(H_v)\subseteq B$ holds. Note that a canonical cut $(A,B)$ in graph $G_1$ naturally defines a cut $(A',B')$ of the same value on graph $G$, where a vertex $v\in V(G)$ is added to $A'$ if $V(H_v)\subseteq A$, and it is added to $B'$ otherwise. Lastly, note that the optimal drawing $\phi$ of $G$ defines a drawing $\phi_1$ of $G_1$ with the same number of crossings. In order to obtain drawing $\phi_1$ of $G_1$, we start with the drawing $\phi$ of $G$, and then inflate the image of every vertex $v\in V(G)$, so that it becomes a disc $D(v)$. We place another smaller disc $D'(v)$ inside $D(v)$, so that the boundaries of both discs are disjoint. We then place the standard drawing of the grid $H_v$ inside disc $D'(v)$, so that vertices $x_1(v),\ldots,x_{d_v}(v)$ appear on the boundary of the disc $D'(v)$ in this counter-clock-wise order. By slightly extending the images of the edges $e_1(v),\ldots,e_{d_v}(v)$ inside $D(v)\setminus D'(v)$, we can ensure that the image of each such edge $e_i(v)$ terminates at the image of the vertex $x_i(v)$. Once all vertices of $V(G)$ are processed in this manner, we obtain a drawing $\phi_1$ of graph $G_1$, in which the number of crossings is bounded by $\cro(\phi)=\optcro(G)$.

In order to obtain the final graph $G'$, we start with $G'=G_1$, and we denote $V(G_1)=X$. Next, for every {\bf external} edge $e'\in E(G_1)$, we subdivide the edge with a new vertex $u_{e'}$. In other words, if $e'=(x_i(v),x_j(v'))$, then we replace the edge with a path consisting of two edges: $(x_i(v),u_{e'})$, and $(u_{e'},x_j(v'))$. We denote this new set of vertices representing the external edges of $G_1$ by $U=\set{u_{e'}\mid e\in E(G)}$. Note that drawing $\phi_1$ of graph $G_1$ can be easily transformed into a drawing of this new graph, without increasing the number of crossings. Denote the resulting drawing by $\phi_2$. In our last step, for every crossing point $p$ between a pair $a,a'$ of edges in drawing $\phi_2$, we replace point $p$ with a new vertex $y_p$. In other words, if $a=(s,t)$ and $a'=(s',t')$, then we add a new vertex $y_p$ to the graph. We then replace edge $a=(s,t)$ with two new edges, $(s,y_p)$ and $(y_p,t)$, and we similarly replace edge $a'$ with two new edges, $(s',y_p)$ and $(y_p,t')$.  We continue processing every crossing point in drawing $\phi_2$ one by one in this manner, until no more crossings remain. We denote this new set of vertices, that represent all crossing points in the original drawing $\phi_2$, by $Y$. Note that $|Y|=\optcro(G)$. This completes the definition of the graph $G'$. Observe that $V(G')=X\cup Y\cup U$, and so:
\[|V(G')|=\sum_{v\in V(G)}(d_v)^2+m+\optcro(G)\leq \Delta\cdot \sum_{v\in V(G)}d_v+m+\optcro(G)\leq 3\Delta m+\optcro(G). \]
It is also immediate to verify that $G'$ is a simple planar graph, and that the maximum vertex degree in $G'$ is at most $4$. We now assign weights $w_v$ to vertices $v\in V(G')$, as follows: every vertex $u_{e'}\in U$ is assigned weight $1/m$, and all other vertices are assigned weight $0$. It is immediate to verify that $\sum_{v\in V(G')}w_v=1$.

From \Cref{thm: weighted_planar_separator}, there is a partition $(A,B,C)$ of $V(G')$, such that no edge connects a vertex of $A$ to a vertex in $B$; $\sum_{v\in A}w_v,\sum_{v\in B}w_v\leq 2/3$; and $|C|\leq \sqrt{8|V(G')|}\leq \sqrt{24\Delta m+8\optcro(G)}$.
We  convert this partition of vertices of $G'$ into a $(3/4)$-balanced cut in graph $G$ in three steps. In the first step, we use the partition $(A,B,C)$ of $V(G')$ in order to construct a cut $(A_1,B_1)$ in graph $G_1$. In the second step, we transform this cut into a canonical cut $(A_2,B_2)$ in graph $G_1$. Lastly, in the third step, we use this canonical cut in order to define the final cut $(A^*,B^*)$ in graph $G$.
We now describe each of the steps in turn.

\paragraph{Step 1: Cut in Graph $G_1$.}

We define a cut $(A_1,B_1)$ in graph $G_1$ as follows. 
We start with $A_1=B_1=\emptyset$, and then process every vertex $v\in V(G)$ one by one. When vertex $v\in V(G)$ is processed, we consider every vertex $x\in V(H_v)$. If $x\in A\cup C$, then we add $x$ to $A_1$, and otherwise we add $x$ to $B_1$. 

Consider now the resulting cut $(A_1,B_1)$ in graph $G_1$. We first claim that $|E_{G_1}(A_1,B_1)|\leq 4|C|$.
In order to prove this, we assign, to every edge $e\in E_{G_1}(A_1,B_1)$, some vertex $x\in C$ that is \emph{responsible} for $e$, and we will ensure that every vertex of $C$ is responsible for at most $4$ edges of $ E_{G_1}(A_1,B_1)$.
 Consider some edge $e\in E_{G_1}(A_1,B_1)$. 
If either of the endpoints of $e$ lies in set $C$, then we assign $e$ to that endpoint. Otherwise, there must be some vertex $x$ of graph $G'$ that subdivided the edge $e$ (so either $x\in U$ or $x\in Y$ holds), and $x\in C$. In this case, we assign $e$ to this vertex $x$. Since the degree of every vertex in $G'$ is at most $4$, every vertex of $C$ may be assigned to at most $4$ edges of $E_{G_1}(A_1,B_1)$, and so we conclude that $|E_{G_1}(A_1,B_1)|\leq 4|C|\leq 4\cdot \sqrt{24\Delta m+8\optcro(G)}$.

Next, we bound the total number of external edges in $E_{G_1}(A_1)$ and in $E_{G_1}(B_1)$. For convenience, denote by $E'$ the set of all external edges in graph $G_1$. Consider some external edge $e'=(x_i(v ),x_j(v'))\in E'$. Let $P(e')$ be the path that replaced the edge $e'$ in graph $G'$. Recall that path $P(e')$ is a path connecting $x_i(v)$ to $x_j(v')$, it contains the vertex $u_{e'}$ representing the edge $e'$, and possibly additional vertices representing the crossing points of edge $e'$ with other edges. 
We claim that either vertex $y_{e'}$ lies in $A$, or some vertex of $P(e')$ (including possibly $x_i(v)$ or $x_j(v')$) must lie in $C$. Indeed, assume that $y_{e'}\not \in A$. If none of the vertices of $P(e')$ lie in $C$, then $y_{e'}\in B$, while $x_i(v),x_j(v')\in A$ must hold. This is impossible since there are no edges connecting vertices of $A$ to vertices of $B$. Therefore, either $y_{e'}\in A$, or at least one vertex on $P(e')$ lies in $C$. Since every vertex of $G'$ has degree at  most $4$, every vertex of $C$ may lie on at most $4$ paths in $\set{P(e')\mid e\in E(G)}$. Since the weight of every vertex in $U$ is $1/m$, we get that: 
\[|E'\cap E_{G_1}(A_1)|\leq |U\cap A|+4|C|\leq m\cdot \sum_{v\in A}w_v+4|C|\leq 2m/3+  4\cdot \sqrt{24\Delta m+8\optcro(G)}.\]
Using similar reasoning,  $|E'\cap E_{G_1}(B_1)|\leq 2m/3+  4\cdot \sqrt{24\Delta m+8\optcro(G)}$.

\paragraph{Step 2: Canonical Cut in Graph $G_1$.}

In this step we construct a cut $(A_2,B_2)$ in graph $G_1$ that is canonical, by gradually modifying the cut $(A_1,B_1)$. For every vertex $v\in V(G)$, we denote $n_A(v)=|X(v)\cap A_1|$, and $n_B(v)=|X(v)\cap B_1|$. We use the following simple observation.

\begin{observation}\label{obs: cut in grid}
	For every vertex $v\in V(G)$, $|E(H_v)\cap E_{G_1}(A_1,B_1)|\geq\min\set{n_A(v),n_B(v)}$.
\end{observation}
\begin{proof}
	We partition the columns of the grid $H_v$ into two subsets, $\wset',\wset''$, as follows. For $1\leq i\leq d_v$, the $i$th column of the grid is added to set $\wset'$ if vertex $x_i(v)$ (the vertex of the $i$th column that lies on the first row of the grid) lies in $A_1$. Otherwise, the $i$th column is added to $\wset''$.
	
	We now consider three cases. The first case happens if, for every row $R$ of the grid $H_v$, at least one edge of $R$ lies in $E_{G_1}(A_1,B_1)$. Clearly, in this case, $|E(H_v)\cap E_{G_1}(A_1,B_1)|\geq d_v\geq \min\set{n_A(v),n_B(v)}$. The second case happens if, for every column $W\in \wset'$, at least one edge of $W$ lies in $E_{G_1}(A_1,B_1)$. In this case, $|E(H_v)\cap E_{G_1}(A_1,B_1)|\geq n_A(v)\geq \min\set{n_A(v),n_B(v)}$. Lastly, the third case happens if,  for every column $W\in \wset''$, at least one edge of $W$ lies in $E_{G_1}(A_1,B_1)$. In this case, $|E(H_v)\cap E_{G_1}(A_1,B_1)|\geq n_B(v)\geq \min\set{n_A(v),n_B(v)}$.
	
	We now claim that at least one of the above three cases has to happen. Indeed, assume otherwise. Then there is some row $R$ of the grid, and two columns $W\in \wset'$, $W'\in \wset''$, such that no edge of $E(R)\cup E(W')\cup E(W'')$ lies in $E_{G_1}(A_1,B_1)$. Assume that $W'$ is the $i$th column and $W''$ is the $j$th column of the grid $H_v$. Since $x_i(v)\in A_1$, $x_j(v)\in B_1$, and $R\cup W'\cup W''$ is a connected graph, this is impossible.
\end{proof}

We say that a vertex $v\in V(G)$ is \emph{indecisive} iff $n_A,n_B\geq d_v/32$; otherwise we say that vertex $v$ is \emph{decisive}. We start with $(A_2,B_2)=(A_1,B_1)$, and then gradually modify this cut, by processing the decisive vertices one by one. When such a vertex $v$ is processed, if $n_A<d_v/32$, then we move the vertices of $V(H_v)\cap A_2$ from $A_2$ to $B_2$. Notice that this transformation adds up to $n_A$ new external edges to cut $E_{G_1}(A_2,B_2)$ -- the edges incident to the vertices of $X(v)\cap A_2$. However, from \Cref{obs: cut in grid}, at least $n_A=|X(v)\cap A_2|$ edges of $H_v$ contributed to the cut $E_{G_1}(A_2,B_2)$ before this transformation, and they no longer contribute to the cut after the transformation. Therefore, $|E_{G_1}(A_2,B_2)|$ does not increase as the result of this transformation. If $n_A\geq d_v/32$, then $n_B<d_v/32$ must hold, and we move the vertices of $V(H_v)\cap B_2$ from $B_2$ to $A_2$. Using the same arguments as before, $|E_{G_1}(A_2,B_2)|$ does not increase.

Consider the cut $(A_2,B_2)$ of $G_1$ that is obtained after all decisive vertices are processed. From the above discussion,  $|E_{G_1}(A_2,B_2)|\leq |E_{G_1}(A_1,B_1)|\leq 4\cdot \sqrt{24\Delta m+8\optcro(G)}$. Moreover, the total number of 
new edges of $E'$ that were added to $E_{G_1}(A_2)$ is bounded by $\sum_{v\in V(G)}d_v/32\leq m/16$. 
Since $|E'\cap E_{G_1}(A_1)|\leq 2m/3+  4\cdot \sqrt{24\Delta m+8\optcro(G)}$, if $\optcro(G)\le m^2/2^{40}$ and $\Delta\leq m/2^{40}$, then
\[|E'\cap E_{G_1}(A_2)|\leq m/16+2m/3+ 4\cdot \sqrt{24\Delta m+8\optcro(G)}\leq 3m/4.\]
Using the same reasoning, if $\optcro(G)\le m^2/2^{40}$, then $|E'\cap E_{G_1}(B_2)|\leq 3m/4$.

Next, we construct a canonical cut $(A_3,B_3)$ in graph $G_1$, by starting with $(A_3,B_3)=(A_2,B_2)$, and processing every indecisive vertex $v\in V(G)$ one by one. When vertex $v$ is processed, if $|E'\cap E_{G_1}(A_3)|\leq |E'\cap E_{G_1}(B_3)|$, then we move all vertices of $H_v\cap B_3$ from $B_3$ to $A_3$, and otherwise we move all vertices of $H_v\cap A_3$ from $A_3$ to $B_3$. Note that in either case, the total number of external edges that are added to cut $(A_3,B_3)$ is bounded by $d_v$. From \Cref{obs: cut in grid}, the number of edges of $H_v$ that contributed to the cut $(A_3,B_3)$ before this transformation is at least $\min\set{n_A,n_B}$. Since vertex $v$ is indecisive, $n_A,n_B\geq d_v/32$. Once every indecisive vertex of $G$ is processed, we obtain the final cut $(A_3,B_3)$ in graph $G_1$, that must be canonical. From the above discussion:
\[|E_{G_1}(A_3,B_3)|\leq 32\cdot |E_{G_1}(A_2,B_2)|\leq 32|E_{G_1}(A_1,B_1)|
\leq  O\left(\sqrt{\Delta m+\optcro(G)}\right ).\]
We claim that $|E'\cap E_{G_1}(A_3)|,|E'\cap E_{G_1}(B_3)|\leq 3m/4$. We prove this for $A_3$, as the proof for $B_3$ is symmetric. If no new vertices were added to set $A_3$, then $|E'\cap E_{G_1}(A_3)|\leq |E'\cap E_{G_1}(A_2)|\leq 3m/4$ holds. Assume now that new vertices were added to $A_3$, and let $v$ be the last indecisive vertex that was processed by the algorithm, for which vertices of $H_v$ were added to $A_3$. Then, before vertex $v$ was processed, $|E'\cap E_{G_1}(A_3)|\leq |E'\cap E_{G_1}(B_3)|$ held; therefore, $|E'\cap E_{G_1}(A_3)|\leq m/2$. Note that moving the vertices of $V(H_v)\cap B_3$ from $B_3$ to $A_3$ could have added at most $d_v\leq \Delta\leq m/2^{40}$ new edges to $E'\cap E_{G_1}(A_3)$, and so $|E'\cap E_{G_1}(A_3)|\leq 3m/4$ must hold at the end of this iteration. Since the iteration when $v$ was processed is the last iteration when vertices were added to $A_3$, we conclude that $|E'\cap E_{G_1}(A_3)|\leq 3m/4$ holds at the end of the algorithm. Using the same reasoning, $|E'\cap E_{G_1}(B_3)|\leq 3m/4$ holds as well.

\paragraph{Step 3: Balanced Cut in $G$.}
We are now ready to define the final cut $(A^*,B^*)$ in graph $G$. We add to $A^*$ every vertex $v\in V(G)$ with $V(H_v)\subseteq A_3$, and we add all remaining vertices of $V(G)$ to $B^*$. It is easy to verify that $|E_G(A^*,B^*)|=|E_{G_1}(A_3,B_3)|\leq O\left(\sqrt{\Delta m+\optcro(G)}\right )$. Additionally, $|E_G(A^*)|\leq |E'\cap E_{G_1}(A_3)|\leq 3m/4$, and similarly $|E_G(B^*)|\leq |E'\cap E_{G_1}(B_3)|\leq 3m/4$. We conclude that $(A^*,B^*)$ is a $(3/4)$-edge-balanced cut in graph $G$, whose value is at most $O(\sqrt{\optcro(G)+\Delta\cdot m})$.


\subsection{Proof of \Cref{obs: grid 1st row well-linked}}
\label{apd: Proof grid 1st row well-linked}

The proof is practically identical to the proof of \Cref{obs: cut in grid}. Consider a cut $(A,B)$ in graph $H$, and denote $n_A=|S\cap A|$ and $n_B=|S\cap B|$. It is enough to show that $|E(A,B)|\geq \min\set{n_A,n_B}$.

	We partition the columns of the grid graph $H$ into two subsets, $\wset',\wset''$, as follows. For $1\leq i\leq r$, the $i$th column of the grid is added to set $\wset'$ if the unique vertex of $S$ lying in the $i$th column belongs to $A$. Otherwise, the $i$th column is added to $\wset''$.

We now consider three cases. The first case happens if, for every row $R$ of the grid $H$, at least one edge of $R$ lies in $E(A,B)$. Clearly, in this case, $|E(A,B)|\geq r\geq \min\set{n_A,n_B}$. The second case happens if, for every column $W\in \wset'$, at least one edge of $W$ lies in $E(A,B)$. In this case, $|E(A,B)|\geq n_A\geq \min\set{n_A,n_B}$. Lastly, the third case happens if,  for every column $W\in \wset''$, at least one edge of $W$ lies in $E(A,B)$. In this case, $| E(A,B)|\geq n_B\geq \min\set{n_A,n_B}$.

We now claim that at least one of the above three cases has to happen. Indeed, assume otherwise. Then there is some row $R$ of the grid, and two columns $W\in \wset'$, $W'\in \wset''$, such that no edge of $E(R)\cup E(W')\cup E(W'')$ lies in $E(A,B)$. But the unique vertex of $S\cap V(W)$ lies in $A$, the unique vertex of $S\cap V(W')$ lies in $B$, and $R\cup W'\cup W''$ is a connected graph, a contradiction.


\subsection{Proof of Theorem~\ref{thm: bandwidth_means_boundary_well_linked}}
\label{apd: Proof of bandwidth_means_boundary_well_linked}

We construct an $s$-$t$ flow network, as follows. We start with the graph $G$, and then add a new source vertex $s$, that connects to every vertex in $T_1$ with an edge. We also add a new destination vertex $t$, and connect every vertex of $T_2$ to $t$ with an edge. 
Denote the resulting graph by $H$. For every edge $e\in E(H)$, if $e$ is incident to $s$ or to $t$, then we set its capacity $c(e)=1$, and otherwise we set $c(e)=\ceil{1/\alpha}$.
Note that the capacity of every edge in the resulting flow network is integral. 

We show below that the value of the maximum $s$-$t$ flow in the resulting flow network is $k=|T_1|$. From the integrality of flow, we can then compute an integral $s$-$t$ flow $f$ of value $k$ in $H$. Let $\pset$ be the set of all $s$-$t$ paths in graph $H$. Since flow $f$ is integral, and since the capacity of every edge incident to $s$ and to $t$ is $1$, for every path $P\in \pset$, $f(P)=0$ or $f(P)=1$ holds. Moreover, if $\pset'\subseteq \pset$ is the set of all paths $P$ with $f(P)=1$, then $|\pset'|=k$. 
Since the capacity of every edge in $\set{(s,x)\mid x\in T_1}$, and the capacity of every edge in $\set{(y,x)\mid y\in T_1}$ is $1$, each such edge belongs to exactly one path in $\pset'$. Therefore,
set $\pset'$ of paths naturally defines a one-to-one routing $\qset$ of vertices of $T_1$ to vertices of $T_2$ in graph $G$, with $\cong_G(\qset)\leq \ceil{1/\alpha}$. In order to complete the proof of the theorem, it is now enough to show that the value of the maximum $s$-$t$ flow in graph $H$ is at least $k$.

Assume for contradiction that this is not the case. Consider a minimum $s$-$t$ cut $(A,B)$ in graph $H$. From our assumption, the value of the cut is less than $k$. We partition the set $T_1$ of vertices into two subsets: set $T_1'=T_1\cap A$ and set $T_1''=T_1\cap B$. Note that, for every vertex $x\in T_1''$, its corresponding edge $(s,x)$ belongs to the cut $E(A,B)$. We denote by $E_1=\set{(s,x)\mid x\in T_1''}$ the corresponding set of edges. We also partition the set  $T_2$ of vertices into two subsets: set $T_2'=T_2\cap B$ and set $T_2''=T_2\cap A$. Note that, for every vertex $y\in T_2''$, its corresponding edge $(y,t)$ belongs to the cut $E(A,B)$. We denote by $E_2=\set{(y,t)\mid y\in T_2''}$ the corresponding set of edges.

Lastly, we denote by $E'=E(A,B)\setminus (E_1\cup E_2)$ the set of the remaining edges in the cut $(A,B)$. Note that each edge in $E'$ has capacity $\ceil{1/\alpha}$, while each edge in $E_1\cup E_2$ has capacity $1$. Since we have assumed that the value of the cut $(A,B)$ is less than $k$, we get that:
\begin{equation}\label{eq: bound the cut}
|T_1''|+|T_2''|+|E'|\cdot \ceil{1/\alpha}= |E_1|+|E_2|+|E'|\cdot \ceil{1/\alpha}=\sum_{e\in E_H(A,B)}c(e)<k 
\end{equation}
We define a cut $(A',B')$ in graph $G$ using cut $(A,B)$ as follows: $A'=A\setminus \set{s}$ and $B'=B\setminus \set{t}$. Notice that $|E_G(A',B')|=|E'|$. From \Cref{eq: bound the cut}, we then get that:
\[|E_G(A',B')|=|E'| < \alpha\cdot (k-|T_1''|-|T_2''|) \]
Since $|T_1'|=k-|T_1''|$ and $|T_2'|=k-|T_2''|$, we get that: 
\[|E_G(A',B')| < \alpha\cdot \min\set{|T_1'|,|T_2'|}.\]
Lastly, since $T_1'\subseteq A'$ and $T_2'\subseteq B'$, we get that $|E_G(A',B')| < \alpha\cdot \min\set{|T\cap A'|,|T\cap B'|}$, contradicting the fact that the set $T$ of vertices is $\alpha$-well-linked in $G$.


\subsection{Proof of Theorem~\ref{thm:well_linked_decomposition}}
\label{apd: Proof of well_linked_decomposition}

Assume first that $0<\alpha\le 1/m$. Then we simply let $\rset=\set{S}$. Since $\alpha\le 1/m$, and $S$ is a connected graph, it is easy to verify that it has the $\alpha$-bandwidth property in graph $G$. For each edge $e\in \delta_G(S)$, we simply let $P(e)$ be the path that contains the single edge $e$. It is easy to verify that cluster set $\rset=\set{S}$, and the set $\pset(S)=\set{P(e)\mid e\in \delta_G(S)}$ of paths have all required properties.

We assume from now on that $\frac 1 m <\alpha< \min\set{\frac 1 {64\alphasc(m)\cdot \log m},\frac 1 {48\log^2 m}}$ holds.

Our algorithm maintains a collection $\rset$ of clusters of $S$, that is initialized to $\rset=\set{S}$. Throughout the algorithm, we ensure that the following invariants hold:

\begin{properties}{I}
	\item all clusters in $\rset$ are mutually disjoint; \label{inv1: disjointness}
	\item  $\bigcup_{R\in \rset}V(R)=V(S)$; and \label{inv2: partition}
	\item for every cluster $R\in \rset$, $|\delta_G(R)|\leq |\delta_G(S)|$. \label{inv3: small boundary}
\end{properties}

 For a given collection $\rset$ of clusters with the above properties, we define a \emph{budget} $b(e)$ for every edge $e\in E(G)$, as follows. If $e\in \delta_G(S)$, and the endpoint of $e$ that lies in $S$ belongs to a cluster $R\in\rset$, then we set the budget $b(e)=1+8\alpha  \cdot \alphasc(m)\cdot \log_{3/2}(|\delta_G(R)|)$. If edge $e$ has its endpoints in two distinct clusters $R,R'\in \rset$, then we set $b(e)=2+8\alpha\cdot \alphasc(m)\cdot \log_{3/2}(|\delta_G(R)|)+8\alpha \cdot \alphasc(m)\cdot \log_{3/2}(|\delta_G(R')|)$. Otherwise, we set $b(e)=0$.
Notice that, for every edge $e\in E(G)$, $b(e)\leq 3$ always holds. Additionally, for every edge $e\in \bigcup_{R\in \rset}\delta_G(R)$, $b(e)\geq 2$ if the endpoints of $e$ lie in two different clusters of $\rset$, and $b(e)\geq 1$ if $e\in \delta_G(S)$. Therefore, if we denote by $B=\sum_{e\in E(G)}b(e)$ the total budget in the system, then, throughout the algorithm, $B\geq \sum_{R\in \rset}|\delta_G(R)|$ holds. Lastly, observe that, at the beginning of the algorithm, $B\leq |\delta_G(S)|\cdot (1+O(\alpha\cdot \alphasc(m)\cdot \log m))$. Throughout the algorithm, we will modify the clusters in set $\rset$, leading to changes in the budgets of the edges of $G$. We will ensure however that the total budget $B$ never increases, and so, if $\rset$ is the final set of clusters that we obtain, then $\sum_{R\in \rset}|\delta_G(R)|\leq B\leq (1+O(\alpha\cdot \alphasc(m)\cdot \log m))=(1+O(\alpha\cdot \log^{1.5} m))$ holds.

Throughout the algorithm, we maintain a partition of the set $\rset$ of clusters into two subsets: set $\rset^A$ of \emph{active} clusters, and set $\rset^I$ of \emph{inactive} clusters. We will ensure that the following additional invariant holds:

\begin{properties}[3]{I}
\item every cluster $R\in \rset^I$ has the $\alpha$-bandwidth property.\label{inv: last - bw}
\end{properties}
 
Additionally, we will store, with every inactive cluster $R\in \rset^I$, a set $\pset(R)=\set{P(e)\mid e\in \delta_G(R)}$ of paths in graph $G$, (that we refer to as \emph{witness set of paths for $R$}), such that $\cong_G(\pset(R))\leq 100$, and, for every edge $e\in \delta_G(R)$, path $P(e)$ has $e$ as its first edge and some edge of $\delta_G(S)$ as its last edge, and all inner vertices of $P(e)$ lie in $V(S)\setminus V(R)$. At the beginning of the algorithm, we set $\rset^A=\rset=\set{S}$ and $\rset^I=\emptyset$. Clearly, all invariants hold at the beginning of the algorithm. We then proceed in iterations, as long as $\rset^A\neq \emptyset$.

In order to execute an iteration, we select an arbitrary cluster $R\in \rset^A$ to process. We will  either establish that $R$ has the $\alpha$-bandwidth property in graph $G$ and compute a witness set $\pset(R)$ of paths for $R$ (in which case $R$ is moved from $\rset^A$ to $\rset^I$); or we will modify the set $\rset$ of clusters so that the total budget of all edges decreases by at least $1/m$. An iteration that processes a cluster $R\in \rset^A$ consists of two steps. The purpose of the first step is to either establish the $\alpha$-bandwidth property of cluster $R$, or to replace it with a collection of smaller clusters in $\rset$. The purpose of the second step is to either compute the witness set $\pset(R)$ of paths for cluster $R$, or to modify the set $\rset$ of clusters so that the total budget of all edges decreases. We now describe each of the two steps in turn.

\paragraph{Step 1: Bandwidth Property.}
Let $R\in \rset^A$ be any active cluster, and
let $R^+$ be the augmentation of $R$ in graph $G$. Recall that $R^+$ is a graph that is obtained from $G$ through the following process. First, we subdivide every edge $e\in \delta_G(R)$ with a vertex $t_e$, and we let $T=\set{t_e\mid e\in \delta_G(R)}$ be the resulting set of vertices. We then let $R^+$ be the subgraph of the resulting graph induced by vertex set $V(R)\cup T$. We apply  Algorithm  \algsc  for computing approximate sparsest cut to graph $R^+$, with the set $T$ of vertices, to obtain a $\alphasc(m)$-approximate sparsest cut $(X,Y)$ in graph $R^+$ with respect to vertex set $T$. 
We now consider two cases. The first case happens if $|E(X,Y)|\geq \alpha\cdot \alphasc(m)\cdot \min\set{|X\cap T|,|Y\cap T|}$. In this case, we are guaranteed that the minimum sparsity of any $T$-cut in graph $R^+$ is at least $\alpha$, or equivalently, set $T$ of vertices is $\alpha$-well-linked in $R^+$. From \Cref{obs: wl-bw}, cluster $R$ has the $\alpha$-bandwidth property in graph $G$. In this case, we proceed to the second step of the algorithm.

Assume now that $|E(X,Y)|< \alpha\cdot \alphasc(m)\cdot \min\set{|X\cap T|,|Y\cap T|}$. 
Since $\alpha\leq \min\set{\frac 1 {64\alphasc(m)\cdot \log m},\frac 1 {48\log^2 m}}$, we get that the sparsity of the cut $(X,Y)$ is less than $1$.  Consider now any vertex $t\in T$, and let $v$ be the unique neighbor  of $t$ in $R^+$. We can assume w.l.o.g. that either $t,v$ both lie in $X$, or they both lie in $Y$. Indeed, if $t\in X$ and $v\in Y$, then moving vertex $t$ from $X$ to $Y$ does not increase the sparsity of the cut $(X,Y)$. This is because, for any two real numbers $1\leq a<b$, $\frac{a-1}{b-1}\leq \frac a b$. Similarly, if $t\in Y$ and $v\in X$, then moving $t$ from $Y$ to $X$ does not increase the sparsity of the cut $(X,Y)$. Therefore, we assume from now on, that for every vertex $t\in T$, if $v$ is the unique neighbor of $t$ in $R^+$, then either both $v,t\in X$, or both $v,t\in Y$.

Consider now the partition $(X',Y')$ of $V(R)$, where $X'=X\setminus T$ and $Y'=Y\setminus T$. It is easy to verify that $|\delta_G(R)\cap \delta_G(X')|=|X\cap T|$, and $|\delta_G(R)\cap \delta_G(Y')|=|Y\cap T|$. Let $E'=E_G(X',Y')$, and assume w.l.o.g. that $|\delta_G(R)\cap \delta_G(X')|\leq |\delta_G(R)\cap \delta_G(Y')|$. Then $|E'|< \alpha\cdot \alphasc(m)\cdot |\delta_G(R)\cap \delta_G(X')|$ must hold. We remove cluster $R$ from sets $\rset$ and $\rset^A$, and we add instead every connected component of graphs $G[X']$ and $G[Y']$ to both sets. It is immediate to verify that $\rset$ remains a collection of disjoint clusters of $G$, and that $\bigcup_{R'\in \rset}V(R')=V(G)$. Since $|E'|<\min\set{|\delta_G(R)\cap \delta_G(X')|,|\delta_G(R)\cap \delta_G(Y')|}$, we get that for every cluster $C$ that we just added to $\rset$, $|\delta_G(C)|\leq |\delta_G(R)|\leq |\delta_G(S)|$ (from Invariant \ref{inv3: small boundary}). Therefore, all invariants continue to hold. We now show that the total budget $B$ decreases by at least $1/m$ as the result of this operation.

Note that the only edges whose budgets may change as the result of this operation are edges of $\delta_G(R)\cup E'$. Observe that, for each edge $e\in \delta_G(R)\cap \delta_G(Y')$, its budget $b(e)$ may not increase. Since we have assumed that $|\delta_G(R)\cap \delta_G(X')|\leq |\delta_G(R)\cap \delta_G(Y')|$, and since $|E'|<|\delta_G(R)|/8$, we get that $|\delta_G(X')|\leq 2|\delta_G(R)|/3$. Therefore, for every edge $e\in \delta_G(X')\cap \delta_G(R)$, its budget $b(e)$ decreases by at least $8\alpha \cdot \alphasc(m)\cdot\log_{3/2}(|\delta_G(R)|)-8\alpha \cdot \alphasc(m)\cdot\log_{3/2}(|\delta_G(X')|)$. Since $|\delta_G(X')|\leq 2|\delta_G(R)|/3$, we get that $ \log_{3/2}(|\delta_G(R)|)\leq \log_{3/2}(3|\delta_G(X')|/2)\leq 1+\log_{3/2}(|\delta_G(X')|$. We conclude that the budget $b(e)$ of each edge $e\in \delta_G(X')\cap \delta_G(R)$ decreases by at least $8\alpha\cdot \alphasc(m)$.
On the other hand, the budget of every edge $e\in E'$ increases by at most $3$. Since $|E'|\leq \alpha\cdot \alphasc(m)\cdot |\delta_G(R)\cap \delta_G(X')|$, we get that the decrease in the budget $B$ is at least:
\[
\begin{split}
&8\alpha\cdot \alphasc(m)\cdot |\delta_G(X')\cap \delta_G(R)|-3|E'|\\&\hspace{3cm}\geq 8\alpha\cdot \alphasc(m)\cdot |\delta_G(X')\cap \delta_G(R)|- 3\alpha\cdot \alphasc(m)\cdot |\delta_G(R)\cap \delta_G(X')| 
\\&\hspace{3cm}\geq 5 \alpha\cdot \alphasc(m)\cdot |\delta_G(R)\cap \delta_G(X')|\\&\hspace{3cm}>1/m,\end{split}\]
since $\alpha\geq 1/m$.
To conclude, if $|E(X,Y)|< \alpha\cdot \alphasc(m)\cdot \min\set{|X\cap T|,|Y\cap T|}$, then we have modified the set $\rset$ of clusters, so that all invariants continue to hold, and the total budget $B$ decreases by at least $1/m$. In this case, we terminate the current iteration.

From now on we assume that $|E(X,Y)|> \alpha\cdot \alphasc(m)\cdot \min\set{|X\cap T|,|Y\cap T|}$, which, as observed already, implies that cluster $R$ has the $\alpha$-bandwidth property. We now proceed to describe the second step of the algorithm.

\paragraph{Step 2: Witness Set of Paths.}
In the second step, we attempt to compute a witness set $\pset(R)$ of paths for cluster $R$. If we succeed in doing so, we will move cluster $R$ from $\rset^A$ to $\rset^I$. Otherwise, we will further modify the set $\rset$ of clusters, so that all invariants  continue to hold, and the total budget decreases by at least $1/m$.

We construct the following flow network. Starting from graph $G$, we contract all vertices of $R$ into a source vertex $s$, and we contract all vertices of $V(G)\setminus V(S)$ into a destination vertex $t$. Denote the resulting graph by $H$, and observe that $\delta_H(s)=\delta_G(R)$, and $\delta_H(t)=\delta_G(S)$. We set the capacity $c(e)$ of every edge incident to $s$ to $1$, and the capacity of every other edge in graph $H$ to $100$. We then compute the maximum $s$-$t$ flow $f$ in the resulting flow network.

We consider two cases. The first case is when the value of the flow $f$ is $|\delta_H(s)|$. Since all edge capacities are integral, we can assume that flow $f$ is integral as well. Note that in this case, for every path $P$ connecting $s$ to $t$, either $f(P)=0$ or $f(P)=1$ must hold, as the capacities of all edges incident to $s$ are $1$. Therefore, flow $f$ naturally defines a collection $\pset'(R)$ of $s$-$t$ paths, with $\cong_H(\pset'(R))\leq 100$, where each edge $e\in \delta_G(s)$ serves as the first edge of exactly one such path. Set $\pset'(R)$ of paths then naturally defines a witness  set $\pset(R)=\set{P(e)\mid e\in \delta_G(R)}$ of paths for cluster $R$ in graph $G$, with $\cong_G(\pset(R))\leq 100$, where, for every edge $e\in \delta_G(R)$, path $P(e)$ has $e$ as its first edge and some edge of $\delta_G(S)$ as its last edge, with all inner vertices of $P(e)$ lying in $V(S)\setminus V(R)$. We then move cluster $R$ from $\rset^A$ to $\rset^I$ and terminate the current iteration. It is easy to verify that all invariants continue to hold, and the total budget $B$ does not change.

It remains to consider the second case, where the value of the flow $f$ in $H$ is less than $|\delta_H(s)|$. We compute a minimum $s$-$t$ cut $(A',B')$ in graph $H$, whose value is less than $|\delta_H(s)|$. We partition the set $E(A',B')$ of edges into two subsets: set $E'=E(A',B')\cap \delta_H(s)$, and set $E''=E(A',B')\setminus E'$. Recall that the capacity of every edge in $E'$ is $1$, while the capacity of every edge in $E''$ is $100$. Therefore, $|E'|+100|E''|<|\delta_G(R)|$.

Observe that cut $(A',B')$ in $H$ naturally defines cut $(A,B)$ of graph $S$: we let $A=(A'\setminus \set{s})\cup V(R)$, and $B=V(S)\setminus A$. Notice that $\delta_G(A)= E_H(A',B')$. 

Let $\aset$ denote the set of all connected components of graph $S[A]=G[A]$.
Let $\xset$ denote the set of all clusers $R'\in \rset$ with $R'\cap A\neq \emptyset$. For each such cluster $R'\in \xset$, let $\yset(R')$ be the set of all connected components of $R'\setminus A$. We need the following observation.

\begin{observation}\label{obs: small boundary for new clusters}
For every cluster $C\in \aset$, $|\delta_G(C)|\leq |\delta_G(R)|$. Additionally,
for every cluster $R'\in \xset$, and every cluster $R''\in \yset(R')$, $|\delta_G(R'')|\leq |\delta_G(R')|$.  
\end{observation}
\begin{proof}
	Consider first some cluster $C\in \aset$. Clearly, $\delta_G(C)\subseteq \delta_G(A)\subseteq E_H(A',B')$. Since $|E_{H}(A',B')|<|\delta_H(s)|=|\delta_G(R)|$, we get that  $|\delta_G(C)|\leq |\delta_G(R)|$.
	
	Consider now some cluster $R'\in \xset$. Denote by $R'_A$ the subgraph of $R'$ induced by $V(R')\cap A$, and denote by $R'_B=R'\setminus A$. Let $E_1=\delta_G(R')\cap \delta_G(R'_A)$, $E_2=\delta_G(R')\cap \delta_G(R'_B)$, and $\hat E=E_G(R'_A,R'_B)$. We show below that $|\hat E|\leq |E_1|$ must hold. Assume for now that this is true, and consider any cluster $R''\in \yset(R')$. Since $R''$ is a connected component of $R'_B$, we get that $\delta_G(R'')\subseteq E_2\cup \hat E$. Therefore, if $|\hat E|\leq |E_1|$, then $|\delta_G(R'')|\leq |E_2|+|\hat E|\leq |E_1|+|E_2|=|\delta_G(R')|$ holds.
	
	It now remains to prove that $|\hat E| \leq |E_1|$. Consider the cut $(A',B')$ in graph $H$, and recall that it is a minimum $s$-$t$ cut. From the definition of the cut $(A,B)$, the edges of $\hat E$ belong to the edge set $E_H(A',B')$. Since none of these edges is incident to $s$, the capacity of every edge in $\hat E$ is $100$. Consider now a new $s$-$t$ cut $(A'',B'')$ in graph $H$, where $A''=A'\setminus V(R'_A)$ and $B''=B'\cup V(R'_B)$. Note that the edges of $\hat E$ no longer contribute to this cut, and the only new edges that were added to this cut are the edges of $E_1$, each of which has a capacity that is either $1$ or $100$. Therefore, $\sum_{e\in E_H(A'',B'')}c(e)\leq \sum_{e\in E_H(A',B')}c(e)-100|\hat E|+100|E_1|$. Since $(A',B')$ is a minimum $s$-$t$ cut in graph $H$, $|\hat E|\leq |E_1|$ must hold.
\end{proof}

We perform the following modifications to the sets $\rset,\rset^I$ and $\rset^A$ of clusters. First, we remove cluster $R$ from $\rset$ and from $\rset^A$, and we add every cluster of $\aset$ to both sets instead. Next, we consider every cluster $R'\in \xset$ one by one. We remove each such cluster $R'$ from $\rset$, and we add instead every cluster in $\yset(R')$ to $\rset$. We also remove cluster $R'$ from the cluster set in $\set{\rset^I,\rset^A}$ to which it belongs, and we add every cluster of $\yset(R')$ to  set $\rset^I$. This completes the description of the modification of the sets $\rset,\rset^I,\rset^A$ of clusters. Note that all clusters in $\rset$ remain disjoint, and $\bigcup_{R''\in \rset'}V(R'')=V(S)$ continues to hold. 
Moreover, from \Cref{obs: small boundary for new clusters}, combined with Invariant \ref{inv3: small boundary}, for every cluster $R'\in \rset$, $|\delta_G(R')|\leq |\delta_G(S)|$ continues to hold.
It remains to show that the total budget $B$ decreases by at least $1/m$.

Consider any edge $e\in E(G)\setminus E''$, whose budget, at the end of the current step, is non-zero. Assume first that the budget of $e$ was non-zero at the beginning of the current step. Then, from \Cref{obs: small boundary for new clusters}, the budget of $e$ could not have increased as the result of the current step. If the budget of an edge $e$ was $0$ at the beginning of the current step and is non-zero at the end of the current step, then $e\in E''$ must hold.
Therefore, the only edges whose budget may have increased  as the result of the current step are edges of $E''$. 

For each edge $e\in E''$, its budget may have grown from $0$ to at most $3$, while the number of all such edges is $|E''|<(|\delta_G(R)|-|E'|)/100$. Therefore, the total increase in the budget $B$ due to the edges of   $E''$ is at most $(|\delta_G(R)|-|E'|)/30$. We show that this increase is compensated by the decrease in the budgets of the edges of $\delta_G(R)\setminus E'$. Consider any edge $e\in \delta_G(R)\setminus E'$. Edge $e$ had budget at least $1$ originally, but after the current iteration, since the endpoints of $e$ both lie in $A$, its budget becomes $0$. Therefore, the decrease in the budget $B$ due to the edges of $\delta_G(R)\setminus E'$ is at least $|\delta_G(R)\setminus E'|$. Overall, we get that the decrease in the budget $B$ is at least:
\[ |\delta_G(R)\setminus E'|- (|\delta_G(R)|-|E'|)/30\geq 1/2.\] 
This concludes the description of an iteration. The algorithm terminates when $\rset^I=\emptyset$ holds, at which point we obtain the final set $\rset$ of clusters, together with the witness sets $\set{\pset(R)}_{R\in \rset}$ of paths, that, from the invariants, have all required properties. 
In particular, as observed above, since  $B\geq\sum_{R\in \rset}|\delta_G(R)|$, and $B$ never increases over the course of the algorithm, $\sum_{R\in \rset}|\delta_G(R)|\leq B\leq (1+O(\alpha\cdot \alphasc(m)\cdot \log m))=(1+O(\alpha\cdot \log^{1.5} m))$ holds.
It remains to prove that the algorithm is efficient. Clearly, the algorithm for executing every iteration is efficient. We now show that the number of iterations is bounded by $O(m^3)$. Consider any iteration $i$ of the algorithm. Recall that, as the result of iteration $i$, either the budget $B$ decreased by at least $1/m$ (in which case we say that $i$ is a type-$1$ iteration); or budget $B$ did not change, but the number of clusters in set $\rset^I$ decreases by $1$ (in which case we say that $i$ is a type-$2$ iteration). It is then immediate to see that the number of type-$1$ iterations, over the course of the algorithm, is bounded by $O(m^2)$. Since every cluster of $\rset$ must contain at least one vertex, and $|V(S)|\leq m$ (because $S$ is a connected graph), the number of type-$2$ iterations executed between every consecutive pair of type-$1$ iterations is bounded by $O(m)$. Therefore, the total number of iterations of the algorithm is $O(m^3)$, and so the algorithm is efficient.

\subsection{Proof of \Cref{thm: layered well linked decomposition}}
\label{sec: layered well linked}

Note that by letting $c$ be a large enough constant, we can ensure that  $\alpha < \min\set{\frac 1 {64\alphasc(m)\cdot \log m},\frac 1 {48\log^2 m}}$ holds.

The algorithm starts with layer $\lset_0$ containing a single subgraph of $H$ -- the subgraph $C$, and then performs iterations. The input to iteration $i$ is layers $\lset_0,\lset_1,\ldots,\lset_{i-1}$, each of which is a collection of disjoint clusters of $H$. We ensure that all clusters in $\bigcup_{i'=0}^{i-1}\lset_{i'}$ are mutually disjoint, and each cluster  $W\in \bigcup_{i'=1}^{i-1}\lset_{i'}$  has $\alpha$-bandwidth property. We let $S_i$ be the subgraph of $H$ induced by vertex set $\bigcup_{i'=0}^{i-1}\bigcup_{W\in \lset_{i'}}V(W)$, and we let $E_i=\delta_H(S_i)$. In subsequent iterations, we will create layers $\lset_i,\lset_{i+1},\ldots$, each of which will contain clusters that are disjoint from $S_i$. Notice that, for all $1\leq i'<i$, for every cluster $W\in \lset_{i'}$, the partition of the edges of $\delta_H(W)$ into $\delta^{\down}(W)$ and $\delta^{\up}(W)$ is now settled, since the layers $\lset_0,\ldots,\lset_{i'-1}$ will not undergo any changes in subsequent iterations, and the edges of $E_i\cap \delta_H(W)$ are guaranteed to lie in $\delta^{\up}(W)$. We will ensure that, for all $1\leq i'<i$, every cluster in $\lset_{i'}$ has properties \ref{condition: layered well linked}, \ref{condition: layered decomp each cluster prop} and \ref{condition: layered decomp edge ratio}. We will also ensure that $|E_i|\leq |\delta_H(C)|/2^{i-1}$. The algorithm terminates once $S_i=H$. Since we ensure that for all $i$, $|E_i|\leq |\delta_H(C)|/2^{i-1}$, the number of iterations is bounded by $\log m$.


We now describe the execution of the $i$th iteration. We start by considering the subgraph $S'_i=H\setminus V(S_i)$ of $H$. We apply the algorithm from \Cref{thm:well_linked_decomposition} to graph $H$, its subgraph $S=S'_i$, and the parameter $\alpha=\frac{1}{c\log^{2.5}m}$. As observed already, $\alpha < \min\set{\frac 1 {64\alphasc(m)\cdot\log m},\frac 1 {48\log^2 m}}$ holds. (If  graph $S'_i$ is not connected, then we apply the algorithm to every connected component of $S'_i$ separately). Let $\wset_i$ be the collection of clusters that the algorithm returns.
Recall that we are guaranteed that the sets $\set{V(W)}_{W\in \wset_i}$ of vertices partition $V(S'_i)$, and for each such cluster $W\in \wset_i$,  $|\delta_H(W)|\le |\delta_H(S'_i)|=|\delta_H(S_i)|\leq |\delta_H(C)|$. We are also guaranteed that each cluster $W\in \wset_i$ has the $\alpha$-bandwidth property, and that  $\sum_{W\in \wset_i}|\delta_H(W)|\le |\delta_H(S'_i)|\cdot\left(1+O(\alpha\cdot \log^{3/2} m)\right)=|E_i|\cdot \left(1+O(\alpha\cdot \log^{3/2} m)\right)$.
Since  $\alpha=\frac{1}{c\log^{2.5}m}$ for a large enough constant $c$, we can ensure that $\sum_{W\in \wset_i}|\delta_H(W)|\le |E_i|\cdot\left (1+\frac{1}{1000\log m}\right )$.
%
If there is a cluster $W\in \wset_i$ with $|E_H(W)|<|\delta_H(W)|/(64\log m)$, then remove $W$ from $\wset_i$ and add each of its vertices as a separate cluster to $\wset_i$. Clearly, we have increased the sum $\sum_{W\in \wset_i}|\delta_H(W)|$ by a factor of at most $\left (1+1/(32\log m)\right )$ (since each edge appears in at most two sets of $\set{\delta_H(W)}_{W\in \wset_i}$). Therefore, for the resulting set $\wset_i$, we get that:
 
\[\sum_{W\in \wset_i}|\delta_H(W)|\le |E_i|\cdot\left (1+\frac{1}{1000\log m}\right )\cdot\left (1+\frac{1}{32\log m}\right )\le |E_i|\cdot\left (1+\frac{1}{16\log m}\right).\]
Recall that the algorithm from  \Cref{thm:well_linked_decomposition} also computes, for each cluster $W\in \wset$, a set $\pset'(W)$ of paths in graph $H$ routing the edges of $\delta_H(W)$ to edges of $\delta_H(S'_i)=E_i$, such that the paths of  
$\pset'(W)$ avoid $W$ and cause congestion at most $100$ in $H$.

We partition the set $\wset_i$ of clusters into two subsets: set $\wset_i'$ contains all clusters $W\in \wset_i$, such that $|\delta_H(W)\setminus E_i|< |\delta_H(W)\cap E_i|/\log m$, and set $\wset_i''$ contains all remaining clusters. We then set $\lset_i=\wset_i'$. This finishes the description of the iteration. Recall that we define $S_{i+1}$ to be the subgraph of $H$ induced by the set $V(S_i)\cup \left (\bigcup_{W\in \lset_i}V(W)\right )$ of vertices, and $E_{i+1}=\delta_H(S_{i+1})$. We now analyze the iteration.
First, from the algorithm, every cluster $W\in \lset_i$ satisfies Property \ref{condition: layered decomp each cluster prop}, since the first inequality is guaranteed by  \Cref{thm:well_linked_decomposition}, and we have replaced every cluster that does not satisfies the second inequality of Property \ref{condition: layered decomp each cluster prop} by single-vertex clusters.
We now prove the following claim.

\begin{claim} $|E_{i+1}|\leq |E_i|/2$.
\end{claim}
\begin{proof}
We partition the set $E_{i+1}$ of edges into two subsets. The first set, $E'_{i+1}$, contains all edges of $E_{i+1}$ that lie in the sets  $\set{\delta_H(W)\setminus E_i}_{W\in \wset'_i}$. Since, for each cluster $W\in \wset'_i$, $|\delta_H(W)\setminus E_i|\leq |\delta_H(W)\cap E_i|/\log m$, we get that $|E_{i+1}'|\leq |E_i|/\log m$. The second set, $E''_{i+1}$, contains all remaining  edges of $E_{i+1}$. It is easy to verify that every edge of $E''_{i+1}$ belongs to set $\delta_H(W)\cap E_i$ of edges for some cluster $W\in \wset_i''$.
	
Recall that	$\sum_{W\in \wset_i}|\delta_H(W)|\le  |E_i|\cdot\left (1+\frac{1}{16\log m}\right )$. Therefore, since $E_i\subseteq \bigcup_{W\in \wset_i}\delta_H(W)$, we get that $\sum_{W\in \wset''_i}|\delta_H(W)\setminus E_i|\leq \frac{|E_i|}{16\log m}$. For every cluster $W\in \wset''_i$, $|\delta_H(W)\setminus E_i|\geq \frac{|\delta_H(W)\cap E_i|}{\log m}$ from the definition of set $\wset''_i$. Therefore:
\[|E_{i+1}''|=\sum_{W\in \wset''_i}|\delta_H(W)\cap E_i|\leq (\log m)\cdot \sum_{W\in \wset''_i}|\delta_H(W)\setminus E_i|\leq \frac{|E_i|}{16}.\]
Altogether, $|E_{i+1}|=|E'_{i+1}|+|E''_{i+1}|\leq |E_{i}|/2$.
\end{proof}

Recall that we have already established that, for every cluster $W\in \lset_{i}$, $|\delta_H(W)|\leq |\delta_H(C)|$. Consider any such cluster $W\in \lset_i$. Since layers $\lset_0,\ldots,\lset_i$ will remain unchanged in the remainder of the algorithm, the partition of the edge set $\delta_H(W)$ into  $\delta^{\up}(W)$ and  $\delta^{\down}(W)$ is now settled, and moreover $\delta^{\down}(W)=\delta_H(W)\cap E_i$, while $\delta^{\up}(W)=\delta_H(W)\setminus E_i$. From the definition of cluster set $\wset'_i=\lset_i$, we get that, for every cluster $W\in \lset_i$, $|\delta^{\up}(W)|\leq |\delta^{\down}(W)|/\log m$ holds. Therefore, property \ref{condition: layered decomp edge ratio} holds for every cluster $W\in \lset_i$. Recall that we have already established property \ref{condition: layered well linked} for each such cluster as well.

The algorithm terminates once $S_i=H$. Let $r$ denote the index of the last iteration. Since, for all $i$, $|E_{i+1}|\leq |E_i|/2$, $r\leq \log m$ holds. We let $\wset=\bigcup_{i=1}^r\lset_i$ be the final collection of clusters. We now claim that $(\wset,(\lset_1,\ldots,\lset_r))$ is a valid layered $\alpha$-well-linked decomposition of $H$ with respect to $C$. Note that our algorithm immediately guarantees Property \ref{condition: layered decomposition is partition}, and we have already established Properties \ref{condition: layered well linked} -- \ref{condition: layered decomp edge ratio} of the decomposition. In order to establish property \ref{condition: layered decomposition few edges}, observe that for all $1\leq i\leq r$:
\[\sum_{W\in \lset_i}|\delta_H(W)|=\sum_{W\in \lset_i}(|\delta^{\up}(W)|+|\delta^{\down}(W)|)\leq \sum_{W\in \lset_i}2|\delta^{\down}(W)|= \sum_{W\in \lset_i}2|\delta_H(W)\cap E_i|=2|E_i|.\]
Therefore,
\[\sum_{W\in \wset}|\delta_H(W)|\leq 2 \sum_{1\le i\le r} |E_i|\leq 4|E_1|=4|\delta_H(C)|.\]
We conclude that property \ref{condition: layered decomposition few edges} holds for the decomposition. Lastly, it remains to establish property \ref{condition: layered decomposition routing}. We do so using the following claim.

\begin{claim}\label{claim: routing the frontier}
	For all $0\leq i< r$, there is a collection $\rset_{i+1}=\set{R(e)\mid e\in E_{i+1}}$ of paths such that, for every edge $e\in E_{i+1}$, path $R(e)$ has $e$ as its first edge, and some edge $e'\in E_i$ as its last edge. Moreover, every edge $e'\in E_i$ participates in at most one path of $\rset_{i+1}$, the paths in $\rset_{i+1}$ cause congestion at most $\ceil{1/\alpha}$, and for each path $R(e)\in \rset_{i+1}$, all inner vertices on $R(e)$ lie in $S_{i+1}\setminus S_i$.
\end{claim}

\begin{proof}
	We partition the set $E_{i+1}$ of edges into two subsets: $E'_{i+1}=E_i\cap E_{i+1}$, and $E''_{i+1}=E_{i+1}\setminus E_i$. For each edge $e\in E'_{i+1}$, the path $R(e)$ consists of a single edge -- the edge $e$.
	
	Observe that $E''_{i+1}\subseteq \left (\bigcup_{W\in \lset_{i}}\delta_H(W)\right )\setminus E_i$. For every cluster $W\in \lset_{i}$, we let $\hat E(W)=\delta_H(W)\cap E_{i+1}''$. Clearly, $\hat E(W)\subseteq \delta^{\up}(W)$. Recall that $\delta^{\down}(W)=\delta_H(W)\cap E_i$, and $|\delta^{\down}(W)|>|\delta^{\up}(W)|$. From \Cref{cor: bandwidth_means_boundary_well_linked}, there is a set $\rset(W)$ of paths, that is a one-to-one routing of edges in $\delta^{\up}(W)$ to a subset of edges in $\delta^{\down}(W)$, such that, for each path $Q\in\rset(W)$, all its edges, except for the first and the last,  belong to $E(W)$, and the paths in $\rset(W)$ cause congestion at most $\ceil{1/\alpha}$. For each edge $e\in \hat E(W)\subseteq \delta^{\up}(W)$, we let $R(e)\in \rset(W)$ be the unique path whose first edge is $e$. We have  now defined, for each edge $e\in E_{i+1}$, a path $R(e)$, whose first edge is $e$, last edge lies in $E_{i}$, and all inner vertices lie in $S_{i+1}\setminus S_i=\bigcup_{W\in \lset_i}V(W)$. It is immediate to verify that the resulting set $\rset_{i+1}$ of paths causes congestion at most $\ceil{1/\alpha}$, and that each edge of $E_i$ participates in at most one such path.
\end{proof}

We obtain the following immediate corollary of \Cref{claim: routing the frontier}.

\begin{corollary}\label{cor: routing the frontier}
	For all $0\leq i< r$, there is a collection $\rset'_{i+1}=\set{R'(e)\mid e\in E_{i+1}}$ of paths such that, for every edge $e\in E_{i+1}$, path $R'(e)$ has $e$ as its first edge,  some edge of $\delta_H(C)$ as its last edge, and all its inner vertices are contained in $S_{i+1}\setminus C$. Moreover, every edge in $\delta_H(C)$ may participate in at most one path in $\rset'_{i+1}$, and  the paths in $\rset'_{i+1}$ cause congestion at most $\ceil{1/\alpha}$.
\end{corollary}

\begin{proof}
	The proof is by induction on $i$. For $i=0$, we let the set $\rset'_1$ of paths contain, for each edge $e\in \delta_H(C)=\delta_H(S_1)$, a path $R'(e)$ that consists of a single edge - the edge $e$. 
	
	Assume now that we have defined the sets $\rset'_1,\ldots,\rset'_{i}$ of paths. In order to define the set $\rset'_{i+1}=\set{R'(e)\mid e\in E_{i+1}}$ of paths, consider any edge $e\in E_{i+1}$, and let $R(e)\in \rset_{i+1}$ be the unique path that has $e$ as its first edge. Denote by $e'\in E_i$ the last edge on path $R(e)$, and consider the path $R'(e')\in \rset_i$, connecting $e'$ to an edge of $\delta_H(C)$. We obtain the path $R'(e)$ by concatenating $R(e)$ and $R'(e')$. It is easy to verify that the resulting set $\rset'_{i+1}=\set{R'(e)\mid e\in E_{i+1}}$ of paths has all required properties.  
\end{proof}

We are now ready to establish Property \ref{condition: layered decomposition routing} of the decomposition. Consider some layer $\lset_i$, for $1\leq i\leq r$, and some cluster $W\in \lset_i$.  Recall that, when the algorithm from  \Cref{thm:well_linked_decomposition} was applied to cluster $S'_{i}$ in iteration $i$, it returned a set $\pset'(W)$ of paths in graph $H$, routing the edges of $\delta_H(W)$ to edges of $\delta_H(S'_i)=E_i$, such that the paths of  
$\pset'(W)$ avoid $W$ and cause congestion at most $100$ in $H$. We can assume without loss of generality that, if an edge of $E_i$ lies on a path of $\pset'(W)$, then it is the last edge on that path. Equivalently, no path of $\pset'(W)$ may contain a vertex of $S_i$ as its inner vertex. Consider now some edge $e\in \delta_H(W)$, and let $P'(e)\in \pset'(W)$ be the path whose first edge is $e$. Let $e'\in E_i$ be the last edge on path $P'(e)$, and let $R'(e')$ be the unique path in $\rset'_i$ that has $e'$ as its first edge. Recall that the last edge of $\rset'_i$ lies in $\delta_H(C)$. Moreover, since all inner vertices on path $R'(e')$ lie in $S_i\setminus C$, no inner vertex of path $R'(e')$ may lie in $W$. By concatenating the paths $P'(e)$ and $R'(e')$, we obtain a path $P(e)$, whose first edge is $e$, and last edge lies in $\delta_H(C)$. From the above discussion, no inner vertex of $P(e)$ lies in $W$. We then let $\pset(W)=\set{P(e)\mid e\in \delta_H(W)}$. We have now obtained a set of paths routing the edges of $\delta_H(W)$ to  edges of $\delta_H(C)$, such that the paths in $\pset(W)$ avoid $W$. It now remains to analyze the congestion that this set of paths causes in graph $H$. As the set $\pset'(W)$ of paths causes congestion at most $100$, every edge $e'\in E_i$ may participate in at most   $100$ such paths. Since the congestion caused by the set $\rset'_i$ of paths is at most $\ceil{1/\alpha}$, and since no vertex of $S_i$ may serve as an inner vertex on a path of $\pset'(W)$, the total congestion caused by paths in $\pset(W)$ is at most $100\cdot \ceil{\frac 1 {\alpha}}\leq \frac{200}{\alpha}$.

\subsection{Proof of Claim~\ref{claim: embed expander}}

\label{apd: Proof of embed expander}

For convenience, we sometimes refer to vertices of $T$ as \emph{terminals}.
We use the cut-matching game of Khandekar, Rao and Vazirani~\cite{khandekar2009graph}, defined as follows. The game is played between two players, called the cut player and the matching player. The input to the game is an even integer $N$.  
The game is played in iterations. We start with a graph $W$, whose vertex set $V$ has cardinality $N$, and the edge set is empty. In every iteration, some edges are added to $W$. The game ends when $W$ becomes a $\half$-expander. 
The goal of the cut player is to construct a $\half$-expander in as few iterations as
possible, whereas the goal of the matching player is to prevent the construction of the expander for as long as possible. The iterations proceed as follows.  In every iteration $j$, the cut player chooses a partition $(Z_j, Z'_j)$ of $V$ with $|Z_j| = |Z'_j|$, and the
matching player chooses a perfect matching $M_j$ that matches the nodes of $Z_j$ to the nodes of $Z'_j$.  The edges of $M_j$ are then
added to $W$.  Khandekar, Rao, and Vazirani~\cite{khandekar2009graph} showed that there is an efficient  randomized algorithm for the cut player (that is, an algorithm that, in every iteration $j$, given the current graph $W$, computes a partition $(Z_j,Z'_j)$ of $V$ with $|Z_j| = |Z'_j|$), that guarantees that after
$O(\log^2{N})$ iterations, with high probability, graph $W$ is a $(1/2)$-expander, regardless of the specific matchings chosen by the matching player. 

We use the above cut-matching game in order to compute an expander $W$  with vertex set $T$, and to embed it into $G$, using standard techniques.
If $|T|$ is an even integer, then we start with the graph $W$ containing the vertices of $T$; otherwise, we let $t\in T$ be an arbitrary vertex, and we start with $V(W)=T\setminus\set{t}$. Initially, $E(W)=\emptyset$. We then perform iterations. In the $i$th iteration, we apply the algorithm of the cut player to the current graph $W$, and obtain a partition  $(Z_i,Z'_i)$ of its vertices with $|Z_j| = |Z'_j|$. Using the algorithm from \Cref{thm: bandwidth_means_boundary_well_linked},  we compute a collection $\pset_i$ of paths in graph $G$, routing vertices of $Z_i$ to vertices of $Z'_i$, so that every vertex of $Z_i\cup Z'_i$ is an endpoint of exactly one path in $\pset_i$, and the paths in $\pset_i$ cause congestion at most $\ceil{1/\alpha}\leq 2/\alpha$ in $G$. Let $M_i$ be the perfect matching between vertices of $Z_i$ and vertices of $Z'_i$ defined by the set $\pset_i$ of paths: that is, we add to $M_i$ a pair $(t,t')$ of vertices iff some path in $\pset_i$ has endpoints $t$ and $t'$. We then treat $M_i$ as the response of the matching player, and add the edges of $M_i$ to $W$, completing the current iteration of the game.

Let $W$ be the graph obtained after $i^*=O(\log^2k)$ iterations, that is guaranteed to be a $1/2$-expander with high probability. We then set $\pset=\bigcup_{i=1}^{i^*}\pset_i$. It is immediate to verify that $\pset$ is an embedding of $W$ into $G$. Since each set $\pset_i$ of paths causes congestion $O(1/\alpha)$, and $i^*\leq O(\log^2k)$, the paths in $\pset$ cause congestion $O((\log^2 k)/\alpha)$. If $|T|$ is even, then we have constructed the desired expander and its embedding into $G$ as required. If $|T|$ is odd, then we add the terminal $t$ to the graph $W$. Let $P$ be any path in graph $G$, connecting $t$ to any terminal $t'\in T\setminus\set{t}$; such a path must exist since the set of terminals is $\alpha$-well-linked in $G$. We then add edge $(t,t')$ to graph $W$, and we let its embedding be $P(e)=P$; we add path $P$ to $\pset$. It is easy to verify that this final graph $W$ is $1/4$ expander, provided that the original graph $W$ obtained at the end of the cut-matching game was a $1/2$-expander. We have also obtained an embedding of $W$ into $G$ with congestion $O((\log^2k)/\alpha)$.
Lastly, since the number of iterations in the cut-matching game is $O(\log^2k)$, and the set of edges that is added to $W$ in every iteration is a maching, we get that the maximum vertex degree in $W$ is $O(\log^2k)$.


\subsection{Proof of Observation~\ref{obs: cr of exp}}

\label{apd: Proof of cr of exp}

We assume that $c>2^{120}$ is a large enough constant. Let $\Delta$ denote the maximum vertex degree in $W$. Since $c$ is a large enough constant, we can assume that $\Delta\leq  c^{1/8}\log^2k<k/2^{40}$. Assume for contradiction that $\optcro(W)<k^2/(c\log^8k)$. From \Cref{lem:min_bal_cut}, there is a $(3/4)$-edge-balanced cut $(A,B)$ in $W$, with:
\[|E_W(A,B)|\leq O(\sqrt{\optcro(W)+\Delta\cdot |E(W)|})\leq O(\sqrt{\optcro(W)+c^{1/4}\cdot k\cdot \log^4k})\leq O\left (\frac{k}{ c^{1/4}\log^4k}\right ).\]
(We have used the fact that, since $k>c$, and $c$ is a large enough constant, $\log^4k<\frac{k}{c^{3/4}\log^8k}$.)

Since cut $(A,B)$ is a $(3/4)$-edge-balanced cut, $|E_W(A)|\leq 3|E(W)|/4$. Therefore, $|E_W(B)|\geq |E(W)|/4-|E_W(A,B)|\geq |E(W)|/8$. Since the degree of every vertex in $W$ is at most $\Delta\leq c^{1/8}\log^2k$, we get that $|B|\geq \frac{|E(W)|}{8\Delta}\geq \frac{|E(W)|}{8c^{1/8}\log^2k}$. Using the same reasoning,  $|A|\geq \frac{|E(W)|}{8c^{1/8}\log^2k}$. Since graph $W$ is a $\frac 1 4$-expander, $|E_W(A,B)|\geq \frac 1 4\cdot \min\set{|A|,|B|}\geq  \frac 1 4 \cdot \frac{|E(W)|}{8c^{1/8}\log^2k}>\frac{k}{32c^{1/8}\log^2k}$ must hold, a contradiction.


\subsection{Proof of Corollary~\ref{cor: routing well linked vertex set}}

\label{apd: Proof of routing well linked vertex set}

The proof relies on known results for routing on expanders, that are summarized in the next claim, that is well-known, and follows  immediately from the results of \cite{leighton1999multicommodity}. A proof can be found, e.g. in \cite{chuzhoy2012routing}.

\begin{claim}[Corollary C.2 in \cite{chuzhoy2012routing}]\label{claim: routing on expander}
	There is an efficient randomized algorithm that, given as input an $n$-vertex $\alpha$-expander $H$, and any partial matching $M$ over the vertices of $H$, computes, for every pair $(u,v)\in M$, a path $P(u,v)$ connecting $u$ to $v$ in $H$, such that with high probability, the set $\set{P(u,v)\mid (u,v)\in M}$ of paths causes congestion $O(\log^2 n/\alpha)$ in $H$. 
\end{claim}

We start by computing a graph $W$ with $V(W)=T$, and its embedding $\pset=\set{P(e)\mid e\in E(W)}$ into $G$ with congestion $O((\log^2k)/\alpha)$ using the algorithm from \Cref{claim: embed expander} (recall that, with high probability, $W$ is an $(1/4)$-expander). Next,
we use the algorithm from \Cref{claim: routing on expander} to compute a collection $\rset'=\set{R'(u,v)\mid (u,v)\in M}$ of paths in graph $W$, where for all $(u,v)\in M$, path $R(u,v)$ connects $u$ to $v$ in $W$, such that the congestion of the set $\rset'$ of paths in $W$ is $O(\log^2k)$ with high probability. 

Lastly, we consider the paths $R'(u,v)\in \rset'$ one by one. We transform each such path $R'(u,v)$ into a path $R(u,v)$ connecting $u$ to $v$ in graph $G$ by replacing, for every edge $e\in R'(u,v)$, the edge $e$ with the path $P(e)\in \pset$ embedding the edge $e$ into $G$. Since the paths in $\rset'$ with high probability cause congestion at most $O(\log^2k)$ in $W$, while the paths in $\pset$ cause congestion $O(\log^2k/\alpha)$ in $G$, we get that with high probability, the paths in the resulting set $\rset=\set{R(u,v)\mid (u,v)\in M}$ cause congestion $O(\log^4k/\alpha)$ in $G$.


\subsection{Proof of Corollary~\ref{cor: embed complete graph}}

\label{apd: Proof of embed complete graph}

We partition the set $E(K)$ of edges into $3z$ matchings $M_1,\ldots,M_{3z}$, and then use \Cref{cor: routing well linked vertex set} to compute, for each $1\leq i\leq 3z$, a set $\tilde \rset_i=\set{\tilde P(e)\mid e\in M_i}$ of paths in graph $G$, where for all $e=(t,t')\in M_i$, path $\tilde P(e)$ connects $t$ to $t'$, and with high probability, the paths in $\tilde \pset_i$ cause edge-congestion $O((\log^4z)/\alpha)$ in graph $G$. Let $\tilde \pset=\bigcup_{i=1}^{3z}\tilde \pset_i$. Then $\tilde \pset$ is an embedding of $K_z$ into $G$, and with high probability, the congestion of this embedding is $O((z\log^4z)/\alpha)$.

\subsection{Proof of \Cref{lem: find reordering}}
\label{subsec: compute reordering}

Suppose we are given a graph $G$, and, for every vertex $v\in V(G)$, an oriented circular ordering $(\oset_v,b_v)$ of edges in $\delta_G(v)$. 
We say that a drawing $\phi$ of $G$ \emph{obeys the oriented orderings $\set{(\oset_v,b_v)}_{v\in V}$ at the vertices of $G$} if, for every vertex $v\in V(G)$, the oriented circular order in which the images of the edges of $\delta_G(v)$ enter $v$ in $\phi$ is $(\oset_v,b_v)$. 
We use the following theorem from~\cite{pelsmajer2011crossing}.
\begin{theorem}[Corollary 5.6 of~\cite{pelsmajer2011crossing}]
	\label{thm: compute_reordering_curves}
	There is an efficient algorithm, that, given a two-vertex loopless multigraph $G$ (so $V(G)=\set{v,v'}$ and $E(G)$ only contains parallel edges connecting $v$ to $v'$), and, for each vertex $v\in V(G)$, an oriented ordering  $(\oset_v,b_v)$ of its incident edges, computes a drawing $\phi$ of $G$ that obeys the given oriented orderings, such that $\cro(\phi)$ is at most twice the minimum number of crossings of any drawing of $G$ that obeys the given oriented orderings.
\end{theorem}

The proof of Lemma~\ref{lem: find reordering} easily follows \Cref{thm: compute_reordering_curves}. Recall that we are given a pair $(\oset,b)$, $(\oset',b')$ of oriented orderings of a collection $U$ of elements. We construct a two-vertex loopless graph $G$ with oriented ordering on its vertices, as follows. Denote $U=\set{u_1,\ldots,u_r}$.
The vertex set of $G$ is $\set{v,v'}$. The edge set of $G$ consists of $r$ parallel edges connecting $v$ to $v'$, that we denote by $e_{u_1}, e_{u_2},\ldots e_{u_r}$, respectively. 
The oriented ordering $(\oset,b)$ of the elements of $U$ naturally defines an oriented ordering $(\hat \oset,b)$ of the edges of $G$, and similarly, the oriented ordering $(\oset,b')$ of the elements of $U$ defines an oriented ordering $(\hat \oset,b')$ of the edges of $G$. 
We define the oriented orderings for the vertices of $G$ as follows: $(\oset_v,b_v)=(\hat \oset,-b)$ and $(\oset_{v'},b_{v'})=(\hat \oset',b')$.

Consider any drawing $\phi$ of $G$ on the sphere that obeys the oriented orderings for $v,v'$ defined above. Let $D'=D_{\phi}(v')$ be a tiny $v'$-disc. For all $1\leq i\leq r$, we denote the unique point on the image of edge $e_{u_i}$ that lies on the boundary of $D'$ by by $p'_i$. 
Similarly, we let $\hat D=D_{\phi}(v)$ be a tiny $v$-disc, and, for  all $1\leq i\leq r$, we denote the unique point on the image of edge $e_{u_i}$ that lies on the boundary of $\hat D$ by $p_i$. Let $D$ be the disc whose boundary is the same as the boundary of $\hat D$, but whose interior is disjoint from that of $\hat D$. Then $D'\subseteq D$, and the boundaries of the two discs are disjoint. Furthermore, points $p_1,\ldots,p_r$ appear on the boundary of $D$ according to the oriented ordering $(\oset,b)$, and points $p'_1,\ldots,p'_r$ appear on the boundary of $D'$ according to the oriented ordering $(\oset',b')$.
For all $1\leq i\leq r$, let $\gamma_i$ be the segment of the image of the edge $e_{u_i}$ between points $p_i$ and $p'_i$. Then $\set{\gamma_i\mid 1\leq i\leq r}$ is a set of reordering curves for the orderings $(\oset,b)$ and $(\oset',b')$, and moreover, the cost of this curve set is exactly the number of crossings in $\phi$.

Using a similar reasoning, any set $\Gamma$ of reordering curves for the orderings $(\oset,b)$ and $(\oset',b')$ can be converted into a drawing of graph $G$ that obeys the oriented orderings at vertices $v$ and $v'$, in which the number of crossings is exactly the cost of $\Gamma$. Therefore, there is a drawing of $G$ that obeys the oriented orderings at $v$ and $v'$, whose number of crossings is bounded by $\dist((\oset,b),(\oset',b'))$.  We apply the algorithm from Theorem~\ref{thm: compute_reordering_curves} to graph $G$, and then compute a set of reordering curves from the resulting drawing of $G$ as described above. From the above discussion, the cost of the resulting set of curves is at most $2\cdot \dist((\oset,b),(\oset',b'))$.


\subsection{Proof of \Cref{lem: ordering modification}}

\label{apd: Proof of find reordering}

The proof easily follows from \Cref{lem: find reordering}.
We denote $\delta_G(v)=\set{e_1,\ldots,e_r}$, and, for all $1\leq i\leq r$, we let $p_i$ be the unique point of $\phi(e_i)$ lying on the boundary of the disc $D$. Let $\sigma_i$ be the segment of $\phi(e_i)$ that is disjoint from the interior of the disc $D$. In other words, if $e_i=(v,u_i)$, then $\sigma_i$ is a curve connecting $\phi(u_i)$ to $p_i$. 
Note that the points $p_1,\ldots,p_r$ appear on the boundary of $D$ according to the circular ordering $\oset'_v$ of their corresponding edges.
We assume w.l.o.g. that the orientation of the ordering is $b_v'=-1$.

Let $D'$ be another disc, that is contained in $D$, with $\phi(v)$ lying in the interior of $D'$, such that the boundaries of $D$ and $D'$ are disjoint. 
We place points $p_1',\ldots,p'_r$ on the boundary of the disc $D'$, so that all resulting points are distinct, and they appear on the boundary of $D'$ in the order $\oset_v$ of their corresponding edges, using a positive orientation of the ordering. For all $1\leq i\leq r$, we can compute a simple curve $\gamma_i$, connecting $\phi(v)$ to $p'_i$, such that $\gamma_i$ is contained in $D'$ and only intersects the boundary of $D'$ at its endpoint $p'_i$. We also ensure that all resulting curves $\gamma_1,\ldots,\gamma_r$ are mutually internally disjoint. Using the algorithm from \Cref{lem: find reordering}, we compute a collection $\Gamma=\set{\gamma_1',\ldots,\gamma'_r}$ of {reordering curves}, where for $1\leq i\leq r$, curve $\gamma'_i$ connects $p_i$ to $p'_i$, is contained in $D$, and is disjoint from the interior of $D'$. Note that the total number of crossings between the curves in $\Gamma$ is at most $2\cdot\dist{((\oset_v,1),(\oset'_v,-1))}$. For all $1\leq i\leq r$, we define a new image of the edge $e_i$ to be the concatenation of the curves $\sigma_i,\gamma'_i$, and $\gamma_i$. The images of all remaining edges and vertices of $G$ remain unchanged. Denote the resulting drawing of the graph $G$ by $\phi'$. It is immediate to verify that the edges of $\delta_G(v)$ enter the image of $v$ in the order $\oset_v$ in $\phi'$, and that the drawings of $\phi$ and $\phi'$ are identical except for the segments of the images of the edges in $\delta_G(v)$ that lie inside the disc $D$. It is also immediate to verify that $\cro(\phi')\leq \cro(\phi)+
2\cdot\dist{((\oset_v,1),(\oset'_v,-1))}$.

We  repeat the same algorithm again, only this time the points $p_1',\ldots,p'_r$  are placed on the boundary of disc $D'$ in the order $\oset_v$ of their corresponding edges, using a negative orientation of the ordering. The remainder of the algorithm remains unchanged, and produces a drawing $\phi''$ of $G$. As before, the edges of $\delta_G(v)$ enter the image of $v$ in the order $\oset_v$ in $\phi''$, and  the drawings of $\phi$ and $\phi''$ are identical except for the segments of the images of the edges in $\delta_G(v)$ that lie inside the disc $D$. Moreover, $\cro(\phi'')\leq \cro(\phi)+
2\cdot\dist{((\oset_v,-1),(\oset'_v,-1))}$. Let $\phi^*$ be the drawing with smaller number of crossings, among $\phi'$ and $\phi''$. Our algorithm returns the drawing $\phi^*$ as its final outcome. From the above discussion, $\cro(\phi^*)\leq \cro(\phi)+
2\cdot\dist{(\oset_v,\oset'_v)}$.

\subsection{Proof of Theorem~\ref{thm: type-1 uncrossing}}
\label{apd: type-1 uncrossing}


Let $Z$ be the set of crossings points between the curves of $\Gamma$. For each such crossing point $z\in Z$, we consider a tiny disc $D_z$, that contains the point $z$ in its interior. We select the discs $D_z$ to ensure that all such discs are disjoint, and, moreover, if $z$ is a crossing point between curves $\gamma_1,\gamma_2$, then for every curve $\gamma\in \Gamma\setminus\set{\gamma_1,\gamma_2}$, $\gamma\cap D_z=\emptyset$, while for every curve $\gamma\in \set{\gamma_1\cup \gamma_2}$,  $\gamma\cap D_z$ is a simple open curve whose endpoints lie on the boundary of $D_z$.

\begin{figure}[h]
	\centering
	\subfigure[Before: Curve $\gamma_1$ is shown in blue and curve $\gamma_2$ is shown in red. The disc on the left is $D_z$, and the disc on the right is $D_{z'}$.]{\scalebox{0.1}{\includegraphics{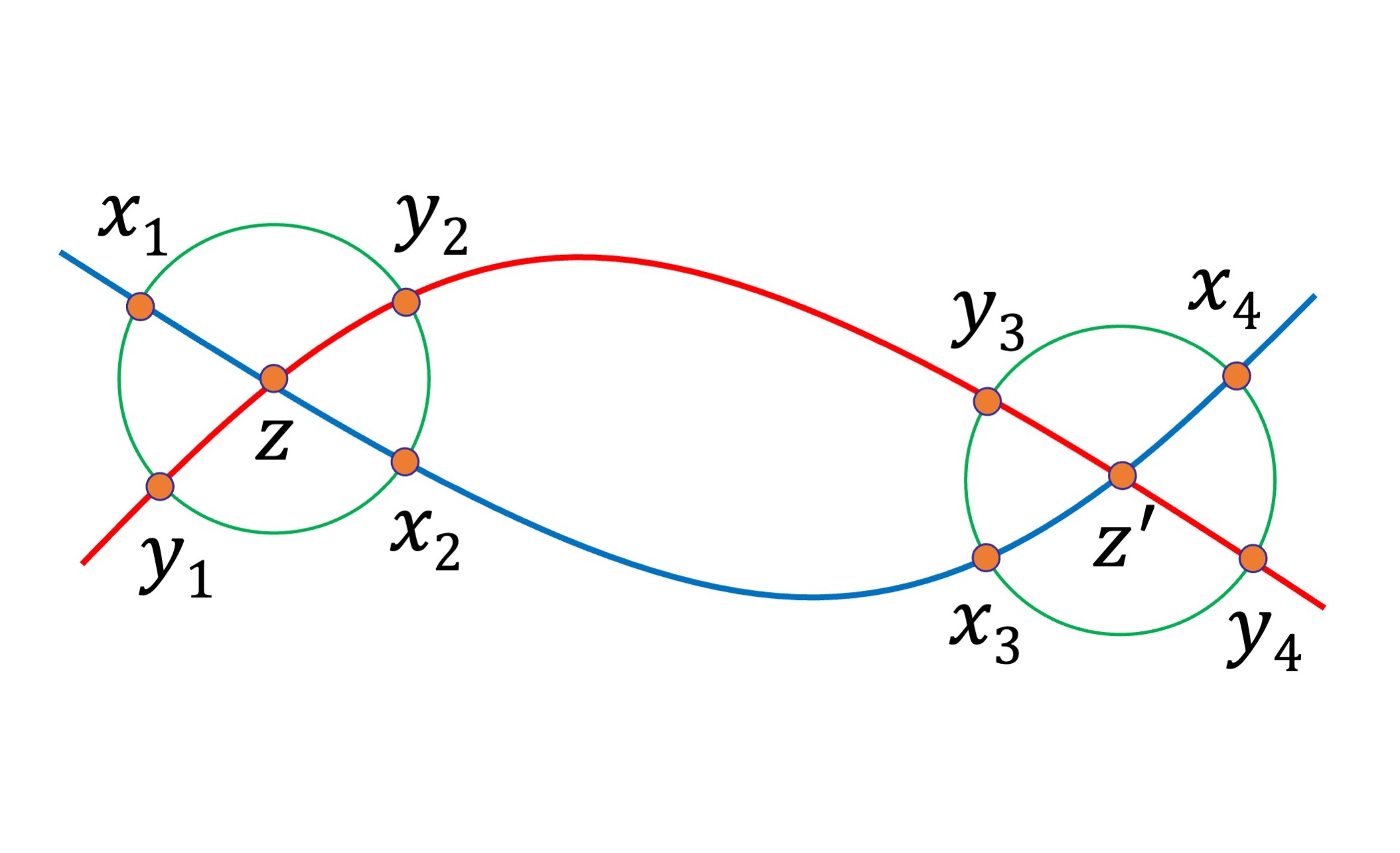}} \label{fig:non_crossing_representation1}
	}
	\hspace{0.5cm}
	\subfigure[After: Curve $\gamma'_1$ is shown in blue and curve $\gamma'_2$ is shown in red.]{
		\scalebox{0.1}{\includegraphics{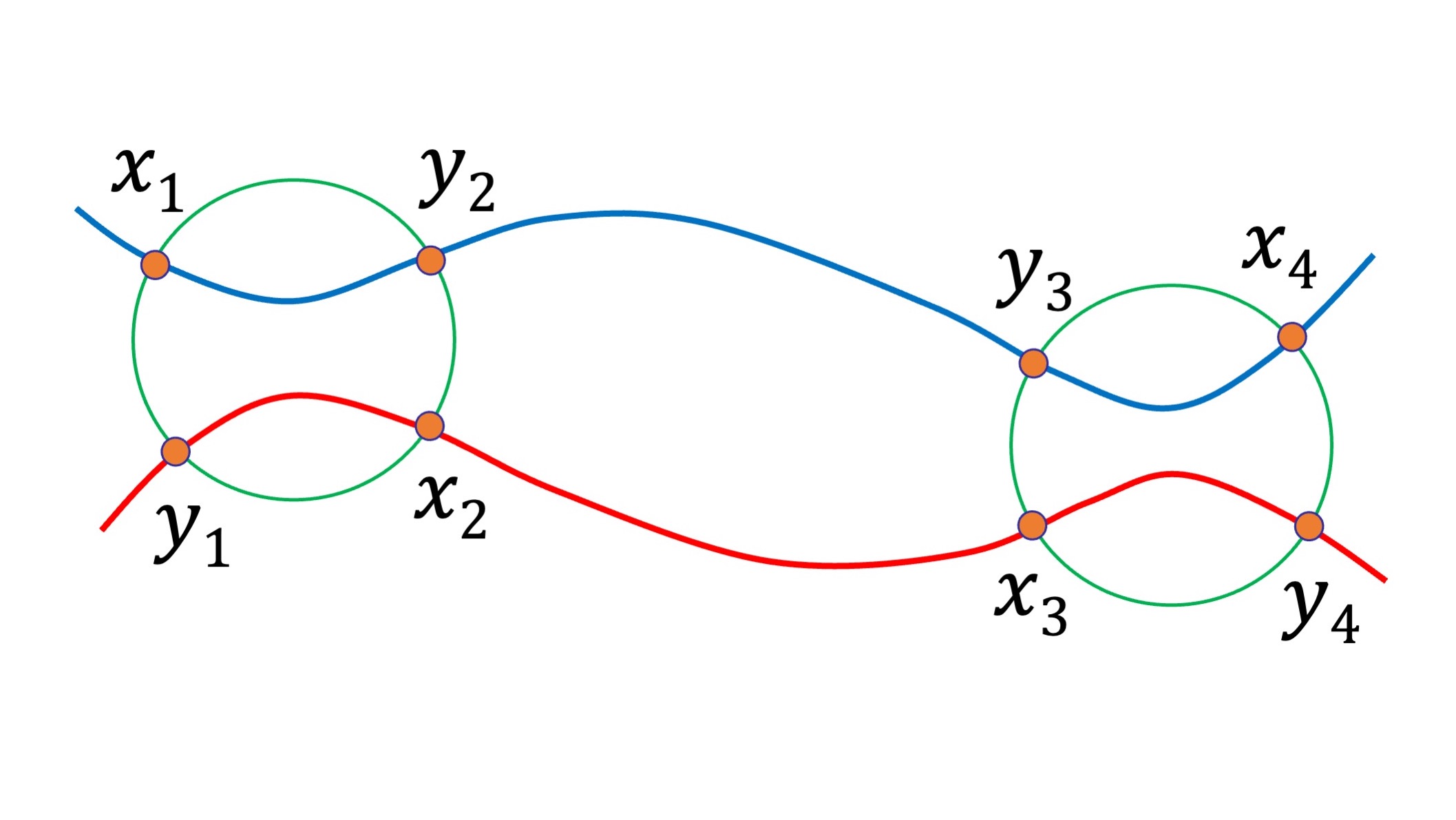}}\label{fig:non_crossing_representation2}}
	\caption{An iteration of the algorithm for performing a type-1 uncrossing.}\label{fig: type_1_uncross_proof}
\end{figure}

We start with $\Gamma_1'=\Gamma_1$, and then iteratively modify the curves in $\Gamma_1'$, as long as there is a pair of distinct curves $\gamma_1,\gamma_2\in \Gamma_1'$ that cross more than once. 
Each iteration is executed as follows. Let $\gamma_1,\gamma_2\in \Gamma_1'$ be a pair of curves that cross more than once, and let  $z,z'$ be two crossing points between $\gamma_1,\gamma_2$, that appear consecutively on $\gamma_1$. In other words, no other point that appears between $z$ and $z'$ on $\gamma_1$ may belong to $\gamma_2$.
We denote by $s_1,t_1$ the endpoints of $\gamma_1$, such that $z$ appears closer to $s_1$ than $z'$ on $\gamma_1$.
Similarly, we denote by $s_2,t_2$ the endpoints of  $\gamma_2$, such that $z$ appears closer to $s_2$ than $z'$ on $\gamma_2$.
We denote by $x_1,x_2$ the two points of $\gamma_1$ that lie on the boundary of disc $D_z$, and denote by $x_3,x_4$ the two points of $\gamma_1$ lying on the boundary of disc $D_{z'}$,
such that the points $x_1,z,x_2,x_3,z',x_4$ appear on $\gamma_1$ in this order.
We define points $y_1,y_2,y_3,y_4$ on $\gamma_2$ similarly (see \Cref{fig:non_crossing_representation1}).

In order to execute the iteration, we slightly modify the curves $\gamma_1, \gamma_2$, by ``swapping'' their segments between points $x_2,x_3$ and $y_2,y_3$, respectively, and slightly nudging them inside the discs $D_z,D_{z'}$, as show in \Cref{fig: type_1_uncross_proof}. We now describe the construction of the new curves $\gamma_1',\gamma_2'$ more formally.
Note that the points $x_1,x_2,y_1,y_2$ appear on the boundary of $D_z$ clockwise in either the order  $(x_1,y_1,x_2,y_2)$ or the order $(x_1,y_2,x_2,y_1)$.
Therefore, we can find two disjoint simple curves $\eta_1$ and $\eta_2$ that are contained in disc $D_z$, with $\eta_1$ connecting $x_1$ to $y_2$, and $\eta_2$ connecting $y_1$ to $x_2$.
Similarly, we compute two disjoint simple curves curves $\eta'_1,\eta'_2$, that are contained in disc $D_{z'}$, with $\eta_1'$ connecting $y_3$ to $x_4$, and $\eta_2'$ connecting $x_3$ to $y_4$ (see \Cref{fig:non_crossing_representation2}). 

We let $\gamma_1'$ be a curve, that is constructed by concatenating the following five curves: (1) the segment of $\gamma_1$ from $s_1$ to $x_1$; (2) curve $\eta_1$; (3) the segment of $\gamma_2$ from $y_2$ to $y_3$; (4) curve $\eta'_1$; and (5) the segment of $\gamma_1$ from $x_4$ to $t_1$.  Similarly, let $\gamma_2'$ be a  curve, that is constructed by concatenating the following five curves: (1) the segment of $\gamma_2$ from $s_2$ to $y_1$; (2) curve $\eta_2$; (3) the segment of $\gamma_1$ from $x_2$ to $x_3$; (4) curve $\eta'_2$; and (5) the segment of $\gamma_2$ from $y_4$ to $t_2$. We then remove any self loops from the two curves, to obtain the final curves $\gamma'_1,\gamma_2'$, that replace the curves $\gamma_1,\gamma_2$ in $\Gamma_1'$. Note that $\gamma'_1$ has the same endpoints as $\gamma_1$, and the same is true for $\gamma'_2$ and $\gamma_2$. It is also easy to verify that, at the end of the iteration, the number of crossings between the curves of $\Gamma_1'\cup \Gamma_2$ strictly decreases, and the number of crossings between the curves of $\Gamma_1'$ and the curves of $\Gamma_2$, that we denoted by $\chi(\Gamma_1',\Gamma_2)$, does not grow. Moreover, for every curve $\gamma\in \Gamma_2$, the number of crossings of $\gamma$ with the curves in $\Gamma_1'$ may not grow either. Once the algorithm terminates, we obtain the desired set $\Gamma_1'$ of curves, in which every pair of distinct curves crosses at most once. From the above discussion, it is immediate to verify that the curves in $\Gamma_1'$ have all required properties. Since the curves in $\Gamma$ are in general position, the number of iterations is bounded by the number of crossing points between the curves.


\subsection{Proof of \Cref{claim: curves in a disc}}
\label{apd: Proof of curves in a disc}

We start by constructing a collection $\Gamma'=\set{\gamma'_i\mid 1\leq i\leq k}$ of curves that are in general position, such that for all $1\leq i\leq k$, curve $\gamma'_i$ has $s_i$ and $t_i$ as its endpoints, and is contained in disc $D$. In order to construct the set $\Gamma'$ of curves, we let $p$ be any point in the interior of the disc $D$, and $r>0$ be some real number, such that a radius-$r$ circle centered at point $p$ is contained in the disc $D$. For all $1\leq i\leq k$, let $\ell_i$ be a straight line, connecting point $s_i$ to $p$, and $\ell'_i$ a straight line, connecting point $t_i$ to $p$. We can assume that both lines are contained in the disc $D$, by stretching the disc as needed. For all $1\leq i\leq k$, we choose a radius $0<r_i<r$, so that $0<r_1<\cdots<r_k<r$ holds. For an index $1\leq i\leq k$, we let $C_i$ be the boundary of a radius-$r_i$ circle centered at point $p$, and we let $q_i,q'_i$ be the points on lines $\ell_i$ and $\ell'_i$, respectively, that lie on $C_i$. We let curve $\gamma'_i$ be a concatenation of three curves: the segment of $\ell_i$ from $s_i$ to $q_i$; a segment of $C_i$ between $q_i$ and $q'_i$; and the segment of $\ell'_i$ from $q'_i$ to $t_i$. Consider the resulting set 
 $\Gamma'=\set{\gamma'_i\mid 1\leq i\leq k}$ of curves. Clearly, for all $1\leq i\leq k$, curve $\gamma'_i$ has $s_i$ and $t_i$ as its endpoints, and is contained in disc $D$. 
 It is also easy to verify that curves of $\Gamma'$ are in general position. 
 
Next, we use the algorithm from \Cref{thm: type-1 uncrossing} to perform a type-1 uncrossing of the curves in $\Gamma'$. Specifically, we set $\Gamma_1=\Gamma'$ and $\Gamma_2=\emptyset$. We denote by $\Gamma=\Gamma_1'=\set{\gamma_i\mid 1\leq i\leq k}$ the set of curves that the algorithm outputs. Recall that, for all $1\leq i\leq k$, curve $\gamma_i$ has $s_i$ and $t_i$ as its endpoints; the curves in $\Gamma$ are in general position; and every pair of curves in $\Gamma$ cross at most once. From the description of the type-1 uncrossing operation, it is easy to verify that all curves in $\Gamma$ are contained in the disc $D$.  

Consider now two pairs $(s_i,t_i),(s_j,t_j)$ of points, with $i\neq j$. Note that curve $\gamma_i$ partitions the disc $D$ into two regions, that we denote by $F$ and $F'$. If the two pairs $(s_i,t_i),(s_j,t_j)$ cross, then $s_j,t_j$ may not lie on the boundary of the same region, and so curve $\gamma_j$ must cross curve $\gamma_i$ exactly once. If the two pairs do not cross, then $s_j,t_j$ either both lie on the boundary of $F$, or they both lie on the boundary of $F'$. It is then impossible that curves $\gamma_i,\gamma_j$ cross exactly once, and, since every pair of curves cross at most once, they cannot cross.

\subsection{Proof of Theorem~\ref{thm: new type 2 uncrossing}}
\label{apd: new type 2 uncrossing}

We start with an initial set $\Gamma'=\set{\gamma'(Q)\mid Q\in \qset}$ of curves, where, for each path $Q\in \qset$, $\gamma'(Q)$ is the image of the path $Q$ in $\phi$. In other words, $\gamma'(Q)$ is the concatenation of the images of the edges of $Q$ in $\phi$. Note however that the resulting set $\Gamma'$ of curves may not be in general position. This is since a vertex $v\in V(G)$ may serve as an inner vertex on more than two paths of $\qset$, and in such a case its image $\phi(v)$ serves as an inner point of more than two curves in $\Gamma'$.
Let $V'\subseteq V(G)$ be the set of all vertices $v\in V(G)$, such that more than two paths in $\qset$ contain $v$.

In our first step, we transform the set $\Gamma'$ of curves so that the resulting curves are in general position, while ensuring that the endpoints of each curve $\gamma'(Q)$ remain unchanged, and each such curve $\gamma'(Q)$ remains aligned with the graph $\bigcup_{Q'\in \qset}Q'$. We do so by performing a \emph{nudging operation} around every vertex $v\in V'$, as follows.

Consider any vertex $v\in V'$, and let $\qset(v)\subseteq \qset$ be the set of all paths containing vertex $v$. Note that $v$ must be an inner vertex on each such path. For convenience, we denote $\qset(v)=\set{Q_1,\ldots,Q_z}$. Consider the tiny $v$-disc $D=D_{\phi}(v)$. For all $1\leq i\leq z$, denote by $a_i$ and $b_i$ the two points on curve $\gamma(Q_i)$ that lie on the boundary of disc $D$. We use the algorithm from \Cref{claim: curves in a disc} to compute a collection $\set{\sigma_1,\ldots,\sigma_z}$ of curves, such that, for all $1\leq i\leq z$, curve $\sigma_i$ connects $a_i$ to $b_i$, and the interior of the curve is contained in the interior of $D$. Recall that every pair of resulting curves crosses at most once, and every point in the interior of $D$ may be contained in at most two curves.
For all $1\leq i\leq z$, we modify the curve $\gamma(Q_i)$, by replacing the segment of the curve that is contained in disc $D$ with $\sigma_i$.
Once every vertex $v\in V'$ is processed, we obtain a collection $\Gamma''=\set{\gamma''(Q)\mid Q\in \qset}$ of curves, where for every path $Q\in \qset$, curve $\gamma''(Q)$ connects $\phi(s(Q))$ to $\phi(t(Q))$. Moreover, it is easy to verify that each resulting curve $\gamma''(Q)\in \Gamma''$ is aligned with the drawing of the graph $\bigcup_{Q'\in \qset}Q'$ induced by $\phi$, and that the curves in $\Gamma''$ are in general position.
 
We let $S$ be the multiset of points that contains, for every curve $\gamma''(Q)\in \Gamma''$ its first endpoint $\phi(s(Q))$, and we let $T$ be the multiset of points contianing the last endpoint of each such curve.

We initially let, for each path $Q\in \qset$, $\gamma(Q)$ be the curve obtained by deleting all loops from $\gamma''(Q)$, and we denote by $\Gamma=\set{\gamma(Q)\mid Q\in \qset}$ the resulting set of curves. We gradually modify the curves in $\Gamma$ in order to eliminate all crossings between them. Throughout the algorithm, we ensure that for each path $Q\in \qset$, curve $\gamma(Q)$ originates at point $\phi(s(Q))$, and moreover, if $e_1(Q)$ is the first edge on path $Q$, then segment $\phi(e_1(Q))\cap D_{\phi}(s(Q))$ is contained in $\gamma(Q)$. We also ensure that the multiset containing the last point on every curve of $\Gamma$ remains unchanged throughout the algorithm. Let $P$ be the collection of all points $p$, such that at least two curves of $\Gamma$ contain $p$ as an inner point. 
We perform iterations, as long as $P\neq\emptyset$. Each iteration is executed as follows. Let $p\in P$ be any point, and let $Q,Q'\in \qset$ be two paths whose corresponding curves $\gamma(Q),\gamma(Q')$ contain the point $p$. Let $x,x'$ be the two points of $\gamma(Q)$ that lie on the boundary of the tiny $p$-disc $D(p)$, with $x$ appearing before $x'$ on $\gamma(Q)$. Let $y,y'$ be the two points of $\gamma(Q')$, that lie on the boundary of $D(p)$, with $y$ appearing before $y'$ on $\gamma(Q')$. We now consider two cases. In the first case, the circular clock-wise ordering of points $x,x',y,y'$ on the boundary of $D(p)$ is either $(x,y,x',y')$, or $(x,y',x',y)$. In this case, the two pairs $(x,y')$ and $(y,x')$ of points on the boundary of $D(p)$ do not cross. Therefore, from \Cref{claim: curves in a disc}, we can construct two disjoint curves $\sigma,\sigma'$ that are contained in $D(p)$, with $\sigma$ connecting $x$ to $y'$ and $\sigma'$ connecting $y$ to $x'$. We let $\gamma'(Q)$ be a curve that is obtained by concatenating the segment of $\gamma(Q)$ from its first endpoint to $x$; the curve $\sigma$; and the segment of $\gamma(Q')$ from $y'$ to its last endpoint. Similarly, we let $\gamma'(Q')$ be a curve that is obtained by concatenating the segment of $\gamma(Q')$ from its first endpoint to $y$; the curve $\sigma'$; and the segment of $\gamma(Q)$ from $x'$ to its last endpoint. We then replace $\gamma(Q)$ with $\gamma'(Q)$ and $\gamma(Q')$ with $\gamma'(Q')$ in $\Gamma$. In the second case, the circular clock-wise ordering of points $x,x',y,y'$ on the boundary of $D(p)$ must be either $(x,x',y,y')$, or $(x',x,y,y')$, or $(x,x',y',y)$, or $(x',x,y',y)$. In either case,  the two pairs $(x,x')$ and $(y,y')$ of points on the boundary of $D(p)$ do not cross. Therefore, from \Cref{claim: curves in a disc}, we can construct two disjoint curves $\sigma,\sigma'$ that are contained in $D(p)$, with $\sigma$ connecting $x$ to $x'$ and $\sigma'$ connecting $y$ to $y'$. We modify curve $\gamma(Q)$ by replacing its segment that is contained in $D(p)$ with $\sigma$, and we similarly replace the segment of $\gamma(Q')$ that is contained in $D(p)$ with $\sigma'$. If either of the new curves $\gamma(Q),\gamma(Q')$ has loops, we turn the corresponding curve into a simple one by removing all loops from it. We also update the set $P$ of points, by removing from it points that no longer belong to two curves in $\Gamma$.  This finishes the description of an iteration. It is easy to verify that no new crossing points between curves in $\Gamma$ are created, the curves in $\Gamma$ remain in general position, and each such curve is aligned with the drawing of the graph $\bigcup_{Q''\in \qset}Q''$ induced by $\phi$. It is also easy to verify that the invariants continue to hold.

Consider the set $\Gamma$ of curves that we obtain at the end of the algorithm. Clearly, the curves in $\Gamma$ do not cross with each other, and they are aligned with the drawing of the graph $\bigcup_{Q\in \qset}Q$ induced by $\phi$. It is also immediate to verify that they have all remaining required properties.
Since $|P|$ strictly decreases from iteration to iteration, the number of iterations is bounded by the number of crossings of the drawing $\phi$ of $G$, which is in turn bounded by the input size. Each iteration can be executed in time polynomial in the input size, so the algorithm is efficient.


\subsection{Proof of \Cref{cor: new type 2 uncrossing}}
\label{apd: cor new type 2 uncrossing}

For each edge $e\in E(G)\setminus E(C)$, we let $n_e=\cong_G(\qset,e)$. Let $H$ be a new graph, with $V(H)=V(G)$, whose edge set consists of the set $E(C)$ of edges, and, for each edge $e\in E(G)\setminus E(C)$, a set $J(e)$ of $n_e$ parallel copies of the edge $e$. Note that the drawing $\phi$ of graph $G$ naturally defines a drawing $\phi'$ of graph $H$. In order to obtain the drawing $\phi'$ of $H$, we start with the drawing $\phi$ of $G$, and then, for every edge $e\in E(G)\setminus E(C)$ with $n_e>0$, we draw the edges of $J(e)$ in parallel to $\phi(e)$, very close to it. We also delete the images of all edges $e\in E(G)\setminus E(C)$ with $n_e=0$. Note that, for every edge $e\in E(C)$, the number of crossings between $\phi'(e)$ and the images of the edges of $E(H)\setminus E(C)$ in the drawing $\phi'$ is at most $\sum_{e'\in  E(G)\setminus E(C)}\chi(e,e')\cdot \cong_G(\qset,e')$, where $\chi(e,e')$ is the number of crossings between $\phi(e)$ and $\phi(e')$.

The set $\qset$ of paths in graph $G$ naturally defines a set $\qset' $ of edge-disjoint paths in graph $H$, where, for each edge $e\in E(G)\setminus E(C)$, for every path $Q\in \qset$ containing the edge $e$, we replace $e$ with a distinct edge of $J(e)$ on path $Q$. In particular, the multisets $S(\qset)$, $S(\qset')$ of vertices containing the first vertex of every path in set $\qset$ and $\qset'$, respectively, remain unchanged, and the same is true regarding the multisets $T(\qset),T(\qset')$ of paths, containing the last vertex of every path in set $\qset$ and $\qset'$, respectively. 

We apply the algorithm from \Cref{thm: new type 2 uncrossing} to graph $H$, the drawing $\phi'$ of $H$, and the set $\qset'$ of edge-disjoint paths in $H$. Let $\Gamma'=\set{\gamma'(Q')\mid Q'\in \qset'}$ be the resulting set of curves. Recall that, for every path $Q\in \qset$, there is a distinct path $Q'\in \qset'$, that is obtained from $Q$ by replacing each edge $e\in E(Q)$ with one of its copies. For each path $Q\in \qset$, we then let $\gamma(Q)=\gamma'(Q')$, and we consider the resulting set  $\Gamma=\set{\gamma(Q)\mid Q\in \qset}$ of curves.  The algorithm from \Cref{thm: new type 2 uncrossing}  ensures that the curves in $\Gamma$ do not cross each other, and that,  for every path $Q\in \qset$, $s(\gamma(Q))=\phi(s(Q))$. It also guarantees that the multiset $T(\Gamma)$ is precisely the multiset $\set{\phi(t(Q))\mid Q\in \qset}$.
Lastly, consider any edge $e\in E(C)$. Since the curves in set $\Gamma'$ are aligned with the drawing of the graph $\bigcup_{Q'\in \qset'}Q'$ induced by $\phi'$, the number of crossings between $\phi'(e)=\phi(e)$ and the curves in set $\Gamma'=\Gamma$ is bounded by the number of crossings between $\phi'(e)$ and the images of the edges of $E(H)\setminus E(C)$ in drawing $\phi'$ of $H$, which is, in turn, bounded by 
$\sum_{e'\in E(G)\setminus E(C)}\chi(e,e')\cdot \cong_G(\qset,e')$.

\subsection{Proof of Claim~\ref{clm: contracted_graph_well_linkedness}}

\label{apd: Proof of contracted_graph_well_linkedness}

Consider any $T$-cut $(A,B)$ in graph $G$, and denote $T_A=T\cap A$ and $T_B=T\cap B$. Assume without loss of generality that $|T_A|\le |T_B|$. It is enough to show that $|E_G(A,B)|\ge (\alpha_1\alpha_2)\cdot |T_A|$. Assume for contradiction that this is not the case.

Denote $H=G_{|\cset}$.  We partition the set $V(H)$ of vertices into two subsets: set $V'=V(H)\cap V(G)$ of regular vertices, and set $V''=\set{v_C\mid C\in \cset}$ of supernodes.
Note that $T\subseteq V'$ must hold. We use the cut $(A,B)$ in $G$, in order to construct a cut $(A',B')$ in graph $H$, with $A'\cap T=T_A$ and $B'\cap T=T_B$, such that $|E_H(A',B')|<\alpha_2\cdot |T_A|$, contradicting the fact that vertex set $T$ is $\alpha_2$-well-linked in graph $G_C$.

In order to construct the cut $(A',B')$ in $H$, we first process every vertex $v\in V'$ one by one. For each such vertex $v$, if $v\in A$, then we add $v$ to $A'$, and otherwise we add it to $B'$. Notice that this process guarantees that $T_A\subseteq A'$ and $T_B\subseteq B'$.

Next, we process every cluster $C\in \cset$ one by one. Notice that partition $(A,B)$ of $V(G)$ naturally defines a partition $(A_C,B_C)$ of $V(C)$, where $A_C=A\cap V(C)$ and $B_C=B\cap V(C)$. We denote $E'_C=E_G(A_C,B_C)$, $E_1(C)=\delta_G(A_C)\setminus E'_C$, and $E_2(C)=\delta_G(B_C)\setminus E'_C$. If $|E_1(C)|\leq |E_2(C)|$, then we add supernode $v_C$ to $B'$, and otherwise we add it to $A'$. Assume w.l.o.g. that $v_C$ was added to $B'$, so $|E_1(C)|\leq |E_2(C)|$ holds. From the $\alpha_1$-bandwidth property of $C$, we get that $|E'_C|\geq \alpha_1\cdot |E_1(C)|$. Notice that the edges of $E'_C$ lie in the cut $(A,B)$ in graph $G$, but they do not contribute to the cut $(A',B')$ in graph $H$. On the other hand, edges of $E_1(C)$ may lie in $E_H(A',B')\setminus E_G(A,B)$. We \emph{charge} the edges of $E_1(C)$ to the edges of $E'$. Since $|E'|\geq \alpha_1\cdot |E_1(C)|$, every edge of $E'$ pays at most $1/\alpha_1$ units for the edges of $E_1(C)$, so the total charge to the edges of $E'$ is $|E_1(C)|$. 

Once every cluster $C\in \cset$ is processed, we obtain the final cut $(A',B')$ in graph $H$. For every edge $e\in E_H(A',B')$, either $e\in E_G(A,B)$, or $e$ is charged to some edges of $E_G(A,B)\setminus E_H(A',B')$. Since the charge to every edge of $E_G(A,B)\setminus E_H(A',B')$ is at most $1/\alpha_1$, we get that $|E_H(A',B')|\leq |E_G(A,B)|/\alpha_1$. Since we have assumed that $|E_G(A,B)|< (\alpha_1\alpha_2)\cdot |T_A|$, we get that $|E_H(A',B')|<\alpha_2\cdot |T_A|=\alpha_2\cdot |T\cap A'|$, contradicting the fact that vertex set $T$ is $\alpha_2$-well-linked in $H$.


\subsection{Proof of \Cref{cor: contracted_graph_well_linkedness}}

\label{apd: Proof of cor contracted_graph_well_linkedness}

Let $G^+$ be the graph obtained from $G$ by subividing each edge $e\in \delta_G(R)$ with a new vertex $t_e$. 
Denote $T=\set{t_e\mid e\in \delta_G(R)}$.
Recall that the augmentation $R^+$ of cluster $R$ in $G$ is defined to be the subgraph of $G^+$ induced by vertex set $V(R)\cup T$.
It is immediate to verify that every cluster $C\in \cset$ has the $\alpha_1$-bandwidth property in graph $R^+$. Furthermore, from \Cref{obs: wl-bw}, the set $T$ of vertices is $\alpha_2$-well-linked in graph $R^+_{|\cset}$. By applying \Cref{clm: contracted_graph_well_linkedness} to graph $R^+$, vertex set $T$ and collection $\cset$ of clusters, we get that $T$ is $(\alpha_1\cdot\alpha_2)$-well-linked in graph $R^+$. From \Cref{obs: wl-bw}, cluster $R$ has the $(\alpha_1\cdot\alpha_2)$-bandwidth property in $G$.


\subsection{Proof of \Cref{claim: routing in contracted graph}}
\label{apx: contracted graph routing}

For convenience, we denote $|T|=k$.
We assume w.l.o.g. that the paths in $\pset$ are simple, and we direct each such path towards $x$. We then graduately modify the paths in $\pset$, by processing the clusters of $\cset$ one by one. 

Consider any cluster $C\in \cset$, and let $\pset(C)\subseteq \pset$ be the subset of paths that contain the supernode $v_C$. For each path $P\in \pset(C)$, let $e_P(C)$ and $e_P'(C)$ denote the edges appearing immediately before and immediately after $v_C$ on $P$. We denote $E_1(C)=\set{e_P(C)\mid P\in \pset(C)}$ and $E_2(C)=\set{e'_P(C)\mid P\in \pset(C)}$. We use the algorithm from \Cref{cor: bandwidth_means_boundary_well_linked}, to compute a collection $\rset(C)$  of paths that is a one-to-one routing of the edges of $E_1(C)$ to the edges of $E_2(C)$, such that all inner vertices on the paths of $\rset(C)$ lie in $C$, and every edge in $E(C)$ participates in at most $\ceil{1/\alpha}$ such paths. We modify the paths in set $\pset(C)$ as follows. First, for each path $P\in \pset(C)$, we delete the vertex $v_C$ from $P$, together with its two incident edges. Let $P_1,P_2$ be the two resulting subpaths of $P$. We then let $\pset_1(C)=\set{P_1\mid P\in \pset(C)}$, and $\pset_2(C)=\set{P_2\mid P\in \pset(C)}$. Lastly, let $\pset'(C)$ be the set of paths obtained by concatenating the paths in $\pset_1(C),\rset(C)$ and $\pset_2(C)$. We  delete from $\pset$ the paths that belong to $\pset(C)$, and add the paths of $\pset'(C)$ instead. It is easy to verify that the resulting set $\pset$ of paths still routes the vertices of $T$ to $x$.

Once we process every cluster $C\in \cset$, we obtain a collection $\pset'$ of $k$ paths, routing the vertices of $T$ to  vertex $x$ in graph $G$. Since the paths of $\pset$ at the beginning of the algorithms are edge-disjoint, for each edge $e\in E(G)\setminus\left( \bigcup_{C\in \cset}E(C)\right )$, $\cong_G(\pset',e)\le 1$. From our construction, for each edge $e\in \bigcup_{C\in \cset}E(C)$, $\cong_{G}(\pset',e)\leq \ceil{1/\alpha}\leq 2/\alpha$. Lastly, we apply the algorithm from \Cref{claim: remove congestion}, to graph $G$ and the set $\pset'$ of paths, to obtain  a collection $\pset''$ of at least $\alpha k/2$ edge-disjoint paths in graph $G$, where each path in $\pset''$ connects a distinct vertex of $T$ to $x$. 


\subsection{Proof of \Cref{lem: crossings in contr graph}}
\label{apd: Proof of crossings in contr graph}

Let $\phi^*$ be an optimal solution to instance $I$ of \cnwrs. Let $G'$ be the graph that is obtained from $G$ by subdividing every edge $e\in \bigcup_{C\in \cset}\delta_G(C)$ with a vertex $t_e$, and let $T=\set{t_e\mid e\in \bigcup_{C\in \cset}\delta_G(C)}$ be the resulting set of new vertices. For every cluster $C\in \cset$, we denote by $T_C=\set{t_e\mid e\in\delta_G(C)}$, and we let $C^+$ be the subgraph of $G'$ induced by vertex set $V(C)\cup T_C$. From \Cref{obs: wl-bw}, vertex set $T_C$ is $\alpha$-well-linked in $C^+$. Observe that drawing $\phi^*$ of $G$ naturally defines a drawing $\phi'$ of graph $G'$, with $\cro(\phi')=\cro(\phi^*)$.
We denote $\cset=\set{C_1,\ldots,C_r}$, where the clusters are indexed arbitrarily. For $1\leq i\leq r$, we let $\cset_i=\set{C_1,\ldots,C_i}$, and we let $G'_i=G'_{|\cset_i}$. We also denote $G'_0=G'$. We perform $r$ iterations. The input to the $i$th iteration is a drawing $\phi'_{i-1}$ of the graph $G'_{i-1}$, and the output is a drawing $\phi'_i$ of the graph $G'_i$. We set $\phi'_0=\phi'$.

We now describe the $i$th iteration, for $1\leq i\leq r$. For convenience, we denote $C_i=C$.
\Cref{cor: simple guiding paths} guarantees that there is a distribution $\dset(C)$ over the set $\Lambda(C)$ of internal $C$-routers, such that, for every edge $e\in E(C)$, $\expect[\qset(C)\sim \dset(C)]{\cong(\qset(C),e)}\leq  O((\log |\delta_G(C)|)^4/\alpha)\leq O((\log^4m)/\alpha)$. We let $\qset(C)=\set{Q(e)\mid e\in \delta_G(C)}$ be an internal $C$-router sampled from the distribution $\dset(C)$, and we denote by $u(C)$ the center of the router $\qset(C)$. Note that the set $\qset(C)$ of paths in graph $G$ naturally defines a set of paths in graph $C^+$, routing the vertices of $T_C$ to vertex $u(C)$. Abusing the notation, we denote this set of paths by $\qset(C)$ as well.

Applying the algorithm from \Cref{cor: new type 2 uncrossing} to graph $G'_{i-1}$, its drawing $\phi'_{i-1}$, subgraph $G'_{i-1}\setminus C^+$, and the set $\qset(C)$ of paths, we obtain a collection $\Gamma(C)=\set{\gamma(e)\mid e\in \delta_{G'_{i-1}}(C)}$ of curves, such that, for every edge $e\in \delta_{G'_{i-1}}(C)$, curve $\gamma(e)$ originates at the image of the endpoint of $e$ that lies in $T_C$, and terminates at the image of $u(C)$. Furthermore, the curves in $\Gamma$ do not cross each other, and, for every edge $e\in E(G'_{i-1})\setminus E(C^+)$, the number of crossings between $\phi'_{i-1}(e)$ and the curves in $\Gamma(C)$ is bounded by 
		$\sum_{e'\in E(C^+)}\chi(e,e')\cdot \cong_{G'}(\qset(C),e')$, where $\chi(e,e')$ is the number of crossings between $\phi'_{i-1}(e)$ and $\phi'_{i-1}(e')$. For every edge $e'\in E(C^+)$, and every crossing $(e,e')_p$ between $e$ and $e'$ in $\phi'_{i-1}$, we \emph{charge} this crossing $\cong_G(\qset(C),e')$ units, and we say that crossing $(e,e')_p$ is \emph{responsible} for $\cong_{G'}(\qset(C),e')$ new crossings between the edge $e$ and the curves in $\Gamma(C)$. Therefore, the total charge to all crossings between $e$ and the edges of $E(C^+)$ is at least the total number of crossings between $\phi'_{i-1}(e)$ and the curves in $\Gamma(C)$. 
We obtain a drawing $\phi'_i$ of the graph $G_i$ as follows. We start from the drawing $\phi'_{i-1}$ of graph $G'_{i-1}$, and delete all edges and vertices of $C_i^+\setminus T_C$ from it. We place the image of the supernode $v_{C_i}$ at the image of the vertex $u(C_i)$ in $\phi'_{i-1}$.
For every edge $e\in \delta_{G'_i}(v_{C_i})$, we let $\gamma(e)\in \Gamma$ be the new image of the edge $e$. This concludes the definition of the drawing $\phi'_i$ of graph $G'_i$. Note that:
\[ \cro(\phi'_i)-\cro(\phi'_{i-1})\leq \sum_{e\in E(G'_{i-1})\setminus E(C_i^+)}\sum_{e'\in E(C_i^+)} \chi(e,e')\cdot \cong_{G'}(\qset(C),e'). \]
Once every cluster $C\in \cset$ is processed in this manner, we obtain the final drawing $\phi$ of $G_{|\cset}$, by suppressing the images of the vertices of $T$ in the drawing $\phi'_r$ of the graph $G'_r$. Note that for every vertex $x\in V(G_{|\cset})\cap V(G)$, we did not modify the images of the edges of $\delta_G(x)$ inside the tiny $\phi^*$-disc $D_{\phi^*}(x)$, so the order in which these edges enter the image of $x$ continues to be $\oset_x$. It now remains to bound the number of crossings in the drawing $\phi$. 
We only bound the number of new crossings that were added due to the transformations that we perform.

Consider some crossing $(e,e')_p$ in the drawing $\phi'$ of graph $G'$. If neither of the edges $e,e'$  lie in $\bigcup_{C\in \cset}E(C^+)$, then no new crossings between these edges were introduced, and this crossing was not charged for any new crossings.  Assume next that $e\in E(C_i^+)$, for some cluster $C_i\in \cset$, and $e'\not\in \bigcup_{C\in \cset}E(C^+)$. Then crossing $(e,e')_p$ may be responsible for up to $\cong_{G'}(\qset(C_i),e)$ new crossings. Each of these new crossings is between the image of $e'$ and the images of the edges of $\delta_{G'_i}(v_{C_i})$, and so they cannot be responsible for any additional new crossings. Since $\expect{\cong_{G'}(\qset(C_i),e)}\leq   O((\log^4m)/\alpha)$, the total expected number of crossings for which crossing $(e,e')_p$ is responsible is at most $O((\log^4m)/\alpha)$. 

Lastly, assume that $e\in E(C_i^+)$ and $e'\in E(C_j^+)$, for $C_i,C_j\in \cset$. If $i=j$, then crossing $(e,e')_p$ is not responsible for any new crossings. Assume now without loss of generality that $i<j$. After cluster $C_i$ is processed, crossing $(e,e')_p$ may be responsible for at most 
 $\cong_{G'}(\qset(C_i),e)$ new crossings. 
 All these new crossings are between the images of the edges of $\delta_{G'_i}(v_{C_i})$ and the image of edge $e'$. Once cluster $C_j$ is processed, each of the resulting crossings may in turn be responsible for at most  $\cong_{G'}(\qset(C_j),e')$ new crossings. 
 Each of these new crossings is between images of edges in  $\delta_{G'_j}(v_{C_i})$ and images of edges in $\delta_{G'_j}(v_{C_j})$, so they in turn will not be responsible for any new crossing.
 Therefore, overall, crossing 
 $(e,e')_p$ may be responsible for up to $\cong_{G'}(\qset(C_i),e)\cdot \cong_{G'}(\qset(C_j),e')$ new crossings. Since $\cong_{G'}(\qset(C_i),e)$ and $ \cong_{G'}(\qset(C_j),e')$ are independent random variables, and the expected value of each of these variables is at most $O((\log^4m)/\alpha)$, the expected number of crossings for which crossing $(e,e')_p$ is responsible is at most $O((\log^8m)/\alpha^2)$. We conclude that $\expect{\cro(\phi)}\leq O((\log^8m)/\alpha^2)\cdot \cro(\phi')\leq O((\log^8m)/\alpha^2)\cdot \optcrors(I)$. Therefore, there exists  a drawing $\phi$ of the contracted graph $G_{|\cset}$, with $\cro(\phi)\leq O((\optcrors(I)\cdot \log^8m)/\alpha^2)$, in which, for every regular vertex $x\in V(G_{|\cset})\cap V(G)$, the ordering of the edges of $\delta_G(x)$ as they enter $x$ in $\phi$ is consistent with the rotation $\oset_x\in \Sigma$.

\section{Proofs Omitted from \Cref{sec: guiding paths orderings basic disengagement}}
\label{sec: appendix: proofs from basic disengagement}

\subsection{Proof of \Cref{lem: basic disengagement combining solutions}}
\label{appx: proof of basic disengagement combining solutions} 
	In this proof, we assume that all drawings are on the sphere.
	For every cluster $C\in \lset$, denote by $\wset(C)\subseteq \lset$ the set of all child clusters of $C$, and by $\wset^*(C)\subseteq \lset$ the set of all descendant clusters of $C$. We define a new instance $I'_C=(G'_C,\Sigma'_C)$ of \cnwrs associated with cluster $C$, as follows. If $C=G$, then $G'_C=G$ and $\Sigma'_C=\Sigma$. Otherwise, graph $G'_C$ is obtained from graph $G$, by contracting all vertices of $V(G)\setminus V(C)$ into a supernode $v^*$. Rotation system $\Sigma'_C$ is defined as follows. Note that $\delta_{G'_C}(v^*)=\delta_G(C)$. We define the rotation $\oset_{v^*}\in \Sigma'_C$ to be $\oset(C)$. For every other vertex $v\in V(G'_C)$, $\delta_{G'_C}(v)=\delta_G(v)$ holds, and its rotation $\oset_v\in \Sigma'_C$ remains the same as in $\Sigma$. This completes the definition of instance $I'_C$.
	Notice that instance $I_C$ can be obtained from instance $I'_C$ by contracting, for each cluster $C'\in \wset(C)$, the vertices of $C'$ into a supernode $v_{C'}$, and then setting the rotation of the edges incident to this supernode to $\oset(C')$.
	
	We prove by induction that there is an efficient algorithm, that, given a cluster $C\in \lset$, and solutions $\set{\phi(I_{C'})}_{C'\in \wset^*(C)}$ to instances associated with the descendant clusters of $C$, computes a solution $\phi'(I'_C)$ to instance $I'_C$, of cost at most $\sum_{C'\in \wset^*(C)}\phi(I_{C'})$. Since $I'_G=I$, this will complete the proof of the lemma.
	
	The proof is by induction of the length of the longest path in the partitioning tree $\tau(\lset)$ between $v(C)$ and its descendant. The base of the induction is when cluster $C$ is the leaf of the tree $\tau(\lset)$. In this case, $I'_C=I_C$ holds, and we let $\phi'(I'_C)=\phi(I_C)$.
	
	For the induction step, we consider some cluster $C\in \lset$, whose corresponding vertex $v(C)$ is not a leaf vertex of the tree $\tau(\lset)$.
	Assume that $\wset(C)=\set{C_1,\ldots,C_r}$. For convenience, for each $1\leq i\leq r$, we denote the supernode $v_{C_i}$ representing cluster $C_i$ in graph $G_C$ by $v_i$. By applying the induction hypothesis to every cluster $C_i\in \wset(C)$, we obtain a solution $\phi'_i=\phi'(I'_{C_i})$ to instance $I'_{C_i}$ of \cnwrs, whose cost is $\cro(\phi'_i)\leq \sum_{C'\in \wset^*(C_i)}\cro(\phi(I_{C'}))$. It is now enough to show an efficient algorithm that constructs a solution $\phi'(I'_C)$ to instance $I'_C$, whose cost is at most $\cro(\phi(I_C))+\sum_{i=1}^r\cro(\phi'_i)\leq \sum_{C'\in \wset^*(C)}\cro(\phi(I_{C'}))$.

	We start with the solution $\tilde \phi=\phi(I_C)$ to instance $I_C$, and we process the clusters $C_1,\ldots,C_r$ one by one, gradually modifying the drawing $\tilde \phi$. We now describe the iteration when cluster $C_i$ is processed. We denote $\delta_{G'_C}(v_i)=\delta_G(C_i)=\set{e^i_1,\ldots,e^i_{q_i}}$, where the edges are indexed according to their ordering in $\oset(C_i)$. 
	For all $1\leq j\leq q_i$, we denote $e^i_j=(x_j,y_j)$, where $x_j\in C_i$.
	Let $D_i=D_{\tilde \phi}(v_i)$ be a tiny $v_i$-disc in the drawing $\tilde \phi$. For all $1\leq j\leq q_i$, we denote by $p^i_j$ the unique point on the image of edge $e^i_j$ that lies on the boundary of the disc $D_i$, and we let $\gamma(e^i_j)$ denote the segment of the image of $e^i_j$ that is disjoint from the interior of $D_i$. Therefore, $\gamma(e^i_j)$ connects the image of vertex $y_j$ to point $p^i_j$. Notice that points $p^i_1,\ldots,p^i_{q_i}$ appear on the boundary of $D_i$ in this circular order. If the orientation of this ordering is positive, then we say that vertex $v_i$ is positive, and otherwise we say that it is negative. We erase the parts of the images of all edges in the interior of disc $D_i$, and we erase the image of the vertex $v_i$ from the current drawing. We place another disc $D'_i$ inside $D_i$, so that $D'_i\subseteq D_i$, and the boundaries of both discs are disjoint.
	
	Next, we consider the drawing $\phi'_i$ of the graph $G'_{C_i}$. We let $\hat D_i=D_{\phi'_i}(v^*)$ be a tiny $v^*$-disc in this drawing. Recall that $\delta_{G'_{C_i}}(v^*)=\delta_G(C_i)$. For all $1\leq j\leq q_i$, we denote by $\hat p^i_j$ the unique point on the image of the edge $e^i_j$ in $\phi'_i$ that lies on the boundary of the disc $\hat D_i$. Note that points $\hat p^i_1,\ldots,\hat p^i_{q_i}$ must appear on the boundary of the disc $\hat D_i$ in this circular order, from the definition of the rotation $\oset_{v^*}\in \Sigma'_{C_i}$. We assume w.l.o.g. that, if vertex $v_i$ is positive, then the orientation of this ordering is negative, and otherwise it is positive (if this is not the case then we simply flip the drawing $\phi'_i$). Let $\hat D'_i$ be the disc that has the same boundary as $\hat D_i$ but whose interior is disjoint from that of $\hat D_i$ (so $\hat D'_i$ is the complement of disc $\hat D_i$; recall that the drawing $\phi'_i$ is on the sphere). For all $1\leq j\leq q_i$, we denote by $\gamma'(e^i_j)$ the segment of the image of edge $e^i_j$ that lies inside $\hat D'_i$. Therefore,  $\gamma'(e^i_j)$ connects the image of vertex $x_j$ to point $\hat p^i_j$.
	
	We copy the disc $\hat D'_i$, together with its contents (in $\phi'_i$), to the current drawing $\tilde \phi$, so that the boundaries and the interiors of the discs $\hat D'_i$ and $D'_i$ coincide. 
	Assume w.l.o.g. that vertex $v_i$ is positive. Then points $p^i_1,\ldots,p^i_{q_i}$ appear on the boundary of disc $D'_i$ in this counter-clock-wise order, while points $\hat p^i_1,\ldots,\hat p^i_{q_i}$ appear on the boundary of disc $\hat D'_i$ in this counter-clock-wise order. Therefore, we can compute a collection $\set{\sigma_1,\ldots,\sigma_{q_i}}$ of mutually disjoint curves, where for all $1\leq j\leq q_i$, curve $\sigma_j$ has endpoints $p_j^i$ and $\hat p_j^i$, and all inner points of $\sigma_j$ lie in $D_i\setminus D'_i$, and are disjoint from the boundary of $D_i$. For all $1\leq j\leq q_i$, we now define the image of the edge $e_j=(x_j,y_j)$ to be the concatenation of the curves $\gamma^i_j,\sigma_j$, and $\hat \gamma^i_j$.
	
	Once every cluster $C_i\in \wset(C)$ is processed in this manner, we obtain a solution $\phi'(I'_C)$ to instance $I'_C$ of \cnwrs. It is immediate to verify that the total number of crossings in this solution is at most $\cro(\phi(I_C))+\sum_{i=1}^r\cro(\phi'_i)\leq \sum_{C'\in \wset^*(C)}\cro(\phi(I_{C'}))$. The lemma follows by letting $\phi$ be the solution $\phi'(I_{C'})$ that we construct for instance $I_{C'}$, where $C'=G$.


\subsection{Proof of \Cref{lem: disengagement final cost}}
\label{subsec: appx basic diseng opt bounds}

Throughout the proof, we denote $|E(G)|=m$.

Consider first a cluster $C\in \lset^{\gd}$. Recall that, in order to define instance $I_C$, we used the distribution $\dset(C)$ over the internal $C$-routers, where $C$ is $\beta$-light with respect to $\dset(C)$.
We selected a router $\qset(C)$ from the distribution $\dset(C)$ at random, whose center vertex is denoted by $u(C)$. We then used the algorithm from \Cref{lem: non_interfering_paths} to compute a non-transversal set $\tilde \qset(C)$ of paths, routing all edges of $\delta_G(C)$ to vertex $u(C)$, so $\tilde \qset(C)$ is also an internal $C$-router. The set $\tilde \qset(C)$ of paths was used in order to define the ordering $\oset(C)$ of the edges of $\delta_G(C)$, which was in turned used in order to define instance $I_C$.

Consider now a cluster $C\in \lset^{\bad}$. We apply the algorithm from \Cref{cor: simple guiding paths} to $C$, obtaining a distribution $\dset(C)$ over the set $\Lambda(C)$ of internal $C$-routers, such that, for every edge $e\in E(C)$, $\expect[\qset\sim \dset]{\cong(\qset(C),e)}\leq  O(\log^4m/\alpha_0)\leq O(\log^{16}m)$. We then select a router $\qset(C)$ from the distribution $\dset(C)$ at random, and denote by $u(C)$ its center vertex. We view the paths of $\qset(C)$ as being directed towards  $u(C)$.
Next, we use the algoritm from \Cref{lem: non_interfering_paths} to compute a non-transversal set $\tilde \qset(C)$ of paths, routing all edges of $\delta_G(C)$ to vertex $u(C)$, so $\tilde \qset(C)$ is also an internal $C$-router. The algorithm ensures that, for every edge $e\in E(G)$,  $\cong_G(\tilde \qset(C),e)\leq \cong_G(\qset(C),e)$.

Consider an optimal solution $\phi^*$ to instance $I$ of \cnwrs. For every cluster $C\in \lset$, denote by $\chi(C)$ the set of all crossings $(e,e')_p$ in the drawing $\phi^*$, where at least one of the edges $e,e'$ lies in $E(C)\cup \delta_G(C)$.
Recall that $\iset=\set{I_C\mid C\in \lset}$. The proof of the lemma follows from the following claim.

\begin{claim}\label{claim: cost of cluster instance}
	For every cluster $C\in \lset$, $\expect{\optcrors(I_C)}\leq O(\beta^2\cdot(|\chi(C)|+|E(C)|))$.
\end{claim}

Indeed, for all $1\leq i\leq \dep(\lset)$, let $\lset_i\subseteq \lset$ be the set of all clusters that lie at level $i$ of the laminar family. Note that all clusters in $\lset_i$ are mutually disjoint. Therefore, every crossing $(e,e')_p$ of the drawing $\phi^*$ may contribute to the sets $\chi(C)$ of at most four clusters of $\lset_i$: at most two clusters $C$ with $e\in E(C)\cup \delta_G(C)$, and at most  two clusters $C'$ with $e\in E(C')\cup \delta_G(C')$. Therefore:

\[
\sum_{C\in \lset_i}\expect{\optcrors(I_C)}\leq \sum_{C\in \lset_i} O(\beta^2\cdot(|\chi(C)|+|E(C)|)) \leq  O(\beta^2\cdot (\optcrors(I)+|E(G)|)). 
\]

Summing this up over all  $1\leq i\leq \dep(\lset)$, we get that $\expect{\sum_{I'\in \iset}\optcrors(I')}\leq O(\dep(\lset)\cdot\beta^2\cdot (\optcrors(I)+|E(G)|))$. In order to complete the proof of \Cref{lem: disengagement final cost}, it is now enough to prove \Cref{claim: cost of cluster instance}, which we do next.

In the remainder of this proof, we fix a cluster $C\in \lset$. Recall that there is a distribution $\dset'(C)$ over the set $\Lambda'(C)$ of external $C$-routers, such that for every edge $e$ of $E(G\setminus C)$, $$\expect[\qset'(C)\sim\dset'(C)]{\cong_G(\qset'(C),e)}\leq \beta.$$ 
We sample an external $C$-router $\qset'(C)$ from the distribution $\dset'(C)$. We view the paths of $\qset'(C)$ as being directed towards vertex $u'(C)$, that is the center of the router. We apply the algorithm from \Cref{lem: non_interfering_paths} to obtain a collection $\tilde \qset'(C)$ of non-transversal paths, routing the edges of $\delta_G(C)$ to $u'(C)$, such that,  for every edge $e\in E(G)$, $\cong_G(\tilde \qset'(C),e)\leq \cong_G(\qset'(C),e)$. In particular, $\tilde \qset'(C)$ is also an external $C$-router with center vertex $u'(C)$.

In order to simplify the notation, we denote $\tilde \qset'(C)$ by $\qset'$, and $\tilde \qset(C)$ by $\qset$. To summarize, $\qset'$ is an external $C$-router, and we are guaranteed that, for every edge $e\in E(G\setminus C)$, $\expect{\cong_G(\qset',e)}\leq \beta$. Set $\qset$ of paths is an internal $C$-router. If $C\in \lset^{\bad}$, then for every edge $e\in E(C)$, 
$\expect{\cong(\qset,e)}\leq  \beta$, while, if $C\in \lset^{\gd}$ then, for every edge $e\in E(C)$, $\expect{(\cong_G(\qset,e))^2}\leq \beta$.
We denote the center vertex of router $\qset$ by $u$, and the center vertex of router $\qset'$ by $u'$.

We denote by $\wset=\set{C_1,\ldots,C_r}$ the set of child-clusters of $C$. We partition $\wset$ into two subsets: $\wset^{\bad}=\wset\cap \lset^{\bad}$, and $\wset^{\gd}=\wset\cap \lset^{\gd}$. For convenience, for all $1\leq i\leq r$, we denote the internal $C_i$-router $\tilde \qset(C_i)$ that we have constructed by $\qset_i$, and its center vertex by $u_i$. Recall that, if $C_i\in \lset^{\bad}$, then  for every edge $e\in E(C_i)$, 
$\expect{\cong(\qset_i,e)}\leq  \beta$, while, if $C_i\in \lset^{\gd}$ then, for every edge $e\in E(C_i)$, $\expect{(\cong_G(\qset_i,e))^2}\leq \beta$.

It will be convenient for us to also define another cluster $C_0$ to be the connected component of $G\setminus C$ containing the vertex $u'$ -- the center vertex of the external $C$-router $\qset'$. It is immediate to verify that the set $\qset'$ of paths is an internal $C_0$-router, and for consistency we denote it by $\qset_0$. We also denote the center vertex of this router by $u_0=u'$. Recall that each one of the sets $\qset_0,\qset_1,\ldots,\qset_r$ of paths is non-transversal with respect to $\Sigma$.
For convenience, we denote $R=C\setminus\left(\bigcup_{i=1}^rC_i\right )$, and $\qset^*=\bigcup_{i=0}^r\qset_i$.

Lastly, it will be convenient for us to assume that no edge of $G$ has its endpoints in two distinct clusters of $C_0,C_1,\ldots,C_r$. For each such edge $e$, we subdivide the edge with a new vertex $v_e$, that is added to graph $R$ as an isolated vertex.
Note that, for all $0\leq i<j\leq r$, the only vertices that may be shared by paths in $\qset_i$ and paths in $\qset_j$ are vertices of $V(R)$, which must serve as endpoints of those paths.

The remainder of the proof of \Cref{claim: cost of cluster instance} consists of three steps. In the first step, we will define a new graph $H$ by slightly modifying graph $G$, and compute its drawing $\psi$. In the second step, we use graph $H$ and its drawing $\psi$ in order to construct an initial drawing $\phi'$ of graph $G_C$ (associated with instance $I_C=(G_C,\Sigma_C)\in \iset$ of \cnwrs). This drawing, however, may not obey all rotations in $\Sigma_C$. In the third and the last step, we modify drawing $\phi'$ to obtain the final drawing $\phi''$ of $G_C$, that is a valid solution to instance $I_C$ of \cnwrs. We now describe each of the three steps in turn.

\subsubsection{Step 1: Graph $H$}

 We assign, to every edge $e\in E(G)$, an integer $n_e\geq 0$, as follows. For every edge $e\in E(R)\cup \delta_G(R)$, we let $n_e=1$. For every other edge $e\in E(G)\setminus E(R)$, we let $n_e=\cong_G(\qset^*,e)$.

In order to construct graph $H$, we start with $V(H)=V(G)$. For every edge $e=(u,v)\in E(G)$ with $n_e>0$, we add a set $J(e)$ of $n_e$ parallel edges $(u,v)$ to graph $H$, and we call the edges of $J(e)$ \emph{copies of edge $e$}. This completes the definition of the graph $H$. Note that graph $H$ is a random graph, as values $\set{n_e}_{e\in E(G)}$ are random variables.

Consider the optimal solution $\phi^*$ to instance $I$ of \cnwrs. We use $\phi^*$ in order to define a drawing $\psi$ of the graph $H$, in a natural way: for every vertex $v\in V(H)$, its image in $\psi$ remains the same as in $\phi^*$. For every edge $e\in E(G)$ with $n(e)\geq 1$, we draw the edges of $J(e)$ in parallel to the image of $e$ in $\phi^*$, immediately next to it, so that their images do not cross.
Consider the resulting drawing $\psi$ of graph $H$, and let $(e'_1,e'_2)_p$ be a crossing of $\psi$. Assume that $e'_1\in J(e_1)$ and $e'_2\in J(e_2)$. Then the images of the edges $e_1,e_2$ cross in $\phi^*$, at some point $p'$ that is very close to point $p$. We say that crossing $(e_1,e_2)_{p'}$ of $\phi^*$ is \emph{responsible} for the crossing $(e_1',e_2')_p$ of $\psi$.

Next, we classify the crossings of the drawing $\psi$ into three types, and we bound the expected number of crossings of some of the types. Consider some crossing $(e'_1,e'_2)_p$ of $\psi$, and let $(e_1,e_2)_{p'}$ be the crossing of $\phi^*$ that is responsible for $(e_1',e_2')_p$. We say that crossing $(e'_1,e'_2)_p$ is a \emph{type-1 crossing} if there is some cluster $C_i\in \wset^{\light}$, with $e_1,e_2\in E(C_i)$. We say that it is a \emph{type-2 crossing} if there is an index $0\leq i\leq r$ with $e_1,e_2\in E(C_i)$, and either $i=0$ or $C_i\in\wset^{\bad}$ holds. We say that $(e_1',e_2')_p$ is a \emph{type-3 crossing} otherwise. 

We now bound the expected number of type-1 and type-3 crossings. We do not bound the number of type-2 crossings, as all such crossings will eventually be eliminated.

\paragraph{Type-1 crossings.} Consider some cluster $C_i\in \wset^{\light}$, and some crossing $(e_1,e_2)_{p'}$ of $\phi^*$, such that $e_1,e_2\in E(C_i)$. Notice that crossing $(e_1,e_2)_{p'}$ lies in $\chi(C_i)$. The number of type-1 crossings in $\phi$ that this crossing is responsible for is $n_{e_1}\cdot n_{e_2}\leq n_{e_1}^2+n_{e_2}^2$. Observe that, for an edge $e\in E(C_i)$, $n_e=\cong_G(\qset_i,e)$, and so  $\expect{n_e^2}=\expect{(\cong_G(\qset_i,e))^2}\leq \beta$. We conclude that the total expected number of type-1 crossings in $\psi$ is bounded by:

\[ \sum_{C_i\in \wset^{\light}} \sum_{(e_1,e_2)_{p'}\in \chi(C_i)}\expect{n_{e_1}^2+n_{e_2}^2 }\leq   \sum_{C_i\in \wset^{\light}} O(\beta \cdot |\chi(C_i)|)\leq O(\beta\cdot |\chi(C)|).
\]

\paragraph{Type-3 crossings.} Consider some crossing $(e_1,e_2)_{p'}$ of $\phi^*$. If edges $e_1,e_2$ lie in the same cluster $C_i$, for $0\leq i\leq r$, then this crossing may not be responsible for any type-3 crossings in $\psi$. Assume now that this is not the case. Then the total number of type-3 crossings that $(e_1,e_2)_{p'}$ is responsible for is at most $n_{e_1}\cdot n_{e_2}$. Furthermore $n_{e_1},n_{e_2}$ are independent random variables, each of which has expectation at most $\beta$. Therefore, the expected number of type-3 crossings for which crossing $(e_1,e_2)_{p'}$ is responsible is at most $\beta^2$. Note that, at least one of the edges $e_1,e_2$ must lie in $E(C)\cup \delta_G(C)$, so crossing $(e_1,e_2)_{p'}$  must lie in $\chi(C)$. We conclude that the total expected number of type-3 crossings in $\phi$ is at most $|\chi(C)|\cdot \beta^2$.

Consider now an index $0\leq i\leq r$, and let $E_i=\delta_G(C_i)$. From our definition, for every edge $e\in E_i$, $n_e=1$. Recall that we have defined a set $\qset_i=\set{Q(e)\mid e\in E_i}$ of paths in graph $G$, routing all edges of $E_i$ to the vertex $u_i\in V(C_i)$. The paths in $\qset_i$ are non-transversal with respect to $\Sigma$, and all their inner vertices are contained in $C_i$. We will now define a corresponding set $\hat \qset_i=\set{\hat Q(e)\mid e\in E_i}$ of paths in graph $H$, routing the edges of $E_i$ to the same vertex $u_i$, such that the paths in $\hat \qset_i$ are edge-disjoint. In order to do so, we assign, to every path $Q(e)\in \qset_i$, for every edge $e'\in E(Q(e))\setminus\set{e}$, a copy of the edge $e'$ from $J(e')$, such that every copy of edge $e'$ is assigned to a distinct path. We will then obtain path $\hat Q(e)$ from path $Q(e)$ by replacing every edge $e'\in E(Q(e))\setminus\set{e}$ with its copy that was assigned to $Q(e)$.

Consider any edge $e'\in E(C_i)$. If $e'$ is not incident to the vertex $u_i$, then we assign each copy of $e'$ to a distinct path in $\qset_i$ that contains $e'$ arbitrarily. Assume now that edge $e'$ is incident to vertex $u_i$, and that $C_i\not\in \wset^{\gd}$. In this case, as before, we assign  each copy of $e'$ to a distinct path in $\qset_i$ that contains $e'$ arbitrarily. It now remains to consider the case where $C_i\in \wset^{\gd}$, and edges that are incident to vertex $u_i$. We need to assign  copies of such edges to paths in $\qset_i$ more carefully. The goal of this more careful assignment is to achieve the following property: if we denote $E_i=\set{e_1,\ldots,e_{h_i}}$, where the edges are indexed according to the ordering $\oset(C_i)$, and, for each such edge $e_j$, we denote by $e'_j$ be the last edge on path $\hat Q(e_j)$ (that we are trying to construct), then the images of the edges $\set{e'_1,\ldots,e'_{h_i}}$ enter the image of $u_i$ in the drawing $\psi$ of graph $H$ in this circular order. We now describe the procedure for assigning copies of edges of $\delta_G(u_i)$ to paths in $\qset_i$.

We start by revisiting the definition of the  ordering $\oset(C_i)$ of the edges of $E_i=\delta_G(C_i)$, which is an ordering that is guided by the set $\qset_i$ of paths and the rotation system $\Sigma$. 
Denote $\delta_G(u_i)=\set{a^i_1,\ldots,a^i_{z_i}}$, where the edges are indexed according to their circular ordering $\oset_{u_i}\in \Sigma$. We assume w.l.o.g. that the orientation of this ordering in the drawing $\phi^*$ of $G$ is negative (or clock-wise). For all $1\leq j\leq z_i$, let $\qset^j_i\subseteq \qset_i$ the set of paths in $\qset_i$ whose last edge is $a^i_j$. We defined an ordering $\hat \oset_i$ of the paths in $\qset_i$, where the paths in sets $\qset_i^1,\ldots,\qset_i^{z_i}$ appear in the natural order of their indices, and for all $1\leq j\leq z_i$, the ordering of the paths in set $\qset_i^j$ is arbitrary. We denote $\qset_i^j=\set{Q(e_1^{i,j}),Q(e_2^{i,j}),\ldots,Q(e_{m_{i,j}}^{i,j})}$, and assume that these paths are indexed according to the ordering that we have chosen when defining $\hat \oset_i$.

Ordering $\hat\oset_i$ of the paths in $\qset_i$ was then used to define the ordering $\oset(C_i)$ of the edges in $E_i$: we obtain the ordering $\oset(C_i)$ from $\hat \oset_i$ by replacing, for every path $Q(e_{\ell}^{i,j})\in \qset_i$, the path $Q(e_{\ell}^{i,j})$ in $\hat \oset$ with the edge $e_{\ell}^{i,j}$ (the first edge of $Q(e_{\ell}^{i,j})$).

Consider now some edge $a^i_j\in \delta_G(u_i)$. Recall that we have defined a set $J(a^i_j)$ of  $n_{a^i_j}= m_{i,j}$ copies of the edge $a^i_j$. We denote these copies by $\hat a^{i,j}_1,\ldots,\hat a^{i,j}_{m_{i,j}}$, where the copies are indexed according to the order in which their images enter the image of vertex $u_i$ in the drawing $\psi$ of $H$, in the clock-wise direction. For all $1\leq \ell\leq m_{i,j}$, we assign the copy $\hat a^{i,j}_{\ell}$ of edge $a^i_j$ to the path $Q(e_{\ell}^{i,j})$. 
This completes the assignment of the copies of the edges incident to vertex $u_i$ to the paths of $\qset_i$.

We now define  a set $\hat \qset_i=\set{\hat Q(e)\mid e\in E_i}$ of paths in graph $H$, routing the edges of $E_i$ to vertex $u_i$, as follows. For every edge $e\in E_i$, path $\hat Q(e)$ is obtained from the path $Q(e)\in \qset_i$ by replacing every edge $e'\in E(Q(e))\setminus \set{e}$ with the copy of $e'$ that was assigned to path $Q(e)$. The following observation summarizes the properties of the path set $\hat \qset_i$, that follow immediately from our construction.  

\begin{observation}\label{obs: properties of new path set}
	Paths in set $\hat \qset_i=\set{\hat Q(e)\mid e\in E_i}$  route the edges of $E_i$ to vertex $u_i$ in graph $H$, and all inner vertices on all paths in $\hat \qset_i$ lie in $V(C_i)$. Moreover, the paths of $\hat \qset_i$ are edge-disjoint. Additionally, if $C_i\in \wset^{\light}$, then the following holds. Denote $E_i=\set{e_1,\ldots,e_{h_i}}$, where the edges are indexed according to the ordering $\oset(C_i)$. For each such edge $e_j$, let $e'_j$ be the last edge on path $\hat Q(e_j)$. Then the images of the edges $e'_1,\ldots,e'_{h_i}$ enter the image of $u_i$ in the drawing $\psi$ of graph $H$ in the circular order of their indices.
\end{observation}

\subsubsection{Step 2: Initial Drawing of Graph $G_C$}

In this step we exploit the drawing $\psi$ of graph $H$ that we have constructed in the first step, in order to construct an initial drawing $\phi'$ of graph $G_C$. 

In order to construct the drawing $\phi'$ of graph $G_C$, we start with the drawing $\psi$ of graph $H$, and then gradually modify it. We  place the image of the vertex $v^*$ in $\phi'$ at point $\psi(u_0)$, and, for all $1\leq i\leq r$, we place the image of the vertex $v_{C_i}$ at point $\psi(u_i)$. Intuitively, the images of the vertices and the edges of $R$ will remain unchanged. For all $0\leq i\leq r$, we will utilize the images of the paths of $\hat\qset_i$ in $\psi$ in order to draw the edges of $E_i$. There are two issues with this approach. First, we did not bound the expected number of type-2 crossings in $\psi$, so there may be many crossings between pairs of edges lying on paths of $\qset_i$, where $i=0$, or $C_i\in \wset^{\bad}$. We take care of this issue by performing a type-2 uncrossing for each such path set $\hat \qset_i$, to obtain the drawings of the edges in $E_i$. The second problem then remains for indices $i$ with $C_i\in \wset^{\gd}$. Since several paths from $\hat \qset_i$ may share the same vertex, there could be points that lie on images of multiple paths of $\hat \qset_i$. We take care of this latter issue by employing a nudging procedure. We now describe each of these two operations in turn.

\paragraph{Uncrossing.}
We consider indices $i$ for which either $i=0$ or $C_i\in \wset^{\bad}$ holds one by one. Consider any such index $i$. We view every path of $\hat \qset_i$ as being directed towards the vertex $u_i$. We use the algorithm from \Cref{thm: new type 2 uncrossing} in order to compute a type-2 uncrossing, that produces, for every edge $e\in E_i$, a directed curve $\gamma(e)$, that connects the image of the endpoint of $e$ that lies in $R$ to the image of $u_i$ in $\psi$. Recall that we are guaranteed that the curves in the resulting set $\Gamma_i=\set{\gamma(e)\mid e\in E_i}$ do not cross each other, and each such curve is aligned with the drawing of graph $\bigcup_{\hat Q(e)\in \hat \qset_i}\hat Q(e)$ induced by $\psi$.

Let $\psi'$ be a drawing obtained from $\psi$ as follows. For every index $i$ with $i=0$ or  $C_i\in \wset^{\bad}$, we delete the images of all vertices of $V(C_i)$ and all edges with at least one endpoint in $V(C_i)$ from the drawing. If $i=0$, then we place the image of vertex $v^*$ at point $\psi(u_0)$, and otherwise we place the image of vertex $v_{C_i}$ at point $\psi(u_i)$. For every edge $e\in E_i$, we then let $\gamma(e)\in \Gamma_i$ be the image of the edge $e$.

Note that this uncrossing step has eliminated all type-2 crossings, and every crossing in the resulting drawing $\psi'$ corresponds to a distinct type-1 or type-3 crossing of $\psi$. Therefore, the expected number of crossings of $\psi'$ is bounded by $O(\beta^2\cdot |\chi(C)|)$.
We call all crossings that are currently present in drawing $\psi'$ \emph{primary crossings}.

\paragraph{Nudging.}
We now consider the indices $i$ with $C_i\in \wset^{\gd}$ one by one. When  such an index $i$ is considered, we delete the images of all vertices of $V(C_i)$ and all edges with at least one endpoint in $V(C_i)$ from the current drawing $\psi'$. We then place the image of vertex $v_{C_i}$ at point $\psi(u_i)$. For every edge $e\in E_i$, we initially let $\gamma(e)$ be the image of the path $\hat Q(e)\in \hat \qset_i$ in $\psi$, and we add $\gamma(e)$ to the current drawing as the image of the edge $e$. Note that the curves in $\set{\gamma(e)\mid e\in E_i}$ enter the image of $v_{C_i}$ in the order $\oset(C_i)$ of their corresponding edges in $E_i$,  from \Cref{obs: properties of new path set}. However, it is possible that, for some vertex $x\in V(C_i)$, point $\psi(x)$ lies on more than two curves from 
$\set{\gamma(e)\mid e\in E_i}$.

We process each vertex $x\in V(C_i)\setminus \set{u_i}$ one by one. Consider any such vertex $x$, and let $\qset^x\subseteq \hat\qset_i$ be the set of all paths containing vertex $x$. Note that $x$ must be an inner vertex on each such path. For convenience, we denote $\qset^x=\set{Q(e_1),\ldots,Q(e_z)}$. Consider the tiny $x$-disc $D=D_{\psi}(x)$. For all $1\leq j\leq z$, denote by $s_j$ and $t_j$ the two points on the curve $\gamma(e_j)$ that lie on the boundary of disc $D$. We use the algorithm from \Cref{claim: curves in a disc} to compute a collection $\set{\sigma_1,\ldots,\sigma_z}$ of curves, such that, for all $1\leq j\leq z$, curve $\sigma_j$ connects $s_j$ to $t_j$, and the interior of the curve is contained in the interior of $D$. Recall that every pair of resulting curves crosses at most once, and every point in the interior of $D$ may be contained in at most two curves.
Consider now a pair $\sigma_{\ell},\sigma_{\ell'}$ of curves, and assume that these two curves cross. Recall that, from \Cref{claim: curves in a disc}, this may only happen if the two pairs $(s_{\ell},t_{\ell})$, $(s_{\ell'},t_{\ell'})$ of points cross. Denote by $e_1,e_2$ the two edges that lie on path $\hat Q(e_{\ell})$ immediately before and immediately after vertex $x$, and denote by $e_1',e_2'$ the two edges that lie on path $\hat Q(e_{\ell'})$ immediately before and immediately after vertex $x$. We assume that edges $e_1,e_2$ are copies of edges $\hat e_1,\hat e_2$ of $G$, and similarly, $e_1',e_2'$ are copies of edges $\hat e_1',\hat e_2'$ of $G$, respectively. Assume first that there are four distinct edges in set $\set{\hat e_1,\hat e_1',\hat e_2,\hat e_2'}$. From the fact that the two pairs  $(s_{\ell},t_{\ell})$, $(s_{\ell'},t_{\ell'})$ of points cross, we get that these four edges must appear in the rotation $\oset_x\in \Sigma$ in the order $(\hat e_1,\hat e_1',\hat e_2,\hat e_2')$. Since the paths of $\qset_i$ are non-transversal with respect to $\Sigma$, this is impossible. Therefore, we conclude that paths $Q(e_{\ell}),Q(e_{\ell'})$ must share an edge that is incident to $x$. If $e^*$ is an edge incident to $x$ that the two paths share, then we say that $e^*$ is \emph{responsible for the crossing between $\sigma_{\ell}$ and $\sigma_{\ell'}$}. 

For all $1\leq j\leq z$, we modify the curve $\gamma(e_j)$, by replacing the segment of the curve that is contained in disc $D$ with $\sigma_j$.
Once every vertex $x\in V(C_i)\setminus \set{u_i}$ is processed, we obtain the final set $\Gamma'_i=\set{\gamma'(e)\mid e\in E_i}$ of curves, which are now guaranteed to be in general position. For every edge $e\in E_i$, we modify the image of edge $e$ in the current drawing, by replacing it with the new curve $\gamma'(e)$. As before, curves of $\Gamma'_i$ enter the image of $v_{C_i}$ according to the ordering $\oset(C_i)$.

Once every index $i$ with $C_i\in \wset^{\gd}$ is processed, we obtain a valid drawing $\phi'$ of the graph $G_C$. In this drawing, for every index $i$ with $C_i\in \wset^{\gd}$, the images of the edges in set $E_i=\delta_{G_C}(v_{C_i})$ enter the image of vertex $v_{C_i}$ according to the ordering $\oset_{v_{C_i}}\in \Sigma_C$, which is precisely $\oset(C_i)$. However, for indices $i$ with $C_i\in \wset^{\bad}$, this property may not hold, and the edges incident to $v^*$ may enter the image of $v^*$ in an arbitrary order. For every other vertex $v$ of $G_C$, the rotation $\oset_v\in \Sigma_{C}$ is identical to the rotation $\oset_v\in \Sigma$, and is obeyed by the current drawing $\phi'$. We modify the drawing $\phi'$ to obtain a drawing that is consistent with the rotation system $\Sigma_{C}$ in the third step. Notice however that the nudging operation may have introduced some new crossings. Each such new crossing must be contained in a disc $D_{\psi}(x)$, for some vertex $x$ that must lie in some cluster $C_i\in \wset^{\gd}$. We call all such new crossings \emph{secondary crossings}. We now bound the total number of secondary crossings.

Fix an index $i$ with $C_i\in \wset^{\gd}$, and consider some vertex $x\in V(C_i)$. Every secondary crossing that is contained in $\dset_{\psi}(x)$ is a crossing between a pair $\sigma_{\ell},\sigma_{\ell'}$ of curves that we have defined when processing vertex $x$, and each such crossing was charged to an edge of $G$ that is incident to $x$, whose copies lie on the corresponding two paths $\hat Q(e_{\ell}),\hat Q(e_{\ell'})\in \hat \qset_i$. If $e$ is an edge that is incident to $x$ in $G$, then there are at most $(\cong_G(\qset_i,e))^2$ pairs of paths in $\qset_i$ that contain $e$, and each such pair of paths may give rise to a single secondary crossing in $D_{\psi}(x)$ that is charged to edge $e$. Therefore, the total expected number of secondary crossings that are contained in discs $D_{\psi}(x)$ for vertices $x\in V(C_i)$ is bounded by:

\[ \sum_{e\in E(C_i)}O(\expect{(\cong_G(\qset_i,e))^2})\leq O(\beta \cdot |E(C_i)|),  \]

since $C_i\in \wset^{\gd}$.

We conclude that the total expected number of secondary crossings in $\phi'$ is at most $\sum_{C_i\in\wset^{\gd}}O(\beta \cdot |E(C_i)|)\leq O(\beta\cdot |E(C)|)$, and the total number of all crossings in $\phi'$ is at most  $O(\beta^2\cdot (|\chi(C)|+|E(C)|))$.

\subsubsection{Step 3: the Final Drawing}

So far we have obtained a drawing $\phi'$ of graph $G_C$, that obeys the rotations $\oset_v\in \Sigma_C$ for all vertices $v\in V(G_C)$, except possibly for vertex $v^*$, and vertices $v_{C_i}$, for $C_i\in \wset^{\bad}$. We now fix this drawing to obtain a final drawing $\phi''$ of $G_C$ that obeys the rotation system $\Sigma_C$.

Let $U=\set{v^*}\cup \set{v_{C_i}\mid C_i\in \wset^{\bad}}$. For each vertex $x\in U$, we denote by $\hat \oset(x)=\oset_x\in \Sigma_C$ the rotation associated with vertex $x$ in the rotation system $\Sigma_C$, and by $\hat \oset'(x)$ the circular order in which the edges of $\delta_{G_C}(x)$ enter the image of $x$ in the current drawing $\phi'$. Note that, for a vertex $x=v_{C_i}$, where $C_i\in \wset^{\bad}$, if we denote by $\Sigma(C_i)$ the rotation system induced by $\Sigma$ for cluster $C$, then the following must hold:
\[\dist(\hat \oset(x),\hat \oset'(x))\leq |\delta_{G_C}(x)|^2=|\delta_G(C_i)|^2\leq \beta\cdot (\optcrors(C_i,\Sigma(C_i))+|E(C_i)|)\leq \beta(|\chi(C_i)|+|E(C_i)|).\]
(we have used the fact that cluster $C_i$ is a $\beta$-bad cluster).
We use the following claim, whose proof appears in Section \ref{subsec: appx  bounding distance between rotations}, in order to bound $\dist(\hat \oset(v^*),\hat \oset'(v^*))$.

\begin{claim}\label{claim: bounding distance between rotations}
	$\expect{\dist(\hat \oset(v^*),\hat \oset'(v^*))}\leq \beta^2\cdot (|\chi(C)|+|E(C)|)$.
\end{claim}

In order to compute the final drawing $\phi''$ of graph $G_C$, we process the vertices $x\in U$ one by one. When vertex $x$ is processed, we apply the algorithm from \Cref{lem: ordering modification} to it. The algorithm modifies the current drawing of graph $G_C$ within the tiny $x$-disc $D(x)$ to ensure that the images of the edges of $\delta_{G_C}(x)$ enter the image of $x$ in the circular order $\hat \oset(x)$. This modification increases the number of crossings in the current drawing by at most $2\cdot\dist(\hat \oset(x),\hat \oset'(x))$. Once every vertex of $U$ is processed, we obtain the final drawing $\phi''$ of graph $G_C$, which obeys the rotation system $\Sigma_C$. Moreover, $\cro(\phi'')\leq \cro(\phi')+\sum_{x\in U}2\cdot\dist(\hat \oset(x),\hat \oset'(x))$. Recall that $\expect{\cro(\phi')}\leq O(\beta^2)\cdot (|\chi(C)|+|E(C)|)$, and that, for every vertex $x=v_{C_i}$ with $C_i\in \wset^{\bad}$, $\dist(\hat \oset(x),\hat \oset'(x))\leq \beta(|\chi(C_i)|+|E(C_i)|)$. Combining this with \Cref{claim: bounding distance between rotations}, we get that:
\[\expect{\cro(\phi'')}\leq O(\beta^2)\cdot (|\chi(C)|+|E(C)|)+\sum_{C_i\in \wset^{\bad}}  O(\beta)\cdot (|\chi(C_i)|+|E(C_i)|)\leq O(\beta^2\cdot (|\chi(C)|+|E(C)|)). \]
This completes the proof of \Cref{claim: cost of cluster instance}.

\subsection{Proof of \Cref{claim: bounding distance between rotations}}
\label{subsec: appx  bounding distance between rotations}

Clearly, $\dist(\hat \oset(v^*),\hat \oset'(v^*))\leq |\delta_{G_C}(v^*)|^2=|\delta_G(C)|^2$. If $C\in \lset^{\bad}$, then cluster $C$ is $\beta$-bad, and so $\dist(\hat \oset(v^*),\hat \oset'(v^*))\leq |\delta_G(C)|^2\leq \beta\cdot (|\chi(C)|+|E(C)|)$ from the definition of $\beta$-bad clusters. Therefore, we assume from now on that $C\in \lset^{\gd}$.

For convenience of notation, we denote $v^*$ by $u'$, and we denote $\hat \oset(v^*)$ and $\hat \oset'(v^*)$ by $\oset$ and $\oset'$, respectively. We also denote $E'=\delta_G(C)=\delta_{G_C}(u')$.
In order to prove the claim, we will construct a collection $\Gamma=\set{\gamma(e)\mid e\in E'}$ of curves in the plane, all of which connect two points $p$ and $q$. We will ensure that the order in which the curves enter the point $p$ is precisely $\oset'$, and the order in which they enter the point $q$ is $\oset$. We will also ensure that the curves of $\Gamma$ are in general position. By showing that the expected number of crossings between the curves in $\Gamma$ is relatively small, we will obtain the desired bound on $\expect{\dist(\oset,\oset')}$.

In order to construct the curves in $\Gamma$, we consider again the intance $I$ of \cnwrs and its optimal solution $\phi^*$. Recall that we have computed an internal $C$-router $\qset=\set{Q(e)\mid e\in E'}$, where for each edge $e\in E'$, path $Q(e)$ originates with edge $e$, terminates at vertex $u$, and all its inner vertices lie in $C$. The paths in $\qset$ are non-transversal with respect to $\Sigma$, and, for every edge $e'\in E(C)$, $\expect{(\cong_G(\qset,e'))^2}\leq \beta$. We have also constructed an  external $C$-router $\qset'=\set{Q'(e)\mid e\in E'}$, where for each edge $e\in E'$, path $Q(e)$ originates with edge $e$, terminates at vertex $u'$, and all its inner vertices are disjoint from $C$. The paths in $\qset'$ are non-transversal with respect to $\Sigma$, and, for every edge $e'\in E(G\setminus C)$, $\expect{\cong_G(\qset',e')}\leq \beta$. We note that path set $\qset'$ is exactly the same as path set $\qset_0$ -- the internal router for $C_0$, that we used in the first step of the algorithm. Intuitively, we would like to let $p$ be the image of vertex $u'$ and $q$ the image of vertex $u$ in $\phi^*$. For every edge $e\in E'$, we would like to use the concatenation of the images of paths $Q(e)$ and $Q'(e)$ in $\phi^*$ in order to construct the curve $\gamma(e)$. This approach has several problems. First, the paths in sets $\qset$ and $\qset'$ may share edges and vertices, and so the resulting curves may not be in a general position. Second, there could be many crossings between edges lying on the paths of $\qset'$, which may lead to many crossings between curves of $\Gamma$. We take care of all these issues in the following three steps. In the first step, we take care of the congestion issue by constructing a graph $H'$ and its drawing $\psi'$. The construction is somewhat similar to the construction of graph $H$, in that we make several copies of some of the edges of $G$, in a way that allows us to define edge-disjoint paths in graph $H'$ replacing the path sets $\qset$ and $\qset'$. In the second step, we perform uncrossing of curves corresponding to the paths in $\qset'$, in order to eliminate some of the crossings. In the third step we perform nudging of curves corresponding to the paths in $\qset$. We now describe each of these steps in turn.

\paragraph{Graph $H'$.}
For each edge $e\in E'$, we set $n'_e=1$. For an edge $e\in E(C)$, we set $n'_e=\cong_G(\qset,e)$, and for an edge $e\in E(G\setminus C)$, we set $n'_e=\cong_G(\qset',e)$. For every other edge $e$, we set $n'_e=0$. Note that for each edge $e\in E(G\setminus C)$, $n'_e=n_e$ holds, where $n_e$ is the parameter that we have used in the construction of graph $H$ in Step 1 of the algorithm.

In order to construct graph $H'$, we start with $V(H')=V(G)$. For every edge $e=(x,y)\in E(G)$ with $n'_e\neq 0$, we add a new set $J'(e)$ of $n'_e$ parallel edges connecting $x$ to $y$ to graph $H'$, that we view as \emph{copies of edge $e$}.  As before, we use the drawing $\phi^*$ of $G$ in order to compute a drawing $\psi'$ of graph $H'$. For every vertex $x\in V(H')$, we let its image in $\psi'$ be $\phi^*(x)$. For every edge $e\in E(G)$ with $n'_e\neq 0$, we draw the edges of $J'(e)$ in parallel to the image of $e$ in $\phi^*$. Recall that for each edge $e\in E(G\setminus C)$, $n'_e=n_e$ holds, and so set $J'(e)$ of copies of $e$ can be thought of as being identical to the set $J(e)$ of copies of $e$ that we have constructed for graph $H$. We ensure that the specific drawing of the edges of $J'(e)$ in $\psi'$ is identical to the drawing of these edges in $\psi$.

We now bound the expected number of crossings in the resulting drawing $\psi'$ of graph $H'$. Consider any such crossing $(e_1',e_2')_{p'}$, and assume that $e_1'\in J'(e_1)$, $e_2'\in J'(e_2)$ holds for some edges $e_1,e_2\in E(G)$. Then there must be some crossing $(e_1,e_2)_p$ in the drawing $\phi^*$ of $G$, with point $p$ lying very close to point $p'$. We say that crossing $(e_1,e_2)_p$ of $\phi^*$ is \emph{responsible} for the crossing $(e_1',e_2')_{p'}$ of $\psi'$. It is immediate to verify that every crossing $(e_1,e_2)_p$ of $\phi^*$ may be responsible for at most $n'_{e_1}\cdot n'_{e_2}$ crossings of $\psi'$.

We classify the crossings of $\psi'$ into three types. Let $(e_1',e_2')_{p'}$ be a crossing of $\psi'$, and let  $(e_1,e_2)_p$ be the crossing of $\phi^*$ responsible for it. We say that $(e_1',e_2')_{p'}$ is a \emph{type-1 crossing} if $e_1,e_2\in E(C)$. We say that it is a \emph{type-2 crossing} if $e_1,e_2\in E(G)\setminus(E(C)\cup \delta_G(C))$. Otherwise, we say that it is a type-3 crossing. We now bound the expected number of type-1 and type-3 crossings; type-2 crossings will eventually be eliminated.

In order to bound the expected number of type-1 crossings, consider any crossing $(e_1,e_2)_p$ of $\phi^*$  with $e_1,e_2\in E(C)$. Recall that this crossing may be responsible for at most $n'_{e_1}\cdot n'_{e_2}\leq (n'_{e_1})^2+(n'_{e_2})^2$ type-1 crossings of $\psi'$, and moreover, $(e_1,e_2)_p\in \chi(C)$ must hold. Since we have assumed that $C\in \lset^{\gd}$, for every edge $e\in E(C)$, $\expect{(n'_e)^2}=\expect{(\cong_G(\qset,e))^2} \leq \beta$. Therefore, the expected number of type-1 crossings of $\psi'$ for which $(e_1,e_2)_p$ is responsible for is at most $\expect{(n'_{e_1})^2+(n'_{e_1})^2}\leq 2\beta$.
 Overall, the total expected number of type-1 crossings in $\psi'$ is then bounded by $O(\beta\cdot |\chi(C)|)$.

In order to bound the expected number of type-3 crossings, consider any crossing $(e_1,e_2)_p$ of $\phi^*$, and assume that neither  $e_1,e_2\in E(C)$ nor $e_1,e_2\in E(G)\setminus (E(C)\cup \delta_G(C))$ holds. Recall that this crossing may be repsonsible for at most $n'_{e_1}\cdot n'_{e_2}$ type-3 crossings of $\psi'$. Moreover, $n'_{e_1},n'_{e_2}$ are independent random variables, and the expected value of each such variable is at most $\beta$. Therefore, the expected number of type-3 crossings of $\psi'$ for which crossing $(e_1,e_2)_p$ of $\phi^*$ is responsible for is at most $\beta^2$.
Notice that one of the edges $e_1,e_2$ lies in $E(C)\cup \delta_G(C)$, and so
$(e_1,e_2)_p\in \chi(C)$ must hold. Overall, the total expected number of type-3 crossings in $\psi'$ is then bounded by $O(\beta^2\cdot |\chi(C)|)$.

We conclude that the total expected number of type-1 and type-3 crossings in $\psi'$ is at most $O(\beta^2\cdot |\chi(C)|)$.

Next, we construct a set $\hat \qset'=\set{\hat Q'(e)\mid e\in E'}$ of edge-disjoint paths in graph $H'$, routing the edges of $E'$ to vertex $u'$. In order to do so, we assign, for every path $Q'(e)\in \qset'$, for every edge $e'\in E(Q'(e))\setminus\set{e}$, a copy of edge $e'$ to path $Q'(e)$. Observe that path set $\qset'$ in graph $G$ was denoted by $\qset_0$ in Step 1 of the algorithm, and, as observed before, for every edge $e'\in E(G)\setminus (E(C)\cup \delta_G(C))$, $n_{e'}=n'_{e'}$ and so $J(e')=J'(e')$. For each such edge $e'\in E(G)\setminus (E(C)\cup \delta_G(C))$, we assign a distinct copy of $e'\in J'(e')$ to every path in $\qset'$ that contains $e'$. We ensure that this assignment is exactly the same as the assignment done in Step 1 of the algorithm. For each edge $e\in E'$, we then obtain a path $\hat Q'(e)$ in graph $H'$ from path $Q'(e)$ by replacing each edge $e'\in E(Q'(e))\setminus\set{e}$ with the copy of $e'$ that was assigned to $e'$. We then denote $\hat \qset'=\set{\hat Q'(e)\mid e\in E'}$. From our construction, $\hat \qset'$ is a set of edge-disjoint paths in graph $H'$, routing the edges of $E'$ to vertex $u'$, and all inner vertices on the paths in $\hat \qset'$ are disjoint from $C$. Moreover, from our construction, $\hat \qset'=\hat \qset_0$ -- the set of edge-disjoint paths in $H$ that we have constructed in Step 1 of the algorithm.

We also construct a set $\hat \qset=\set{\hat Q(e)\mid e\in E'}$ of edge-disjoint paths in graph $H'$, routing the edges of $E'$ to vertex $u$. In order to do so, we assign, for every path $Q(e)\in \qset$, for every edge $e'\in E(Q(e))\setminus\set{e}$, a copy of edge $e'$ to path $Q(e)$. 
Consider now any edge $e'\in E(C)$. If edge $e'$ is not incident to vertex $u$, then we assign every copy of $e$ in $J'(e')$ to a distinct path of $\qset$ containing $e'$ arbitrarily. If edge $e'$ is incident to vertex $u$, then we perform the assignment more carefully, using the same procedure that we used for every cluster $C_i\in \wset^{\gd}$ in order to assign, for each edge $e'$ incident to $u_i$, copies of $e'$ to paths in $\qset_i$; we do not repeat the description of the procedure there. For each edge $e\in E'$, we  obtain a path $\hat Q(e)$ in graph $H'$ from path $Q(e)$ by replacing each edge $e'\in E(Q(e))\setminus\set{e}$ with the copy of $e'$ that was assigned to $e'$. We then denote $\hat \qset=\set{\hat Q(e)\mid e\in E'}$. 
Recall that $\oset=\oset(C)$ is a circular ordering of the edges of $E'=\delta_G(C)$ that is guided by the set $\qset$ of paths and the rotation system $\Sigma$. As in Step 1 of the algorithm, our assignment of edges incident to vertex $u$ ensures the following crucial property. Denote $E'=\set{e_1,\ldots,e_k}$, where the edges are indexed according to the ordering $\oset$. For each such edge $e_j$, let $e'_j$ be the last edge on path $\hat Q(e_j)$. Then the images of edges $e'_1,\ldots,e'_k$ enter the image of the vertex $u$ in the drawing $\psi'$ of $H'$ in the order of their indices. 

\paragraph{Uncrossing.}

We view every path of $\hat \qset'$ as being directed towards the vertex $u'$. We use the algorithm from \Cref{thm: new type 2 uncrossing} in order to compute a type-2 uncrossing, that produces, for every edge $e_j\in E'$, a directed curve $\hat \gamma'(e_j)$, that connects the image of the endpoint of $e_j$ lying in $C$ to $u'$. Recall that we are guaranteed that the curves in the resulting set $\hat \Gamma'=\set{\hat \gamma'(e_j)\mid e_j\in E'}$ do not cross each other, and each such curve is aligned with the drawing of graph $\bigcup_{\hat Q'(e_{\ell})\in \hat \qset'}\hat Q'(e_{\ell})$ induced by $\psi'$ (which is identical to the drawing induced by $\psi$).
Notice that the steps that we have followed in constructing the set $\hat \Gamma'$ of curves are identical to those we followed in order to construct the set $\Gamma_0$ of curves, and so the resulting two sets of curves are identical. In particular, the order in which the curves of $\hat \Gamma$ enter the image of vertex $u'$ in $\psi'$ is exactly $\oset'$. 

We now construct another set of curves, $\hat \Gamma=\set{\hat \gamma(e_j)\mid e_j\in E'}$, by letting, for each edge $e_j\in E'$, $\hat \gamma(e_j)$ be the image of the path $\hat Q(e_j)\in \hat \qset$ in the drawing $\psi'$ of $H'$. From our construction of the set $\hat \qset$ of paths, the order in which the curves of $\hat \Gamma$ enter the image of vertex $u$ in $\psi$ is exactly $\oset$. For every edge $e_j\in E'$, we then let $\gamma(e_j)$ be a curve, connecting the images of $u$ and $u'$ in $\psi'$, obtained by combining the curves $\hat \gamma(e_j)$ and $\hat \gamma'(e_j)$. In order to combine the two curves, let $y_j$ be the endpoint of $e_j$ lying in $C$, and let $p_j$ be the unique point on $\psi'(e_j)$ that lies on the boundary of the tiny $y_j$-disc $D_{\psi'}(y_j)$. We truncate curve $\hat\gamma'(e_j)$ so it connects point $p_j$ to the image of vertex $u'$, and we truncate the curve $\hat \gamma(e_j)$, so it connects point $p_j$ to the image of vertex $u$. We then concatenate the resulting two curves to obtain the curve $\gamma(e_j)$.

Consider the resulting set $\Gamma=\set{\gamma(e_j)\mid e_j\in E'}$ of curves. From the above discussion, the curves enter the image of $u'$ in $\psi'$ according to the ordering $\oset'$, and they enter the image of $u$ according to the ordering $\oset$. The total number of crossings between the curves in $\Gamma$ is bounded by $\cro(\psi')$. We call all such crossings \emph{primary crossings}. Recall that the expected number of primary crossings is at most $O(\beta^2\cdot |\chi(C)|)$. However, the curves in $\Gamma$ may not be in  general position. This is since some vertices $x\in V(C)$ may lie on a number of paths in $\hat \qset$. In the next step we perform ``nudging'' around such vertices, to ensure that the resulting curves are in general position.

\paragraph{Nudging.}
The nudging procedure and its analysis are identical to those from Step 2 of the algorithm. We only need to perform nudging of the curves in $\Gamma$ around vertices $x\in V(C)\setminus\set{u}$.

We process each vertex $x\in V(C)\setminus \set{u}$ one by one. Consider any such vertex $x$, and let $\qset^x\subseteq \hat\qset$ be the set of all paths containing vertex $x$. Note that $x$ must be an inner vertex on each such path. For convenience, we denote $\qset^x=\set{\hat Q(e_1),\ldots,\hat Q(e_z)}$. Consider the tiny $x$-disc $D=D_{\psi'}(x)$. For all $1\leq j\leq z$, denote by $s_j$ and $t_j$ the two points on the curve $\gamma(e_j)$ that lie on the boundary of disc $D$. We use the algorithm from \Cref{claim: curves in a disc} to compute a collection $\set{\sigma_1,\ldots,\sigma_z}$ of curves, such that, for all $1\leq j\leq z$, curve $\sigma_j$ connects $s_j$ to $t_j$, and the interior of the curve is contained in the interior of $D$. Recall that every pair of resulting curves crosses at most once, and every point in the interior of $D$ may be contained in at most two curves.
Consider now a pair $\sigma_{\ell},\sigma_{\ell'}$ of curves, and assume that these two curves cross. Recall that, from \Cref{claim: curves in a disc}, this may only happen if the two pairs $(s_{\ell},t_{\ell})$, $(s_{\ell'},t_{\ell'})$ of points cross. Denote by $e_1,e_2$ the two edges that lie on path $\hat Q(e_{\ell})$ immediately before and immediately after vertex $x$, and denote by $e_1',e_2'$ the two edges that lie on path $\hat Q(e_{\ell'})$ immediately before and immediately after vertex $x$. We assume that edges $e_1,e_2$ are copies of edges $\hat e_1,\hat e_2$ of $G$, and similarly, $e_1',e_2'$ are copies of edges $\hat e_1',\hat e_2'$ of $G$, respectively. Assume first that there are four distinct edges in set $\set{\hat e_1,\hat e_1',\hat e_2,\hat e_2'}$. From the fact that the two pairs  $(s_{\ell},t_{\ell})$, $(s_{\ell'},t_{\ell'})$ of points cross, we get that these four edges must appear in $\oset_x\in \Sigma$ in the order $(\hat e_1,\hat e_1',\hat e_2,\hat e_2')$. Since the paths of $\qset$ are non-transversal with respect to $\Sigma$, this is impossible. Therefore, we conclude that paths $Q(e_{\ell}),Q(e_{\ell'})$ must share an edge that is incident to $x$. If $e^*$ is an edge incident to $x$ that the two paths share, then we say that $e^*$ is \emph{responsible for the crossing between $\sigma_{\ell}$ and $\sigma_{\ell'}$}. 

For all $1\leq j\leq z$, we modify the curve $\gamma(e_j)$, by replacing the segment of the curve that is contained in disc $D$ with $\sigma_j$.
Once every vertex $x\in V(C)\setminus \set{u}$ is processed, we obtain the final set $\Gamma^*=\set{\gamma^*(e)\mid e\in E'}$ of curves, which are now guaranteed to be in general position. Notice that as before, the curves in $\Gamma^*$ enter the image of $u'$ according to the ordering $\oset'$, and they enter the image of $u$ according to the ordering $\oset$. It now only remains to bound the expected number of crossings between the curves of $\Gamma^*$.

Notice that the nudging operation may have introduced some new crossings. Each such new crossing must be contained in a disc $D_{\psi'}(x)$, for some vertex $x\in V(C)\setminus\set{u}$. We call all such new crossings \emph{secondary crossings}. We now bound the expected number of secondary crossings.

Consider some vertex $x\in V(C)\setminus\set{u}$. Every secondary crossing that is contained in $\dset_{\psi'}(x)$ is a crossing between a pair $\sigma_{\ell},\sigma_{\ell'}$ of curves that we have defined when processing vertex $x$, and each such crossing was charged to an edge of $G$ that is incident to $x$. If $e$ is an edge of $G$ that is incident to $x$, then there are at most $(\cong_G(\qset,e))^2$ pairs of paths in $\qset^x$ that contain copies of $e$, and each such pair of paths may give rise to a single secondary crossing in $D_{\psi'}(x)$ that is charged to edge $e$. Therefore, the total expected number of secondary crossings is bounded by:
\[ \sum_{e\in E(C)}O(\expect{(\cong_G(\qset,e))^2})\leq O(\beta \cdot|E(C)|),\]
since we have assumed that $C\in \wset^{\gd}$.

Overall, the expected number of crossings between the curves in $\Gamma^*$ is at most  $O(\beta^2\cdot (|\chi(C)|+|E(C)|))$, proving that $\expect{\dist(\oset,\oset')}\leq  O(\beta^2\cdot (|\chi(C)|+|E(C)|))$.

\section{Proofs Omitted from \Cref{sec: routing within a cluster}}
\subsection{Proof of \Cref{thm: basic decomposition of a graph}}
\label{sec: appx-decomposition-good-bad-other}
	Note that, since graph $G$ is connected, $|V(G)|\leq m+1\leq 2m$ must hold. 
Throughout, we use a parameter $\eta'=c\eta\log_{3/2}m\log_2m$, were $c$ is a large enough constant, whose value we set later.
	
	The algorithm maintains a collection $\cset$ of disjoint clusters of $G\setminus T$, such that $\bigcup_{C\in \cset}V(C)=V(G)\setminus T$. Set $\cset$ of clusters is partitioned into two subsets: set $\cset^A$ of active clusters and set $\cset^I$ of inactive clusters. We will ensure that every cluster $C\in \cset^I$ has the $\alpha'$-bandwidth property. Set $\cset^I$ of inactive clusters is, in turn, partitioned into three subsets, $\cset_1^I,\cset_2^I$, and $\cset_3^I$. For every cluster $C\in \cset_3^I$, we will define  a vertex $u(C)\in V(C)$, and an internal $C$-router $\qset(C)$, whose center vertex is $u(C)$, such that the paths in $\qset(C)$ are edge-disjoint. For every cluster 
	$C\in \cset^I_1$, we will ensure that $|E(C)|\leq O(\eta^4\log^8m)\cdot |\delta_G(C)|$ holds. 
	Lastly, for every cluster $C\in \cset^I_2$, we will ensure that 
$\optcro(C)\geq \Omega(|E(C)|^2/(\eta^2\poly\log m))$, and $|E(C)|> \Omega(\eta^4 |\delta_G(C)|\log^8m)$. We start with $\cset^I=\emptyset$, and $\cset^A$ containing a single cluster $G\setminus T$ (note that graph $G\setminus T$ is connected since $G$ is connected and every terminal has degree $1$). The algorithm terminates once $\cset^A=\emptyset$, and once this happens, we return $\cset^I$ as the algorithm's outcome.
	
	In order to bound the number of edges in the contracted graph $E(G_{|\cset})$, we will use edge budgets and vertex budgets, that are defined as follows.

\paragraph{Edge Budgets.}	
	If an edge $e$ belongs to the boundary $\delta_G(C)$ of a cluster $C\in \cset$, then, if $C\in \cset^I$, we set the budget $B_C(e)=1$, and otherwise we set it to be $B_C(e)=\log_{3/2}(|\delta_G(C)|)$. If cluster $C$ is the unique cluster with $e\in \delta_G(C)$, then we set $B(e)=B_C(e)$. If there are two clusters $C\neq C'\in \cset$ with $e\in \delta_G(C)$ and $e\in \delta_G(C')$, then we set $B(e)=B_C(e)+B_{C'}(e)$. Lastly, if no cluster $C\in \cset$ with $e\in \delta_G(C)$ exists, then we set $B(e)=0$.
	
\paragraph{Vertex Budgets.}
	Vertex budgets are defined as follows. For every cluster $C\in \cset^A$, for every vertex $v\in V(C)$, we set the budget $B(v)=\frac{c\deg_C(v)\log_{3/2}m\cdot \log_{2}(|E(C)|)}{8\eta'}$, where $c$ is the constant used in the definition of $\eta'$. The budgets of all other vertices are set to $0$.

\paragraph{Cluster Budgets and Total Budget.}
	For a cluster $C\in\cset$, we define its edge-budget $B^E(C)=\sum_{e\in \delta_G(C)}B_C(e)$, and its vertex-budget $B^V(C)=\sum_{v\in V(C)}B(v)$. The total budget of a cluster $C\in \cset$ is $B(C)=B^E(C)+B^V(C)$, and the total budget in the system is $B^*=\sum_{C\in \cset}B(C)=\sum_{e\in E(G)}B(e)+\sum_{v\in V(G\setminus T)}B(v)$.

	Notice that at the beginning of the algorithm, the budget of every vertex $v\in V(G)\setminus T$ is bounded by:
	 $$\frac{c\cdot \deg_{G\setminus T}(v)\cdot \log_{3/2}m\cdot \log_2|E(G)|}{8\eta'}\leq \frac{\deg_G(v)}{8\eta},$$ 
	 the budget of every edge incident to a vertex in $T$ is at most $\log_{3/2}(|T|)\leq 16\log m$, while the budget of every other edge is $0$. Therefore, the total budget $B^*$ in the system at the beginning of the algorithm is:
\[	\frac{m}{4\eta}+ 16k\log m\leq \frac{m}{\eta},\]
since $k\leq \frac{m}{16\eta\log m}$ from the statement of \Cref{thm: basic decomposition of a graph}.

We will ensure that, throughout the algorithm, the total budget $B^*$ never increases. Since, from the definition, $B^*\geq \sum_{C\in \cset}|\delta_G(C)|$, this ensures that, when the algorithm terminates,  $|E(G_{|\cset})|\leq m/\eta$, so the set $\cset^I$ of clusters and its partition $(\cset_1^I,\cset_2^I,\cset_3^I)$ is a valid output of the algorithm.

As mentioned above, the algorithm starts with $\cset^I=\emptyset$, and $\cset^A$ contains a single cluster -- cluster $G\setminus T$. As long as $\cset^A\neq \emptyset$, we perform iterations, where in each iteration we select an arbitrary cluster $C\in \cset^A$ to process.
We now describe the execution of an iteration in which cluster $C\in \cset^A$ is processed.
The algorithm for processing cluster $C$ consists of three steps, that we describe next.

\paragraph{Step 1: Bandwidth Property.}
In this step we will either establish that $C$ has the $\alpha'$-bandwidth property, or we will partition it into smaller clusters that will replace $C$ in set $\cset^A$.
Let $C^+$ be the augmentation of cluster $C$. Recall that $C^+$ is a graph that is obtained as follows. We start with the graph $G$, and we subdivide every edge $e\in \delta_G(C)$ with a vertex $t_e$, letting $T(C)=\set{t_e\mid e\in \delta_G(C)}$ be this new set of vertices. We then let $C^+$ be the subgraph of the resulting graph induced by $V(C)\cup T(C)$. From \Cref{obs: wl-bw}, cluster $C$ has the $\alpha'$-bandwidth property iff the set $T(C)$ of vertices is $\alpha'$-well-linked in $C^+$. We apply the algorithm  \algsc to graph $C^+$ and terminal set $T(C)$, to obtain an  $\alphasc(m)=O(\sqrt{\log m})$-approximate sparsest cut $(X,Y)$ in graph $C^+$ with respect to the set $T(C)$ of terminals.

We can assume without loss of generality that, for every vertex $t_e\in T(C)$, if $t_e\in X$, and $e'=(t_e,v)$ is the unique edge that is incident to $t_e$ in $C^+$, then $v\in X$ as well (as otherwise we can move $t_e$ to $Y$, making the cut only sparser). Similarly, if $t_e\in Y$, then $v\in Y$ as well.
We assume w.l.o.g. that $|X\cap T(C)|\leq |Y\cap T(C)|$.
 We then consider two cases. First, if $|E(X,Y)|\geq \alpha'\cdot \alphasc(m)\cdot |X\cap T(C)|$, then we are guaranteed that vertex set $T(C)$ is $\alpha'$-well-linked in $C^+$, and therefore cluster $C$ has the $\alpha'$-bandwidth property. Assume now that  $|E(X,Y)|< \alpha'\cdot \alphasc(m)\cdot |X\cap T(C)|$.

 Let $X'=X\setminus T(C)$ and $Y'=Y\setminus T(C)$, so $(X',Y')$ is a partition of $C$. Note that $|T(C)\cap X|=|\delta_G(C)\cap \delta_G(X')|$ and similarly $|T(C)\cap Y|=|\delta_G(C)\cap \delta_G(Y')|$. 
 
 We remove cluster $C$ from $\cset^A$, and we add all connected components of $C[X']$ and $C[Y']$ to $\cset^A$ instead. Observe that we are still guaranteed that $\bigcup_{C'\in \cset}V(C')=V(G)\setminus T$. We now show that the total budget in the system does not increase as the result of this step. 

Since $|X'|,|Y'|<|V(C)|$, it is immediate to verify that, for every vertex $v$ of $C$, its budget may only decrease. The only edges whose budget may increase are the edges of $E_C(X',Y')$. The number of such edges is bounded by $\alpha'\cdot \alphasc(m)\cdot |X\cap T(C)|=\alpha'\cdot \alphasc(m)\cdot |\delta_G(C)\cap \delta_G(X')|$, and the budget of each of them increases by at most $2\log_{3/2}m\leq 8\log m$, so the total increase in the budget of all edges due to this step is bounded by:
\[8\alpha'\cdot \alphasc(m)\cdot |\delta_G(C)\cap \delta_G(X')|\cdot \log m\leq |\delta_G(C)\cap \delta_G(X')|,\]
since $\alpha'=\frac{1}{16\alphasc(m)\cdot \log m}$.

Consider now some edge $e\in \delta_G(C)\cap \delta_G(X')$. Since we have assumed that $|X\cap T(C)|\leq |Y\cap T(C)|$, it is easy to verify that $|\delta_G(X')|\leq 2|\delta_G(C)|/3$. Therefore, if an endpoint of an edge $e\in \delta_G(C)$ belongs to a new cluster $C'\subseteq G[X']$, then the new budget $B_{C'}(e)$ becomes at most:
\[\log_{3/2}(|\delta_G(X')|)\leq \log_{3/2}(|\delta_G(C)|)-1. \]
The total decrease in the global budget due to the edges of 
$\delta_G(X')\cap \delta_G(C)$ is then at least $|\delta_G(C)\cap \delta_G(X')|$. We conclude that overall the global budget does not increase.

We assume from now on that algorithm \algsc returned a cut $(X,Y)$ of $C^+$ with $|E(X,Y)|\geq \alpha'\cdot \alphasc(m)\cdot |X\cap T(C)|$, and so cluster $C$ has the $\alpha'$-bandwidth property.

Assume now that $|E(C)|\leq (\eta')^4|\delta_G(C)|$. From the definition of $\eta'$, we are then guaranteed that $|E(C)|\leq O(\eta^4\log^8m)\cdot |\delta_G(C)|$.
We then remove cluster $C$ from $\cset^A$ and add it to the set $\cset^I$ of inactive clusters, and to the set $\cset^I_1$ of clusters. Therefore, we assume from now on that 
$|E(C)|> (\eta')^4|\delta_G(C)|$.

\paragraph{Step 2: Sparse Balanced Cut.}

In this step, we apply the algorithm from \Cref{cor: approx_balanced_cut} to graph $C$ with parameter $\hat c=3/4$, to obtain a $\hat c'$-edge-balanced cut $(Z,Z')$ of $C$ (where $1/2<\hat c'<1$), whose value is at most 
$O(\alphasc(m))$ times the value of a minimum $3/4$-edge-balanced cut  of $C$. We say that this step is \emph{successful} if $|E_G(Z,Z')|<|E(C)|/\eta'$. Assume first that the step was successful. Then we remove cluster $C$ from set $\cset^A$, and add all connected components of graphs $C[Z],C[Z']$ to set $\cset^A$ insead. We now show that the total budget in the system does not increase as the result of this step. Observe that the budget of every vertex may only decrease, and the same is true for the budget of every edge, except for the edges in set $\delta_G(C)\cup E_G(Z,Z')$. The budget of each such edge may increase by at most $2\log_{3/2}m$, so the total increase in the budgets of all edges is bounded by $(|\delta_G(C)|+|E_G(Z,Z')|)\cdot 2\log_{3/2}m \leq \frac{4|E(C)|\cdot \log_{3/2}m}{\eta'}$ (we have used the fact that $|E(C)|> (\eta')^4|\delta_G(C)|$). We now show that this increase in total budget is compensated by the decrease in the budgets of the vertices of $Z$.

Assume without loss of generality that $|E(Z)|\leq |E(Z')|$.
Recall that for every vertex $v\in Z$, its original  budget is:
$B(v)=\frac{c\deg_C(v)\log_{3/2}m\cdot \log_{2}(|E(C)|)}{8\eta'}$.
From our assumption that $|E(Z)|\leq |E(Z')|$, $\log_{2}(|E(Z)|)\leq \log_{2}(|E(C)|)-1$.
The new budget of vertex $v$ is:
$$B'(v)=\frac{c\deg_Z(v)\log_{3/2}m\cdot \log_{2}(|E(Z)|)}{8\eta'}\leq \frac{c\deg_Z(v)\log_{3/2}m\cdot (\log_{2}(|E(C)|)-1)}{8\eta'}.$$
Therefore, for every vertex $v\in Z$, its budget decreases by at least 
$\frac{c\deg_Z(v)\log_{3/2}m}{8\eta'}$. 

 From the definition of a $(3/4)$-edge-balanced cut, $|E(Z')|\leq \hat c'|E(C)|$, for some universal constant $\hat c'$. In particular: 
\[
\sum_{v\in Z}\deg_Z(v)\geq  |E(C)|-|E(Z')|\geq (1-\hat c')|E(C)|. \label{eq: many edges in Z}
\]

Overall, the budget of the vertices in $Z$ decreases by at least:
\[\frac{c\log_{3/2}m}{8\eta'}\cdot \sum_{v\in Z}\deg_Z(v)\geq \frac{c\log_{3/2}m}{8\eta'}\cdot (1-\hat c')\cdot |E(C)|.\]
Since $\hat c'<1$, and we can set $c$ to be a large enough constant, we can ensure that this is at least $\frac{4|E(C)|\cdot \log_{3/2}m}{\eta'}$, so the overall budget in the system does not increase.

If the current step is successful, then we replace cluster $C$ with the connected components of $C[Z]$ and $C[Z']$ in set $\cset^A$ and continue to the next iteration. Therefore, we assume from now on that the current step was not successful. In other words, the algorithm from  \Cref{cor: approx_balanced_cut} returned a cut$(Z,Z')$ with $|E_G(Z,Z')|\geq |E(C)|/\eta'$. Since this cut is within factor $\alphasc(m)=O(\sqrt{\log m})$ from the minimum $3/4$-edge-balanced cut, we conclude that the value of the minimum $3/4$-edge-balanced cut in $C$ is at least $\rho=\frac{|E(C)|}{\eta'\cdot  \alphasc(m)}$.

From \Cref{lem:min_bal_cut}, if the maximum vertex degree $\Delta$ in graph $C$ is at most $|E(C)|/2^{40}$ and $\optcro(C)\le |E(C)|^2/2^{40}$,
then graph $C$ has a $(3/4)$-edge-balanced cut of value at most $\tilde c\cdot\sqrt{\optcro(C)+\Delta\cdot|E(C)|}$ where $\tilde c>2^{40}$ is some universal constant.
As the size of the minimum $3/4$-balanced cut in $C$ is at least $\rho$, we conclude that either $\Delta\geq |E(C)|/2^{40}$, or $\optcro(C)> |E(C)|^2/2^{40}$, or 
$\sqrt{\optcro(C)+\Delta\cdot|E(C)|}\geq \rho/\tilde c$. The latter can only happen if either $\optcro(C)\geq \rho^2/\tilde c^2$, or $\Delta\geq \rho^2/(\tilde c^2\cdot |E(C)|)$. Substituting the value of $\rho=\frac{|E(C)|}{\eta'\cdot  \alphasc(m)}$, we conclude that either $\optcro(C)\geq \frac{|E(C)|^2}{(\tilde c \eta'\alphasc(m))^2} $, or $\Delta\geq \frac{|E(C)|}{(\tilde c\eta'\alphasc(m))^2}$, or $\Delta\geq \frac{|E(C)|}{2^{40}}$. Note that we can check whether the last two conditions hold efficiently. If they do not hold, then we are guaranteed that $\optcro(C)\geq \frac{|E(C)|^2}{(\tilde c\eta'\alphasc(m))^2 }\geq \frac{|E(C)|^2}{\eta^2\poly\log m }$. Recall that we are also guaranteed that $|E(C)|> (\eta')^4\cdot |\delta_G(C)|=\Omega(\eta^4|\delta^G(C)|\log^8m)$. We then remove cluster $C$ from the set $\cset^A$ of active clusters and add it to set $\cset^I$ of inactive clusters, where it joins the set $\cset^I_2$ of clusters.

From now on we can assume that 
$|E(C)|>\Omega(\eta^4|\delta^G(C)|\log^8m)$, and that either $\Delta\geq \frac{|E(C)|}{(\tilde c \eta'\alphasc(m))^2}$, or $\Delta\geq \frac{|E(C)|}{2^{40}}$ hold. Note that, since $\eta'=c\eta\log_{3/2}m\log_2 m$, and since we can set $c$ to be a large enough constant, we can ensure that $\frac{|E(C)|}{2^{40}}\geq \frac{|E(C)|}{\tilde c^2(\eta')^2\alphasc(m)}$. Therefore, from now on we assume that there is some vertex $v^*$ in graph $C$, whose degree in $C$ is at least $\frac{|E(C)|}{(\tilde c \eta'\alphasc(m))^2}$.
Since we have assumed that $|E(C)|\geq (\eta')^4|\delta_G(C)|$, we get that
$\deg_G(v^*)\geq 8|\delta_G(C)|\eta'$.

\paragraph{Step 3: routing to a vertex}

We consider again the graph $C^+$ that we have defined in Step 1, with the corresponding set $T(C)=\set{t_e\mid e\in \delta_G(C)}$ of terminal vertices.
Using standard Maximum Flow computation, we compute a maximum-cardinality set $\pset$ of edge-disjoint paths, where each path connects a distinct vertex of $T(C)$ to to vertex $v^*$. We now consider two cases. In the first case, $|\pset|=|\delta_G(C)|$. In this case, there is a set $\qset(C)$ of edge-disjoint paths in cluster $C$, routing edges of $\delta_e(C)$ to vertex $v^*$, which can be easily obtained from $\pset$. We then remove cluster $C$ from the set $\cset^A$ of active clusters and add it to set $\cset^I$ of inactive clusters, where it joins cluster set $\cset^I_3$.

Assume now that $|\pset|<|\delta_G(C)|$. From the maximum flow / minimum cut theorem, there is a collection $E'$ of at most $|\delta_G(C)|$ edges in graph $C^+$, such that, in graph $C^+\setminus E'$, there is no path connecting an edge of $\delta_G(C)$ to vertex $u^*$. Let $C'$ be the connected component of $C^+\setminus E'$ containing $v^*$, so $C'\subseteq C$. Note that $\delta_G(C')\subseteq E'$, so there is a set $\qset(C')$ of edge-disjoint paths routing the edges of $\delta_G(C')$ to vertex $v^*$, with all inner vertices of the paths in $\qset(C')$ lying in $C'$; in other words, $\qset(C')$ is an internal $C'$-router. We add cluster $C'$ to set $\cset^I$ and $\cset^I_3$. Next, we delete cluster $C$ from $\cset^A$, and we add every connected component of $C\setminus C'$ as a new cluster to set $\cset^A$.

It now remains to prove that the total budget in the system does not increase as the result of these changes. Note that the budget of every vertex may only decrease, and the budget of every edge, except for the edges of $\delta_G(C)\cup E'$, may also only decrease. The increase in the budget of every edge in $\delta_G(C)\cup E'$ is bounded by $2\log_{3/2}m$. Therefore, the total increase in the budget is bounded by $(|\delta_G(C)|+|E'|)\cdot 2\log_{3/2}m\leq 4|\delta_G(C)|\cdot  \log_{3/2}m$. Note that the budget of vertex $v^*$ was initially at least $\frac{c\deg_C(v)\log_{3/2}m\cdot \log_{2}(|E(C)|)}{8\eta'}$, and it becomes $0$ at the end of this step. Therefore, the decrease in the budget is at least:
\[\frac{c\deg_C(v^*)\log_{3/2}m\cdot \log_{2}(|E(C)|)}{8\eta'}\geq 4|\delta_G(C)|\cdot  \log_{3/2}m,\]
since $\deg_C(v^*)\geq 8|\delta_G(C)|\eta'$. Overall, the budget does not grow.
The algorithm terminates once $\cset^A$ becomes empty, and it returns the set $\cset^I$ of clusters. From our invariants, it is immediate to verify that this set of clusters has all required properties. It remains to establish that the algorithm is efficient. In every iteration of the algorithm, we either add a cluster to set $\cset^I$, or we split a single cluster of $\cset^A$ into at least two clusters. Once a cluster is added to $\cset^I$, it remains there until the end of the algorithm. It is then easy to verify that the number of iterations is bounded by $O(|V(G)|)\le O(m)$, and every iteration can be executed efficiently.

\subsection{Proof of \Cref{obs: opt is small}}
\label{subsec: proof of obs opt is small}

Let $\phi^*$ be the optimal solution to instance $(\tilde C,\Sigma_{\tilde C})$. We can assume w.l.o.g. that every pair of edges cross at most once in $\phi^*$. Denote by $\chi$ the collection of all pairs $e,e'\in E(\tilde C)$ of edges, such that the images of $e$ and $e'$ cross in $\phi^*$.

Assume for now that, for every cluster $W\in \wset^{\light}$, the routers $\qset(W)$ and $\hat \qset(W)$ are fixed, and so the rotation system $\hat \Sigma$ for graph $H$ is fixed as well. For every edge $e\in E(\tilde C)$, we define an integer $N(e)$ as follows. If there is a cluster $W\in \wset^{\light}$ with $e\in E(W)$, then we let $N(e)$ be the number of paths in $\hat \qset(W)$ containing $e$. Therefore, $N(e)=\cong_G(\hat \qset(W),e)\leq \cong_G(\qset(W),e)$. Otherwise, we set $N(e)=1$.
We will prove the following observation:

\begin{observation}\label{obs: opt is small fixed choice}
	Suppose the routers $\hat \qset(W)$ for all clusters $W\in \wset^{\light}$ are fixed, and let $\hat \Sigma$ be the corresponding rotation system for $H$. Then:
	
	$$\optcrors(H, \hat \Sigma)\leq O\left(\sum_{(e,e')\in \chi}N(e)\cdot N(e')\right )+O\left(\sum_{e\in E(\tilde C)}(N(e))^2\right )$$.
\end{observation}

The proof of \Cref{obs: opt is small} immediately follows from \Cref{obs: opt is small fixed choice}. Indeed:

\[
\begin{split}
 \expect{\optcrors(H, \hat \Sigma)}&\leq \expect{\sum_{(e,e')\in \chi}O(N(e)\cdot N(e'))+\sum_{e\in E(\tilde C)}O((N(e))^2)}\\
 &\leq O\left (\sum_{(e,e')\in \chi}\expect{N(e)\cdot N(e')}\right )+O\left( \sum_{e\in E(\tilde C)}\expect{(N(e))^2}  \right )\\
 &\leq O\left (\sum_{(e,e')\in \chi}\left(\expect{(N(e))^2}+\expect{ (N(e'))^2}\right )\right )+O\left( \sum_{e\in E(\tilde C)}\expect{N(e))^2}  \right ).
\end{split} \]

Consider now some edge $e\in E(\tilde C)$. Assume first that there is some cluster $W\in \wset^{\light}$ with $e\in E(W)$. Then, as observed above, $N(e)\leq \cong_G(\qset(W),e)$. Since $\qset(W)$ is a router of $\Lambda_G(W)$ that is drawn from the distribution $\dset(W)$, and since cluster $W$ is $\beta_i$-light with respect to $\dset(W)$, we get that: 

$$\expect[\qset(W)\sim \dset(W)]{(N(e))^2}\leq \expect[\qset(W)\sim\dset(W)]{(\cong_{G}(\qset(W),e))^2}\leq \beta_{i}.$$

Otherwise, $e\in E(\tilde C)\setminus\left(\bigcup_{W\in \wset^{\light}}E(W)\right)$, and so $N(e)=1$ holds. Overall, for every edge $e\in E(\tilde C)$, $\expect{(N(e))^2}\leq \beta_i$. Therefore, we get that:

\[ \expect{\optcrors(H, \hat \Sigma)}\leq|\chi|\cdot O(\beta_i)+|E(\tilde C)|\cdot O(\beta_i)= O\left(\beta_i\cdot\left(\optcrors(\tilde C,\Sigma_{\tilde C})+|E(\tilde C)|\right )\right ).\]

In order to complete the proof of \Cref{obs: opt is small}, it is now enough to prove \Cref{obs: opt is small fixed choice}.

\newpage
\begin{proofof}{\Cref{obs: opt is small fixed choice}}
The proof uses arguments that are very similar to those used in the proof of \Cref{lem: disengagement final cost}, and more specifically in the proof of \Cref{claim: cost of cluster instance}. Similar argument are used in several places throughout this paper, so we only provide a proof sketch here. We start with the optimal solution $\phi^*$ to instance $(\tilde C,\Sigma_{\tilde C})$, and then gradually transform it to obtain a solution $\hat\psi$ to instance $(H,\hat \Sigma)$.

Let $C^*$ be the graph that is obtained from $\tilde C$ as follows. We let $V(C^*)=V(\tilde C)$. For every edge $e=(u,v)\in E(\tilde C)$ with $N(e)>0$, we add a set $J(e)$ of $N(e)$ parallel edges $(u,v)$ to graph $C^*$. We call the edges of $J(e)$ \emph{copies of edge $e$}. We can modify the solution $\phi^*$ to instance $\tilde C$ to obtain a drawing $\psi$ of graph $C^*$ in a natural way: the images of all vertices of $\tilde C$ remain unchanged. For every edge $e\in E(\tilde C)$, we draw the images of the edges of $J(e)$ in $\psi$ in parallel very close to the original image of edge $e$ in $\phi^*$. It is immediate to verify that the number of crossings in the resulting drawing $\psi$ of graph  $C^*$ is bounded by $\sum_{(e,e')\in \chi}N(e)\cdot N(e')$.

Consider now some cluster $W\in \wset^{\light}$. We will now define, for every edge $e\in \delta_{\tilde C}(W)$, a curve $\gamma^*(e)$, that will serve as the image of $e$ in the solution $\hat \psi$ to instance $(H,\hat \Sigma)$ that we construct. Note that for each edge $e\in \delta_{\tilde C}(W)$, $N(e)=1$, and so set $J(e)$ of edges contains exactly one copy of edge $e$. We do not distinguis between edge $e$ and its unique copy in $J(e)$.

We use the internal $W$-router $\hat \qset(W)$ in graph $G$, in order to define a collection $\hat \qset'(W)=\set{\hat Q'(e)\mid e\in \delta_{\tilde C}(W)}$ of edge-disjoint paths in graph $C^*\cup \delta_{\tilde C}(W)$, such that, for every edge $e\in \delta_{\tilde C}(W)$, path $\hat Q'(e)$ originates at edge $e$, terminates at vertex $u(W)$ (the center vertex of the router $\hat \qset(W)$), and all internal vertices of $\hat Q'(e)$ lie in $V(W)$. In order to obtain the collection $\hat \qset'(W)$ of paths from the router $\hat \qset(W)$, for every edge $e\in E(W)$ with $N(e)>0$, we assign every copy of $e$ in $J(e)$ to a distinct path of $\hat \qset(W)$ that contains the edge $e$.

Consider any edge $e\in \delta_{\tilde C}(W)$, and denote $e=(x_e,y_e)$, where $x_e\in V(W)$. Initially, we let $\gamma(e)$ be the image of the path $\hat Q'(e)$ in the drawing $\psi$ of $C^*$. Let $\Gamma(W)=\set{\gamma(e)\mid e\in \delta_{\tilde C}(W)}$ be the resulting collection of curves. Note that, for every edge $e\in \delta_{\tilde C}(W)$, curve $\gamma(e)$ connects the image of vertex $y_e$ in drawing $\phi^*$ to the image of vertex $u(W)$ in the same drawing.

We are now ready to construct an initial drawing $\psi'$ of the graph $H$. For regular every vertex $v\in V(H)\cap V(\tilde C)$, the image of $v$ in $\psi'$ remains the same as in $\phi^*$ (and in $\psi$). For every edge $e$ of $H$ whose both endpoints are regular vertices, the image of $e$ remains the same as in $\phi^*$ (and the same as in $\psi$). Consider now some cluster $W\in \wset^{\light}$. The image of the supernode $v_W$ in drawing $\psi'$ is the image of vertex $u(W)$ in $\phi^*$ (and in $\psi$). For every edge $e\in \delta_{\tilde C}(W)$, the initial image of edge $e$ is the curve $\gamma(e)$. Lastly, since the degree of every terminal $t\in T$ is $1$ in $H$, we can add these terminals and their adjacent edges to the current drawing without increasing the number of crossings. We note that the resulting drawing $\psi'$ of graph $H$ may not be a valid drawing. This is since, whenever there is a cluster $W\in \wset^{\light}$ and a vertex $x\in V(W)$ that lies on more than two paths of $\hat \qset(W)$, then point $p=\phi^*(x)$ belongs to more than two curves of $\Gamma(W)$, and hence more than two edges cross at point $p$. In order to overcome this difficulty, for every cluster $W\in \wset^{\light}$, for every vertex $x\in V(W)$ that lies on at least two paths of $\hat \qset(W)$, we perform a nudging operatio of the curves in $\Gamma(W)$ in the vicinity of vertex $x$ (see \Cref{sec: curves in a disc}). The curves are modified locally inside the tiny $x$-disc $D_{\phi^*}(x)$ to ensure that every point of  $D_{\phi^*}(x)$ lies on at most two curves. Consider now a pair $e_1,e_2\in \delta_{\tilde C}(W)$ of edges. Let $\hat Q(e_1),\hat Q(e_2)$ be the paths of $\hat \qset(W)$ that originate at $e_1$ and $e_2$, respectively, and assume that both paths contain vertex $x$. From the definition of the nudging procedure (see also \Cref{claim: curves in a disc}), and since the paths in $\hat \qset(W)$ are non-transversal with respect to $\Sigma_{\tilde C}$, the curves $\gamma(e_1),\gamma(e_2)$ that are obtained at the end of the nudging operation may only cross inside disc $D_{\phi^*}(x)$ if some edge $e^*\in\delta_{\tilde C}(x)$ lies on both $\hat Q(e_1)$ and $\hat Q(e_2)$. We say that edge $e^*$ is \emph{responsible} for this crossing of $\gamma(e_1)$ and $\gamma(e_2)$.

For every edge $e\in \delta_{\tilde C}(W)$, let $\gamma'(e)$ be the curve that is obtained after the nudging operation is performed for every vertex $x\in V(W)$ that belongs to at least two paths of $\hat \qset(W)$, and denote $\Gamma'(W)=\set{\gamma'(e)\mid e\in \delta_{\tilde C}(W)}$. For every edge $e\in \delta_{\tilde C}(W)$, we replace the image of edge $e$ in the current drawing with the curve $\gamma'(e)$.

Consider the drawing $\psi'$ of graph  $H$ that is obtained after all clusters $W\in \wset^{\light}$ are processed. It is now easy to verify that $\psi'$ is a valid drawing of graph $H$. We partition the crossigns of drawing $\psi'$ into two types: a crossing is of type 1 if it is present in drawing $\psi$, and it is of type 2 otherwise. Equivalently, a crossing $(e,e')_{p}$ of $\psi'$ is of type 2 if there is a cluster $W\in \wset^{\light}$, and a vertex $x\in V(W)$, such that the crossing point $p$ lies in $D_{\phi^*}(x)$.

The number of type-1 crossings in $\psi'$ remains bounded by $\cro(\psi)\leq \sum_{(e,e')\in \chi}N(e)\cdot N(e')$. In order to bound the number of type-2 crossings, observe that for every cluster $W\in \wset^{\light}$, vertex $x\in V(W)$ and edge $e\in \delta_{\tilde C}(x)$, the number of type-2 crossings for which edge $e$ may be responsible is at most $(\cong_{\tilde C}(\hat \qset(W),e))^2=(N(e))^2$. Overall, we get that $\cro(\psi')\leq \sum_{(e,e')\in \chi}N(e)\cdot N(e')+\sum_{e\in E(\tilde C)}(N(e))^2$.

Observe that for every regular vertex $x\in V(H)\cap V(\tilde C)$, drawing $\psi'$ of $H$ obeys the rotation $\oset_x\in \hat \Sigma$ (which is identical to the rotation of $x$ in $\Sigma_{\tilde C}$). However, it is possible that for some supernodes $v_W$, the rotation $\oset_{v_W}\in \hat \Sigma$ is not obeyed by $\psi'$. We fix this issue by introducing at most $O\left (\sum_{e\in E(\tilde C)}(N(e))^2\right )$ additional crossings, as follows.

Consider a cluster $W\in \wset^{\light}$, and denote $\delta_{\tilde C}(u(W))=\set{a_1,a_2,\ldots,a_r}$, where the edges are indexed according to their order in the rotation $\oset_{u(W)}\in \Sigma_{\tilde C}$. For all $1\leq i\leq r$, let $\qset_i\subseteq \hat \qset(W)$ be the set of paths whose last edge is $a_i$. Denote by $A_i\subseteq \delta_{\tilde C}(W)$ the set of edges $e$, for which the unique path $\hat Q(e)\in \hat \qset(W)$ that originates at $e$ terminates at edge $a_i$; in other words, $\hat Q(e)\in \qset_i$. Denote by $\Gamma'_i\subseteq \Gamma'(W)$ the set of the images of the edges of $A_i$ in the current drawing $\psi'$. Let $\oset'$ be that the circular order of the edges of $\delta_{\tilde C}(W)=A_1\cup A_2\cup\cdots\cup A_r$, in which their images in $\psi'$ enter the image of $v_W$.
Then for all $1\leq i\leq r$, the edges of $A_i$ appear consecutively in $\oset'$ in some arbitrary order, while the edges lying in different groups of $A_1,A_2,\ldots,A_r$ appear in $\oset'$ in the natural order of the indices of these groups. 

Recall that the rotation $\oset_{v_W}\in \hat \Sigma$ is a circular ordering of the edges of $\delta_{\tilde C}(W)=A_1\cup A_2\cup\cdots\cup A_r$ that is guided by the paths of $\hat \qset(W)$. In this ordering, 
for all $1\leq i\leq r$, the edges of $A_i$ appear consecutively  in some arbitrary order, while the edges lying in different groups of $A_1,A_2,\ldots,A_r$ appear in $\oset_{v_W}$ in the natural order of the indices of these groups. In other words, the only difference between the orderings $\oset'$ and $\oset_{v_W}$ is that, for all $1\leq i\leq r$, the edges of $A_i$ may appear in different order in the two orderings. Therefore, $\dist(\oset',\oset_{v_W})\leq \sum_{i=1}^r|A_i|^2=\sum_{i=1}^r(\cong_{\tilde C}(\hat \qset(W),e_i))^2\leq \sum_{i=1}^r(N(e_i))^2$. For all $1\leq i\leq r$, we slightly modify the curves of $\Gamma'_i$ inside the tiny $v_W$-disc $D_{\psi'}(v_W)$ to ensure that the images of the edges of $\delta_{\tilde C}(W)$ enter the image of vertex $v_W$ in the circular order $\oset_{v_W}$. From the above discussion, this can be done by introducing at most $\sum_{i=1}^r(N(e_i))^2$ new crossings. Once every cluster $W\in \wset^{\light}$ is processed, we obtain a valid solution $\hat \psi$ to instance $(H,\hat \Sigma)$, with $\cro(\hat \psi)\leq O\left(\sum_{(e,e')\in \chi}N(e)\cdot N(e')+\sum_{e\in E(\tilde C)}(N(e))^2\right )$.

\end{proofof}

\section{Proofs Omitted from \Cref{sec: main disengagement}}

\subsection{Proof of \Cref{claim: path length in decomposition tree}} 
\label{subsec: bounding tree height}
Consider any cluster $R\in \lset$. Let $N(R)$ denote the number of clusters $C\in \cset$ with $C\subseteq R$. Clearly, $N(R)\leq |\cset|\leq m$ must hold.

Assume first that $R\neq G$, and vertex $v(R)$ is unmarked in the tree $\tau(\lset)$. Let $R'$ be the parent-cluster of $R$, that is $v(R')$ is the parent vertex of $v(R)$ in $\tau(\lset)$. In this case, the algorithm from  \Cref{thm: construct one level of laminar family}, when applied to the graph $G'$ corresponding to cluster $R'$, returned a type-2 legal clustering $\rset$ of $G'$, with $R\in \rset$, and moreover, $R$ is not the distinguished cluster $R^*$. Let $\cset'$ denote the set of all clusters $C\in \cset$ with $C\subseteq R'$, so that $N(R')=|\cset'|$. Recall that, from the definition of type-2 legal clustering, the distinguished cluster $R^*$ must contain at least $\floor{\left(1-1/2^{(\log m)^{3/4}}\right )|\cset'|}$ clusters of $\cset'$. Therefore, if $N(R')\geq  2^{(\log m)^{3/4}}$, then $R$ may contain at most $2N(R')/2^{(\log m)^{3/4}}$ clusters of $\cset'$, that is, $N(R)\leq 2N(R')/2^{(\log m)^{3/4}}$. Otherwise, $N(R)\leq 1$ must hold. Consider now any root-to-leaf path $P$ in three $\tau(\lset)$, and assume that $R_1,R_2,\ldots,R_z$ is the sequence of unmarked clusters whose corresponding vertices appear on the path in this order. Then, for all $1\leq i<z$, $N(R_{i+1})\leq \ceil{2N(R_i)/2^{(\log m)^{3/4}}}$, and so $z\leq O\left (\log^{3/4}m\right )$ must hold. 

Next, we consider a cluster $R\in \lset\setminus\set{G}$, whose corresponding vertex $v(R)$ in tree $\tau(\lset)$ is marked. Let $R'$ be the parent-cluster of $R$. Note that two cases are possible. The first case is that the algorithm from \Cref{thm: construct one level of laminar family} was applied to the graph corresponding to $R'$, and it returned a type-1 legal clustering $\rset$ with $R\in \rset$. In this case, the theorem guarantees that $N(R)\leq $\floor{\left(1-1/2^{(\log m)^{3/4}}\right )N(R')}$ $. In the second case, there is a parent-cluster $R''$ of cluster $R'$, to which the algorithm from \Cref{thm: construct one level of laminar family} was applied, and it returned a type-2 legal clustering $\rset$, with $R'\in \rset$, such that $R'=R^*$ is the distinguished cluster of the decomposition, and cluster $R$ lies in the type-1 legal clustering $\rset'$ of the graph corresponding to cluster $R'$. In this latter case, from the definition of type-2 legal clustering, we are guaranteed that $N(R)\leq \floor{\left(1-1/2^{(\log m)^{3/4}}\right )N(R'')}$. Therefore, if we consider any root-to-leaf path $P$ in the tree $\tau(\lset)$, and we let $R_1',R_2',\ldots, R_{y}'$ be the sequence of marked clusters whose corresponding vertices lie on $P$ in this order, then for all $1< i<\floor{y/2}$, $N(R'_{2i})\leq \left(1-1/2^{(\log m)^{3/4}}\right )N(R'_{2i-2})$. Since $N(G)\leq m$, we get that $y\leq O\left(2^{(\log m)^{3/4}}\cdot \log m\right )\leq 2^{O((\log m)^{3/4})}$. 	

\subsection{Proof of \Cref{claim: compose distributions}}
\label{subsec: external routers}
The proof is by induction on the distance from $v(R)$ to the root of the tree $\tau(\lset)$. Recall that, for $R=G$, we set $\dset''(R)$ to assign probability $1$ to an empty set of paths. 

If $v(R)$ is the child vertex of $v(G)$, then consider the graph $G'$, that is obtained from $G$ by adding a new special vertex $v^*$ to it, that connects with an edge to some arbitrary vertex $v_0$. Recall that, when we applied the algorithm from \Cref{thm: construct one level of laminar family} to this graph $G'$, it computed a legal clustering $\rset$ of $G'$, with $R\in \rset$, together with a distribution $\dset'(R)$ over the sets of paths in $\Lambda'_{G'}(R)$, such that, for every edge $e\in E(G')\setminus E(R)$, $\expect[\qset'(R)\sim\dset'(R)]{\cong_{G'}(\qset'(R),e)}\leq \beta$.
Consider any external router $\qset'(R)\in \Lambda'_{G'}(R)$ that is assigned a non-zero probabilty. Let $u$ be the vertex at which every path of $\qset'(R)$ terminates. If $u\neq v^*$, then, since vertex $v^*$ has degree $1$ in $G'$, no path in $\qset'(R)$ contains the vertex $v^*$, and so the paths of $\qset'(R)$ lie in $G$. Otherwise, by removing the last vertex from each path in $\qset'(R)$, we obtain a new set $\qset''(R)$ of paths in $\Lambda'_G(R)$, such that, for every edge $e\in E(G)$, $\cong_G(\qset''(R),e)\leq \cong_G(\qset'(R),e)$. Therefore, we can transform $\dset'(R)$ into a distribution $\dset''(R)$ over the family $\Lambda'_{G}(R)$ of external $R$-routers, such that, for every edge $e\in E(G)\setminus E(R)$, $\expect[\qset'(R)\sim\dset''(R)]{\cong_{G}(\qset'(R),e)}\leq \beta$.

Next, we consider any cluster $R\in \lset$, such that the distance from $v(R)$ to $v(G)$ in tree $\tau(\lset)$ is greater than $1$. Let $R'$ be the parent-cluster of $R$, and let $G'$ be the graph obtained from $G$ by contracting all vertices of $V(G)\setminus V(R')$ into a special vertex $v^*$.
Recall that, from  \Cref{thm: construct one level of laminar family}, we have obtained a distribution  $\dset'(R)$ over external routers in $\Lambda'_{G'}(R)$, such that, for every edge $e\in E(G')\setminus E(R)$, $\expect[\qset'(R)\sim\dset'(R)]{\cong_{G'}(\qset'(R),e)}\leq \beta$. Recall that, if $v(R)$ is a marked vertex, every router $\qset'(R)\in \Lambda'_{G'}(R)$, to which $\dset'(R)$ assigns a non-zero probability, is careful with respect to $v^*$, that is, the paths in $\qset'(R)$ cause congestion at most $1$ on every edge $e\in \delta_{G'}(v^*)$.

Let $i$ denote the number of unmarked vertices on the path connecting $v(R')$ to the root of $\tau(\lset)$. From the induction hypothesis, we have computed a distribution  $\dset''(R')$ over the external routers in $\Lambda'_G(R')$, such that, for every edge $e\in E(G)\setminus E(R')$, $\expect[\qset'(R')\sim\dset''(R')]{\cong_{G}(\qset'(R'),e)}\leq \beta^{i+1}$.

We now compute the desired distribution a distribution  $\dset''(R)$ over the external routers in $\Lambda'_G(R)$, such that, for every edge $e\in E(G)\setminus E(R)$, $\expect[\qset'(R)\sim\dset''(R)]{\cong_{G}(\qset'(R),e)}\leq \beta^{j+1}$, where $j=i$ if vertex $v(R)$ is marked, and $j=i+1$ otherwise. We provide the distribution implicitly, by providing an efficient algorithm for drawing a set $\tilde \qset'(R)$ of paths from the distribution. The algorithm for drawing an external router from the distribution $\dset''(R)$ proceeds as follows.

First, the algorithm draws an external router $\qset'(R)\in \Lambda'_{G'}(R)$ for cluster $R$ in graph $G'$. If the paths in $\qset'(R)$ do not contain the vertex $v^*$, then this is the set of paths that we return. We now assume that at least one path in set $\qset'(R)$ contains vertex $v^*$. We denote by $u'$ the vertex of $G'$ that serves as the last vertex on every path in $\qset'(R)$.

Next, the algorithm draws a router $\qset'(R')\in \Lambda'_{G}(R')$ from the distribution $\dset''(R')$ that we have constructed by the induction hypothesis. We denote by $u''$ the vertex of $G$ that serves as the last vertex on every path in $\qset'(R')$. The final set $\tilde \qset'(R)$ of paths that the algorithm returns is constructed by combining the sets $\qset'(R)$ and $\qset'(R')$ of paths, as follows.

We consider two cases. The first case is when $u'=v^*$. In this case, for every edge $e\in \delta_{G'}(R)=\delta_G(R)$, the unique path $Q(e)\in \qset'(R)$ that as $e$ as its first edge terminates at vertex $v^*$. We denote by $e'$ the last edge on path $Q(e)$. Note that edge $e'$, that is incident to vertex $v^*$ in graph $G'$, corresponds to an edge of $\delta_G(R')$ in graph $G$; we do not distinguish between the two edges. Therefore, there is some path $Q'(e)\in \qset'(R')$, whose first edge is $e'$, and last vertex is $u''$. By concatenating the paths $Q(e)$ and $Q'(e)$, we obtain  path $Q^*(e)$ in graph $G$, connecting edge $e$ to vertex $u''$. We then let $\tilde \qset'(R)=\set{Q^*(e)\mid e\in \delta_G(R)}$ be the final set of paths that the algorithm outputs. 

The second case is when $u'\neq v^*$. In this case, some paths in $\qset'(R)$ may contain vertex $v^*$ as an inner vertex. Consider any path $Q\in \qset'(R)$ that contains the vertex $v^*$, and let $e\in \delta_{G'}(v^*)$ be any edge that is incident to $v^*$ that the path contains. Recall that $e\in \delta_G(R)$, and so there is some path $Q'(e)\in \qset'(R')$, whose first edge is $e$, and last vertex is $u''$. We replace edge $e$ with path $Q'(e)$ on path $Q$. Note that originally path $Q$ must have contained two edges that are incident to $v^*$; denote them by $e$ and $e'$. We have replaced edge $e$ with a path connecting $e$ to vertex $u''$ in graph $G$, and we replace edge $e'$ with a path connecting $e'$ to $u''$ in graph $G$, but we reverse the direction of the path. In this way, we can glue the two paths to each other via the vertex $u''$. Once every path of $\qset'(R)$ containing $v^*$ is processed in this manner, we obtain the final set $\tilde \qset'(R)$ of paths in graph $G$.

This completes the definition of the distribution $\dset''(R)$ over the set $\Lambda'_G(R)$ of external routers for $R$ in $G$. It now remains to analyze the expected congestion on each edge of $E(G)\setminus E(R)$. Fix any edge $e\in E(G)\setminus E(R)$. First, if edge $e$ lies in graph $R'\cup \delta_G(R')$,  then $\cong_{G}(\tilde \qset'(R),e)=\cong_{G'}(\qset'(R),e)$, and so, from the definition of helpful clustering, $\expect[\tilde \qset'(R)\sim\dset''(R)]{\cong_{G}(\tilde \qset'(R),e)}\leq \expect[ \qset'(R)\sim\dset'(R)]{\cong_{G'}( \qset'(R),e)} \leq \beta$.

Assume now that $e\in E(G)\setminus (E(R')\cup \delta_G(R'))$. Note that, if the set $\qset'(R)\in \Lambda'_{G'}(R)$ that was drawn from distribution $\dset'(R)$ is careful with respect to vertex $v^*$, then every edge of $\delta_{G'}(v^*)$ may lie on at most one path of $\qset'(R)\in \Lambda'_{G'}(R)$, and so every path in the set $\qset'(R')$ that was drawn from distribution $\dset''(R')\in \Lambda'_G(R')$ is used by at most one path in $\tilde \qset'(R)$. Therefore:
$\expect[\tilde \qset'(R)\sim\dset''(R)]{\cong_{G}(\tilde \qset'(R),e)}\leq \expect[ \qset'(R')\sim\dset''(R')]{\cong_{G}( \qset'(R'),e)} \leq \beta^{i+1}$ in this case.

Recall that, if vertex $v(R)$ is marked, then every router $\qset'(R)$ that has non-zero probability to be drawn from distribution $\dset'(R)$ is careful with respect to $v^*$, while the number of unmarked vertices on the unique path connecting $v(R)$ to $v(G)$ in tree $\tau(\lset)$ is $i$.

It remains to consider the case where vertex $v(R)$ is unmarked. 
Consider the following two-step process for drawing a router $\tilde \qset'(R)\in \Lambda_G(R)$ from the distribution $\dset''(R)$, that is equivalent to the one described above. In the first step, we select a router $\qset'(R')\in \Lambda_G(R')$ from the distribution $\dset''(R')$. Then, in the second step, we select a router $\qset'(R)\in \Lambda_{G'}(R)$ from distribution  $\dset'(R)$. Lastly, composing the two sets of paths as described above, we obain the final router $\tilde \qset'(R)$.

Fix an edge $e\in E(G)\setminus (E(R')\cup \delta_G(R'))$, and assume that the set of paths $\qset'(R')\in \Lambda_G(R')$ that was chosen from the distribution $\dset''(R')$ causes congestion $z$ on edge $e$. We denote $\qset'(R')=\set{Q(e')\mid e'\in \delta_G(R')}$, where path $Q(e')$ originates at edge $e'$ and terminates at vertex $u''$. Let $E'\subseteq \delta_G(R')$ be the set of all edges $e'$, whose corresponding path $Q(e')$ contains the edge $e$, so  $|E'|=z$. Denoting $E'=\set{e_1,\ldots,e_z}$, and assuming that the path set $\qset'(R')$ is fixed, we can now write:
\[
\expect[\qset'(R)\sim \dset'(R)]{\cong_{G}(\tilde \qset'(R),e)}=\expect[\qset'(R)\sim \dset'(R)]{\sum_{i=1}^z\cong_{G'}(\qset'(R),e_i)}  \leq \beta z.
\]
Recall that $z$ is the congestion caused by the set $\qset'(R')$ of paths on edge $e$. Therefore, overall:
\[\begin{split}
\expect[\tilde \qset'(R) \sim \dset''(R)]{\cong_G(\tilde \qset'(R),e)}
&\leq \beta\cdot \expect[\qset'(R')\sim \dset''(R')]{\cong_G(\qset'(R'),e)}\leq \beta^{i+2},
\end{split}
\]

from the induction hypothesis. Since, in the case that $v(R)$ is unmarked, the number of unmarked vertices on the path connecting $v(R)$ to $v(G)$ in tree $\tau(\lset)$ is $i+1$, this completes the proof of the claim.

\subsection{Proof of \Cref{claim: few edges}}
\label{subsec:appx few edges}
Consider some cluster $R\in \lset$. Let $\rset$ be the set of child-clusters of $R$. Consider the graph $\tilde R=R\setminus\left(\bigcup_{R'\in \rset}R'\right )$. Note that, from the definition of the laminar family, every edge of $E(G)$ may lie in at most one graph in the collection $\set{\tilde R\mid R\in \lset}$.

Observe that collection $\iset_1$ of instances can be defined as $\iset_1=\set{I(R)\mid R\in \lset}$, where $I(R)=(G(R),\Sigma(R))$ is the instance associated with cluster $R$. Graph $G(R)$ is obtained from graph $G$, by first contracting the vertices of $V(G)\setminus V(R)$ into a supernode $v^*$, and then contracting each child cluster $R'$ of $R$ into a supernode $v(R')$. We partition the set of edges of $G(R)$ into two subsets: the first subset, that we call \emph{internal edges}, and denote by $E_1(R)$, is the edge set $ E(\tilde R)$. The second subset, that we call \emph{external edges}, and denote by $E_2(R)$, is the set of all edges that are incident to the supernodes of $G(R)$. From the above discussion, $\sum_{R\in \lset}|E_1(R)|\leq |E(G)|$, as every edge may serve as an internal edge for at most one graph $G(R)$. It now remains to bound the total number of external edges in all graphs in $\set{G(R)\mid R\in \lset}$.

For all $1\leq i\leq \dep(\lset)$, we denote by $\lset_i\subseteq \lset$ the set of all clusters $R\in \lset$, such that vertex $v(R)$ lies at distance exactly $i$ from the root of the tree $\tau(\lset)$. Note that for each cluster $R\in \lset_i$, every external edge $e\in E_2(R)$ corresponds to some edge of the original graph $G$ that has at least one endpoint in cluster $R$. Moreover, since every basic cluster $C\in \cset$ is either contained in $R$ or is disjoint from $R$, each such edge must lie in $\Eout(\cset)$. Since the clusters in set $\lset_i$ are disjoint from each other, we get that $\sum_{R\in \lset_i}|E_2(R)|\leq \sum_{C\in \cset}|\delta_G(C)|\leq  |E(G)|/\mu^{0.1}$, from the statement of \Cref{thm: advanced disengagement get nice instances}.

Since, from  \Cref{claim: path length in decomposition tree}, $\dep(\lset)\leq 2^{O((\log m)^{3/4})}$, while 
$\mu\geq 2^{c^*(\log m )^{7/8}\log\log m}$ for a large enough constant $c^*$, we get that, overall:

\[\sum_{R\in \lset}|E_2(R)|\leq \dep(\lset)\cdot \frac{|E(G)|}{\mu^{0.1}}\leq \frac{2^{O((\log m)^{3/4})}\cdot |E(G)|}{2^{0.1\cdot c^*(\log m)^{7/8}\log\log m}}\leq |E(G)|.
\]


\subsection{Proof of \Cref{obs: subtree to cluster}}
\label{appx: subtree to cluster}
Let $u'$ be the parent-vertex of $u$ in $\tau$. Denote $U=V(\tau_u)$, and $U'=V(\hat H)\setminus U$. Recall that we have denoted by $S$ the cluster of $H$ that is defined by vertex set $U\subseteq V(\hat H)$. 

Let $\rset'\subseteq \rset$ contain all clusters $R$ with $v_R\in U$. Notice that, from \Cref{thm: GH tree properties}, $(U,U')$ is the minimum cut in graph $\hat H$ separating $u$ from $u'$. Let $E'=E_{\hat H}(U,U')$. Observe that, equivalently, $E'=\delta_H(S)$. From the properties of minimum cut, there is a set $\pset$ of edge-disjoint paths in graph $\hat H$, routing the edges of $E'=\delta_{\hat H}(U)$ to vertex $u$, such that all internal vertices on every path of $\pset$ lie in $U$. Similarly, there is a set $\pset'$ of edge-disjoint paths in graph $\hat H$, routing the edges of $E'=\delta_{\hat H}(U')$ to vertex $u'$, such that all internal vertices on every path of $\pset'$ lie in $U'$.

The existence of the set $\pset$ of paths in $\hat H[U]$ immediately implies that cluster $\hat H[U]$ has the $1$-bandwidth property in graph $\hat H$.  Since graph $\hat H[U]$ is precisely the contracted graph of $S$ with respect to cluster set $\rset'$, that is, $\hat H[U]=S_{|\rset'}$, and since every cluster in $\rset'$ has the $\alpha$-bandwidth property, from \Cref{cor: contracted_graph_well_linkedness}, cluster $S$ has the $\alpha$-bandwidth property in graph $H$.

Next, we show an algorithm to construct the desired distribution $\dset'(S)$ over the external routers in $\Lambda'_{H}(S)$. We start with the set $\pset'$ of paths routing the edges of $E'$ to vertex $u'$ in graph $\hat H[U']$. 

Assume first that $u'$ is not a supernode.
Let $\hat H'$ be the graph obtained as follows: we first subdivide every edge $e\in E'$ with a terminal vertex $t_e$, and we let $T=\set{t_e\mid e\in E'}$ be the resulting set of terminals. We then let $\hat H'$ be the subgraph of the resulting graph induced by vertex set $U'\cup T$. Let $\rset''\subseteq \rset$ be the set of all clusters $R$ with $v_R\in U'$. 
We  apply the algorithm from \Cref{claim: routing in contracted graph} to graph $\hat H'$, cluster set $\rset''$, and set $\pset'$ of paths, to obtain a set $\pset''$ of paths in graph $H$, such that, for every edge $e\in \bigcup_{R\in \rset''}E(R)$, the paths of $\pset''$ cause congestion at most $ \ceil{1/\alpha}$, and for every edge $e\in E(H\setminus U)\setminus \left(\bigcup_{R\in \rset''}E(R)\right )$, the paths of $\pset''$ cause congestion at most $1$. 
We are also guaranteed that the paths in $\pset''$ route the edges of $\delta_H(S)$ to vertex $u'$, and all internal vertices on every path in $\pset''$ are disjoint from $U$. Lastly, since the edges incident to the special vertex $v^*$ do not lie in the clusters of $\rset''$, the set $\pset''$ of paths is careful with respect to $v^*$. From the above discussion, $\pset''\in \Lambda'_H(S)$. We then let distribution $\dset'(S)$ assign probability $1$ to the set $\pset''$ of paths.

Assume now that vertex $u'$ is a supernode, and $u'=v_R$ for some cluster $R\in \rset$. We repeat the algorithm from above, except that, when we apply the algorithm from  \Cref{claim: routing in contracted graph} to graph $\hat H'$, we use cluster set $\rset''\setminus \set{R}$ instead of $\rset''$. The resulting set of paths $\pset''$ then routes the edges of $E'$ to the edges of $\delta_G(R)$, and they remain internally disjoint from cluster $S$. As before, the set $\pset''$ of edges is careful with respect to $v^*$, and it causes edge-congestion at most $\ceil{1/\alpha}$. Moreover, every edge of $\delta_G(R)$ participates in at most one path in $\pset''$. We then use the algorithm from \Cref{lem: simple guiding paths} in order to compute a distribution $\dset(R)$ over the internal  $R$-routers in $\Lambda_H(R)$, such that, for every edge $e\in E(R)$, $\expect[\qset\in \dset(R)]{\cong(\qset,e)}\leq  O(\log^4m/\alpha)$ (in order to use the lemma we subdivide every edge in $\delta_G(R)$ with a terminal, and apply the lemma to the augmented cluster $R^+$ together with the resulting set of terminals). In order to define the distribution $\dset'(S)$, we first select a set $\qset\in \Lambda_G(R)$ of paths from the distribution $\dset(R)$, and then concatenate the paths in $\pset''$ with the paths in $\qset$. It is immediate to verify that the resulting distribution $\dset'(S)$ is supported over the external $S$-routers in $\Lambda'_H(S)$, which are careful with respect to $v^*$, and that $\expect[\qset'(S)\sim\dset'(S)]{\cong_{H}(\qset'(S),e)}\leq O(\log^4m/\alpha)$.

\subsection{Proof of \Cref{obs:J wl}}
\label{subsec:J-clusters well-linked} 
Consider some cluster $J\in \jset$, and let $u_0$ be the center node of cluster $J$. It is enough to show that there is a set $\pset$ of paths (internal $J$-router) in graph $\hat H'$, routing all edges of $\delta_{\hat H'}(J)$ to vertex $u_0$, so that every inner vertex on every path in $\pset$ lies in $J$, and the congestion of $\pset$ is at most $O(\log m)$.
Let $\pset^*$ be the set of all paths $P$ in  graph $\hat H'$, such that the first edge on $P$ lies in $\delta_{\hat H'}(J)$, the last vertex of $P$ is $u_0$, and all inner vertices of $P$ lie in  $J$. 

Recall that a flow $f$ defined over a set $\pset'$ of (directed) paths is an assigment of a flow value $f(P)$ to every path $P\in \pset'$. Given such a flow $f$, we say that an edge $e$ sends one flow unit iff the total amount of flow $f(P)$, for all paths $P\in \pset'$ that originate at edge $e$, is $1$. The congestion caused by flow $f$ is the maximum, over all edges $e'$, of $\sum_{\stackrel{P\in \pset':}{e\in P}}f(P)$. 

We show below that there is a flow $f$, defined over the set $\pset^*$ of paths, in which every edge $e\in \delta_{H'}(J)$ sends one flow unit, and the flow causes congestion at most $O(\log m)$. From the integrality of flow, it then follows that  that there is a set $\pset$ of paths routing the edges of $\delta_{\hat H'}(J)$ to vertex $u_0$, so that for every path of $\pset$, every inner vertex on the path lies in $J$, and the congestion of $\pset$ is at most $O(\log m)$, and so cluster $J$ has $\Omega(1/\log m)$-bandwidth property. From now on we focus on defining the flow $f$.

For $1\leq i\leq h$, let $L'_i=L_i\cap V(J)$, and let $J_i$ be the subgraph of $J$ induced by vertex set $\set{u_0}\cup L'_1\cup\cdots\cup L'_{i}$. We also let $J_0$ be the graph that consists of a single vertex -- vertex $u_0$. For all $0\leq i\leq h$, we denote $E_i=\delta_{\hat H'}(J_i)$, and we denote by $\pset^*_i$ the set of  all paths $P$ in graph $\hat H'$, such that the first edge of $P$ lies in $E_i$, the last vertex of $P$ is $u_0$, and all inner vertices of $P$ are contained in $J_i$. 
Additionally, we let $\tilde E_i\subseteq E_i$ be the set of all the edges $e\in E_i$, such that, for some vertex $v\in V(J_i)\setminus\set{u_0}$, $e\in \delta^{\up}(v)$. We let $\tilde \pset^*_i\subseteq \pset^*_i$ be the set of all paths whose first edge lies in $\tilde E_i$. Note that any flow $f_i$ defined over the set $\pset^*_i$ of paths immediately defines a flow $f'_i$ over the set $\tilde \pset^*_i$ of paths, by setting, for every path $P\in \tilde \pset^*_i$, $f'(P)=f(P)$, and setting the flow on all other paths to $0$. We call $f'_i$ the \emph{restriction of flow $f_i$ to the set $\tilde \pset^*_i$ of paths}.
We prove the following claim.

\begin{claim}\label{claim: route level by level}
	For all $1\leq i\leq h$, there is a flow $f_i$, defined over the set $\pset^*_i$ of paths, in which every edge of $E_i$ sends one flow unit, and the total congestion is bounded by $2^{512}\cdot i$. Moreover, if we let $f'_i$ be the restriction of $f_i$ to the set $\tilde \pset^*_i$ of paths, then the congestion caused by $f'_i$ is at most $\left (1+\frac{256}{\log m}\right)^i$.
\end{claim}

Notice that the proof of  \Cref{obs:J wl} immediately follows from \Cref{claim: route level by level}, since $J_h=J$, and flow $f_h$ defines the desired flow in graph $J$, that causes congestion at most $O(h)\leq O(\log m)$. It now remains to prove \Cref{claim: route level by level}.

\begin{proofof}{\Cref{claim: route level by level}}
	The proof is by induction on $i$. The base is when $i=0$. In this case, $\delta_{\hat H'}(J_0)=\delta_{\hat H'}(u_0)$. The set $\pset^*_0$ of paths contains, for every edge $e\in \delta_{\hat H'}(u_0)$, a path $P(e)$ that only consists of the edge $e$ itself. We obtain flow $f_0$ by sending one flow unit on each such path $P(e)$. Note that $\tilde E_0=\emptyset$ in this case, and the resulting flow has congestion $1$.
	
	Assume now that the claim holds for some $0\leq i< h$. We now prove it for $i+1$. 
	
	We partition the set $E_{i+1}$ of edges into two subsets. The first subset, $E'_{i+1}$ is $E_{i+1}\cap E_i$. The second subset, $E''_{i+1}$ contains all remaining edges of $E_{i+1}$. It is easy to verify that, for every edge $e\in E''_{i+1}$, there is some vertex $v\in L'_{i+1}$, with $e\in \delta_{\hat H'}(v)\setminus E_i$.

	
	For every edge $e\in E'_{i+1}$, the flow on the paths that originate from $e$ remains unchanged from $f_i$. In other words, for every path $P\in \pset^*_i$ that starts with edge $e\in E'_{i+1}$ (and hence $P\in \pset^*_{i+1}$), we set $f_{i+1}(P)=f_i(P)$.
	
	Consider now some vertex $v\in L'_i$. Note that, when vertex $v$ was added to cluster $J$, the number of edges connecting $v$ to vertices that belonged to $J$ at that time was at least $|\delta_{\hat H'}(v)|/128$. From the definition of layered well-linked decomposition, $|\delta^{\up}(v)|<|\delta^{\down}(v)|/\log m\leq |\delta_{\hat H'}(v)|/\log m$. Therefore, at the time when $v$ was added to $J$, there were at least $|\delta_{\hat H'}(v)|/256$ edges in $\delta^{\down}(v)$, that connected $v$ to the vertices of $J$. It is immediate to verify that each such edge must lie in $E_i$, and moreover, it must lie in $\tilde E_i$. To conclude, there is a set $E'(v)\subseteq \delta^{\down}(v)$ of at least $|\delta_{\hat H'}(v)|/256$ edges, that lie in $\tilde E_i$. Note that none of these edges may lie in $E_{i+1}$. We now define the flow $f_{i+1}$ that originates at the edges of $\delta_{\hat H'}(v)\cap E''_{i+1}$.
	
	Every edge $e\in \delta_{\hat H'}(v)\setminus E'(v)$ that lies in $E''_{i+1}$, spreads one unit of flow evenly among the edges of $E'(v)$, and then uses the flow that each of these edges sends in $f_i$, in order to reach $u_0$.  In other words, for every edge $e\in \delta_{\hat H'}(v)\cap E_{i+1}''$, for every edge $e'\in E'(v)$, and for every path $P\in \pset^*_i$ whose first edge is $e'$, we consider the path $P'\in \pset^*_{i+1}$, that is obtained by appending the edge $e$ at the beginning of path $P$, and we set $f_{i+1}(P')=f_i(P)/|E'(v)|$. Since each edge $e\in E'(v)$ sends one flow unit in $f_i$, each edge $e\in \delta_{\hat H'}(v)\cap E_{i+1}''$ now sends one flow unit in $f_{i+1}$.
	This completes the definition of the flow $f_{i+1}$. We now analyze the congestion caused by this flow. 
	
	First, the flow on the paths originating at the edges of $E'_{i+1}$ remains unchanged, and causes congestion at most $2^{512}i$. Next, we consider on flow on paths originating at edges of $E''_{i+1}$. Consider again some vertex $v\in L_{i+1}'$, and recall that $|E'(v)|\geq |\delta_{\hat H'}(v)|/256$. Observe that edges of $E'(v)$ must lie in edge set $\tilde E_i$. Since every edge $e\in \delta_{\hat H'}(v)\cap E''_{i+1}$ spreads one unit of flow evenly among the edges of $E'(v)$, each edge of $E'(v)$ is responsible for sending at most $\frac{|\delta_{\hat H'}(v)|}{|E'(v)|}\leq 256$ flow units. In other words, for each edge $e\in E'(v)$, the flow originating at $e$ in $f_i$ is scaled by at most factor $256$ in order to obtain flow $f_{i+1}$. Therefore, the flow $f_{i+1}$ originating at edges of $E''_{i+1}$ causes congestion at most $256\cdot \left (1+\frac{256}{\log m}\right)^i$. Overall, flow $f_i$ causes congestion at most $2^{512}i+256\cdot \left (1+\frac{256}{\log m}\right)^i\leq 2^{512}\cdot (i+1)$, since $i+1\leq h\leq \log m$.
	
	Lastly, we bound the congestion of the flow $f'_{i+1}$, which is the restriction of the flow $f_{i+1}$ to the set $\tilde \pset_i^*$ of paths. We partition the edges of $\tilde E_{i+1}$ into two subsets: $\tilde E'_{i+1}=\tilde E_i\cap \tilde E_{i+1}$, and $\tilde E_{i+1}''$ containing all remaining edges. Observe that for each edge $e\in \tilde E_{i+1}''$, there is some vertex $v\in L'_{i+1}$ with $e\in \delta^{\up}(v)$.
	
	The flow $f'_{i+1}$ that originates at the edges of $\tilde E'_{i+1}$ remains unchanged from $f'_i$, and causes congestion at most $\left (1+\frac{256}{\log m}\right)^i$. In order to bound the congestion caused by the flow $f'_{i+1}$ originating from edges of $\tilde E'_{i+1}$, consider some vertex $v\in L'_{i+1}$. Recall that $|E'(v)| \geq |\delta_{\hat H'}(v)|/256$, while 
	$|\delta^{\up}(v)|<|\delta_{\hat H'}(v)|/\log m\leq 256|E'(v)| /\log m$. Since every edge $e\in \delta^{\up}(v)\setminus E'(v)$ spreads one unit of flow evenly among the edges of $E'(v)$, each edge of $E'(v)$ is responsible for sending at most $\frac{|\delta^{\up}(v)|}{|E'(v)|}\leq \frac{256}{\log m}$ flow units. In other words, for each edge $e\in E'(v)$, the flow originating at $e$ in $f'_i$ is scaled by at most factor $256/\log m$ in order to obtain flow $f'_{i+1}$. Therefore, the flow $f'_{i+1}$ originating at edges of $\tilde E''_{i+1}$ causes congestion at most $\frac{256}{\log m}\cdot \left (1+\frac{256}{\log m}\right)^i$. Overall, flow $f'_{i+1}$ causes congestion at most $\left (1+\frac{256}{\log m}\right)^i+ \frac{256}{\log m}\cdot \left (1+\frac{256}{\log m}\right)^i\leq \left (1+\frac{256}{\log m}\right )^{i+1}$.
\end{proofof}


\subsection{Proof of \Cref{claim: simplifying cluster is enough}}
\label{subsec: simplifying cluster is enough}
We denote by $E'=\delta_{\cH}(S)$. We define a cluster $S'$ in graph $G$, corresponding to cluster $S$, as usual: First, we define vertex set $V(S')$, and then we let $S'$ be the subgraph of $G$ induced by $V(S')$. Vertex set $V(S')$ contains every regular vertex of $S$. Additionally, for every $R$-node $v_{R}\in S$, it contains all vertices of $R$, and for every $J$-node $v_{J'}\in S$, it contains all vertices of the cluster $J'\in \jset'$. 
We start with the following simple observation.

\begin{observation}\label{obs: compute external paths for S'}
	There is an efficient algorithm to compute a distribution $\dset'(S')$ over the set $\Lambda'_G(S')$ of external $S'$-routers in $G$, such that, for every edge $e\in E(G)\setminus E(S')$, 
	$\expect[\qset'(S')\sim \dset'(S')]{\cong(\qset'(S'),e)}\leq  O(\log^{14.5}m)$.
\end{observation}
\begin{proof}
Recall that every cluster
$R\in \rset$ has the $\alpha_1=1/\log^6m$-bandwidth property in graph $G$, and every cluster $J'\in \jset'$ has the $\Omega(1/\log^{9.5} m)$-bandwidth property in $G$. Let $x$ be the vertex of graph $\cH$ that serves as the last vertex on every path of $\pset(S)$. Assume first that $x$ is a regular vertex, that is, $x\in V(G)$. Then we can use the algorithm from \Cref{claim: routing in contracted graph} (by first subdividing every edge of $E'$ with a terminal) to obtain a collection $\pset(S')$ of paths in graph $G$, that is an external $S'$-router, such that the paths in $\pset(S')$ cause congestion $O\left (\log^{10.5}m\right )$. We define a distribution $\dset'(S')$ over the set $\Lambda'_G(S')$ of external $S'$-routers, that assigns probability $1$ to the router $\pset(S')$.

Assume now that $x$ is a supernode, that corresponds to some cluster $A\in \rset''\cup\jset'$. Again, applying the algorithm from  \Cref{claim: routing in contracted graph} (but this time replacing the graph $G$ with the graph that is obtained from $G$ by contracting cluster $A$ into a supernode $x$), we obtain  a collection $\pset(S')$ of paths in graph $G$, routing the edges of $E'=\delta_G(S')$ to the edges of $\delta_G(A)$, with congestion $O( \log^{10.5}m )$, such that the paths in $\pset(S')$ are internally disjoint from both $S'$ and $A$. Moreover, from \Cref{claim: routing in contracted graph} the paths in $\pset(S')$ cause congestion at most $\beta'=O(\log m)$ on the edges of $\delta_G(A)$.
Recall that cluster $A$ must have the $\Omega(1/\log^{9.5} m)$-bandwidth property in $G$. By applying the algorithm from \Cref{lem: simple guiding paths} to cluster $A$, we obtain a distribution $\dset(A)$ over the set $\Lambda_G(A)$ of internal $A$-routers, such that, for every edge $e\in E(A)$, $\expect[\qset(A)\sim \dset(A)]{\cong(\qset(A),e)}\leq  O((\log^4m)\cdot (\log^{9.5}m))=O(\log^{13.5}m)$. We now define the distribution $\dset'(S')$ over the set $\Lambda'_G(S')$ of external $S'$-routers, by providing an algorithm to draw a set of paths from the distribution. In order to do so, we first choose a set $\qset(A)\in \Lambda_G(A)$ of paths from the distribution $\dset(A)$.
Denote $\qset(A)=\set{Q(e)\mid e\in \delta_G(A)}$, where path $Q(e)$ has $e$ as its first edge. Let $\qset'$ be a multi-set of paths, in which, for every edge $e\in \Lambda_G(A)$,  the path $Q(e)$ is included   $\cong_G(\pset(S'),e)\leq O(\log m)$ times. 
We then concatenate the paths in $\pset(S')$ with the paths in $\qset'$, obtaining an external $S'$-router $\qset'(S')\in \Lambda_{G}(S')$. From the above discussion, it is immediate to verify that, for every edge $e\in E(G)\setminus E(S')$, 
$\expect[\qset'(S')\sim \dset'(S')]{\cong(\qset'(S'),e)}\leq  O(\log^{14.5}m)$.
\end{proof}

We now consider three cases, depending on whether cluster $S$ contains any $J$-node, and whether the cluster  $J'\in \jset$ corresponding to the $J$-node has a cluster of $\cset$ or of $\wset'$ as its center cluster.

\begin{figure}[h]
	\centering
	\includegraphics[scale=0.12]{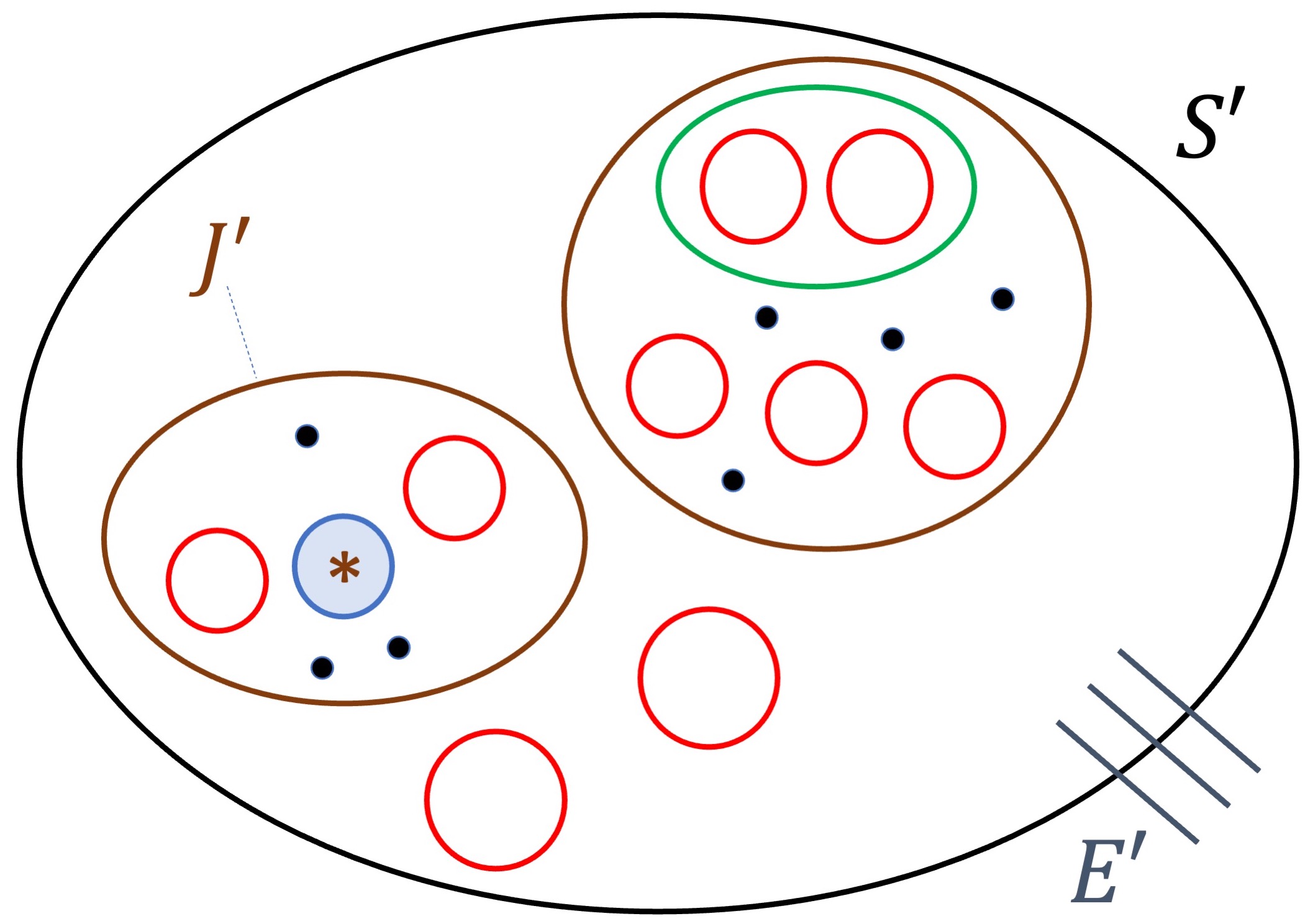}
	\caption{Case $1$: $S'$ contains at least one $J'$-cluster (shown in brown) whose center cluster $C^*\in \cset$ is marked with $*$. Clusters of $\rset^*$ are shown in red.}\label{fig: NF6}
\end{figure}

\paragraph{Case 1.} The first case happens if there is at least one $J$-node $v_{J'}\in S$, such that the center-cluster of the cluster $J'\in \jset$ is a cluster of $\cset$, that we denote by $C^*$ (see \Cref{fig: NF6}).
Let $\cset^*\subseteq \cset$ be the set of all clusters $C\in \cset$ with $C\subseteq S'$, and let $\rset^*\subseteq \rset$ be the set of all clusters $R\in \rset$ with $R\subseteq S'$. Observe that cluster $C^*$ may not be contained in any cluster of $\rset$, from the definition of cluster set $\jset$. We will modify the set $\rset$ of clusters, by deleting the clusters of $\rset^*$ from it, and adding a new set $\rset^{**}$ of clusters instead, so that the resulting cluster set $\tilde \rset=(\rset\setminus\rset^*)\cup \rset^{**}$ is a helpful clustering that is better than $\rset$.

In order to define the new set $\rset^{**}$ of clusters, we start with the augmented cluster $X=(S')^+$; that is, we subdivide every edge $e\in E'$ in graph $G$ with a terminal $t_e$, and we let $T=\set{t_e\mid e\in E'}$ be the resulting set of terminals. We then let $X$ be the subgraph of the resulting graph induced by $T\cup V(S')$. Next, we obtain a graph $X'$ from $X$, by contracting every basic cluster $C\in \cset^*$ into a supernode $v_C$, so $X'=X_{|\cset^*}$. We apply the algorithm from \Cref{thm:well_linked_decomposition} to graph $X'$, its cluster $X'\setminus T$, and parameter $\alpha_0=1/\log^3m$ (so the requirement that $\alpha_0< \min\set{\frac 1 {64\alphasc(m)\cdot\log m},\frac 1 {48\log^2 m}}$ is satisfied). The algorithm computes a collection $\yset$ of clusters of $X'\setminus T$ (the well-linked decomosition), such that the vertex sets $\set{V(Y)}_{Y\in \yset}$ partition $V(X')\setminus T$, and every cluster $Y\in \tset$ has the $\alpha_0$-bandwidth property, with $|\delta_{X'}(Y)|\leq |T|=|E'|$. We are also guaranteed that:

\begin{equation}\label{eq: few edges in WLD}
\sum_{Y\in \yset}|\delta_{X'}(Y)|\le |T|\cdot\left(1+O(\alpha_0\cdot \log^{1.5} m)\right)= |E'|\cdot \left(1+O(\alpha_0\cdot \log^{1.5} m)\right).
\end{equation}

Recall that, additionally, the algorithm computes, 
for every cluster $Y\in \yset$, a set $\pset(Y)=\set{P(e)\mid e\in \delta_{X'}(Y)}$ of paths, such that, for every edge $e\in \delta_{X'}(Y)$, path $P(e)$ has $e$ as its first edge, and some terminal of $T$ as its last vertex, with all inner vertices of $P(e)$ lying in $V(X')\setminus V(Y)$. We are also guaranteed that, for every cluster $Y\in \yset$, the set $\pset(Y)$ of paths causes edge-congestion at most $100$ in $X'$.

We now define the set $\rset^{**}$ of clusters, that will be added to $\rset$ instead of the clusters of $\rset^*$. For every cluster $Y\in \yset$, we let $R_Y$ be the cluster of graph $G$ corresponding to cluster $Y$. Intuitively, $R_Y$ is obtained from $Y$ by uncontracting all basic clusters whose corresponding supernode lies in $Y$. Formally, we denote by $\cset(Y)\subseteq \cset$ the set of all basic clusters $C$ whose corresponding supernode $v_C$ lies in $Y$. We define vertex set $V(R_Y)$ to contain all regular vertices lying in $Y$ (that is, vertices of $V(Y)\cap V(G)$), and all vertices of $\bigcup_{C\in \cset(Y)}V(C)$. We then let $R_Y$ be the subgraph of $G$ induced by $V(R_Y)$. Note that $Y=(R_Y)_{|\cset(Y)}$. Since every cluster in $\cset$ has the $\alpha_0$-bandwidth property, and every cluster $Y\in \yset$ has the $\alpha_0$-bandwidth property, from \Cref{cor: contracted_graph_well_linkedness}, cluster $R_Y$ has the $\alpha_0^2=\alpha_1$-bandwidth property. Consider now the set $\pset(Y)$ of paths in graph $X'$, routing the edges of $\delta_{X'}(Y)$ to the vertices of $T$, with congestion at most $100$, and recall that the paths of $\pset(Y)$ are internally disjoint from $Y$. Since $Y=(R_Y)_{|\cset(Y)}$, and  every cluster in $\cset$ has the $\alpha_0$-bandwidth property, we can use the algorithm from \Cref{claim: routing in contracted graph}, to obtain a collection $\pset(R_Y)$ of paths in graph $G$, routing the edges of $\delta_{G}(R_Y)$ to the edges of $E'$, with congestion at most $200/\alpha_0$, such that the paths in $\pset(R_Y)$ are internally disjoint from $R_Y$, and they cause congestion at most $100$ on edges of $E'$. 
We are now ready to define a distribution $\dset'(R_Y)$ over the set $\Lambda'_G(R_Y)$ of external $R_Y$-routers in $G$, by providing an algorithm to draw an external router $\qset'(R_Y)$ from the distribution.  In order to draw a set $\qset'(R_Y)$ of paths from distribution $\dset'(R_Y)$, we start by drawing  an external $S'$-router $\qset'(S')\in \Lambda_G(S')$ from the distribution $\dset'(S')$ given by \Cref{obs: compute external paths for S'}. Recall that paths in $\qset'(S')$ route the edges of $E'$ to some vertex $u\not\in S'$, they are internally disjoint from $S'$, and 	 $\expect[\qset'(S')\sim \dset'(S')]{\cong(\qset'(S'),e)}\leq  O(\log^{14.5}m)$. 
We denote $\qset'(S)=\set{Q(e)\mid e\in \delta_G(S')}$, where for all $e\in \delta_G(S')$, path $Q(e)$ has $e$ as its first edge.
We let $\qset''$ be a multi-set of paths obtained by including, for each edge $e\in \delta_G(S')$, exactly $\cong_G(\pset(R_Y),e)\leq 100$ copies of the path $Q(e)$.
By concatenating the paths in $\pset(R_Y)$ with the paths in $\qset''$,  we obtain a set $\qset'(R_Y)$ of paths (an external $R_Y$-router), routing the edges of $\delta_G(R_Y)$ to vertex $u$ in $G$, such that the paths in the set are inernally disjoint from $R_Y$. For every edge in $E(S')$, the congestion caused by the paths in $\qset'(R_Y)$ is at most $200/\alpha_0\leq O(\log^3m)$, while for every edge $e\in E(G\setminus S')$, $\cong_G(\qset'(R_Y),e)\leq 100\cong_G(\qset'(S'),e)$. This completes the definition of the distribution $\dset'(R_Y)$ over the set $\Lambda'_G(R_Y)$ of external $R_Y$-routers. From the above discussion, for every edge $e\in E(G)\setminus E(R_Y)$, $\expect[\qset'(R_Y)\sim \dset'(R_Y)]{\cong(\qset'(R_Y),e)}\leq  O(\log^{15.5}m)\leq \beta$, since $\beta=\log^{18}m$. 
We set $\rset^{**}=\set{R_Y\mid Y\in \yset}$, and we define a new clustering $\tilde \rset=(\rset\setminus \rset^*)\cup \rset^{**}$. 

It is immediate to verify that all clusters in $\tilde \rset$ are disjoint from each other and every cluster $R\in \tilde \rset$ has $\alpha_1$-bandwidth property. From the construction of the graph $X'$, we are guaranteed that, for every basic cluster $C\in \cset$, and for every cluster $R\in \rset$, either $C\subseteq R$, or $C\cap R=\emptyset$ must hold. We have also defined, for every cluster $R\in \rset$, a distribution  	
$\dset'(R)$ over external $R$-routers in $\Lambda'_{G}(R)$, such that, for every edge $e\in E(G)\setminus E(R)$, $\expect[\qset'(R)\sim\dset'(R)]{\cong_{G}(\qset'(R),e)}\leq \beta$. Since we have ensured that vertices $v^*,u^*$ do not lie in set $S$, we are guaranteed that $R^*\in \rset\setminus \rset^*$, and so $R^*\in \tilde \rset$. Similarly, the special vertex $v^*$ does not lie in any cluster of $\tilde \rset$. Therefore, $(\tilde \rset,\set{\dset'(R)_{R\in \tilde \rset}})$ is a {helpful clustering} of $G$ with respect to $v^*$ and $\cset$, with $R^*\in \tilde \rset$. It now only remains to show that it is better than the original clustering $\rset$. 

Observe that the only basic clusters that may be contained in the clusters of $\rset^*$ are the clusters of $\cset^*$. From the definition of the set $\jset$ of clusters, since cluster $C^*\in \cset^*$ is a center-cluster of some cluster $J'\in \jset'$, we are guaraneed that cluster $C^*$ is not contained in any cluster of $\rset$. But, since the vertices of $\set{V(Y)}_{Y\in \yset}$ partition vertex set $V(X')\setminus T$, we are guaranteed that every cluster of $\cset^*$ is contained in some cluster of $\rset^{**}$. Therefore, the number of basic clusters of $\cset$ contained in $G\setminus\left(\bigcup_{R\in \rset}R\right )$ is strictly greater than the number of 
basic clusters of $\cset$ contained in $G\setminus\left(\bigcup_{R\in \tilde\rset}R\right )$. We conclude that clustering $\tilde \rset$ is a better clustering than $\rset$.

\paragraph{Case 2.}
We now consider the second case, where cluster $S$ does not contain any $J$-node, so every vertex of $S$ is either an $R$-node or a regular vertex. In this case, we proceed exactly as before: we define the augmented cluster $X=(S')^+$ and its contracted version $X'=X_{|\cset^*}$. We then compute a well-linked decomposition $\yset$ of $X'\setminus T$, and the corresponding set $\rset^{**}=\set{R_Y\mid Y\in \yset}$ of clusters of $G$ exactly as before. We also define a new clustering $\tilde \rset$ and the distributions $\dset'(R_Y)$ over sets of $R_Y$-routers for clusters $R_Y\in \rset^{**}$ exactly as before. Using the same reasoning as in Case 1, the final clustering $(\tilde \rset,\set{\dset'(R)_{R\in \tilde \rset}})$ is a {helpful clustering} of $G$ with respect to $v^*$ and $\cset$, with $R^*\in \tilde \rset$. However, since we are no longer guaranteed that $S$ contains a $J$-node (which in turn contains a cluster of $\cset$ as a center cluster), we need to employ a different argument in order to prove that $\tilde \rset$ is a better clustering than $\rset$. Since $S$ does not contain any $J$-node, from the definition of a simplifying cluster,  $|E_{\cH}(S)|\geq |\delta_{\cH}(S)|/\log m=|E'|/\log m$ must hold. Notice that every edge of $E_{\cH}(S)$ corresponds to an edge of $S'_{|\rset^*}$, and so $|E(S'_{|\rset^*})|\geq |E'|/\log m$. On the other hand, from \Cref{eq: few edges in WLD}, $|E(S'_{|\rset^{**}})|\leq \sum_{Y\in \yset}|\delta_{X'}(Y)|-|E'|\le O(|E'|\alpha_0\cdot \log^{1.5} m)< |E'|/\log m$, since $\alpha_0= 1/\log^3m$. 
From our definition of the set $\tilde \rset$ of clusters, $|E(G_{|\tilde \rset})|=|E(G_{|\rset})|-|E(S'_{|\rset^*})|+|E(S'_{|\rset^{**}})|<|E(G_{|\rset})|$.
We conclude that the new helpful clustering $\tilde \rset$ is better than $\rset$.

\paragraph{Case 3.}
It remains to consider the third case, where $S$ contains at least one $J$-node, and for each such $J$-node $v_{J'}$, the center-cluster of the cluster $J'\in \jset'$ lies in $\wset'$. We fix any $J$-node $v_{J'}\in V(S)$, and we denote by $W'\in \wset'$ the center-cluster of $J'$. 

There is a difficulty with following the approach used in Cases 1 and 2 in this case: it is possible that every cluster of $\cset^*$ is contained in some cluster of $\rset^*$, and, additionally, it is possible that the total number of edges in $S'_{|\rset^*}$ is quite small compared to $|E'|$. While we could still obtain a new helpful clustering $\tilde \rset$ in the same way as in Cases 1 and 2, it may no longer be the case that $\tilde \rset$ is a better clustering than $\rset$. In order to overcome this difficulty, we will replace cluster $S'$ of $G$ with a different cluster $\tilde S'\subseteq S'$ that has similar properties to cluster $S'$, except that, if we denote by $\tilde \rset^*$ the set of all clusters $R\in \rset^*$ with $R\subseteq \tilde S'$, then $|E(S'_{|\tilde \rset^*})|$ is sufficiently large compared to $|E_G(\tilde S')|$. We construct the cluster $\tilde S'$ using the following claim.

\begin{claim} \label{claim: case 3}
	There is an efficient algorithm, that, if Case 3 happenned, computes a cluster $\tilde S'\subseteq S'$ in graph $G$, such that for every cluster $R\in \rset^*$, either $R\subseteq \tilde S'$ or $R\cap \tilde S'=\emptyset$ holds, and similarly, for every cluster $C\in \cset^*$, either $C\subseteq \tilde S'$ or $C\cap \tilde S'=\emptyset$ holds. Moreover, if we denote by $\tilde \rset^*$ the set of all clusters $R\in \rset^*$ with $R\subseteq \tilde S'$, then $|E(\tilde S'_{|\tilde \rset^*})|\geq |\delta_G(\tilde S')|/(64\log m)$. Additionally, the algorithm computes a distribution $\dset'(\tilde S')$ over the set $\Lambda'_G(\tilde S')$ of external $\tilde S'$-routers in graph $G$, such that, for every edge $e\in E(G\setminus \tilde S')$, 
	$\expect[\qset'(\tilde S')\sim \dset'(\tilde S')]{\cong(\qset'(\tilde S'),e)}\leq  O(\log^{14.5}m)$.
\end{claim}

We provide the proof of \Cref{claim: case 3} below, after completing the proof of	\Cref{claim: simplifying cluster is enough} using it. We employ the algorithm from Cases 1 and 2, except that we apply it to cluster $\tilde S'$ of $G$ instead of $S'$, and we replace the set $\rset^*$ of clusters with $\tilde \rset^{*}$. Let $\tilde \cset^*\subseteq \cset^*$ be the set of all basic clusters $C\in \cset^*$ with $C\subseteq \tilde S'$. Let $\tilde \rset^{**}$ be the set of clusters that the algorithm from Cases 1 and 2 computes (that were denoted by $\rset^{**}$ before), when applied to cluster $\tilde S'$. For every cluster $R\in \tilde \rset^{**}$, the algorithm obtains a distribution $\dset'(R)$ over the set $\Lambda'_G(R)$ of external $R$-routers.
Let $\tilde \rset=(\rset\setminus \tilde\rset^*)\cup \tilde \rset^{**}$. Using the same arguments as in Cases 1 and 2, $(\tilde \rset,\set{\dset'(R)}_{R\in \tilde \rset})$ is a helpful clustering in $G$ with respect to $v^*$ and $\cset$, with $R^*\in \tilde \rset$. It now only remains to show that $\tilde \rset$ is a better clustering than $\rset$. 

As in Case 2, the only basic clusters of $\cset$ that the clusters of $\tilde \rset^*$ may contain are the clusters of $\tilde \cset^*$. As in Case 2, each such cluster is guaranteed to be contained in some cluster of $\tilde \rset^{**}$. Therefore, the number of clusters of $\cset$ that are contained in $G\setminus(\bigcup_{R\in\rset}R)$ is greater than or equal to the number of clusters of $\cset$ that are contained in $G\setminus (\bigcup_{R\in \tilde \rset}R)$. In order to prove that $\tilde \rset$ is a better clustering than $\rset$, it is now enough to prove that $|E(G_{|\tilde \rset})|<|E(G_{|\rset})|$.

We denote $\delta_G(\tilde S')$ by $E''$. On the one hand, 
from \Cref{claim: case 3}, $|E(\tilde S'_{|\tilde \rset^*})|\geq |E''|/(64\log m)$. On the other hand, from \Cref{eq: few edges in WLD}, $|E(\tilde S'_{|\tilde \rset^{**}})|\leq 
\sum_{Y\in \yset}|\delta_{X'}(Y)|-|E''| \le |E''|\cdot O(\alpha_0\cdot \log^{1.5} m))<|E''|/(64\log m)$, since $\alpha_0=1/\log^3m$.
As in Case 2, $|E(G_{|\tilde \rset})|=|E(G_{|\rset})|-|E(S'_{|\tilde \rset^*})|+|E(S'_{|\tilde \rset^{**}})|<|E(G_{|\rset})|$.
 Therefore, $\tilde \rset$ is a better clustering than $\rset$. It now remains to complete the proof of \Cref{claim: case 3}, which we do next.

\begin{proofof}{\Cref{claim: case 3}}
Recall that we have fixed a $J$-node $v_{J'}\in V(S)$, and a center-cluster $W'\in \wset'$ of the cluster $J'\in \jset'$.
We let $\tilde S'$ be a vertex-induced subgraph of $S'$ with the following properties:
	
	\begin{itemize}
		\item $W'\subseteq \tilde S'$;
		\item for every cluster $R\in \rset^*$, either $R\subseteq \tilde S'$, or $R\cap \tilde S'=\emptyset$;
		\item for every cluster $C\in \cset^*$, either $C\subseteq \tilde S'$ or $C\cap \tilde S'=\emptyset$; and
		\item $|E(\tilde S')|$ is minimized among all graphs $\tilde S'$ for which the above conditions hold.
	\end{itemize}
	
	Such a graph $\tilde S'$ can be computed via standard minimum cut computation: we start with graph $G$, and we let $G_1$ be the graph obtained from $G$ by contracting the cluster $W'$ into a source $s$, and contracting $G\setminus S'$ into a destination $t$. Next, we obtain a graph $G_2$ from $G$, by contracting every cluster $R\in \rset$ with $R\subseteq S'\setminus W'$ into a supernode $v_R$, and similarly contracting every basic cluster $C\in \cset'$ with $C\subseteq S'\setminus W'$ into a supernode $v_C$. Let $(Z,Z')$ be the minimum $s$-$t$ cut in $G_2$, and denote $E''=E_{G_2}(Z,Z')$. Observe that, from the max-flow / min-cut theorem, there is a collection $\qset$ of edge-disjoint path in graph $G_2$, routing the edges of $E''$ to $t$, such that all paths in $\qset$ are internally disjoint from $Z$. We let $\tilde S'\subseteq S'$ be the subgraph of $\tilde S'$ that is defined by $Z$ (that is, we un-contract every cluster of $\rset\cup\cset$ whose corresponding supernode lies in $Z$). Note that $\delta_G(\tilde S')=E''$, and $W'\subseteq \tilde S'$. 
	Let $W\in \wset$ be the $W$-cluster in graph $\hat H=G_{|\cset'\cup \rset}$ that corresponds to $W'$; in other words, cluster $W'$ was obtained from $W$ by un-contracting its supernodes. 
	Let $\tilde \rset^*$ the set of all clusters $R\in \rset^*$ with $R\subseteq \tilde S'$.
	Clearly, every edge of $W$ is an edge of $S'_{|\tilde \rset^*}$. From the definition of a valid set of $W$-clusters, 
	$|E(S'_{|\tilde \rset^*})|\geq |E_{\hat H}(W)|\geq |\delta_{\hat H}(W)|/(64\log m)\geq |E''|/(64\log m)$ (we have used the fact that $|E(Z,Z')|\leq |\delta_G(W')|=|\delta_{\hat H}(W)|$ must hold, from the definition of minimum cut).


	It now remains to define the distribution $\dset'(\tilde S')$ over the set $\Lambda'_G(\tilde S')$ of external $\tilde S'$-routers in graph $G$. As before, we will provide an efficient algorithm to draw a set $\qset'(S')$ of paths from the distribution. As observed already, there is a set $\qset$ of edge-disjoint paths in graph $G_2$ routing the edges of $\delta_{G_2}(Z)$ to vertex $t$, so that the paths are internally disjoint from cluster $Z$. Note that every edge $e\in E''$ has exactly one path $Q(e)\in \qset$ whose first edge is $e$, and the last edge of $Q(e)$ lies in $E'$. Since every cluster of $\rset$ has the $\alpha_1=1/\log^6m$-bandwidth property, and every cluster $C\in \cset$ has the $\alpha_0=1/\log^3m$-bandwidth property, we can use the algorithm from \Cref{claim: routing in contracted graph} to compute a collection $\pset=\set{P(e) \mid e\in E''}$ of paths in graph $G_1$, where for every edge $e\in E''$, path $P(e)$ has $e$ as its first edge and some edge of $E'$ as its last edge. Moreover, the paths in $ \pset$ cause edge-congestion at most $2/\alpha_1\leq 2/\log^6m$ in graph $G_1$, and congestion at most $1$ on edges of $E'$, and they are internally disjoint from $\tilde S'$. Note that the paths of $ \pset$ are also contained in the original graph $G$. We are now ready to define the distribution $\dset'(\tilde S')$. In order to draw an external $\tilde S'$-router $\qset'(\tilde S')\in \Lambda'_G(\tilde S')$ from the distribution, we first
	draw  a set $\qset'(S')\in \Lambda_G(S')$ of paths from the distribution $\dset'(S')$, given by \Cref{obs: compute external paths for S'}. Recall that paths in $\qset'(S')$ route the edges of $E'$ to some vertex $u\not\in S'$, they are internally disjoint from $S'$, and 	 $\expect[\qset'(S')\sim \dset'(S')]{\cong(\qset'(S'),e)}\leq  O(\log^{14.5}m)$. By concatenating the paths in $\pset$ with the paths in $\qset'(S')$, we obtain a set $\qset'(\tilde S')$ of paths, routing the edges of $\delta_G(\tilde S')$ to vertex $u$ in $G$, such that the paths in the set are inernally disjoint from $\tilde S'$. Moreover, for every edge of $ S'\cup E'$, the congestion caused by the paths in $\qset'(\tilde S')$ is at most $O(\log^6m)$, while for every edge of $G\setminus  S'$, $\cong_G(\qset'(\tilde S'),e)\leq \cong_G(\qset'(S'),e)$. This completes the definition of the distribution $\dset'(\tilde S')$ over the set $\Lambda'_G(\tilde S')$ of external $\tilde S'$-routers. Clearly, for every edge $e\in E(G\setminus \tilde S')$, $\expect[\qset'(\tilde S')\sim \dset'(\tilde S')]{\cong(\qset'(\tilde S'),e)}\leq  O(\log^{14.5}m)$.
\end{proofof}




\subsection{Proof of \Cref{obs: left and right down-edges}}
\label{subsec: left and right down-edges}
Consider any vertex $v\in V(\cH)\setminus\left(\bigcup_{i=1}^rS'_i\right )$, and assume that it lies in $S_i\setminus S'_i$, for some $1\leq i\leq r$.

Recall that $|\delta^{\up}(v)|\leq |\delta_{\cH}(v)|/\log m$ must hold, and, since $v$ was not added to $S'_i$, $|\delta^{\down,\straight''}(v)|\leq |\delta_{\cH}(v)|/128$.
Therefore, $|\delta^{\down,\rig}(v)|+|\delta^{\down,\lef}(v)|+| \delta^{\down,\straight'}(v)|\geq 63|\delta_{\cH}(v)|/64$ holds.

\begin{figure}[h]
\centering
\includegraphics[scale=0.15]{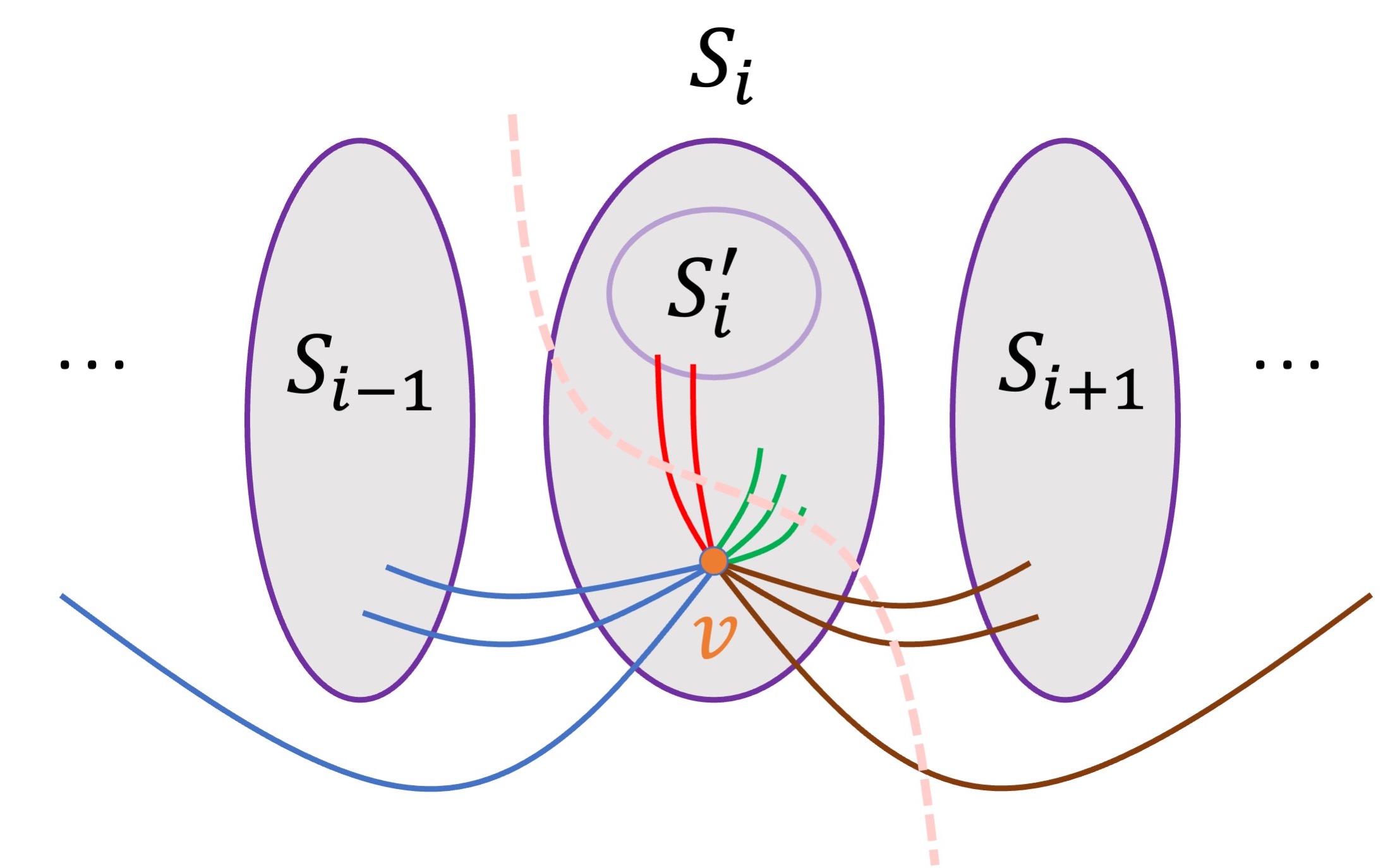}
\caption{Illustration for the proof of \Cref{obs: left and right down-edges}. The edges of $\delta^{\down,\lef}(v)$ are shown in blue, the edges of $\delta^{\down,\straight''}(v)$ are shown in red, the edges of $\delta^{\down,\straight'}(v)$ are shown in green, and  the edges of $\delta^{\down,\rig}(v)$ are shown in brown. The blue edges lie in  $E(A,B)\setminus E(A',B')$.  The  edges crossing the pink dashed line lie in $E(A',B')\setminus E(A,B)$. Additionally, edges of $\delta^{\up}(v)$ may lie in set $E(A',B')\setminus E(A,B)$.}\label{fig: NF7}
\end{figure}

Assume now that $i>1$, and consider the Gomory-Hu tree $\tau$ of the graph $\cH$, and the two connected components of the graph obtained from $\tau$ after the edge $(u_{i-1},u_i)$ is removed from $\tau$. Denote by $A$ the set of all vertices lying in the connected component containing $u_{i-1}$, and by $B$ the set of all vertices lying in the other connected component. From the definition of cluster $S_i$, $v\in B$, and moreover, $(A,B)$ is the minimum $u_{i-1}$-$u_i$ cut in graph $\cH$. Therefore, if we let $A'=A\cup \set{v}$ and $B'=B\setminus\set{v}$, then $|E_{\cH}(A',B')|\geq |E_{\cH}(A,B)|$ must hold. Observe that the only difference between the edge sets $E_{\cH}(A',B')$ and $E_{\cH}(A,B)$ is that the edges of $\delta^{\down,\lef}(v)$ contribute to $E_{\cH}(A,B)$ but not to $E_{\cH}(A',B')$; the edges of $\delta^{\down,\rig}(v)\cup \delta^{\down,\straight'}(v)\cup \delta^{\down,\straight''}(v)$ contribute to $E_{\cH}(A',B')$ but not to $E_{\cH}(A,B)$; and the edges of $\delta^{\up}(v)$ may contribute to either cut (see \Cref{fig: NF7}).
Therefore, it must be the case that:

$$|\delta^{\down,\lef}(v)|\leq |\delta^{\down,\rig}(v)|+| \delta^{\down,\straight'}(v)|+|\delta^{\down,\straight''}(v)|+|\delta^{\up}(v)|.$$ 

From the definition of the layers $L_1,\ldots,L_h$, $|\delta^{\up}(v)|\leq |\delta(v)|/\log m$, and, since $v$ was not added to $S'_i$, $|\delta^{\down,\straight''}(v)|\leq |\delta(v)|/128$. Therefore:

$$|\delta^{\down,\lef}(v)|\leq |\delta^{\down,\rig}(v)|+| \delta^{\down,\straight'}(v)|+|\delta(v)|/64.$$ 

If we assume, for contradiction, that $|\delta^{\down,\lef}(v)|> 2(|\delta^{\down,\rig}(v)|+| \delta^{\down,\straight'}(v)|)$, then, combining this with the above inequality, we will get that $|\delta^{\down,\rig}(v)|+| \delta^{\down,\straight'}(v)|<|\delta(v)|/64$, and therefore, $|\delta^{\down,\lef}(v)|<|\delta(v)|/32$, contradicting the fact that 
$|\delta^{\down,\lef}(v)|+|\delta^{\down,\rig}(v)|+| \delta^{\down,\straight'}(v)|\geq 63|\delta(v)|/64$.
We conclude that $|\delta^{\down,\lef}(v)|\leq 2(|\delta^{\down,\rig}(v)|+| \delta^{\down,\straight'}(v)|)$ must hold. 

We now show that $S_r=S'_r$. Assume for contradiction that this is not the case, and let $v\in S_r\setminus S'_r$ be a vertex that minimizes the index $j$ for which $v\in L'_j$. Notice that, from the definition, $\delta^{\down,\rig}(v)$, $\delta^{\down,\straight'}(v)=\emptyset$ must hold. 
Since we have established that $|\delta^{\down,\lef}(v)|\leq 2(|\delta^{\down,\rig}(v)|+| \delta^{\down,\straight'}(v)|)$, we get that $\delta^{\down,\lef}=\emptyset$ as well. Altogether, we get that $\delta^{\down}(v)=\delta^{\down,\straight''}(v)$. Since $|\delta^{\up}(v)|\leq |\delta_{\cH}(v)|/\log m$, we get that the number of edges connecting $v$ to vertices of $S'_r$, $|\delta^{\down,\straight''}(v)|>|\delta_{\cH}(v)|/2$, and so $v$ should have been added to set $S'_r$. Therefore, $S_r=S'_r$ must hold.

The proof that $|\delta^{\down,\rig}(v)|\leq 2(|\delta^{\down,\lef}(v)|+|\delta^{\down,\straight'}(v)|)$ is similar, except that now we need to consider the cut obtained by deleting the edge $(u_i,u_{i+1})$ from the tree $\tau$. The proof that $S_1=S'_1$ is symmetric to the proof that $S_r=S'_r$ (in fact, if we reverse the order of the vertices $u_1,\ldots,u_r$ on path $P^*$, by rooting the tree $\tau$ at vertex $u_r=u^*$, then the definitions of sets $S_i,S'_i$ will remain unchanged, and we can simply repeat the proof from above).



\subsection{Proof of \Cref{obs: left and right mappings}}
\label{subsec: left and right mappings}

Consider any vertex $v\in V(\cH)\setminus\left(\bigcup_{i=1}^rS'_i\right )$. From \Cref{obs: left and right down-edges}, $|\delta^{\down,\lef}(v)|\leq 2(|\delta^{\down,\rig}(v)|+|\delta^{\down,\straight'}(v)|)$. Therefore:

$$|\delta^{\down,\lef}(v)|+|\delta^{\down,\rig}(v)|+|\delta^{\down,\straight'}(v)|\leq  3(|\delta^{\down,\rig}(v)|+|\delta^{\down,\straight'}(v)|).$$

On the other hand, since $|\delta^{\down,\straight''}(v)|\leq |\delta_{\cH}(v)|/128$, and $|\delta^{\up}(v)|\leq |\delta_{\cH}(v)|/\log m$, we get that:

$$|\delta^{\down,\lef}(v)|+|\delta^{\down,\rig}(v)|+|\delta^{\down,\straight'}(v)|\geq \left (\frac {127}{128} -\frac 1 {\log m}\right )\cdot |\delta_{\cH}(v)|.$$

By combining the two inequalities, we get that:

\[ |\delta^{\down,\rig}(v)|+|\delta^{\down,\straight'}(v)|\geq \left (\frac {127} {384} -\frac 1 {3\log m}\right )\cdot |\delta_{\cH}(v)|>|\delta^{\down,\straight''}(v)|+|\delta^{\up}(v)|.  \]

Therefore, we can define a mapping $f^{\rig}(v)$, that maps every edge of $\delta^{\down,\straight''}(v)\cup\delta^{\up}(v)$ to a distinct edge of $\delta^{\down,\rig}(v)\cup \delta^{\down,\straight'}(v)$.

Using exactly the same reasoning, we get that:

\[ |\delta^{\down,\lef}(v)|+|\delta^{\down,\straight'}(v)|>|\delta^{\down,\straight''}(v)|+|\delta^{\up}(v)|.  \]

Therefore, we can define a mapping $f^{\lef}(v)$, that maps every edge of $\delta^{\down,\straight''}(v)\cup\delta^{\up}(v)$ to a distinct edge of $\delta^{\down,\lef}(v)\cup \delta^{\down,\straight'}(v)$.

\subsection{Constructing the Monotone Paths -- proof of \Cref{lem: prefix and suffix path}}
\label{subsubsec: monotone paths}

For all $1\leq j\leq h$, we define two sets of edges, $E^{\lef}_j$ and $E^{\rig}_j$, as follows. Start with $E^{\lef}_j=E^{\rig}_j=\emptyset$. For all $1\leq i\leq r$, for every vertex $v\in U_{i,j}\setminus \set{S'_i}$, we add all edges of $\delta^{\down,\lef}(v)\cup \delta^{\down,\straight'}(v)$ to $E^{\lef}_j$, and similarly, we add all edges of $\delta^{\down,\rig}(v)\cup \delta^{\down,\straight'}(v)$ to $E^{\rig}_j$. We prove the following claim.

\begin{claim}\label{claim: level by level routing}
	There is an efficient algorithm to construct, for all $1\leq j\leq h$, a set $\pset^{\lef}_j=\set{P^{\lef}(e)\mid e\in E^{\lef}_j}$ of edge-disjoint left-monotone paths, and a set $\pset^{\rig}_j=\set{P^{\rig}(e)\mid e\in E^{\rig}_j}$ of edge-disjoint right-monotone paths, such that, for every edge $e\in E^{\lef}_j$, path $P^{\lef}(e)$ starts with edge $e$, and similarly, for every  edge $e'\in E^{\rig}_j$, path $P^{\rig}(e')$ starts with edge $e'$. 
\end{claim}

\begin{proof}
	The proof is by induction on $j$. The base is when $j=1$. Consider some vertex $v\in U_{i,1}\setminus \set{S_i'}$, for some $1\leq i\leq r$. Note that every edge in $\delta^{\down}(v)$ must connect $v$ to a vertex of $L_0'$, and every vertex of $L_0'$ lies in $\set{u_1,\ldots,u_r}$. Consider now some edge $e=(v,u)$ that is incident to $v$, that lies in $E^{\lef}_1$. Note that $e\not \in \delta^{\down,\straight'}(v)$, as all edges connecting $u$ to $u_i$ lie in $\delta^{\down,\straight''}(v)$, and no edge of $\delta^{\down}(v)$ may connect $v$ to a vertex outside  $\set{u_1,\ldots,u_r}$. Therefore, $e\in \delta^{\lef}(v)$ must hold. We then let $P^{\lef}(e)=(e)$. It is immediate to verify that it is a left-monotone path. We define paths $P^{\rig}(e)$ for every edge $e\in \delta_{\cH(v)}\cap E^{\rig}_1$ similarly.

We assume now that the claim holds for some index $1\leq j<h$, and we prove it for index $j+1$.
Consider some vertex $v\in U_{i,j+1}\setminus \set{S_i'}$, for some $1\leq i\leq r$, and let $e=(v,u)$ be any edge that is incident to $v$ and lies in $E^{\lef}_j$. In this case, $e\in \delta^{\down,\lef}(v)\cup \delta^{\down,\straight'}(v)$ must hold. If we denote by $1\leq i'\leq r$, $1\leq j'\leq h$ the indices for which $u\in U_{i',j'}$, then $i'\leq i$ and $j'\leq j$ must hold. 

Assume first that $u\in S'_1\cup\cdots\cup S'_{r}$. In this case, since $e\not\in \delta^{\down,\straight''}(v)$, $e\in \delta^{\down,\lef}(v)$ must hold, and $i'<i$. In this case, we let $P^{\lef}(e)=(e)$. It is easy to see that this path is a valid left-monotone path. Otherwise, $u\not \in S'_1\cup\cdots\cup S'_{r}$, and $e\in \delta^{\up}(u)$. We then consider the edge $e'\in \delta^{\down,\lef}(u)\cup \delta^{\down,\straight'}(u)$ to which edge $e$ is mapped by $f^{\lef}(u)$. We then let $P^{\lef}(e)$ be the path obtained by concatenating the edge $e$ and the path $P^{\lef}(e')\in \pset^{\lef}_{j'}$.
Since path $P^{\lef}(e')$ is left-monotone, and since $j'\leq j$ and $i\leq i$, path $P^{\lef}(e)$ is also left-monotone.
 We note that edge $e$, by the definition, may not lie on any path in $\pset^{\lef}_1\cup\cdots\cup \pset^{\lef}_j$. Moreover, edge $e$ is the only edge that is mapped to edge $e'$ by map $f^{\lef}(u)$. This ensures that all paths in the resulting set $\pset^{\lef}_j=\set{P^{\lef}(e'')\mid e''\in E^{\lef}_j}$ are edge-disjoint.
 
 The construction of the set $\pset^{\rig}_{j+1}$ of paths is symmetric.
\end{proof}

We are now ready to complete the proof of \Cref{lem: prefix and suffix path}. Consider an edge $e=(u,v)\in \hat E$, with $u\in S_i$, $v\in S_{i'}$, and $i<i'$. 
We describe the construction of path $P(e,u)$; the construction of path $P(e,v)$ is symmetric. If $u\in S'_i$, then path $P(e,u)$ consists only of the vertex $u$. It is a left-monotone path by definition. Therefore, we assume now that $u\not\in S'_i$.
We assume that $u$ lies in layer $L'_j$, for some $1\leq j \leq h$, and $v$ lies in layer $L'_{j'}$, for some $0\leq j'\geq h$ (vertex $u$ may not lie in $L'_0$, since we have assumed that $u\not\in S'_i$). 

We now consider two cases. The first case is when $j\leq j'$. In this case, $e\in \delta^{\up}(u)$. We let $e'\in \delta^{\down,\lef}(u)\cup \delta^{\down,\straight'}(u)$ be the edge to which $e$ is mapped by $f^{\lef}(u)$. Since $e'\in E^{\lef}_j$, there is a left-monotone path $P^{\lef}(e')\in \pset^{\lef}_j$. We then let $P(e,u)$ be the path obtained by concatenating edge $e$ with path $P^{\lef}(e')$. It is easy to verify that path $P(e,u)$ is left-monotone.

The second case is when $j>j'$. In this case, $e\in \delta^{\rig}(u)$. Recall that, from \Cref{obs: left and right down-edges}, 
 $|\delta^{\down,\rig}(u)|\leq 2(|\delta^{\down,\lef}(u)|+|\delta^{\down,\straight'}(u)|)$. Therefore, we can define another mapping $g^{\lef}(u)$, that maps the edges of $\delta^{\down,\rig}(u)$ to edges of $\delta^{\down,\lef}(u)\cup\delta^{\down,\straight'}(u)$, such that at most two edges are mapped to every edge of $\delta^{\down,\lef}(u)\cup\delta^{\down,\straight'}(u)$. We then let $e'$ be the edge to which $e$ is mapped by $g^{\lef}(u)$. As before, $e'\in E^{\lef}_j$ must hold, and so there is a left-monotone path $P^{\lef}(e')\in \pset^{\lef}_j$. We then let $P(e,u)$ be the path obtained by concatenating edge $e$ with path $P^{\lef}(e')$ as before. 
This finishes the definition of the path $P(e,u)$. Path $P(e,v)$ is defined symmetrically. Since the paths in $\left (\bigcup_{j=1}^h\pset_j^{\rig}\right )\cup \left(\bigcup_{j=1}^h\pset_j^{\rig}\right )$ cause congestion $O(\log m)$, it is easy to verify that the set $\set{P(e,v),P(e,u)\mid e=(u,v)\in \hat E}$ of paths causes congestion $O(\log m)$.

\renewcommand{\tE}{\tilde E}

\subsection{Proof of \Cref{obs: bad inded structure}}
\label{subsec: edges between Sis}

Fix an index $1<i<r$. Consider the two connected components of the Gomory-Hu tree $\tau$ that are obtained after the edge $(u_{i-1},u_i)$ is deleted from it. Denote the sets of vertices of the two components by $A$ and $B$, where $u_{i-1}\in A$. Recall that $(A,B)$ is the minimum cut separating $u_{i-1}$ from $u_{i}$ in $\cH$. Observe that $A=V(S_1)\cup \cdots\cup V(S_{i-1})$, while $B=V(S_i)\cup \cdots\cup V(S_r)$. Note also that:

\[|E(A,B)|\geq |E_{i-1}'|+|\tE_i^{\lef}|+|\tE_{i+1}^{\lef}|+|\tE_i^{\through}|\]

\begin{figure}[h]
	\centering
	\includegraphics[scale=0.12]{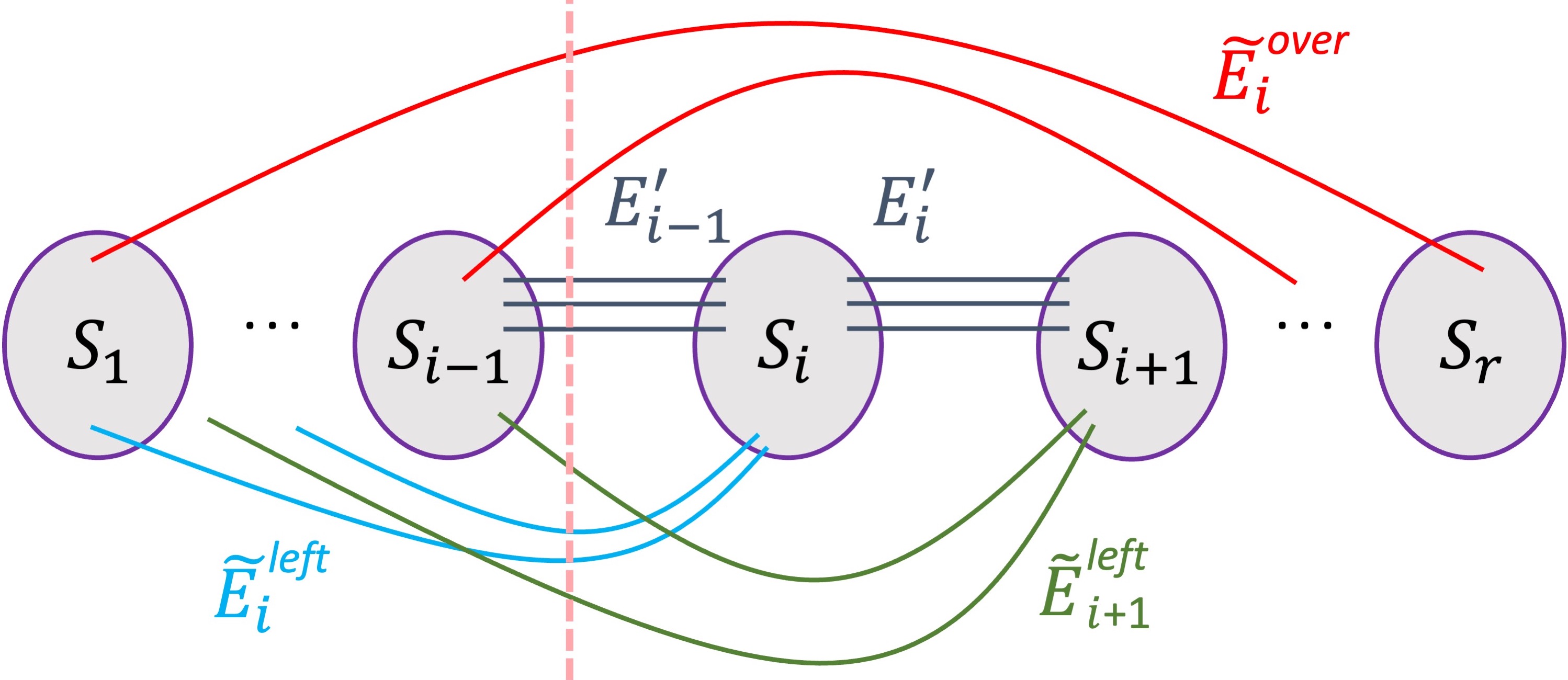}
	\caption{Illustration for the Proof of \Cref{obs: bad inded structure}. Cut $(A,B)$ is shown in a pink dashed line.}\label{fig: NF10a}
\end{figure}

(see \Cref{fig: NF10a}).

Next, we consider another $u_{i-1}$--$u_i$ cut $(X,Y)$ in $\cH$, where $Y=V(S_i)$, and $Xß=V(\cH)\setminus Y$. Observe that:

\[|E(X,Y)|= |E_{i-1}'|+|\tE_i^{\lef}|+ |\tE_i^{\rig}|+|E_i'|\]

\begin{figure}[h]
	\centering
	\includegraphics[scale=0.12]{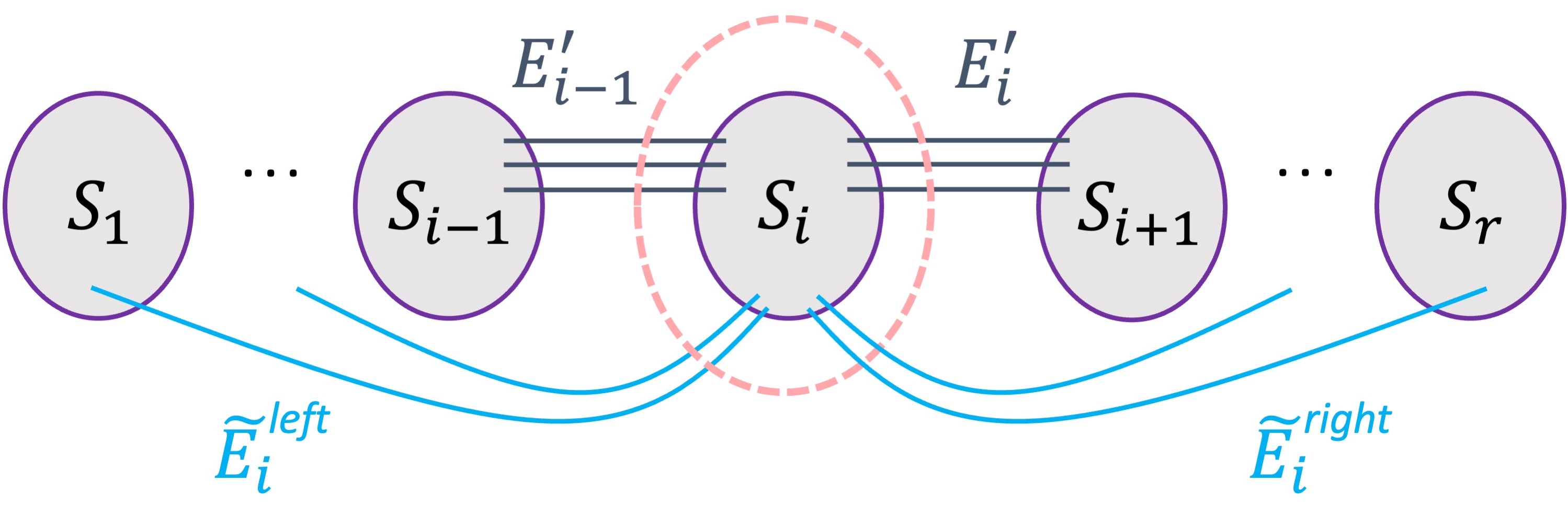}
	\caption{Illustration for the Proof of \Cref{obs: bad inded structure}. Cut $(X,Y)$ is shown in a pink dashed line.}\label{fig: NF10b}
\end{figure}

(see \Cref{fig: NF10b}).

Since $|E(A,B)|\leq |E(X,Y)|$ must hold, we conclude that:

\begin{equation}\label{eq: bound from left}
|\tE_{i+1}^{\lef}|+|\tE_i^{\through}|\leq  |\tE_i^{\rig}|+|E_i'|.
\end{equation}

We now repeat the same reasoning with cuts separating $u_{i+1}$ from $u_{i+2}$. 
Consider the two connected components of the Gomory-Hu tree $\tau$ that are obtained after the edge $(u_{i+1},u_{i+2})$ is deleted from it. Denote the sets of vertices of the two components by $A'$ and, $B'$, where $u_{i+1}\in A'$. Recall that $(A',B')$ is the minimum cut separating $u_{i+1}$ from $u_{i+2}$ in $\cH$. Note that $A'=V(S_1)\cup \cdots\cup V(S_{i+1})$, and:

\[|E(A',B')|\geq |E_{i+1}'|+|\tE_{i+1}^{\rig}|+|\tE_{i}^{\rig}|+|\tE_i^{\through}|\]

\begin{figure}[h]
	\centering
	\includegraphics[scale=0.12]{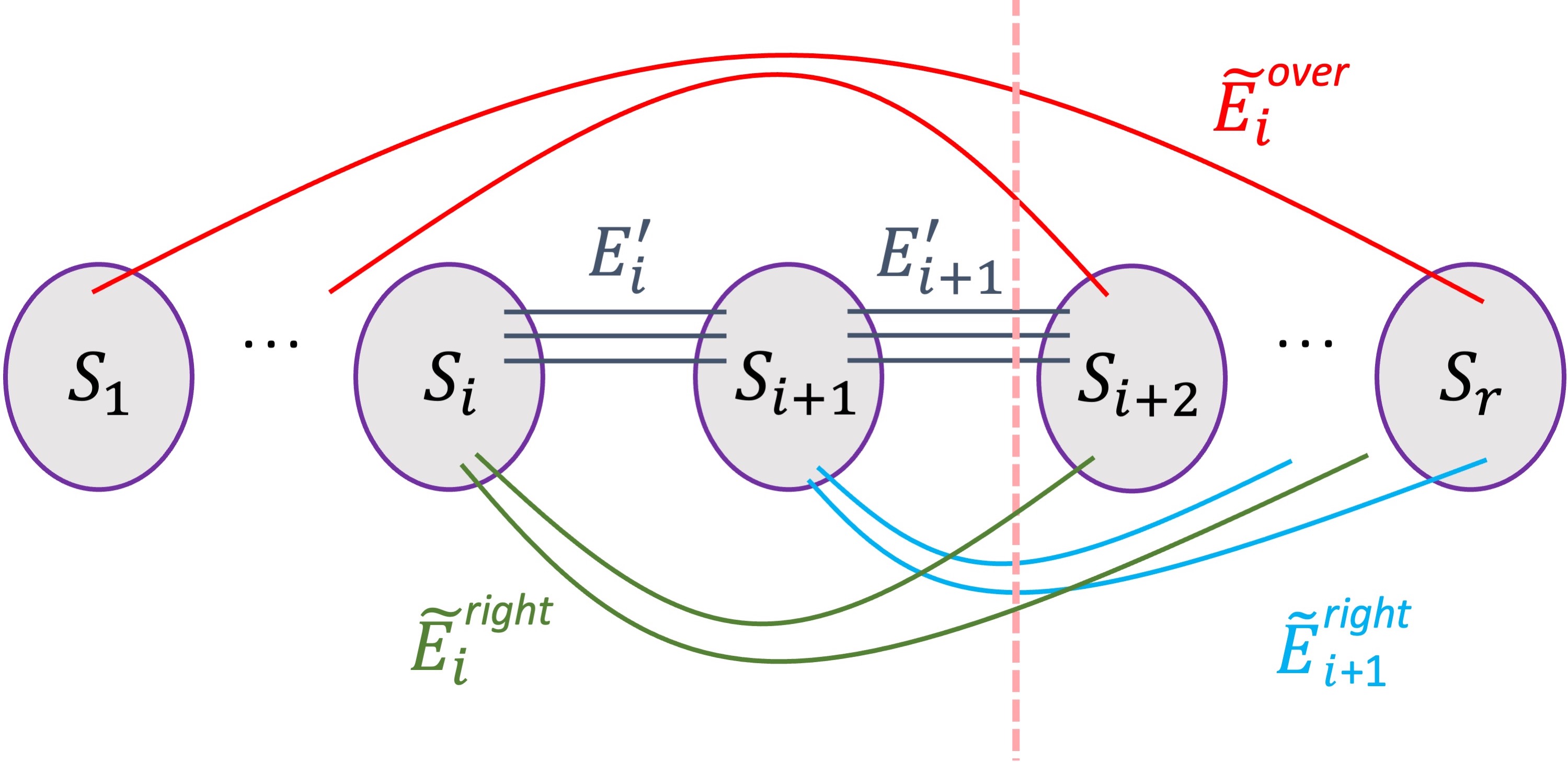}
	\caption{Illustration for the proof of \Cref{obs: bad inded structure}, with cut $(A,B)$ shown in a pink dashed line.
	}\label{fig: NF11a}
\end{figure}

(see \Cref{fig: NF11a}).

We now consider another $u_{i+1}$--$u_{i+2}$ cut $(X',Y')$ in $\cH$, where $X'=V(S_{i+1})$, and $Y'=V(\cH)\setminus X'$. Observe that:

\[|E(X',Y')= |E_{i+1}'|+|\tE_{i+1}^{\rig}|+|E_{i}'|+|\tE_{i+1}^{\lef}|\]

\begin{figure}[h]
	\centering
	\includegraphics[scale=0.12]{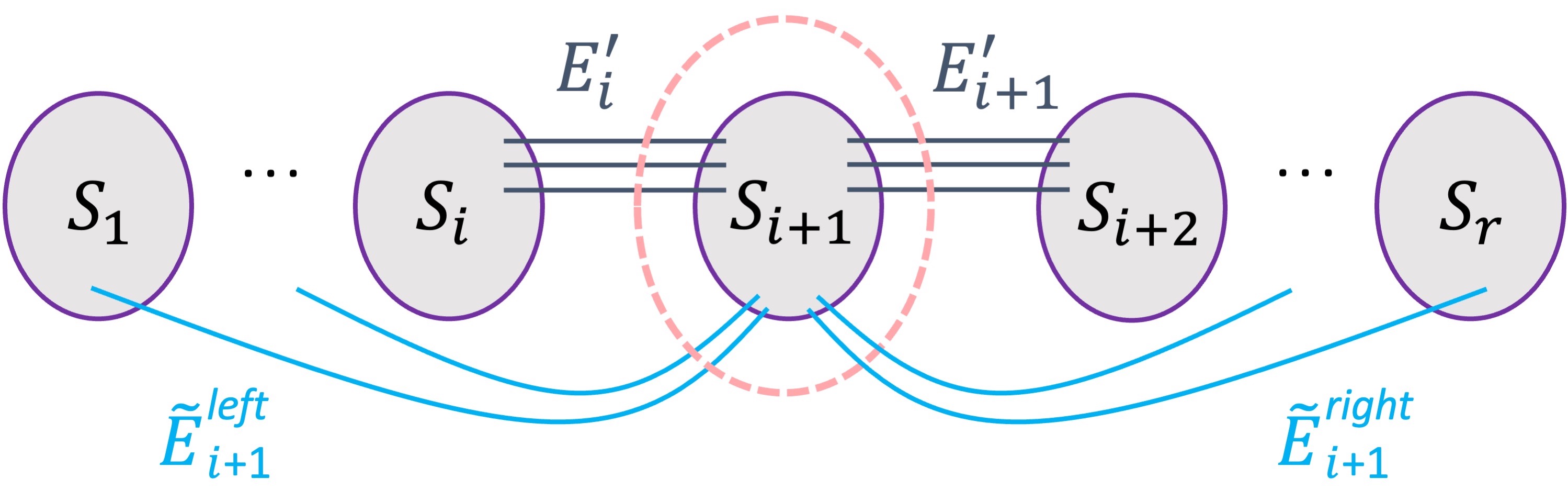}
	\caption{Illustration for the proof of \Cref{obs: bad inded structure}, with cut $(X,Y)$ shown in a pink dashed line.}\label{fig: NF11b}
\end{figure}

(see \Cref{fig: NF11b}).

Since $|E(A',B')|\leq |E(X',Y')|$ must hold, we conclude that:

\begin{equation}\label{eq: bound from right}
|\tE_{i}^{\rig}|+|\tE_i^{\through}|\leq 
|E_{i}'|+|\tE_{i+1}^{\lef}|.
\end{equation}

By adding \Cref{eq: bound from left} and \Cref{eq: bound from right}, we conclude that $|\tE_i^{\through}|\leq 
|E_{i}'|$. We also immediately get that $|\tE_{i+1}^{\lef}|\leq 
|E_{i}'|+|\tE_{i}^{\rig}|$ from \Cref{eq: bound from left} and $|\tE_{i}^{\rig}|\leq 
|E_{i}'|+|\tE_{i+1}^{\lef}|$ from \Cref{eq: bound from right}.


\subsection{Proof of \Cref{claim: bound S' to S'' edges}}
\label{subsec: bound S' to S'' edges}
Fix an index $1<i<r$. We prove that $|\downedges{i}|\leq 0.1|\rightedges{i}|$, and show the existence of the set $\pset^{\rig}$ of paths. The proof that $|\downedges{i}|\leq 0.1|\leftedges{i}|$ and the proof of the existence of the set $\pset^{\lef}$ of paths is symmetric.

Let $\pset$ be the set of all paths $P$ in graph $\cH$, such that the first edge of $P$ lies in $\downedges{i}$, the last edge lies in $\rightedges{i}$, and all inner vertices of $P$ lies in $S''_i$. We show a flow $f$, defined over the set $\pset$ of paths, in which every edge in $\downedges{i}$ sends one flow unit, every edge in $\rightedges{i}$ receives at most $1/10$ flow unit, and each edge of $E(S''_i)$ carries at most one flow unit. The existence of such a flow will then prove that $|\downedges{i}|\leq 0.1|\rightedges{i}|$, and will imply the existence of the path set $\pset^{\rig}$ with the required properties from the integrality of flow. 

In order to define the flow $f$, we consider the vertices of $S''_i$ in the decreasing order of their levels $L'_h,\ldots,L'_1$. When we consider level $L'_j$, for $1\leq j\leq h$, we assume that, for every vertex $v\in L'_j\cap S''_i$, for every edge $e\in \delta^{\up}(v)$, connecting $v$ to another vertex of $S''_i$, the flow on edge $e$ is fixed already, and that the only edges that are incident to $v$ that may carry non-zero flow are edges of $\delta^{\up}(v)\cup\delta^{\down,\straight''}(v)$. Initially, for every edge $e\in \downedges{i}$, we set $f(e)=1$. Note that for every edge $e\in \downedges{i}$, if $v$ is the endpoint of $e$ lying in $S''_i$, then $e\in \delta^{\down,\straight''}(v)$.

We start with the level $L'_h$. Consider any vertex $v\in L'_h$. From the definition, $\delta^{\up}(v)=\emptyset$. Recall that, from \Cref{obs: left and right down-edges}, $|\delta^{\down,\rig}(v)|+|\delta^{\down,\lef}(v)|+| \delta^{\down,\straight'}(v)|\geq 63|\delta(v)|/64$, and additionally, $|\delta^{\down,\lef}(v)|\leq 2(|\delta^{\down,\rig}(v)|+|\delta^{\down,\straight'}(v)|)$. 
Therefore, $|\delta^{\down,\rig}(v)|+|\delta^{\down,\straight'}(v)|\geq 21|\delta(v)|/64$ must hold. At the same time, $|\delta^{\down,\straight''}(v)|\leq |\delta(v)|/128$. We now consider two cases. The first case is when $|\delta^{\down,\straight'}(v)|\geq  |\delta(v)|/64$. In this case, $|\delta^{\down,\straight''}(v)|\leq |\delta^{\down,\straight'}(v)|$ holds. We assign, to each edge $e\in \delta^{\down,\straight''}(v)$, a distinct edge $e'\in \delta^{\down,\straight'}(v)$, and we set $f(e')=f(e)$. Consider now the second case, where $|\delta^{\down,\rig}(v)|\geq |\delta(v)|/8$ must hold. In this case, $|\delta^{\down, \straight''}(v)|\leq |\delta^{\down,\straight'}(v)|/10$. We can now assign, to each edge $e\in  \delta^{\down,\straight''}(v)$ ten distinct edges $e_1,\ldots,e_{10}\in \delta^{\down,\rig}(v)$, such that each edge of $\delta^{\down,\rig}(v)$ is assigned to at most one edge of $\delta^{\down,\straight''}(v)$. If edge $e'\in \delta^{\down,\rig}(v)$ is assigned to edge $e\in  \delta^{\down,\straight''}(v)$, then we set $f(e')=f(e)/10$.

Assume now that we have processed levels $L'_h,\ldots,L'_{j+1}$, and consider some level $L'_j$. Let $v\in L'_j\cap S''_i$ be any vertex. The processing of vertex $v$ is similar to the one above, except that now some edges of $\delta^{\up}(v)$ may carry flow, and we need to forward this flow to edges in $\delta^{\down}(v)$. As before, from \Cref{obs: left and right down-edges}, $|\delta^{\down,\rig}(v)|+|\delta^{\down,\lef}(v)|+| \delta^{\down,\straight'}(v)|\geq 63|\delta(v)|/64$, and additionally, $|\delta^{\down,\lef}(v)|\leq 2(|\delta^{\down,\rig}(v)|+|\delta^{\down,\straight'}(v)|)$. 
Therefore, $|\delta^{\down,\rig}(v)|+|\delta^{\down,\straight'}(v)|\geq 21|\delta(v)|/64$ must hold. At the same time, $|\delta^{\down,\straight''}(v)|\leq |\delta(v)|/128$, and $\delta^{\up}(v)\leq \delta^{\down}(v)/\log m$. We again consider two cases. The first case is when $|\delta^{\down,\straight'}(v)|\geq  |\delta(v)|/64$. In this case, $|\delta^{\down''}(v)|+|\delta^{\up}(v)|\leq |\delta^{\down,\straight'}(v)|$ holds. We assign, to each edge $e\in \delta^{\down,\straight''}(v)\cup \delta^{\up}(v)$, a distinct edge $e'\in \delta^{\down,\straight'}(v)$, and we set $f(e')=f(e)$. Consider now the second case, where $|\delta^{\down,\rig}(v)|\geq |\delta(v)|/8$ must hold. In this case, $|\delta^{\down, \straight''}(v)|+|\delta^{\up}(v)|\leq |\delta^{\down,\straight'}(v)|/10$. As before, we can  assign, to each edge $e\in  \delta^{\down,\straight''}(v)\cup \delta^{\up}(v)$ ten distinct edges $e_1,\ldots,e_{10}\in \delta^{\down,\rig}(v)$, such that each edge of $\delta^{\down,\rig}(v)$ is assigned to at most one edge of $\delta^{\down,\straight''}(v)\cup \delta^{\up}(v)$. If edge $e'\in \delta^{\down,\rig}(v)$ is assigned to edge $e\in  \delta^{\down,\straight''}(v)\cup \delta^{\up}(v)$, then we set $f(e')=f(e)/10$.

Notice that, when vertices $v\in L'_1$ are processed, we are guaranteed that $\delta^{\down'}(v)=\emptyset$, and so eventually all flow originating at the edges of $\downedges{i}$ reaches the edges of $\rightedges{i}$. This completes the definition of the flow $f$. It is immediate to verify that $f$ is defined over the set $\pset$ of paths;  every edge in $\downedges{i}$ sends one flow unit; every edge in $\rightedges{i}$ receives at most $1/10$ flow units; and each edge of $E(S''_i)$ carries at most one flow unit. 



\subsection{Proof of \Cref{claim: bound left and right for S''}}
\label{subsec: bound left and right for S''}
\begin{figure}[h]
	\centering
	\includegraphics[scale=0.12]{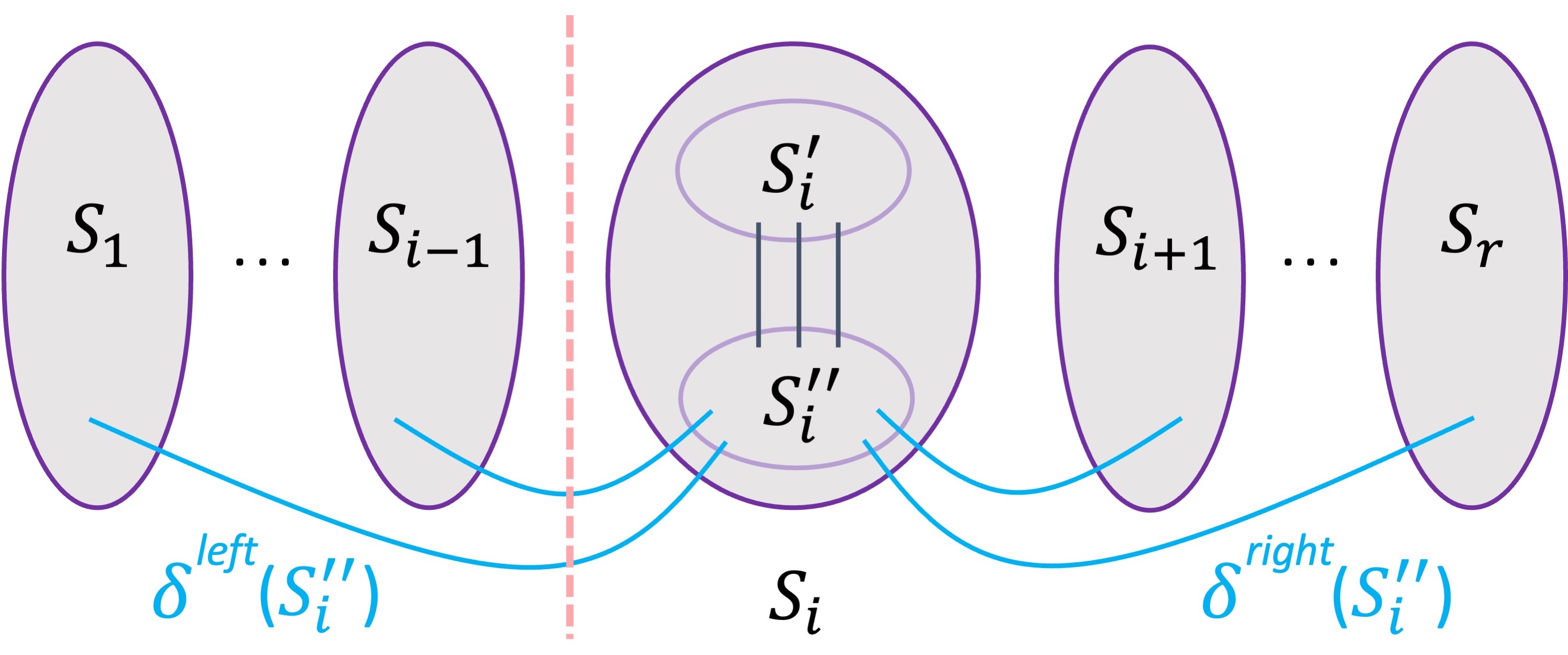}
	\caption{Illustration for the proof of \Cref{claim: bound left and right for S''}, with cut $(A,B)$ shown in  a pink dashed line.}\label{fig: NF14a}
\end{figure}

We prove that $|\leftedges{i}|\leq 1.1|\rightedges{i}|$; the proof that $|\rightedges{i}|\leq 1.1|\leftedges{i}|$ is symmetric.
Consider the cut $(A,B)$ in graph $\cH$, where $A=V(S_1)\cup\cdots\cup V(S_{i-1})$, and $B=V(S_i)\cup\cdots\cup V(S_r)$ (see \Cref{fig: NF14a}).

\begin{figure}[h]
	\centering
	\includegraphics[scale=0.12]{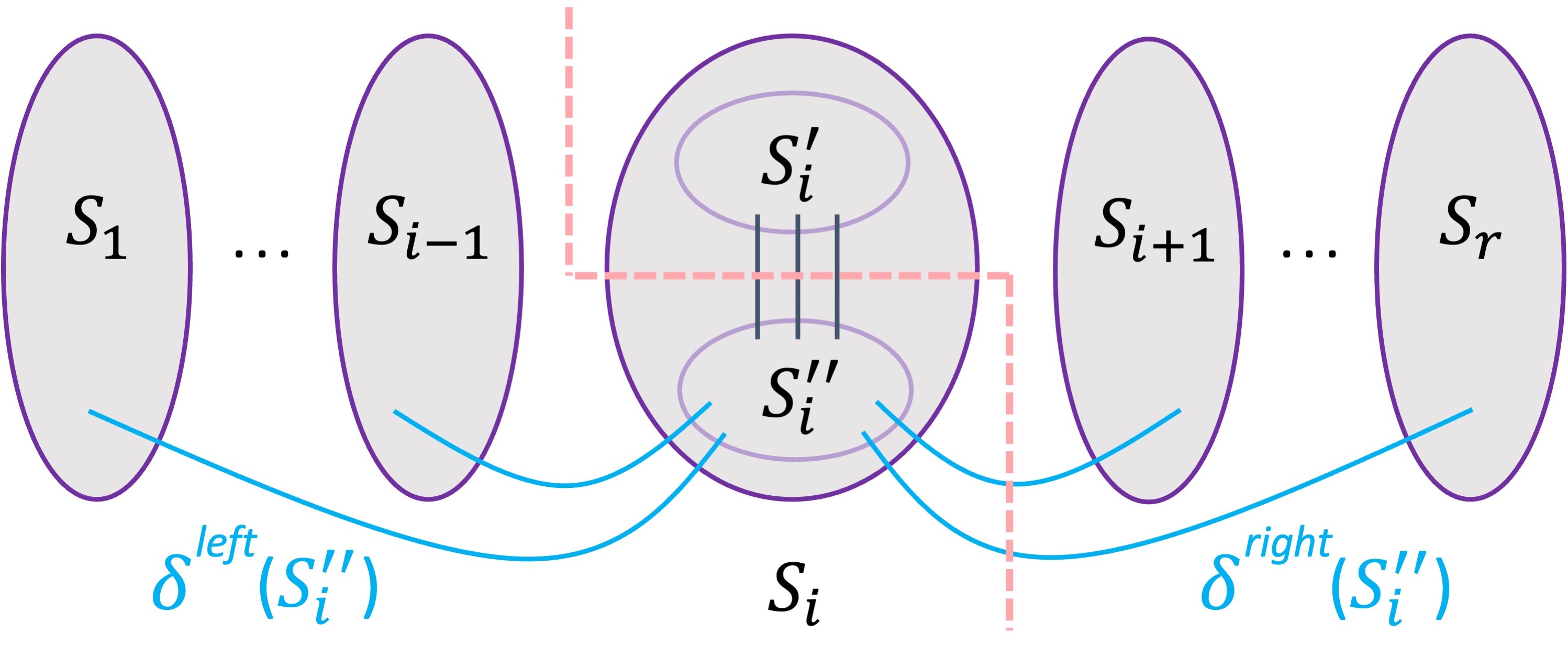}
	\caption{Illustration for the proof of \Cref{claim: bound left and right for S''}, with cut $(A',B')$ shown in  a pink dashed line. Edges of $\delta^{\down}(S''_{i})$ are shown in black.}\label{fig: NF14b}
\end{figure}

Note that $A$ and $B$ are precisely the sets of vertices of the two connected components of the graph $\tau\setminus\set{(u_{i-1},u_i)}$, where $\tau$ is the Gomory-Hu tree of graph $\cH$. Therefore, $(A,B)$ is the minimum $u_{i-1}$-$u_i$ cut in $\cH$. We now consider another $u_{i-1}$-$u_i$ cut $(A',B')$ in $\cH$, where $A'=A\cup V(S''_i)$, and $B'=B\setminus V(S''_i)$ (see \Cref{fig: NF14b}).

Observe that edges of $\leftedges{i}$ contribute to $E(A,B)$ but not to $E(A',B')$, while edges of $\rightedges{i}\cup \downedges{i}$ contribute to $E(A',B')$ but not to $E(A,B)$, and this is the only difference between the two edge sets. Since $(A,B)$ is the minimum $u_{i-1}$-$u_i$ cut, we get that $|\leftedges{i}|\leq |\rightedges{i}|+|\downedges{i}|$ must hold. Since, from \Cref{claim: bound S' to S'' edges}, $|\downedges{i}|\leq 0.1|\rightedges{i}|$, we conclude that $|\leftedges{i}|\leq 1.1|\rightedges{i}|$. 

\subsection{Proof of \Cref{claim left edges for S' and S''}}
\label{subsec: left edges for S' and S''}
Fix an index $1<i<r$. We start by proving that $|\rightedges{i}|\leq 1.3|E_{i}|+1.3|\rightCedges{i}|$.

\begin{figure}[h]
	\centering
	\includegraphics[scale=0.12]{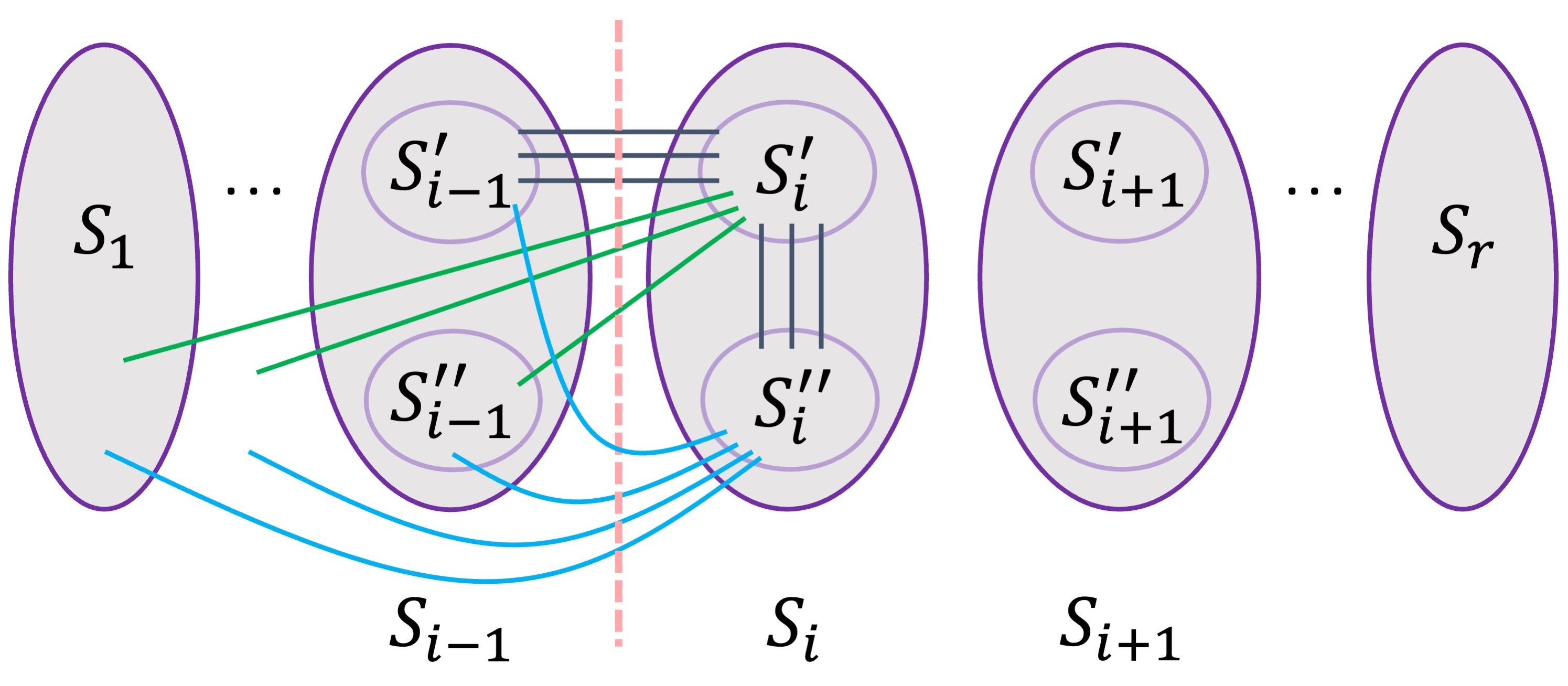}
	\caption{Illustration for the proof of \Cref{claim left edges for S' and S''}, with cut $(A,B)$ shown in a pink dashed line. Edges of $\delta^{\lef}(S'_{i})$ are shown in light green, and edges of $\delta^{\lef}(S''_{i})$ are shown in blue. 
	}\label{fig: NF16a}
\end{figure}

\begin{figure}[h]
	\centering
	\includegraphics[scale=0.12]{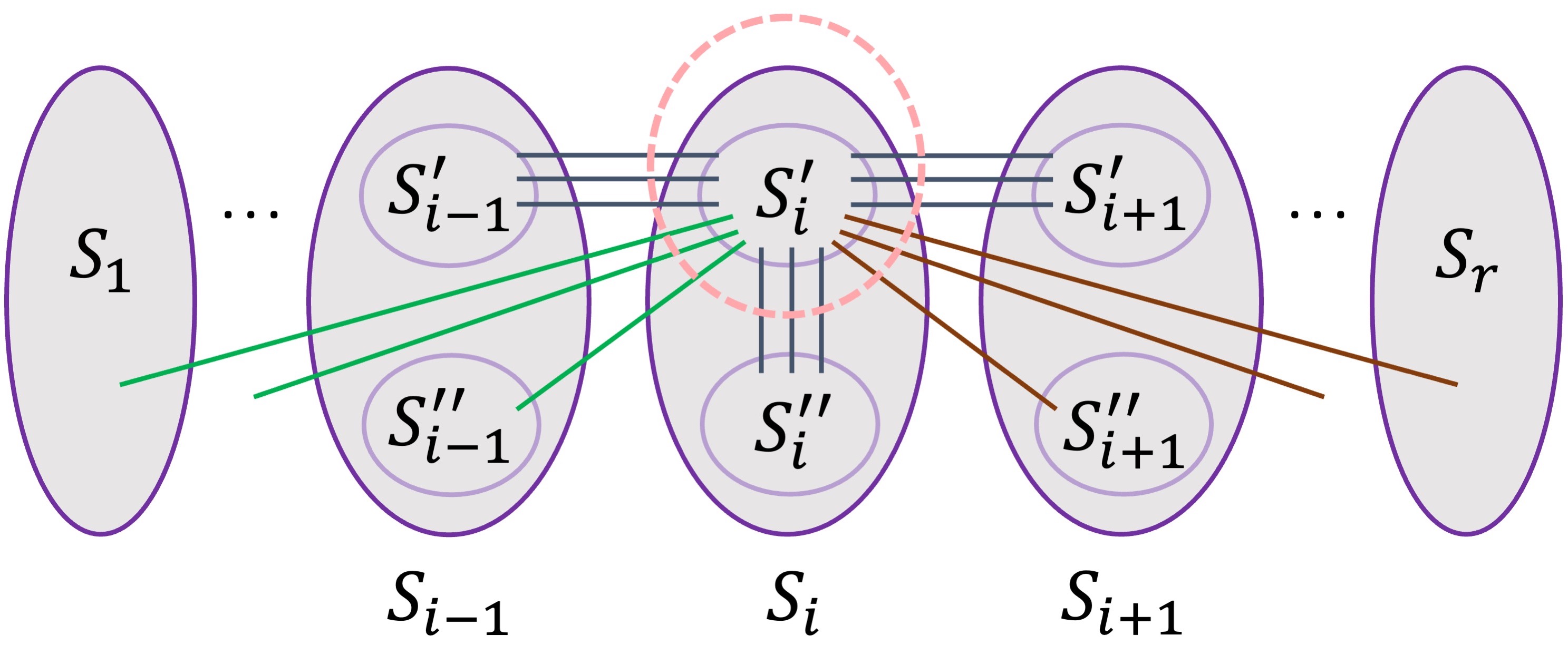}
	\caption{Illustration for the proof of \Cref{claim left edges for S' and S''}, with cut $(A',B')$ shown in a pink dashed line. Edges of $\delta^{\rig}(S'_{i})$ are shown in brown, and edges of $\delta^{\lef}(S'_{i})$ are shown in green.}\label{fig: NF16b}
\end{figure}

As before, we consider the cut $(A,B)$ in graph $\cH$, where $A=V(S_1)\cup\cdots\cup V(S_{i-1})$, and $B=V(S_i)\cup\cdots\cup V(S_r)$  (see \Cref{fig: NF16a}). As before, $A$ and $B$ are precisely the sets of vertices of the two connected components of the graph $\tau\setminus\set{(u_{i-1},u_i)}$, where $\tau$ is the Gomory-Hu tree of graph $\cH$, and so $(A,B)$ is the minimum $u_{i-1}$-$u_i$ cut in $\cH$. We now consider another $u_{i-1}$-$u_i$ cut $(A',B')$ in $\cH$, where $B'=V(S'_i)$, and $A'=V(\cH)\setminus V(S'_i)$ (see  \Cref{fig: NF16b}).

Note that $|E(A,B)|\geq |E_{i-1}|+|\leftCedges{i}|+|\leftedges{i}|$, while $|E(A',B')|=|E_{i-1}|+|\leftCedges{i}|+|\downedges{i}|+|E_{i}|+|\rightCedges{i}|$. From the fact that $(A,B)$ is the minimum $u_{i-1}$-$u_i$ cut, it must be the case that:

\[
|\leftedges{i}|\leq |\downedges{i}|+|E_{i}|+|\rightCedges{i}|
\]

Since, from \Cref{claim: bound left and right for S''}, $|\rightedges{i}|\leq 1.1|\leftedges{i}|$, and, from \Cref{claim: bound S' to S'' edges},  $|\downedges{i}|\leq 0.1|\rightedges{i}|$, we get that:

\[
\begin{split}
|\rightedges{i}|&\leq 1.1|\leftedges{i}|\\ &\leq 1.1|\downedges{i}|+1.1|E_{i}|+1.1|\rightCedges{i}|\\ &\leq 0.11|\rightedges{i}|+1.1|E_{i}|+1.1|\rightCedges{i}|,
\end{split}
\]

and so $|\rightedges{i}|\leq 1.3|E_{i}|+1.3|\rightCedges{i}|$.

\begin{figure}[h]
	\centering
	\includegraphics[scale=0.12]{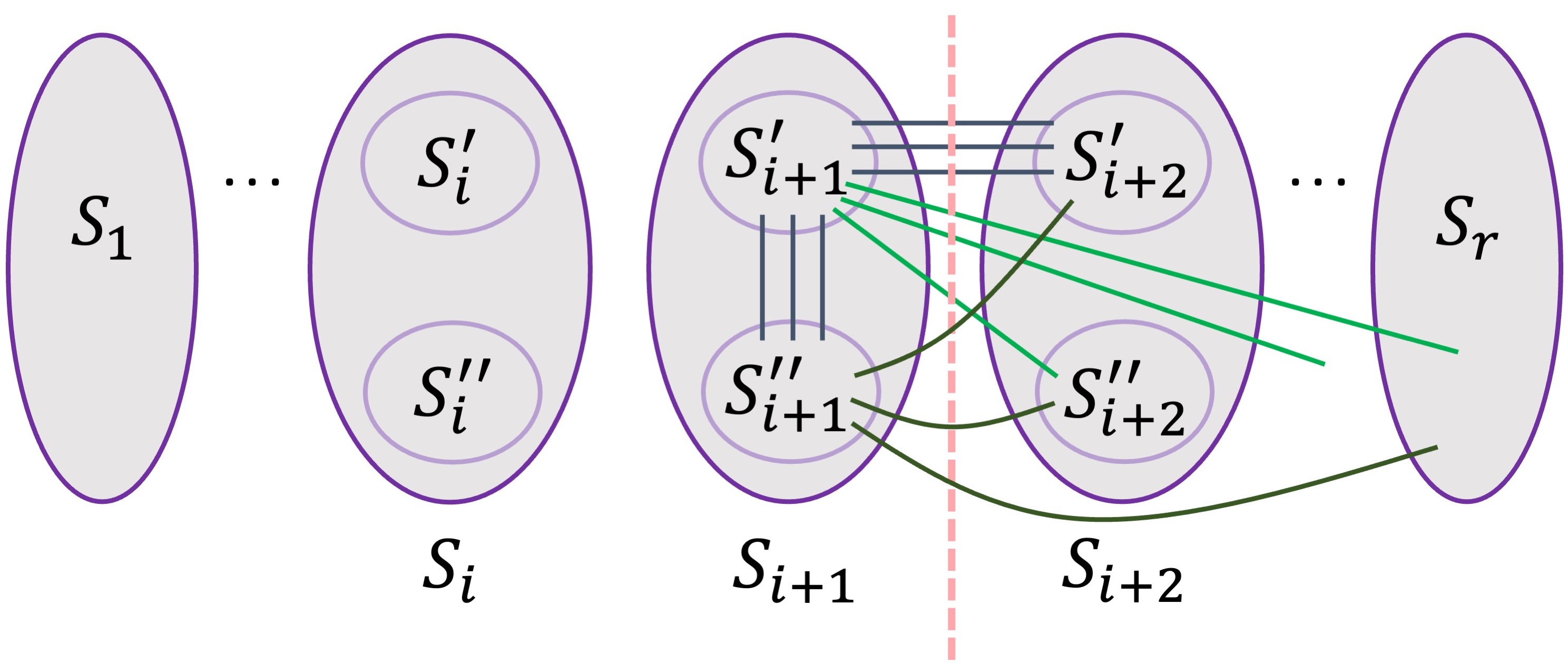}
	\caption{Illustration for the proof of \Cref{claim left edges for S' and S''}, with cut $(X,Y)$ shown in a pink dashed line. Edges of $\delta^{\rig}(S'_{i+1})$ are shown in light green, and edges of $\delta^{\rig}(S''_{i+1})$ are shown in dark green.}\label{fig: NF22a}
\end{figure}

\begin{figure}[h]
	\centering
	\includegraphics[scale=0.12]{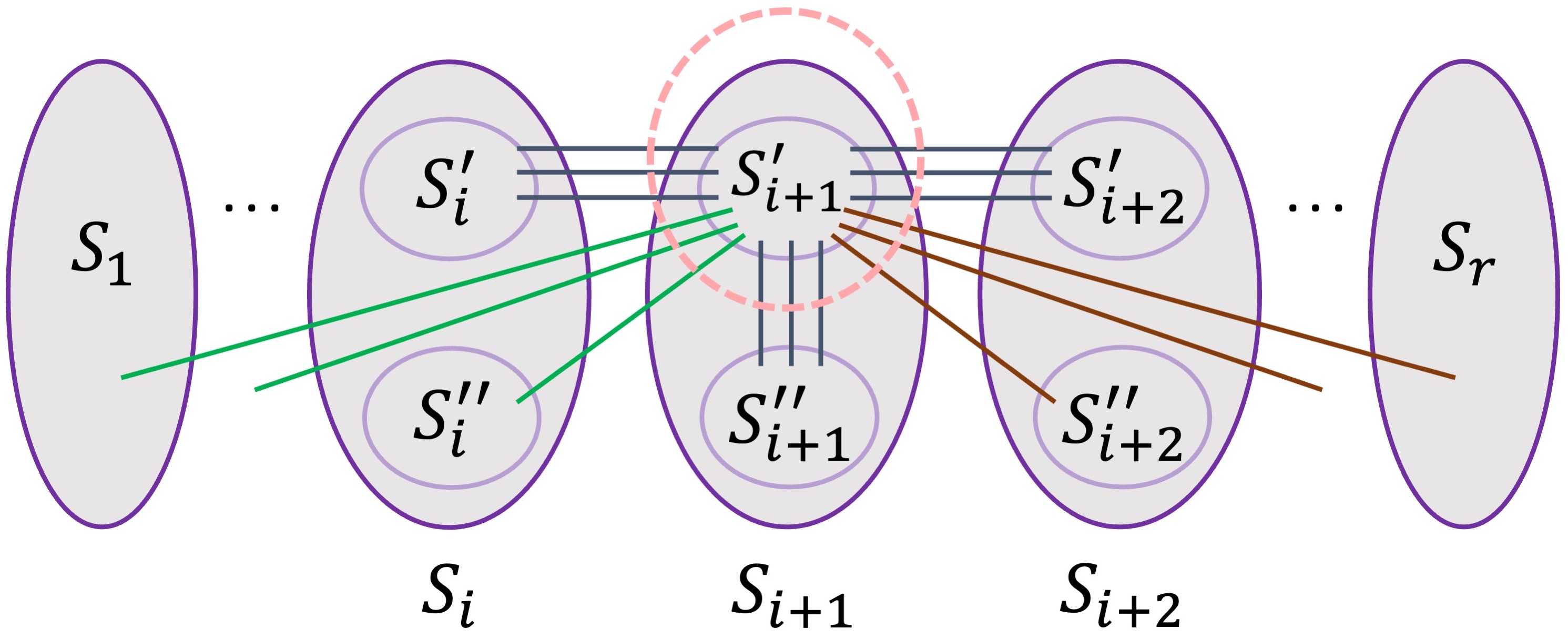}
	\caption{Illustration for the proof of \Cref{claim left edges for S' and S''}, with cut $(X',Y')$ shown in a pink dashed line. Edges of $\delta^{\rig}(S'_{i+1})$ are shown in brown, and edges of $\delta^{\lef}(S'_{i+1})$ are shown in green.}\label{fig: NF22b}
\end{figure}

The proof that $|\leftedges{i+1}|\leq 1.3|E_{i}|+1.3|\leftCedges{i+1}|$ is symmetric. 
We consider the cut $(X,Y)$ in graph $\cH$, where $X=V(S_1)\cup\cdots\cup V(S_{i+1})$, and $Y=V(S_{i+2})\cup\cdots\cup V(S_r)$ (see  \Cref{fig: NF22a}). Note that $X$ and $Y$ are precisely the sets of vertices of the two connected components of the graph $\tau\setminus\set{(u_{i+1},u_{i+2})}$, where $\tau$ is the Gomory-Hu tree of graph $\cH$, and so $(X,Y)$ is the minimum $u_{i+1}$-$u_{i+2}$ cut in $\cH$. We now consider another $u_{i+1}$-$u_{i+2}$ cut $(X',Y')$ in $\cH$, where $X'=V(S'_{i+1})$, and $Y'=V(\cH)\setminus V(S'_{i+1})$ (see \Cref{fig: NF22b}).

Note that $|E(X,Y)|\geq |E_{i+1}|+|\rightCedges{i+1}|+|\rightedges{i+1}|$, while $|E(X',Y')|=|E_{i+1}|+|\rightCedges{i+1}|+|\leftCedges{i+1}|+|\downedges{i+1}|+|E_{i}|$. From the fact that $(X,Y)$ is the minimum $u_{i}$-$u_{i+1}$ cut, it must be the case that:

\[
|\rightedges{i+1}|\leq |\leftCedges{i+1}|+|\downedges{i+1}|+|E_{i}|
\]

Since, from \Cref{claim: bound left and right for S''}, $|\leftedges{i+1}|\leq 1.1|\rightedges{i+1}|$, and, from \Cref{claim: bound S' to S'' edges},  $|\downedges{i+1}|\leq 0.1|\leftedges{i+1}|$, we get that:

\[
\begin{split}
|\leftedges{i+1}|&\leq 1.1|\rightedges{i+1}|\\ &\leq 1.1|\downedges{i+1}|+1.1|E_{i}|+1.1|\leftCedges{i+1}|\\ &\leq 0.11|\leftedges{i+1}|+1.1|E_{i}|+1.1|\leftCedges{i}|,
\end{split}
\]

and so $|\leftedges{i+1}|\leq 1.3|E_{i}|+1.3|\leftCedges{i+1}|$.



\subsection{Proof of \Cref{claim left edges for S' only}}
\label{subsec: left edges for S' only}

\begin{figure}[h]
	\centering
	\includegraphics[scale=0.12]{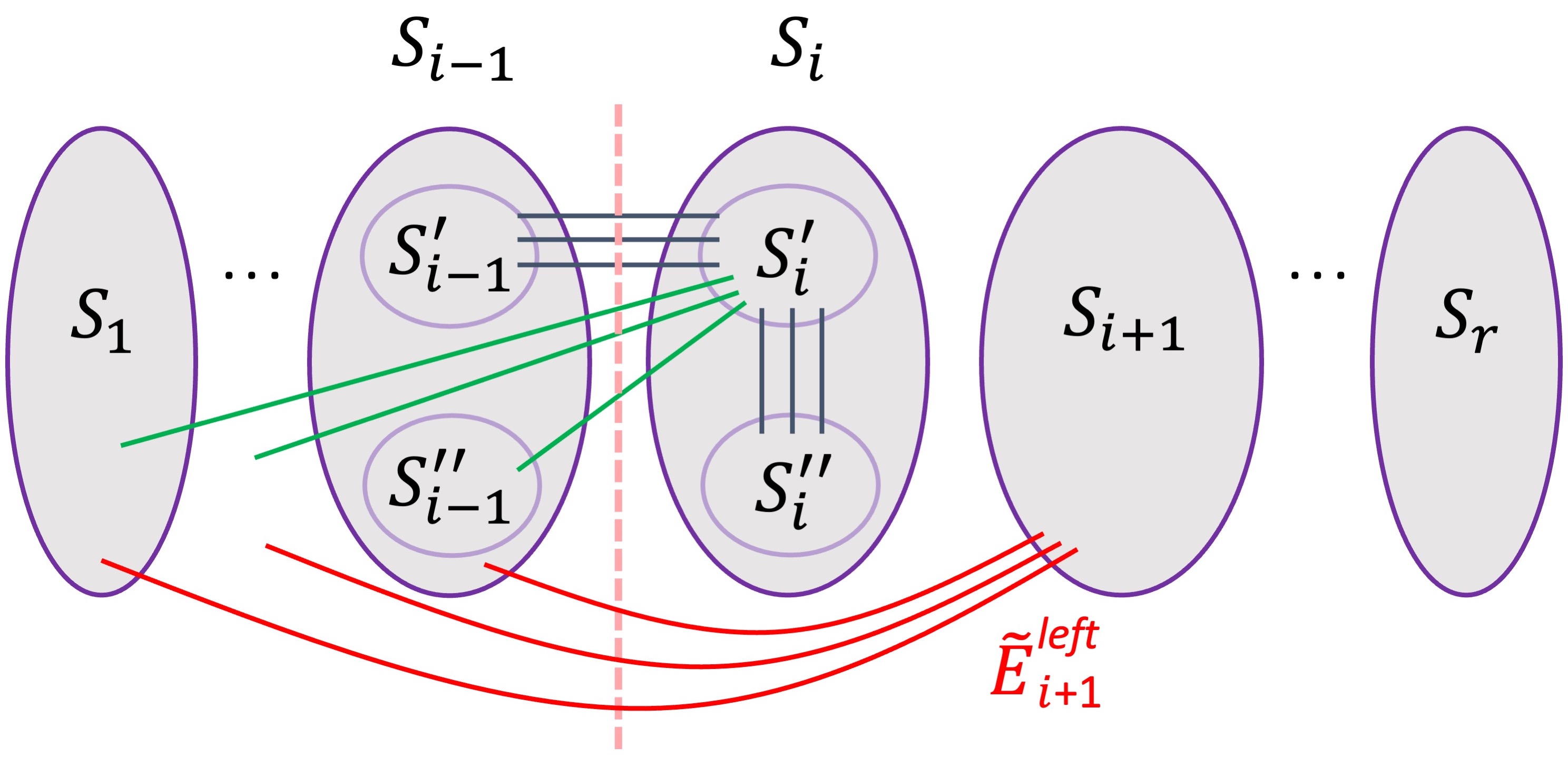}
	\caption{Illustration for the proof of \Cref{claim left edges for S' only}, with cut $(A,B)$ shown in pink dashed line. Edges of $\delta^{\lef}(S'_{i})$ are shown in green.}\label{fig: NF17a}
\end{figure}

Fix an index $1\leq i<r$. 
As before, we consider the cut $(A,B)$ in graph $\cH$, where $A=V(S_1)\cup\cdots\cup V(S_{i-1})$, and $B=V(S_i)\cup\cdots\cup V(S_r)$. As before, $A$ and $B$ are precisely the sets of vertices of the two connected components of the graph $\tau\setminus\set{(u_{i-1},u_i)}$, where $\tau$ is the Gomory-Hu tree of graph $\cH$, and so $(A,B)$ is the minimum $u_{i-1}$-$u_i$ cut in $\cH$ (see \Cref{fig: NF17a}).

\begin{figure}[h]
	\centering
	\includegraphics[scale=0.12]{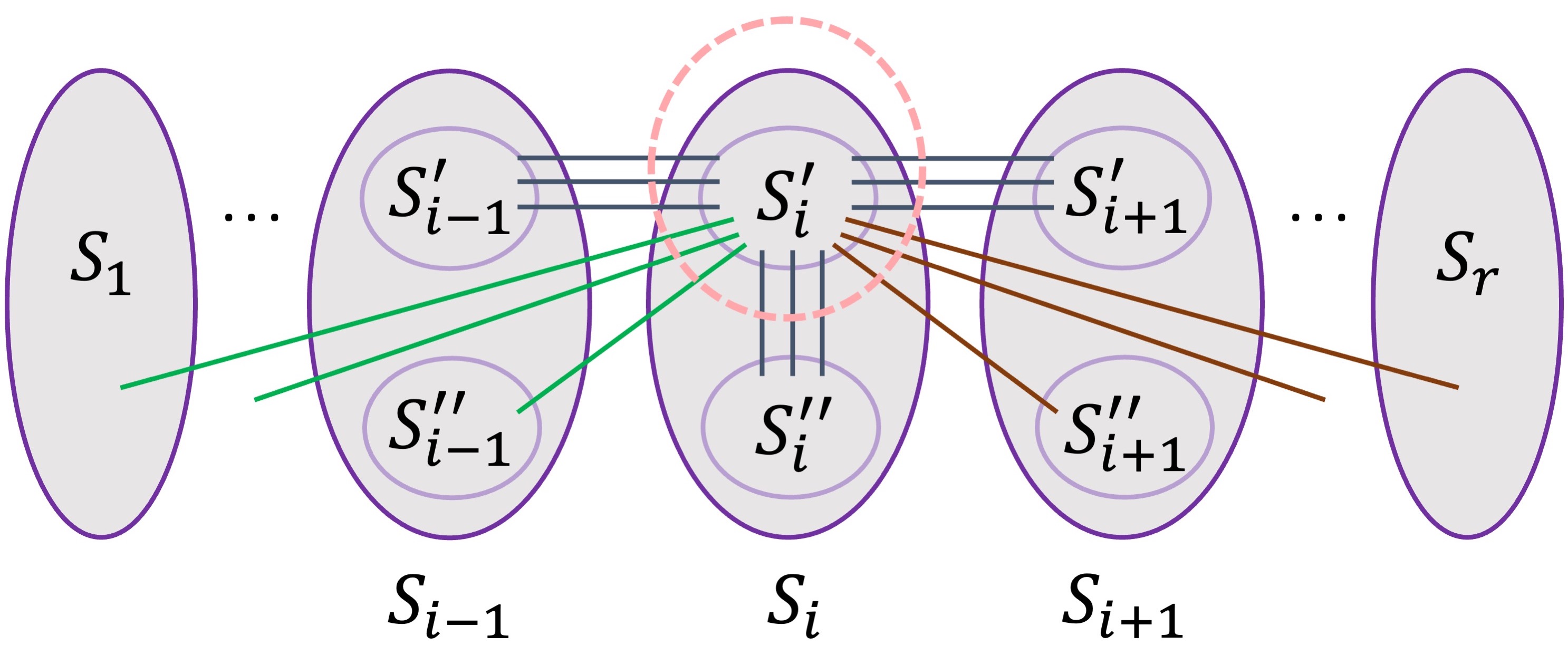}
	\caption{Illustration for the proof of \Cref{claim left edges for S' only}, with cut $(A',B')$ shown in pink dashed line. Edges of $\delta^{\lef}(S'_{i})$ are shown in green, and edges of $\delta^{\rig}(S'_{i})$ are shown in brown.}\label{fig: NF17b}
\end{figure}

We now consider another $u_{i-1}$-$u_i$ cut $(A',B')$ in $\cH$, where $B'=V(S'_i)$, and $A'=V(\cH)\setminus V(S'_i)$ (see \Cref{fig: NF17b}).

Note that $|E(A,B)|\geq |E_{i-1}|+|\leftCedges{i}|+|\tilde E_{i+1}^{\lef}|$, while $|E(A',B')|=|E_{i-1}|+|\leftCedges{i}|+|\downedges{i}|+|E_{i}|+|\rightCedges{i}|$. From the fact that $(A,B)$ is the minimum $u_{i-1}$-$u_i$ cut, it must be the case that:

\[
|\tilde E_{i+1}^{\lef}|\leq |\downedges{i}|+|E_{i}|+|\rightCedges{i}|
\]

Recall that, from \Cref{claim: bound S' to S'' edges}, $|\downedges{i}|\leq 0.1|\rightedges{i}|$, and, from \Cref{claim left edges for S' and S''}, $|\rightedges{i}|\leq 1.3|E_{i}|+1.3|\rightCedges{i}|$. 
Therefore, $\downedges{i}\leq 0.13|E_i|+0.13|\rightCedges{i}|$, and:

\begin{equation}\label{eq: left bound }
|\tilde E_{i+1}^{\lef}|\leq 1.13|E_{i}|+1.13|\rightCedges{i}|.
\end{equation}

\begin{figure}[h]
	\centering
	\includegraphics[scale=0.12]{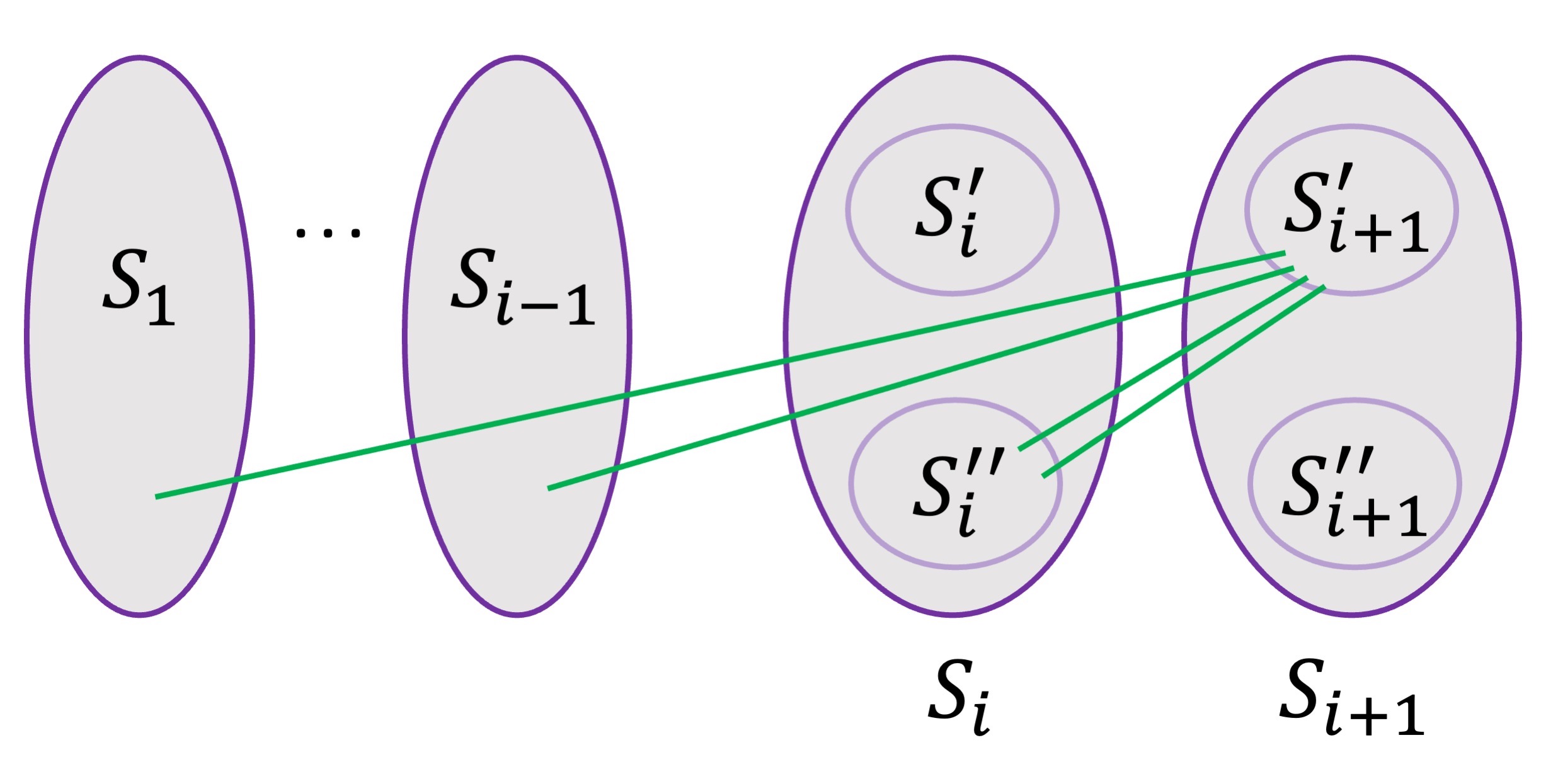}
	\caption{Set $\delta^{\lef}(S'_{i+1})$ of edges (shown in green).}\label{fig: NF18}
\end{figure}

Consider now the set $\leftCedges{i+1}$ of edges (see \Cref{fig: NF18}). 
Recall that this set contains every edge $e=(u,v)$ with $u\in V(S'_{i+1})$, and $v$ either lying in $V(S_1)\cup\cdots\cup V(S_{i-1})$, or in $V(S''_i)$. In the former case, $e\in \tilde E_{i+1}^{\lef}$, while in the latter case, $e\in \rightedges{i}$. Therefore, $|\leftCedges{i+1}|\leq |\tilde E_{i+1}^{\lef}|+|\rightedges{i}|$. From \Cref{claim left edges for S' and S''}: 	$|\rightedges{i}|\leq 1.3|E_{i}|+1.3|\rightCedges{i}|$.
Combining this with \Cref{eq: left bound }, we get that $|\leftCedges{i+1}|\leq |\tilde E_{i+1}^{\lef}|+|\rightedges{i}|\leq 2.5|E_{i}|+2.5|\rightCedges{i}|$, as required.

\begin{figure}[h]
	\centering
	\includegraphics[scale=0.12]{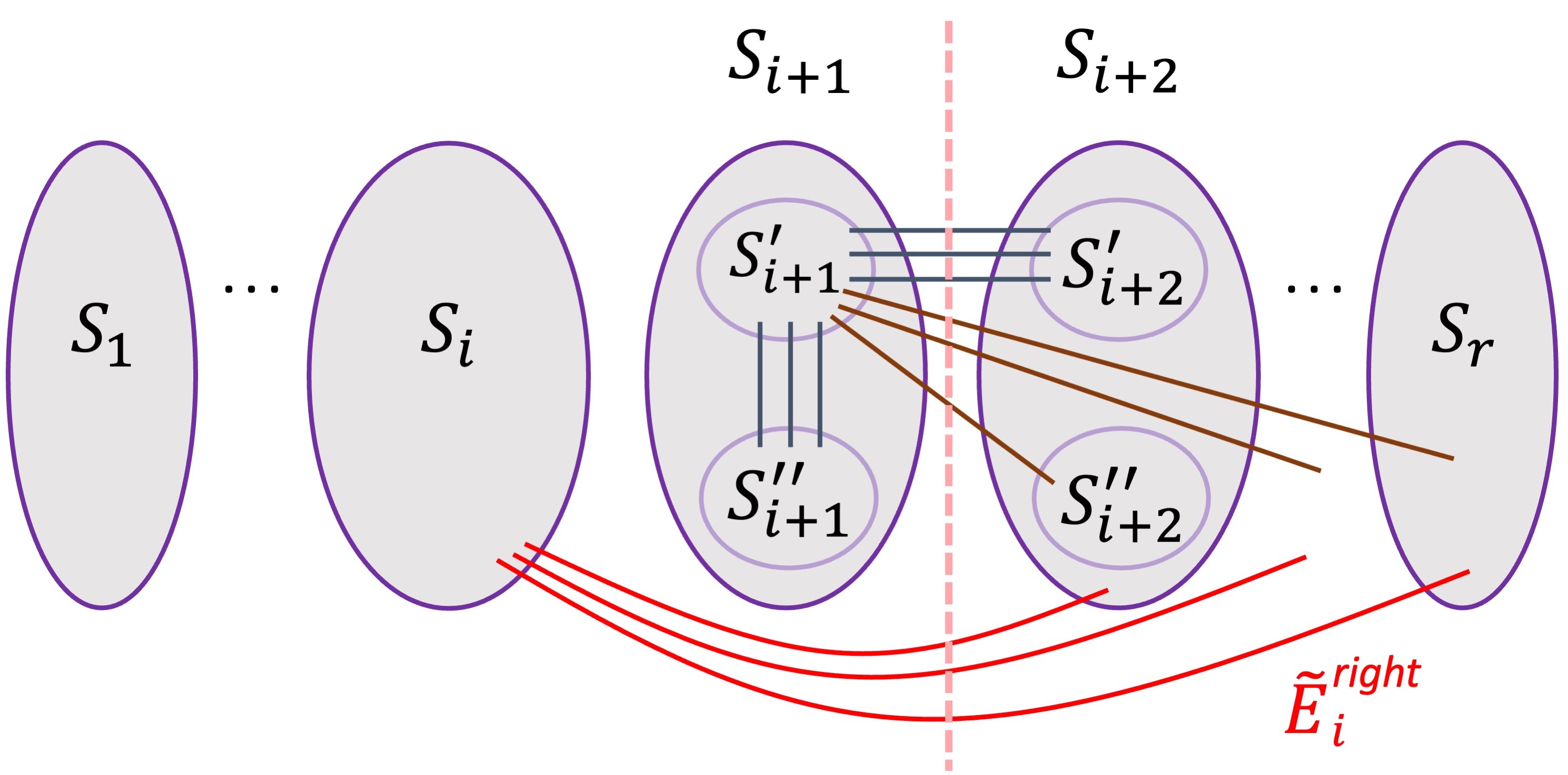}
	\caption{Illustration for the proof of \Cref{claim left edges for S' only}, with cut $(X,Y)$ shown in pink dashed line. Edges of $\delta^{\rig}(S'_{i+1})$ are shown in brown.}\label{fig: NF19a}
\end{figure}

We now employ a symmetric argument in order to bound $|\rightCedges{i}|$: consider the cut $(X,Y)$ in graph $\cH$, where $X=V(S_1)\cup\cdots\cup V(S_{i+1})$, and $Y=V(S_{i+2})\cup\cdots\cup V(S_r)$ (see \Cref{fig: NF19a}).

\begin{figure}[h]
	\centering
	\includegraphics[scale=0.12]{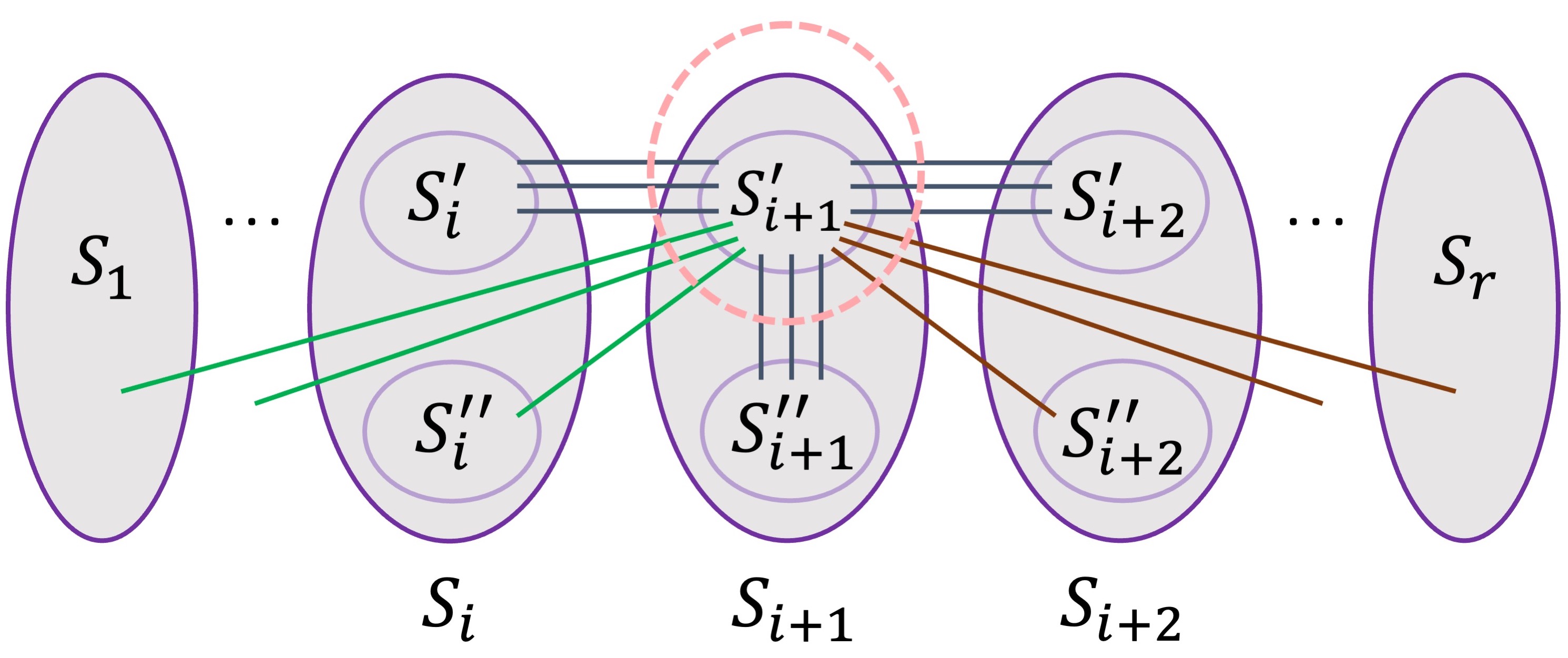}
	\caption{Illustration for the proof of \Cref{claim left edges for S' only}, with cut $(X',Y')$ shown in pink dashed line. Edges of $\delta^{\lef}(S'_{i+1})$ are shown in green, and edges of $\delta^{\rig}(S'_{i+1})$ are shown in brown.}\label{fig: NF19b}
\end{figure}

As before, $X$ and $Y$ are precisely the sets of vertices of the two connected components of the graph $\tau\setminus\set{(u_{i+1},u_{i+2})}$, where $\tau$ is the Gomory-Hu tree of graph $\cH$, and so $(X,Y)$ is the minimum $u_{i+1}$-$u_{i+2}$ cut in $\cH$. We now consider another $u_{i+1}$-$u_{i+2}$ cut $(X',Y')$ in $\cH$, where $X'=V(S'_{i+1})$, and $Y'=V(\cH)\setminus V(S'_{i+1})$ 
(see \Cref{fig: NF19b}).
Note that $|E(X,Y)|\geq |E_{i+1}|+|\rightCedges{i+1}|+|\tilde E_i^{\rig}|$, while $|E(X',Y')|=|E_{i+1}|+|\rightCedges{i+1}|+|E_{i}|+|\leftCedges{i+1}|+|\downedges{i+1}|$. From the fact that $(X,Y)$ is the minimum $u_{i+1}$-$u_{i+2}$ cut, it must be the case that:
\[
|\tilde E_i^{\rig}|\leq |E_{i}|+|\leftCedges{i+1}|+|\downedges{i+1}|.
\]

As before, from 	\Cref{claim: bound S' to S'' edges}, $|\downedges{i+1}|\leq 0.1\cdot |\leftedges{i+1}|$, and, from \Cref{claim left edges for S' and S''}, $|\leftedges{i+1}|\leq 1.3|E_{i}|+1.3|\leftCedges{i+1}|$.
Therefore, $|\downedges{i}|\leq 0.13 |E_{i}|+0.13|\leftCedges{i+1}|$, and:

\begin{equation}\label{eq: right bound}
|\tilde E_i^{\rig}|\leq 1.13|E_{i}|+1.13|\leftCedges{i+1}|.
\end{equation}

Consider now edge set $\rightCedges{i}$. Recall that it contains every edge $e=(u,v)$ with  $u\in V(S_i)$, such that either $v\in V(S_{i+2})\cup\cdots \cup V(S_r)$, or $v\in V(S''_{i+1})$. In the former case, $e\in \tilde E_i^{\rig}$, and in the latter case, $v\in \leftedges{i+1}$ holds. Therefore, $|\rightCedges{i}|\leq 	|\tilde E_i^{\rig}|+|\leftedges{i+1}|$.	From \Cref{claim left edges for S' and S''}: $|\leftedges{i+1}|\leq 1.3|E_{i}|+1.3|\leftCedges{i+1}|$. Therefore, altogether, $|\rightCedges{i}|\leq 	|\tilde E_i^{\rig}|+|\leftedges{i+1}|\leq 2.5|E_{i}|+2.5|\leftCedges{i+1}|$.

\subsection{Proof of \Cref{claim: non-J-node-boundary size}}
\label{subsec: non-J-node boundary size}
Fix some index $1<i<r$, and recall how graph $S'_i\subseteq S_i$ was constructed. Initially, we set $V(S'_i)=\set{u_i}$. As long as there was any vertex $x\in S_i\setminus S'_i$, such that the number of edges connecting $x$ to vertices of $V(S'_i)$ was at least $|\delta(x)|/128$, we added $x$ to $V(S'_i)$. Recall that $|\delta^{\up}(x)|\leq |\delta(x)|/\log m$. Therefore, at least $|\delta(x)|\left(\frac 1 {128}-\frac{1}{\log m}\right )$ edges that connect $x$ to vertices of $V(S'_i)$ lie in $\delta^{\down}(x)$. We then let $S'_i$ be the subgraph of $\check H$ induced by $V(S'_i)$.

Denote $V(S'_i)=\set{u_i=x_0,x_1,\ldots,x_z}$, where the vertices $x_1,\ldots,x_z$ were added to $S'_i$ in the order of their indices. 
Notice that for all $1\leq a\leq z$, if $x_a\in L'_{j_a}$, then, from the above discussion, at least one vertex in $\set{x_0,\ldots,x_{a-1}}$ must lie in $L'_1\cup L'_2\cup\cdots\cup L'_{j_a-1}$. It is then easy to see (by induction on $a$), that, if  $u_i\in L'_j$, for some $1\leq j\leq h$, then every vertex of $S'_i\setminus\set{u_i}$ lies in $L'_{j+1}\cup\cdots\cup L'_h$.

For all $0\leq a\leq z$, we consider the vertex set $X_a=\set{x_0,\ldots,x_a}$, and we define a weight $w_a(e)$ of every edge $e\in E(\cH)$ with respect to $X_a$ as follows. First, we let $\delta^{\up}(X_a)$ be the set of all edges $e=(x,y)$ with $x\in X_a$, $y\not\in X_a$, such that $e\in \delta^{\up}(x)$. Every edge $e\in \delta^{\up}(X_a)$ is assigned weight $w_a(e)=130$. Every edge $e\in \left ( \bigcup_{v\in X_a}\delta(v)\right )\setminus \delta^{\up}(X_a)$ is assigned weight $w_a(e)=1$. All other edges $e\in E(\cH)$ are assigned weight $w_a(e)=0$. We denote $W_a=\sum_{e\in E(\cH)}w_a(e)$. Observe that $W_0=|\delta^{\down}(u_i)|+130|\delta^{\up}(u_i)|\leq |\delta(u_i)|\cdot \left(1+\frac{130}{\log m}\right )$. Additionally, $W_z\geq |\bigcup_{v\in X_z}\delta(v)|=|\bigcup_{v\in S'_i}\delta(v)|$. We now show that for all $1\leq a\leq z$, $W_a\leq W_{a-1}$. Notice that, if this is the case, then $|\delta(S'_i)|\leq W_z\leq W_0\leq |\delta(u_i)|\cdot \left(1+\frac{130}{\log m}\right )$. Therefore, in order to complete the proof of \Cref{claim: non-J-node-boundary size}, it is enough to prove the following observation.

\begin{observation}
	For all $1\leq a\leq z$, $W_a\leq W_{a-1}$.
\end{observation}
\begin{proof}
	Consider some index $1\leq a\leq z$. Recall that $X_a=X_{a-1}\cup \set{x_a}$. Recall that, as observed above, at least $|\delta(x_a)|\left(\frac 1 {128}-\frac{1}{\log m}\right )$ edges that connect $x_a$ to vertices of $X_{a-1}$ lie in $\delta^{\down}(x_a)$. Denote this edge set by $E^*\subseteq \delta^{\up}(X_a)$. Each edge $e\in E^*$ has $w_{a-1}(e)=130$, and $w_a(e)=1$. Additionally, every edge $e\in \delta^{\down}(x_a)\setminus E^*$ has $w_{a-1}(e)=0$ and $w_a(e)=1$. Finally, each edge $e\in \delta^{\up}(x_a)$ has $w_{a-1}(e)=0$ and $w_a(e)=130$. For all other edges $e'\in E(\cH)$, $w_{a-1}(e')=w_a(e')$.
	Therefore, altogether, we get that:
	
	\[ \begin{split}
	W_{a}&=W_{a-1}-129|E^*|+|\delta^{\down}(x_a)\setminus E^*| +130|\delta^{\up}(x_a)|\\
	&=W_{a-1}-130|E^*|+|\delta^{\down}(x_a)|+130|\delta^{\up}(x_a)|\\
	&\leq W_{a-1}-130\cdot |\delta(x_a)|\left(\frac 1 {128}-\frac{1}{\log m}\right )+|\delta(x_a)|\cdot \left(1 +\frac{130}{\log m}\right )\\
	&\leq W_{a-1}-\frac{|\delta(x_a)|}{64}+\frac{|\delta(x_a)|\cdot 260}{\log m}\\
	&\leq W_{a-1}.
	\end{split} \]
	
	(we have used the fact that $|E^*|\geq |\delta(x_a)|\left(\frac 1 {128}-\frac{1}{\log m}\right )$ and that $m$ is sufficiently large).	
\end{proof}

\subsection{Proof of \Cref{claim: many edges left right large}}
\label{subsec:many edges left right large}
Assume that there is an index $1\leq a\leq r$, such that at least $|\delta(u_{i^*})|/16$ edges connect $u_{i^*}$ to vertices of $S''_a$. We start by proving that $|\rightedges{a}|\geq |\delta(u_{i^*})|\cdot\frac {\log m}{256}$. 
Denote by $E'$ the set of all edges connecting $u_{i^*}$ to vertices of $S''_a$, so $|E'|\geq |\delta(u_{i^*})|/16$. We denote by $E''=E'\cap \delta^{\down}(u_{i^*})$.
Since $|\delta^{\up}(u_{i^*})|\leq |\delta(u_{i^*})|/\log m$, $|E''|\geq |\delta(u_{i^*})|/32$.

Consider now a set $\pset$ of paths, defined as follows: $\pset$ contains every path $P$ of $\cH$, whose first edge lies in $E''$, last edge lies in $\rightedges{a}$, and all inner vertices lie in $S''_a$. We will show a flow $f$ defined over the paths in $\pset$, in which every edge in $E''$ sends $\frac{\log m}{8}$ flow units, every edge in $\rightedges{a}$ receives at most one flow unit, and every edge $e\in E(S''_a)$ carries at most one flow unit. Clearly, this will prove that $|\rightedges{a}|\geq |E''|\cdot  \frac{\log m}{8}\geq |\delta(u_{i^*})|\cdot\frac {\log m}{256}$.

We now focus on defining the flow $f$. Initially, we set the flow on every edge $e\in E''$ to $\frac{\log m}{8}$, and for every edge of $e'\in E(\cH)\setminus E''$, we set $f(e')=0$. Note that, if $e=(u_{i^*},v)\in E''$, then $e\in \delta^{\up}(v)$ must hold, since $e\in \delta^{\down}(u_{i^*})$ from the definition of $E''$. Next, we consider indices $j=h,h-1,\ldots,1$ one by one. We assume that, when index $j$ is considered, for every vertex $v\in L'_j\cap S''_a$, for every edge $e\in \delta^{\up}(v)$, the flow $f(e)$ is fixed, and $f(e)\leq \frac{\log m}{8}$. During the iteration when index $j$ is processed we will finalize the flow values $f(e')$ for every edge $e'\in \set{\delta^{\down}(v)\mid v\in L'_j\cap S''_a}$.

We now describe the iteration when index $j$ is processed. Consider any vertex $v\in L'_j\cap S''_a$. Recall that $|\delta^{\up}(v)|\leq |\delta(v)|/\log m$, and so the total flow that the edges of $\delta^{\up}(v)$ carry is bounded by $\frac{|\delta(v)|}{\log m}\cdot \frac{\log m}{8}\leq \frac{|\delta(v)|}{8}$.
Recall that, from \Cref{obs: left and right down-edges},
$|\delta^{\down,\rig}(v)|+|\delta^{\down,\lef}(v)|+| \delta^{\down,\straight'}(v)|\geq 63|\delta(v)|/64$, and 
$|\delta^{\down,\lef}(v)|\leq 2(|\delta^{\down,\rig}(v)|+|\delta^{\down,\straight'}(v)|)$.
By combining the two inequalities, we get that $|\delta^{\down,\rig}(v)|+|\delta^{\down,\straight'}(v)|\geq 21|\delta(v)|/64$.
We now define the flow on every edge $e'\in \delta^{\down}(v)$, as follows. If $e'\in \delta^{\down,\lef}(v)\cup\delta^{\down,\straight''}(v)$, then we set $f(e')=0$. Otherwise, $e'\in \delta^{\down,\rig}(v)\cup \delta^{\down,\straight'}(v)$. We then set $f(e')=\frac{\sum_{e\in \delta^{\up}(v)}f(e)}{|\delta^{\down,\rig}(v)|+|\delta^{\down,\straight'}(v)|}$. From the above discussion, we are guaranteed that for every edge $e\in \delta^{\down}(v)$, $f(e)\leq 1$. This completes the description of the interation where index $j$ is processed. Once al indices $j=r,r-1,\ldots,1$ are processed, we obtain a final flow $f$.
From the construction of the flow $f$, flow conservation constraints hold for every vertex $v\in S''_a$. The only edges that carry non-zero flow are edges of $E''\cup E(S''_a)\cup \rightedges{a}$. Moreover, each edge in $E''$ carries $\frac{\log m}{8}$ flow units, and every edge in $\rightedges{a}$ carries at most one flow unit. We can then apply standard flow-paths decomposition of the flow $f$, to obtain a flow that is defined over the set $\pset$ of paths, where every edge of $E''$ sends  $\frac{\log m}{8}$ flow units, and every edge in $\rightedges{a}$ receives at most one flow unit. We conclude that $|\rightedges{a}|\geq |E''|\cdot  \frac{\log m}{8}\geq |\delta(u_{i^*})|\cdot\frac {\log m}{256}$.
The proof that $|\leftedges{a}|\geq  |\delta(u_{i^*})|\cdot\frac {\log m}{256}$ is symmetric.

\begin{figure}[h]
	\centering
	\includegraphics[scale=0.12]{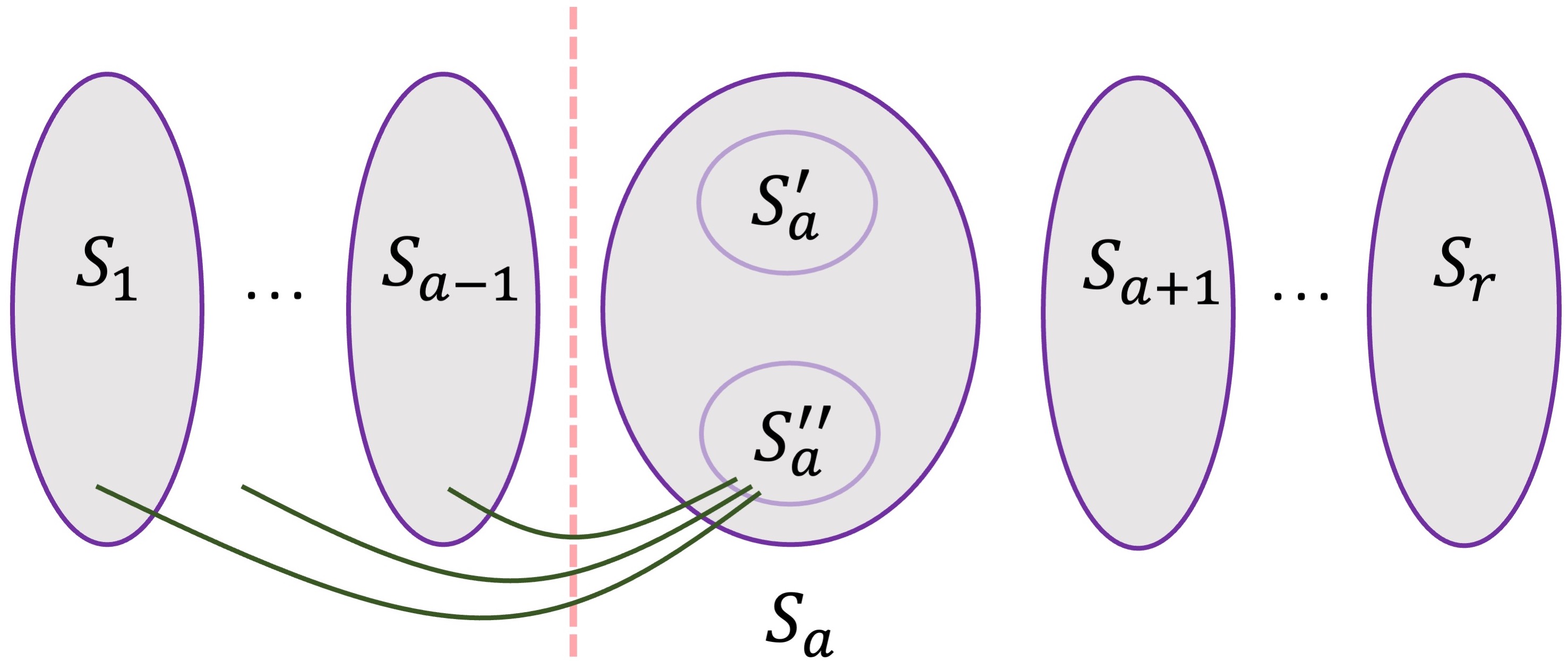}
	\caption{Illustration for the Proof of \Cref{claim: many edges left right large}, with cut $(A,B)$ shown in pink dashed line. Edges of $\delta^{\lef}(S''_{a})$ are shown in green.}\label{fig: NF24a}
\end{figure}

Lastly, we prove that $u_a$ is a $J$-node. Observe first that, since $S''_a\neq\emptyset$, from \Cref{obs: left and right down-edges}, $a\not\in\set{1,r}$. Conisder a cut $(A,B)$ in graph $\cH$, where $A=S_1\cup\cdots\cup S_{a-1}$, and $B=S_a\cup S_{a+1}\cup\cdots\cup S_r$ (see \Cref{fig: NF24a}). From the construction of the Gomory-Hu tree $\tau$, $(A,B)$ is a minimum $u_{a-1}$-$u_a$ cut in graph $\cH$. Since $\leftedges{a}\subseteq E(A,B)$, we get that $|E(A,B)|\geq |\leftedges{a}|\geq |\delta(u_{i^*})|\cdot\frac {\log m}{256}$.

\begin{figure}[h]
 	\centering
 	\includegraphics[scale=0.12]{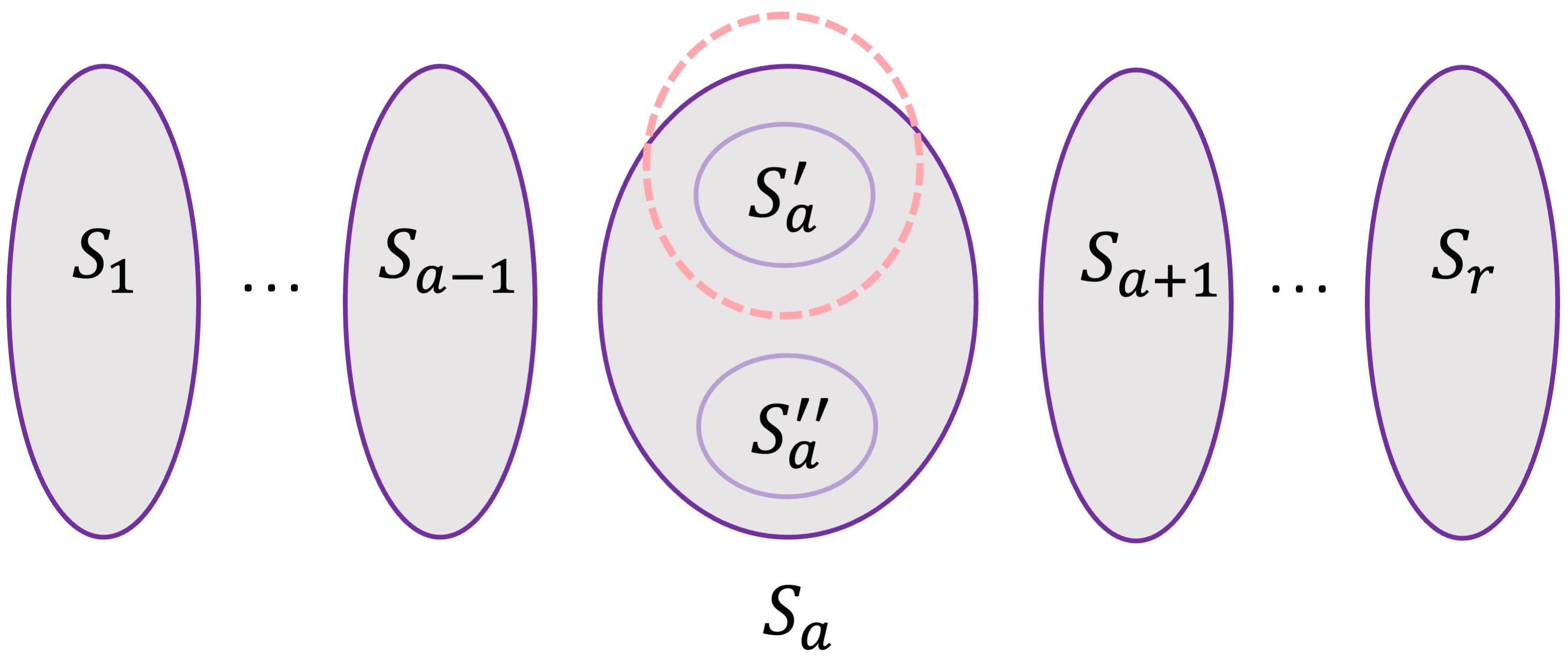}
 	\caption{Illustration for the Proof of \Cref{claim: many edges left right large}, with cut $(A',B')$ shown in pink dashed line.}\label{fig: NF24b}
 \end{figure}

 Consider another $u_{a-1}$-$u_a$ cut $(A',B')$ in graph $\cH$, where $B'=S'_a$ and $A'=V(\cH)\setminus S'_a$ (see \Cref{fig: NF24b}). Then $|E(A',B')|\geq |E(A,B)|\geq |\delta(u_{i^*})|\cdot\frac {\log m}{256}$.

 Assume now for contradiction that $u_a$ is not a $J$-node. Then from \Cref{claim: non-J-node-boundary size}, $|E(A',B')|=|\delta(S'_a)|\leq  \left(1+\frac{130}{\log m}\right )|\delta(u_a)|\leq 2|\delta(u_a)|$. Therefore, we conclude that $|\delta(u_a)|\geq \frac{|E(A',B')|}{2}\geq 
|\delta(u_{i^*})|\cdot\frac {\log m}{512}>|\delta(u_{i^*})|$, contradicting the choice of the index $i^*$ (we have used the fact that $m$ is sufficiently large).

\subsection{Proof of \Cref{claim: simplifying cluster case 1}}
\label{subsec: simplifying cluster Case 1}
We consider two cuts in graph $\cH$. 
The first cut, $(A_1,B_1)$ is defined as follows: 
$A_1=V(S_1)\cup\cdots\cup V(S_{i})$, $B_1=V(S_{i+1})\cup \cdots\cup V(S_r)$. 
From the defintion of the Gomory-Hu tree $\tau$, and from \Cref{cor: G-H tree_edge_cut}, $(A_1,B_1)$ is a minimum $u_i$--$u_{i+1}$ cut in graph $\cH$, and moreover, there is a set $\pset_1=\set{P(e)\mid e\in E(A_1,B_1)}$ of edge-disjoint paths in graph $\cH$, where, for each edge $e\in E(A_1,B_1)$, path $P(e)$ has $e$ as its first edge, vertex $u_{i}$ as its last vertex, and is internally disjoint from $B_1$, and hence from $S^*$. 
We can define another $u_i$--$u_{i+1}$ cut $(A'_1,B'_1)$ in graph $\cH$, where $B'_1=V(S'_{i+1})$, and $A'_1=V(\cH)\setminus B'_1$. Since $(A_1,B_1)$ is a minimum $u_i$--$u_{i+1}$ cut, we get that: 

\[|E(A_1,B_1)|\leq |E(A'_1,B_1')|=|\delta(S'_{i+1})|\leq 2|\delta(u_{i+1})|\leq 2|\delta(u_i)|\]

(we have used \Cref{claim: non-J-node-boundary size} for the penultimate inequality, and the definition of the index $i=i^*$ for the last inequality).

Similarly, we consider a second cut  $(A_2,B_2)$, that is defined as follows: $A_2=V(S_1)\cup\cdots\cup V(S_{i+2})$, $B_2=V(S_{i+3})\cup \cdots\cup V(S_r)$. As before, from the defintion of the Gomory-Hu tree $\tau$, and from \Cref{cor: G-H tree_edge_cut}, $(A_2,B_2)$ is a minimum $u_{i+2}$--$u_{i+3}$ cut in graph $\cH$, and moreover, there is a set $\pset_2=\set{P'(e)\mid e\in E(A_2,B_2)}$ of edge-disjoint paths in graph $\cH$, where, for each edge $e\in E(A_2,B_2)$, path $P'(e)$ has $e$ as its first edge, vertex $u_{i+3}$ as its last vertex, and is internally disjoint from $A_2$, and hence from $S^*$. 
As before, we can define another $u_{i+2}$--$u_{i+3}$ cut $(A'_2,B'_2)$ in graph $\cH$, setting $A'_2=V(S'_{i+2})$, and $B'_2=V(\cH)\setminus A'_2$. Since $(A_2,B_2)$ is a minimum $u_{i+2}$--$u_{i+3}$ cut, we get that: 

\[|E(A_2,B_2)|\leq |E(A'_2,B_2')|=|\delta(S'_{i+2})|\leq 2|\delta(u_{i+2})|\leq 2|\delta(u_i)|.\]

\begin{figure}[h]
	\centering
	\includegraphics[scale=0.15]{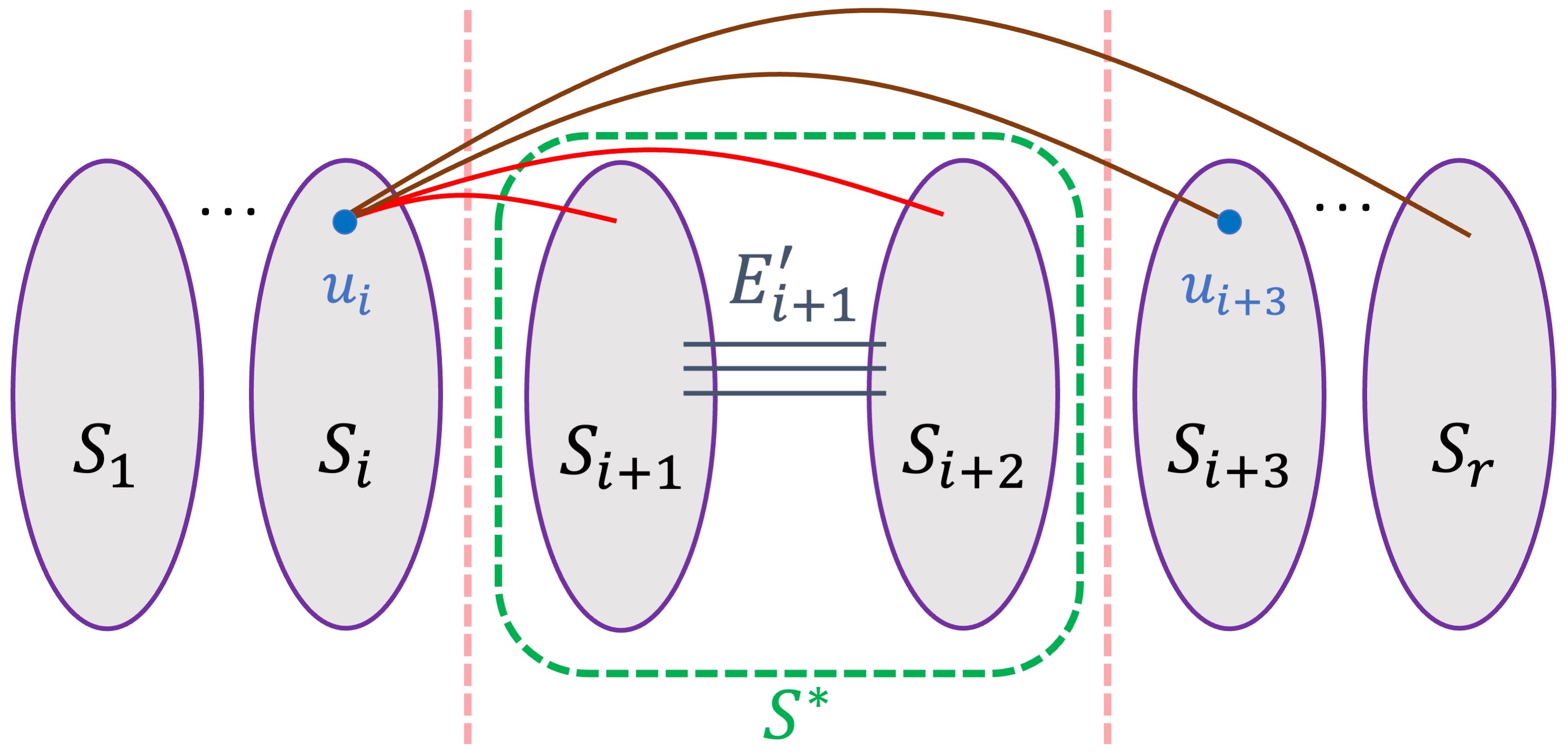}
	\caption{Illustration for the proof of \Cref{claim: simplifying cluster case 1}. The edges of $E^*$  are shown in red and brown. At least $|\delta(u_i)|/16$ edges of $E^*$ (shown in brown) have an endpoint in $S_{i+1}\cup\ldots,\cup S_r$, and these edges belong to set $E(A_2,B_2)$.}\label{fig: NF25}
\end{figure}

Observe that $\delta(S^*)=E(A_1,B_1)\cup E(A_2,B_2)$  (see \Cref{fig: NF25}). Therefore, from the above discussion, $|\delta(S^*)|\leq 4|\delta(u_i)|$. On the other hand, all 
edges connecting $u_{i}$ to vertices of $\bigcup_{a>i+2}S_a$ lie in set 
$\tilde E^{\through}_{i+1}$ (see \Cref{fig: NF25}), and so, from \Cref{obs: bad inded structure},  $|E_{i+1}'|\geq |\tE_{i+1}^{\through}|\geq |\delta(u_{i})|/16$. Recall that
$E'_{i+1}=E(S_{i+1},S_{i+2})$, and so in particular, $E'_{i+1}\subseteq E(S^*)$. We conclude that $|E(S^*)|\geq |E'_{i+1}|\geq |\delta(u_{i})|/16\geq |\delta(S^*)|/64$.

In order to prove that $S^*$ is a simplifying cluster, it is now enough to show a collection $\pset^*=\set{P^*(e)\mid e\in \delta(S^*)}$ of paths in graph $\cH$, that cause congestion at most $\beta'=O(\log m)$, such that, for every edge $e\in \delta(S^*)$, path $P^*(e)$ has $e$ as its first edge, vertex $u_{i+3}$ as its last vertex, and it is internally disjoint from $S^*$.

Recall that we have defined a set
$\pset_2=\set{P'(e)\mid e\in E(A_2,B_2)}$ of edge-disjoint paths in graph $\cH$, where, for each edge $e\in E(A_2,B_2)$, path $P'(e)$ has $e$ as its first edge, vertex $u_{i+3}$ as its last vertex, and is internally $S^*$. For each edge $e\in E(A_2,B_2)$, we set $P^*(e)=P(e)$. Since $\delta(S^*)=E(A_1,B_1)\cup E(A_2,B_2)$, it remains to define the paths $P^*(e)$ for edges $e\in E(A_1,B_1)$.

As observed above, $|E(A_1,B_1)|\leq 2|\delta(u_i)|$. 
From \Cref{obs: many through edges}, 
at least $ |\delta(u_{i})|/16$ edges connect $u_{i}$ to vertices of $\bigcup_{a>i+2}S_a$. Denote this set of edges by $\tilde E^*$. Clearly, $\tilde E^*\subseteq E(A_2,B_2)$ (see \Cref{fig: NF25}). 
Since $|E(A_1,B_1)|\leq 2|\delta(u_i)| \leq 32|\tilde E^*|$, we can define a mapping $M$ from the edges of 
$E(A_1,B_1)$ to edges of $\tilde E^*$, such that, for every edge $e'\in \tilde E^*$, at most $32$ edges of $E(A_1,B_1)$ are mapped to it. Consider now some edge $e\in E(A_1,B_1)$. The final path $P^*(e)$ is a concatenation of two paths. The first path is $P(e)\in \pset_1$, that originates at $e$, terminates at $u_i$, and it is internally disjoint from $S^*$. Let $e'=M(e)$ be the edge of $\tilde E^*$ to which edge $e$ is mapped (recall that $e'$ must be incident to $u_i$). The second path is $P'(e')\in \pset_2$, that starts at edge $e'$ and terminates at vertex $u_{i+3}$. This completes the definition of the set $\pset^*=\set{P^*(e)\mid e\in \delta(S^*)}$ of paths. From the construction, for every edge $e\in \delta(S^*)$, path $P^*(e)$ has $e$ as its first edge, vertex $u_{i+3}$ as its last vertex,  and it is internally disjoint from $S^*$. It now remains to bound the congestion of the path set $\pset^*$. Recall that each of the path sets $\pset_1,\pset_2$ causes congestion $1$. Each path in $\pset_1$ is used once, and each path in $\pset_2$ may be used by up to $33$ paths. Therefore, the total congestion caused by paths in $\pset^*$ is at most $34$. We conclude that $S^*$ is a simplfying cluster.

\subsection{Proof of \Cref{claim: simplifying cluster case 2}}
\label{subsec: simplifying cluster Case 2}

Since $u_{i+1}$ is a $J$-node, in order to prove that $S^*$ is a simplifying cluster, it is enough to show a collection $\pset^*=\set{P^*(e)\mid e\in \delta(u_{i+1})}$ 
of paths in graph $\cH$, causing congestion at most $\beta'=O(\log m)$, where for each edge $e\in \delta(u_{i+1})$, path $P^*(e)$ has $e$ as its first edge, vertex $u_{i+2}$ as its last vertex, and is internally disjoint from $S^*$. 
As before, we define two cuts in graph $\cH$: cut $(A_1,B_1)$, with $A_1=V(S_1)\cup \cdots\cup V(S_i)$ and $B_1=V(\cH)\setminus A_1$, and cut $(A_2,B_2)$, with $A_2=V(S_1)\cup\cdots\cup V(S_{i+1})$ and $B_2=V(\cH)\setminus A_2$. As before, from our construction, $(A_1,B_1)$ is a minimum $u_i$--$u_{i+1}$ cut in $\cH$, and so there is a set $\pset_1=\set{P(e)\mid e\in E(A_1,B_1)}$ of edge-disjoint paths, that are internally disjoint from $B_1$, where for each edge $e\in E(A_1,B_1)$, path $P(e)$ has $e$ as its first edge and vertex $u_i$ as its last vertex. Similarly, $(A_2,B_2)$ 
is a minimum $u_{i+1}$--$u_{i+2}$ cut in $\cH$, and so there is a set $\pset_2=\set{P'(e)\mid e\in E(A_2,B_2)}$ of edge-disjoint paths, that are internally disjoint from $A_2$, where for each edge $e\in E(A_2,B_2)$, path $P'(e)$ has $e$ as its first edge and vertex $u_{i+2}$ as its last vertex.

\begin{figure}[h]
	\centering
	\includegraphics[scale=0.15]{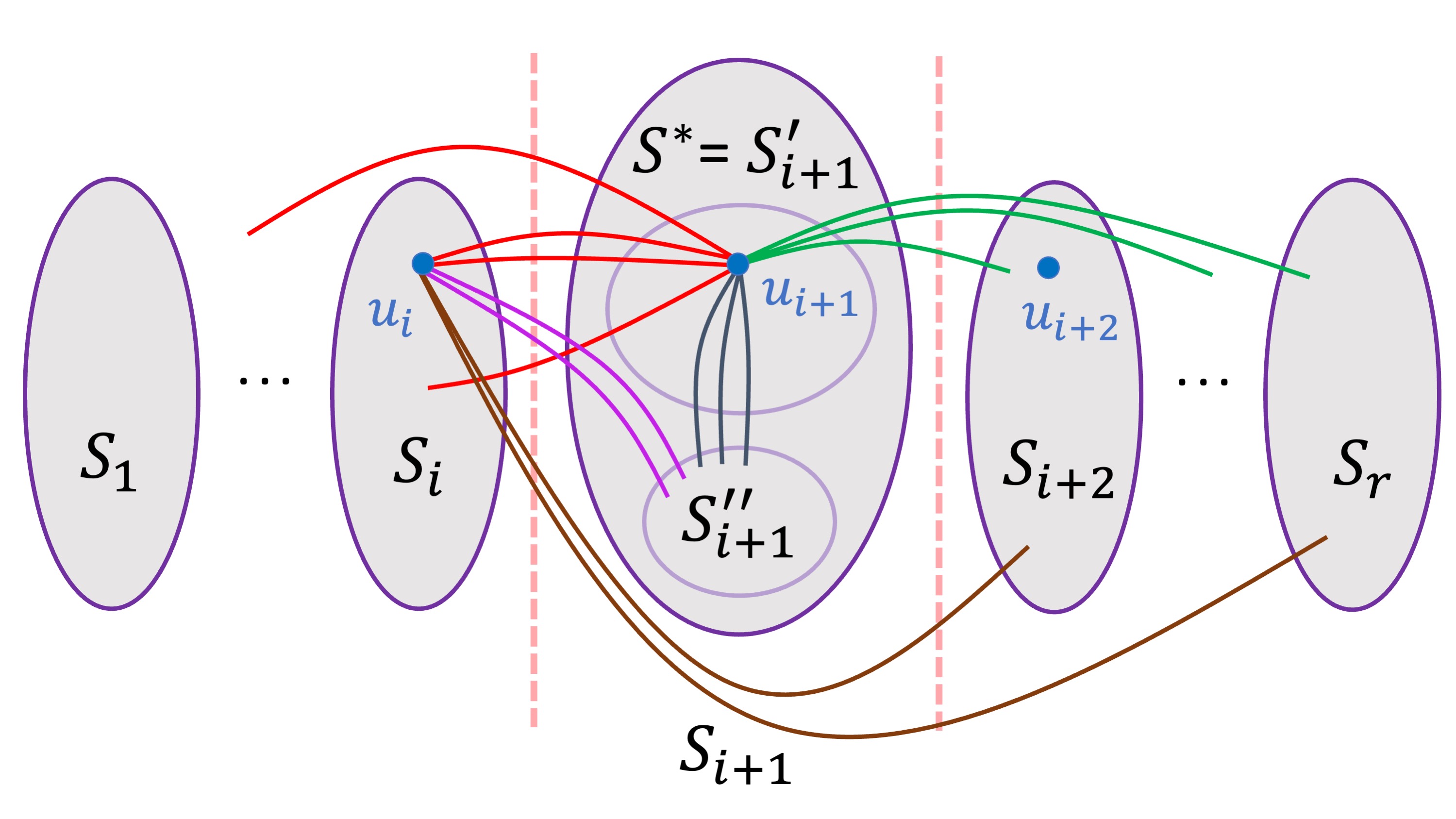}
	\caption{Partition of the set $\delta(S^*)$ of edges into three subsets: set $\delta_1$ (shown in green), set $\delta_2$ (shown in black), and set $\delta_3$ (shown in red). The pink dashed line on the left shows cut $(A_1,B_1)$, and the pink dashed line on the right shows cut $(A_2,B_2)$. 
	}\label{fig: NF26}
\end{figure}

We partition the edge set $\delta(S^*)$ into three subsets (see \Cref{fig: NF26}). 
The first subset, that we denote by $\delta_1$, contains all edges of $\delta(S^*)$ that lie in the set $E(A_2,B_2)$. The second set, that we denote by $\delta_2$, contains all edges of $\delta(S^*)$ that lie in $\delta^{\down}(S_{i+1}'')$ -- that is, they connect $u_{i+1}$ to vertices of $S''_{i+1}$. The third set $\delta_3$ contains all remaining edges. Note that every edge of $\delta_3$ must lie in $E_i\cup \leftCedges{i+1}\subseteq E(A_1,B_1)$.
We now consider each of the three sets of edges in turn.

First, for every edge $e\in \delta_1$, we let $P^*(e)=P'(e)$, where $P'(e)$ is the path of $\pset_2$, that has $e$ as its first edge, vertex $u_{i+2}$ as its last vertex, and is internally disjoint from $S^*$. We set $\pset^*_1=\set{P^*(e)\mid e\in \delta_1}$. Clearly, $\pset^*_1\subseteq \pset_2$, and the paths in $\pset^*_1$ are edge-disjoint. 

Next, we consider the edges of $\delta_2=\delta^{\down}(S_{i+1}'')$. Recall that, from \Cref{claim: bound S' to S'' edges}, there is a set $\pset^{\rig}=\set{P^{\rig}(e)\mid e\in \delta_2}$ of edge-disjoint paths in $\cH$, where, for each edge  $ e\in \delta_2$, path $P^{\rig}(e)$ has $e$ as its first edge, some edge of $\rightedges{i+1}$ as its last edge, and all inner vertices of $P^{\rig}(e)$ are contained in $S''_{i+1}$, so that the paths of $\pset^{\rig}$ are internally disjoint from $S^*$. Consider an edge $e\in \delta_2$, and let $e'\in \rightedges{i+1}$ be the last edge on the path $P^{\rig}(e)$. Then $e'\in E(A_2,B_2)$. We let $P^*(e)$ be the path obtained by concatenating the path $P^{\rig}(e)$ with the path $P'(e')\in \pset_2$. Clearly, path $P^*(e)$ has $e$ as its first edge, vertex $u_{i+2}$ as its last vertex, and it is internally disjoint from $S^*$. We set $\pset^*_2=\set{P^*(e)\mid e\in \delta_2}$. It is easy to verify that the paths of $\pset^*_2$ are edge-disjoint.

Lastly, we consider the edges of $\delta_3\subseteq E(A_1,B_1)$. Observe that a cut $(A_1',B_1')$, where $A_1'=\set{u_i}$, and $B_1'=V(\cH)\setminus A_1'$ is a $u_i$--$u_{i+1}$-cut in graph $\cH$, and so $|\delta_3|\leq |E(A_1,B_1)|\leq |E(A_1',B_1')|\leq  |\delta(u_i)|$. We denote by $\tilde E'$ the set of all edges connecting $u_i$ to vertices of $S''_{i+1}$, and we denote by $\tilde E''$ the set of all edges connecitng $u_i$ to vertices of $S_{i+2}\cup\cdots\cup S_r$.

Recall that set $E^*$ of edges contains all edges connecting $u_i$ to vertices of $\bigcup_{a>i}S_a$. 
Let $\check E_i\subseteq E_i$ be the set of edges connecting $u_i$ to $u_{i+1}$. Since $S'_{i+1}=\set{u_{i+1}}$, $E^*=\tilde E'\cup \tilde E''\cup \check E_i$. From our assumtion,  $|E^*|\geq |\delta(u_i)|/4$. 
Furthermore, since vertex $u_i$ was not added to the $J$-cluster corresponding to vertex $u_{i+1}$, $|\check E_i|\leq |\delta(u_i)|/128$. 
Therefore, either $|\tilde E'|\geq |\delta(u_i)|/16$, or $|\tilde E''|\geq |\delta(u_i)|/16$ must hold. 

We assume first that it is the latter. Since $|\delta_3|\leq |\delta(u_i)|$, we can define a mapping $M$ from the edges of $\delta_3$ to edges of $\tilde E''$, where, for each edge $e'\in \tilde E''$, at most $16$ edges of $\delta_3$ are mapped to $e'$. Observe that $\tilde E''\subseteq E(A_2,B_2)$.  Consider now some edge $e\in \delta_3$. We obtain the path $P^*(e)$ by concatenating two paths: path $P(e)\in \pset_1$, connecting $e$ to vertex $u_i$, and the path $P'(e')\in \pset_2$, where $e'$ is the edge of $\tilde E''$ to which $e$ is mapped. Recall that $P'(e')$ has $e'$ as its first edge and $u_{i+2}$ as its last vertex; edge $e'$ is incident to $u_i$. Therefore, path $P^*(e)$ has $e$ as its first edge, vertex $u_{i+2}$ as its last vertex, and it is internally disjoint from $S^*$. We then set  $\pset^*_3=\set{P^*(e)\mid e\in \delta_3}$. It is easy to verify that the paths of $\pset^*_3$ cause edge-congestion at most $17$.

Lastly, we assume that $|\tilde E'|\geq |\delta(u_i)|/16$. As before, we define a mapping $M$ from edges of $\delta_3$ to edges of $\tilde E'$, where, for each edge $e'\in \tilde E'$, at most $16$ edges of $\delta_3$ are mapped to $e'$. 
Consider now some edge $e'=(u_i,v)\in \tilde E'$, and recall that $v\in S''_{i+1}$. Recall that the algorithm from \Cref{lem: prefix and suffix path} provided a construction of a right-monotone path $P(e',v)$. This path starts with edge $e'$, and it must terminate at some vertex of $V(S'_{i+2})\cup  V(S'_{i+3})\cup\cdots\cup V(S'_r)$. All inner vertices on path $P(e',v)$ must lie in $V(S''_{i+1})\cup V(S''_{i+2})\cup\cdots\cup V(S''_r)$. Therefore, if $e''$ is the first edge of $P(e',v)$ that is not contained in $S''_{i+1}$, then $e''\in E(A_2,B_2)$. We denote by $\tilde P(e')$ the subpath of $P(e',v)$ that starts with edge $e'$ and ends with edge $e''$. From \Cref{lem: prefix and suffix path}, we are guaranteed that all paths in set $\set{\tilde P(e')\mid e'\in \tilde E'}$ cause congestion $O(\log m)$. Consider now some edge $e\in \delta_3$. We let $P^*(e)$ be a path that is a concatenation of three paths. The first path is the path $P(e)\in \pset_1$, that connects $e$ to $u_i$. Let $e'\in \tilde E'$ be the edge to which $e$ is mapped by $M$. The second path is $\tilde P(e')$, connecting $e'$ to some edge $e''\in E(A_2,B_2)$. The third path is the path $P'(e'')\in\pset_2$, connecting $e''$ to vertex $u_{i+2}$. It is immediate to verify that the resulting path $P^*(e)$ has $e$ as its first edge, $u_{i+2}$ as its last vertex, and it is internally disjoint from $S^*$. 
We then set  $\pset^*_3=\set{P^*(e)\mid e\in \delta_3}$. It is easy to verify that the paths of $\pset^*_3$ cause congestion at most $O(\log m)$.

Finally, we set $\pset^*=\pset^*_1\cup \pset^*_2\cup \pset^*_3$. From our construction, the set $\pset^*$ of paths routes the edges of $\delta(S^*)$ to vertex $u_{i+2}$, the paths of $\pset^*$ are internally disjoint from  $S^*$, and they cause edge-congestion $O(\log m)$ as required. We conclude that $S^*$ is a simplifying cluster.


\subsection{Proof of \Cref{claim: simplifying cluster case 3}}
\label{subsec: simplifying cluster Case 3}

Since $u_{i+2}$ is a $J$-node, in order to prove that $S^*$ is a simplifying cluster, it is enough to show a collection $\pset^*=\set{P^*(e)\mid e\in \delta(u_{i+2})}$ 
of paths in graph $\cH$, where for each edge $e\in \delta(u_{i+2})$, path $P^*(e)$ has $e$ as its first edge, vertex $u_{i+3}$ as its last vertex, and is internally disjoint from $S^*$. 
The construction of the set $\pset^*$ of paths is almost identical to that from Case 2.

\begin{figure}[h]
	\centering
	\includegraphics[scale=0.15]{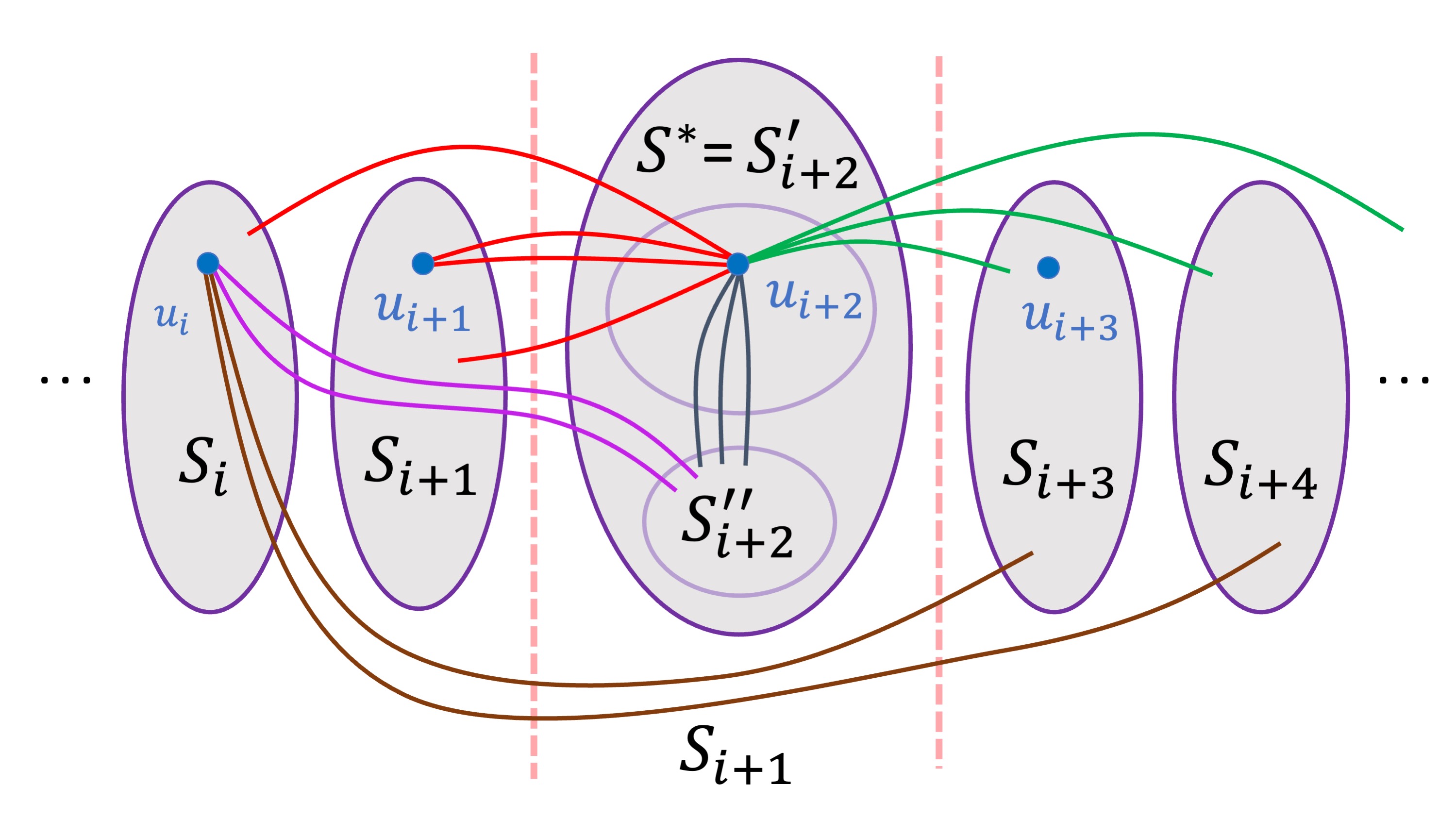}
	\caption{Illustration for the proof of \Cref{claim: simplifying cluster case 3}. Edges of $\delta_1$ are shown in green, edges of $\delta_2$ are shown in black, and edges of $\delta_3$ are shown in red. The pink dashed line on the left shows cut $(A_1,B_1)$ and the pink dashed line on the right shows cut $(A_2,B_2)$. 
	}\label{fig: NF27}
\end{figure}

As before, we define two cuts in graph $\cH$:
cut $(A_1,B_1)$, with $A_1=V(S_1)\cup \cdots\cup V(S_{i+1})$ and $B_1=V(\cH)\setminus A_1$, and cut $(A_2,B_2)$, with $A_2=V(S_1)\cup\cdots\cup V(S_{i+2})$ and $B_2=V(\cH)\setminus A_2$ 
(see \Cref{fig: NF27}).
As before, from our construction, $(A_1,B_1)$ is a minimum $u_{i+1}$--$u_{i+2}$ cut in $\cH$, and so there is a set $\pset_1=\set{P(e)\mid e\in E(A_1,B_1)}$ of edge-disjoint paths, that are internally disjoint from $B_1$, where for each edge $e\in E(A_1,B_1)$, path $P(e)$ has $e$ as its first edge and vertex $u_{i+1}$ as its last vertex. Similarly, $(A_2,B_2)$ 
is a minimum $u_{i+2}$--$u_{i+3}$ cut in $\cH$, and so there is a set $\pset_2=\set{P'(e)\mid e\in E(A_2,B_2)}$ of edge-disjoint paths, that are internally disjoint from $A_2$, where for each edge $e\in E(A_2,B_2)$, path $P'(e)$ has $e$ as its first edge and vertex $u_{i+3}$ as its last vertex.

As before, we partition the edge set $\delta(S^*)$ into three subsets. The first subset, that we denote by $\delta_1$, contains all edges of $\delta(S^*)$ that lie in the set $E(A_2,B_2)$. The second set, that we denote by $\delta_2$, contains all edges of $\delta(S^*)$ that lie in $\delta^{\down}(S_{i+2}'')$ -- that is, they connect $u_{i+2}$ to vertices of $S''_{i+2}$. The third set $\delta_3$ contains all remaining edges. As before, $\delta_3$ must lie in $E_{i+1}\cup \leftCedges{i+2}\subseteq E(A_1,B_1)$.
We consider each of the three sets of edges in turn.

The constructions of the path sets  $\pset^*_1=\set{P^*(e)\mid e\in \delta_1}$  and $\pset^*_2=\set{P^*(e)\mid e\in \delta_2}$ remain exactly the same as in Case 2. We now focus on constructing the set $\pset^*_3=\set{P^*(e)\mid e\in \delta_3}$ of paths.

Consider the edges of $\delta_3\subseteq E(A_1,B_1)$. Recall that we have assumed that $u_{i+1}$ is not a $J$-node. From the choice of the index $i^*=i$, $|\delta(u_{i+1})|\leq |\delta(u_i)|$. As before, we consider another $u_{i+1}$--$u_{i+2}$ cut $(A_1',B_1')$, where $A_1'=\set{u_{i+1}}$, and $B_1'=V(\cH)\setminus A_1'$. As before, $|\delta_3|\leq |E(A_1,B_1)|\leq |E(A_1',B_1')|\leq |\delta(u_{i+1})|\leq |\delta(u_i)|$.

We denote by $\tilde E'$ the set of all edges connecting $u_i$ to vertices of $S''_{i+2}$, and by $\tilde E''$ the set of all edges connecitng $u_i$ to vertices of $S_{i+3}\cup\cdots\cup S_r$.
Since $u_{i+2}$ is a $J$-node, 
$S'_{i+2}=\set{u_{i+2}}$. As before, since vertex $u_i$ was not added to the $J$-cluster corresponding to vertex $u_{i+2}$, the number of edges connecting $u_i$ to $u_{i+2}$ is bounded by $|\delta(u_{i})|/128$.
Recall that we have established that at least $|\delta(u_i)|/8$ edges connect  $u_i$ to vertices of  $\bigcup_{a>i+1}V(S_a)$.  Each such edge either lies in $\tilde E'\cup \tilde E''$, or it connects $u_i$ to $u_{i+2}$. Therefore, either $|\tilde E'|\geq |\delta(u_i)|/32$, or $|\tilde E''|\geq |\delta(u_i)|/32$ must hold. 
The remainder of the construction of the paths in $\pset^*_3$ is very similar to that for Case 2. 

We assume first that $|\tilde E''|\geq |\delta(u_i)|/32$. Since $|\delta_3|\leq |\delta(u_i)|$, we can define a mapping $M$ from the edges of $\delta_3$ to edges of $\tilde E''$, where, for each edge $e'\in \tilde E''$, at most $32$ edges of $\delta_3$ are mapped to $e'$. Observe that $\tilde E''\subseteq E(A_1,B_1)$ and $\tilde E''\subseteq E(A_2,B_2)$.  Consider now some edge $e\in \delta_3$. We obtain the path $P^*(e)$ by concatenating three paths. The first path is path $P(e)\in \pset_1$, connecting $e$ to vertex $u_{i+1}$.
Denote by $e'$ the edge of $\tilde E''$ to which edge $e$ is mapped. The second path is path $P(e')\in \pset_1$ (which we reverse), connecting vertex $u_{i+1}$ to edge $e'$. The third path is path $P'(e')\in \pset_2$, connecting edge $e'$ to vertex $u_{i+3}$. Clearly, path $P^*(e)$ has $e$ as its first edge, vertex $u_{i+3}$ as its last vertex, and it is internally disjoint from $S^*$. We then set  $\pset^*_3=\set{P^*(e)\mid e\in \delta_3}$. It is easy to verify that the paths of $\pset^*_3$ cause edge-congestion at most $O(1)$.

Finally, we assume that $|\tilde E'|\geq |\delta(u_i)|/32$.  Recall that the edges of $\tilde E'$ connect vertex $u_i$ to vertices of $S''_{i+2}$, and in particular $\tilde E'\subseteq E(A_1,B_1)$. 
As before, we define a mapping $M$ from edges of $\delta_3$ to edges of $\tilde E'$, where, for each edge $e'\in \tilde E'$, at most $32$ edges of $\delta_3$ are mapped to $e'$. 

Consider now some edge $e'=(u_i,v)\in \tilde E'$, and recall that $v\in S''_{i+1}$. Recall that the algorithm from \Cref{lem: prefix and suffix path} provides a construction of a right-monotone path $P(e',v)$. This path starts with edge $e'$, and it must terminate at some vertex of $V(S'_{i+3})\cup  V(S'_{i+4})\cup\cdots\cup V(S'_r)$. All inner vertices on path $P(e',v)$ must lie in $V(S''_{i+2})\cup V(S''_{i+3})\cup\cdots\cup V(S''_r)$. Therefore, if $e''$ is the first edge of $P(e',v)$ that is not contained in $S''_{i+2}$, then $e''\in E(A_2,B_2)$. We denote by $\tilde P(e')$ the subpath of $P(e',v)$ that starts with edge $e'$ and ends with edge $e''$. From \Cref{lem: prefix and suffix path}, we are guaranteed that all paths in set $\set{\tilde P(e')\mid e'\in \tilde E'}$ cause congestion $O(\log m)$. Consider now some edge $e\in \delta_3$. We let $P^*(e)$ be a path that is a concatenation of four paths. The first path is the path $P(e)\in \pset_1$, that connects $e$ to $u_{i+1}$. Let $e'\in \tilde E'$ be the edge to which $e$ is mapped by $M$, and recall that $e'\in E(A_1,B_1)$. The second path is $P(e')\in \pset_1$, which we reverse, so the path now connects vertex $u_{i+1}$ to edge $e'$. 
The third path is
$\tilde P(e')$, connecting $e'$ to some edge $e''\in E(A_2,B_2)$. The fourth and the last path is the path $P'(e'')\in\pset_2$, connecting $e''$ to vertex $u_{i+3}$. It is immediate to verify that the resulting path $P^*(e)$ has $e$ as its first edge, $u_{i+3}$ as its last vertex, and it is internally disjoint from $S^*$. 
We then set  $\pset^*_3=\set{P^*(e)\mid e\in \delta_3}$. From the above discussion, the paths of $\pset^*_3$ cause congestion at most $O(\log m)$.

Lastly, we set $\pset^*=\pset^*_1\cup \pset^*_2\cup \pset^*_3$. It is easy to verify that the set $\pset^*$ of paths routes the edges of $\delta(S^*)$ to vertex $u_{i+3}$, the paths are internally disjoint from $S^*$, and they cause congestion $O(\log m)$ as required. We conclude that $S^*$ is a simplifying cluster.


\subsection{Proof of \Cref{claim: simplifying cluster last case}}
\label{subsec: simplifying cluster last}

Since $u_{i+1}$ is a $J$-node,  it is enough to show a collection $\pset^*=\set{P^*(e)\mid e\in \delta(u_{i+1})}$ 
of paths in graph $\cH$, where for each edge $e\in \delta(u_{i+1})$, path $P^*(e)$ has $e$ as its first edge, vertex $u_{i+2}$ as its last vertex, and is internally disjoint from $S^*$. 
As before, we define two cuts in graph $\cH$: cut $(A_1,B_1)$, with $A_1=V(S_1)\cup \cdots\cup V(S_i)$ and $B_1=V(\cH)\setminus A_1$, and cut $(A_2,B_2)$, with $A_2=V(S_1)\cup\cdots\cup V(S_{i+1})$ and $B_2=V(\cH)\setminus A_2$. As before, from our construction, $(A_1,B_1)$ is a minimum $u_i$--$u_{i+1}$ cut in $\cH$, and so there is a set $\pset_1=\set{P(e)\mid e\in E(A_1,B_1)}$ of edge-disjoint paths, that are internally disjoint from $B_1$, where for each edge $e\in E(A_1,B_1)$, path $P(e)$ has $e$ as its first edge and vertex $u_i$ as its last vertex. Similarly, $(A_2,B_2)$ 
is a minimum $u_{i+1}$--$u_{i+2}$ cut in $\cH$, and so there is a set $\pset_2=\set{P'(e)\mid e\in E(A_2,B_2)}$ of edge-disjoint paths, that are internally disjoint from $A_2$, where for each edge $e\in E(A_2,B_2)$, path $P'(e)$ has $e$ as its first edge and vertex $u_{i+2}$ as its last vertex.

\begin{figure}[h]
	\centering
	\includegraphics[scale=0.15]{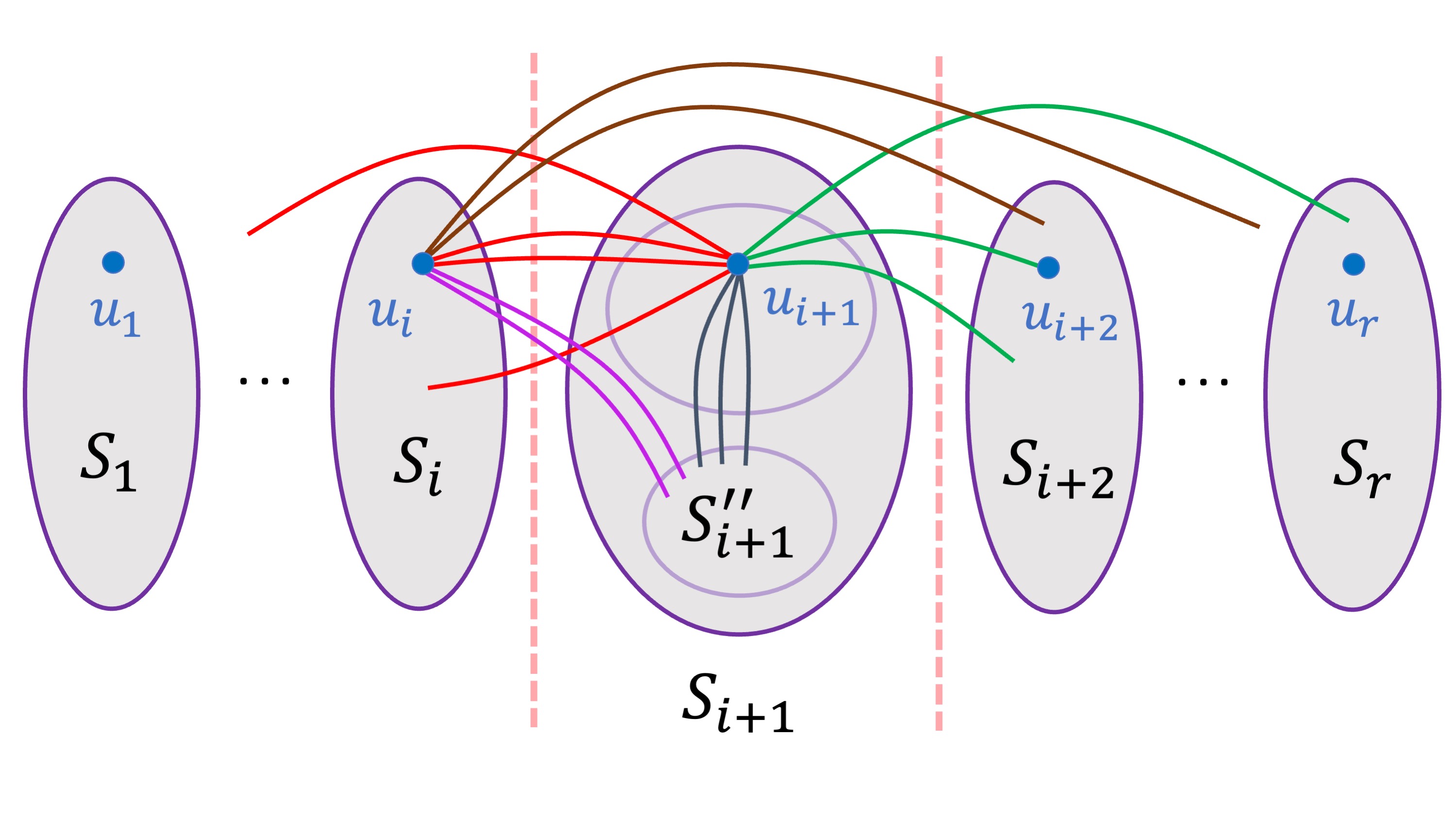}
	\caption{Illustration for the proof of \Cref{claim: simplifying cluster last case}. Edges of $\delta_1$ are shown in green, edges of $\delta_2$ are shown in black, and edges of $\delta_3$ are shown in red. Additionally, edges of $\tilde E'$ are shown in purple and edges of $\tilde E''$ are shown in brown. The pink dashed line on the left shows cut $(A_1,B_1)$ and the pink dashed line on the right shows cut $(A_2,B_2)$. 
	}\label{fig: NF28}
\end{figure}

As before, we partition the set $\delta(S^*)$ of edges into three subsets (see \Cref{fig: NF28}). 
The first subset, that we denote by $\delta_1$, contains all edges of $\delta(S^*)$ that lie in the set $E(A_2,B_2)$. The second set, that we denote by $\delta_2$, contains all edges of $\delta(S^*)$ that lie in $\delta^{\down}(S_{i+1}'')$ -- that is, they connect $u_{i+1}$ to vertices of $S''_{i+1}$. The third set, $\delta_3$, contains all remaining edges. As before, very edge of $\delta_3$ must lie in $E_i\cup \leftCedges{i+1}\subseteq E(A_1,B_1)$.
We now consider each of the three sets of edges in turn.

First, for every edge $e\in \delta_1$, we let $P^*(e)=P'(e)$, where $P'(e)$ is the path of $\pset_2$, that has $e$ as its first edge, vertex $u_{i+2}$ as its last vertex, and is internally disjoint from $S^*$. We set $\pset^*_1=\set{P^*(e)\mid e\in \delta_1}$. Clearly, $\pset^*_1\subseteq \pset_2$, and the paths in $\pset^*_1$ are edge-disjoint. 

Next, we consider the edges of $\delta_2=\delta^{\down}(S_{i+1}'')$. Recall that, from \Cref{claim: bound S' to S'' edges}, there is a set $\pset^{\rig}=\set{P^{\rig}(e)\mid e\in \delta_2}$ of edge-disjoint paths in $\cH$, where, for each edge  $ e\in \delta_2$, path $P^{\rig}(e)$ has $e$ as its first edge, some edge of $\rightedges{i+1}$ as its last edge, and all inner vertices of $P^{\rig}(e)$ are contained in $S''_{i+1}$, so that the paths of $\pset^{\rig}$ are internally disjoint from $S^*$. Consider an edge $e\in \delta_2$, and let $e'\in \rightedges{i+1}$ be the last edge on the path $P^{\rig}(e)$. Then $e'\in E(A_2,B_2)$. We let $P^*(e)$ be the path obtained by concatenating the path $P^{\rig}(e)$ with the path $P'(e')\in \pset_2$. Clearly, path $P^*(e)$ has $e$ as its first edge, vertex $u_{i+2}$ as its last vertex, and it is internally disjoint from $S^*$. We set $\pset^*_2=\set{P^*(e)\mid e\in \delta_2}$. It is easy to verify that the paths of $\pset^*_2$ are edge-disjoint.

Lastly, we consider the edges of $\delta_3\subseteq E(A_1,B_1)$.
Clearly, $|\delta_3|\leq |E_i|+|\hat E_i|\leq 8|E_i|+7|\rightCedges{i}|+7|\leftCedges{i+1}|$, from \Cref{eq: bound on hat E}.
Note that, if $i=1$, then, from \Cref{obs: left and right down-edges},
$S'_1=S_1$, and so $\leftCedges{2}=\emptyset$. Otherwise, from  
\Cref{claim left edges for S' only}, $|\leftCedges{i+1}|\leq 2.5|E_{i}|+2.5|\rightCedges{i}|$.
In either case, we get that: 

$$|\delta_3|\leq  30|E_i|+29|\rightCedges{i}|\leq 30|\rightCedges{i}|,$$

since have assumed that 
$|\rightCedges{i}|>64|E_i|$.
We partition the 
set $\rightCedges{i}$ into two subsets: set $\tilde E'$, containing all edges $(u_i,v)$, with $v\in S_{i+1}''$, and set $\tilde E''$, containing all remaining edges. Note that, for each edge $(u_i,v)\in \tilde E''$, $v\in S_{i+2}\cup\cdots\cup S_{r}$ must hold. 
The remainder of the construction of the set $\pset^*_3=\set{P^*(e)\mid e\in \delta_3}$ of paths is very similar to the analysis of Case 2 in the proof of \Cref{lem: each ui is a J-node}. Since $|\delta_3|\leq 30|\rightCedges{i}|$, either $|\tilde E'|\geq |\delta_3|/60$ or $|\tilde E''|\geq |\delta_3|/60$ must hold. Assume first that it is the latter. Then we can define a map $M$ from the edges of $\delta_3$ to edges of $\tilde E''$, where, for each edge $e'\in \tilde E''$, at most $60$ edges of $\delta_3$ are mapped to $e'$. Observe that $\tilde E''\subseteq E(A_2,B_2)$.  Consider now some edge $e\in \delta_3$. We obtain the path $P^*(e)$ by concatenating two paths: path $P(e)\in \pset_1$, connecting $e$ to vertex $u_i$, and the path $P'(e')\in \pset_2$, where $e'$ is the edge of $\tilde E''$ to which $e$ is mapped. Recall that path $P'(e')$ has $e'$ as its first edge and $u_{i+2}$ as its last vertex, and that edge $e'$ is incident to $u_i$. Therefore, path $P^*(e)$ has $e$ as its first edge, vertex $u_{i+2}$ as its last vertex, and it is internally disjoint from $S^*$. We then set  $\pset^*_3=\set{P^*(e)\mid e\in \delta_3}$. It is easy to verify that the paths of $\pset^*_3$ cause edge-congestion at most $60$.

Lastly, we assume that $|\tilde E'|\geq |\delta_3|/60$. As before, we define a mapping $M$ from edges of $\delta_3$ to edges of $\tilde E'$, where, for each edge $e'\in \tilde E'$, at most $60$ edges of $\delta_3$ are mapped to $e'$. 
Consider now some edge $e'=(u_i,v)\in \tilde E'$, and recall that $v\in S''_{i+1}$. Recall that the algorithm from \Cref{lem: prefix and suffix path} provided a construction of a right-monotone path $P(e',v)$. This path starts with edge $e'$, and it must terminate in some vertex of $V(S'_{i+2})\cup  V(S'_{i+3})\cup\cdots\cup V(S'_r)$. All inner vertices on path $P(e',v)$ must lie in $V(S''_{i+1})\cup V(S''_{i+2})\cup\cdots\cup V(S''_r)$. Therefore, if $e''$ is the first edge of $P(e',v)$ that is not contained in $S''_{i+1}$, then $e''\in E(A_2,V_2)$. We denote by $\tilde P(e')$ the subpath of $P(e',v)$ that starts with edge $e'$ and ends with edge $e''$. From \Cref{lem: prefix and suffix path}, we are guaranteed that all paths in set $\set{\tilde P(e')\mid e'\in \tilde E'}$ cause congestion $O(\log m)$. Consider now some edge $e\in \delta_3$. We let $P^*(e)$ be a path that is a concatenation of three paths. The first path is the path $P(e)\in \pset_1$, that connects $e$ to $u_i$. Let $e'\in \tilde E'$ be the edge to which $e$ is mapped by $M$. The second path is $\tilde P(e')$, connecting $e'$ to some edge $e''\in E(A_2,B_2)$. The third path is the path $P'(e'')\in\pset_2$, connecting $e''$ to vertex $u_{i+2}$. It is immediate to verify that the resulting path $P^*(e)$ has $e$ as its first edge, $u_{i+2}$ as its last vertex, and it is internally disjoint from $S^*$. 
We then set  $\pset^*_3=\set{P^*(e)\mid e\in \delta_3}$. From the above discussion, the paths of $\pset^*_3$ cause congestion at most $O(\log m)$.

Finally, we set $\pset^*=\pset^*_1\cup \pset^*_2\cup \pset^*_3$. It is easy to verify that the set $\pset^*$ of paths routes the edges of $\delta(S^*)$ to vertex $u_{i+2}$   with congestion $O(\log m)$, and all paths of $\pset^*$ are internally disjoint from $S^*$. We conclude that $S^*$ is a simplifying cluster, a contradiction.


\renewcommand{\tE}{\textbf{E}'}

\subsection{Proof of \Cref{claim: computing out-paths}}
\label{subsec: computing out-paths}

We provide the construction of the set $\pset^{\out,\lef}$ of paths; the construction of the set $\pset^{\out,\rig}$ of paths is symmetric.

We maintain a set $\rset=\set{R(e)\mid e\in \hat E}$ of paths, that we gradually modify over the course of the algorithm. We will ensure that, throughout the algorithm, $\rset$ is a collection of simple paths that contains, for each edge $e\in \hat E$, a path $R(e)$ that originates at $e$ and is a left-monotone path. Additionally, for every vertex $v\in V'$, the number of paths of $\rset$ terminating at $v$ is $n_1(v)$ throughout the algorithm. 
Initially, for each edge $e\in \hat E$, we let $R(e)$ be the path obtained by appending edge $e$ at the beginning of the path $P^1(e)\in \hat\pset(e)$, and we set $\rset=\set{R(e)\mid e\in \hat E}$.
Clearly, all invariants hold for the initial set $\rset$ of paths.

We perform the algorithm as long as there are two paths $R(e),R(e')$ in $\rset$, and some vertex $v$ that lies on both paths, such that the intersection of $R(e)$ and $R(e')$ at  $v$ is transversal. Note that $v$ must be an inner vertex on both $R(e)$ and $R(e')$. Assume that path $R(e)$ terminates at some vertex $u\in V'$, while path $R(e')$ terminates at vertex $u'\in V'$. We perform splicing of paths $R(e),R(e')$ at vertex $v$ (see \Cref{subsec: non-transversal paths and splicing}), obtaining two new paths: path $\tilde R(e)$, whose first edge is $e$ and last vertex is $u'$; and path $\tilde R(e')$, whose first edge is $e'$ and last vertex is $u$. If any of the resulting paths $\tilde R(e),\tilde R(e')$ is non-simple, we remove cycles from it, until it becomes a simple path. We then update the set $\rset$ of paths by replacing $R(e)$ and $R(e')$ with paths $\tilde R(e)$ and $\tilde R(e')$, respectively. It is easy to verify that, if  $R(e),R(e')$  were both left-monotone paths, then so are paths  $\tilde R(e)$ and $\tilde R(e')$. It is also immediate to verify that all other invariants hold. Clearly, when the algorithm terminates, the final set $\pset^{\out,\lef}=\rset$ of paths has all required properties.

It now remains to show that the algorithm is efficient. From \Cref{obs: splicing}, after every iteration, either (i) $\sum_{R\in \rset}|E(R)|$ decreases, or (ii) $|\Pi^T(\rset)|$ decreases, and $\sum_{R\in \rset}|E(R)|$ remains fixed. It is then immediate to verify that the algorithm terminates after $\poly(|E(G)|)$ iterations.

\subsection{Proof of \Cref{claim: enough segments}}
\label{subsec: enough segments}
In order to prove the claim, we use the following observation.

\begin{observation}\label{obs: intervals}
	Let $\iset$ be a collection of $k$ intervals of non-zero length, where for each interval $I\in \iset$, $I\subseteq [0,r]$, interval $I$ is closed on the left and open on the right, and the endpoints of $I$ are integers in $\set{0,\ldots,r}$. 
	Let $\iset'$ be another collection of $k$ intervals of non-zero length, where for each interval $I'\in \iset'$, $I'\subseteq [0,r]$, interval $I'$ is closed on the left and open on the right, and the endpoints of $I'$ are integers in $\set{0,\ldots,r}$. Assume further that for every integer $0\leq i\leq r$, the number of intervals of $\iset$ for which $i$ serves as the left endpoint is equal to the number of intervals of $\iset'$ for which $i$ serves as the left endpoint, and similarly, the number of intervals of $\iset$ for which $i$ serves as the right endpoint is equal to the number of intervals of $\iset'$ for which $i$ serves as the right endpoint. Then for every integer $p\in [0,r)$, the total number of intervals in $\iset$ containing $p$ is equal to the total number of intervals in $\iset'$ containing $p$.
\end{observation}

\begin{proof}
	Let $p$ be any integer in $[0,r)$.
	Consider any interval $I=(a,b)\in \iset\cup \iset'$. Clearly, $p\in I$ iff $a\leq p<b$. 
	
	Let $\iset_1\subseteq \iset$ be the set of all intervals $I=[a,b)\in \iset$ with $a\leq p$, and let $\iset_2\subseteq \iset$ be the set of all intervals $I=[a,b)\in \iset$ with $b\leq p$. Clearly, $\iset_2\subseteq \iset_1$, and an interval $I\in \iset$ contains the point $p$ iff $I\in \iset_1\setminus \iset_2$. We define subsets $\iset_1',\iset_2'$ of intervals of $\iset'$ similarly. As before, an interval $I'\in \iset'$ contains the point $p$ iff $I'\in \iset'_1\setminus \iset'_2$. Since, for every integer $0\leq i\leq p$, the number of intervals in $\iset$ whose left endpoint is $i$ is equal to the number of intervals in $\iset'$ whose left endpoint is $i$, we get that $|\iset_1|=|\iset_1'|$. Similarly, since, for every integer $0\leq i\leq p$, the number of intervals in $\iset$ whose right endpoint is $i$ is equal to the number of intervals in $\iset'$ whose right endpoint is $i$, we get that $|\iset_2|=|\iset'_2|$. 
	Therefore, $|\iset_1\setminus \iset_2|=|\iset_2'\setminus \iset_2'|$. Since $\iset_1\setminus \iset_2$ is precisely the set of all  intervals $I\in \iset$ that contain $p$, and $\iset_1'\setminus \iset_2'$ is precisely the set of all intervals $I'\in \iset'$ that contiain $p$, the observation follows.
\end{proof}

We construct two collections of intervals, $\iset=\set{I(e)\mid e\in \hat E}$, and $\iset'=\set{I'(e)\mid e\in \hat E}$, as follows. Consider an edge $e\in \hat E$. Assume that $\spann'(e)=\set{i',i'+1,\ldots,j'-1}$, and that $\spann''(e)=\set{i'',i''+1,\ldots,j''-1}$. We then let $I(e)=[i',j')$ and $I'(e)=[i'',j'')$.

Let $L$ be the multiset of integers, that serve as the left endpoint of every interval in $\iset$. Consider an integer $1\leq i\leq r$. The number of times that $i$ appears in set $L$ is equal to the number of edges $e\in \hat E$, such that $i$ is the first element of $\spann'(e)$; equivalently, the prefix path $P^1(e)$ must terminate at a vertex of $S_i$. Therefore, the number of times that integer $i$ appears in $L$ is $\sum_{v\in V(S_i)}n_1(v)$. Note that, if $i$ is the left endpoint of some interval $I'(e)\in \iset'$, then path $P^{\out}(e)$ must originate at a vertex of $S_i$. Therefore, path $P^{\out,\lef}(e)$ that was constructed by the algorithm from \Cref{claim: computing out-paths} must terminate at a vertex of $S_i$. From \Cref{claim: computing out-paths}, for every vertex $v\in V'$, the number of paths of $\pset^{\out,\lef}$ terminating at $v$ is exactly $n_1(v)$. Therefore, the number of intervals of $\iset'$ for which $i$ serves as the left endpoint is equal to $\sum_{v\in V(S_i)}n_1(v)$, which is exactly the number of intervals of $\iset$, for which $i$ serves as the left endpoint.

Using similar reasoning, for every integer $1\leq i\leq r$,  
the number of intervals of $\iset'$ for which $i$ serves as the right endpoint is equal to  the number of intervals of $\iset$, for which $i$ serves as the right endpoint.

Note that for each integer $1\leq t\leq r$, the number of intervals of $\iset$ containing $t$ is precisely $N_t$, while the number of intervals of $\iset'$ containing $t$ is precisely $N'_t$. We conclude that $N_t=N'_t$.

In order to prove the second assertion, observe that, for every index $1\leq t<r$, for every edge $e\in \hat E $ with $t\in \spann'(e)$, the mid-segment $P^2(e)$ of the nice guiding path $P(e)\in \hat \pset$ must contain some edge of $E_t$. Since the paths in $\hat \pset$ cause congestion at most $O(\log^{18}m)$, we get that $N_t\leq O(\log^{18}m)\cdot |E_t|$.


\subsection{Proof of \Cref{obs: bound N' values}}
\label{subsec:bound N' values}

Consider an edge $e\in E(G)$.  Recall that, if $e$ is a primary edge for an index $1\leq z\leq r$ (that is, $e\in E(\tilde S_z)\cup \delta_G(\tilde S_z)$), then $N'_z(e)=1$. Otherwise, $N'_z(e)$ is the number of paths in set: $$\set{Q(e')\mid e'\in E_{z-1}\cup E_z^{\lef}}\cup \set{Q'(e')\mid e'\in E_z\cup E_z^{\rig}}$$ that contain $e$. Here, for an edge  $e'\in E_{z-1}\cup E_z^{\lef}$, $Q(e')$ is the unique path of the internal router $\qset(U_{z-1})$ that originates at $e'$, and for an edge $e'\in E_z\cup E_z^{\rig}$, $Q'(e')$ is the unique path of the external router $\qset'(U_z)$ that originates at $e'$. 
If a secondary edge $e\in E(S_{z-1})\cup E(S_{z+1})$, then, from \Cref{obs: bound on num of copies}, $\expect{N'_z(e)}\leq \expect{N_z(e)}\leq \hat \eta$.
Otherwise, $N'_z(e)$ is the total number of auxiliary cycles in set $\set{W(e')\mid e'\in  E_z^{\lef}\cup E_z^{\rig}}$
that contain the edge $e$. Consider now some edge $e'\in \hat E$, and assume that $\spann(e')=\set{i,\ldots,j-1}$. Then $e'$ may lie in set $E_z^{\lef}\cup E_z^{\rig}$ only for $z\in \set{i,j}$. It may also be a primary edge only for indices $z\in \set{i,j}$. 
Since, from \Cref{obs: bound congestion of cycles}, edge $e$ may appear on at most $O(\log^{34}m)$ cycles in $\wset$, and since there are at most $O(1)$ indices $z$, for which $e\in E_{z-1}\cup E_z\cup E_z^{\lef}\cup E_z^{\rig}$, or $e$ is a primary edge, we get that, overall,

\[\expect{\sum_{z=1}^rN'_z(e)}\leq O(\hat \eta)+O(\log^{34}m)\leq O(\hat \eta).\]
\subsection{Proof of \Cref{obs: bound Pi triples}}
\label{subsec: proof of obs bound Pi triples}

Recall that, for $1\leq z\leq r$, $\Pi^T_z$ is the set of all triples $(e,e',v)$, where $e\in E_z^{\rig}$, $e'\in \hat E_z$, and $v$ is a vertex that lies on both $W(e)$ and $W(e')$, such that cycles $W(e)$ and $W(e')$ have a transversal intersection at $v$.  Note that $E_z^{\rig}\subseteq \hat E_z$. 
Recall that, from \Cref{obs: auxiliary cycles non-transversal at at most one}, for every pair $e,e'\in \hat E_z$ of edges, there is at most 
one vertex $v$, such that $W(e)$ and $W(e')$ have a transversal intersection at vertex $v$. We say that a triple $(e,e',v)\in \Pi^T_z$ is a \emph{type-1 triple} if the cycles $W(e), W(e')$ share an edge. Let $(e,e',v)$ be a type-1 triple of $\Pi^T_z$, and  let $e^*$ be an arbitrary edge shared by $W(e)$ and $W(e')$. We say that edge $e^*$ is \emph{responsible} for the triple $(e,e',v)$. If the cycles $W(e),W(e')$ do not share edges, then we say that triple $(e,e',v)$ is a type-2 triple.

We now bound the total number of type-1 triples in $\bigcup_{z=1}^r\Pi^T_z$. Consider an edge $e^*\in E(G)$. From \Cref{obs: bound congestion of cycles}, edge $e^*$ may appear on at most $O(\log^{34}m)$ cycles in $\wset$. Consider now any such pair $W(e),W(e')$ of cycles. Assume that $\spann(e)=\set{i,\ldots,j-1}$, and $\spann(e')=\set{i',\ldots,j'-1}$. Recall that a triple $(e,e',v)$ may only lie in a set $\Pi^T_z$ if $e\in E_z^{\rig}$, so $z=i$ must hold. Therefore, every pair $e,e'\in \hat E$ of edges, for which $e^*\in E(W(e))\cap E(W(e'))$, contributes at most $O(1)$ triples to set $\bigcup_{z=1}^r\Pi^T_z$. Overall, edge $e^*$ may be responsible for at most $O(\log^{68}m)$ triples in  $\bigcup_{z=1}^r\Pi^T_z$, and the total number of type-1 triples in $\bigcup_{z=1}^r\Pi^T_z$ is at most $|E(G)|\cdot O(\log^{68}m)$.

Next, we consider a type-2 triple $(e,e',v)\in \Pi^T_z$. Observe that $v$ is the only vertex at which $W(e)$ and $W(e')$ have a transversal intersection, and cycles $W(e),W(e')$ do not share any edges. It is then easy to see that, in the drawing $\phi^*$ of graph $G$, there must be a crossing between an edge of $W(e)$ and an edge of $W(e')$. We say that this crossing is responsible for the triple $(e,e',v)$.

Consider now some crossing $(e_1,e_2)\in \chi^*$. As before, from \Cref{obs: bound congestion of cycles} edge $e_1$ lies on at most $O(\log^{34}m)$ cycles of $\wset$, and the same bound holds for edge $e_2$. Therefore, there are at most $O(\log^{68}m)$ pairs $(e,e')\in \hat E$ of edges, with $e_1\in W(e)$ and $e_2\in W(e')$. As before, there is at most one index $z$ for which $e_1\in E_z^{\rig}$, and at most one index $z$, for which $e_2\in E_z^{\rig}$. Therefore, crossing $(e_1,e_2)$ may be responsible for at most $O(\log^{68}m)$ triples in $\bigcup_{z=1}^r\Pi^T_z$, and the total number of type-2 triples in $\bigcup_{z=1}^r\Pi^T_z$ is at most $|\chi^*|\cdot O(\log^{68}m)$. 

\subsection{Proof of \Cref{obs: bound transversal pairs}}
\label{subsec: proof of bound transversal pairs}

From now on, we assume for contradiction that 
we assume that $E(H(e_1))\cap E(H(e_2))=\emptyset$, and, additionally, there is no pair of edges $\tilde e_1\in H(e_1)$, $\tilde e_2\in H(e_2)\cup W''(e_2)$, whose images cross in $\phi^*$, and similarly, there is no pair of edges $\tilde e_1'\in H(e_1)\cup W''(e_1)$, $\tilde e_2'\in H(e_2)$,  whose images cross in  $\phi^*$. 
From \Cref{obs: transversal pairs property}, edges $\hat a_{e_1}',\hat a_{e_2}',a'_{e_1},a'_{e_2}$ appear in this order in the rotation $\oset_{u_{z-1}}\in \Sigma$.

\begin{figure}[h]
	\centering
	\includegraphics[scale=0.12]{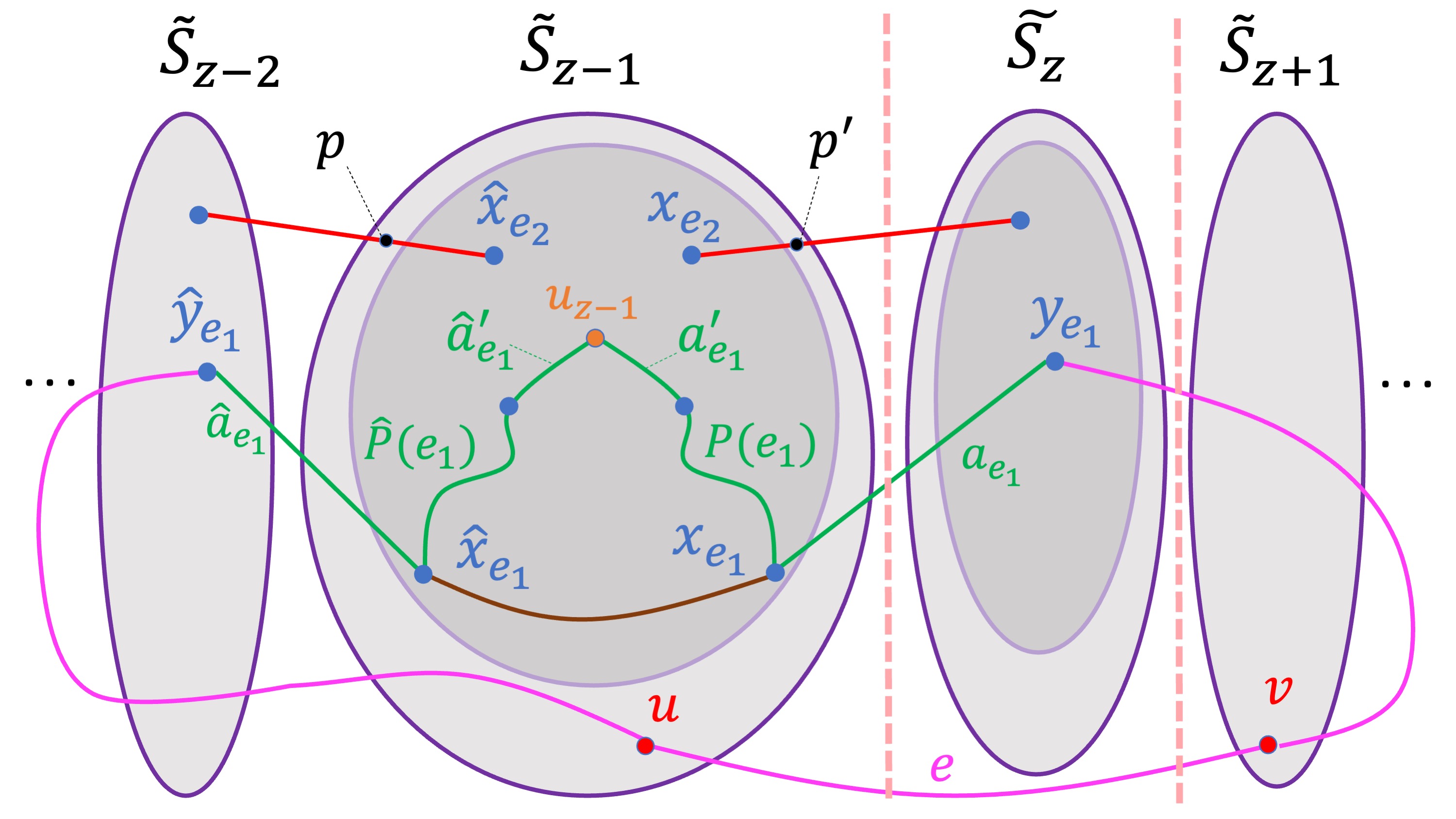}
	\caption{Illustration for the proof of \Cref{obs: bound transversal pairs}. Edges $\hat a_{e_1}$ and $a_{e_2}$ are shown in red, with points $p$ and $p'$ marked. Path $W'(e_1)$ is the concatenation of the brown path and edges $a_{e_1}, \hat a_{e_1}$. Path $W''(e_1)$ is shown in pink. 
	}\label{fig: NN8}
\end{figure}

We now define two points in the drawing $\phi^*$ of $G$: point $p$, that is an internal point on the image of edge $\hat a_{e_2}$, that is very close to the image of the vertex $\hat x_{e_2}$, such that the segment of the image of edge $a_{e_2}$ between $p$ and the image of $\hat x_{e_2}$ does not participate in any crossings. The second point, $p'$, is defined similarly on the image of edge $a_{e_2}$, very close to the image of vertex $x_{e_2}$ 
(see \Cref{fig: NN8}).

\begin{figure}[h]
	\centering
	\includegraphics[scale=0.1]{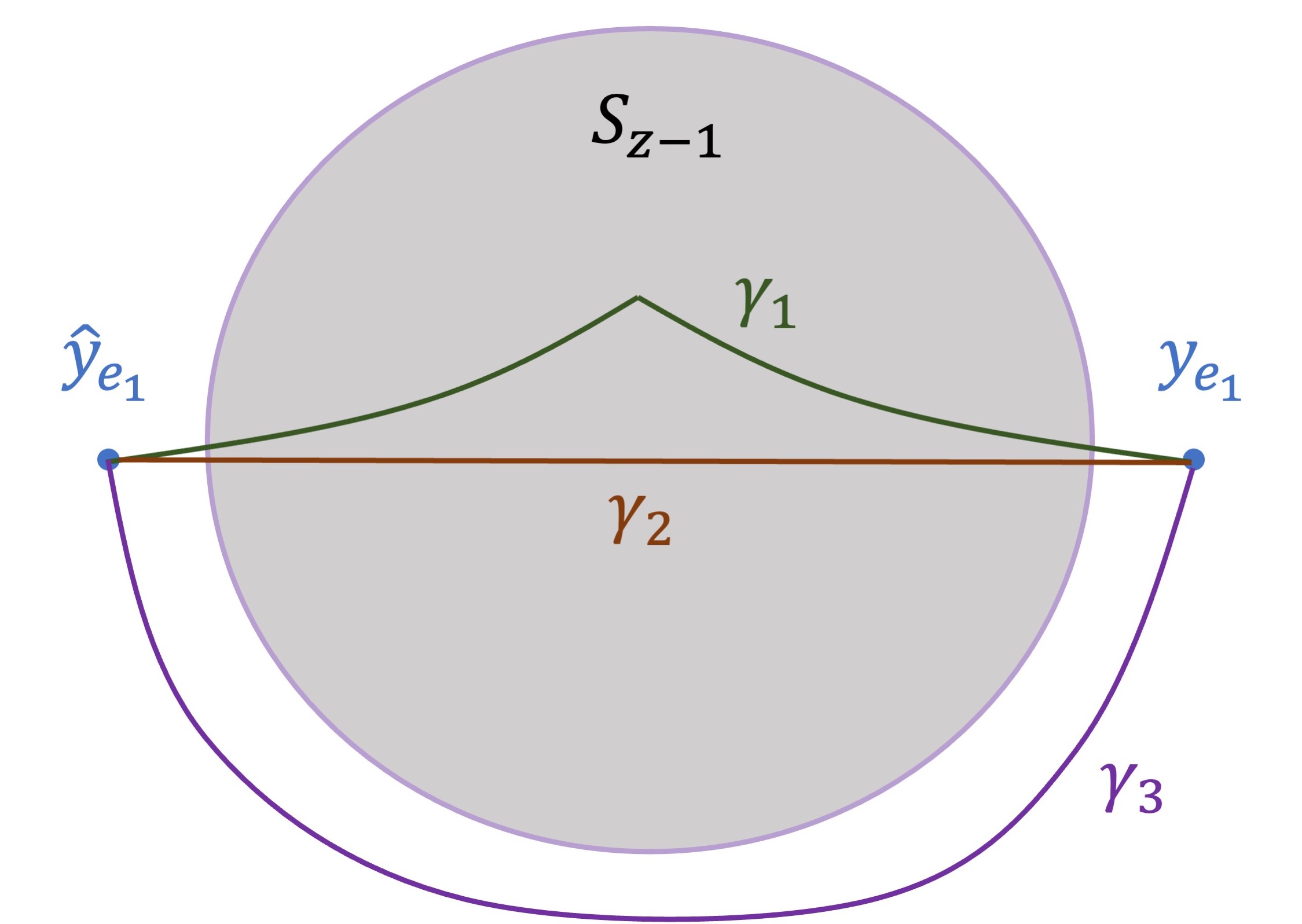}
	\caption{Curves $\gamma_1,\gamma_2$, and  $\gamma_3$.}\label{fig: NN9}
\end{figure}

Next, we consider three curves in the drawing $\phi^*$ of $G$. The first curve, $\gamma_1$, is the union of the images of paths  $\hat P(e_1)$ and $ P(e_1)$ in $\phi^*$. The second curve, $\gamma_2$, is the image of the path $W'(e_1)$, and the third curve, $\gamma_3$, is the image of the path $W''(e_1)$ in $\phi^*$ 
(see \Cref{fig: NN9}).
Observe that the endpoints of each of the three curves are the images of vertices $y_{e_1}$ and $\hat y_{e_1}$, and that points $p$ and $p'$ may not lie on any of these curves, as we have assumed that $E(H(e_1))\cap E(H(e_2))=\emptyset$. We will next show that the closed curve obtained by the union of curves $\gamma_1$ and $\gamma_2$ may not separate points $p$ and $p'$; the closed curve obtained by the union of the curves $\gamma_1$ and $\gamma_3$ must separate points $p$ and $p'$; and the closed curve obtained by the union of the curves $\gamma_2$ and $\gamma_3$ may not separate points $p$ and $p'$. We will then show that this is impossible, reaching a contradiction.

\begin{observation}\label{obs: first two curves}
	Let $\tilde \gamma_1$ be the closed curve obtained by the union of the curves $\gamma_2$ and $\gamma_3$. The points $p$ and $p'$ are not separated by $\tilde \gamma_1$. In other words, if we consider the open regions into which curve $\tilde \gamma_1$ partitions the sphere, then points $p,p'$ lie in the same region.
\end{observation}
\begin{proof}
	Assume otherwise. Consider another curve $\gamma^*$, which is obtained from the image of the path $W'(e_2)$, by truncating it so it connects point $p$ to point $p'$.
	Notice that path $W'(e_2)$ may not share any vertices with $W''(e_1)$ (except possibly for the endpoints of path $W'(e_2)$, whose images do not appear on curve $\gamma^*$). 
	Moreover, cycles $W(e_1)$ and $W(e_2)$ may not have transversal intersections at a vertex of $V(S_{z-1})$ (from \Cref{obs: auxiliary cycles non-transversal at at most one}). 
	Since we have assumed that no edge of $W'(e_2)\subseteq H(e_2)$ may cross an edge of $H(e_1)\cup W''(e_1)$, we get that curve $\gamma^*$ may not cross curve $\tilde \gamma_1$, and so points $p$ and $p'$ may not be separated by curve $\tilde \gamma_1$. 
\end{proof}

From \Cref{obs: first two curves}, we can define a simple curve $\zeta$, whose endpoints are $p$ and $p'$, such that neither of the curves $\gamma_2,\gamma_3$ crosses $\zeta$.

\begin{observation}\label{obs: second two curves}
	Let $\tilde \gamma_2$ be the closed curve obtained by the union of the curves $\gamma_1$ and $\gamma_2$. Then points $p$ and $p'$ are not separated by $\tilde \gamma_2$. In other words, if we consider the open regions into which curve $\tilde \gamma_2$ partitions the sphere, then points $p,p'$ lie in the same region.
\end{observation}
\begin{proof}
	Assume otherwise. Consider another curve $\gamma^*$, which is obtained from the image of the path $W''(e_2)$, by appending to it the image of edge $\hat a_{e_2}$ between the image of vertex $\hat y_{e_2}$ and point $p$, and the image of edge $a_{e_2}$ between the image of vertex $y_{e_2}$ and point $p'$. 
	
	Notice that path $W''(e_2)$ may not share any vertices with paths $W'(e_1)$, $P(e_1)$, and $P(e_2)$ (except for possibly the endpoints of $W''(e_2)$). Observe also that paths $W'(e_1),P(e_1)$ both contain the edge $a_{e_1}=(x_{e_1},y_{e_1})$, so, even if vertex $y_{e_1}\in W''(e_2)$, curve $\gamma^*$ may not cross curve $\tilde \gamma_2$ at the image of vertex $y_{e_1}$. Similarly, paths $W'(e_1),\hat P(e_1)$ both contain the edge $\hat a_{e_1}=(\hat x_{e_1},\hat y_{e_1})$. Therefore, even if vertex $\hat y_{e_1}\in W''(e_2)$, curve $\gamma^*$ may not cross curve $\tilde \gamma_2$ at the image of vertex $\hat y_{e_1}$. Since we have assumed that no edge of $W''(e_2)$ may cross an edge of $H(e_1)$, we conclude that curve $\gamma^*$ may not cross curve $\gamma_2$. Therefore, points $p$ and $p'$ must lie in the same region defined by $\tilde \gamma_2$.
\end{proof}

From \Cref{obs: second two curves}, we can now define a simple curve $\zeta'$, whose endpoints are $p$ and $p'$, such that neither of the curves $\gamma_1,\gamma_2$ crosses $\zeta'$.
Let $\zeta^*$ be the curve obtained from the union of the curves $\zeta,\zeta'$. Recall that curve $\gamma_2$, that connects the images of the vertices $y_{e_1}$ and $\hat y_{e_1}$, may not cross the curve $\zeta^*$. Therefore, if we denote by $q$ and $q'$ the images of the vertices $y_{e_1}$ and $\hat y_{e_1}$, respectively, then points $q$ and $q'$ do not lie on curve $\zeta^*$, and they are not separated by curve $\zeta^*$. Lastly, we need the following observation.

\begin{observation}\label{obs: third two curves}
	Let $\tilde \gamma_3$ be the closed curve obtained by the union of the curves $\gamma_1$ and $\gamma_3$. Then points $p$ and $p'$ are separated by $\tilde \gamma_3$. In other words, if we consider the set $\fset$ open regions into which curve $\tilde \gamma_3$ partitions the sphere, then points $p,p'$ lie in different regions.
\end{observation}

\begin{proof}
	Recall that we have assumed that $E(H(e_1))\cap E(H(e_2))=\emptyset$, and that no edge of $H(e_1)\cup W''(e_1)$ may cross an edge of $P(e_2)\cup \hat P(e_2)\subseteq H(e_2)$. Additionally, path $W''(e_1)$ is internally disjoint from paths $P(e_2)$ and $\hat P(e_2)$, and, 
	since paths in $\qset(S_{z-1})$ are non-transversal with respect to $\Sigma$, paths $P(e_1),\hat P(e_1),P(e_2),\hat P(e_2)$ do not have transversal intersections. Therefore, the image of path $P(e_2)$  beween point $p'$ and the image of vertex $u_{z-1}$ may not cross the curve $\tilde \gamma_3$, and it must be contained in a single region of $\fset$, that we denote by $F$. Similarly, the image of path $\hat P(e_2)$  beween point $p$ and the image of vertex $u_{z-1}$ may not cross the curve $\tilde \gamma_3$, and it must be contained in a single region of $\fset$, that we denote by $F'$. Lastly, recall that we have established that the images of edges  $\hat a_{e_1}',\hat a_{e_2}',a'_{e_1},a'_{e_2}$ enter the image of edge $u_{z-1}$ in this circular order. Recall that edges $\hat a_{e_1}',a'_{e_1}$ lie on paths $\hat P(e_1),P(e_1)$, respectively, while edges  $\hat a_{e_2}',a'_{e_2}$ lie on paths $\hat P(e_2),P(e_2)$, respectively. Since curve $\tilde \gamma_3$ is a closed curve, it must be the case that $F\neq F'$.
\end{proof}

To summarize, so far we have defined two points $p,p'$, and two curves $\zeta,\zeta'$ connecting them. We have also denoted by $\zeta^*$ the closed curve obtained by taking the union of these two curves. We also defined two points $q$ and $q'$, and two curves $\gamma_1,\gamma_3$ connecting them, and we denoted by $\tilde \gamma_3$ the closed curve obtained by taking the union of these two curves. We have established that curve $\zeta^*$ may not separate $q$ and $q'$, while curve $\tilde \gamma_3$ must separate $p$ and $p'$. Lastly, from the definition, curve $\gamma_1$ may not cross $\zeta'$, while curve $\gamma_3$ may not cross $\zeta$. We now show that this is impossible to achieve. We denote by $D$ a disc in the plane, whose boundary lies on curve $\zeta^*$, and whose interior contains the points $q$ and $q'$, and does not contain any point of $\zeta^*$. Such a disc must exist, since points $q$ and $q'$ are not separated by $\zeta^*$. Note that points $p$ and $p'$ do not lie in the interior of the disc $D$, and yet they are separated by curve $\tilde \gamma_3$. this may only happen if curve $\tilde \gamma_3$ intersects both $\zeta$ and $\zeta'$. 

Consider curve $\tilde \gamma$ that is obtained from $\tilde \gamma_3$ by deleting the point $q$ from it. On this curve, we can mark two points $a$ and $b$, such that $a$ lies on $\zeta$, $b$ lies on $\zeta'$, and no point of $\tilde \gamma$ that lies between $a$ and $b$ belongs to $\zeta^*$. Denote by $\tilde \gamma'$ the segment of $\tilde \gamma$ between $a$ and $b$. Note that this segment is disjoint from the interior of $D$, so it may not contain the point $q'$. Therefore, either $\tilde \gamma\subseteq \gamma_1$, or $\tilde \gamma\subseteq \gamma_3$. In the former case, we get that $\gamma_1$ crosses $\zeta'$, while in the latter case, we get that $\gamma_3$ crosses $\zeta$, a contradiction.

\section{Proofs Omitted from \Cref{sec: many paths}}
\label{Proofs Omitted from many paths}

\subsection{Proof of \Cref{claim: find potential augmentors}}
\label{subsec: proof of finding potential augmentors}

Consider the $\jset$-contracted instance $\hat I =(\hat G ,\hat \Sigma )$. Since it is a wide instance, there is a high-degree vertex $v^*\in V(\hat G )$, a partition $(E_1,E_2)$ of the edges of $\delta_{\hat G }(v^*)$, such that the edges of $E_1$ appear consequently in the rotation $\oset_{v^*}\in \hat \Sigma $, and a collection $\rset$  of at least $\floor{|E(\hat G)|/\mu^{50}}=\floor{\hat m(I)/\mu^{50}}$ simple edge-disjoint cycles in $\hat G$, such that every cycle $P\in \rset$ contains one edge of $E_1$ and one edge of $E_2$.
Note that we can compute the vertex $v^*$, the partition $(E_1,E_2)$ of the edges of $\delta_{\hat G }(v^*)$, and the set $\rset$ of cycles with the above properties efficiently.


Assume first that $v^*=v_J$. In this case, $\delta_{\hat G}(v^*)=\delta_G(J)$, so $(E_1,E_2)$ is also a partition of the edges of $\delta_G(J)$. From the definition of a $\jset$-contracted instance, the rotation $\oset_{v^*}\in \hat \Sigma$ is identical to the ordering $\oset(J)$ of the edges of  $\delta_{\hat G}(v^*)=\delta_G(J)$. Therefore, edges of $E_1$ appear consecutively in the ordering $\oset(J)$.
The set $\rset$ of cycles in graph $\hat G$ then naturally defines a set $\pset$ of simple paths in graph $G$, where every path $P'\in \pset$ has an edge of $E_1$ as its first edge, an edge of $E_2$ as its last edge, and it is internally disjoint from $J$. We discard arbitrary paths from $\pset$ until $|\pset|=\floor{\frac{\hat m(I)}{\mu^{50}}}$ holds, obtaining the desired set of promising paths.

From now on we assume that $v^*\neq v_J$, that is, $v^*$ is a vertex of $G$. From the definition of a high-degree vertex, 
$\deg_G(v)\geq \frac{\hat m(I)}{\mu^4}$. Therefore, there is a collection $\qset$ of at least $\ceil{\frac{2\hat m(I)}{\mu^{50}}}$ edge-disjoint paths in $G$ connecting $v^*$ to vertices of $J$, and we can compute such a collection of paths efficiently using standard Maximum $s$-$t$ Flow. We can assume w.l.o.g. that every path in $\qset$ is internally disjoint from $J$, and we view the paths in $\qset$ as being directed away from $v^*$. We can then compute a partition $(E_1',E_2')$ of the edges of $\delta_G(J)$, such that the edges of $E_1'$ appear consecutively in the ordering $\oset(J)$, and there are two subsets $\qset_1,\qset_2\subseteq \qset$ of paths of cardinality $\floor{\frac{\hat m(I)}{\mu^{50}}}$ each, such that the last edge of every path in $\qset_1$ lies in $E_1'$, and the last edge of every path in $\qset_2$ lies in $E_2'$. By arbitrarily matching the paths in $\qset_1$ to the paths in $\qset_2$ and concatenating the  pairs of matched paths, we obtain a  collection $\pset$ of edge-disjoint paths, each of which has an edge of $E_1'$ as its first edge, an edge of $E_2'$ as its last edge, and is internally disjoint from $J$. We discard paths from $\qset$ until $|\qset|=\floor{\frac{\hat m(I)}{\mu^{50}}}$ holds, obtaining the desired promising set of paths.

\subsection{Proof of \Cref{obs: combine solutions for split}}
\label{subsec: combine solutions for split}

We denote the input instances by $I=(G,\Sigma)$,  $I_{1}=(G_{1},\Sigma_{1})$, and $I_{2}=(G_{2},\Sigma_{2})$. We denote the core structure by $\jset=(J,\set{b_u}_{u\in V(J)},\rho_J, F^*(\rho_J))$, and the $\jset$-enhancement structure by  $\aset=\set{P,\set{b_u}_{u\in V(J')},\rho'}$, where $J'=J\cup P$.
We let $(\jset_1,\jset_2)$ be the split of the core structure $\jset$ via the enhancement structure $\aset$, and we denote  $\jset_1=(J_1,\set{b_u}_{u\in V(J_1)},\rho_{J_1}, F_1)$ and $\jset_2=(J_2,\set{b_u}_{u\in V(J_2)},\rho_{J_2}, F_2)$.

Let $E^{\del}=E(G)\setminus(E(G_1)\cup E(G_2))$, let $G'=G\setminus E^{\del}$, and let $\Sigma'$ be the rotation system for $G'$ that is induced by $\Sigma$. We first compute a $\jset$-clean solution $\phi'$ to instance $I'=(G',\Sigma')$ of \cnwrs with $\cro(\phi')\leq \cro(\phi_1)+\cro(\phi_2)$, and then insert the edges of $E^{\del}$ into this drawing, to obtain the final solution $\phi$ to instance $I$.

We now describe the construction of the solution $\phi'$ to instance $I'$. 
Consider the drawing $\rho'$ of graph $J'$, and the faces $F_1,F_2$ of this drawing that were used to define the split $(\jset_1,\jset_2)$ of the core structure $\jset$.
Recall that the drawing of graph $J_1$ induced by $\rho'$ is precisely $\rho_{J_1}$, and the drawing of graph $J_2$ induced by $\rho'$ is precisely $\rho_{J_2}$.

Consider now the $\jset_1$-clean drawing $\phi_1$ of graph $G_1$ on the sphere. The drawing of $J_1$ induced by $\phi_1$  is identical to $\rho_{J_1}$, and the images of all edges and vertices of $G_1$ are contained in region $F_1$ of this drawing. We plant the drawing $\phi_1$ of $G_1$ into the face $F_1$ of drawing $\rho'$, so that the images of the edges and the vertices of $J_1$ in both drawings coincide, and the images of all vertices and edges of $G_1$ appear in face $F_1$. We similarly plant drawing $\phi_2$ of $G_2$ inside face $F_2$ of $\rho'$, obtaining a solution $\phi'$ to instance $I'$, whose cost is bounded by $\cro(\phi_1)+\cro(\phi_2)$. Since the images of all edges and vertices of $G'$ are contained in region $F^*_{\rho_J}=F_1\cup F_2$, this drawing is $\jset$-clean.

In order to complete the construction of the solution $\phi$ to instance $I$, it remains to ``insert'' the images of the edges of $E^{\del}$ into $\phi'$. 
We do so using \Cref{lem: edge insertion}.
There is, however, one subtlety in using this lemma directly in order to insert the edges of $E^{\del}$ into the drawing $\phi'$: we need to ensure that the drawing remains $\jset$-clean, so the images of the newly inserted edges may not cross the images of the edges of $J$. This is easy to achieve, for example, by first contracting core $J$ into a supernode and modifying the drawing $\phi$ to obtain a drawing of the resulting graph in a natural way. We then insert the edges of $E^{\del}$ into this drawing of the contracted instance using the algorithm from  \Cref{lem: edge insertion}, and then un-contract the supernode $v_J$. We obtain a $\jset$-clean solution $\phi$ to instance $I$, whose number of crossings is bounded by $\cro(\phi_1)+\cro(\phi_2)+|E^{\del}|\cdot |E(G)|$.

\subsection{Proof of \Cref{claim: curves orderings crossings}}
\label{subsec: curves orderings crossings}

The proof of \Cref{claim: curves orderings crossings} is similar to the proof of Claim 9.9 in \cite{chuzhoy2020towards}.
For all $1\leq i\leq 4k+2$, we denote by $\gamma_i$ the image of path $P_i$ in $\phi$, that is, $\gamma_i$ is the concatenation of the images of the edges of $P_i$. Clearly, all curves in the resulting set $\Gamma=\set{\gamma_1,\ldots,\gamma_{4k+2}}$ connect $\phi(u)$ to $\phi(v)$, and they enter the image of $u$ in $\phi$ in the order of their indices.
Notice that the curves $\gamma_i\in \Gamma$ are not necessarily simple: if a pair of edges lying on path $P_i$ cross at some point $p$, then curve $\gamma_i$ crosses itself at point $p$. 
In such a case, point $p$ may not lie on any other curve in $\Gamma$.

Next, we will slightly modify the curves in $\Gamma$ by ``nudging'' them in the vicinity of their common vertices. In order to do so, we consider every vertex $x\in V(G)\setminus\set{u,v}$ that belongs to at least two paths of $\pset$ one by one.

We now describe an iteration when a vertex $x\in V(G)\setminus\set{u,v}$ is processed.
Let $\pset^x\subseteq \pset$ be the set of all paths containing vertex $x$. Note that $x$ must be an inner vertex on each such path. For convenience, we denote $\qset^x=\set{P_{i_1},\ldots,P_{i_z}}$. Consider the tiny $x$-disc $D(x)=D_{\phi}(x)$. For all $1\leq j\leq z$, denote by $s_j$ and $t_j$ the two points on the curve $\gamma_{i_j}$ that lie on the boundary of disc $D(x)$. We use the algorithm from \Cref{claim: curves in a disc} to compute a collection $\set{\sigma_1,\ldots,\sigma_z}$ of curves, such that, for all $1\leq j\leq z$, curve $\sigma_j$ connects $s_j$ to $t_j$, and the interior of the curve is contained in the interior of $D(x)$. Recall that every pair of resulting curves crosses at most once, and every point in the interior of $D(x)$ may be contained in at most two curves. Moreover, a pair $\sigma_r,\sigma_q$ of such curves may only cross if the two pairs $(s_r,t_r),(s_q,t_q)$ of points on the boundary of $D(x)$ cross. This, in turn, may only happen if paths $P_{i_r},P_{i_q}$ have a transversal intersection at vertex $x$, which is impossible. Therefore, the curves $\sigma_1,\ldots,\sigma_z$ do not cross each other. 
For all $1\leq j\leq z$, we modify the curve $\gamma_{i_j}$, by replacing the segment of the curve that is contained in disc $D(x)$ with $\sigma_j$.

Once every vertex $x\in V(G)\setminus\set{u,v}$ is processed, we obtain the final set $\Gamma'=\set{\gamma'_1,\ldots,\gamma'_{4k+2}}$ of curves.
Notice that for any pair $1\leq j<j'\leq 4k+2$ of indices, curves $\gamma'_j,\gamma'_{j'}$ cross if and only if there is a crossing $(e,e')_{p}$ in $\phi$ with $e\in E(P_j)$ and $e'\in E(P_{j'})$.

It is now enough to prove that curves $\gamma'_1$ and $\gamma'_{2k+1}$ do not cross. Assume for contradiction that the two curves cross.  Among all crossing points between these two curves, let $p'$ be the point that is closest to $\phi(u)$ on $\gamma'_1$. Let $\lambda$ be the segment of $\gamma'_1$ from $\phi(u)$ to $p'$, and let $\lambda'$ be defined similarly for $\gamma'_{2k+1}$. We modify curves $\lambda$ and $\lambda'$ to remove all their self-loops, so the curves become simple. Note that curves $\lambda$ and $\lambda'$ both originate at $\phi(u)$ and terminate at $p'$, and they do not cross. Let $\lambda^*$ be the simple closed curve obtained from the union of $\lambda$ and $\lambda'$. Curve $\lambda^*$ partitions the sphere into two internally disjoint discs, that we denote by $D$ and $D'$. 
Since the curves $\gamma'_1,\ldots,\gamma'_{4k+2}$ enter the image of $u$ in $\phi$ in the order of their indices, either (i) for all $1< i<2k+1$, the intersection of $\gamma'_i$ with the tiny $u$-disc $D_{\phi}(u)$ lies in $D$, and for all  $2k+1< i\leq 4k+2$, the intersection of $\gamma'_i$ with $D_{\phi}(u)$ lies in $D'$, or (ii) the opposite is true. We assumme without loss of generality  that it is the former.
Note that point $\phi(v)$ must lie in the interior of one of these discs -- we assume without loss of generality that it is $D'$. 

Consider now some index $1<i<2k+1$.  Recall that the segment of $\gamma'_i$ that is contained in $D_{\phi}(u)$ is contained in $D$, while point $\phi(v)$, that is an endpoint of $\gamma'_i$, lies in the interior of $D'$. Therefore, there must be some point $r$ that lies on $\gamma_i$ and on $\lambda^*$, such that the two curves have a transversal intersection at $r$. 
This point may not be $p'$, since curves $\gamma'_1$ and $\gamma'_{2k+1}$ have a crossing at $p'$, and this crossing corresponds to a crossing between an edge of $P_1$ and an edge of $P_{2k+1}$ in $\phi$. Therefore, $\gamma_i'$ has a transversal crossing with either $\lambda$ or $\lambda'$. In the former case, there is a  crossing  between an edge of $P_1$ and an edge of $P_i$, while in the latter case there is a  crossing between an edge of $P_{2k+1}$ and an edge of $P_i$. We conclude that for all $1<i<2k+1$, some edge of $P_i$ must cross an image of an edge of $P_1$ or of $P_{2k+1}$ in $\phi$. Since the edges of $P_1$ participate in at most $k$ crossings, and so do the edges of $P_{2k+1}$, and since we have assumed that an edge of $P_1$ crosses an edge of $P_{2k+1}$, this is impossible.

\subsection{Proof of \Cref{claim: bound unlucky paths}}
\label{sec: bound unlucky paths}

Assume for contradiction that there is a vertex $x\in V(G)\setminus V(J)$, and a set $\qset\subseteq \pset^*$ of $\ceil{\frac{512\mu^{13b}\cro(\phi)}{m}}$ good paths that are unlucky with respect to $x$. We denote $\qset=\set{Q_1,\ldots, Q_{\lambda}}$, where ${\lambda}=\ceil{\frac{512\mu^{13b}\cro(\phi)}{m}}$. Let ${\lambda}'=4\cdot  \ceil{{\frac{4\mu^{13b}\cro(\phi)}{m}}}$, so ${\lambda}/{\lambda}'\geq 16$. For all $1\leq i\leq {\lambda}$, we denote by $\hat e_i$ the first edge on path $Q_i$ that is incident to vertex $x$, and by $\hat e_i'$ the edge following $\hat e_i$ on path $Q_i$. We assume that the paths in $\qset$ are indexed so that the edges $\hat e_1,\ldots,\hat e_\lambda$ appear in the rotation $\oset_x\in \Sigma$ in the order of their indices. 

 Recall that we are given a partition $(E_1,E_2)$ of the edges of $\delta_G(J)$, such that the edges of $E_1$ appear consecutively in the ordering $\oset(J)$, and every path in $\pset$ has an edge of $E_1$ as its first edge, and an edge of $E_2$ as its last edge. From our construction, every path in $\pset^*$, and hence in $\qset$, has an edge of $E_1$ as its first edge, and an edge of $E_2$ as its last edge. Consider the  solution $\phi$ to instance $I$, that is $\jset$-valid. Let $\phi'$ be the drawing that is obtained from $\phi$ after we delete all edges and vertices from it, except for those lying in $J$, and on the paths of $\qset$. Since all paths in $\qset$ are good, there are no crossings in $\phi'$ in which the edges of $J$ are involved. Let $D(J)$ be the disc associated with core $J$ in $\phi$. This disc contains the image of $J$ in $\phi'$ in its interior, and its boundary follows closely the drawing of $J$. 
 Notice that the image of every path $Q\in \qset$ in $\phi$ must intersect the interior of the region $F^*$, from the definition of a valid core structure (see \Cref{def: valid core 2}), and since each such path contains edges incident to vertices of $J$. Therefore, the image of every path $Q\in \qset$ in $\phi'$ is contained in the region $F^*$. We can then ensure that no crossing points of $\phi'$ are contained in disc $D(J)$; the only vertices and edges whose images in $\phi'$ are contained in $D(J)$ are the vertices and edges of $J$; and the only other edges whose images intersect $D(J)$ in $\phi'$ are the edges of $\delta_G(J)$ that lie on the paths of $\qset$. Moreover, for each such edge $e$, the intersection of $\phi'(e)$ and $D(J)$ is a simple curve. 
 
\begin{figure}[h]
	\centering
	\includegraphics[scale=0.13]{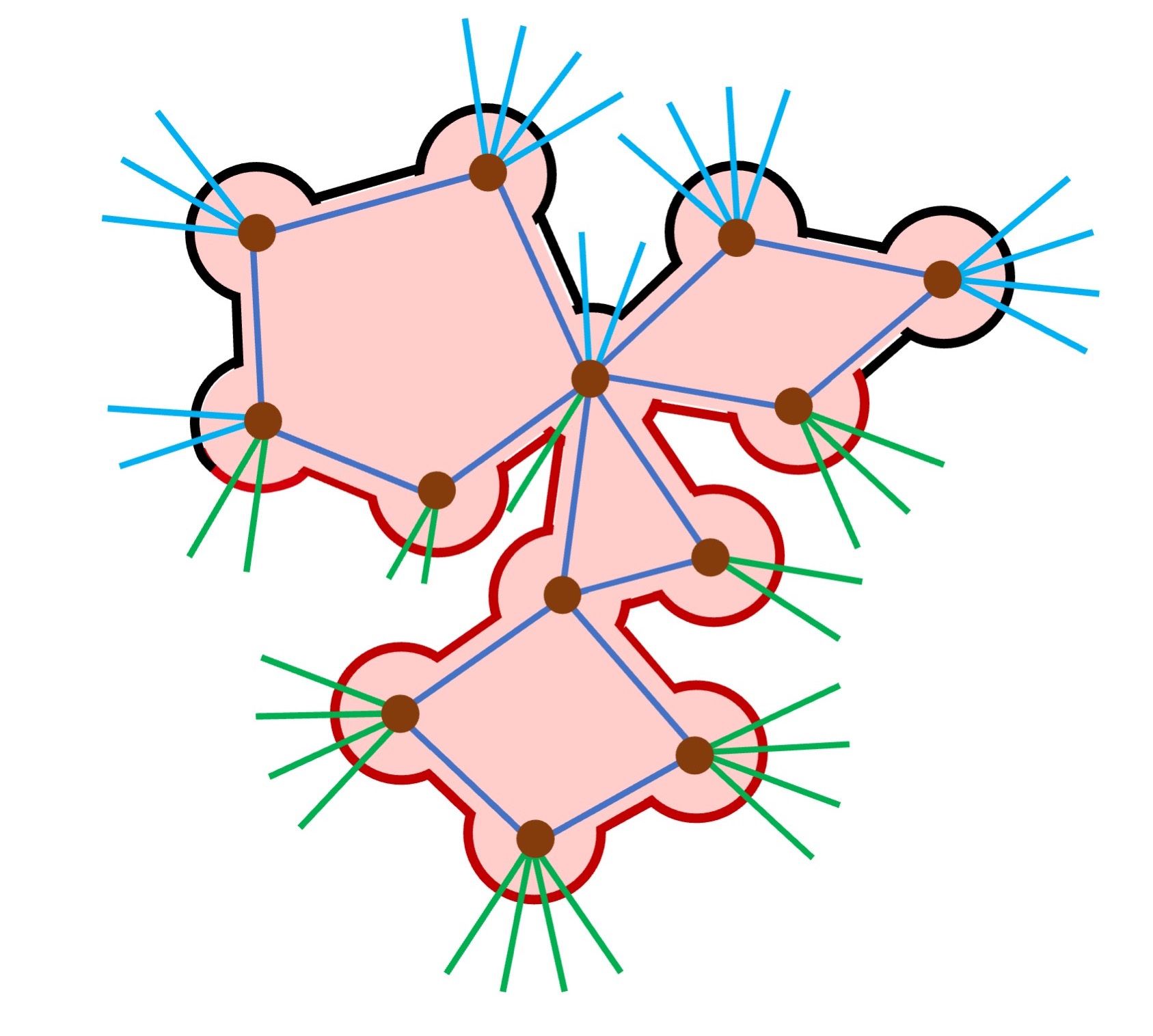}
	\caption{Partitioning the boundary of disc $D(J)$ into segments $\sigma$ (black) and $\sigma'$ (red).
		Edges of $E_1$ are shown in  light blue and edges of $E_2$ are shown in  green.}\label{fig: proof520_3} 
\end{figure}

 We partition the boundary of disc $D(J)$ into two segment $\sigma$ and $\sigma'$, such that $\sigma$ contains all intersection points of the boundary of $D(J)$ with the images of the edges in $E_1$ in $\phi$, while $\sigma'$ contains all intersection points of the boundary of $D(J)$ with the images of the edges in $E_2$ in $\phi$ (see \Cref{fig: proof520_3}).

Let $\phi''$ be the drawing that is obtained from $\phi'$ by deleting the images of the vertices and the edges of $J$ from it, and deleting the segment of the image of every edge $e\in \delta_G(J)$ that is contained in the interior of $D(J)$. We then contract segment $\sigma$ into a point $p$, and segment $\sigma'$ into a point $q$, so that the image of every path $Q\in \qset$ in the resulting drawing connects $p$ to $q$, and for every edge $e\in \bigcup_{Q\in \qset}E(Q)$, the number of crossings in which $e$ participates in $\phi''$ is bounded by the number of crossings in which $e$ participates in $\phi'$.

For all $1\leq i\leq {\lambda}$, we let $R_i$ be the subpath of $Q_i$ from its first vertex to $x$, so $R_i$ contains an edge of $E_1$, and let $\rset=\set{R_i\mid 1\leq i\leq {\lambda}}$.
We also denote $\qset'=\set{Q_{i\cdot {\lambda}'}\mid 1\leq i\leq 16}$, and, for all $1\leq i\leq 16$, we let $\tilde R_i=R_{i\cdot {\lambda}'}$ -- the subpath of path $Q_{i\lambda'}$, from its first endpoint to vertex $x$. Let $\tilde \rset=\set{\tilde R_1,\ldots,\tilde R_{16}}$.

Recall that all paths in $\qset$ are good, and so each such path participates in at most $\frac{\cro(\phi)\cdot \mu^{12b}}{m}$ crossings. Since for every pair $P,P'\in \pset^*$ of paths, for every vertex $v\in V(P)\cap V(P')$ with $v\not\in V(J)$, the intersection of $P$ and $P'$ at $v$ is non-transversal with respect to $\Sigma$, the paths in set $\rset$ are non-transversal with respect to $\Sigma$, so are the paths in $\tilde \rset$. Therefore, from \Cref{claim: curves orderings crossings}, for any pair $2\cdot  \ceil{\frac{\cro(\phi)\cdot \mu^{12b}}{m}}< i<j\leq {\lambda}$ of indices with $j-i> 2\cdot\ceil{ \frac{\cro(\phi)\cdot \mu^{12b}}{m}}$, there is no crossing in $\phi''$ between an edge of $R_i$ and an edge of $R_j$. In particular, since  ${\lambda}'=4\cdot  \ceil{{\frac{4\mu^{13b}\cro(\phi)}{m}}}$, there are no crossings in $\phi''$ between pairs of edges lying in distinct paths of $\tilde \rset$.

For $1\leq i\leq 16$, we let $\gamma_i$ be the curve that is obtained from the image of path $\tilde R_i$ in $\phi''$, after removing all self-loops. Consider the resulting collection $\Gamma=\set{\gamma_1,\ldots,\gamma_{16}}$ of curves. All curves in $\Gamma$ originate at point $p$ and terminate at the image of $x$ in $\phi''$. The curves do not cross themselves or each other, and they enter the image of $x$ in the order of their indices. The curves in $\Gamma$ partition the sphere into $16$ regions, that we denote by $\tilde F_1,\ldots,\tilde F_{16}$. Region $\tilde F_1$ has curves $\gamma_1$ and $\gamma_{16}$ as its boundaries, and for $1<i\leq 16$, region $\tilde F_i$ has curves $\gamma_{i-1}$ and $\gamma_i$ as its boundaries. Note that point $q$ must be contained in the interior of one of these regions. We assume without loss of generality that it is $\tilde F_1$ (as otherwise we could re-index the paths of $\qset$ accordingly).

Next, we consider the path $Q^*=Q_{8\lambda'+1}$, and we prove that path $Q^*$ is not unlucky for vertex $x$, reaching a contradiction. 

\begin{observation}\label{obs: not unlucky}
	Path $Q^*$ is not unlucky for vertex $x$.
\end{observation}

\begin{proof}
	Let $e^*$, $e^{**}$ be the edges of $Q^*$ that are incident to $x$, with $e^*$ lying before $e^{**}$ on the path. 
	 Let  $\hat E_1(x)\subseteq \delta_G(x)$ be the set of edges $\hat e\in \delta_G(x)$, such that $\hat e$ lies between $e^*$ and $e^{**}$ in the rotation $\oset_x\in \Sigma$ (in clock-wise orientation), and $\hat e$ lies on some good path of $\pset^*$. Let $\hat E_2(x)\subseteq \delta_G(x)$ be the set of edges $\hat e\in \delta_G(x)$, such that $\hat e$ lies between $e^{**}$ and $e^{*}$ in the rotation $\oset_x\in \Sigma$ (in clock-wise orientation), and $\hat e$ lies on some good path of $\pset^*$. It is enough to prove that $|\hat E_1(x)|,|\hat E_2(x)|\geq \frac{\cro(\phi)\mu^{13b}}{m}$.

	For all $2\leq i\leq 16$, let $\qset_i=\set{Q_j\mid (i-1){\lambda}'<j< i\lambda'}$. Recall that ${\lambda}'=4\cdot  \ceil{\frac{4\mu^{13b}\cro(\phi)}{m}}$, and every path $\tilde R_i\in \tilde \rset$ participates in at most  $ \frac{\cro(\phi)\cdot \mu^{12b}}{m}$  crossings in $\phi''$, since the paths in $\qset$ are good. Therefore, there must be a subset $\qset'_i\subseteq \qset_i$ of at least $2\cdot  \ceil{\frac{4\mu^{13b}\cro(\phi)}{m}}$ paths, such that for every path $Q_j\in \qset'_i$, the image of $Q_j$ in $\phi$ does not cross the curves $\gamma_{i-1}$ and $\gamma_i$. In particular, the image of every path in set $\set{R_j\mid Q_j\in \qset'_i}$ is contained in region $\tilde F_i$ in $\phi''$.

	Recall that we have defined a set $\qset'_4\subseteq \qset_4$ of at least $2\cdot \ceil{\frac{4\mu^{13b}\cro(\phi)}{m}}$ paths, where for each path $Q_j\in \qset'_4$, the image of the corresponding path $R_j$ is contained in $\tilde F_{4}$. Recall that, for every path $Q_j\in \qset$, we denoted by 
	$\hat e_j$ the first edge on path $Q_j$ that is incident to $x$. 
	We denote by $E^L=\set{\hat e_j\mid Q_j\in \qset'_4}$. Clearly, for every edge $\hat e_j\in E^L$, the image of $\hat e_j$ in $\phi''$ lies in region $\tilde F_4$.
	
	Similarly, we have defined a set 
	$\qset'_{14}\subseteq \qset_{14}$ of at least $2\cdot \ceil{\frac{4\mu^{13b}\cro(\phi)}{m}}$ paths, where for each path $Q_j\in \qset'_{14}$, the image of the corresponding path $R_j$ is contained in $\tilde F_{14}$. 
	We denote by $E^R=\set{\hat e_j\mid Q_j\in \qset'_{14}}$. Clearly, for every edge $\hat e_j\in E^R$, the image of $\hat e_j$ in $\phi''$ lies in region $\tilde F_{14}$.
	
	It is also easy to see that the imgage of edge $e^*$ must be contained in $\tilde F_7\cup \tilde F_8\cup \tilde F_9\cup \tilde F_{10}$ (as otherwise path $Q^*$ would need to cross more than $\frac{\cro(\phi)\cdot \mu^{12b}}{m}$ other paths in $\qset$, for example, the paths of $\qset'_7$, or the paths of $\qset'_{10}$).
	
	Lastly, we show that the intersection of the image of edge $e^{**}$ and the tiny $x$-disc $D_{\phi''}(x)$, that we denote by $\sigma(e^{**})$, must be contained in $\tilde F_2\cup \tilde F_1\cup \tilde F_{16}$. Note that, if this is the case, then either (i) $E^L\subseteq \hat E_1(x)$ and $E^R\subseteq \hat E_2(x)$; or (ii) $E^R\subseteq \hat E_1(x)$ and $E^L\subseteq \hat E_2(x)$ hold. In either case, since $|E^R|,|E^L|\geq 
	2\cdot \ceil{\frac{4\mu^{13b}\cro(\phi)}{m}}$, path $Q^*$ is not unlucky for $x$.
	
	It now remains to prove that $\sigma(e^{**})$ is contained in $\tilde F_2\cup \tilde F_1\cup \tilde F_{16}$. Assume otherwise. From the definition of tiny $x$-disc, $\sigma(e^{**})$ must be contained in some region $\tilde F_i$, for $3\leq i\leq 15$. Recall that we have defined a subset $\qset'_2\subseteq \qset_2$ of at least $2\cdot \ceil{\frac{4\mu^{13b}\cro(\phi)}{m}}$ paths, such that, for every path $Q_j\in \qset'_2$, the image of the corresponding path $R_j\in \rset$ is contained in region $\tilde F_2$. We have also defined a collection $\qset'_{16}\subseteq \qset_{16}$ of at least 
	$2\cdot \ceil{\frac{4\mu^{13b}\cro(\phi)}{m}}$ paths, such that, for every path $Q_j\in \qset'_{16}$, the image of the corresponding path $R_j\in \rset$ is contained in region $\tilde F_{16}$. 
	
	Consider now the segment $\gamma^*$ of the image of path $Q^{*}$ in $\phi''$ from vertex $x$ to point $q$. Since $\sigma(e^{**})$ is contained in $\bigcup_{3\leq i\leq 15}\tilde F_i$, while point $q$ is contained in $\tilde F_1$, curve $\gamma^*$ has to either cross the image of every path in $\set{R_j\mid Q_j\in \qset'_2}$, or it has to cross the image of every path in $\set{R_j\mid Q_j\in \qset'_{16}}$. In either case, since the paths of $\qset$ are non-transversal with respect to $\Sigma$, the edges of path $Q^*$ must participate in at least $2\cdot \ceil{\frac{4\mu^{13b}\cro(\phi)}{m}}$ crossings, contradicting the fact that $Q^*$ is a good path.
\end{proof}

\subsection{Proof of \Cref{claim: new drawing}}
\label{sec:getting new drawing}

We start with $\phi'$ being the drawing of graph $G'$ that is induced by $\phi$.
Since bad event $\event_1$ did not happen, drawing $\phi'$ does not contain crossings between edges of $E(P^*)$ and edges of $E(J)$, but it may contain crossings between pairs of edges in $E(P^*)$. We next show how to modify this drawing in order to eliminate all such crossings.
Let $\gamma$ be the image of the path $P^*$ in $\phi'$. Notice that $ \gamma$ is either a closed or an open curve, that may cross itself in a number of points.
Recall that path $P^*$ is internally disjont from $J$, and there are no crossings in $\phi'$ between edges of $P^*$ and edges of $J$. Moreover, from the definition of valid core structure (see \Cref{def: valid core 2}), and since $\phi$ is a $\jset$-valid drawing of $G$, $\gamma$ must intersect the interior of the region $F^*$. Therefore, $\gamma\subseteq F^*$ must hold. In order to obtain the desired final drawing of graph $G'$, we will only modify the images of the edges and vertices that lie on $P^*$, with the new images contained in $F^*$, so that the resulting drawing of $G'$ is  $\phi$-compatible.

 We can partition the curve $\gamma$ into a collection $\Gamma$ of curves, for which the following hold. First, there is a single special curve $\gamma^*\in \Gamma$, which is either a simple closed or a simple open curve. In the former case, $\gamma^*$ contains the image of exactly one vertex of $J$, and in the latter case, the endpoints of $\gamma^*$ are images of two distinct vertices of $J$. All other curves in $\Gamma$ are simple closed curves. For every pair $\gamma_1,\gamma_2\in \Gamma$ of distinct curves, $\gamma_1$ and $\gamma_2$ may share at most one point, and that point must be a crossing point between a pair of edges of $E(P^*)$ in $\phi'$. We ensure that every point of $ \gamma$ lies on at least one curve of $\Gamma$. Since all curves in $\Gamma$ are simple, no curve in $\Gamma$ may cross itself. 
We need the following observation.

\begin{observation}\label{obs: no heavy vertices on loops}
	If neither of the Events $\event_1,\event_3$ happenned, then for every curve $\gamma'\in \Gamma\setminus\set{\gamma^*}$, every vertex $x$ with $\phi'(x)\in \gamma'$ is a light vertex.
\end{observation}

\begin{proof}
	Assume otherwise. Let  $\gamma'\in \Gamma\setminus\set{\gamma^*}$ be a curve, and $x$ a vertex with  $\phi'(x)\in \gamma'$, such that $x$ is a heavy vertex. Denote $\phi'(x)$ by $p$, and notice that point $p$ may not lie on any other curves in $\Gamma$ (since every point shared by a pair of curves in $\Gamma$ is a crossing point between a pair of edges from $E(P^*)$.)

	Let $D=D_{\phi'}(x)$ be a tiny $x$-disc. For every edge $\hat e$ that is incident to $x$, we denote by $\sigma(\hat e)$ the intersection of $\phi'(\hat e)$ with $D$. Let $e,e'$ be the two edges of $P^*$ that are incident to $x$.
	Denote by  $\hat E_1(x)\subseteq \delta_G(x)$ the set of edges $\hat e\in \delta_G(x)$, such that $\hat e$ lies strictly between $e$ and $e'$ in the rotation $\oset_x\in \Sigma$ (in clock-wise orientation), and $\hat e$ lies on some good path of $\pset^*$. Let $\hat E_2(x)\subseteq \delta_G(x)$ be the set of edges $\hat e\in \delta_G(x)$, such that $\hat e$ lies strictly between $e'$ and $e$ in the rotation $\oset_x\in \Sigma$ (in clock-wise orientation), and $\hat e$ lies on some good path of $\pset^*$. Since Event $\event_3$ did not happen, path $P$ may not be unlucky with respect to $x$, so  $|\hat E_1(x)|,|\hat E_2(x)|\geq \frac{\cro(\phi)\mu^{13b}}{m}$ holds.
	
	Curve $\gamma'$ partitions the sphere into two regions, that we denote by $\tilde F$ and $\tilde F'$. Since the curves $\sigma(e),\sigma(e')$ are contained in $\gamma'$, it must be the case that either (i) for every edge $\hat e\in \hat E_1(x)$, $\sigma(\hat e)\subseteq \tilde F$, and for every edge $\hat e\in \hat E_2(x)$, $\sigma(\hat e)\subseteq \tilde F'$, or (ii) the opposite is true. We assume w.l.o.g. that it is the former. Note that, since $\gamma$ is disjoint from the image of the core $J$ in $\phi'$ (except for its endpoints), the image of $J$ in $\phi'$ must be contained either in   the interior of $\tilde F$, or in the interior of $\tilde F'$; we assume w.l.o.g. that it is the former. Consider now some edge $\hat e\in \hat E_2(x)$, and let $\hat P\in \pset^*$ be a path that contains $\hat e$. Observe that $\sigma(\hat e)\subseteq \tilde F'$, while both endpoints of $\hat P$ belong to $J$, whose image lies in the interior of $\tilde F$. Therefore, the image of path $\hat P$ in $\phi'$ must cross the curve $\gamma'$. Let $q$ be a point of $\gamma'$ that lies on the image of $\hat P$, such that the image of $\hat P$ and $\gamma'$ have a transversal intersection at $q$. Note that $q$ may not be the image of a vertex of $G$, since for every vertex $v\in V(G)\setminus V(J)$, for every pair $P,P'\in \pset^*$ of paths containing $v$, the intersection of $P$ and $P'$ at $v$ is non-transversal  with respect to $\Sigma$. Therefore, $q$ is a crossing point between an edge of $\hat P$ and an edge of $P^*$. We conclude that the edges of $P^*$ participate in at least $|\hat E_2(x)|\geq \frac{\cro(\phi)\mu^{13b}}{m}$ crossings. But since we have assumed that Event $\event_1$ did not happen, path $P^*$ must be good, a contradiction. (Note that it is possible that, for some good path $\hat P\in \pset^*$, both edges of $\hat P$ that are incident to $x$ lie in $\hat E_2(x)$. But in that case, the image of $\hat P$ must cross $\gamma$ twice, since both endpoints of $\hat P$ lie in $J$).
\end{proof}

Let $p_1,\ldots,p_z$ denote the points on the curve $\gamma^*$ that correspond to crossing points between pairs of edges of $P^*$, and assume that these points appear on $\gamma^*$ in this order. For all $1\leq i\leq z$, let $e_i,e'_i$ be the pair of edges of $P^*$ that cross at point $p_i$, with edge $e_i$ appearing before edge $e'_i$ on path $P^*$. 
For all $1\leq i\leq z$, let $Q_i$ be the subpath of path $P^*$ from edge $e_i$ to edge $e'_{i}$, and we denote by $y_i,y'_i$ the second and the penultimate vertices of $Q_i$, respectively. Let $Q'_i$ be the subpath of $Q_i$ connecting $y_i$ to $y'_i$. From \Cref{obs: no heavy vertices on loops}, if Events $\event_1,\event_3$ did not happen, the every vertex of $Q'_i$ is a light vertex. Since we have deleted all edges of $E'$ from $G$ to obtain graph $G'$, every vertex of $Q'_i$ is incident to exactly two edges in $G'$, and these edges lie on path $P^*$.

\begin{figure}[h]
	\centering
	\subfigure[Before: tiny $p_i$-disc $D_i$ is shown in gray, and the original image of path $Q_i$ is shown in blue.]{
		\scalebox{0.42}{\includegraphics[scale=0.25]{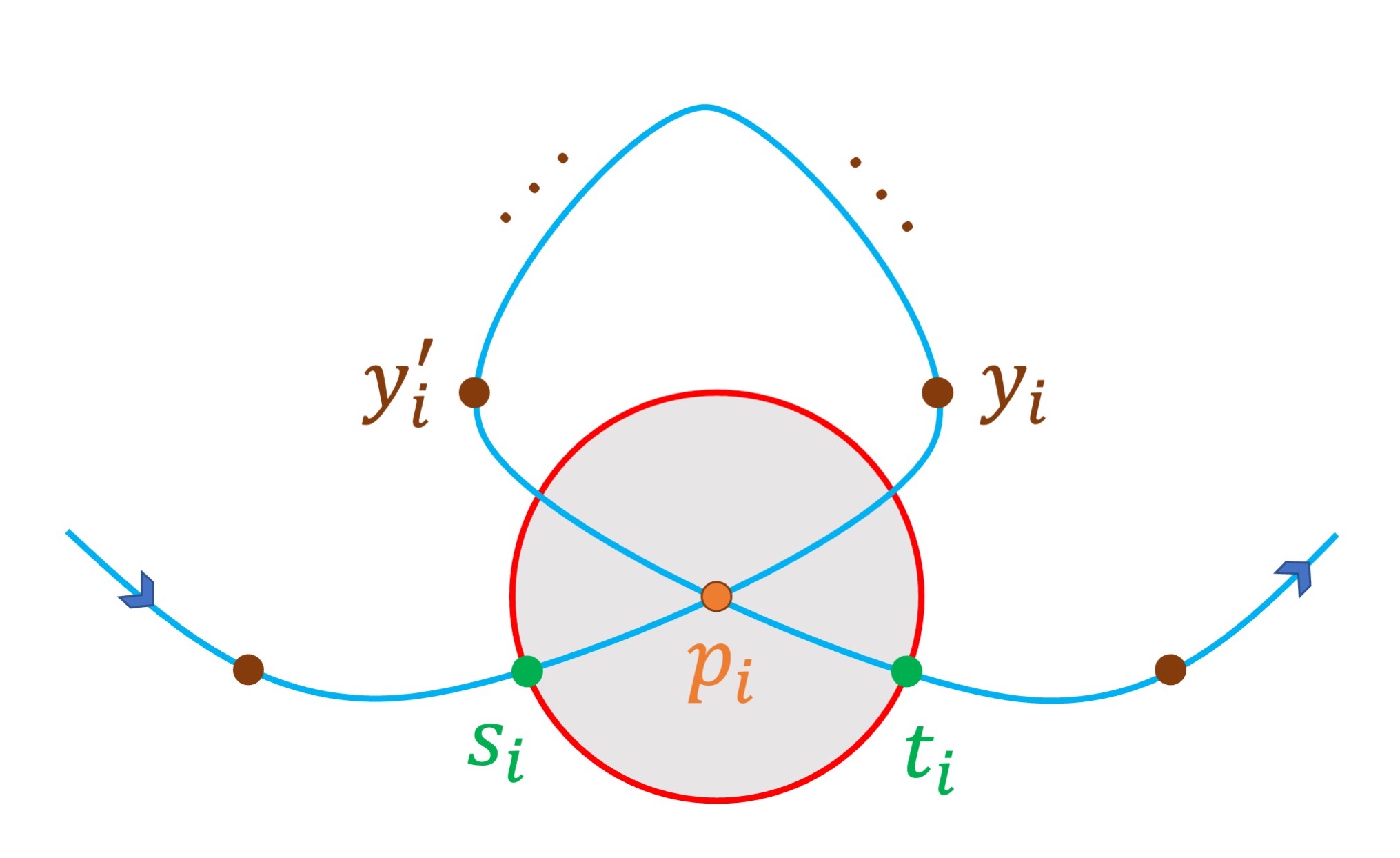}}}
	\hspace{0.5cm}
	\subfigure[After: the new images of vertices $y_i,y'_i$ are shown in brown, and the new image of path $Q_i$ is shown in blue.]{\scalebox{0.42}	{\includegraphics[scale=0.25]{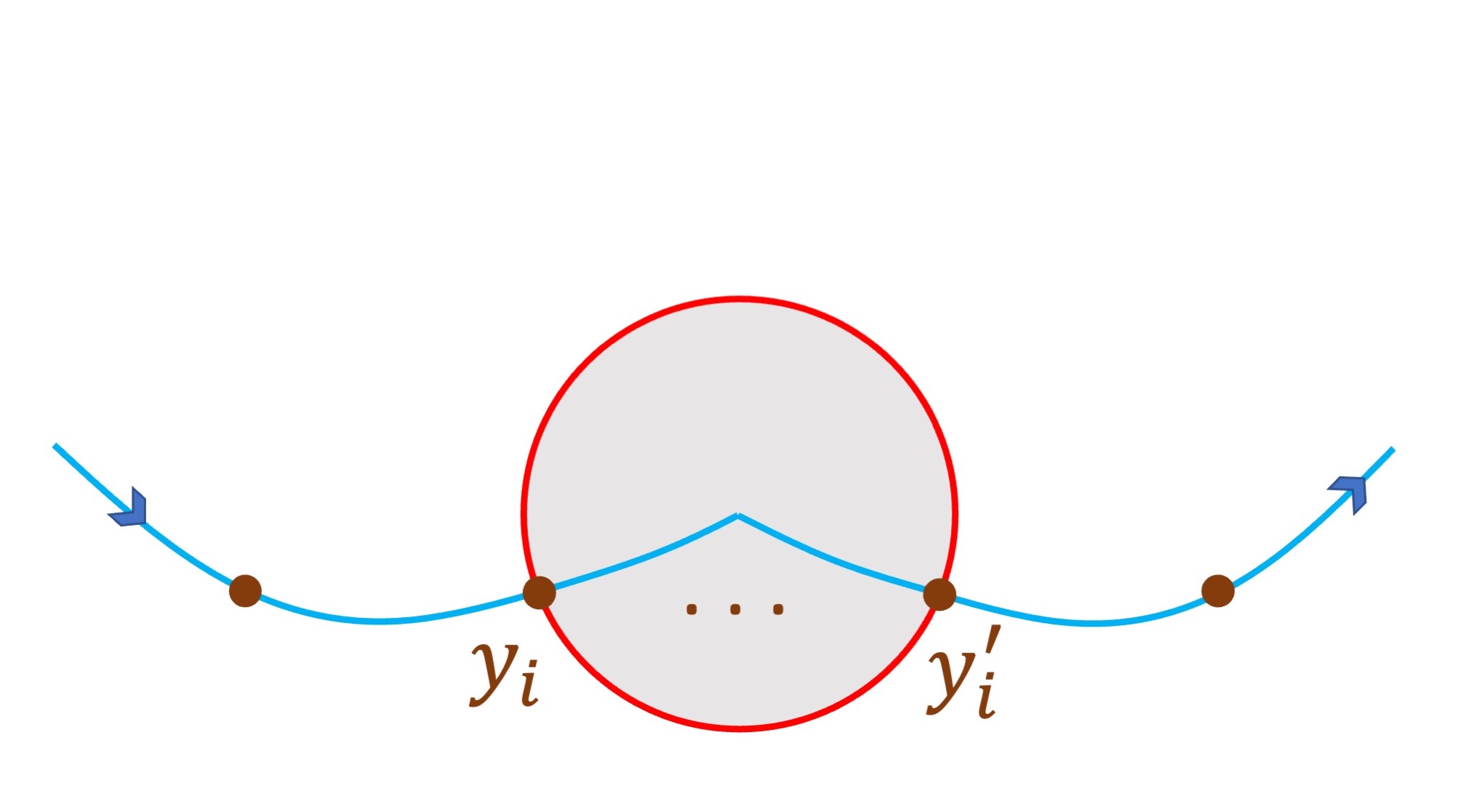}}
	}
	\caption{Modifying the image of path $Q_i$. 
	}\label{fig: proof522}
\end{figure}

We let $D_i$ be a tiny $p_i$-disc in the drawing $\phi'$. Let $s_i$ be the point on the image of edge $e_i$ lying on the boundary of $D_i$, and let $t_i$ be the point on the image of edge $e'_i$ lying on the boundary of $D_i$. We modify the drawing $\phi'$ in order to ``straighten'' the loop corresponding to the image of the path $Q'_i$, as follows. First, we truncate the images of the edges $e_i$ and $e'_i$, by deleting the segment of $\phi'(e_i)$ between $s_i$ and $\phi'(y_i)$, and similarly deleting the segment of $\phi'(e'_i)$ between $t_i$ and $\phi'(y'_i)$. We then delete the images of all vertices and edges of $Q'_i$ from $\phi'$. We place the new image of $y_i$ at point $s_i$, and the new image of $y'_i$ at point $t_i$. We then add an image of the path $Q'_i$ as a simple curve with endpoints $s_i$ and $t_i$, that is contained in $D_i$, so that the image of $Q'_i$ is contained in $\gamma^*\cap D_i$ (see \Cref{fig: proof522}). 

Clearly, this modification does not increase the number of crossings, and it is local to region $F^*$. Once every point $p_1,\ldots,p_z$ is processed, we obtain the final solution $\phi'$ to instance $I'$ that is compatible with $\phi$, with $\cro(\phi')\leq \cro(\phi)$. Note that our transformation step does not introduce any new crossings. Therefore, if $(e_1,e_2)_p$ is a crossing in drawing $\phi'$, then there is a crossing between edges $e_1$ and $e_2$ at point $p$ in drawing $\phi$.


\subsection{Proof of \Cref{claim: cut set small case2}}
\label{subsec: small cut set in case 2}
Assume for contradiction that Event $\event$ did not  happen, but $|E''|> \frac{2\cro(\phi)\cdot \mu^{12b}}{m}+|\chi^{\dirty}(\phi)|$. From the Maximum Flow - Minimum Cut Theorem, there is a collection $\qset$ of $\ceil{ \frac{2\cro(\phi)\cdot \mu^{12b}}{m}}+|\chi^{\dirty}(\phi)|$ edge-disjont paths connecting $s$ to $t$ in $H$. Note that every edge in $H$ corresponds to some distinct edge in graph $G$. We do not distinguish between these edges. We  show that, for every path $Q\in \qset$, there is a crossing between an edge of $Q$ and an edge of $E(P^*)\cup E(\thecore)$ in $\phi'$. From \Cref{claim: new drawing}, and since the number of crossings in which the edges of $\thecore$ may participate is bounded by $|\chi^{\dirty}(\phi)|$, it then follows that the edges of $P^*$ participate in at least $\ceil{ \frac{2\cro(\phi)\cdot \mu^{12b}}{m}}$ crossings in $\phi$. But, since we have assumed that Event $\event_1$ did not happen, path $P^*$ is good, so its edges may particpate in at most $\frac{\cro(\phi)\cdot \mu^{12b}}{m}$ crossings, a contradiction. It now remains to prove that, for every path $Q\in \qset$, there is a crossing between an edge of $Q$ and an edge of $E(P^*)\cup E(\thecore)$ in $\phi'$.

Consider any path $Q\in \qset$. This path naturally defines a path $Q'$ in graph $G$, whose first edge, denoted by $e(Q)$, lies in $\tilde E_1$, and last edge, denoted by  $e'(Q)$, lies in $\tilde E_2$. Then the image of edge $e(Q)$ in $\phi'$ must intersect the interior of region $F_1$, while the image of edge $e'(Q)$ in $\phi'$ must intersect the interior of region $F_2$.  Therefore, the image of the path $Q'$ in $\phi'$ must cross the boundary of the face $F_1$.
Since path $Q'$ is internally disjoint from $V(J')$, the image of some edge on path $Q'$ must cross the image of some edge of $E(J')=E(P^*)\cup E(\thecore)$ in $\phi'$.

\subsection{Proof of \Cref{obs: few edges in split graphs case2}}
\label{subsec: few edges in split Case 2}
We start by recalling how the enhancement path $P^*$ was selected. Recall that initial set $\pset^*$ of paths had cardinality 
$k\geq \frac{15m}{16\mu^{b}}$.
We denoted by $E^*_1=\set{e_1,\ldots,e_{k}}\subseteq E_1$ the subset of edges that belong to the paths of $\pset^*$, where the edges are indexed so that $e_1,\ldots,e_{k}$ appear consecutively, in the order of their indices in the ordering $\oset(J)$. For all $1\leq j\leq k$, we denote by $P_j\in \pset^*$ the unique path originating at the edge $e_j$. We then selected an index $\floor{k/3}<j^*<\ceil{2k/3}$ uniformly at random, and we let $P^*=P_{j^*}$. 
Let $e'\in E_2$ be the edge of $P^*$ lying in $E_2$. Let $\tilde E_1$ be the set of edges lying between $e_{j^*}$ and $e'$ in  $\oset(\thecore)$, and let $\tilde E_2$ be the set of edges lying between $e'$ and $e_{j^*}$  in $\oset(\thecore)$. Then one of the sets $\tilde E_1, \tilde E_2$ of edges must contain all edges in $\set{e_1,\ldots,e_{\floor{k/3}-1}}$, while the other must contain all edges in $\set{e_{\ceil{2k/3}+1},\ldots,e_{k}}$. Therefore, $|\tilde E_1|, |\tilde E_2|\geq \frac {k} 6\geq  \frac{m}{12\mu^{b}}$. 
 
 Notice that graph $G_1$ may only contain edges from one of the sets $\tilde  E_1, \tilde E_2$, and so $|E(G_1)|\leq m-\frac{m}{32\mu^{b}}$. Using similar reasoning, $|E(G_2)|\leq  m-\frac{m}{32\mu^{b}}$.

\subsection{Proof of \Cref{thm: wld all paths congestion}}
\label{subsec: proof of wld cor}

Our algorithm consists of a number of phases. For all $j\geq 1$, the input to phase $j$ consists of a collection $\rset_j$ of disjoint clusters of $S$, and, for every cluster $R\in \rset_j$,  two sets $\pset_1(R),\pset_2(R)$ of paths in graph $G$. We require that $\pset_1(R)=\set{P_1(e)\mid e\in \delta_G(R)}$, where for every edge $e\in \delta_G(R)$, path $P(e)$ has $e$ as its first edge and some edge of $\delta_G(S)$ as its last edge, and all inner vertices of $P(e)$ lie in $V(S)\setminus V(R)$. Additionally, $\cong_G(\pset_1(R))\leq 400/\alpha$. We also require that there is a subset $\hat E_R\subseteq \delta_G(R)$ of at least $\floor{|\delta_G(R)|/64}$ edges, such that $\pset_2(R)=\set{P_2(e)\mid e\in \hat E_R}$, where for every edge $e\in \hat E_R$, path $P(e)$ has $e$ as its first edge and some edge of $\delta_G(S)$ as its last edge, and all inner vertices of $P(e)$ lie in $V(S)\setminus V(R)$.
We denote by $S_j$ the subgraph of $G$ induced by the set $V(S)\setminus\left (\bigcup_{R\in \rset_j}V(R)\right )$ of vertices.
We will ensure that the following ivariants hold:

\begin{properties}{P}
	\item for every cluster $R\in\rset_j$, $|\delta_G(R)|\le |\delta_G(S)|$; \label{prop: small boundary each cluster}

	\item every cluster $R\in\rset_j$ has the $\alpha$-bandwidth property in graph $G$;  \label{prop: bw prop each cluster}

	\item $\sum_{R\in \rset_j}|\delta_G(R)|\le 2|\delta_G(S)|\cdot \sum_{j'=0}^{j-1}\frac{1}{2^{j'}}$; \label{prop: total boundary sum}
	
	\item the congestion caused by the set $\bigcup_{R\in \rset_j}\pset_2(R)$ of paths is at most $400j/\alpha$;  \label{prop: bound on congestion}
	
	\item $|\delta_G(S_j)|\leq |\delta_G(S)|/16^{j-1}$, and there is a set $\qset_j$ of paths in graph $G$, routing the edges of $\delta_G(S_j)$ to edges of $\delta_G(S)$, such that for every path in $\qset_j$, all inner vertices on the path lie in $V(S)\setminus V(S_j)$, and the paths in $\qset_j$ cause congestion at most $2/\alpha$. \label{prop: boundary goes down}
\end{properties}

The algorithm terminates once $\bigcup_{R\in \rset_j}V(R)=V(S)$. 
Notice that, if we ensure that the above properties hold after each phase, the number of phases of the algorithm is $z\leq \ceil{\log m}$ (since $|\delta_G(S_z)|\geq 1$ must hold). Once the algorithm terminates, we return the final set $\rset_z$ of clusters. It is then 
easy to verify that this set of clusters has all required properties.

The input to the first phase is $\rset_1=\emptyset$, so $S_1=S$. The set $\qset_1$ of paths contains, for every edge $e\in \delta_G(S)$, a path $Q(e)$ consisting of the edge $e$ only.
It is easy to verify that all invariants hold for this input.

We now assume that we are given an input $\rset_j$ to phase $j$, for which Properties \ref{prop: small boundary each cluster} -- \ref{prop: boundary goes down} hold. We now describe the algorithm for executing the $j$th phase.

The algorithm consists of two steps. In the first step, we apply the algorithm from \Cref{thm:well_linked_decomposition} to graphs $G$ and $S_j$ (if $S_j$ is not connected, then we apply the algorithm to every connected component of $S_j$). We let $\rset'$ be the set of clusters that the algorithm returns. We start by setting $\rset_{j+1}=\rset_j\cup \rset'$ (but eventually we will discard some clusters from $\rset_{j+1}$ in the second step). Before we continue to the second step, we verify that Invariants \ref{prop: small boundary each cluster} -- \ref{prop: total boundary sum} hold for the current set $\rset_{j+1}$ of clusters.

Recall that, from Invariant \ref{prop: boundary goes down}, $|\delta_G(S_j)|\leq \delta_G(S)/16^{j-1}$. The algorithm from \Cref{thm:well_linked_decomposition} ensures that, for every cluster $R\in \rset'$, $|\delta_G(R)|\leq |\delta_G(S_j)|\leq |\delta_G(S)|$. It also ensures that every cluster $R\in \rset'$ has the $\alpha$-bandwidth property in $G$, and that: 

\[\sum_{R\in \rset'}|\delta_G(R)|\leq 
|\delta_G(S_{j})|\cdot\left(1+O(\alpha\cdot \log^{1.5} m)\right)\leq 2|\delta_G(S_{j})|\leq 2|\delta_G(S)|/16^{j-1}.
\]

(we have used the fact that $\alpha< \frac 1 {c\log^2 m}$ for a large enough constant $c$). Therefore, Invariants \ref{prop: small boundary each cluster} -- \ref{prop: total boundary sum} hold for the current set $\rset_{j+1}$ of clusters.  Notice that currently $V(S)=\bigcup_{R\in \rset_{j+1}}V(R)$ holds. 

We now describe the second step of the algorithm. Our goal is to discard some clusters from $\rset_{j+1}$, and to define the sets $\pset_1(R)$, $\pset_2(R)$ of paths for each cluster that remains in $\rset_{j+1}$, so that  Invariants  \ref{prop: bound on congestion} and \ref{prop: boundary goes down} hold.

In order to do so, we construct a flow network $H$, as follows. We start from graph $G$, and contract all vertices of $V(G)\setminus V(S)$ into a destination vertex $t$. We also contract every cluster $R\in \rset_{j+1}$ into a vertex $u(R)$. Additionally, we add a source vertex $s$. For every cluster $R\in \rset'$, we connect $s$ to vertex $u(R)$ with an edge of capacity $|\delta_G(R)|$. All remaining edges of $H$ have capacity $64$. This completes the definition of the flow network $H$.

Next, we compute a minimum $s$-$t$ cut $(X,Y)$ in $H$. We partition the edges of $E_H(X,Y)$ into two subsets: set $E'$ containing all edges incident to $s$, and set $E''$ containing all remaining edges. Recall that the capacity of every edge in $E''$ is $64$. Clearly, the value of the minimum $s$-$t$ cut in $H$ is at most: 

$$\sum_{R\in \rset'}|\delta_G(R)|\leq 2|\delta_G(S)|/16^{j-1},$$

as we could set $X=\set{s}$ and $Y=V(G)\setminus X$. Therefore, the total capacity of all edges in $E''$ is at most $2|\delta_G(S)|/16^{j-1}$, and $|E''|\leq \frac{2|\delta_G(S)|}{16^{j-1}\cdot 64}\leq \frac{|\delta_G(S)|}{16^j}$.

We discard from set $\rset_{j+1}$ all clusters $R$ with $u(R)\in X$, obtaining the final set $\rset_{j+1}$ of clusters. Clearly, Invariants 
\ref{prop: small boundary each cluster} -- \ref{prop: total boundary sum} continue to hold for this final set of clusters. 
Let $S_{j+1}$ the subgraph of $S$ induced by $V(S)\setminus\left (\bigcup_{R\in \rset_{j+1}}V(R)\right )$. Then $\delta_G(S_{j+1})= E''$, and so $|\delta_G(S_{j+1})|\leq |\delta_G(S)|/16^{j}$.

We now show that there exists  a set $\qset_{j+1}$ of paths in graph $G$, routing the edges of $\delta_G(S_{j+1})$ to edges of $\delta_G(S)$, such that all inner vertices on every path lie in $V(S)\setminus V(S_{j+1})$, and the paths of $\qset_{j+1}$ cause congestion at most $2/\alpha$. In order to do so, we first consider the flow network $H$. Let $\pset^*$ be the set of all paths $P$ in $H$, such that the first edge on $P$ lies in $E''$, the last vertex of $P$ is $t$, and all inner vertices of $P$ lie in $Y$. 
From the maximum flow / minimum cut theorem, there is a flow $f$ in $H$, defined over the set $\pset^*$ of paths, where every edge of $E''$ sends $64$ flow units. Note that the edges of $E'$ (and in particular, all edges incident to $s$) do not carry any flow in $f$. Let $H'$ be the graph obtained from $H$ after we contract all vertices of $X$ into a supernode $s^*$, and delete all edges incident to the orginal source $s$ from this graph. From the above discussion, there is an $s^*$-$t$ flow in the resulting graph, in which every edge of $E''$ carries one flow unit, and all other edges of $H'$ carry at most one flow unit each (the flow is obtained by scaling flow $f$ by factor $64$). Therefore, there is a collection $\qset$ of $|E''|$ edge-disjoint paths in graph $H'$, routing the edges of $E''$ to vertex $t$. Let $\rset^*$ be the set of all clusters $R$ whose corresponding supernode $u(R)$ lies in $Y$. Since every cluster in $\rset^*$ has $\alpha$-bandwidth property, from \Cref{claim: routing in contracted graph}, there is a collection $\qset_{j+1}$ of paths in graph $G$, routing the edges of $E''=\delta_G(S_{j+1})$ to edges of $\delta_G(S)$, with congestion at most $2/\alpha$, such that all inner vertices on every path in $\qset_{j+1}$ lie in $V(S)\setminus V(S_{j+1})$. 
This establishes Property \ref{prop: boundary goes down}.

For every cluster $R\in \rset_{j+1}\cap \rset_j$, we leave the sets $\pset_1(R)$ and $\pset_2(R)$ of paths unchanged. This ensures that the congestion caused by the set $\pset_1(R)$ of paths is at most $400/\alpha$, and that the total congestion caused by the set $\bigcup_{R\in  \rset_{j+1}\cap \rset_j}\pset_2(R)$ of paths is at most $400j/\alpha$. Next, we define the sets $\pset_1(R)$ and $\pset_2(R)$ of paths for clusters $R\in \rset_{j+1}\setminus \rset_j$.

Consider some cluster $R\in \rset_{j+1}\setminus \rset_j$, and recal that $\rset_{j+1}\setminus \rset_j\subseteq \rset'$.
Recall that the algorithm from  \Cref{thm:well_linked_decomposition} returned a set $\pset'(R)=\set{P'(e)\mid e\in \delta_G(R)}$ of paths in graph $G$ with $\cong_G(\pset'(R))\leq 100$, such that, for every edge $e\in \delta_G(R)$, path $P'(e)$ has $e$ as its first edge and some edge of $\delta_G(S_j)$ as its last edge, and all inner vertices of $P'(e)$ lie in $V(S_j)\setminus V(R)$. We combine these paths with the set $\qset_j$ of paths given by Invariant \ref{prop: boundary goes down} to obtain the desired set $\pset_1(R)=\set{P_1(e)\mid e\in \delta_G(R)}$ of paths, where for every edge $e\in \delta_G(R)$, path $P(e)$ has $e$ as its first edge and some edge of $\delta_G(S)$ as its last edge, and all inner vertices of $P(e)$ lie in $V(S)\setminus V(R)$. Since $\cong_G(\pset'(R))\leq 100$, while $\cong_G(\qset_j)\leq 2/\alpha$, it is easy to verify that $\cong_G(\pset_1(R))\leq 400/\alpha$.

It now remains to define the sets $\pset_2(R)$ of paths for all clusters $R\in \rset_{j+1}\setminus \rset_j$. In order to do so, we
consider again the flow network $H$. Recall that for every cluster $R\in \rset_{j+1}\setminus \rset_j$, $u(R)\in Y$ holds, and moreover, there is an edge $(s,u(R))$ of capacity $|\delta_G(R)|$, that belongs to $E'\subseteq E_H(X,Y)$. 
Let $\pset^{**}$ be the set of all paths $P$ that connect $s$ to $t$ in $H$, such that the first edge on $P$ lies in $E'$.
From the maximum flow/minimum cut theorem, there is a flow $f$ in $H$ over the set $\pset^{**}$ of paths, in which every edge $e=(s,u(R))\in E'$ sends $|\delta_G(R)|$ flow units (the capacity of the edge $e$). Scaling this flow down by factor $64$, using the integrality of flow, and deleting the first edge from every flow-path, we obtain a collection $\qset'$ of edge-disjoint paths in graph $H$, such that, for every cluster $R\in \rset_{j+1}\setminus \rset_j$, at least $\floor{|\delta_G(R)|/64}$
paths in $\qset'$ originate at edges of $\delta_H(u(R))$, and all paths in $\qset'$ terminate at vertex $t$. As before, we use the algorithm from \Cref{claim: routing in contracted graph} in order to obtain a collection $\qset''$ of paths in graph $G$, such that, for every cluster $R\in \rset_{j+1}\setminus\rset_j$, there is a subset $\qset''(R)\subseteq \qset''$ of at least $\floor{|\delta_G(R)|/64}$ paths that originate at edges of $\delta_G(R)$, and all paths in $\qset''(R)$ terminate at edges of $\delta_G(S)$. 
The algorithm ensures that, for every edge $e\in \bigcup_{R\in \rset_{j+1}\setminus\rset_j}\delta_G(R)$, at most one path of $\qset''$ uses $e$, and the total congestion caused by the paths of $\qset''$ is at most $2/\alpha$. Consider now a cluster $R\in \rset_{j+1}\setminus\rset_j$, and the corresponding set $\qset''(R)$ of paths. Let $Q\in \qset''(R)$ be any such path. Observe that path $Q$ may not be internally disjoint from $R$. We let $e$ be the last edge on $Q$ that belongs to $\delta_G(R)$, and we truncate path $Q$, so that it now originates at edge $e$ and terminates at some edge of $\delta_G(S)$. This ensures that path $Q$ is internally disjoint from $R$. We let $\pset_2(R)$ be  the resulting set of paths, obtained after every path of $\qset''(R)$ was processed. From the above discussion, the set $\pset_2(R)$ routes a subset $\hat E_R\subseteq \delta_G(R)$ of at least $\floor{|\delta_G(R)|/64}$ edges to edges of $\delta_G(S)$; all paths in $\pset_2(R)$ are internally disjoint from $R$; and the total congestion caused by the paths in $\bigcup_{R\in \rset_{j+1}\setminus \rset_j}\pset_2(R)$ is at most $2/\alpha$. Altogether, the paths in $\bigcup_{R\in \rset_{j+1}}\pset_2(R)$ cause congestion at most $400(j+1)/\alpha$, establising Property \ref{prop: bound on congestion}.

\subsection{Proof of \Cref{claim: avoid guiding curves}}
\label{subsec: proof of claim avoind guiding curves}

Consider a pair of indices $0\leq j\leq r$ and $0\leq a<2^{r-j}$, and recall that there is a level-$j$ curve $\lambda_{j,a}\in \Lambda_j$ connecting point $p_{a\cdot 2^j}$ to point $p_{(a+1)\cdot 2^j}$. Recall that we have defined a segment $\sigma_{j,a}$ of the boundary of disc $D$, whose endpoints are $p_{a\cdot 2^j}$ and $p_{(a+1)\cdot 2^j}$, where $\sigma_{j,a}$ does not contain the point $p_{(a+1)\cdot 2^j+1}$. It will be convenient for us to view the segment $\sigma_{j,a}$ as closed on one side and open on another side, that is, $p_{a\cdot 2^j}\in \sigma_{j,a}$, and $p_{(a+1)\cdot 2^j}\not\in \sigma_{j,a}$. We let $T_a^j$ be the set of all anchor vertices whose images lie on the curve $\sigma_{j,a}$.

Notice that, for each level $j$, the collections $\set{T_a^j\mid 0\leq a<2^{r-j}}$ of vertices are disjoint from each other. For every pair $0\leq j<j'\leq r$ of levels and indices $0\leq a< 2^{r-j}$ and $0\leq a'< 2^{r-j'}$, either $\sigma_{j,a}\cap \sigma_{j',a'}=\emptyset$, or $\sigma_{j,a}\subseteq \sigma_{j',a'}$. In the former case, $T_a^j\cap T_{a'}^{j'}=\emptyset$, while in the latter case, $T_a^j\subseteq T_{a'}^{j'}$. 

For all $0\leq j\leq r$ and $0\leq a<2^{r-j}$, we let $X^j_a$ be the subset of vertices of $G'$ with the following properties:

\begin{itemize}
	\item $X^j_a\cap A=T^j_a$;
	\item $|\delta_{G'}(X^j_a)|$ is minimized among all sets $X^j_a$ with the above property; and
	\item $|X^j_a|$ is minimized among all sets $X^j_a$ with the above two properties.
\end{itemize}

In other words, we let $(X^j_a,V(G')\setminus X^j_a)$ be a minimum cut separating vertices of $T^j_a$ from the remaining vertices of $A$, that minimizes the number of vertices in $X^j_a$. Note that, from \Cref{obs: small boundary cuts}, $|\delta_{G'}(X^j_a)|\leq 4\cm'/\mu^{2b}$.

The following simple observation follows immediately from submodularity of cuts.

\begin{observation}\label{obs: laminar}
	For every pair $0\leq j\leq j'\leq r$ of levels and indices $0\leq a< 2^{r-j}$ and $0\leq a'< 2^{r-j'}$, if $T_a^j\cap T_{a'}^{j'}=\emptyset$ then  $X_a^j\cap X_{a'}^{j'}=\emptyset$, and if $T_a^j\subseteq T_{a'}^{j'}$, then $X_a^j\subseteq X_{a'}^{j'}$.
	\end{observation}

\begin{proof}
	Consider a pair  $0\leq j\leq j'\leq r$ of levels, and indices $0\leq a< 2^{r-j}$ and $0\leq a'< 2^{r-j'}$. Assume first that $T_a^j\cap T_{a'}^{j'}=\emptyset$, but $X_a^j\cap X_{a'}^{j'}\neq \emptyset$. Let $Y=X_a^j\setminus X_{a'}^{j'}$ and $Y'=X_{a'}^{j'}\setminus X_{a}^{j}$. Since $X_a^j\cap A=T_a^j$ and $X_{a'}^{j'}\cap A=T_{a'}^{j'}$, we get that $Y\cap A=T_a^j$ and $Y'\cap A=T_{a'}^{j'}$. Since $X_a^j\cap X_{a'}^{j'}\neq \emptyset$, $|Y|< |X_{a}^j|$ and $|Y'|<|X_{a'}^{j'}|$ holds. Lastly, from submodularity of cuts:
	\[|\delta_{G'}(Y)|+|\delta_{G'}(Y')|\leq |\delta_{G'}(X_{a}^j)|+ |\delta_{G'}(X_{a'}^{j'})|. \]
	Since $|\delta_{G'}(X_{a}^j)|$ minimizes the number of edges in a cut separating the vertices of $T_a^j$ from the remaining vertices of $A$, and similarly $|\delta_{G'}(X_{a'}^{j'})|$ minimizes the number of edges in a cut separating the vertices of $T_{a'}^{j'}$ from the remaining vertices of $A$, $|\delta_{G'}(Y)|=|\delta_{G'}(X_{a}^j)|$ and $|\delta_{G'}(Y')|\leq |\delta_{G'}(X_{a'}^{j'})|$ must hold, a contradiction.
	
	Assume now that $T_a^j\subseteq T_{a'}^{j'}$, but $X_a^j\not\subseteq X_{a'}^{j'}$.
	Let $Y=X_a^j\cap X_{a'}^{j'}$, and let $Y'=X_{a'}^{j'}\cup X_{a}^{j}$. It is immediate to verify that $Y\cap A=T_a^j$, $Y'\cap A=T_{a'}^{j'}$, and $|Y|<|X_{a}^{j}|$. From submodularity of cuts:
 	\[|\delta_{G'}(Y)|+|\delta_{G'}(Y')|\leq |\delta_{G'}(X_{a}^j)|+ |\delta_{G'}(X_{a'}^{j'})|. \] 
 	Using the same argument as before, $|\delta_{G'}(Y)|=|\delta_{G'}(X_{a}^j)|$ and $|\delta_{G'}(Y')|= |\delta_{G'}(X_{a'}^{j'})|$ must hold. This contradicts the minimality of the cut $X_{a}^{j}$, as $|Y|< |X_{a}^{j}|$.
\end{proof}

We denote, for all $0\leq j\leq r$, $\xset^j=\set{X^j_a\mid 0\leq a<2^{r-j}}$, and $\xset=\bigcup_{j=0}^r\xset^j$. For simplicity, we will refer to the sets of vertices in $\xset$ as \emph{clusters} (each such vertex set $X^j_a$ indeed naturally defines a cluster $G'[X^j_a]$ of graph $G'$). Note that the set $\xset$ of clusters is laminar.

We can naturally associate a partitioning tree $\tau$ with the set $\xset$ of clusters. The set of vertices of the tree $\tau$ is $\set{u(X)\mid X\in \xset\cup \set{V(G')}}$. The root of the tree is vertex $u(X)$ where $X=V(G')$. This vertex has one child vertex -- $u(X^r_0)$, corresponding to the unique cluster in $\xset^r$. For every non-root vertex $u(X^j_a)$, there are exactly two level-$(j-1)$ clusters that are contained in $X^j_a$: clusters $X^{j-1}_{a'}$ and $X^{j-1}_{a''}$, where $a'=2a$ and $a''=2a+1$. Vertices $u(X^{j-1}_{a'})$ and $u(X^{j-1}_{a''})$ become child-vertices of $u(X^j_a)$ in the tree; we refer to clusters $X^{j-1}_{a'}$ and $X^{j-1}_{a''}$ as \emph{child-clusters} of $X^j_a$, where $X^{j-1}_{a'}$ is the left child and $X^{j-1}_{a''}$ is the right child. We also say that $X^j_a$ is a \emph{parent-cluster} of $X^{j-1}_{a'}$ and $X^{j-1}_{a''}$. The leaves of the tree $\tau$ are vertices in set $\set{u(X)\mid X\in \xset^0}$.

It will be convenient for us to subdivide some of the edges of $G'$, in order to ensure the following two properties:

\begin{properties}{P}
	\item for every cluster $X\in \xset$,  if $e=(x,y)\in \delta_{G'}(X)$, with $x\in X$, then vertex $y$ lies in the parent-cluster of $X$, and neither $x$ nor $y$ are anchor vertices;  \label{prop: subdivision1}

	\item for every cluster $X\in \xset$, if $X'$ and $X''$ are the two child-clusters of $X$, and we denote $Y=X\setminus (X'\cup X'')$, then for every pair $e,e'\in \delta_{G'}(Y)$ of edges, the two edges $e,e'$ do not share endpoints. \label{prop: subdivision2}

\end{properties}

In order to achieve the above properties, we will subdivide some edges of $G'$, and we will update the clusters in $\xset$ accordingly. We will ensure that the clusters remain laminar, and that, for all $0\leq j\leq r$ and $0\leq a<2^{r-j}$,  $X^j_a$ remains the smallest cut separating the vertices of $T^j_a$ from the remaining vertices of $A$, with $|\delta_{G'}(X^j_a)|$ remaining unchanged.

In order to perform this transformation, we process the clusters of $\xset$ in sets $\xset^0,\xset^1,\ldots,\xset^r$ in this order of the sets.

Consider an iteration when some cluster $X^j_a$ is processed, for $0\leq j\leq r$ and $0\leq a<2^{r-j}$. Consider any edge $e=(x,y)\in \delta_{G'}(X^j_a)$, and assume that $x\in X^j_a$. We subdivide edge $e$ with two new vertices, replacing it with a path $(x,t_e,t'_e,y)$. We add vertex $t_e$ to both $X^j_a$ and all its ancestor clusters, and we add vertex $t'_e$ to all ancestor clusters of $X^j_a$ (but not to $X^j_a$). Notice that, after this subdivision step, the updated set $\xset$ of clusters remains laminar, and for every cluster $X\in \xset$, $|\delta_{G'}(X)|$ does not grow. Once we process every edge $e\in \delta_{G'}(X^j_a)$, we complete the processing of cluster $X^j_a$. Once every cluster in $\xset$ is processed, we obtain an updated graph $G'$, with the updated family $\xset$ of clusters, for which properties \ref{prop: subdivision1} and \ref{prop: subdivision2} hold.
We update the input drawing $\phi$ of graph $G'$ by subdiving the images of its edges appropriately, to obtain a drawing of the current graph $G'$, that we also denote by $\phi$. Note that for all $0\leq j\leq r$ and $0\leq a<2^{r-j}$, every edge $e$ in the current set $\delta_{G'}(X^j_a)$ is obtained by subdiving some edge in the original graph $G'$, and both endpoints of $e$ are new vertices that were used for the subdivision. We can then place the images of the endpoints of $e$ close enough to each other, so that the image of the edge $e$ does not participate in any crossings in the new drawing $\phi$. We will assume from now on that for all $0\leq j\leq r$ and $0\leq a<2^{r-j}$, the edges of $\delta_{G'}(X^j_a)$  do not participate in crossings in $\psi$.

For all $0\leq j\leq r$ and $0\leq a<2^{r-j}$, we define a graph $H_{j,a}$ associated with cluster $X^j_a$, as follows. If $j=0$, then $H_{j,a}=G'[X^j_a]$. Otherwise, let $X^{j-1}_{a'},X^{j-1}_{a''}$ be the two child clusters of cluster $X^j_a$, where $X^{j-1}_{a'}$ is the left child cluster. We let $H_{j,a}$ be the subgraph of $G'$ induced by vertex set $X^j_a\setminus \left (X^{j-1}_{a'}\cup X^{j-1}_{a''}  \right )$. We also define three subsets of vertices of $H_{j,a}$: set $T^{\parent}_{j,a}$ contains all vertices of $H_{j,a}$ that serve as endpoints of the edges of $\delta_{G'}(X^j_a)$; set $T^{\lchild}_{j,a}$ contains all vertices of $H_{j,a}$ that serve as endpoints of the edges of $\delta_{G'}(X^{j-1}_{a'})$; and set $T^{\rchild}_{j,a}$ contains all vertices of $H_{j,a}$ that serve as endpoints of the edges of $\delta_{G'}(X^{j-1}_{a''})$. From Property 
\ref{prop: subdivision2}, these three sets of vertices are mutually disjoint. 
We denote by $T(H_{j,a})=T^{\parent}_{j,a}\cup T^{\lchild}_{j,a}\cup T^{\rchild}_{j,a}$.

For $j=0$, for all $0\leq a< 2^r$, we define the set $T^{\parent}_{j,a}$ of vertices of graph $H_{j,a}$ similarly. We do not define the sets $T^{\lchild}_{j,a},T^{\rchild}_{j,a}$ of vertices, but instead we use the set $T^j_a$ of anchor vertices that we defined already. From Property \ref{prop: subdivision1}, vertex sets $T^{\parent}_{j,a}$ and $T^j_a$ are disjoint. 
We denote $T(H_{j,a})=T^{\parent}_{j,a}\cup T^j_a$.

Lastly, we define a graph $H^*$ -- a subgraph of $G'$ induced by vertex set $V(G')\setminus X^r_0$. We let $T^*$ be the set of all anchor vertices in $A\setminus T^r_0$, and we let $T^{**}$ be the set of all vertices of $H^*$ that serve as endpoints of the edges of $\delta_{G'}(X^r_0)$. We denote $T(H^*)=T^*\cup T^{**}$.

Let $\hset=\set{H^*}\cup \set{H_{j,a}\mid 0\leq j\leq r, 0\leq a<2^{r-j}}$ be the resulting collection of subgraphs of $G'$. Note that the graphs in $\hset$ are mutually disjoint from each other and $\bigcup_{H\in \hset}V(H)=V(G')$.

Next, for every graph $H\in \hset$, we will define a disc $D(H)$, and we will also define an ordering of the vertices in $T(H)$. We will then modify the current drawing $\phi$ of graph $G'$, so that, for every graph $H\in \hset$, the image of $H$ lies in disc $D(H)$, with the vertices of $T(H)$ lying on the disc boundary, in the pre-specified order. We will ensure that, for all $0\leq j\leq r$ and $0\leq a< 2^{r-j}$, the only edges whose images cross the curve $\lambda^j_a$ are the edges of $\delta_{G'}(X^j_a)$. Since, as observed above, $|\delta_{G'}(X^j_a)|\leq 4\cm'/\mu^{2b}$, this will ensure that at most $4\cm'/\mu^{2b}$ edges cross each curve $\lambda\in \Lambda$ in the final drawing.
We now consider every graph $H\in \hset$ in turn, define the corresponding disc $D(H)$, and the ordering of the vertices of $T(H)$.

Consider some pair of indices $0\leq j< r$ and $0\leq a< 2^{r-j}$. Recall that $(X^j_a,V(G')\setminus X^j_a)$ is a minimum cut in the current graph $G'$ separating vertices of $T^j_a$ from the remaining vertices of $A$. Therefore, there is a set $\qset_{j,a}=\set{Q_{j,a}(e)\mid e\in \delta_{G'}(X^j_a)}$ of edge-disjoint paths, that are internally disjoint from $X^j_a$, such that, for every edge $e\in \delta_{G'}(X^j_a)$, path $Q_{j,a}(e)$ originates at edge $e$, and terminates at some vertex of $A\setminus T^j_a$. Similarly, there is a set $\qset'_{j,a}=\set{Q'_{j,a}(e)\mid e\in \delta_{G'}(X^j_a)}$ of edge-disjoint paths, whose inner vertices are contained in $X^j_a$, such that, for every edge $e\in \delta_{G'}(X^j_a)$, path $Q'_{j,a}(e)$ originates at edge $e$, and terminates at some vertex of $T^j_a$. From \Cref{lem: non_interfering_paths}, we can assume w.l.o.g. that the paths in set $\qset_{j,a}$ are non-transversal with respect to the rotation system $\Sigma'$, and the same is true regarding the paths in $\qset'_{j,a}$. We define an oriented ordering $\oset_{j,a}$ of the edges in set $\delta_{G'}(X^j_a)$, as follows. For every edge $e\in \delta_{G'}(X^j_a)$, let $v_e$ be the last vertex on path $Q'_{j,a}(e)$, that must lie in $T^j_a$. From our construction, it is easy to verify that every vertex of $A$ has degree $1$ in $G'$, so all vertices in set $\set{v_e\mid e\in \delta_{G'}(X^j_a)}$ are distinct. We define the oriented ordering $\oset_{j,a}$ of the edges of $\delta_{G'}(X^j_a)$ to be the order in which their corresponding vertices $v_e$ are encountered along the boundary of the disc $D$, as we traverse it in counter-clock-wise direction. We use this ordering in order to define an oriented ordering $\oset(T^{\parent}_{j,a})$ of the set $T^{\parent}_{j,a}$ of vertices of graph $H^j_a$: recall that the vertices of $T^{\parent}_{j,a}$ are the endpoints of the edges of $\delta_{G'}(X^j_a)$ that lie in $X^j_a$, and every edge in $\delta_{G'}(X^j_a)$ is incident to a distinct vertex of $T^{\parent}_{j,a}$. We let the oriented ordering $\oset(T^{\parent}_{j,a})$  of the vertices of $T^{\parent}_{j,a}$  be identical to the oriented ordering $\oset_{j,a}$ of the edges of $\delta_{G'}(X^j_a)$, except that we reverse the orientation. In other words, in order to obtain the ordering $\oset(T^{\parent}_{j,a})$, we replace, in ordering $\oset_{j,a}$ every edge $e\in \delta_{G'}(X^j_a)$ with its endpoint that lies in $T^{\parent}_{j,a}$, and then reverse the orientation of the resulting ordering. 

Assume now that cluster $X^j_a$ is the child cluster of some other cluster $X^{j'}_{a'}$. We assume w.l.o.g. that it is the left child cluster; the other case is dealt with similarly. Recall that the set $T^{\lchild}_{j',a'}$ of vertices contains all endpoints of the edges of $\delta_{G'}(X^j_a)$ that lie in $H_{j',a'}$. We define an ordering $\oset(T^{\lchild}_{j',a'})$ of the vertices of $T^{\lchild}_{j',a'}$ to be identical to the oriented ordering $\oset_{j,a}$ of the edges of $\delta_{G'}(X^j_a)$. In other words, in order to obtain the ordering $\oset(T^{\lchild}_{j',a'})$, we replace, in ordering $\oset_{j,a}$ every edge $e\in \delta_{G'}(X^j_a)$ with its endpoint that lies in $T^{\lchild}_{j',a'}$. If $j=r$ and $a=0$, then cluster $X^r_0$ is the child cluster of $V(G')$. The latter cluster, in turn, corresponds to graph $H^*$. In this case, the set $T^{**}$ of vertices of $H^*$ contains all endpoints of the edges of $\delta_{G'}(H_{r,0})$ that lie in $V(H^*)$. We define an ordering $\oset(T^{**})$ of the vertices of $T^{**}$ similarly: it is identical to the ordering  $\oset_{r,0}$ of the edges of $\delta_{G'}(X^r_0)$.

\begin{figure}[h]
	\centering
	\subfigure[Disc $D(H*)$.]{
		\scalebox{0.32}{\includegraphics[scale=0.25]{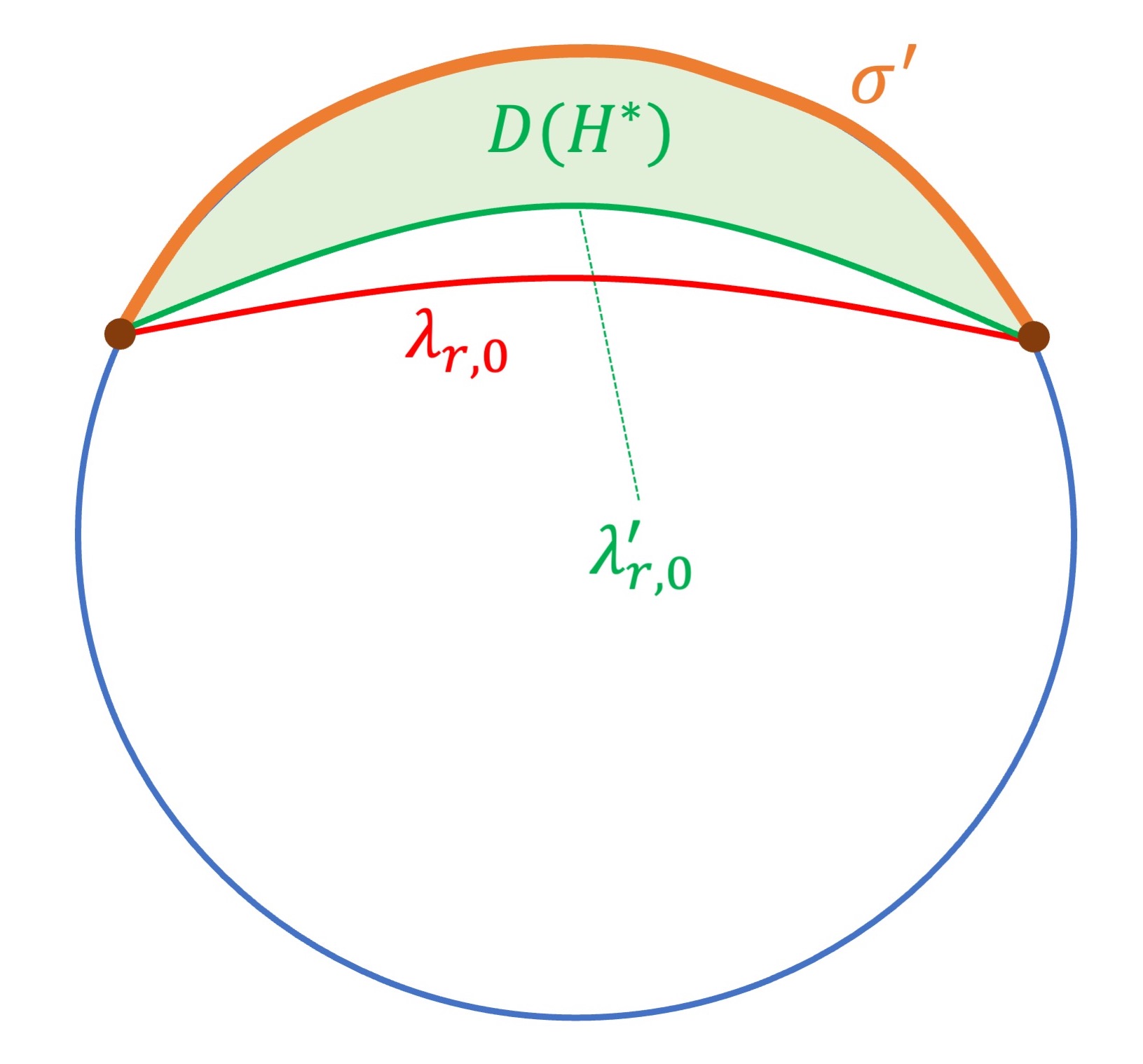}}\label{fig: discD_0}}
		\hspace{0.2cm}
	\subfigure[Disc $D(H_{j,a})$ for $0<j\le r$.]{
		\scalebox{0.32}{\includegraphics[scale=0.25]{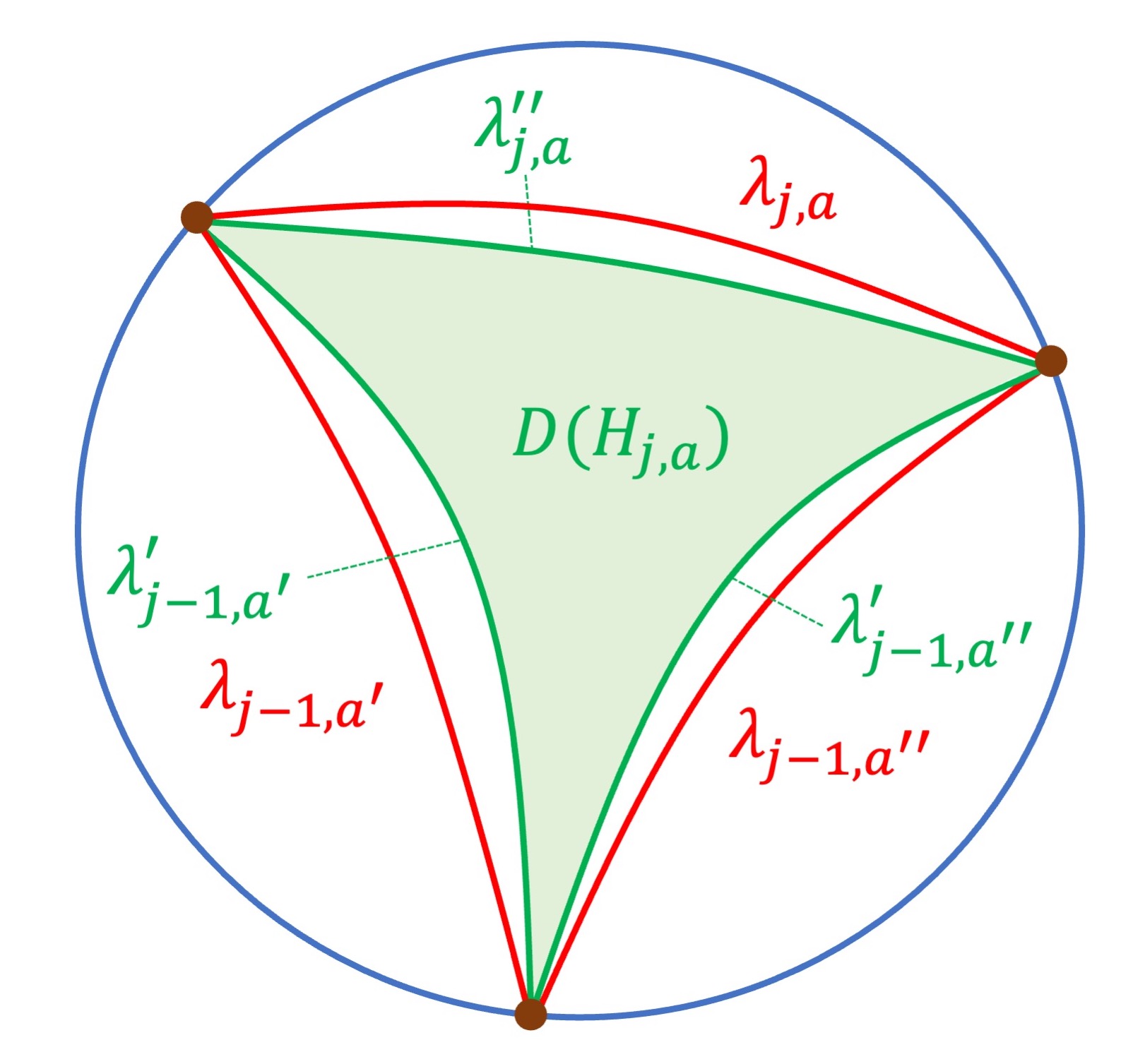}}\label{fig: discD_1}}
	\hspace{0.2cm}
	\subfigure[Disc $D(H_{0,a})$.]{\scalebox{0.32}	{\includegraphics[scale=0.25]{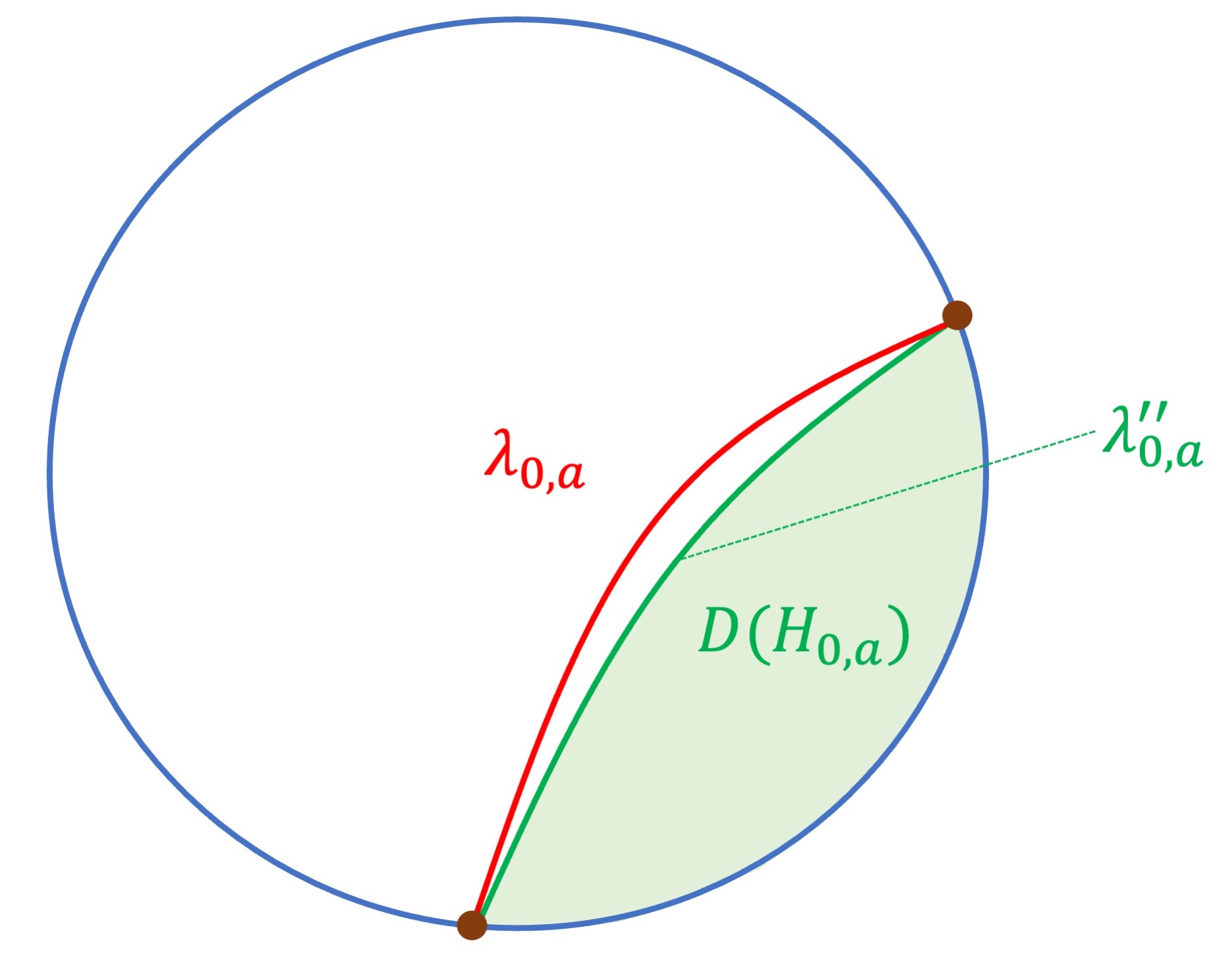}}\label{fig: discD_2}
	}
	\caption{Discs $D(H)$ for graphs $H\in \hset$}\label{fig: discs}
\end{figure}

Next, we define, for every graph $H\in \hset$, a corresponding disc $D(H)$. Consider first the graph $H^*$. Let $\lambda'_{r,0}$ be a curve that has the same endpoints as $\lambda_{r,0}$, is internally disjoint from $\lambda_{r,0}$, and follows the curve $\lambda_{r,0}$ closely outside the disc $D^r_0$. Let $\sigma'$ be the segment of the boundary of disc $D$ that connects the two endpoints of $\lambda_{r,0}$, and is internally disjoint from the boundary of disc $D^r_0$. We let $D(H^*)$ be the disc that is contained in $D$, whose boundary is the concatenation of the curves $\lambda'_{r,0}$ and $\sigma'$ (see \Cref{fig: discD_0}).

Consider now indices $0<j\leq r$ and $0\leq a<2^{r-j}$, and the graph $H_{j,a}$. Let $X_{j-1,a'}$ and $X_{j-1,a''}$ be the left and the right child clusters of $X_{j,a}$, respectively. We let $\lambda''_{j,a}$ be a curve whose endpoints are the same as those of $\lambda_{j,a}$, so that $\lambda''_{j,a}$  follows the curve  $\lambda_{j,a}$ closely inside disc $D^j_a$. We let $\lambda'_{j-1,a'}$ be a curve whose endpoints are the same as those of $\lambda_{j-1,a'}$, so that $\lambda'_{j-1,a'}$ follows the curve $\lambda_{j-1,a'}$ closely, and is internally disjoint from disc $D^{j-1}_{a'}$. Similarly, we let $\lambda'_{j-1,a''}$ be a curve whose endpoints are the same as those of $\lambda_{j-1,a''}$, so that $\lambda'_{j-1,a''}$ follows the curve $\lambda_{j-1,a''}$ closely, and is internally disjoint from disc $D^{j-1}_{a''}$.
The concatenation of the curves $\lambda''_{j,a}, \lambda'_{j-1,a'}$ and $\lambda'_{j-1,a''}$ is a simple closed curve that is contained in disc $D^j_a$. We let $D(H^j_a)$ be the disc that is contained in $D$, whose boundary is the concatenation of $\lambda''_{j,a}, \lambda'_{j-1,a'}$ and $\lambda'_{j-1,a''}$. 
(see \Cref{fig: discD_1}).

Lastly, we consider the index $j=0$, and any index $0\leq a<2^r$. 
We let $\lambda''_{0,a}$ be a curve whose endpoints are the same as those of $\lambda_{0,a}$, so that $\lambda''_{0,a}$  follows the curve  $\lambda_{0,a}$ closely inside disc $D^0_a$. Recall that $\sigma^0_a$ is a segment of the boundary of disc $D$, whose endpoints are the same as those of $\lambda_{0,a}$, with point $p_{a+1}$ not lying on $\sigma^0_a$. We let $D(H_{0,a})$ be the disc that is contained in $D$, whose boundary is the concatenation of $\lambda''_{0,a}$ and $\sigma^0_a$
(see \Cref{fig: discD_2}). 

We note that every anchor vertex in $A$ must belong to one of the graphs in $\set{H^*}\cup\set{H_{0,a}\mid 0\leq a< 2^r}$.
For every graph $H\in \hset$, we define a set $\chi(H)$ of crossings as follows. For $H=H^*$, $\chi(H)$ contains all crossings in the drawing $\phi$. For a pair of indices $0\leq j\leq r$ and $0\leq a< 2^{r-j}$, $\chi(H_{j,a})$ is the set of all crossings in $\phi$ in which the edges of $G'[X^j_a]$ participate.
The following claim is central to the proof of \Cref{claim: avoid guiding curves}.

\begin{claim}\label{claim: move to discs2}
	Consider any graph $H\in \hset$,  let $\Sigma_H$ be the rotation system for $H$ induced by $\Sigma'$, and let $I_H=(H,\Sigma_H)$ be the resulting instance of \cnwrs. There is a solution $\psi(H)$ to instance $I_H$ with $\cro(\psi(H))\leq O(|\chi(H)|)$, where the image of the graph $H$ is contained in disc $D(H)$. Moreover, the following  hold:
	
	\begin{itemize}
		\item If $H=H^*$, then the images of the vertices of $T^*=V(H^*)\cap A$ appear on segment $\sigma'$ of the boundary of $D(H^*)$, in the same locations as in $\phi$, and the images of the vertices of $T^{**}$ appear on segment $\lambda'_{r,0}$ of the boundary of $D(H^*)$, in the same order as in  $\oset(T^{**})$, incuding the orientation (that is defined with respect to disc $D(H^*)$).
		
		\item If $H=H_{j,a}$ for $j=0$, then the images of the vertices of $T^j_a=V(H_{j,a})\cap A$ appear on the segment $\sigma^j_a$ of the boundary of disc $D(H_{j,a})$, in the same locations as in $\phi$, and the images of the vertices of $T^{\parent}_{j,a}$ appear on the segment $\lambda''_{j,a}$ of the boundary of disc $D(H_{j,a})$, in the same order as in $\oset(T^{\parent}_{j,a})$, including the orientation (that is defined with respect  disc $D(H_{j,a})$).
		
		\item If $H=H_{j,a}$ for $j>0$, then the images of the vertices of $T^{\parent}_{j,a}$ appear on the segment $\lambda''_{j,a}$ of the boundary of disc $D(H_{j,a})$, in the same order as in $\oset(T^{\parent}_{j,a})$, including the orientation, and similarly, the images of the vertices in sets $T^{\lchild}_{j,a}$ and $T^{\rchild}_{j,a}$  appear on the segments $\lambda'_{j-1,a'}$ and $\lambda'_{j-1,a''}$ of the boundary of the disc $D(H_{j,a})$, respectively, where $X^{j-1}_{a'}$ is the left child of $X^{j}_{a}$ and $X^{j-1}_{a''}$ is its right child. The ordering of the images of the vertices of $T^{\lchild}_{j,a}$  on  $\lambda'_{j-1,a'}$ is identical to $\oset(T^{\lchild}_{j,a})$, including orientation, and the ordering of the images of the vertices of $T^{\rchild}_{j,a}$  on  $\lambda'_{j-1,a''}$ is identical to $\oset(T^{\rchild}_{j,a})$, including orientation. The orientations of all orderings are with respect to disc $D(H_{j,a})$.
	\end{itemize}
\end{claim}

We provide the proof of \Cref{claim: move to discs2} in the following subsection, after we complete the proof of \Cref{claim: avoid guiding curves} using it.

In order to construct the solution $\psi'$ to instance $I'$ we start by planting, for every graph $H\in \hset$, the image $\psi(H)$ of $H$ into the disc $D(H)$. From \Cref{claim: move to discs2}, and since every vertex of $A$ lies in either $T^*$ or in $\bigcup_{a=0}^{2^r-1}T^0_a$, the images of the vertices of $A$ remain the same as in $\phi$. In order to complete the drawing of graph $G'$, we need to insert the edges of $\delta_{G'}(X^j_a)$ for all $0\leq j\leq r$ and $0\leq a<2^{r-j}$ into the current drawing.
Observe that the endpoints of all such edges have degree $2$ in $G'$ from our construction of graph $G'$.

We now fix an index $0\leq j< r$ and $0\leq a<2^{r-j}$. Assume that cluster $X^j_a$ is a child cluster of some cluster $X^{j+1}_{a'}$, and assume w.l.o.g. that it is a left child cluster (the other case is dealt with similarly). Consider the set $E'=\delta_{G'}(X^j_a)$ of edges. Recall that these edges define a perfect matching between the sets $T^{\lchild}_{j+1,a'}$ and $T^{\parent}_{j,a}$ of vertices. The images of the vertices of $T^{\lchild}_{j+1,a'}$ appear on curve $\lambda'_{j,a}$, while the images of the vertices of $T^{\parent}_{j,a}$ appear on curve $\lambda''_{j,a}$. Let $D^*_{j,a}$ be the disc that is contained in $D$, whose boundary is the concatenation of the curves $\lambda'_{j,a}$ and $\lambda''_{j,a}$. Denote $E'=\set{e_1,e_2,\ldots,e_q}$, where the edges are indexed according to the ordering $\oset_{j,a}$. For all $1\leq i\leq q$, let $e_i=(x_i,y_i)$, where $x_i\in T^{\lchild}_{j+1,a'}$ and $y_i\in T^{\parent}_{j,a}$. Then the images of vertices $x_1,\ldots,x_q$ appear on curve $\lambda'_{j,a}$ in the order of their indices, and the images of vertices $y_1,\ldots,y_q$ appear on curve $\lambda''_{j,a}$ in the order of their indices, but the orientations of the two orderings are different (the orientation of the first ordering is with respect to $D(H_{j+1,a'})$, while the orientation of the second ordering is with respect to $D(H_{j,a})$). Therefore, the images of vertices $x_1,x_2,\ldots,x_q,y_q,\ldots,y_1$ appear on the boundary of disc $D^*_{j,a}$ in this circular order. We can then define, for all $1\leq i\leq q$, a curve $\gamma_i$ that is contained in disc $D^*_{j,a}$ connecting the image of $x_i$ to the image of $y_i$. We can ensure that no two such curves cross each other, and each curve crosses $\lambda_{j,a}$ exactly once. We then let, for all $1\leq i\leq q$, curve $\gamma_i$ be the image of edge $e_i$. Since, as observed above, $|E'|=|\delta_{G'}(X^j_a)|\leq 4\cm'/\mu^{2b}$, we introduce at most   $4\cm'/\mu^{2b}$ crossings between images of edges of $G'$ and curve $\lambda_{j,a}$.
It now only remains to take care of edge set $\delta_{G'}(X^r_0)$. We insert these edges exactly as before. The only difference is that vertex set $T^{\lchild}_{j+1,a}$ is replaced with $T^{**}$.

We have now obtained a solution $\psi'$ to instance $I'$, in which the images of all vertices and edges of $G'$ lie in disc $D$, and the images of all anchor vertices remain the same as in $\phi$. For every curve $\lambda\in \Lambda$, for every vertex $v\in V(G')$, the image of $v$ in $\psi'$ does not lie on an inner point of $\lambda$, and for every edge $e\in E(G')$, the image of $e$ in $\psi'$ may intersect $\lambda$ in at most one point. For every curve $\lambda\in \Lambda$,  the total number of edges in $E(G')$ whose images intersect $\lambda$ is at most $4\cm'/\mu^{2b}$.

It now remains to bound the number of crossings in $\psi'$. Since the insertion of the edges of $\delta_{G'}(X^j_a)$ for all $0\leq j\leq r$ and $0\leq a<2^{r-j}$ did not introduce any crossings, from  \Cref{claim: move to discs2}, $\cro(\psi')\leq \sum_{H\in \hset}O(|\chi(H)|)$.
Recall  $|\chi(H^*)|=\cro(\phi)$, and, for all $0\leq j\leq r$ and $0\leq a< 2^{r-j}$, $|\chi(H^j_a)|$ is number of  crossings in $\phi$ in which the edges of $G'[X^j_a]$ participate.
 Observe that  vertex sets in $\set{X^j_a \mid 0\leq j\leq r, 0\leq a< 2^{r-j}}$ define a laminar family of depth $O(\log \cm')$. Therefore, every edge $e$ may belong to at most $O(\log \cm')$ graphs in set $\set{G'[X^j_a] \mid 0\leq j\leq r, 0\leq a< 2^{r-j}}$. We conclude that every crossing $(e,e')_p$ of $\phi$ may belong to at most $O(\log \cm')$ sets $\set{\chi(H)\mid H\in \hset}$. Therefore, overall, $\cro(\psi')\leq \sum_{H\in \hset}O(|\chi(H)|)\leq \cro(\phi)\cdot O (\log \cm')$.
 In order to complete the proof of \Cref{claim: avoid guiding curves}, it remains to prove \Cref{claim: move to discs2}, which we do next.

\subsection{Proof of \Cref{claim: move to discs2}}


Fix  a pair of indices $0\leq j\leq r$ and $0\leq a< 2^{r-j}$.
Recall that we have defined two sets of paths associated with the edges of $\delta_{G'}(X^j_a)$. The first set of paths is a set $\qset_{j,a}=\set{Q_{j,a}(e)\mid e\in \delta_{G'}(X^j_a)}$ of edge-disjoint paths, that are internally disjoint from $X^j_a$, such that, for every edge $e\in \delta_{G'}(X^j_a)$, path $Q_{j,a}(e)$ originates at edge $e$, and terminates at some vertex of $A\setminus T^j_a$. The second set of paths is a set $\qset'_{j,a}=\set{Q'_{j,a}(e)\mid e\in \delta_{G'}(X^j_a)}$ of edge-disjoint paths, that are contained in $G'[X^j_a]$, such that, for every edge $e\in \delta_{G'}(X^j_a)$, path $Q'_{j,a}(e)$ originates at edge $e$, and terminates at some vertex of $T^j_a$. Both sets of paths are non-transversal with respect to $\Sigma'$. We now define two sets of curves: $\Gamma_{j,a}$ (corresponding to the paths in $\qset_{j,a}$), and $\Gamma'_{j,a}$ (corresponding to the paths in $\qset'_{j,a}$) that will be used in constructing the new drawings for graphs in $\hset$. We denote by $\sigma'_{j,a}$ the segment of the boundary of disc $D$ that is the complement of $\sigma_{j,a}$. In other words, if the boundary of $D$ is denoted by $\beta$, then $\sigma'_{j,a}=\beta\setminus\sigma_{j,a}$. For every edge $e\in \delta_{G'}(X^j_a)$, we denote $e=(x_e,y_e)$, where $x_e\in X^{j}_a$. 

We start with the set $\Gamma'_{j,a}$ of curves. Initially, for every edge $e\in \delta_{G'}(X^j_a)$, we let $\gamma'_{j,a}(e)$ be the image of the path $Q'_{j,a}(e)$. 
Note that curve $\gamma'_{j,a}(e)$ connects the image of $y_e$ to some point on $\sigma_{j,a}$, and it contains $\phi(e)$. Let $\Gamma'_{j,a}=\set{\gamma'_{j,a}(e)\mid e\in \delta_{G'}(X^j_a)}$.
From the definition of the ordering $\oset_{j,a}$, the oriented ordering $\oset_{j,a}$ of the edges in $\delta_{G'}(X^j_a)$ is identical to the order of the endpoints of their corresponding curves $\gamma'_{j,a}(e)$ along the boundary of the disc $D$. We assume w.l.o.g. that the orientation of the ordering $\oset_{j,a}$ is counter-clock-wise.
 The curves of $\Gamma'_{j,a}$ may not be in general position: if a vertex of $V(G')\setminus X^j_a$ lies on more than two paths in $\qset'_{j,a}$, then its image belongs to more than two curves of $\Gamma'_{j,a}$. We perform a nudging procedure by modifying the curves in $\Gamma'_{j,a}$ locally within tiny discs $D_{\phi}(v)$ of vertices $v\in X^j_a$  that belong to at least two paths of $\qset'_{j,a}$, using the algorithm from \Cref{claim: curves in a disc} (see also \Cref{sec: curves in a disc} for the definition of a nuding procedure). Since that paths in $\qset'_{j,a}$ are non-transversal with respect to $\Sigma'$, this nudging procedure does not introduce any new crossings between the curves in $\Gamma'_{j,a}$. We summarize the properties of the resulting set $\Gamma'_{j,a}$ of curves, that are immediate form our definitions and construction, and the fact that the vertices of $A$ have degree $1$ in $G'$, in the following observation.

\begin{observation}\label{obs: inner curves}
	Consider the final set $\Gamma'_{j,a}=\set{\gamma'_{j,a}(e)\mid e\in \delta_{G'}(X^j_a)}$ of curves. For every edge $e\in \delta_{G'}(X^j_a)$, curve $\gamma'_{j,a}(e)$ connects the image of vertex $y_e$ in $\phi$ to some point on $\sigma_{j,a}$, and it contains $\phi(e)$. 
The number of crossings between the curves in $\Gamma'_{j,a}$ is at most $|\chi(H_{j,a})|$, and the number of crossings between the curves in $\Gamma'_{j,a}$ and the edges of $G'\setminus X^j_a$ is at most $|\chi(H_{j,a})|$. The oriented ordering $\oset_{j,a}$ of the edges of $\delta_{G'}(X^j_a)$ is identical to the oriented ordering of the endpoints of the corresponding curves $\gamma'_{j,a}(e)$ on the boundary of disc $D$; we assume that the orientation of the ordering is counter-clock-wise.
\end{observation}

The construction of the set $\Gamma_{j,a}$ of curves is very similar, except that we also perform a type-2 uncrossing for them. In order to obtain the set $\Gamma_{j,a}$ of curves, we simply
 apply the algorithm from \Cref{thm: new type 2 uncrossing} to perform a type-2 uncrossing of the set $\qset_{j,a}$ of paths. We denote the resulting set of curves by $\Gamma_{j,a}=\set{\gamma_{j,a}(e)\mid e\in \delta_{G'}(X^j_a)}$. We summarize the properties of the resulting set of curves in the following observation, that follows immediately from the discussion so far and \Cref{thm: new type 2 uncrossing}.

\begin{observation}\label{obs: outer curves}
	Consider the  set $\Gamma_{j,a}=\set{\gamma_{j,a}(e)\mid e\in \delta_{G'}(X^j_a)}$ of curves. For every edge $e\in \delta_{G'}(X^j_a)$, curve $\gamma_{j,a}(e)$ connects the image of vertex $y_e$ in $\phi$ to some point on $\sigma'_{j,a}$. 
	There are no crossings between the curves in $\Gamma'_{j,a}$, and the number of crossings between the curves in $\Gamma_{j,a}$ and the edges of $G'[X^j_a]$ is at most $|\chi(H_{j,a})|$. The number of crossings between the curves in $\Gamma_{j,a}$ and the curves in $\Gamma'_{j,a}$ is bounded by $|\chi(H_{j,a})|$.
\end{observation}

(The last assertion follows from \Cref{obs: inner curves} and the fact that the curves in $\Gamma_{j,a}$ are aligned with the graph $\bigcup_{Q\in \qset_{j,a}}Q$.)

We let $\oset'_{j,a}$ be the oriented ordering of the edges in $\delta_{G'}(X^j_a)$ defined by the oriented ordering of the endpoints of the corresponding curves in $\set{\gamma_{j,a}(e)\mid e\in \delta_{G'}(X^j_a)}$ on the boundary of disc $D$. We assume w.l.o.g. that the orientation of the ordering is counter-clock-wise. 
We need the following obervation:

\begin{observation}\label{obs: bound distance between orderings}
	$\dist(\oset_{j,a},\oset'_{j,a})\leq O(|\chi(H_{j,a})|)$.
\end{observation}

Note that the above observation bounds the distance between two {\bf oriented} orderings.

\begin{proof}
	For every edge $e\in \delta_{G'}(X^j_a)$, let $\gamma^*(e)$ be the curve obtained by  concatenating curves $\gamma_{j,a}(e)$ and $\gamma'_{j,a}(e)$. Let $\Gamma^*=\set{\gamma^*(e)\mid e\in \delta_{G'}(X^j_a)}$ be the resulting set of curves. It is immediate to verify that $\Gamma^*$ is a valid reordering set of curves for the oriented orderings $\oset_{j,a},\oset'_{j,a}$. From Observations \ref{obs: inner curves} and \ref{obs: outer curves}, the number of crossings between the curves of $\Gamma$ is at most $O(|\chi(H_{j,a})|)$.
\end{proof}

We are now ready to define a solution $\psi(H)$ for every instance $I_H$ with $H\in \hset$. We start with a graph $H=H_{j,a}$, where $0<j\leq r$ and $0\leq a< 2^{r-j}$. 
Assume that $X^{j-1}_{a'}$ and $X^{j-1}_{a''}$ are the left and the right child clusters of $X^j_a$ respectively.
Recall that the boundary of the disc $D(H_{j,a})$ is the concatenation of the curves $\lambda''_{j,a},\lambda'_{j-1,a'}$ and $\lambda'_{j-1,a''}$. We let $\lambda^*_{j,a}$ be a curve with the same endpoints as $\lambda''_{j,a}$, that is internally disjoint from $\lambda''_{j,a}$, and is contained in disc $D(H_{j,a})$. We let $D'(H_{j,a})$ be the disc that is contained in $D(H_{j,a})$, whose boundary is the concatenation of the curves $\lambda^*_{j,a},\lambda'_{j-1,a'}$ and $\lambda'_{j-1,a''}$. 

We obtain an initial drawing $\psi(H_{j,a})$ of graph $H_{j,a}$ as follows. We start with the drawing of the graph $H_{j,a}$ that is induced by $\phi$. Conside now some vertex $t\in T^{\parent}_{j,a}$, and let $e_t$ be the unique edge of $H_{j,a}$ incident to $t$. We replace the current image of $e_t$ with the concatenation of $\phi(e_t)$ and the curve $\gamma_{j,a}(e_t)\in \Gamma_{j,a}$, and we move the image of $t$ to the endpoint of this curve that lies on segment $\sigma'_{j,a}$ of the boundary of disc $D$. Consider now some vertex $t\in T^{\lchild}_{j,a}$, and let $e_t$ be the unique edge of $H_{j,a}$ incident to $t$. We replace the current image of $e_t$ with the curve $\gamma'_{j-1,a'}(e_t)\in \Gamma'_{j-1,a'}$, and we place the image of $t$ on the endpoint of the curve that lies in the segment $\sigma'_{j-1,a'}$ of the boundary of $D$. Lastly, we consider vertices $t\in  T^{\rchild}_{j,a}$, and for each such vertex, we let $e_t$ be the unique edge of $H_{j,a}$ incident to $t$. We replace the current image of $e_t$ with the curve $\gamma'_{j-1,a''}(e_t)\in \Gamma'_{j-1,a''}$, and we place the image of $t$ on the endpoint of the curve that lies in the segment $\sigma'_{j-1,a'}$ of the boundary of $D$. We plant the resulting drawing of graph $H_{j,a}$ into the disc $D'(H_{j,a})$, so that the segments $\sigma'_{j,a},\sigma_{j-1,a'}$ and $\sigma_{j-1,a''}$ of the boundary of $D$ coincide with segments $\lambda^*_{j,a},\lambda'_{j-1,a'}$ and $\lambda'_{j-1,a''}$ of the boundary of $D'(H_{j,a})$, respectively. We denote the resulting drawing of $H_{j,a}$ by $\psi'_{j,a}$. It is easy to verify that it is a valid solution to instance $I_{H_{j,a}}$. Observe that, from our construction, the images of the vertices of $T^{\parent}_{j,a}$ appear on curve $\lambda^*_{j,a}$ in drawing $\psi'_{j,a}$, and their (oriented) ordering on this curve (with respect to disc $D'(H_{j,a})$) is identical to the ordering $\oset'_{j,a}$ of their corresponding edges in $\delta_{G'}(X_{j,a})$ that we have defined above. The vertices  
of $T^{\lchild}(j,a)$ appear on curve $\lambda'_{j-1,a'}$, and their (oriented) ordering on this curve (with respect to disc $D'(H_{j,a})$) is identical to $\oset(T^{\lchild}_{j,a})$. Similarly, the vertices  
of $T^{\rchild}(j,a)$ appear on curve $\lambda'_{j-1,a''}$, and their (oriented) ordering on this curve (with respect to disc $D'(H_{j,a})$) is identical to $\oset(T^{\rchild}_{j,a})$.

We now bound the number of crossings in drawing $\psi'_{j,i}$. For convenience, denote $H'= H_{j,a}\setminus \left (T^{\parent}_{j,a}\cup T^{\lchild}_{j,a}\cup T^{\rchild}_{j,a}\right )$.
Recall that $\chi(H_{j,a})$ is the set of all crossings in drawing $\psi'$ of $G'$ in which the edges of $G'[X^j_a]$ participate.
First, the total number of crossings between the edges of $E(H')$ is clearly bounded by $|\chi(H_{j,a})|$. There are no crossings between the curves in $\Gamma_{j,a}$. The number of crossings between the curves in $\Gamma_{j,a}$ and the edges of  $E(H')$ is bounded by $|\chi(H_{j,a})|$ from \Cref{obs: outer curves}. The number of crossings between the curves in $\Gamma'_{j-1,a'}$ and the edges of $E(H')$ is bounded by $|\chi(H_{j,a})|$, and the number of crossings between the curves in $\Gamma'_{j-1,a'}$ is also bounded by $|\chi(H_{j,a})|$  from \Cref{obs: inner curves}. Similarly, the number of crossings  between the curves in $\Gamma'_{j-1,a''}$, and the number of crossings between the curves in $\Gamma'_{j-1,a''}$ and the edges of $E(H')$ is bounded by $|\chi(H_{j,a})|$. It now remains to bound the number of crossings between the curves of $\Gamma'_{j-1,a'}$ and the curves of $\Gamma'_{j-1,a''}$. Notice that each such crossing corresponds to a unique crossing between an edge of $G'[X^{j-1}_{a'}]$ and an edge of $G'[X^{j-1}_{a''}]$. Since $G'[X^{j-1}_{a'}], G'[X^{j-1}_{a''}]\subseteq G'[X^j_a]$, the number of crossings between the curves of $\Gamma'_{j-1,a'}$ and the curves of $\Gamma'_{j-1,a''}$ is also bounded by $|\chi(H_{j,a})|$. Overall, the total number of crossings in $\phi_{j,a}$ is bounded by $O(|\chi(H_{j,a})|)$.

Finally, we need to ``fix'' the current drawing of graph $H_{j,a}$ by reordering the images of the vertices of $T^{\parent}_{j,a}$. Recall that every vertex $t\in T^{\parent}_{j,a}$ is an endpoint of a distinct edge in $\delta_{G'}(X^j_a)$. We have defined two oriented orderings of the edges of $\delta_{G'}(X^j_a)$: ordering $\oset_{j,a}$ and ordering $\oset'_{j,a}$. Each of these orderings naturally defines an oriented ordering of the vertices of $T^{\parent}_{j,a}$: we have denoted by  $\oset(T^{\parent}_{j,a})$ the ordering of the vertices of  $T^{\parent}_{j,a}$ defined by $\oset_{j,a}$, after we reverse the ordering. We denote the oriented ordering of the vertices of $T^{\parent}_{j,a}$ corresponding to  $\oset'_{j,a}$ by $\oset'(T^{\parent}_{j,a})$. Note that the images of the vertices of $T^{\parent}_{j,a}$ appear on the curve $\lambda^*_{j,a}$ in the ordering $\oset'(T^{\parent}_{j,a})$, as we traverse the boundary of the disc $D'(H_{j,a})$ in the clockwise direction. But we are required to ensure that the imges of the vertices of $T^{\parent}_{j,a}$ appear on the curve $\lambda''_{j,a}$ in the ordering $\oset(T^{\parent}_{j,a})$, as we traverse the boundary of the disc $D'(H_{j,a})$ in the clockwise direction. Let $D^*$ be the disc that is contained in $D(H_{j,a})$, whose boundary is the concatenation of the curves $\lambda''_{j,a}$ and $\lambda^*_{j,a}$. 

We place the images of the vertices of $T^{\parent}_{j,a}$ on curve $\lambda''_{j,a}$, so that they are encountered in the order $\oset(T^{\parent}_{j,a})$, as we traverse the boundary of $D^*$ in the clockwise direction. Notice that the previous images of the vertices of $T^{\parent}_{j,a}$ appeared on curve $\lambda^*_{j,a}$, and they are encountered on that curve in the order $\oset'(T^{\parent}_{j,a})$, as we traverse the boundary of $D^*$ in the counter-clockwise direction. Recall that, from \Cref{obs: bound distance between orderings}, the distance between the oriented orderings $\oset(T^{\parent}_{j,a})$ and $\oset'(T^{\parent}_{j,a})$ is
$\dist(\oset_{j,a},\oset'_{j,a})\leq O(|\chi(H_{j,a})|)$. Therefore, we can define, for every vertex $t\in T^{\parent}_{j,a}$ a curve $\gamma^*_t$, that is contained in disc $D^*$, and connects the original image of $t$ to the new image of $t$. The total number of crossings between the curves in $\set{\gamma^*_t\mid t\in T^{\parent}_{j,a}}$ is bounded by $O(|\chi(H_{j,a})|)$. In order to obtain the final solution $\psi(H_{j,a})$ to instance $I_{H_{j,a}}$, we start with solution $\psi'_{j,a}$ to the same instance. For every vertex $t\in T^{\parent}_{j,a}$, we consider the unique edge $e_t$ of $H_{j,a}$ that is incident to $t$. We extend the image of $e_t$ by appending the curve $\gamma^*_t$ to it, and we move the image of $t$ to its new location on curve $\lambda''_{j,a}$. This completes the construction of solution  $\psi(H_{j,a})$ to instance $I_{H_{j,a}}$, for $j>0$.

Next, we consider a graph $H_{j,a}\in \hset$ with $j=0$. We first define a curve $\lambda^*_{j,a}$ exactly as before. We then let  $D'(H_{j,a})$ be the disc that is contained in $D(H_{j,a})$, whose boundary is the concatenation of the curves $\lambda^*_{j,a}$ and $\sigma_{j,a}$. In order to construct the initial solution $\psi'_{j,a}$ to instance $H_{j,a}$, we start with the drawing of graph $H_{j,a}$ that is induced by $\phi$. We then process every vertex $t\in T^{\parent}_{j,a}$ exactly as before, replacing the image of the unique edge $e_t$ incident to $t$ with the curve $\gamma_{j,a}(e_t)$, and moving the image of $t$ to segment $\sigma'_{j,a}$ of the boundary of $D$. We plant the resulting drawing of graph $H_{j,a}$ inside disc $D'(H_{j,a})$, so that the segments $\sigma_{j,a}$ in disc $D$ and $D'(H_{j,a})$ coincide,  and the images of the anchor vertices in set $T_{j,a}$ remain unchanged. We also ensure that segment $\sigma'_{j,a}$ on the bondary of disc $D$ coincides with segment $\lambda^*_{j,a}$ on the boundary of $D'(H_{j,a})$. We then modify the images of the edges incident to the vertices $ T^{\parent}_{j,a}$, and update the images of the vertices of $ T^{\parent}_{j,a}$ exactly as before.

The solution $\psi(H^*)$ to the instance $I_{H^*}$ associated with graph $H^*$ is computed very similarly. The main difference is that this time we do not need to define the curve $\lambda^*_{j,a}$, and instead we can plant the initial image of $H^*$ directly into the disc $D(H^*)$, making sure that the images of the anchor vertices that belong to $H^*$ (the vertices of $T^*$) remain unchanged. We no longer need to take care of the set $T^{\parent}_{j,a}$ of vertices, and the vertices of $T^{**}$ are treated like the vertices of $T^{\lchild}_{j,a}$ or $T^{\rchild}_{j,a}$ in the case where $j>0$.
\subsection{Proof of \Cref{obs: tunnels}}
\label{subsec: appx tunnels}

We start with the following simple observation.

\begin{observation}\label{obs: nice tunnel}
	Let $p_i,p_{i'}$ be a pair of distinct points of $\Pi$, and assume that, for some integers $0\leq j\leq r$ and $0\leq a\leq 2^{r-j}$, $i'=a\cdot 2^j$. Assume further that $|i'-i|\leq 2^j$. Then there is a tunnel of length at most $j+1$ connecting $p_i$ to $p_{i'}$.
\end{observation}

\begin{proof}
	We assume w.l.o.g. that $i'<i$; the other case is symmetric. The proof is by induction on $j$. If $j=0$, then $i=i'+1$, and we can let tunnel $L$ consist of a single level-$0$ curve $\lambda_{0,i'}$, that connects $p_{i'}$ to $p_i$.
	
Consider now some integer $j>0$, and assume that the claim holds for all integers $0\leq \tilde j<j$. We prove that the claim holds for $j$. If $i'-i=2^j$, then there is a single level-$j$ curve $\lambda_{j,a}$ that connects $p_{i'}$ to $p_i$, and we let the tunnel $L$ consist of this one curve. We assume from now on that $i'-i<2^j$. Let $j'$ be the largest integer so that $2^{j'}\leq i'-i$, so $0\leq j'<j$ holds. Then there must be a level-$j'$ curve $\lambda_{j',a'}$, whose endpoints are $p_{i'}$ and $p_{i''}$, where $i''=i'+2^{j'}$. Notice that $i'<i''\leq i$ must hold, and moreover, $|i-i''|\leq 2^{j'}<2^j$ must hold.  If $p_{i''}=p_i$, then we let the tunnel $L$ consist of the curve $\lambda_{j',a'}$. Otherwise, from the induction hypothesis, there is a tunnel $L'$, of length at most $j'<j$, that connects $p_{i''}$ to $p_i$. We let $L$ be a tunnel that is obtained by appending curve $\lambda_{j',a'}$ at the beginning of tunnel $L'$.
\end{proof}

We are now ready to complete the proof of \Cref{obs: tunnels}. Consider  a pair $p_i,p_{i'}$ of distinct points of $\Pi$, and assume w.l.o.g. that $i'<i$. Let $j$ be the largest integer, so that at least two points lie in $\set{p_{a\cdot 2^j}\mid 0\leq a<2^{r-j}}\cap \set{p_{i''}\mid i'\leq i''\leq i}$.  Then there must be a pair of points $p_x,p_y$, with $i'\leq x<y\leq i$, such that $x=2^j\cdot a$ for some integer $a$, and $y=2^j\cdot (a+1)$. Note that there is a level-$j$ curve $\lambda_{j,a}$ in $\Lambda$ connecting $p_x$ and $p_y$. Moreover, $i-y\leq 2^j$ and $x-i'\leq 2^j$. From \Cref{obs: nice tunnel}, there is a tunnel $L_1$ of length at most $j$ connecting $p_{i'}$ to $p_x$, and a tunnel $L_2$ of length at most $j$ connecting $p_y$ to $p_{i}$. We then let $L$ be a tunnel obtained by concatenating tunnel $L_1$, curve $\lambda_{j,a}$, and tunnel $L_2$. Note that tunnel $L$ connects $p_{i'}$ to $p_{i}$, and its length is at most $2j+3\leq 2r+3\leq O(\log \cm')$.

\section{Proofs Omitted from \Cref{sec: computing the decomposition}}

\subsection{Proof of \Cref{lem: decomposition into small clusters}}
\label{sec: appx-decomposition-small-clusters}

We use a parameter $\tau'=c\tau\log_{3/2}m\log_2m$, were $c$ is a large enough constant, whose value we set later.

The algorithm maintains a collection $\cset$ of disjoint clusters of $H\setminus T$, with $\bigcup_{C\in \cset}V(C)=V(H)\setminus T$. 
Set $\cset$ of clusters is partitioned into two subsets: set $\cset^A$ of active clusters and set $\cset^I$ of inactive clusters. We will ensure that every cluster $C\in \cset^I$ has the $\alpha'$-bandwidth property, and $|E(C)|\leq m/\tau$. We start with $\cset^I=\emptyset$, and $\cset^A$ containing all connected components of $H\setminus T$. The algorithm terminates once $\cset^A=\emptyset$, and, when this happens, we return $\cset^I$ as the algorithm's outcome.

In order to bound the number of edges in $|\bigcup_{C\in \cset}\delta_H(C)|$, we use edge budgets and vertex budgets, that we define next.

\paragraph{Edge budgets.}
Consider a cluster $C\in \cset$ and an edge $e\in \delta_H(C)$. If $C\in \cset^I$, we set the budget $B_C(e)=1$, and otherwise we set it to be $B_C(e)=\log_{3/2}(2|\delta_H(C)|)$. If cluster $C$ is the unique cluster with $e\in \delta_H(C)$, then we set the budget of the edge $e$ to be $B(e)=B_C(e)$. If there are two clusters $C\neq C'\in \cset$ with $e\in \delta_H(C)$ and $e\in \delta_H(C')$, then we set the budget of the edge $e$ to be $B(e)=B_C(e)+B_{C'}(e)$. Lastly, if no cluster $C\in \cset$ with $e\in \delta_H(C)$ exists, then we set $B(e)=0$.

\paragraph{Vertex budgets.}
Vertex budgets are defined as follows. For every cluster $C\in \cset^A$, for every vertex $v\in V(C)$, we set the budget $B(v)=\frac{c\deg_C(v)\log_{3/2}m\cdot \log_{2}(|E(C)|)}{8\tau'}$, where $c$ is the constant used in the definition of $\tau'$. The budgets of all other vertices are set to $0$.

\paragraph{Cluster budgets and total budget.}
For a cluster $C\in\cset$, we define its edge-budget $B^E(C)=\sum_{e\in \delta_H(C)}B_C(e)$, and its vertex-budget $B^V(C)=\sum_{v\in V(C)}B(v)$. The total budget of a cluster $C\in \cset$ is $B(C)=B^E(C)+B^V(C)$, and the total budget in the system is $B^*=\sum_{C\in \cset}B(C)= 2\cdot\sum_{e\in E(G)}B(e)+\sum_{v\in V(G)}B(v)$.

\paragraph{Initial budget.}
At the beginning of the algorithm, the budget of every vertex $v\in V(H)\setminus T$ is at most:
$$\frac{c\deg_{H\setminus T}(v) \log_{3/2}m\log_2|E(H)|}{8\tau'}\leq \frac{\deg_H(v)}{8\tau},$$ 
the budget of every edge incident to a vertex in $T$ is at most $\log_{3/2}(2|T|)\leq 16\log m$, while the budget of every other edge is $0$. Therefore, the total budget $B^*$ in the system at the beginning of the algorithm is at most:
\[	\frac{m}{4\tau}+ 2\cdot 16k\log m\leq \frac{m}{\tau},\]
since $k\leq m/(64\tau\log m)$.

We will ensure that, throughout the algorithm, the total budget $B^*$ never increases. Since, from the definition, $B^*\geq \sum_{C\in \cset}|\delta_H(C)|$, this ensures that, when the algorithm terminates,  $\sum_{C\in \cset}|\delta_H(C)|\leq m/\tau$, so the set $\cset^I$ of clusters is a valid output of the algorithm.

\paragraph{Algorithm execution.}
As mentioned above, the algorithm starts with $\cset^I=\emptyset$, and $\cset^A$ containing all connected components of $H\setminus T$, with $\cset=\cset^A\cup \cset^I$. As long as $\cset^A\neq \emptyset$, we perform iterations, where in each iteration we select an arbitrary cluster $C\in \cset^A$ to process.
We now describe the execution of an iteration in which a cluster $C\in \cset^A$ is processed.
The algorithm for processing cluster $C$ consists of two parts, that we describe next.

\paragraph{Part 1: bandwidth property.}
In this step we either establish that $C$ has the $\alpha'$-bandwidth property, or we partition it into smaller clusters that will replace $C$ in set $\cset^A$.
Recall that an augmentation $C^+$ of cluster $C$ is a graph that is obtained from graph $H$, by subdividing every edge $e\in \delta_H(C)$ with a vertex $t_e$, letting $T(C)=\set{t_e\mid e\in \delta_H(C)}$ be this new set of vertices, and then letting $C^+$ be the subgraph of the resulting graph induced by $V(C)\cup T(C)$. Recall that cluster $C$ has the $\alpha'$-bandwidth property iff the set $T(C)$ of vertices is $\alpha'$-well-linked in $C^+$. We apply the algorithm  \algsc to graph $C^+$ and terminal set $T(C)$, to obtain an  $\alphasc(m)$-approximate sparsest cut $(X,Y)$ in graph $C^+$ with respect to the set $T(C)$ of terminals. We assume w.l.o.g. that $|X\cap T(C)|\leq |Y\cap T(C)|$.

We consider two cases. First, if $|E(X,Y)|\geq \alpha'\cdot \alphasc(m)\cdot |X\cap T(C)|$, then we are guaranteed that the set $T(C)$ of vertices is $\alpha'$-well-linked in $C^+$, and therefore cluster $C$ has the $\alpha'$-bandwidth property. In this case, we continue to Part 2 of the algorithm. Assume now that  $|E(X,Y)|< \alpha'\cdot \alphasc(m)\cdot |X\cap T(C)|$. 

We can assume without loss of generality that, for every vertex $t_e\in T(C)$, if $t_e\in X$, and $e'=(t_e,v)$ is the unique edge that is incident to $t_e$ in $C^+$, then $v\in X$ as well (since otherwise we can move vertex $t_e$ to $Y$, only making the cut sparser). Similarly, if $t_e\in Y$, then $v\in Y$ as well.

Let $X'=X\setminus T(C)$ and $Y'=Y\setminus T(C)$, so $(X',Y')$ is a partition of $C$. Note that $|T(C)\cap X|=|\delta_H(C)\cap \delta_H(X')|$ and similarly $|T(C)\cap Y|=|\delta_H(C)\cap \delta_H(Y')|$. 

We remove cluster $C$ from $\cset^A$ and from $\cset$, and we all connected components of  $C[X']$ and $C[Y']$ to $\cset^A$ and to $\cset$ instead. Observe that we are still guaranteed that $\bigcup_{C'\in \cset}V(C')=V(H)\setminus T$. We now show that the total budget in the system does not increase as the result of this step. 

Since every cluster that was newly added to $\cset^A$ is contained in $C$, it is immediate to verify that, for every vertex $v$ of $C$, its budget may only decrease. The only edges whose budget may increase are the edges of $E_C(X',Y')$. The number of such edges is bounded by $\alpha'\cdot \alphasc(m)\cdot |X\cap T(C)|=\alpha'\cdot \alphasc(m)\cdot |\delta_H(C)\cap \delta_H(X')|$, and the budget of each such edge increases by at most $\log_{3/2}(2m)\leq 4\log m$, so the total increase in the budget of all edges due to this step is bounded by:
\[4\alpha'\cdot \alphasc(m)\cdot |\delta_H(C)\cap \delta_H(X')|\cdot \log m\leq |\delta_H(C)\cap \delta_H(X')|,\]
since $\alpha'=\frac{1}{16\alphasc(m)\cdot \log m}$.

Consider now some edge $e\in \delta_H(C)\cap \delta_H(X')$. Since we have assumed that $|X\cap T(C)|\leq |Y\cap T(C)|$, it is easy to verify that $|\delta_H(X')|\leq 2|\delta_H(C)|/3$. Therefore, for every edge  $e\in \delta_H(X')\cap \delta_H(C)$, if $C'\subseteq H[X']$ is the new cluster with $e\in \delta(C')$, then:
\[B_{C'}(e)=\log_{3/2}(2|\delta_H(C')|)\leq \log_{3/2}(2|\delta_H(X')|)\leq \log_{3/2}(2|\delta_H(C)|)-1. \]
Therefore, the total decrease in the global budget due to the edges of 
$\delta_H(X')\cap \delta_H(C)$ is at least $|\delta_H(C)\cap \delta_H(X')|$. We conclude that overall the budget $B^*$ does not increase.

We assume from now on that algorithm \algsc returned a cut $(X,Y)$ of $C^+$ with $|E(X,Y)|\geq \alpha'\cdot \alphasc(m)\cdot |X\cap T(C)|$, and so cluster $C$ has the $\alpha'$-bandwidth property.

If $|E(C)|\leq m/\tau$, then we remove cluster $C$ from $\cset^A$, and add it to the set $\cset^I$ of inactive clusters. It is easy to verify that total budget $B^*$ may not increase as the result of this step. Therefore, we assume from now on that 
$|E(C)|>  m/\tau$.

\paragraph{Part 2: sparse balanced cut.}

In this step, we apply the algorithm from \Cref{cor: approx_balanced_cut} to graph $C$ with parameter $\hat c=3/4$, to obtain a $\hat c'$-edge-balanced cut $(Z,Z')$ of $C$ (where $1/2<\hat c'<1$), whose size is at most 
$O(\alphasc(m))$ times the size of a minimum $3/4$-edge-balanced cut  of $C$. We say that this step is \emph{successful} if $|E_H(Z,Z')|<|E(C)|/\tau'$. Assume first that the step was successful. Then we remove cluster $C$ from $\cset^A$ and from $\cset$, and add all connected components of  $C[Z],C[Z']$ to $\cset^A$ and to $\cset$ instead. We now show that the total budget in the system does not increase as the result of this step. Observe that the budget of every vertex may only decrease, and the same is true for the budget of every edge, except for the edges in set $\delta_H(Z)\cup \delta_H(Z')$.

We first bound the increase in the budgets of the edges of $\delta_H(Z)\cup \delta_H(Z')$. We consider two cases. The first case happens if $|\delta_H(C)|\leq \frac{8 |E(C)|}{\tau'}$. Since the budget of every edge is always bounded by  $\log_{3/2}(2m)$,
and since $|E_H(Z,Z')|\leq \frac{|E(C)|}{\tau'}$, 
 so the total increase in the budgets of all edges is bounded by $\frac{10|E(C)|\cdot \log_{3/2}(2m)}{\tau'}$.

Consider now the second case, where $|\delta_H(C)|> \frac{8 |E(C)|}{\tau'}$, and assume without loss of generality that $|\delta_H(Z)|\leq |\delta_H(Z')|$. In this case, since $|E_H(Z,Z')|\leq \frac{|E(C)|}{\tau'}$, $|\delta_H(Z)|\leq |\delta_H(C)|$, so the budgets of the edges in $\delta_H(Z)\cap \delta_H(C)$ may not grow. As before, the budgets of the edges of $E_H(Z,Z')$ may grow by at most $|E_H(Z,Z')|\cdot\log_{3/2}(2m)\leq \frac{|E(C)|\cdot \log_{3/2}(2m)}{\tau'}$. Lastly, for every edge $e\in \delta_H(Z')\cap \delta_H(C)$, the original budget $B_C(e)$ is $\log_{3/2}(2|\delta_H(C)|)$, and, if $C'\subseteq H[Z']$ is the new cluster with $e\in \delta_H(C')$, then the new budget $B_{C'}(e)= 
\log_{3/2}(2|\delta_H(C')|)$. Since $|\delta_H(C')|\leq |\delta_H(C)|+|E_H(Z,Z')|$, we get that the increase in the budget of $e$ is bounded by $\log_{3/2}\left (\frac{|\delta_H(C)|+|E_H(Z,Z')|}{|\delta_H(C)|}\right )=\log_{3/2}\left (1+\frac{|E_H(Z,Z')|}{|\delta_H(C)|}\right )$.

Since we have assumed that $|\delta_H(C)|> \frac{8 |E(C)|}{\tau'}$, while $|E_H(Z,Z')|<\frac{|E(C)|}{\tau'}$, we get that $\frac{|E_H(Z,Z')|}{|\delta_H(C)|}<1/2$. Since for all $\eps\in (0,1/2)$, $\ln(1+\eps)\leq \eps$, we get that the increase in the budget of $e$ is bounded by $\frac{|E_H(Z,Z')|}{|\delta_H(C)|\cdot \ln(3/2)}\leq \frac{4|E_H(Z,Z')|}{|\delta_H(C)|}$. Overall, we get that the budget of the edges of $\delta_H(Z')\cap \delta_H(C)$ increases by at most:

\[|\delta_H(Z')\cap \delta_H(C)|\cdot \frac{4|E_H(Z,Z')|}{|\delta_H(C)|}\leq  4|E_H(Z,Z')|\leq \frac{4|E(C)|}{\tau'}.  \]

To summarize, regardless of which of the above two cases happened, the total increase in the budgets of all edges is bounded by $\frac{10|E(C)|\cdot \log_{3/2}(2m)}{\tau'}$. Next, we show that the total decrease in the budgets of the vertices is high enough to compensate for this increase.

Assume without loss of generality that $|E(Z)|\leq |E(Z')|$. From the definition of edge-balanced cut, $|E(Z')|\leq \hat c'|E(C)|$, for some universal constant $\hat c'$. In particular: 
\begin{equation}
\sum_{v\in Z}\deg_Z(v)\geq 2(|E(C)|-|E(Z')|-|E(Z,Z')|)\geq 2(1-\hat c'-1/\tau')\cdot |E(C)|. \label{eq: many edges in Z 2}
\end{equation}
%

On the other hand, from our assumption that $|E(Z)|\leq |E(Z')|$, $\log_{2}(|E(Z)|)\leq \log_{2}(|E(C)|)-1$.
Recall that for every vertex $v\in Z$, its original vertex budget is:
$B(v)=\frac{c\deg_C(v)\log_{3/2}m\cdot \log_{2}(|E(C)|)}{8\tau'}$, and its new budget is: 
$$B'(v)=\frac{c\deg_Z(v)\log_{3/2}m\cdot \log_{2}(|E(Z)|)}{8\tau'}\leq \frac{c\deg_Z(v)\log_{3/2}m\cdot (\log_{2}(|E(C)|)-1)}{8\tau'}.$$
Therefore, for every vertex $v\in Z$, its budget decreases by at least 
$\frac{c\deg_Z(v)\log_{3/2}m}{8\tau'}$. Overall, the budget of the vertices in $Z$ decreases by at least:
\[\frac{c\log_{3/2}m}{8\tau'}\cdot \sum_{v\in Z}\deg_Z(v)\geq \frac{c\log_{3/2}m}{4\tau'}\cdot (1-\hat c'-1/\tau')\cdot |E(C)|\]
(from Equation \ref{eq: many edges in Z 2}.) Since $\tau'=c\tau\log_{3/2}m\log_2m$, and since we can set $c$ to be a large enough constant, we can ensure that this is at least $\frac{16|E(C)|\cdot \log_{3/2}(2m)}{\tau'}$, so the overall budget in the system does not increase.

We assume from now on that the current step was not successful. In other words, the algorithm from  \Cref{cor: approx_balanced_cut} returned a cut $(Z,Z')$ with $|E_H(Z,Z')|\geq |E(C)|/\tau'$. Since the size of this cut is within factor $O(\alphasc(m))$ from the minimum $3/4$-edge-balanced cut, we conclude that the value of the minimum $3/4$-edge-balanced cut in $C$ is at least $\rho=\Omega\left(\frac{|E(C)|}{\tau'\cdot  \alphasc(m)}\right )$.

Recall that, from \Cref{lem:min_bal_cut}, if the maximum vertex degree $\Delta$ in graph $C$ is at most $|E(C)|/2^{40}$, and $\optcro(C)\le |E(C)|^2/2^{40}$, then graph $C$ must contain a $(3/4)$-edge-balanced cut of value at most $\tilde c\cdot\sqrt{\optcro(C)+\Delta\cdot|E(C)|}$ where $\tilde c$ is some universal constant.
As the size of the minimum $3/4$-balanced cut in $C$ is at least $\rho$, we conclude that either $\Delta\geq |E(C)|/2^{40}$, or  $\optcro(C)> |E(C)|^2/2^{40}$, or
$\sqrt{\optcro(C)+\Delta\cdot|E(C)|}\geq \rho/\tilde c$ must hold. 
The latter can only happen if either $\optcro(C)\geq \frac{\rho^2}{2\tilde c^2}$, or $\Delta\geq \frac{\rho^2}{2\tilde c^2\cdot |E(C)|}$. Substituting the value of $\rho=\Omega\left(\frac{|E(C)|}{\tau'\cdot  \alphasc(m)}\right )$,
and recalling that $|E(C)|>m/\tau$, while $\tau'=c\tau\log_{3/2}m\log_2m$,
 we conclude that either (i) $\optcro(C)\geq \Omega\left(\frac{|E(C)|^2}{2(\tilde c \tau'\alphasc(m))^2}\right ) \geq \Omega \left( \frac{|E(C)|^2}{ \tau^2\log^5 m} \right ) \geq \Omega \left( \frac{m^2}{ \tau^4\log^5 m} \right )$; or (ii) $\Delta\geq \Omega\left(\frac{|E(C)|}{2(\tilde c\tau'\alphasc(m))^2}\right )\geq  \Omega\left(\frac{m}{\tau^3\log^5m}\right ) $; or (iii) $\Delta\geq \frac{|E(C)|}{2^{40}}\geq \frac{m}{2^{40}\tau} $. 
However, since we are guaranteed that $\Delta\leq \frac{m}{c^*\tau^3\log^5m}$ for a large enough constant $c^*$, we can rule out the latter two options, and conclude that $\optcro(H)\geq \optcro(C)\geq \Omega \left( \frac{m^2}{ \tau^4\log^5 m} \right )$.



If Phase 2 is unsuccessful, then we terminate the algorithm and declare that $\optcro(H)\geq  \Omega \left( \frac{|E(H)|^2}{ \tau^4\log^5 m} \right )$.

\subsection{Proof of \Cref{lem: solution to split to solution to original}}
\label{apd: Proof of solution to split to solution to original}

	
	Let $\phi'$ be a solution to instance $I'$. 
	Throughout the proof, we denote by $D=D_{\phi'}(u^*)$ the tiny $u^*$-disc in drawing $\phi'$.
	Recall that 
	$\delta_{G'}(u^*)=\set{a'_{1,1},\ldots,a'_{1,\hat q_1},a'_{2,1},\ldots,a'_{2,\hat q_2},\ldots,a'_{k,1},\ldots,a'_{k,\hat q_k}}$. Moreover, the edges of $\delta_{G'}(u^*)$ appear in this circular order in the rotation $\oset_{u^*}'\in \Sigma'$. 
	We assume w.l.o.g. that the orientation of this ordering in $\phi'$ is positive. In other words, the edges are encountered in this order as we traverse the boundary of $D$ so that the interior of $D$ always lies to our left (see, e.g. \Cref{fig: positive}, in which the orientation of the ordering $\oset_{u^*}'$  is positive).
	
	Consider now vertex $u_i$, for some $1\leq i\leq k$. Recall that 
	the set $\delta_{G'}(u_i)$ of edges is the union of two subsets: set
	$A'_i=\set{a'_{i,1},\ldots,a'_{i,\hat q_i}}$ of parallel edges connecting $u_i$ to $u^*$, and set $A_i=\set{a_{i,1},\ldots,a_{i, q_i}}$ of edges corresponding to the edge set $E_i\subseteq E(G)$. Recall that the ordering of the edges in $\delta_{G'}(u_i)$ in the rotation system $\Sigma'$ is: $(a'_{i,1},a'_{i,2},\ldots,a'_{i, \hat q_i}, a_{i, q_i}, a_{i, q_i-1},\ldots, a_{i,1})$. We say that vertex $u_i$ is \emph{synchronized} with $u^*$, if the orientation of the above ordering in $\phi'$ is negative  (see \Cref{fig: syn_example}). In other words, if we traverse the boundary of the tiny $u_i$-disc $D_{\phi'}(u_i)$ so that its interior always lies to our right, then we will encounter the edges of $\delta_{G'}(u_i)$ in the order $(a'_{i,1},a'_{i,2},\ldots,a'_{i, \hat q_i}, a_{i, q_i}, a_{i, q_i-1},\ldots, a_{i,1})$. 
	We need the following simple observation.

	\begin{observation}\label{obs: not synchronized}
		If, for some index $1\leq i\leq k$, vertex $u_i$ is not synchronized with $u^*$, then there are at least $\hat q_i^2/8$ crossings $(e,e')$ in $\phi'$ with $e,e'\in A_i'$. 
	\end{observation}

	\begin{proof}
		We delete from drawing $\phi'$ all vertices and edges except for vertices $u^*,u_i$ and edges of $A'_i$. For all $1\leq j\leq \hat q_i$, let $s_j$ be the point on the boundary of $D$ that lies on the image of edge $e_{i,j}$, let $t_j$ be the point on the boundary of $D_{\phi'}(u_i)$ that lies on the image of $e_{i,j}$, and let $\gamma_{j}$ be the segment of the image of edge $e_{i,j}$ between $s_j$ and $t_j$. 
		We assume without loss of generality that $\gamma_j$ does not cross itself; if it does, then we remove self loops until $\gamma_j$ does not cross itself. Denote $\Gamma=\set{\gamma_1,\ldots,\gamma_{\hat q_j}}$ the resulting set of curves.
	
\begin{figure}[h]
	\centering
	\subfigure[The ordering $\oset'_{u^*}$ with positive orientation. 
	]{		\scalebox{0.45}{\includegraphics[scale=0.42]{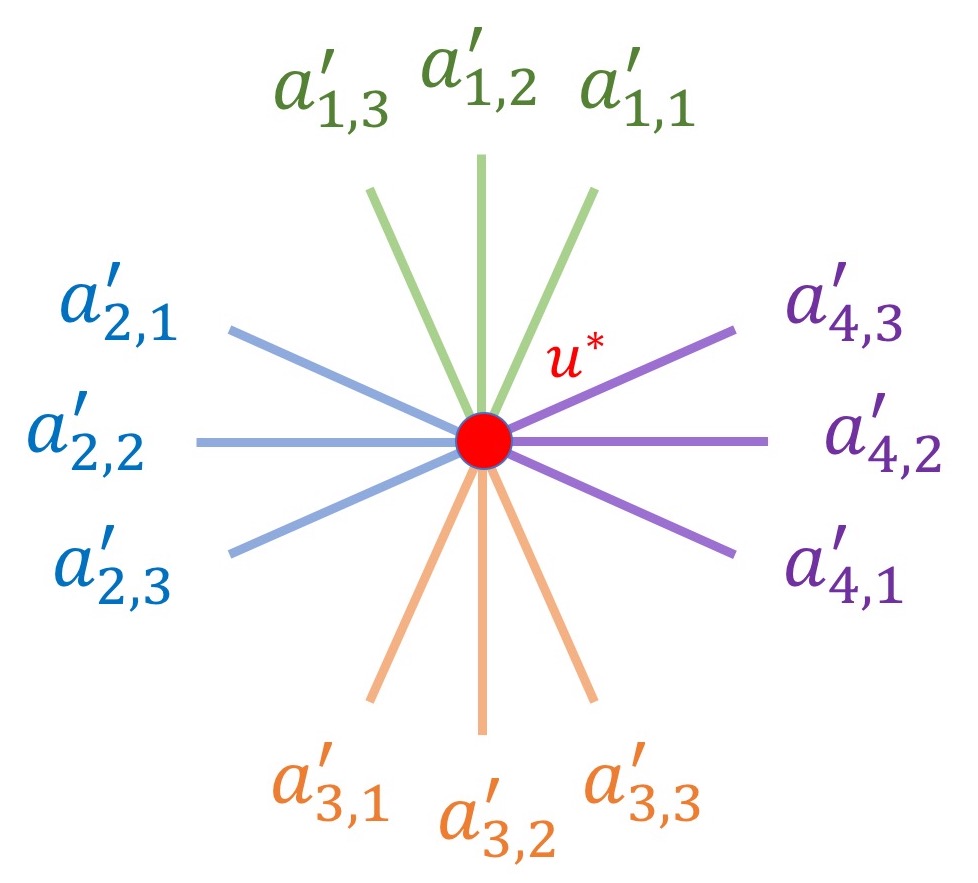}}\label{fig: positive}
	}
	\hspace{0.4cm}
	\subfigure[Vertex $u_i$ is synchronized with $u^*$. Note that edges of $A'_i$ may cross each other but they intersect the boundaries of the discs $D_{\phi'}(u^*)$ and $D_{\phi'}(u_i)$  in the order indicated above.]{
		\scalebox{0.45}{\includegraphics[scale=0.25]{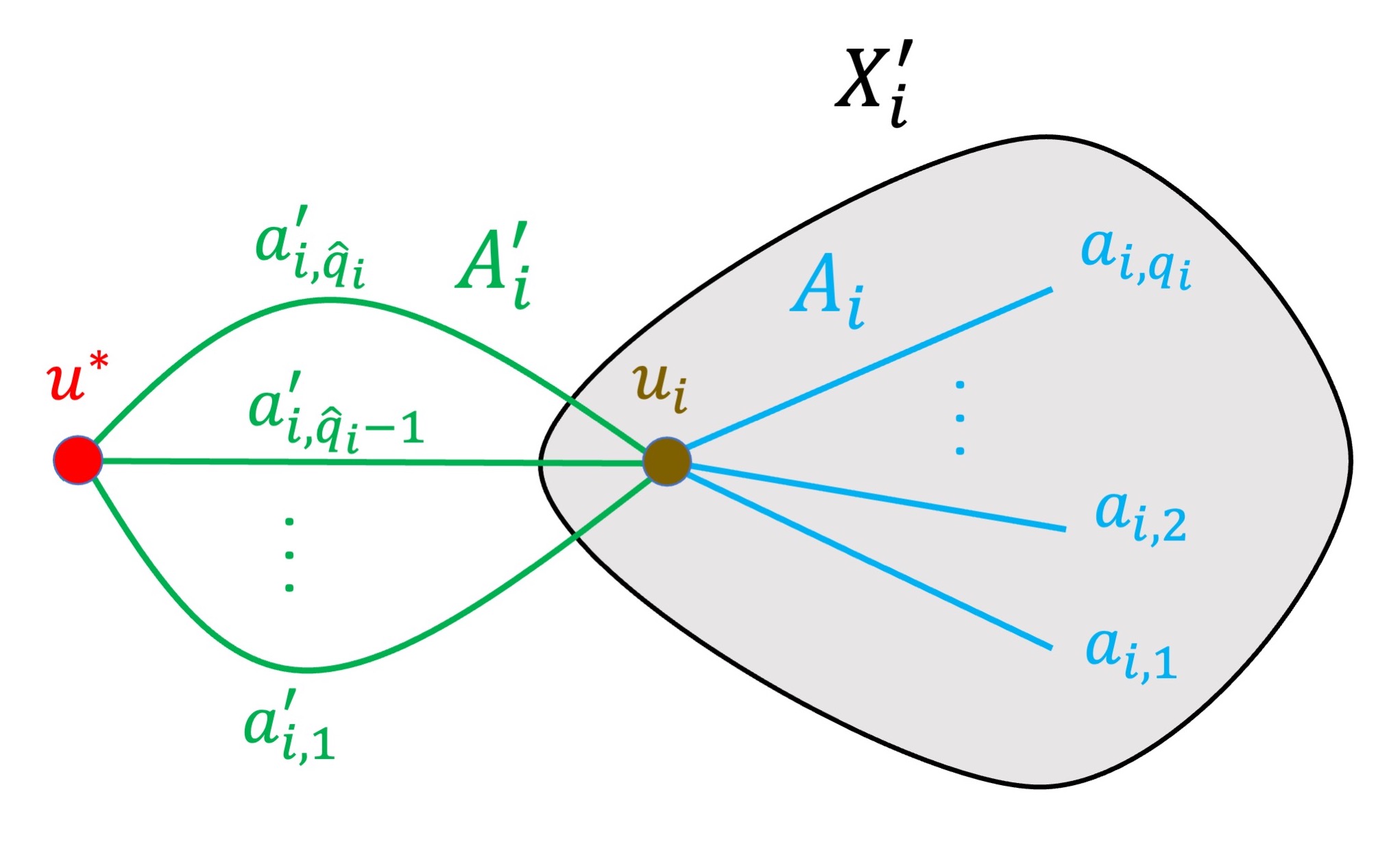}}\label{fig: syn_example}}
	\caption{An illustration of the positive orientation of $\oset'_{u^*}$ and the oriented rotation of the synchronized vertex $u_i$. 
	}
\end{figure}
	
		From our assumptions, points $s_1,\ldots,s_{\hat q_i}$ appear in this order on the boundary of $D$, when we traverse it so that the interior of $D$ lies to our left, while points $t_1,\ldots,t_{\hat q_i}$ appear in this order on the boundary of $\eta_i$, when we traverse it so that the interior of the disc is to our left (see \Cref{fig: synchronize_curve}).

		\begin{figure}[h]
			\centering
			\subfigure[Points $\set{s_k,t_k}_{1\le k\le  \hat q_i}$ and curves $\gamma_1, \gamma',\gamma''$. 
			]{\scalebox{0.42}	{\includegraphics[scale=0.35]{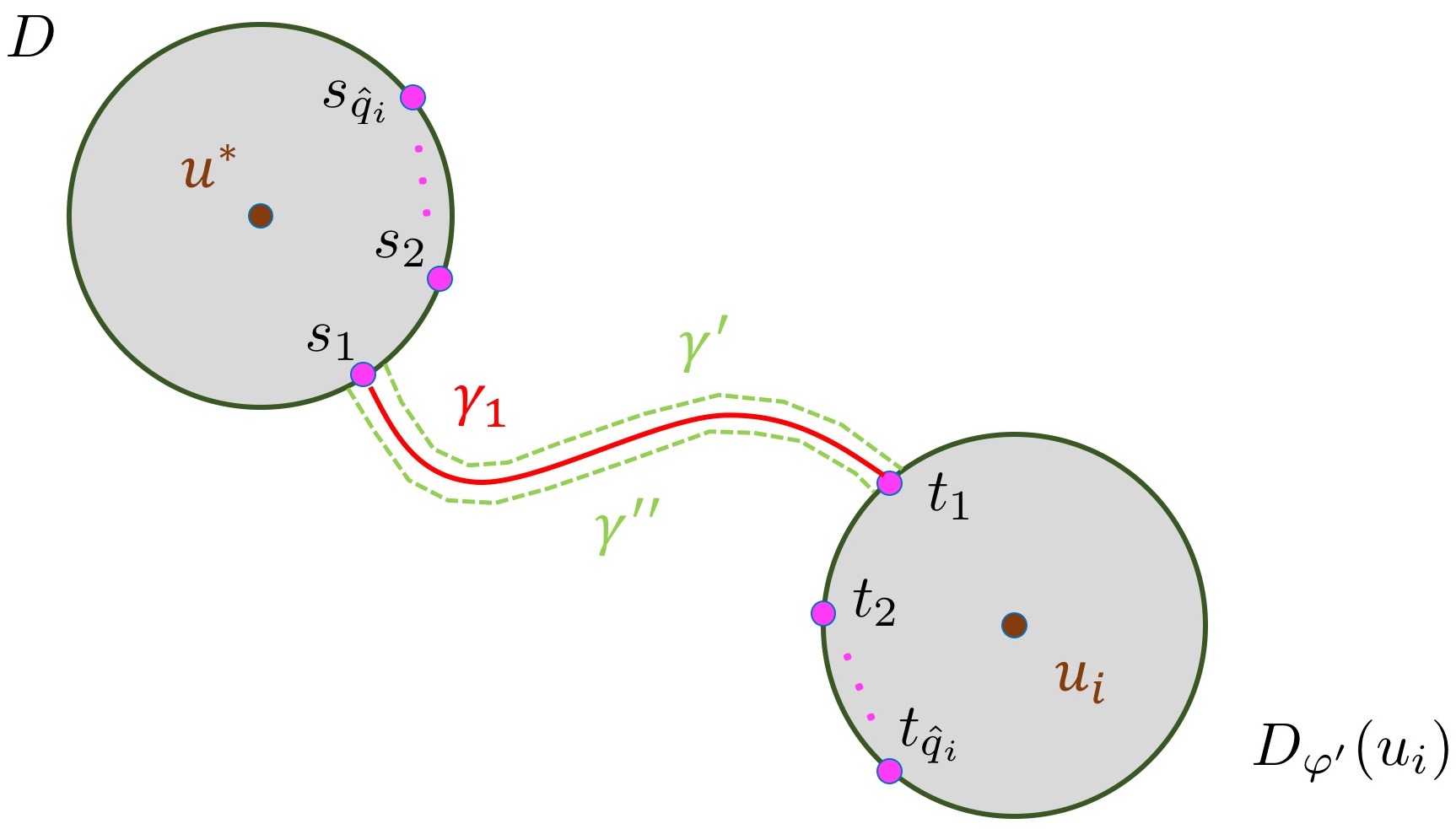}\label{fig: synchronize_curve}}
			}
			\hspace{0.2cm}
			\subfigure[Curves $\sigma,\gamma'',\sigma', \gamma'$ and disc $\tilde D$. 
			]{
				\scalebox{0.35}{\includegraphics[scale=0.3]{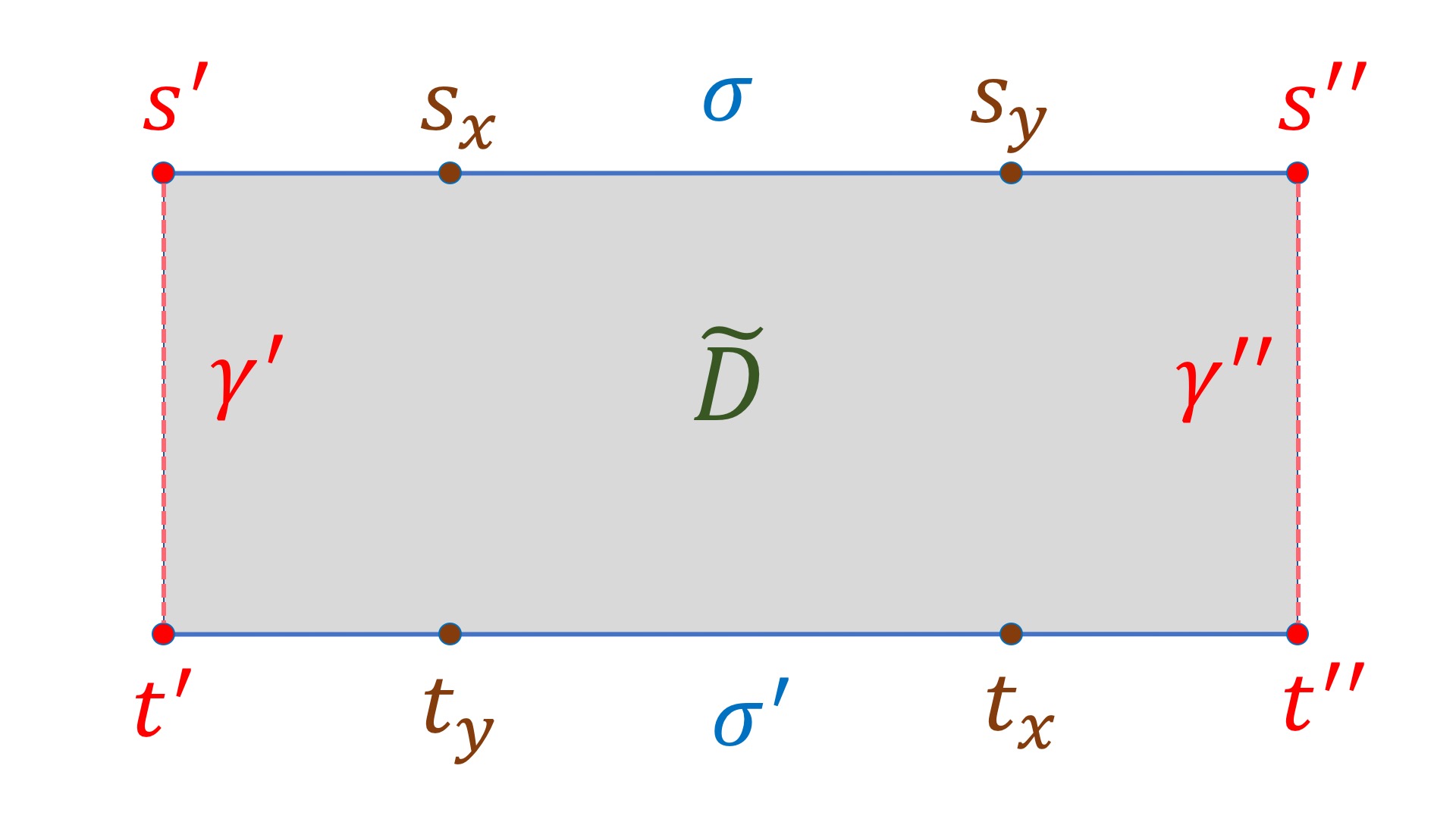}}\label{fig: closed_curve}}
			\caption{Illustrations for the proof of \Cref{obs: not synchronized}.}
		\end{figure}
		
		Assume for contradiction that there are fewer than $\hat q_i^2/8$ crossings $(e,e')$ in $\phi'$ with $e,e'\in A_i'$. Then there is some curve $\gamma_j\in \Gamma$, whose image crosses fewer than $\hat q_i/8$ curves in $\Gamma$. Assume w.l.o.g. that this curve is $\gamma_1$.
		
		Let $\Gamma'\subseteq \Gamma\setminus\set{\gamma_1}$ be the set of curves that do not cross $\gamma_1$, so $|\Gamma'|\geq |\Gamma|/2$. We next show that every pair distinct of curves in $\Gamma'$ must cross, leading to a conradiction. 	Indeed, consider any pair $\gamma_x,\gamma_y\in \Gamma'$ of distinct curves in $\Gamma'$, and assume without loss of generality that $x<y$. 
		Let $\gamma'$ and $\gamma''$ be two curves that follow curve $\gamma_1$ immediately to the left and immediately to the right, respectively. Let $s',s''$ be the endpoints of curves $\gamma'$ and $\gamma''$ lying on the boundary of $D$, respectively, and let $t',t''$ be  be the endpoints of curves $\gamma'$ and $\gamma''$ lying on the boundary of $D_{\phi'}(u_i)$, respectively. Notice that points $s',s''$ partition the boundary of $D$ into two segments, whose endpoints are $s'$ and $s''$; we let $\sigma$ be the segment that does not contain $s_1$. Similarly,  points $t',t''$ partition the boundary of $D_{\phi'}(u_i)$ into two segments, whose endpoints are $t'$ and $t''$; we let $\sigma'$ be the segment that does not contain $t_1$. Let $\lambda$ be the closed curve obtained by concatenating curves $\sigma,\gamma'',\sigma'$, and $\gamma'$ (see \Cref{fig: closed_curve}), and let $\tilde D$ be the disc whose boundary is $\lambda$, that contains the images of the curves $\gamma_x$ and $\gamma_y$. Then points $s_x,s_y,t_x,t_y$ appear on the boundary of $\eta^*$ in this order. Therefore, curves $\gamma_x,\gamma_y$ must cross. We conclude that every pair of curves in $\Gamma'$ must cross, a contradiction.
	\end{proof}

	

	In order to transform the drawing $\phi'$ of $G'$ into a drawing $\phi$ of $G$, we start by considering the tiny $u^*$-disc $D$ in the drawing $\phi'$. For each edge $a'_{i,j}\in \delta_{G'}(u^*)$, the image of the edge in $\phi'$ intersects the boundary of $D$ at exactly one point, that we denote by $p_{i,j}$.
	Recall that, from our assumptions, points $ p_{1,1},\ldots, p_{1, \hat q_1}, p_{2,1},\ldots, p_{2, \hat q_2},\ldots, p_{k,1},\ldots, p_{k, \hat q_k}$ appear in this order on the boundary of $D$, if we traverse it so that the interior of $D$ lies to our left.

\begin{figure}[h]
		\centering
		\includegraphics[scale=0.25]{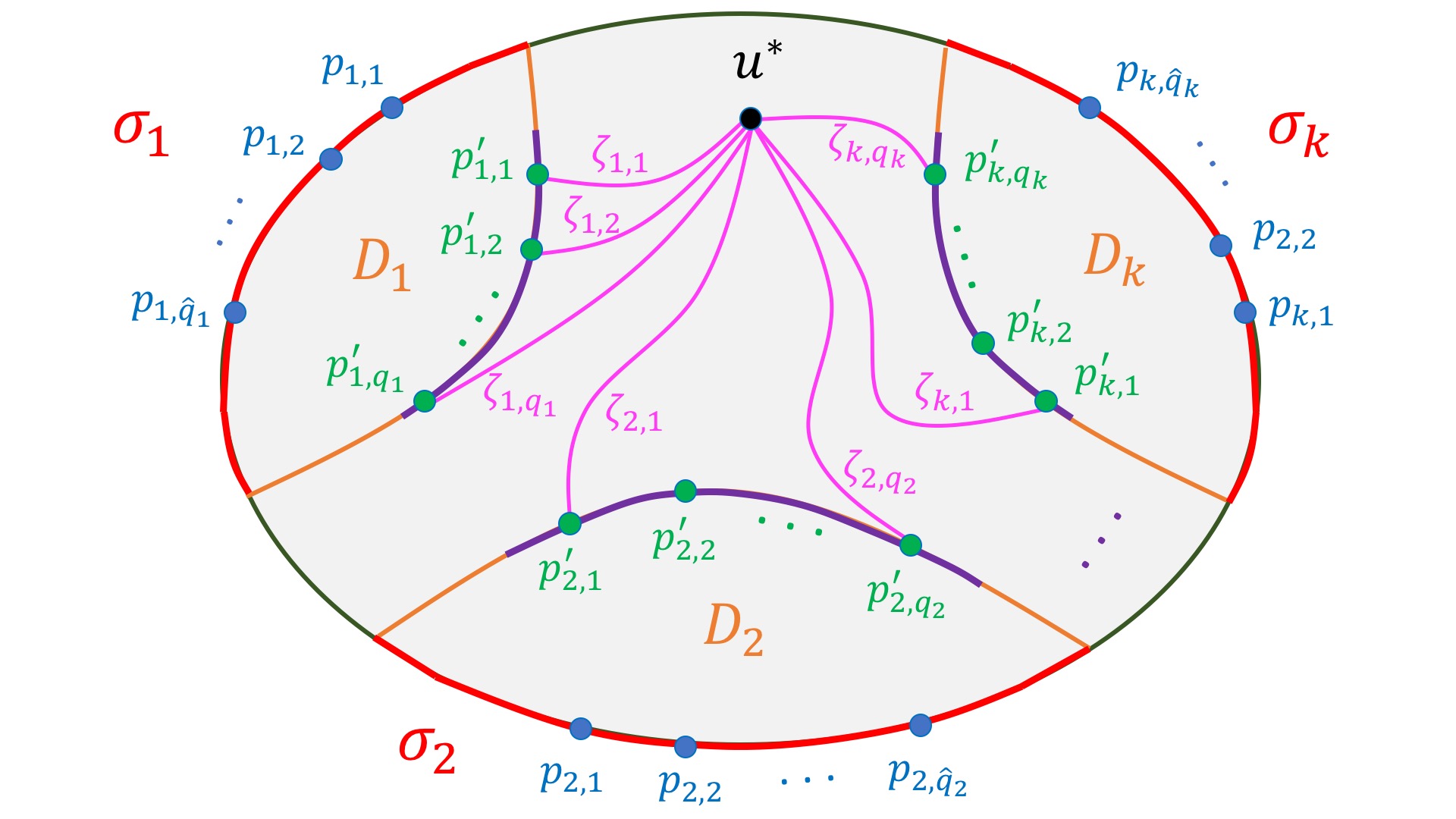}
		\caption{The interior of the disc $D$. Segments $\sigma'_1,\ldots,\sigma'_k$ are shown in purple. 
	 }\label{fig: segments}
	\end{figure}

	For all $1\leq i\leq k$, we define a segment $\sigma_i$ of the boundary of $D$, that contains all points $p_{i,1},\ldots, p_{i,\hat q_i}$. Observe that these segments can be defined so that they are mutually disjoint, and they appear on the boundary of $D$ in their natural order $ \sigma_1,\ldots, \sigma_k$, as we traverse the boundary  of $D$ so that its interior lies to our left. Next, for each $1\leq i\leq k$, we define a disc $D_i$, that is contained in $D$, such that the intersection of the boundary of $D$ and the boundary of $D_i$ is precisely $\sigma_i$, the image of $u^*$ lies outside $D_i$, and all discs $D_1,\ldots,D_k$ are mutually disjoint. 
	From the above discussion, for all $1\leq i\leq k$, the points $ p_{i,1},\ldots, p_{i,\hat q_i}$ appear in this order on segment $ \sigma_i$ of the boundary of $D_i$, as we traverse this boundary so that the interior of the disc $D_i$ lies to our left (see \Cref{fig: segments}).

	Consider now some index $1\leq i\leq k$.	
	Let $\sigma'_i$ be any segment of non-zero length on the boundary of disc $D_i$, that is disjoint from segment $\sigma_i$.
	Let $p'_{i,1},\ldots,p'_{i,q_i}$ be an arbitrary collection of distinct points on $\sigma'_i$, that appear on $\sigma'_i$ in this order, as we traverse the boundary of $D_i$ so that its interior lies to our right  (see \Cref{fig: segments}). We can then define, for each $1\leq i\leq k$ and $1\leq j\leq q_i$, a curve $\zeta_{i,j}$, that originates at the image of $u^*$ and terminates at point $p'_{i,j}$, such that all curves in set $\set{\zeta_{i,j}\mid 1\leq i\leq k, 1\leq j\leq q_i}$ are mutually internally disjoint. From our definitions so far, the circular order in which these curves enter the image of $u^*$ is:
	$(\zeta_{1,1},\ldots,\zeta_{1,q_1},\zeta_{2,1},\ldots,\zeta_{2,q_2},\ldots,\zeta_{k,1},\ldots,\zeta_{k,q_k})$ (see \Cref{fig: segments}). For all $1\leq i\leq k$ and $1\leq j\leq q_i$, we will use the curve $\zeta_{i,j}$ in order to draw the edge $e_{i,j}\in E_i$; in fact we will refer to $\zeta_{i,j}$ as the \emph{first segment of the drawing of edge $e_{i,j}$}. We will later define a second segment of the drawing of this edge, and then eventually stitch the two segments together to complete the drawing of the edge.

\begin{figure}[h]
	\centering
 	\includegraphics[scale=0.17]{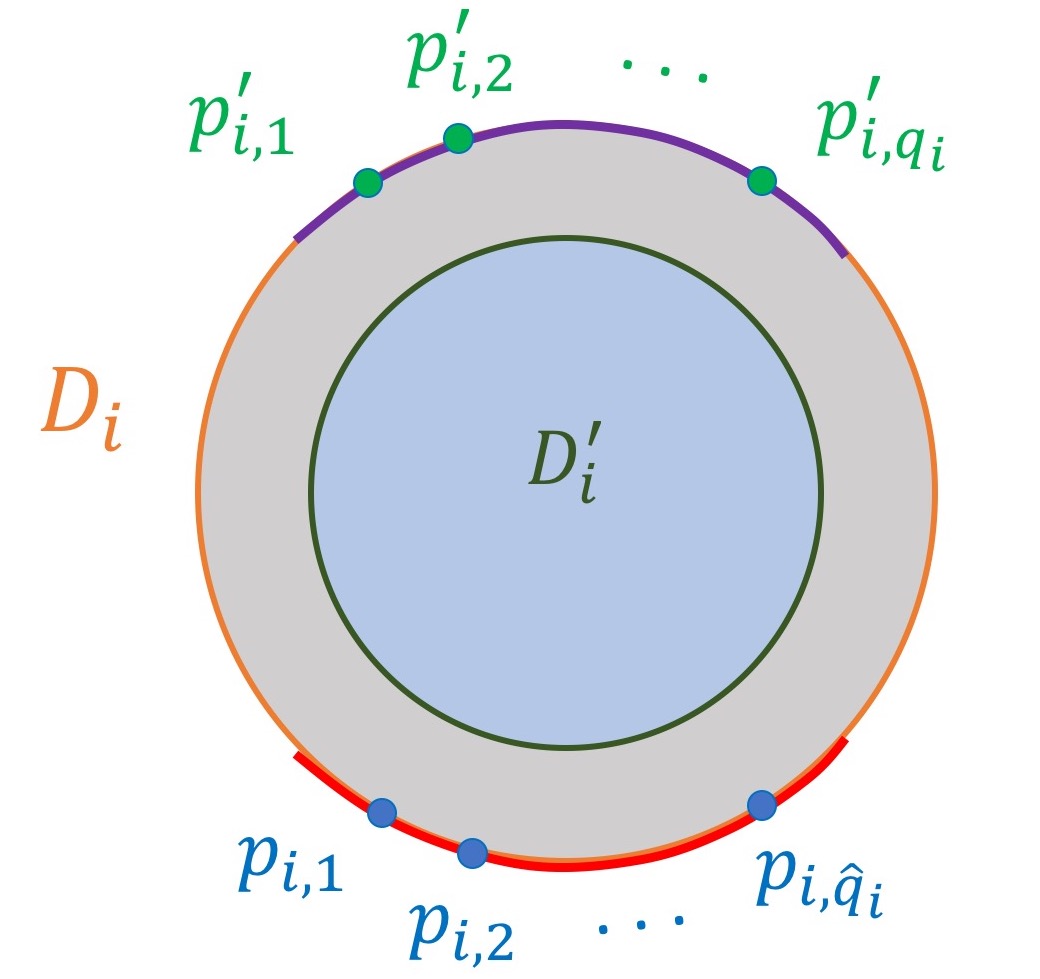}
	\caption{Discs $D_i$ and $D'_i$. Segments $\sigma_i$ and $\sigma'_i$ on the boundary of $D_i$ are shown in red and purple, respectively.}\label{fig: two discs}
\end{figure}

	Notice that so far, for all $1\leq i\leq k$, we have defined a collection $\set{p'_{i,1}, p'_{i,2},\ldots,p'_{i,q_i},p_{i,\hat q_i},p_{i,\hat q_i-1},\ldots, p_{i,1}}$  of points on the boundary of $D_i$, that appear on it in this order, as we traverse the boundary so that the interior of $D_i$ lies to our right (see \Cref{fig: two discs}).
	Lastly, for all $1\leq i\leq k$, we define another disc  $D'_i\subseteq D_i$, whose boundary is disjoint from the boundary of $D_i$ (see \Cref{fig: two discs}).

	In the remainder of the algorithm, we process each index $1\leq i\leq k$ one by one. We let $G_0=G'$, and for all $1\leq i\leq k$, we let $G_i$ be the graph obtained from $G'$ by contracting vertices $u_1,\ldots,u_i$ into the vertex $u^*$; we delete self-loops but keep parallel edges. Note that graph $G_k$ is identical to graph $G$, except that some of the edges of $G$ are subdivided in $G_k$. Therefore, a drawing of graph $G_k$ immediately gives a drawing of graph $G$. For each $1\leq i\leq k$, the input to the $i$th iteration is a drawing $\phi_{i-1}$ of graph $G_{i-1}$, in which, for all $1\leq i'\leq i-1$, all vertices and edges of $X_{i'}$ are drawn inside the disc $D'_{i'}$. The goal of the $i$th iteration is to produce a drawing $\phi_i$ of graph $G_i$, in which,  for all $1\leq i'\leq i$, all vertices and edges of $X_{i'}$ are drawn inside the disc $D'_i$. The final drawing $\phi_k$ of graph $G_k$, obtained at the end of the last iteration immediately provides a drawing of graph $G$. Let $\phi_0=\phi'$ be the given drawing of graph $G_0$, that is a solution to instance $I'$ of \cnwrs. For all $1\leq i\leq k$, we denote by $\cro_i$ the total number of crossings in $\phi_0$, in which edges of $E(X'_i)\cup A'_i\cup \hat A_i$ participate. Clearly, $\sum_{i=1}^k\cro_i\leq 2\cro(\phi)$.
	We will ensure that the following invariants hold, for all $1\leq i\leq k$:

	\begin{properties}{Inv}
		\item  over the course of iteration $i$, we may only change the images of the vertices and edges of $X'_i$, and the images of the edges of $\delta_G(X_i)\cup \delta_{G'}(X'_i)$; the images of the remaining edges and vertices of the graph remain unchanged;
		\label{inv: only drawing for Xi changes}
		
		\item for every edge $e\in E(G_{i-1})\setminus (E(X'_i)\cup A'_i\cup\hat A_i)$, the number of crossings in which edge $e$ participates in $\phi_i$ is bounded by the number of crossings in which edge $e$ participated in $\phi_{i-1}$; and \label{inv: num of crossings does not increase}
		
		 \label{inv: number of crossings does not increase}
		 
		\item $\cro(\phi_i)\leq \cro(\phi)+O(\cro_i)$. \label{inv: small increase in crossings}
	\end{properties}

From the above invariants, it is immediate to see that the final drawing $\phi_k$ of graph $G_k$ has at most $O(\cro(\phi))$ crossings.

In order to execute the $i$th iterations, we use the two sets $\qset_i,\qset'_i$ of paths that we have defined, in order to define two sets $\Gamma_i,\Gamma'_i$ of curves, that will serve as ``guiding curves'' for the transformation of the drawing $\phi_{i-1}$. We also use the current drawing $\phi_{i-1}$ in order to compute a ``nice'' drawing $\psi_i$ of graph $X_i$, together with a partial drawing of edges incident to vertices of $X_i$ in $G$. We then ``plant'' this drawing inside the disc $D'_i$, and then complete the drawings of the edges of $\delta_G(X_i)$.

	The input to the first iteration is the initial drawing $\phi_0=\phi'$ of graph $G_0=G'$ on the sphere. 
	We now fix a single index $1\leq i\leq k$, and describe the iteration in which index $i$ is processed. Our starting point is a drawing $\phi_{i-1}$ of graph $G_{i-1}$.
	Note that, from Invariant \ref{inv: number of crossings does not increase}, the total number of crossings in which the edges of $E(X'_i)\cup A'_i\cup \hat A_i$ participate in drawing $\phi_{i-1}$ is at most $\cro_i$. We will use this fact later.
	
	The algorithm for processing index $i$ consists of three stages. In the first stage, we use the set $\qset_i$ of paths in order to define the first set $\Gamma_i$, of ``guiding'' curves. In the second stage, we use the set $\qset'_i$ of paths in order to define the second set $\Gamma'_i$ of ``guiding'' curves, and we compute a drawing $\psi_i$ of $X_i$, together with a partial drawing of edges of $\delta_{G}(X_i)$. In the third and the final stage stage, we ``plant'' this drawing inside the disc $D'_i$, and complete the drawing $\phi_i$. We now describe each of the three stages in turn.

	\subsubsection{Stage 1: First Set of Guiding Curves, and Partial Drawing of Edges of $\hat E_i$}

	Recall that we have defined a collection $\qset_i=\set{Q(e)\mid e\in \hat E_i}$ of edge-disjoint paths in graph $G$, where for each edge $e\in \hat E_i$, path $Q(e)$ has $e$ as its first edge and $u^*$ as its last vertex, and all its inner vertices are contained in $X_i$. From the definition of graph $X'_i$, and from the fact that there are $|\hat E_i|=\hat q_i$ edges connecting $u_i$ to $u^*$ in $G'$ (the edges of $A'_i$), it is immediate to see that there must be a set $\hat \qset_i=\set{\hat Q(\hat a)\mid \hat a\in \hat A_i}$ of edge-disjoint paths in graph $G'_{i-1}$, where for each edge $\hat a\in \hat A_i$, path $\hat Q(\hat a)$ contains $\hat a$ as its first edge and terminates at vertex $u^*$, such that all inner vertices of $\hat Q(\hat a)$ are contained in $X'_i$.
	
	We apply the algorithm from \Cref{thm: new type 2 uncrossing} to perform a type-2 uncrossing of the paths in $\hat \qset_i$. The input to this algorithm is graph $G_{i-1}$ and its drawing $\phi_{i-1}$ on the sphere, and the set $\hat \qset_i$ of paths, which we view as being directed away from $u^*$. Let $\Gamma_i=\set{\gamma(\hat a)\mid \hat a\in \hat A_i}$ denote the set of curves that the algorithm outputs, that are aligned with the graph $\bigcup_{Q\in \hat Q_i}Q$. For each edge $\hat a\in \hat A_i$, if $y(\hat a)$ is the endpoint of $\hat a$ that does not lie in $X'_i$, then curve $\gamma(\hat a)$ originates at the image of $u^*$ and terminates at the image of $y(\hat a)$. Moreover, the curves in set $\Gamma_i$ do not cross each other.

	Recall that, for every path $\hat Q\in \hat \qset_i$, the first edge of $\hat Q$ (the edge incident to $u^*$) must be an edge of $A'_i$. From the definition of aligned curves, the theorem guarantees that, for every edge $a'_i\in A'_i$, there is a unique curve $\gamma(\hat a)\in \Gamma_i$ that contains the segment of the image of $a'_i$ that lies inside disc $D$. In particular, for all $1\leq j\leq \hat q_i$, there is a unique curve $\gamma(\hat a)\in \Gamma_i$ containing the point $p_{i,j}$ on the boundary of $D_i$. We denote $\hat A_i=\set{\hat a_{i,1},\ldots,\hat a_{i,\hat q_i}}$, where for all $1\leq j\leq \hat q_i$, edge $\hat a_{i,j}$ is the unique edge whose corresponding curve $\gamma(\hat a_{i,j})$ contains the point $p_{i,j}$. From the definition of aligned curves, each such curve $\gamma(\hat e_{i,j})$ intersects the boundary of $D$ at a unique point - point $p_{i,j}$. For all $1\leq j\leq \hat q_i$, we denote $\hat a_{i,j}=(\hat x_{i,j},\hat y_{i,j})$, where $x_{i,j}\in X'_i$. For all $1\leq j\leq \hat q_j$, we let $\hat \zeta_{i,j}$ be the segment of curve $\gamma(\hat a_{i,j})$ from the image of $y_{i,j}$ to point $p_{i,j}$ on the boundary of disc $D$. We denote the resulting set of curves by $\hat Z_{i}=\set{\hat \zeta_{i,j}\mid 1\leq j\leq \hat q_i}$.

	Consider now the drawing $\phi_{i-1}$ of graph $G_{i-1}$. We slightly modify this drawing, as follows. First, we delete from $\phi_{i-1}$ the images of all vertices of $X'_i$ and all edges of $E(X'_i)\cup \hat A_i\cup A'_i$. Next, we add to this drawing the set $Z_i=\set{\zeta_{i,j}\mid 1\leq j\leq q}$ of curves; recall that for all $1\leq j\leq q$, curve $\zeta_{i,j}$ is contained in disc $D$, is internally disjoint from disc $D_i$, and connects the image of $u^*$ to point $p'_{i,j}$ on the boundary of disc $D_i$ (see \Cref{fig: segments}). Recall that we called curve $\zeta_{i,j}$ the first segment in the drawing of edge $e_{i,j}$. Additionally, we add to the current drawing the set $\hat Z_i=\set{\hat \zeta_{i,j}\mid 1\leq j\leq \hat q_i}$ of curves. Recall that, for all $1\leq j\leq \hat q_i$, curve $\hat \zeta_{i,j}$ is internally disjoint from disc $D$, and it connects the image of vertex $y_{i,j}$ (the endpoint of edge $\hat a_{i,j}\in \hat A_i$ lying outside $X'_i$) to point $p_{i,j}$ on the boundary of disc $D_i$ (see also \Cref{fig: two discs}). We refer to curve $\hat \zeta_{i,j}$ as \emph{the first segment in the drawing of edge $\hat a_{i,j}$}.
	We denote the resulting drawing by $\phi'_{i-1}$. Note that, since the curves in $\Gamma_i$ are aligned with the graph $\bigcup_{Q\in \hat \qset_i}Q$, and, since the paths in set $\hat \qset_i$ are edge-disjoint, the total number of crossings in drawing $\phi'_{i-1}$ (including crossings between curves representing edges that were not erased and curves in sets $Z_{i}$ and $\hat Z_i$) is bounded by $\cro(\phi_{i-1})$. Moreover, for every edge $e\in E(G_{i-1})\setminus (E(X'_i)\cup A_i'\cup \hat A_i$), the number of crossings in which $e$ participates in $\phi'_{i-1}$ is bounded by the number of crossings in which $e$ participates in $\phi_{i-1}$, and the image of $e$ is disjoint from disc $D_i$.	
	This completes the first stage of the algorithm.	
	
	

\subsubsection{Stage 2: Second Set of Guiding Curves and Drawing of $X_i$}
	
	In this stage we consider again drawing $\phi_{i-1}$ of graph $G_{i-1}$. We will start by defining another set $\Gamma'_i$ of guiding curves in this graph (which we will eventually use in order to draw a second segment of each edge in $\hat A_i$). We then exploit the drawing $\phi_{i-1}$ and the curves in $\Gamma'_i$ in order to compute a drawing $\psi_i$ of graph $X_i$, and, for each edge $e\in \delta_{G}(X_i)$, a drawing of a segment of $e$ that is incident to its endpoint that lies in $X_i$. We will also define a new collection $\Gamma^*_i$ of curves, that will be useful for us in Stage 3.

\paragraph{Set $\Gamma'_i$ of guiding curves.}
We start by defining a set $\Gamma'_i$ of guiding curves. Consider the drawing $\phi_{i-1}$ of $G_{i-1}$.

Recall that petal $X_i$ is routable in graph $G$. Therefore, there is a set $\qset'_i$ of paths routing the edges of $\hat E_i$ to vertex $u^*$ in graph $G$, such that the paths in $\qset'_i$ are internally disjoint from $X_i$, and cause congestion at most $3000$. 
Consider any path $Q=Q(\hat e)\in \qset'_i$, whose first edge is $\hat e \in \hat E_i$. Recall that we have subdivided each such edge $\hat e\in \hat E_i$ with a vertex in graph $G'$. Let $e',e''$ denote the two edges that we obtained from subdividing edge $\hat e$, and assume that $e'\in \hat A_i$. We then replace edge $\hat e$ with $\hat e''$ on path $Q$. Assume now that the last edge of $Q$ is $e_{i',j}\in E_{i'}$. Since path $Q$ is internally disjoint from $X_i$, $i'\neq i$ must hold. If $i'>i$, then, by replacing edge $e_{i',j}$ with the corresponding edge $a_{i',j}$, we obtain a new path $Q'$ in graph $G_{i-1}$, whose first vertex is an endpoint of an edge of $\hat A_i$, and last vertex is $u_{i'}$. If $i'<i$, then we set $Q'=Q$. Let $\qset''_i=\set{Q'\mid Q\in \qset'_i}$ be the resulting set of paths in graph $G_{i-1}$.

 Consider now any vertex $u_{i'}$, for $i< i'\leq k$. Every path $Q'\in \qset''_i$ that terminates at $u_{i'}$ must contain an edge of $\hat A_{i'}$. Therefore, the number of paths in $\qset''_i$ terminating at $u_{i'}$ is bounded by $3000\hat q_{i'}$. Since, for all $1\leq i''\leq k$, $|A'_{i''}|=\hat q_{i''}$, we can then extend all such paths, using the edges of $A'_{i'}$, to ensure that they terminate at vertex $u^*$, such that  all such paths cause congestion at most $3000$ in $G_{i-1}$. Therefore, we have established that there is a set $\qset'''_i=\set{Q'''_{i,j} \mid 1\leq j\leq \hat q_i}$ of paths in graph $G_{i-1}$ that cause congestion at most $3000$, such that, for all $1\leq j\leq \hat q_i$, path $Q'''_{i,j}$ originates at vertex $\hat y_{i,j}$ (the endpoint of the edge $\hat a_{i,j}\in \hat A_i$ that does not lie in $X'_i$), terminates at vertex $u^*$, and is internally disjoint from $X'_i$.

In order to construct the set $\Gamma'_i$ of curves, we consider a graph $H$, that is obtained as follows. We start with $H=G_{i-1}$. We delete from this graph all edges $e\in E(G_{i-1})\setminus (E(X'_i)\cup \hat A_i\cup A'_i)$ that do not participate in the paths of $\qset'''_i$. For all remaining edges  $e\in E(G_{i-1})\setminus (E(X'_i)\cup \hat A_i\cup A'_i)$, we replace $e$ with  $\cong_{G_{i-1}}(\qset'''_i,e)$ parallel copies. Lastly, we delete all isolated vertices from the resulting graph. Note that drawing $\phi_{i-1}$ of graph $G_{i-1}$ naturally defines a drawing $\phi'$ of graph $H$: after deleting all edges of $E(G_{i-1})\setminus E(H)$ and all vertices of $V(G_{i-1})\setminus V(H)$ from the drawing, for every remaining edge $e\in E(H)\setminus (E(X'_i)\cup \hat A_i\cup A'_i)$, we draw the parallel copies of $e$ along the original image of edge $e$ in $\phi_{i-1}$. 
We can now use the set $\qset'''_i$ paths in graph $G_{i-1}$ in order to define a set $\hat \qset'_i=\set{\hat Q'_{i,j}\mid 1\leq j\leq \hat q_i}$ of edge-disjoint paths in graph $H$, where for all $1\leq j\leq \hat q_i$, path $\hat Q_{i,j}$ originates at vertex $\hat y_{i,j}$, terminates at vertex $u^*$, and is disjoint from $X_i'$.

We use the algorithm from \Cref{thm: new type 2 uncrossing} in order to compute a type-2 uncrossing of the paths in $\hat \qset'_i$. Specifically, the algorithm is applied to graph $H$, its drawing $\phi'$, and the set $\hat \qset'_i$ of edge-disjoint paths.
The algorithm returns a set $\hat \Gamma=\set{\hat \gamma_{i,j}\mid 1\leq j\leq \hat q_i}$ 
of internally disjoint curves, where, for all $1\leq j\leq \hat q_i$, curve $\hat \gamma_{i,j}$ connects the image of $\hat y_j$ to the image of vertex $u^*$ in drawing $\phi'$, and all curves in $\hat \Gamma$ are aligned with the graph $\bigcup_{\hat Q\in \hat \qset'_i}Q$. We will also consider the curves in set $\hat \Gamma$ in the drawing $\phi_{i-1}$ of $G_{i-1}$. As before, each curve $\hat \gamma_{i,j}$ connects the image of $\hat y_{i,j}$ to the image of vertex $u^*$ in drawing $\phi_{i-1}$.
Since the paths in the original set $\qset'$ caused congestion at most $3000$, it is immediate to verify that the number of crossings between the images of the edges of $E(X'_i)\cup A'_i\cup \hat A_i$ in $\phi_{i-1}$ and the  curves of $\hat \Gamma$ is at most $3000\cdot\cro_i$.

Since the curves in $\hat \Gamma$ are aligned with the graph $\bigcup_{\hat Q\in \hat \qset'_i}Q$, each curve $\hat \gamma_{i,j}\in \hat \Gamma$ intersects the boundary of the tiny $u^*$-disc $D$ in a single point, that we denote by $z_j$. Sine the paths in $\hat \qset'$ may not use the edges of $A'_i$, we are guaranteed that each such point $z_j$ may not lie on the segment $\sigma_i$ (the segment containing the points $p_{i,1},\ldots,p_{i,\hat q_i}$; point $p_{i,j'}$ is the intersection point of the image of edge $a'_{i,j'}$ and the boundary of $D$, see \Cref{fig: segments}).
For convenience, we re-index the points in set $\set{z_j}_{1\leq j\leq \hat q_i}$, so that points $z_1,z_2,\ldots,z_{\hat q_i},p_{i,\hat q_i},\ldots,p_{i,1}$ appear on the boundary of $D$ in this order. For each $1\leq j\leq \hat q_i$, we denote by $\ell(j)$ the unique index such that curve $\hat \gamma_{i,j}$ contains the point $z_{\ell(j)}$.  For all $1\leq j\leq \hat q_i$, we let $\gamma'_{i,j}$ be the segment of curve $\hat \gamma_{i,j}$ from the image of vertex $\hat y_{i,j}$ to point $z_{\ell(j)}$. We then set $\Gamma'_i=\set{\gamma'_{i,j}\mid 1\leq j\leq \hat q_i}$. From the above discussion, the total number of crossings 
 between the images of the edges of $E(X'_i)\cup A'_i\cup \hat A_i$ in $\phi_{i-1}$ and the  curves of $\Gamma'_i$ is at most $3000\cdot\cro_i$.

\paragraph{Set $\Gamma^*_i$ of Auxiliary curves.}
We need to define another set of curves, that we will use in Stage 3. Recall that we have defined a set of points $z_1,z_2,\ldots,z_{\hat q_i},p_{i,\hat q_i},\ldots,p_{i,1}$ that appear on the boundary of disc $D$ in this order. We can consider two orderings of elements of $\set{1,\ldots,\hat q_i}$: the first ordering is their natural ordering, while the second ordering is $\ell(1),\ell(2),\ldots,\ell(\hat q_i)$ -- ordering that is defined by the curves in $\Gamma'_i$. In Stage 3 of our algorithm, we will need to show that the distance between these two orderings is small, in order to combine different segments of the drawings of the edges of $A_i$ to complete their drawing. The set $\Gamma^*_i$ of curves, that we define in the next observation, will be used in order to do so.

\begin{observation}\label{obs: curves of gamma star}
	There is an efficient algorithm to construct a collection $\Gamma^*_i=\set{\gamma^*_{i,j}\mid 1\leq j\leq \hat q_i}$ of curves, such that, for all $1\leq j\leq \hat q_i$, curve $\gamma^*_{i,j}$ connects point $p_{i,j}$ on the boundary of disc $D$ to point $z_{\ell(j)}$, and it is internally disjoint from disc $D$. Moreover, the total number of crossings between the curves of $\Gamma^*_i$ is $O(\cro_i)$.
\end{observation}

\begin{proof}
	Consider an index $1\leq j\leq \hat q_i$. Recall that we have defined a curve $\hat \zeta_{i,j}$, which is a sub-curve of some curve of $\Gamma_i$, connecting point $p_{i,j}$ to the image of vertex $\hat y_{i,j}$ in $\phi_{i-1}$. We concatenate this curve with curve $\gamma'_{i,j}\in \Gamma'_i$, connecting the image of vertex $\hat y_{i,j}$ to point $z_{\ell(j)}$, obtaining the curve $\gamma^*_{i,j}$, that connects $p_{i,j}$ to $z_{\ell(j)}$. From the construction of curves in $\Gamma_i$ and $\Gamma'_i$, and the alignemnt properties of each such curve, we are guaranteed that each resulting curve is internally disjoint from disc $D$.
	
	In order to bound the number of crossings between the curves of $\Gamma^*$, recall that the total number of crossings 
	between the images of the edges of $E(X'_i)\cup A'_i\cup \hat A_i$ in $\phi_{i-1}$ and the  curves of $\Gamma'_i$ is at most $3000\cro_i$. Since the curves of $\Gamma_i$ are aligned with graph $\bigcup_{Q\in \hat \qset_i}Q$, and the paths of $\qset_i$ are edge-disjoint and contained in $X'_i\cup A'_i\cup \hat A_i$, we get that the total number of crossings between the curves of $\Gamma^*_i$ is $O(\cro_i)$.
\end{proof}

\paragraph{Partial drawing of edges of $\hat A_i$.}
Next, we define a collection $\hat Z_i'=\set{\hat \zeta'_{i,j}\mid 1\leq j\leq \hat q_i}$, of curves, that we will use in order to obtain partial drawing of the edges of $\hat A_i$. For all $1\leq j\leq \hat q_i$, curve $\hat \zeta'_{i,j}$ will connect the image of vertex $\hat x_{i,j}$ (the endpoint of edge $\hat a_{i,j}$ lying in $X'_i$) to a point on the boundary of the tiny $u_i$-disc in $\phi_{i-1}$. We will then view curve $\hat \zeta'_{i,j}$ as the \emph{second segment in the drawing of edge $\hat a_{i,j}$}, and we will add all such curves to the drawing $\psi_i$ that we compute in this stage. 
	
In order to compute the set $\hat Z_i'$ of curves, we will utilize the cuves of $\Gamma'_i$, the images of the edges of $\hat A_i\cup A'_i$ in drawing $\phi_{i-1}$, and another set of curves that we define next.

\begin{figure}[h]
	\centering
	\includegraphics[scale=0.15]{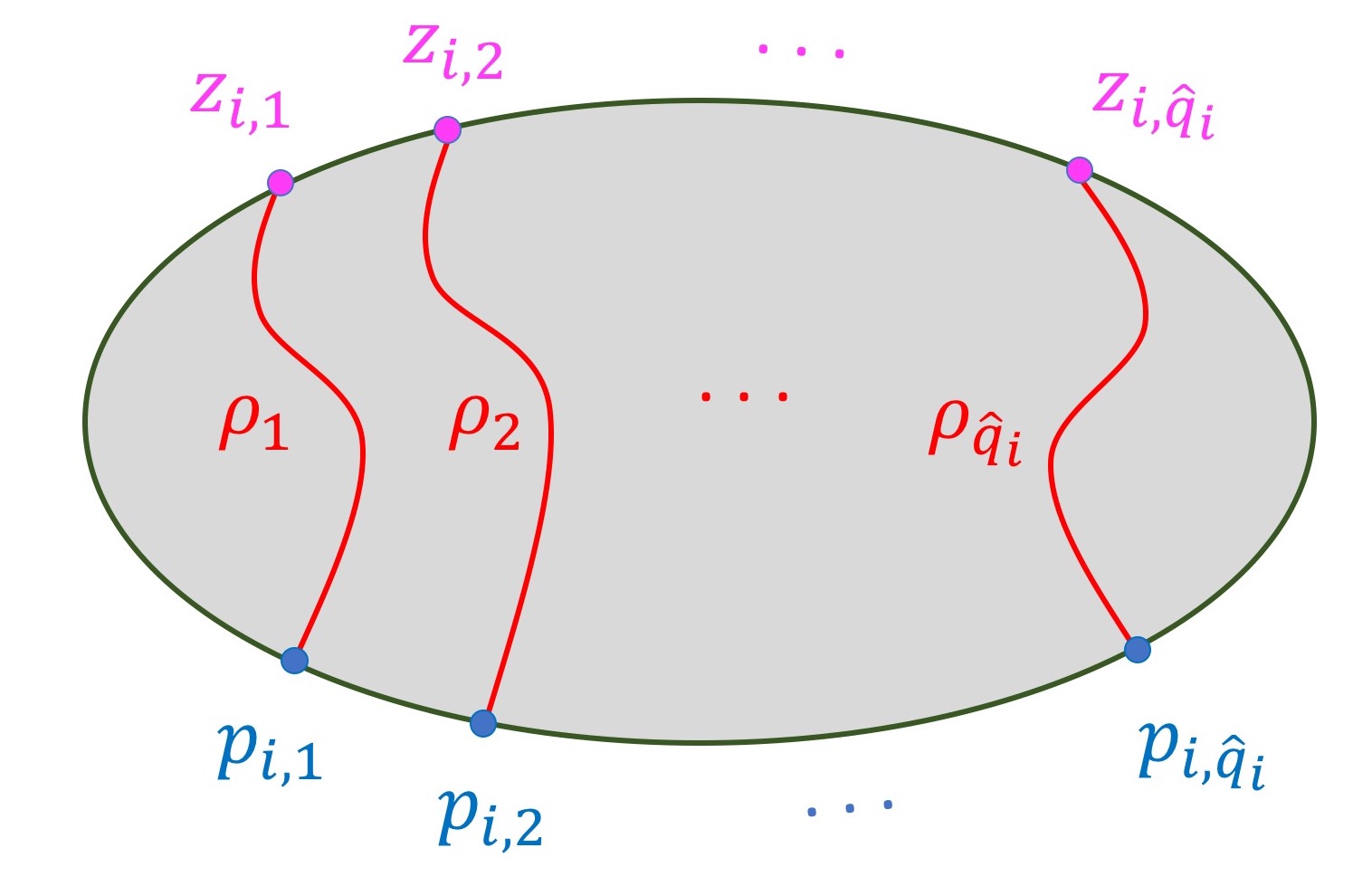}
	\caption{An illustration of curves $\rho_1,\ldots,\rho_{\hat q_i}$ in disc $D$.}\label{fig: rho curves}
\end{figure}

Recall that we have defined a set $z_1,z_2,\ldots,z_{\hat q_i},p_{i,\hat q_i},\ldots,p_{i,1}$ of points on the boundary of disc $D$, that appear on the boundary of $D$ in this order. We define, for all $1\leq j\leq \hat q_i$ a curve $\rho_j$, connecting $z_j$ to $p_{i,j}$, such that all curves in $\set{\rho_1,\ldots,\rho_{\hat q_i}}$ are contained in disc $D$ and are disjoint from each other (see \Cref{fig: rho curves}).

\begin{figure}[h]
	\centering
	\includegraphics[scale=0.2]{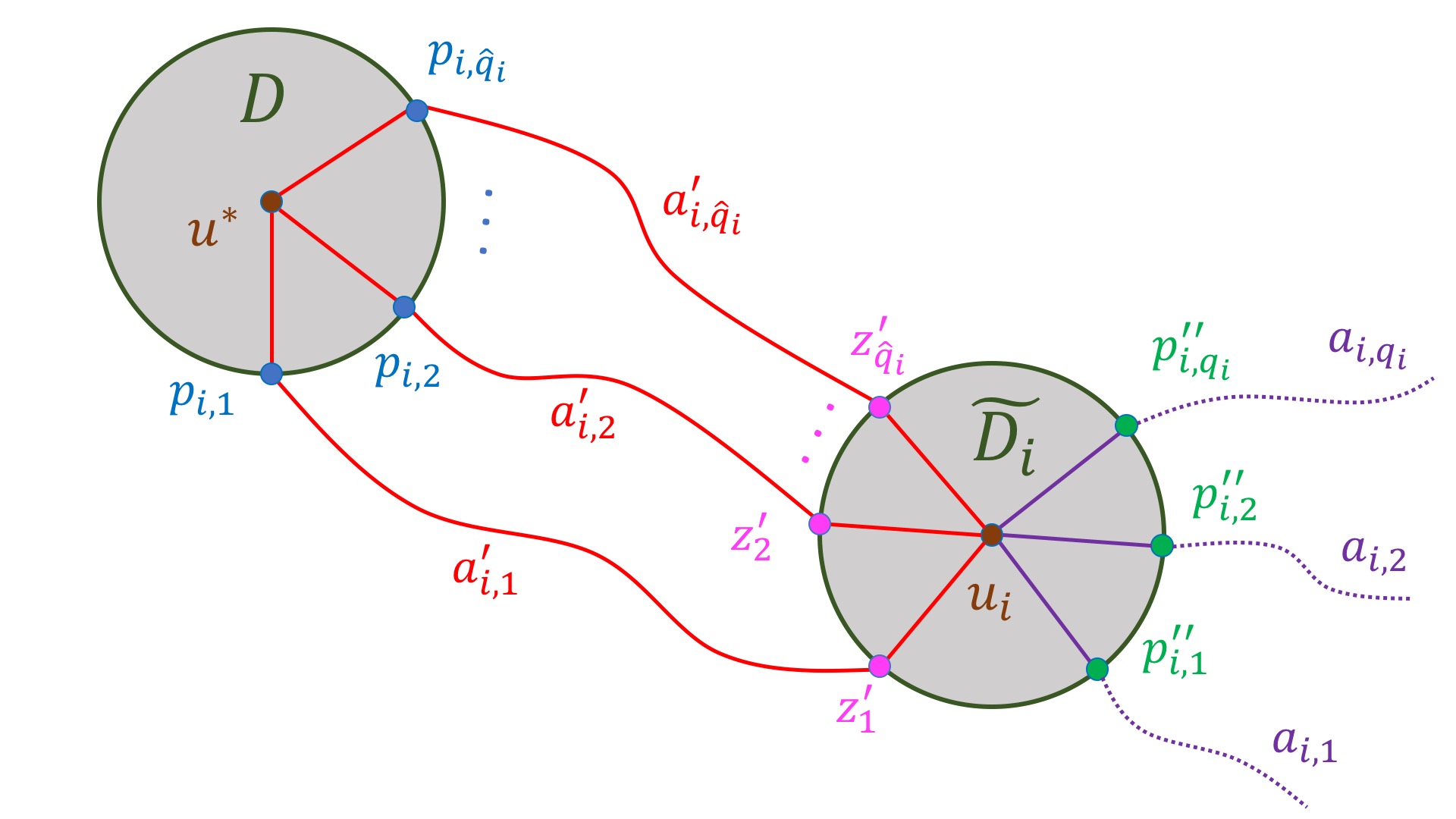}
	\caption{Schematic view of images of edges of $A'_i$ and $A_i$ and relevant points on the boundary of $D$ and $\tilde D_i$. Images of edges of $A_i$ are shown in red. Image of each edge $a_{i,j}\in A_i$ is shown in purple, with segments $\zeta_{i,j}$ dotted. 
	}\label{fig: image construction}
\end{figure}

We denote by $\tilde D_i=D_{\phi_{i-1}}(u_i)$ the tiny $u_i$-disc in $\phi_{i-1}$. Consider the images of the edges in set $A'_i=\set{a'_{i,j}\mid 1\leq j\leq \hat q_i}$ in drawing $\phi_{i-1}$ (these are the parallel edges connecting $u^*$ to $u_i$). From our definition, for all $1\leq j\leq \hat q_i$, point $p_{i,j}$ is the unique point on the boundary of disc $D$ that lies on the image of edge $a'_{i,j}$. We denote the unique point of the image of $a'_{i,j}$ lying on the boundary of disc $\tilde D_i$ by $z'_{j}$. Recall that, from our assumptions, points $p_{i,1},\ldots,p_{i,\hat q_i},z_{\hat q_i},\ldots,z_1$ appear in this order on the boundary of $D$, as we traverse it so that the interior of the disc lies to our left (see \Cref{fig: image construction}). If $u_i$ is synchronized with $u^*$, then, from the definition of the rotation $\oset'_{u_i}\in \Sigma'$, points $z'_{1},\ldots,z'_{\hat q_i}$ appear in this order on the boundary of disc $\tilde D_i$, as we traverse it so that the interior of the disc lies to our right; if $u_i$ is not synchronized with $u^*$, then this order is reversed (see  \Cref{fig: image construction}).

We are now ready to define the curves of $\hat Z_i'$. Consider some index $1\leq j\leq \hat q_i$. Curve $\hat \zeta'_{i,j}$ is the concatenation of four curves: (i) the image of the edge $\hat a_i=(\hat x_{i,j},\hat y_{i,j})$ in $\phi_{i-1}$; (ii) curve $\gamma'_{i,j}\in \Gamma'_i$ (that connects the image of $\hat y_{i,j}$ to point $z_\ell(j)$ on the boundary of $D$); (iii) curve $\rho_j$ (that connects $z_\ell(j)$ to $p_{i,j}$ and is contained in $D$); and (iv) a segment of the image of the edge $a'_{i,j}$ in $\phi_{i-1}$, from point $p_{i,j}$ on the boundary of disc $D$, to point $z'_{\ell'(j)}$ on the boundary of disc $\tilde D_i$, where $1\leq \ell'(j)\leq \hat q_i$. Note that, if $u_i$ is synchronized with $u^*$, then $\ell'(j)=\ell(j)$ must hold, while otherwise $\ell'(j)=\hat q_i-\ell(j) +1$. We refer to the resulting curve $\hat \zeta'_{i,j}$ as the \emph{second segment in the drawing of edge $\hat a_{i,j}$}, and we denote $\hat Z_i'=\set{\hat \zeta'_{i,j}\mid 1\leq j\leq \hat q_i}$.

\paragraph{Computing the drawing $\psi_i$.}	
	Consider the current drawing $\phi_{i-1}$ of graph $G_{i-1}$. We slightly modify this drawing in order to obtain a drawing $\psi_i$ of graph $X_i$, and, for each edge $e\in \delta_{G_i}(X_i)\setminus \delta_{G_i}(u^*)$, a drawing of a segment of $e$ that is incident to its endpoint that lies in $X_i$.

	In order to do so, we start with the drawing $\phi_{i-1}$ of graph $G_{i-1}$, and we delete from it the images of all edges and vertices, except for those lying in $X'_i$. We will also make use of the disc $\tilde D_{i-1}$ that we have defined. Next, for all $1\leq j\leq q_i$, we consider the image of edge $a_{i,j}$ in the current drawing. Recall that this image intersects the boundary of $\tilde D_i$ at a single point, that we denote by $p''_{i,j}$. From the definition of the rotation $\oset'_{u_i}\in \Sigma'$, points $z'_{1},\ldots,z'_{\hat q_i},p''_{i,q_i},\ldots,p''_{i,2},p''_{i,1}$ appear on the boundary of disc $\tilde D_i$ in this order, and, if $u_i$ is synchronized with  $u^*$, then they are encountered in this order as we traverse the boundary of $\tilde D_i$ so its interior lies on our right; see \Cref{fig: image construction}.

	 Consider now some edge $a_{i,j}\in A_i$, for $1\leq j\leq q_i$, and assume that $a_{i,j}=(u_i,x_{i,j})$, where $x_{i,j}\in V(X_i)$. We denote by $\zeta'_{i,j}$ the segment of the image of edge $a_{i,j}$ from the image of $x_{i,j}$ to point $p''_{i,j}$ (see \Cref{fig: image construction}), and we refer to $\zeta'_{i,j}$ \emph{the second segment in the drawing of edge $e_{i,j}$}. We delete, from the current drawing, the portion of the image of $a_{i,j}$ lying in the interior of the disc $\tilde D_i$; in other words, we replace the image of $a_{i,j}$ with the curve $\zeta'_{i,j}$.
	
	Lastly, we add the curves in set $\hat Z'_i=\set{\hat \zeta'_{i,j}\mid 1\leq j\leq \hat q_i}$ to the current drawing. Recall that, for each 
	$1\leq j\leq \hat q$, curve $\hat \zeta'_{i,j}$ connects the image of vertex $\hat x_{i,j}$ (the endpoint of edge $\hat a_{i,j}\in \hat E_i$ that lies in $X_i$) to point $z'_{\ell'(j)}$ on the boundary of $\tilde D_i$, where $1\leq \ell'(j)\leq \hat q_i$.

	This completes the drawing $\psi_i$. 
	We now bound the number of crossings in this drawing.
	In order to bound the number of crossings, recall that every curve $\zeta'_{i,j}\in Z'_{i,j}$ is a segment of the image of edge $a_{i,j}\in A_{i,j}$, and every curve $\hat \zeta'_{i,j}\in \hat Z'_i$ is a concatenation of four curves: the image of the edge $\hat a_{i,j}\in \hat A_i$; the curve $\gamma'_{i,j}\in \Gamma'_i$; the curve $\rho_j$, and the image of the edge $a'_{i,\ell(j)}\in A'_i$. Since the curves in $\Gamma'_i$ are disjoint from each other, and the curves $\rho_1,\ldots,\rho_{\hat q_i}$ are disjoint from each other and are contained in disc $D$, the total number of crossings in $\psi_i$ is bounded by (i) the number of crossings between pairs of edges in $E(X'_i)\cup \hat A_i\cup A'_i$ (which is bounded by $\cro_i$ by definition); and (ii) the number of crossings between edges of $E(X'_i)\cup \hat A_i\cup A'_i$ and curves of $\Gamma'_i$ (which is bounded by $O(\cro_i)$ from the discussion above. We conclude that drawing $\psi_i$ has $O(\cro_i)$ crossings.

	We will view the interior of the disc $\tilde D_i$ as the ``outer face'' of the drawing $\psi_i$.  If we denote by $\tilde D'_i$ the disc in the sphere whose boundary is the same as the boundary of $\tilde D_i$, but its interior is disjoint from the interior of $\tilde D_i$, then the current drawing $\psi_i$ is contained in $\tilde D'_i$. To summarize, drawing $\psi_i$ consists of: (i) the drawing of all edges and vertices of $X_i\setminus\set{u^*}$; (ii) for every edge $a_{i,j}=(x_{i,j},u^*)\in E_i$, a curve $\zeta'_{i,j}$, connecting the image of $x_{i,j}$ to point $p''_{i,j}$ on the boundary of $\tilde D'_i$; and (iii) for every edge $\hat a_{i,j}=(\hat x_{i,j},\hat y_{i,j})\in \hat A_i$ with $\hat x_{i,j}\in X_i$, a curve $\hat \zeta'_{i,j}$, connecting the image of $\hat x_{i,j}$ to point $z'_{\ell'(j)}$ on the boundary of $\tilde D_i'$, where $1\leq \ell'(j)\leq \hat q_i$. Note that the points $p''_{i,1},\ldots,p''_{i,q_i},z'_{\hat q_i},\ldots,z'_{2},z'_{1}$ appear on the bounary of disc $\tilde D_i'$ in this order, and, if $u_i$ is synchronized with  $u^*$, then they are encountered in this order as we traverse the boundary of $\tilde D'_i$ so its interior lies to our right.

	\subsubsection{Stage 3: Computing the Drawing $\phi_i$ of $G_i$}
	
\begin{figure}[h]
	\centering
	\includegraphics[scale=0.105]{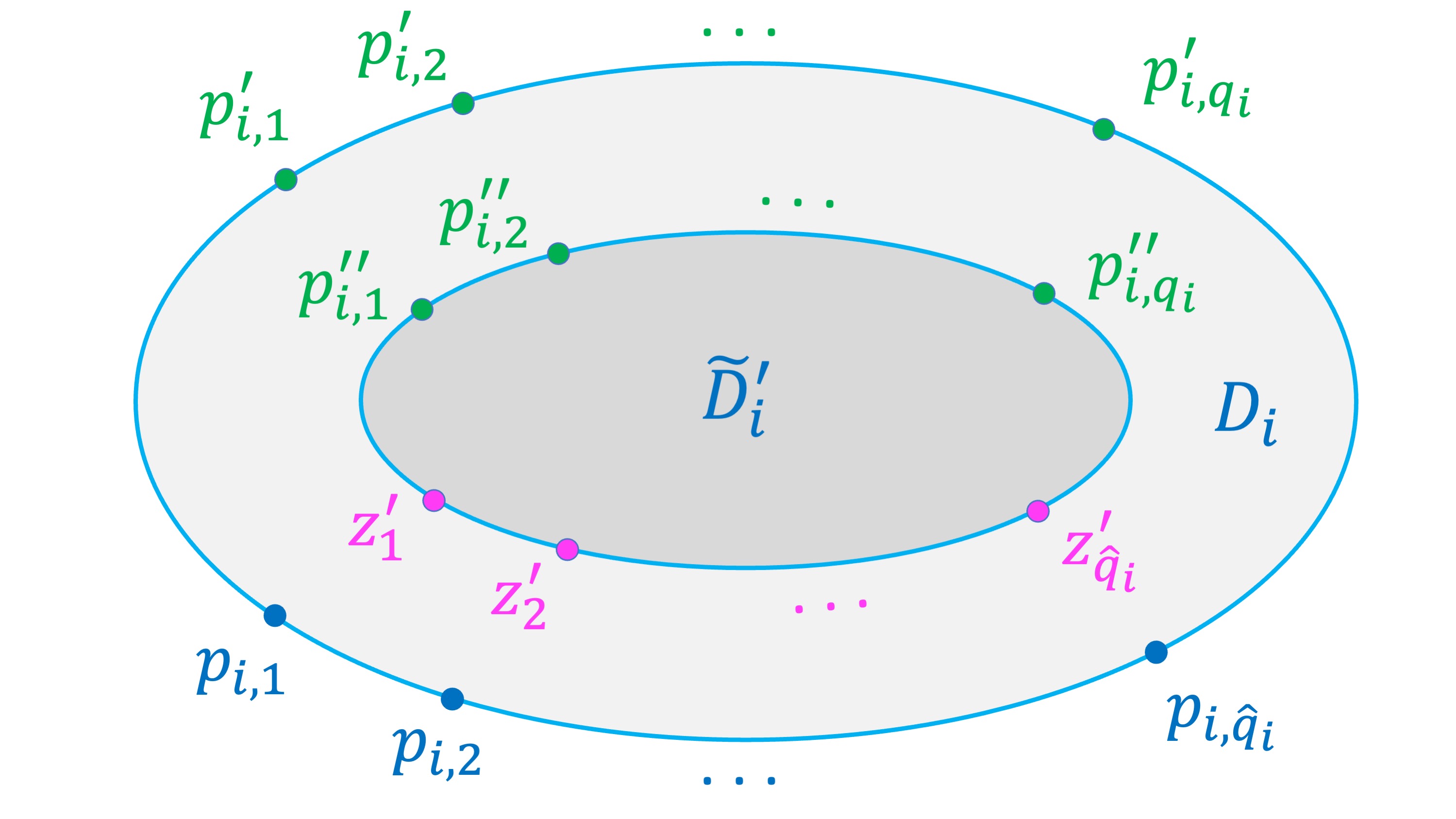}
	\caption{Planting disc $\tilde D'_i$ inside $D'_i$.
	}\label{fig: too_many_points_1}
\end{figure}

	We start with the drawing $\phi_{i-1}'$ that we computed in Stage 1. 
	We consider two cases. The first case is when vertex $u_i$ is synchronized with vertex $u^*$. In this case, we place the drawing $\psi_i$ that we computed in Stage 2 of the algorithm inside the disc $D'_i$, so that the discs $D'_i$ and $\tilde D'_i$ coincide. In the second case, vertex $u_i$ is not synchronized with vertex $u^*$. In this case, we place a mirror image of the drawing $\phi_i$ inside the disc $D'_i$, so that the boundaries of the discs $D'_i$ and (the mirror image of) disc $\tilde D'_i$ coincide (see \Cref{fig: too_many_points_1}). In either case, we obtain two disjoint segments on the boundary of $D'_i$: segment $\tilde \sigma_i$, containing the points $p''_{i,1},\ldots,p''_{i,q_i}$, and segment $\tilde  \sigma'_i$, containing the points $z'_1,\ldots,z'_{\hat q_i}$. Moreover, we are now guaranteed that points $p''_{i,1},\ldots,p''_{i,q_i},z'_{\hat q_i},\ldots,z'_{2},z'_{1}$  are encountered in this order as we traverse the boundary of $D_i'$, so that the interior of $D_i'$ lies to our 
	right (see \Cref{fig: too_many_points_1}). 
	Recall that we have defined a collection of points $\set{p'_{i,1}, p'_{i,2},\ldots,p'_{i, q_i},p_{i,\hat q_i},p_{i,\hat q_i-1},\ldots,  p_{i,1}}$ on the boundary of $D_i$, that appear in this order on the boundary of $D_i$, as we traverse it so that the interior of $D_i$ lies to our right (see Figures \ref{fig: segments} and \ref{fig: image construction}). Consider now some edge $e_{i,j}=(u^*,x_{i,j})\in E_i$, for $1\leq i\leq q_i$. Recall that we have already defined a curve $\zeta_{i,j}$, that serves as the first segment of the drawing of $e_{i,j}$, and connects the image of $u^*$ to point $p'_{i,j}$ on the boundary of $D_i$, such that curve $\zeta_{i,j}$ is internally disjoint from $D_i$. We have also defined a curve $\zeta'_{i,j}$ inside the disc $D'_i$, that connects the image of vertex $x_{i,j}$ to point $p''_{i,j}$ on the boundary of $D'_{i,j}$. 

\begin{figure}[h]
	\centering
	\includegraphics[scale=0.105]{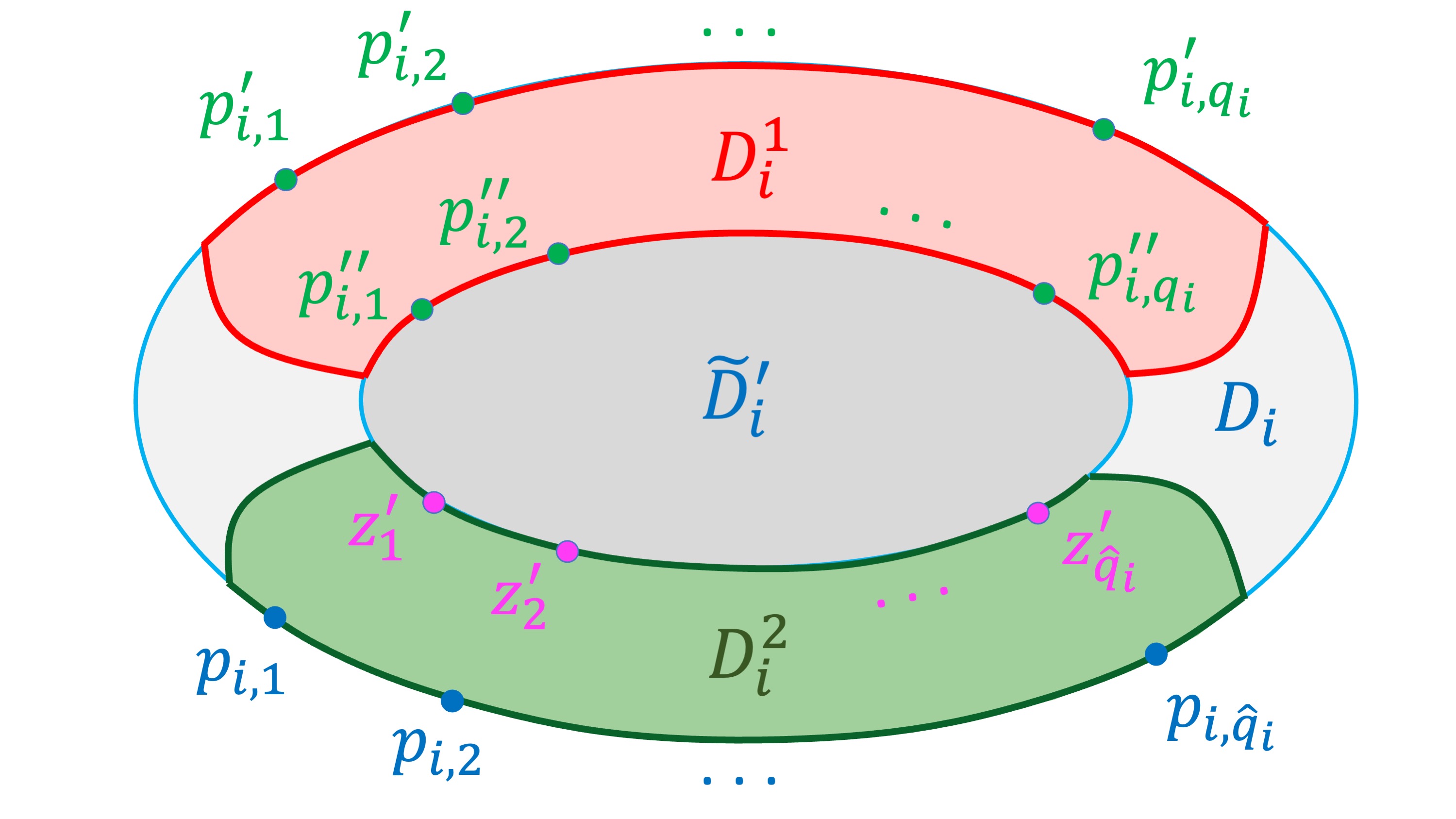}
	\caption{An illustration of discs $D^1_i$ and $D^2_i$.}\label{fig: too_many_points_2}
\end{figure}

	Recall that we have also defined another segment $\sigma'_i$ on the boundary of $D_i$, that contains points $p'_{i,1},\ldots, p'_{i, q_i}$, and a segment $\hat \sigma'_{i}$ on the boundary of $D'_i$, containing the points in $z'_1,\ldots,z'_{\hat q_i}$. We can then define two disjoint discs  that are both contained in $D_i\setminus D'_i$: one disc, $D^1_i$, with segments $\sigma_i$ and $\tilde \sigma_i$ on its boundary, and another disc, $D^2_i$, with segments $\tilde \sigma'_i,\sigma_i'$ on its boundary (see \Cref{fig: too_many_points_2}).

\begin{figure}[h]
	\centering
	\subfigure[An illustration of curves $\zeta_{i,1}.\ldots,\zeta_{i,q_i}$.]{
		\scalebox{0.42}{\includegraphics[scale=0.37]{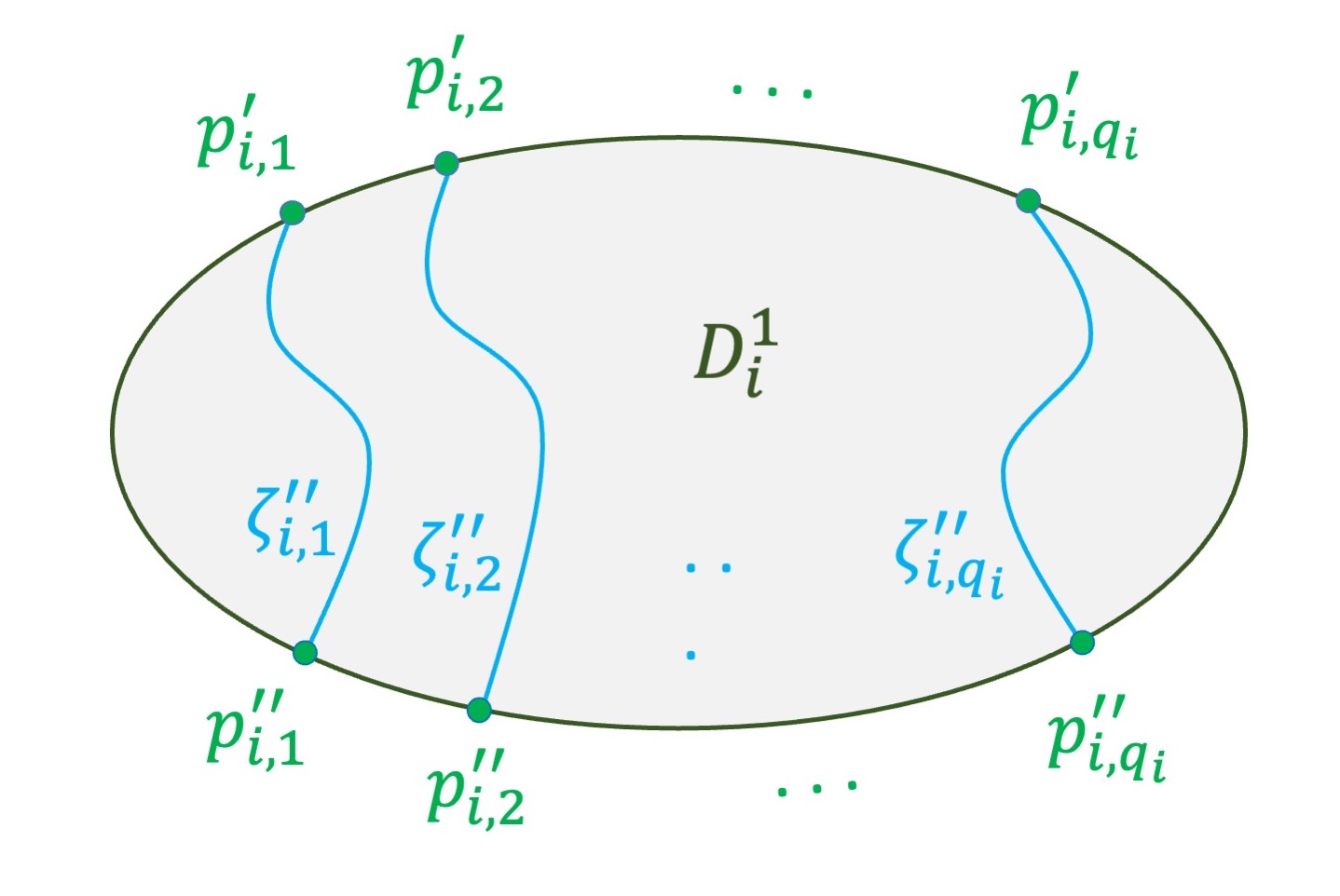}}\label{fig: pp_and_ppp_curves}}
	\hspace{0.5cm}
	\subfigure[An illustration of disc $D^2_i$ and points on its boundary.]{\scalebox{0.42}	{\includegraphics[scale=0.37]{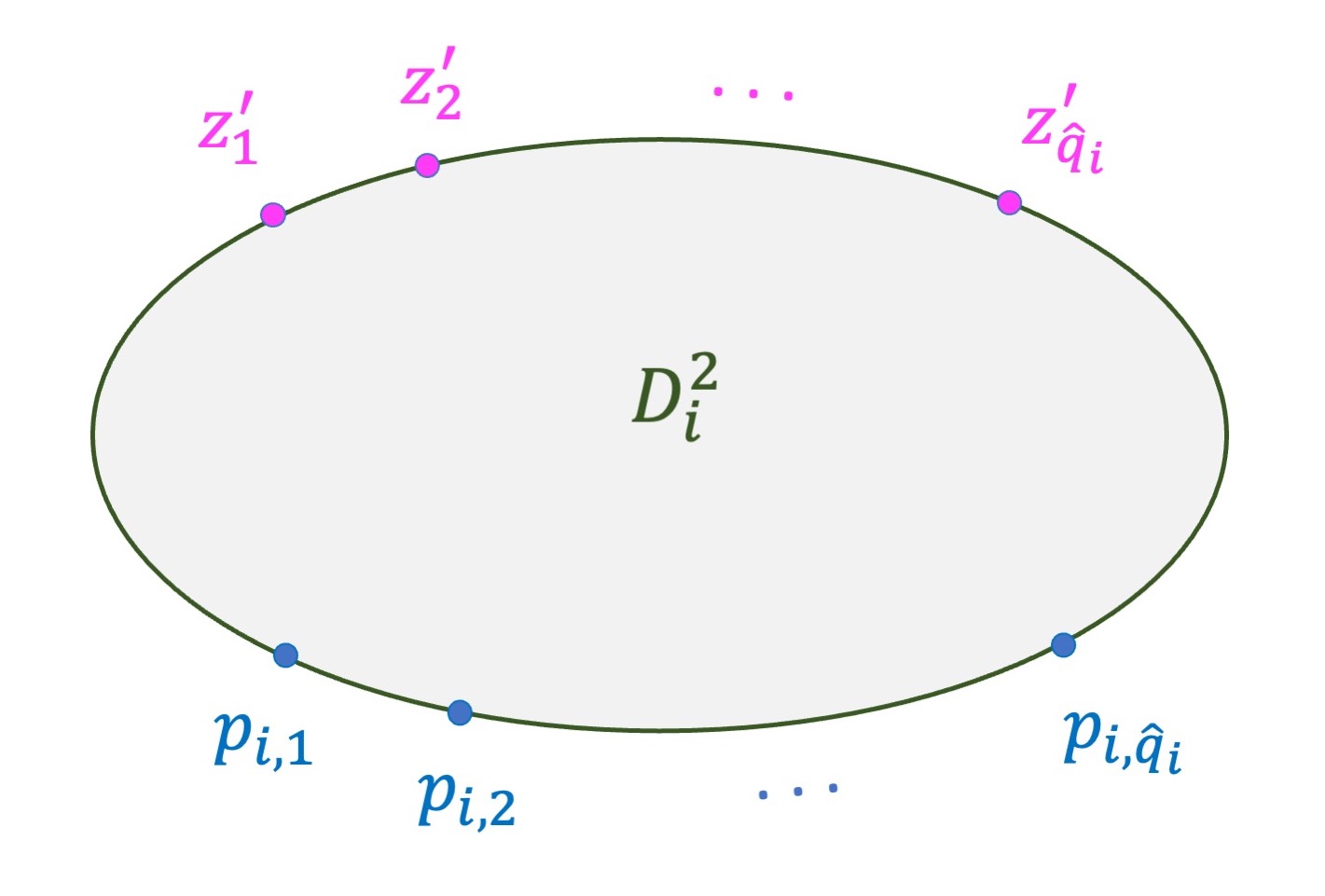}\label{fig: p_and_zp_curves}}
	}
\caption{Stitching the images of the edgs of $E_i$ and $\hat E_i$.
}
\end{figure}

	Observe that points $\set{p'_{i,1}, p'_{i,2},\ldots,p'_{i,q_i},p''_{i,q_i},\ldots,p''_{i,1}}$ appear in this circular order on the boundary of disc $D^1_i$. Therefore, we can define, for all $1\leq j\leq q_i$, a curve $\zeta''_{i,j}$, connecting point $p'_{i,j}$ to point $p''_{i,j}$, such that the interior  of the curve is contained in disc $D^1_i$, and all curves in $\set{\zeta''_{i,j}\mid 1\leq j\leq q_i}$ are mutually disjoint (see \Cref{fig: pp_and_ppp_curves}).
	For all $1\leq j\leq q_i$, we then let the image of the edge $e_{i,j}$ be the concatenation of the curves $\zeta_{i,j}$, $\zeta''_{i,j}$, and $\zeta'_{i,j}$.
	
	Lastly, it remains to complete the drawing of the edges of $\hat A_i$.
	Recall that for every edge $\hat a_{i,j}=(\hat x_{i,j},\hat y_{i,j})$ (where $\hat x_{i,j}\in X_i$), we have already defined two segments of the drawing of $\hat a_{i,j}$. The first segment, $\hat \zeta_{i,j}$, is internally disjoint from disc $D$, and connects the image of $\hat y_{i,j}$ to point $p_{i,j}$ on the boundary of $D_i$. The second segment, $\hat \zeta'_{i,j}$, is contained in disc $D'_i$, and connect the image of $\hat x_{i,j}$ to point $z'_{\ell'(j)}$ on the boundary of disc $D'_{i}$.  In order to complete the drawing of edge $\hat a_{i,j}$, we will define a third curve, $\hat \zeta''_{i,j}$, that is contained in disc $D_i^2$, and connects points $p_{i,j}$ and $z'_{\ell'(j)}$ to each other. See \Cref{fig: p_and_zp_curves} for an illustration.

	We will use the following observation in order to complete the drawing.
	
	\begin{observation}\label{obs: completing the drawing}
		There is an efficient algorithm to compute a collection $\set{\hat \zeta''_{i,j}\mid 1\leq j\leq \hat q_i}$ of curves that are contained in disc $D_i^2$, such that, for all $1\leq j\leq \hat q_i$, curve $\zeta''_{i,j}$ connects points $p_{i,j}$ and $z'_j$, and the number of crossings between the curves in $\set{\hat \zeta''_{i,j}\mid 1\leq j\leq \hat q_i}$ is at most $O(\cro_i)$.
	\end{observation}

\begin{proof}
	We consider two cases. The first case is when $u^*$ and $u_i$ are not synchonized. In this case, we let  $\set{\hat \zeta''_{i,j}\mid 1\leq j\leq \hat q_i}$ of curves that are contained in disc $D_i^2$, such that, for all $1\leq j\leq \hat q_i$, curve $\zeta''_{i,j}$ connects points $p_{i,j}$ and $z'_j$, and every pair of curves cross at most once. In this case, the number of crossings between the curves in $\set{\hat \zeta''_{i,j}\mid 1\leq j\leq \hat q_i}$ is at most $\hat q_i^2$. Since, from \Cref{obs: not synchronized}, there were at least $\hat q_i^2/8$ crossings $(e,e')$ in $\phi'$ with $e,e'\in A_i'$, we get that $\cro_i\geq \Omega(\hat q_i^2)$, and so the  the number of crossings between the curves in $\set{\hat \zeta''_{i,j}\mid 1\leq j\leq \hat q_i}$ is at most $O(\cro_i)$ as required.
	
	In the second case, $u^*$ and $u_i$ are synchronized. 
Recall that in Stage 2 of the algorithm, in \Cref{obs: curves of gamma star}, we have constructed a collection $\Gamma^*_i=\set{\gamma^*_{i,j}\mid 1\leq j\leq \hat q_i}$ of curves, such that, for all $1\leq j\leq \hat q_i$, curve $\gamma^*_{i,j}$ connects point $p_{i,j}$ on the boundary of disc $D$ to point $z_{\ell(j)}$, and it is internally disjoint from disc $D$; the total number of crossings between the curves of $\Gamma^*_i$ is $O(\cro_i)$. Recall that points  $z_1,z_2,\ldots,z_{\hat q_i},p_{i,\hat q_i},\ldots,p_{i,1}$ that appear on the boundary of disc $D$ in this order, while points $z'_1,z'_2,\ldots,z'_{\hat q_i},p_{i,\hat q_i},\ldots,p_{i,1}$ appear on the boundary of disc $D^2_i$ in this order. The key point is that, as observed in Stage 2 of the algorithm, if vertex $u_i$ is synchronized with vertex $u^*$, then for all $1\leq j\leq \hat q_i$, $\ell(j)=\ell(j')$. Therefore, we can copy the collection $\Gamma^*_i$ of curves to the interior of the disc $D^2_i$, such that, for all $1\leq j\leq \hat q_i$, one of the resulting curves, that we denote by $\hat \zeta''_{i,j}$ connects $p_{i,j}$ to $z'_{\ell'(j)}=z'_{\ell(j)}$.
\end{proof}

\subsubsection{Analysis}

We now show that, assuming that Invariants \ref{inv: only drawing for Xi changes}--\ref{inv: small increase in crossings} hold at the beginning of the $i$th iteration, they continue to hold at the end of the iteration. Indeed, it is immediate to see that we only change the images of vertices and edges of $X'_i$, and $A'_i\cup \hat A_i$, which establishes Invariant \ref{inv: only drawing for Xi changes}. Consider now some edge 
$e\in E(G_{i-1})\setminus (E(X'_i)\cup A'_i\cup\hat A_i)$, and some other edge $e'$ that crosses $e$ in $\phi_i$. If $e'$ is not an edge of $E(X_i)\cup\hat A_i$, then the image of $e'$ was not changed in the current iteration, and the crossing lies in $\phi_{i-1}$ as well. Notice that $e$ may not cross edges of $E(X_i)$, as for each such edge $e'$, either $e'$ is drawn inside disc $D'_i$, or $e'\in E_i$, so the first segemnt of $e'$ is some curve $\zeta_{i,j}$ (that is contained in $D$), and the remainder of the image of $e'$ is contained in $D'_i$. Assume now that $e'\in \hat A_i$. In this case, only the first segment of the drawing $e'$, which is a segment of some curve in $\Gamma_i$, may cross edge $e$, as the remainder of the image of $e'$ lies in disc $D$. Overall, the number of crossings in which edge $e$ participates in drawing $\phi_i$ is bounded by the number of crossings in which edge $e$ participates in drawing $\phi'_{i-1}$ (that was defined in Stage 1), which is in turn bounded by the number of crossings in which $e$ participates in $\phi_{i-1}$. We conclude that Invariant \ref{inv: num of crossings does not increase} continues to hold. Lastly, it remains to establish Invariant \ref{inv: small increase in crossings}. From the above discussion, for each edge $e\in E(G_{i-1})\setminus (E(X'_i)\cup A'_i\cup\hat A_i)$, the number of crossings in which $e$ participates in drawing $\phi_i$ is bounded by the number of crossings in which $e$ participates in drawing $\phi_{i-1}$. From our analysis of Stage 2, the number of crossings in $\psi_i$ is bounded by $O(\cro_i)$. Recall that the curves of $\Gamma_i$ cannot cross each other, and they are internally disjoint from disc $D$. The only additional crossings that we introduced are the crossings between the curves of 
$\set{\hat \zeta''_{i,j}\mid 1\leq j\leq \hat q_i}$ that were computed in \Cref{obs: completing the drawing}; from the observation, the number of such crossings is  at most $O(\cro_i)$.
This establishes Invariant \ref{inv: small increase in crossings}.

Overall, after $k$ iterations, we obtain a drawing $\phi_k$ of graph $G_k$ with $O(\cro(\phi')+\sum_{i=1}^k\cro_i)\leq O(\cro(\phi'))$ crossings. Since graph $G_k$ can be obtained from graph $G$ by subdividing some of its edges, this immediately provides a drawing of graph $G$ with $O(\cro(\phi'))$ crossings. From our construction it is immediate to verify that the resulting drawing obeys the rotation system $\Sigma$, so we obtain a feasible solution to instance $I$ of \cnwrs.

\section{Proofs Omitted from \Cref{sec: guiding paths}}
\label{sec: appx guiding paths}


\subsection{Proof of \Cref{lem: routing path extension}}
\label{apd: Proof of routing path extension}

Denote $z=\ceil{\frac{|T|}{|T'|}}$. We arbitrarily partition the vertices of $T\setminus T'$ into $(z-1)$ subsets $T_1,\ldots,T_{z-1}$ of cardinality at most $|T'|$ each.  
Consider some index $1\leq i\leq z-1$. Since vertices of $T$ are $\alpha$-well-linked in $G$, using the algorithm from \Cref{thm: bandwidth_means_boundary_well_linked}, we can compute a collection $\qset'_i$ of paths in graph $G$, routing vertices of $T_i$ to vertices of $T'$, such that $\cong_G(\qset'_i)\le \ceil{1/\alpha}$, and each vertex of $T'\cup T_i$ is the endpoint of at most one path in $\qset'_i$. 
By concatenating the paths in $\qset'_i$ with paths in $\pset$, we obtain a collection $\qset_i$ of paths in graph $G$ routing vertices of $T_i$ to vertex $x$. For every edge $e\in E(G)$, $\cong_G(\qset_i,e)\leq \ceil{1/\alpha}+\cong_G(\pset,e)$. 
Lastly, we set $\pset'=\bigcup_{i=1}^{z-1}\qset_i$.
It is clear that the paths in $\pset$ route the vertices of $T$ to $x$.
Moreover, for every edge $e\in E(G)$,
\[
\cong_G(\pset',e) \le \sum_{i=1}^z\cong_G(\qset_i,e)\\
\le \ceil{\frac{|T|}{|T'|}}\bigg(\cong_G(\pset,e)+\ceil{1/\alpha}\bigg).
\]

\subsection{Proof of \Cref{lem: splitting}}
\label{sec: splitting}


We denote $\talpha'=\frac{\talpha}{c\log^2m}$, where $c$ is a large enough constant whose value we set later.
Throughout the algorithm, we will maintain a collection $\wset$ of disjoint clusters of $G\setminus T$. We will ensure that this collection $\wset$ of clusters has some useful properties, that are summarized in the following definition.

\begin{definition}
	A collection $\wset$ of disjoint clusters of $G\setminus T$ is a \emph{legal clustering of $G$} if the following hold:
	\begin{itemize}
		\item $\bigcup_{W\in \wset}V(W)=V(G)\setminus T$;
		\item every cluster $W\in \wset$ has the $\talpha'$-bandwidth property; and
		\item for every cluster $W\in \wset$, $|\delta_G(W)|\leq \talpha k/64$.
	\end{itemize}
\end{definition}

Given a legal clustering $\wset$ of $G$, we associate with it a contracted graph $\hat G=G_{|\wset}$; recall that $\hat G$ is obtained from graph $G$ by contracting every cluster $W\in \wset$ into a supernode $v_W$; we keep parallel edges but delete self loops.  Observe that from the definition of a legal clustering, the only regular (non-supernode) vertices of $\hat G$ are the vertices of $T$. 
Given a legal clustering $\wset$ of $G$, we denote by $\heout(\wset)$ the set of all edges $(u,v)$ of $G$, where $u$ and $v$ belong to different clusters of $\wset$; equivalently, $\heout(\wset)=E(\hat G\setminus T)$. We will not distinguish between the edges of $\heout(\wset)$ and the edges of $\hat G\setminus T$.

We need the following simple claim, whose analogues were proved in 
\cite{chuzhoy2012routing,chuzhoy2012polylogarithmic,chekuri2016polynomial,chuzhoy2016improved}.

\begin{claim}\label{claim: contracted graph lots of edges}
	Let $\wset$ be a legal clustering of $G$. If the set $T$ of terminals is $\talpha$-well-linked in $G$, then 
$|\heout(\wset)|\geq \talpha k/4$.
\end{claim}

\begin{proof}
	For every cluster $W\in \wset$, let $T_W\subseteq T$ be the set containing every terminal $t\in T$, such that, if $e=(t,v)$ is the unique edge incident to $t$ in $G$, then $v\in W$. Denote $n_W=|T_W|$. Since we are guaranteed that for every cluster $W\in \wset$, $|\delta_G(W)|\leq \talpha k/64$, $n_W\leq \talpha k/64$ must hold. Note that there is a partition $\wset_1,\wset_2$ of $\wset$, such that $\sum_{W\in \wset_1}n_W,\sum_{W\in \wset_2}n_W\geq k/4$. Indeed, we can compute such a partition using a simple greedy algorithm: start with $\wset_1,\wset_2=\emptyset$, and process the clusters $W\in \wset$ one by one. When cluster $W\in \wset$ is processed, we add it to $\wset_1$ if $\sum_{W\in \wset_1}n_W<\sum_{W\in \wset_2}n_W$, and we add it to $\wset_2$ otherwise. We are guaranteed that at the end of this procedure, $\left |\sum_{W\in \wset_1}n_W-\sum_{W\in \wset_2}n_W\right |\leq \max_{W\in \wset}\set{n_W}\leq \talpha k/64$ holds. It is then immediate to verify that $\sum_{W\in \wset_1}n_W,\sum_{W\in \wset_2}n_W\geq k/4$.
	
	We construct a partition $(X,Y)$ of $V(G)$, where $X=\bigcup_{W\in \wset_1}(V(W)\cup T_W)$, and $Y=\bigcup_{W\in \wset_2}(V(W)\cup T_W)$. Then $|X\cap T|,|Y\cap T|\geq k/4$, and, since we have assumed that the set $T$ of terminals is $\talpha$-well-linked in $G$, $|E_G(X,Y)|\geq \talpha k/4$. Notice that every edge in $E_G(X,Y)$ connects a pair of vertices lying in different clusters of $\wset$, so $|\heout(\wset)|\geq |E_G(X,Y)|\geq \talpha k/4$.
\end{proof}

The following lemma is key to the proof of \Cref{lem: splitting}.

\begin{lemma}\label{lem: better clustering}
There is an efficient algorithm, that, given a legal clustering $\wset$ of $G$ with $|\heout(\wset)|\geq \talpha k/4$, either produces another legal clustering $\wset'$ of $G$ with $|\heout(\wset')|<|\heout(\wset)|$, or it computes two disjoint subgraphs $C_1,C_2$ of $G$, each of which has the $\talpha'$-bandwidth property, such that, for all $i\in \set{1,2}$, there is a set $\rset_i$ of at least $\Omega(\talpha^2k/\log^2m)$ edge-disjoint paths in $G$, routing a subset $T_i\subseteq T$ of terminals to the edges of $\delta_G(C_i)$.
\end{lemma}

We prove \Cref{lem: better clustering} in the following subsection, after we complete the proof of  \Cref{lem: splitting} using it.
Throughout the algorithm, we maintain a legal clustering $\wset$ of $G$. If, at any point in the algorithm's execution, we obtain a legal clustering $\wset$ with $|\heout(\wset)|<\talpha k/4$, then, from \Cref{claim: contracted graph lots of edges}, the set $T$ of terminals is not $\talpha$-well-linked in $G$. We then terminate the algorithm and return FAIL. Therefore, from now on we assume that every legal clustering $\wset$ that the algorithm obtains has $|\heout(\wset)|\geq \talpha k/4$.

We start with an initial collection $\wset$ of clusters of $V(G)\setminus T$, where for every vertex $v\in V(G)\setminus T$, we add a cluster $\set{v}$ to $\wset$. It is easy to verify that $\wset$ is a legal clustering of $G$, since the degree of every 
vertex in $G$ is guaranteed to be at most $\talpha k/64$. We then perform a number of iterations. In every iteration, we apply the algorithm from \Cref{lem: better clustering} to the current legal clustering $\wset$. If the outcome of the algorithm is another legal clustering $\wset'$ of $G$ with $|\heout(\wset')|<|\heout(\wset)|$, then we replace $\wset$ with $\wset'$ and continue to the next iteration. Assume now that the outcome of the algorithm from \Cref{lem: better clustering} is a pair $C_1,C_2$ of disjoint subgraphs of $G$, each of which has the $\talpha'$-bandwidth property, such that for all $i\in \set{1,2}$, there is a set $\rset_i$ of at least $\Omega(\talpha^2k/\log^2m)$ edge-disjoint paths in $G$, routing a subset $T_i\subseteq T$ of terminals to the edges of $\delta_G(C_i)$.
In this case, we let $(X,Y)$ be a partition of $V(G)$ with $V(C_1)\subseteq X$ and $V(C_2)\subseteq Y$, that minimizes $|E_G(X,Y)|$ among all such partitions. Notice that such a partition $(X,Y)$ can be computed via a standard minimum $s$-$t$ cut computation. Moreover, from the Maximum Flow / Minimum Cut theorem, we are guaranteed that there is a collection $\qset$ of $|E_G(X,Y)|$ edge-disjoint paths, each of which connects a vertex of $V(C_1)$ to a vertex of $V(C_2)$; we can assume w.l.o.g. that each path in $\qset$ is simple and it does not contain vertices of $V(C_1)\cup V(C_2)$ as inner vertices. Notice that every path in $\qset$ must contain exactly one edge of the set $E'=E_G(X,Y)$, and every edge of $E'$ must lie on exactly one such path. For every path $Q\in \qset$, we denote by $Q_1$ the subpath of $Q$ from its endpoint that lies in $C_1$ to an edge of $E'$, and we denote by $Q_2$ the subpath of $Q$ from an edge of $E'$ to a vertex of $C_2$. Let $\qset_1=\set{Q_1\mid Q\in \qset}$ and $\qset_2=\set{Q_2\mid Q\in \qset}$. Then $\qset_1$ is a set of edge-disjoint paths routing the edges of $E'$ to edges of $\delta_G(C_1)$, in graph $G[X]\cup E'$, and similarly, $\qset_2$ is a set of edge-disjoint paths routing the edges of $E'$ to edges of $\delta_G(C_2)$ in graph $G[Y]\cup E'$.
We now show that the partition $(X,Y)$ of $V(G)$ has all required properties.

Assume w.l.o.g. that $|X\cap T|\geq |Y\cap T|$.
 Recall that there is a set $\rset_2$ of  at least $\Omega(\talpha^2k/\log^2m)$ edge-disjoint paths in $G$, routing a subset $T_2\subseteq T$ of terminals to the edges of $\delta_G(C_2)$. Assume first that $|T_2\cap X|\geq |T_2|/2$. Let $\rset_2'\subseteq \rset_2$ be the set of paths whose endpoint lies in $T_2\cap X$, so $|\rset_2'|\geq |\rset_2|/2\geq \Omega(\talpha^2k/\log^2m)$. Then each path  $R\in \rset_2'$ connects a vertex of $T_2\cap X$ to a vertex of $Y$, so it must contain an edge of $E_G(X,Y)$. By suitably truncating each such path $R\in \rset_2'$, we obtain a collection $\rset$ of $\Omega(\talpha^2k/\log^2m)$ edge-disjoint paths, routing the terminals of $T_2\cap X$ to the edges of $E_{G}(X,Y)$.
 
 Assume now that $|T_2\cap X|<|T_2|/2$. Let $h=\ceil{|T_2|/2}$. Then $|X\cap T|,|Y\cap T|\geq h$ must hold. We apply the algorithm from \Cref{thm: bandwidth_means_boundary_well_linked} to graph $G$ and two arbitrary subsets $T_1'\subseteq X\cap T, T'_2\subseteq Y\cap T$ of terminals, of cardinality $h$ each. 
 If the set $T$ of terminals is $\talpha$-well-linked in $G$, the algorithm must return a collection $\rset'$ of paths in graph $G$, such that $\rset'$ is an one-to-one routing of vertices of $T_1'$ to vertices of $T_2'$, and $\cong_G(\rset')\leq \ceil{1/\talpha}$. If the algorithm fails to return such a collection of paths, then we are guaranteed that the set $T$ of terminals is not $\talpha$-well-linked in $G$. We then terminate the algorithm and return FAIL. Therefore, we assume from now on that the algorithm from   \Cref{thm: bandwidth_means_boundary_well_linked} returned a collection $\rset'$ of paths with  $\cong_G(\rset')\leq \ceil{1/\talpha}$, such that $\rset'$ is an one-to-one routing of $T_1'$ to $T_2'$. From \Cref{claim: remove congestion}, there is a collection $\rset''$ of at least $\Omega(h\talpha)=\Omega(\talpha^3k/\log^2m)$ edge-disjont paths in graph $G$, routing a subset  $T''_1\subseteq T'_1$ of terminals to a subset $T''_2\subseteq T'_2$ of terminals. Each path $R\in \rset''$ must then contain an edge of $E_G(X,Y)$. By suitably truncating each such path, we obtain a collection $\rset$ of $\Omega(\talpha^3k/\log^2m)$ edge-disjoint paths, routing the terminals of $T''_1$ to the edges of $E_G(X,Y)$.

It is now enough to prove that each of the clusters $G[X],G[Y]$ has the $\talpha'/2$-bandwidth property, which we do in the following claim.

\begin{claim}
	Each of the clusters $G[X],G[Y]$ has the $\talpha'/2$-bandwidth property.
\end{claim}

\begin{proof}
We show this for $G[X]$; the proof for $G[Y]$ is symmetric.

Assume for contradiction that $G[X]$ does not have the $\talpha'/2$-bandwidth property. Then there must be a partition $(A,B)$ of $X$, such that $|E_G(A,B)|<\talpha'\cdot \min\set{|\delta_G(X)\cap \delta_G(A)|,|\delta_G(X)\cap \delta_G(B)|}/2$. We assume w.l.o.g. that $|\delta_G(X)\cap \delta_G(A)|\leq |\delta_G(X)\cap \delta_G(B)|$, and we denote $|\delta_G(X)\cap \delta_G(A)|$ by $r$.

Note that partition $(A,B)$ of $X$ naturally defines a partition $(A',B')$ of $V(C_1)$, with $A'=A\cap V(C_1)$ and $B'=B\cap V(C_1)$. Since cluster $C_1$ has the $\talpha'$-bandwidth property, while $|E_{C_1}(A',B')|\leq |E_G(A,B)|< \talpha' r/2$,
either $|\delta_G(A')\cap \delta_G(C)|<r/2$, or $|\delta_G(B')\cap \delta_G(C_1)|<r/2$ must hold. Assume w.l.o.g. that it is the former. Recall that $|\delta_G(X)\cap \delta_G(A)|= r$, and there is a set $\qset_1$ is a set of edge-disjoint paths routing the edges of $E'=\delta_G(X)$ to edges of $\delta_G(C_1)$. Let $\qset'\subseteq \qset_1$ be the paths that originate at edges of $|\delta_G(X)\cap \delta_G(A)|$, so $|\qset'|\geq r$. Recall that each path in $\qset'$ terminates at an edge of $\delta_G(C_1)$. However, since  $|\delta_G(A')\cap \delta_G(C_1)|<r/2$, at least $r/2$ of the paths in $\qset'$ must contain a vertex of $B$. Therefore, at least $r/2$ of the paths in $\qset'$ contain an edge of $E_G(A,B)$, and so $|E_G(A,B)|\geq r/2$, a contradiction.
\end{proof}

\subsection*{Proof of \Cref{lem: better clustering}}
We need the following claim, which is a constructive version of Lemma 5.8 from \cite{chuzhoy2016improved}; the proof is almost identical to that in  \cite{chuzhoy2016improved} and is included here for completeness.

\begin{claim}\label{claim: partition into two}
There is an efficient algorithm that, given any graph $G'$ with maximum vertex degree at most $\Delta$, computes a partition $(A,B)$ of $V(G')$, with $|E(A)|,|E(B)|\geq \frac{|E(G')|}{4}-\Delta$.
\end{claim}

\begin{proof}
	For every vertex $v\in V(G')$, let $d(v)$ denote its degree in $G'$. For a subset $S\subseteq V(G')$ of vertices, let $\vol(S)=\sum_{v\in S}d(v)$. 
	We start by computing an initial partition $(A,B)$ of $V(G')$, with $|\vol(A)-\vol(B)|\leq \Delta$, using a simple greedy algorithm: start with $A=B=\emptyset$, and process the vertices of $G'$ one-by-one. When $v$ is processed, add it to $A$ if $\vol(A)<\vol(B)$ currently holds, and add it to $B$ otherwise. It is easy to see that at the end of this procedure, we obtain a partition $(A,B)$ of $V(G')$ with $|\vol(A)-\vol(B)|\leq \Delta$.
	
	We then iterate. The input to iteration $i$ is a partition $(A_i,B_i)$ of $V(G')$ with $|\vol(A_i)-\vol(B_i)|\leq 2\Delta$, where the input to the first iteration is the partition $(A_1,B_1)=(A,B)$ that we have just computed. We assume w.l.o.g. that $\vol(A_i)\geq \vol(B_i)$ holds, so $|E(A_i)|\geq |E(B_i)|$. If $|E(B_i)|<  \frac{|E(G')|}{4}-\Delta$, then the outcome of the $i$th iteration is a partition 
	$(A_{i+1},B_{i+1})$ of $V(G')$ with $|\vol(A_{i+1})-\vol(B_{i+1})|\leq 2\Delta$, and $|E(A_{i+1},B_{i+1})|<|E(A_{i},B_i)|$; otherwise, the algorithm terminates. In the latter case, we get that $|E(A_i)|\geq|E(B_i)|\geq \frac{|E(G')|}{4}-\Delta$, as required.

	We now describe the execution of the $i$th iteration, whose input is  is a partition $(A_i,B_i)$ of $V(G')$ with $|\vol(A_i)-\vol(B_i)|\leq 2\Delta$, such that $\vol(A_i)\geq \vol(B_i)$  and $|E(B_i)|<  \frac{|E(G')|}{4}-\Delta$ hold.
%
%
	
	For every vertex $v\in A_i$, let $d_1(v)$ be the number of edges incident to $v$ whose other endpoint belongs to $A_i$, and let $d_2(v)$ be the number of edges incident to $v$ whose other endpoint belongs to $B_i$. As we show later, there must exist a vertex $v\in A_i$ with $d_1(v)< d_2(v)$. Let $v$ be any such vertex. We then define a new partition $(A_{i+1},B_{i+1})$ of $V(G')$ as follows: $A_{i+1}=A_i\setminus\set{v}$ and $B_{i+1}=B_i\cup\set{v}$. It is easy to verify that $|E(A_{i+1},B_{i+1})|<|E(A_{i},B_i)|$, while:
	 $$|\vol(A_{i+1})-\vol(B_{i+1})|\leq \max\set{|\vol(A_{i})-\vol(B_{i})|,2d(v)}\leq 2\Delta.$$ 
	 
	 We then output the partition $(A_{i+1},B_{i+1})$ of $V(G')$ and terminate the iteration.
	
	It now remains to show that, if $|E(B_i)|<  \frac{|E(G')|}{4}-\Delta$, there must exist a vertex $v\in A_i$ with $d_1(v)< d_2(v)$.
	Indeed, assume for contradiction that for every vertex $v\in A_i$, $d_1(v)\geq d_2(v)$. 
	
	Then $|E(A_i)|= \half\sum_{v\in A_i}d_1(v)\geq \half\sum_{v\in A_i}d_2(v)=\half |E(A_i,B_i)|$.
	Altogether, $|E(G')|=|E(A_i)|+|E(B_i)|+|E(A_i,B_i)|\leq 4|E(A_i)|$, and so $|E(A_i)|\geq |E(G')|/4$.
	
	On the other hand: 
	
	\[|E(B_i)|= \frac {\vol(B_i)-|E(A_i,B_i)|} 2 \geq \frac {\vol(A_i)-2\Delta-|E(A_i,B_i)|} 2 \geq \frac{2|E(A_i)|-2\Delta}2\geq \frac{|E(G')|}{4}-\Delta,\]

	a contradiction to our assumption that $|E(B_i)|<\frac{|E(G')|}4-\Delta$.
	
	The algorithm terminates once we obtain a partition $(A_i,B_i)$ of $V(G')$ with $|E(A_i)|,|E(B_i)|\geq \frac{|E(G')|}{4}-\Delta$. From the above discussion, this is guaranteed to happen after at most $|E(G')|$ iterations. Since each iteration can be executed efficiently, the claim follows.
\end{proof}

We apply the algorithm from \Cref{claim: partition into two} to graph $\hat G'=\hat G\setminus T$. Recall that, since for every cluster $W\in \wset$, $|\delta_G(W)|\leq \talpha k/64$, every vertex in graph $\hat G'$ has degree at most $\talpha k/64$. Moreover, 
$|E(\hat G')|=|\heout(\wset)|\geq \talpha k/4$. Therefore, we obtain a partition $(A,B)$ of $V(\hat G')$ with $|E_{\hat G'}(A)|,|E_{\hat G'}(B)|\geq |E(\hat G')|/4-\talpha k/64\geq |E(\hat G')|/8\geq \talpha k/32$.

Let $\wset_1$ be the set of all clusters $W\in \wset$ with $v_W\in A$, and let $\wset_2$ be the set of all clusters $W\in \wset$ with $v_W\in B$. Clearly, $(\wset_1,\wset_2)$ is a partition of $\wset$. Our next step is summarized in the following claim.

\begin{claim}\label{claim: processing one side}
	There is an efficient algorithm, that, given any subset $\cset\subseteq \wset$ of clusters, such that the total number of edges $e=(u,v)\in E(G)$ where $u$ and $v$ lie in distinct clusters of $\cset$ is at least $|E(\hat G')|/8$, outputs one of the following:
	
	\begin{itemize}
		\item either a legal clustering $\wset'$ of $G$ with $|\heout(\wset')|<|\heout(\wset)|$; or
		\item a cluster $C$ with $V(C)\subseteq \bigcup_{C'\in \cset}V(C')$, such that $C$ has the $\talpha'$-bandwidth property, and there exists a collection $\rset$ of at least $\talpha \cdot \talpha' k/256$ edge-disjoint paths in $G$ routing a subset of terminals to the edges of $\delta_G(C)$.
	\end{itemize}
\end{claim}

Observe that \Cref{claim: processing one side} finishes the proof of \Cref{lem: better clustering}, as follows. 
Let $A'=\bigcup_{W\in \wset_1}V(W)$, and let $B'=\bigcup_{W\in \wset_2}V(W)$; clearly, $A'\cap B'=\emptyset$.
We aply the algorithm from \Cref{claim: processing one side} to the set $\wset_1$ of clusters, and then separately to the set $\wset_2$ of clusters. If the outcome of any of the two algorithms is a legal  clustering $\wset'$ of $G$ with $|\heout(\wset')|<|\heout(\wset)|$, then we return this legal clustering $\wset'$ and terminate the algorithm. Therefore, we assume from now on that the outcome of the algorithm from  \Cref{claim: processing one side} when applied to cluster set $\wset_1$ is a cluster $C_1$ with $V(C_1)\subseteq A'$, such that $C_1$ has the $\talpha'$-bandwidth property, and there exists a collection $\rset_1$ of  at least $\talpha \cdot \talpha' k/256=\Omega(\talpha^2k/\log^2m)$ edge-disjoint paths in $G$ routing some subset $T_1\subseteq T$ of terminals to the edges of $\delta_G(C_1)$. Similarly, the outcome of the algorithm from  \Cref{claim: processing one side} when applied to cluster set $\wset_2$ is a cluster $C_2$ with $V(C_2)\subseteq B'$, such that $C_2$ has the $\talpha'$-bandwidth property, and there exists a collection $\rset_2$ of  at least $\talpha \cdot \talpha' k/256=\Omega(\talpha^2k/\log^2m)$ edge-disjoint paths routing some subset $T_2\subseteq T$ of terminals to the edges of $\delta_G(C_2)$. We then return $C_1$ and $C_2$. From the above discussion, $C_1$ and $C_2$ are both disjoint and have the required properties. In order to complete the proof of 
\Cref{lem: better clustering}, it is now enough to prove \Cref{claim: processing one side}, which we do next.

\begin{proofof}{\Cref{claim: processing one side}}
Let $S\subseteq V(G)$ be a set of vertices that contains, for every cluster $W\in \cset$, the vertices of $W$. Since every edge that is incident to a vertex of $S$ either has a terminal of $T$ as its other endpoint, or belongs to $\heout(\wset)$, from \Cref{claim: contracted graph lots of edges}, we get that $|\delta_G(S)|\leq k+|E(\hat G')|\leq 8|E(\hat G')|/\talpha$, since, from \Cref{claim: contracted graph lots of edges}, $|E(\hat G')|=|\heout(\wset)|\geq \talpha k/4$. 

We apply the algorithm from \Cref{thm:well_linked_decomposition} to graph $G$ and its cluster $G[S]$, with parameter $\alpha=\talpha'$ (recall that $\tilde \alpha'=\frac{\tilde{\alpha}}{c\log^2 m}$ for a large enough constant $c$,  so the requirement that $\tilde \alpha'< \min\set{\frac 1 {64\alphasc(m)\log m},\frac 1 {48\log^2 m}}$ is satisfied), to obtain a collection $\cset'$ of disjoint clusters of $G[S]$ (if grah $G[S]$ is not connected, then we apply the algorithm from \Cref{thm:well_linked_decomposition} to each connected component of $G[S]$ separately; this does not affect the remainder of the analysis). 
Recall that $\set{V(C')\mid C'\in \cset'}$ partitions $S$; every cluster $C'\in \cset'$ has the $\talpha'$-bandwidth property, and:

\[\begin{split}
\sum_{C'\in \cset'}|\delta_G(C')| & \le |\delta_G(S)|\cdot\left(1+O(\talpha'\cdot \log^{3/2} m)\right)\\
&\leq |\delta_G(S)|+O\left(\frac{8|E(\hat G')|\talpha'\log^{3/2}m}{\talpha}\right )\\
&\leq  |\delta_G(S)|+O\left(\frac{|E(\hat G')|}{c\log^{1/2}m}\right )   \\
&\leq |\delta_G(S)|+\frac{|E(\hat G')|}{64},
\end{split} \]

since
$\talpha'=\talpha/c\log^2m$, and $c$ is a large enough constant.

We consider now a new clustering $\wset'$ of $G$, that is obtained as follows: start from $\wset'=\wset\setminus \cset$, and then add the clusters of $\cset'$ to $\wset'$. It is easy to verify that the clusters in $\wset'$ are all mutually disjoint, and that $\bigcup_{W'\in \wset}V(W)=V(G)\setminus T$. Moreover, every clsuter $W\in \wset$ has the $\tilde \alpha'$-bandwidth property. Next, we show that $|\heout(\wset')|< |\heout(\wset)|$.

Indeed, we can partition the edge set $\heout(\wset)$ into three subsets: set $E_1$ contains all edges $e=(u,v)$, where $u$ and $v$ lie in different clusters of $\wset\setminus \cset$; set $E_2$ contains all edges $e=(u,v)$, where $u$ lies in a cluster of $\wset\setminus \cset$, and $v$ lies in a cluster of $\cset$; lastly, $E_3$ contains all edges $e=(u,v)$, where $u$ and $v$ lie in different clusters of $\cset$.

We can partition $\heout(\wset')$ into three subsets $E_1',E_2',E_3'$ similarly, using cluster set $\cset'$ instead of $\cset$. Clearly, $E_1=E_1'$, and $E_2=E_2'=\delta_G(S)$. From the statement of \Cref{claim: processing one side}, $|E_3|\geq |E(\hat G')|/8$. On the other hand, since we have established that $\sum_{C\in \cset'}|\delta_G(C)|\leq |\delta_G(S)|+\frac{|E(\hat G')|}{64}$, and since $E_2=\delta_G(S)\subseteq \bigcup_{C\in \cset'}\delta_G(C)$, we get that $|E_3'|\leq \frac{|E(\hat G')|}{64}$. We conclude that $|E_3'|<|E_3|$, and $|\heout(\wset')|<|\heout(W)|$.  Note however that $\wset'$ may not be a legal clustering of $G$, since we do not guarantee that for every cluster $C\in \cset'$, $\delta_G(C)\leq \talpha k/64$ (this property is guaranteed to hold for every cluster of $\wset\setminus \cset'$ though, since $\wset$ is a legal clustering). In the remainer of the algorithm, we will attempt to ``fix'' the clustering $\wset'$ so it becomes a legal clustering of $G$, and, if we fail to do so, we will produce the desired cluster $C$.

In the remainder of the algorithm, we will maintain a set $\wset^*$ of clusters, starting with $\wset^*=\wset'$, and we will ensure that the following invariants hold for $\wset^*$ at all times:

\begin{properties}{I}
	\item all clusters in $\wset^*$ are disjoint from each other, and $\bigcup_{W\in \wset^*}V(W)=V(G)\setminus T$; \label{inv: partition}
	\item every cluster $W\in \wset^*$ has the $\talpha'$-bandwidth property; \label{inv: bandwidth prop}
	\item $|\heout(\wset^*)|< |\heout(\wset)|$; \label{inv: boundaries dont grow} and
	\item if $|\delta_G(W)|>\talpha k/64$ for some cluster $W\in \wset^*$, then $V(W)\subseteq S$. \label{inv: large cluster in S}
\end{properties}

Note that all these invariants hold for the initial setting $\wset^*=\wset'$. The algorithm performs a number of iterations, as long as there is some cluster $W\in \wset^*$ with $|\delta_G(W)|>\talpha k/64$. We now describe the execution of a single iteration.

Let $W\in \wset^*$ be any cluster with $|\delta_G(W)|>\talpha k/64$. Using the standard max-flow computation, we compute a maximum-cardinality set $\rset$ of edge-disjoint paths in graph $G$,  where each path in $\rset$ connects a distinct terminal of $T$ to a distinct edge of $\delta_G(W)$. 

We now consider two cases. The first case happens if $|\rset|\geq \talpha \talpha' k/256$. In this case, we terminate the algorithm and return the cluster $W$. Notice that, from Invariant \ref{inv: large cluster in S}, we are guaranteed that $V(W)\subseteq S=\bigcup_{C'\in \cset}V(C)$, and from Invariant \ref{inv: bandwidth prop}, $W$ has the $\talpha'$-bandwidth property.

Assume now that the second case happens, that is, $|\rset|<\talpha \talpha' k/256$. From the maximum flow / minimum cut theorem, there is a partition $(A,B)$ of $V(G)$, with $V(W)\subseteq A$, $T\subseteq B$, and $|E_G(A,B)|\leq |\rset|<\talpha \talpha' k/256$.

We slightly modify the cut $(A,B)$ in graph $G$, to compute a new cut $(A',B')$ with $W\subseteq A'$, $T\subseteq V(B')$, such that for every cluster $C\in \wset^*$, either $V(C)\subseteq A'$, or $V(C)\subseteq B'$ holds. In order to do so, we process every cluster $C\in \wset^*\setminus\set{W}$ one by one. Consider an iteration when cluster $C$ is processed. If $V(C)\subseteq A$, or $V(C)\subseteq B$, then no further updates for cluster $C$ are necessary. Otherwise, we denote by $E'(C)=E_G(A,B)\cap E(C)$ -- the set of edges that cluster $C$ contributes to $E_G(A,B)$. 
We partition the set $\delta_G(C)$ of edges into two subsets: set $\delta^A(C)$, $\delta^B(C)$, as follows. Let $e=(u,v)\in \delta_G(C)$ be any such edge, and assume that $v\in V(C)$. If $v\in A$, then we add $e$ to $\delta^A(C)$, and otherwise we add $e$ to $\delta^B(C)$. If $|\delta^A(C)|<|\delta^B(C)|$, then we move all vertices of $V(C)\cap A$ to $B$. Notice that in this case, since $C$ has the $\talpha'$-bandwidth property, $|\delta^A(C)|\leq |E'(C)|/\talpha'$. Once the vertices of $V(C)\cap A$ are moved to $B$, the edges of $E'(C)$ no longer lie in the cut $E_G(A,B)$, and the only new edges that may have been added to the cut $E_G(A,B)$ are the edges of $\delta^A(C)$. Since $|\delta^A(C)|\leq |E'(C)|/\talpha'$, we charge every edge of $E'(C)$ $1/\talpha'$ units for the edges of $\delta^A(C)$. Note that the edges of $E'(C)$ will never be charged again by the algorithm. Otherwise, $|\delta^B(C)|\leq |\delta^A(C)|$ holds, and we move the vertices of $B\cap V(C)$ to $A$. As before, $|\delta^B(C)|\leq |E'(C)|/\talpha'$ must hold, and the edges of $E'(C)$ no longer belong to the cut $E_G(A,B)$. The only new edges that may have been added to the cut are the edges of $\delta^B(C)$. As before, we charge every edge of $E'(C)$ $1/\talpha'$ units for the edges of $\delta^B(C)$.

Once every cluster in $\wset^*\setminus \set{W}$ is processed, we obtain a new partition $(A',B')$ of $V(G)$, with $W\subseteq A'$, and $T\subseteq V(B')$, such that for every cluster $C\in \wset^*$, either $V(C)\subseteq A'$, or $V(C)\subseteq B'$ holds. Moreover, since every edge in $E_G(A',B')\setminus E_G(A,B)$ is charged to some edge of $E_G(A,B)\setminus E_G(A',B')$, and the charge to each edge of $E_G(A,B)$ is at most $1/\talpha'$, we get that $|E_G(A',B')|\leq |E_G(A,B)|/\talpha'\leq \talpha k/256$.

We modify the clustering $\wset^*$ in two steps. In the first step, we remove from $\wset^*$ all clusters $W'$ with $V(W')\subseteq A'$, and we add to it cluster $G[A']$  (notice that we can assume w.l.o.g. that graph $G[A']$ is connected, since otherwise, we can move the sets of vertices corresponding to all connected components of $G[A']$ that are disjoint from $W$ to $B'$). Let $\wset^{**}$ denote this new clustering. It is immediate to verify that Invariants \ref{inv: partition} and \ref{inv: large cluster in S} continue to hold in $\wset^{**}$, since $|\delta_G(A')|<\talpha k/64$.
Note also that the edges of $\delta_G(W)\setminus \delta_G(A')$ no longer belong to $\heout(\wset^{**})$, while no new edges were added to 
$\heout(\wset^{**})$.
 Since $|\delta_G(W)|>\talpha k/64$, while $|\delta_G(A')|=|E_G(A',B')|\leq \talpha k/256$, we get that $|\heout(\wset^{**})|\leq |\heout(\wset^{*})|-\talpha k/64$.

Note that Invariant \ref{inv: bandwidth prop} is guaranteed to hold for every cluster of $\wset^{**}$ except for possibly $G[A']$. In our last step, 
we apply the algorithm from \Cref{thm:well_linked_decomposition} to compute a well-linked decomposition of the cluster $G[A']$ with parameter $\talpha'$. Recall that the algorithm computes a collection $\cset^*$ of clusters, such that vertex sets $\set{V(C')\mid C'\in \cset^*}$ partition $A'$, each cluster $C'\in \cset^*$ has the $\talpha'$-bandwidth property, and $\delta_G(C')\leq \delta_G(A')<\talpha k/64$ for all $C'\in \cset^*$. Moreover, we are guaranteed that:

\[\sum_{C'\in \cset^*}|\delta_G(C')|\le |\delta_G(A')|\cdot\left(1+O(\talpha'\cdot \log^{3/2} m)\right)\leq 2|\delta_G(A')|\leq \talpha k/128.
\]

We let $\wset^{***}$ be obtained from $\wset^{**}$ by replacing $G[A']$ with the collection $\cset^*$ of clusters. From the above discussion, it is immediate to verify that Invariants \ref{inv: partition}, \ref{inv: bandwidth prop} and \ref{inv: large cluster in S} hold for $\wset^{***}$. Moreover, $\heout(\wset^{***})\subseteq \heout(\wset^{**})\cup \left(\bigcup_{C'\in \cset^*}\delta_G(C')\right )$. Therefore, $|\heout(\wset^{***})|\leq |\heout(\wset^{**})|+\sum_{C'\in \cset^*}|\delta_G(C')|\leq |\heout(\wset^*)|$. This establishes Invariant \ref{inv: boundaries dont grow} for $\wset^{***}$. We then set $\wset^*=\wset^{***}$, and continue to the next iteration.

If the algorithm never terminates with a cluster $W$ and a collection $\rset$ of at least $\talpha\talpha'k/256$ edge-disjoint paths routing a subset of terminals to the edges of $\delta_G(W)$, then the algorithm terminates once every cluster $W'\in \wset^*$ has $|\delta_G(W')|\leq \talpha k/64$. We are then guaranteed that $\wset^*$ is a legal clustering of $G$, and moreover, $|\heout(\wset^*)|<|\heout(\wset)|$. We return the clustering $\wset^*$ as the output of the algorithm.
\end{proofof}

\subsection{Proof of \Cref{lem: high opt or lots of paths}}
\label{sec: few paths high opt}

We assume for contradiction that for every regular vertex $x\in V(\hH_1)$, $|\jset(x)|<\tk'$, and yet $\optcrors(I)<\frac{(\tk\talpha \alpha')^2}{c'\eta'\log^{20}m}$, where $c'$ is the constant from the definition of $\tilde k'$. 
The high-level idea of the proof is the following. 
We use the algorithm from \Cref{claim: embed expander} in order to embed an expander $W$ over the set $\tT$ of terminals into $\hat H_1$, which, from  \Cref{obs: cr of exp} has a high crossing number. On the other hand, from \Cref{lem: crossings in contr graph}, there is a drawing of the contracted graph $\hat H_1$ with relatively few crossings. We exploit this latter drawing of $\hat H_1$, the embedding of the expander $W$ into $\hat H_1$, and the fact that any set $\jset'(x)$ of edge-disjoint paths routing a subset of terminals to a vertex $x$ of $\hat H_1\cap V(H_1)$ must have a small cardinality, in order to obtain a drawing of the expander $W$ with relatively few crosings, reaching a contradiction. We now proceed with a formal proof. 

By applying \Cref{claim: embed expander} to graph $\hH_1$ and the set $\tT$ of terminals (that is $\talpha$-well-linked in $\hH_1$ from Property \ref{prop after step 1: terminals in H1}), we conclude that there exist a graph $W$ with $V(W)=\tT$ and maximum vertex degree at most $\cCMG\log^2\tk$, and an embedding $\hpset$ of $W$ into $\hH_1$ with congestion at most $\frac{\cCMG\log^2\tk}{\talpha}$, such that $W$ is a $1/4$-expander.
Moreover, from \Cref{obs: cr of exp}, the crossing number of $W$ is at least $\tk^2/(\tilde c\log^8\tk)\geq \tk^2/(\tilde c\log^8m)$, for some constant $\tilde c$. 
Recall that we have assumed for contradiction that $\optcrors(I)< \left(\frac{(\tk\talpha \alpha')^2}{c'\eta'\log^{20}m}\right )$ for some large enough constant $c'$. We will exploit this fact in order to show that there exists a drawing of $W$ with fewer than $\tk^2/(\tilde c\log^8m)$
crossings, reaching a contradiction.

Recall that, from \Cref{lem: crossings in contr graph}, there is a drawing $\phi$ of the contracted graph $\hH_1$, whose number of crossings is bounded by:

\[O\left(\frac{\optcrors(I)\cdot \log^8m}{(\alpha')^2}\right)\leq 
O\left(\frac{\tk^2\talpha^2}{c'\eta'\log^{12}m}\right),\]

from our assumption that $\optcrors(I)< \left(\frac{(\tk\talpha \alpha')^2}{c'\eta'\log^{20}m}\right )$.

In the remainder of the proof, we gradually modify the drawing $\phi$ of $\hH_1$, to transform it into a drawing of the expander $W$ with fewer than $\tk^2/(\tilde c\log^8m)$ crossings, leading to a contradiction.
The drawing of $W$ is obtained from the drawing $\phi$ of $\hH_1$ by exploiting the embedding $\hpset$ of $W$ into $\hH_1$. Intuitively, we would like to use the images of the paths in $\hpset$ in $\phi$ in order to draw the edges of the graph $W$. Unfortunately, the curves representing the images of these paths are not in general position. This is since that paths in $\hpset$ may share edges, and they may also share vertices other than their endpoints. We modify the drawing $\phi$ in three stages. In the first stage, we create a number of copies of every edge in $\hH_1$, so that the paths in $\hpset$ no longer share edges, and in the second stage, we modify the drawing of the curves corresponding to the images of the paths in $\hpset$ in the vicinity of the vertices they share, using a nudging operation. Then in the third and the last stage we define the final drawing of the expander $W$.

\paragraph{Stage 1: shared edges.}

For every edge $e\in E(\hH_1)$, let $N_e$ be the total number of paths in $\hpset$ containing $e$; recall that $N_e\leq  O((\log^2\tk)/\talpha)\leq O((\log^2m)/\talpha)$ must hold. Let $\hH'$ be the graph obtained from graph $\hH_1$ by deleting from it all edges that do not participate in paths in $\hpset$, and, for each remaining edge $e$, creating $N_e$ parallel copies of edge $e$. Drawing $\phi$ of graph $\hH_1$ can be easily extended to a drawing $\phi'$ of graph $\hH'$, by deleting the images of all edges $e$ with $N_e=0$, and, for every edge $e$ with $N_e>1$, drawing the parallel copies of $e$ in parallel to the original image of $e$, at a very small distance from it, so that the images of the copies of $e$ do not cross. Since, for every edge $e$, $N_e\leq  O((\log^2m)/\talpha)$, every crossing in the drawing $\phi$ may give rise to at most  $O((\log^4m)/\talpha^2)$ crossings in the drawing $\phi'$, and so the number of crossings in $\phi'$ is bounded by:

\[\cro(\phi)\cdot O\left(\frac{\log^4m}{\talpha^2}\right )\leq O\left(\frac{\tk^2\talpha^2}{c'\eta'\log^{12}m}\right )\cdot O\left(\frac{\log^4m}{\talpha^2}\right )\leq O\left(\frac{\tk^2}{c'\eta'\log^8m}\right ).\] 

Notice that the set $\hpset$ of paths in graph $\hH_1$ naturally defines a set $\hpset'$ of edge-disjoint paths in graph $\hH'$, embedding the expander $W$ into $\hat H'$ (where for every edge $e\in E(\hH)$, every path in $\hpset$ containing $e$ now uses a different copy of $e$).

\paragraph{Stage 2: shared vertices.}

We process every vertex $x\in V(\hH')\setminus \tT$ one by one.
Let $\hpset'(x)=\set{P_1,\ldots,P_r}$ be the set of all paths in $\hpset'$ containing $x$. For each such path $P_i(x)$, let $e(P_i,x)$ and $e'(P_i,x)$ be the edges immediately preceding and immediately following $x$ on path $P_i$. Let $D(x)$ be a very small disc containing the image $x$ in the drawing $\phi'$ in its interior. Consider now some path $P_i\in \hpset'(x)$, and let $q_i,q'_i$ be the points where the images of $e(P_i,x)$ and $e'(P_i,x)$ intersect the boundary of $D(x)$, respectively. We define a curve $\gamma(P_i,x)$ inside disc $D(x)$, connecting $q_i$ and $q'_i$, such that, for every pair $P_i,P_j\in \hpset'(x)$ of distinct paths, the two corresponding curves $\gamma(P_i,x)$, $\gamma(P_j,x)$ cross at most once, and every point of $D(X)$ lies on at most two such curves.

\paragraph{Stage 3: final drawing of $W$.}
We are now ready to define the final drawing $\phi''$ of the expander $W$. Recall that $V(W)=\tT$. For every terminal $t\in \tT$, the image of $t$ in $\phi''$ remains the same as the image of $t$ in $\phi'$. 

Consider now some edge $\hat e=(t,t')\in E(W)$, and let $P(\hat e)\in \hpset'$ be its embedding path. Denote $P=(e_1,e_2,\ldots,e_{\ell})$, and denote the vertices of $P$ by $t=x_0,x_1,x_2,\ldots,x_{\ell-1},x_{\ell}=t'$ in the order of their appearance on $P$. For each edge $e_i$, let $\gamma(e_i)$ be its image in the drawing $\phi'$. If $i>1$, then we delete the portion of $\gamma(e_{i})$ that lies inside the disc $D(x_{i-1})$, and similarly, if $i<\ell$, then we delete the portion of $\gamma(e_i)$ that lies inside the disc $D(x_{i})$. The image of the edge $\hat e$ is obtained by concatenating the curves $\gamma(e_1),\gamma(P,x_1),\gamma(e_2),\ldots,\gamma(P,x_{\ell-1}),\gamma(e_{\ell})$.

This completes the definition of the drawing $\phi''$ of the expander $W$. Our last step is to show that this drawing contains few crossings, leading to a contradiction. The following claim will then complete the proof of \Cref{lem: high opt or lots of paths}.

\begin{claim}
	$\cro(\phi'')<\tk^2/(\tilde c\log^8m)$.
\end{claim}

\begin{proof}
Recall that drawing $\phi'$ of $\hH'$ contained at most $O\left(\frac{\tk^2}{c'\eta'\log^8m}\right )$ crossings.
For every vertex $u\in V(\hat H_1)$, let $N_u$ de note the number of paths in $\hat \pset$ containing $u$. 
 It is then easy to verify that the total number of crossings in $\phi''$ is at most: 
 
 \[\cro(\phi')+ \sum_{u\in V(\hH_1)}N_u^2\leq O\left(\frac{\tk^2}{c'\eta'\log^8m}\right )+ \sum_{u\in V(\hH_1)}N_u^2. \]  

Assuming that $c'$ is a large enough constant, the above expression is bounded by $
\tk^2/(2\tilde c\log^8m)+ \sum_{u\in V(\hH_1)}N_u^2$. Therefore, it is now enough to prove that $\sum_{u\in V(\hH_1)}N_u^2\leq \tk^2/(2\tilde c\log^8m)$.

We partition the vertices of $\hH_1$ into two subsets: the set $U=\set{v_C\mid C\in \cset_X}$ of supernodes, and the set $U'=V(\hH_1)\setminus U$ of regular vertices, that lie in $H_1$.

Consider first a supernode $v=v_C$. Since the paths in $\hpset$ cause edge-congestion at most $O(\log^2m/\talpha)$ in graph $\hH_1$, $N_v\leq O(\deg_{\hH_1}(v)\log^2m/\talpha)$, and so $N_v^2\leq O(|\delta_{H_1}(C)|^2\log^4m/\talpha^2)$. 
Recall that from Property \ref{prop after step 1: small squares of boundaries}:

\[\sum_{C\in \cset}|\delta_H(C)|^2\leq \frac{(\tk\talpha\alpha')^2}{c_1\log^{20}m}.\]  

Therefore, we conclude that:

\[
\begin{split}
\sum_{v\in U}N_v^2&\leq \sum_{C\in \cset}|\delta_H(C)|^2 \cdot O\left(\frac{\log^4m}{\talpha^2}\right)\\
& \leq    \frac{(\tk\talpha\alpha')^2}{c_1\log^{20}m}
   \cdot O\left(\frac{\log^{4}m}{\talpha^2}\right)\\
&\leq \frac{\tk^2}{4\tilde c\log^8m},
\end{split}\]

since we have assumed that $c_1$ is a large enough constant.
Lastly, it remains to show that $\sum_{v\in U'}N_v^2\leq \frac{\tk^2}{4\tilde c\log^8m}$. We do so using the following claim.


\begin{claim}\label{claim: small degree for non super nodes}
	For every vertex $v\in U'$, $N_v< \frac{8\cCMG^2\tk'\log^4m}{\talpha}$.
\end{claim}

\begin{proof}
	Assume for contradiction that there is some vertex $v\in U'$, with $N_v\geq \frac{8\cCMG^2\tk'\log^4m}{\talpha}$. Then there is a set $\pset\subseteq \hpset$ of at least $\frac{8\cCMG^2\tk'\log^4m}{\talpha}$ paths in the embedding $\hpset$ of $W$ into $\hat H_1$, containing $v$. Recall that for every path $P\in\hpset$, both endpoints of $P$ are terminals in $\tT$, and that, since the maximum vertex degree in $W$ is  
	at most $\cCMG\log^2 \tk\leq \cCMG\log^2 m$, every terminal may serve as an endpoint of at most $\cCMG\log^2m$ paths in $\pset$. 
	
	Consider any path $P\in \pset$, and let $t,t'$ be its two endpoints. Let $P'$ be the subpath of $P$ between $t$ and $v$. We then set $\pset_1=\set{P'\mid P\in \pset}$. Furthermore, since every terminal may serve as an endpoint of at most $\cCMG\log^2m$ paths in $\pset_1$, there is a subset $\pset_2\subseteq \pset_1$ of at least $\frac{|\pset|}{\cCMG\log^2m}$ paths, each of which originates at a distinct terminal of $\tT$, and terminates at $v$. Recall that the paths in $\hpset$ cause edge-congestion at most $\frac{\cCMG\log^2m}{\talpha}$ in $\hH_1$.
	Lastly, from \Cref{claim: remove congestion}, there is a collection $\pset_3$ of edge-disjoint paths in $\hH_1$, routing a subset of terminals to $v$, such that: 
	
\[|\pset_3|\geq \frac{\talpha\cdot |\pset_2|}{2\cCMG\log^2m}\geq \frac{\talpha |\pset|}{2\cCMG^2\log^4m}>\tk',\]
 contradicting the fact that the largest number of edge-disjoint paths routing terminals of $\tT$ to $v$ is bounded by $\tk'$.
\end{proof}

	We group the vertices of $U'$ into groups $S_1,\ldots,S_q$, where $q=\ceil{\log \left (\frac{8\cCMG^2\tk'\log^4m}{\talpha}\right )}+1$. Set $S_i$ contains all vertices $v\in U'$, with $2^{i-1}\leq N_v<2^i$.
	
	Since paths in $\hpset$ cause edge-congestion $O((\log^2m)/\talpha)$, for all $1\leq i\leq q$, for every vertex $v\in S_i$, $\deg_{\hH}(v)\geq \Omega(\talpha N_v/\log^2m)\geq \Omega(\talpha \cdot 2^i/\log^2m)$. Since the total number of edges in graph $\hH_1$ is 
	$|E(\hat H_1)|\leq O(\tk\cdot \eta \log^8m/\alpha^3)$ from Property \ref{prop after step 1: few edges}, 
	while $\talpha=\Omega(\alpha/\log^4m)$ from Property \ref{prop after step 1: terminals in H1},
	we get that, for all $1\leq i\leq q$,
	
	\[\sum_{v\in S_i}N_v\leq \sum_{v\in S_i}O\left (\frac{\deg_{\hH}(v)\log^2m}{\talpha}\right )
	\leq O\left(\frac{|E(\hat H_1)| \log^6m}{\alpha}\right )\leq O\left(\frac{\tk\eta\log^{14}m}{\alpha^4}\right).  \]
	
	Therefore:
	
\[\sum_{v\in S_i}N_v^2\leq  2^{i+1}\cdot \sum_{v\in S_i}N_v\leq  O\left(\frac{2^i\cdot \tk\eta\log^{14}m}{\alpha^4}\right).\]
	
	Summing up over all sets $S_1,\ldots,S_q$, we get that:
	
	\[
	\begin{split}
	\sum_{v\in U'}N_v^2& \leq 
	 O\left(\frac{2^q\tk\eta\log^{14}m}{\alpha^4}\right) \\
	 &\leq O\left(\frac{\cCMG^2\tk'\tk\eta\log^{18}m}{\alpha^4\talpha}\right)\\
	 &\leq O\left(\frac{\cCMG^2\tk^2}{c'\log^8m}\right).
	 \end{split}\]

since	
$q=\ceil{\log \left (\frac{8\cCMG^2\tk'\log^4m}{\talpha}\right )}+1$,
$\talpha=\Omega(\alpha/\log^4m)$, and
$\tk'=\tk\alpha^5/(c'\eta\log^{36}m)$.

 Lastly, since we can fix $c'$ to be a large enough constant, we get that $\sum_{v\in U'}N_v^2< \frac{k^2}{4\tilde c\log^8m}$.

To conclude, we have shown that $\cro(\phi'')\leq \frac{\tk^2}{\tilde c\log^8m}$, a contradiction.
\end{proof}	

\subsection{Proof of \Cref{lem: find ordering of terminals}}
\label{sec:ordering of terminals}

We start by defining a new expanded graph $H'$, whose construction is similar to that of $H^*$, except that now we expand every vertex $v\in V(H)\setminus (\tT\cup \set{x})$ into a grid. Specifically, 
we start with $H'=\emptyset$, and then process every vertex $u\in V(H)\setminus (\tT\cup \set{x})$ one by one. We denote by $d(u)$ the degree of the vertex $u$ in graph $H$. We now describe an iteration when a vertex $u\in V(H)\setminus (\tT\cup \set{x})$ is processed. Let $e_1(u),\ldots,e_{d(u)}(u)$ be the edges that are incident to $u$ in $H$, indexed according to their ordering in $\oset_u\in  \Sigma$. We let $\Pi(u)$ be the $(d(u)\times d(u))$ grid, and we denote the vertices on the first row of this grid by $s_1(u),\ldots,s_{d(u)}(u)$ indexed in their natural left-to-right order. 
We add the vertices and the edges of the grid $\Pi(u)$ to graph $H'$. As before, we refer to the edges in the resulting grids $\Pi(u)$ as inner edges.
Once every vertex $u\in V(H)\setminus (\tT\cup \set{x})$ is processed, we add 
the vertices of $\tT$ to the graph $H'$. Recall that every terminal $t\in \tT$ has degree $2$ in $H$. We denote $s_1(t)=s_2(t)=t$, and we arbitrarily denote the two edges incident to $t$ by $e_1(t)$ and $e_2(t)$.
We also add vertex $x$ to $H'$. We denote $s_1(x)=\cdots=s_{d(x)}(x)=x$, and we denote the edges incident to $x$ by $e_1(x),\ldots,e_{d(x)}(x)$, indexed consistently with the circular ordering $\oset_x\in \Sigma$, where $\Sigma$ is the rotation system for graph $H$.
Next, we add 
a collection of outer to graph $H'$, exactly as before. Consider any edge $e=(u,v)\in E(H)$. Assume that $e$ is the $i$th edge of $u$ and the $j$th edge of $v$, that is, $e=e_i(u)=e_j(v)$. Then we add an edge $e'=(s_i(u),s_j(v))$ to graph $H'$, and we view this edge as the copy of the edge $e\in E(H)$. This completes the definition of graph $H'$.

The partition $(X,Y)$ of the vertices of $V(H)\setminus \tT$ naturally defines a partition $(X',Y')$ of the vertices of $V(H')\setminus \tT$, as follows: $X'=\left ( \bigcup_{u\in X\setminus\set{x}}V(\Pi(u))\right )\cup \set{x}$ and $Y'= \bigcup_{u\in Y}V(\Pi(u))$.
We denote $H'_1=H'[X'\cup \tT]$ and $H'_2=H'[Y\cup \tT]$.
Let $\wset_X=\set{\Pi(u)\mid u\in X\setminus\set{x}}$. Then $\wset_X$ is a collection of disjoint clusters in graph $H'_1$. Moreover, since, for every vertex $u\in X\setminus\set{x}$, the set of vertices on the first row of a grid $\Pi(u)$ is $1$-well-linked in $\Pi(u)$ (from \Cref{obs: grid 1st row well-linked}), the corresponding cluster $\Pi(u)$ has the $1$-bandwidth property.

\begin{observation}\label{obs: routing terminals to $x$}
	There is a collection $\jset'$ of paths in graph $H_1'$, routing all terminals of $\tT$ to $x$, with $\cong_{H_1'}(\jset')\leq O\left(\frac{\eta\log^{36}m}{\alpha^6(\alpha')^2}\right )$.
\end{observation}

\begin{proof}
Recall that there exists a set $\jset$ of edge-disjoint paths in graph $\hat H_1$, routing a subset $\tT_0\subseteq \tT$ of terminals to $x$, with $|\jset|\geq \tk'$. Since every cluster in $\cset$ has the $\alpha'$-bandwidth property, from
\Cref{claim: routing in contracted graph}, there is a collection $\jset_0$ of edge-disjoint paths in graph $H_1$, routing a subset $\tilde T_0'\subseteq \tilde T_0$ to $x$, where $|\tilde T'_0|\geq \alpha'\tk'/2$.  Using the $1$-bandwidth property of the clusters $\Pi(u)$ in graph $H'_1$, it is easy to verify that there is a collection $\jset'_0$ of edge-disjoint paths in graph $H_1'$, routing the terminals of $\tT'_0$ to vertex $x$. 
Recall that $\tk'=\tk\alpha^5/(c'\eta\log^{36}m)$, where $c'$ is a constant whose value was fixed in the proof of \Cref{lem: high opt or lots of paths}.

From Property \ref{prop after step 1: terminals in H1}, the set $\tT$ of terminals is $\talpha$-well-linked in $\hat H_1$, and so, from \Cref{clm: contracted_graph_well_linkedness}, it is $(\talpha\alpha')$-well-linked in $H_1$. Moreover, since $H_1=(H_1')_{|\wset_X}$, and since each cluster in $\wset_X$ has the $1$-bandwidth property, from \Cref{clm: contracted_graph_well_linkedness}, the set $\tT$ of terminals is $(\talpha\alpha')$-well-linked in $H_1'$.

From \Cref{lem: routing path extension}, there is a set $\jset'$ of paths in graph $H'_1$, routing the terminals in $\tilde T$ to vertex $x$, with:

\[\cong_{H_1'}(\jset')\leq O\left(\frac {|\tilde T|}{|\tilde T_0'|}\cdot\frac{1}{\alpha\alpha'}\right )
\leq O\left(\frac {\tk}{\tk'\alpha(\alpha')^2}\right )
\leq O\left(\frac{\eta\log^{36}m}{\alpha^6(\alpha')^2}\right ).\]

\end{proof}

For convenience, we denote by $\rho=O\left(\frac{\eta\log^{36}m}{\alpha^6(\alpha')^2}\right )$ the bound on $\cong_{H_1'}(\jset')$. Note that we can compute such a collection $\jset'$ of paths efficiently using standard maximum $s$-$t$ flow computation. In order to compute the ordering $\tilde \oset$ of the terminals in $\tT$, we need one additional property from the paths in $\jset'$: we need them to be \emph{confluent}. In order to define confluent paths, we need to assign each path a direction, so that one of its endpoint becomes the first vertex on the path.
If $P$ is a simple directed path, whose first endpoint is $s$ and last endpoint is $t$, a \emph{suffix} of $P$ is any subpath $P'\subseteq P$ that contains the vertex $t$.

\begin{definition}[Confluent Paths]
	Let $\pset$ be a collection of directed paths. We say that the paths in $\pset$ are \emph{confluent} iff for every pair $P_1,P_2\in \pset$ of paths, either $P_1\cap P_2=\emptyset$, or $P_1\cap P_2$ is a suffix of both $P_1$ and $P_2$.
\end{definition}

The following claim, that easily follows from the work of \cite{confluent},
allows us to transform the set $\jset'$ of paths into a confluent one.

\begin{claim}\label{claim: confluent paths}
	There is an efficient algorithm that computes a set $\jset''$ of confluent paths in graph $H_1'$, routing the vertices of $\tT$ to $x$, (where every path is directed towards $x$), with $\cong_{H'_1}(\jset'')\leq O(\rho \log m)$.
\end{claim}

\begin{proof}
	We use the notion of confluent flows from \cite{confluent}. The following definitions are from \cite{confluent}. Let $G=(V,E)$ be a directed graph, and let $\dem:V\rightarrow \reals^+$ be a demand function for vertices $v\in V$. Let $S\subseteq V$ denote a collection of sink vertices. We assume that every edge incident to a sink vertex $s\in S$ is directed towards $s$. A flow $f: E\rightarrow \reals^+$ is a \emph{valid flow} if it satisfies, for every vertex $v\in V\setminus S$:
	
	\[\sum_{e=(v,w)\in E}f(e)-\sum_{e=(u,v)\in E}f(e)=\dem(v). \]
	
	In other words, the total amount of flow leaving $v$ is equal to the demand on $v$ plus the total amount of flow entering $v$. The \emph{congestion} on vertex $v$ is defined to be as:
	
	\[ \sum_{e=(u,v)\in E}f(e)+\dem(v), \]
	
	the total amount of flow entering $v$ plus the demand at $v$. The congestion of $f$ is the maximum, over all vertices $v\in V$, of the congestion of $f$ at $v$. 
	
	We say that a flow $f$ is \emph{confluent} if for every vertex $v\in V$, there is at most one edge $(v,u)$ with $f(v,u)>0$. A confluent flow therefore defines a subgraph of $G$ (induced by edges carrying non-zero flow), consisting of disjoint components $\set{T_1,\ldots,T_k}$, where each $T_i$ is an arborescence directed towards the root $s_i\in S$. In each such arborescence $T_i$, the maximum vertex congestion occurs at the sink $s_i$, and is equal to the total demand of all vertices of $T_i$.
	The following result was proved in \cite{confluent}.
	
	\begin{theorem}[Theorem 3 in \cite{confluent}]\label{thm: confluent flow}
There is an efficient algorithm, that, given a directed $n$-vertex graph $G$ with a collection $S\subseteq V(G)$ of sinks, and demands $\dem(v)\geq 0$ for vertices $v\in V(G)$, such that there exists a (regular, splittable) flow $f$ with node congestion $1$ satisfying the demands in graph $G$, computes a confluent flow satisfying all demands, with congestion $O(\log n)$.
\end{theorem}

We construct a flow network from graph $H_1'$, as follows. First, we subdivide every edge $e$ that is incident to $x$ with a new sink vertex $s_e$, setting $S=\set{s_e\mid e\in \delta_{H_1'}(x)}$, and delete the vertex $x$ from the graph. Next, we bi-direct every edge of the resulting graph (by replacing it with two anti-parallel edges), except that for every sink vertex $s_e\in S$, the unique edge incident to $s_e$ is directed towards $s_e$. For every terminal $t\in \tT$, we set its demand $\dem(t)=1/(4\rho)$, and we set the demands of all other vertices to $0$. Let $G$ be the resulting flow network. Denote $n=|V(G)|$, and observe that $|V(G)|=O\left(\sum_{u\in V(H_1)}(\deg_{H_1}(u))^2\right )\leq O(m^2)$. Note that the collection $\jset'$ of paths in graph $H_1'$ immediately defines a collection $\tilde \jset$ of paths, routing the set $\tT$ of terminals to the vertices of $S$, in graph $G$, with edge-congestion at most $\rho$. Since the degree of every vertex in $G$ is at most $4$, by sending $1/(4\rho)$ flow units on every path in $\jset''$, we obtain a valid flow from vertices of $\tT$ to vertices of $S$, satisfying all demands, with vertex-congestion at most $1$.  From \Cref{thm: confluent flow}, there is a confluent flow $f$ in graph $G$ with vertex-congestion $O(\log n)\leq O(\log m)$, satisfying all demands. We can use standard flow-decomposition of $f$ to obtain a collection $\tilde \jset'$ of flow-paths, routing the terminals in $\tT$ to vertices of $S$, such that the paths in $\jset''$ are confluent. Each such flow-paths carries $1/(4\rho)$ flow units in the flow $f$, so the total edge-congestion caused by paths in $\tilde \jset'$ is at most $O(\rho\log m)$. Moreover, every sink vertex $s_e\in S$ is an endpoint of at most $O(\rho\log m)$ paths. The set $\tilde \jset'$ of paths naturally define a set $\jset''$ of confluent paths in graph $H_1'$, routing the set $\tT$ of terminals to vertex $x$, with edge-congestion $O(\rho \log m)$.
\end{proof}

We will use the set $\jset''$ of confluent paths to both compute the desired ordering $\tilde \oset$ of the terminals, and to show that there exists the desired drawing $\phi$ of graph $H^*$.

For convenence, let $d$ denote the degree of vertex $x$ in $H$ and the edges incident to $x$ by $e_1(x),\ldots,e_{d}(x)$, indexed consistently with the circular ordering $\oset_x\in \Sigma$, where $\Sigma$ is the rotation system for graph $H$. For all $1\leq i\leq d$ let $\pset_i\subseteq\jset''$ be the subset of paths whose last edge is $e_i$, so $(\pset_1,\ldots,\pset_d)$ is a partition of $\jset''$. We define a circular ordering of the paths in $\jset''$ as follows: for each $1\leq i\leq d$, the paths in $\pset_i$ appear consecutively in this ordering, in an arbitrary order, and paths belonging to different subsets appear in the natural ordering $\pset_1,\ldots,\pset_d$ of these subsets. Denote $\jset''=\set{P_1,\ldots,P_{\tk}}$, where the paths are indexed by the ordering that we have just defined. For all $1\leq j\leq |\tT|$, let $t_j$ be the terminal of $\tT$ that serves as an endpoint of path $P_j$. We have then defined an ordering $t_1,\ldots,t_{\tk}$ of the terminals in $\tT$, that we denote by $\tilde \oset$. For all $1\leq i\leq \tk$, let $e_i$ denote the edge $(x,t_i)$ in graph $H^*$.

It is now enough to show that there is a drawing $\phi$ of graph $H^*$, in which the inner edges do not participate in crossings, and the images of edges $e_1,\ldots,e_{\tk}$ enter the image of $x$ in this order, such that $\cro(\phi)\leq O\left(\optcrors(I)\cdot\frac{\eta^2\log^{74}m}{\alpha^{12}(\alpha')^4}\right ) + \left (
\frac{\tk \eta\log^{37}m}{\alpha^6(\alpha')^2}\right )$. The following observation will then finish the proof of \Cref{lem: find ordering of terminals}.

\begin{observation}
	There is a drawing $\phi$ of graph $H^*$ with at most $O\left(\optcrors(I)\cdot\frac{\eta^2\log^{74}m}{\alpha^{12}(\alpha')^4}\right ) + O\left (
	\frac{\tk \eta\log^{37}m}{\alpha^6(\alpha')^2}\right )$ crossings, in which all crossings are between pairs of outer edges, and the images of edges $e_1,\ldots,e_{\tk}$ enter the image of $x$ in this circular order.
\end{observation}

\begin{proof}
Let $\phi_1$ be the optimal solution to instance $I$ of \CNwRS, and denote by $\chi_1=\optcrors(I)$ its number of crossings. We can easily transform drawing $\phi_1$ to a drawing $\phi_2$ of graph $H'$, with the same number of crossings, such that every crossing in $\phi_2$ is between a pair of outer edges. In order to do so, we consider, for every vertex  $v\in V(H)\setminus(\tT\cup \set{x})$, the tiny $v$-disc $D_{\phi}(v)$. We place a drawing of the grid $\Pi(v)$ inside the disc, using the natural layout of the grid (depending on the orientation of vertex $v$ in $\phi$, we may need to flip the image of $\Pi(v)$).

Next, we slightly modify the graph $H'$, as follows. 
First, for every edge $e_i$ that is incident to $x$, we subdivide $e_i$ with a new vertex $s_i$, and then delete $x$ from the graph. Let $S=\set{s_1,\ldots,s_d}$ be the resulting set of new vertices. We modify the paths in $\jset''$ so that each path now connects a distinct vertex of $\tT$ to some vertex of $S$, and the paths remain confluent. As before, we denote by $\pset_i\subseteq \jset''$ the set of paths that terminate at vertex $s_i$.

For every edge $e$ of the resulting graph, we let $N_e$ the number of paths in $\jset''$ in which edge $e$ participates; recall that $N_e\leq O(\rho\log m)$ must hold. For every edge $e\in E(H_1')$, if $N_e=0$, then we delete edge $e$, and otherwise we replace $e$ with $N_e$ parallel copies. 
We denote the resulting graph by $H''$, and we denote by $H_1''$ the subgraph of $H''$ corresponding to $H_1'$, that is, if we let $X''=V(H_1')\cap V(H'')$, then
 $H_1''$ is the subgraph of $H''$ that is induced by vertices of $X''\cup S$. Observe that graph $H_1''$ can be viewed as consisting of disjoint trees $\tau_1,\ldots,\tau_d$ (though some of the trees may consist of a single vertex $s_i$), where for all $1\leq i\leq d$, the root of $\tau_i$ is the vertex $s_i$, but these trees may have parallel edges. If $v\in V(\tau_i)\setminus\set{s_i}$, and the subtree rooted at $v$ contains $n_v$ terminal vertices, then the edge connecting $v$ to its parent has exactly $n_v$ parallel copies, and $n_v\leq O(\rho\log m)$. We further modify the paths set $\jset''$, so that the paths become edge-disjoint in graph $H_1''$, that is, we ensure that for every edge $e\in E(H_1')$, each path in $\jset''$ containing $e$
uses a different copy of the edge $e$.

We can modify the drawing $\phi_2$ of graph $H'$ to obtain a drawing $\phi_3$ of graph $H''$, as follows. First, consider the tiny $x$-disc $D=D_{\phi_2}(x)$. Recall that, that for every edge $e_i$, the intersection of the image of $e_i$ in $\phi_2$ and the boundary of the disc $D$ is a single point, that we denote by $p_i$. We place the image of the new vertex $s_i$ on point $p_i$, and we erase the portion of the image of $e_i$ that lies in disc $D$. We also delete the image of $x$. We delete images of edges and vertices as needed, and then, for every edge $e\in E(H_1'')$ with $N_e>0$, we create $N_e$ copies of $e$, all of which are drawn in parallel to the original image of $e$, very close to it, so that the images of these copies of $e$ do not cross each other. Let $\phi_3$ denote this resulting drawing of the graph $H''$. Since, for every edge $e$, $N_e\leq O(\rho\log m)$, it is easy to verify that $\cro(\phi_3)\leq \cro(\phi_2)\cdot O(\rho^2\log^2m)\leq O(\optcrors(I)\cdot \rho^2\log^2m)$.

For every vertex $v\in V(H'')$, this drawing $\phi_3$ of $H''$ naturally defines a circular ordering of the edges of $\delta_{H''}(v)$, that we denote by $\oset^3_v$ -- the order in which the images of the edges of $\delta_{H''}(v)$ enter the image of $v$ in this drawing. (Note that the drawing $\phi_3$ depends on the optimal solution to instance $I$ of \CNwRS, so we cannot compute the drawing or the resulting orderings $\oset^3(v)$ for $v\in V(H'')$ efficiently; we only use their existence here).
Let $\Sigma^3=\set{\oset_v^3}_{v\in V(H'')}$ be the resulting rotation system for graph $H''$. 

Recall that for all $1\leq i\leq d$, $\pset_i\subseteq \jset''$ is a set of paths routing a subset of the terminals of $\tT$ to $s_i$, and for $1\leq i\neq j\leq d$, no vertex may belong to a path in $\pset_i$ and to a path in $\pset_j$.
For all $1\leq i\leq d$, let $\tT_i\subseteq \tT$ be the set of terminals that serve as endpoints of paths in $\pset_i$.

 We apply \Cref{lem: non_interfering_paths} to each such path set $\pset_i$ separately, to obtain a path set $\pset'_i$, that is non-transversal with respect to the rotation system $\Sigma^3$. The lemma ensures that the paths in $\pset'_i$ route terminals of $\tT_i$ to vertex $s_i$ in $H''$; the paths are edge-disjoint, and moreover, an edge $e\in E(H'')$ may only belong to a path of $\pset'_i$ if it belonged to a path of $\pset_i$. Let $\jset'''=\bigcup_{i=1}^d\pset'_i$.
 
 We are now ready to define a drawing $\phi$ of the graph $H^*$. Notice that graph $H''$ can be viewed as the union of graph $H^*\setminus\set{x}$, and the paths in $\jset'''$. As discussed above, there is a drawing $\phi_3$ of $H''$ with at most $O(\optcrors(I)\cdot \rho^2\log^2m)$ crossings. We have defined a rotation system $\Sigma^3$ for graph $H''$, such that the drawing $\phi_3$ is consistent with this rotation system. Moreover, from the definition of the drawing $\phi_3$, there is a disc $D$ (that originally contained the image of the vertex $x$), whose interior is disjoint from the drawing $\phi_3$, with vertices $s_1,\ldots,s_d$ appearing on the boundary of $D$ in this circular order. Lastly, the paths in sets $\pset_1',\ldots,\pset_d'$ are all non-transversal with respect to $\Sigma_3$, and for all $1\leq i\neq j\leq d$, paths in $\pset_i$ and paths in $\pset_j$ cannot share vertices with each other.
 
 In order to obtain the drawing $\phi$ of $H^*$, we slightly modify the drawing $\phi_3$, as follows. First, we place an image of vertex $x$ in the interior of the disc $D$. 
 For every edge $e_j=(t_j,x)$ of $H^*$, let $P_j\in \jset'''$ be the path whose endpoint is $t_j$, and let $s_{i_j}$ be its other endpoint. Let $\gamma_j$ be the image of the path $P_j$ in the drawing $\phi_3$. Since all paths in set $\jset'''$ are non-transversal with respect to $\Sigma_3$ we can apply the nudging algorithm from \Cref{claim: curves in a disc} to the image of every vertex of $H''$ that lies on some path in $\jset'''$, in order to compute, for each $1\leq j\leq \tk$, a curve $\gamma_j'$ connecting the image of $t_j$ to the image of $s_{i_j}$, so that, if we delete from $\phi_3$ the images of all inner vertices and of all edges participating in the paths in $\jset'''$, and add instead the curves $\gamma'_1,\ldots,\gamma'_{\tk}$, then the number of crossings does not increase. For all $1\leq j\leq d$, curve $\gamma'_j$ is obtained from $\gamma_j$ by ``nudging'' it in the vicinity of every vertex $v\in V(P_j)$ using the algorithm from  \Cref{claim: curves in a disc}.
 
 Note that for all $1\leq i\leq d$, for every terminal $t_j\in \tT_i$, the curve $\gamma'_j$ terminates at the image of the vertex $s_i$, and moreover, the images of the vertices $s_1,\ldots,s_d$ appear in this circular order on the boundary of the disc $D$. However, the order in which the curves in $\Gamma'_i=\set{\gamma'_j\mid t_j\in \tT_i}$ enter the image of $s_i$ may be different from the ordering of the corresponding terminals in $\tT$. We need to reorder the curves in $\Gamma'_i$ in the vicinity of $s_i$, so that they enter the image of $s_i$ in the order consistent with $\tilde \oset$. Since for all $1\leq j\leq d$, $|\Gamma_j|\leq O(\rho\log m)$, we can perform these reorderings while introducing crossings whose number is bounded by:
 
 \[\sum_{i=1}^d|\Gamma'_i|^2\leq \sum_{i=1}^d|\Gamma'_i|\cdot O(\rho\log m)\leq O(\tk\rho \log m).  \]

For all $1\leq j\leq \tk$, let $\gamma''_j$ be the curve obtained form $\gamma'_j$ after the reordering, so that $\gamma''_j$ connects the image of $t_j$ to the image of $s_{i_j}$. By slighty extending this curve inside the disc $D$, we can ensure that it terminates at the image of $x$. This can be done for all $1\leq j\leq \tk$ without introducing any new crossings, while ensuring that the resulting curves $\gamma''_1,\ldots,\gamma''_{\tk}$ enter the image of $x$ in this order. We then let, for all $1\leq j\leq \tk$, $\gamma''_j$ be the image of the edge $(t_j,x)$, obtaining a drawing $\phi$ of $H^*$. It is immediate to verify that every crossing in $\phi$ is between a pair of outer edges, and from our construction, the edges $(t_1,x),\ldots,(t_{\tk},x)$ enter the image of $x$ in this circular order. From the above discussion, the total number of crossings in $\phi$ is at most:

\[O(\optcrors(I)\cdot \rho^2\log^2m)+O(\tk\rho \log m) \leq O\left(\optcrors(I)\cdot\frac{\eta^2\log^{74}m}{\alpha^{12}(\alpha')^4}\right ) + \left (
\frac{\tk \eta\log^{37}m}{\alpha^6(\alpha')^2}\right ).
 \]

\end{proof}

\subsection{Proof of \Cref{obs: bounds on opt}}
\label{subsec: proof of obs on bounds on opt}

Assume that
$\optcrors(I)<\frac{(\tilde k\tilde \alpha\alpha')^2}{c_1\eta'\log^{20}m}$. Then, from \Cref{lem: find ordering of terminals}:

\begin{equation}\label{eq: bound on cr}
\begin{split}
\cro(\phi)\leq &O\left(\frac{(\tilde k\tilde \alpha\alpha')^2\eta^2\log^{54}m}{c_1\eta'\alpha^{12}(\alpha')^4}\right ) + O\left (
\frac{\tk \eta\log^{37}m}{\alpha^6(\alpha')^2}\right ) \\
&\leq O\left(\frac{\tilde k^2\log^{46}m}{c_1\eta^7\alpha^{10}(\alpha')^2}\right ) + O\left (
\frac{\tk \eta\log^{37}m}{\alpha^6(\alpha')^2}\right ),
\end{split}
\end{equation}
since $\talpha=\Theta(\alpha/\log^4m)$ and $\eta'\geq \eta^{13}$ (from the statement of \Cref{thm: find guiding paths}).

Recall that, since we have assumed that Special Case 4 did not happen, $k\geq \eta^6$, and from Property \ref{prop after step 1: number of pseudoterminals},
  $\tilde k \geq \Omega(\alpha^3k/\log^8m)$. Therefore, $\tk\geq \Omega(\eta^6\alpha^3/\log^8m)$, and $\eta\leq O\left (\frac{\tk\log^8m}{\eta^5\alpha^3}\right )$. We can now bound the second term in Equation \ref{eq: bound on cr} as follows:

  \[ O\left (
  \frac{\tk \eta\log^{37}m}{\alpha^6(\alpha')^2}\right )
\leq O\left (
\frac{\tk^2\log^{45}m}{\eta^5\alpha^9(\alpha')^2}\right )
\]

Recall that we have assumed that $\log m>c_0'$, for some large enough constant $c_0'$, whose value we can set to be greater than $c_1$.
Therefore, $\cro(\phi)\leq 
O\left(\frac{\tilde k^2\log^{46}m}{c_1\eta^6\alpha^{10}(\alpha')^2}\right )$ must hold.

Lastly, from the conditions of \Cref{thm: find guiding paths}, $\eta\geq c^*\log^{46}m/(\alpha^{10}(\alpha')^2)$. Since we can assume that $c_1$ is a sufficiently large constant, we conclude that, if $\optcrors(I)<\frac{(\tilde k\tilde \alpha\alpha')^2}{c_1\eta'\log^{20}m}$, then $\cro(\phi)\leq \frac{\tilde k^2}{c_2\eta^5}$ holds, where $c_2$ is an arbitrarily large constant whose value we can set later.

\subsection{Proof of \Cref{obs: transform paths 2}}
\label{subsec: transform paths 2}

In order to obtain the distribution $\dset'$, for every router $\qset\in \Lambda'$, whose probability value in $\dset$ is $p(\qset)>0$, we construct a router $\qset'\in \Lambda(H,\tT)$, as follows. Let $y'$ be the center vertex of $\qset$, and assume that $y'\in \Pi(y)$ for some vertex $y\in V(H)$. For every terminal $t\in \tT$, let $Q_t\in \qset$ be the unique path connecting $t$ to $y'$ in $H''$. By suppressing all inner edges on path $Q_t$, we obtain a path $Q'_t$ in graph $H$, connecting $t$ to $y$. We then set $\qset'=\set{Q'_t\mid t\in \tT}$. It is easy to verify that paths in $\qset'$ route $\tT$ to $y$ in graph $H$, so $\qset'\in \Lambda(H,\tT)$. Moreover, for every edge $e\in E(H)$, $\cong_H(\qset',e)\leq \cong_{H''}(\qset,e)$. We assign to the router $\qset'\in \Lambda(H,\tT)$ the same probability value $p(\qset)$. Let $\dset'$ be the resulting distribution over the routers of $\Lambda(H,\tT)$. Since every edge $e\in E(H)$ is an outer edge of $H''$, it is immediate to verify that $\expect[\qset'\sim \dset']{(\cong_{H}(\qset',e))^2}\leq \beta$.

\newpage

\bibliography{REF}

\end{document}